\documentclass{aastex61}

\usepackage{amsmath}
\usepackage{tikz}
\usepackage{color}
\usepackage{soul}
\usepackage{array}
\usepackage{multirow}
\usepackage{graphicx}
\usepackage{longtable}
\usepackage{rotating}
\usepackage{natbib}
\usepackage{geometry}
\usepackage{graphicx}
\usepackage{verbatim}
\usepackage{tikz}
\usepackage{subfigure}
\usepackage{longtable}
\usepackage{geometry}
\usepackage{bm}
\newcounter{RomanNumber}
\newcommand{\MyRoman}[1]{\setcounter{RomanNumber}{#1}\Roman{RomanNumber}}
\newcommand{\tabincell}[2]{\begin{tabular}{@{}#1@{}}#2\end{tabular}}

\shorttitle{Statistics on GRBs}
\shortauthors{Wang et al.}

\begin{document}


\title{A comprehensive statistical study on gamma-ray bursts}


\correspondingauthor{Yuan-Chuan Zou}
\email{zouyc@hust.edu.cn}

\author{Feifei Wang}
\affil{School of Physics \\
Huazhong University of Science and Technology \\
Wuhan 430074, China}
\affil{School of Mathematics and Physics \\
Qingdao University of Science and Technology \\
Qingdao 266061, China}

\author{Yuan-Chuan Zou}
\affiliation{School of Physics \\
Huazhong University of Science and Technology \\
Wuhan 430074, China}

\author{Fuxiang Liu}
\affiliation{College of Science and TGMRC \\
China Three Gorges University \\
Yichang 443002, China}

\author{Bin Liao}
\affiliation{School of Physics \\
Huazhong University of Science and Technology \\
Wuhan 430074, China}

\author{Yu Liu}
\affiliation{School of Physics \\
Huazhong University of Science and Technology \\
Wuhan 430074, China}

\author{Yating Chai}
\affiliation{School of Physics \\
Huazhong University of Science and Technology \\
Wuhan 430074, China}

\author{Lei Xia}
\affiliation{School of Physics \\
Huazhong University of Science and Technology \\
Wuhan 430074, China}




\begin{abstract}
In order to obtain an overview of the gamma-ray bursts (GRBs), we need a full sample. In this paper, we collected 6289 GRBs  (from GRB 910421 to GRB 160509A) from the literature, including prompt emission, afterglow and host galaxy properties. We hope to use this large sample to reveal the intrinsic properties of GRB. We have listed all the data in machine readable tables, including the properties of the GRBs, correlation coefficients and linear regression results of two arbitrary parameters, and linear regression results of any three parameters. These machine readable tables could be used as a data reservoir for further studies on the classifications or correlations. One may find some intrinsic properties from these statistical results. With this comprehensive table, it is possible to find relations between different parameters, and to classify the GRBs into different kinds of sub-groups. With the completion, it may reveal the nature of GRBs and may be used as tools like pseudo-redshift indicators, standard candles, etc. All the machine readable data and statistical results are available on the website of the journal. 
\end{abstract}


\section{Introduction} \label{sec:intro}

Gamma-ray bursts (GRBs) were  firstly detected in 1967 \citep{Klebesadel1973} by the Vela Satellite Network. They have been intensely studied since 1990s, especially after the operation of Burst and Transient Source Experiment (BATSE) onboard Compton Gamma Ray Observatory (CGRO) \citep{Meegan1992}. For a single event, the fluence is between $10^{\rm -7}$ and $10^{\rm -5}$ $\rm ergs ~ cm^{\rm -2}$, and the isotropic energy is from about $10^{\rm 48}$ to $10^{\rm 55}$ $\rm ergs$ \citep{Nakar2007,Zhang2011,Gehrels2013,Berger2014,Kumar2015}. So GRBs are the most violent and energetic stellar explosions known to humankind in distant galaxies \citep{Kumar2015}. The nature of GRBs is still under discovered even though there are many researches have been going on for many years. Fireball model is one of historical models to explain the mechanism of GRBs \citep{1992MNRAS.258P..41R,1993MNRAS.263..861P,Wijers1997,Meszaros1998,Meszaros2006}, which implies GRBs are produced by highly relativistic and collimated jets. Interaction of blobs in the jet is believed to produce the prompt emission, and the interaction of the jet with the ambient material produces the multi-wavelength afterglow (X-ray, optical and sometimes also radio). For a clear understanding of the current situation for GRBs, some recent reviews could be read, such as \citet{1999PhR...314..575P, 2004RvMP...76.1143P, 2004IJMPA..19.2385Z, Meszaros2006, 2007ChJAA...7....1Z, Kumar2015} for the general physics, \citet{Nakar2007, Berger2014} for short GRBs, \citet{2009ARA&A..47..567G} concentrating on the observations, \citet{Woosley2006, 2012RvMP...84...25M} on the progenitors, and \citet{2000PhR...325...83L} on the central engines.

In order to explain the physics of GRBs,  many attempts have been made to describe the spectra in different physical frameworks, and significant progress is achieved. There are two main frameworks. One is internal or external shocks, the emissions are assumed to be non-thermal intrinsically \citep{Katz1994,Rees1994,Tavani1996,Sari1998,Gao2015B}. Another is photospheric emission, which is predicted to occur in the  ``fireball" model \citep{Meszaros2000,Rees2005,Peer2007,Thompson2007}. \citet{Racz2017} found the synchrotron radiation is significant in \textit{Fermi}/Gamma-ray Burst Monitor (GBM) spectra. Sub-photospheric dissipation \citep{Rees2005,Peer2006,Giannios2006,Chhotray2015} was suggested as the emission mechanism in some bursts which can broaden the spectrum. \citet{Ahlgren2015} firstly provided a full physical model, DREAM (Dissipation with Radiative Emission as A table Model), which is based on sub-photospheric dissipation, and gave acceptable fits to GRB 090618 and GRB 100724B for more details. If we take sub-photospheric dissipation and/or high latitude effects \citep{Lundman2013}, the photospheric model can account for a large diversity of spectra. Recently, the spectra observations of GRBs show a mixture of thermal and non-thermal spectra \citep{Ryde2005,Ryde2009,Guiriec2010,Nappo2017}, which implies that there is an interplay between different emission mechanisms. The exist of GRBs subgroups might be the consequence of different emission mechanisms \citep{Begue2016}. \citet{Acuner2017} searched the full \textit{Fermi}/GBM catalogue using Gaussian Mixture Models to cluster bursts according to their low energy photon index of Band model \citep{Band1993} ($\alpha_{\rm band}$), high energy photon index of Band model ($\beta_{\rm band}$), spectral peak energy of Band model ($E_{\rm p,band}$), fluence ($F_{\rm g}$) and duration of 5\% to 95\% $\gamma$-ray fluence ($T_{\rm 90}$), in order to divide bursts into photospheric origin and synchrotron origin. They thought both emission from the photosphere and optically-thin synchrotron radiation are operating but different emissions dominate differently for individual GRBs. They found 1/3 of the bursts are consistent with synchrotron radiation, and 2/3 of all bursts are consistent with photospheric emission. Beside the two main groups, they also found subgroups. It is maybe due to the dissipation pattern in the jet, alternatively due to whether the jet is dominated by thermal or magnetic energy, or due to the viewing angle \citep{Acuner2017}. For the afterglows of GRBs, we can also find thermal and non-thermal components. Up to now, there are totally 16 GRBs reported to have thermal components in the X-ray light curves (LCs) \citep{Campana2006,Page2011,Thone2011,Starling2011,Sparre2012,Starling2012,Friis2013,Nappo2017}. \citet{Valan2017} reported six detections of additional thermal components in the GRBs' early X-ray afterglows. A cocoon breaking out from a thick wind or late prompt emission maybe the explanation for the additional thermal components, which may be hidden by bright afterglows in the majority of GRBs. That is why small part of the afterglows detected with additional thermal components \citep{Valan2017}. Some other possible explanations are that shock break-out is appealing for GRBs with associated supernova \citep{Campana2006}, or the cocoon that surrounds the jet \citep{Meszaros2001,Peer2006B,Ghisellini2007,Starling2012}, or the jet itself \citep{Friis2013,Irwin2016,Nappo2017}.

With accumulated data, statistical study may reveal the underlying physics. It can be roughly split into classification and correlation seeking.
Nowadays, the classification of GRBs is still in suspense. Traditionally, they are grouped into short GRBs (SGRBs) and long GRBs (LGRBs) depending on $T_{\rm 90}$ smaller or greater than 2 s, where $T_{\rm 90}$ is the time difference between the 95th and 5th percentile of the total counts, which is often taken as the typical duration of a GRB. \citet{Kouveliotou1993} analysed the $\log T_{\rm 90}$ distribution of 222 BATSE GRBs. They found that the distribution is a bimodal form, and the dividing line is about 2 $\rm s$. 
However, $T_{\rm 90}$ is a function of energy band. Consequently, this value is detector dependent \citep{Bromberg2013,Resmi2017}. \citet{Bromberg2013} calculated a useful threshold duration that separates collapsars (long) from non-collapsars (short), $T_{\rm 90} = 3.1 \pm 0.5$ s in BATSE, $T_{\rm 90} = 1.7_{-0.6}^{+0.4}$ s in \textit{Fermi} GBM, $T_{\rm 90} = 0.8 \pm 0.3$ s in \textit{Swift}. LGRBs are rich in low-energy photons. They are thought to be originated from the gravitational collapse of massive stars, owing to the observational evidences of some LGRBs associated with Type \MyRoman{1}c supernovae (SNe) \citep{Galama1998,Hjorth2003,Stanek2003,Woosley2006,Hjorth2012A,Xu2013}. The host galaxies of LGRBs have low metallicity, sometimes interacting with other galaxies \citep{Sahu1997,Bloom1998,Bloom2002,Chary2002,Christensen2004,Savaglio2009,KruhlerAA2015}, high star-forming rate and small host galaxy offset \citep{Bloom2002,Fruchter2006,Peter2016}. Comparing to LGRBs, the SGRBs are rich in higher energetic photons. They are thought to be the product of compact binary mergers with at least one neutron star, such as a black hole and a neutron star (BH-NS), or two neutron stars (NS-NS) \citep{Eichler1989,Paczynski1991,Narayan1992,Nakar2007,Zhang2009}. Although some studies preferred other compact merger models, like an ONeMg with a CO white dwarf (WD) merger model \citep{Lyutikov2017}. The presence of kilonova emission \citep{Berger2013,Tanvir2013} and the locations of SGRBs \citep{Berger2013} provide evidences to compact merger model. Therefore SGRBs maybe the candidate sources of gravitational waves (GWs) (e.g., \citet{Eichler1989,Nakar2007,Piro2012,Berger2014}). Until now, advanced LIGO (Laser Interferometer Gravitational Wave Observatory) have detected some GWs, such as GW 150914 \citep{Abbott2016B}, GW151226 \citep{Abbott2016C}, GW170104 \citep{Abbott2017C} and GW170817 \citep{Abbott2017A}. The observation of the association of GW170817 and GRB 170817A confirms the NS-NS as a progenitor of SGRBs \citep{Abbott2017A,Xiao2017,Pozanenko2017,Goldstein2017,Abbott2017B,Zou2017,Granot2017}. The host galaxies of SGRBs include late-type and early-type galaxies, and have large offset which is about 5 times larger than LGRBs \citep{Fong2010A}. Some SGRBs are putatively associated with r-process-powered ``kilonovae/macronovae", like GRB 130603B, GRB 060614, and probably GRB 080503, GRB 050709 as well \citep{Li1998,Metzger2010,Tanvir2013,Berger2013,Yang2015,Gao2015A,Jin2016}.

With an increasing number of instruments applied for detecting GRBs, such as $Beppo$SAX/GRBM (40 - 700 $\rm keV$) \citep{1997A&AS..122..299B}, Konus-\textit{Wind} (10 keV - 10 MeV) \citep{1995SSRv...71..265A}, HETE-2 (6 - 400 $\rm keV$) \citep{2003AIPC..662....3R}, \textit{INTEGRAL}  (15 - 200 $\rm keV$) \citep{2003A&A...411L.291M}, \textit{Swift}/XRT (0.2 - 10 $\rm keV$) \citep{2004ApJ...611.1005G,2005SSRv..120..165B}, \textit{Swift}/BAT (15 - 150 $\rm keV$) \citep{2004ApJ...611.1005G,2005SSRv..120..143B}, \textit{Polar} (10 - 500 $\rm keV$) \citep{2018NIMPA.877..259P}, \textit{HXMT} (10 - 500 $\rm keV$) \citep{2014SPIE.9144E..21Z}, \textit{Fermi}/LAT (20 MeV - 300 $\rm GeV$) \citep{1999APh....11..277G,2012ApJS..203....4A}, \textit{Fermi}/GBM (8 keV - 40 $\rm MeV$) \citep{1999APh....11..277G,2009ApJ...702..791M}, the Alpha Magnetic Spectrometer (AMS-02) (0.5 GeV - 2 TeV) \citep{2012IJMPE..2130005K}, the DArk Matter Particle Explorer (DAMPE, or called Wukong) (5 GeV -10 TeV) \citep{2017APh....95....6C}, we can have much better quality data for analyzing. Some researches argued that there maybe exits a third class of GRBs \citep[e.g.][]{De2011}. \citet{Tsutsui2013A} identified subclasses of LGRBs with cumulative light curve of the prompt emission. Different subclasses have different Fundamental Planes, which is a correlation between $L_{\rm pk}$, $E_{\rm pk}$ and $T_{\rm L}$ ($T_{\rm L}=E_{\rm iso}/L_{\rm pk}$) \citep{Tsutsui2011}. Forthermore, \citet{Tsutsui2014} confirmed that there do exist the third class in GRBs in addition to SGRBs and LGRBs. The classification method is based on two properties both quantified with light curve shapes of the
prompt emission: the absolute deviation from the constant luminosity of their cumulative light curve, and the ratio of the mean counts to the maximum counts \citep{Tsutsui2013A, Tsutsui2014}. A new class of GRBs with thousands of seconds of duration known as ultra-long bursts have been discovered \citep{Stratta2013,Levan2014,Cucchiara2015,Horesh2015,Gompertz2017}. The possible candidates for ultra-long bursts origin are blue supergiant collapsars, magnetars, and white dwarf tidal disruption events caused by massive black holes \citep{Stratta2013,Greiner2015,Ioka2016,Perets2016,Gompertz2017}. However, \citet{Evans2014} considered and rejected the possibility that ultra-long GRB 130925A was form of tidal disruption event, and instead showed that if the circumburst density around ultra-long GRB 130925A is low, the long duration of the burst and faint external shock emission are naturally explained. On the contrary, \citet{Gompertz2017} suggested a high density circumburst environment for ultra-long GRB 111209A, which was powered by the spin-down of a highly magnetized millisecond pulsar. \citet{Norris2006} found the existence of an intermediate class (IC) or SGRBs with Extended Emission (SGRBsEE), which shows mixed properties between SGRBs and LGRBs. Some models have been proposed to explain SGRBsEE, like fallback accretion onto a newborn magnetar \\citep{Rowlinson2012,Rowlinson2014,Rea2015,Gibson2017,Stratta2018}. \citet{Horvath1998} and \citet{Horvath2002}  investigated 797 and 1929 BASTE GRBs respectively, and found three-Gaussian (3-G) distribution is better than two-Gaussian (2-G) statistically. Similar results can be found in the analysis of $Beppo$SAX \citep{Horvath2009}, \textit{Swift} /BAT \citep{Horvath2008,Horvath2016}, \textit{Fermi}/GBM \citep{Tarnopolski2015} etc. The analyses method is  $\chi^{2}$ fitting \citep{Horvath1998,Tarnopolski2015} or Maximum Likelihood method \citep{Horvath2002,Horvath2008,Horvath2009}, and the rejection probability is less than 0.5$\%$. \citet{Zitouni2015} investigated 248 \textit{Swift}/BAT GRBs and also prefers to 3-G distribution, but prefers to 2-G distribution for BATSE bursts. It was suggested by \citet{Zitouni2015} that, because the distribution of envelope masses of the progenitors is non-symmetrical, the duration distribution corresponding to the collapsar scenario might not be symmetrical. \citet{Chattopadhyay2017} even found five kinds of GRBs in the BATSE catalog using Gaussian Mixture model. However, some recent works still insisted on two components \citep{Tarnopolski2015,Tarnopolski2016,Zhang2016,Yang2016A,Kulkarni2017,Bhave2017}, and showed the intermediate GRB class is unlikely. The plausible explanation of the duration bimodal distribution is that the two-class GRB duration distributions are intrinsically non-symmetrical \citep{Tarnopolski2016}. This indicates that we need more GRB parameters to evaluate the optimum number of components, and $T_{\rm 90}$ should not be the unique criterion. Because we have found some SGRBs have the properties of LGRBs, and LGRBs have the properties of SGRBs, duration criterion is not enough to reveal the physical origin of the GRBs. For example, GRB 060505, GRB 060614 and GRB 111005A have $T_{\rm 90}$ greater than 2 $\rm s$, but we have no detection of supernova associated with these three GRBs \citep{Dong2017}. They are also named as long-short GRBs or SN-less LGRBs \citep{Wang2017}. GRB 060614 is also more like an SGRB from the temporal lag and peak luminosity \citep{Gehrels2006}. Several models were proposed to explain GRB 060614, including the merger of an NS and a massive WD \citep{King2007}, the tidal disruption of a star by an intermediate-mass black hole \citep{Lu2008}, and a stellar-mass BH and a WD \citep{Dong2017}. A near-infrared bump was discovered in the afterglow, which probably arose from a Li-Paczy$\acute{\rm n}$ski macronova \citep{Li1998,Dong2017}, which supports the compact binary merger model for GRB 060614 \citep{Yang2015}. On the contrary, the $T_{\rm 90}$ of GRB 090426 is 1.24 $\rm s$, but it lies in a blue, star-forming and interacting host galaxy. The afterglow located at a small offset from the center of its host galaxy, and it is in the LGRB region of $E_{\rm p,rest}$-$E_{\rm iso}$ plot \citep{Antonelli2009,Levesque2010A}, where $E_{\rm p,rest}$ is spectral peak energy in rest-frame, $E_{\rm iso}$ is istropic $\gamma$-ray energy. All these properties indicate that its origin might be collapsing of massive star. \citet{Zhang2009} suggested that we need multi-wavelength criteria to decide the physical origin of individual GRBs. \citet{Li2016ApJS} gave a comparative overlapping properties study of SGRBs and LGRBs, and found that the three best parameters for the classification purpose are $T_{\rm 90}$, $f_{\rm eff}$ and $F_{\rm light}$, where $f_{\rm eff}$ is the effective amplitude parameter \citep{Lv2014B} and $F_{\rm light}$ is surface brightness fraction. \citet{Ruffini2016} divided LGRBs and SGRBs further into two sub-classes, depending on whether a BH is formed in the merger or in the hypercritical accretion process exceeding the NS critical mass, and they indicated two additional progenitor systems: WD-NS and BH-NS. Therefore, with the complex of the GRBs, the muti-dimensional analysis based on more complete data is likely to reveal the true classification, which needs a comprehensive data collection.

The other method is to seeking the underlying correlation between different properties. The most quoted relations are Amati relation \citep{Amati2002, Sakamoto2004, Lamb2004, Amati2006, Amati2009, Virgili2012, Demianski2017} and Ghirlanda relation \citep{Ghirlanda2004}. \citet{Amati2002} found a correlation between $E_{\rm iso}$ in the rest-frame 1-$10^{4}$ $\rm keV$ energy band and spectral peak energy ($E_{\rm pk}$) with 12 $Beppo$SAX GRBs. It is maybe due to larger initial Lorentz factor $\Gamma_{0}$ in GRBs also have larger energy and peak energy \citep{Ghirlanda2012}, or an optically thin synchrotron shock model \citep{Lloyd2000A}. \citet{Amati2008} updated the samples. They used 70 LGRBs and XRFs (X-ray flashes), and analyzed the correlation between rest-frame peak energy ($E_{\rm pk,i}$) and $E_{\rm iso}$. They also pointed that $E_{\rm pk,i}$-$E_{\rm iso}$ correlation is not affected by significant selection effects. However, this relation is also challenged by other works \citep{2005MNRAS.360L..73N,2005ApJ...627..319B}. \citet{Ghirlanda2004} used 40 GRBs to derive a correlation between collimation-corrected energy and $E_{\rm pk}$.  They gave a further test for this correlation later \citep{Ghirlanda2007}. \citet{Yonetoku2004} also gave a reliable relation between $E_{\rm pk}$ and peak luminosity $L_{\rm pk}$ (Yonetoku correlation) using the data of $Beppo$SAX and BATSE. Then they estimated redshifts and GRB formation rate for some GRBs without known distances. Afterwards, a growing number of samples are used to investigate Yonetoku correlation \citep{Ghirlanda2004A,Ghirlanda2005,Yonetoku2010,Lu2010,Tsutsui2013C}. \citet{Yonetoku2010} reanalyzed Amati and Yonetoku relation with 101 GRBs, and examined how the truncation of the detector sensitivity affects the correlations, and concluded they are surely intrinsic properties of GRBs. \citet{Firmani2006} reported that adding $T_{\rm R45}$ \citep{Reichart2001} can reduce the dispersion of Yonetoku correlation. Therefore, they discovered a correlation between $L_{\rm pk}$, $E_{\rm pk}$ and $T_{\rm R45}$. However, the studies of \citet{Rossi2008} and \citet{Collazzi2008} did not confirm this relation. \citet{Tsutsui2009} investigated the correlation between the residuals of $L_{\rm pk}$ and $E_{\rm iso}$ from the best function, and found that the luminosity time ($T_{\rm L}=E_{\rm iso}/L_{\rm pk}$) can improve the Amati and Yonetoku relations. Later, \citet{Tsutsui2011} gave a new correlation between $T_{\rm L}$, $L_{\rm pk}$ and $E_{\rm pk}$ with considering the systematic errors. \citet{Tsutsui2010} pointed that the intrinsic dispersion of correlations among $E_{\rm pk}$, $L_{\rm pk}$ and $E_{\rm iso}$ depends on the quality of data set. \citet{Tsutsui2013C} analysed Amati and Yonetoku relations for 13 SGRBs, and the correlations are dimmer than those of LGRBs for the same $E_{\rm pk}$. \citet{Zhang2017} used the $E_{\rm pk}$ and $L_{\rm pk}$ correlation with 16 SGRBs to study the luminosity function and formation rate of SGRBs. Isotropic luminosity ($L_{\rm iso}$) and $E_{\rm pk}$  are also found to have good correlation \citep{Schaefer2003A,Frontera2012,Nava2012}. \citet{Schaefer2001} and \citet{Schaefer2003A} pointed that the $E_{\rm pk}$ and $L_{\rm iso}$ correlation is due to their dependence on $\Gamma_{0}$. With this correlation, we can further study the structure of the ultra relativistic outflow, the shock acceleration and the magnetic field generation \citep{Lloyd2002A}. \citet{Dichiara2016} found a highly significant anti-correlation between $E_{\rm pk,i}$ and power density spectra (PDS) slope $\alpha$  from 123 LGRBs. They put forward a model based on magnetic reconnection for this phenomenon. The $E_{\rm pk,i}$ and $\alpha$ are linked to the ejecta magnetization at the dissipation site, so that more magnetised outflows would produce more variable GRB light curves at short timescales, shallower PDS and higher values of $E_{\rm pk,i}$. Addition to the relations extracted from the spectra, some relations are obtained from the GRB light curves. \citet{Norris2000} found an anti-correlation between $L_{\rm pk}$ and spectral lag with 174 BATSE GRBs. \citet{Tsutsui2008} got a new spectral lag-$L_{\rm pk}$ relation with 565 BATSE GRBs, which is different with the result of \citet{Norris2000}. This anti-correlation might contain indirect connections to $\Gamma_{0}$, and it has been confirmed by several studies \citep{Salmonson2000,Schaefer2001,Daigne2003,Zhang2006}. The interpretation of this relation might be kinematic effect \citep{Salmonson2000}, or energy formation affecting the development of the pulse much more than dissipation \citep{Norris2000}. The correlation between variability and $L_{\rm pk}$ can estimate GRB redshift and luminosities \citep{Fenimore2000,Reichart2001}. The origin of this correlation may be relativistically shocked jets \citep{Schaefer2007}. \citet{Mallozzi1995} found a correlation between $E_{\rm pk}$ and peak photon flux ($P_{\rm pk}$) in 256 ms time bin of 50-300 $\rm keV$. Then \citet{Lloyd2000} simulated 1000 GRBs  in 50-300 $\rm keV$ energy band, and found a similar strong correlation between $E_{\rm pk}$ and $F_{\rm g}$. \citet{Goldstein2010} used this correlation to classify LGRBs and SGRBs, and confirmed the presence of two GRB classes. There are many similar works to confirm this correlation \citep{Borgonovo2001,Ghirlanda2010,Guiriec2010,Ghirlanda2011,Lu2012A}. With the estimated $\Gamma_{0}$, some related correlations are also found, such as  $\Gamma_{0}-E_{\rm iso}$ \citep{Liang2010},  $\Gamma_{0}-L_{\rm iso}$ \citep{Lv2012} and $L_{\rm iso} - E_{\rm pk,i} - \Gamma_{0}$ \citep{Liang2015}. \citet{Willingale2007A} showed that a source frame characteristic-photon-energy/peak luminosity ratio, $K_{\rm z}$, can be constructed, and it is constant within a factor of 2 for all bursts. The existence of $K_{\rm z}$ indicates that the mechanism responsible for the prompt emission from all GRBs is probably predominantly thermal \citep{Willingale2007A}. \citet{Willingale2010} analyzed the individual prompt pulses of a GRB. They showed that the luminosity of pulses is correlated with the peak energy of the pulse spectrum, and anti-correlated with the time since ejection of the pulse.

Besides the individual study on  GRB prompt emission, the statistics on the afterglows is also helpful in understanding the nature. A correlation was found between the early optical/UV luminosity (measured at rest-frame 200 s) and average decay rate (measured from 200 s) \citep{Oates2009, Oates2012}. The luminosity-decay correlation also exits in X-ray band, and is consistent with optical/UV \citep{Oates2015, Oates2016, Racusin2016}. At the same time, the early optical/UV luminosity (measured at rest-frame 200 s) correlations with isotropic energy $E_{\rm iso}$ and rest-frame peak spectral energy $E_{\rm pk}$ \citep{Oates2015, Davanzo2012, Margutti2013}. \citet{Evans2007} and \citet{Evans2009} provided the methods and results of an automatic analysis of a complete sample of \textit{Swift}-XRT observations of GRBs. \citet{Liang2008A} and \citet{Liang2008B} gave more detailed analysis, like the jet breaks \citep{Liang2008B}, and the early shallow decay to late jet like decay phases \citep{Liang2008A}. \citet{Yi2016ApJS} gave a comprehensive study of the X-ray flares from GRBs observed by \textit{Swift}. They analyzed the ten-year X-ray flare data of \textit{Swift}/XRT until the end of March 2015, and studied the distributions of energy, duration, waiting time, rise time, decay time, peak time and peak flux. After that, \citet{Yi2017} statistically studied GRB optical flares from the \textit{Swift}/UVOT catalog. They found optical flares and X-ray flares may share the similar physical origin and both of them are possibly related to central engine activities. \citet{Jia2016} gave a statistical study of GRB X-ray flares to prove the ubiquitous bulk acceleration in the emission region. With the recent development of networks of robotic telescopes, we are able to follow up the early optical emission of GRBs and to find a peak in the optical afterglow light curves ($t_{\rm pkOpt}$) of some GRBs. It is due to the dynamics of the fireball deceleration. So $t_{\rm pkOpt}$ can provide the $\Gamma_{0}$ of the fireball before deceleration, and $\Gamma_{0}$ represents the maximum value attained by the outflow during this dynamical evolution. Bulk Lorentz factor $\Gamma_{0}$ of GRBs is very important. We can use it to compute GRBs' comoving frame properties shedding light on their physics \citep{Ghirlanda2017}. We can use $t_{\rm pkOpt}$ to derive the distribution of $\Gamma_{0}$ \citep{Rykoff2009,Ghirlanda2012,Ghirlanda2017}, also other methods to estimate $\Gamma_{0}$ \citep{Sari1999,Molinari2007,2010MNRAS.402.1854Z,Ghisellini2010,2011ApJ...726L...2Z,Lv2012,Nava2013,Nappo2014,2015ApJ...800L..23Z}. \citet{Geng2016A} developed a numerical method to calculate the dynamic of the system consisting of a forward shock and a reverse shock. They found that the steep optical re-brightenings would be caused by the fall-back accretion of black holes, while the shallow optical re-brightenings are the consequence of the injection of the electron-positron-pair wind from the central magnetar. \citet{Dainotti2008} examined the X-ray decay curves of all the GRBs measured by the \textit{Swift} that available. They found a correlation between $\log [T_{\rm a}/(1+z)]$ ( the time in the X-ray at the end of the plateau in rest-frame) and $\log [L_{\rm X}(T_{\rm a})]$ (X-ray luminosity at the time $T_{\rm a}$), hereafter also referred as LT. The slope is $-0.74_{-0.19}^{+0.2}$. This LT anti-relation shows that the shorter the plateau duration, the more luminous the plateau. \citet{Dainotti2008} believed that this LT anti-correlation is a further tool towards the standardization of GRBs as a distance indicator. This result is confirmed by \citet{Ghisellini2009} and \citet{Yamazaki2009}. A physical subsample of LGRBs with a significant LT anti-correlation in the GRB rest-frame is discovered \citep{Dainotti2010}. \citet{Dainotti2016} revealed that the subsample of LGRBs associated with SNe presents a very high LT anti-relation. This analysis may open new perspectives in future theoretical investigations of the GRBs with plateau emission and associated with SNe \citep{Dainotti2017}. Many researches have been carried on the LT relation, like expanding the sample \citep{Dainotti2011A, Mangano2012, Dainotti2015B}, or the selection biases influence on the slope of the relation \citep{Dainotti2013A, Dainotti2018A, Dainotti2018B}.There are also many correlations between the prompt and the afterglow. For the LT relation, \citet{Xu2012} added $E_{\rm iso}$ as a third parameter to get a tighter three-parameter correlation. When adding peak luminosity in the prompt emission $L_{\rm pk}$, \citet{Dainotti2016} also found a good correlation. Similarly, \citet{Si2018} analyzed the optical light curves of 50 GRBs. They calculated the break time of optical plateaus and the break luminosity. When they added $E_{\rm iso}$, a significantly tighter correlation was found. \citet{Dainotti2015A} demonstrate that $L_{\rm pk}$ (the peak luminosity in the prompt emission) and $L_{\rm prompt}$ (the averaged prompt luminosity) have intrinsic correlations with $L_{\rm X}(T_{\rm a})$ respectively. \citet{2005ApJ...633..611L} derived a three parameters correlation, which is between $E_{\rm \gamma,iso}$, spectral peak energy in the rest-frame and rest-frame break time of the optical afterglow light curves. \citet{Zaninoni2016} gave a comprehensive statistical analysis of \textit{Swift} X-ray light curves of GRBs collected from December 2004 to June 2014, and found a three parameter correlation between $E_{\rm iso}$ in the rest-frame 1-$10^{4}$ $\rm keV$ energy band, $E_{\rm pk,i}$ and $E_{\rm X,iso}$ (X-ray energy emitted in the rest frame 0.3-30 $\rm keV$) \citep{Margutti2013}.

There are also some researches on GRB host galaxies. The LGRBs host galaxy offsets (the distance from the site of the GRBs to the center of its host galaxies) are consistent with the expected distribution of massive stars \citep{Bloom2002}, and SGRBs host galaxy offsets are in good agreement with neutron star binary mergers \citep{Fong2010A,Church2011}. SGRBsEE seems to be the subgroup of SGRBs, because SGRBsEE mostly have smaller projected physical offsets \citep{Troja2008} and occur closer to their host galaxies in denser interstellar environments \citep{Malesani2007}. SGRBs without extended emission are opposite. It also implies that SGRBs possibly have two distinct populations. SGRBsEE are due to NS-BH mergers, and SGRBs without extended emission are due to NS-NS mergers \citep{Troja2008}. Furthermore, X-ray absorption column densities and SGRBs host galaxy offsets correlation gives another evidence that SGRBs possibly have two distinct populations \citep{Kopac2012}.
\citet{Japelj2016} studied the host galaxies of a complete sample of bright LGRBs to investigate the impact of the environment on GRB formation. \citet{Arabsalmani2018} studied the mass-metallicity (MZ) relation from 33 GRB hosts spanning a redshift range between  $\sim$ 0.3 and $\sim$ 3.4 and a mass range from $10^{8.2}$ $M_{\bigodot}$ to $10^{11.1}$ $M_{\bigodot}$. They found that GRB hosts track the the MZ relation of the general star-forming galaxy population with an average offset of 0.15$\pm$0.15 dex below the MZ relation of the general population, and metallicity measurements can influence the relation result \citep{Arabsalmani2018}. The offset may be the result of the different methods used to select their respective galaxy populations \citep{Kocevski2011}. There are also some relations between host galaxy and afterglow emission (or prompt emission). \citet{Zhang2017B} carried out some statical analysis and found possible correlations between SGRBs afterglow luminosities and their host galaxy offset. This may be due to the number density of circum-burst medium should decrease with the distance to the host galaxy center. However there are some other uncertainties related to the correlation, including the angle between the line of sight, the host galaxy disk and SGRBs occurred in globular cluster \citep{Zhang2017B}. Selection effects usually occur when the sample observed is not representative of the true population itself \citep{Dainotti2017}.

All the previous statistical studies are all based on the data directly from the original data or the catalogs. For each individual study, one needs to collect the data from the beginning. Here we are trying to collect all the data and let the data as a reservoir for future studies. The data are all manually collected from the literature, including almost all the properties belongs to GRBs, i.e., prompt emission, afterglows and the host galaxies. Comparing with the automatic gathering data by machine, the manual collection is slow. However, it is hard to let the machine to recognize the variety of symbols, and many of data are expressed in different ways, which makes the task even harder. Before the advance of the machine algorithm, for the precision of the data gathering, we can only manually collect them in the present stage.
This paper is organized as follows. In Section \ref{sec:sample}, we introduce the data preparation, including data gathering from published papers. The table is in a machine readable table, while a sample of the table is shown in Table \ref{tab:bigtable}. Error imputation for the data with central values but without error bars is introduced in Section \ref{sec:imputation}. In Section \ref{sec:method}, we introduce our statistical methods. We give the distributions for each parameter in observer-frame and some parameters in rest-frame. All the histograms are shown in Fig. Set 1. We put two histograms in Figure \ref{fig:distribution}. We gave scatter plots between two parameters with at least 5 samples. All the plots are shown in Fig. Set 2 and two scatter plots in Figure \ref{fig:scatter}. We  analyzed the linear coefficient and non-linear correlation ratio between two parameters with at least 5 GRBs. We excluded the  trivial results, like $T_{\rm 90}$ and $T_{\rm 90,i}$ correlation, which are obviously correlated but do not give extra information. A small part of all the results are shown in Table \ref{tab:coefficient}, and the comprehensive results are shown in machine the readable table. We gave all the linear regression results between two and three parameters with at least 5 GRBs. A small part of all the results are shown in Table \ref{tab:linear2} and Table \ref{tab:linear3}, and all the results are given in two machine readable tables. Conclusions are given in Section \ref{sec:tworesult} and Section \ref{sec:threeresult}, we gave more detailed analysis for some good results, and gave some reasonable explanation for these good results. Discussion is given in Section \ref{sec:discuss} and conclusion is given in Section \ref{sec:conclusion}. A concordance cosmology with parameters $H_{\rm 0}=67.8 \pm 0.9 \rm km ~ s^{\rm -1} ~ Mpc^{\rm -1}$, $\Omega_{\rm M}=0.308 \pm 0.012$ \citep{Planck2016} is adopted in all part of this work.

\section{Samples} \label{sec:sample}

We collected all the possible data for 6289 GRBs from GRB 910421 (April 21st, 1994) to GRB 160509A (May 9th, 2016). There are 46 parameters in this catalog, which includes the basic information, the prompt emission, the afterglow and the host galaxy.

For basic information of each GRB, we recorded the trigger time, instrument, trigger number, coordinate and position error. Most of basic information is from Gamma-ray Coordinates Network (GCN) and published papers, while most of BATSE GRBs information is from BATSE website \footnote{\url{https://gammaray.msfc.nasa.gov/batse/grb/catalog/current/}}. The basic information is quite important. It can help us to make sure weather some GRBs with the same names are different GRBs or not. Because for some GRBs detected by different instruments have the same GRB name, we changed the names for a small part of GRBs. For example, there is a \textit{Fermi} GRB 100911,  while there is also a MAXI GRB 100911. They have same name in different papers, however, the trigger time is different. Therefore, they must be different GRBs. We made the \textit{Fermi} GRB name as GRB 100911A and the MAXI GRB name as GRB 100911B, following the trigger time. Besides, some GRBs have different names in different papers. For all the GRBs that have different names or are changed the names by us, we use the flag ``A" to remark. Not all the basic information are available for every GRB. When we did not find the information, we let the place empty.

The prompt emission properties include $z$ (redshift), $D_{\rm L}$ (luminosity distance, $\rm 10^{\rm 28} ~ cm$), $T_{\rm 90}$, $T_{\rm 50}$, $T_{\rm R45}$ \citep[defined in][]{Reichart2001}, $variability_{\rm 1}$ (variability of \citet{Fenimore2000} definition), $variability_{\rm 2}$ (variability of \citet{Reichart2001} definition), $variability_{\rm 3}$ (variability of \citet{Schaefer2007} definition), $F_{\rm g}$ (fluence in 20-2000 $\rm keV$ energy band, in unit of $\rm 10^{\rm -6} ~ ergs ~ cm^{\rm -2}$), HR (hardness ratio between 100-2000 $\rm keV$ and 20-100 $\rm keV$),$E_{\rm iso}$ (isotropic $\gamma$-ray energy in rest-frame 1-$10^{4}$ $\rm keV$ energy band,  in unit of $\rm 10^{\rm 52} ~ ergs$), $L_{\rm pk}$ (peak luminosity of 1 $\rm s$ time bin in rest-frame 1-$10^{4}$ $\rm keV$ energy band,  in unit of $\rm 10^{\rm 52} ~ erg ~ s^{\rm -1}$), $F_{\rm pk1}$ (peak energy flux of 1 $\rm s$ time bin in rest-frame 1-$10^{4}$ $\rm keV$ energy band, in unit of  $\rm 10^{\rm -6} ~ ergs ~ cm^{\rm -2} ~ s^{\rm -1}$), $F_{\rm pk2}$ (peak energy flux of 64 $\rm ms$ time bin in rest-frame 1-$10^{4}$ $\rm keV$ energy band,  in unit of $\rm 10^{\rm -6} ~ ergs ~ cm^{\rm -2} ~ s^{\rm -1}$), $F_{\rm pk3}$ (peak energy flux of 256 $\rm ms$ time bin in rest-frame 1-$10^{4}$ $\rm keV$ energy band,  in unit of $\rm 10^{\rm -6} ~ ergs ~ cm^{\rm -2} ~ s^{\rm -1}$), $F_{\rm pk4}$ (peak energy flux of 1024 $\rm ms$ time bin in rest-frame 1-$10^{4}$ $\rm keV$ energy band,  in unit of $\rm 10^{\rm -6} ~ ergs ~ cm^{\rm -2} ~ s^{\rm -1}$), $P_{\rm pk1}$ (peak photon flux of 64 $\rm ms$ time bin in 10-1000 $\rm keV$, in unit of  $\rm photons ~ cm^{\rm -2} ~ s^{\rm -1}$), $P_{\rm pk2}$ (peak photon flux of 256 $\rm ms$ time bin in 10-1000 $\rm keV$,  in unit of $\rm photons ~ cm^{\rm -2} ~ s^{\rm -1}$), $P_{\rm pk3}$ (peak photon flux of 1024 $\rm ms$ time bin in 10-1000 $\rm keV$,  in unit of $\rm photons ~ cm^{\rm -2} ~ s^{\rm -1}$), $P_{\rm pk4}$ (peak photon flux of 1 $\rm s$ time bin in 10-1000 $\rm keV$,  in unit of $\rm photons ~ cm^{\rm -2} ~ s^{\rm -1}$). $\alpha_{\rm band}$ (low energy spectral index of band model), $\beta_{\rm band}$ (high energy spectral index of band model), $E_{\rm p,band}$ (spectral peak energy of band model, $\rm keV$), $\alpha_{\rm cpl}$ (low energy spectral index of CPL model), $E_{\rm p,cpl}$ (spectral peak energy of CPL model,  in unit of $\rm keV$), $\alpha_{\rm spl}$ (spectral index of SPL model), $\theta_{\rm j}$ (jet open angle,  in unit of $\rm rad$), spectral time lag (in unit of $\rm ms ~ MeV^{\rm -1}$), $\Gamma_{0}$ (initial Lorentz factor). We removed the spectral time lag for GRB 060218, because it is an clear outlier \citep{Foley2008}. $P_{\rm pk}$ and $F_{\rm pk}$ have four time bins: 64 $\rm ms$, 256 $\rm ms$, 1024 $\rm ms$ and 1 $\rm s$. We make them as different parameters. As the data are from different instruments, while different instruments have different energy bands and different sensitivities. In order to avoid the influence of different energy band, we corrected the $P_{\rm pk}$ into observer-frame in 10-1000 $\rm keV$ energy band, $F_{\rm pk}$ into rest-frame 1-$10^{4}$ $\rm keV$ energy band using the method of \citet{Schaefer2007} of four time bins respectively. At the same time, using the same method, we changed all the $F_{\rm g}$ into 20-2000 $\rm keV$, HR into 100-2000/20-100 $\rm keV$, $E_{\rm iso}$ into rest-frame 1-$10^{\rm 4}$ $\rm keV$ energy band, $L_{\rm pk}$ into rest-frame 1-$10^{\rm 4}$ $\rm keV$ energy band. Besides, we calculated a small part of $E_{\rm iso}$ also using the method of \citet{Schaefer2007} with $F_{\rm g}$ and $z$. We gave different flags for different data calculation and correction.

The afterglow properties include $\log t_{\rm burst}$ (central engine active duration in logarithm,  in unit of $\rm s$) \citep[defined in][]{Zhang2014}, $t_{\rm pkX}$ (peak time in X-ray LC,  in unit of $\rm s$), $t_{\rm pkOpt}$ (peak time in optical LC, $\rm s$), $F_{\rm X11hr}$ (flux density in X-ray band at 11 hours related to the trigger time,  in unit of $\rm Jy$), $\beta_{\rm X11hr}$ (index in X-ray band at 11 hours related to the trigger time), $F_{\rm Opt11hr}$ (flux density in optical band at 11 hours after the trigger time, $\rm Jy$), $t_{\rm radio,pk}$ (peak time in radio band,  in unit of $\rm s$), $F_{\rm radio,pk}$ (peak flux density in radio band at 8.46 $\rm GHz$,  in unit of $\rm Jy$).

The host galaxy properties include host galaxy offset (the distance from GRB location to the centre of its host galaxy, in unit of $\rm kpc$), metallicity ($12+\log {\rm [O/H]}$), Mag (absolute magnitude in AB system at rest 3.6 $\mu m$ wavelength), $N_{\rm H}$ (column density of hydrogen,  in unit of $\rm 10^{\rm 21} ~ cm^{\rm -2}$), $A_{\rm V}$ (dust extinction), SFR (star formation rate,  in unit of $\rm M_{\bigodot} ~ yr^{\rm -1}$), $\log SSFR$ (specific star formation rate in logarithm,  in unit of $\rm Gyr^{\rm -1}$), Age ( in unit of $\rm Myr$), $\log Mass$ (stellar mass in logarithm,  in unit of $M_{\bigodot}$). For metallicity, we converted different metallicity calibrations into \citet{Kobulnicky2004} calibrator using the method given in \citet{Kewley2008}.

We also changed 10 parameters from observer-frame to rest-frame, including $T_{\rm 90}$, $T_{\rm 50}$, $T_{\rm R45}$, $E_{\rm p,band}$, $E_{\rm p,cpl}$, spectral time lag, $\log t_{\rm burst}$, $t_{\rm pkX}$, $t_{\rm pkOpt}$ and $t_{\rm radio,pk}$. We use label ``i" to remark the parameters in rest-frame. We put the 10 rest-frame parameters as new parameters joining the statistics. But we did not show the results for the same parameter just in different frames, such as $T_{\rm 90}$ and $T_{\rm 90,i}$. In the table, we also did not include the 10 rest-frame parameters as they can be easily obtained. When we collected the data from the literatures, some data have different units for the same parameter. We convert them into the same. For example, the unit of $D_{\rm L}$ (luminosity distance) includes $\rm 10^{\rm 28} ~ cm$, $\rm kpc$ and others. In order to have an easy conversion in $\rm ergs$ for $E_{\rm iso}$ calculation, we choose the unit $\rm 10^{\rm 28} ~ cm$ for $D_{\rm L}$. The units of the parameters are all given in table \ref{tab:bigtable}.

All the information above of the 6289 GRBs are collected in the machine readable table. We just put the first 5 samples in Table \ref{tab:bigtable} as an example. In the machine readable table, every GRB has the basic information and 46 parameter values.
If the basic information is not available, we let it as blank. If the parameter value is not available, we use ``..." to remark. For every available parameter value, we also put the relevant reference or flag (or both). For different situations, we use different flags. We introduce every flag one by one in the following. ``a" means the errors are imputed by MICE algorithm. The details of error imputation are shown in Section \ref{sec:imputation}. ``b" means the errors in the original papers were in 90\% confidence level, and we change the errors to 1 $\sigma$ confidence level by multiplying 0.995/1.645. ``c" means the values are calculated using the spectral values in order to change the energy band to a uniform band. Because for some parameters, the energy band is different for different instruments. In order to avoid the influence of different energy band, we use the spectral values to correct the energy band of the 4 time bins $P_{\rm pk}$, 4 time bins $F_{\rm pk}$, $F_{\rm g}$, HR, $E_{\rm iso}$ and $L_{\rm pk}$. ``d" means the unit is different from the original papers. For example, the unit for $D_{\rm L}$ in \citet{Cano2014} is $\rm Mpc$, we change it to $\rm 10^{\rm 28} ~ cm$. ``e" means the error is estimated as the central value multiplying 0.1. One can see Section \ref{sec:imputation} for more details. When coming to spectral values, sometimes the spectral index cannot be constrained very well. We use the BASTE $\alpha$ peak value -1.1 as common with flag ``f", and the BASTE $\beta$ peak value -2.2 as common with flag ``g". Some values are in rest-frame, then we change the values from rest-frame to observer-frame with flag ``h". ``i" means the values are converted into logarithm or from logarithm into the normal form. ``j" means the values are calculated using other parameters. For example, almost half of HRs are calculated from $F_{\rm g}$ and spectral values. ``k" means we converted different metallicity calibrations into \citet{Kobulnicky2004} calibrator using the method in \citet{Kewley2008}. We found the metallicity is different with different calibrators, so this step is necessary. ``m" means the $D_{\rm L}$ is calculated using $z$ with cosmology parameters in \citet{Planck2016}. There is also a description in Table \ref{tab:bigtable}.

The data are from different literatures, GCN,  websites \footnote{such as \url{http://www.mpe.mpg.de/~jcg/grbgen.html}} and calculations if it is not directly available and can be derived. Principle of data collection is the following. We collect variables as many as possible. We only take the certain values, which are mainly taken from literature manually. Every data should be directed reversely to the original reference. The data should include error bars in 1 $\sigma$. If the error is in other confidence, we convert the error into 1 $\sigma$.  If it is not available, we use the imputation, which is shown in Section \ref{sec:imputation}. For those data are shown in different literature, we choose the values in the following order (higher to lower priority): 1st, published papers (catalogs, data gatherings, other articles); 2nd, GCNs; 3rd, other websites; 4th self-calculation if they are able to obtain. For those data shown in ordinary published papers, the paper from the official team is in higher priority. Otherwise, the newer paper is of higher priority.

We excluded \textit{Fermi} GRBs 120222A, 110920A, 110517A and 101214A, because they have the same GRB names while two different trigger numbers \citep{Bhat2016}. This might be a mistake for GRB names.  For GRB 091024A, it triggered \textit{Fermi}/GBM twice and triggered \textit{Swift} once \citep{Bhat2016,Ukwatta2012}. We set it as one GRB for the analysis. During the statistical study, we deleted some values with central values smaller than error bars (which means the values in the lower side are not physical). Because for some parameters, the lower limit must be greater than 0, such as $T_{\rm 90}$. If the central value is smaller than error bar for $T_{\rm 90}$, the lower limit is smaller than 0. For example, in \citet{Von2014},  the $T_{\rm 90}$ of GRB 080719A is $16.128 \pm 17.887$ s, we did not include this datum. Notice we still put this kind of data in the table, and just did not include them in the statistics.

For the spectral parameters, the spectra are mainly fitted by three models: Band model, cutoff power law (CPL) model and simple power law (SPL) model \citep{Li2016ApJS}. Band model is a smoothly joint broken power law with the definition \citep{Band1993}:
\begin{equation}
\label{eq:Band}
N(E)=
\left.
\Big \{
\begin{array}{lr}
A\left(\frac{E}{100\ {\rm keV}}\right)^{\alpha} e ^{-\frac{E}{E_0}},& E < (\alpha-\beta)E_0, \\
A\left(\frac{E}{100\ {\rm keV}}\right)^{\beta} \left[ \frac{(\alpha-\beta)E_0}{100\ {\rm keV}}\right]^{\alpha-\beta} e^{\beta-\alpha},& E \ge (\alpha-\beta)E_0,
\end{array}
\right.
\end{equation}
where $\alpha$ is low energy photon index, $\beta$ is high energy photon index, $A$ is the coefficient for normalization, $E$ is the energy of the photons,
and $E_0$ is the break energy. Mostly we used $E_{\rm p}$ instead of $E_0$. $E_{\rm p}$ is the peak energy in spectrum $E^2N$, and $E_{\rm p}=(2+\alpha)E_0$. 
In the big table, we use $\alpha_{\rm band}$, $\beta_{\rm band}$ and $E_{\rm p,band}$ as Band function spectral parameters. $\alpha_{\rm cpl}$ and $E_{\rm p,cpl}$ to mark CPL model spectrum parameters. The formula of SPL model is $N(E) = A E^{\alpha_{\rm spl}}$.

Most of the data are taken from the following literatures: (1) \textit{Swift} lookup website \footnote{\url{https://swift.gsfc.nasa.gov/archive/grb_table/index.php}} with the number of selected data  1003; (2) ``How long does a burst burst?" \citep{Zhang2014} with the number of selected data  354; (3) ``Cosmic evolution of long gamma-ray burst luminosity" \citep{Deng2016} with the number of selected data  177. (4) ``Effect of gamma-ray burst (GRB) spectra on the empirical luminosity correlations and the GRB Hubble diagram" \citep{Lin2016} with the number of selected data  219; (5) ``On the classification of GRBs and their occurrence rates" \citep{Ruffini2016} with the number of selected data  571; (6) ``The third Fermi GBM gamma-ray burst catalog: the first six years" \citep{Bhat2016} with the number of selected data  3702; (7) ``A comparative study of long and short GRBs. \MyRoman{1}. overlapping properties"  \citep{Li2016ApJS} with the number of selected data  786; (8) ``GRB hosts through cosmic time. VLT/X-Shooter emission-line spectroscopy of 96 {$\gamma$}-ray-burst-selected galaxies at $0.1 < z < 3.6$"  \citep{KruhlerAA2015} with the number of selected data  107; (9) ``Uncovering the intrinsic variability of gamma-ray bursts"  \citep{Golkhou2014} with the number of selected data  323; (10) ``The E$_{peak}$ - E$_{iso}$ relation revisited with Fermi GRBs. Resolving a long-standing debate"  \citep{Heussaff2013} with the number of selected data  671; (11) ``The Second Fermi GBM Gamma-Ray Burst Catalog: The First Four Years"  \citep{Von2014} with the number of selected data  2019; (12) ``The spectral catalogue of INTEGRAL gamma-ray bursts. results of the joint IBIS/SPI spectral analysis"  \citep{Bosnjak2014} with the number of selected data  138; (13) ``The Swift GRB Host Galaxy Legacy Survey. \MyRoman{2}. Rest-frame Near-IR Luminosity Distribution and Evidence for a Near-solar Metallicity Threshold"  \citep{Perley2016A} with the number of selected data  132; (14) ``The dark bursts population in a complete sample of bright Swift long gamma-ray bursts"  \citep{Melandri2012} with the number of selected data  107; (15) ``A Radio-selected Sample of Gamma-Ray Burst Afterglows"  \citep{Chandra2012} with the number of selected data  347; (16) ``Statistical Analysis of the Observable Data of Gamma-Ray Bursts"  \citep{Ripa2011} with the number of selected data  260; (17) ``Spectral properties of 438 GRBs detected by Fermi/GBM"  \citep{Nava2011} with the number of selected data  694; (18) ``Possible Origins of Dispersion of the Peak Energy-Brightness Correlations of Gamma-Ray Bursts"  \citep{Yonetoku2010} with the number of selected data  254; (19) ``The Cosmic Rate, Luminosity Function and Intrinsic Correlations of Long Gamma-Ray Bursts"  \citep{Butler2010} with the number of selected data  874; (20) ``The updated spectral catalogue of INTEGRAL gamma-ray bursts"  \citep{Vianello2009} with the number of selected data  121; (21) ``Redshift Catalog for Swift Long Gamma-ray Bursts"  \citep{Xiao2011} with the number of selected data  415; (22) ``Search for gamma-ray burst classes with the RHESSI satellite"  \citep{Ripa2009} with the number of selected data  323; (23) ``Statistical studies of optically dark gamma-ray bursts in the Swift era"  \citep{Zheng2009} with the number of selected data  271; (24) ``Correlations of Prompt and Afterglow Emission in Swift Long and Short Gamma-Ray Bursts"  \citep{Gehrels2008} with the number of selected data  209; (25) ``A Complete Catalog of Swift Gamma-Ray Burst Spectra and Durations: Demise of a Physical Origin for Pre-Swift High-Energy Correlations"  \citep{Butler2007} with the number of selected data  1069; (26) ``Intrinsic properties of a complete sample of HETE-2 gamma-ray bursts. A measure of the GRB rate in the Local Universe"  \citep{Pelangeon2008} with the number of selected data  272; (27) ``Global characteristics of X-ray flashes and X-ray rich GRBs observed by HETE-2"  \citep{Sakamoto2005} with the number of selected data  101; (28) ``Konus catalog of SGRBs"  \citep{Mazets2002} with the number of selected data  128; (29) ``The Gamma-Ray Burst Catalog Obtained with the Gamma-Ray Burst Monitor Aboard BeppoSAX"  \citep{Frontera2009} with the number of selected data  1619; (30) ``Spectral catalogue of bright gamma-ray bursts detected with the BeppoSAX/GRBM"  \citep{Guidorzi2011} with the number of selected data  248; (31) ``BASTE current gamma-ray burst catalog \footnote{\url{https://gammaray.nsstc.nasa.gov/batse/grb/catalog/current/index.html}} with the number of selected data  8983; (32) ``Hardness as a spectral peak estimator for gamma-ray bursts"  \citep{Shahmoradi2010} with the number of selected data  2128; (33) ``The BATSE 5B Gamma-Ray Burst Spectral Catalog"  \citep{Goldstein2013} with the number of selected data  4764; (34) ``Are There Any Redshift $>8$ Gamma-Ray Bursts in the BATSE Catalog"  \citep{Ashcraft2007} with the number of selected data  110; (35) ``Short versus long gamma-ray bursts: spectra, energetics and luminosities"  \citep{Ghirlanda2009} with the number of selected data  139; (36) ``The Fourth BATSE Gamma-Ray Burst Catalog (Revised)"  \citep{Paciesas1999} with the number of selected data  1049; (37) ``On the Spectral Energy Dependence of Gamma-Ray Burst Variability"  \citep{Lloyd2002} with the number of selected data  159; (38) ``The Third BATSE Gamma-Ray Burst Catalog"  \citep{Meegan1996} with the number of selected data  1380. A full reference list is given following Table \ref{tab:bigtable}.

\section{Error Imputation} \label{sec:imputation}

In the table, there are many incomplete data. Especially there are many data just having central values without error bars. However, when we perform the statistics, we need to consider the uncertainty for each data. Therefore, we used the R package $mice$ imputes incomplete multivariate data by multiple imputation \citep{Rubin1987,Rubin1996} by chained equations (MICE). Multiple imputation is the method to deal with the problems of incomplete data. In the big table, many parameters have missing errors. For the first step of generating multiple imputation, we used the predictive mean matching \citep{Little1988} (PMM) method. It is a general purpose semi-parametric imputation method for numeric scale type. We did 5 times iterations for every missing error, and we set a threshold 0.25 for imputation. It means the error of each parameter  should have at least 25\% percentage information. Because if the information is not enough, the imputation maybe not reliable. Then we use three indictors to asses the imputation results: RIV (relative increased variance), FMI (fraction of missing information) and RE (relative efficiency) \citep{Rubin1987,Rubin1996}, One can see the results in Table \ref{tab:imputation}. However, there are some parameters that we cannot impute the missing errors. Then we use the 10$\%$ of the absolute central value as the error bars \citep{Schaefer2007}. The errors of $t_{\rm pkX}$, $F_{\rm Opt11hr}$, Mag, $\log SSFR$, Age, $t_{\rm pkX,i}$ used the 10$\%$ of the absolute central value as the error bars. We do not impute the errors of $z$ and $D_{\rm L}$, as the errors are relatively very small.

\begin{enumerate}
\item Relative increase in variance due to missing data $r_{\rm m}$ (RIV). It is the ratio of between-imputation variance and within-imputation variance of the 5 data sets, then multiplying the imputation time m. It stands for the increase fraction in variance due to missing data, the influence of the missing data is bigger when $r_{\rm m}$ is bigger. When $r_{\rm m}$ is smaller, it indicates influence of the change of m is smaller, this is to say that missing data has smaller influence to the whole data parameters, the imputation results are more stable, the imputations are better. $r_{\rm m}$ is defined as
\begin{equation} \label{eq:RIV}
r_{m}=\frac{(1+\frac{1}{m}) {\sigma_{\rm B}}^{2}}{{\sigma_{\rm W}}^{2}},
\end{equation}
where ${\sigma_{\rm W}}^{2}=\frac{1}{m} \sum_{i=1}^m {\sigma_{\rm i}}^{2}$ is within-imputation variance, ${\sigma_{\rm B}}^{2}=\frac{1}{m-1} \sum_{i=1}^m {(\widehat{\theta}_{\rm i}-\widehat{\theta})}^{2}$ is between-imputation variance, and $\widehat{\theta}_{\rm i}$ is the mean of every complete data set, $\widehat{\theta}=\frac{1}{m}\sum_{i=1}^m \widehat{\theta}_{\rm i}$.
$\sigma_{\rm W}$ stands for the mean of the variance of m data sets, and $\sigma_{\rm B}$ stands for the variance of the mean of m data sets.

\item Fraction of missing information $\gamma_{\rm m}$ (FMI). It stands for the influence of the missing data for the whole parameters (e.g. mean), FMI is smaller, the imputation results are more stable.
\begin{equation} \label{eq:FMI}
\gamma_{\rm m}=\frac{r_{\rm m}+\frac{2}{v_{\rm m}+3}}{r_{\rm m}+1},
\end{equation}
where $v_{\rm m}=(m-1)(1+\frac{1}{r_{\rm m}}^{2})$ is freedom degree.

\item Relative efficiency (RE). RE is a comprehensive analysis of RIV and FMI, it stands for the imputation fraction for missing information by MICE. RE is bigger, the results are better.
\begin{equation} \label{eq:RE}
{\rm RE}={(1+\frac{\gamma_{\rm m}}{m})}^{-1},
\end{equation}
\end{enumerate}

For analyzing imputed data and pooling analysis results, we use the mean of every imputed error bar, because we also need to calculate some values and plot scatter plots with error bars, we can not use 5 values for one error bar, we must get a best estimate for every imputed error bar, so we use the mean.

The imputation results are shown in Table \ref{tab:imputation}. From the results, we can see that RIV and FMI are very close to 0, which means the imputation is stable. RE is very close to 1, which means the imputation efficiency is very high. We almost imputed all the missing information. Therefore, the imputation is reliable.

\section{Statistical methods} \label{sec:method}

\subsection{Distributions for each parameter} \label{subsec:distribution}

\begin{figure*}[!b]
\centering
\includegraphics[width=0.45\textwidth]{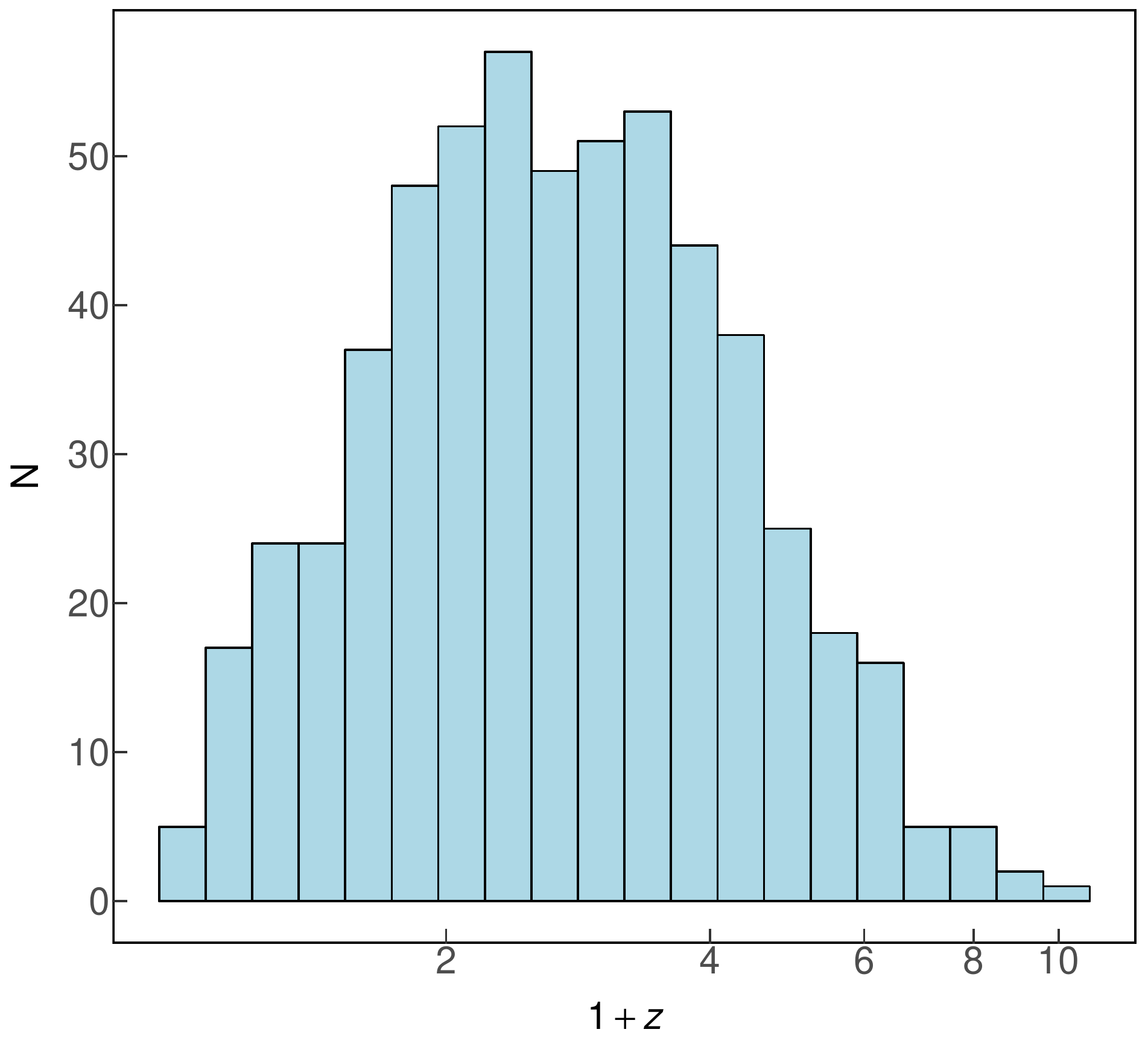}

\caption{
Histograms for $1+z$ and $D_{\rm L}$ as examples.  All of the histograms for the collected parameters are available in the Figure. Set 1. The description of each parameter is shown in Section \ref{sec:sample}.}
\label{fig:distribution}
\end{figure*}

We did the histograms for every parameter in observer-frame and some parameters in rest-frame. Two histograms are shown in Figure \ref{fig:distribution}. We made the figure set for all the figures, which are in total 58 figures. Readers can see all the figures in electronic version.

\figsetstart
\figsetnum{1}
\figsettitle{Distributions for every parameter}

\figsetgrpstart
\figsetgrpnum{1.1}
\figsetplot{./figset/distribution/1.pdf}
\figsetgrpend

\figsetgrpstart
\figsetgrpnum{1.2}
\figsetplot{./figset/distribution/2.pdf}
\figsetgrpend

\figsetgrpstart
\figsetgrpnum{1.3}
\figsetplot{./figset/distribution/3.pdf}
\figsetgrpend

\figsetgrpstart
\figsetgrpnum{1.4}
\figsetplot{./figset/distribution/4.pdf}
\figsetgrpend

\figsetgrpstart
\figsetgrpnum{1.5}
\figsetplot{./figset/distribution/5.pdf}
\figsetgrpend

\figsetgrpstart
\figsetgrpnum{1.6}
\figsetplot{./figset/distribution/6.pdf}
\figsetgrpend

\figsetgrpstart
\figsetgrpnum{1.7}
\figsetplot{./figset/distribution/7.pdf}
\figsetgrpend

\figsetgrpstart
\figsetgrpnum{1.8}
\figsetplot{./figset/distribution/8.pdf}
\figsetgrpend

\figsetgrpstart
\figsetgrpnum{1.9}
\figsetplot{./figset/distribution/9.pdf}
\figsetgrpend

\figsetgrpstart
\figsetgrpnum{1.10}
\figsetplot{./figset/distribution/10.pdf}
\figsetgrpend

\figsetgrpstart
\figsetgrpnum{1.11}
\figsetplot{./figset/distribution/11.pdf}
\figsetgrpend

\figsetgrpstart
\figsetgrpnum{1.12}
\figsetplot{./figset/distribution/12.pdf}
\figsetgrpend

\figsetgrpstart
\figsetgrpnum{1.13}
\figsetplot{./figset/distribution/13.pdf}
\figsetgrpend

\figsetgrpstart
\figsetgrpnum{1.14}
\figsetplot{./figset/distribution/14.pdf}
\figsetgrpend

\figsetgrpstart
\figsetgrpnum{1.15}
\figsetplot{./figset/distribution/15.pdf}
\figsetgrpend

\figsetgrpstart
\figsetgrpnum{1.16}
\figsetplot{./figset/distribution/16.pdf}
\figsetgrpend

\figsetgrpstart
\figsetgrpnum{1.17}
\figsetplot{./figset/distribution/17.pdf}
\figsetgrpend

\figsetgrpstart
\figsetgrpnum{1.18}
\figsetplot{./figset/distribution/18.pdf}
\figsetgrpend

\figsetgrpstart
\figsetgrpnum{1.19}
\figsetplot{./figset/distribution/19.pdf}
\figsetgrpend

\figsetgrpstart
\figsetgrpnum{1.20}
\figsetplot{./figset/distribution/20.pdf}
\figsetgrpend

\figsetgrpstart
\figsetgrpnum{1.21}
\figsetplot{./figset/distribution/21.pdf}
\figsetgrpend

\figsetgrpstart
\figsetgrpnum{1.22}
\figsetplot{./figset/distribution/22.pdf}
\figsetgrpend

\figsetgrpstart
\figsetgrpnum{1.23}
\figsetplot{./figset/distribution/23.pdf}
\figsetgrpend

\figsetgrpstart
\figsetgrpnum{1.24}
\figsetplot{./figset/distribution/24.pdf}
\figsetgrpend

\figsetgrpstart
\figsetgrpnum{1.25}
\figsetplot{./figset/distribution/25.pdf}
\figsetgrpend

\figsetgrpstart
\figsetgrpnum{1.26}
\figsetplot{./figset/distribution/26.pdf}
\figsetgrpend

\figsetgrpstart
\figsetgrpnum{1.27}
\figsetplot{./figset/distribution/27.pdf}
\figsetgrpend

\figsetgrpstart
\figsetgrpnum{1.28}
\figsetplot{./figset/distribution/28.pdf}
\figsetgrpend

\figsetgrpstart
\figsetgrpnum{1.29}
\figsetplot{./figset/distribution/29.pdf}
\figsetgrpend

\figsetgrpstart
\figsetgrpnum{1.30}
\figsetplot{./figset/distribution/30.pdf}
\figsetgrpend

\figsetgrpstart
\figsetgrpnum{1.31}
\figsetplot{./figset/distribution/31.pdf}
\figsetgrpend

\figsetgrpstart
\figsetgrpnum{1.32}
\figsetplot{./figset/distribution/32.pdf}
\figsetgrpend

\figsetgrpstart
\figsetgrpnum{1.33}
\figsetplot{./figset/distribution/33.pdf}
\figsetgrpend

\figsetgrpstart
\figsetgrpnum{1.34}
\figsetplot{./figset/distribution/34.pdf}
\figsetgrpend

\figsetgrpstart
\figsetgrpnum{1.35}
\figsetplot{./figset/distribution/35.pdf}
\figsetgrpend

\figsetgrpstart
\figsetgrpnum{1.36}
\figsetplot{./figset/distribution/36.pdf}
\figsetgrpend

\figsetgrpstart
\figsetgrpnum{1.37}
\figsetplot{./figset/distribution/37.pdf}
\figsetgrpend

\figsetgrpstart
\figsetgrpnum{1.38}
\figsetplot{./figset/distribution/38.pdf}
\figsetgrpend

\figsetgrpstart
\figsetgrpnum{1.39}
\figsetplot{./figset/distribution/39.pdf}
\figsetgrpend

\figsetgrpstart
\figsetgrpnum{1.40}
\figsetplot{./figset/distribution/40.pdf}
\figsetgrpend

\figsetgrpstart
\figsetgrpnum{1.41}
\figsetplot{./figset/distribution/41.pdf}
\figsetgrpend

\figsetgrpstart
\figsetgrpnum{1.42}
\figsetplot{./figset/distribution/42.pdf}
\figsetgrpend

\figsetgrpstart
\figsetgrpnum{1.43}
\figsetplot{./figset/distribution/43.pdf}
\figsetgrpend

\figsetgrpstart
\figsetgrpnum{1.44}
\figsetplot{./figset/distribution/44.pdf}
\figsetgrpend

\figsetgrpstart
\figsetgrpnum{1.45}
\figsetplot{./figset/distribution/45.pdf}
\figsetgrpend

\figsetgrpstart
\figsetgrpnum{1.46}
\figsetplot{./figset/distribution/46.pdf}
\figsetgrpend

\figsetgrpstart
\figsetgrpnum{1.47}
\figsetplot{./figset/distribution/47.pdf}
\figsetgrpend

\figsetgrpstart
\figsetgrpnum{1.48}
\figsetplot{./figset/distribution/48.pdf}
\figsetgrpend

\figsetgrpstart
\figsetgrpnum{1.49}
\figsetplot{./figset/distribution/49.pdf}
\figsetgrpend

\figsetgrpstart
\figsetgrpnum{1.50}
\figsetplot{./figset/distribution/50.pdf}
\figsetgrpend

\figsetgrpstart
\figsetgrpnum{1.51}
\figsetplot{./figset/distribution/51.pdf}
\figsetgrpend

\figsetgrpstart
\figsetgrpnum{1.52}
\figsetplot{./figset/distribution/52.pdf}
\figsetgrpend

\figsetgrpstart
\figsetgrpnum{1.53}
\figsetplot{./figset/distribution/53.pdf}
\figsetgrpend

\figsetgrpstart
\figsetgrpnum{1.54}
\figsetplot{./figset/distribution/54.pdf}
\figsetgrpend

\figsetgrpstart
\figsetgrpnum{1.55}
\figsetplot{./figset/distribution/55.pdf}
\figsetgrpend

\figsetgrpstart
\figsetgrpnum{1.56}
\figsetplot{./figset/distribution/56.pdf}
\figsetgrpend

\figsetgrpstart
\figsetgrpnum{1.57}
\figsetplot{./figset/distribution/57.pdf}
\figsetgrpend

\figsetgrpstart
\figsetgrpnum{1.58}
\figsetplot{./figset/distribution/58.pdf}
\figsetgrpend

\figsetend

\subsection{scatter plots between two arbitrary parameters} \label{subsec:scatter}

\begin{figure}
\plottwo{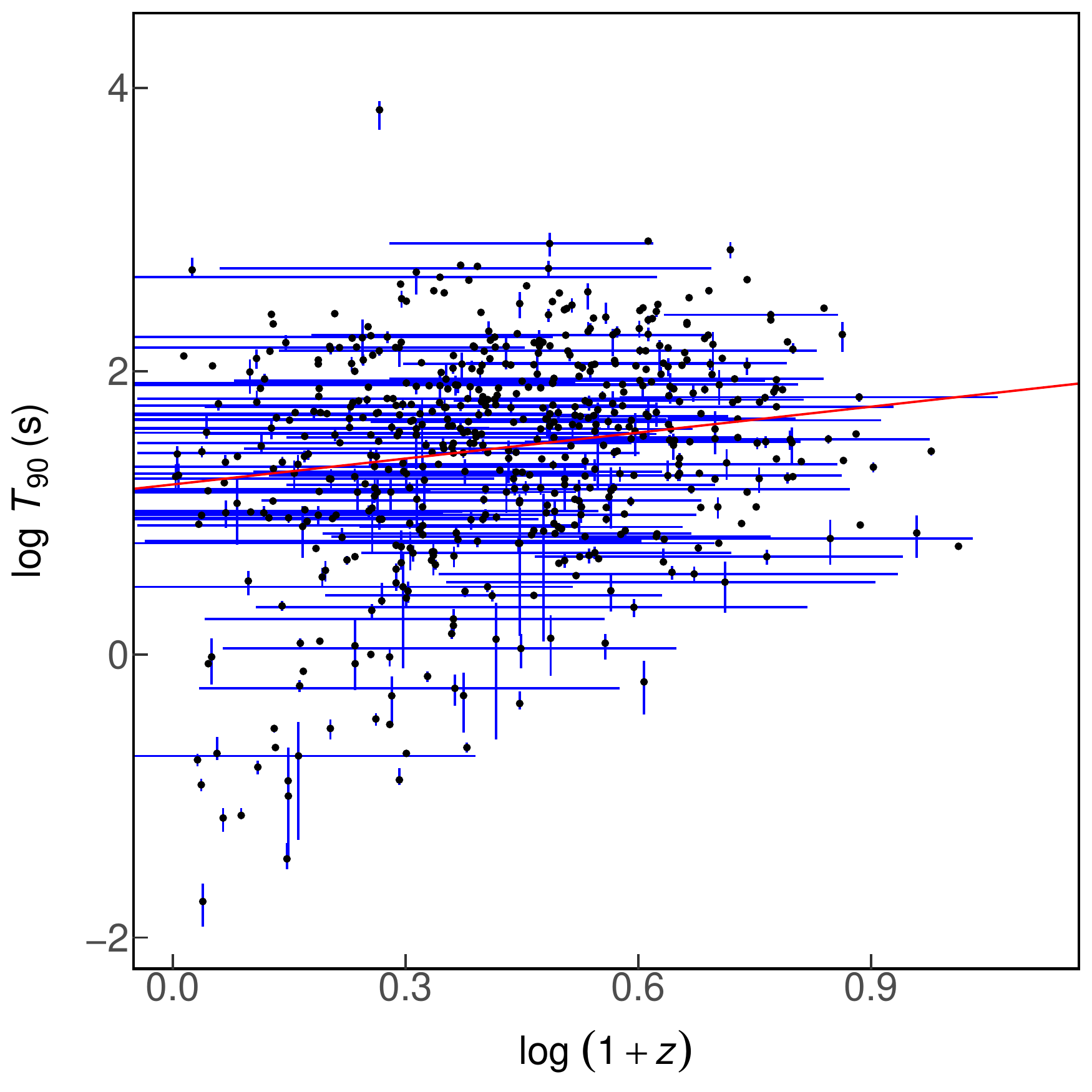}{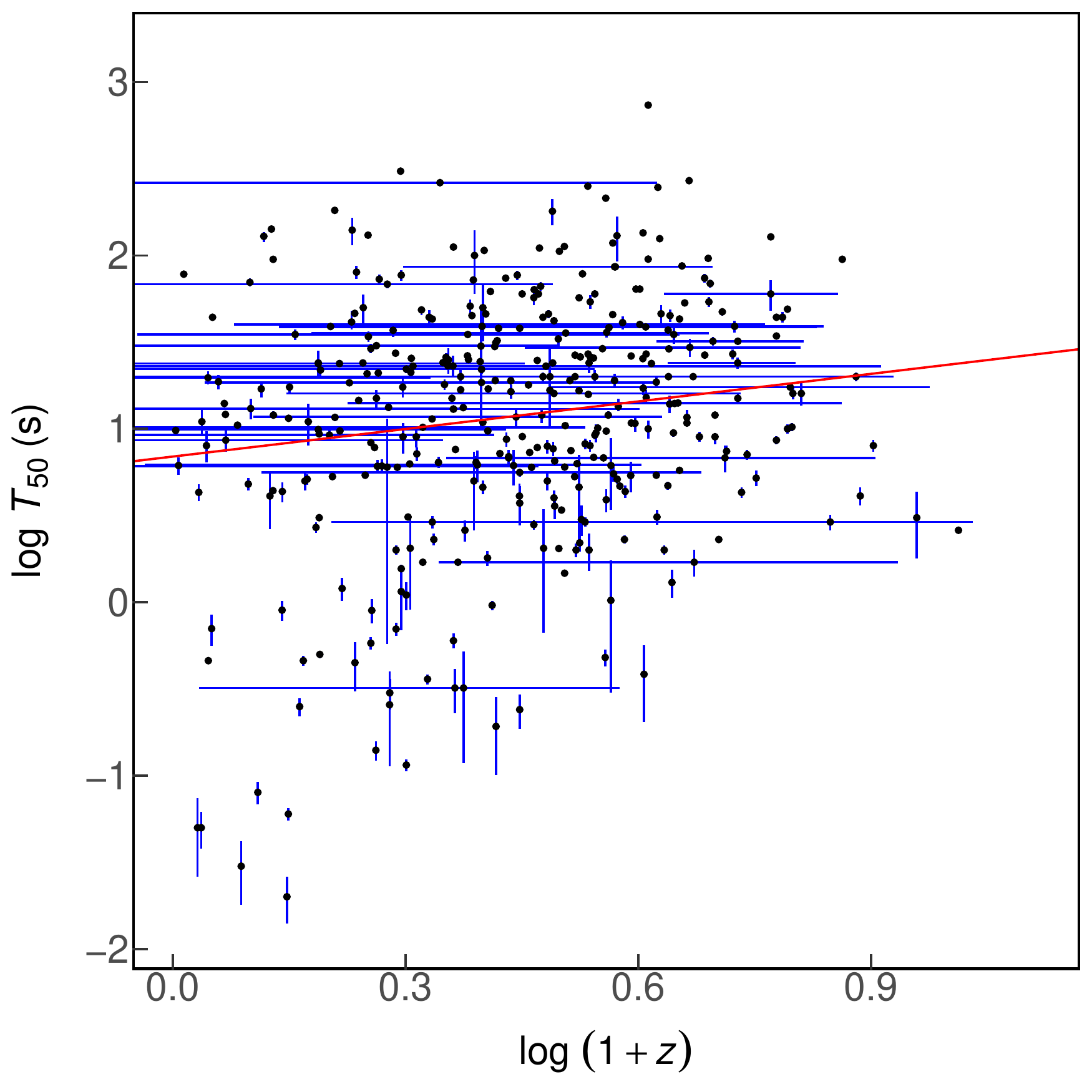}
\caption{
Scatter plots for two arbitrary parameters. One can see all the 1468 plots in Figure. Set 2 \label{fig:scatter}. The description of every parameter is defined in Section \ref{sec:sample}.}
\end{figure}

We plotted the scatter plots between two arbitrary parameters when the points is not less than 5. We can find that, with this large samples, we can also keep the relationships that found before. We removed the spectral lag value of GRB 060218, because it is an outlier \citep{Foley2008}. The two results are shown in Figure \ref{fig:scatter}. We also made figure set for all the scatter plots. In total there are 1468 plots.

\figsetstart
\figsetnum{2}
\figsettitle{Scatter plots for two arbitrary parameters}

\figsetgrpstart
\figsetgrpnum{2.1}
\figsetplot{./figset/scatterchangexy/1.pdf}
\figsetgrpend

\figsetgrpstart
\figsetgrpnum{2.2}
\figsetplot{./figset/scatterchangexy/2.pdf}
\figsetgrpend

\figsetgrpstart
\figsetgrpnum{2.3}
\figsetplot{./figset/scatterchangexy/3.pdf}
\figsetgrpend

\figsetgrpstart
\figsetgrpnum{2.4}
\figsetplot{./figset/scatterchangexy/4.pdf}
\figsetgrpend

\figsetgrpstart
\figsetgrpnum{2.5}
\figsetplot{./figset/scatterchangexy/5.pdf}
\figsetgrpend

\figsetgrpstart
\figsetgrpnum{2.6}
\figsetplot{./figset/scatterchangexy/6.pdf}
\figsetgrpend

\figsetgrpstart
\figsetgrpnum{2.7}
\figsetplot{./figset/scatterchangexy/7.pdf}
\figsetgrpend

\figsetgrpstart
\figsetgrpnum{2.8}
\figsetplot{./figset/scatterchangexy/8.pdf}
\figsetgrpend

\figsetgrpstart
\figsetgrpnum{2.9}
\figsetplot{./figset/scatterchangexy/9.pdf}
\figsetgrpend

\figsetgrpstart
\figsetgrpnum{2.10}
\figsetplot{./figset/scatterchangexy/10.pdf}
\figsetgrpend

\figsetgrpstart
\figsetgrpnum{2.11}
\figsetplot{./figset/scatterchangexy/11.pdf}
\figsetgrpend

\figsetgrpstart
\figsetgrpnum{2.12}
\figsetplot{./figset/scatterchangexy/12.pdf}
\figsetgrpend

\figsetgrpstart
\figsetgrpnum{2.13}
\figsetplot{./figset/scatterchangexy/13.pdf}
\figsetgrpend

\figsetgrpstart
\figsetgrpnum{2.14}
\figsetplot{./figset/scatterchangexy/14.pdf}
\figsetgrpend

\figsetgrpstart
\figsetgrpnum{2.15}
\figsetplot{./figset/scatterchangexy/15.pdf}
\figsetgrpend

\figsetgrpstart
\figsetgrpnum{2.16}
\figsetplot{./figset/scatterchangexy/16.pdf}
\figsetgrpend

\figsetgrpstart
\figsetgrpnum{2.17}
\figsetplot{./figset/scatterchangexy/17.pdf}
\figsetgrpend

\figsetgrpstart
\figsetgrpnum{2.18}
\figsetplot{./figset/scatterchangexy/18.pdf}
\figsetgrpend

\figsetgrpstart
\figsetgrpnum{2.19}
\figsetplot{./figset/scatterchangexy/19.pdf}
\figsetgrpend

\figsetgrpstart
\figsetgrpnum{2.20}
\figsetplot{./figset/scatterchangexy/20.pdf}
\figsetgrpend

\figsetgrpstart
\figsetgrpnum{2.21}
\figsetplot{./figset/scatterchangexy/21.pdf}
\figsetgrpend

\figsetgrpstart
\figsetgrpnum{2.22}
\figsetplot{./figset/scatterchangexy/22.pdf}
\figsetgrpend

\figsetgrpstart
\figsetgrpnum{2.23}
\figsetplot{./figset/scatterchangexy/23.pdf}
\figsetgrpend

\figsetgrpstart
\figsetgrpnum{2.24}
\figsetplot{./figset/scatterchangexy/24.pdf}
\figsetgrpend

\figsetgrpstart
\figsetgrpnum{2.25}
\figsetplot{./figset/scatterchangexy/25.pdf}
\figsetgrpend

\figsetgrpstart
\figsetgrpnum{2.26}
\figsetplot{./figset/scatterchangexy/26.pdf}
\figsetgrpend

\figsetgrpstart
\figsetgrpnum{2.27}
\figsetplot{./figset/scatterchangexy/27.pdf}
\figsetgrpend

\figsetgrpstart
\figsetgrpnum{2.28}
\figsetplot{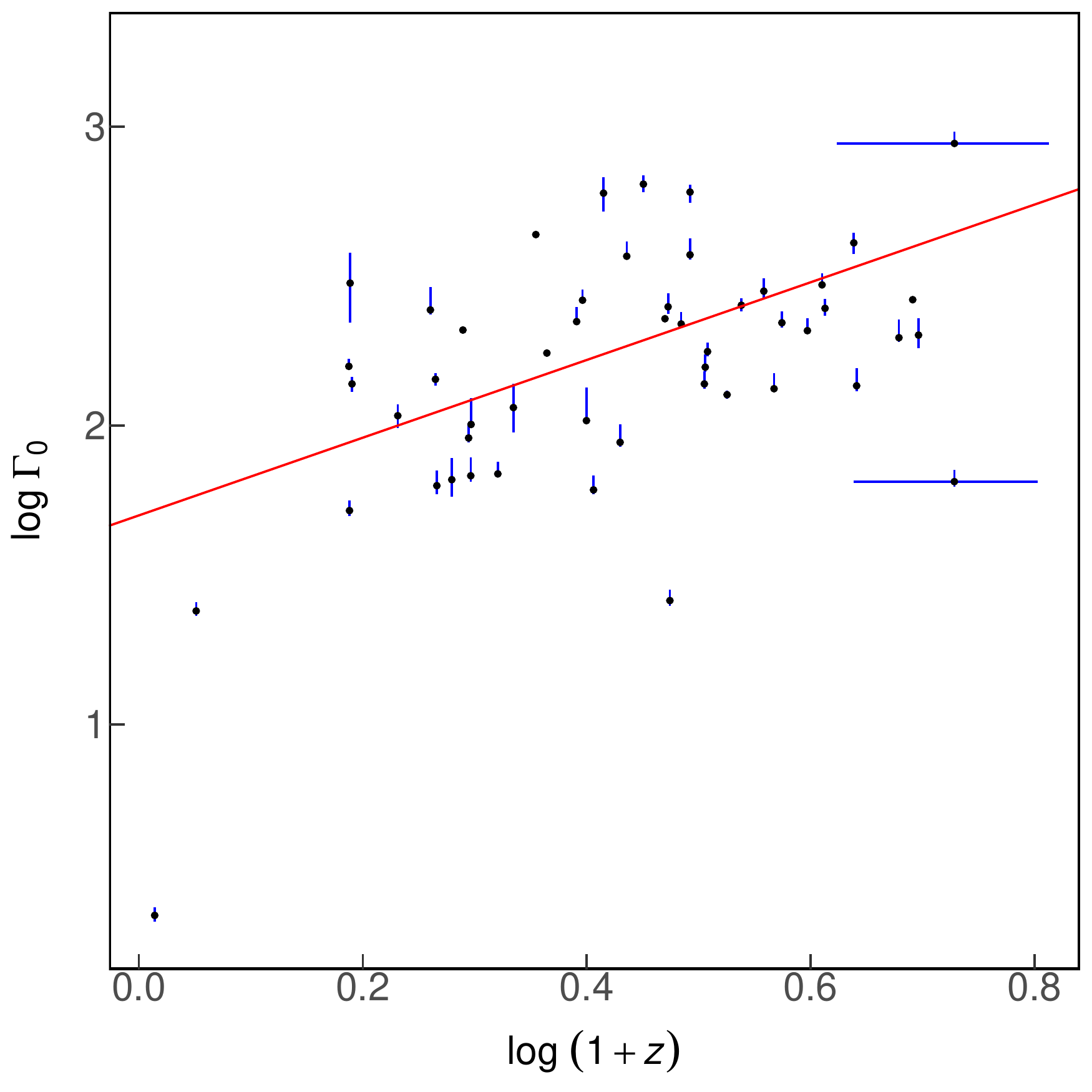}
\figsetgrpend

\figsetgrpstart
\figsetgrpnum{2.29}
\figsetplot{./figset/scatterchangexy/29.pdf}
\figsetgrpend

\figsetgrpstart
\figsetgrpnum{2.30}
\figsetplot{./figset/scatterchangexy/30.pdf}
\figsetgrpend

\figsetgrpstart
\figsetgrpnum{2.31}
\figsetplot{./figset/scatterchangexy/31.pdf}
\figsetgrpend

\figsetgrpstart
\figsetgrpnum{2.32}
\figsetplot{./figset/scatterchangexy/32.pdf}
\figsetgrpend

\figsetgrpstart
\figsetgrpnum{2.33}
\figsetplot{./figset/scatterchangexy/33.pdf}
\figsetgrpend

\figsetgrpstart
\figsetgrpnum{2.34}
\figsetplot{./figset/scatterchangexy/34.pdf}
\figsetgrpend

\figsetgrpstart
\figsetgrpnum{2.35}
\figsetplot{./figset/scatterchangexy/35.pdf}
\figsetgrpend

\figsetgrpstart
\figsetgrpnum{2.36}
\figsetplot{./figset/scatterchangexy/36.pdf}
\figsetgrpend

\figsetgrpstart
\figsetgrpnum{2.37}
\figsetplot{./figset/scatterchangexy/37.pdf}
\figsetgrpend

\figsetgrpstart
\figsetgrpnum{2.38}
\figsetplot{./figset/scatterchangexy/38.pdf}
\figsetgrpend

\figsetgrpstart
\figsetgrpnum{2.39}
\figsetplot{./figset/scatterchangexy/39.pdf}
\figsetgrpend

\figsetgrpstart
\figsetgrpnum{2.40}
\figsetplot{./figset/scatterchangexy/40.pdf}
\figsetgrpend

\figsetgrpstart
\figsetgrpnum{2.41}
\figsetplot{./figset/scatterchangexy/41.pdf}
\figsetgrpend

\figsetgrpstart
\figsetgrpnum{2.42}
\figsetplot{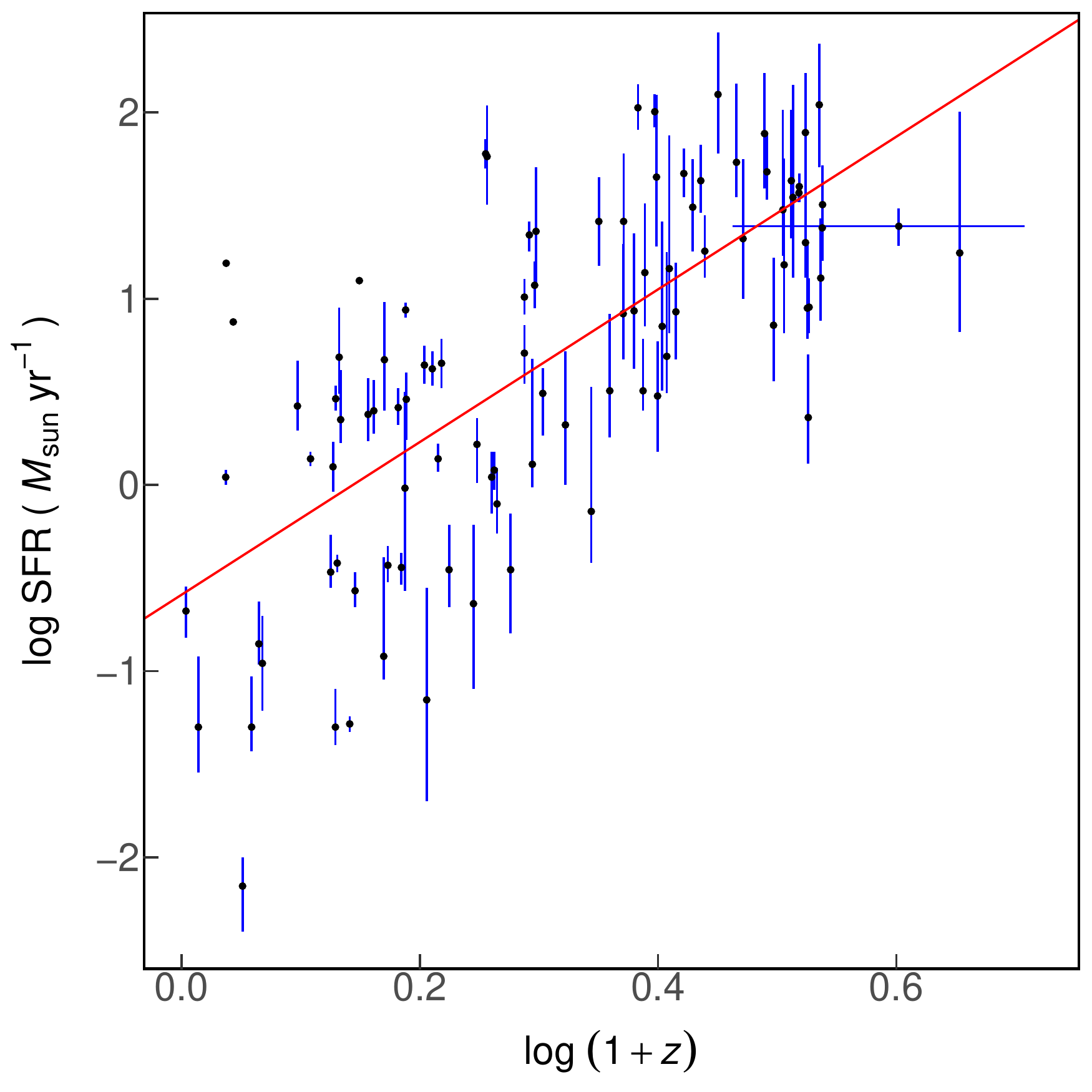}
\figsetgrpend

\figsetgrpstart
\figsetgrpnum{2.43}
\figsetplot{./figset/scatterchangexy/43.pdf}
\figsetgrpend

\figsetgrpstart
\figsetgrpnum{2.44}
\figsetplot{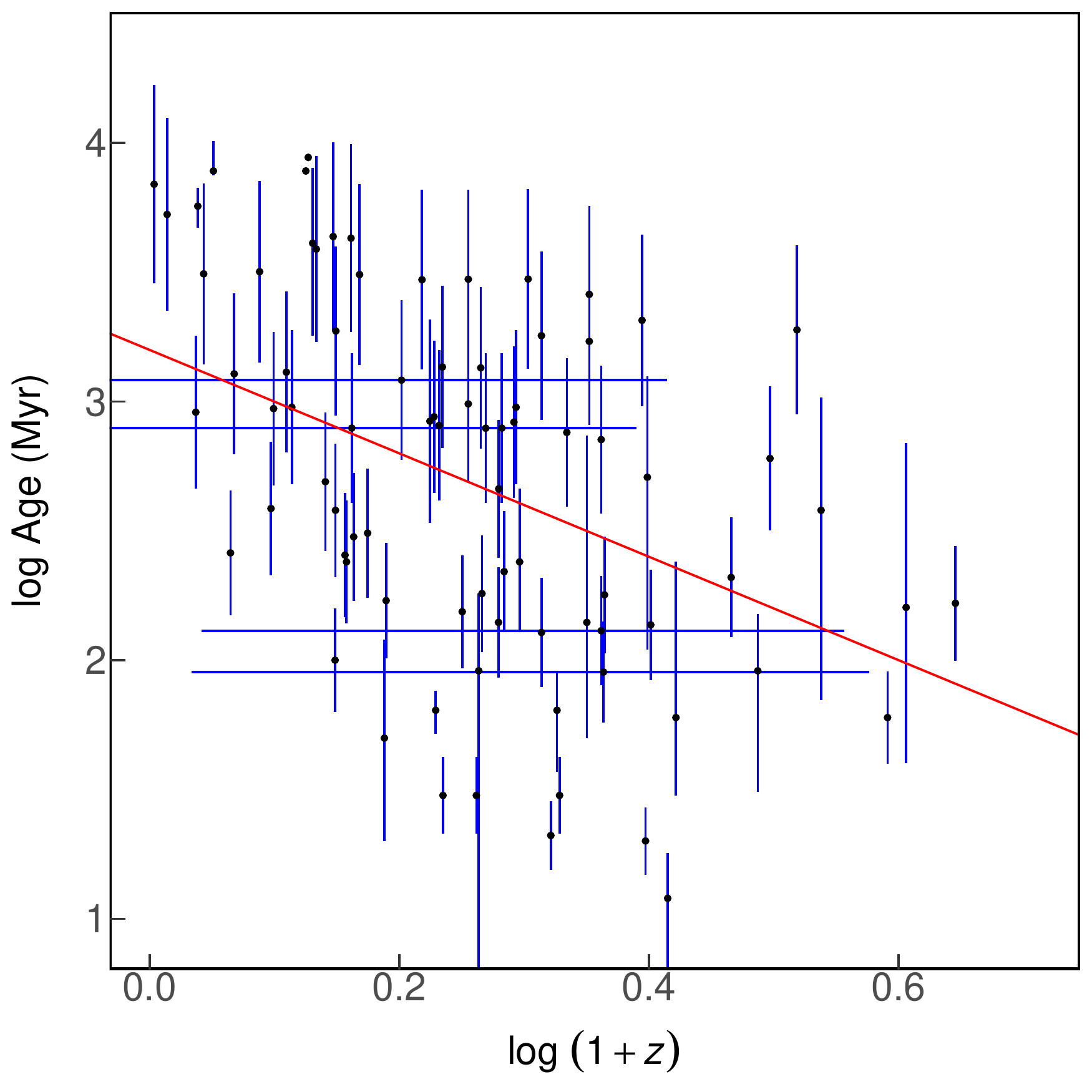}
\figsetgrpend

\figsetgrpstart
\figsetgrpnum{2.45}
\figsetplot{./figset/scatterchangexy/45.pdf}
\figsetgrpend

\figsetgrpstart
\figsetgrpnum{2.46}
\figsetplot{./figset/scatterchangexy/46.pdf}
\figsetgrpend

\figsetgrpstart
\figsetgrpnum{2.47}
\figsetplot{./figset/scatterchangexy/47.pdf}
\figsetgrpend

\figsetgrpstart
\figsetgrpnum{2.48}
\figsetplot{./figset/scatterchangexy/48.pdf}
\figsetgrpend

\figsetgrpstart
\figsetgrpnum{2.49}
\figsetplot{./figset/scatterchangexy/49.pdf}
\figsetgrpend

\figsetgrpstart
\figsetgrpnum{2.50}
\figsetplot{./figset/scatterchangexy/50.pdf}
\figsetgrpend

\figsetgrpstart
\figsetgrpnum{2.51}
\figsetplot{./figset/scatterchangexy/51.pdf}
\figsetgrpend

\figsetgrpstart
\figsetgrpnum{2.52}
\figsetplot{./figset/scatterchangexy/52.pdf}
\figsetgrpend

\figsetgrpstart
\figsetgrpnum{2.53}
\figsetplot{./figset/scatterchangexy/53.pdf}
\figsetgrpend

\figsetgrpstart
\figsetgrpnum{2.54}
\figsetplot{./figset/scatterchangexy/54.pdf}
\figsetgrpend

\figsetgrpstart
\figsetgrpnum{2.55}
\figsetplot{./figset/scatterchangexy/55.pdf}
\figsetgrpend

\figsetgrpstart
\figsetgrpnum{2.56}
\figsetplot{./figset/scatter/56.pdf}
\figsetgrpend

\figsetgrpstart
\figsetgrpnum{2.57}
\figsetplot{./figset/scatter/57.pdf}
\figsetgrpend

\figsetgrpstart
\figsetgrpnum{2.58}
\figsetplot{./figset/scatter/58.pdf}
\figsetgrpend

\figsetgrpstart
\figsetgrpnum{2.59}
\figsetplot{./figset/scatter/59.pdf}
\figsetgrpend

\figsetgrpstart
\figsetgrpnum{2.60}
\figsetplot{./figset/scatter/60.pdf}
\figsetgrpend

\figsetgrpstart
\figsetgrpnum{2.61}
\figsetplot{./figset/scatter/61.pdf}
\figsetgrpend

\figsetgrpstart
\figsetgrpnum{2.62}
\figsetplot{./figset/scatter/62.pdf}
\figsetgrpend

\figsetgrpstart
\figsetgrpnum{2.63}
\figsetplot{./figset/scatter/63.pdf}
\figsetgrpend

\figsetgrpstart
\figsetgrpnum{2.64}
\figsetplot{./figset/scatter/64.pdf}
\figsetgrpend

\figsetgrpstart
\figsetgrpnum{2.65}
\figsetplot{./figset/scatter/65.pdf}
\figsetgrpend

\figsetgrpstart
\figsetgrpnum{2.66}
\figsetplot{./figset/scatter/66.pdf}
\figsetgrpend

\figsetgrpstart
\figsetgrpnum{2.67}
\figsetplot{./figset/scatter/67.pdf}
\figsetgrpend

\figsetgrpstart
\figsetgrpnum{2.68}
\figsetplot{./figset/scatter/68.pdf}
\figsetgrpend

\figsetgrpstart
\figsetgrpnum{2.69}
\figsetplot{./figset/scatter/69.pdf}
\figsetgrpend

\figsetgrpstart
\figsetgrpnum{2.70}
\figsetplot{./figset/scatter/70.pdf}
\figsetgrpend

\figsetgrpstart
\figsetgrpnum{2.71}
\figsetplot{./figset/scatter/71.pdf}
\figsetgrpend

\figsetgrpstart
\figsetgrpnum{2.72}
\figsetplot{./figset/scatter/72.pdf}
\figsetgrpend

\figsetgrpstart
\figsetgrpnum{2.73}
\figsetplot{./figset/scatter/73.pdf}
\figsetgrpend

\figsetgrpstart
\figsetgrpnum{2.74}
\figsetplot{./figset/scatter/74.pdf}
\figsetgrpend

\figsetgrpstart
\figsetgrpnum{2.75}
\figsetplot{./figset/scatter/75.pdf}
\figsetgrpend

\figsetgrpstart
\figsetgrpnum{2.76}
\figsetplot{./figset/scatter/76.pdf}
\figsetgrpend

\figsetgrpstart
\figsetgrpnum{2.77}
\figsetplot{./figset/scatter/77.pdf}
\figsetgrpend

\figsetgrpstart
\figsetgrpnum{2.78}
\figsetplot{./figset/scatter/78.pdf}
\figsetgrpend

\figsetgrpstart
\figsetgrpnum{2.79}
\figsetplot{./figset/scatter/79.pdf}
\figsetgrpend

\figsetgrpstart
\figsetgrpnum{2.80}
\figsetplot{./figset/scatter/80.pdf}
\figsetgrpend

\figsetgrpstart
\figsetgrpnum{2.81}
\figsetplot{./figset/scatter/81.pdf}
\figsetgrpend

\figsetgrpstart
\figsetgrpnum{2.82}
\figsetplot{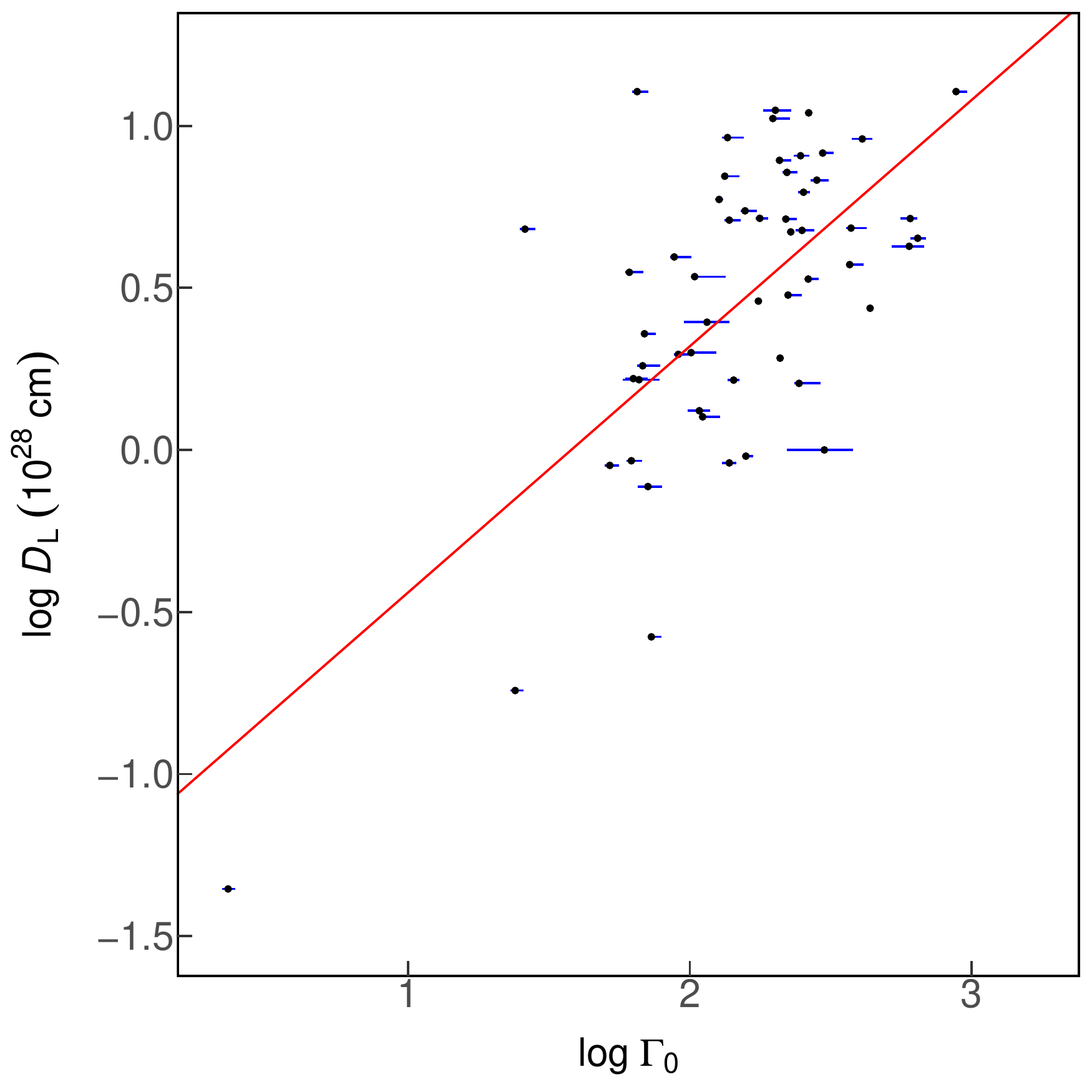}
\figsetgrpend

\figsetgrpstart
\figsetgrpnum{2.83}
\figsetplot{./figset/scatter/83.pdf}
\figsetgrpend

\figsetgrpstart
\figsetgrpnum{2.84}
\figsetplot{./figset/scatter/84.pdf}
\figsetgrpend

\figsetgrpstart
\figsetgrpnum{2.85}
\figsetplot{./figset/scatter/85.pdf}
\figsetgrpend

\figsetgrpstart
\figsetgrpnum{2.86}
\figsetplot{./figset/scatter/86.pdf}
\figsetgrpend

\figsetgrpstart
\figsetgrpnum{2.87}
\figsetplot{./figset/scatter/87.pdf}
\figsetgrpend

\figsetgrpstart
\figsetgrpnum{2.88}
\figsetplot{./figset/scatter/88.pdf}
\figsetgrpend

\figsetgrpstart
\figsetgrpnum{2.89}
\figsetplot{./figset/scatter/89.pdf}
\figsetgrpend

\figsetgrpstart
\figsetgrpnum{2.90}
\figsetplot{./figset/scatter/90.pdf}
\figsetgrpend

\figsetgrpstart
\figsetgrpnum{2.91}
\figsetplot{./figset/scatter/91.pdf}
\figsetgrpend

\figsetgrpstart
\figsetgrpnum{2.92}
\figsetplot{./figset/scatter/92.pdf}
\figsetgrpend

\figsetgrpstart
\figsetgrpnum{2.93}
\figsetplot{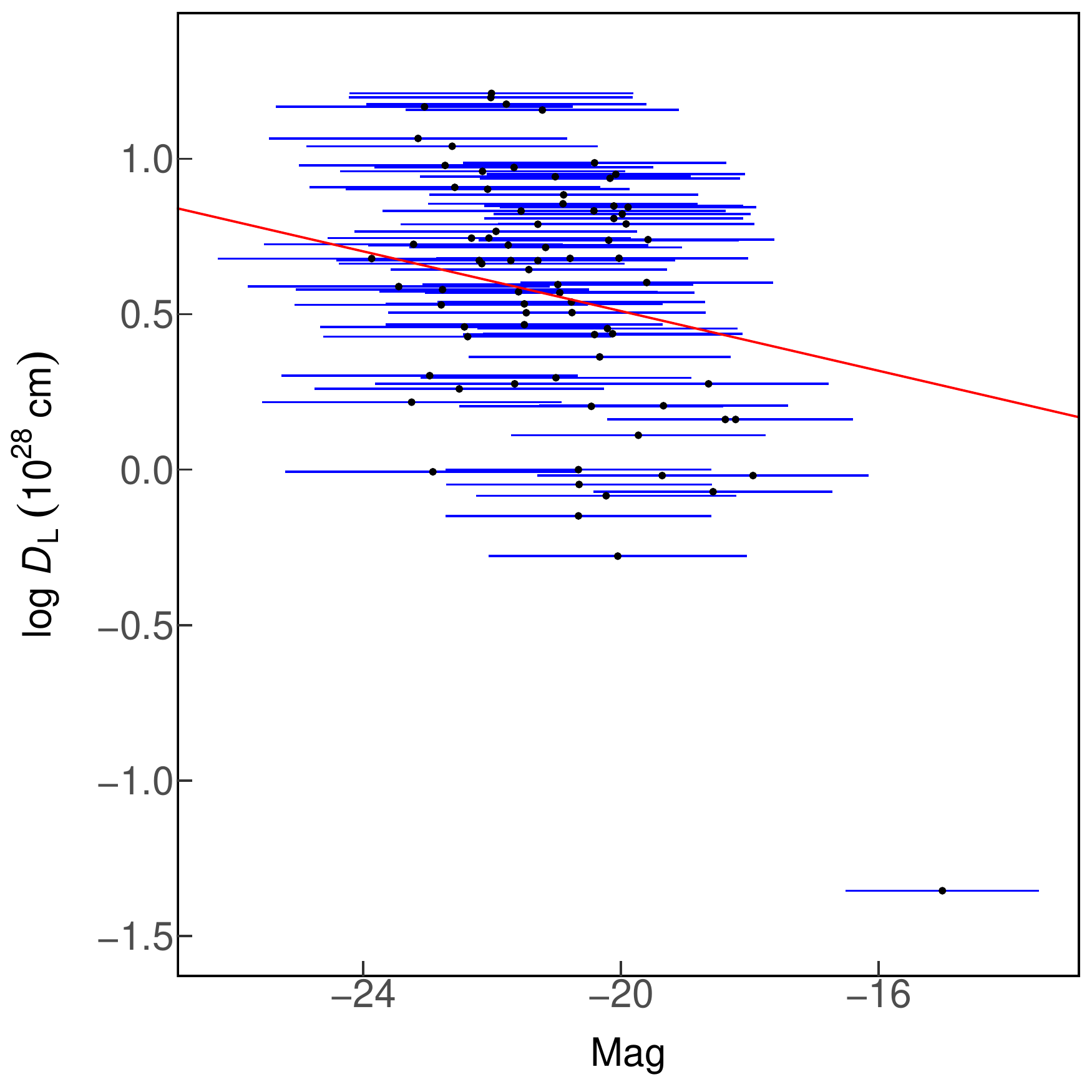}
\figsetgrpend

\figsetgrpstart
\figsetgrpnum{2.94}
\figsetplot{./figset/scatter/94.pdf}
\figsetgrpend

\figsetgrpstart
\figsetgrpnum{2.95}
\figsetplot{./figset/scatter/95.pdf}
\figsetgrpend

\figsetgrpstart
\figsetgrpnum{2.96}
\figsetplot{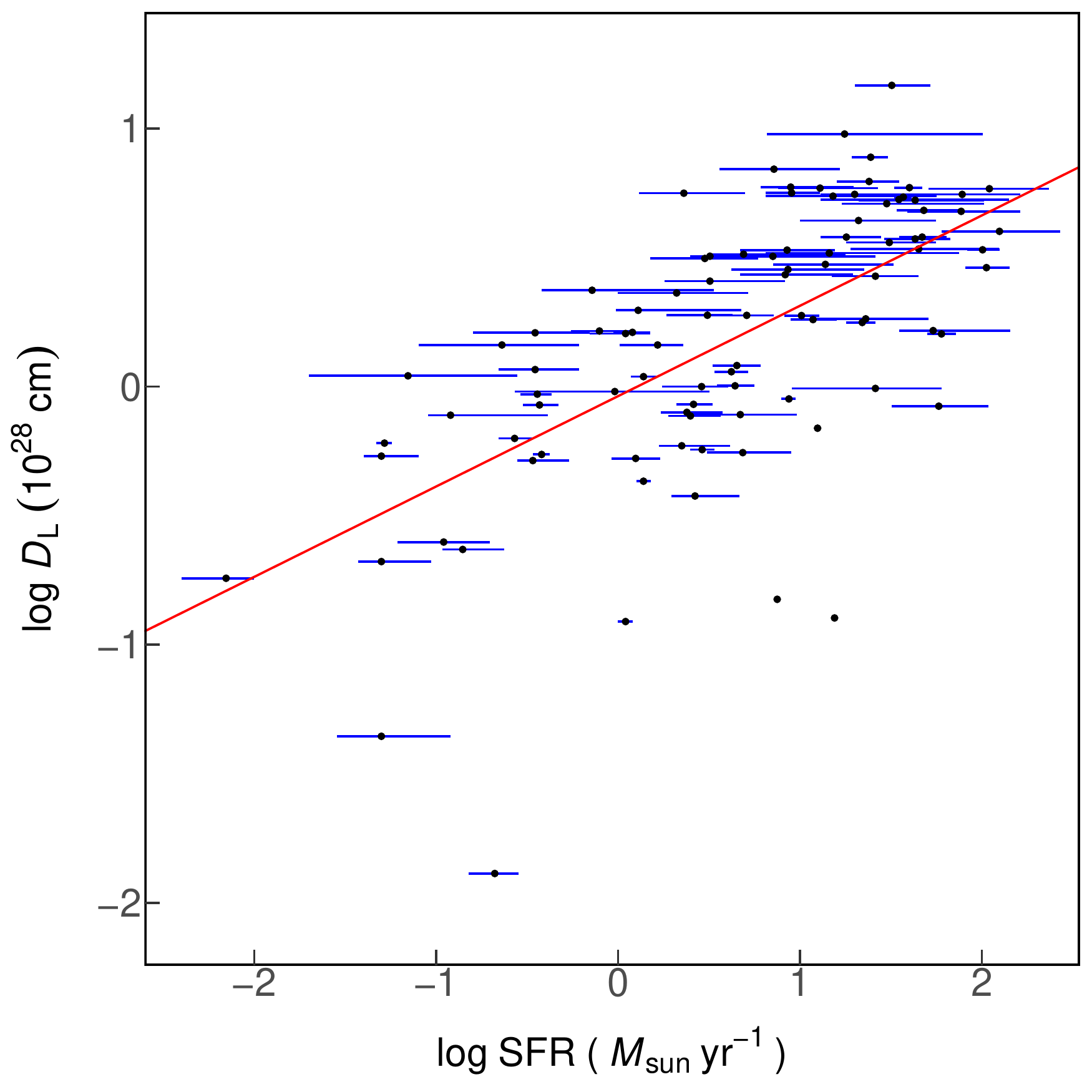}
\figsetgrpend

\figsetgrpstart
\figsetgrpnum{2.97}
\figsetplot{./figset/scatter/97.pdf}
\figsetgrpend

\figsetgrpstart
\figsetgrpnum{2.98}
\figsetplot{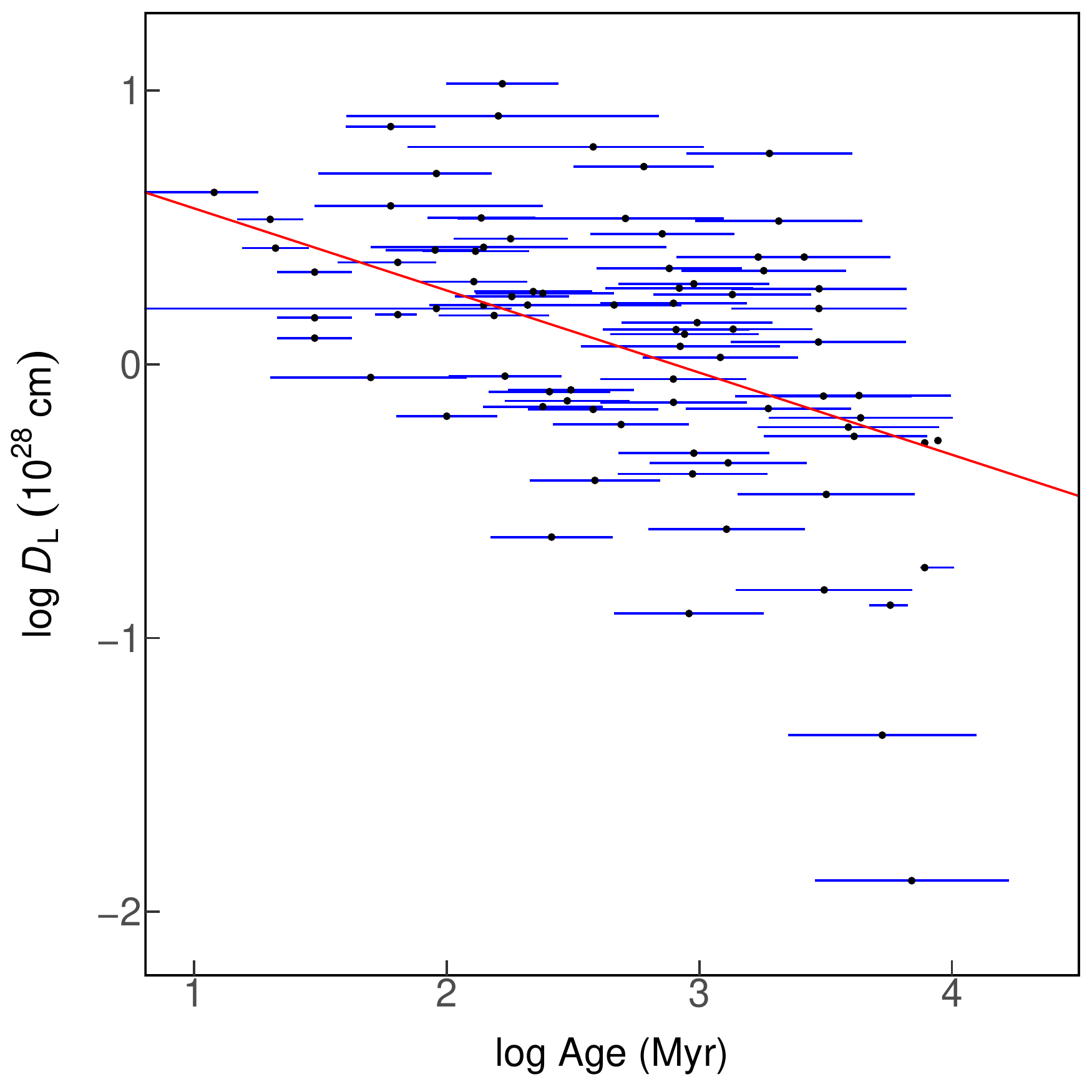}
\figsetgrpend

\figsetgrpstart
\figsetgrpnum{2.99}
\figsetplot{./figset/scatter/99.pdf}
\figsetgrpend

\figsetgrpstart
\figsetgrpnum{2.100}
\figsetplot{./figset/scatter/100.pdf}
\figsetgrpend

\figsetgrpstart
\figsetgrpnum{2.101}
\figsetplot{./figset/scatter/101.pdf}
\figsetgrpend

\figsetgrpstart
\figsetgrpnum{2.102}
\figsetplot{./figset/scatter/102.pdf}
\figsetgrpend

\figsetgrpstart
\figsetgrpnum{2.103}
\figsetplot{./figset/scatter/103.pdf}
\figsetgrpend

\figsetgrpstart
\figsetgrpnum{2.104}
\figsetplot{./figset/scatter/104.pdf}
\figsetgrpend

\figsetgrpstart
\figsetgrpnum{2.105}
\figsetplot{./figset/scatter/105.pdf}
\figsetgrpend

\figsetgrpstart
\figsetgrpnum{2.106}
\figsetplot{./figset/scatter/106.pdf}
\figsetgrpend

\figsetgrpstart
\figsetgrpnum{2.107}
\figsetplot{./figset/scatter/107.pdf}
\figsetgrpend

\figsetgrpstart
\figsetgrpnum{2.108}
\figsetplot{./figset/scatter/108.pdf}
\figsetgrpend

\figsetgrpstart
\figsetgrpnum{2.109}
\figsetplot{./figset/scatter/109.pdf}
\figsetgrpend

\figsetgrpstart
\figsetgrpnum{2.110}
\figsetplot{./figset/scatter/110.pdf}
\figsetgrpend

\figsetgrpstart
\figsetgrpnum{2.111}
\figsetplot{./figset/scatter/111.pdf}
\figsetgrpend

\figsetgrpstart
\figsetgrpnum{2.112}
\figsetplot{./figset/scatter/112.pdf}
\figsetgrpend

\figsetgrpstart
\figsetgrpnum{2.113}
\figsetplot{./figset/scatter/113.pdf}
\figsetgrpend

\figsetgrpstart
\figsetgrpnum{2.114}
\figsetplot{./figset/scatter/114.pdf}
\figsetgrpend

\figsetgrpstart
\figsetgrpnum{2.115}
\figsetplot{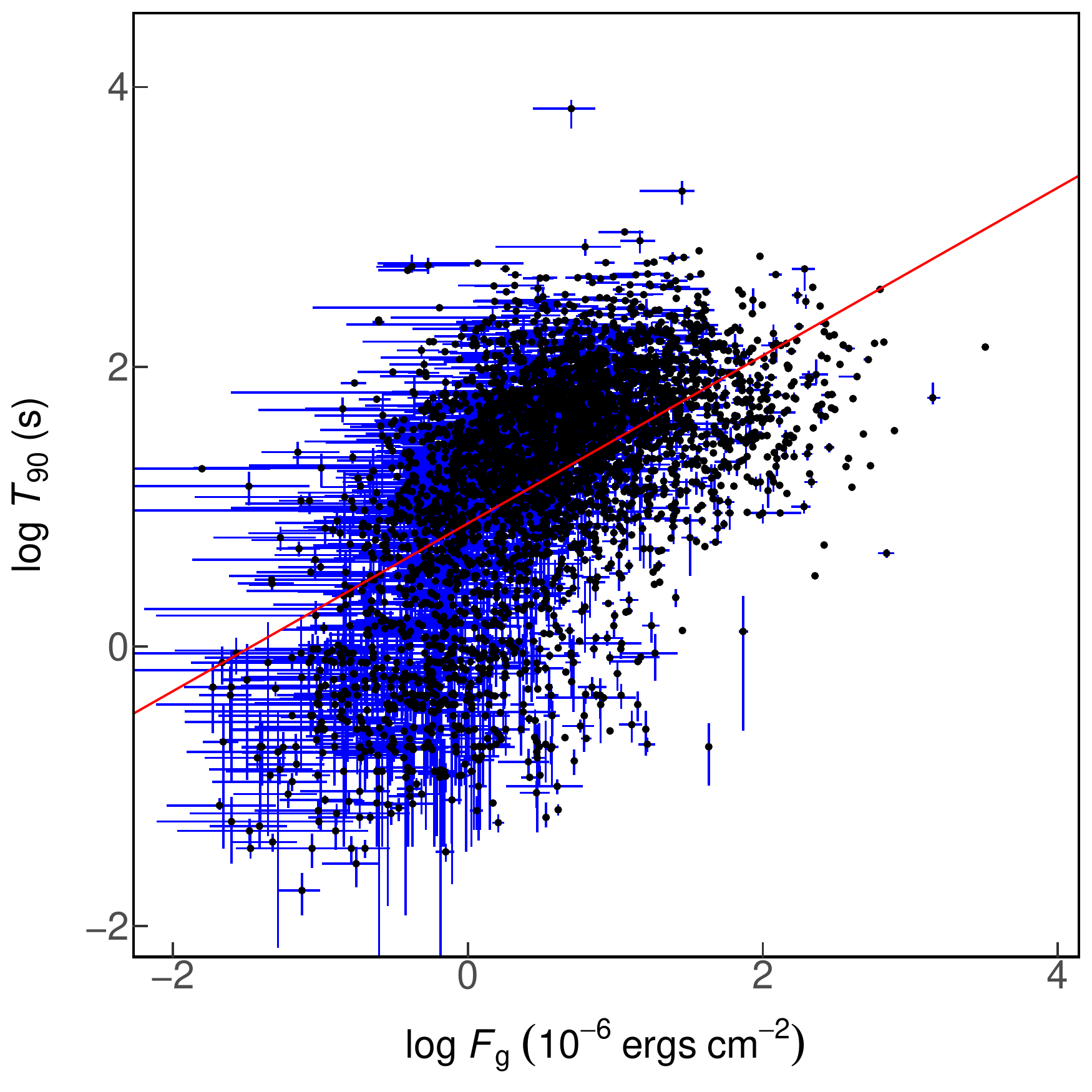}
\figsetgrpend

\figsetgrpstart
\figsetgrpnum{2.116}
\figsetplot{./figset/scatter/116.pdf}
\figsetgrpend

\figsetgrpstart
\figsetgrpnum{2.117}
\figsetplot{./figset/scatter/117.pdf}
\figsetgrpend

\figsetgrpstart
\figsetgrpnum{2.118}
\figsetplot{./figset/scatter/118.pdf}
\figsetgrpend

\figsetgrpstart
\figsetgrpnum{2.119}
\figsetplot{./figset/scatter/119.pdf}
\figsetgrpend

\figsetgrpstart
\figsetgrpnum{2.120}
\figsetplot{./figset/scatter/120.pdf}
\figsetgrpend

\figsetgrpstart
\figsetgrpnum{2.121}
\figsetplot{./figset/scatter/121.pdf}
\figsetgrpend

\figsetgrpstart
\figsetgrpnum{2.122}
\figsetplot{./figset/scatter/122.pdf}
\figsetgrpend

\figsetgrpstart
\figsetgrpnum{2.123}
\figsetplot{./figset/scatter/123.pdf}
\figsetgrpend

\figsetgrpstart
\figsetgrpnum{2.124}
\figsetplot{./figset/scatter/124.pdf}
\figsetgrpend

\figsetgrpstart
\figsetgrpnum{2.125}
\figsetplot{./figset/scatter/125.pdf}
\figsetgrpend

\figsetgrpstart
\figsetgrpnum{2.126}
\figsetplot{./figset/scatter/126.pdf}
\figsetgrpend

\figsetgrpstart
\figsetgrpnum{2.127}
\figsetplot{./figset/scatter/127.pdf}
\figsetgrpend

\figsetgrpstart
\figsetgrpnum{2.128}
\figsetplot{./figset/scatter/128.pdf}
\figsetgrpend

\figsetgrpstart
\figsetgrpnum{2.129}
\figsetplot{./figset/scatter/129.pdf}
\figsetgrpend

\figsetgrpstart
\figsetgrpnum{2.130}
\figsetplot{./figset/scatter/130.pdf}
\figsetgrpend

\figsetgrpstart
\figsetgrpnum{2.131}
\figsetplot{./figset/scatter/131.pdf}
\figsetgrpend

\figsetgrpstart
\figsetgrpnum{2.132}
\figsetplot{./figset/scatter/132.pdf}
\figsetgrpend

\figsetgrpstart
\figsetgrpnum{2.133}
\figsetplot{./figset/scatter/133.pdf}
\figsetgrpend

\figsetgrpstart
\figsetgrpnum{2.134}
\figsetplot{./figset/scatter/134.pdf}
\figsetgrpend

\figsetgrpstart
\figsetgrpnum{2.135}
\figsetplot{./figset/scatter/135.pdf}
\figsetgrpend

\figsetgrpstart
\figsetgrpnum{2.136}
\figsetplot{./figset/scatter/136.pdf}
\figsetgrpend

\figsetgrpstart
\figsetgrpnum{2.137}
\figsetplot{./figset/scatter/137.pdf}
\figsetgrpend

\figsetgrpstart
\figsetgrpnum{2.138}
\figsetplot{./figset/scatter/138.pdf}
\figsetgrpend

\figsetgrpstart
\figsetgrpnum{2.139}
\figsetplot{./figset/scatter/139.pdf}
\figsetgrpend

\figsetgrpstart
\figsetgrpnum{2.140}
\figsetplot{./figset/scatter/140.pdf}
\figsetgrpend

\figsetgrpstart
\figsetgrpnum{2.141}
\figsetplot{./figset/scatter/141.pdf}
\figsetgrpend

\figsetgrpstart
\figsetgrpnum{2.142}
\figsetplot{./figset/scatter/142.pdf}
\figsetgrpend

\figsetgrpstart
\figsetgrpnum{2.143}
\figsetplot{./figset/scatter/143.pdf}
\figsetgrpend

\figsetgrpstart
\figsetgrpnum{2.144}
\figsetplot{./figset/scatter/144.pdf}
\figsetgrpend

\figsetgrpstart
\figsetgrpnum{2.145}
\figsetplot{./figset/scatter/145.pdf}
\figsetgrpend

\figsetgrpstart
\figsetgrpnum{2.146}
\figsetplot{./figset/scatter/146.pdf}
\figsetgrpend

\figsetgrpstart
\figsetgrpnum{2.147}
\figsetplot{./figset/scatter/147.pdf}
\figsetgrpend

\figsetgrpstart
\figsetgrpnum{2.148}
\figsetplot{./figset/scatter/148.pdf}
\figsetgrpend

\figsetgrpstart
\figsetgrpnum{2.149}
\figsetplot{./figset/scatter/149.pdf}
\figsetgrpend

\figsetgrpstart
\figsetgrpnum{2.150}
\figsetplot{./figset/scatter/150.pdf}
\figsetgrpend

\figsetgrpstart
\figsetgrpnum{2.151}
\figsetplot{./figset/scatter/151.pdf}
\figsetgrpend

\figsetgrpstart
\figsetgrpnum{2.152}
\figsetplot{./figset/scatter/152.pdf}
\figsetgrpend

\figsetgrpstart
\figsetgrpnum{2.153}
\figsetplot{./figset/scatter/153.pdf}
\figsetgrpend

\figsetgrpstart
\figsetgrpnum{2.154}
\figsetplot{./figset/scatter/154.pdf}
\figsetgrpend

\figsetgrpstart
\figsetgrpnum{2.155}
\figsetplot{./figset/scatter/155.pdf}
\figsetgrpend

\figsetgrpstart
\figsetgrpnum{2.156}
\figsetplot{./figset/scatter/156.pdf}
\figsetgrpend

\figsetgrpstart
\figsetgrpnum{2.157}
\figsetplot{./figset/scatter/157.pdf}
\figsetgrpend

\figsetgrpstart
\figsetgrpnum{2.158}
\figsetplot{./figset/scatter/158.pdf}
\figsetgrpend

\figsetgrpstart
\figsetgrpnum{2.159}
\figsetplot{./figset/scatter/159.pdf}
\figsetgrpend

\figsetgrpstart
\figsetgrpnum{2.160}
\figsetplot{./figset/scatter/160.pdf}
\figsetgrpend

\figsetgrpstart
\figsetgrpnum{2.161}
\figsetplot{./figset/scatter/161.pdf}
\figsetgrpend

\figsetgrpstart
\figsetgrpnum{2.162}
\figsetplot{./figset/scatter/162.pdf}
\figsetgrpend

\figsetgrpstart
\figsetgrpnum{2.163}
\figsetplot{./figset/scatter/163.pdf}
\figsetgrpend

\figsetgrpstart
\figsetgrpnum{2.164}
\figsetplot{./figset/scatter/164.pdf}
\figsetgrpend

\figsetgrpstart
\figsetgrpnum{2.165}
\figsetplot{./figset/scatter/165.pdf}
\figsetgrpend

\figsetgrpstart
\figsetgrpnum{2.166}
\figsetplot{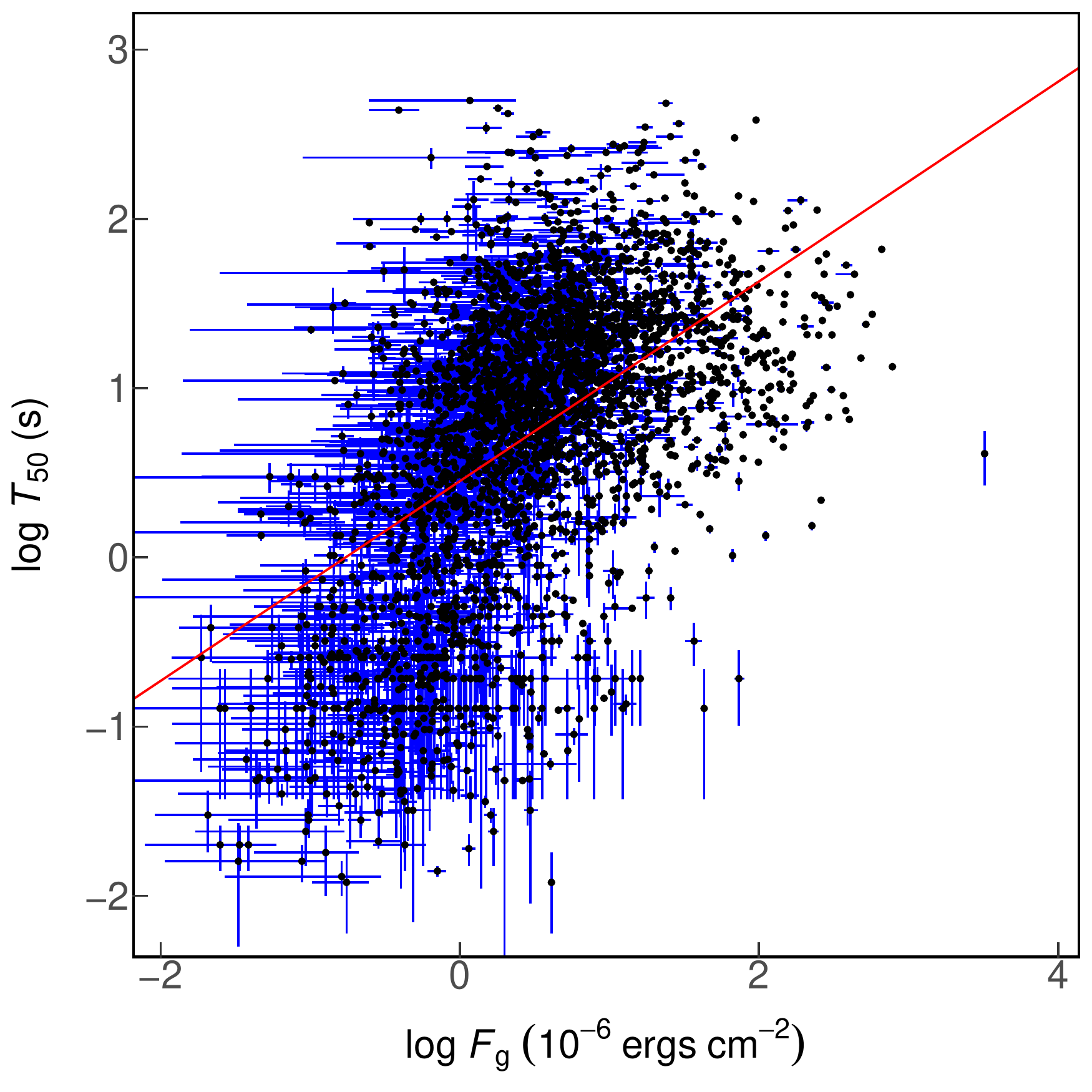}
\figsetgrpend

\figsetgrpstart
\figsetgrpnum{2.167}
\figsetplot{./figset/scatter/167.pdf}
\figsetgrpend

\figsetgrpstart
\figsetgrpnum{2.168}
\figsetplot{./figset/scatter/168.pdf}
\figsetgrpend

\figsetgrpstart
\figsetgrpnum{2.169}
\figsetplot{./figset/scatter/169.pdf}
\figsetgrpend

\figsetgrpstart
\figsetgrpnum{2.170}
\figsetplot{./figset/scatter/170.pdf}
\figsetgrpend

\figsetgrpstart
\figsetgrpnum{2.171}
\figsetplot{./figset/scatter/171.pdf}
\figsetgrpend

\figsetgrpstart
\figsetgrpnum{2.172}
\figsetplot{./figset/scatter/172.pdf}
\figsetgrpend

\figsetgrpstart
\figsetgrpnum{2.173}
\figsetplot{./figset/scatter/173.pdf}
\figsetgrpend

\figsetgrpstart
\figsetgrpnum{2.174}
\figsetplot{./figset/scatter/174.pdf}
\figsetgrpend

\figsetgrpstart
\figsetgrpnum{2.175}
\figsetplot{./figset/scatter/175.pdf}
\figsetgrpend

\figsetgrpstart
\figsetgrpnum{2.176}
\figsetplot{./figset/scatter/176.pdf}
\figsetgrpend

\figsetgrpstart
\figsetgrpnum{2.177}
\figsetplot{./figset/scatter/177.pdf}
\figsetgrpend

\figsetgrpstart
\figsetgrpnum{2.178}
\figsetplot{./figset/scatter/178.pdf}
\figsetgrpend

\figsetgrpstart
\figsetgrpnum{2.179}
\figsetplot{./figset/scatter/179.pdf}
\figsetgrpend

\figsetgrpstart
\figsetgrpnum{2.180}
\figsetplot{./figset/scatter/180.pdf}
\figsetgrpend

\figsetgrpstart
\figsetgrpnum{2.181}
\figsetplot{./figset/scatter/181.pdf}
\figsetgrpend

\figsetgrpstart
\figsetgrpnum{2.182}
\figsetplot{./figset/scatter/182.pdf}
\figsetgrpend

\figsetgrpstart
\figsetgrpnum{2.183}
\figsetplot{./figset/scatter/183.pdf}
\figsetgrpend

\figsetgrpstart
\figsetgrpnum{2.184}
\figsetplot{./figset/scatter/184.pdf}
\figsetgrpend

\figsetgrpstart
\figsetgrpnum{2.185}
\figsetplot{./figset/scatter/185.pdf}
\figsetgrpend

\figsetgrpstart
\figsetgrpnum{2.186}
\figsetplot{./figset/scatter/186.pdf}
\figsetgrpend

\figsetgrpstart
\figsetgrpnum{2.187}
\figsetplot{./figset/scatter/187.pdf}
\figsetgrpend

\figsetgrpstart
\figsetgrpnum{2.188}
\figsetplot{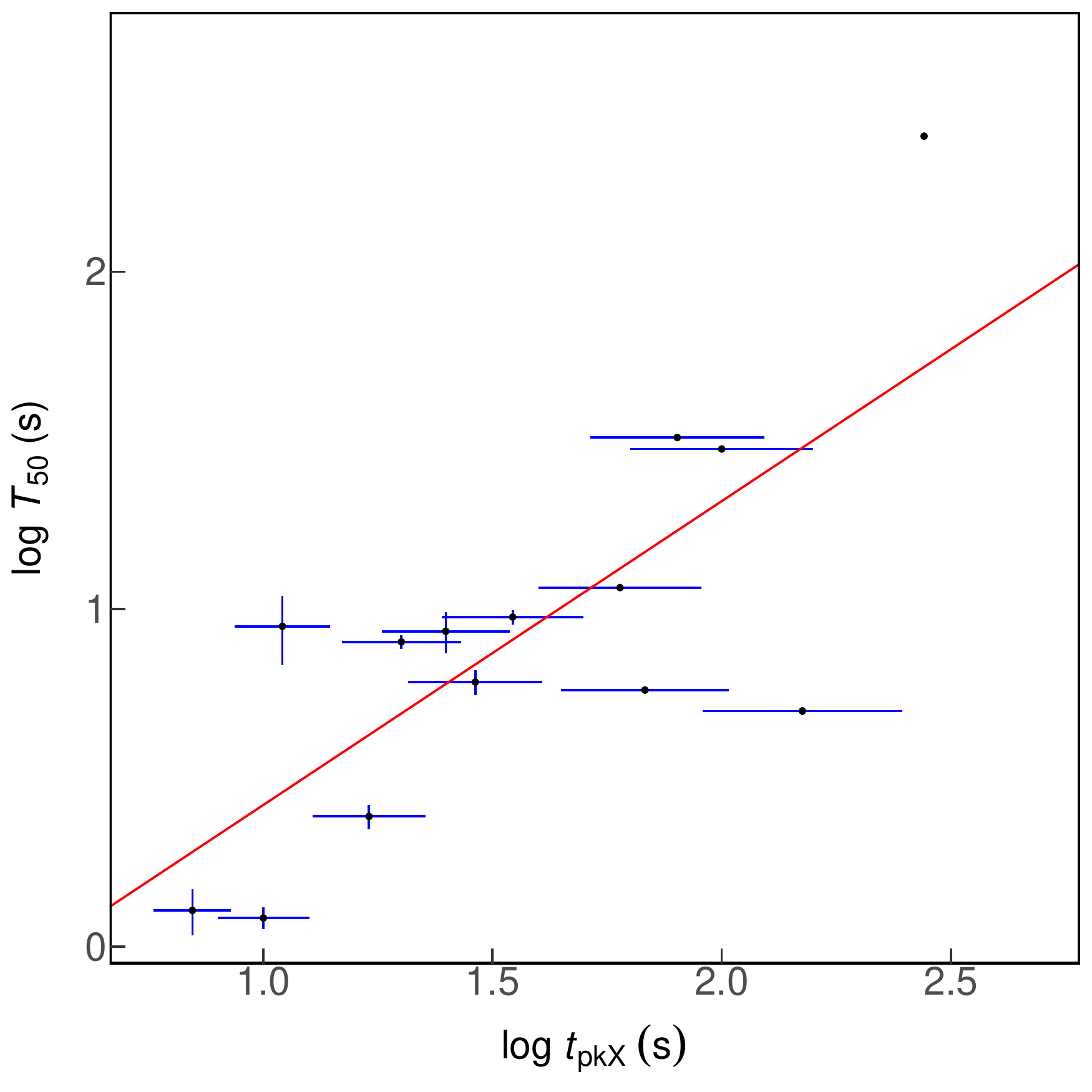}
\figsetgrpend

\figsetgrpstart
\figsetgrpnum{2.189}
\figsetplot{./figset/scatter/189.pdf}
\figsetgrpend

\figsetgrpstart
\figsetgrpnum{2.190}
\figsetplot{./figset/scatter/190.pdf}
\figsetgrpend

\figsetgrpstart
\figsetgrpnum{2.191}
\figsetplot{./figset/scatter/191.pdf}
\figsetgrpend

\figsetgrpstart
\figsetgrpnum{2.192}
\figsetplot{./figset/scatter/192.pdf}
\figsetgrpend

\figsetgrpstart
\figsetgrpnum{2.193}
\figsetplot{./figset/scatter/193.pdf}
\figsetgrpend

\figsetgrpstart
\figsetgrpnum{2.194}
\figsetplot{./figset/scatter/194.pdf}
\figsetgrpend

\figsetgrpstart
\figsetgrpnum{2.195}
\figsetplot{./figset/scatter/195.pdf}
\figsetgrpend

\figsetgrpstart
\figsetgrpnum{2.196}
\figsetplot{./figset/scatter/196.pdf}
\figsetgrpend

\figsetgrpstart
\figsetgrpnum{2.197}
\figsetplot{./figset/scatter/197.pdf}
\figsetgrpend

\figsetgrpstart
\figsetgrpnum{2.198}
\figsetplot{./figset/scatter/198.pdf}
\figsetgrpend

\figsetgrpstart
\figsetgrpnum{2.199}
\figsetplot{./figset/scatter/199.pdf}
\figsetgrpend

\figsetgrpstart
\figsetgrpnum{2.200}
\figsetplot{./figset/scatter/200.pdf}
\figsetgrpend

\figsetgrpstart
\figsetgrpnum{2.201}
\figsetplot{./figset/scatter/201.pdf}
\figsetgrpend

\figsetgrpstart
\figsetgrpnum{2.202}
\figsetplot{./figset/scatter/202.pdf}
\figsetgrpend

\figsetgrpstart
\figsetgrpnum{2.203}
\figsetplot{./figset/scatter/203.pdf}
\figsetgrpend

\figsetgrpstart
\figsetgrpnum{2.204}
\figsetplot{./figset/scatter/204.pdf}
\figsetgrpend

\figsetgrpstart
\figsetgrpnum{2.205}
\figsetplot{./figset/scatter/205.pdf}
\figsetgrpend

\figsetgrpstart
\figsetgrpnum{2.206}
\figsetplot{./figset/scatter/206.pdf}
\figsetgrpend

\figsetgrpstart
\figsetgrpnum{2.207}
\figsetplot{./figset/scatter/207.pdf}
\figsetgrpend

\figsetgrpstart
\figsetgrpnum{2.208}
\figsetplot{./figset/scatter/208.pdf}
\figsetgrpend

\figsetgrpstart
\figsetgrpnum{2.209}
\figsetplot{./figset/scatter/209.pdf}
\figsetgrpend

\figsetgrpstart
\figsetgrpnum{2.210}
\figsetplot{./figset/scatter/210.pdf}
\figsetgrpend

\figsetgrpstart
\figsetgrpnum{2.211}
\figsetplot{./figset/scatter/211.pdf}
\figsetgrpend

\figsetgrpstart
\figsetgrpnum{2.212}
\figsetplot{./figset/scatter/212.pdf}
\figsetgrpend

\figsetgrpstart
\figsetgrpnum{2.213}
\figsetplot{./figset/scatter/213.pdf}
\figsetgrpend

\figsetgrpstart
\figsetgrpnum{2.214}
\figsetplot{./figset/scatter/214.pdf}
\figsetgrpend

\figsetgrpstart
\figsetgrpnum{2.215}
\figsetplot{./figset/scatter/215.pdf}
\figsetgrpend

\figsetgrpstart
\figsetgrpnum{2.216}
\figsetplot{./figset/scatter/216.pdf}
\figsetgrpend

\figsetgrpstart
\figsetgrpnum{2.217}
\figsetplot{./figset/scatter/217.pdf}
\figsetgrpend

\figsetgrpstart
\figsetgrpnum{2.218}
\figsetplot{./figset/scatter/218.pdf}
\figsetgrpend

\figsetgrpstart
\figsetgrpnum{2.219}
\figsetplot{./figset/scatter/219.pdf}
\figsetgrpend

\figsetgrpstart
\figsetgrpnum{2.220}
\figsetplot{./figset/scatter/220.pdf}
\figsetgrpend

\figsetgrpstart
\figsetgrpnum{2.221}
\figsetplot{./figset/scatter/221.pdf}
\figsetgrpend

\figsetgrpstart
\figsetgrpnum{2.222}
\figsetplot{./figset/scatter/222.pdf}
\figsetgrpend

\figsetgrpstart
\figsetgrpnum{2.223}
\figsetplot{./figset/scatter/223.pdf}
\figsetgrpend

\figsetgrpstart
\figsetgrpnum{2.224}
\figsetplot{./figset/scatter/224.pdf}
\figsetgrpend

\figsetgrpstart
\figsetgrpnum{2.225}
\figsetplot{./figset/scatter/225.pdf}
\figsetgrpend

\figsetgrpstart
\figsetgrpnum{2.226}
\figsetplot{./figset/scatter/226.pdf}
\figsetgrpend

\figsetgrpstart
\figsetgrpnum{2.227}
\figsetplot{./figset/scatter/227.pdf}
\figsetgrpend

\figsetgrpstart
\figsetgrpnum{2.228}
\figsetplot{./figset/scatter/228.pdf}
\figsetgrpend

\figsetgrpstart
\figsetgrpnum{2.229}
\figsetplot{./figset/scatter/229.pdf}
\figsetgrpend

\figsetgrpstart
\figsetgrpnum{2.230}
\figsetplot{./figset/scatter/230.pdf}
\figsetgrpend

\figsetgrpstart
\figsetgrpnum{2.231}
\figsetplot{./figset/scatter/231.pdf}
\figsetgrpend

\figsetgrpstart
\figsetgrpnum{2.232}
\figsetplot{./figset/scatter/232.pdf}
\figsetgrpend

\figsetgrpstart
\figsetgrpnum{2.233}
\figsetplot{./figset/scatter/233.pdf}
\figsetgrpend

\figsetgrpstart
\figsetgrpnum{2.234}
\figsetplot{./figset/scatter/234.pdf}
\figsetgrpend

\figsetgrpstart
\figsetgrpnum{2.235}
\figsetplot{./figset/scatter/235.pdf}
\figsetgrpend

\figsetgrpstart
\figsetgrpnum{2.236}
\figsetplot{./figset/scatter/236.pdf}
\figsetgrpend

\figsetgrpstart
\figsetgrpnum{2.237}
\figsetplot{./figset/scatter/237.pdf}
\figsetgrpend

\figsetgrpstart
\figsetgrpnum{2.238}
\figsetplot{./figset/scatter/238.pdf}
\figsetgrpend

\figsetgrpstart
\figsetgrpnum{2.239}
\figsetplot{./figset/scatter/239.pdf}
\figsetgrpend

\figsetgrpstart
\figsetgrpnum{2.240}
\figsetplot{./figset/scatter/240.pdf}
\figsetgrpend

\figsetgrpstart
\figsetgrpnum{2.241}
\figsetplot{./figset/scatter/241.pdf}
\figsetgrpend

\figsetgrpstart
\figsetgrpnum{2.242}
\figsetplot{./figset/scatter/242.pdf}
\figsetgrpend

\figsetgrpstart
\figsetgrpnum{2.243}
\figsetplot{./figset/scatter/243.pdf}
\figsetgrpend

\figsetgrpstart
\figsetgrpnum{2.244}
\figsetplot{./figset/scatter/244.pdf}
\figsetgrpend

\figsetgrpstart
\figsetgrpnum{2.245}
\figsetplot{./figset/scatter/245.pdf}
\figsetgrpend

\figsetgrpstart
\figsetgrpnum{2.246}
\figsetplot{./figset/scatter/246.pdf}
\figsetgrpend

\figsetgrpstart
\figsetgrpnum{2.247}
\figsetplot{./figset/scatter/247.pdf}
\figsetgrpend

\figsetgrpstart
\figsetgrpnum{2.248}
\figsetplot{./figset/scatter/248.pdf}
\figsetgrpend

\figsetgrpstart
\figsetgrpnum{2.249}
\figsetplot{./figset/scatter/249.pdf}
\figsetgrpend

\figsetgrpstart
\figsetgrpnum{2.250}
\figsetplot{./figset/scatter/250.pdf}
\figsetgrpend

\figsetgrpstart
\figsetgrpnum{2.251}
\figsetplot{./figset/scatter/251.pdf}
\figsetgrpend

\figsetgrpstart
\figsetgrpnum{2.252}
\figsetplot{./figset/scatter/252.pdf}
\figsetgrpend

\figsetgrpstart
\figsetgrpnum{2.253}
\figsetplot{./figset/scatter/253.pdf}
\figsetgrpend

\figsetgrpstart
\figsetgrpnum{2.254}
\figsetplot{./figset/scatter/254.pdf}
\figsetgrpend

\figsetgrpstart
\figsetgrpnum{2.255}
\figsetplot{./figset/scatter/255.pdf}
\figsetgrpend

\figsetgrpstart
\figsetgrpnum{2.256}
\figsetplot{./figset/scatter/256.pdf}
\figsetgrpend

\figsetgrpstart
\figsetgrpnum{2.257}
\figsetplot{./figset/scatter/257.pdf}
\figsetgrpend

\figsetgrpstart
\figsetgrpnum{2.258}
\figsetplot{./figset/scatter/258.pdf}
\figsetgrpend

\figsetgrpstart
\figsetgrpnum{2.259}
\figsetplot{./figset/scatter/259.pdf}
\figsetgrpend

\figsetgrpstart
\figsetgrpnum{2.260}
\figsetplot{./figset/scatter/260.pdf}
\figsetgrpend

\figsetgrpstart
\figsetgrpnum{2.261}
\figsetplot{./figset/scatter/261.pdf}
\figsetgrpend

\figsetgrpstart
\figsetgrpnum{2.262}
\figsetplot{./figset/scatter/262.pdf}
\figsetgrpend

\figsetgrpstart
\figsetgrpnum{2.263}
\figsetplot{./figset/scatter/263.pdf}
\figsetgrpend

\figsetgrpstart
\figsetgrpnum{2.264}
\figsetplot{./figset/scatter/264.pdf}
\figsetgrpend

\figsetgrpstart
\figsetgrpnum{2.265}
\figsetplot{./figset/scatter/265.pdf}
\figsetgrpend

\figsetgrpstart
\figsetgrpnum{2.266}
\figsetplot{./figset/scatter/266.pdf}
\figsetgrpend

\figsetgrpstart
\figsetgrpnum{2.267}
\figsetplot{./figset/scatter/267.pdf}
\figsetgrpend

\figsetgrpstart
\figsetgrpnum{2.268}
\figsetplot{./figset/scatter/268.pdf}
\figsetgrpend

\figsetgrpstart
\figsetgrpnum{2.269}
\figsetplot{./figset/scatter/269.pdf}
\figsetgrpend

\figsetgrpstart
\figsetgrpnum{2.270}
\figsetplot{./figset/scatter/270.pdf}
\figsetgrpend

\figsetgrpstart
\figsetgrpnum{2.271}
\figsetplot{./figset/scatter/271.pdf}
\figsetgrpend

\figsetgrpstart
\figsetgrpnum{2.272}
\figsetplot{./figset/scatter/272.pdf}
\figsetgrpend

\figsetgrpstart
\figsetgrpnum{2.273}
\figsetplot{./figset/scatter/273.pdf}
\figsetgrpend

\figsetgrpstart
\figsetgrpnum{2.274}
\figsetplot{./figset/scatter/274.pdf}
\figsetgrpend

\figsetgrpstart
\figsetgrpnum{2.275}
\figsetplot{./figset/scatter/275.pdf}
\figsetgrpend

\figsetgrpstart
\figsetgrpnum{2.276}
\figsetplot{./figset/scatter/276.pdf}
\figsetgrpend

\figsetgrpstart
\figsetgrpnum{2.277}
\figsetplot{./figset/scatter/277.pdf}
\figsetgrpend

\figsetgrpstart
\figsetgrpnum{2.278}
\figsetplot{./figset/scatter/278.pdf}
\figsetgrpend

\figsetgrpstart
\figsetgrpnum{2.279}
\figsetplot{./figset/scatter/279.pdf}
\figsetgrpend

\figsetgrpstart
\figsetgrpnum{2.280}
\figsetplot{./figset/scatter/280.pdf}
\figsetgrpend

\figsetgrpstart
\figsetgrpnum{2.281}
\figsetplot{./figset/scatter/281.pdf}
\figsetgrpend

\figsetgrpstart
\figsetgrpnum{2.282}
\figsetplot{./figset/scatter/282.pdf}
\figsetgrpend

\figsetgrpstart
\figsetgrpnum{2.283}
\figsetplot{./figset/scatter/283.pdf}
\figsetgrpend

\figsetgrpstart
\figsetgrpnum{2.284}
\figsetplot{./figset/scatter/284.pdf}
\figsetgrpend

\figsetgrpstart
\figsetgrpnum{2.285}
\figsetplot{./figset/scatter/285.pdf}
\figsetgrpend

\figsetgrpstart
\figsetgrpnum{2.286}
\figsetplot{./figset/scatter/286.pdf}
\figsetgrpend

\figsetgrpstart
\figsetgrpnum{2.287}
\figsetplot{./figset/scatter/287.pdf}
\figsetgrpend

\figsetgrpstart
\figsetgrpnum{2.288}
\figsetplot{./figset/scatter/288.pdf}
\figsetgrpend

\figsetgrpstart
\figsetgrpnum{2.289}
\figsetplot{./figset/scatter/289.pdf}
\figsetgrpend

\figsetgrpstart
\figsetgrpnum{2.290}
\figsetplot{./figset/scatter/290.pdf}
\figsetgrpend

\figsetgrpstart
\figsetgrpnum{2.291}
\figsetplot{./figset/scatter/291.pdf}
\figsetgrpend

\figsetgrpstart
\figsetgrpnum{2.292}
\figsetplot{./figset/scatter/292.pdf}
\figsetgrpend

\figsetgrpstart
\figsetgrpnum{2.293}
\figsetplot{./figset/scatter/293.pdf}
\figsetgrpend

\figsetgrpstart
\figsetgrpnum{2.294}
\figsetplot{./figset/scatter/294.pdf}
\figsetgrpend

\figsetgrpstart
\figsetgrpnum{2.295}
\figsetplot{./figset/scatter/295.pdf}
\figsetgrpend

\figsetgrpstart
\figsetgrpnum{2.296}
\figsetplot{./figset/scatter/296.pdf}
\figsetgrpend

\figsetgrpstart
\figsetgrpnum{2.297}

\figsetplot{./figset/scatter/297.pdf}

\figsetgrpend

\figsetgrpstart
\figsetgrpnum{2.298}

\figsetplot{./figset/scatter/298.pdf}

\figsetgrpend

\figsetgrpstart
\figsetgrpnum{2.299}

\figsetplot{./figset/scatter/299.pdf}

\figsetgrpend

\figsetgrpstart
\figsetgrpnum{2.300}

\figsetplot{./figset/scatter/300.pdf}

\figsetgrpend

\figsetgrpstart
\figsetgrpnum{2.301}

\figsetplot{./figset/scatter/301.pdf}

\figsetgrpend

\figsetgrpstart
\figsetgrpnum{2.302}

\figsetplot{./figset/scatter/302.pdf}

\figsetgrpend

\figsetgrpstart
\figsetgrpnum{2.303}

\figsetplot{./figset/scatter/303.pdf}

\figsetgrpend

\figsetgrpstart
\figsetgrpnum{2.304}

\figsetplot{./figset/scatter/304.pdf}

\figsetgrpend

\figsetgrpstart
\figsetgrpnum{2.305}

\figsetplot{./figset/scatter/305.pdf}

\figsetgrpend

\figsetgrpstart
\figsetgrpnum{2.306}

\figsetplot{./figset/scatter/306.pdf}

\figsetgrpend

\figsetgrpstart
\figsetgrpnum{2.307}

\figsetplot{./figset/scatter/307.pdf}

\figsetgrpend

\figsetgrpstart
\figsetgrpnum{2.308}

\figsetplot{./figset/scatter/308.pdf}

\figsetgrpend

\figsetgrpstart
\figsetgrpnum{2.309}

\figsetplot{./figset/scatter/309.pdf}

\figsetgrpend

\figsetgrpstart
\figsetgrpnum{2.310}

\figsetplot{./figset/scatter/310.pdf}

\figsetgrpend

\figsetgrpstart
\figsetgrpnum{2.311}

\figsetplot{./figset/scatter/311.pdf}

\figsetgrpend

\figsetgrpstart
\figsetgrpnum{2.312}

\figsetplot{./figset/scatter/312.pdf}

\figsetgrpend

\figsetgrpstart
\figsetgrpnum{2.313}

\figsetplot{./figset/scatter/313.pdf}

\figsetgrpend

\figsetgrpstart
\figsetgrpnum{2.314}

\figsetplot{./figset/scatter/314.pdf}

\figsetgrpend

\figsetgrpstart
\figsetgrpnum{2.315}

\figsetplot{./figset/scatter/315.pdf}

\figsetgrpend

\figsetgrpstart
\figsetgrpnum{2.316}

\figsetplot{./figset/scatter/316.pdf}

\figsetgrpend

\figsetgrpstart
\figsetgrpnum{2.317}

\figsetplot{./figset/scatter/317.pdf}

\figsetgrpend

\figsetgrpstart
\figsetgrpnum{2.318}

\figsetplot{./figset/scatter/318.pdf}

\figsetgrpend

\figsetgrpstart
\figsetgrpnum{2.319}

\figsetplot{./figset/scatter/319.pdf}

\figsetgrpend

\figsetgrpstart
\figsetgrpnum{2.320}

\figsetplot{./figset/scatter/320.pdf}

\figsetgrpend

\figsetgrpstart
\figsetgrpnum{2.321}

\figsetplot{./figset/scatter/321.pdf}

\figsetgrpend

\figsetgrpstart
\figsetgrpnum{2.322}

\figsetplot{./figset/scatter/322.pdf}

\figsetgrpend

\figsetgrpstart
\figsetgrpnum{2.323}

\figsetplot{./figset/scatter/323.pdf}

\figsetgrpend

\figsetgrpstart
\figsetgrpnum{2.324}

\figsetplot{./figset/scatter/324.pdf}

\figsetgrpend

\figsetgrpstart
\figsetgrpnum{2.325}

\figsetplot{./figset/scatter/325.pdf}

\figsetgrpend

\figsetgrpstart
\figsetgrpnum{2.326}

\figsetplot{./figset/scatter/326.pdf}

\figsetgrpend

\figsetgrpstart
\figsetgrpnum{2.327}

\figsetplot{./figset/scatter/327.pdf}

\figsetgrpend

\figsetgrpstart
\figsetgrpnum{2.328}

\figsetplot{./figset/scatter/328.pdf}

\figsetgrpend

\figsetgrpstart
\figsetgrpnum{2.329}

\figsetplot{./figset/scatter/329.pdf}

\figsetgrpend

\figsetgrpstart
\figsetgrpnum{2.330}

\figsetplot{./figset/scatter/330.pdf}

\figsetgrpend

\figsetgrpstart
\figsetgrpnum{2.331}

\figsetplot{./figset/scatter/331.pdf}

\figsetgrpend

\figsetgrpstart
\figsetgrpnum{2.332}

\figsetplot{./figset/scatter/332.pdf}

\figsetgrpend

\figsetgrpstart
\figsetgrpnum{2.333}

\figsetplot{./figset/scatter/333.pdf}

\figsetgrpend

\figsetgrpstart
\figsetgrpnum{2.334}

\figsetplot{./figset/scatter/334.pdf}

\figsetgrpend

\figsetgrpstart
\figsetgrpnum{2.335}

\figsetplot{./figset/scatter/335.pdf}

\figsetgrpend

\figsetgrpstart
\figsetgrpnum{2.336}

\figsetplot{./figset/scatter/336.pdf}

\figsetgrpend

\figsetgrpstart
\figsetgrpnum{2.337}

\figsetplot{./figset/scatter/337.pdf}

\figsetgrpend

\figsetgrpstart
\figsetgrpnum{2.338}

\figsetplot{./figset/scatter/338.pdf}

\figsetgrpend

\figsetgrpstart
\figsetgrpnum{2.339}

\figsetplot{./figset/scatter/339.pdf}

\figsetgrpend

\figsetgrpstart
\figsetgrpnum{2.340}

\figsetplot{./figset/scatter/340.pdf}

\figsetgrpend

\figsetgrpstart
\figsetgrpnum{2.341}

\figsetplot{./figset/scatter/341.pdf}

\figsetgrpend

\figsetgrpstart
\figsetgrpnum{2.342}

\figsetplot{./figset/scatter/342.pdf}

\figsetgrpend

\figsetgrpstart
\figsetgrpnum{2.343}

\figsetplot{./figset/scatter/343.pdf}

\figsetgrpend

\figsetgrpstart
\figsetgrpnum{2.344}

\figsetplot{./figset/scatter/344.pdf}

\figsetgrpend

\figsetgrpstart
\figsetgrpnum{2.345}

\figsetplot{./figset/scatter/345.pdf}

\figsetgrpend

\figsetgrpstart
\figsetgrpnum{2.346}

\figsetplot{./figset/scatter/346.pdf}

\figsetgrpend

\figsetgrpstart
\figsetgrpnum{2.347}

\figsetplot{./figset/scatter/347.pdf}

\figsetgrpend

\figsetgrpstart
\figsetgrpnum{2.348}

\figsetplot{./figset/scatter/348.pdf}

\figsetgrpend

\figsetgrpstart
\figsetgrpnum{2.349}

\figsetplot{./figset/scatter/349.pdf}

\figsetgrpend

\figsetgrpstart
\figsetgrpnum{2.350}

\figsetplot{./figset/scatter/350.pdf}

\figsetgrpend

\figsetgrpstart
\figsetgrpnum{2.351}

\figsetplot{./figset/scatter/351.pdf}

\figsetgrpend

\figsetgrpstart
\figsetgrpnum{2.352}

\figsetplot{./figset/scatter/352.pdf}

\figsetgrpend

\figsetgrpstart
\figsetgrpnum{2.353}

\figsetplot{./figset/scatter/353.pdf}

\figsetgrpend

\figsetgrpstart
\figsetgrpnum{2.354}

\figsetplot{./figset/scatter/354.pdf}

\figsetgrpend

\figsetgrpstart
\figsetgrpnum{2.355}

\figsetplot{./figset/scatter/355.pdf}

\figsetgrpend

\figsetgrpstart
\figsetgrpnum{2.356}

\figsetplot{./figset/scatter/356.pdf}

\figsetgrpend

\figsetgrpstart
\figsetgrpnum{2.357}

\figsetplot{./figset/scatter/357.pdf}

\figsetgrpend

\figsetgrpstart
\figsetgrpnum{2.358}

\figsetplot{./figset/scatter/358.pdf}

\figsetgrpend

\figsetgrpstart
\figsetgrpnum{2.359}

\figsetplot{./figset/scatter/359.pdf}

\figsetgrpend

\figsetgrpstart
\figsetgrpnum{2.360}

\figsetplot{./figset/scatter/360.pdf}

\figsetgrpend

\figsetgrpstart
\figsetgrpnum{2.361}

\figsetplot{./figset/scatter/361.pdf}

\figsetgrpend

\figsetgrpstart
\figsetgrpnum{2.362}

\figsetplot{./figset/scatter/362.pdf}

\figsetgrpend

\figsetgrpstart
\figsetgrpnum{2.363}

\figsetplot{./figset/scatter/363.pdf}

\figsetgrpend

\figsetgrpstart
\figsetgrpnum{2.364}

\figsetplot{./figset/scatter/364.pdf}

\figsetgrpend

\figsetgrpstart
\figsetgrpnum{2.365}

\figsetplot{./figset/scatter/365.pdf}

\figsetgrpend

\figsetgrpstart
\figsetgrpnum{2.366}

\figsetplot{./figset/scatter/366.pdf}

\figsetgrpend

\figsetgrpstart
\figsetgrpnum{2.367}

\figsetplot{./figset/scatter/367.pdf}

\figsetgrpend

\figsetgrpstart
\figsetgrpnum{2.368}

\figsetplot{./figset/scatter/368.pdf}

\figsetgrpend

\figsetgrpstart
\figsetgrpnum{2.369}

\figsetplot{./figset/scatter/369.pdf}

\figsetgrpend

\figsetgrpstart
\figsetgrpnum{2.370}

\figsetplot{./figset/scatter/370.pdf}

\figsetgrpend

\figsetgrpstart
\figsetgrpnum{2.371}

\figsetplot{./figset/scatter/371.pdf}

\figsetgrpend

\figsetgrpstart
\figsetgrpnum{2.372}

\figsetplot{./figset/scatter/372.pdf}

\figsetgrpend

\figsetgrpstart
\figsetgrpnum{2.373}

\figsetplot{./figset/scatter/373.pdf}

\figsetgrpend

\figsetgrpstart
\figsetgrpnum{2.374}

\figsetplot{./figset/scatter/374.pdf}

\figsetgrpend

\figsetgrpstart
\figsetgrpnum{2.375}

\figsetplot{./figset/scatter/375.pdf}

\figsetgrpend

\figsetgrpstart
\figsetgrpnum{2.376}

\figsetplot{./figset/scatter/376.pdf}

\figsetgrpend

\figsetgrpstart
\figsetgrpnum{2.377}

\figsetplot{./figset/scatter/377.pdf}

\figsetgrpend

\figsetgrpstart
\figsetgrpnum{2.378}

\figsetplot{./figset/scatter/378.pdf}

\figsetgrpend

\figsetgrpstart
\figsetgrpnum{2.379}

\figsetplot{./figset/scatter/379.pdf}

\figsetgrpend

\figsetgrpstart
\figsetgrpnum{2.380}

\figsetplot{./figset/scatter/380.pdf}

\figsetgrpend

\figsetgrpstart
\figsetgrpnum{2.381}

\figsetplot{./figset/scatter/381.pdf}

\figsetgrpend

\figsetgrpstart
\figsetgrpnum{2.382}

\figsetplot{./figset/scatter/382.pdf}

\figsetgrpend

\figsetgrpstart
\figsetgrpnum{2.383}

\figsetplot{./figset/scatter/383.pdf}

\figsetgrpend

\figsetgrpstart
\figsetgrpnum{2.384}

\figsetplot{./figset/scatter/384.pdf}

\figsetgrpend

\figsetgrpstart
\figsetgrpnum{2.385}

\figsetplot{./figset/scatter/385.pdf}

\figsetgrpend

\figsetgrpstart
\figsetgrpnum{2.386}

\figsetplot{./figset/scatter/386.pdf}

\figsetgrpend

\figsetgrpstart
\figsetgrpnum{2.387}

\figsetplot{./figset/scatter/387.pdf}

\figsetgrpend

\figsetgrpstart
\figsetgrpnum{2.388}

\figsetplot{./figset/scatter/388.pdf}

\figsetgrpend

\figsetgrpstart
\figsetgrpnum{2.389}

\figsetplot{./figset/scatter/389.pdf}

\figsetgrpend

\figsetgrpstart
\figsetgrpnum{2.390}

\figsetplot{./figset/scatter/390.pdf}

\figsetgrpend

\figsetgrpstart
\figsetgrpnum{2.391}

\figsetplot{./figset/scatter/391.pdf}

\figsetgrpend

\figsetgrpstart
\figsetgrpnum{2.392}

\figsetplot{./figset/scatter/392.pdf}

\figsetgrpend

\figsetgrpstart
\figsetgrpnum{2.393}

\figsetplot{./figset/scatter/393.pdf}

\figsetgrpend

\figsetgrpstart
\figsetgrpnum{2.394}

\figsetplot{./figset/scatter/394.pdf}

\figsetgrpend

\figsetgrpstart
\figsetgrpnum{2.395}

\figsetplot{./figset/scatter/395.pdf}

\figsetgrpend

\figsetgrpstart
\figsetgrpnum{2.396}

\figsetplot{./figset/scatter/396.pdf}

\figsetgrpend

\figsetgrpstart
\figsetgrpnum{2.397}

\figsetplot{./figset/scatter/397.pdf}

\figsetgrpend

\figsetgrpstart
\figsetgrpnum{2.398}

\figsetplot{./figset/scatter/398.pdf}

\figsetgrpend

\figsetgrpstart
\figsetgrpnum{2.399}

\figsetplot{./figset/scatter/399.pdf}

\figsetgrpend

\figsetgrpstart
\figsetgrpnum{2.400}

\figsetplot{./figset/scatter/400.pdf}

\figsetgrpend

\figsetgrpstart
\figsetgrpnum{2.401}

\figsetplot{./figset/scatter/401.pdf}

\figsetgrpend

\figsetgrpstart
\figsetgrpnum{2.402}

\figsetplot{./figset/scatter/402.pdf}

\figsetgrpend

\figsetgrpstart
\figsetgrpnum{2.403}

\figsetplot{./figset/scatter/403.pdf}

\figsetgrpend

\figsetgrpstart
\figsetgrpnum{2.404}

\figsetplot{./figset/scatter/404.pdf}

\figsetgrpend

\figsetgrpstart
\figsetgrpnum{2.405}

\figsetplot{./figset/scatter/405.pdf}

\figsetgrpend

\figsetgrpstart
\figsetgrpnum{2.406}

\figsetplot{./figset/scatter/406.pdf}

\figsetgrpend

\figsetgrpstart
\figsetgrpnum{2.407}

\figsetplot{./figset/scatter/407.pdf}

\figsetgrpend

\figsetgrpstart
\figsetgrpnum{2.408}

\figsetplot{./figset/scatter/408.pdf}

\figsetgrpend

\figsetgrpstart
\figsetgrpnum{2.409}

\figsetplot{./figset/scatter/409.pdf}

\figsetgrpend

\figsetgrpstart
\figsetgrpnum{2.410}

\figsetplot{./figset/scatter/410.pdf}

\figsetgrpend

\figsetgrpstart
\figsetgrpnum{2.411}

\figsetplot{./figset/scatter/411.pdf}

\figsetgrpend

\figsetgrpstart
\figsetgrpnum{2.412}

\figsetplot{./figset/scatter/412.pdf}

\figsetgrpend

\figsetgrpstart
\figsetgrpnum{2.413}

\figsetplot{./figset/scatter/413.pdf}

\figsetgrpend

\figsetgrpstart
\figsetgrpnum{2.414}

\figsetplot{./figset/scatter/414.pdf}

\figsetgrpend

\figsetgrpstart
\figsetgrpnum{2.415}

\figsetplot{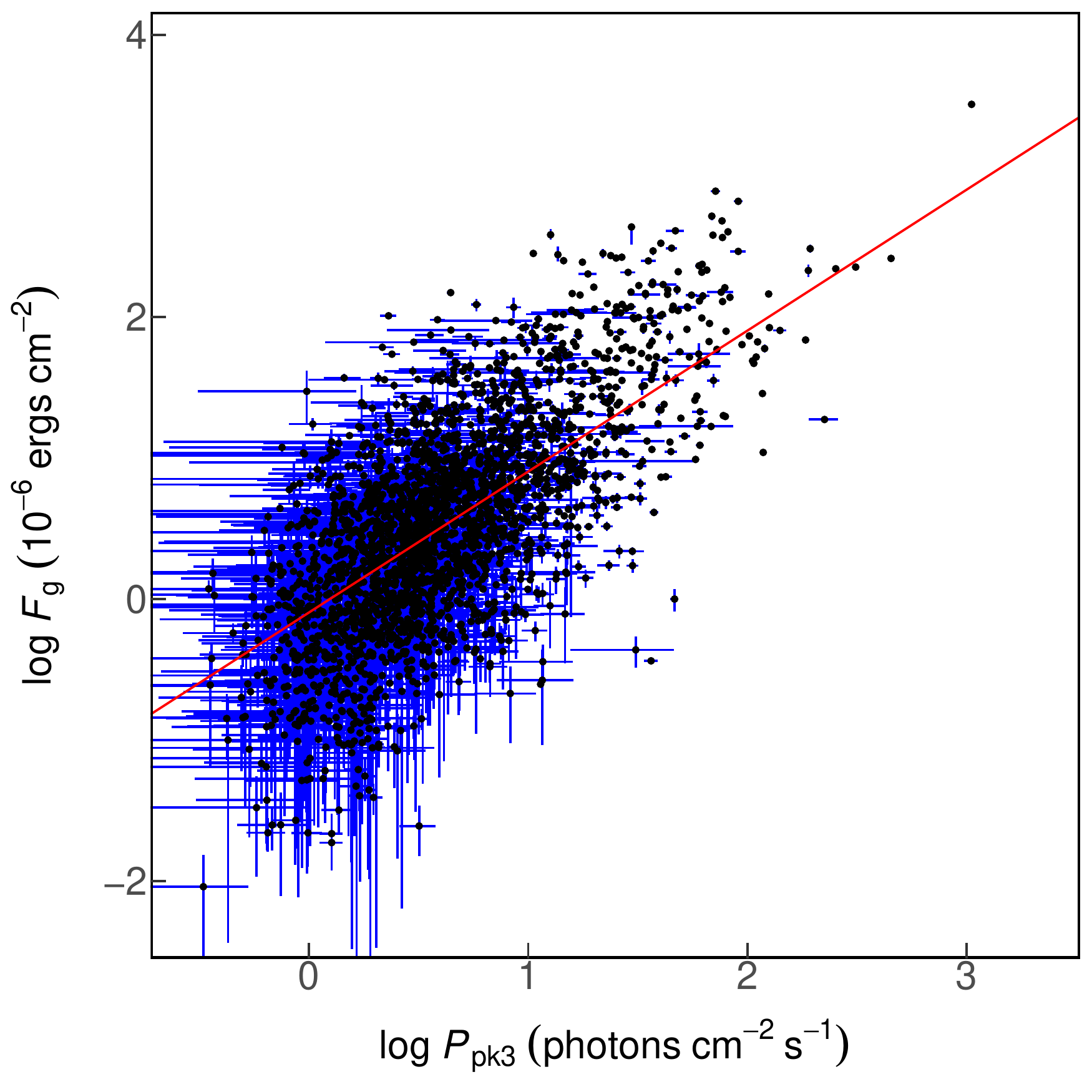}

\figsetgrpend

\figsetgrpstart
\figsetgrpnum{2.416}

\figsetplot{./figset/scatter/416.pdf}

\figsetgrpend

\figsetgrpstart
\figsetgrpnum{2.417}

\figsetplot{./figset/scatter/417.pdf}

\figsetgrpend

\figsetgrpstart
\figsetgrpnum{2.418}

\figsetplot{./figset/scatter/418.pdf}

\figsetgrpend

\figsetgrpstart
\figsetgrpnum{2.419}

\figsetplot{./figset/scatter/419.pdf}

\figsetgrpend

\figsetgrpstart
\figsetgrpnum{2.420}

\figsetplot{./figset/scatter/420.pdf}

\figsetgrpend

\figsetgrpstart
\figsetgrpnum{2.421}

\figsetplot{./figset/scatter/421.pdf}

\figsetgrpend

\figsetgrpstart
\figsetgrpnum{2.422}

\figsetplot{./figset/scatter/422.pdf}

\figsetgrpend

\figsetgrpstart
\figsetgrpnum{2.423}

\figsetplot{./figset/scatter/423.pdf}

\figsetgrpend

\figsetgrpstart
\figsetgrpnum{2.424}

\figsetplot{./figset/scatter/424.pdf}

\figsetgrpend

\figsetgrpstart
\figsetgrpnum{2.425}

\figsetplot{./figset/scatter/425.pdf}

\figsetgrpend

\figsetgrpstart
\figsetgrpnum{2.426}

\figsetplot{./figset/scatter/426.pdf}

\figsetgrpend

\figsetgrpstart
\figsetgrpnum{2.427}

\figsetplot{./figset/scatter/427.pdf}

\figsetgrpend

\figsetgrpstart
\figsetgrpnum{2.428}

\figsetplot{./figset/scatter/428.pdf}

\figsetgrpend

\figsetgrpstart
\figsetgrpnum{2.429}

\figsetplot{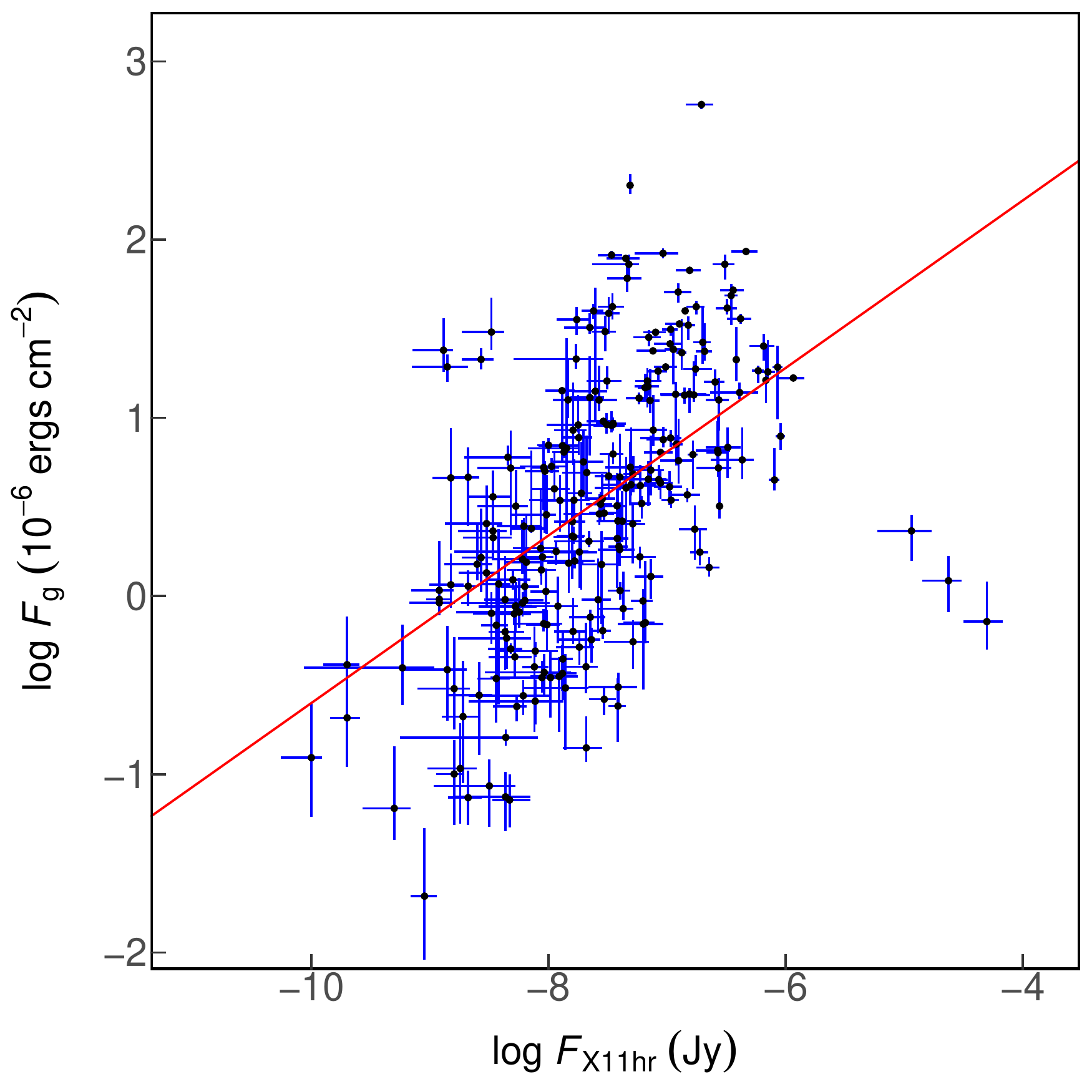}

\figsetgrpend

\figsetgrpstart
\figsetgrpnum{2.430}

\figsetplot{./figset/scatter/430.pdf}

\figsetgrpend

\figsetgrpstart
\figsetgrpnum{2.431}

\figsetplot{./figset/scatter/431.pdf}

\figsetgrpend

\figsetgrpstart
\figsetgrpnum{2.432}

\figsetplot{./figset/scatter/432.pdf}

\figsetgrpend

\figsetgrpstart
\figsetgrpnum{2.433}

\figsetplot{./figset/scatter/433.pdf}

\figsetgrpend

\figsetgrpstart
\figsetgrpnum{2.434}

\figsetplot{./figset/scatter/434.pdf}

\figsetgrpend

\figsetgrpstart
\figsetgrpnum{2.435}

\figsetplot{./figset/scatter/435.pdf}

\figsetgrpend

\figsetgrpstart
\figsetgrpnum{2.436}

\figsetplot{./figset/scatter/436.pdf}

\figsetgrpend

\figsetgrpstart
\figsetgrpnum{2.437}

\figsetplot{./figset/scatter/437.pdf}

\figsetgrpend

\figsetgrpstart
\figsetgrpnum{2.438}

\figsetplot{./figset/scatter/438.pdf}

\figsetgrpend

\figsetgrpstart
\figsetgrpnum{2.439}

\figsetplot{./figset/scatter/439.pdf}

\figsetgrpend

\figsetgrpstart
\figsetgrpnum{2.440}

\figsetplot{./figset/scatter/440.pdf}

\figsetgrpend

\figsetgrpstart
\figsetgrpnum{2.441}

\figsetplot{./figset/scatter/441.pdf}

\figsetgrpend

\figsetgrpstart
\figsetgrpnum{2.442}

\figsetplot{./figset/scatter/442.pdf}

\figsetgrpend

\figsetgrpstart
\figsetgrpnum{2.443}

\figsetplot{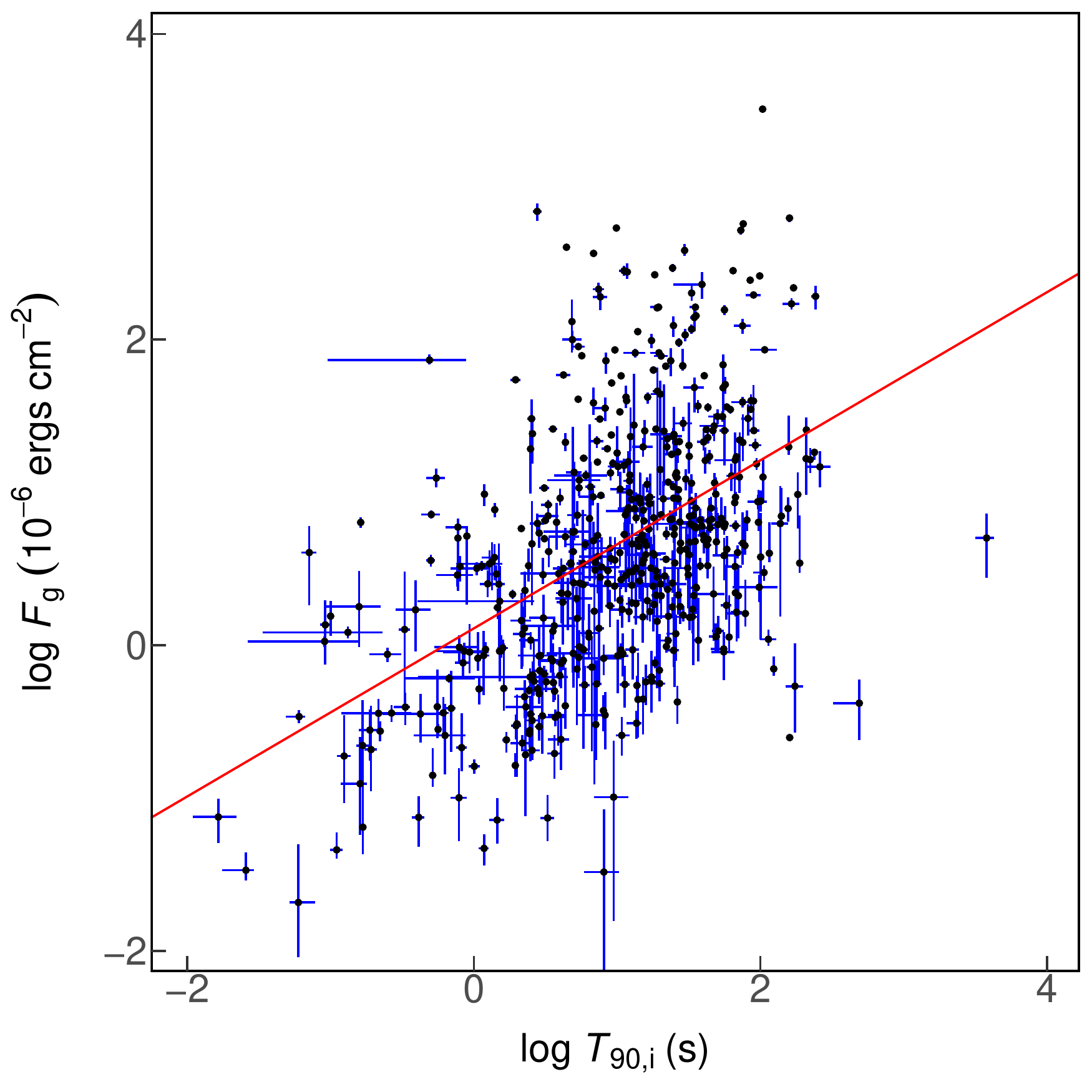}

\figsetgrpend

\figsetgrpstart
\figsetgrpnum{2.444}

\figsetplot{./figset/scatter/444.pdf}

\figsetgrpend

\figsetgrpstart
\figsetgrpnum{2.445}

\figsetplot{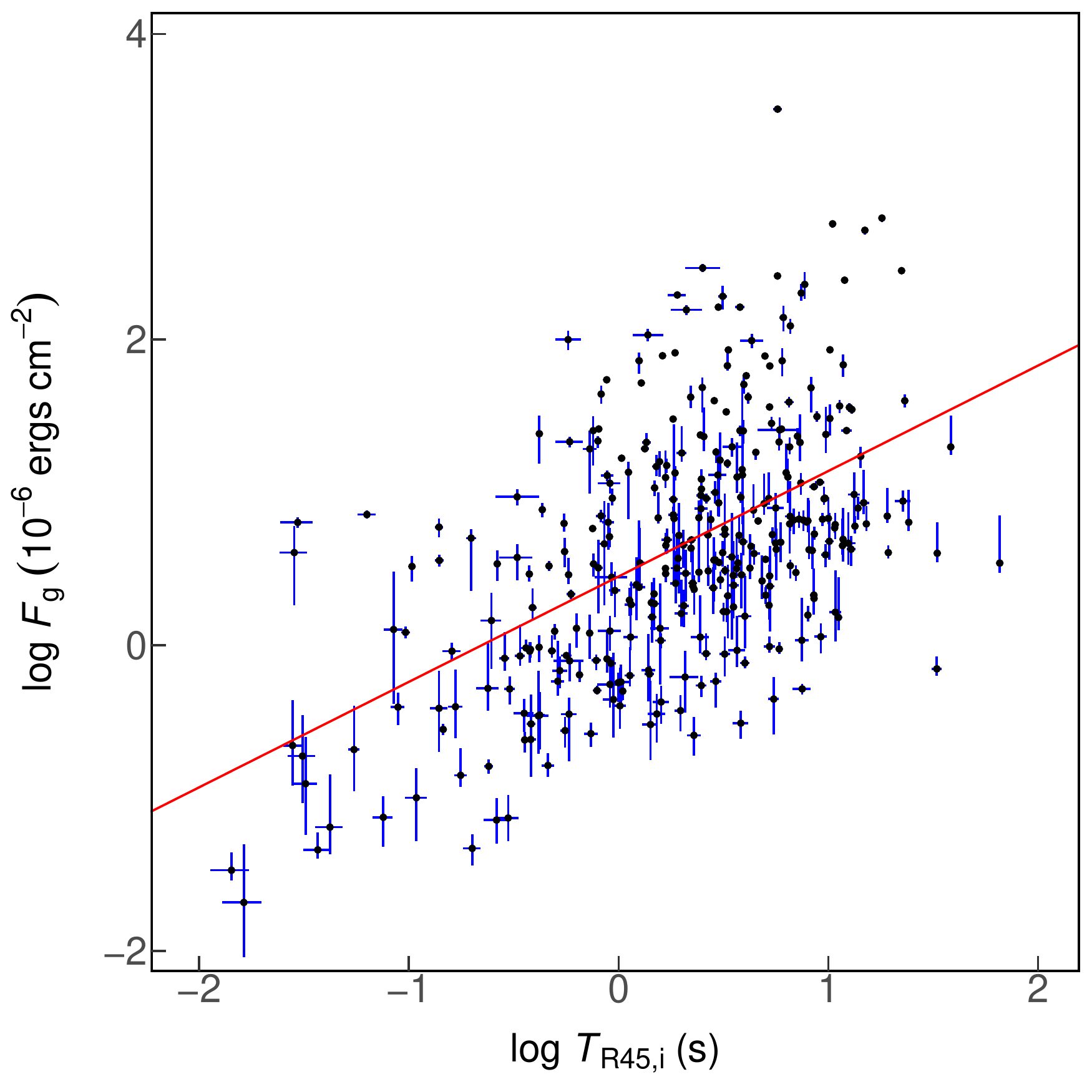}

\figsetgrpend

\figsetgrpstart
\figsetgrpnum{2.446}

\figsetplot{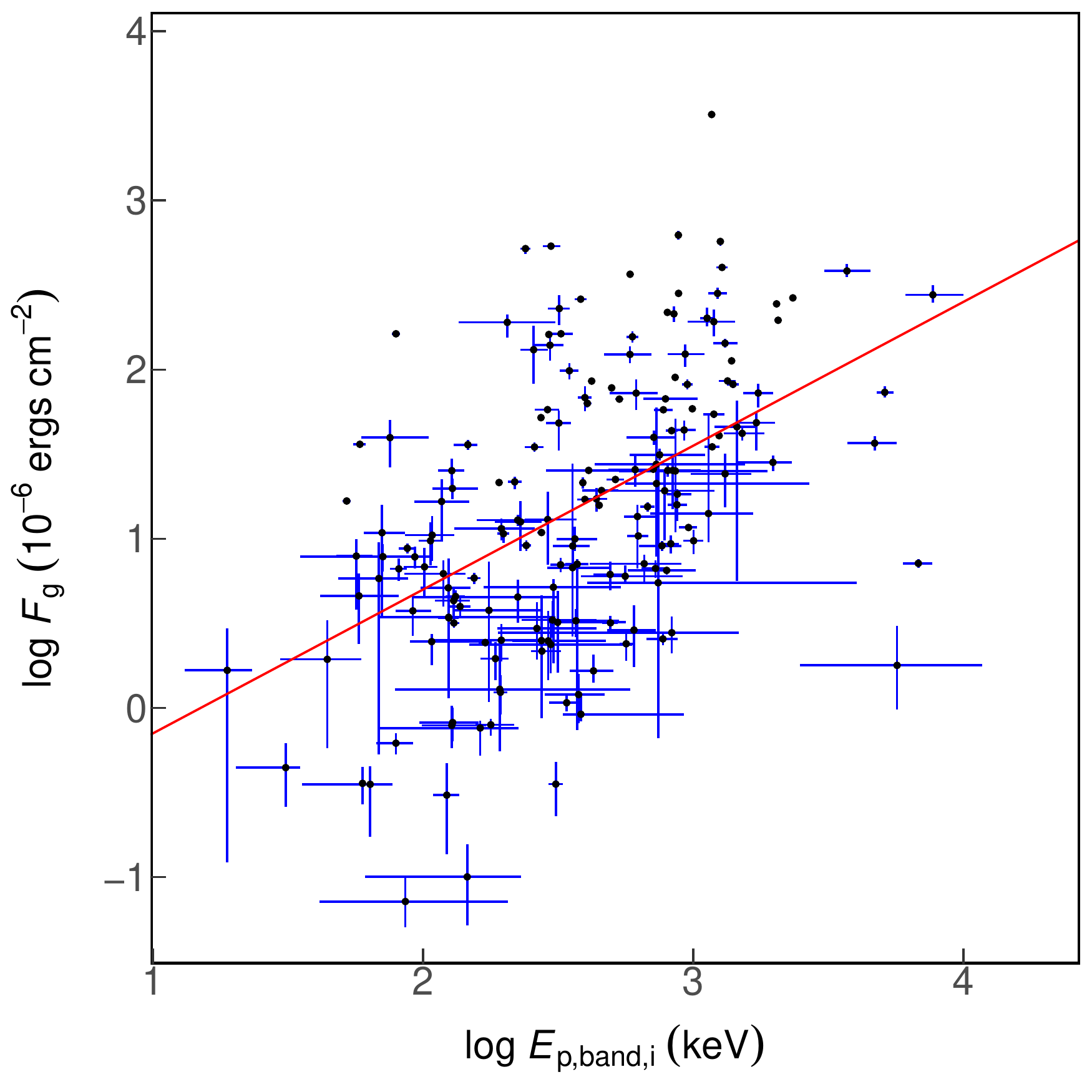}

\figsetgrpend

\figsetgrpstart
\figsetgrpnum{2.447}

\figsetplot{./figset/scatter/447.pdf}

\figsetgrpend

\figsetgrpstart
\figsetgrpnum{2.448}

\figsetplot{./figset/scatter/448.pdf}

\figsetgrpend

\figsetgrpstart
\figsetgrpnum{2.449}

\figsetplot{./figset/scatter/449.pdf}

\figsetgrpend

\figsetgrpstart
\figsetgrpnum{2.450}

\figsetplot{./figset/scatter/450.pdf}

\figsetgrpend

\figsetgrpstart
\figsetgrpnum{2.451}

\figsetplot{./figset/scatter/451.pdf}

\figsetgrpend

\figsetgrpstart
\figsetgrpnum{2.452}

\figsetplot{./figset/scatter/452.pdf}

\figsetgrpend

\figsetgrpstart
\figsetgrpnum{2.453}

\figsetplot{./figset/scatter/453.pdf}

\figsetgrpend

\figsetgrpstart
\figsetgrpnum{2.454}

\figsetplot{./figset/scatter/454.pdf}

\figsetgrpend

\figsetgrpstart
\figsetgrpnum{2.455}

\figsetplot{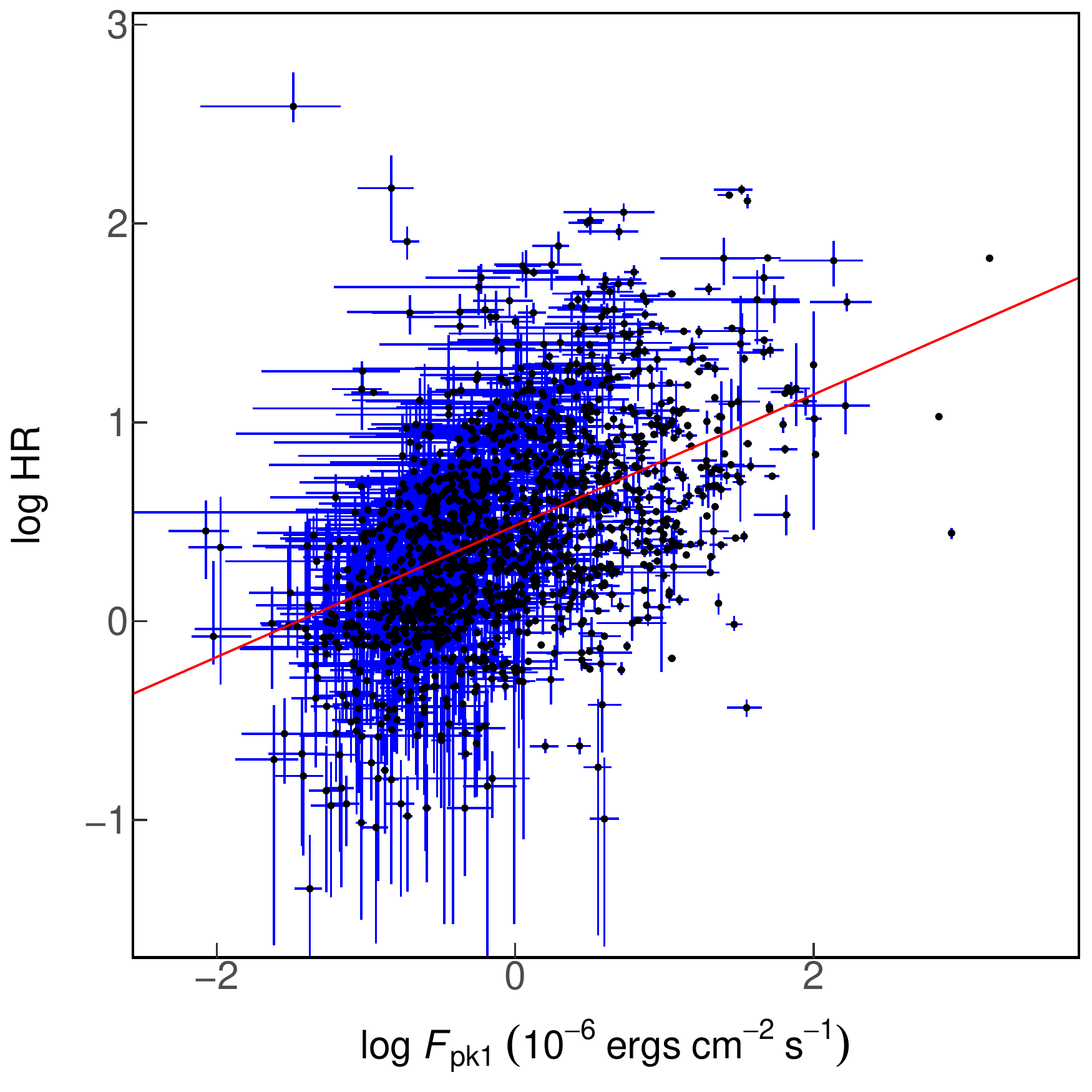}

\figsetgrpend

\figsetgrpstart
\figsetgrpnum{2.456}

\figsetplot{./figset/scatter/456.pdf}

\figsetgrpend

\figsetgrpstart
\figsetgrpnum{2.457}

\figsetplot{./figset/scatter/457.pdf}

\figsetgrpend

\figsetgrpstart
\figsetgrpnum{2.458}

\figsetplot{./figset/scatter/458.pdf}

\figsetgrpend

\figsetgrpstart
\figsetgrpnum{2.459}

\figsetplot{./figset/scatter/459.pdf}

\figsetgrpend

\figsetgrpstart
\figsetgrpnum{2.460}

\figsetplot{./figset/scatter/460.pdf}

\figsetgrpend

\figsetgrpstart
\figsetgrpnum{2.461}

\figsetplot{./figset/scatter/461.pdf}

\figsetgrpend

\figsetgrpstart
\figsetgrpnum{2.462}

\figsetplot{./figset/scatter/462.pdf}

\figsetgrpend

\figsetgrpstart
\figsetgrpnum{2.463}

\figsetplot{./figset/scatter/463.pdf}

\figsetgrpend

\figsetgrpstart
\figsetgrpnum{2.464}

\figsetplot{./figset/scatter/464.pdf}

\figsetgrpend

\figsetgrpstart
\figsetgrpnum{2.465}

\figsetplot{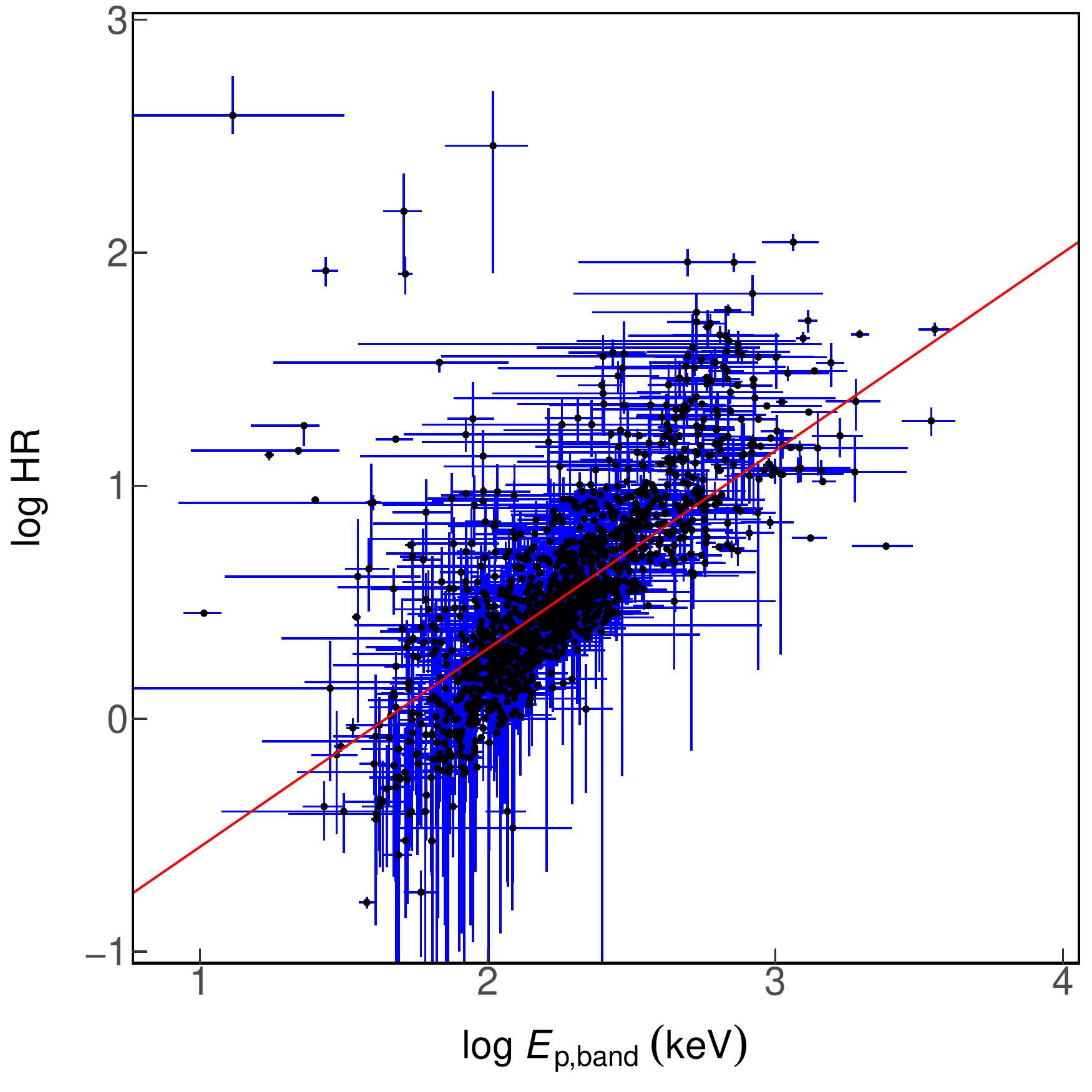}

\figsetgrpend

\figsetgrpstart
\figsetgrpnum{2.466}

\figsetplot{./figset/scatter/466.pdf}

\figsetgrpend

\figsetgrpstart
\figsetgrpnum{2.467}

\figsetplot{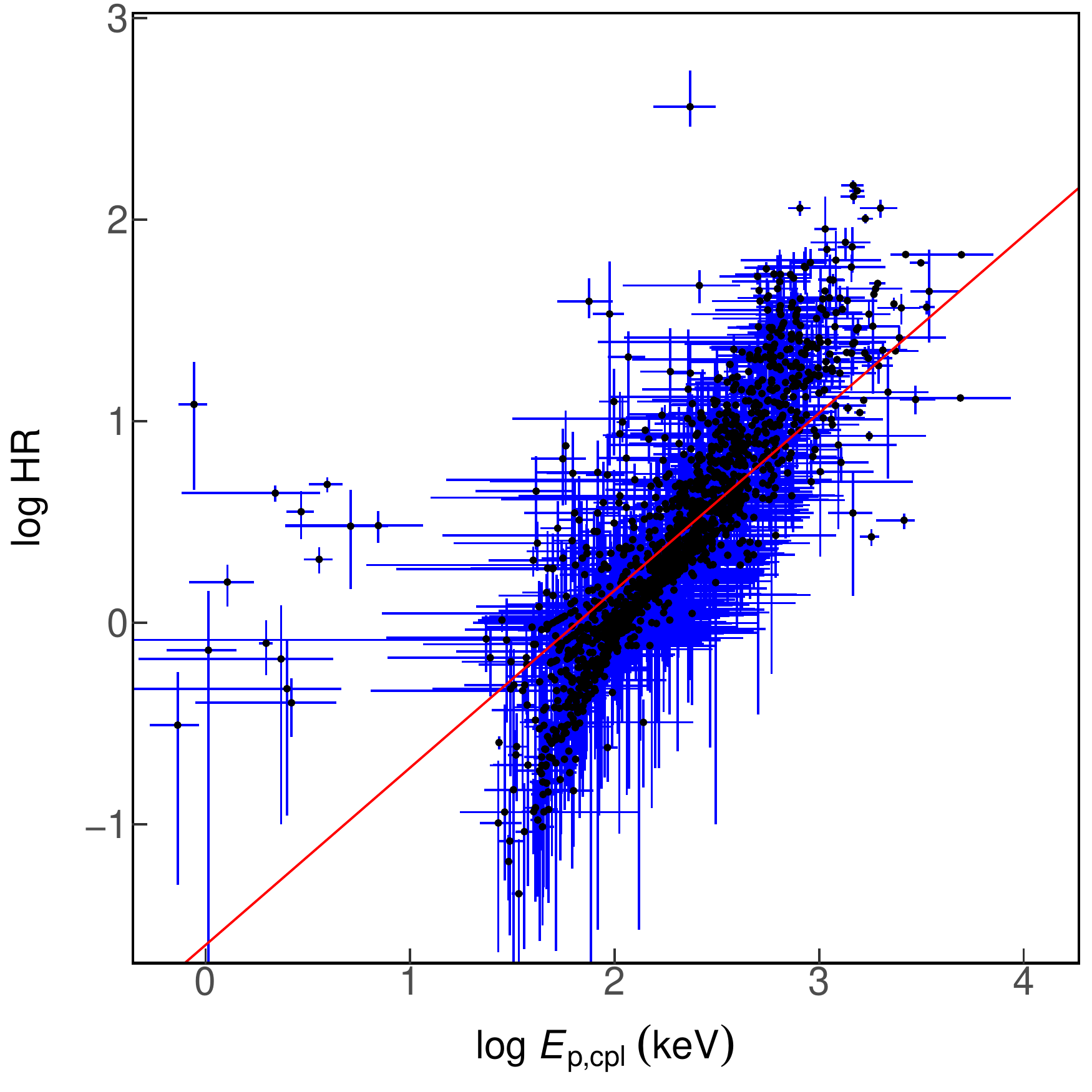}

\figsetgrpend

\figsetgrpstart
\figsetgrpnum{2.468}

\figsetplot{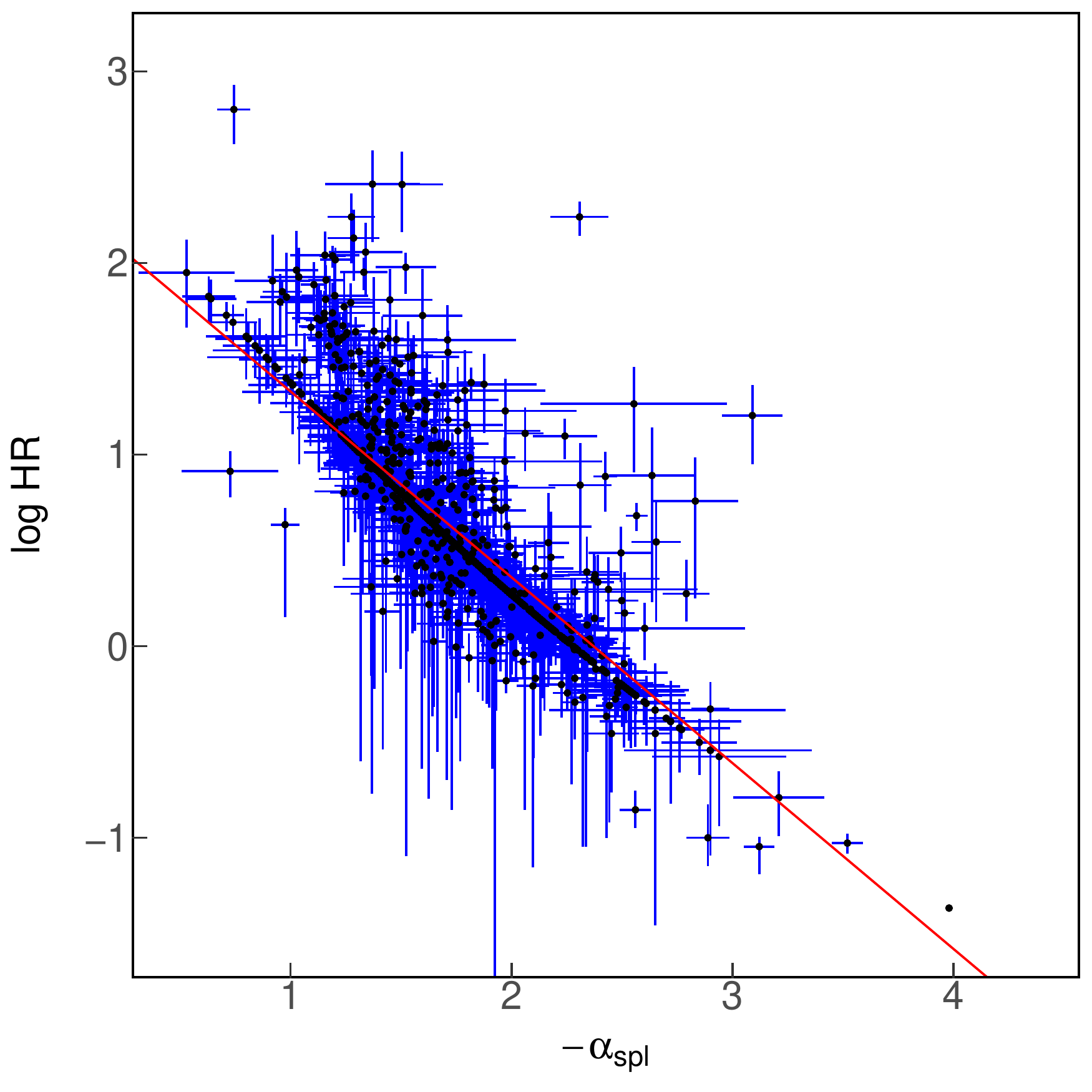}

\figsetgrpend

\figsetgrpstart
\figsetgrpnum{2.469}

\figsetplot{./figset/scatter/469.pdf}

\figsetgrpend

\figsetgrpstart
\figsetgrpnum{2.470}

\figsetplot{./figset/scatter/470.pdf}

\figsetgrpend

\figsetgrpstart
\figsetgrpnum{2.471}

\figsetplot{./figset/scatter/471.pdf}

\figsetgrpend

\figsetgrpstart
\figsetgrpnum{2.472}

\figsetplot{./figset/scatter/472.pdf}

\figsetgrpend

\figsetgrpstart
\figsetgrpnum{2.473}

\figsetplot{./figset/scatter/473.pdf}

\figsetgrpend

\figsetgrpstart
\figsetgrpnum{2.474}

\figsetplot{./figset/scatter/474.pdf}

\figsetgrpend

\figsetgrpstart
\figsetgrpnum{2.475}

\figsetplot{./figset/scatter/475.pdf}

\figsetgrpend

\figsetgrpstart
\figsetgrpnum{2.476}

\figsetplot{./figset/scatter/476.pdf}

\figsetgrpend

\figsetgrpstart
\figsetgrpnum{2.477}

\figsetplot{./figset/scatter/477.pdf}

\figsetgrpend

\figsetgrpstart
\figsetgrpnum{2.478}

\figsetplot{./figset/scatter/478.pdf}

\figsetgrpend

\figsetgrpstart
\figsetgrpnum{2.479}

\figsetplot{./figset/scatter/479.pdf}

\figsetgrpend

\figsetgrpstart
\figsetgrpnum{2.480}

\figsetplot{./figset/scatter/480.pdf}

\figsetgrpend

\figsetgrpstart
\figsetgrpnum{2.481}

\figsetplot{./figset/scatter/481.pdf}

\figsetgrpend

\figsetgrpstart
\figsetgrpnum{2.482}

\figsetplot{./figset/scatter/482.pdf}

\figsetgrpend

\figsetgrpstart
\figsetgrpnum{2.483}

\figsetplot{./figset/scatter/483.pdf}

\figsetgrpend

\figsetgrpstart
\figsetgrpnum{2.484}

\figsetplot{./figset/scatter/484.pdf}

\figsetgrpend

\figsetgrpstart
\figsetgrpnum{2.485}

\figsetplot{./figset/scatter/485.pdf}

\figsetgrpend

\figsetgrpstart
\figsetgrpnum{2.486}

\figsetplot{./figset/scatter/486.pdf}

\figsetgrpend

\figsetgrpstart
\figsetgrpnum{2.487}

\figsetplot{./figset/scatter/487.pdf}

\figsetgrpend

\figsetgrpstart
\figsetgrpnum{2.488}

\figsetplot{./figset/scatter/488.pdf}

\figsetgrpend

\figsetgrpstart
\figsetgrpnum{2.489}

\figsetplot{./figset/scatter/489.pdf}

\figsetgrpend

\figsetgrpstart
\figsetgrpnum{2.490}

\figsetplot{./figset/scatter/490.pdf}

\figsetgrpend

\figsetgrpstart
\figsetgrpnum{2.491}

\figsetplot{./figset/scatter/491.pdf}

\figsetgrpend

\figsetgrpstart
\figsetgrpnum{2.492}

\figsetplot{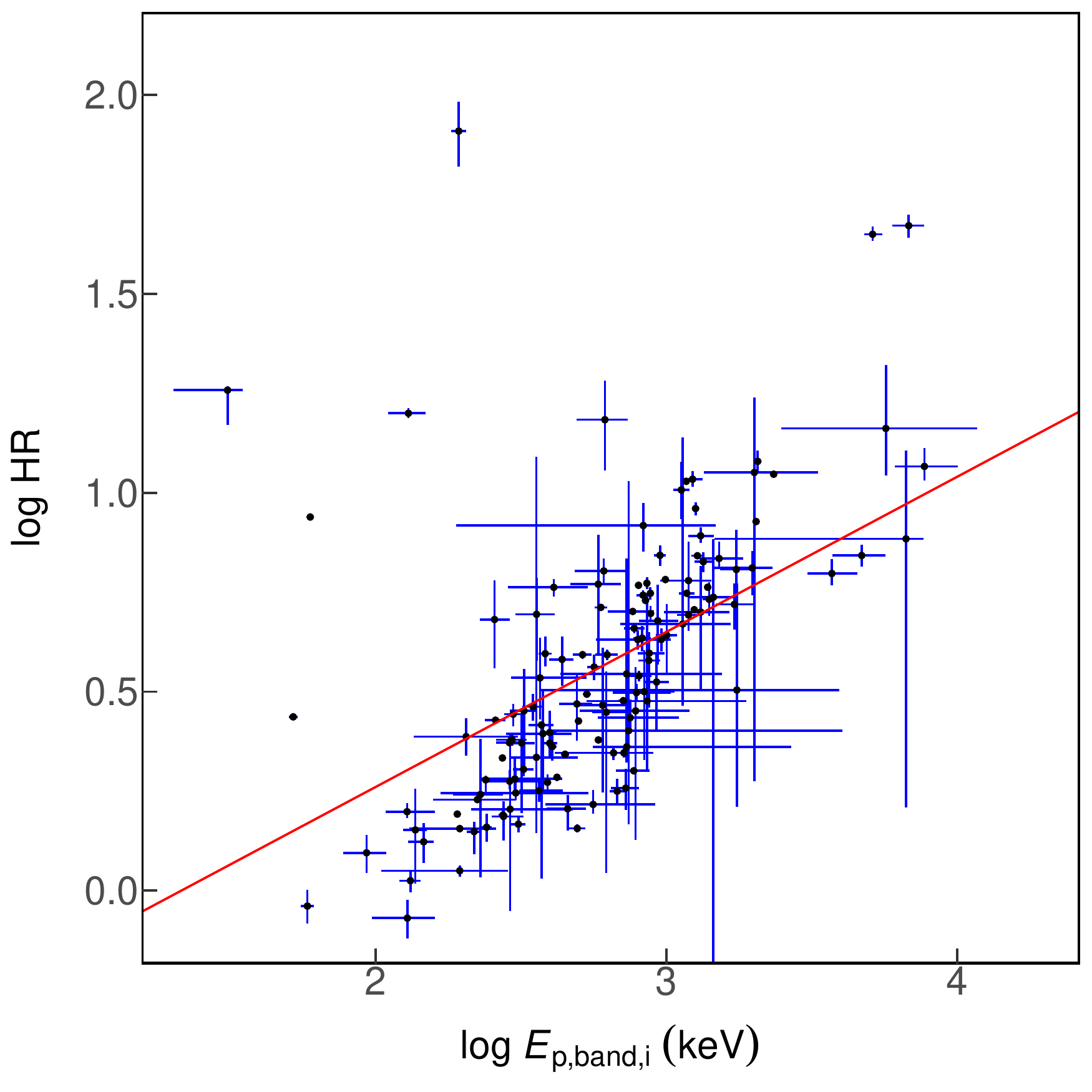}

\figsetgrpend

\figsetgrpstart
\figsetgrpnum{2.493}

\figsetplot{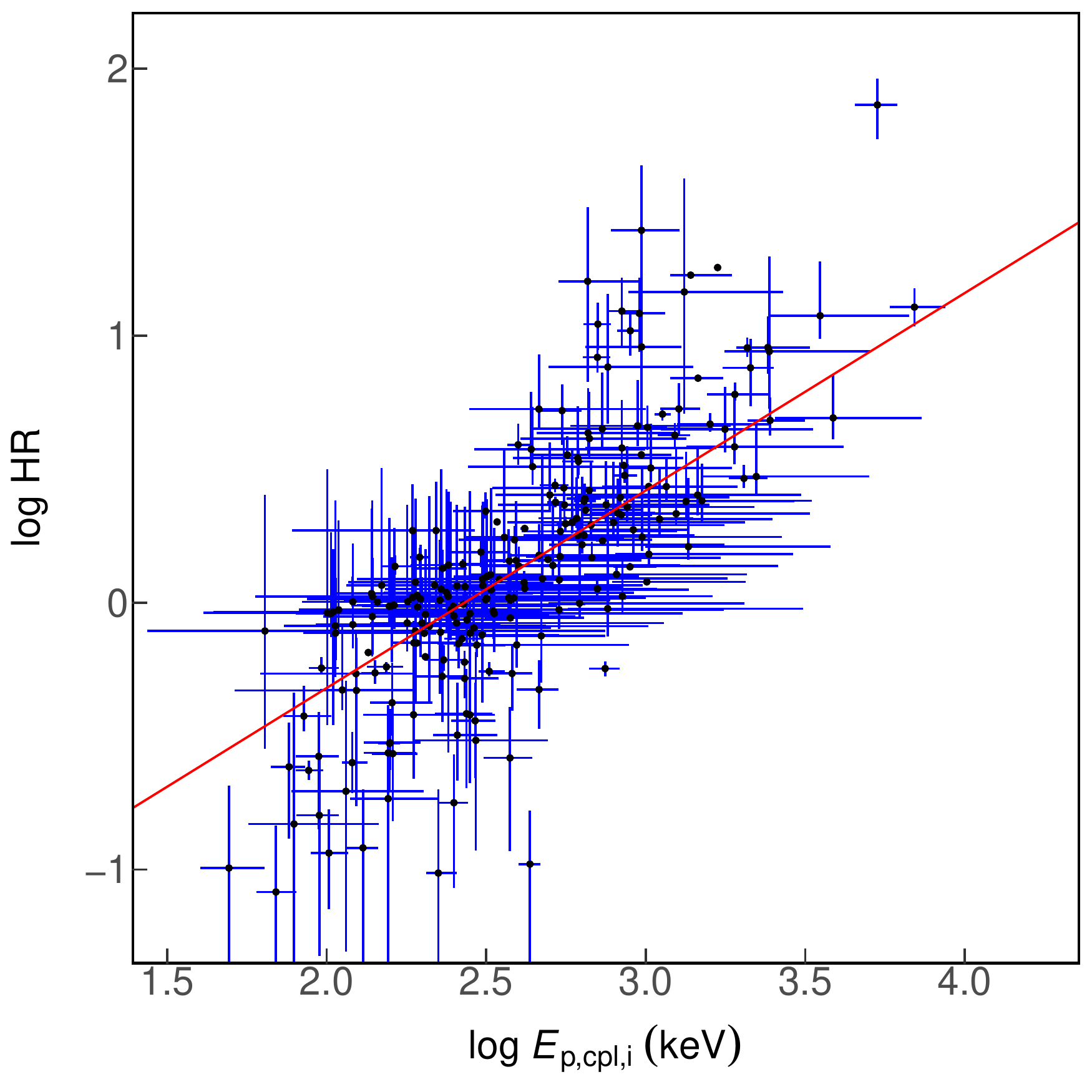}

\figsetgrpend

\figsetgrpstart
\figsetgrpnum{2.494}

\figsetplot{./figset/scatter/494.pdf}

\figsetgrpend

\figsetgrpstart
\figsetgrpnum{2.495}

\figsetplot{./figset/scatter/495.pdf}

\figsetgrpend

\figsetgrpstart
\figsetgrpnum{2.496}

\figsetplot{./figset/scatter/496.pdf}

\figsetgrpend

\figsetgrpstart
\figsetgrpnum{2.497}

\figsetplot{./figset/scatter/497.pdf}

\figsetgrpend

\figsetgrpstart
\figsetgrpnum{2.498}

\figsetplot{./figset/scatter/498.pdf}

\figsetgrpend

\figsetgrpstart
\figsetgrpnum{2.499}

\figsetplot{./figset/scatter/499.pdf}

\figsetgrpend

\figsetgrpstart
\figsetgrpnum{2.500}

\figsetplot{./figset/scatter/500.pdf}

\figsetgrpend

\figsetgrpstart
\figsetgrpnum{2.501}

\figsetplot{./figset/scatter/501.pdf}

\figsetgrpend

\figsetgrpstart
\figsetgrpnum{2.502}

\figsetplot{./figset/scatter/502.pdf}

\figsetgrpend

\figsetgrpstart
\figsetgrpnum{2.503}

\figsetplot{./figset/scatter/503.pdf}

\figsetgrpend

\figsetgrpstart
\figsetgrpnum{2.504}

\figsetplot{./figset/scatter/504.pdf}

\figsetgrpend

\figsetgrpstart
\figsetgrpnum{2.505}

\figsetplot{./figset/scatter/505.pdf}

\figsetgrpend

\figsetgrpstart
\figsetgrpnum{2.506}

\figsetplot{./figset/scatter/506.pdf}

\figsetgrpend

\figsetgrpstart
\figsetgrpnum{2.507}

\figsetplot{./figset/scatter/507.pdf}

\figsetgrpend

\figsetgrpstart
\figsetgrpnum{2.508}

\figsetplot{./figset/scatter/508.pdf}

\figsetgrpend

\figsetgrpstart
\figsetgrpnum{2.509}

\figsetplot{./figset/scatter/509.pdf}

\figsetgrpend

\figsetgrpstart
\figsetgrpnum{2.510}

\figsetplot{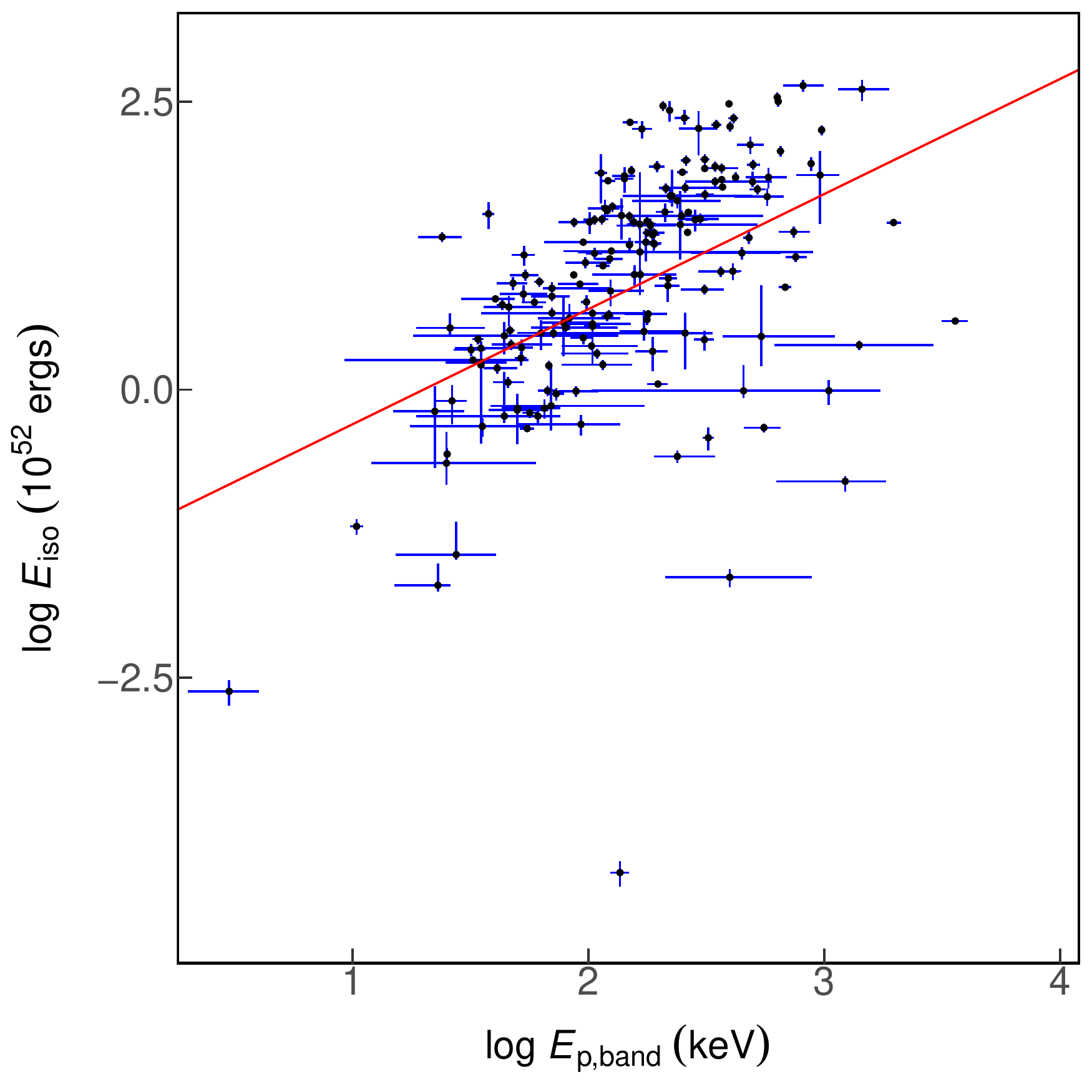}

\figsetgrpend

\figsetgrpstart
\figsetgrpnum{2.511}

\figsetplot{./figset/scatter/511.pdf}

\figsetgrpend

\figsetgrpstart
\figsetgrpnum{2.512}

\figsetplot{./figset/scatter/512.pdf}

\figsetgrpend

\figsetgrpstart
\figsetgrpnum{2.513}

\figsetplot{./figset/scatter/513.pdf}

\figsetgrpend

\figsetgrpstart
\figsetgrpnum{2.514}

\figsetplot{./figset/scatter/514.pdf}

\figsetgrpend

\figsetgrpstart
\figsetgrpnum{2.515}

\figsetplot{./figset/scatter/515.pdf}

\figsetgrpend

\figsetgrpstart
\figsetgrpnum{2.516}

\figsetplot{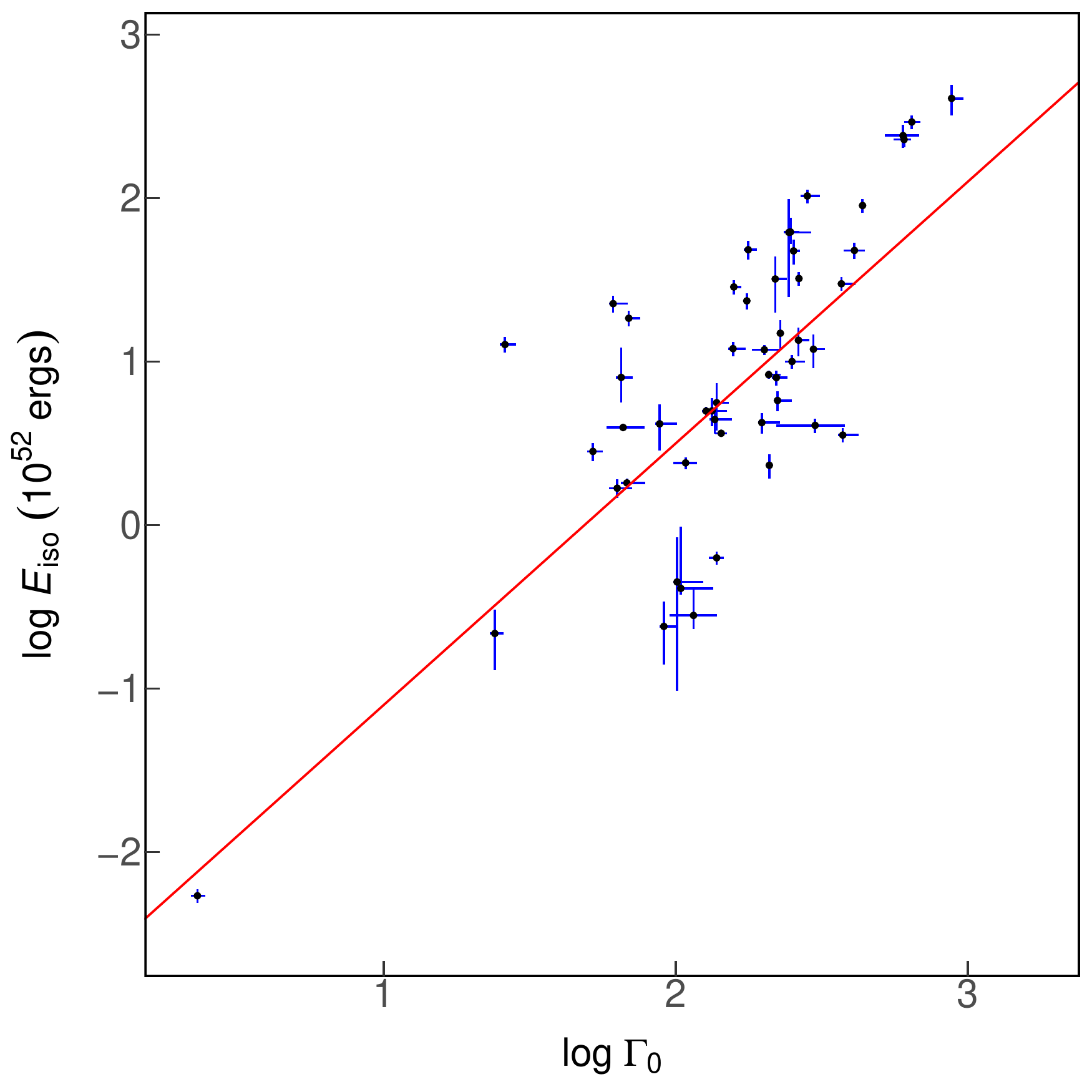}

\figsetgrpend

\figsetgrpstart
\figsetgrpnum{2.517}

\figsetplot{./figset/scatter/517.pdf}

\figsetgrpend

\figsetgrpstart
\figsetgrpnum{2.518}

\figsetplot{./figset/scatter/518.pdf}

\figsetgrpend

\figsetgrpstart
\figsetgrpnum{2.519}

\figsetplot{./figset/scatter/519.pdf}

\figsetgrpend

\figsetgrpstart
\figsetgrpnum{2.520}

\figsetplot{./figset/scatter/520.pdf}

\figsetgrpend

\figsetgrpstart
\figsetgrpnum{2.521}

\figsetplot{./figset/scatter/521.pdf}

\figsetgrpend

\figsetgrpstart
\figsetgrpnum{2.522}

\figsetplot{./figset/scatter/522.pdf}

\figsetgrpend

\figsetgrpstart
\figsetgrpnum{2.523}

\figsetplot{./figset/scatter/523.pdf}

\figsetgrpend

\figsetgrpstart
\figsetgrpnum{2.524}

\figsetplot{./figset/scatter/524.pdf}

\figsetgrpend

\figsetgrpstart
\figsetgrpnum{2.525}

\figsetplot{./figset/scatter/525.pdf}

\figsetgrpend

\figsetgrpstart
\figsetgrpnum{2.526}

\figsetplot{./figset/scatter/526.pdf}

\figsetgrpend

\figsetgrpstart
\figsetgrpnum{2.527}

\figsetplot{./figset/scatter/527.pdf}

\figsetgrpend

\figsetgrpstart
\figsetgrpnum{2.528}

\figsetplot{./figset/scatter/528.pdf}

\figsetgrpend

\figsetgrpstart
\figsetgrpnum{2.529}

\figsetplot{./figset/scatter/529.pdf}

\figsetgrpend

\figsetgrpstart
\figsetgrpnum{2.530}

\figsetplot{./figset/scatter/530.pdf}

\figsetgrpend

\figsetgrpstart
\figsetgrpnum{2.531}

\figsetplot{./figset/scatter/531.pdf}

\figsetgrpend

\figsetgrpstart
\figsetgrpnum{2.532}

\figsetplot{./figset/scatter/532.pdf}

\figsetgrpend

\figsetgrpstart
\figsetgrpnum{2.533}

\figsetplot{./figset/scatter/533.pdf}

\figsetgrpend

\figsetgrpstart
\figsetgrpnum{2.534}

\figsetplot{./figset/scatter/534.pdf}

\figsetgrpend

\figsetgrpstart
\figsetgrpnum{2.535}

\figsetplot{./figset/scatter/535.pdf}

\figsetgrpend

\figsetgrpstart
\figsetgrpnum{2.536}

\figsetplot{./figset/scatter/536.pdf}

\figsetgrpend

\figsetgrpstart
\figsetgrpnum{2.537}

\figsetplot{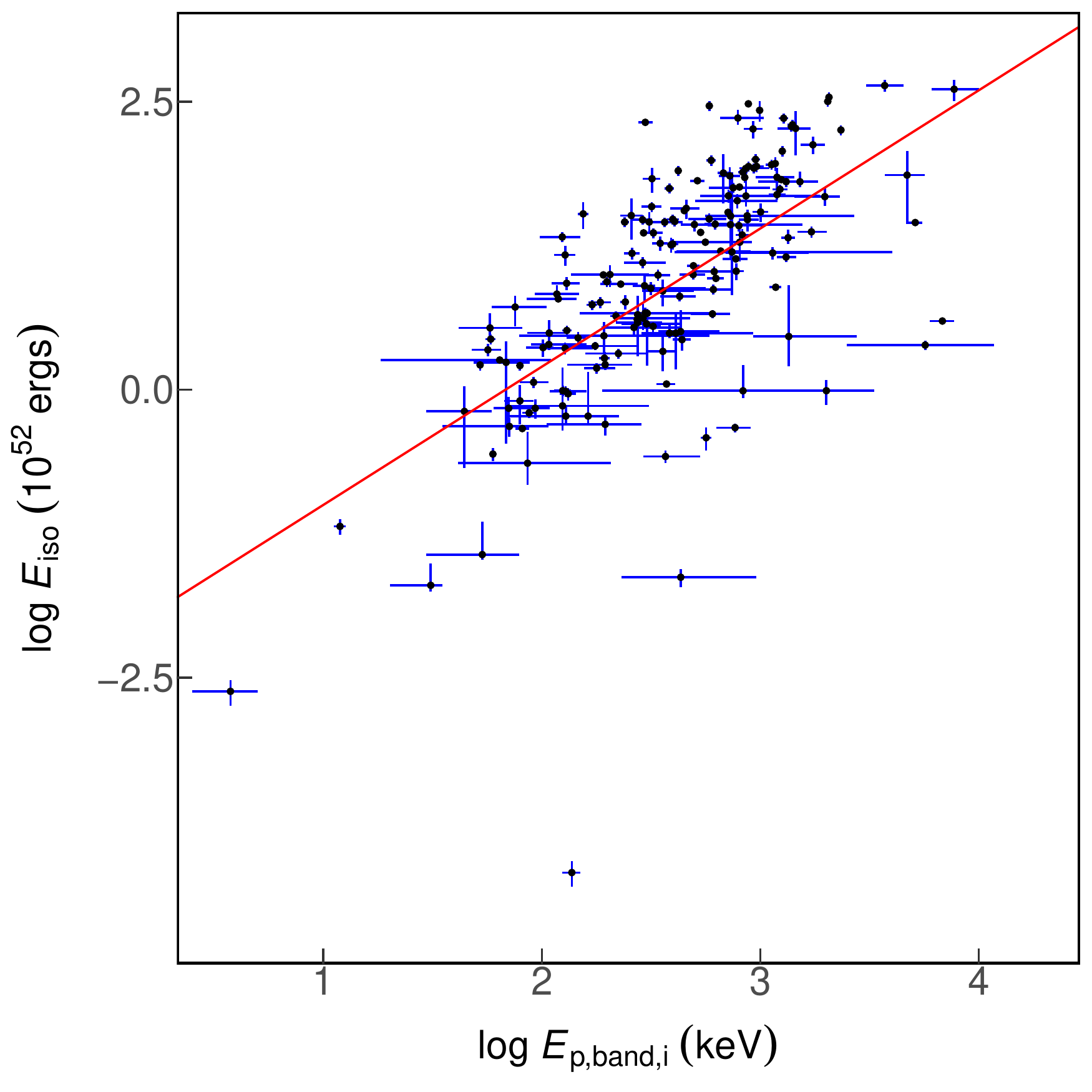}

\figsetgrpend

\figsetgrpstart
\figsetgrpnum{2.538}

\figsetplot{./figset/scatter/538.pdf}

\figsetgrpend

\figsetgrpstart
\figsetgrpnum{2.539}

\figsetplot{./figset/scatter/539.pdf}

\figsetgrpend

\figsetgrpstart
\figsetgrpnum{2.540}

\figsetplot{./figset/scatter/540.pdf}

\figsetgrpend

\figsetgrpstart
\figsetgrpnum{2.541}

\figsetplot{./figset/scatter/541.pdf}

\figsetgrpend

\figsetgrpstart
\figsetgrpnum{2.542}

\figsetplot{./figset/scatter/542.pdf}

\figsetgrpend

\figsetgrpstart
\figsetgrpnum{2.543}

\figsetplot{./figset/scatter/543.pdf}

\figsetgrpend

\figsetgrpstart
\figsetgrpnum{2.544}

\figsetplot{./figset/scatter/544.pdf}

\figsetgrpend

\figsetgrpstart
\figsetgrpnum{2.545}

\figsetplot{./figset/scatter/545.pdf}

\figsetgrpend

\figsetgrpstart
\figsetgrpnum{2.546}

\figsetplot{./figset/scatter/546.pdf}

\figsetgrpend

\figsetgrpstart
\figsetgrpnum{2.547}

\figsetplot{./figset/scatter/547.pdf}

\figsetgrpend

\figsetgrpstart
\figsetgrpnum{2.548}

\figsetplot{./figset/scatter/548.pdf}

\figsetgrpend

\figsetgrpstart
\figsetgrpnum{2.549}

\figsetplot{./figset/scatter/549.pdf}

\figsetgrpend

\figsetgrpstart
\figsetgrpnum{2.550}

\figsetplot{./figset/scatter/550.pdf}

\figsetgrpend

\figsetgrpstart
\figsetgrpnum{2.551}

\figsetplot{./figset/scatter/551.pdf}

\figsetgrpend

\figsetgrpstart
\figsetgrpnum{2.552}

\figsetplot{./figset/scatter/552.pdf}

\figsetgrpend

\figsetgrpstart
\figsetgrpnum{2.553}

\figsetplot{./figset/scatter/553.pdf}

\figsetgrpend

\figsetgrpstart
\figsetgrpnum{2.554}

\figsetplot{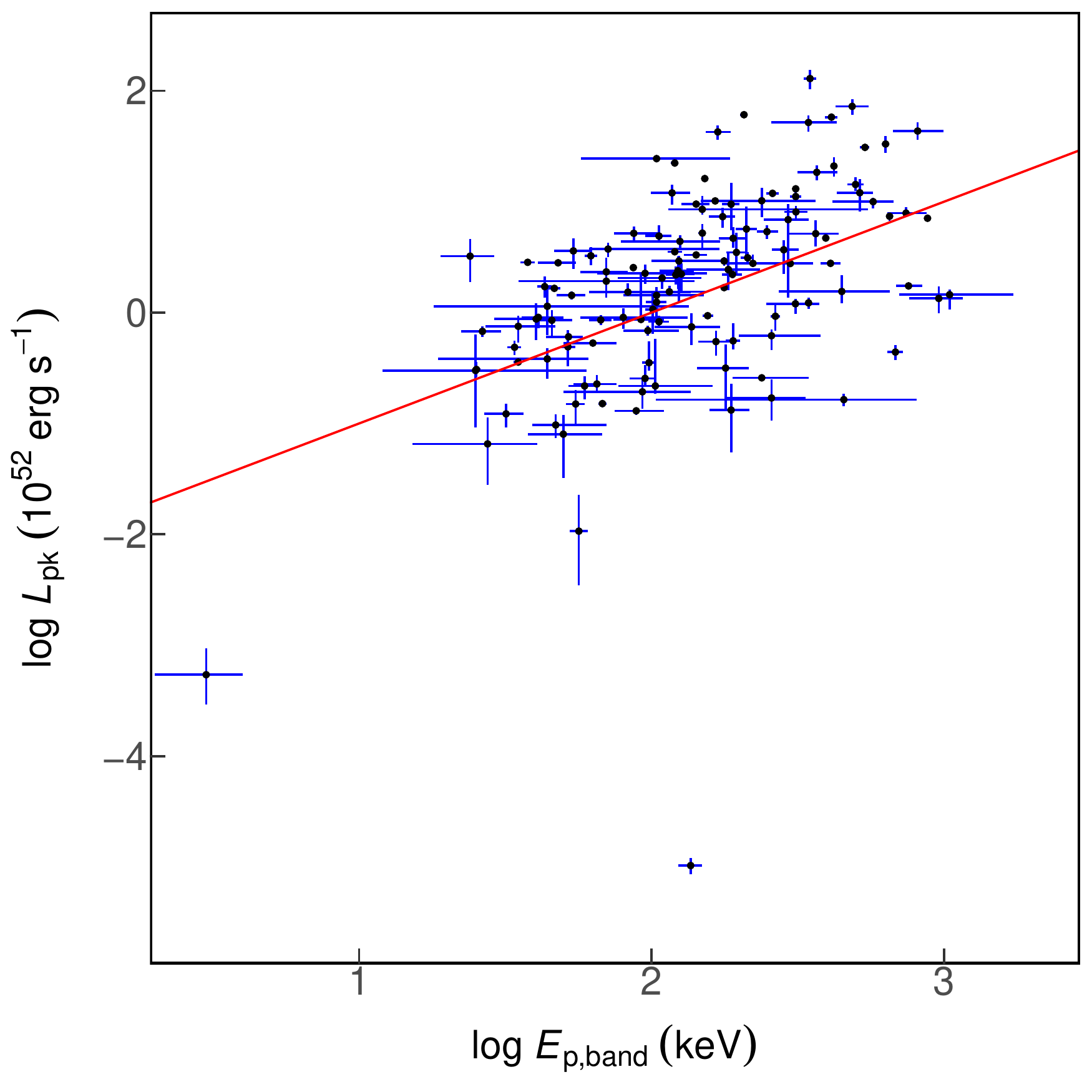}

\figsetgrpend

\figsetgrpstart
\figsetgrpnum{2.555}

\figsetplot{./figset/scatter/555.pdf}

\figsetgrpend

\figsetgrpstart
\figsetgrpnum{2.556}

\figsetplot{./figset/scatter/556.pdf}

\figsetgrpend

\figsetgrpstart
\figsetgrpnum{2.557}

\figsetplot{./figset/scatter/557.pdf}

\figsetgrpend

\figsetgrpstart
\figsetgrpnum{2.558}

\figsetplot{./figset/scatter/558.pdf}

\figsetgrpend

\figsetgrpstart
\figsetgrpnum{2.559}

\figsetplot{./figset/scatter/559.pdf}

\figsetgrpend

\figsetgrpstart
\figsetgrpnum{2.560}

\figsetplot{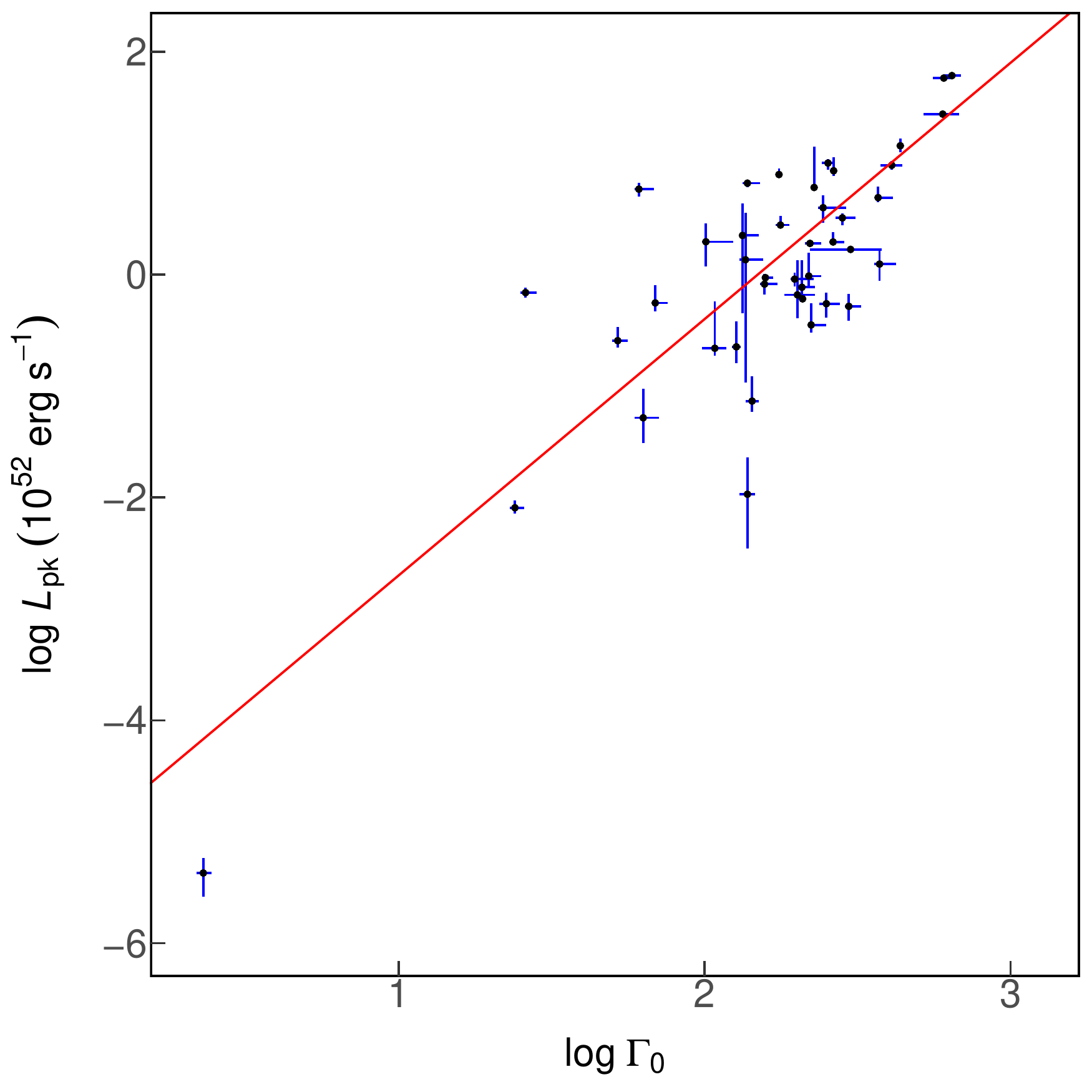}

\figsetgrpend

\figsetgrpstart
\figsetgrpnum{2.561}

\figsetplot{./figset/scatter/561.pdf}

\figsetgrpend

\figsetgrpstart
\figsetgrpnum{2.562}

\figsetplot{./figset/scatter/562.pdf}

\figsetgrpend

\figsetgrpstart
\figsetgrpnum{2.563}

\figsetplot{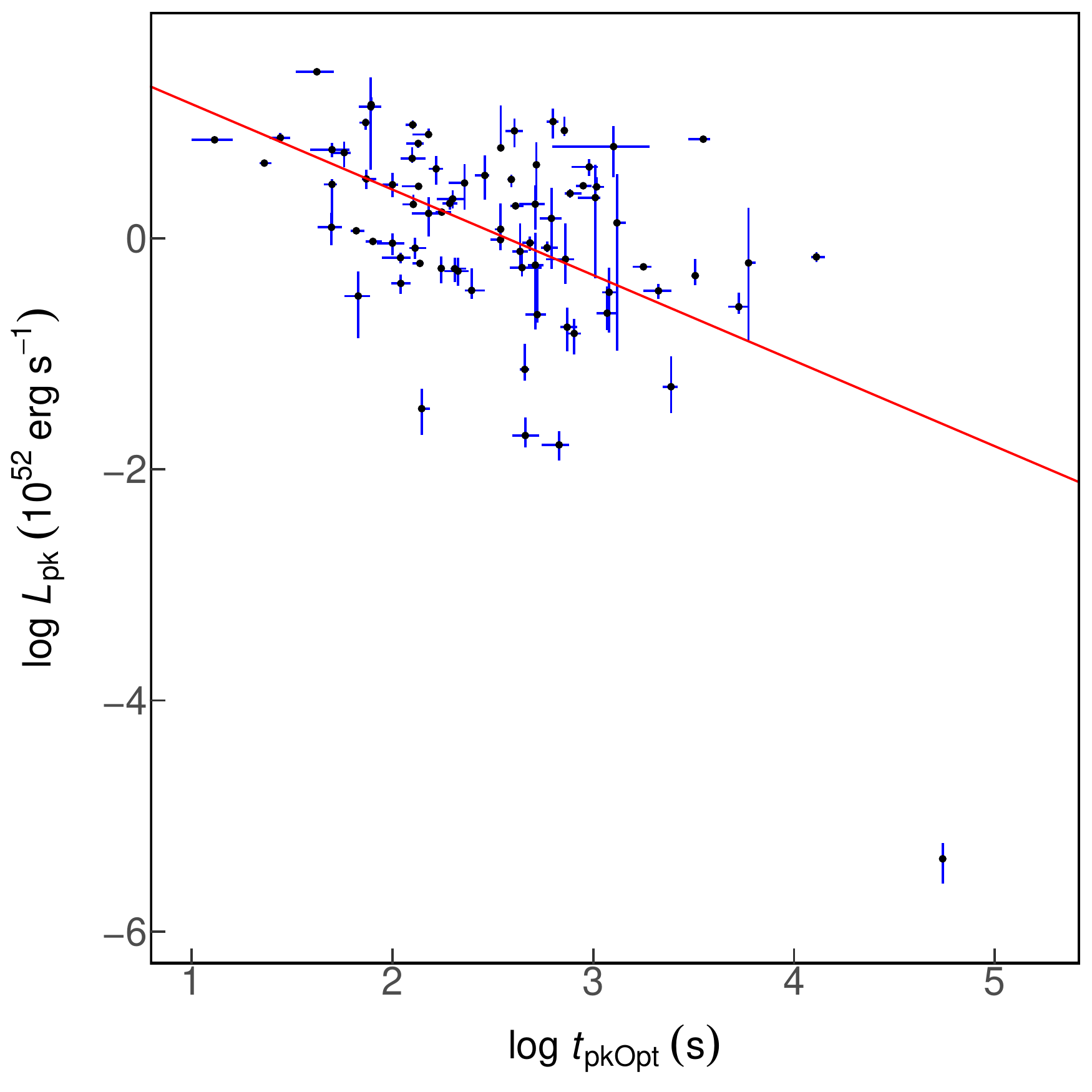}

\figsetgrpend

\figsetgrpstart
\figsetgrpnum{2.564}

\figsetplot{./figset/scatter/564.pdf}

\figsetgrpend

\figsetgrpstart
\figsetgrpnum{2.565}

\figsetplot{./figset/scatter/565.pdf}

\figsetgrpend

\figsetgrpstart
\figsetgrpnum{2.566}

\figsetplot{./figset/scatter/566.pdf}

\figsetgrpend

\figsetgrpstart
\figsetgrpnum{2.567}

\figsetplot{./figset/scatter/567.pdf}

\figsetgrpend

\figsetgrpstart
\figsetgrpnum{2.568}

\figsetplot{./figset/scatter/568.pdf}

\figsetgrpend

\figsetgrpstart
\figsetgrpnum{2.569}

\figsetplot{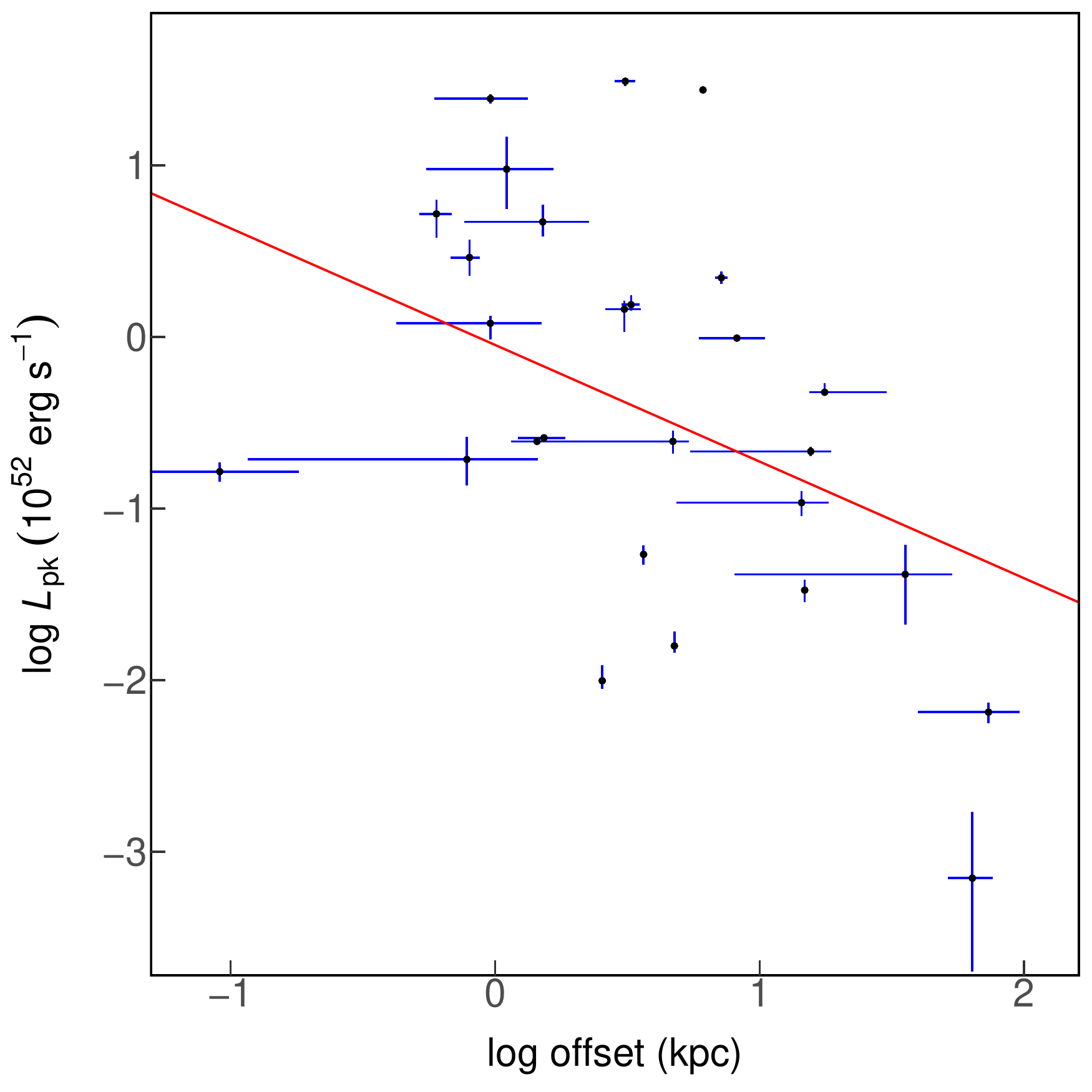}

\figsetgrpend

\figsetgrpstart
\figsetgrpnum{2.570}

\figsetplot{./figset/scatter/570.pdf}

\figsetgrpend

\figsetgrpstart
\figsetgrpnum{2.571}

\figsetplot{./figset/scatter/571.pdf}

\figsetgrpend

\figsetgrpstart
\figsetgrpnum{2.572}

\figsetplot{./figset/scatter/572.pdf}

\figsetgrpend

\figsetgrpstart
\figsetgrpnum{2.573}

\figsetplot{./figset/scatter/573.pdf}

\figsetgrpend

\figsetgrpstart
\figsetgrpnum{2.574}

\figsetplot{./figset/scatter/574.pdf}

\figsetgrpend

\figsetgrpstart
\figsetgrpnum{2.575}

\figsetplot{./figset/scatter/575.pdf}

\figsetgrpend

\figsetgrpstart
\figsetgrpnum{2.576}

\figsetplot{./figset/scatter/576.pdf}

\figsetgrpend

\figsetgrpstart
\figsetgrpnum{2.577}

\figsetplot{./figset/scatter/577.pdf}

\figsetgrpend

\figsetgrpstart
\figsetgrpnum{2.578}

\figsetplot{./figset/scatter/578.pdf}

\figsetgrpend

\figsetgrpstart
\figsetgrpnum{2.579}

\figsetplot{./figset/scatter/579.pdf}

\figsetgrpend

\figsetgrpstart
\figsetgrpnum{2.580}

\figsetplot{./figset/scatter/580.pdf}

\figsetgrpend

\figsetgrpstart
\figsetgrpnum{2.581}

\figsetplot{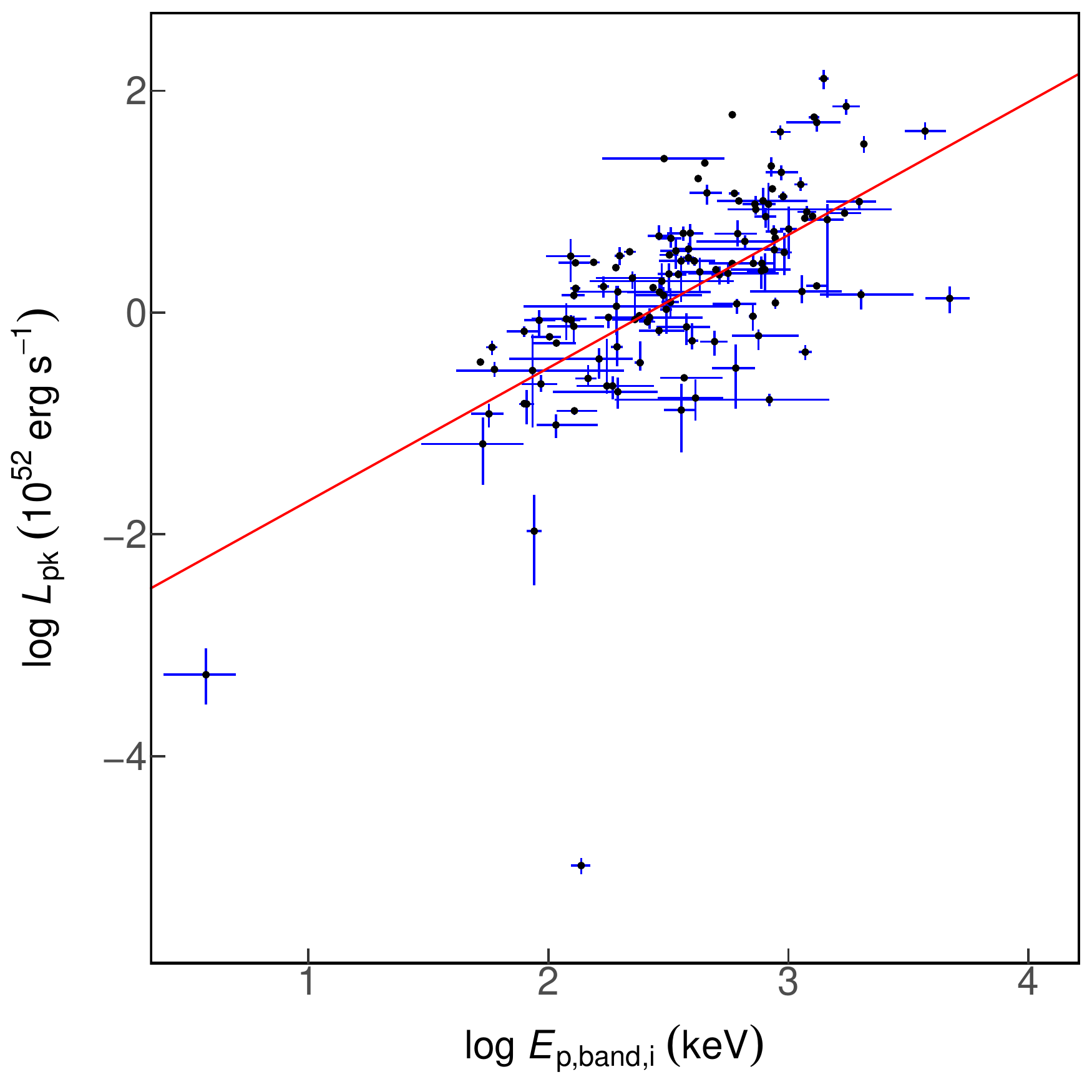}

\figsetgrpend

\figsetgrpstart
\figsetgrpnum{2.582}

\figsetplot{./figset/scatter/582.pdf}

\figsetgrpend

\figsetgrpstart
\figsetgrpnum{2.583}

\figsetplot{./figset/scatter/583.pdf}

\figsetgrpend

\figsetgrpstart
\figsetgrpnum{2.584}

\figsetplot{./figset/scatter/584.pdf}

\figsetgrpend

\figsetgrpstart
\figsetgrpnum{2.585}

\figsetplot{./figset/scatter/585.pdf}

\figsetgrpend

\figsetgrpstart
\figsetgrpnum{2.586}

\figsetplot{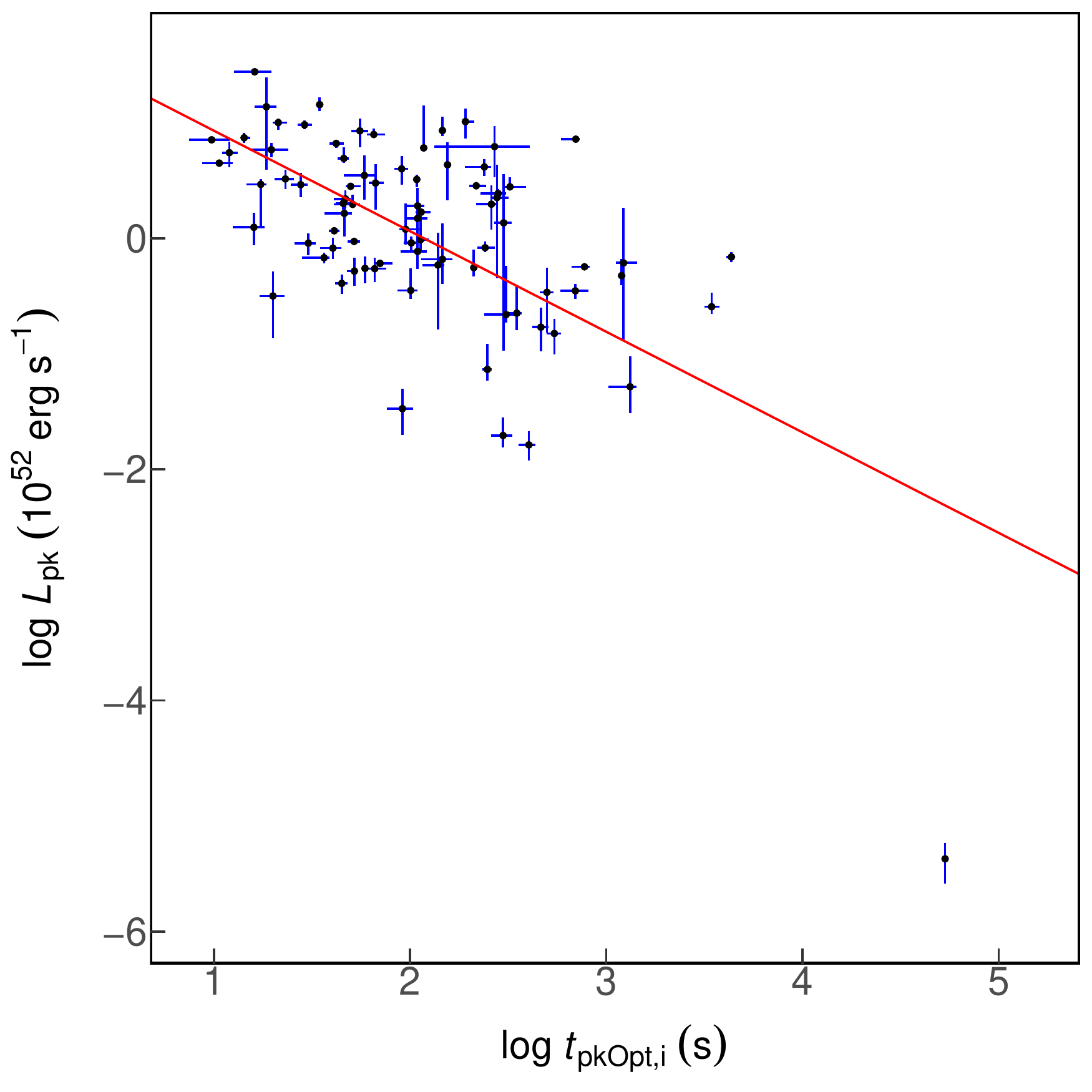}

\figsetgrpend

\figsetgrpstart
\figsetgrpnum{2.587}

\figsetplot{./figset/scatter/587.pdf}

\figsetgrpend

\figsetgrpstart
\figsetgrpnum{2.588}

\figsetplot{./figset/scatter/588.pdf}

\figsetgrpend

\figsetgrpstart
\figsetgrpnum{2.589}

\figsetplot{./figset/scatter/589.pdf}

\figsetgrpend

\figsetgrpstart
\figsetgrpnum{2.590}

\figsetplot{./figset/scatter/590.pdf}

\figsetgrpend

\figsetgrpstart
\figsetgrpnum{2.591}

\figsetplot{./figset/scatter/591.pdf}

\figsetgrpend

\figsetgrpstart
\figsetgrpnum{2.592}

\figsetplot{./figset/scatter/592.pdf}

\figsetgrpend

\figsetgrpstart
\figsetgrpnum{2.593}

\figsetplot{./figset/scatter/593.pdf}

\figsetgrpend

\figsetgrpstart
\figsetgrpnum{2.594}

\figsetplot{./figset/scatter/594.pdf}

\figsetgrpend

\figsetgrpstart
\figsetgrpnum{2.595}

\figsetplot{./figset/scatter/595.pdf}

\figsetgrpend

\figsetgrpstart
\figsetgrpnum{2.596}

\figsetplot{./figset/scatter/596.pdf}

\figsetgrpend

\figsetgrpstart
\figsetgrpnum{2.597}

\figsetplot{./figset/scatter/597.pdf}

\figsetgrpend

\figsetgrpstart
\figsetgrpnum{2.598}

\figsetplot{./figset/scatter/598.pdf}

\figsetgrpend

\figsetgrpstart
\figsetgrpnum{2.599}

\figsetplot{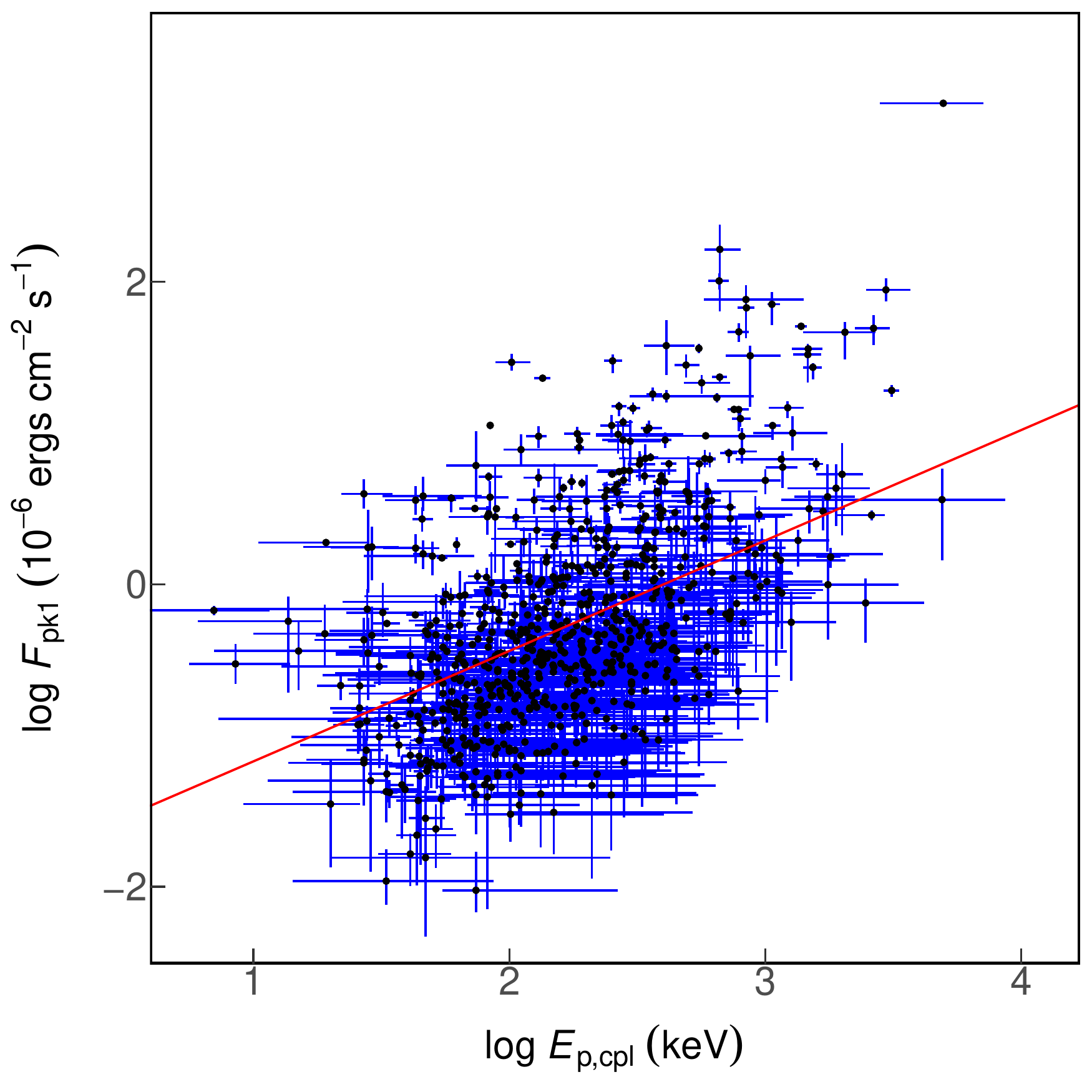}

\figsetgrpend

\figsetgrpstart
\figsetgrpnum{2.600}

\figsetplot{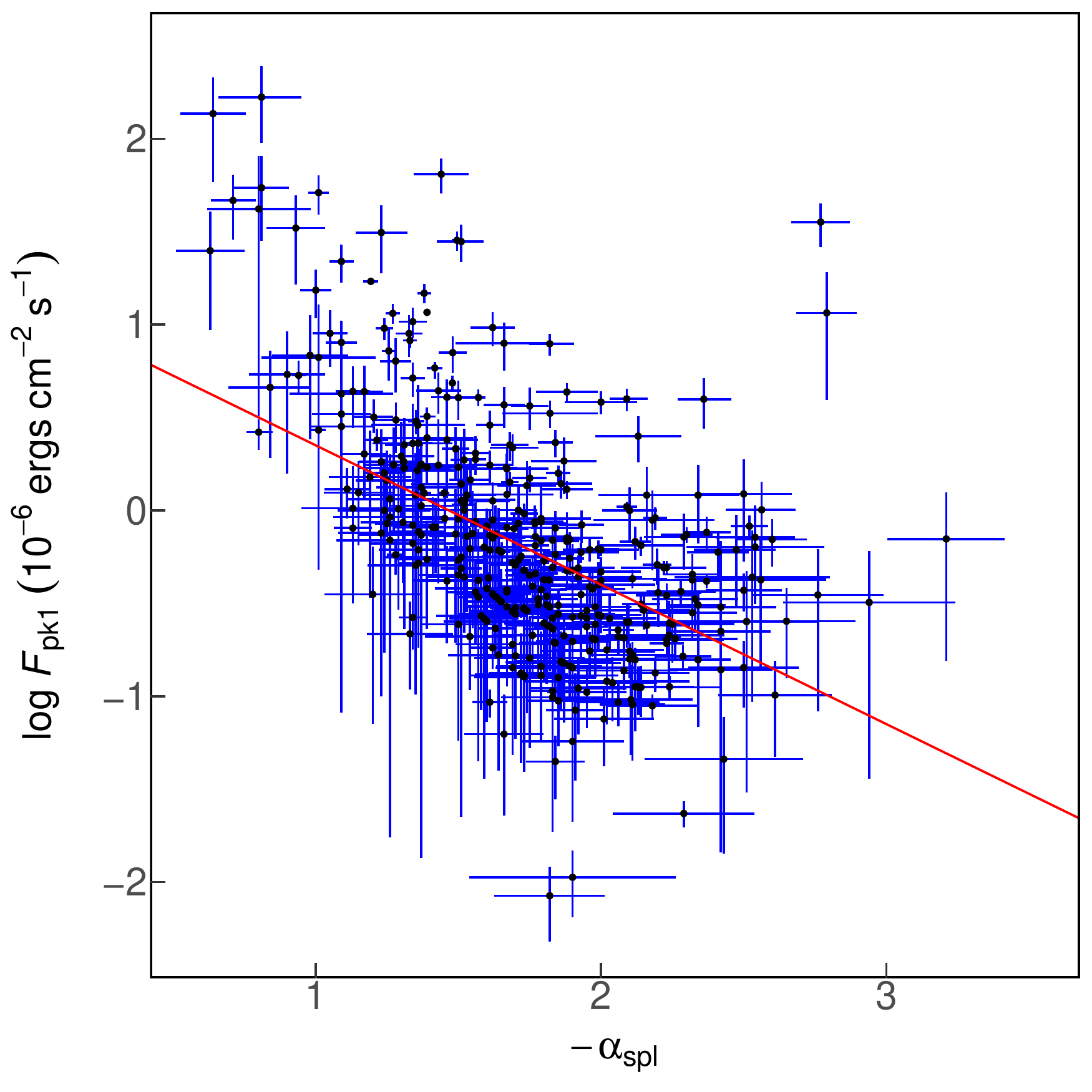}

\figsetgrpend

\figsetgrpstart
\figsetgrpnum{2.601}

\figsetplot{./figset/scatter/601.pdf}

\figsetgrpend

\figsetgrpstart
\figsetgrpnum{2.602}

\figsetplot{./figset/scatter/602.pdf}

\figsetgrpend

\figsetgrpstart
\figsetgrpnum{2.603}

\figsetplot{./figset/scatter/603.pdf}

\figsetgrpend

\figsetgrpstart
\figsetgrpnum{2.604}

\figsetplot{./figset/scatter/604.pdf}

\figsetgrpend

\figsetgrpstart
\figsetgrpnum{2.605}

\figsetplot{./figset/scatter/605.pdf}

\figsetgrpend

\figsetgrpstart
\figsetgrpnum{2.606}

\figsetplot{./figset/scatter/606.pdf}

\figsetgrpend

\figsetgrpstart
\figsetgrpnum{2.607}

\figsetplot{./figset/scatter/607.pdf}

\figsetgrpend

\figsetgrpstart
\figsetgrpnum{2.608}

\figsetplot{./figset/scatter/608.pdf}

\figsetgrpend

\figsetgrpstart
\figsetgrpnum{2.609}

\figsetplot{./figset/scatter/609.pdf}

\figsetgrpend

\figsetgrpstart
\figsetgrpnum{2.610}

\figsetplot{./figset/scatter/610.pdf}

\figsetgrpend

\figsetgrpstart
\figsetgrpnum{2.611}

\figsetplot{./figset/scatter/611.pdf}

\figsetgrpend

\figsetgrpstart
\figsetgrpnum{2.612}

\figsetplot{./figset/scatter/612.pdf}

\figsetgrpend

\figsetgrpstart
\figsetgrpnum{2.613}

\figsetplot{./figset/scatter/613.pdf}

\figsetgrpend

\figsetgrpstart
\figsetgrpnum{2.614}

\figsetplot{./figset/scatter/614.pdf}

\figsetgrpend

\figsetgrpstart
\figsetgrpnum{2.615}

\figsetplot{./figset/scatter/615.pdf}

\figsetgrpend

\figsetgrpstart
\figsetgrpnum{2.616}

\figsetplot{./figset/scatter/616.pdf}

\figsetgrpend

\figsetgrpstart
\figsetgrpnum{2.617}

\figsetplot{./figset/scatter/617.pdf}

\figsetgrpend

\figsetgrpstart
\figsetgrpnum{2.618}

\figsetplot{./figset/scatter/618.pdf}

\figsetgrpend

\figsetgrpstart
\figsetgrpnum{2.619}

\figsetplot{./figset/scatter/619.pdf}

\figsetgrpend

\figsetgrpstart
\figsetgrpnum{2.620}

\figsetplot{./figset/scatter/620.pdf}

\figsetgrpend

\figsetgrpstart
\figsetgrpnum{2.621}

\figsetplot{./figset/scatter/621.pdf}

\figsetgrpend

\figsetgrpstart
\figsetgrpnum{2.622}

\figsetplot{./figset/scatter/622.pdf}

\figsetgrpend

\figsetgrpstart
\figsetgrpnum{2.623}

\figsetplot{./figset/scatter/623.pdf}

\figsetgrpend

\figsetgrpstart
\figsetgrpnum{2.624}

\figsetplot{./figset/scatter/624.pdf}

\figsetgrpend

\figsetgrpstart
\figsetgrpnum{2.625}

\figsetplot{./figset/scatter/625.pdf}

\figsetgrpend

\figsetgrpstart
\figsetgrpnum{2.626}

\figsetplot{./figset/scatter/626.pdf}

\figsetgrpend

\figsetgrpstart
\figsetgrpnum{2.627}

\figsetplot{./figset/scatter/627.pdf}

\figsetgrpend

\figsetgrpstart
\figsetgrpnum{2.628}

\figsetplot{./figset/scatter/628.pdf}

\figsetgrpend

\figsetgrpstart
\figsetgrpnum{2.629}

\figsetplot{./figset/scatter/629.pdf}

\figsetgrpend

\figsetgrpstart
\figsetgrpnum{2.630}

\figsetplot{./figset/scatter/630.pdf}

\figsetgrpend

\figsetgrpstart
\figsetgrpnum{2.631}

\figsetplot{./figset/scatter/631.pdf}

\figsetgrpend

\figsetgrpstart
\figsetgrpnum{2.632}

\figsetplot{./figset/scatter/632.pdf}

\figsetgrpend

\figsetgrpstart
\figsetgrpnum{2.633}

\figsetplot{./figset/scatter/633.pdf}

\figsetgrpend

\figsetgrpstart
\figsetgrpnum{2.634}

\figsetplot{./figset/scatter/634.pdf}

\figsetgrpend

\figsetgrpstart
\figsetgrpnum{2.635}

\figsetplot{./figset/scatter/635.pdf}

\figsetgrpend

\figsetgrpstart
\figsetgrpnum{2.636}

\figsetplot{./figset/scatter/636.pdf}

\figsetgrpend

\figsetgrpstart
\figsetgrpnum{2.637}

\figsetplot{./figset/scatter/637.pdf}

\figsetgrpend

\figsetgrpstart
\figsetgrpnum{2.638}

\figsetplot{./figset/scatter/638.pdf}

\figsetgrpend

\figsetgrpstart
\figsetgrpnum{2.639}

\figsetplot{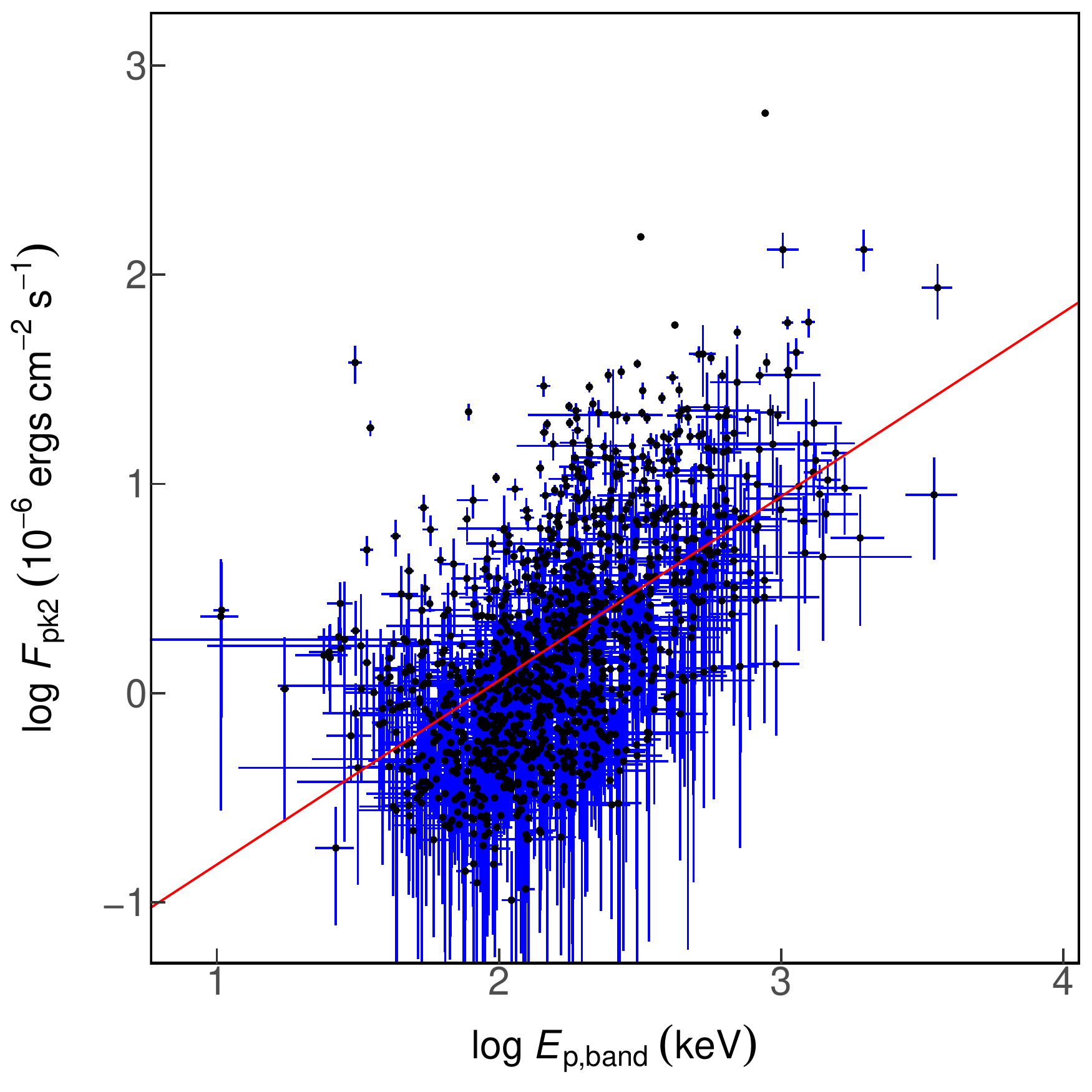}

\figsetgrpend

\figsetgrpstart
\figsetgrpnum{2.640}

\figsetplot{./figset/scatter/640.pdf}

\figsetgrpend

\figsetgrpstart
\figsetgrpnum{2.641}

\figsetplot{./figset/scatter/641.pdf}

\figsetgrpend

\figsetgrpstart
\figsetgrpnum{2.642}

\figsetplot{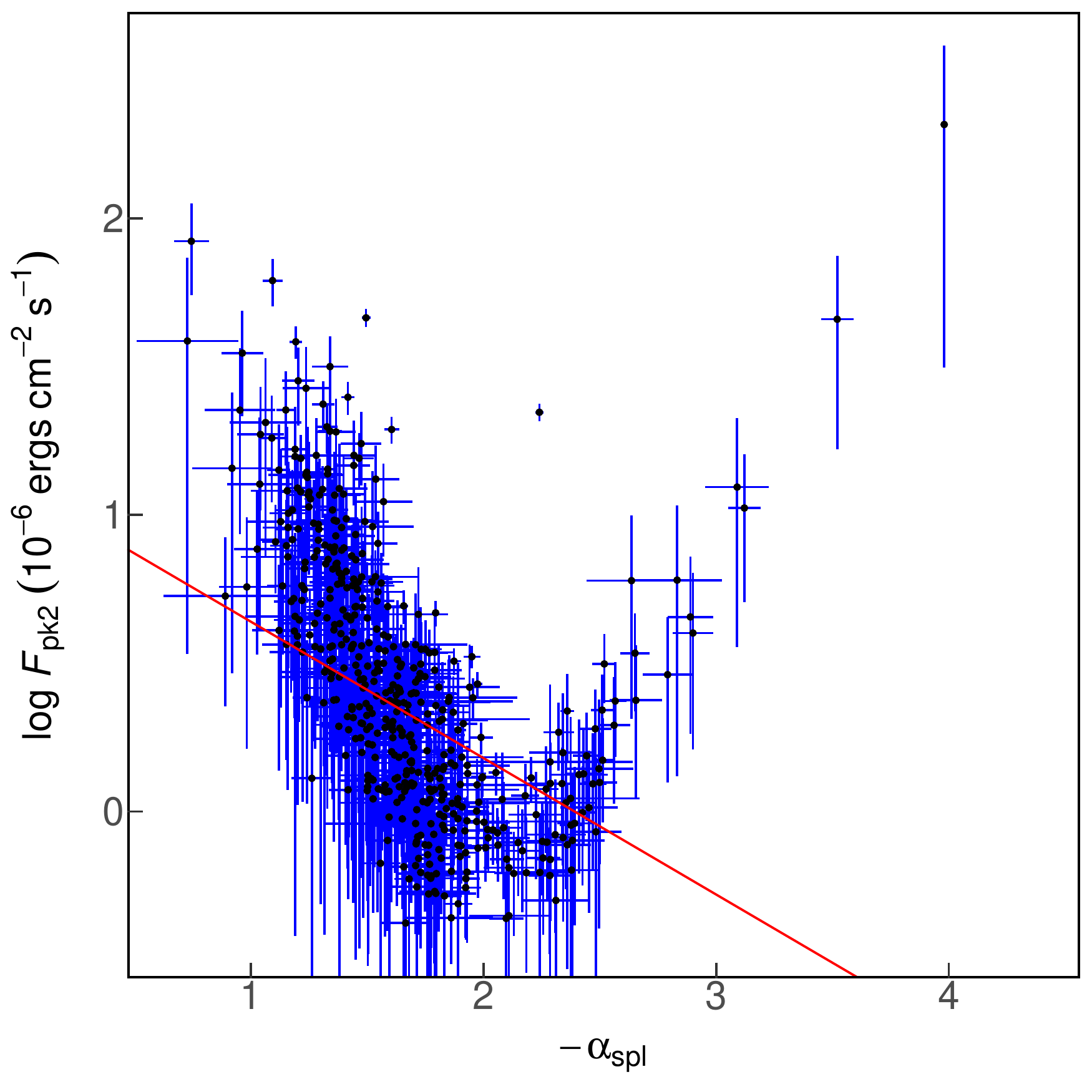}

\figsetgrpend

\figsetgrpstart
\figsetgrpnum{2.643}

\figsetplot{./figset/scatter/643.pdf}

\figsetgrpend

\figsetgrpstart
\figsetgrpnum{2.644}

\figsetplot{./figset/scatter/644.pdf}

\figsetgrpend

\figsetgrpstart
\figsetgrpnum{2.645}

\figsetplot{./figset/scatter/645.pdf}

\figsetgrpend

\figsetgrpstart
\figsetgrpnum{2.646}

\figsetplot{./figset/scatter/646.pdf}

\figsetgrpend

\figsetgrpstart
\figsetgrpnum{2.647}

\figsetplot{./figset/scatter/647.pdf}

\figsetgrpend

\figsetgrpstart
\figsetgrpnum{2.648}

\figsetplot{./figset/scatter/648.pdf}

\figsetgrpend

\figsetgrpstart
\figsetgrpnum{2.649}

\figsetplot{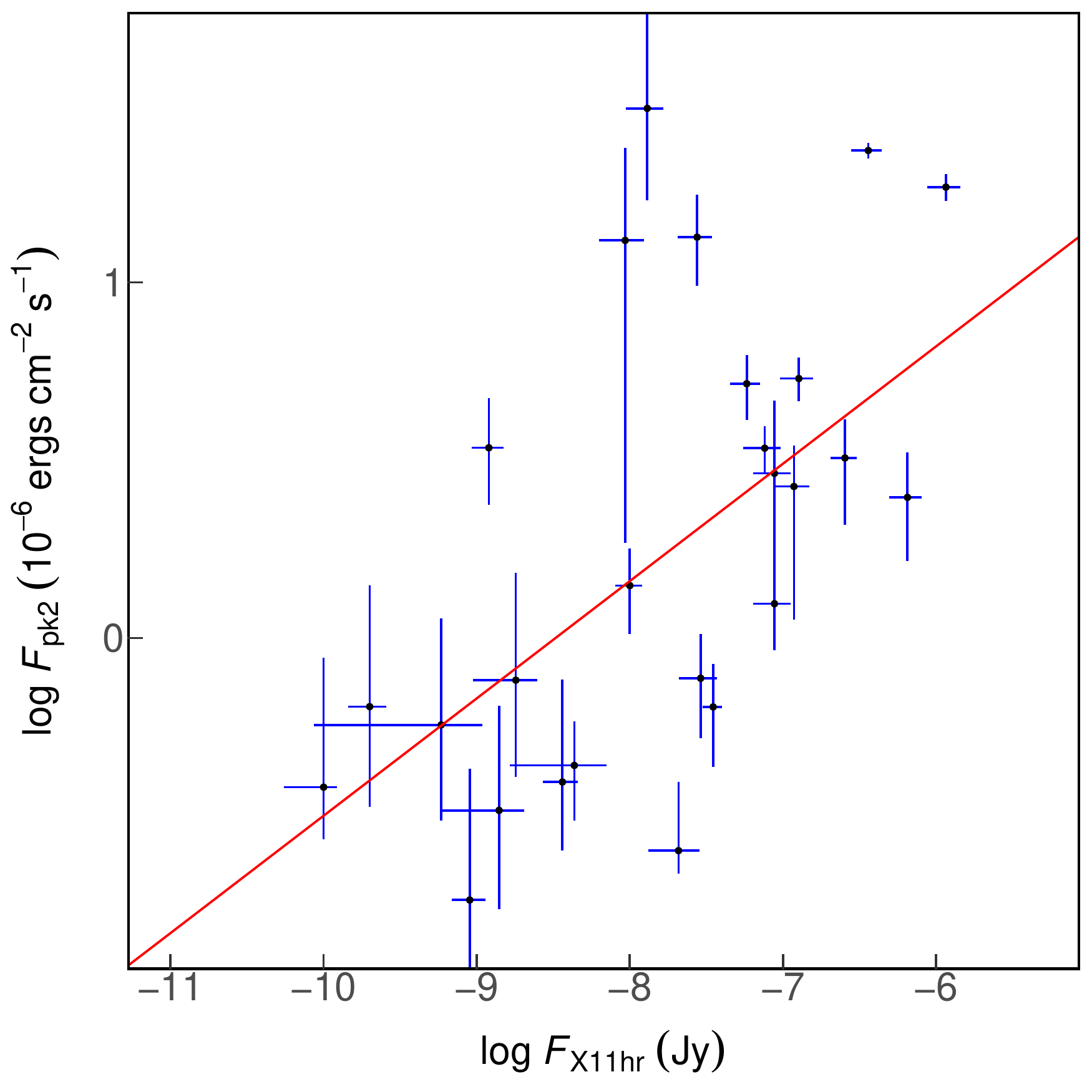}

\figsetgrpend

\figsetgrpstart
\figsetgrpnum{2.650}

\figsetplot{./figset/scatter/650.pdf}

\figsetgrpend

\figsetgrpstart
\figsetgrpnum{2.651}

\figsetplot{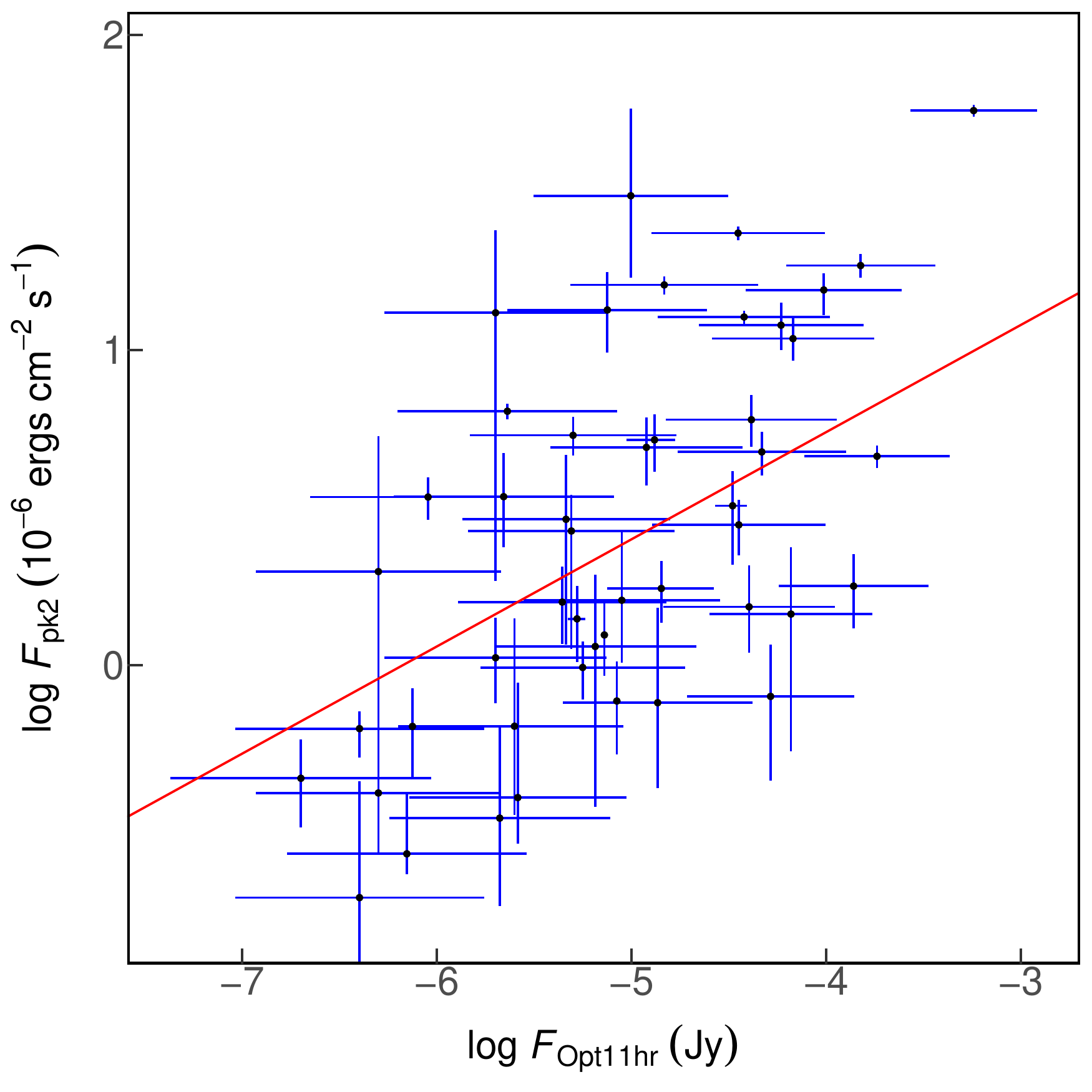}

\figsetgrpend

\figsetgrpstart
\figsetgrpnum{2.652}

\figsetplot{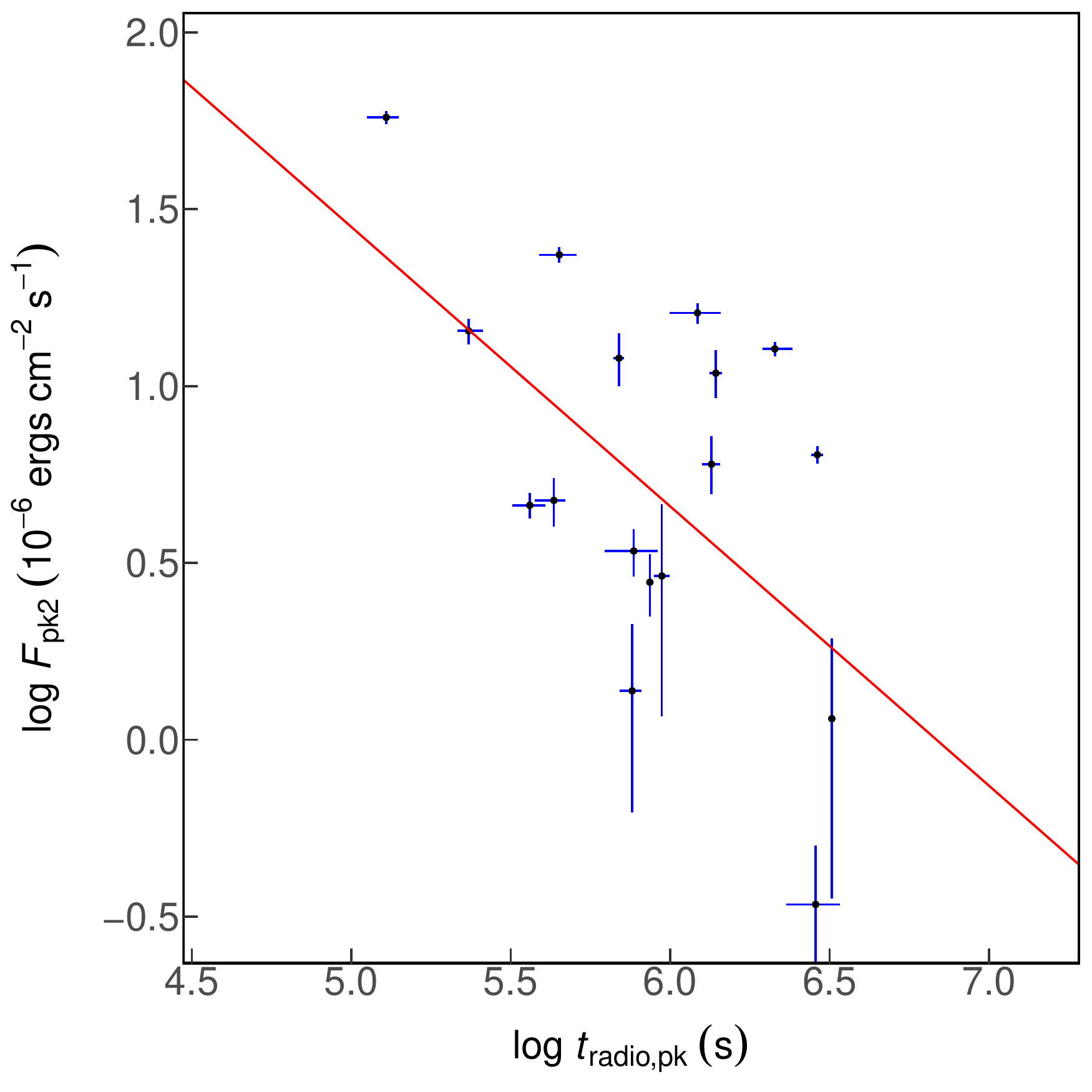}

\figsetgrpend

\figsetgrpstart
\figsetgrpnum{2.653}

\figsetplot{./figset/scatter/653.pdf}

\figsetgrpend

\figsetgrpstart
\figsetgrpnum{2.654}

\figsetplot{./figset/scatter/654.pdf}

\figsetgrpend

\figsetgrpstart
\figsetgrpnum{2.655}

\figsetplot{./figset/scatter/655.pdf}

\figsetgrpend

\figsetgrpstart
\figsetgrpnum{2.656}

\figsetplot{./figset/scatter/656.pdf}

\figsetgrpend

\figsetgrpstart
\figsetgrpnum{2.657}

\figsetplot{./figset/scatter/657.pdf}

\figsetgrpend

\figsetgrpstart
\figsetgrpnum{2.658}

\figsetplot{./figset/scatter/658.pdf}

\figsetgrpend

\figsetgrpstart
\figsetgrpnum{2.659}

\figsetplot{./figset/scatter/659.pdf}

\figsetgrpend

\figsetgrpstart
\figsetgrpnum{2.660}

\figsetplot{./figset/scatter/660.pdf}

\figsetgrpend

\figsetgrpstart
\figsetgrpnum{2.661}

\figsetplot{./figset/scatter/661.pdf}

\figsetgrpend

\figsetgrpstart
\figsetgrpnum{2.662}

\figsetplot{./figset/scatter/662.pdf}

\figsetgrpend

\figsetgrpstart
\figsetgrpnum{2.663}

\figsetplot{./figset/scatter/663.pdf}

\figsetgrpend

\figsetgrpstart
\figsetgrpnum{2.664}

\figsetplot{./figset/scatter/664.pdf}

\figsetgrpend

\figsetgrpstart
\figsetgrpnum{2.665}

\figsetplot{./figset/scatter/665.pdf}

\figsetgrpend

\figsetgrpstart
\figsetgrpnum{2.666}

\figsetplot{./figset/scatter/666.pdf}

\figsetgrpend

\figsetgrpstart
\figsetgrpnum{2.667}

\figsetplot{./figset/scatter/667.pdf}

\figsetgrpend

\figsetgrpstart
\figsetgrpnum{2.668}

\figsetplot{./figset/scatter/668.pdf}

\figsetgrpend

\figsetgrpstart
\figsetgrpnum{2.669}

\figsetplot{./figset/scatter/669.pdf}

\figsetgrpend

\figsetgrpstart
\figsetgrpnum{2.670}

\figsetplot{./figset/scatter/670.pdf}

\figsetgrpend

\figsetgrpstart
\figsetgrpnum{2.671}

\figsetplot{./figset/scatter/671.pdf}

\figsetgrpend

\figsetgrpstart
\figsetgrpnum{2.672}

\figsetplot{./figset/scatter/672.pdf}

\figsetgrpend

\figsetgrpstart
\figsetgrpnum{2.673}

\figsetplot{./figset/scatter/673.pdf}

\figsetgrpend

\figsetgrpstart
\figsetgrpnum{2.674}

\figsetplot{./figset/scatter/674.pdf}

\figsetgrpend

\figsetgrpstart
\figsetgrpnum{2.675}

\figsetplot{./figset/scatter/675.pdf}

\figsetgrpend

\figsetgrpstart
\figsetgrpnum{2.676}

\figsetplot{./figset/scatter/676.pdf}

\figsetgrpend

\figsetgrpstart
\figsetgrpnum{2.677}

\figsetplot{./figset/scatter/677.pdf}

\figsetgrpend

\figsetgrpstart
\figsetgrpnum{2.678}

\figsetplot{./figset/scatter/678.pdf}

\figsetgrpend

\figsetgrpstart
\figsetgrpnum{2.679}

\figsetplot{./figset/scatter/679.pdf}

\figsetgrpend

\figsetgrpstart
\figsetgrpnum{2.680}

\figsetplot{./figset/scatter/680.pdf}

\figsetgrpend

\figsetgrpstart
\figsetgrpnum{2.681}

\figsetplot{./figset/scatter/681.pdf}

\figsetgrpend

\figsetgrpstart
\figsetgrpnum{2.682}

\figsetplot{./figset/scatter/682.pdf}

\figsetgrpend

\figsetgrpstart
\figsetgrpnum{2.683}

\figsetplot{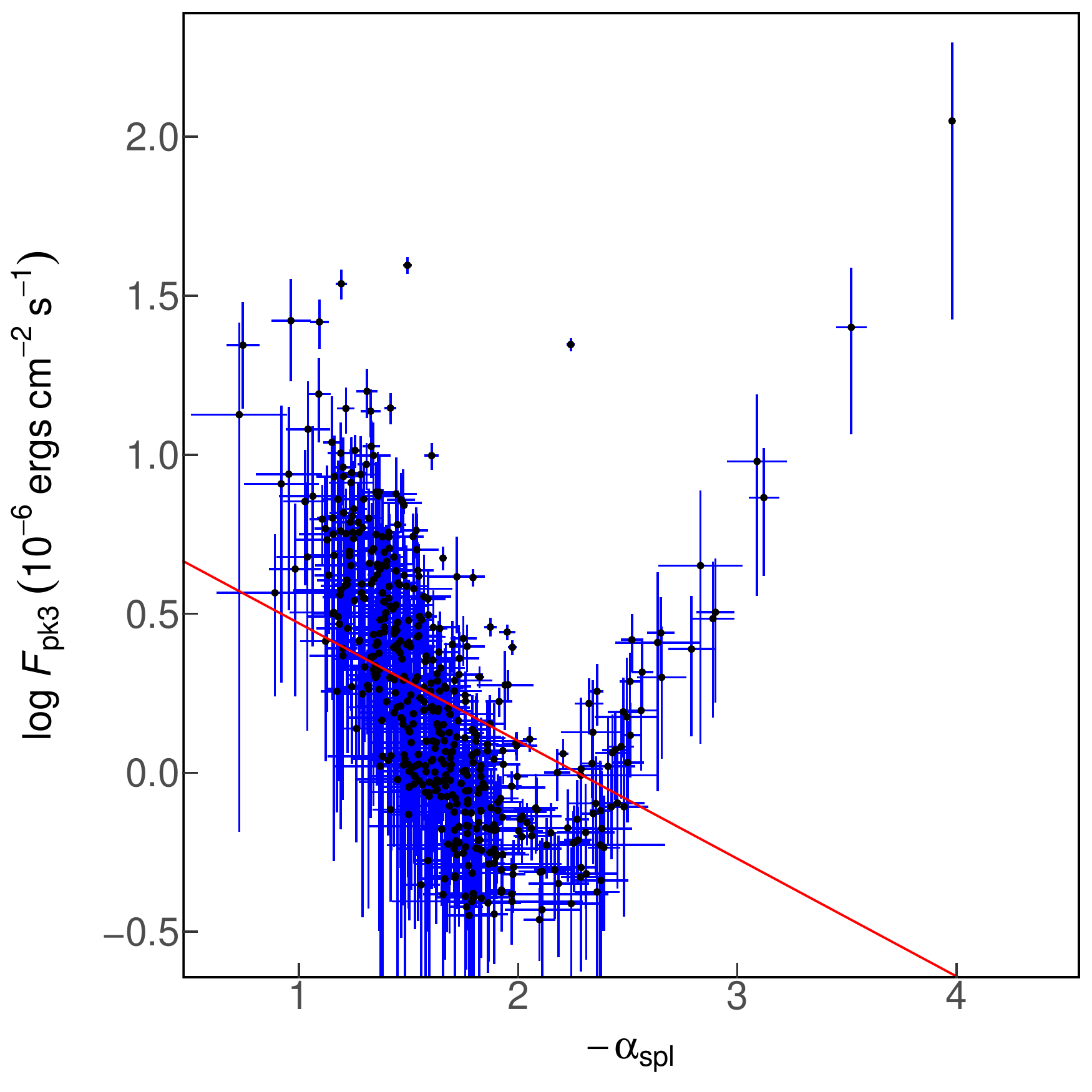}

\figsetgrpend

\figsetgrpstart
\figsetgrpnum{2.684}

\figsetplot{./figset/scatter/684.pdf}

\figsetgrpend

\figsetgrpstart
\figsetgrpnum{2.685}

\figsetplot{./figset/scatter/685.pdf}

\figsetgrpend

\figsetgrpstart
\figsetgrpnum{2.686}

\figsetplot{./figset/scatter/686.pdf}

\figsetgrpend

\figsetgrpstart
\figsetgrpnum{2.687}

\figsetplot{./figset/scatter/687.pdf}

\figsetgrpend

\figsetgrpstart
\figsetgrpnum{2.688}

\figsetplot{./figset/scatter/688.pdf}

\figsetgrpend

\figsetgrpstart
\figsetgrpnum{2.689}

\figsetplot{./figset/scatter/689.pdf}

\figsetgrpend

\figsetgrpstart
\figsetgrpnum{2.690}

\figsetplot{./figset/scatter/690.pdf}

\figsetgrpend

\figsetgrpstart
\figsetgrpnum{2.691}

\figsetplot{./figset/scatter/691.pdf}

\figsetgrpend

\figsetgrpstart
\figsetgrpnum{2.692}

\figsetplot{./figset/scatter/692.pdf}

\figsetgrpend

\figsetgrpstart
\figsetgrpnum{2.693}

\figsetplot{./figset/scatter/693.pdf}

\figsetgrpend

\figsetgrpstart
\figsetgrpnum{2.694}

\figsetplot{./figset/scatter/694.pdf}

\figsetgrpend

\figsetgrpstart
\figsetgrpnum{2.695}

\figsetplot{./figset/scatter/695.pdf}

\figsetgrpend

\figsetgrpstart
\figsetgrpnum{2.696}

\figsetplot{./figset/scatter/696.pdf}

\figsetgrpend

\figsetgrpstart
\figsetgrpnum{2.697}

\figsetplot{./figset/scatter/697.pdf}

\figsetgrpend

\figsetgrpstart
\figsetgrpnum{2.698}

\figsetplot{./figset/scatter/698.pdf}

\figsetgrpend

\figsetgrpstart
\figsetgrpnum{2.699}

\figsetplot{./figset/scatter/699.pdf}

\figsetgrpend

\figsetgrpstart
\figsetgrpnum{2.700}

\figsetplot{./figset/scatter/700.pdf}

\figsetgrpend

\figsetgrpstart
\figsetgrpnum{2.701}

\figsetplot{./figset/scatter/701.pdf}

\figsetgrpend

\figsetgrpstart
\figsetgrpnum{2.702}

\figsetplot{./figset/scatter/702.pdf}

\figsetgrpend

\figsetgrpstart
\figsetgrpnum{2.703}

\figsetplot{./figset/scatter/703.pdf}

\figsetgrpend

\figsetgrpstart
\figsetgrpnum{2.704}

\figsetplot{./figset/scatter/704.pdf}

\figsetgrpend

\figsetgrpstart
\figsetgrpnum{2.705}

\figsetplot{./figset/scatter/705.pdf}

\figsetgrpend

\figsetgrpstart
\figsetgrpnum{2.706}

\figsetplot{./figset/scatter/706.pdf}

\figsetgrpend

\figsetgrpstart
\figsetgrpnum{2.707}

\figsetplot{./figset/scatter/707.pdf}

\figsetgrpend

\figsetgrpstart
\figsetgrpnum{2.708}

\figsetplot{./figset/scatter/708.pdf}

\figsetgrpend

\figsetgrpstart
\figsetgrpnum{2.709}

\figsetplot{./figset/scatter/709.pdf}

\figsetgrpend

\figsetgrpstart
\figsetgrpnum{2.710}

\figsetplot{./figset/scatter/710.pdf}

\figsetgrpend

\figsetgrpstart
\figsetgrpnum{2.711}

\figsetplot{./figset/scatter/711.pdf}

\figsetgrpend

\figsetgrpstart
\figsetgrpnum{2.712}

\figsetplot{./figset/scatter/712.pdf}

\figsetgrpend

\figsetgrpstart
\figsetgrpnum{2.713}

\figsetplot{./figset/scatter/713.pdf}

\figsetgrpend

\figsetgrpstart
\figsetgrpnum{2.714}

\figsetplot{./figset/scatter/714.pdf}

\figsetgrpend

\figsetgrpstart
\figsetgrpnum{2.715}

\figsetplot{./figset/scatter/715.pdf}

\figsetgrpend

\figsetgrpstart
\figsetgrpnum{2.716}

\figsetplot{./figset/scatter/716.pdf}

\figsetgrpend

\figsetgrpstart
\figsetgrpnum{2.717}

\figsetplot{./figset/scatter/717.pdf}

\figsetgrpend

\figsetgrpstart
\figsetgrpnum{2.718}

\figsetplot{./figset/scatter/718.pdf}

\figsetgrpend

\figsetgrpstart
\figsetgrpnum{2.719}

\figsetplot{./figset/scatter/719.pdf}

\figsetgrpend

\figsetgrpstart
\figsetgrpnum{2.720}

\figsetplot{./figset/scatter/720.pdf}

\figsetgrpend

\figsetgrpstart
\figsetgrpnum{2.721}

\figsetplot{./figset/scatter/721.pdf}

\figsetgrpend

\figsetgrpstart
\figsetgrpnum{2.722}

\figsetplot{./figset/scatter/722.pdf}

\figsetgrpend

\figsetgrpstart
\figsetgrpnum{2.723}

\figsetplot{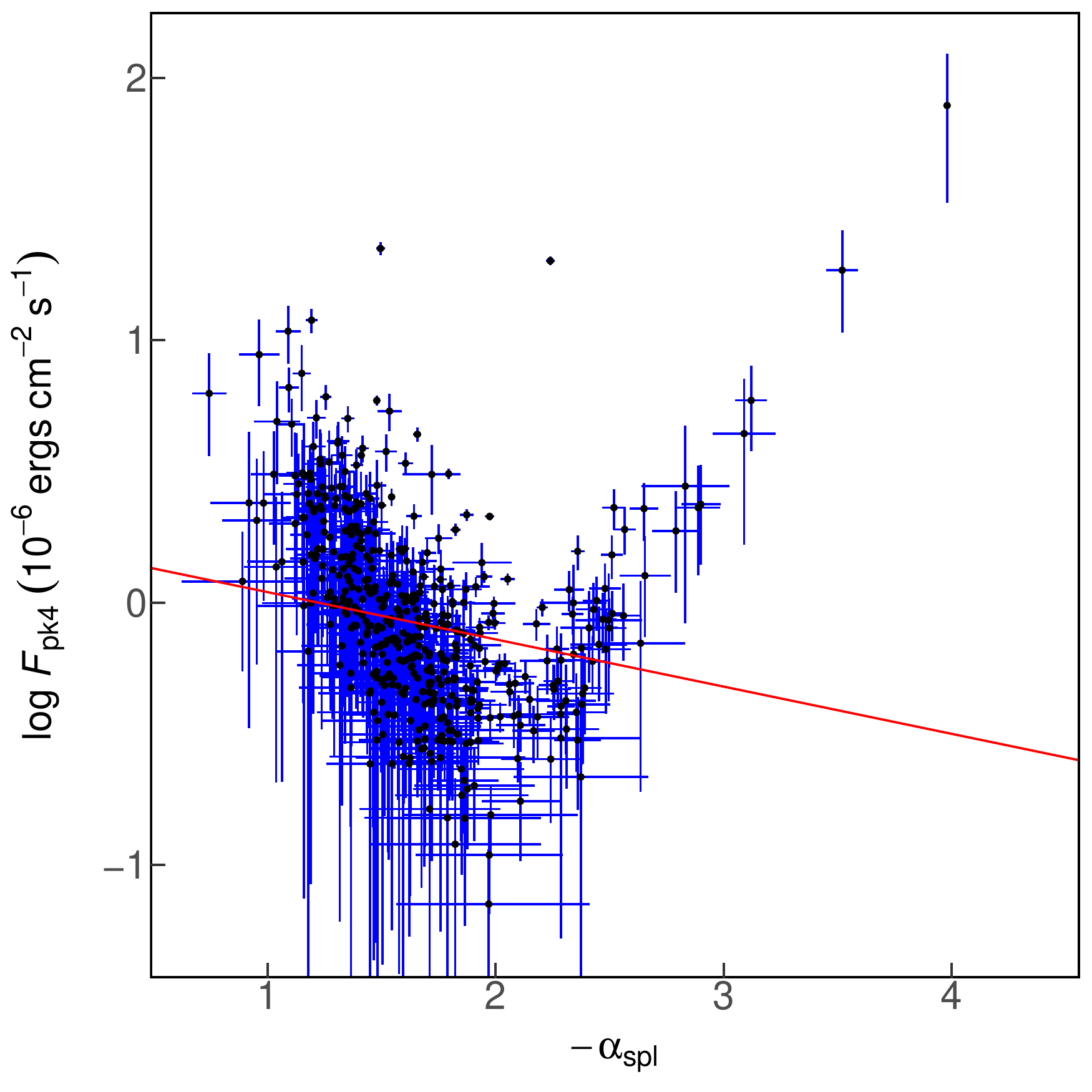}

\figsetgrpend

\figsetgrpstart
\figsetgrpnum{2.724}

\figsetplot{./figset/scatter/724.pdf}

\figsetgrpend

\figsetgrpstart
\figsetgrpnum{2.725}

\figsetplot{./figset/scatter/725.pdf}

\figsetgrpend

\figsetgrpstart
\figsetgrpnum{2.726}

\figsetplot{./figset/scatter/726.pdf}

\figsetgrpend

\figsetgrpstart
\figsetgrpnum{2.727}

\figsetplot{./figset/scatter/727.pdf}

\figsetgrpend

\figsetgrpstart
\figsetgrpnum{2.728}

\figsetplot{./figset/scatter/728.pdf}

\figsetgrpend

\figsetgrpstart
\figsetgrpnum{2.729}

\figsetplot{./figset/scatter/729.pdf}

\figsetgrpend

\figsetgrpstart
\figsetgrpnum{2.730}

\figsetplot{./figset/scatter/730.pdf}

\figsetgrpend

\figsetgrpstart
\figsetgrpnum{2.731}

\figsetplot{./figset/scatter/731.pdf}

\figsetgrpend

\figsetgrpstart
\figsetgrpnum{2.732}

\figsetplot{./figset/scatter/732.pdf}

\figsetgrpend

\figsetgrpstart
\figsetgrpnum{2.733}

\figsetplot{./figset/scatter/733.pdf}

\figsetgrpend

\figsetgrpstart
\figsetgrpnum{2.734}

\figsetplot{./figset/scatter/734.pdf}

\figsetgrpend

\figsetgrpstart
\figsetgrpnum{2.735}

\figsetplot{./figset/scatter/735.pdf}

\figsetgrpend

\figsetgrpstart
\figsetgrpnum{2.736}

\figsetplot{./figset/scatter/736.pdf}

\figsetgrpend

\figsetgrpstart
\figsetgrpnum{2.737}

\figsetplot{./figset/scatter/737.pdf}

\figsetgrpend

\figsetgrpstart
\figsetgrpnum{2.738}

\figsetplot{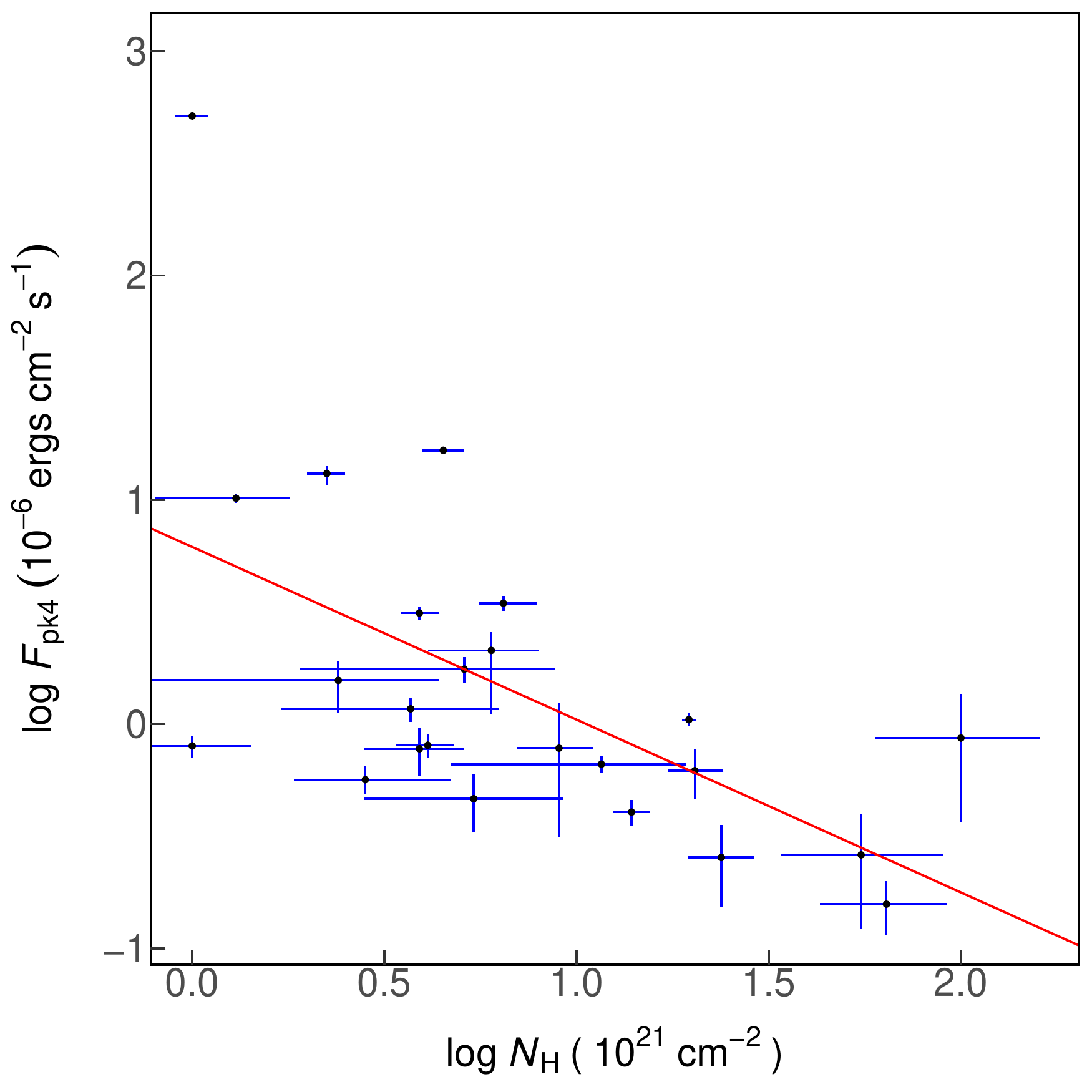}

\figsetgrpend

\figsetgrpstart
\figsetgrpnum{2.739}

\figsetplot{./figset/scatter/739.pdf}

\figsetgrpend

\figsetgrpstart
\figsetgrpnum{2.740}

\figsetplot{./figset/scatter/740.pdf}

\figsetgrpend

\figsetgrpstart
\figsetgrpnum{2.741}

\figsetplot{./figset/scatter/741.pdf}

\figsetgrpend

\figsetgrpstart
\figsetgrpnum{2.742}

\figsetplot{./figset/scatter/742.pdf}

\figsetgrpend

\figsetgrpstart
\figsetgrpnum{2.743}

\figsetplot{./figset/scatter/743.pdf}

\figsetgrpend

\figsetgrpstart
\figsetgrpnum{2.744}

\figsetplot{./figset/scatter/744.pdf}

\figsetgrpend

\figsetgrpstart
\figsetgrpnum{2.745}

\figsetplot{./figset/scatter/745.pdf}

\figsetgrpend

\figsetgrpstart
\figsetgrpnum{2.746}

\figsetplot{./figset/scatter/746.pdf}

\figsetgrpend

\figsetgrpstart
\figsetgrpnum{2.747}

\figsetplot{./figset/scatter/747.pdf}

\figsetgrpend

\figsetgrpstart
\figsetgrpnum{2.748}

\figsetplot{./figset/scatter/748.pdf}

\figsetgrpend

\figsetgrpstart
\figsetgrpnum{2.749}

\figsetplot{./figset/scatter/749.pdf}

\figsetgrpend

\figsetgrpstart
\figsetgrpnum{2.750}

\figsetplot{./figset/scatter/750.pdf}

\figsetgrpend

\figsetgrpstart
\figsetgrpnum{2.751}

\figsetplot{./figset/scatter/751.pdf}

\figsetgrpend

\figsetgrpstart
\figsetgrpnum{2.752}

\figsetplot{./figset/scatter/752.pdf}

\figsetgrpend

\figsetgrpstart
\figsetgrpnum{2.753}

\figsetplot{./figset/scatter/753.pdf}

\figsetgrpend

\figsetgrpstart
\figsetgrpnum{2.754}

\figsetplot{./figset/scatter/754.pdf}

\figsetgrpend

\figsetgrpstart
\figsetgrpnum{2.755}

\figsetplot{./figset/scatter/755.pdf}

\figsetgrpend

\figsetgrpstart
\figsetgrpnum{2.756}

\figsetplot{./figset/scatter/756.pdf}

\figsetgrpend

\figsetgrpstart
\figsetgrpnum{2.757}

\figsetplot{./figset/scatter/757.pdf}

\figsetgrpend

\figsetgrpstart
\figsetgrpnum{2.758}

\figsetplot{./figset/scatter/758.pdf}

\figsetgrpend

\figsetgrpstart
\figsetgrpnum{2.759}

\figsetplot{./figset/scatter/759.pdf}

\figsetgrpend

\figsetgrpstart
\figsetgrpnum{2.760}

\figsetplot{./figset/scatter/760.pdf}

\figsetgrpend

\figsetgrpstart
\figsetgrpnum{2.761}

\figsetplot{./figset/scatter/761.pdf}

\figsetgrpend

\figsetgrpstart
\figsetgrpnum{2.762}

\figsetplot{./figset/scatter/762.pdf}

\figsetgrpend

\figsetgrpstart
\figsetgrpnum{2.763}

\figsetplot{./figset/scatter/763.pdf}

\figsetgrpend

\figsetgrpstart
\figsetgrpnum{2.764}

\figsetplot{./figset/scatter/764.pdf}

\figsetgrpend

\figsetgrpstart
\figsetgrpnum{2.765}

\figsetplot{./figset/scatter/765.pdf}

\figsetgrpend

\figsetgrpstart
\figsetgrpnum{2.766}

\figsetplot{./figset/scatter/766.pdf}

\figsetgrpend

\figsetgrpstart
\figsetgrpnum{2.767}

\figsetplot{./figset/scatter/767.pdf}

\figsetgrpend

\figsetgrpstart
\figsetgrpnum{2.768}

\figsetplot{./figset/scatter/768.pdf}

\figsetgrpend

\figsetgrpstart
\figsetgrpnum{2.769}

\figsetplot{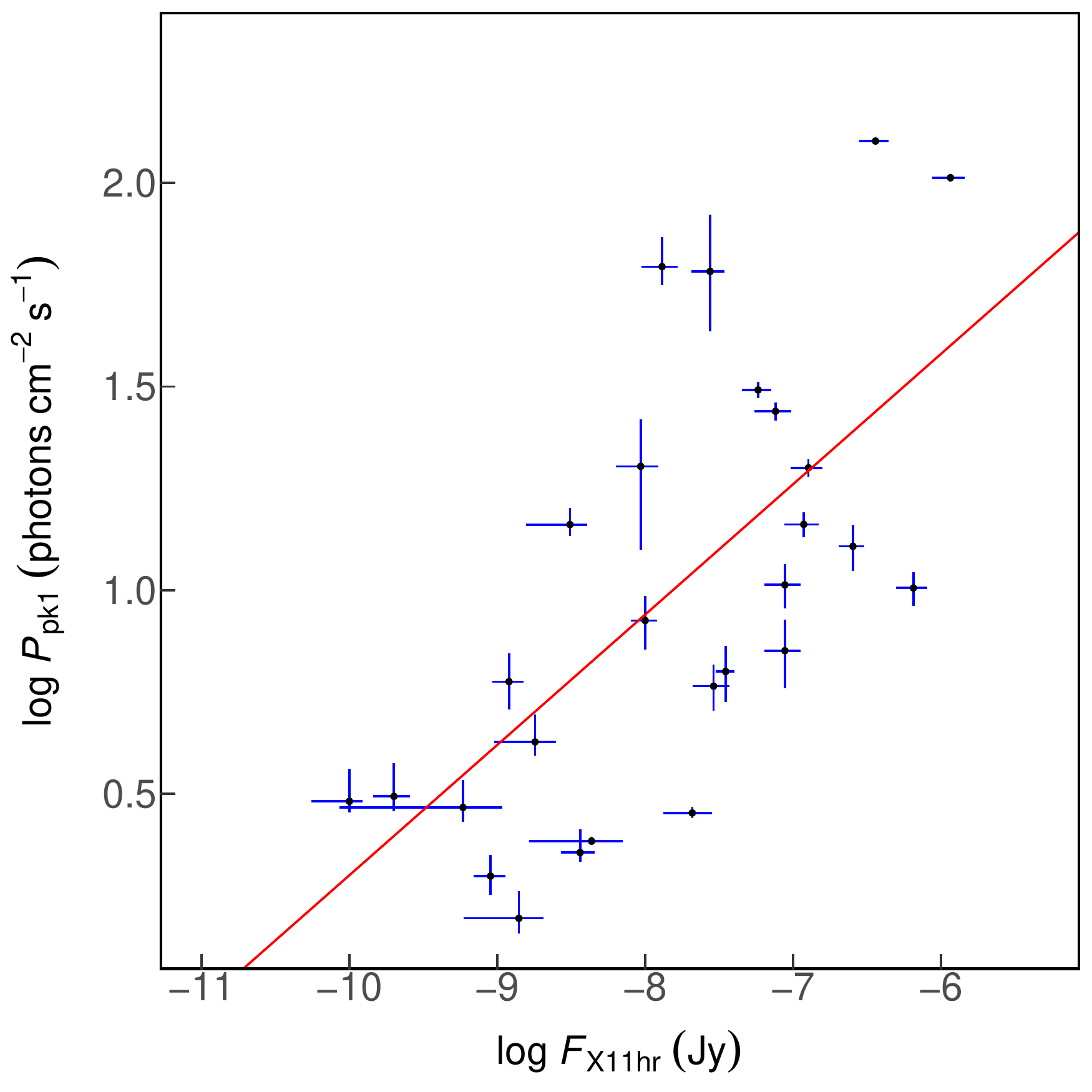}

\figsetgrpend

\figsetgrpstart
\figsetgrpnum{2.770}

\figsetplot{./figset/scatter/770.pdf}

\figsetgrpend

\figsetgrpstart
\figsetgrpnum{2.771}

\figsetplot{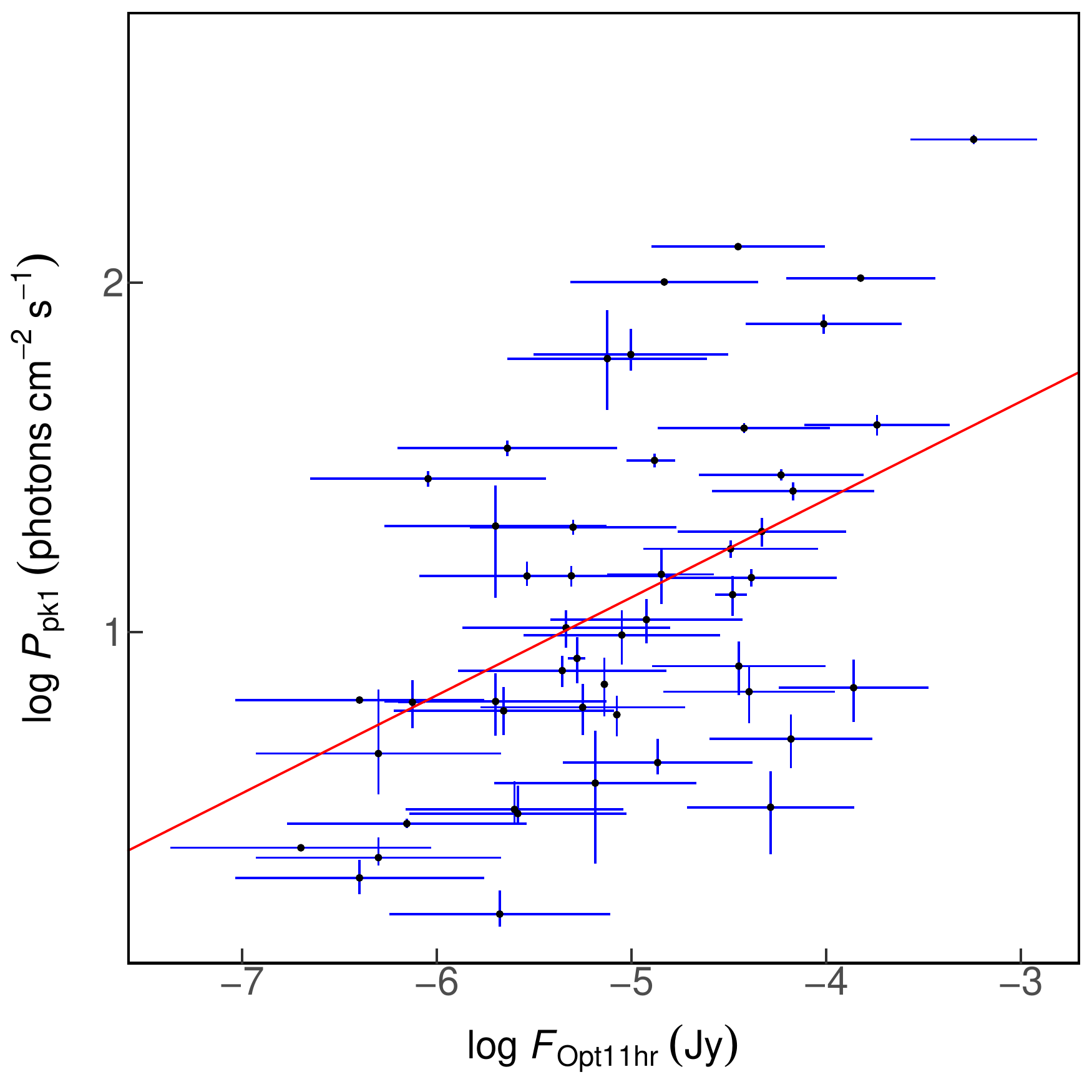}

\figsetgrpend

\figsetgrpstart
\figsetgrpnum{2.772}

\figsetplot{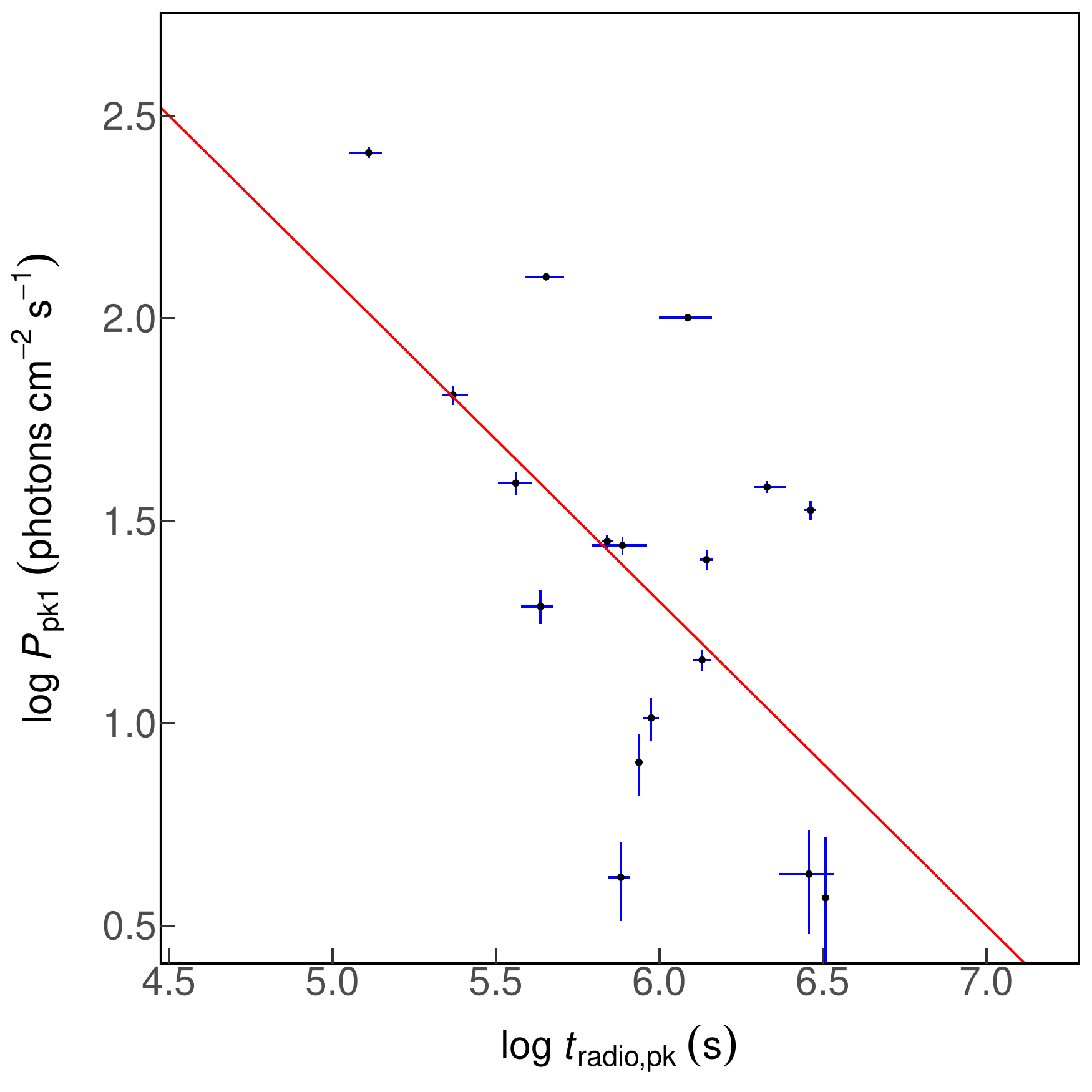}

\figsetgrpend

\figsetgrpstart
\figsetgrpnum{2.773}

\figsetplot{./figset/scatter/773.pdf}

\figsetgrpend

\figsetgrpstart
\figsetgrpnum{2.774}

\figsetplot{./figset/scatter/774.pdf}

\figsetgrpend

\figsetgrpstart
\figsetgrpnum{2.775}

\figsetplot{./figset/scatter/775.pdf}

\figsetgrpend

\figsetgrpstart
\figsetgrpnum{2.776}

\figsetplot{./figset/scatter/776.pdf}

\figsetgrpend

\figsetgrpstart
\figsetgrpnum{2.777}

\figsetplot{./figset/scatter/777.pdf}

\figsetgrpend

\figsetgrpstart
\figsetgrpnum{2.778}

\figsetplot{./figset/scatter/778.pdf}

\figsetgrpend

\figsetgrpstart
\figsetgrpnum{2.779}

\figsetplot{./figset/scatter/779.pdf}

\figsetgrpend

\figsetgrpstart
\figsetgrpnum{2.780}

\figsetplot{./figset/scatter/780.pdf}

\figsetgrpend

\figsetgrpstart
\figsetgrpnum{2.781}

\figsetplot{./figset/scatter/781.pdf}

\figsetgrpend

\figsetgrpstart
\figsetgrpnum{2.782}

\figsetplot{./figset/scatter/782.pdf}

\figsetgrpend

\figsetgrpstart
\figsetgrpnum{2.783}

\figsetplot{./figset/scatter/783.pdf}

\figsetgrpend

\figsetgrpstart
\figsetgrpnum{2.784}

\figsetplot{./figset/scatter/784.pdf}

\figsetgrpend

\figsetgrpstart
\figsetgrpnum{2.785}

\figsetplot{./figset/scatter/785.pdf}

\figsetgrpend

\figsetgrpstart
\figsetgrpnum{2.786}

\figsetplot{./figset/scatter/786.pdf}

\figsetgrpend

\figsetgrpstart
\figsetgrpnum{2.787}

\figsetplot{./figset/scatter/787.pdf}

\figsetgrpend

\figsetgrpstart
\figsetgrpnum{2.788}

\figsetplot{./figset/scatter/788.pdf}

\figsetgrpend

\figsetgrpstart
\figsetgrpnum{2.789}

\figsetplot{./figset/scatter/789.pdf}

\figsetgrpend

\figsetgrpstart
\figsetgrpnum{2.790}

\figsetplot{./figset/scatter/790.pdf}

\figsetgrpend

\figsetgrpstart
\figsetgrpnum{2.791}

\figsetplot{./figset/scatter/791.pdf}

\figsetgrpend

\figsetgrpstart
\figsetgrpnum{2.792}

\figsetplot{./figset/scatter/792.pdf}

\figsetgrpend

\figsetgrpstart
\figsetgrpnum{2.793}

\figsetplot{./figset/scatter/793.pdf}

\figsetgrpend

\figsetgrpstart
\figsetgrpnum{2.794}

\figsetplot{./figset/scatter/794.pdf}

\figsetgrpend

\figsetgrpstart
\figsetgrpnum{2.795}

\figsetplot{./figset/scatter/795.pdf}

\figsetgrpend

\figsetgrpstart
\figsetgrpnum{2.796}

\figsetplot{./figset/scatter/796.pdf}

\figsetgrpend

\figsetgrpstart
\figsetgrpnum{2.797}

\figsetplot{./figset/scatter/797.pdf}

\figsetgrpend

\figsetgrpstart
\figsetgrpnum{2.798}

\figsetplot{./figset/scatter/798.pdf}

\figsetgrpend

\figsetgrpstart
\figsetgrpnum{2.799}

\figsetplot{./figset/scatter/799.pdf}

\figsetgrpend

\figsetgrpstart
\figsetgrpnum{2.800}

\figsetplot{./figset/scatter/800.pdf}

\figsetgrpend

\figsetgrpstart
\figsetgrpnum{2.801}

\figsetplot{./figset/scatter/801.pdf}

\figsetgrpend

\figsetgrpstart
\figsetgrpnum{2.802}

\figsetplot{./figset/scatter/802.pdf}

\figsetgrpend

\figsetgrpstart
\figsetgrpnum{2.803}

\figsetplot{./figset/scatter/803.pdf}

\figsetgrpend

\figsetgrpstart
\figsetgrpnum{2.804}

\figsetplot{./figset/scatter/804.pdf}

\figsetgrpend

\figsetgrpstart
\figsetgrpnum{2.805}

\figsetplot{./figset/scatter/805.pdf}

\figsetgrpend

\figsetgrpstart
\figsetgrpnum{2.806}

\figsetplot{./figset/scatter/806.pdf}

\figsetgrpend

\figsetgrpstart
\figsetgrpnum{2.807}

\figsetplot{./figset/scatter/807.pdf}

\figsetgrpend

\figsetgrpstart
\figsetgrpnum{2.808}

\figsetplot{./figset/scatter/808.pdf}

\figsetgrpend

\figsetgrpstart
\figsetgrpnum{2.809}

\figsetplot{./figset/scatter/809.pdf}

\figsetgrpend

\figsetgrpstart
\figsetgrpnum{2.810}

\figsetplot{./figset/scatter/810.pdf}

\figsetgrpend

\figsetgrpstart
\figsetgrpnum{2.811}

\figsetplot{./figset/scatter/811.pdf}

\figsetgrpend

\figsetgrpstart
\figsetgrpnum{2.812}

\figsetplot{./figset/scatter/812.pdf}

\figsetgrpend

\figsetgrpstart
\figsetgrpnum{2.813}

\figsetplot{./figset/scatter/813.pdf}

\figsetgrpend

\figsetgrpstart
\figsetgrpnum{2.814}

\figsetplot{./figset/scatter/814.pdf}

\figsetgrpend

\figsetgrpstart
\figsetgrpnum{2.815}

\figsetplot{./figset/scatter/815.pdf}

\figsetgrpend

\figsetgrpstart
\figsetgrpnum{2.816}

\figsetplot{./figset/scatter/816.pdf}

\figsetgrpend

\figsetgrpstart
\figsetgrpnum{2.817}

\figsetplot{./figset/scatter/817.pdf}

\figsetgrpend

\figsetgrpstart
\figsetgrpnum{2.818}

\figsetplot{./figset/scatter/818.pdf}

\figsetgrpend

\figsetgrpstart
\figsetgrpnum{2.819}

\figsetplot{./figset/scatter/819.pdf}

\figsetgrpend

\figsetgrpstart
\figsetgrpnum{2.820}

\figsetplot{./figset/scatter/820.pdf}

\figsetgrpend

\figsetgrpstart
\figsetgrpnum{2.821}

\figsetplot{./figset/scatter/821.pdf}

\figsetgrpend

\figsetgrpstart
\figsetgrpnum{2.822}

\figsetplot{./figset/scatter/822.pdf}

\figsetgrpend

\figsetgrpstart
\figsetgrpnum{2.823}

\figsetplot{./figset/scatter/823.pdf}

\figsetgrpend

\figsetgrpstart
\figsetgrpnum{2.824}

\figsetplot{./figset/scatter/824.pdf}

\figsetgrpend

\figsetgrpstart
\figsetgrpnum{2.825}

\figsetplot{./figset/scatter/825.pdf}

\figsetgrpend

\figsetgrpstart
\figsetgrpnum{2.826}

\figsetplot{./figset/scatter/826.pdf}

\figsetgrpend

\figsetgrpstart
\figsetgrpnum{2.827}

\figsetplot{./figset/scatter/827.pdf}

\figsetgrpend

\figsetgrpstart
\figsetgrpnum{2.828}

\figsetplot{./figset/scatter/828.pdf}

\figsetgrpend

\figsetgrpstart
\figsetgrpnum{2.829}

\figsetplot{./figset/scatter/829.pdf}

\figsetgrpend

\figsetgrpstart
\figsetgrpnum{2.830}

\figsetplot{./figset/scatter/830.pdf}

\figsetgrpend

\figsetgrpstart
\figsetgrpnum{2.831}

\figsetplot{./figset/scatter/831.pdf}

\figsetgrpend

\figsetgrpstart
\figsetgrpnum{2.832}

\figsetplot{./figset/scatter/832.pdf}

\figsetgrpend

\figsetgrpstart
\figsetgrpnum{2.833}

\figsetplot{./figset/scatter/833.pdf}

\figsetgrpend

\figsetgrpstart
\figsetgrpnum{2.834}

\figsetplot{./figset/scatter/834.pdf}

\figsetgrpend

\figsetgrpstart
\figsetgrpnum{2.835}

\figsetplot{./figset/scatter/835.pdf}

\figsetgrpend

\figsetgrpstart
\figsetgrpnum{2.836}

\figsetplot{./figset/scatter/836.pdf}

\figsetgrpend

\figsetgrpstart
\figsetgrpnum{2.837}

\figsetplot{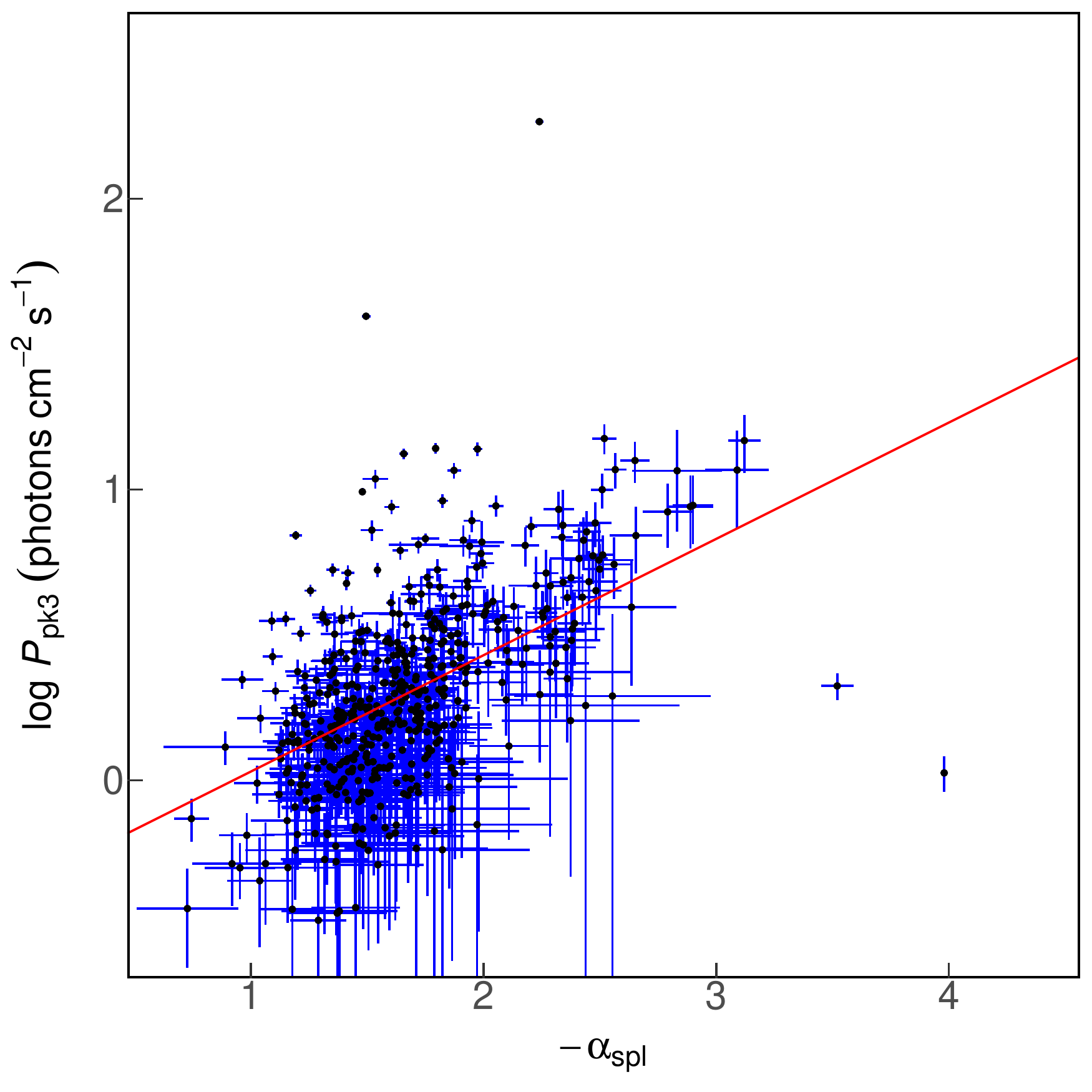}

\figsetgrpend

\figsetgrpstart
\figsetgrpnum{2.838}

\figsetplot{./figset/scatter/838.pdf}

\figsetgrpend

\figsetgrpstart
\figsetgrpnum{2.839}

\figsetplot{./figset/scatter/839.pdf}

\figsetgrpend

\figsetgrpstart
\figsetgrpnum{2.840}

\figsetplot{./figset/scatter/840.pdf}

\figsetgrpend

\figsetgrpstart
\figsetgrpnum{2.841}

\figsetplot{./figset/scatter/841.pdf}

\figsetgrpend

\figsetgrpstart
\figsetgrpnum{2.842}

\figsetplot{./figset/scatter/842.pdf}

\figsetgrpend

\figsetgrpstart
\figsetgrpnum{2.843}

\figsetplot{./figset/scatter/843.pdf}

\figsetgrpend

\figsetgrpstart
\figsetgrpnum{2.844}

\figsetplot{./figset/scatter/844.pdf}

\figsetgrpend

\figsetgrpstart
\figsetgrpnum{2.845}

\figsetplot{./figset/scatter/845.pdf}

\figsetgrpend

\figsetgrpstart
\figsetgrpnum{2.846}

\figsetplot{./figset/scatter/846.pdf}

\figsetgrpend

\figsetgrpstart
\figsetgrpnum{2.847}

\figsetplot{./figset/scatter/847.pdf}

\figsetgrpend

\figsetgrpstart
\figsetgrpnum{2.848}

\figsetplot{./figset/scatter/848.pdf}

\figsetgrpend

\figsetgrpstart
\figsetgrpnum{2.849}

\figsetplot{./figset/scatter/849.pdf}

\figsetgrpend

\figsetgrpstart
\figsetgrpnum{2.850}

\figsetplot{./figset/scatter/850.pdf}

\figsetgrpend

\figsetgrpstart
\figsetgrpnum{2.851}

\figsetplot{./figset/scatter/851.pdf}

\figsetgrpend

\figsetgrpstart
\figsetgrpnum{2.852}

\figsetplot{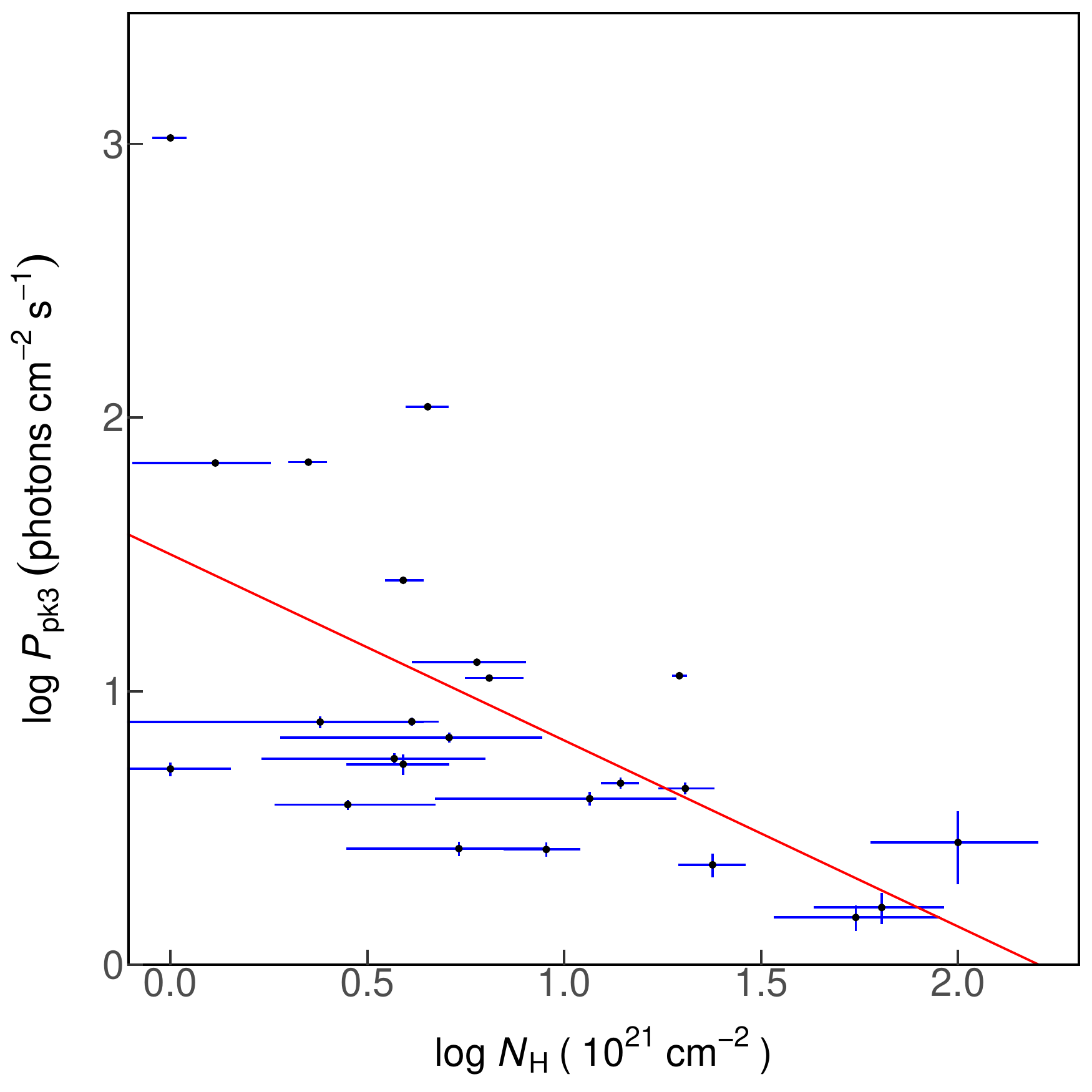}

\figsetgrpend

\figsetgrpstart
\figsetgrpnum{2.853}

\figsetplot{./figset/scatter/853.pdf}

\figsetgrpend

\figsetgrpstart
\figsetgrpnum{2.854}

\figsetplot{./figset/scatter/854.pdf}

\figsetgrpend

\figsetgrpstart
\figsetgrpnum{2.855}

\figsetplot{./figset/scatter/855.pdf}

\figsetgrpend

\figsetgrpstart
\figsetgrpnum{2.856}

\figsetplot{./figset/scatter/856.pdf}

\figsetgrpend

\figsetgrpstart
\figsetgrpnum{2.857}

\figsetplot{./figset/scatter/857.pdf}

\figsetgrpend

\figsetgrpstart
\figsetgrpnum{2.858}

\figsetplot{./figset/scatter/858.pdf}

\figsetgrpend

\figsetgrpstart
\figsetgrpnum{2.859}

\figsetplot{./figset/scatter/859.pdf}

\figsetgrpend

\figsetgrpstart
\figsetgrpnum{2.860}

\figsetplot{./figset/scatter/860.pdf}

\figsetgrpend

\figsetgrpstart
\figsetgrpnum{2.861}

\figsetplot{./figset/scatter/861.pdf}

\figsetgrpend

\figsetgrpstart
\figsetgrpnum{2.862}

\figsetplot{./figset/scatter/862.pdf}

\figsetgrpend

\figsetgrpstart
\figsetgrpnum{2.863}

\figsetplot{./figset/scatter/863.pdf}

\figsetgrpend

\figsetgrpstart
\figsetgrpnum{2.864}

\figsetplot{./figset/scatter/864.pdf}

\figsetgrpend

\figsetgrpstart
\figsetgrpnum{2.865}

\figsetplot{./figset/scatter/865.pdf}

\figsetgrpend

\figsetgrpstart
\figsetgrpnum{2.866}

\figsetplot{./figset/scatter/866.pdf}

\figsetgrpend

\figsetgrpstart
\figsetgrpnum{2.867}

\figsetplot{./figset/scatter/867.pdf}

\figsetgrpend

\figsetgrpstart
\figsetgrpnum{2.868}

\figsetplot{./figset/scatter/868.pdf}

\figsetgrpend

\figsetgrpstart
\figsetgrpnum{2.869}

\figsetplot{./figset/scatter/869.pdf}

\figsetgrpend

\figsetgrpstart
\figsetgrpnum{2.870}

\figsetplot{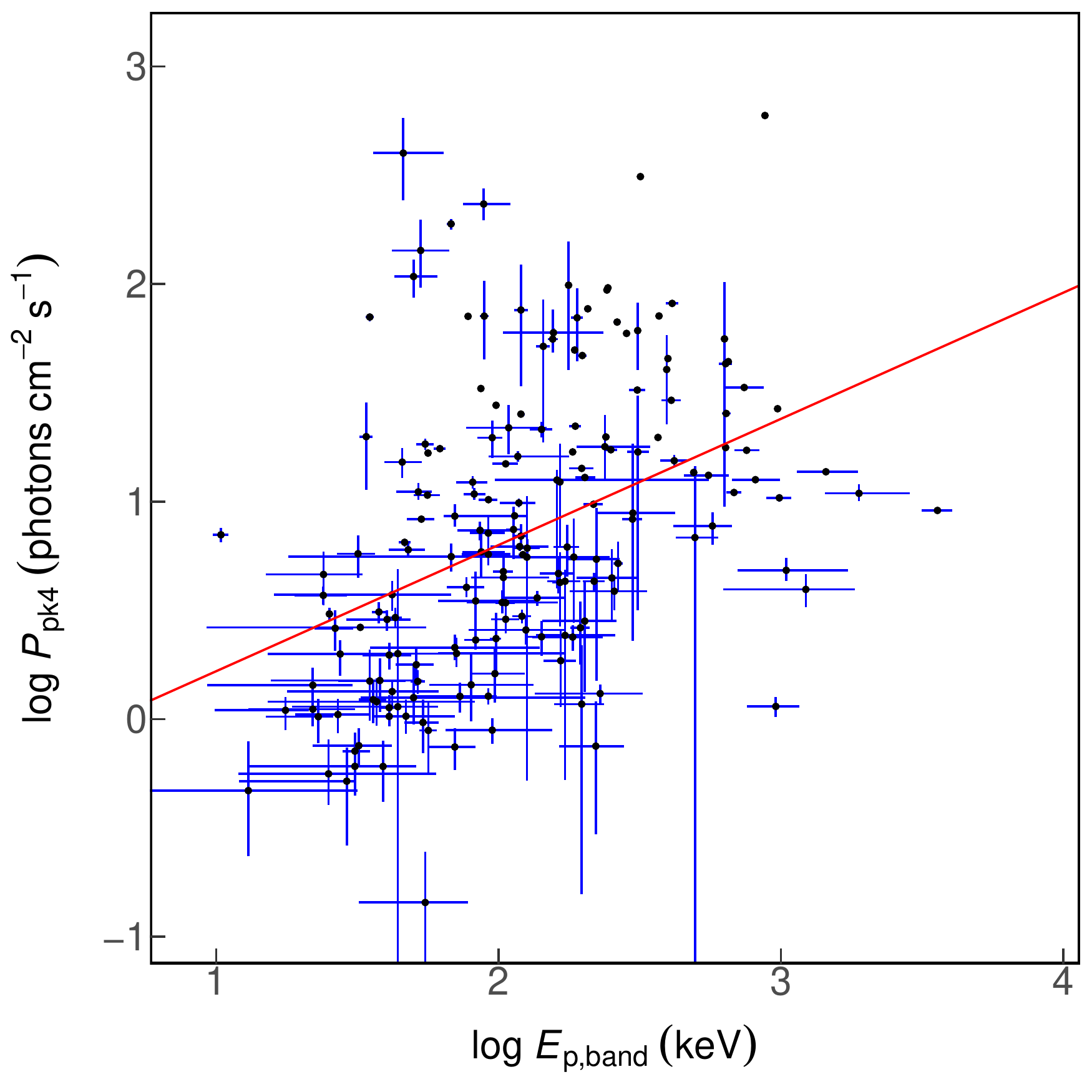}

\figsetgrpend

\figsetgrpstart
\figsetgrpnum{2.871}

\figsetplot{./figset/scatter/871.pdf}

\figsetgrpend

\figsetgrpstart
\figsetgrpnum{2.872}

\figsetplot{./figset/scatter/872.pdf}

\figsetgrpend

\figsetgrpstart
\figsetgrpnum{2.873}

\figsetplot{./figset/scatter/873.pdf}

\figsetgrpend

\figsetgrpstart
\figsetgrpnum{2.874}

\figsetplot{./figset/scatter/874.pdf}

\figsetgrpend

\figsetgrpstart
\figsetgrpnum{2.875}

\figsetplot{./figset/scatter/875.pdf}

\figsetgrpend

\figsetgrpstart
\figsetgrpnum{2.876}

\figsetplot{./figset/scatter/876.pdf}

\figsetgrpend

\figsetgrpstart
\figsetgrpnum{2.877}

\figsetplot{./figset/scatter/877.pdf}

\figsetgrpend

\figsetgrpstart
\figsetgrpnum{2.878}

\figsetplot{./figset/scatter/878.pdf}

\figsetgrpend

\figsetgrpstart
\figsetgrpnum{2.879}

\figsetplot{./figset/scatter/879.pdf}

\figsetgrpend

\figsetgrpstart
\figsetgrpnum{2.880}

\figsetplot{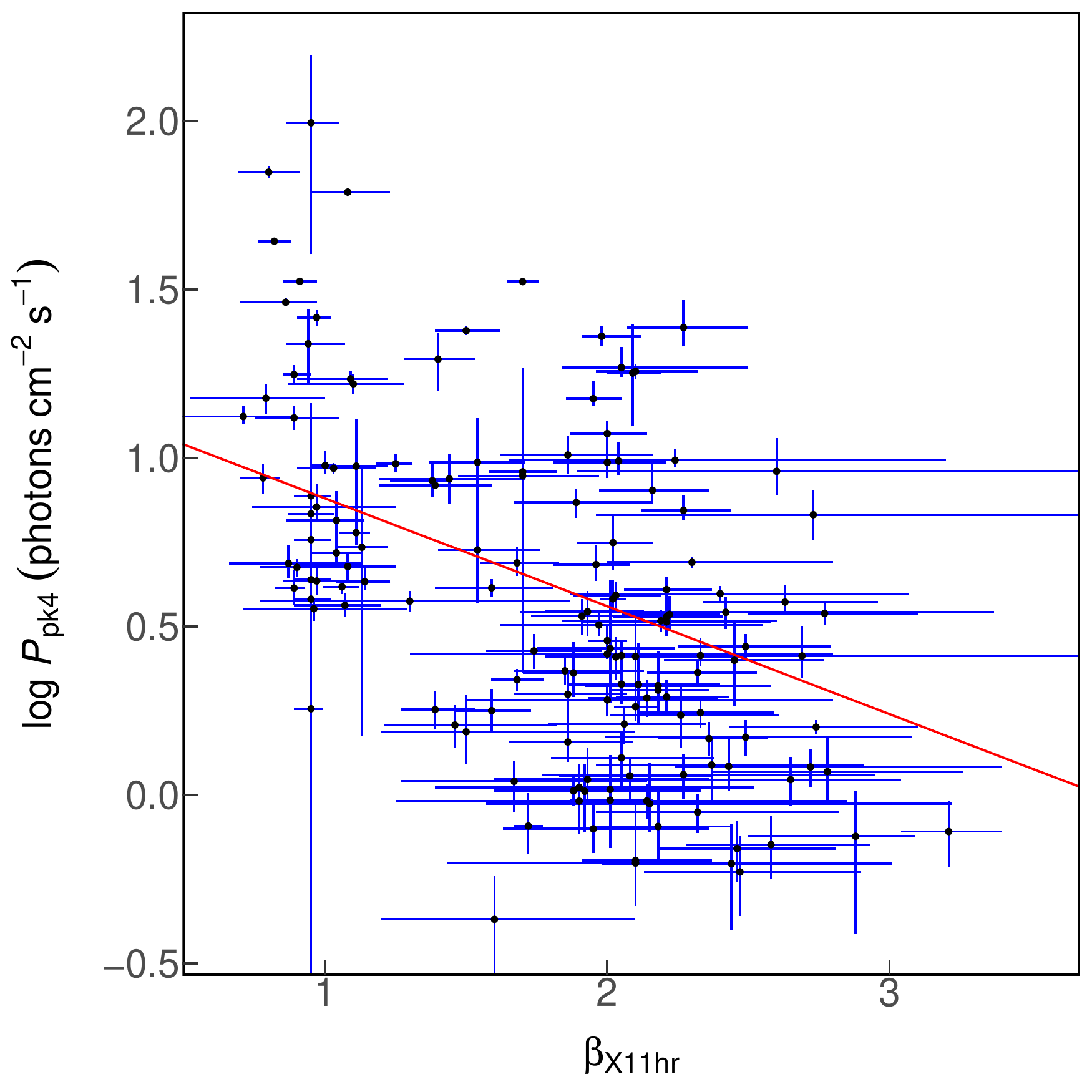}

\figsetgrpend

\figsetgrpstart
\figsetgrpnum{2.881}

\figsetplot{./figset/scatter/881.pdf}

\figsetgrpend

\figsetgrpstart
\figsetgrpnum{2.882}

\figsetplot{./figset/scatter/882.pdf}

\figsetgrpend

\figsetgrpstart
\figsetgrpnum{2.883}

\figsetplot{./figset/scatter/883.pdf}

\figsetgrpend

\figsetgrpstart
\figsetgrpnum{2.884}

\figsetplot{./figset/scatter/884.pdf}

\figsetgrpend

\figsetgrpstart
\figsetgrpnum{2.885}

\figsetplot{./figset/scatter/885.pdf}

\figsetgrpend

\figsetgrpstart
\figsetgrpnum{2.886}

\figsetplot{./figset/scatter/886.pdf}

\figsetgrpend

\figsetgrpstart
\figsetgrpnum{2.887}

\figsetplot{./figset/scatter/887.pdf}

\figsetgrpend

\figsetgrpstart
\figsetgrpnum{2.888}

\figsetplot{./figset/scatter/888.pdf}

\figsetgrpend

\figsetgrpstart
\figsetgrpnum{2.889}

\figsetplot{./figset/scatter/889.pdf}

\figsetgrpend

\figsetgrpstart
\figsetgrpnum{2.890}

\figsetplot{./figset/scatter/890.pdf}

\figsetgrpend

\figsetgrpstart
\figsetgrpnum{2.891}

\figsetplot{./figset/scatter/891.pdf}

\figsetgrpend

\figsetgrpstart
\figsetgrpnum{2.892}

\figsetplot{./figset/scatter/892.pdf}

\figsetgrpend

\figsetgrpstart
\figsetgrpnum{2.893}

\figsetplot{./figset/scatter/893.pdf}

\figsetgrpend

\figsetgrpstart
\figsetgrpnum{2.894}

\figsetplot{./figset/scatter/894.pdf}

\figsetgrpend

\figsetgrpstart
\figsetgrpnum{2.895}

\figsetplot{./figset/scatter/895.pdf}

\figsetgrpend

\figsetgrpstart
\figsetgrpnum{2.896}

\figsetplot{./figset/scatter/896.pdf}

\figsetgrpend

\figsetgrpstart
\figsetgrpnum{2.897}

\figsetplot{./figset/scatter/897.pdf}

\figsetgrpend

\figsetgrpstart
\figsetgrpnum{2.898}

\figsetplot{./figset/scatter/898.pdf}

\figsetgrpend

\figsetgrpstart
\figsetgrpnum{2.899}

\figsetplot{./figset/scatter/899.pdf}

\figsetgrpend

\figsetgrpstart
\figsetgrpnum{2.900}

\figsetplot{./figset/scatter/900.pdf}

\figsetgrpend

\figsetgrpstart
\figsetgrpnum{2.901}

\figsetplot{./figset/scatter/901.pdf}

\figsetgrpend

\figsetgrpstart
\figsetgrpnum{2.902}

\figsetplot{./figset/scatter/902.pdf}

\figsetgrpend

\figsetgrpstart
\figsetgrpnum{2.903}

\figsetplot{./figset/scatter/903.pdf}

\figsetgrpend

\figsetgrpstart
\figsetgrpnum{2.904}

\figsetplot{./figset/scatter/904.pdf}

\figsetgrpend

\figsetgrpstart
\figsetgrpnum{2.905}

\figsetplot{./figset/scatter/905.pdf}

\figsetgrpend

\figsetgrpstart
\figsetgrpnum{2.906}

\figsetplot{./figset/scatter/906.pdf}

\figsetgrpend

\figsetgrpstart
\figsetgrpnum{2.907}

\figsetplot{./figset/scatter/907.pdf}

\figsetgrpend

\figsetgrpstart
\figsetgrpnum{2.908}

\figsetplot{./figset/scatter/908.pdf}

\figsetgrpend

\figsetgrpstart
\figsetgrpnum{2.909}

\figsetplot{./figset/scatter/909.pdf}

\figsetgrpend

\figsetgrpstart
\figsetgrpnum{2.910}

\figsetplot{./figset/scatter/910.pdf}

\figsetgrpend

\figsetgrpstart
\figsetgrpnum{2.911}

\figsetplot{./figset/scatter/911.pdf}

\figsetgrpend

\figsetgrpstart
\figsetgrpnum{2.912}

\figsetplot{./figset/scatter/912.pdf}

\figsetgrpend

\figsetgrpstart
\figsetgrpnum{2.913}

\figsetplot{./figset/scatter/913.pdf}

\figsetgrpend

\figsetgrpstart
\figsetgrpnum{2.914}

\figsetplot{./figset/scatter/914.pdf}

\figsetgrpend

\figsetgrpstart
\figsetgrpnum{2.915}

\figsetplot{./figset/scatter/915.pdf}

\figsetgrpend

\figsetgrpstart
\figsetgrpnum{2.916}

\figsetplot{./figset/scatter/916.pdf}

\figsetgrpend

\figsetgrpstart
\figsetgrpnum{2.917}

\figsetplot{./figset/scatter/917.pdf}

\figsetgrpend

\figsetgrpstart
\figsetgrpnum{2.918}

\figsetplot{./figset/scatter/918.pdf}

\figsetgrpend

\figsetgrpstart
\figsetgrpnum{2.919}

\figsetplot{./figset/scatter/919.pdf}

\figsetgrpend

\figsetgrpstart
\figsetgrpnum{2.920}

\figsetplot{./figset/scatter/920.pdf}

\figsetgrpend

\figsetgrpstart
\figsetgrpnum{2.921}

\figsetplot{./figset/scatter/921.pdf}

\figsetgrpend

\figsetgrpstart
\figsetgrpnum{2.922}

\figsetplot{./figset/scatter/922.pdf}

\figsetgrpend

\figsetgrpstart
\figsetgrpnum{2.923}

\figsetplot{./figset/scatter/923.pdf}

\figsetgrpend

\figsetgrpstart
\figsetgrpnum{2.924}

\figsetplot{./figset/scatter/924.pdf}

\figsetgrpend

\figsetgrpstart
\figsetgrpnum{2.925}

\figsetplot{./figset/scatter/925.pdf}

\figsetgrpend

\figsetgrpstart
\figsetgrpnum{2.926}

\figsetplot{./figset/scatter/926.pdf}

\figsetgrpend

\figsetgrpstart
\figsetgrpnum{2.927}

\figsetplot{./figset/scatter/927.pdf}

\figsetgrpend

\figsetgrpstart
\figsetgrpnum{2.928}

\figsetplot{./figset/scatter/928.pdf}

\figsetgrpend

\figsetgrpstart
\figsetgrpnum{2.929}

\figsetplot{./figset/scatter/929.pdf}

\figsetgrpend

\figsetgrpstart
\figsetgrpnum{2.930}

\figsetplot{./figset/scatter/930.pdf}

\figsetgrpend

\figsetgrpstart
\figsetgrpnum{2.931}

\figsetplot{./figset/scatter/931.pdf}

\figsetgrpend

\figsetgrpstart
\figsetgrpnum{2.932}

\figsetplot{./figset/scatter/932.pdf}

\figsetgrpend

\figsetgrpstart
\figsetgrpnum{2.933}

\figsetplot{./figset/scatter/933.pdf}

\figsetgrpend

\figsetgrpstart
\figsetgrpnum{2.934}

\figsetplot{./figset/scatter/934.pdf}

\figsetgrpend

\figsetgrpstart
\figsetgrpnum{2.935}

\figsetplot{./figset/scatter/935.pdf}

\figsetgrpend

\figsetgrpstart
\figsetgrpnum{2.936}

\figsetplot{./figset/scatter/936.pdf}

\figsetgrpend

\figsetgrpstart
\figsetgrpnum{2.937}

\figsetplot{./figset/scatter/937.pdf}

\figsetgrpend

\figsetgrpstart
\figsetgrpnum{2.938}

\figsetplot{./figset/scatter/938.pdf}

\figsetgrpend

\figsetgrpstart
\figsetgrpnum{2.939}

\figsetplot{./figset/scatter/939.pdf}

\figsetgrpend

\figsetgrpstart
\figsetgrpnum{2.940}

\figsetplot{./figset/scatter/940.pdf}

\figsetgrpend

\figsetgrpstart
\figsetgrpnum{2.941}

\figsetplot{./figset/scatter/941.pdf}

\figsetgrpend

\figsetgrpstart
\figsetgrpnum{2.942}

\figsetplot{./figset/scatter/942.pdf}

\figsetgrpend

\figsetgrpstart
\figsetgrpnum{2.943}

\figsetplot{./figset/scatter/943.pdf}

\figsetgrpend

\figsetgrpstart
\figsetgrpnum{2.944}

\figsetplot{./figset/scatter/944.pdf}

\figsetgrpend

\figsetgrpstart
\figsetgrpnum{2.945}

\figsetplot{./figset/scatter/945.pdf}

\figsetgrpend

\figsetgrpstart
\figsetgrpnum{2.946}

\figsetplot{./figset/scatter/946.pdf}

\figsetgrpend

\figsetgrpstart
\figsetgrpnum{2.947}

\figsetplot{./figset/scatter/947.pdf}

\figsetgrpend

\figsetgrpstart
\figsetgrpnum{2.948}

\figsetplot{./figset/scatter/948.pdf}

\figsetgrpend

\figsetgrpstart
\figsetgrpnum{2.949}

\figsetplot{./figset/scatter/949.pdf}

\figsetgrpend

\figsetgrpstart
\figsetgrpnum{2.950}

\figsetplot{./figset/scatter/950.pdf}

\figsetgrpend

\figsetgrpstart
\figsetgrpnum{2.951}

\figsetplot{./figset/scatter/951.pdf}

\figsetgrpend

\figsetgrpstart
\figsetgrpnum{2.952}

\figsetplot{./figset/scatter/952.pdf}

\figsetgrpend

\figsetgrpstart
\figsetgrpnum{2.953}

\figsetplot{./figset/scatter/953.pdf}

\figsetgrpend

\figsetgrpstart
\figsetgrpnum{2.954}

\figsetplot{./figset/scatter/954.pdf}

\figsetgrpend

\figsetgrpstart
\figsetgrpnum{2.955}

\figsetplot{./figset/scatter/955.pdf}

\figsetgrpend

\figsetgrpstart
\figsetgrpnum{2.956}

\figsetplot{./figset/scatter/956.pdf}

\figsetgrpend

\figsetgrpstart
\figsetgrpnum{2.957}

\figsetplot{./figset/scatter/957.pdf}

\figsetgrpend

\figsetgrpstart
\figsetgrpnum{2.958}

\figsetplot{./figset/scatter/958.pdf}

\figsetgrpend

\figsetgrpstart
\figsetgrpnum{2.959}

\figsetplot{./figset/scatter/959.pdf}

\figsetgrpend

\figsetgrpstart
\figsetgrpnum{2.960}

\figsetplot{./figset/scatter/960.pdf}

\figsetgrpend

\figsetgrpstart
\figsetgrpnum{2.961}

\figsetplot{./figset/scatter/961.pdf}

\figsetgrpend

\figsetgrpstart
\figsetgrpnum{2.962}

\figsetplot{./figset/scatter/962.pdf}

\figsetgrpend

\figsetgrpstart
\figsetgrpnum{2.963}

\figsetplot{./figset/scatter/963.pdf}

\figsetgrpend

\figsetgrpstart
\figsetgrpnum{2.964}

\figsetplot{./figset/scatter/964.pdf}

\figsetgrpend

\figsetgrpstart
\figsetgrpnum{2.965}

\figsetplot{./figset/scatter/965.pdf}

\figsetgrpend

\figsetgrpstart
\figsetgrpnum{2.966}

\figsetplot{./figset/scatter/966.pdf}

\figsetgrpend

\figsetgrpstart
\figsetgrpnum{2.967}

\figsetplot{./figset/scatter/967.pdf}

\figsetgrpend

\figsetgrpstart
\figsetgrpnum{2.968}

\figsetplot{./figset/scatter/968.pdf}

\figsetgrpend

\figsetgrpstart
\figsetgrpnum{2.969}

\figsetplot{./figset/scatter/969.pdf}

\figsetgrpend

\figsetgrpstart
\figsetgrpnum{2.970}

\figsetplot{./figset/scatter/970.pdf}

\figsetgrpend

\figsetgrpstart
\figsetgrpnum{2.971}

\figsetplot{./figset/scatter/971.pdf}

\figsetgrpend

\figsetgrpstart
\figsetgrpnum{2.972}

\figsetplot{./figset/scatter/972.pdf}

\figsetgrpend

\figsetgrpstart
\figsetgrpnum{2.973}

\figsetplot{./figset/scatter/973.pdf}

\figsetgrpend

\figsetgrpstart
\figsetgrpnum{2.974}

\figsetplot{./figset/scatter/974.pdf}

\figsetgrpend

\figsetgrpstart
\figsetgrpnum{2.975}

\figsetplot{./figset/scatter/975.pdf}

\figsetgrpend

\figsetgrpstart
\figsetgrpnum{2.976}

\figsetplot{./figset/scatter/976.pdf}

\figsetgrpend

\figsetgrpstart
\figsetgrpnum{2.977}

\figsetplot{./figset/scatter/977.pdf}

\figsetgrpend

\figsetgrpstart
\figsetgrpnum{2.978}

\figsetplot{./figset/scatter/978.pdf}

\figsetgrpend

\figsetgrpstart
\figsetgrpnum{2.979}

\figsetplot{./figset/scatter/979.pdf}

\figsetgrpend

\figsetgrpstart
\figsetgrpnum{2.980}

\figsetplot{./figset/scatter/980.pdf}

\figsetgrpend

\figsetgrpstart
\figsetgrpnum{2.981}

\figsetplot{./figset/scatter/981.pdf}

\figsetgrpend

\figsetgrpstart
\figsetgrpnum{2.982}

\figsetplot{./figset/scatter/982.pdf}

\figsetgrpend

\figsetgrpstart
\figsetgrpnum{2.983}

\figsetplot{./figset/scatter/983.pdf}

\figsetgrpend

\figsetgrpstart
\figsetgrpnum{2.984}

\figsetplot{./figset/scatter/984.pdf}

\figsetgrpend

\figsetgrpstart
\figsetgrpnum{2.985}

\figsetplot{./figset/scatter/985.pdf}

\figsetgrpend

\figsetgrpstart
\figsetgrpnum{2.986}

\figsetplot{./figset/scatter/986.pdf}

\figsetgrpend

\figsetgrpstart
\figsetgrpnum{2.987}

\figsetplot{./figset/scatter/987.pdf}

\figsetgrpend

\figsetgrpstart
\figsetgrpnum{2.988}

\figsetplot{./figset/scatter/988.pdf}

\figsetgrpend

\figsetgrpstart
\figsetgrpnum{2.989}

\figsetplot{./figset/scatter/989.pdf}

\figsetgrpend

\figsetgrpstart
\figsetgrpnum{2.990}

\figsetplot{./figset/scatter/990.pdf}

\figsetgrpend

\figsetgrpstart
\figsetgrpnum{2.991}

\figsetplot{./figset/scatter/991.pdf}

\figsetgrpend

\figsetgrpstart
\figsetgrpnum{2.992}

\figsetplot{./figset/scatter/992.pdf}

\figsetgrpend

\figsetgrpstart
\figsetgrpnum{2.993}

\figsetplot{./figset/scatter/993.pdf}

\figsetgrpend

\figsetgrpstart
\figsetgrpnum{2.994}

\figsetplot{./figset/scatter/994.pdf}

\figsetgrpend

\figsetgrpstart
\figsetgrpnum{2.995}

\figsetplot{./figset/scatter/995.pdf}

\figsetgrpend

\figsetgrpstart
\figsetgrpnum{2.996}

\figsetplot{./figset/scatter/996.pdf}

\figsetgrpend

\figsetgrpstart
\figsetgrpnum{2.997}

\figsetplot{./figset/scatter/997.pdf}

\figsetgrpend

\figsetgrpstart
\figsetgrpnum{2.998}

\figsetplot{./figset/scatter/998.pdf}

\figsetgrpend

\figsetgrpstart
\figsetgrpnum{2.999}

\figsetplot{./figset/scatter/999.pdf}

\figsetgrpend

\figsetgrpstart
\figsetgrpnum{2.1000}

\figsetplot{./figset/scatter/1000.pdf}

\figsetgrpend

\figsetgrpstart
\figsetgrpnum{2.1001}

\figsetplot{./figset/scatter/1001.pdf}

\figsetgrpend

\figsetgrpstart
\figsetgrpnum{2.1002}

\figsetplot{./figset/scatter/1002.pdf}

\figsetgrpend

\figsetgrpstart
\figsetgrpnum{2.1003}

\figsetplot{./figset/scatter/1003.pdf}

\figsetgrpend

\figsetgrpstart
\figsetgrpnum{2.1004}

\figsetplot{./figset/scatter/1004.pdf}

\figsetgrpend

\figsetgrpstart
\figsetgrpnum{2.1005}

\figsetplot{./figset/scatter/1005.pdf}

\figsetgrpend

\figsetgrpstart
\figsetgrpnum{2.1006}

\figsetplot{./figset/scatter/1006.pdf}

\figsetgrpend

\figsetgrpstart
\figsetgrpnum{2.1007}

\figsetplot{./figset/scatter/1007.pdf}

\figsetgrpend

\figsetgrpstart
\figsetgrpnum{2.1008}

\figsetplot{./figset/scatter/1008.pdf}

\figsetgrpend

\figsetgrpstart
\figsetgrpnum{2.1009}

\figsetplot{./figset/scatter/1009.pdf}

\figsetgrpend

\figsetgrpstart
\figsetgrpnum{2.1010}

\figsetplot{./figset/scatter/1010.pdf}

\figsetgrpend

\figsetgrpstart
\figsetgrpnum{2.1011}

\figsetplot{./figset/scatter/1011.pdf}

\figsetgrpend

\figsetgrpstart
\figsetgrpnum{2.1012}

\figsetplot{./figset/scatter/1012.pdf}

\figsetgrpend

\figsetgrpstart
\figsetgrpnum{2.1013}

\figsetplot{./figset/scatter/1013.pdf}

\figsetgrpend

\figsetgrpstart
\figsetgrpnum{2.1014}

\figsetplot{./figset/scatter/1014.pdf}

\figsetgrpend

\figsetgrpstart
\figsetgrpnum{2.1015}

\figsetplot{./figset/scatter/1015.pdf}

\figsetgrpend

\figsetgrpstart
\figsetgrpnum{2.1016}

\figsetplot{./figset/scatter/1016.pdf}

\figsetgrpend

\figsetgrpstart
\figsetgrpnum{2.1017}

\figsetplot{./figset/scatter/1017.pdf}

\figsetgrpend

\figsetgrpstart
\figsetgrpnum{2.1018}

\figsetplot{./figset/scatter/1018.pdf}

\figsetgrpend

\figsetgrpstart
\figsetgrpnum{2.1019}

\figsetplot{./figset/scatter/1019.pdf}

\figsetgrpend

\figsetgrpstart
\figsetgrpnum{2.1020}

\figsetplot{./figset/scatter/1020.pdf}

\figsetgrpend

\figsetgrpstart
\figsetgrpnum{2.1021}

\figsetplot{./figset/scatter/1021.pdf}

\figsetgrpend

\figsetgrpstart
\figsetgrpnum{2.1022}

\figsetplot{./figset/scatter/1022.pdf}

\figsetgrpend

\figsetgrpstart
\figsetgrpnum{2.1023}

\figsetplot{./figset/scatter/1023.pdf}

\figsetgrpend

\figsetgrpstart
\figsetgrpnum{2.1024}

\figsetplot{./figset/scatter/1024.pdf}

\figsetgrpend

\figsetgrpstart
\figsetgrpnum{2.1025}

\figsetplot{./figset/scatter/1025.pdf}

\figsetgrpend

\figsetgrpstart
\figsetgrpnum{2.1026}

\figsetplot{./figset/scatter/1026.pdf}

\figsetgrpend

\figsetgrpstart
\figsetgrpnum{2.1027}

\figsetplot{./figset/scatter/1027.pdf}

\figsetgrpend

\figsetgrpstart
\figsetgrpnum{2.1028}

\figsetplot{./figset/scatter/1028.pdf}

\figsetgrpend

\figsetgrpstart
\figsetgrpnum{2.1029}

\figsetplot{./figset/scatter/1029.pdf}

\figsetgrpend

\figsetgrpstart
\figsetgrpnum{2.1030}

\figsetplot{./figset/scatter/1030.pdf}

\figsetgrpend

\figsetgrpstart
\figsetgrpnum{2.1031}

\figsetplot{./figset/scatter/1031.pdf}

\figsetgrpend

\figsetgrpstart
\figsetgrpnum{2.1032}

\figsetplot{./figset/scatter/1032.pdf}

\figsetgrpend

\figsetgrpstart
\figsetgrpnum{2.1033}

\figsetplot{./figset/scatter/1033.pdf}

\figsetgrpend

\figsetgrpstart
\figsetgrpnum{2.1034}

\figsetplot{./figset/scatter/1034.pdf}

\figsetgrpend

\figsetgrpstart
\figsetgrpnum{2.1035}

\figsetplot{./figset/scatter/1035.pdf}

\figsetgrpend

\figsetgrpstart
\figsetgrpnum{2.1036}

\figsetplot{./figset/scatter/1036.pdf}

\figsetgrpend

\figsetgrpstart
\figsetgrpnum{2.1037}

\figsetplot{./figset/scatter/1037.pdf}

\figsetgrpend

\figsetgrpstart
\figsetgrpnum{2.1038}

\figsetplot{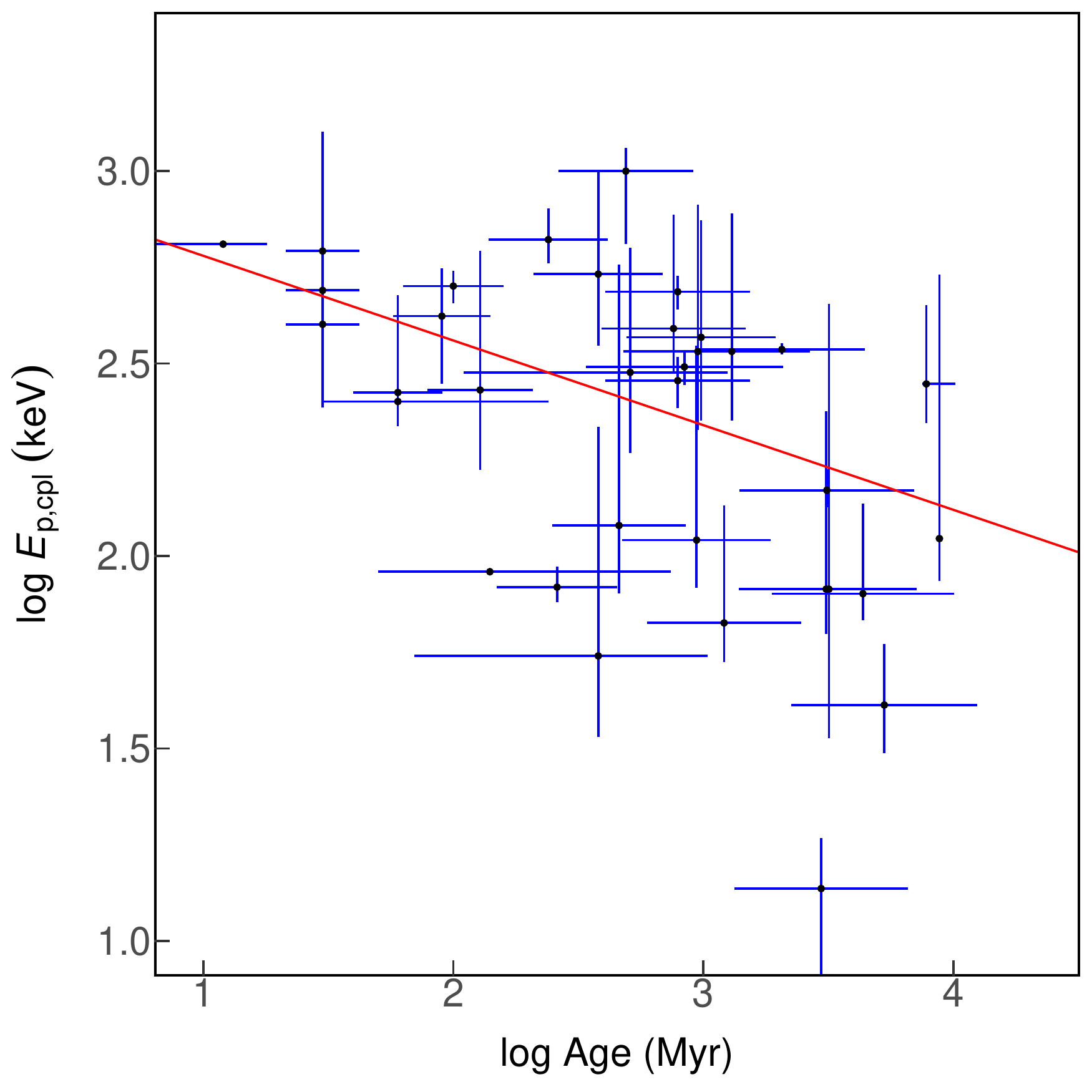}

\figsetgrpend

\figsetgrpstart
\figsetgrpnum{2.1039}

\figsetplot{./figset/scatter/1039.pdf}

\figsetgrpend

\figsetgrpstart
\figsetgrpnum{2.1040}

\figsetplot{./figset/scatter/1040.pdf}

\figsetgrpend

\figsetgrpstart
\figsetgrpnum{2.1041}

\figsetplot{./figset/scatter/1041.pdf}

\figsetgrpend

\figsetgrpstart
\figsetgrpnum{2.1042}

\figsetplot{./figset/scatter/1042.pdf}

\figsetgrpend

\figsetgrpstart
\figsetgrpnum{2.1043}

\figsetplot{./figset/scatter/1043.pdf}

\figsetgrpend

\figsetgrpstart
\figsetgrpnum{2.1044}

\figsetplot{./figset/scatter/1044.pdf}

\figsetgrpend

\figsetgrpstart
\figsetgrpnum{2.1045}

\figsetplot{./figset/scatter/1045.pdf}

\figsetgrpend

\figsetgrpstart
\figsetgrpnum{2.1046}

\figsetplot{./figset/scatter/1046.pdf}

\figsetgrpend

\figsetgrpstart
\figsetgrpnum{2.1047}

\figsetplot{./figset/scatter/1047.pdf}

\figsetgrpend

\figsetgrpstart
\figsetgrpnum{2.1048}

\figsetplot{./figset/scatter/1048.pdf}

\figsetgrpend

\figsetgrpstart
\figsetgrpnum{2.1049}

\figsetplot{./figset/scatter/1049.pdf}

\figsetgrpend

\figsetgrpstart
\figsetgrpnum{2.1050}

\figsetplot{./figset/scatter/1050.pdf}

\figsetgrpend

\figsetgrpstart
\figsetgrpnum{2.1051}

\figsetplot{./figset/scatter/1051.pdf}

\figsetgrpend

\figsetgrpstart
\figsetgrpnum{2.1052}

\figsetplot{./figset/scatter/1052.pdf}

\figsetgrpend

\figsetgrpstart
\figsetgrpnum{2.1053}

\figsetplot{./figset/scatter/1053.pdf}

\figsetgrpend

\figsetgrpstart
\figsetgrpnum{2.1054}

\figsetplot{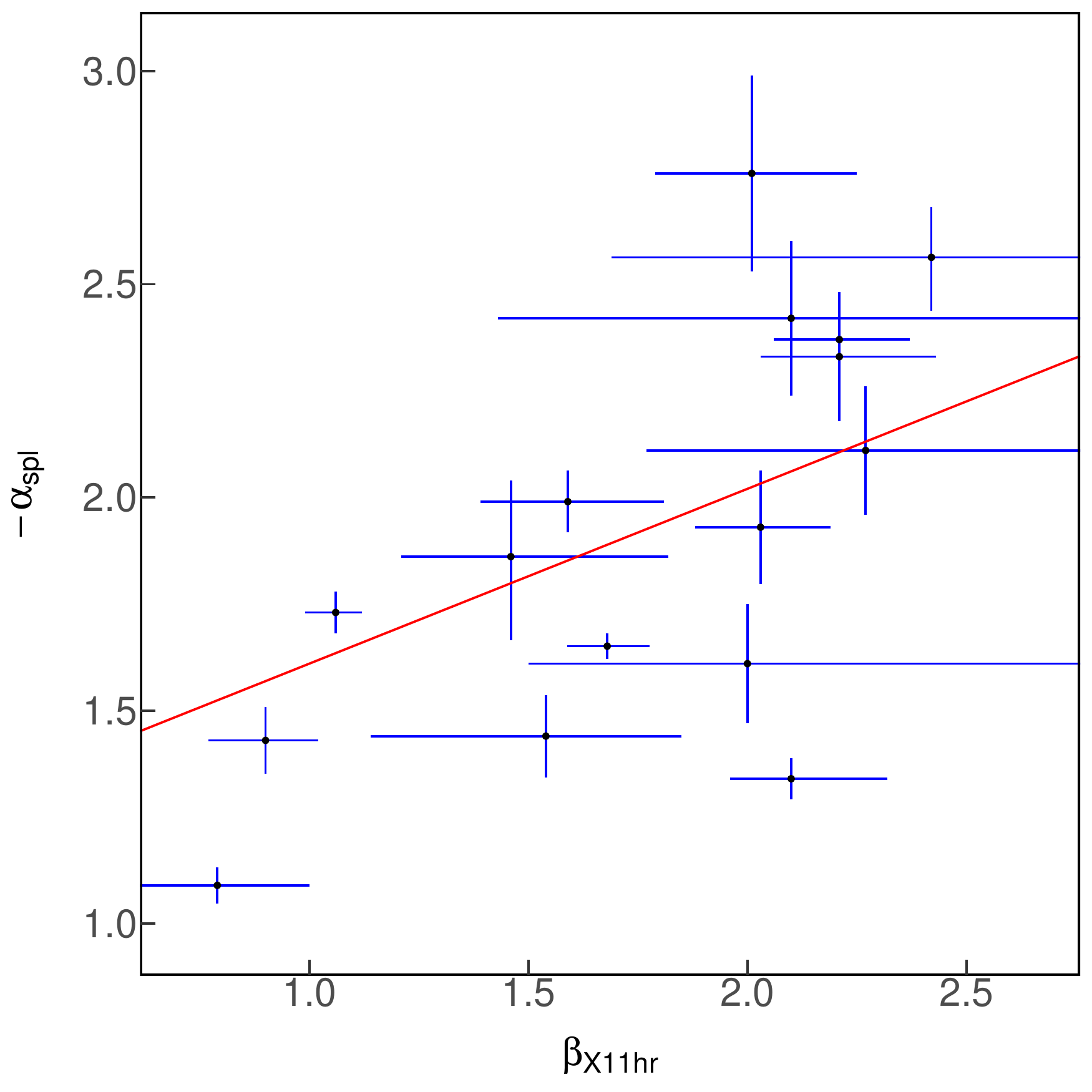}

\figsetgrpend

\figsetgrpstart
\figsetgrpnum{2.1055}

\figsetplot{./figset/scatter/1055.pdf}

\figsetgrpend

\figsetgrpstart
\figsetgrpnum{2.1056}

\figsetplot{./figset/scatter/1056.pdf}

\figsetgrpend

\figsetgrpstart
\figsetgrpnum{2.1057}

\figsetplot{./figset/scatter/1057.pdf}

\figsetgrpend

\figsetgrpstart
\figsetgrpnum{2.1058}

\figsetplot{./figset/scatter/1058.pdf}

\figsetgrpend

\figsetgrpstart
\figsetgrpnum{2.1059}

\figsetplot{./figset/scatter/1059.pdf}

\figsetgrpend

\figsetgrpstart
\figsetgrpnum{2.1060}

\figsetplot{./figset/scatter/1060.pdf}

\figsetgrpend

\figsetgrpstart
\figsetgrpnum{2.1061}

\figsetplot{./figset/scatter/1061.pdf}

\figsetgrpend

\figsetgrpstart
\figsetgrpnum{2.1062}

\figsetplot{./figset/scatter/1062.pdf}

\figsetgrpend

\figsetgrpstart
\figsetgrpnum{2.1063}

\figsetplot{./figset/scatter/1063.pdf}

\figsetgrpend

\figsetgrpstart
\figsetgrpnum{2.1064}

\figsetplot{./figset/scatter/1064.pdf}

\figsetgrpend

\figsetgrpstart
\figsetgrpnum{2.1065}

\figsetplot{./figset/scatter/1065.pdf}

\figsetgrpend

\figsetgrpstart
\figsetgrpnum{2.1066}

\figsetplot{./figset/scatter/1066.pdf}

\figsetgrpend

\figsetgrpstart
\figsetgrpnum{2.1067}

\figsetplot{./figset/scatter/1067.pdf}

\figsetgrpend

\figsetgrpstart
\figsetgrpnum{2.1068}

\figsetplot{./figset/scatter/1068.pdf}

\figsetgrpend

\figsetgrpstart
\figsetgrpnum{2.1069}

\figsetplot{./figset/scatter/1069.pdf}

\figsetgrpend

\figsetgrpstart
\figsetgrpnum{2.1070}

\figsetplot{./figset/scatter/1070.pdf}

\figsetgrpend

\figsetgrpstart
\figsetgrpnum{2.1071}

\figsetplot{./figset/scatter/1071.pdf}

\figsetgrpend

\figsetgrpstart
\figsetgrpnum{2.1072}

\figsetplot{./figset/scatter/1072.pdf}

\figsetgrpend

\figsetgrpstart
\figsetgrpnum{2.1073}

\figsetplot{./figset/scatter/1073.pdf}

\figsetgrpend

\figsetgrpstart
\figsetgrpnum{2.1074}

\figsetplot{./figset/scatter/1074.pdf}

\figsetgrpend

\figsetgrpstart
\figsetgrpnum{2.1075}

\figsetplot{./figset/scatter/1075.pdf}

\figsetgrpend

\figsetgrpstart
\figsetgrpnum{2.1076}

\figsetplot{./figset/scatter/1076.pdf}

\figsetgrpend

\figsetgrpstart
\figsetgrpnum{2.1077}

\figsetplot{./figset/scatter/1077.pdf}

\figsetgrpend

\figsetgrpstart
\figsetgrpnum{2.1078}

\figsetplot{./figset/scatter/1078.pdf}

\figsetgrpend

\figsetgrpstart
\figsetgrpnum{2.1079}

\figsetplot{./figset/scatter/1079.pdf}

\figsetgrpend

\figsetgrpstart
\figsetgrpnum{2.1080}

\figsetplot{./figset/scatter/1080.pdf}

\figsetgrpend

\figsetgrpstart
\figsetgrpnum{2.1081}

\figsetplot{./figset/scatter/1081.pdf}

\figsetgrpend

\figsetgrpstart
\figsetgrpnum{2.1082}

\figsetplot{./figset/scatter/1082.pdf}

\figsetgrpend

\figsetgrpstart
\figsetgrpnum{2.1083}

\figsetplot{./figset/scatter/1083.pdf}

\figsetgrpend

\figsetgrpstart
\figsetgrpnum{2.1084}

\figsetplot{./figset/scatter/1084.pdf}

\figsetgrpend

\figsetgrpstart
\figsetgrpnum{2.1085}

\figsetplot{./figset/scatter/1085.pdf}

\figsetgrpend

\figsetgrpstart
\figsetgrpnum{2.1086}

\figsetplot{./figset/scatter/1086.pdf}

\figsetgrpend

\figsetgrpstart
\figsetgrpnum{2.1087}

\figsetplot{./figset/scatter/1087.pdf}

\figsetgrpend

\figsetgrpstart
\figsetgrpnum{2.1088}

\figsetplot{./figset/scatter/1088.pdf}

\figsetgrpend

\figsetgrpstart
\figsetgrpnum{2.1089}

\figsetplot{./figset/scatter/1089.pdf}

\figsetgrpend

\figsetgrpstart
\figsetgrpnum{2.1090}

\figsetplot{./figset/scatter/1090.pdf}

\figsetgrpend

\figsetgrpstart
\figsetgrpnum{2.1091}

\figsetplot{./figset/scatter/1091.pdf}

\figsetgrpend

\figsetgrpstart
\figsetgrpnum{2.1092}

\figsetplot{./figset/scatter/1092.pdf}

\figsetgrpend

\figsetgrpstart
\figsetgrpnum{2.1093}

\figsetplot{./figset/scatter/1093.pdf}

\figsetgrpend

\figsetgrpstart
\figsetgrpnum{2.1094}

\figsetplot{./figset/scatter/1094.pdf}

\figsetgrpend

\figsetgrpstart
\figsetgrpnum{2.1095}

\figsetplot{./figset/scatter/1095.pdf}

\figsetgrpend

\figsetgrpstart
\figsetgrpnum{2.1096}

\figsetplot{./figset/scatter/1096.pdf}

\figsetgrpend

\figsetgrpstart
\figsetgrpnum{2.1097}

\figsetplot{./figset/scatter/1097.pdf}

\figsetgrpend

\figsetgrpstart
\figsetgrpnum{2.1098}

\figsetplot{./figset/scatter/1098.pdf}

\figsetgrpend

\figsetgrpstart
\figsetgrpnum{2.1099}

\figsetplot{./figset/scatter/1099.pdf}

\figsetgrpend

\figsetgrpstart
\figsetgrpnum{2.1100}

\figsetplot{./figset/scatter/1100.pdf}

\figsetgrpend

\figsetgrpstart
\figsetgrpnum{2.1101}

\figsetplot{./figset/scatter/1101.pdf}

\figsetgrpend

\figsetgrpstart
\figsetgrpnum{2.1102}

\figsetplot{./figset/scatter/1102.pdf}

\figsetgrpend

\figsetgrpstart
\figsetgrpnum{2.1103}

\figsetplot{./figset/scatter/1103.pdf}

\figsetgrpend

\figsetgrpstart
\figsetgrpnum{2.1104}

\figsetplot{./figset/scatter/1104.pdf}

\figsetgrpend

\figsetgrpstart
\figsetgrpnum{2.1105}

\figsetplot{./figset/scatter/1105.pdf}

\figsetgrpend

\figsetgrpstart
\figsetgrpnum{2.1106}

\figsetplot{./figset/scatter/1106.pdf}

\figsetgrpend

\figsetgrpstart
\figsetgrpnum{2.1107}

\figsetplot{./figset/scatter/1107.pdf}

\figsetgrpend

\figsetgrpstart
\figsetgrpnum{2.1108}

\figsetplot{./figset/scatter/1108.pdf}

\figsetgrpend

\figsetgrpstart
\figsetgrpnum{2.1109}

\figsetplot{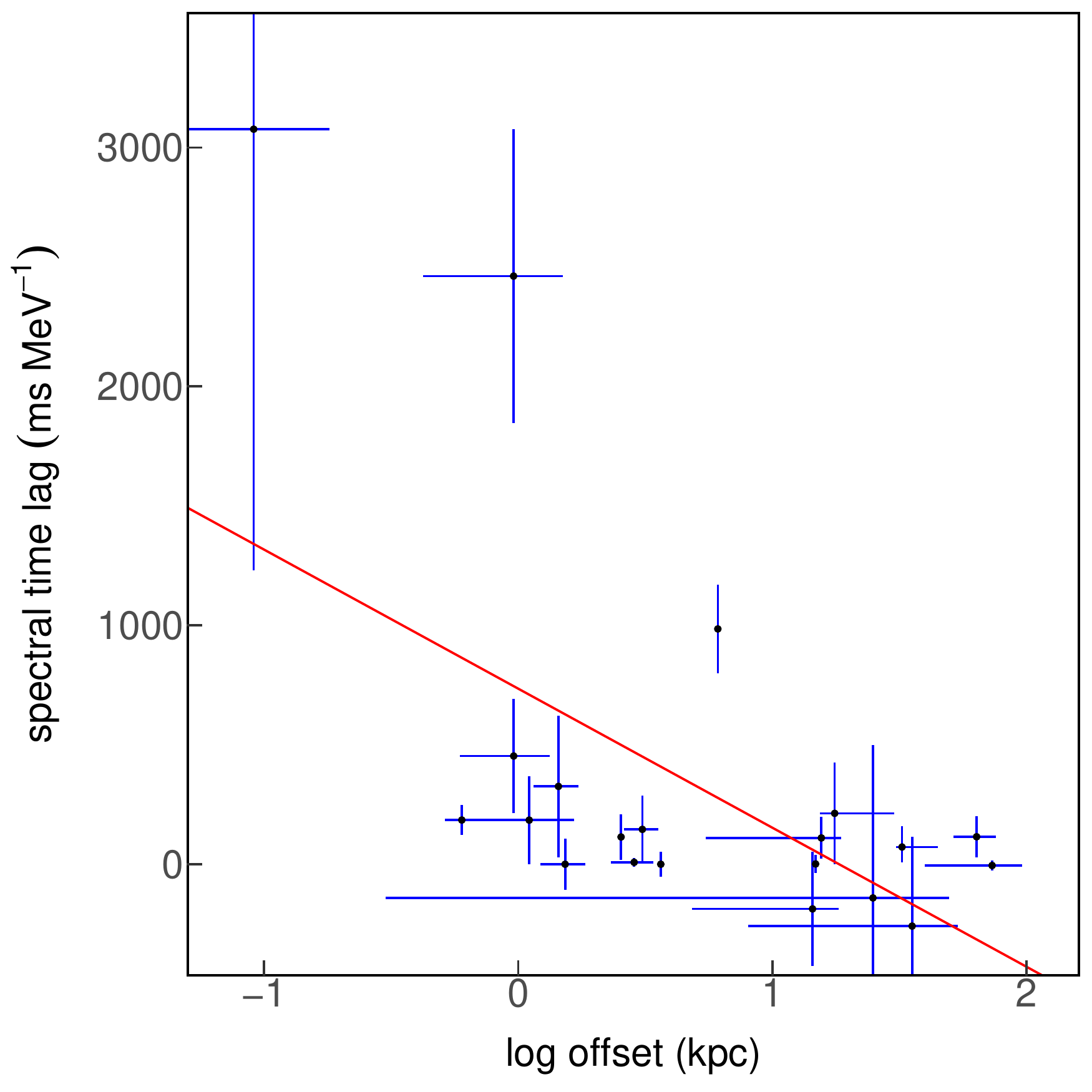}

\figsetgrpend

\figsetgrpstart
\figsetgrpnum{2.1110}

\figsetplot{./figset/scatter/1110.pdf}

\figsetgrpend

\figsetgrpstart
\figsetgrpnum{2.1111}

\figsetplot{./figset/scatter/1111.pdf}

\figsetgrpend

\figsetgrpstart
\figsetgrpnum{2.1112}

\figsetplot{./figset/scatter/1112.pdf}

\figsetgrpend

\figsetgrpstart
\figsetgrpnum{2.1113}

\figsetplot{./figset/scatter/1113.pdf}

\figsetgrpend

\figsetgrpstart
\figsetgrpnum{2.1114}

\figsetplot{./figset/scatter/1114.pdf}

\figsetgrpend

\figsetgrpstart
\figsetgrpnum{2.1115}

\figsetplot{./figset/scatter/1115.pdf}

\figsetgrpend

\figsetgrpstart
\figsetgrpnum{2.1116}

\figsetplot{./figset/scatter/1116.pdf}

\figsetgrpend

\figsetgrpstart
\figsetgrpnum{2.1117}

\figsetplot{./figset/scatter/1117.pdf}

\figsetgrpend

\figsetgrpstart
\figsetgrpnum{2.1118}

\figsetplot{./figset/scatter/1118.pdf}

\figsetgrpend

\figsetgrpstart
\figsetgrpnum{2.1119}

\figsetplot{./figset/scatter/1119.pdf}

\figsetgrpend

\figsetgrpstart
\figsetgrpnum{2.1120}

\figsetplot{./figset/scatter/1120.pdf}

\figsetgrpend

\figsetgrpstart
\figsetgrpnum{2.1121}

\figsetplot{./figset/scatter/1121.pdf}

\figsetgrpend

\figsetgrpstart
\figsetgrpnum{2.1122}

\figsetplot{./figset/scatter/1122.pdf}

\figsetgrpend

\figsetgrpstart
\figsetgrpnum{2.1123}

\figsetplot{./figset/scatter/1123.pdf}

\figsetgrpend

\figsetgrpstart
\figsetgrpnum{2.1124}

\figsetplot{./figset/scatter/1124.pdf}

\figsetgrpend

\figsetgrpstart
\figsetgrpnum{2.1125}

\figsetplot{./figset/scatter/1125.pdf}

\figsetgrpend

\figsetgrpstart
\figsetgrpnum{2.1126}

\figsetplot{./figset/scatter/1126.pdf}

\figsetgrpend

\figsetgrpstart
\figsetgrpnum{2.1127}

\figsetplot{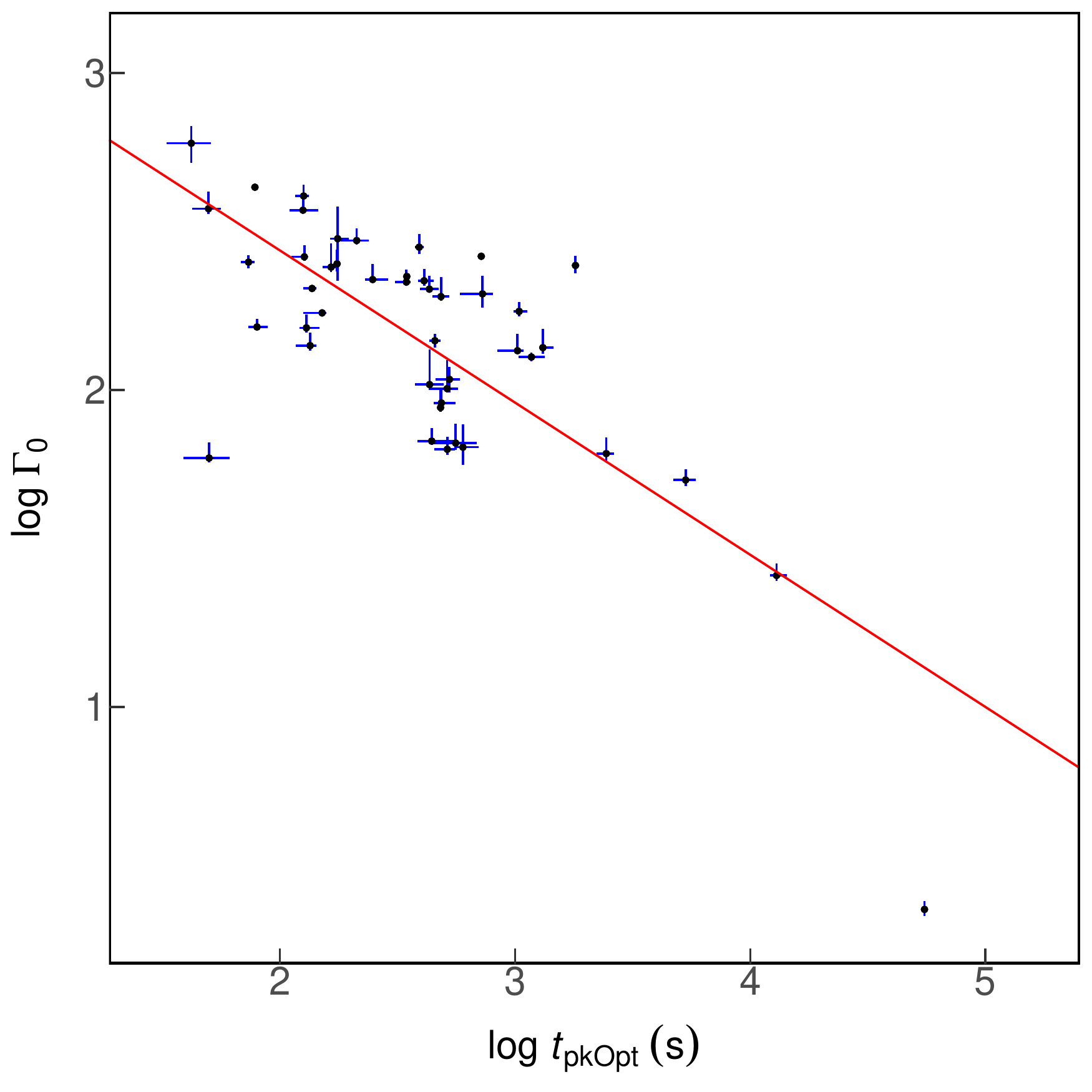}

\figsetgrpend

\figsetgrpstart
\figsetgrpnum{2.1128}

\figsetplot{./figset/scatter/1128.pdf}

\figsetgrpend

\figsetgrpstart
\figsetgrpnum{2.1129}

\figsetplot{./figset/scatter/1129.pdf}

\figsetgrpend

\figsetgrpstart
\figsetgrpnum{2.1130}

\figsetplot{./figset/scatter/1130.pdf}

\figsetgrpend

\figsetgrpstart
\figsetgrpnum{2.1131}

\figsetplot{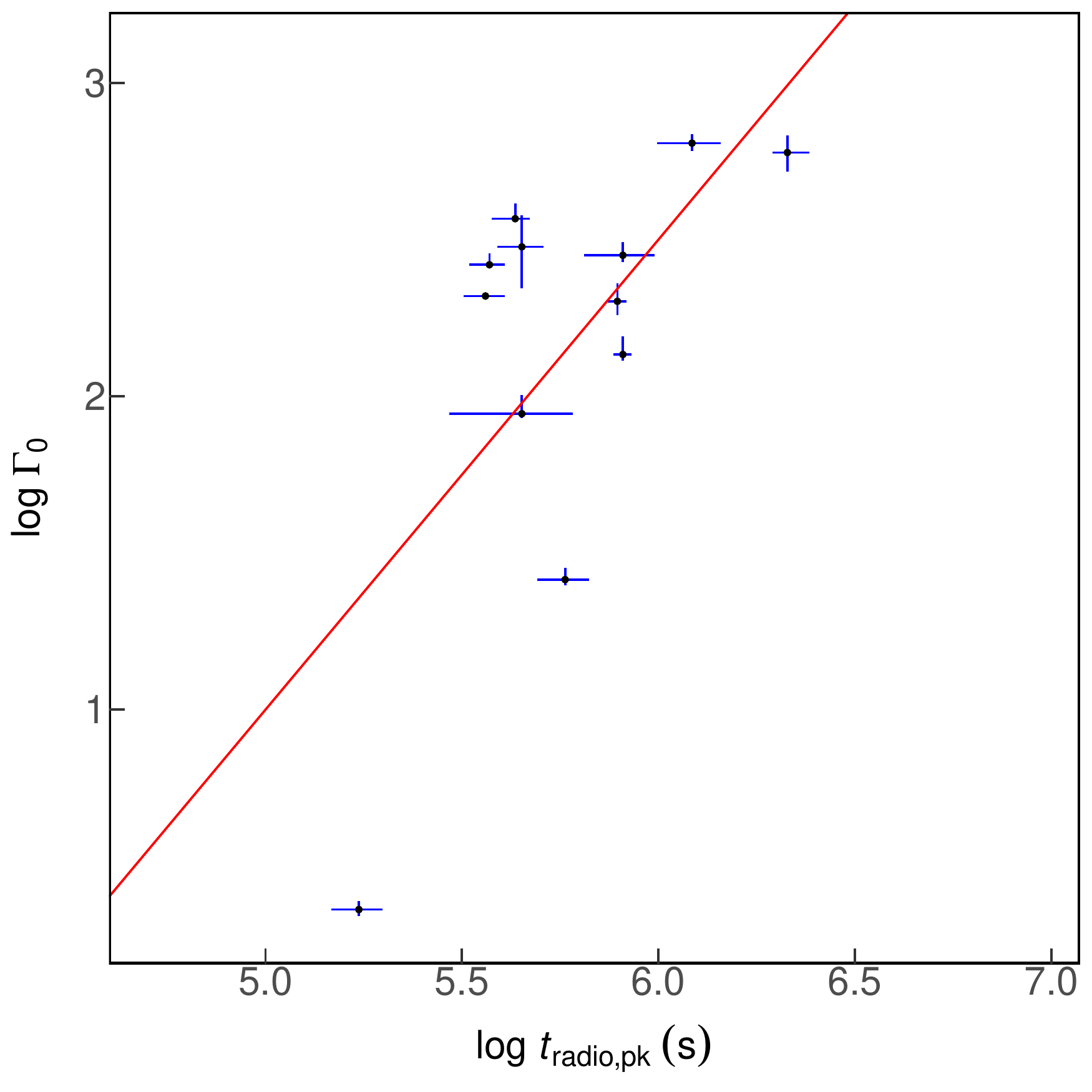}

\figsetgrpend

\figsetgrpstart
\figsetgrpnum{2.1132}

\figsetplot{./figset/scatter/1132.pdf}

\figsetgrpend

\figsetgrpstart
\figsetgrpnum{2.1133}

\figsetplot{./figset/scatter/1133.pdf}

\figsetgrpend

\figsetgrpstart
\figsetgrpnum{2.1134}

\figsetplot{./figset/scatter/1134.pdf}

\figsetgrpend

\figsetgrpstart
\figsetgrpnum{2.1135}

\figsetplot{./figset/scatter/1135.pdf}

\figsetgrpend

\figsetgrpstart
\figsetgrpnum{2.1136}

\figsetplot{./figset/scatter/1136.pdf}

\figsetgrpend

\figsetgrpstart
\figsetgrpnum{2.1137}

\figsetplot{./figset/scatter/1137.pdf}

\figsetgrpend

\figsetgrpstart
\figsetgrpnum{2.1138}

\figsetplot{./figset/scatter/1138.pdf}

\figsetgrpend

\figsetgrpstart
\figsetgrpnum{2.1139}

\figsetplot{./figset/scatter/1139.pdf}

\figsetgrpend

\figsetgrpstart
\figsetgrpnum{2.1140}

\figsetplot{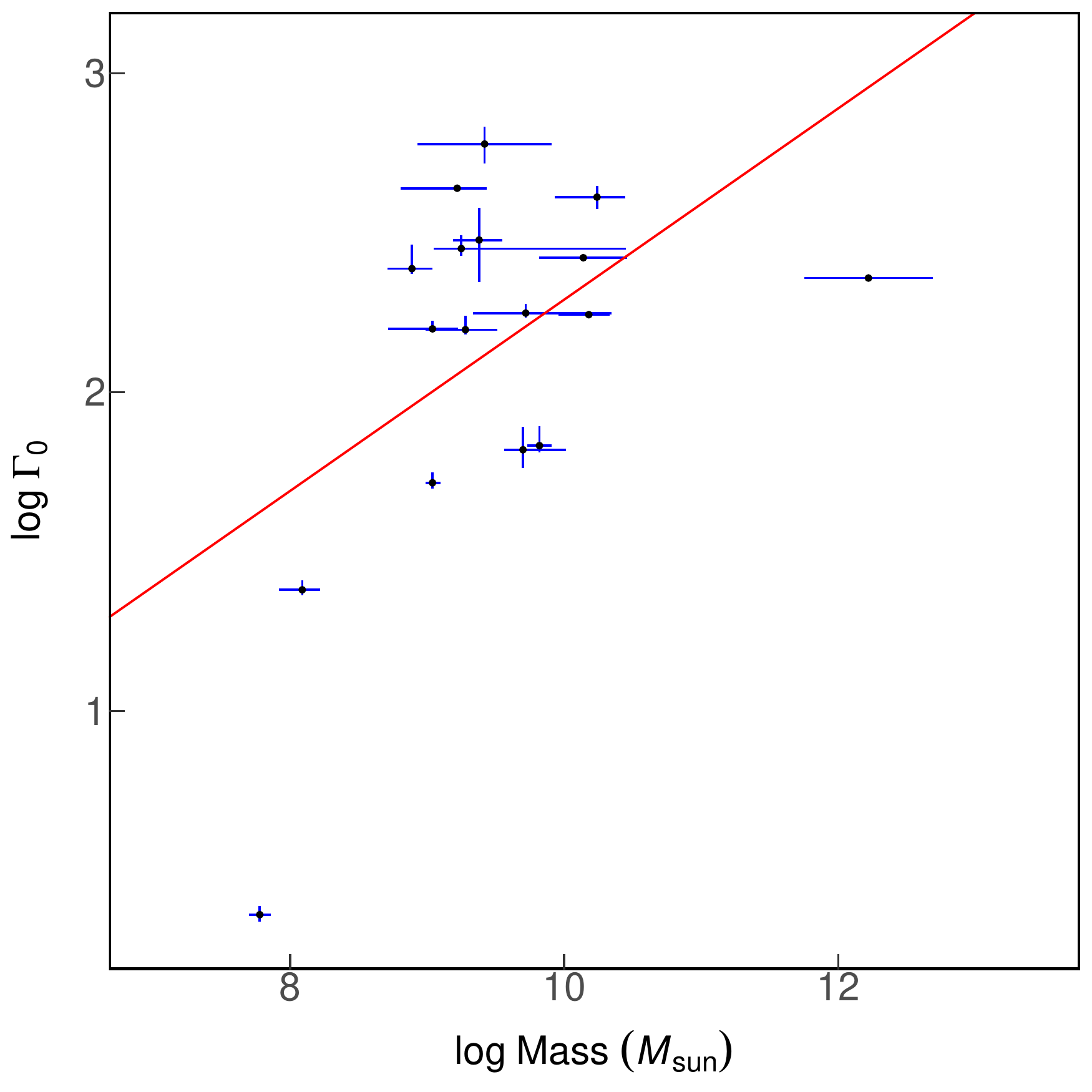}

\figsetgrpend

\figsetgrpstart
\figsetgrpnum{2.1141}

\figsetplot{./figset/scatter/1141.pdf}

\figsetgrpend

\figsetgrpstart
\figsetgrpnum{2.1142}

\figsetplot{./figset/scatter/1142.pdf}

\figsetgrpend

\figsetgrpstart
\figsetgrpnum{2.1143}

\figsetplot{./figset/scatter/1143.pdf}

\figsetgrpend

\figsetgrpstart
\figsetgrpnum{2.1144}

\figsetplot{./figset/scatter/1144.pdf}

\figsetgrpend

\figsetgrpstart
\figsetgrpnum{2.1145}

\figsetplot{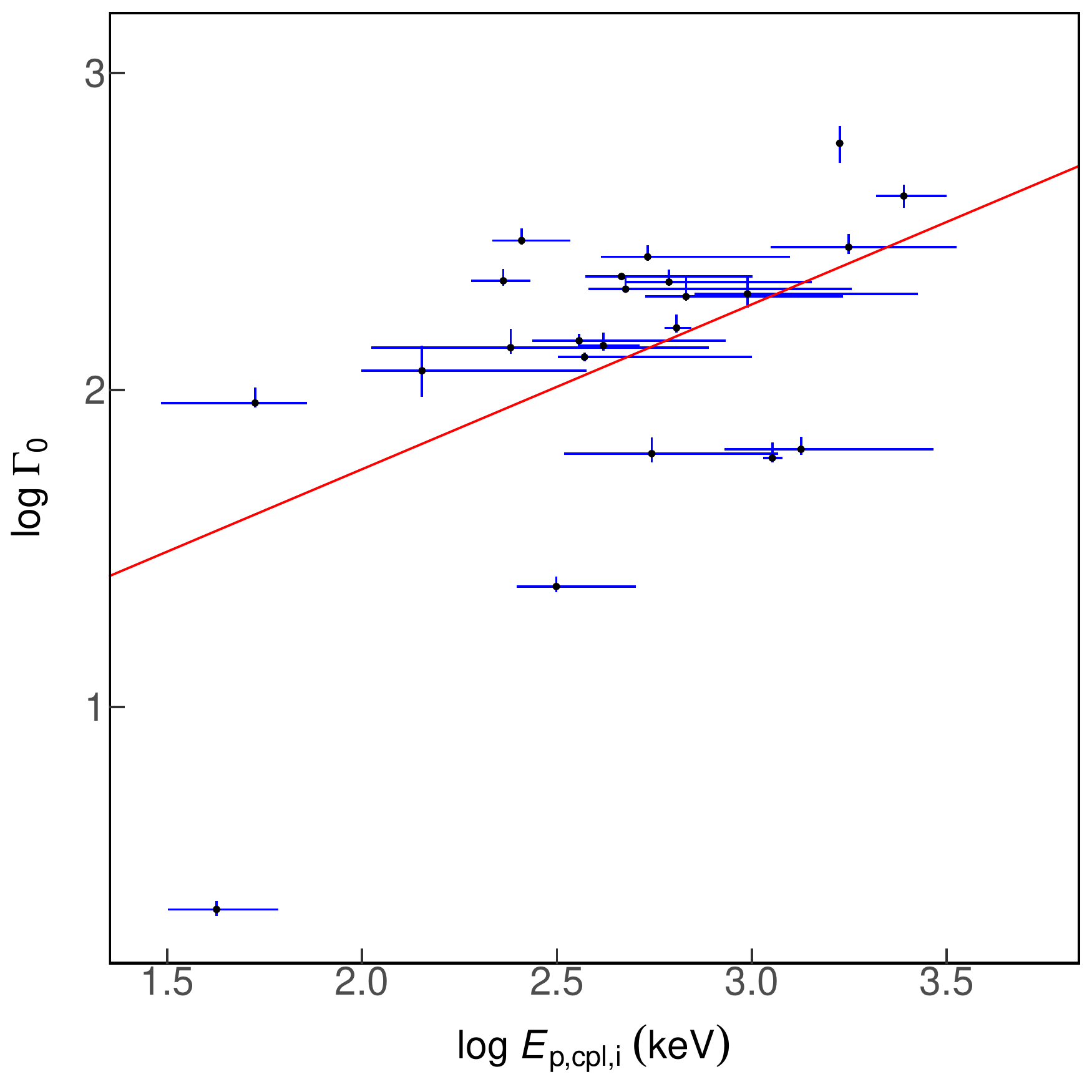}

\figsetgrpend

\figsetgrpstart
\figsetgrpnum{2.1146}

\figsetplot{./figset/scatter/1146.pdf}

\figsetgrpend

\figsetgrpstart
\figsetgrpnum{2.1147}

\figsetplot{./figset/scatter/1147.pdf}

\figsetgrpend

\figsetgrpstart
\figsetgrpnum{2.1148}

\figsetplot{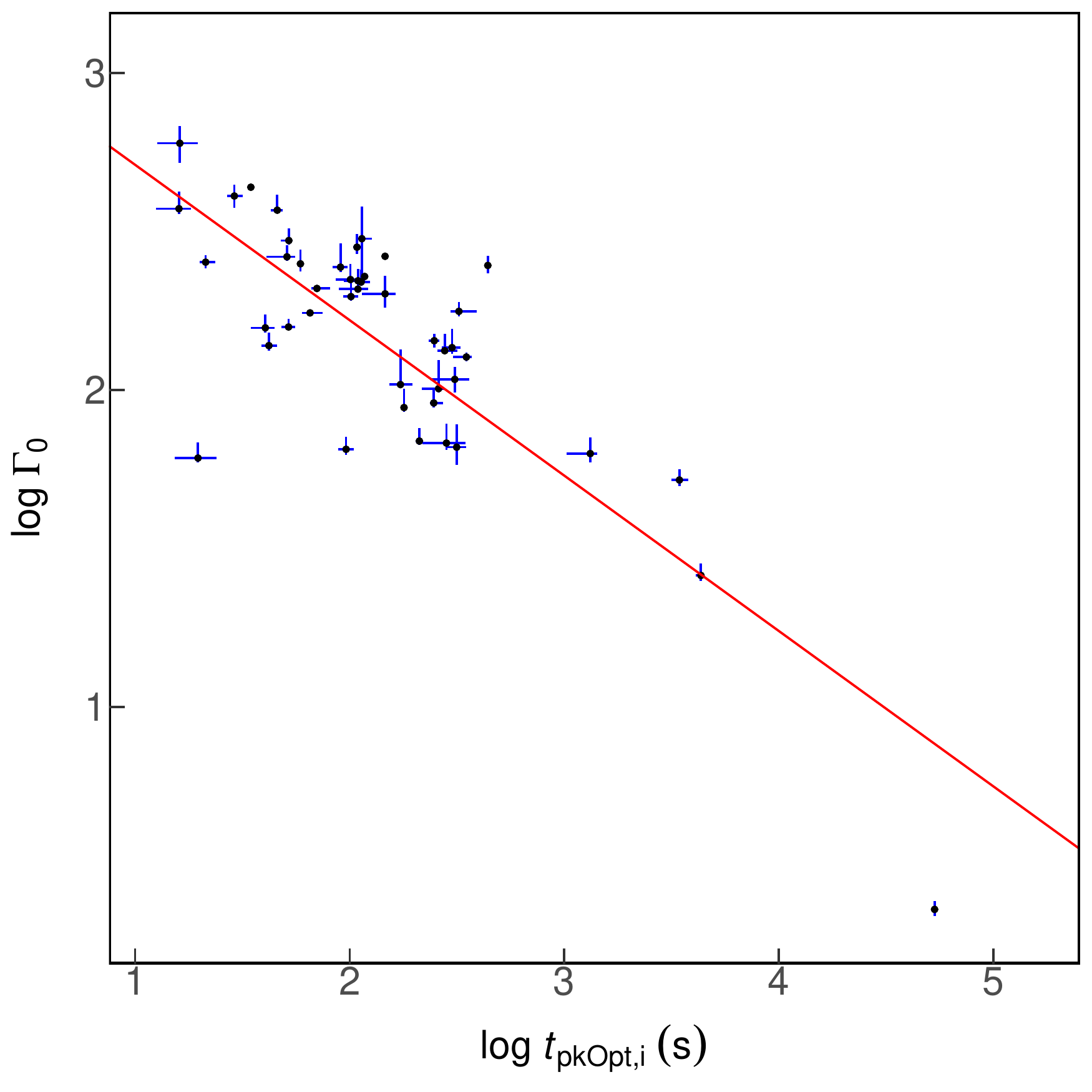}

\figsetgrpend

\figsetgrpstart
\figsetgrpnum{2.1149}

\figsetplot{./figset/scatter/1149.pdf}

\figsetgrpend

\figsetgrpstart
\figsetgrpnum{2.1150}

\figsetplot{./figset/scatter/1150.pdf}

\figsetgrpend

\figsetgrpstart
\figsetgrpnum{2.1151}

\figsetplot{./figset/scatter/1151.pdf}

\figsetgrpend

\figsetgrpstart
\figsetgrpnum{2.1152}

\figsetplot{./figset/scatter/1152.pdf}

\figsetgrpend

\figsetgrpstart
\figsetgrpnum{2.1153}

\figsetplot{./figset/scatter/1153.pdf}

\figsetgrpend

\figsetgrpstart
\figsetgrpnum{2.1154}

\figsetplot{./figset/scatter/1154.pdf}

\figsetgrpend

\figsetgrpstart
\figsetgrpnum{2.1155}

\figsetplot{./figset/scatter/1155.pdf}

\figsetgrpend

\figsetgrpstart
\figsetgrpnum{2.1156}

\figsetplot{./figset/scatter/1156.pdf}

\figsetgrpend

\figsetgrpstart
\figsetgrpnum{2.1157}

\figsetplot{./figset/scatter/1157.pdf}

\figsetgrpend

\figsetgrpstart
\figsetgrpnum{2.1158}

\figsetplot{./figset/scatter/1158.pdf}

\figsetgrpend

\figsetgrpstart
\figsetgrpnum{2.1159}

\figsetplot{./figset/scatter/1159.pdf}

\figsetgrpend

\figsetgrpstart
\figsetgrpnum{2.1160}

\figsetplot{./figset/scatter/1160.pdf}

\figsetgrpend

\figsetgrpstart
\figsetgrpnum{2.1161}

\figsetplot{./figset/scatter/1161.pdf}

\figsetgrpend

\figsetgrpstart
\figsetgrpnum{2.1162}

\figsetplot{./figset/scatter/1162.pdf}

\figsetgrpend

\figsetgrpstart
\figsetgrpnum{2.1163}

\figsetplot{./figset/scatter/1163.pdf}

\figsetgrpend

\figsetgrpstart
\figsetgrpnum{2.1164}

\figsetplot{./figset/scatter/1164.pdf}

\figsetgrpend

\figsetgrpstart
\figsetgrpnum{2.1165}

\figsetplot{./figset/scatter/1165.pdf}

\figsetgrpend

\figsetgrpstart
\figsetgrpnum{2.1166}

\figsetplot{./figset/scatter/1166.pdf}

\figsetgrpend

\figsetgrpstart
\figsetgrpnum{2.1167}

\figsetplot{./figset/scatter/1167.pdf}

\figsetgrpend

\figsetgrpstart
\figsetgrpnum{2.1168}

\figsetplot{./figset/scatter/1168.pdf}

\figsetgrpend

\figsetgrpstart
\figsetgrpnum{2.1169}

\figsetplot{./figset/scatter/1169.pdf}

\figsetgrpend

\figsetgrpstart
\figsetgrpnum{2.1170}

\figsetplot{./figset/scatter/1170.pdf}

\figsetgrpend

\figsetgrpstart
\figsetgrpnum{2.1171}

\figsetplot{./figset/scatter/1171.pdf}

\figsetgrpend

\figsetgrpstart
\figsetgrpnum{2.1172}

\figsetplot{./figset/scatter/1172.pdf}

\figsetgrpend

\figsetgrpstart
\figsetgrpnum{2.1173}

\figsetplot{./figset/scatter/1173.pdf}

\figsetgrpend

\figsetgrpstart
\figsetgrpnum{2.1174}

\figsetplot{./figset/scatter/1174.pdf}

\figsetgrpend

\figsetgrpstart
\figsetgrpnum{2.1175}

\figsetplot{./figset/scatter/1175.pdf}

\figsetgrpend

\figsetgrpstart
\figsetgrpnum{2.1176}

\figsetplot{./figset/scatter/1176.pdf}

\figsetgrpend

\figsetgrpstart
\figsetgrpnum{2.1177}

\figsetplot{./figset/scatter/1177.pdf}

\figsetgrpend

\figsetgrpstart
\figsetgrpnum{2.1178}

\figsetplot{./figset/scatter/1178.pdf}

\figsetgrpend

\figsetgrpstart
\figsetgrpnum{2.1179}

\figsetplot{./figset/scatter/1179.pdf}

\figsetgrpend

\figsetgrpstart
\figsetgrpnum{2.1180}

\figsetplot{./figset/scatter/1180.pdf}

\figsetgrpend

\figsetgrpstart
\figsetgrpnum{2.1181}

\figsetplot{./figset/scatter/1181.pdf}

\figsetgrpend

\figsetgrpstart
\figsetgrpnum{2.1182}

\figsetplot{./figset/scatter/1182.pdf}

\figsetgrpend

\figsetgrpstart
\figsetgrpnum{2.1183}

\figsetplot{./figset/scatter/1183.pdf}

\figsetgrpend

\figsetgrpstart
\figsetgrpnum{2.1184}

\figsetplot{./figset/scatter/1184.pdf}

\figsetgrpend

\figsetgrpstart
\figsetgrpnum{2.1185}

\figsetplot{./figset/scatter/1185.pdf}

\figsetgrpend

\figsetgrpstart
\figsetgrpnum{2.1186}

\figsetplot{./figset/scatter/1186.pdf}

\figsetgrpend

\figsetgrpstart
\figsetgrpnum{2.1187}

\figsetplot{./figset/scatter/1187.pdf}

\figsetgrpend

\figsetgrpstart
\figsetgrpnum{2.1188}

\figsetplot{./figset/scatter/1188.pdf}

\figsetgrpend

\figsetgrpstart
\figsetgrpnum{2.1189}

\figsetplot{./figset/scatter/1189.pdf}

\figsetgrpend

\figsetgrpstart
\figsetgrpnum{2.1190}

\figsetplot{./figset/scatter/1190.pdf}

\figsetgrpend

\figsetgrpstart
\figsetgrpnum{2.1191}

\figsetplot{./figset/scatter/1191.pdf}

\figsetgrpend

\figsetgrpstart
\figsetgrpnum{2.1192}

\figsetplot{./figset/scatter/1192.pdf}

\figsetgrpend

\figsetgrpstart
\figsetgrpnum{2.1193}

\figsetplot{./figset/scatter/1193.pdf}

\figsetgrpend

\figsetgrpstart
\figsetgrpnum{2.1194}

\figsetplot{./figset/scatter/1194.pdf}

\figsetgrpend

\figsetgrpstart
\figsetgrpnum{2.1195}

\figsetplot{./figset/scatter/1195.pdf}

\figsetgrpend

\figsetgrpstart
\figsetgrpnum{2.1196}

\figsetplot{./figset/scatter/1196.pdf}

\figsetgrpend

\figsetgrpstart
\figsetgrpnum{2.1197}

\figsetplot{./figset/scatter/1197.pdf}

\figsetgrpend

\figsetgrpstart
\figsetgrpnum{2.1198}

\figsetplot{./figset/scatter/1198.pdf}

\figsetgrpend

\figsetgrpstart
\figsetgrpnum{2.1199}

\figsetplot{./figset/scatter/1199.pdf}

\figsetgrpend

\figsetgrpstart
\figsetgrpnum{2.1200}

\figsetplot{./figset/scatter/1200.pdf}

\figsetgrpend

\figsetgrpstart
\figsetgrpnum{2.1201}

\figsetplot{./figset/scatter/1201.pdf}

\figsetgrpend

\figsetgrpstart
\figsetgrpnum{2.1202}

\figsetplot{./figset/scatter/1202.pdf}

\figsetgrpend

\figsetgrpstart
\figsetgrpnum{2.1203}

\figsetplot{./figset/scatter/1203.pdf}

\figsetgrpend

\figsetgrpstart
\figsetgrpnum{2.1204}

\figsetplot{./figset/scatter/1204.pdf}

\figsetgrpend

\figsetgrpstart
\figsetgrpnum{2.1205}

\figsetplot{./figset/scatter/1205.pdf}

\figsetgrpend

\figsetgrpstart
\figsetgrpnum{2.1206}

\figsetplot{./figset/scatter/1206.pdf}

\figsetgrpend

\figsetgrpstart
\figsetgrpnum{2.1207}

\figsetplot{./figset/scatter/1207.pdf}

\figsetgrpend

\figsetgrpstart
\figsetgrpnum{2.1208}

\figsetplot{./figset/scatter/1208.pdf}

\figsetgrpend

\figsetgrpstart
\figsetgrpnum{2.1209}

\figsetplot{./figset/scatter/1209.pdf}

\figsetgrpend

\figsetgrpstart
\figsetgrpnum{2.1210}

\figsetplot{./figset/scatter/1210.pdf}

\figsetgrpend

\figsetgrpstart
\figsetgrpnum{2.1211}

\figsetplot{./figset/scatter/1211.pdf}

\figsetgrpend

\figsetgrpstart
\figsetgrpnum{2.1212}

\figsetplot{./figset/scatter/1212.pdf}

\figsetgrpend

\figsetgrpstart
\figsetgrpnum{2.1213}

\figsetplot{./figset/scatter/1213.pdf}

\figsetgrpend

\figsetgrpstart
\figsetgrpnum{2.1214}

\figsetplot{./figset/scatter/1214.pdf}

\figsetgrpend

\figsetgrpstart
\figsetgrpnum{2.1215}

\figsetplot{./figset/scatter/1215.pdf}

\figsetgrpend

\figsetgrpstart
\figsetgrpnum{2.1216}

\figsetplot{./figset/scatter/1216.pdf}

\figsetgrpend

\figsetgrpstart
\figsetgrpnum{2.1217}

\figsetplot{./figset/scatter/1217.pdf}

\figsetgrpend

\figsetgrpstart
\figsetgrpnum{2.1218}

\figsetplot{./figset/scatter/1218.pdf}

\figsetgrpend

\figsetgrpstart
\figsetgrpnum{2.1219}

\figsetplot{./figset/scatter/1219.pdf}

\figsetgrpend

\figsetgrpstart
\figsetgrpnum{2.1220}

\figsetplot{./figset/scatter/1220.pdf}

\figsetgrpend

\figsetgrpstart
\figsetgrpnum{2.1221}

\figsetplot{./figset/scatter/1221.pdf}

\figsetgrpend

\figsetgrpstart
\figsetgrpnum{2.1222}

\figsetplot{./figset/scatter/1222.pdf}

\figsetgrpend

\figsetgrpstart
\figsetgrpnum{2.1223}

\figsetplot{./figset/scatter/1223.pdf}

\figsetgrpend

\figsetgrpstart
\figsetgrpnum{2.1224}

\figsetplot{./figset/scatter/1224.pdf}

\figsetgrpend

\figsetgrpstart
\figsetgrpnum{2.1225}

\figsetplot{./figset/scatter/1225.pdf}

\figsetgrpend

\figsetgrpstart
\figsetgrpnum{2.1226}

\figsetplot{./figset/scatter/1226.pdf}

\figsetgrpend

\figsetgrpstart
\figsetgrpnum{2.1227}

\figsetplot{./figset/scatter/1227.pdf}

\figsetgrpend

\figsetgrpstart
\figsetgrpnum{2.1228}

\figsetplot{./figset/scatter/1228.pdf}

\figsetgrpend

\figsetgrpstart
\figsetgrpnum{2.1229}

\figsetplot{./figset/scatter/1229.pdf}

\figsetgrpend

\figsetgrpstart
\figsetgrpnum{2.1230}

\figsetplot{./figset/scatter/1230.pdf}

\figsetgrpend

\figsetgrpstart
\figsetgrpnum{2.1231}

\figsetplot{./figset/scatter/1231.pdf}

\figsetgrpend

\figsetgrpstart
\figsetgrpnum{2.1232}

\figsetplot{./figset/scatter/1232.pdf}

\figsetgrpend

\figsetgrpstart
\figsetgrpnum{2.1233}

\figsetplot{./figset/scatter/1233.pdf}

\figsetgrpend

\figsetgrpstart
\figsetgrpnum{2.1234}

\figsetplot{./figset/scatter/1234.pdf}

\figsetgrpend

\figsetgrpstart
\figsetgrpnum{2.1235}

\figsetplot{./figset/scatter/1235.pdf}

\figsetgrpend

\figsetgrpstart
\figsetgrpnum{2.1236}

\figsetplot{./figset/scatter/1236.pdf}

\figsetgrpend

\figsetgrpstart
\figsetgrpnum{2.1237}

\figsetplot{./figset/scatter/1237.pdf}

\figsetgrpend

\figsetgrpstart
\figsetgrpnum{2.1238}

\figsetplot{./figset/scatter/1238.pdf}

\figsetgrpend

\figsetgrpstart
\figsetgrpnum{2.1239}

\figsetplot{./figset/scatter/1239.pdf}

\figsetgrpend

\figsetgrpstart
\figsetgrpnum{2.1240}

\figsetplot{./figset/scatter/1240.pdf}

\figsetgrpend

\figsetgrpstart
\figsetgrpnum{2.1241}

\figsetplot{./figset/scatter/1241.pdf}

\figsetgrpend

\figsetgrpstart
\figsetgrpnum{2.1242}

\figsetplot{./figset/scatter/1242.pdf}

\figsetgrpend

\figsetgrpstart
\figsetgrpnum{2.1243}

\figsetplot{./figset/scatter/1243.pdf}

\figsetgrpend

\figsetgrpstart
\figsetgrpnum{2.1244}

\figsetplot{./figset/scatter/1244.pdf}

\figsetgrpend

\figsetgrpstart
\figsetgrpnum{2.1245}

\figsetplot{./figset/scatter/1245.pdf}

\figsetgrpend

\figsetgrpstart
\figsetgrpnum{2.1246}

\figsetplot{./figset/scatter/1246.pdf}

\figsetgrpend

\figsetgrpstart
\figsetgrpnum{2.1247}

\figsetplot{./figset/scatter/1247.pdf}

\figsetgrpend

\figsetgrpstart
\figsetgrpnum{2.1248}

\figsetplot{./figset/scatter/1248.pdf}

\figsetgrpend

\figsetgrpstart
\figsetgrpnum{2.1249}

\figsetplot{./figset/scatter/1249.pdf}

\figsetgrpend

\figsetgrpstart
\figsetgrpnum{2.1250}

\figsetplot{./figset/scatter/1250.pdf}

\figsetgrpend

\figsetgrpstart
\figsetgrpnum{2.1251}

\figsetplot{./figset/scatter/1251.pdf}

\figsetgrpend

\figsetgrpstart
\figsetgrpnum{2.1252}

\figsetplot{./figset/scatter/1252.pdf}

\figsetgrpend

\figsetgrpstart
\figsetgrpnum{2.1253}

\figsetplot{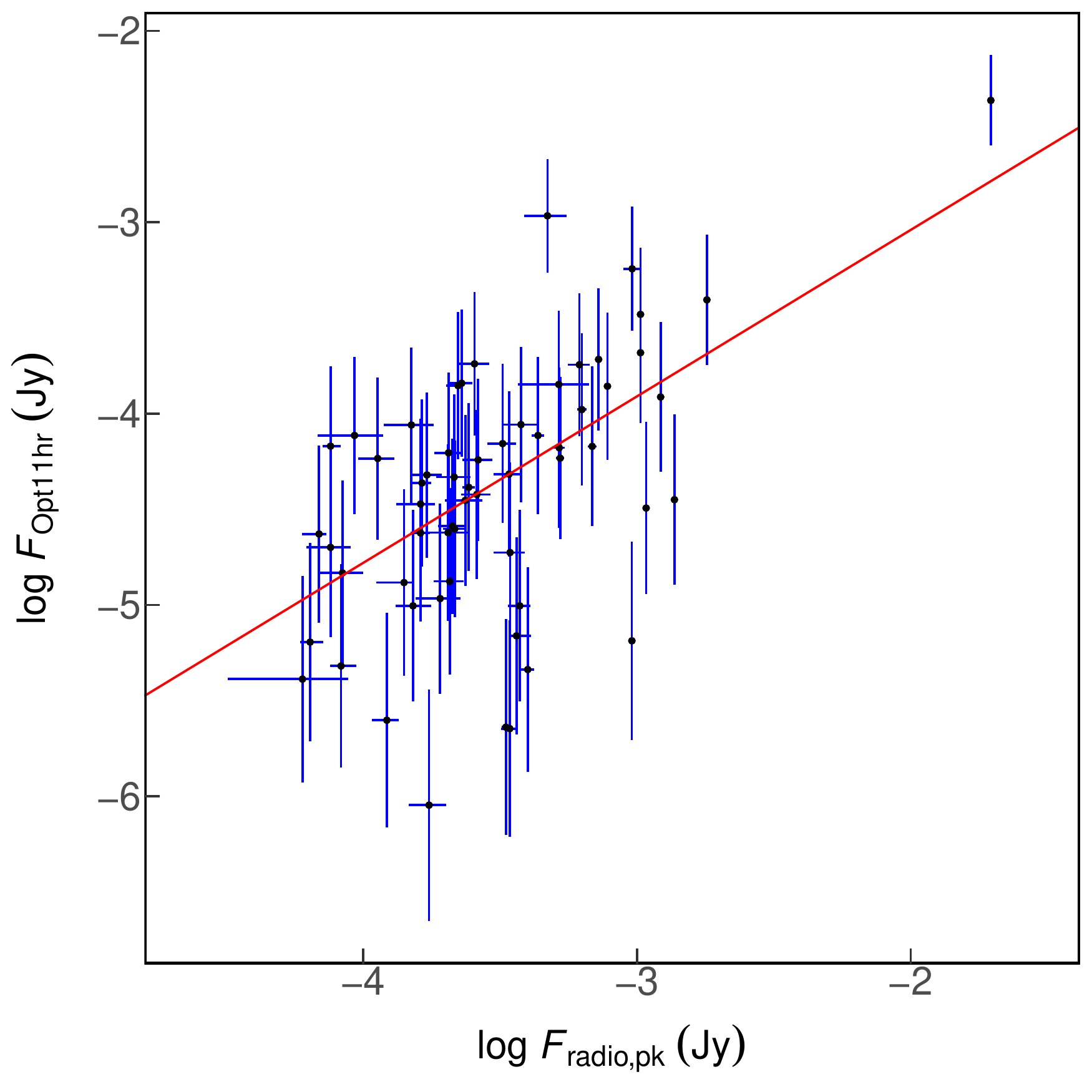}

\figsetgrpend

\figsetgrpstart
\figsetgrpnum{2.1254}

\figsetplot{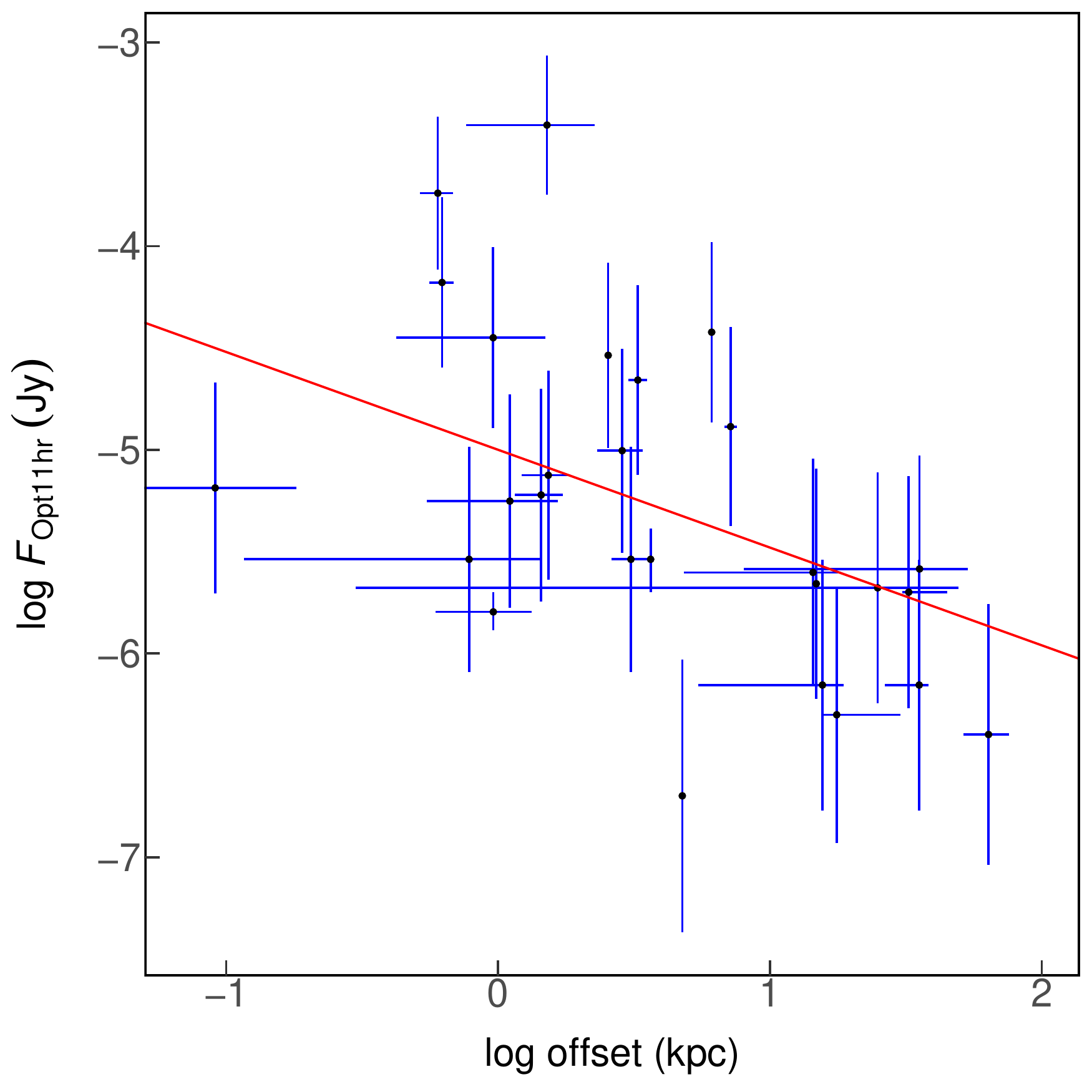}

\figsetgrpend

\figsetgrpstart
\figsetgrpnum{2.1255}

\figsetplot{./figset/scatter/1255.pdf}

\figsetgrpend

\figsetgrpstart
\figsetgrpnum{2.1256}

\figsetplot{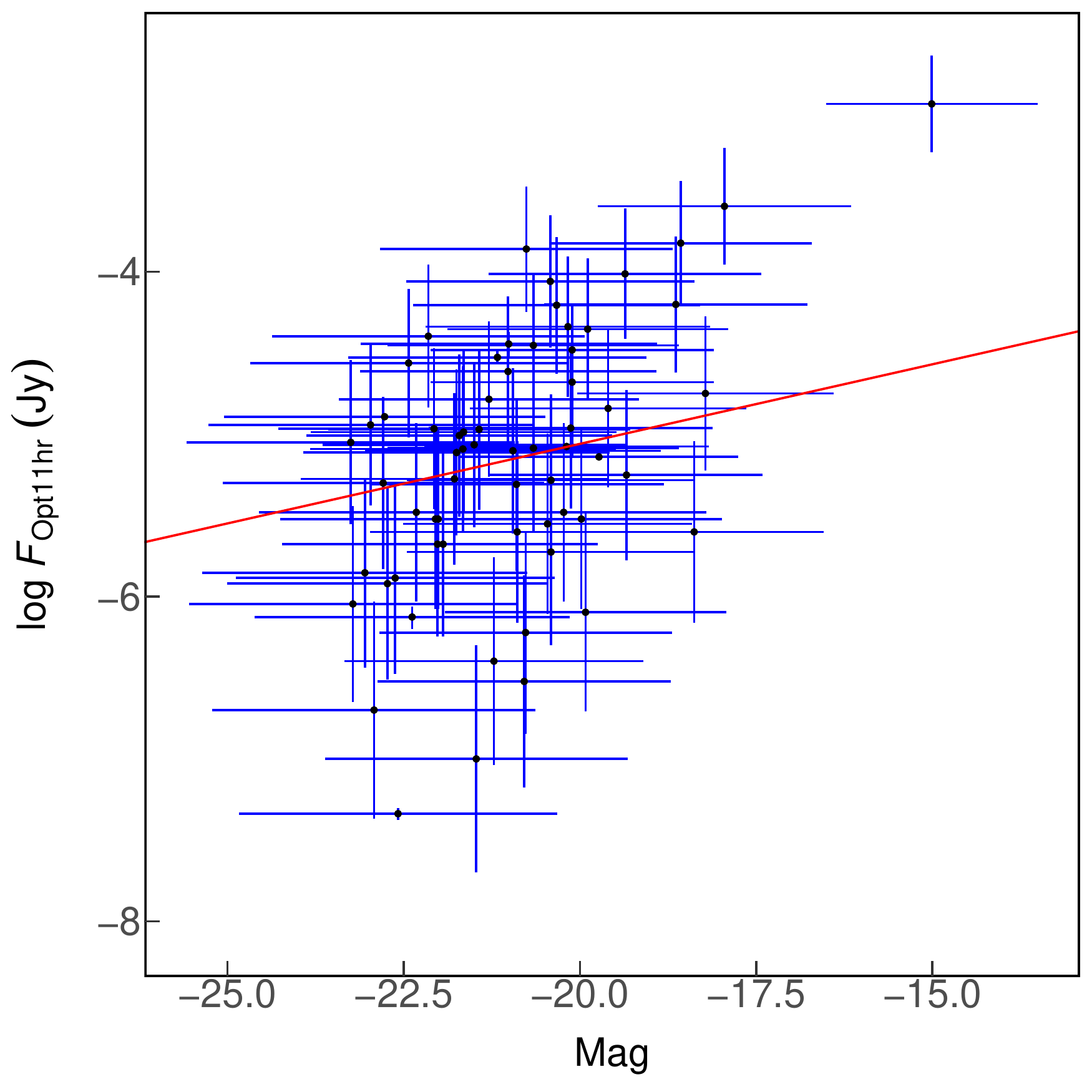}

\figsetgrpend

\figsetgrpstart
\figsetgrpnum{2.1257}

\figsetplot{./figset/scatter/1257.pdf}

\figsetgrpend

\figsetgrpstart
\figsetgrpnum{2.1258}

\figsetplot{./figset/scatter/1258.pdf}

\figsetgrpend

\figsetgrpstart
\figsetgrpnum{2.1259}

\figsetplot{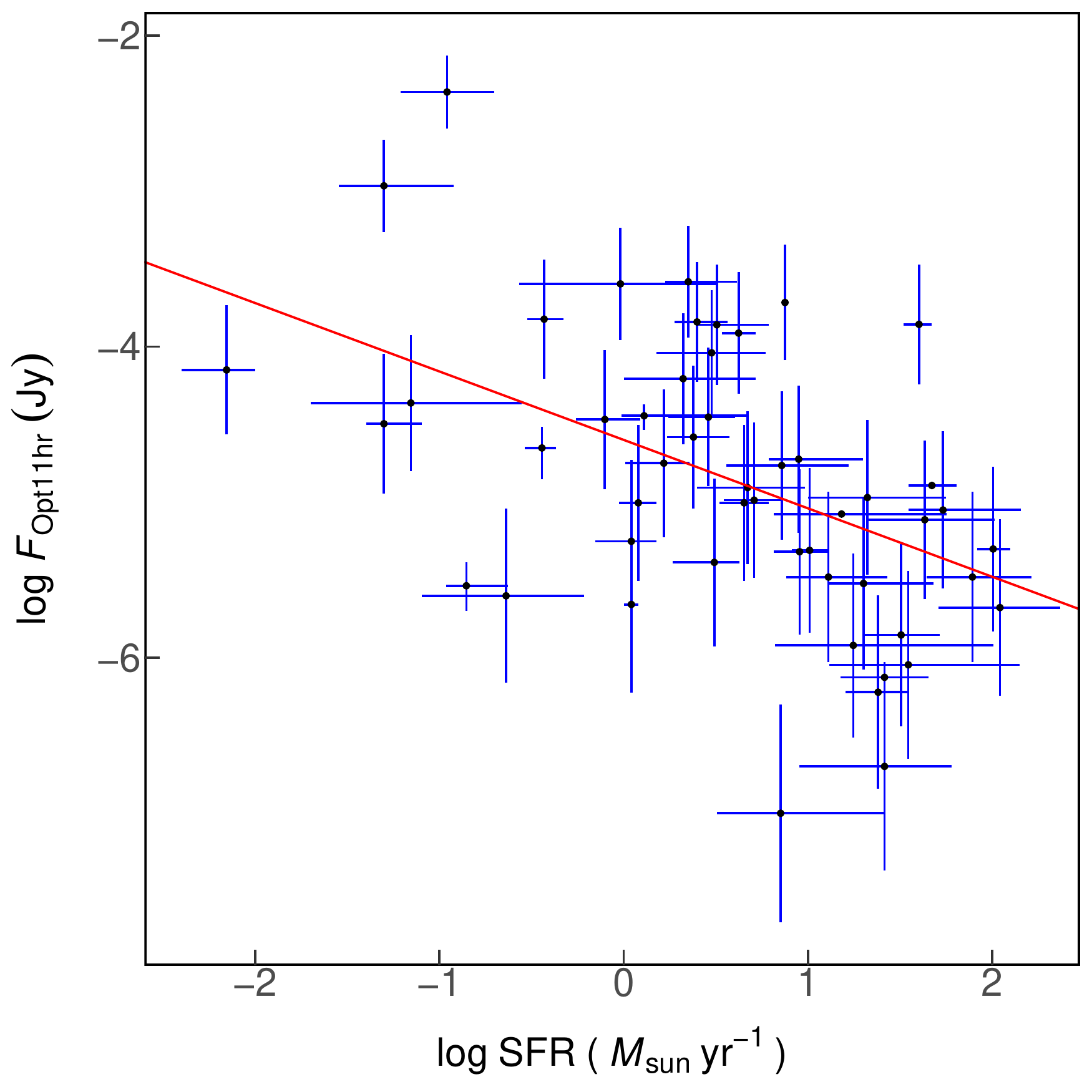}

\figsetgrpend

\figsetgrpstart
\figsetgrpnum{2.1260}

\figsetplot{./figset/scatter/1260.pdf}

\figsetgrpend

\figsetgrpstart
\figsetgrpnum{2.1261}

\figsetplot{./figset/scatter/1261.pdf}

\figsetgrpend

\figsetgrpstart
\figsetgrpnum{2.1262}

\figsetplot{./figset/scatter/1262.pdf}

\figsetgrpend

\figsetgrpstart
\figsetgrpnum{2.1263}

\figsetplot{./figset/scatter/1263.pdf}

\figsetgrpend

\figsetgrpstart
\figsetgrpnum{2.1264}

\figsetplot{./figset/scatter/1264.pdf}

\figsetgrpend

\figsetgrpstart
\figsetgrpnum{2.1265}

\figsetplot{./figset/scatter/1265.pdf}

\figsetgrpend

\figsetgrpstart
\figsetgrpnum{2.1266}

\figsetplot{./figset/scatter/1266.pdf}

\figsetgrpend

\figsetgrpstart
\figsetgrpnum{2.1267}

\figsetplot{./figset/scatter/1267.pdf}

\figsetgrpend

\figsetgrpstart
\figsetgrpnum{2.1268}

\figsetplot{./figset/scatter/1268.pdf}

\figsetgrpend

\figsetgrpstart
\figsetgrpnum{2.1269}

\figsetplot{./figset/scatter/1269.pdf}

\figsetgrpend

\figsetgrpstart
\figsetgrpnum{2.1270}

\figsetplot{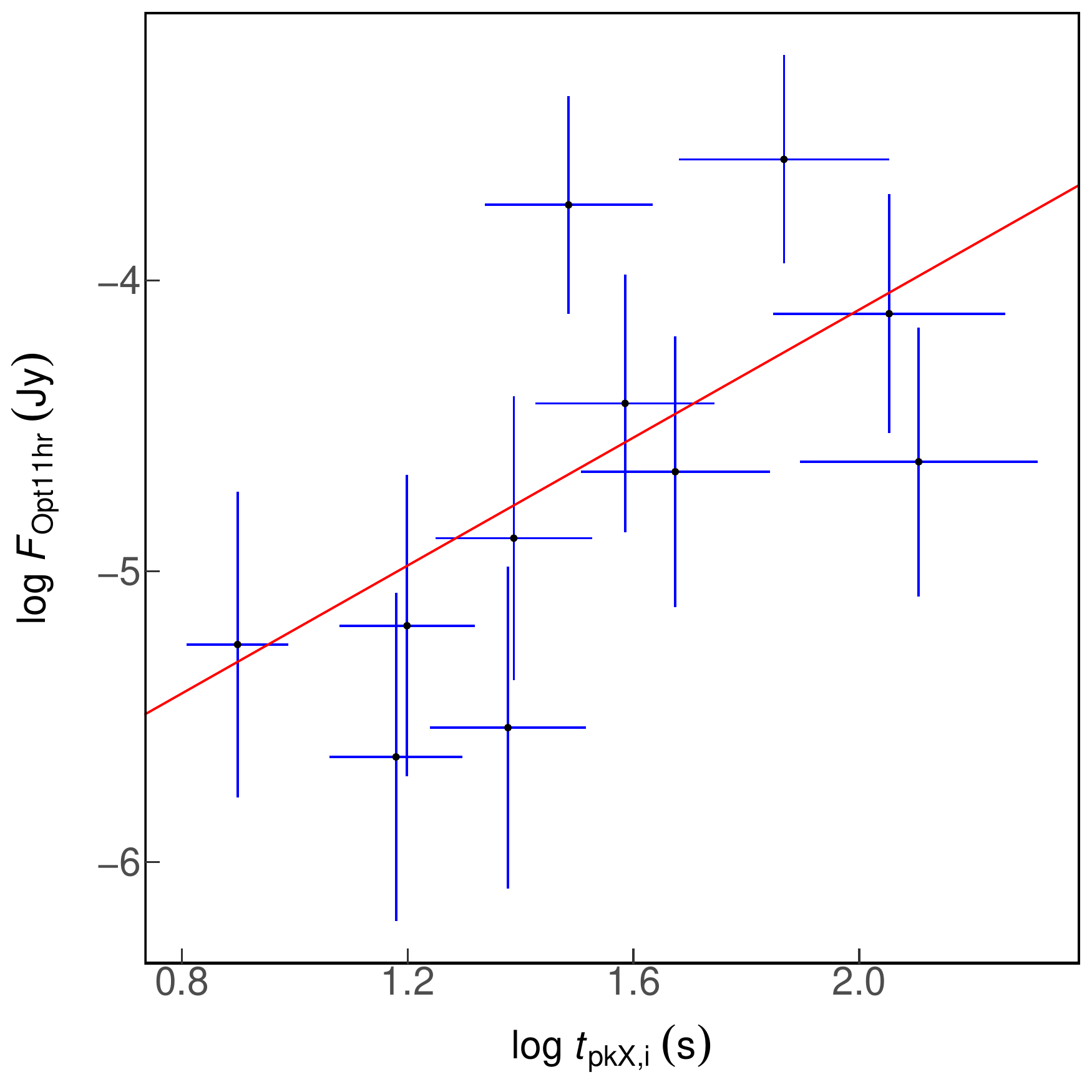}

\figsetgrpend

\figsetgrpstart
\figsetgrpnum{2.1271}

\figsetplot{./figset/scatter/1271.pdf}

\figsetgrpend

\figsetgrpstart
\figsetgrpnum{2.1272}

\figsetplot{./figset/scatter/1272.pdf}

\figsetgrpend

\figsetgrpstart
\figsetgrpnum{2.1273}

\figsetplot{./figset/scatter/1273.pdf}

\figsetgrpend

\figsetgrpstart
\figsetgrpnum{2.1274}

\figsetplot{./figset/scatter/1274.pdf}

\figsetgrpend

\figsetgrpstart
\figsetgrpnum{2.1275}

\figsetplot{./figset/scatter/1275.pdf}

\figsetgrpend

\figsetgrpstart
\figsetgrpnum{2.1276}

\figsetplot{./figset/scatter/1276.pdf}

\figsetgrpend

\figsetgrpstart
\figsetgrpnum{2.1277}

\figsetplot{./figset/scatter/1277.pdf}

\figsetgrpend

\figsetgrpstart
\figsetgrpnum{2.1278}

\figsetplot{./figset/scatter/1278.pdf}

\figsetgrpend

\figsetgrpstart
\figsetgrpnum{2.1279}

\figsetplot{./figset/scatter/1279.pdf}

\figsetgrpend

\figsetgrpstart
\figsetgrpnum{2.1280}

\figsetplot{./figset/scatter/1280.pdf}

\figsetgrpend

\figsetgrpstart
\figsetgrpnum{2.1281}

\figsetplot{./figset/scatter/1281.pdf}

\figsetgrpend

\figsetgrpstart
\figsetgrpnum{2.1282}

\figsetplot{./figset/scatter/1282.pdf}

\figsetgrpend

\figsetgrpstart
\figsetgrpnum{2.1283}

\figsetplot{./figset/scatter/1283.pdf}

\figsetgrpend

\figsetgrpstart
\figsetgrpnum{2.1284}

\figsetplot{./figset/scatter/1284.pdf}

\figsetgrpend

\figsetgrpstart
\figsetgrpnum{2.1285}

\figsetplot{./figset/scatter/1285.pdf}

\figsetgrpend

\figsetgrpstart
\figsetgrpnum{2.1286}

\figsetplot{./figset/scatter/1286.pdf}

\figsetgrpend

\figsetgrpstart
\figsetgrpnum{2.1287}

\figsetplot{./figset/scatter/1287.pdf}

\figsetgrpend

\figsetgrpstart
\figsetgrpnum{2.1288}

\figsetplot{./figset/scatter/1288.pdf}

\figsetgrpend

\figsetgrpstart
\figsetgrpnum{2.1289}

\figsetplot{./figset/scatter/1289.pdf}

\figsetgrpend

\figsetgrpstart
\figsetgrpnum{2.1290}

\figsetplot{./figset/scatter/1290.pdf}

\figsetgrpend

\figsetgrpstart
\figsetgrpnum{2.1291}

\figsetplot{./figset/scatter/1291.pdf}

\figsetgrpend

\figsetgrpstart
\figsetgrpnum{2.1292}

\figsetplot{./figset/scatter/1292.pdf}

\figsetgrpend

\figsetgrpstart
\figsetgrpnum{2.1293}

\figsetplot{./figset/scatter/1293.pdf}

\figsetgrpend

\figsetgrpstart
\figsetgrpnum{2.1294}

\figsetplot{./figset/scatter/1294.pdf}

\figsetgrpend

\figsetgrpstart
\figsetgrpnum{2.1295}

\figsetplot{./figset/scatter/1295.pdf}

\figsetgrpend

\figsetgrpstart
\figsetgrpnum{2.1296}

\figsetplot{./figset/scatter/1296.pdf}

\figsetgrpend

\figsetgrpstart
\figsetgrpnum{2.1297}

\figsetplot{./figset/scatter/1297.pdf}

\figsetgrpend

\figsetgrpstart
\figsetgrpnum{2.1298}

\figsetplot{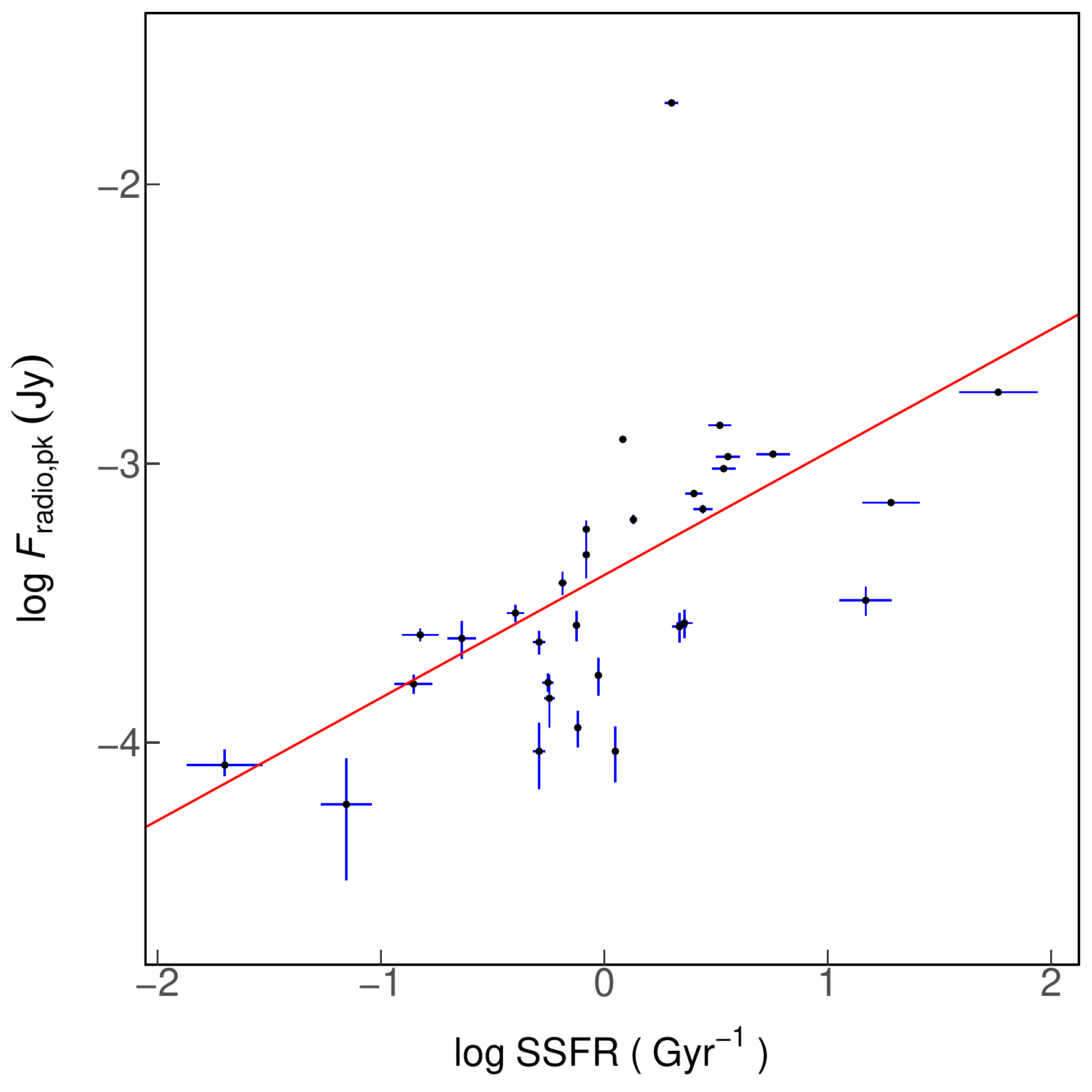}

\figsetgrpend

\figsetgrpstart
\figsetgrpnum{2.1299}

\figsetplot{./figset/scatter/1299.pdf}

\figsetgrpend

\figsetgrpstart
\figsetgrpnum{2.1300}

\figsetplot{./figset/scatter/1300.pdf}

\figsetgrpend

\figsetgrpstart
\figsetgrpnum{2.1301}

\figsetplot{./figset/scatter/1301.pdf}

\figsetgrpend

\figsetgrpstart
\figsetgrpnum{2.1302}

\figsetplot{./figset/scatter/1302.pdf}

\figsetgrpend

\figsetgrpstart
\figsetgrpnum{2.1303}

\figsetplot{./figset/scatter/1303.pdf}

\figsetgrpend

\figsetgrpstart
\figsetgrpnum{2.1304}

\figsetplot{./figset/scatter/1304.pdf}

\figsetgrpend

\figsetgrpstart
\figsetgrpnum{2.1305}

\figsetplot{./figset/scatter/1305.pdf}

\figsetgrpend

\figsetgrpstart
\figsetgrpnum{2.1306}

\figsetplot{./figset/scatter/1306.pdf}

\figsetgrpend

\figsetgrpstart
\figsetgrpnum{2.1307}

\figsetplot{./figset/scatter/1307.pdf}

\figsetgrpend

\figsetgrpstart
\figsetgrpnum{2.1308}

\figsetplot{./figset/scatter/1308.pdf}

\figsetgrpend

\figsetgrpstart
\figsetgrpnum{2.1309}

\figsetplot{./figset/scatter/1309.pdf}

\figsetgrpend

\figsetgrpstart
\figsetgrpnum{2.1310}

\figsetplot{./figset/scatter/1310.pdf}

\figsetgrpend

\figsetgrpstart
\figsetgrpnum{2.1311}

\figsetplot{./figset/scatter/1311.pdf}

\figsetgrpend

\figsetgrpstart
\figsetgrpnum{2.1312}

\figsetplot{./figset/scatter/1312.pdf}

\figsetgrpend

\figsetgrpstart
\figsetgrpnum{2.1313}

\figsetplot{./figset/scatter/1313.pdf}

\figsetgrpend

\figsetgrpstart
\figsetgrpnum{2.1314}

\figsetplot{./figset/scatter/1314.pdf}

\figsetgrpend

\figsetgrpstart
\figsetgrpnum{2.1315}

\figsetplot{./figset/scatter/1315.pdf}

\figsetgrpend

\figsetgrpstart
\figsetgrpnum{2.1316}

\figsetplot{./figset/scatter/1316.pdf}

\figsetgrpend

\figsetgrpstart
\figsetgrpnum{2.1317}

\figsetplot{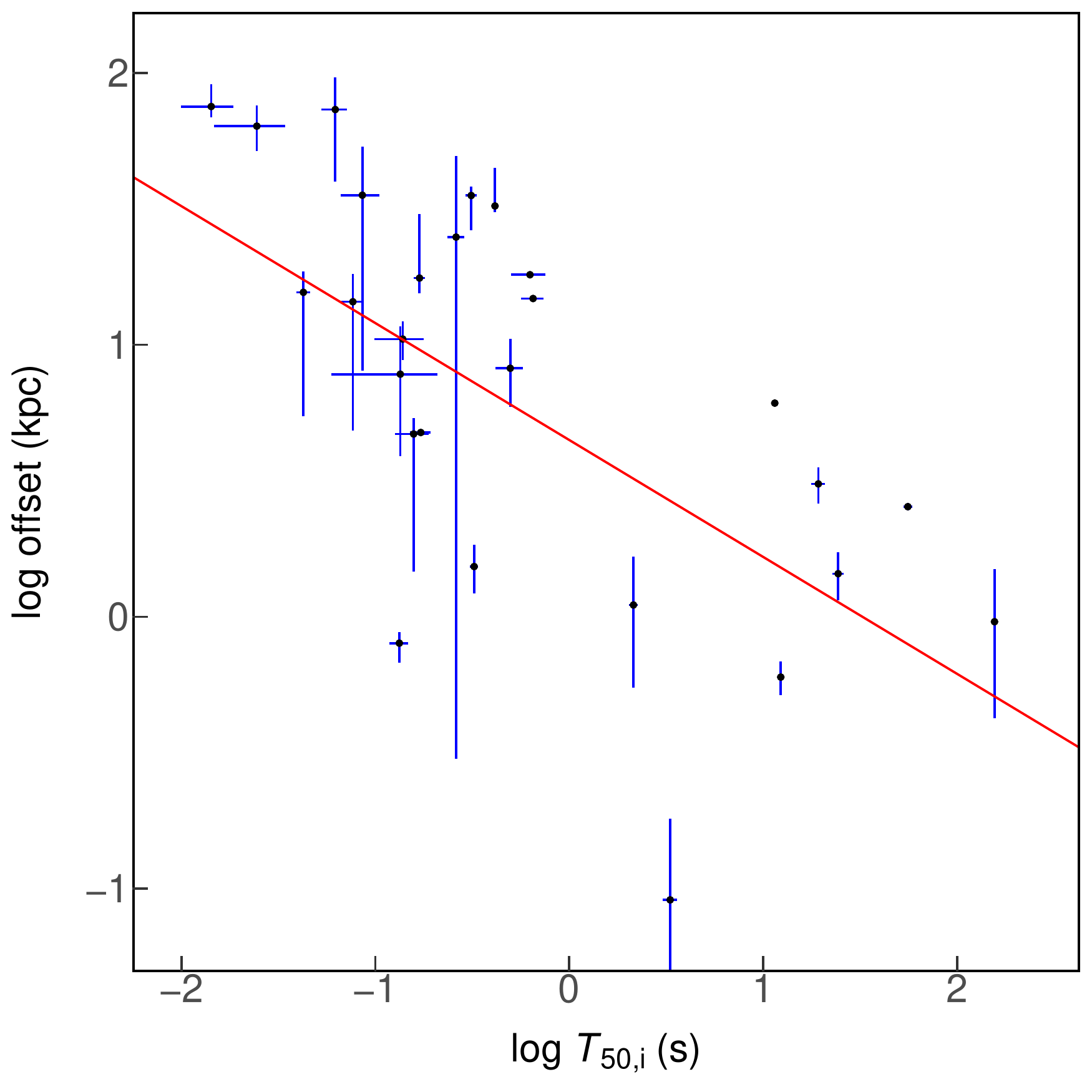}

\figsetgrpend

\figsetgrpstart
\figsetgrpnum{2.1318}

\figsetplot{./figset/scatter/1318.pdf}

\figsetgrpend

\figsetgrpstart
\figsetgrpnum{2.1319}

\figsetplot{./figset/scatter/1319.pdf}

\figsetgrpend

\figsetgrpstart
\figsetgrpnum{2.1320}

\figsetplot{./figset/scatter/1320.pdf}

\figsetgrpend

\figsetgrpstart
\figsetgrpnum{2.1321}

\figsetplot{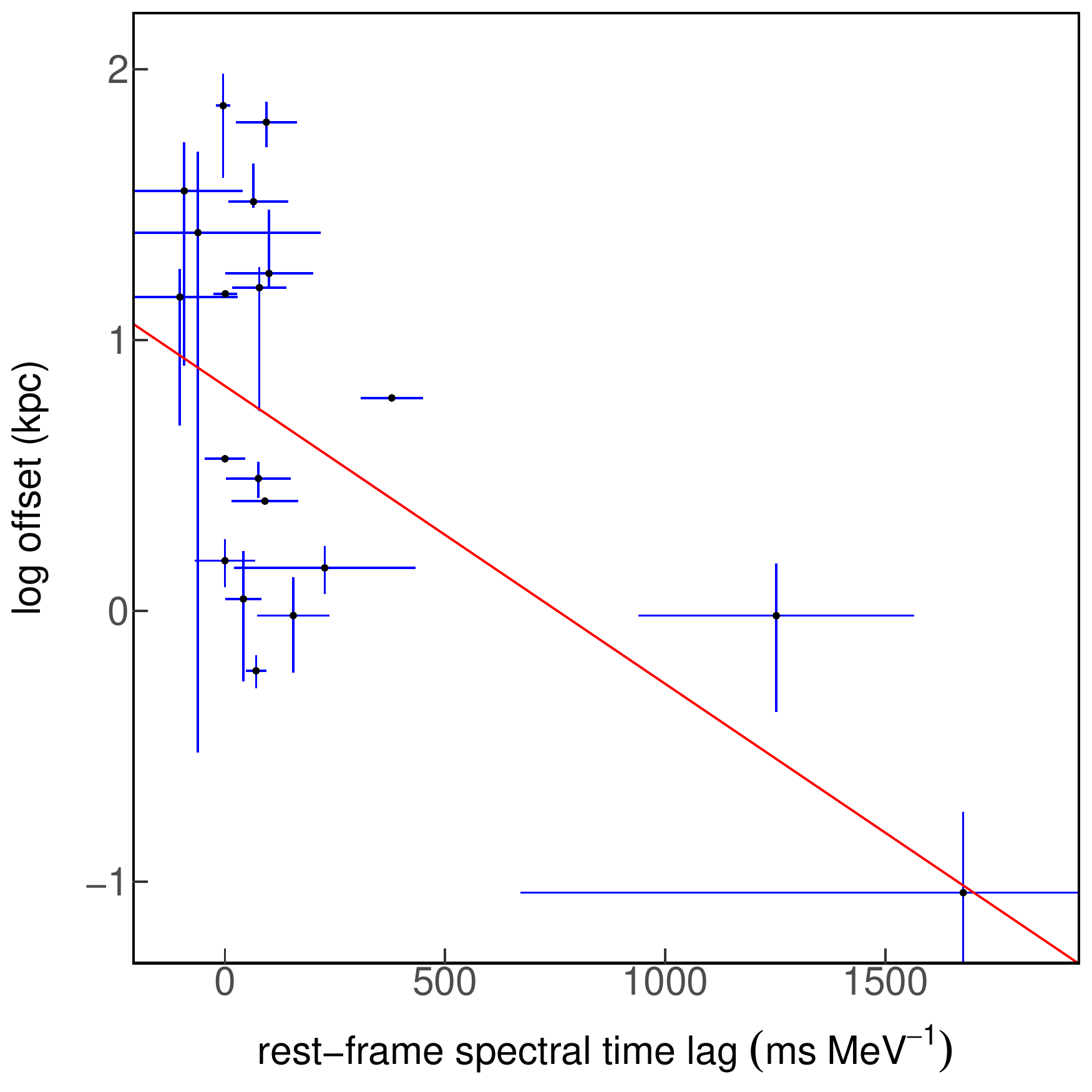}

\figsetgrpend

\figsetgrpstart
\figsetgrpnum{2.1322}

\figsetplot{./figset/scatter/1322.pdf}

\figsetgrpend

\figsetgrpstart
\figsetgrpnum{2.1323}

\figsetplot{./figset/scatter/1323.pdf}

\figsetgrpend

\figsetgrpstart
\figsetgrpnum{2.1324}

\figsetplot{./figset/scatter/1324.pdf}

\figsetgrpend

\figsetgrpstart
\figsetgrpnum{2.1325}

\figsetplot{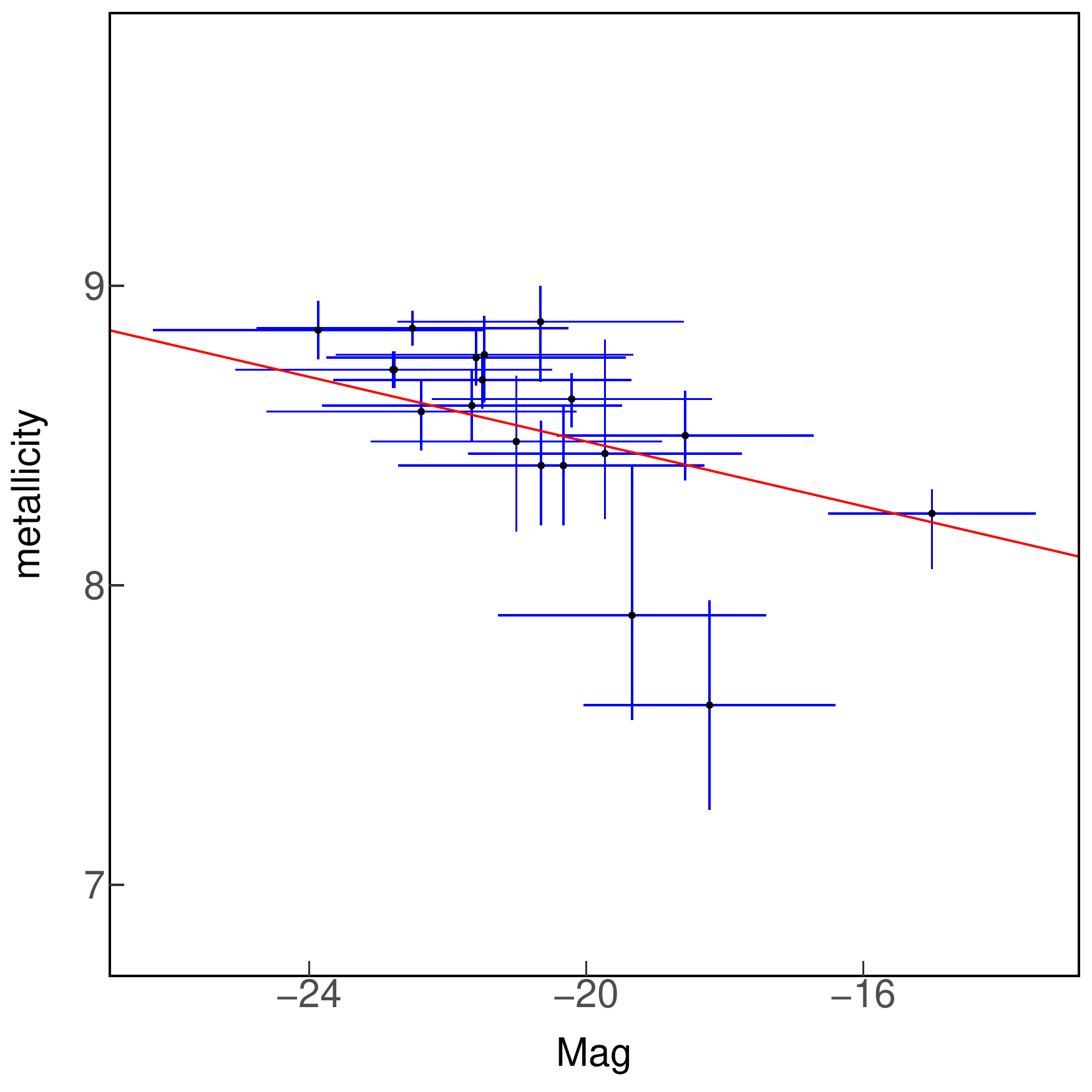}

\figsetgrpend

\figsetgrpstart
\figsetgrpnum{2.1326}

\figsetplot{./figset/scatter/1326.pdf}

\figsetgrpend

\figsetgrpstart
\figsetgrpnum{2.1327}

\figsetplot{./figset/scatter/1327.pdf}

\figsetgrpend

\figsetgrpstart
\figsetgrpnum{2.1328}

\figsetplot{./figset/scatter/1328.pdf}

\figsetgrpend

\figsetgrpstart
\figsetgrpnum{2.1329}

\figsetplot{./figset/scatter/1329.pdf}

\figsetgrpend

\figsetgrpstart
\figsetgrpnum{2.1330}

\figsetplot{./figset/scatter/1330.pdf}

\figsetgrpend

\figsetgrpstart
\figsetgrpnum{2.1331}

\figsetplot{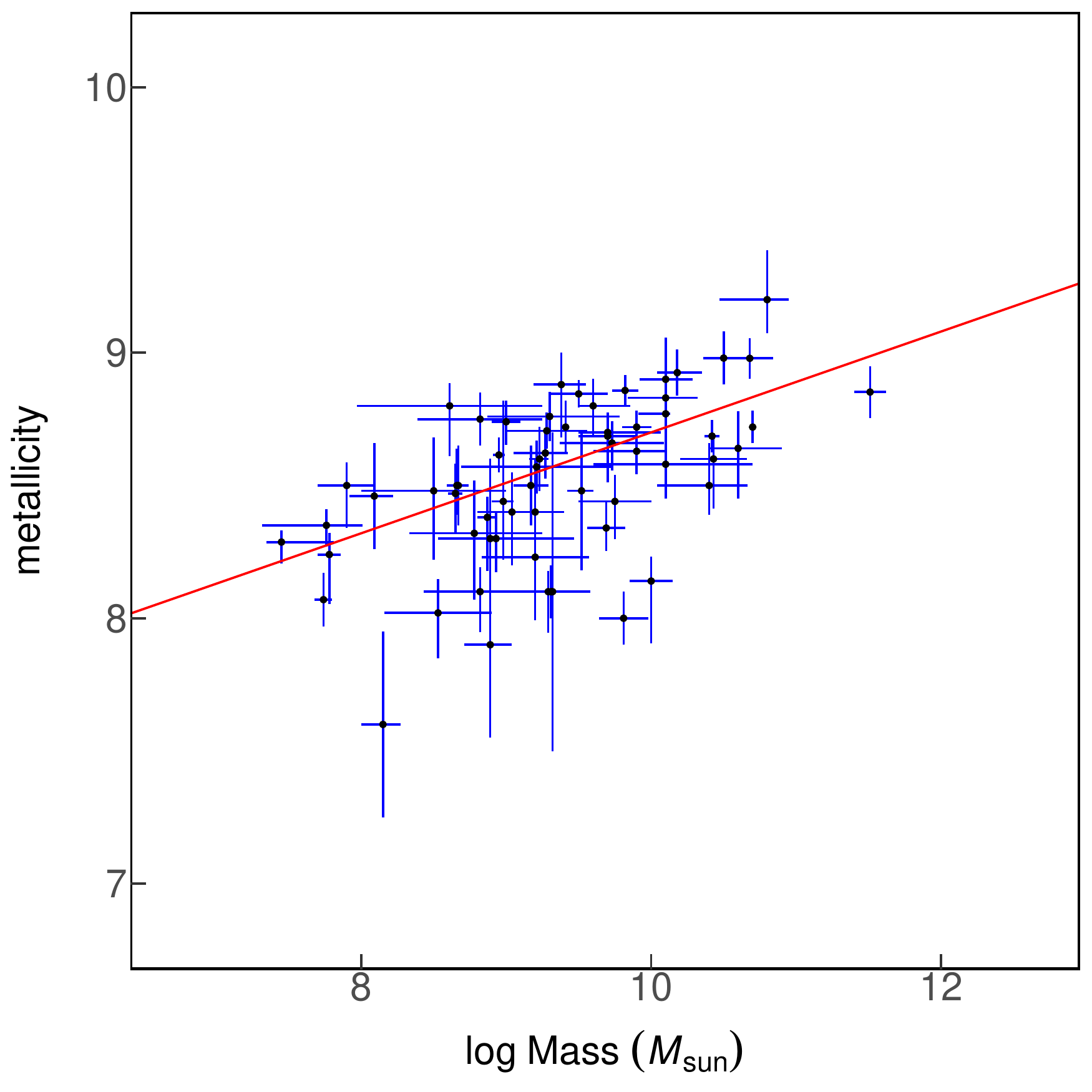}

\figsetgrpend

\figsetgrpstart
\figsetgrpnum{2.1332}

\figsetplot{./figset/scatter/1332.pdf}

\figsetgrpend

\figsetgrpstart
\figsetgrpnum{2.1333}

\figsetplot{./figset/scatter/1333.pdf}

\figsetgrpend

\figsetgrpstart
\figsetgrpnum{2.1334}

\figsetplot{./figset/scatter/1334.pdf}

\figsetgrpend

\figsetgrpstart
\figsetgrpnum{2.1335}

\figsetplot{./figset/scatter/1335.pdf}

\figsetgrpend

\figsetgrpstart
\figsetgrpnum{2.1336}

\figsetplot{./figset/scatter/1336.pdf}

\figsetgrpend

\figsetgrpstart
\figsetgrpnum{2.1337}

\figsetplot{./figset/scatter/1337.pdf}

\figsetgrpend

\figsetgrpstart
\figsetgrpnum{2.1338}

\figsetplot{./figset/scatter/1338.pdf}

\figsetgrpend

\figsetgrpstart
\figsetgrpnum{2.1339}

\figsetplot{./figset/scatter/1339.pdf}

\figsetgrpend

\figsetgrpstart
\figsetgrpnum{2.1340}

\figsetplot{./figset/scatter/1340.pdf}

\figsetgrpend

\figsetgrpstart
\figsetgrpnum{2.1341}

\figsetplot{./figset/scatter/1341.pdf}

\figsetgrpend

\figsetgrpstart
\figsetgrpnum{2.1342}

\figsetplot{./figset/scatter/1342.pdf}

\figsetgrpend

\figsetgrpstart
\figsetgrpnum{2.1343}

\figsetplot{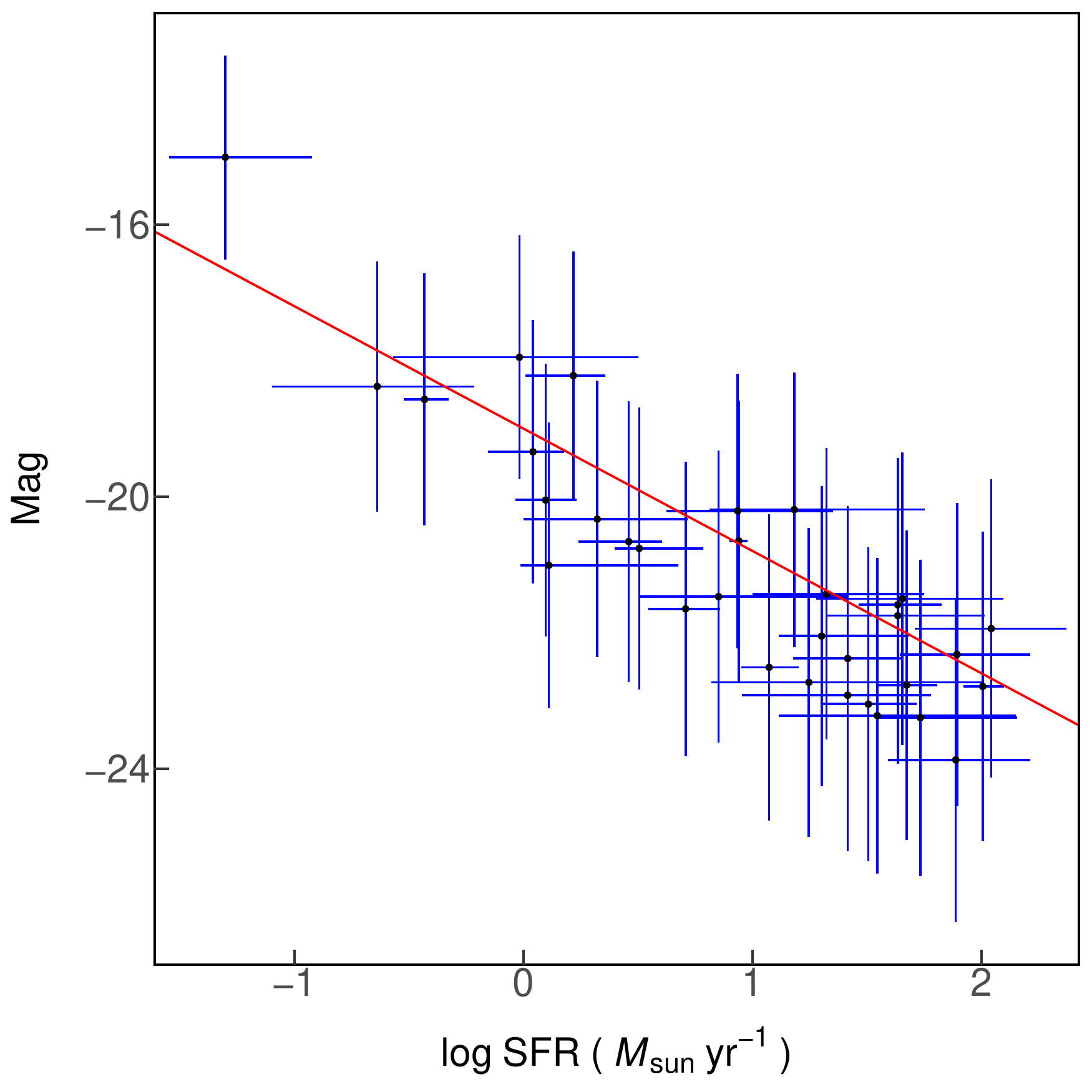}

\figsetgrpend

\figsetgrpstart
\figsetgrpnum{2.1344}

\figsetplot{./figset/scatter/1344.pdf}

\figsetgrpend

\figsetgrpstart
\figsetgrpnum{2.1345}

\figsetplot{./figset/scatter/1345.pdf}

\figsetgrpend

\figsetgrpstart
\figsetgrpnum{2.1346}

\figsetplot{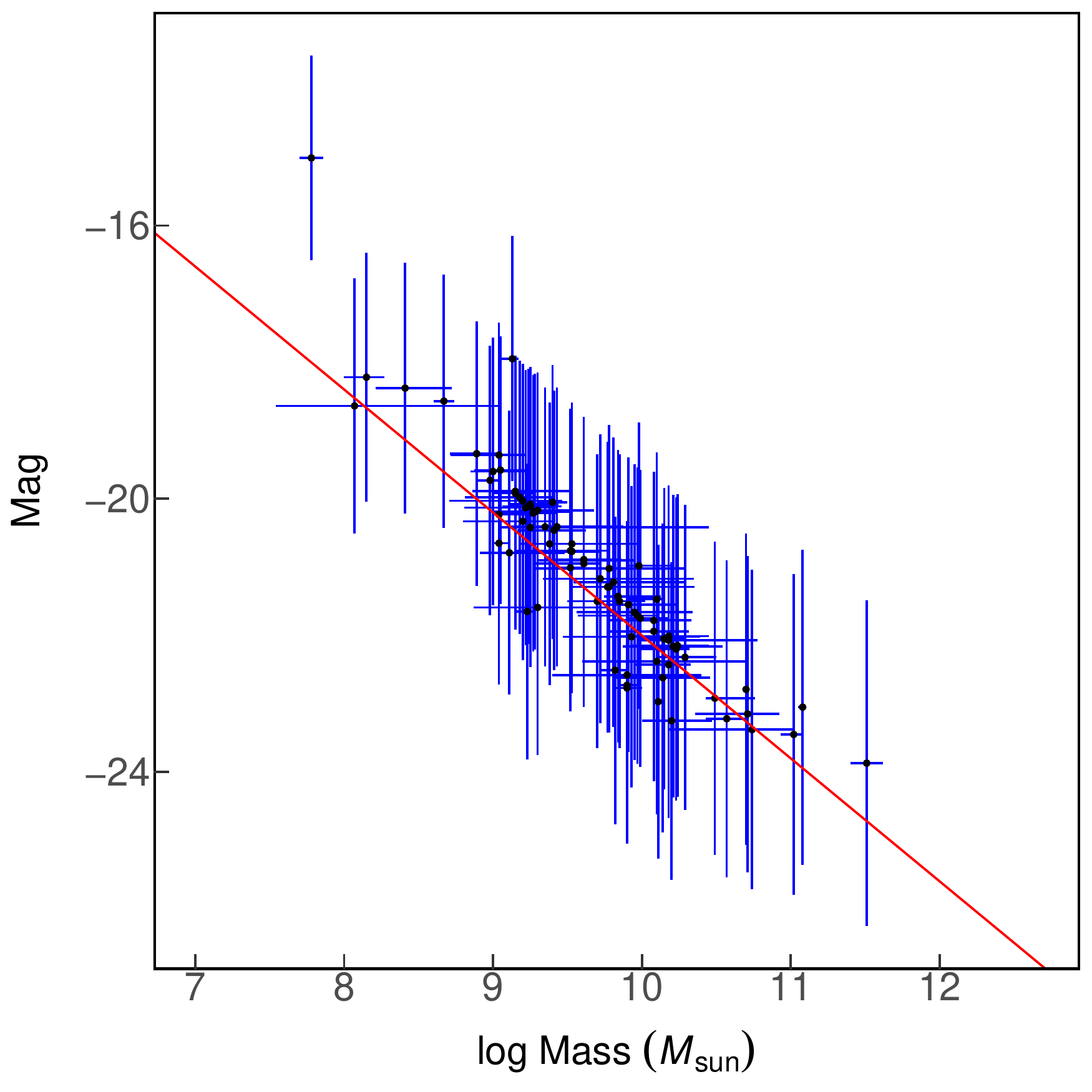}

\figsetgrpend

\figsetgrpstart
\figsetgrpnum{2.1347}

\figsetplot{./figset/scatter/1347.pdf}

\figsetgrpend

\figsetgrpstart
\figsetgrpnum{2.1348}

\figsetplot{./figset/scatter/1348.pdf}

\figsetgrpend

\figsetgrpstart
\figsetgrpnum{2.1349}

\figsetplot{./figset/scatter/1349.pdf}

\figsetgrpend

\figsetgrpstart
\figsetgrpnum{2.1350}

\figsetplot{./figset/scatter/1350.pdf}

\figsetgrpend

\figsetgrpstart
\figsetgrpnum{2.1351}

\figsetplot{./figset/scatter/1351.pdf}

\figsetgrpend

\figsetgrpstart
\figsetgrpnum{2.1352}

\figsetplot{./figset/scatter/1352.pdf}

\figsetgrpend

\figsetgrpstart
\figsetgrpnum{2.1353}

\figsetplot{./figset/scatter/1353.pdf}

\figsetgrpend

\figsetgrpstart
\figsetgrpnum{2.1354}

\figsetplot{./figset/scatter/1354.pdf}

\figsetgrpend

\figsetgrpstart
\figsetgrpnum{2.1355}

\figsetplot{./figset/scatter/1355.pdf}

\figsetgrpend

\figsetgrpstart
\figsetgrpnum{2.1356}

\figsetplot{./figset/scatter/1356.pdf}

\figsetgrpend

\figsetgrpstart
\figsetgrpnum{2.1357}

\figsetplot{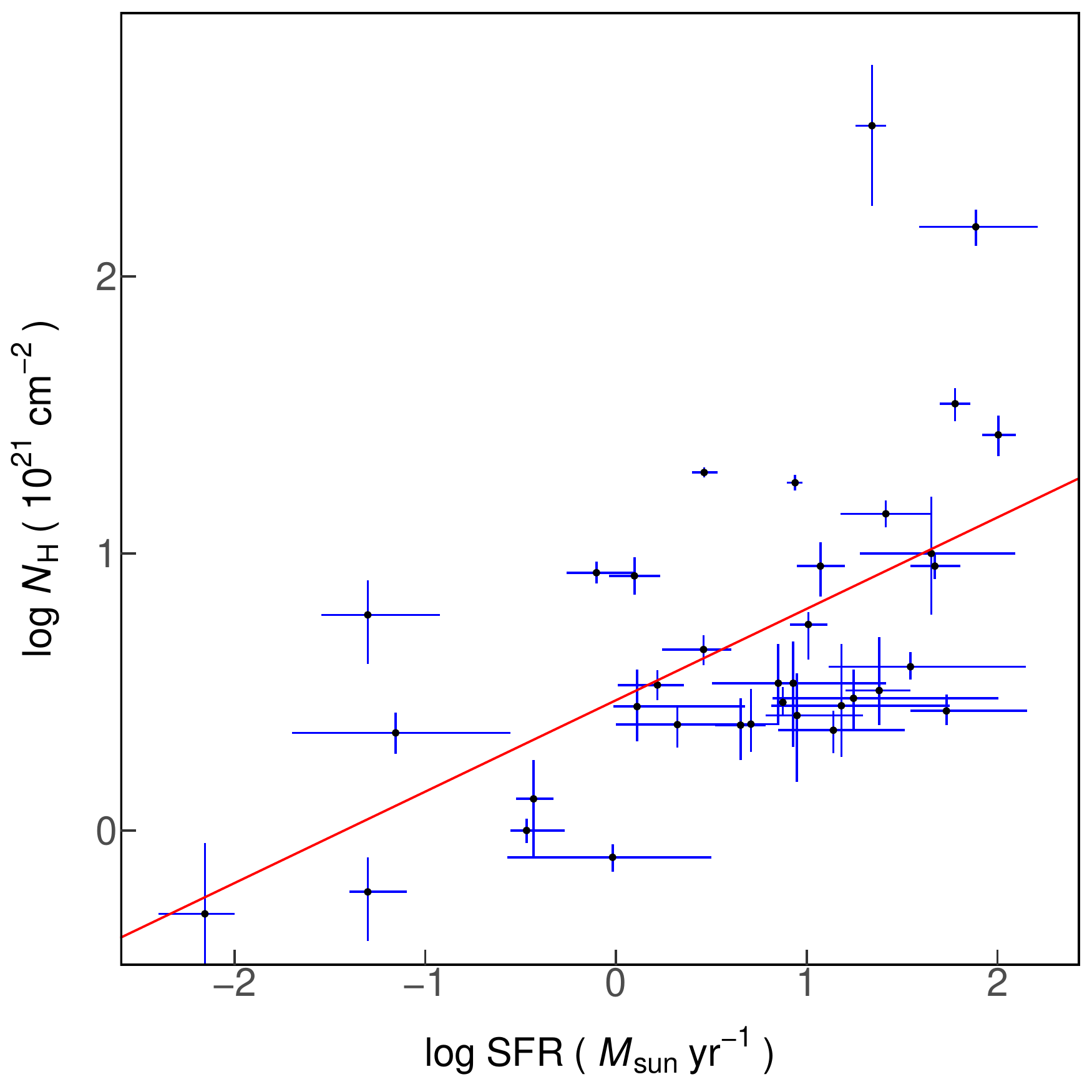}

\figsetgrpend

\figsetgrpstart
\figsetgrpnum{2.1358}

\figsetplot{./figset/scatter/1358.pdf}

\figsetgrpend

\figsetgrpstart
\figsetgrpnum{2.1359}

\figsetplot{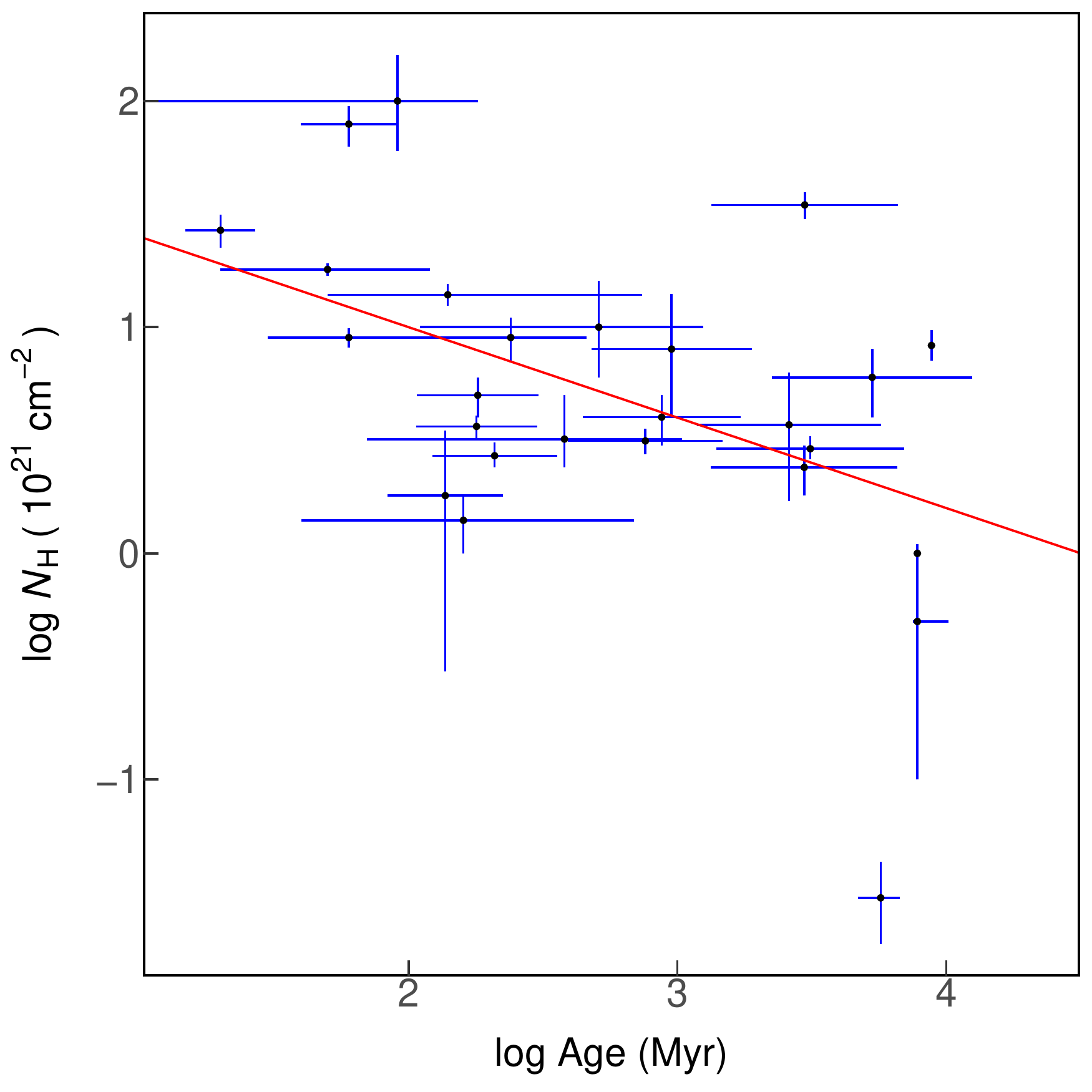}

\figsetgrpend

\figsetgrpstart
\figsetgrpnum{2.1360}

\figsetplot{./figset/scatter/1360.pdf}

\figsetgrpend

\figsetgrpstart
\figsetgrpnum{2.1361}

\figsetplot{./figset/scatter/1361.pdf}

\figsetgrpend

\figsetgrpstart
\figsetgrpnum{2.1362}

\figsetplot{./figset/scatter/1362.pdf}

\figsetgrpend

\figsetgrpstart
\figsetgrpnum{2.1363}

\figsetplot{./figset/scatter/1363.pdf}

\figsetgrpend

\figsetgrpstart
\figsetgrpnum{2.1364}

\figsetplot{./figset/scatter/1364.pdf}

\figsetgrpend

\figsetgrpstart
\figsetgrpnum{2.1365}

\figsetplot{./figset/scatter/1365.pdf}

\figsetgrpend

\figsetgrpstart
\figsetgrpnum{2.1366}

\figsetplot{./figset/scatter/1366.pdf}

\figsetgrpend

\figsetgrpstart
\figsetgrpnum{2.1367}

\figsetplot{./figset/scatter/1367.pdf}

\figsetgrpend

\figsetgrpstart
\figsetgrpnum{2.1368}

\figsetplot{./figset/scatter/1368.pdf}

\figsetgrpend

\figsetgrpstart
\figsetgrpnum{2.1369}

\figsetplot{./figset/scatter/1369.pdf}

\figsetgrpend

\figsetgrpstart
\figsetgrpnum{2.1370}

\figsetplot{./figset/scatter/1370.pdf}

\figsetgrpend

\figsetgrpstart
\figsetgrpnum{2.1371}

\figsetplot{./figset/scatter/1371.pdf}

\figsetgrpend

\figsetgrpstart
\figsetgrpnum{2.1372}

\figsetplot{./figset/scatter/1372.pdf}

\figsetgrpend

\figsetgrpstart
\figsetgrpnum{2.1373}

\figsetplot{./figset/scatter/1373.pdf}

\figsetgrpend

\figsetgrpstart
\figsetgrpnum{2.1374}

\figsetplot{./figset/scatter/1374.pdf}

\figsetgrpend

\figsetgrpstart
\figsetgrpnum{2.1375}

\figsetplot{./figset/scatter/1375.pdf}

\figsetgrpend

\figsetgrpstart
\figsetgrpnum{2.1376}

\figsetplot{./figset/scatter/1376.pdf}

\figsetgrpend

\figsetgrpstart
\figsetgrpnum{2.1377}

\figsetplot{./figset/scatter/1377.pdf}

\figsetgrpend

\figsetgrpstart
\figsetgrpnum{2.1378}

\figsetplot{./figset/scatter/1378.pdf}

\figsetgrpend

\figsetgrpstart
\figsetgrpnum{2.1379}

\figsetplot{./figset/scatter/1379.pdf}

\figsetgrpend

\figsetgrpstart
\figsetgrpnum{2.1380}

\figsetplot{./figset/scatter/1380.pdf}

\figsetgrpend

\figsetgrpstart
\figsetgrpnum{2.1381}

\figsetplot{./figset/scatter/1381.pdf}

\figsetgrpend

\figsetgrpstart
\figsetgrpnum{2.1382}

\figsetplot{./figset/scatter/1382.pdf}

\figsetgrpend

\figsetgrpstart
\figsetgrpnum{2.1383}

\figsetplot{./figset/scatter/1383.pdf}

\figsetgrpend

\figsetgrpstart
\figsetgrpnum{2.1384}

\figsetplot{./figset/scatter/1384.pdf}

\figsetgrpend

\figsetgrpstart
\figsetgrpnum{2.1385}

\figsetplot{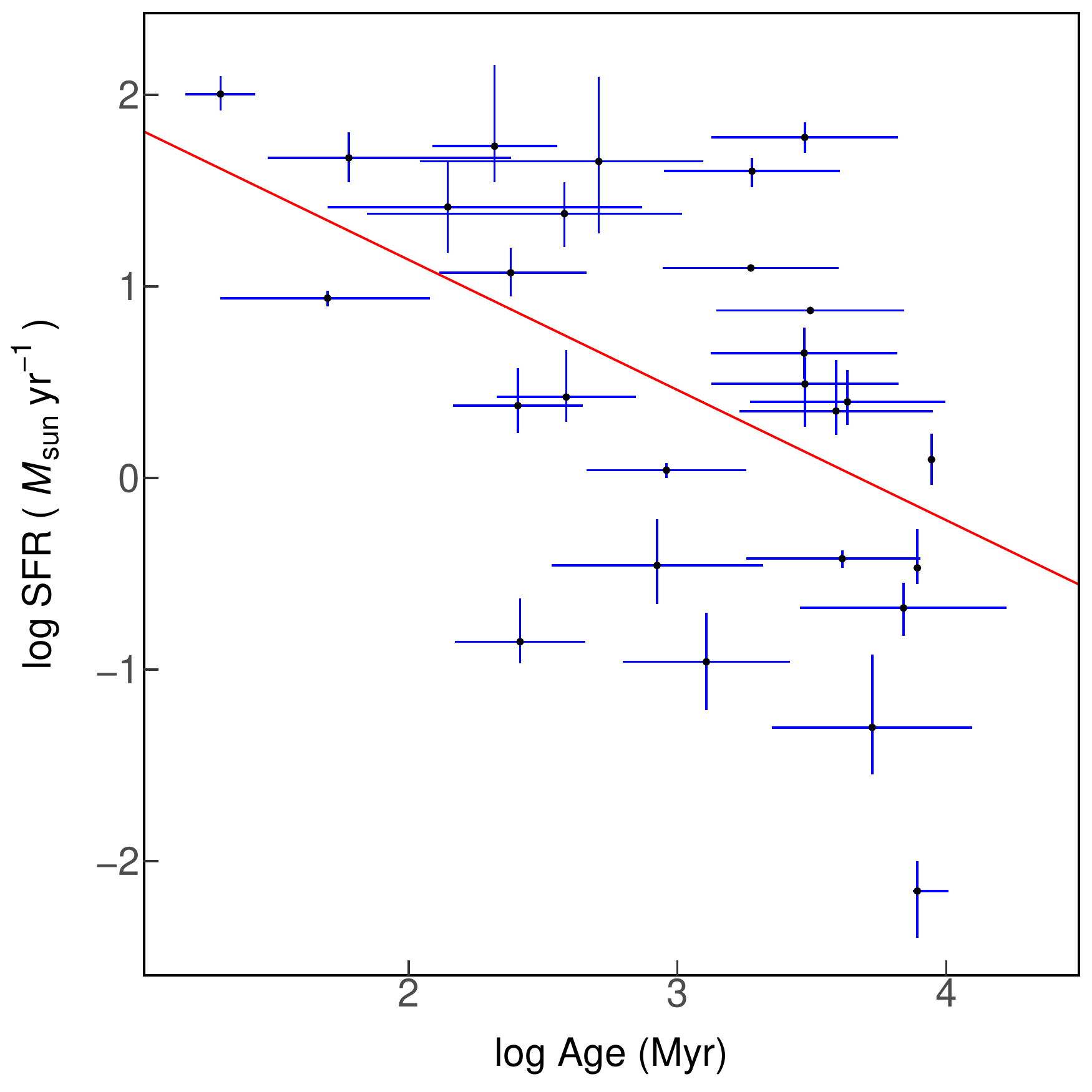}

\figsetgrpend

\figsetgrpstart
\figsetgrpnum{2.1386}

\figsetplot{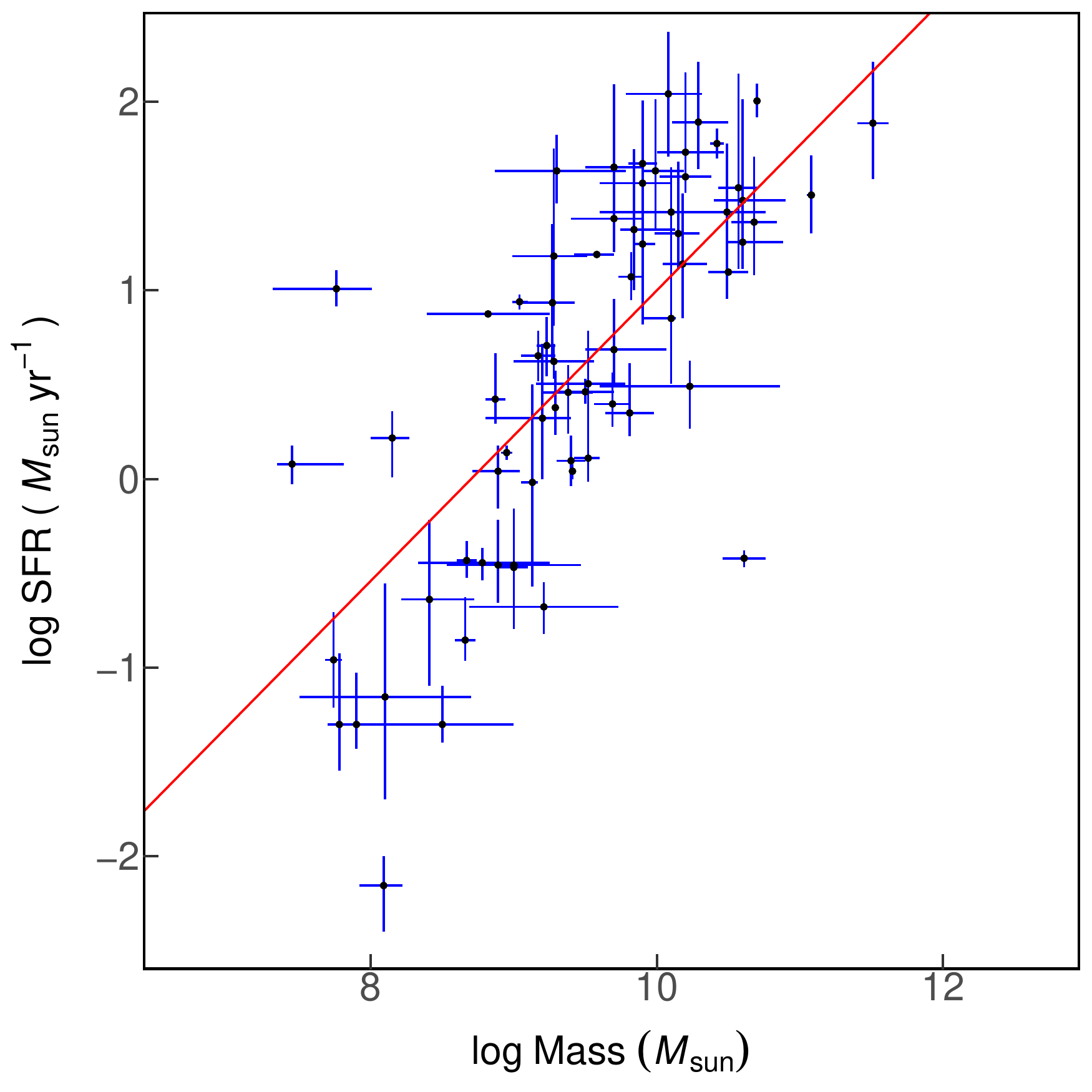}

\figsetgrpend

\figsetgrpstart
\figsetgrpnum{2.1387}

\figsetplot{./figset/scatter/1387.pdf}

\figsetgrpend

\figsetgrpstart
\figsetgrpnum{2.1388}

\figsetplot{./figset/scatter/1388.pdf}

\figsetgrpend

\figsetgrpstart
\figsetgrpnum{2.1389}

\figsetplot{./figset/scatter/1389.pdf}

\figsetgrpend

\figsetgrpstart
\figsetgrpnum{2.1390}

\figsetplot{./figset/scatter/1390.pdf}

\figsetgrpend

\figsetgrpstart
\figsetgrpnum{2.1391}

\figsetplot{./figset/scatter/1391.pdf}

\figsetgrpend

\figsetgrpstart
\figsetgrpnum{2.1392}

\figsetplot{./figset/scatter/1392.pdf}

\figsetgrpend

\figsetgrpstart
\figsetgrpnum{2.1393}

\figsetplot{./figset/scatter/1393.pdf}

\figsetgrpend

\figsetgrpstart
\figsetgrpnum{2.1394}

\figsetplot{./figset/scatter/1394.pdf}

\figsetgrpend

\figsetgrpstart
\figsetgrpnum{2.1395}

\figsetplot{./figset/scatter/1395.pdf}

\figsetgrpend

\figsetgrpstart
\figsetgrpnum{2.1396}

\figsetplot{./figset/scatter/1396.pdf}

\figsetgrpend

\figsetgrpstart
\figsetgrpnum{2.1397}

\figsetplot{./figset/scatter/1397.pdf}

\figsetgrpend

\figsetgrpstart
\figsetgrpnum{2.1398}

\figsetplot{./figset/scatter/1398.pdf}

\figsetgrpend

\figsetgrpstart
\figsetgrpnum{2.1399}

\figsetplot{./figset/scatter/1399.pdf}

\figsetgrpend

\figsetgrpstart
\figsetgrpnum{2.1400}

\figsetplot{./figset/scatter/1400.pdf}

\figsetgrpend

\figsetgrpstart
\figsetgrpnum{2.1401}

\figsetplot{./figset/scatter/1401.pdf}

\figsetgrpend

\figsetgrpstart
\figsetgrpnum{2.1402}

\figsetplot{./figset/scatter/1402.pdf}

\figsetgrpend

\figsetgrpstart
\figsetgrpnum{2.1403}

\figsetplot{./figset/scatter/1403.pdf}

\figsetgrpend

\figsetgrpstart
\figsetgrpnum{2.1404}

\figsetplot{./figset/scatter/1404.pdf}

\figsetgrpend

\figsetgrpstart
\figsetgrpnum{2.1405}

\figsetplot{./figset/scatter/1405.pdf}

\figsetgrpend

\figsetgrpstart
\figsetgrpnum{2.1406}

\figsetplot{./figset/scatter/1406.pdf}

\figsetgrpend

\figsetgrpstart
\figsetgrpnum{2.1407}

\figsetplot{./figset/scatter/1407.pdf}

\figsetgrpend

\figsetgrpstart
\figsetgrpnum{2.1408}

\figsetplot{./figset/scatter/1408.pdf}

\figsetgrpend

\figsetgrpstart
\figsetgrpnum{2.1409}

\figsetplot{./figset/scatter/1409.pdf}

\figsetgrpend

\figsetgrpstart
\figsetgrpnum{2.1410}

\figsetplot{./figset/scatter/1410.pdf}

\figsetgrpend

\figsetgrpstart
\figsetgrpnum{2.1411}

\figsetplot{./figset/scatter/1411.pdf}

\figsetgrpend

\figsetgrpstart
\figsetgrpnum{2.1412}

\figsetplot{./figset/scatter/1412.pdf}

\figsetgrpend

\figsetgrpstart
\figsetgrpnum{2.1413}

\figsetplot{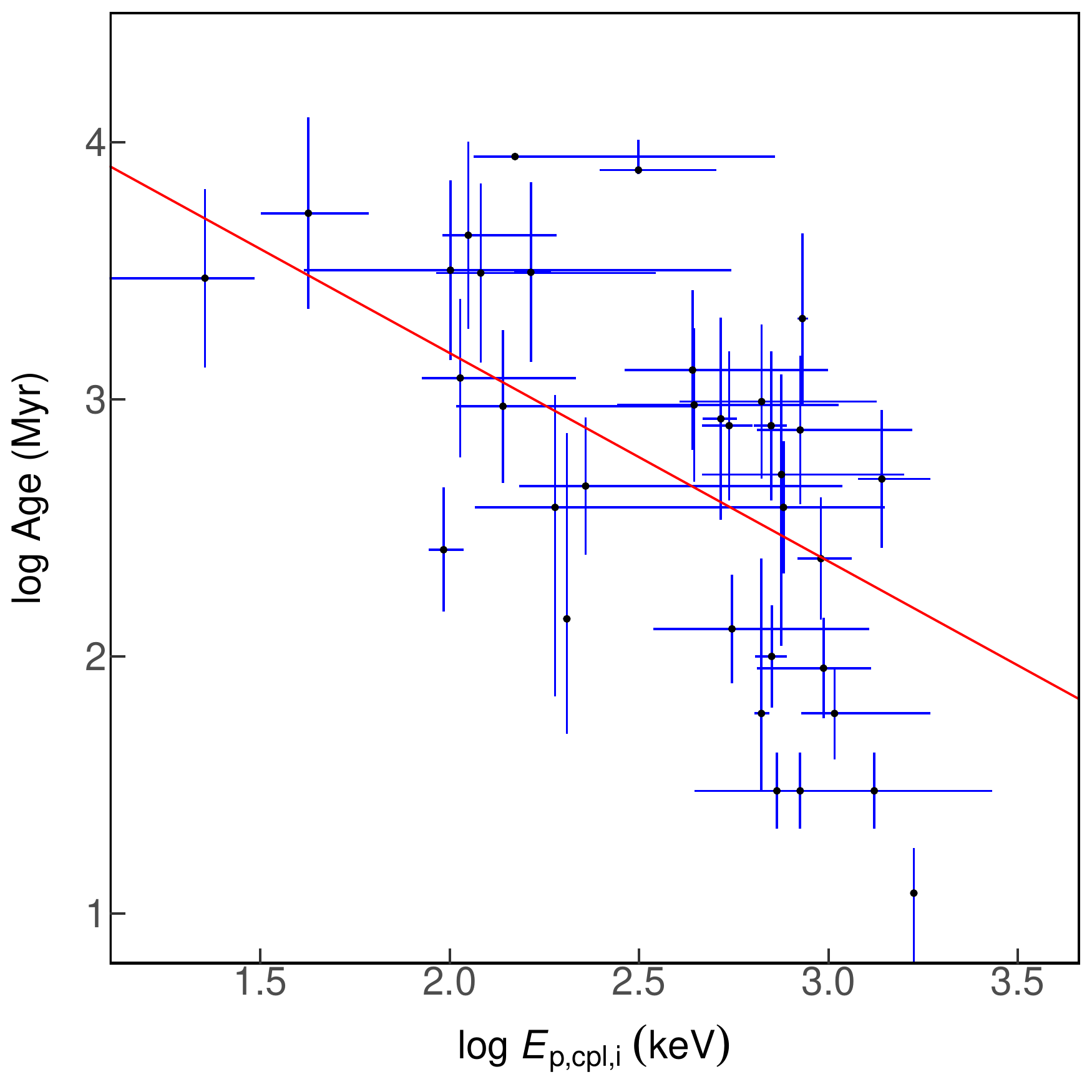}

\figsetgrpend

\figsetgrpstart
\figsetgrpnum{2.1414}

\figsetplot{./figset/scatter/1414.pdf}

\figsetgrpend

\figsetgrpstart
\figsetgrpnum{2.1415}

\figsetplot{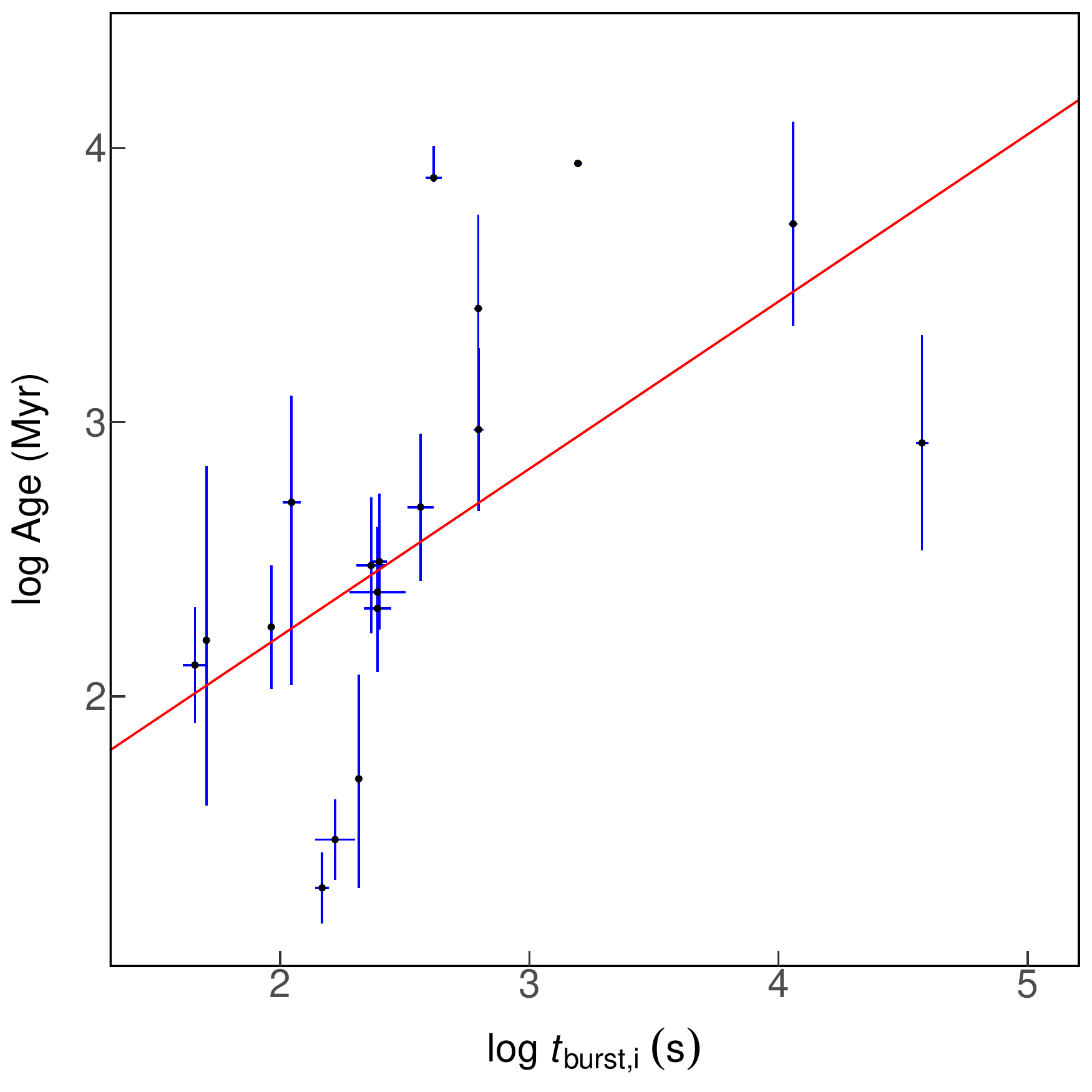}

\figsetgrpend

\figsetgrpstart
\figsetgrpnum{2.1416}

\figsetplot{./figset/scatter/1416.pdf}

\figsetgrpend

\figsetgrpstart
\figsetgrpnum{2.1417}

\figsetplot{./figset/scatter/1417.pdf}

\figsetgrpend

\figsetgrpstart
\figsetgrpnum{2.1418}

\figsetplot{./figset/scatter/1418.pdf}

\figsetgrpend

\figsetgrpstart
\figsetgrpnum{2.1419}

\figsetplot{./figset/scatter/1419.pdf}

\figsetgrpend

\figsetgrpstart
\figsetgrpnum{2.1420}

\figsetplot{./figset/scatter/1420.pdf}

\figsetgrpend

\figsetgrpstart
\figsetgrpnum{2.1421}

\figsetplot{./figset/scatter/1421.pdf}

\figsetgrpend

\figsetgrpstart
\figsetgrpnum{2.1422}

\figsetplot{./figset/scatter/1422.pdf}

\figsetgrpend

\figsetgrpstart
\figsetgrpnum{2.1423}

\figsetplot{./figset/scatter/1423.pdf}

\figsetgrpend

\figsetgrpstart
\figsetgrpnum{2.1424}

\figsetplot{./figset/scatter/1424.pdf}

\figsetgrpend

\figsetgrpstart
\figsetgrpnum{2.1425}

\figsetplot{./figset/scatter/1425.pdf}

\figsetgrpend

\figsetgrpstart
\figsetgrpnum{2.1426}

\figsetplot{./figset/scatter/1426.pdf}

\figsetgrpend

\figsetgrpstart
\figsetgrpnum{2.1427}

\figsetplot{./figset/scatter/1427.pdf}

\figsetgrpend

\figsetgrpstart
\figsetgrpnum{2.1428}

\figsetplot{./figset/scatter/1428.pdf}

\figsetgrpend

\figsetgrpstart
\figsetgrpnum{2.1429}

\figsetplot{./figset/scatter/1429.pdf}

\figsetgrpend

\figsetgrpstart
\figsetgrpnum{2.1430}

\figsetplot{./figset/scatter/1430.pdf}

\figsetgrpend

\figsetgrpstart
\figsetgrpnum{2.1431}

\figsetplot{./figset/scatter/1431.pdf}

\figsetgrpend

\figsetgrpstart
\figsetgrpnum{2.1432}

\figsetplot{./figset/scatter/1432.pdf}

\figsetgrpend

\figsetgrpstart
\figsetgrpnum{2.1433}

\figsetplot{./figset/scatter/1433.pdf}

\figsetgrpend

\figsetgrpstart
\figsetgrpnum{2.1434}

\figsetplot{./figset/scatter/1434.pdf}

\figsetgrpend

\figsetgrpstart
\figsetgrpnum{2.1435}

\figsetplot{./figset/scatter/1435.pdf}

\figsetgrpend

\figsetgrpstart
\figsetgrpnum{2.1436}

\figsetplot{./figset/scatter/1436.pdf}

\figsetgrpend

\figsetgrpstart
\figsetgrpnum{2.1437}

\figsetplot{./figset/scatter/1437.pdf}

\figsetgrpend

\figsetgrpstart
\figsetgrpnum{2.1438}

\figsetplot{./figset/scatter/1438.pdf}

\figsetgrpend

\figsetgrpstart
\figsetgrpnum{2.1439}

\figsetplot{./figset/scatter/1439.pdf}

\figsetgrpend

\figsetgrpstart
\figsetgrpnum{2.1440}

\figsetplot{./figset/scatter/1440.pdf}

\figsetgrpend

\figsetgrpstart
\figsetgrpnum{2.1441}

\figsetplot{./figset/scatter/1441.pdf}

\figsetgrpend

\figsetgrpstart
\figsetgrpnum{2.1442}

\figsetplot{./figset/scatter/1442.pdf}

\figsetgrpend

\figsetgrpstart
\figsetgrpnum{2.1443}

\figsetplot{./figset/scatter/1443.pdf}

\figsetgrpend

\figsetgrpstart
\figsetgrpnum{2.1444}

\figsetplot{./figset/scatter/1444.pdf}

\figsetgrpend

\figsetgrpstart
\figsetgrpnum{2.1445}

\figsetplot{./figset/scatter/1445.pdf}

\figsetgrpend

\figsetgrpstart
\figsetgrpnum{2.1446}

\figsetplot{./figset/scatter/1446.pdf}

\figsetgrpend

\figsetgrpstart
\figsetgrpnum{2.1447}

\figsetplot{./figset/scatter/1447.pdf}

\figsetgrpend

\figsetgrpstart
\figsetgrpnum{2.1448}

\figsetplot{./figset/scatter/1448.pdf}

\figsetgrpend

\figsetgrpstart
\figsetgrpnum{2.1449}

\figsetplot{./figset/scatter/1449.pdf}

\figsetgrpend

\figsetgrpstart
\figsetgrpnum{2.1450}

\figsetplot{./figset/scatter/1450.pdf}

\figsetgrpend

\figsetgrpstart
\figsetgrpnum{2.1451}

\figsetplot{./figset/scatter/1451.pdf}

\figsetgrpend

\figsetgrpstart
\figsetgrpnum{2.1452}

\figsetplot{./figset/scatter/1452.pdf}

\figsetgrpend

\figsetgrpstart
\figsetgrpnum{2.1453}

\figsetplot{./figset/scatter/1453.pdf}

\figsetgrpend

\figsetgrpstart
\figsetgrpnum{2.1454}

\figsetplot{./figset/scatter/1454.pdf}

\figsetgrpend

\figsetgrpstart
\figsetgrpnum{2.1455}

\figsetplot{./figset/scatter/1455.pdf}

\figsetgrpend

\figsetgrpstart
\figsetgrpnum{2.1456}

\figsetplot{./figset/scatter/1456.pdf}

\figsetgrpend

\figsetgrpstart
\figsetgrpnum{2.1457}

\figsetplot{./figset/scatter/1457.pdf}

\figsetgrpend

\figsetgrpstart
\figsetgrpnum{2.1458}

\figsetplot{./figset/scatter/1458.pdf}

\figsetgrpend

\figsetgrpstart
\figsetgrpnum{2.1459}

\figsetplot{./figset/scatter/1459.pdf}

\figsetgrpend

\figsetgrpstart
\figsetgrpnum{2.1460}

\figsetplot{./figset/scatter/1460.pdf}

\figsetgrpend

\figsetgrpstart
\figsetgrpnum{2.1461}

\figsetplot{./figset/scatter/1461.pdf}

\figsetgrpend

\figsetgrpstart
\figsetgrpnum{2.1462}

\figsetplot{./figset/scatter/1462.pdf}

\figsetgrpend

\figsetgrpstart
\figsetgrpnum{2.1463}

\figsetplot{./figset/scatter/1463.pdf}

\figsetgrpend

\figsetgrpstart
\figsetgrpnum{2.1464}

\figsetplot{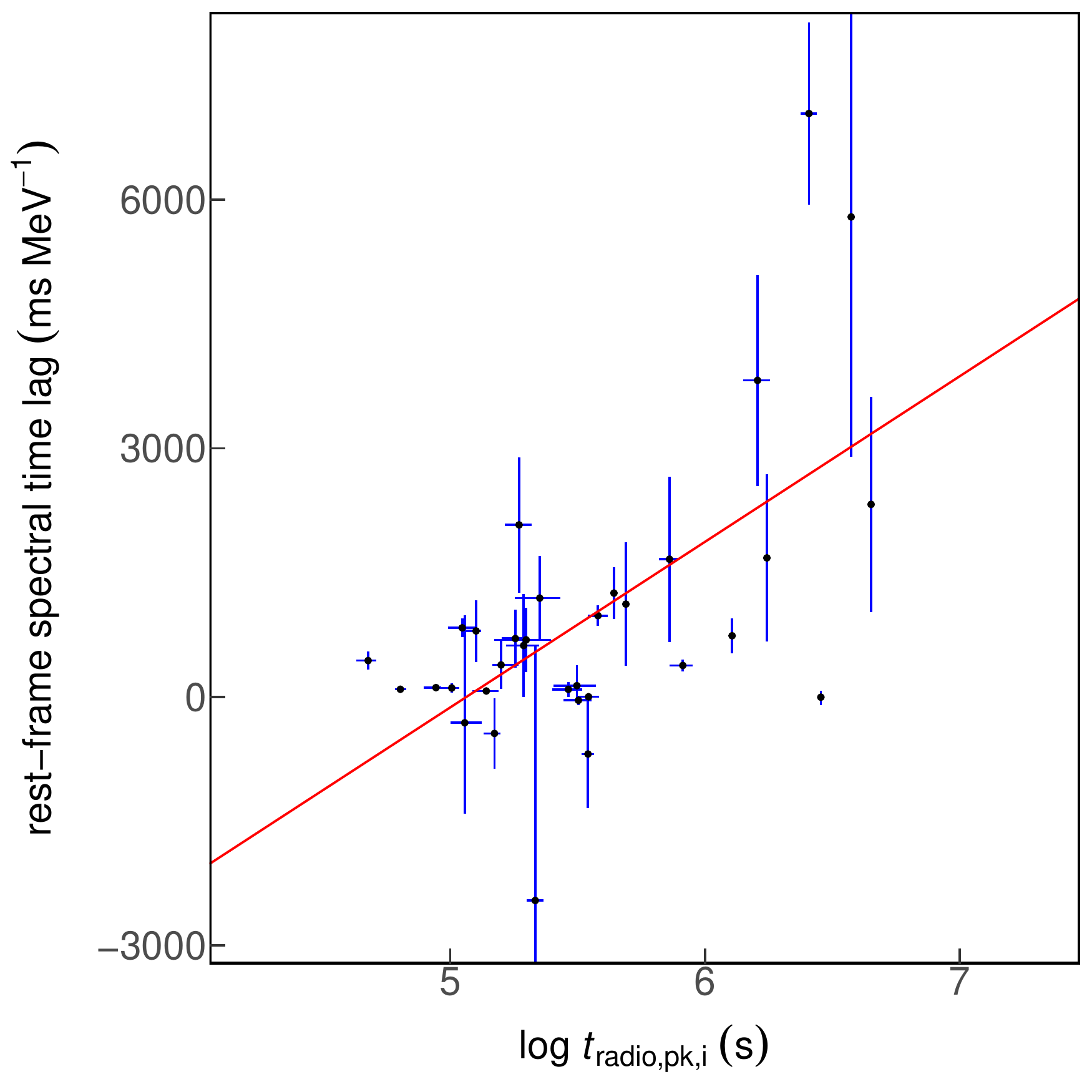}

\figsetgrpend

\figsetgrpstart
\figsetgrpnum{2.1465}

\figsetplot{./figset/scatter/1465.pdf}

\figsetgrpend

\figsetgrpstart
\figsetgrpnum{2.1466}

\figsetplot{./figset/scatter/1466.pdf}

\figsetgrpend

\figsetgrpstart
\figsetgrpnum{2.1467}

\figsetplot{./figset/scatter/1467.pdf}

\figsetgrpend

\figsetgrpstart
\figsetgrpnum{2.1468}

\figsetplot{./figset/scatter/1468.pdf}

\figsetgrpend

\figsetend

\subsection{Correlation coefficients} \label{subsec:coefficient}
Correlation coefficient is to measure the correlation between two parameters, there are 4 methods:

\begin{enumerate}

\item $\rm Pearson$ correlation coefficient \citep{Pearson1895}. Supposed the two variables are $x_{\rm t}$ and $y_{\rm t}$.
\begin{equation} \label{eq:pearson}
\gamma=\frac{\sum_{t=1}^{N}(x_{\rm t}-\bar{x})(y_{\rm t}-\bar{y})}{\sqrt{\sum_{t=1}^{N}(x_{\rm t}-\bar{x})^{2}}\sqrt{\sum_{t=1}^{N}(y_{\rm t}-\bar{y})^{2}}},
\end{equation}
where $\bar{x}$ is the mean of variable $x$, $\bar{y}$ is the mean of variable $y$. $\rm Pearson$ correlation coefficient is based on normal distribution, so when the data is normal distribution, the $\rm Pearson$ correlation coefficient works very well.

\item $\rm Spearman$ correlation coefficient \citep{Spearman1987}. This coefficient does not depend on the data distribution. It is a kind of rank measure.
\begin{equation} \label{eq:spearman}
\rho=\frac{n(\sum_{t=1}^{N}x_{\rm t}y_{\rm t})-(\sum_{t=1}^{N}x_{\rm t})(\sum_{t=1}^{N}y_{\rm t})}{\sqrt{n(\sum_{t=1}^{N}x_{\rm t}^{2})-(\sum_{t=1}^{N}x_{\rm t})^{2}}\sqrt{n(\sum_{t=1}^{N}y_{\rm t}^{2})-(\sum_{t=1}^{N}y_{\rm t})^{2}}},
\end{equation}

\item $\rm Kendall$ $\tau$ correlation coefficient \citep{Kendall1938}. This coefficient compares the order of the two variables, not comparing the values. It is also a kind of rank measure.
\begin{equation} \label{eq:kendall}
\tau=\frac{n_{\rm c}-n_{\rm d}}{1/2n(n-1)},
\end{equation}
where $n$ is the length of the two variables, $n_{\rm c}$ is the number of concordant pairs, $n_{\rm d}$ is the number of discordant pairs.

\item Cosine similarity \citep{Van2012}. Supposed the two variables are $x_{\rm t}$ and $y_{\rm t}$, the cosine similarity definition is
\begin{equation} \label{eq:cos}
cos\theta_{\rm x,y}=\frac{\sum_{t=1}^{N}x_{\rm t}y_{\rm t}}{\sqrt{\sum_{t=1}^{N}x_{\rm t}^{2}\sum_{t=1}^{N}y_{\rm t}^{2}}},
\end{equation}
when $cos\theta_{\rm x,y}=1$, it means $x_{\rm t}$ and $y_{\rm t}$ are completely similar, when $cos\theta_{\rm x,y}$ is more close to 1, the two variables are more similar.

\end{enumerate}

\subsection{Correlation ratio} \label{subsec:ratio}
Pearson correlation coefficients measure the linear correlation between two variables. But when coming to nonlinear correlation, correlation coefficients do not work well. So we need correlation ratio \citep{Fisher1970} to measure the nonlinear correlation between different variables. Correlation ratio is a ratio of the statistical dispersion within individual categories and the dispersion across the whole population or sample, defined as the ratio of the standard deviation of within individual categories and the standard deviation of the the whole population or sample.

Supposed the variables is $y_{\rm xi}$, x means the category, i means the $ith$ value for category x, the number of category x is $n_{\rm x}$,
\begin{equation} \label{eq:ratio1}
\bar{y}_{\rm x}=\frac{\sum_{\rm i}y_{\rm xi}}{n_{\rm x}},
\end{equation}

\begin{equation} \label{eq:ratio2}
\bar{y}=\frac{\sum_{\rm x}n_{\rm x}\bar{y}_{\rm x}}{\sum_{\rm x}n_{\rm x}},
\end{equation}
where $\bar{y}_{\rm x}$ is the mean of category x, $\bar{y}$ is the mean of the whole sample, the definition of correlation ratio $\eta$ is

\begin{equation} \label{eq:ratio3}
\eta^{2}=\frac{\sum_{\rm x}n_{\rm x}(\bar{y}_{\rm x}-\bar{y})^{2}}{\sum_{\rm x,i}(y_{\rm xi}-\bar{y})^{2}},
\end{equation}
it is the ratio of the weighted variance of the category means and the variance of the whole sample. $\eta$ is a value between 0 and 1, when $\eta$ is more close to 1, the nonlinear correlation is stronger.

When the two parameters have at least 5 GRBs, then we gave the correlation coefficients and correlation ratio, and we excluded some trivial  results. It means we removed the correlations between the rest-frame and observer-frame for the same parameters, like the correlation between $T_{\rm 90}$ and $T_{\rm 90,i}$. We have many results, so we just put the first 5 results in Table \ref{tab:coefficient}. One can find all the results in machine readable table. In Table \ref{tab:coefficient}, we gave the results of Pearson, Spearman, Kendall $\tau$ coefficients and the related p-values. We also considered all the error bars using MC method \citep{Zou2017}. The last two rows are correlation ratio and cosine similarity with error bars, also using MC method. We did the statistics between two arbitrary parameters in order to find the remarkable correlations. In Section \ref{sec:tworesult}, we analyzed some interesting results, and gave some reasonable physical explanations. In this paper, we just analyzed the linear correlations. During all the analysis, Pearson, Spearman, Kendall $\tau$ coefficients and the related p-values are the most important references of linear correlations. When the hypothesis testing p-value is smaller than 0.1, it means the correlation has very high probability to be true. The sample selection biases is quite complicated (see for examples of the related discussion, \citep{Lloyd1999, Lloyd2000, Dainotti2013, Dainotti2015A, Dainotti2015B}). One should be caution when considering the results. When the absolute values of Pearson, Spearman, Kendall $\tau$ coefficients are bigger, the correlation is stronger. For cosine similarity, there is no hypothesis testing, so it is just used to confirm the linear correlations. For correlation ratio, we will analyzed the nonlinear correlations in the future with this value.

\subsection{Linear regression of two parameters and three parameters} \label{subsec:regression}

We did the linear regression between two parameters and three parameters arbitrarily when the sample number is not smaller than 5. We also considered all the error bars using MC method \citep{Zou2017}. We excluded some trivial results. As an example, the first 5 linear regression results are shown in Table \ref{tab:linear2} and Table \ref{tab:linear3}, respectively. The full linear regression results are given in two machine readable tables.

In Table \ref{tab:linear2}, $y$ is dependent variable, $x$ is independent variable, $b$ is the intercept of the linear model, and $a$ is the linear coefficient of $x$. Adjusted $R^{2}$ is used to measure the goodness of the regression model. It means the percentage of variance explained considering the parameter freedom. We use the adjusted $R^{2}$ calculated with central values for representation. Because every time of MC, adjusted $R^{2}$ does not change too much. In Table \ref{tab:linear3}, $x_{1}$ and $x_{2}$ are independent variables, $a_{1}$ and $a_{2}$ are the linear coefficients of $x_{1}$ and $x_{2}$ respectively. $y$, $b$ and adjusted $R^{2}$ are same meanings as in Table \ref{tab:linear2}. Readers can see \citet{Hron2012} for more details about linear regression.


\section{Remarkable results between two parameters} \label{sec:tworesult}

In this section, we show some remarkable correlations between two parameters. The strategy of choosing the results are the following. The correlation should have at least 10 GRBs. Adjusted $R^{2}$ should be bigger than 0.2 except Section \ref{subsec:break}. The whole linear model and $a$ have hypothesis testing p-values smaller than 0.05. Pearson, Spearman, Kendall $\tau$ coefficients have at least one hypothesis testing p-value smaller than 0.1. For peak energy flux $F_{\rm pk}$ and peak photon flux $P_{\rm pk}$, both of them have four different time bins. If $F_{\rm pk}$ in all the four time bins correlates with another one same parameter, we just showed the best time bin result. The best time bin result has biggest adjusted $R^{2}$. $P_{\rm pk}$ is the same. Other results are all in machine readable tables. Since some relations have been discovered before, we compared the differences between the current and previous analysis. Readers can also see Table \ref{tab:comparison} for a clear comparison. We just show a part of results in the paper. We provide all the results with figure sets and machine readable tables including our original data. If readers want to analyze with other statistical methods, they can use our original data in machine readable table. One may notice that not only the correlations, but the non-existence of correlations among a certain quantities may also reveal underlying physics of GRBs.

\subsection{Some interesting breaks} \label{subsec:break}

Interestingly, we found there is a break in the $\log F_{\rm pk}$ and $(-\alpha_{\rm spl})$ plots for all 4 time bins.

\begin{figure*}[!b]
\centering
\includegraphics[width=0.45\textwidth]{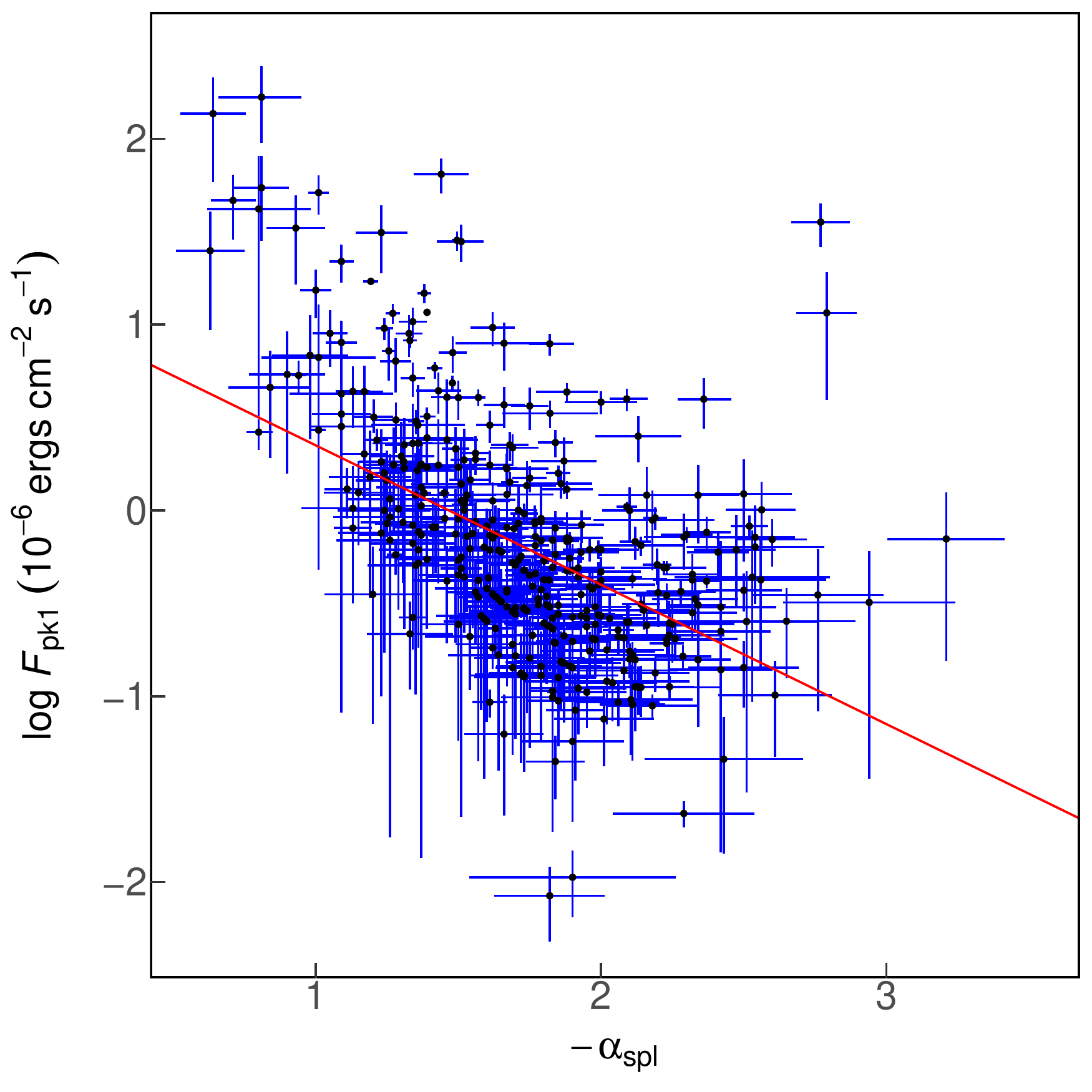}

\caption{
Scatter plot for $\log F_{\rm pk1}$ and $(-\alpha_{\rm spl})$.
The red line is our fit result. The formula of the red line is $\log F_{\rm pk1} = (-0.75 \pm 0.046) \times (-\alpha_{\rm spl}) + (1.1 \pm 0.083)$. The description of every parameter is in Section \ref{sec:sample}.
}
\label{fig:fpk1alpl}
\end{figure*}

The correlation between $\log F_{\rm pk1}$ and $(-\alpha_{\rm spl})$ is:
\begin{equation} \label{eq:fpk1alpl}
\log F_{\rm pk1} = (-0.75 \pm 0.046) \times (-\alpha_{\rm spl}) + (1.1 \pm 0.083),
\end{equation}
where $F_{\rm pk1}$ is peak energy flux of 1 $\rm s$ time bin in rest-frame 1-$10^{4}$ $\rm keV$ energy band and in unit of $\rm 10^{\rm -6} ~ ergs ~ cm^{\rm -2} ~ s^{\rm -1}$. The adjusted $R^{2}$ is 0.3. The Pearson coefficient is $-0.47 \pm 0.028$ with p-value $4.2 \times 10^{-22}$. The Spearman coefficient is $-0.49 \pm 0.024$ with p-value $5.8 \times 10^{-25}$. The Kendall $\tau$ coefficient is $-0.34 \pm 0.018$ with p-value $7.3 \times 10^{-24}$. The correlation ratio is $0.86 \pm 0.0037$. The cosine similarity is $-0.37 \pm 0.013$. The scatter plot is in Figure \ref{fig:fpk1alpl} \footnote{Notice all the figures in this section are also listed in Figure. Set 2.}. The GRB sample number is 385.

\begin{figure*}[!b]
\centering
\includegraphics[width=0.45\textwidth]{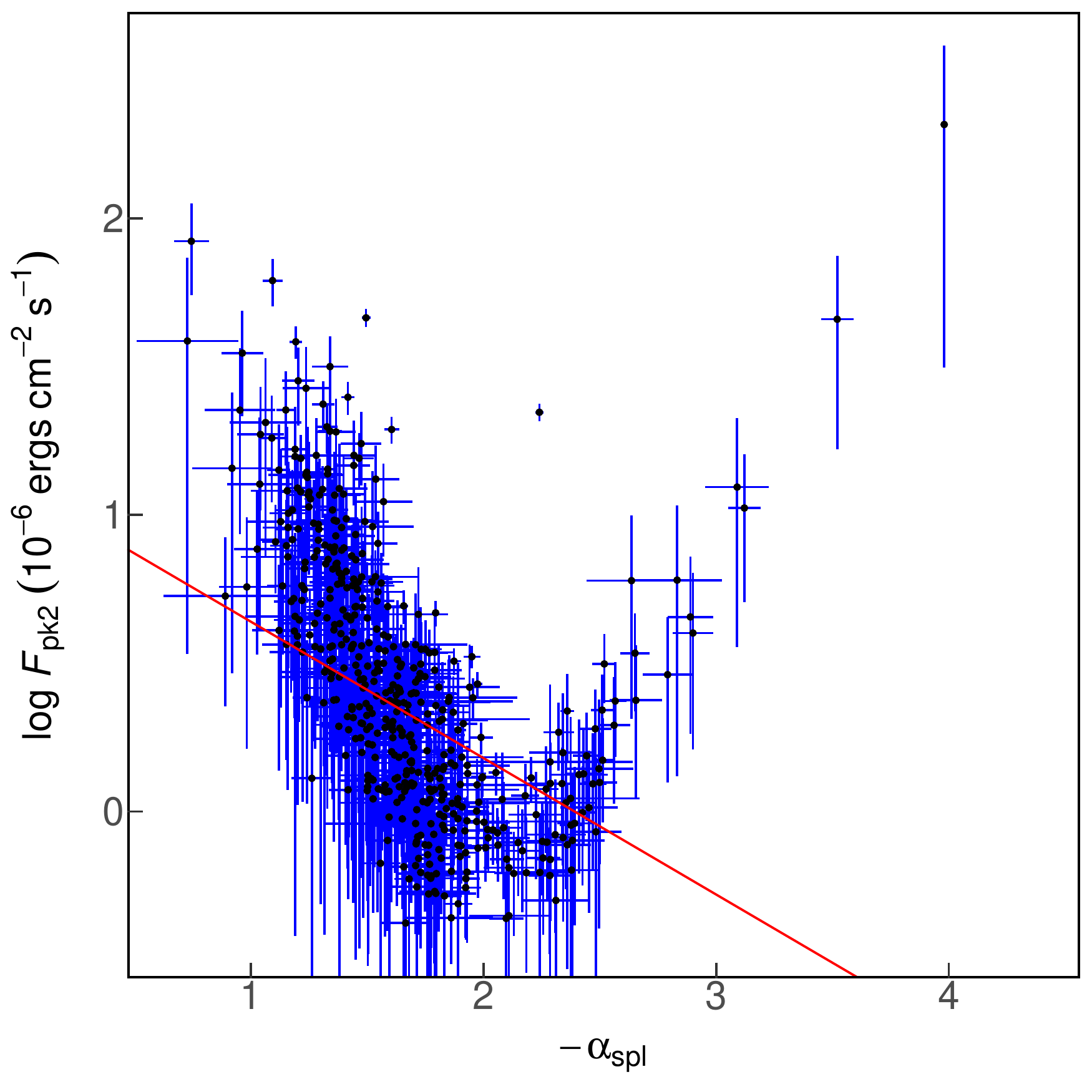}

\caption{
Scatter plot for $\log F_{\rm pk2}$ and $(-\alpha_{\rm spl})$.
The red line is our fit result. The formula of the red line is $\log F_{\rm pk2} = (-0.46 \pm 0.036) \times (-\alpha_{\rm spl}) + (1.1 \pm 0.063)$. The description of every parameter is in Section \ref{sec:sample}.
}
\label{fig:fpk2alpl}
\end{figure*}

The correlation between $\log F_{\rm pk2}$ and $(-\alpha_{\rm spl})$ is:
\begin{equation} \label{eq:fpk2alpl}
\log F_{\rm pk2} = (-0.46 \pm 0.036) \times (-\alpha_{\rm spl}) + (1.1 \pm 0.063),
\end{equation}
where $F_{\rm pk2}$ is peak energy flux of 64 $\rm ms$ time bin in rest-frame 1-$10^{4}$ $\rm keV$ energy band and in unit of $\rm 10^{\rm -6} ~ ergs ~ cm^{\rm -2} ~ s^{\rm -1}$. The adjusted $R^{2}$ is 0.21. The Pearson coefficient is $-0.37 \pm 0.03$ with p-value $1.8 \times 10^{-17}$. The Spearman coefficient is $-0.55 \pm 0.023$ with p-value $8.8 \times 10^{-41}$. The Kendall $\tau$ coefficient is $-0.39 \pm 0.018$ with p-value $3.2 \times 10^{-38}$. The correlation ratio is $0.81 \pm 0.0045$. The cosine similarity is $0.5 \pm 0.018$. The scatter plot is in Figure \ref{fig:fpk2alpl}. The GRB sample number is 490.

\begin{figure*}[!b]
\centering
\includegraphics[width=0.45\textwidth]{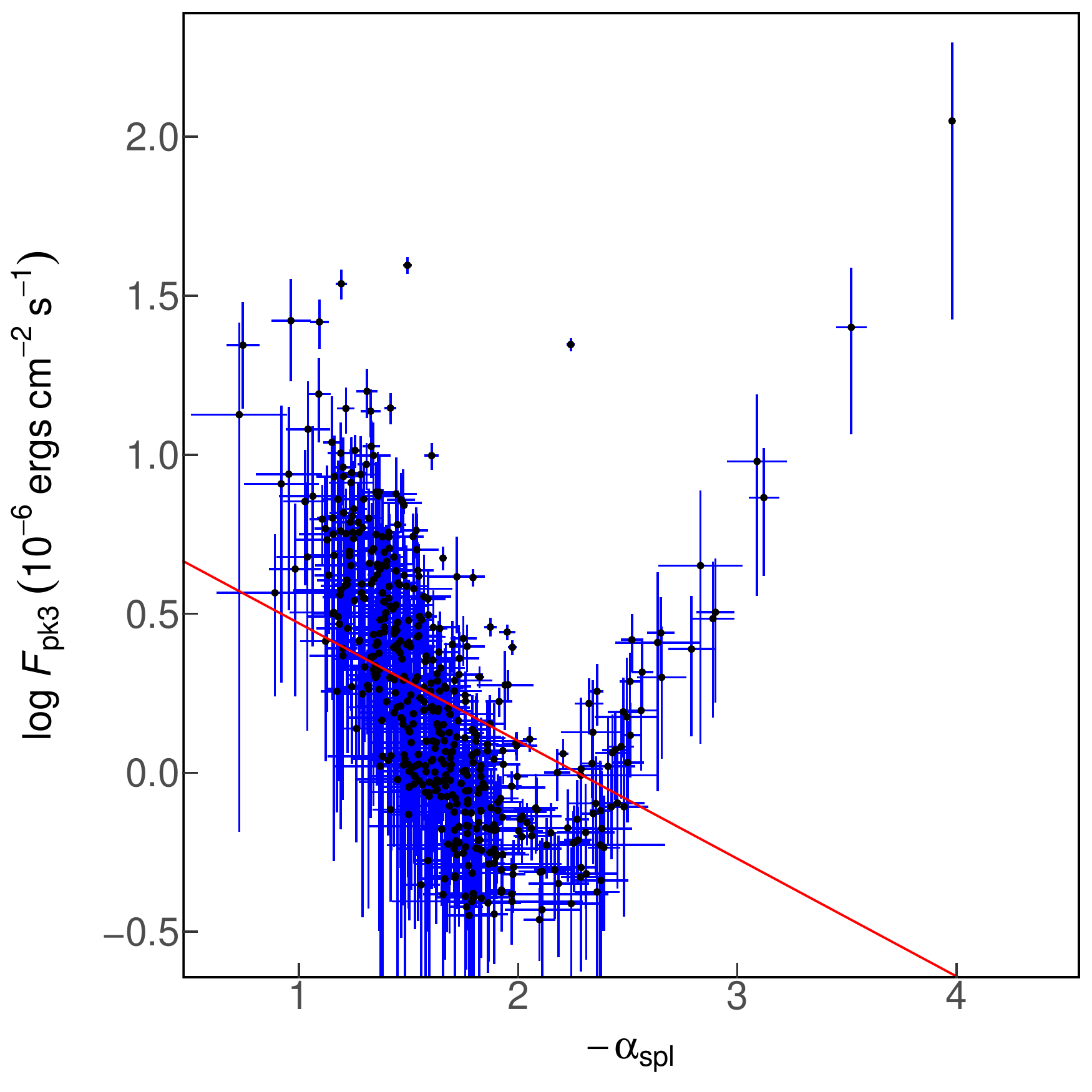}

\caption{
Scatter plot for $\log F_{\rm pk3}$ and $(-\alpha_{\rm spl})$.
The red line is our fit result. The formula of the red line is $\log F_{\rm pk3} = (-0.37 \pm 0.032) \times (-\alpha_{\rm spl}) + (0.84 \pm 0.057)$. The description of every parameter is in Section \ref{sec:sample}.
}
\label{fig:fpk3alpl}
\end{figure*}

The correlation between $\log F_{\rm pk3}$ and $(-\alpha_{\rm spl})$ is:
\begin{equation} \label{eq:fpk3alpl}
\log F_{\rm pk3} = (-0.37 \pm 0.032) \times (-\alpha_{\rm spl}) + (0.84 \pm 0.057),
\end{equation}
where $F_{\rm pk3}$ is peak energy flux of 256 $\rm ms$ time bin in rest-frame 1-$10^{4}$ $\rm keV$ energy band and in unit of $\rm 10^{\rm -6} ~ ergs ~ cm^{\rm -2} ~ s^{\rm -1}$. The adjusted $R^{2}$ is 0.17. The Pearson coefficient is $-0.33 \pm 0.031$ with p-value $4.1 \times 10^{-14}$. The Spearman coefficient is $-0.54 \pm 0.024$ with p-value $1.6 \times 10^{-37}$. The Kendall $\tau$ coefficient is $-0.38 \pm 0.018$ with p-value $6.7 \times 10^{-36}$. The correlation ratio is $0.85 \pm 0.0038$. The cosine similarity is $0.33 \pm 0.021$. The scatter plot is in Figure \ref{fig:fpk3alpl}. The GRB sample number is 487.

\begin{figure*}[!b]
\centering
\includegraphics[width=0.45\textwidth]{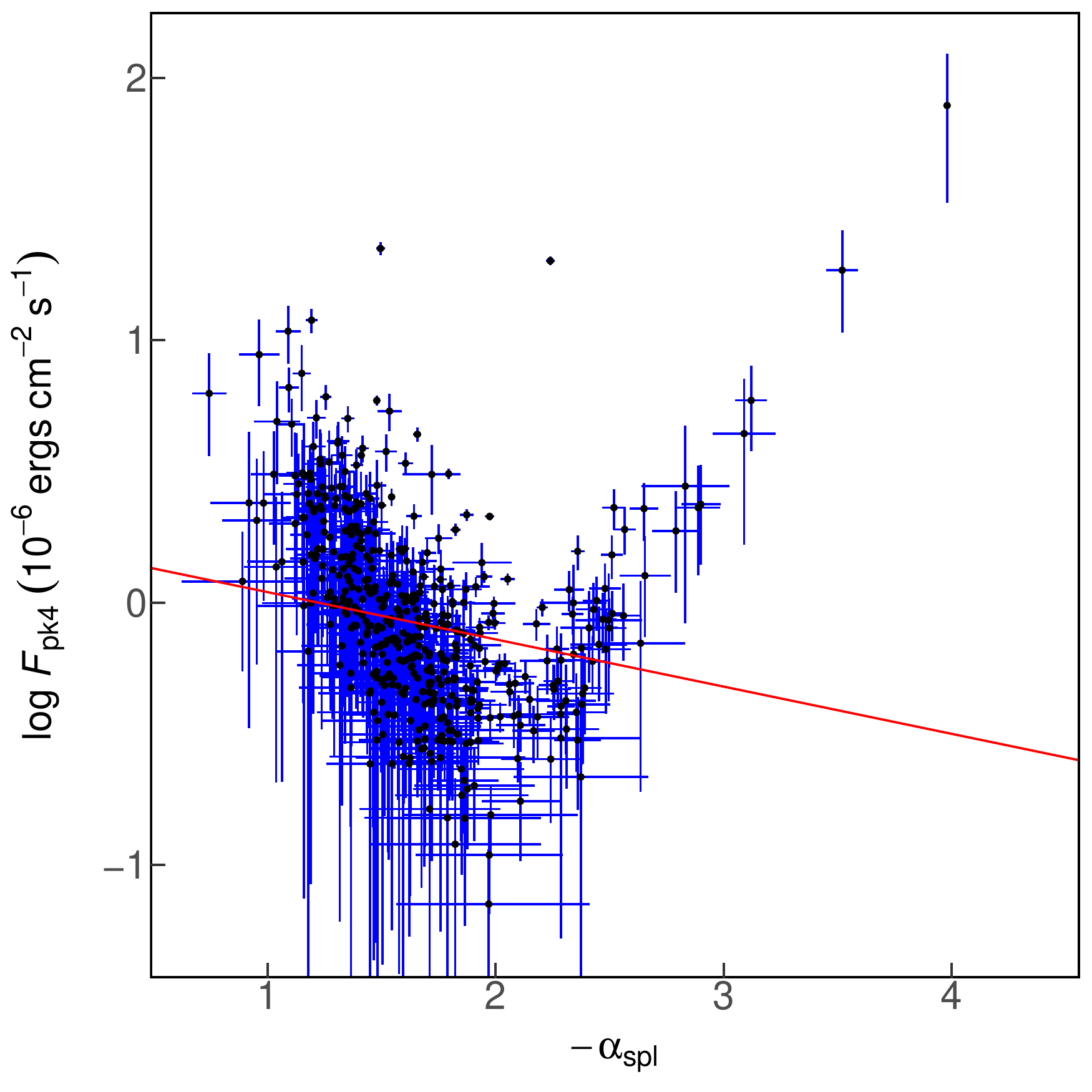}

\caption{
Scatter plot for $\log F_{\rm pk4}$ and $(-\alpha_{\rm spl})$.
The red line is our fit result. The formula of the red line is $\log F_{\rm pk4} = (-0.18 \pm 0.033) \times (-\alpha_{\rm spl}) + (0.22 \pm 0.059)$. The description of every parameter is in Section \ref{sec:sample}.
}
\label{fig:fpk4alpl}
\end{figure*}

The correlation between $\log F_{\rm pk4}$ and $(-\alpha_{\rm spl})$ is:
\begin{equation} \label{eq:fpk4alpl}
\log F_{\rm pk4} = (-0.18 \pm 0.033) \times (-\alpha_{\rm spl}) + (0.22 \pm 0.059),
\end{equation}
where $F_{\rm pk4}$ is peak energy flux of 1024 $\rm ms$ time bin in rest-frame 1-$10^{4}$ $\rm keV$ energy band and in unit of $\rm 10^{\rm -6} ~ ergs ~ cm^{\rm -2} ~ s^{\rm -1}$. The adjusted $R^{2}$ is 0.055. The Pearson coefficient is $-0.16 \pm 0.032$ with p-value $3.4 \times 10^{-4}$. The Spearman coefficient is $-0.39 \pm 0.027$ with p-value $3.3 \times 10^{-19}$. The Kendall $\tau$ coefficient is $-0.28 \pm 0.02$ with p-value $1.3 \times 10^{-19}$. The correlation ratio is $0.89 \pm 0.004$. The cosine similarity is $-0.19 \pm 0.018$. The scatter plot is in Figure \ref{fig:fpk4alpl}. The GRB sample number is 483.

We also found there is a break in $\log HR$-$\log F_{\rm pk2}$-$(-\alpha_{\rm spl})$ plot. In other three time bins of $F_{\rm pk}$, there are also breaks. We just show the best one.

\begin{figure*}[!b]
\centering
\includegraphics[width=0.45\textwidth]{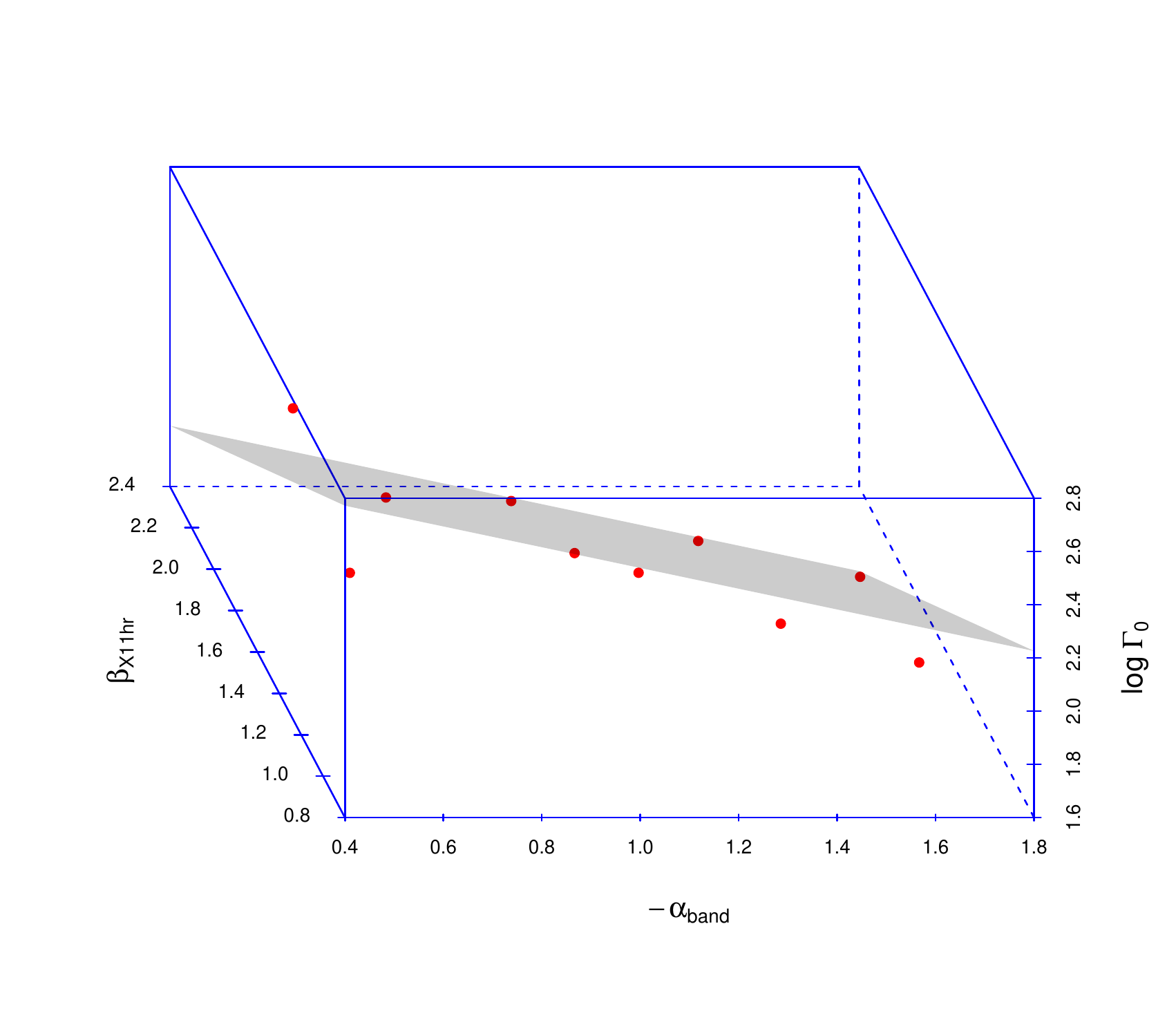}
\includegraphics[width=0.45\textwidth]{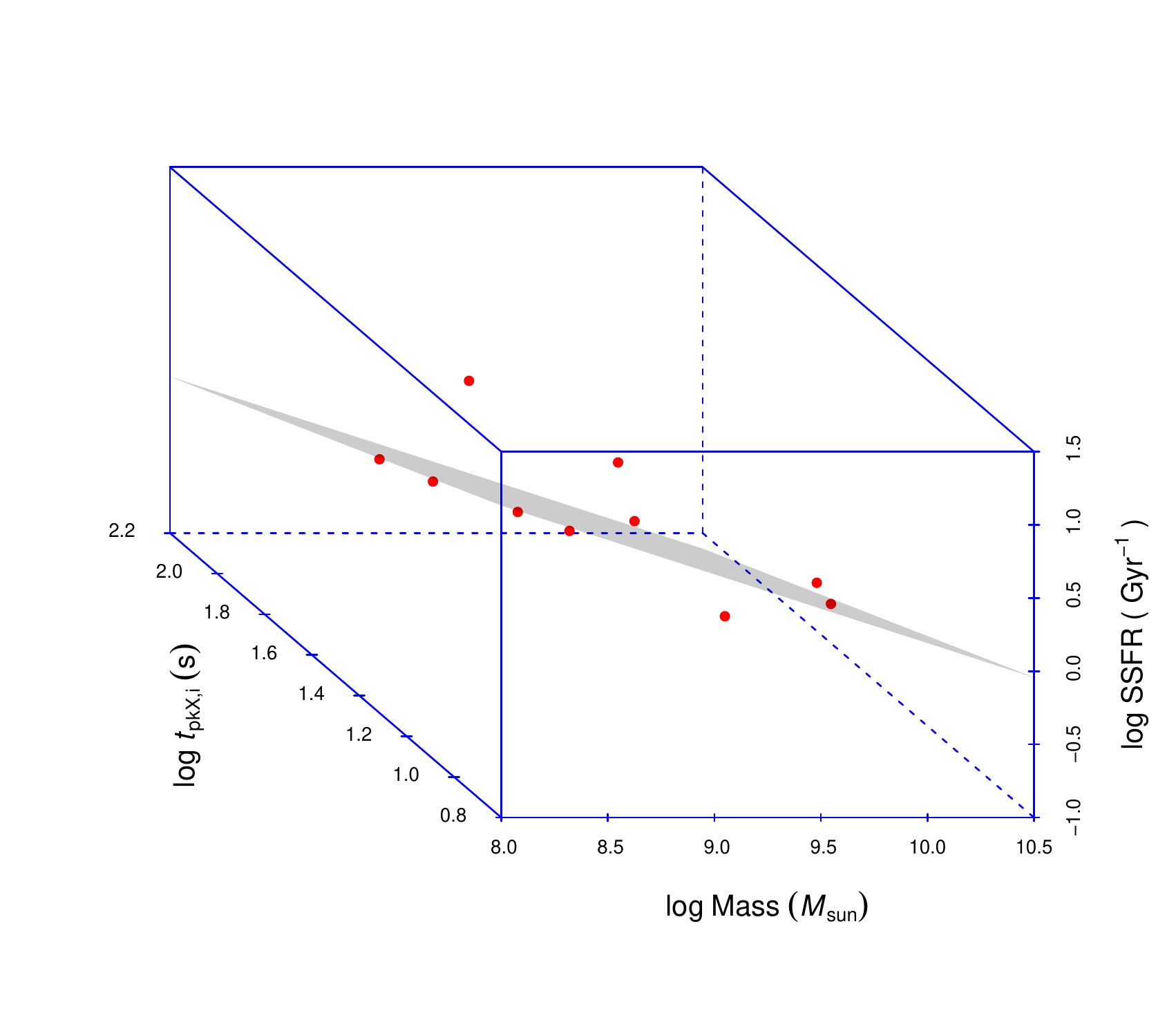}

\includegraphics[width=0.45\textwidth]{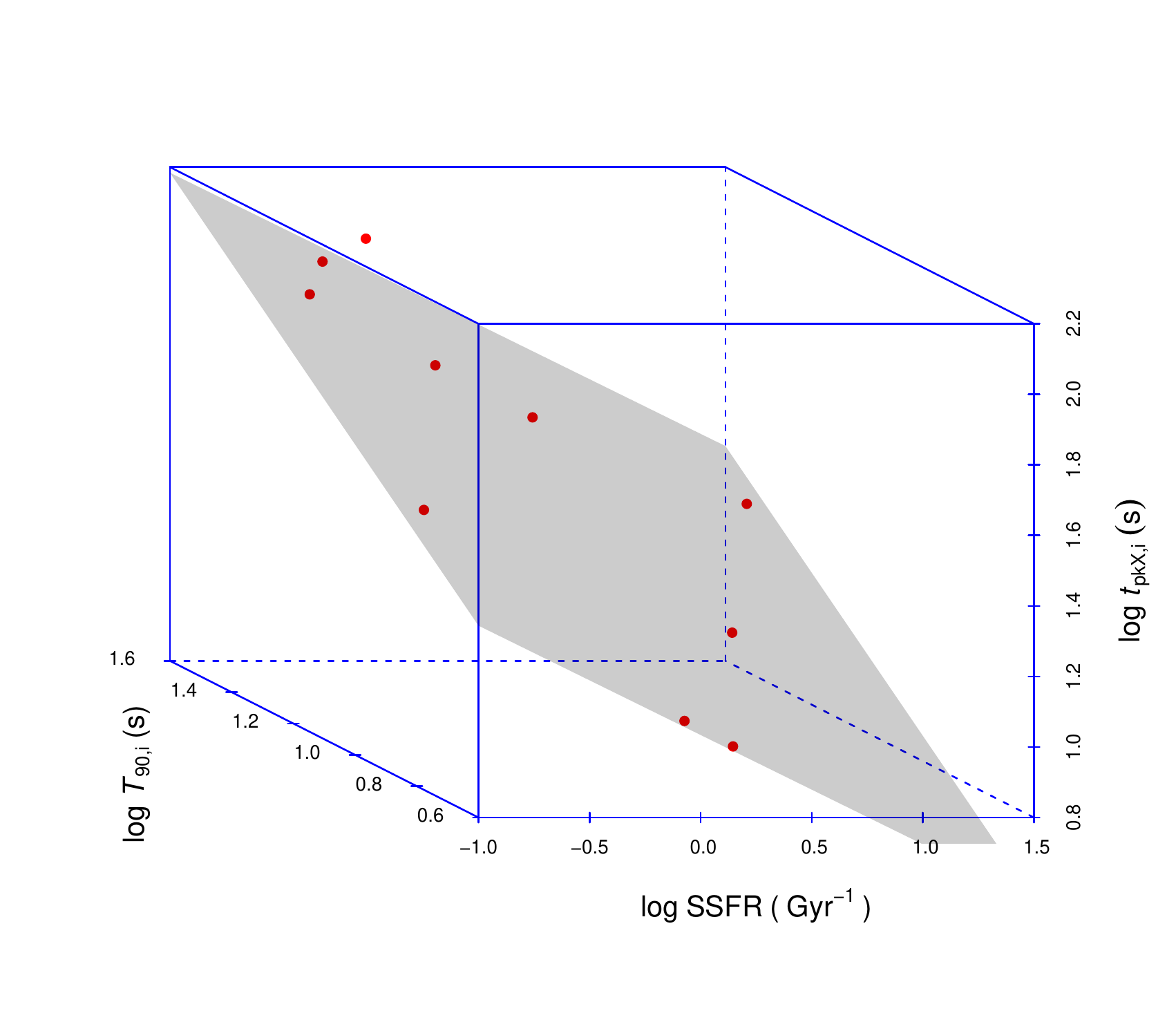}
\includegraphics[width=0.45\textwidth]{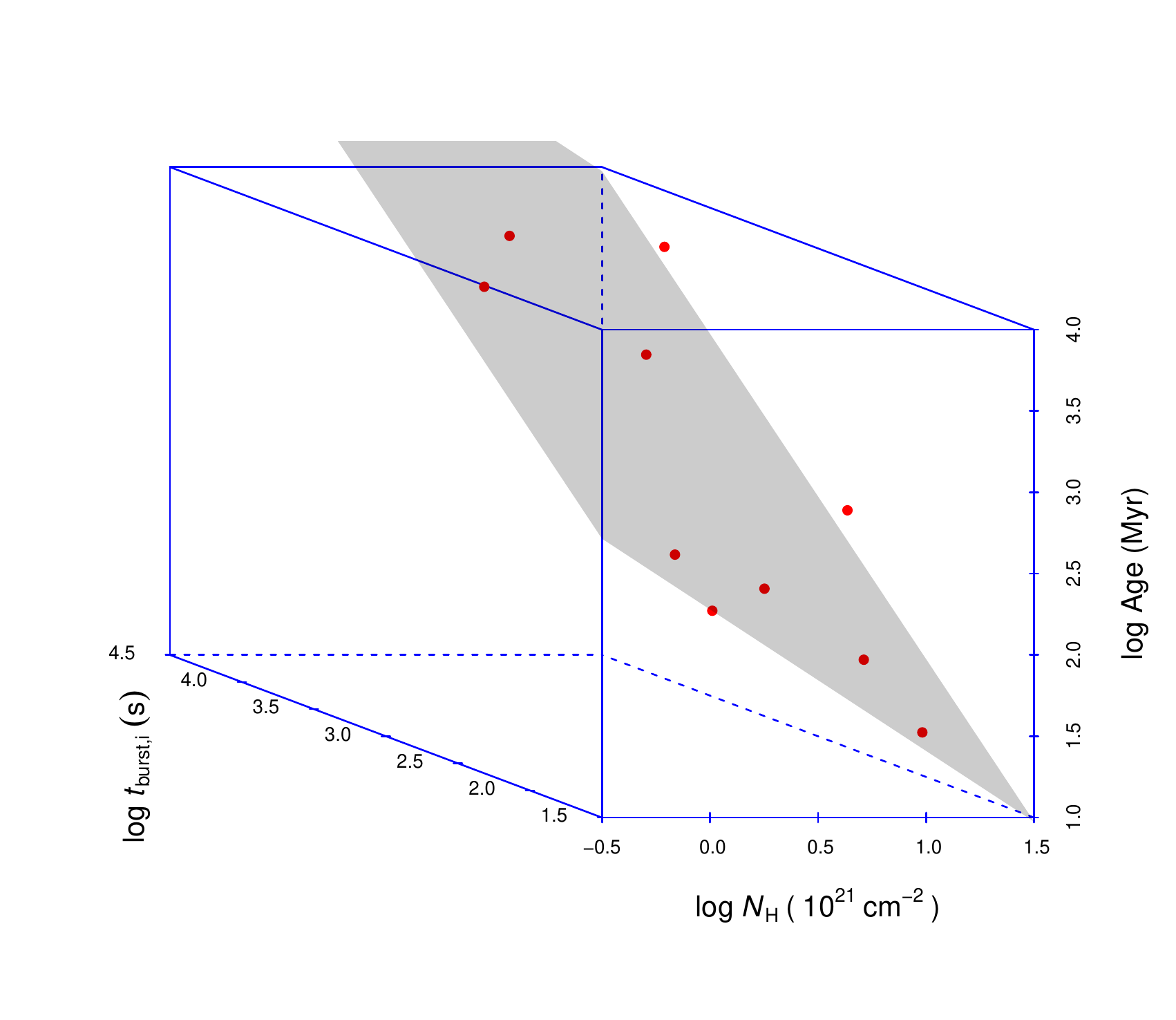}

\includegraphics[width=0.45\textwidth]{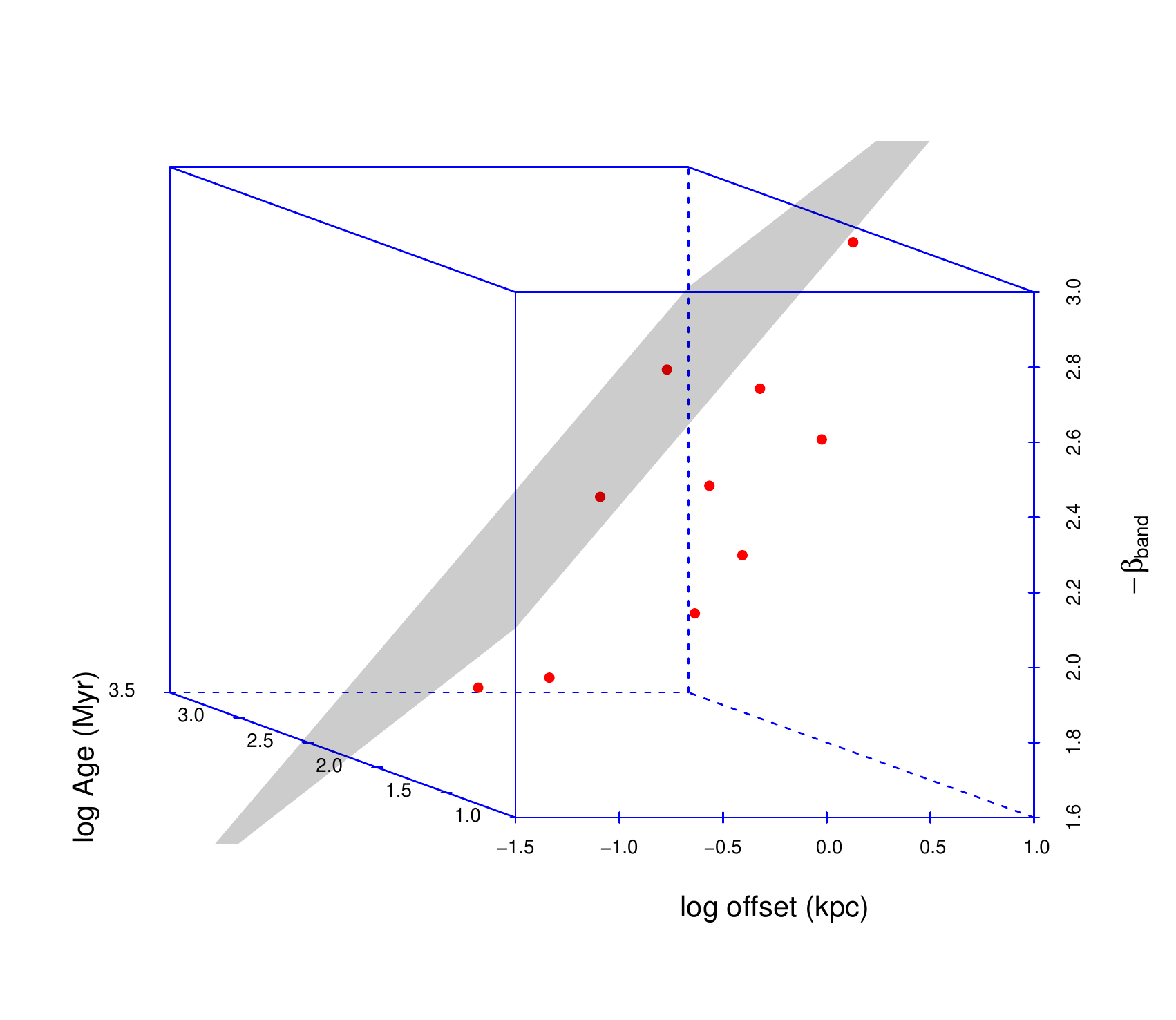}
\includegraphics[width=0.45\textwidth]{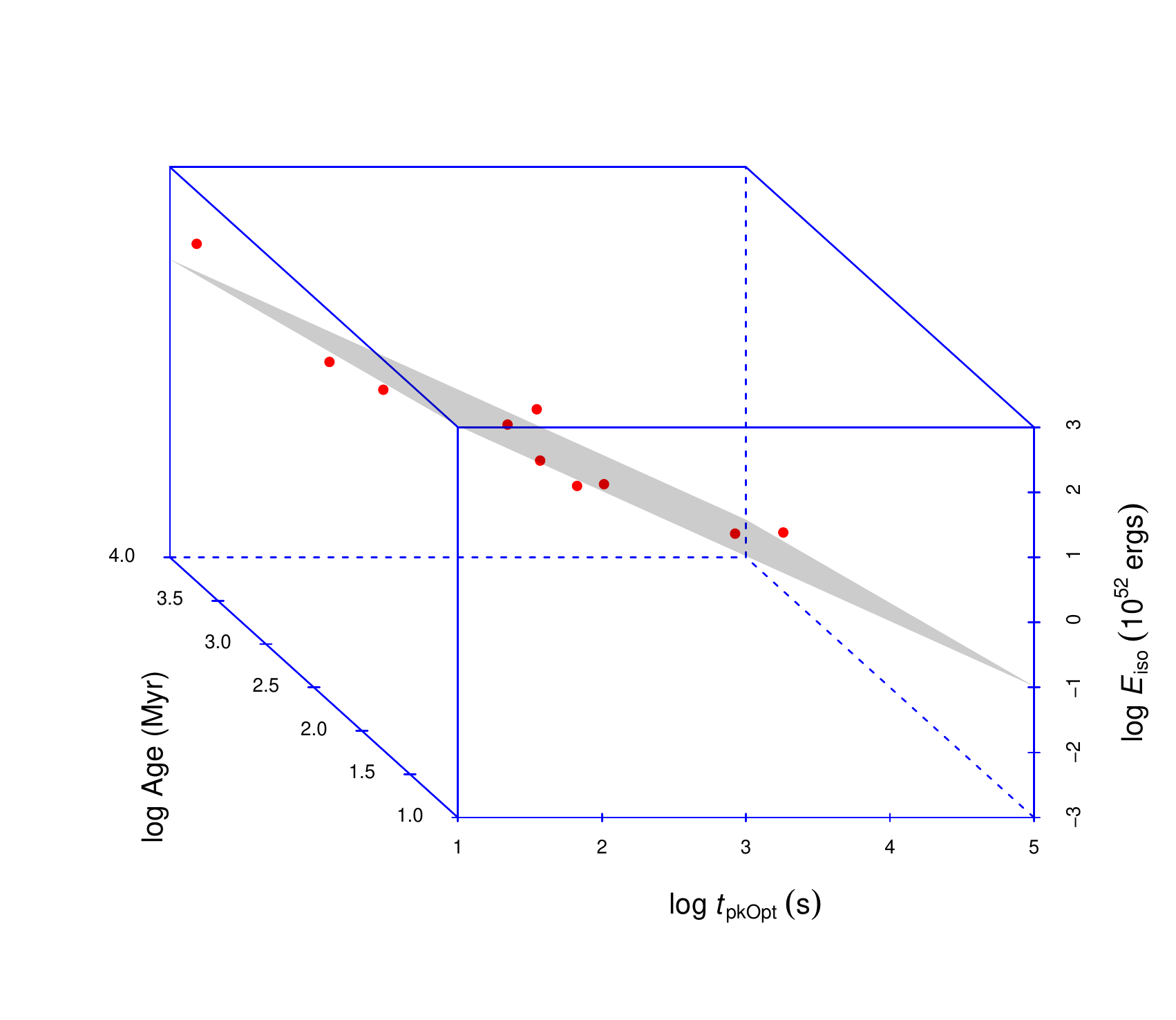}

\caption{
The scatter plots between three parameters. The grey plane in every plot is our fitted result. You can find all the results in machine readable tables. The description of every parameter is in Section \ref{sec:sample}.
}
\label{fig:three}
\end{figure*}

The $\log HR$-$\log F_{\rm pk2}$-$(-\alpha_{\rm spl})$ relation is:
\begin{equation} \label{eq:hfa}
\log HR = (0.17 \pm 0.036) \times \log F_{\rm pk2} + (-0.9 \pm 0.039) \times (-\alpha_{\rm spl}) + (2.2 \pm 0.075),
\end{equation}
where $F_{\rm pk2}$ is peak energy flux of 64 $\rm ms$ time bin in rest-frame 1-$10^{4}$ $\rm keV$ energy band and in unit of $\rm 10^{\rm -6} ~ ergs ~ cm^{\rm -2} ~ s^{\rm -1}$. The adjusted $R^{2}$ is 0.57. One can see \ref{fig:three} for the scatter plot between these three parameters. The GRB sample number is 490.

It is not clear what causes these breaks. There is also weak signals in $P_{\rm pk}-\alpha_{\rm spl}$ plot. 
It might be a selection effect, as there is no such effect in $L_{\rm pk}-\alpha_{\rm spl}$ plot or in $F_{\rm g}-\alpha_{\rm spl}$ plot. 
All the corresponding figures can be found in the figure sets.

\subsection{Amati relation}

Amati relation \citep{Amati2002, Amati2008} is a widely known relation of GRBs.

\begin{figure*}[!b]
\centering
\includegraphics[width=0.45\textwidth]{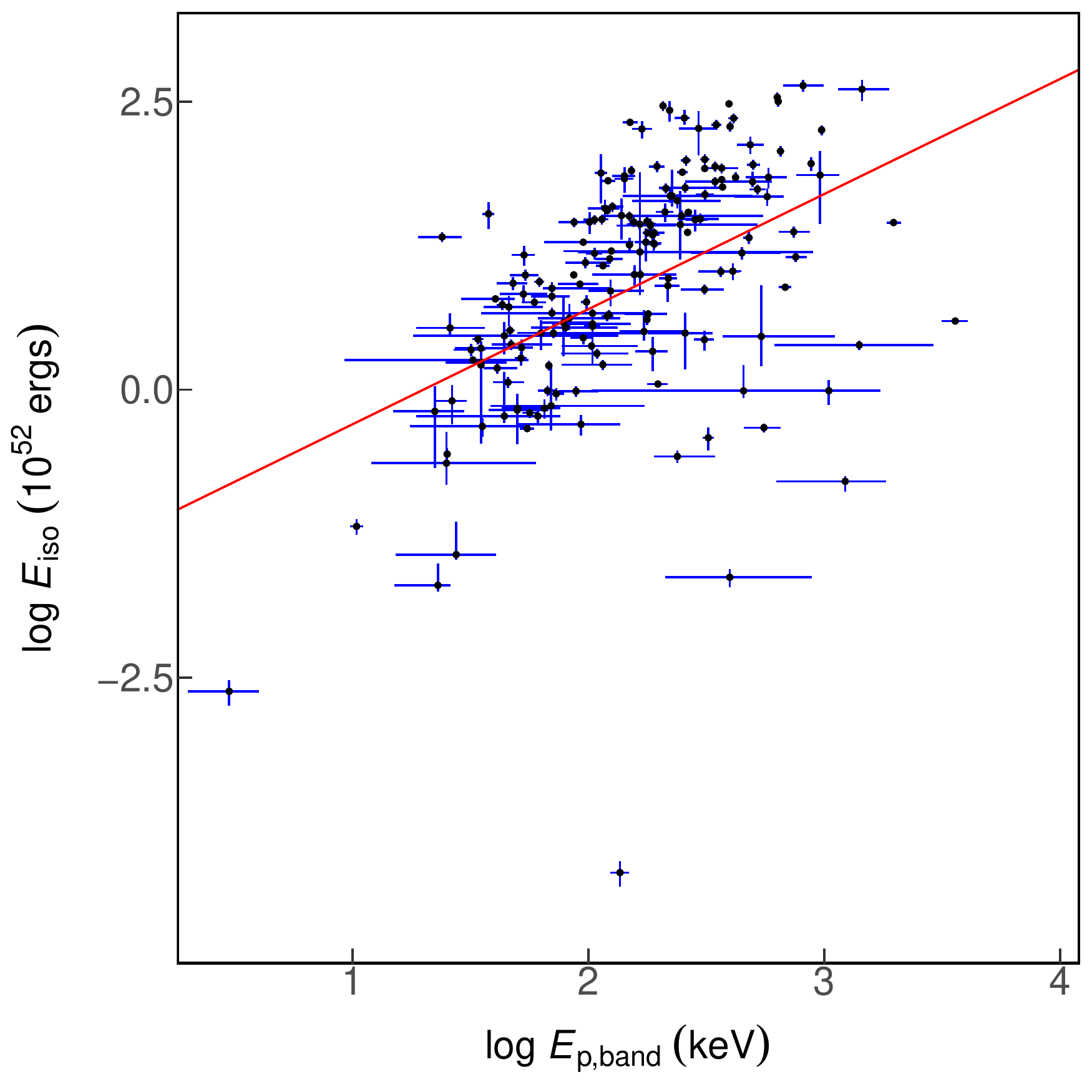}

\caption{
Scatter plot for $\log  E_{\rm p,band}$ and $\log E_{\rm iso}$.
The red line is our fit result. The formula of the red line is $\log  E_{\rm p,band} = (0.24 \pm 0.011) \times \log E_{\rm iso} + (1.9 \pm 0.017)$. The description of every parameter is in Section \ref{sec:sample}.
}
\label{fig:epndeiso}
\end{figure*}

The correlation between $\log  E_{\rm p,band}$ and $\log E_{\rm iso}$ is:
\begin{equation} \label{eq:epndeiso}
\log  E_{\rm p,band} = (0.24 \pm 0.011) \times \log E_{\rm iso} + (1.9 \pm 0.017),
\end{equation}
where $E_{\rm p,band}$ is in unit of $\rm keV$, $E_{\rm iso}$ is in unit of $\rm 10^{\rm 52} ~ ergs$ and in rest-frame 1-$10^{4}$ $\rm keV$ energy band. In \citet{Amati2002} initial data, the slope is $0.52 \pm 0.06$ with 12 GRBs. Here we have 180 GRBs. The adjusted $R^{2}$ is 0.26. The Pearson coefficient is $0.5 \pm 0.021$ with p-value $8.2 \times 10^{-13}$. The Spearman coefficient is $0.56 \pm 0.019$ with p-value $5.2 \times 10^{-16}$. The Kendall $\tau$ coefficient is $0.41 \pm 0.014$ with p-value $2.1 \times 10^{-16}$. The correlation ratio is $0.62 \pm 0.0049$. The cosine similarity is $0.75 \pm 0.0055$. It is maybe due to larger $\Gamma_{0}$ in GRBs also have larger energy and peak energy \citep{Ghirlanda2012}, or an optically thin synchrotron shock model \citep{Lloyd2000A}. One can see Section \ref{sec:gamma0}, $E_{\rm iso}$ and peak energy both correlate with $\Gamma_{0}$. The scatter plot is in Figure \ref{fig:epndeiso}. The GRB sample number is 180.

\begin{figure*}[!b]
\centering
\includegraphics[width=0.45\textwidth]{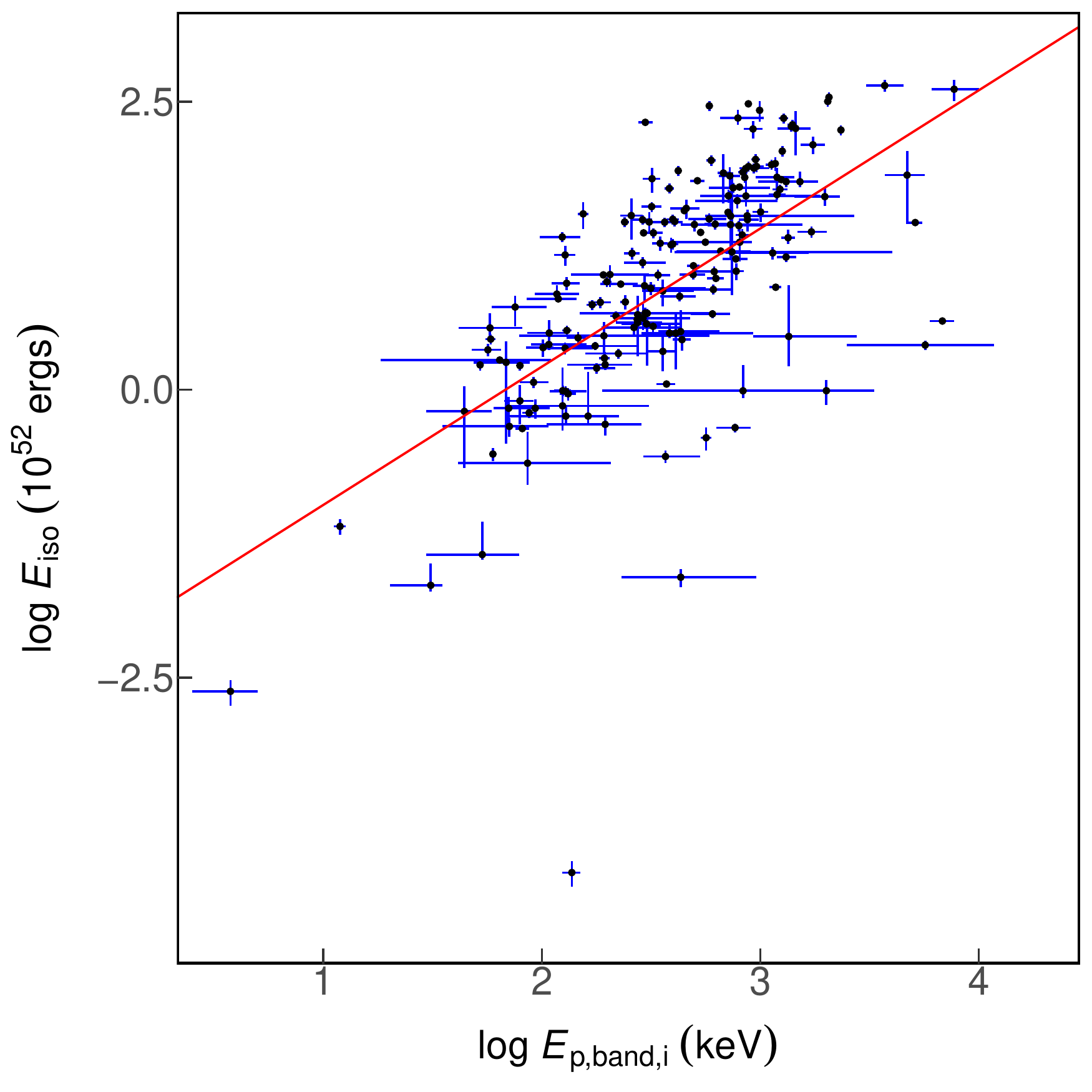}

\caption{
Scatter plot for $\log E_{\rm p,band,i}$ and $\log E_{\rm iso}$.
The red line is our fit result. The formula of the red line is $\log E_{\rm p,band,i} = (0.35 \pm 0.011) \times \log E_{\rm iso} + (2.3 \pm 0.018)$. The description of every parameter is in Section \ref{sec:sample}.
}
\label{fig:epdieiso}
\end{figure*}

The correlation between $\log E_{\rm p,band,i}$ and $\log E_{\rm iso}$ is:
\begin{equation} \label{eq:epdieiso}
\log E_{\rm p,band,i} = (0.35 \pm 0.011) \times \log E_{\rm iso} + (2.3 \pm 0.018),
\end{equation}
In \citet{Amati2008}, the slope is $0.57 \pm 0.01$ with 70 LGRBs and X-ray flares. We have 178 GRBs. where $E_{\rm p,band,i}$ is in unit of $\rm keV$. $E_{\rm iso}$ is in unit of $\rm 10^{\rm 52} ~ ergs$ and in rest-frame 1-$10^{4}$ $\rm keV$ energy band. The adjusted $R^{2}$ is 0.45. The Pearson coefficient is $0.65 \pm 0.019$ with p-value $7.2 \times 10^{-23}$. The Spearman coefficient is $0.69 \pm 0.018$ with p-value $4.9 \times 10^{-26}$. The Kendall $\tau$ coefficient is $0.51 \pm 0.015$ with p-value $3.1 \times 10^{-24}$. The correlation ratio is $0.72 \pm 0.0038$. The cosine similarity is $0.78 \pm 0.0052$. The scatter plot is in Figure \ref{fig:epdieiso}. The GRB sample number is 178.  The better correlation between $\log E_{\rm p,band,i}$ and $\log E_{\rm iso}$ comparing with $\log  E_{\rm p,band}$ and $\log E_{\rm iso}$ suggests this correlation is intrinsic. With enough data, one would investigate this relation within subgroups \citep{2009ApJ...703.1696Z, Qin2013B, Zou2017}. A previous study gave $E_{\rm iso} = 10^{53.00 \pm 0.045} \times [\frac{E_{\rm p,band,i}}{355keV}]^{1.57 \pm 0.099}$ with 101 samples \citep{Yonetoku2010}, and $\log (\frac{E_{\rm iso}}{1 erg}) = 1.75^{+0.18}_{-0.16} \times \log [\frac{E_{\rm p,band,i}}{300 keV}] + (52.53 \pm 0.02)$ with 162 samples \citep{Demianski2017}.

\subsection{Some correlations about $L_{\rm pk}$}

\subsubsection{Yonetoku relation}

\citet{Yonetoku2004} found the correlation between rest-frame peak energy  and peak luminosity. We also did the correlations of  $\log  E_{\rm p,band}$ - $\log L_{\rm pk}$ and $\log  E_{\rm p,band,i}$ - $\log L_{\rm pk}$.

The correlation between $\log  L_{\rm pk}$ and $\log E_{\rm p,band}$ is:
\begin{equation} \label{eq:lpkepnd}
\log  L_{\rm pk} = (1 \pm 0.061) \times \log E_{\rm p,band} + (-2 \pm 0.13),
\end{equation}
where $L_{\rm pk}$ is in unit of $\rm 10^{\rm 52} ~ erg ~ s^{\rm -1}$ and in 1-$10^{4}$ $\rm keV$ energy band. $E_{\rm p,band}$ is in unit of $\rm keV$. The adjusted $R^{2}$ is 0.25. The Pearson coefficient is $0.49 \pm 0.022$ with p-value $3.2 \times 10^{-9}$. The Spearman coefficient is $0.52 \pm 0.025$ with p-value $1.5 \times 10^{-10}$. The Kendall $\tau$ coefficient is $0.37 \pm 0.019$ with p-value $4.9 \times 10^{-10}$. The correlation ratio is $0.8 \pm 0.004$. The cosine similarity is $0.33 \pm 0.014$. The scatter plot is in Figure \ref{fig:lpkepnd}. The GRB sample number is 130.

\begin{figure*}[!b]
\centering
\includegraphics[width=0.45\textwidth]{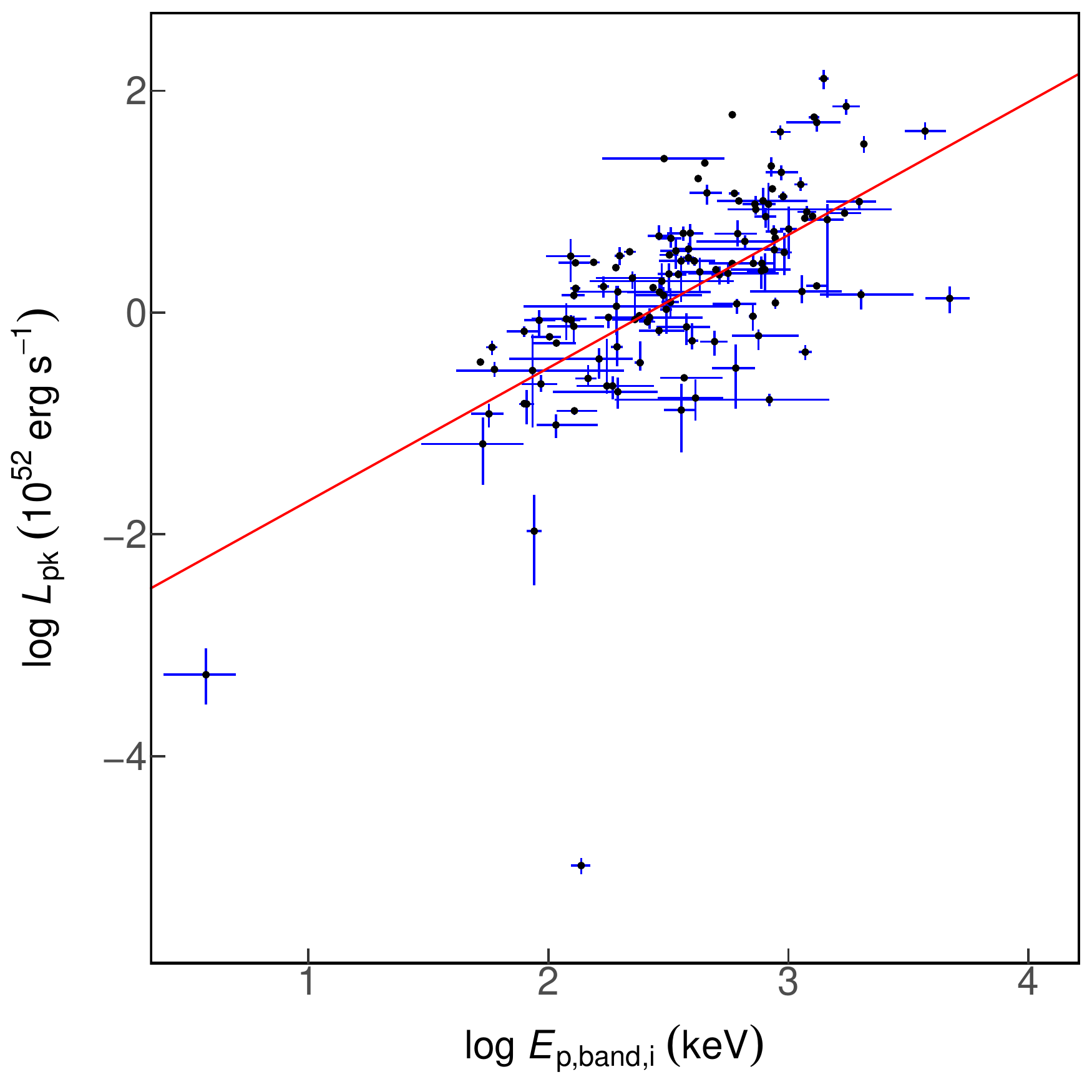}

\caption{
Scatter plot for $\log L_{\rm pk}$ and $\log E_{\rm p,band,i}$.
The red line is our fit result. The formula of the red line is $\log L_{\rm pk} = (1.2 \pm 0.055) \times \log E_{\rm p,band,i} + (-2.9 \pm 0.14)$. The description of every parameter is in Section \ref{sec:sample}. There is an outlier in the plot with lowest $L_{\rm pk}$, this point is BATSE GRB 980425B, which is an well-known low luminosity GRB.
}
\label{fig:lpkepdi}
\end{figure*}

\begin{figure*}[!b]
\centering
\includegraphics[width=0.45\textwidth]{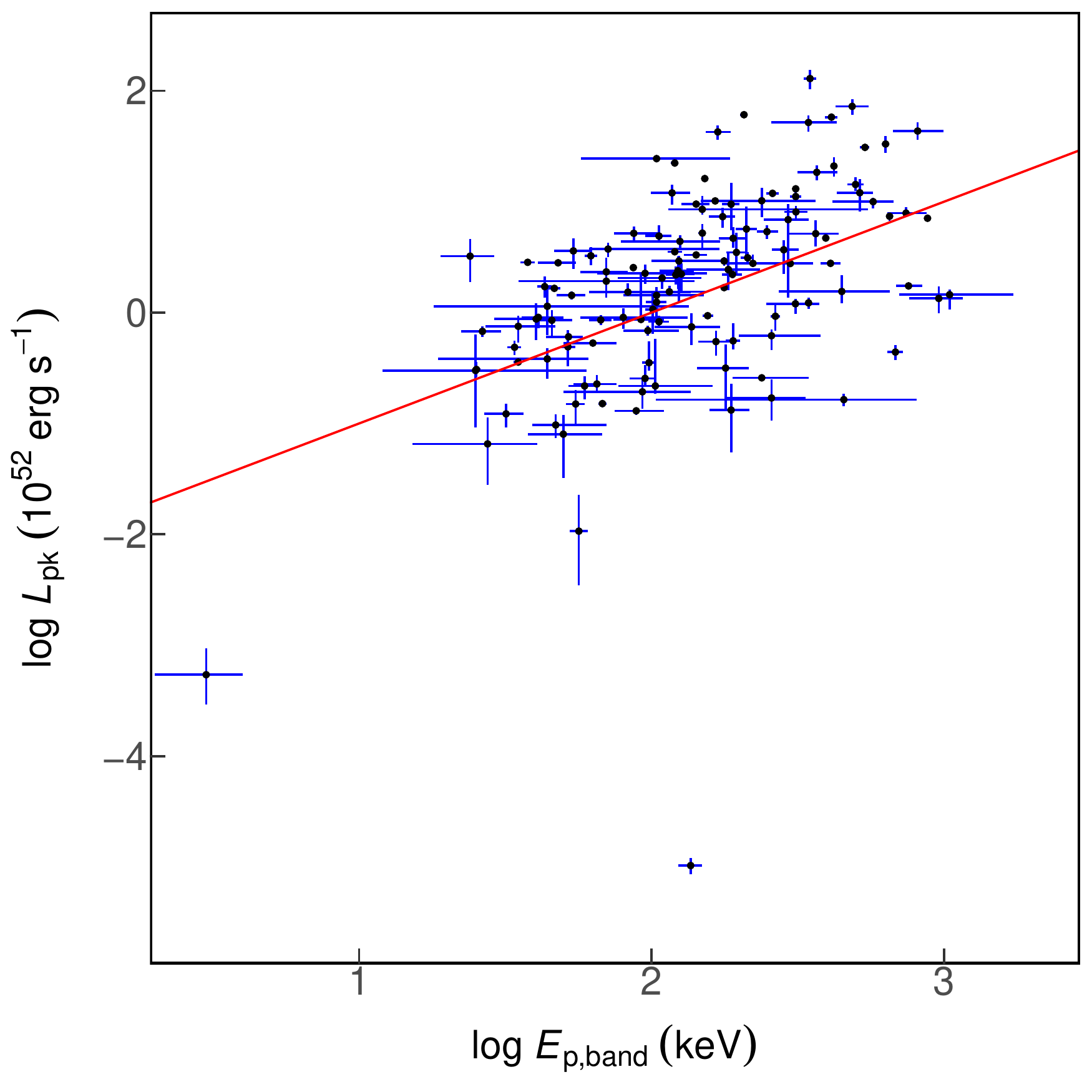}

\caption{
Scatter plot for $\log  L_{\rm pk}$ and $\log E_{\rm p,band}$.
The red line is our fit result. The formula of the red line is $\log  L_{\rm pk} = (1 \pm 0.061) \times \log E_{\rm p,band} + (-2 \pm 0.13)$. The description of every parameter is in Section \ref{sec:sample}. The outlier with lowest $L_{\rm pk}$ is also GRB 980425B as in Figure \ref{fig:lpkepdi}
}
\label{fig:lpkepnd}
\end{figure*}

The correlation between $\log L_{\rm pk}$ and $\log E_{\rm p,band,i}$ is:
\begin{equation} \label{eq:lpkepdi}
\log L_{\rm pk} = (1.2 \pm 0.055) \times \log E_{\rm p,band,i} + (-2.9 \pm 0.14),
\end{equation}
In \citet{Yonetoku2004}, the slope is $1.94 \pm 0.19$ with 11 samples. where $L_{\rm pk}$ is in unit of $\rm 10^{\rm 52} ~ erg ~ s^{\rm -1}$ and in 1-$10^{4}$ $\rm keV$ energy band. $E_{\rm p,band,i}$ is in unit of $\rm keV$. The adjusted $R^{2}$ is 0.42. The Pearson coefficient is $0.63 \pm 0.019$ with p-value $4 \times 10^{-15}$. The Spearman coefficient is $0.64 \pm 0.025$ with p-value $4.3 \times 10^{-16}$. The Kendall $\tau$ coefficient is $0.47 \pm 0.019$ with p-value $7.5 \times 10^{-15}$. The correlation ratio is $0.85 \pm 0.0031$. The cosine similarity is $0.34 \pm 0.013$. The scatter plot is in Figure \ref{fig:lpkepdi}. The GRB sample number is 127. The rest frame peak energy is also better correlated with the luminosity. \citet{Yonetoku2010} reanalyzed 101 GRBs, and examined how the truncation of the detector sensitivity affects the Amati and Yonetoku relations. They conclude they are surely intrinsic properties of GRBs. The result of \citet{Yonetoku2010} is $L_{\rm pk} = 10^{52.43 \pm 0.037} \times [\frac{E_{\rm p,band,i}}{355keV}]^{1.60 \pm 0.082}$ with 101 samples.

\subsubsection{Other correlations about $L_{\rm pk}$}

\begin{figure*}[!b]
\centering
\includegraphics[width=0.45\textwidth]{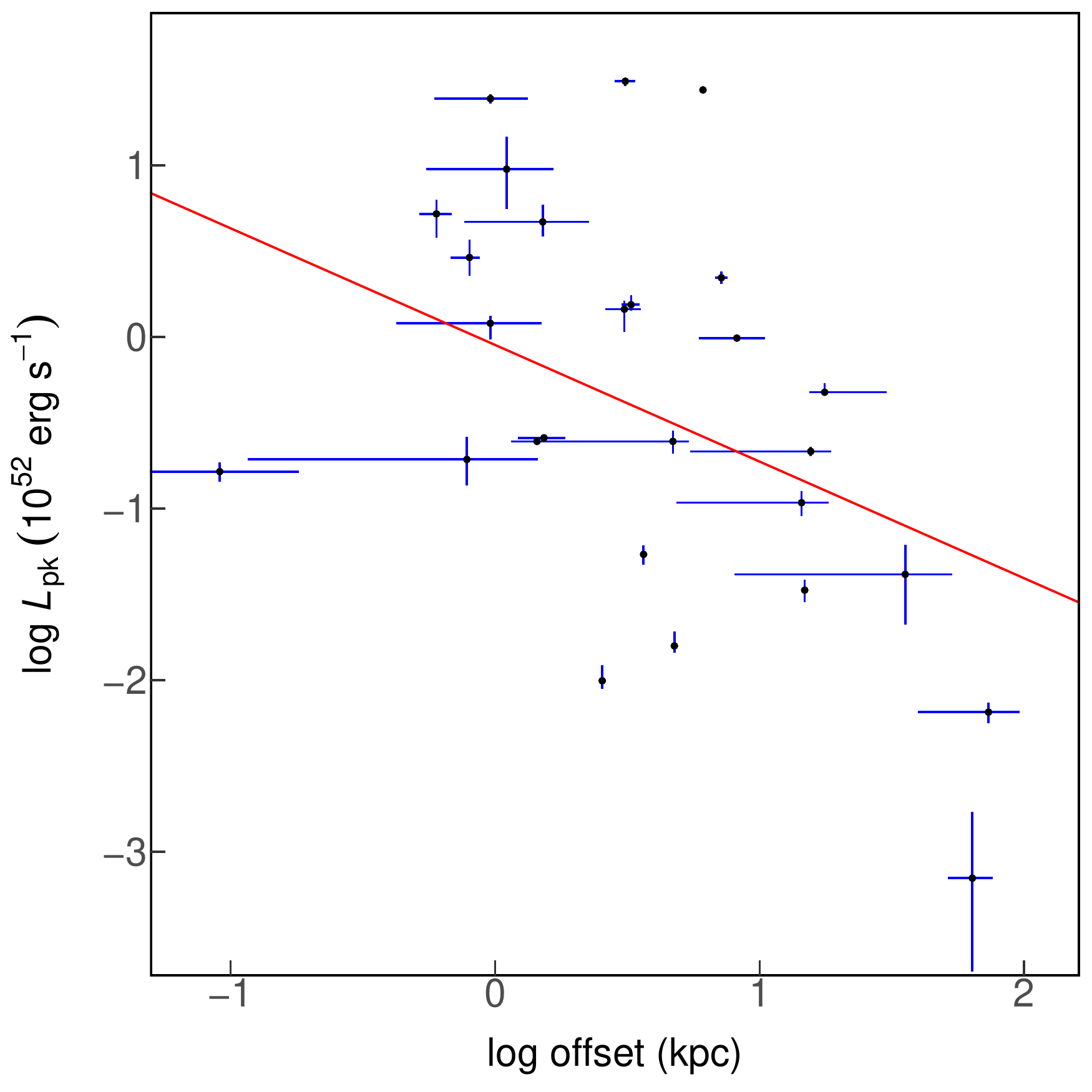}

\caption{
Scatter plot for $\log  L_{\rm pk}$ and $\log offset$.
The red line is our fit result. The formula of the red line is $\log  L_{\rm pk} = (-0.68 \pm 0.24) \times \log offset + (-0.047 \pm 0.15)$. The description of every parameter is in Section \ref{sec:sample}.
}
\label{fig:lpkofet}
\end{figure*}

The correlation between $\log  L_{\rm pk}$ and $\log offset$ is:
\begin{equation} \label{eq:lpkofet}
\log  L_{\rm pk} = (-0.68 \pm 0.24) \times \log offset + (-0.047 \pm 0.15),
\end{equation}
where $L_{\rm pk}$ is in unit of $\rm 10^{\rm 52} ~ erg ~ s^{\rm -1}$ and in 1-$10^{4}$ $\rm keV$ energy band. Host galaxy offset is in unit of $\rm kpc$. The adjusted $R^{2}$ is 0.22. The Pearson coefficient is $-0.42 \pm 0.092$ with p-value $3 \times 10^{-2}$. The Spearman coefficient is $-0.44 \pm 0.041$ with p-value $2.2 \times 10^{-2}$. The Kendall $\tau$ coefficient is $-0.31 \pm 0.035$ with p-value $2.3 \times 10^{-2}$. The correlation ratio is $0.42 \pm 0.043$. The cosine similarity is $-0.5 \pm 0.085$. The scatter plot is in Figure \ref{fig:lpkofet}. The GRB sample number is 27. As the short GRBs locate in the outer regions of the host galaxies, and they are less luminous, this may cause the anti-correlation between $L_{\rm pk} $ and the offset.

\begin{figure*}[!b]
\centering
\includegraphics[width=0.45\textwidth]{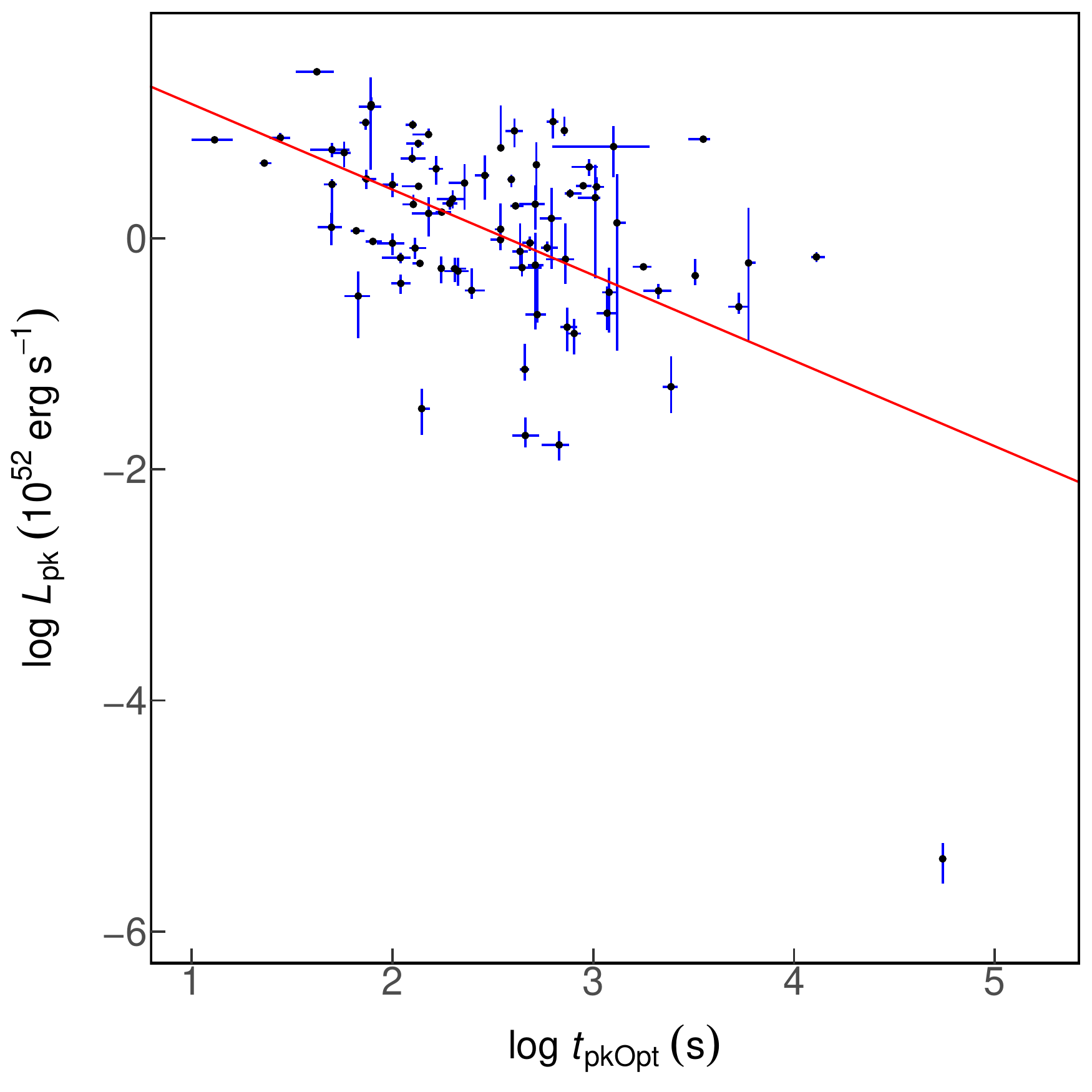}

\caption{
Scatter plot for $\log  L_{\rm pk}$ and $\log t_{\rm pkOpt}$.
The red line is our fit result. The formula of the red line is $\log  L_{\rm pk} = (-0.74 \pm 0.04) \times \log t_{\rm pkOpt} + (1.9 \pm 0.092)$. The description of every parameter is in Section \ref{sec:sample}. The outlier is low luminosity GRB 060218A with lowest $L_{\rm pk}$ in the plot.
}
\label{fig:lpktppt}
\end{figure*}

The correlation between $\log  L_{\rm pk}$ and $\log t_{\rm pkOpt}$ is:
\begin{equation} \label{eq:lpktppt}
\log  L_{\rm pk} = (-0.74 \pm 0.04) \times \log t_{\rm pkOpt} + (1.9 \pm 0.092),
\end{equation}
where $L_{\rm pk}$ is in unit of $\rm 10^{\rm 52} ~ erg ~ s^{\rm -1}$ and in 1-$10^{4}$ $\rm keV$ energy band. $t_{\rm pkOpt}$ is in unit of $\rm s$. The adjusted $R^{2}$ is 0.26. The Pearson coefficient is $-0.51 \pm 0.023$ with p-value $1.5 \times 10^{-6}$. The Spearman coefficient is $-0.39 \pm 0.033$ with p-value $3.5 \times 10^{-4}$. The Kendall $\tau$ coefficient is $-0.27 \pm 0.024$ with p-value $4 \times 10^{-4}$. The correlation ratio is $0.84 \pm 0.0038$. The cosine similarity is $-0.067 \pm 0.026$. The scatter plot is in Figure \ref{fig:lpktppt}. The GRB sample number is 80.

\begin{figure*}[!b]
\centering
\includegraphics[width=0.45\textwidth]{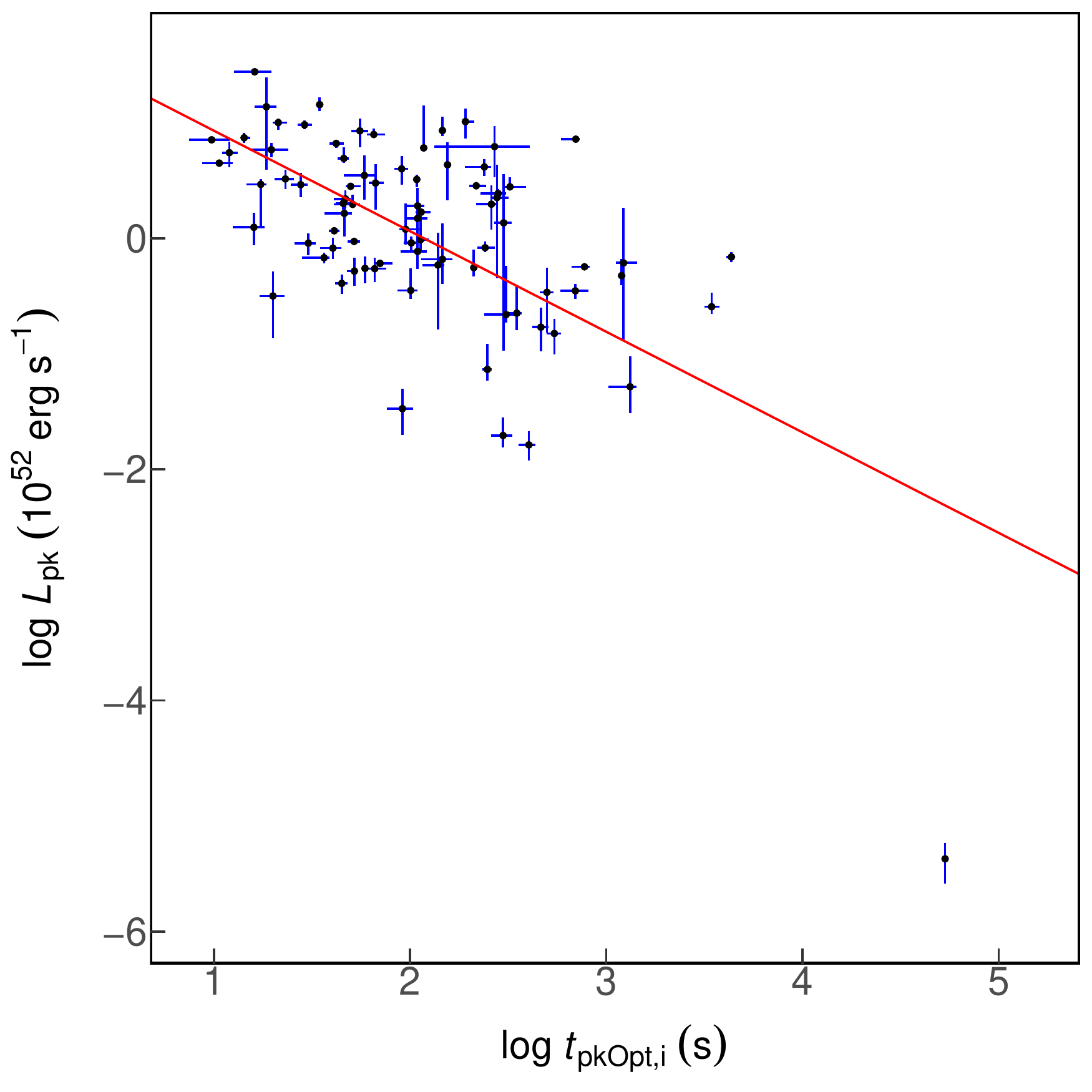}

\caption{
Scatter plot for $\log L_{\rm pk}$ and $\log t_{\rm pkOpt,i}$.
The red line is our fit result. The formula of the red line is $\log L_{\rm pk} = (-0.87 \pm 0.036) \times \log t_{\rm pkOpt,i} + (1.8 \pm 0.069)$. The description of every parameter is in Section \ref{sec:sample}. The outlier is low luminosity GRB 060218A with lowest $L_{\rm pk}$ in the plot.
}
\label{fig:lpktpti}
\end{figure*}

The correlation between $\log L_{\rm pk}$ and $\log t_{\rm pkOpt,i}$ is:
\begin{equation} \label{eq:lpktpti}
\log L_{\rm pk} = (-0.87 \pm 0.036) \times \log t_{\rm pkOpt,i} + (1.8 \pm 0.069),
\end{equation}
where $L_{\rm pk}$ is in unit of $\rm 10^{\rm 52} ~ erg ~ s^{\rm -1}$ and in 1-$10^{4}$ $\rm keV$ energy band. $t_{\rm pkOpt,i}$ is in unit of $\rm s$. The adjusted $R^{2}$ is 0.4. The Pearson coefficient is $-0.62 \pm 0.019$ with p-value $1.1 \times 10^{-9}$. The Spearman coefficient is $-0.48 \pm 0.031$ with p-value $5.6 \times 10^{-6}$. The Kendall $\tau$ coefficient is $-0.33 \pm 0.023$ with p-value $1.3 \times 10^{-5}$. The correlation ratio is $0.78 \pm 0.0049$. The cosine similarity is $-0.13 \pm 0.025$. The scatter plot is in Figure \ref{fig:lpktpti}. The GRB sample number is 80. This anti-correlation is mainly caused by the Lorentz factor, of which higher Lorentz factor corresponds to shorter deceleration time, and consequently smaller $t_{\rm pkOpt,i}$. We can also see the rest frame peak time is better correlated with the luminosity than the observational peak time.

\subsection{Some correlations about host galaxy}

We found there are some correlations between host galaxy parameters and redshift, also $D_{\rm L}$. Especially Mag has correlations with many parameters.

\subsubsection{Some correlations about Mag and other parameters}

\begin{figure*}[!b]
\centering
\includegraphics[width=0.45\textwidth]{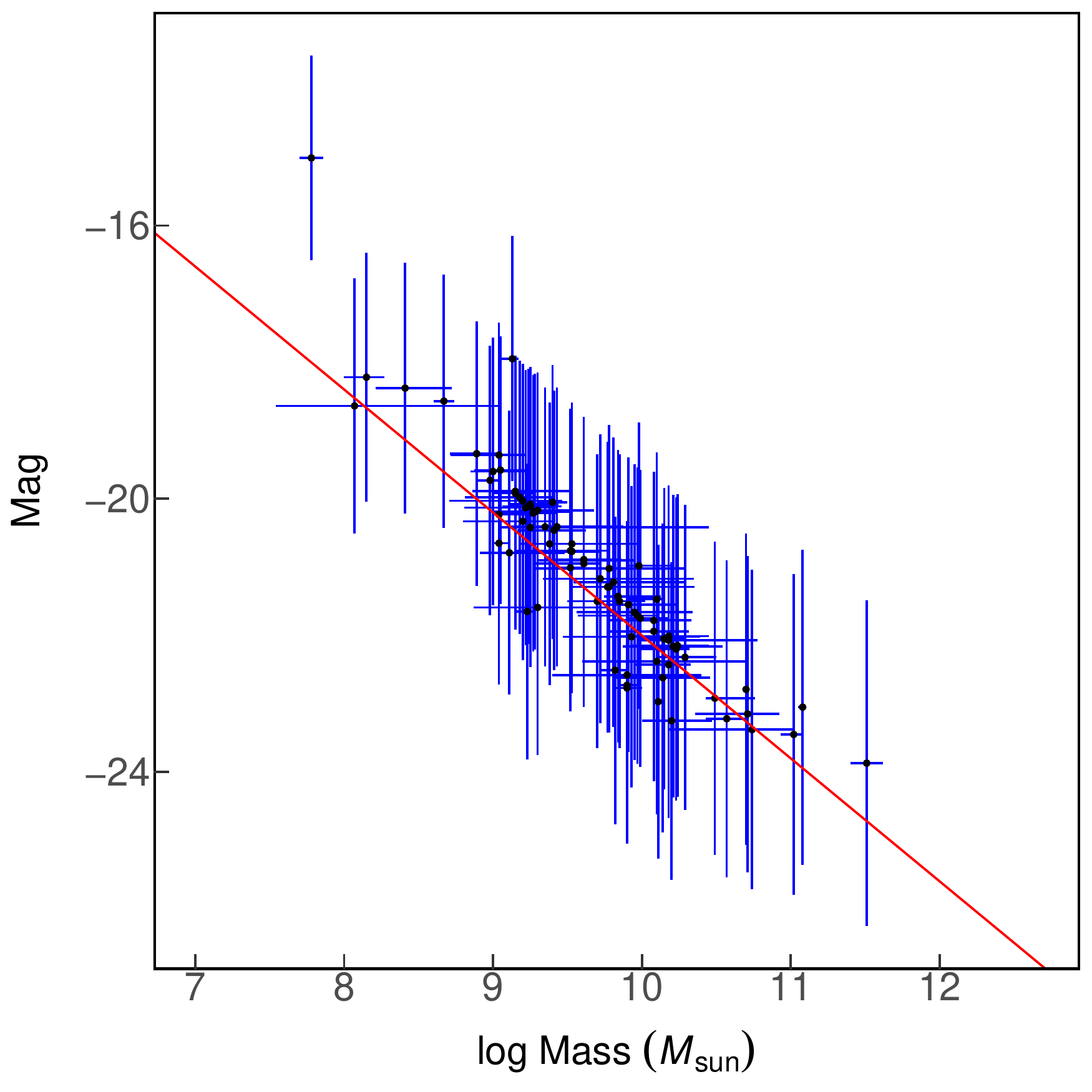}

\caption{
Scatter plot for Mag and $\log Mass$.
The red line is our fit result. The formula of the red line is $Mag = (-1.8 \pm 0.34) \times \log Mass + (-4 \pm 3.3)$. The description of every parameter is in Section \ref{sec:sample}.
}
\label{fig:magmass}
\end{figure*}

The correlation between Mag and $\log Mass$ is:
\begin{equation} \label{eq:magmass}
Mag = (-1.8 \pm 0.34) \times \log Mass + (-4 \pm 3.3),
\end{equation}
where Mass is in unit of $M_{\bigodot}$, and Mag is in unit of magnitude. The adjusted $R^{2}$ is 0.84. The Pearson coefficient is $-0.48 \pm 0.077$ with p-value $4.2 \times 10^{-6}$. The Spearman coefficient is $-0.45 \pm 0.083$ with p-value $1.8 \times 10^{-5}$. The Kendall $\tau$ coefficient is $-0.32 \pm 0.061$ with p-value $2.5 \times 10^{-5}$. The correlation ratio is $0.99 \pm 0.00099$. The cosine similarity is $-0.99 \pm 0.00091$. The scatter plot is in Figure \ref{fig:magmass}. The GRB sample number is 82. The luminosity of the host galaxy is positively related with its mass. It might be worthy to investigate the correlations for different types of galaxies, such as host galaxies of GRBs, of supernovae, radio load galaxies etc.

\begin{figure*}[!b]
\centering
\includegraphics[width=0.45\textwidth]{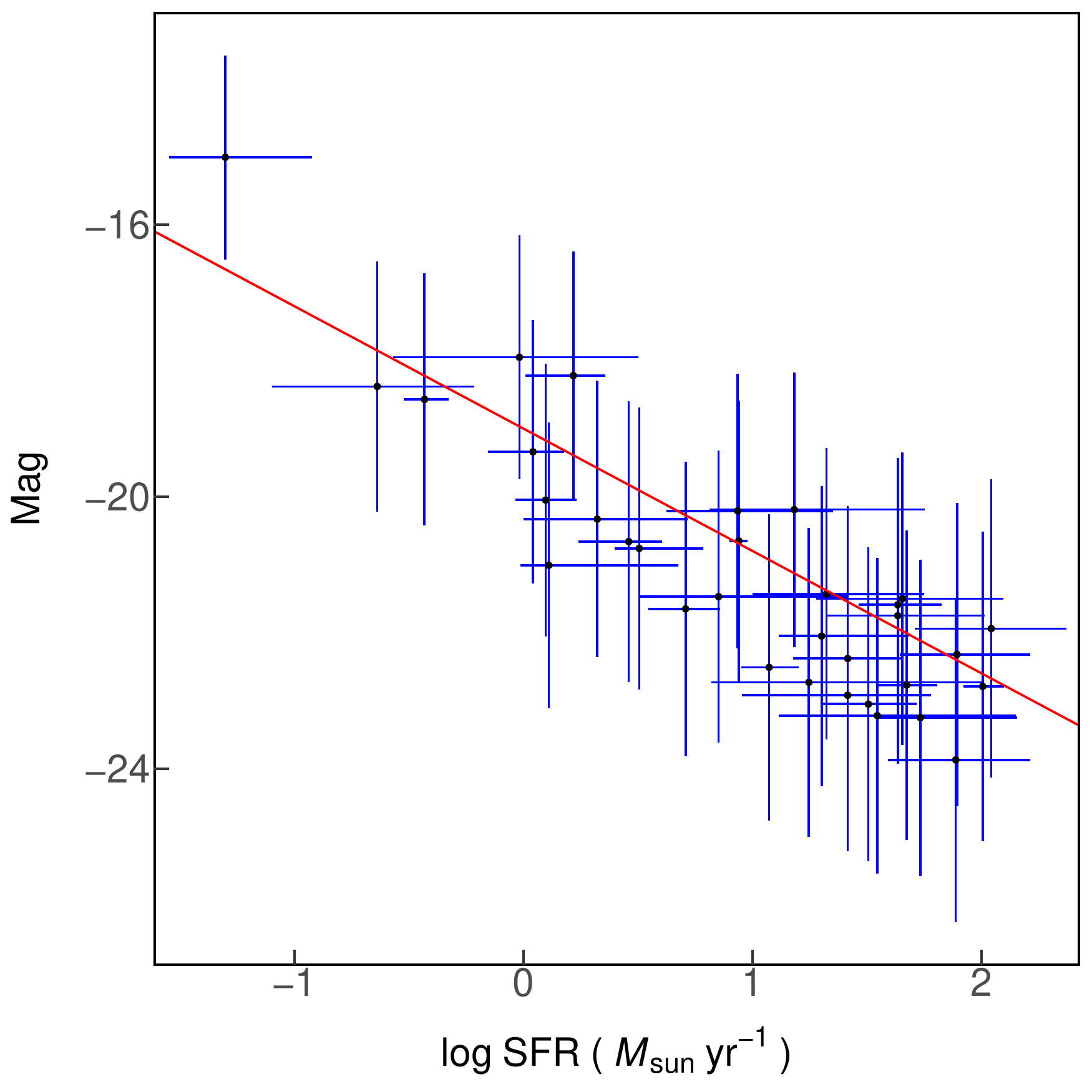}

\caption{
Scatter plot for Mag and $\log SFR$.
The red line is our fit result. The formula of the red line is $Mag = (-1.8 \pm 0.41) \times \log SFR + (-19 \pm 0.51)$. The description of every parameter is in Section \ref{sec:sample}.
}
\label{fig:magsfr}
\end{figure*}

The correlation between Mag and $\log SFR$ is:
\begin{equation} \label{eq:magsfr}
Mag = (-1.8 \pm 0.41) \times \log SFR + (-19 \pm 0.51),
\end{equation}
where SFR is in unit of $\rm M_{\bigodot} ~ yr^{\rm -1}$. Mag is in unit of magnitude. The adjusted $R^{2}$ is 0.78. The Pearson coefficient is $-0.56 \pm 0.1$ with p-value $7.1 \times 10^{-4}$. The Spearman coefficient is $-0.5 \pm 0.12$ with p-value $3.1 \times 10^{-3}$. The Kendall $\tau$ coefficient is $-0.36 \pm 0.091$ with p-value $3.6 \times 10^{-3}$. The correlation ratio is $0.98 \pm 0.0032$. The cosine similarity is $-0.78 \pm 0.027$. The scatter plot is in Figure \ref{fig:magsfr}. The GRB sample number is 33. It might be more interesting to get the correlations for the specific SFR. However, those correlations are not obvious enough.

\begin{figure*}[!b]
\centering
\includegraphics[width=0.45\textwidth]{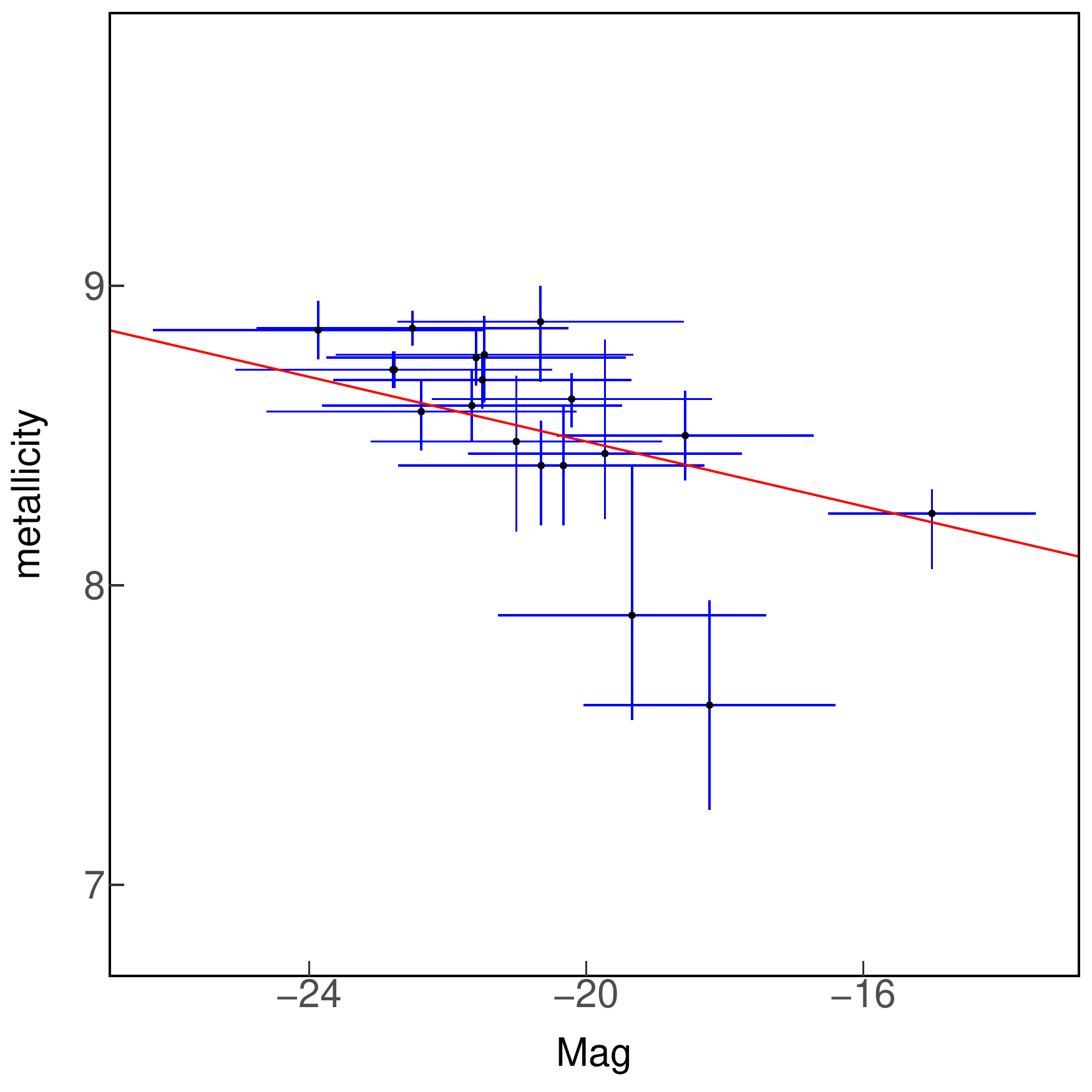}

\caption{
Scatter plot for Mag and metallicity.
The red line is our fit result. The formula of the red line is $Mag = (-3.4 \pm 1.6) \times metallicity + (8.4 \pm 14)$. The description of every parameter is in Section \ref{sec:sample}.
}
\label{fig:magmety}
\end{figure*}

The correlation between Mag and metallicity is:
\begin{equation} \label{eq:magmety}
Mag = (-3.4 \pm 1.6) \times metallicity + (8.4 \pm 14),
\end{equation}
where Mag is in unit of magnitude. Metallicity is the value of $12+\log O/H$. The adjusted $R^{2}$ is 0.41. The Pearson coefficient is $-0.42 \pm 0.15$ with p-value $7.4 \times 10^{-2}$. The Spearman coefficient is $-0.41 \pm 0.17$ with p-value $7.8 \times 10^{-2}$. The Kendall $\tau$ coefficient is $-0.29 \pm 0.13$ with p-value $7.4 \times 10^{-2}$. The correlation ratio is $0.99 \pm 0.0025$. The cosine similarity is $-0.99 \pm 0.0022$. The scatter plot is in Figure \ref{fig:magmety}. The GRB sample number is 19. Metallicity indicates the star formation properties, and consequently the information of the progenitors of GRBs. However, the sample is not big enough, and the correlation is not very tight. One may not find much information from this correlation right now.

\begin{figure*}[!b]
\centering
\includegraphics[width=0.45\textwidth]{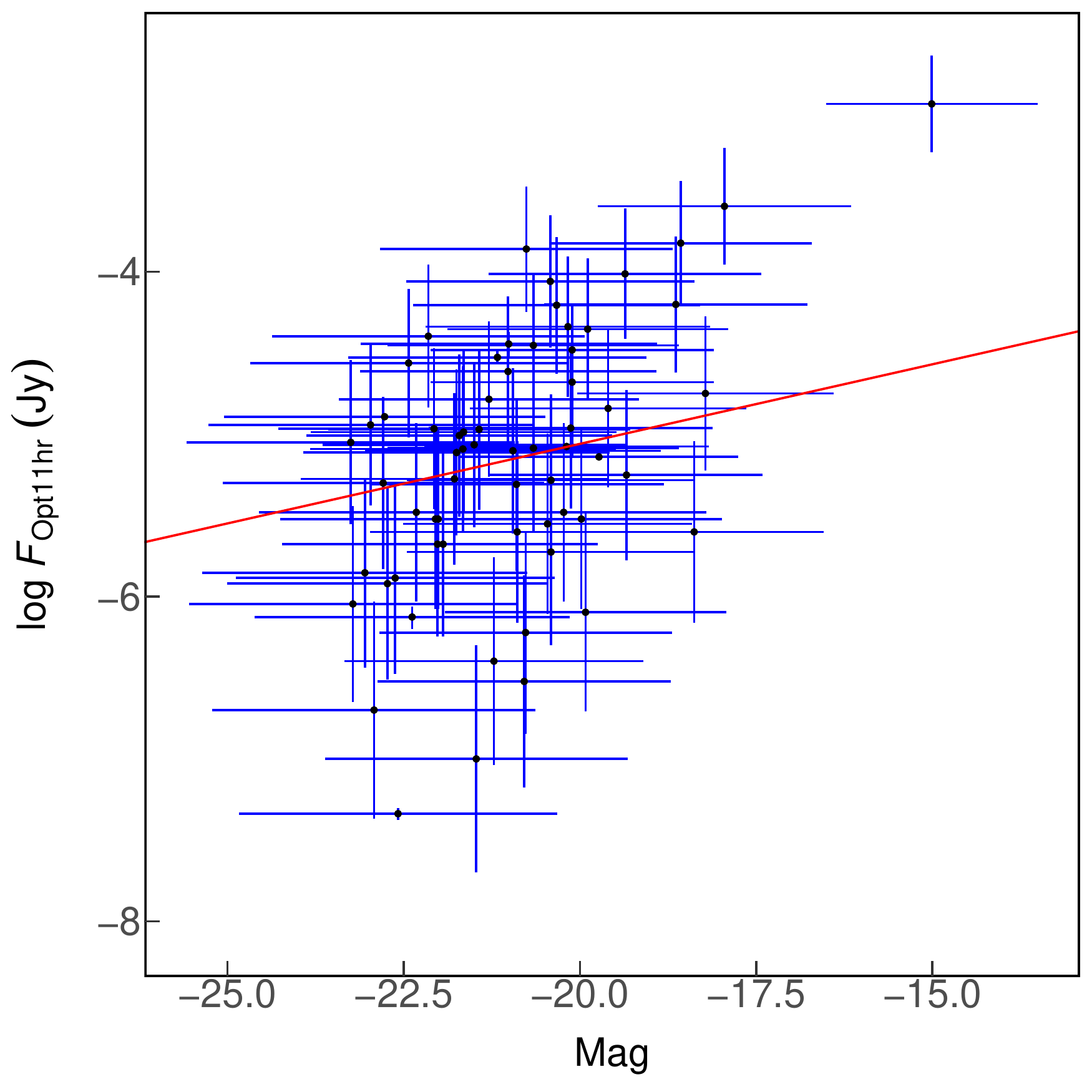}

\caption{
Scatter plot for Mag and $\log  F_{\rm Opt11hr}$.
The red line is our fit result. The formula of the red line is $Mag = (0.73 \pm 0.3) \times \log  F_{\rm Opt11hr} + (-17 \pm 1.5)$. The description of every parameter is in Section \ref{sec:sample}.
}
\label{fig:magfohr}
\end{figure*}

The correlation between Mag and $\log  F_{\rm Opt11hr}$ is:
\begin{equation} \label{eq:magfohr}
Mag = (0.73 \pm 0.3) \times \log  F_{\rm Opt11hr} + (-17 \pm 1.5),
\end{equation}
where $F_{\rm Opt11hr}$ is in unit of $\rm Jy$. Mag is in unit of magnitude. The adjusted $R^{2}$ is 0.27. The Pearson coefficient is $0.27 \pm 0.1$ with p-value $3.3 \times 10^{-2}$. The Spearman coefficient is $0.23 \pm 0.11$ with p-value $7.3 \times 10^{-2}$. The Kendall $\tau$ coefficient is $0.16 \pm 0.075$ with p-value $6.9 \times 10^{-2}$. The correlation ratio is $0.97 \pm 0.004$. The cosine similarity is $0.98 \pm 0.0028$. The scatter plot is in Figure \ref{fig:magfohr}. The GRB sample number is 64. It is interesting that the magnitude of the host galaxy is related to the observed flux density of the afterglow. It is probably that the number density of the smaller galaxy is higher, and the luminosity at 11 hour is brighter.

\begin{figure*}[!b]
\centering
\includegraphics[width=0.45\textwidth]{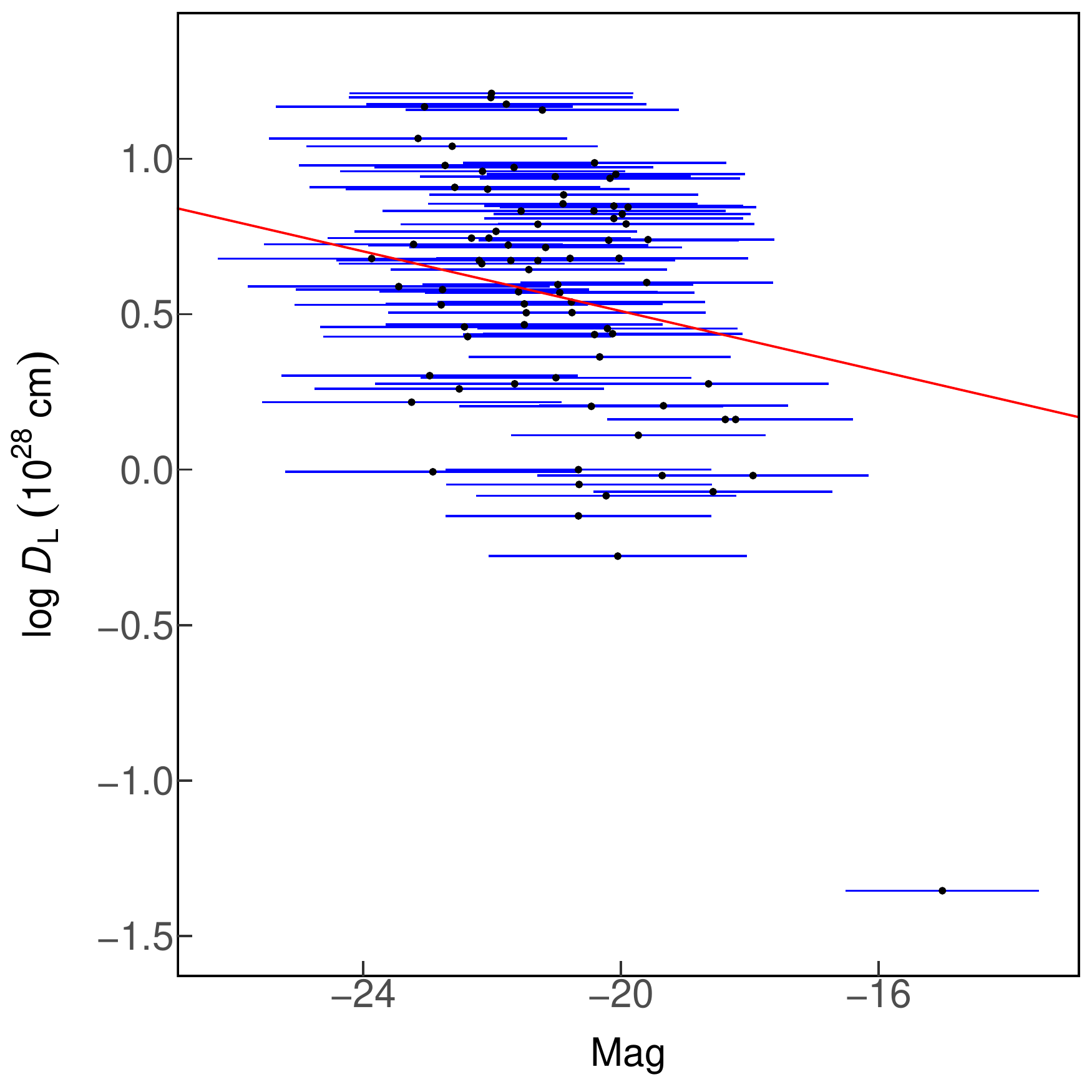}

\caption{
Scatter plot for Mag and $\log  D_{\rm L}$.
The red line is our fit result. The formula of the red line is $Mag = (-1.9 \pm 0.53) \times \log  D_{\rm L} + (-20 \pm 0.37)$. The description of every parameter is in Section \ref{sec:sample}. The outlier is low luminosity GRB 060218A with highest Mag and lowest $D_{\rm L}$ in the plot.
}
\label{fig:magdl}
\end{figure*}

The correlation between Mag and $\log  D_{\rm L}$ is:
\begin{equation} \label{eq:magdl}
Mag = (-1.9 \pm 0.53) \times \log  D_{\rm L} + (-20 \pm 0.37),
\end{equation}
where $D_{\rm L}$ is in unit of $\rm 10^{\rm 28} ~ cm$. Mag is in unit of magnitude. The adjusted $R^{2}$ is 0.26. The Pearson coefficient is $-0.3 \pm 0.081$ with p-value $6.9 \times 10^{-3}$. The Spearman coefficient is $-0.2 \pm 0.091$ with p-value $7.5 \times 10^{-2}$. The Kendall $\tau$ coefficient is $-0.14 \pm 0.064$ with p-value $7.3 \times 10^{-2}$. The correlation ratio is $0.99 \pm 0.002$. The cosine similarity is $-0.82 \pm 0.0058$. The scatter plot is in Figure \ref{fig:magdl}. The GRB sample number is 81. This relation shows the observational selection effect on the data. Because of the flux limit of the observations, only bright galaxies can be observed in the far distance. The observed host galaxies is correlated to the distance.

\subsubsection{Some other correlations between parameters related to host galaxies}

\begin{figure*}[!b]
\centering
\includegraphics[width=0.45\textwidth]{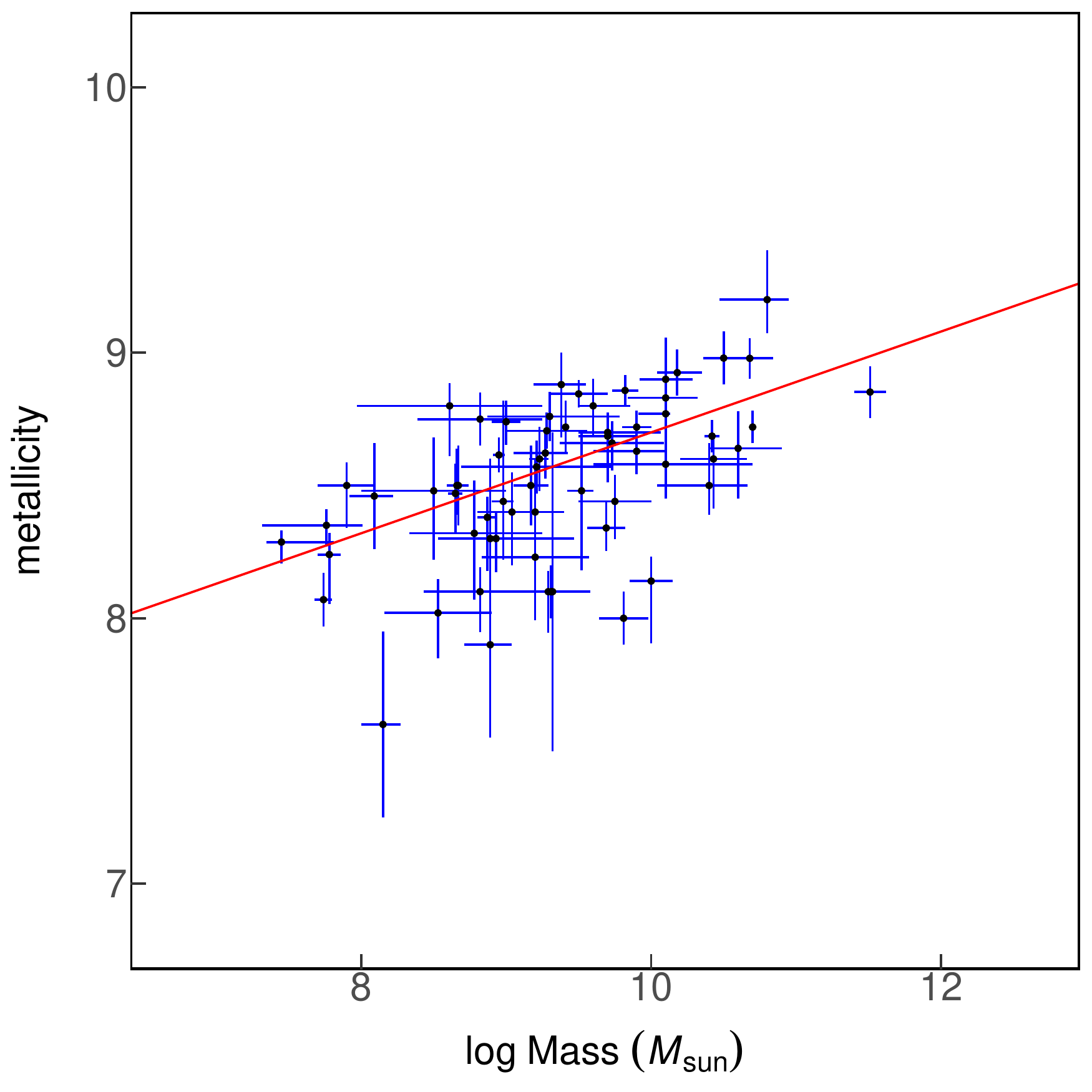}

\caption{
Scatter plot for metallicity and $\log Mass$.
The red line is our fit result. The formula of the red line is $metallicity = (0.19 \pm 0.025) \times \log Mass + (6.8 \pm 0.24)$. The description of every parameter is in Section \ref{sec:sample}.
}
\label{fig:metymass}
\end{figure*}

The correlation between metallicity and $\log Mass$ is:
\begin{equation} \label{eq:metymass}
metallicity = (0.19 \pm 0.025) \times \log Mass + (6.8 \pm 0.24),
\end{equation}
where metallicity is the value of $12+\log O/H$. Mass is in unit of $M_{\bigodot}$. The adjusted $R^{2}$ is 0.3. The Pearson coefficient is $0.47 \pm 0.054$ with p-value $9.3 \times 10^{-5}$. The Spearman coefficient is $0.49 \pm 0.057$ with p-value $3.7 \times 10^{-5}$. The Kendall $\tau$ coefficient is $0.34 \pm 0.043$ with p-value $6.7 \times 10^{-5}$. The correlation ratio is $0.54 \pm 0.022$. The cosine similarity is $1 \pm 0.00033$. The scatter plot is in Figure \ref{fig:metymass}. The GRB sample number is 64. The metallicity is positively correlated with the mass, which may indicate the history of merging of the galaxies. \citet{Arabsalmani2018} studied the mass-metallicity relation for GRB host galaxies, and found GRB-selected galaxies appear to track the mass-metallicity relation of star forming galaxies but
with an offset of 0.15 towards lower metallicities.

\begin{figure*}[!b]
\centering
\includegraphics[width=0.45\textwidth]{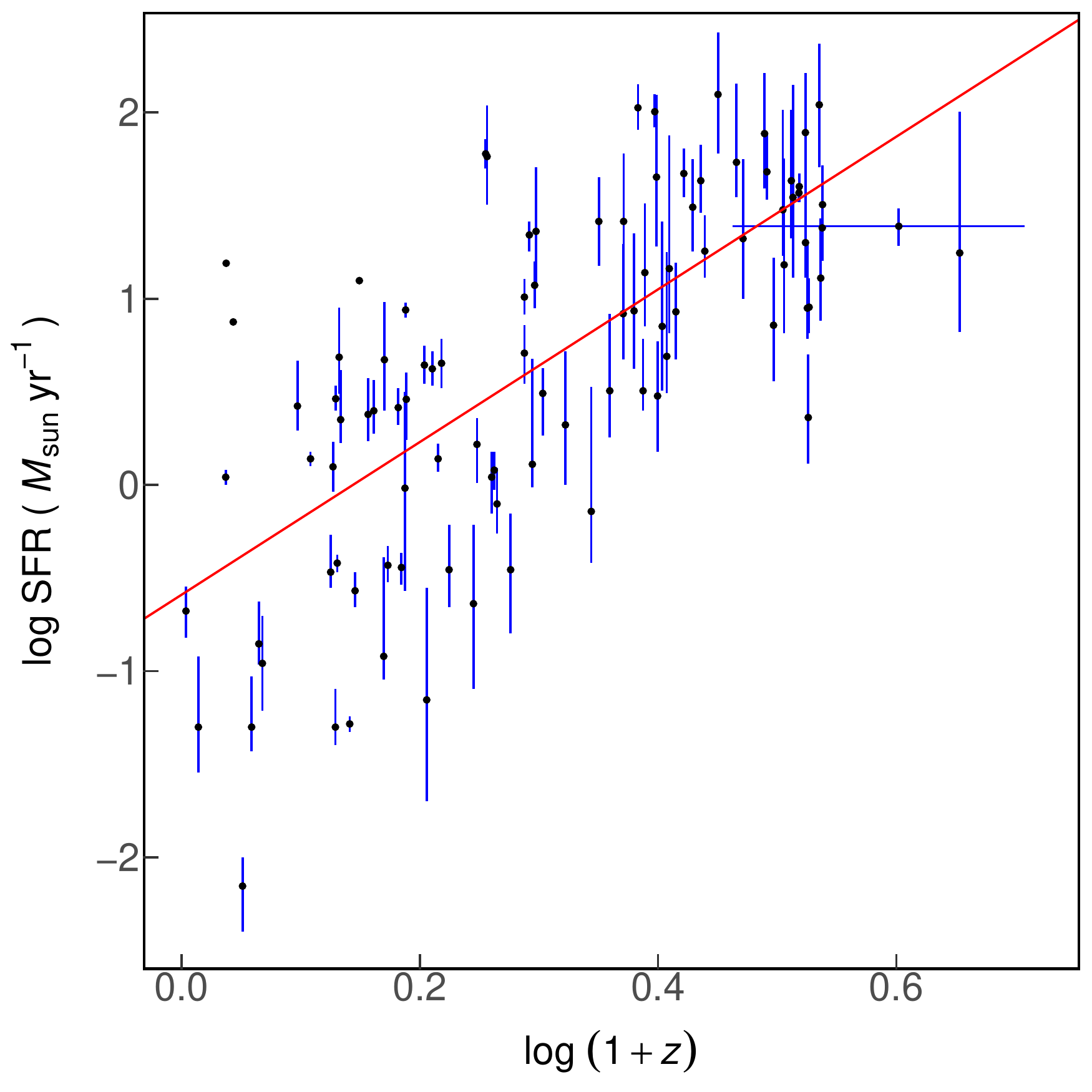}

\caption{
Scatter plot for $\log SFR$ and $\log (1+z)$.
The red line is our fit result. The formula of the red line is $\log SFR = (4.1 \pm 0.19) \times \log (1+z) + (-0.59 \pm 0.051)$. The description of every parameter is in Section \ref{sec:sample}.
}
\label{fig:sfrz}
\end{figure*}

The correlation between $\log SFR$ and $\log (1+z)$ is:
\begin{equation} \label{eq:sfrz}
\log SFR = (4.1 \pm 0.19) \times \log (1+z) + (-0.59 \pm 0.051),
\end{equation}
where SFR is in unit of $\rm M_{\bigodot} ~ yr^{\rm -1}$. The adjusted $R^{2}$ is 0.49. The Pearson coefficient is $0.69 \pm 0.019$ with p-value $1.3 \times 10^{-14}$. The Spearman coefficient is $0.69 \pm 0.02$ with p-value $4.8 \times 10^{-15}$. The Kendall $\tau$ coefficient is $0.49 \pm 0.018$ with p-value $1.6 \times 10^{-12}$. The correlation ratio is $0.24 \pm 0.018$. The cosine similarity is $0.76 \pm 0.016$. The scatter plot is in Figure \ref{fig:sfrz}. The GRB sample number is 96. This relation may indicate the history of the star formation. However, one should be caution about the data selection effect.

\begin{figure*}[!b]
\centering
\includegraphics[width=0.45\textwidth]{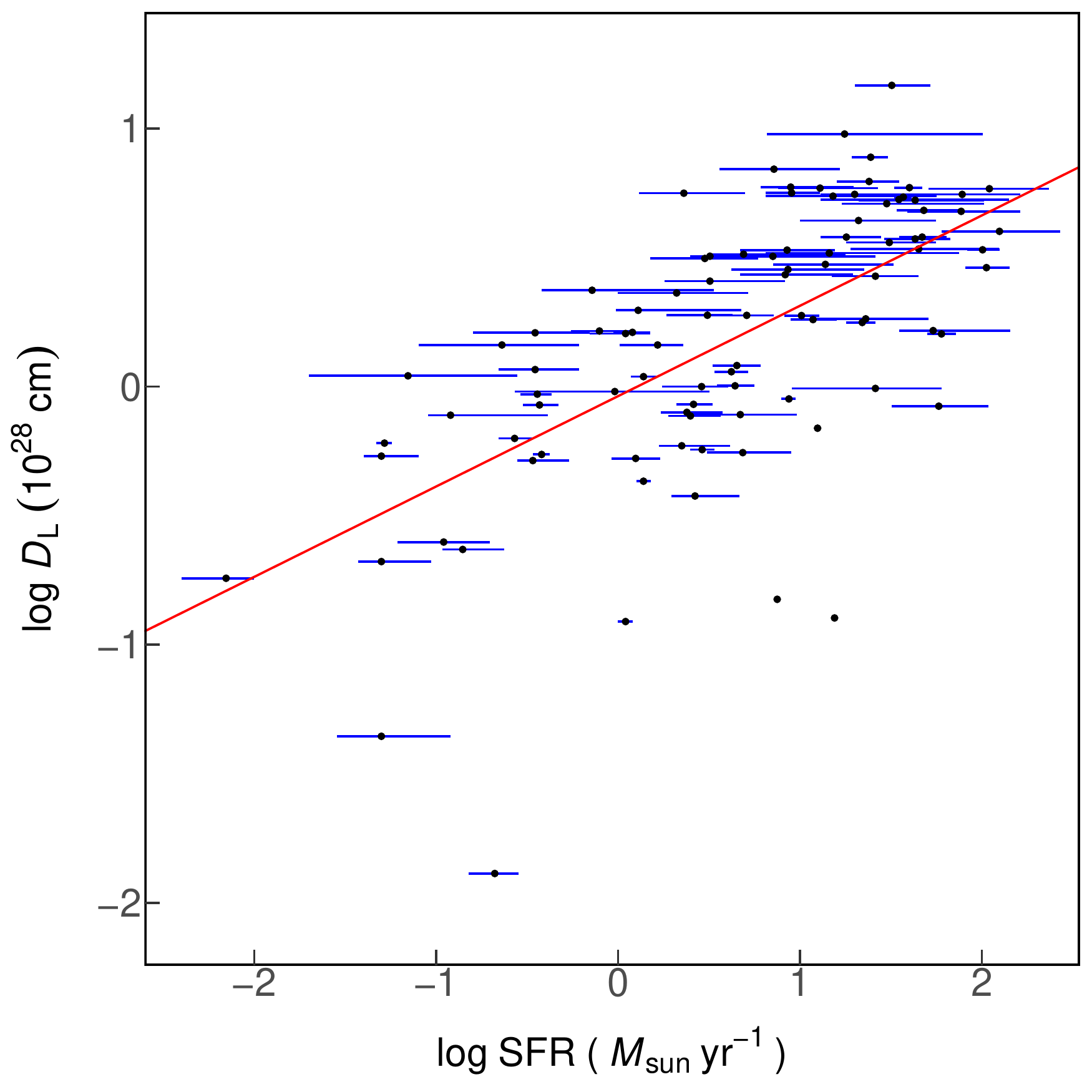}

\caption{
Scatter plot for $\log SFR$ and $\log D_{\rm L}$.
The red line is our fit result. The formula of the red line is $\log SFR = (1.2 \pm 0.047) \times \log D_{\rm L} + (0.43 \pm 0.025)$. The description of every parameter is in Section \ref{sec:sample}.
}
\label{fig:sfrdl}
\end{figure*}

The correlation between $\log SFR$ and $\log D_{\rm L}$ is:
\begin{equation} \label{eq:sfrdl}
\log SFR = (1.2 \pm 0.047) \times \log D_{\rm L} + (0.43 \pm 0.025),
\end{equation}
where SFR is in unit of $\rm M_{\bigodot} ~ yr^{\rm -1}$. $D_{\rm L}$ is in unit of $\rm 10^{\rm 28} ~ cm$. The adjusted $R^{2}$ is 0.42. The Pearson coefficient is $0.64 \pm 0.018$ with p-value $3.2 \times 10^{-12}$. The Spearman coefficient is $0.66 \pm 0.021$ with p-value $2.1 \times 10^{-13}$. The Kendall $\tau$ coefficient is $0.46 \pm 0.019$ with p-value $2.1 \times 10^{-11}$. The correlation ratio is $0.29 \pm 0.016$. The cosine similarity is $0.69 \pm 0.014$. The scatter plot is in Figure \ref{fig:sfrdl}. The GRB sample number is 96. This is similar to the ${\rm SFR}$ and $(1+z)$ relation.

\begin{figure*}[!b]
\centering
\includegraphics[width=0.45\textwidth]{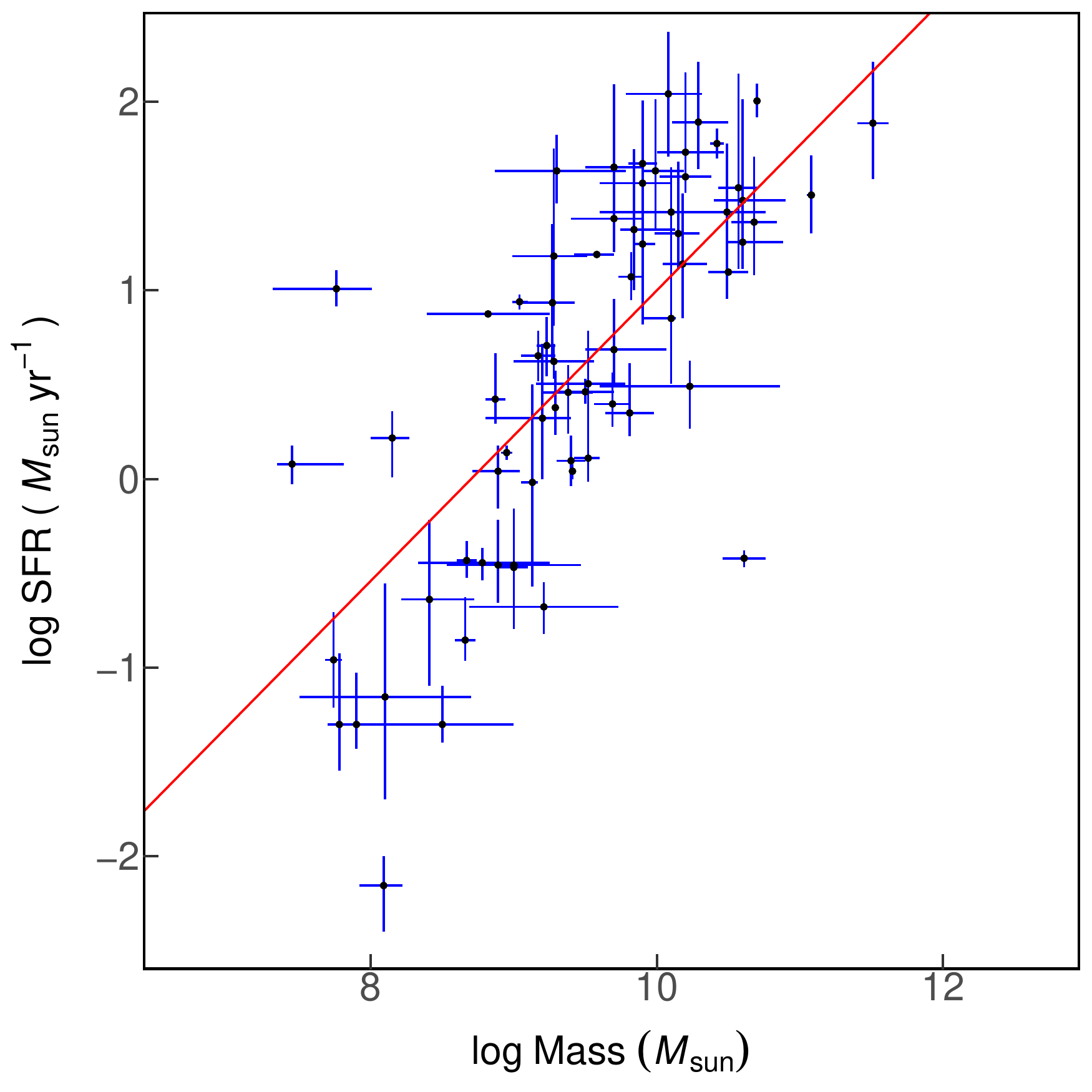}

\caption{
Scatter plot for $\log Mass$ and $\log SFR$.
The red line is our fit result. The formula of the red line is $\log Mass = (0.62 \pm 0.041) \times \log SFR + (9.1 \pm 0.046)$. The description of every parameter is in Section \ref{sec:sample}.
}
\label{fig:masssfr}
\end{figure*}

The correlation between $\log Mass$ and $\log SFR$ is:
\begin{equation} \label{eq:masssfr}
\log Mass = (0.62 \pm 0.041) \times \log SFR + (9.1 \pm 0.046),
\end{equation}
where mass is in unit of $M_{\bigodot}$. SFR is in unit of $\rm M_{\bigodot} ~ yr^{\rm -1}$. The adjusted $R^{2}$ is 0.54. The Pearson coefficient is $0.69 \pm 0.034$ with p-value $5.2 \times 10^{-11}$. The Spearman coefficient is $0.71 \pm 0.034$ with p-value $9.6 \times 10^{-12}$. The Kendall $\tau$ coefficient is $0.52 \pm 0.032$ with p-value $2 \times 10^{-10}$. The correlation ratio is $0.98 \pm 0.0011$. The cosine similarity is $0.57 \pm 0.021$. The scatter plot is in Figure \ref{fig:masssfr}. The GRB sample number is 69. It is natural that the more massive galaxy has more intense total star formation rate. Notice the index is $0.62 \pm 0.041$, which means the less massive galaxies is more effeminate for the star forming per unit mass.

\begin{figure*}[!b]
\centering
\includegraphics[width=0.45\textwidth]{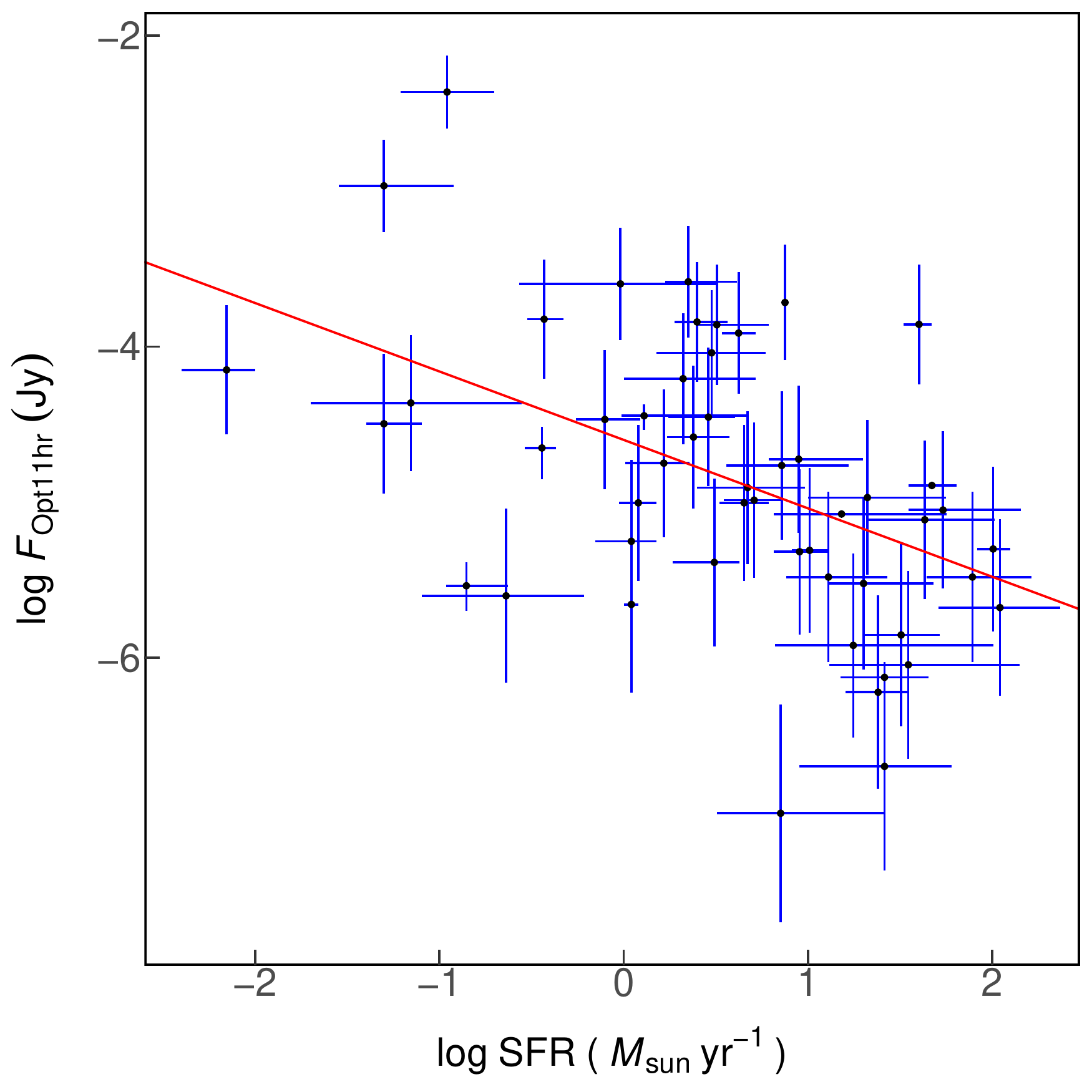}

\caption{
Scatter plot for $\log  F_{\rm Opt11hr}$ and $\log SFR$.
The red line is our fit result. The formula of the red line is $\log  F_{\rm Opt11hr} = (-0.44 \pm 0.073) \times \log SFR + (-4.6 \pm 0.074)$. The description of every parameter is in Section \ref{sec:sample}.
}
\label{fig:fohrsfr}
\end{figure*}

The correlation between $\log  F_{\rm Opt11hr}$ and $\log SFR$ is:
\begin{equation} \label{eq:fohrsfr}
\log  F_{\rm Opt11hr} = (-0.44 \pm 0.073) \times \log SFR + (-4.6 \pm 0.074),
\end{equation}
where $F_{\rm Opt11hr}$ is in unit of $\rm Jy$. SFR is in unit of $\rm M_{\bigodot} ~ yr^{\rm -1}$. The adjusted $R^{2}$ is 0.24. The Pearson coefficient is $-0.43 \pm 0.065$ with p-value $1.7 \times 10^{-3}$. The Spearman coefficient is $-0.43 \pm 0.069$ with p-value $1.6 \times 10^{-3}$. The Kendall $\tau$ coefficient is $-0.29 \pm 0.051$ with p-value $2.3 \times 10^{-3}$. The correlation ratio is $0.94 \pm 0.0042$. The cosine similarity is $-0.57 \pm 0.03$. The scatter plot is in Figure \ref{fig:fohrsfr}. The GRB sample number is 51. This correlation is related to the correlations in eq. (\ref{eq:masssfr}) and in eq. (\ref{eq:magfohr}).

\begin{figure*}[!b]
\centering
\includegraphics[width=0.45\textwidth]{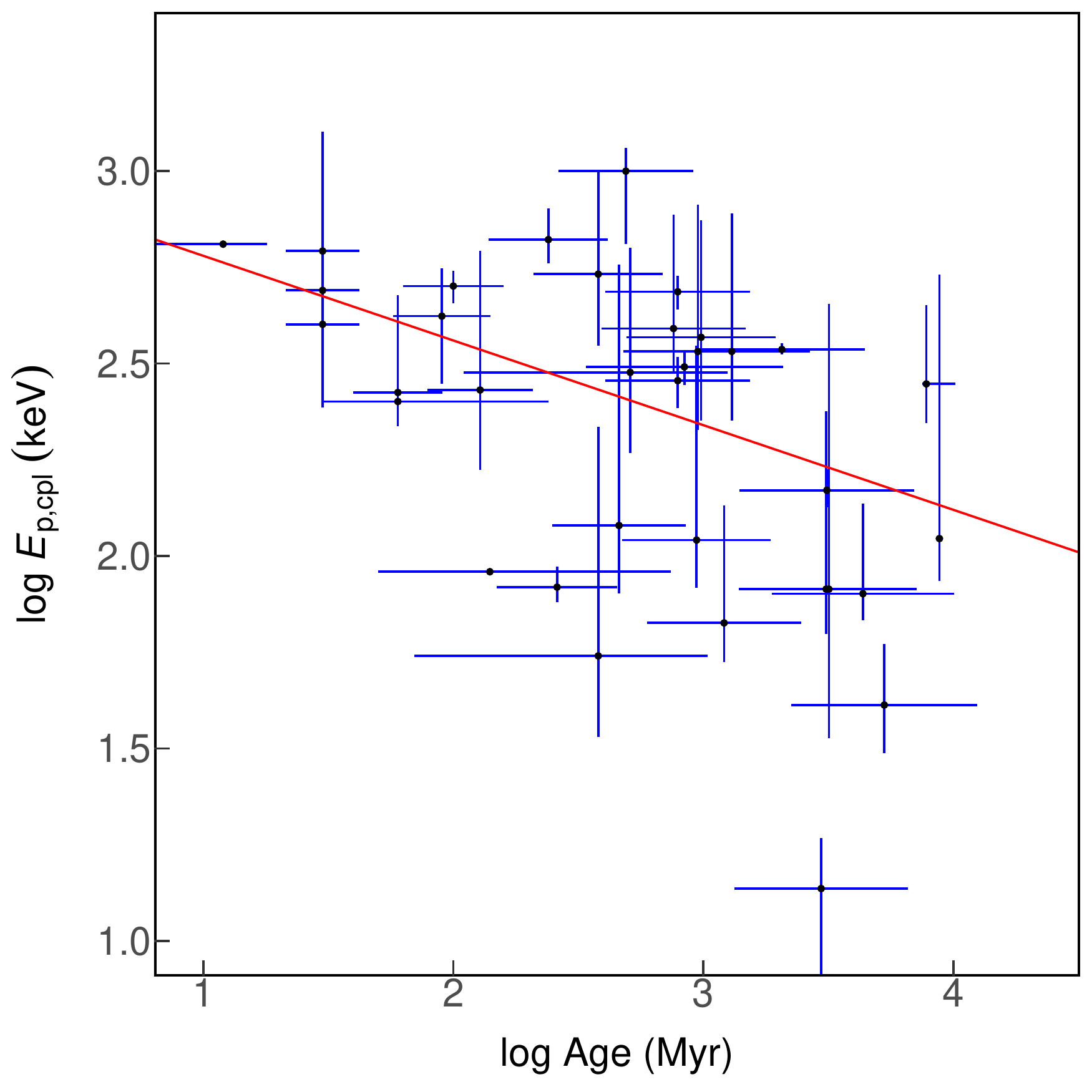}

\caption{
Scatter plot for $\log Age$ and $\log  E_{\rm p,cpl}$.
The red line is our fit result. The formula of the red line is $\log  Age = (-0.69 \pm 0.19) \times \log E_{\rm p,cpl} + (4.3 \pm 0.47)$. The description of every parameter is in Section \ref{sec:sample}.
}
\label{fig:ageepcpl}
\end{figure*}

The correlation between $\log Age$ and $\log  E_{\rm p,cpl}$ is:
\begin{equation} \label{eq:ageepcpl}
\log  Age = (-0.69 \pm 0.19) \times \log E_{\rm p,cpl} + (4.3 \pm 0.47),
\end{equation}
where $E_{\rm p,cpl}$ is in unit of $\rm keV$. Age is in unit of $\rm Myr$. The adjusted $R^{2}$ is 0.24. The Pearson coefficient is $-0.39 \pm 0.1$ with p-value $2.2 \times 10^{-2}$. The Spearman coefficient is $-0.39 \pm 0.11$ with p-value $1.9 \times 10^{-2}$. The Kendall $\tau$ coefficient is $-0.27 \pm 0.077$ with p-value $2.1 \times 10^{-2}$. The correlation ratio is $0.24 \pm 0.049$. The cosine similarity is $0.92 \pm 0.0086$. The scatter plot is in Figure \ref{fig:ageepcpl}. The GRB sample number is 35.

\begin{figure*}[!b]
\centering
\includegraphics[width=0.45\textwidth]{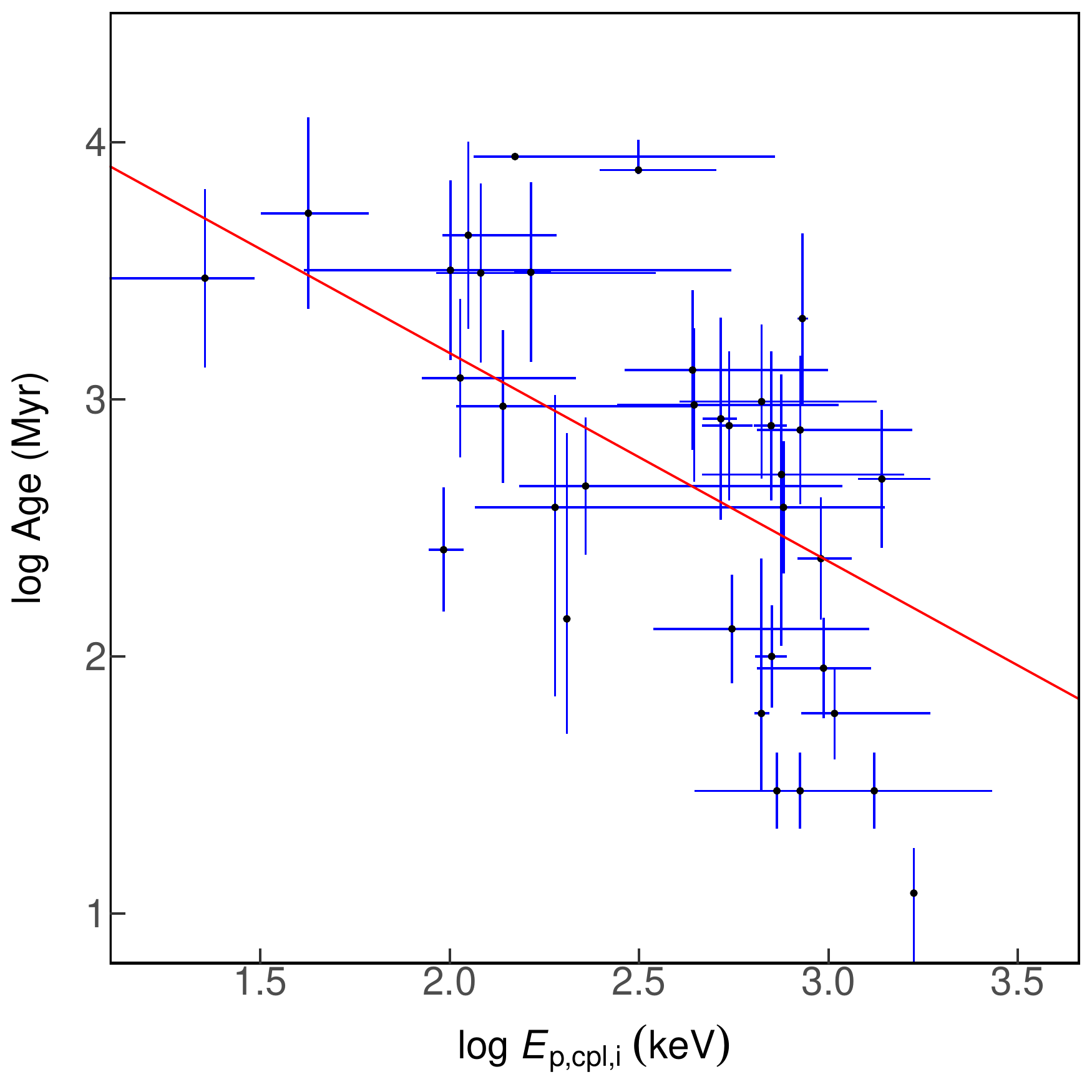}

\caption{
Scatter plot for $\log Age$ and $\log E_{\rm p,cpl,i}$.
The red line is our fit result. The formula of the red line is $\log Age = (-0.81 \pm 0.16) \times \log E_{\rm p,cpl,i} + (4.8 \pm 0.43)$. The description of every parameter is in Section \ref{sec:sample}.
}
\label{fig:ageepcpli}
\end{figure*}

The correlation between $\log Age$ and $\log E_{\rm p,cpl,i}$ is:
\begin{equation} \label{eq:ageepcpli}
\log Age = (-0.81 \pm 0.16) \times \log E_{\rm p,cpl,i} + (4.8 \pm 0.43),
\end{equation}
where Age is in unit of $\rm Myr$. $E_{\rm p,cpl,i}$ is in unit of $\rm keV$. The adjusted $R^{2}$ is 0.39. The Pearson coefficient is $-0.5 \pm 0.087$ with p-value $2.1 \times 10^{-3}$. The Spearman coefficient is $-0.51 \pm 0.094$ with p-value $1.8 \times 10^{-3}$. The Kendall $\tau$ coefficient is $-0.36 \pm 0.071$ with p-value $2.3 \times 10^{-3}$. The correlation ratio is $0.069 \pm 0.042$. The cosine similarity is $0.91 \pm 0.0086$. The scatter plot is in Figure \ref{fig:ageepcpli}. The GRB sample number is 35. This interesting correlation might be an counter evidence for the two origins of the GRBs. Short GRBs are from merger of double compact objects, which needs longer time of evolution, and consequently in older galaxies. And Short GRBs are harder in spectrum, and the peak energy $E_{\rm p,cpl,i}$  should be higher. Therefore, The age should be positively correlated to the peak energy, while eq. (\ref{eq:ageepcpli}) shows it contrarily.

\subsection{Correlations about HR}

\subsubsection{Correlations with HR}

\begin{figure*}[!b]
\centering
\includegraphics[width=0.45\textwidth]{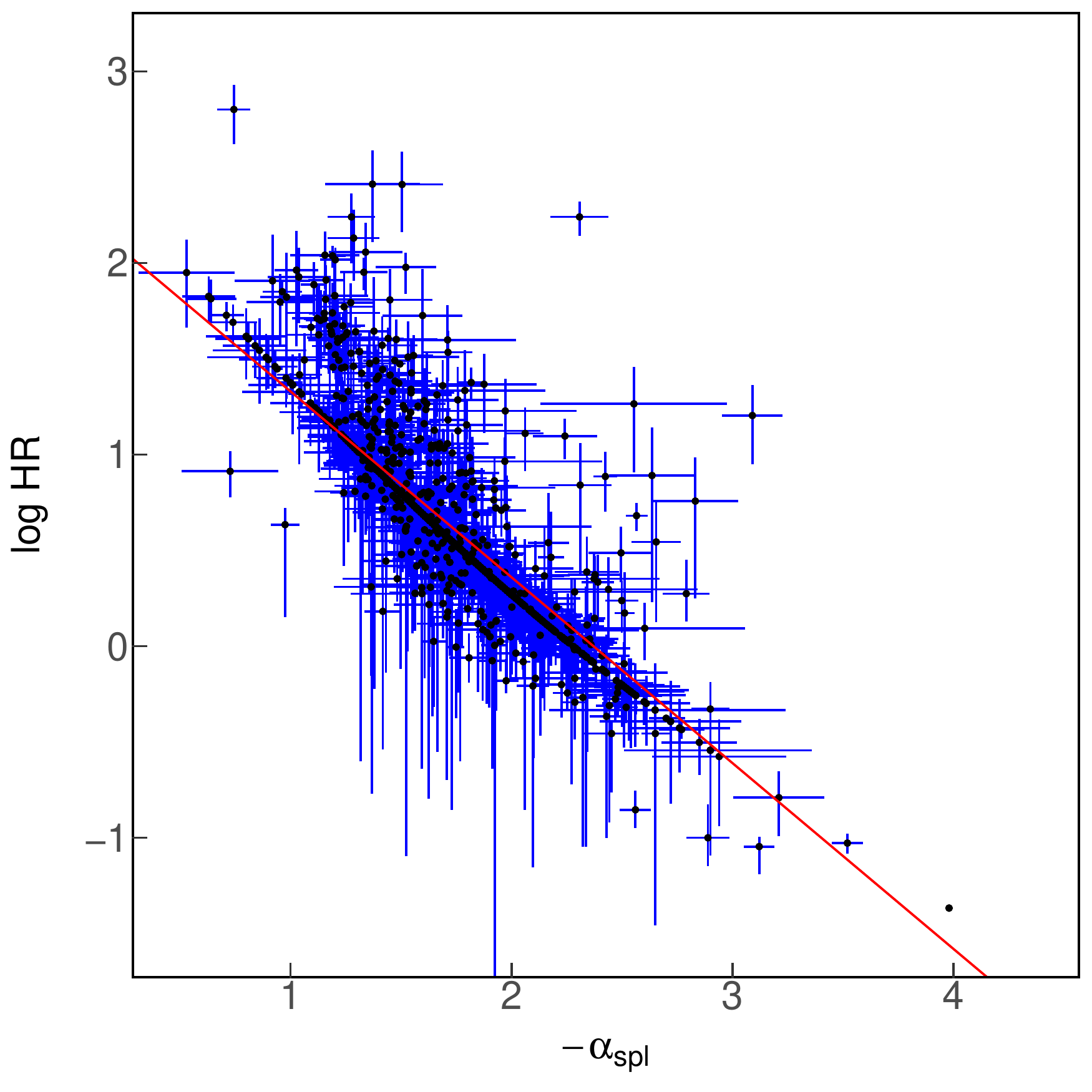}

\caption{
Scatter plot for $(-\alpha_{\rm spl})$ and $\log HR$.
The red line is our fit result. The formula of the red line is $(-\alpha_{\rm spl}) = (-0.55 \pm 0.021) \times \log HR + (2.1 \pm 0.017)$. The description of every parameter is in Section \ref{sec:sample}.
}
\label{fig:alplhr}
\end{figure*}

The correlation between $(-\alpha_{\rm spl})$ and $\log HR$ is:
\begin{equation} \label{eq:alplhr}
(-\alpha_{\rm spl}) = (-0.55 \pm 0.021) \times \log HR + (2.1 \pm 0.017),
\end{equation}
the adjusted $R^{2}$ is 0.69. The Pearson coefficient is $-0.73 \pm 0.016$ with p-value $1.6 \times 10^{-148}$. The Spearman coefficient is $-0.78 \pm 0.011$ with p-value $1.9 \times 10^{-176}$. The Kendall $\tau$ coefficient is $-0.6 \pm 0.01$ with p-value $8.4 \times 10^{-158}$. The correlation ratio is $0.71 \pm 0.0043$. The cosine similarity is $0.59 \pm 0.0093$. The scatter plot is in Figure \ref{fig:alplhr}. The GRB sample number is 874. This just shows the consistency of the spectrum. The outliers and the width show the hardness ratio itself is not good enough to present the information of the spectrum. All the correlations below in this section show the similar information.

\begin{figure*}[!b]
\centering
\includegraphics[width=0.45\textwidth]{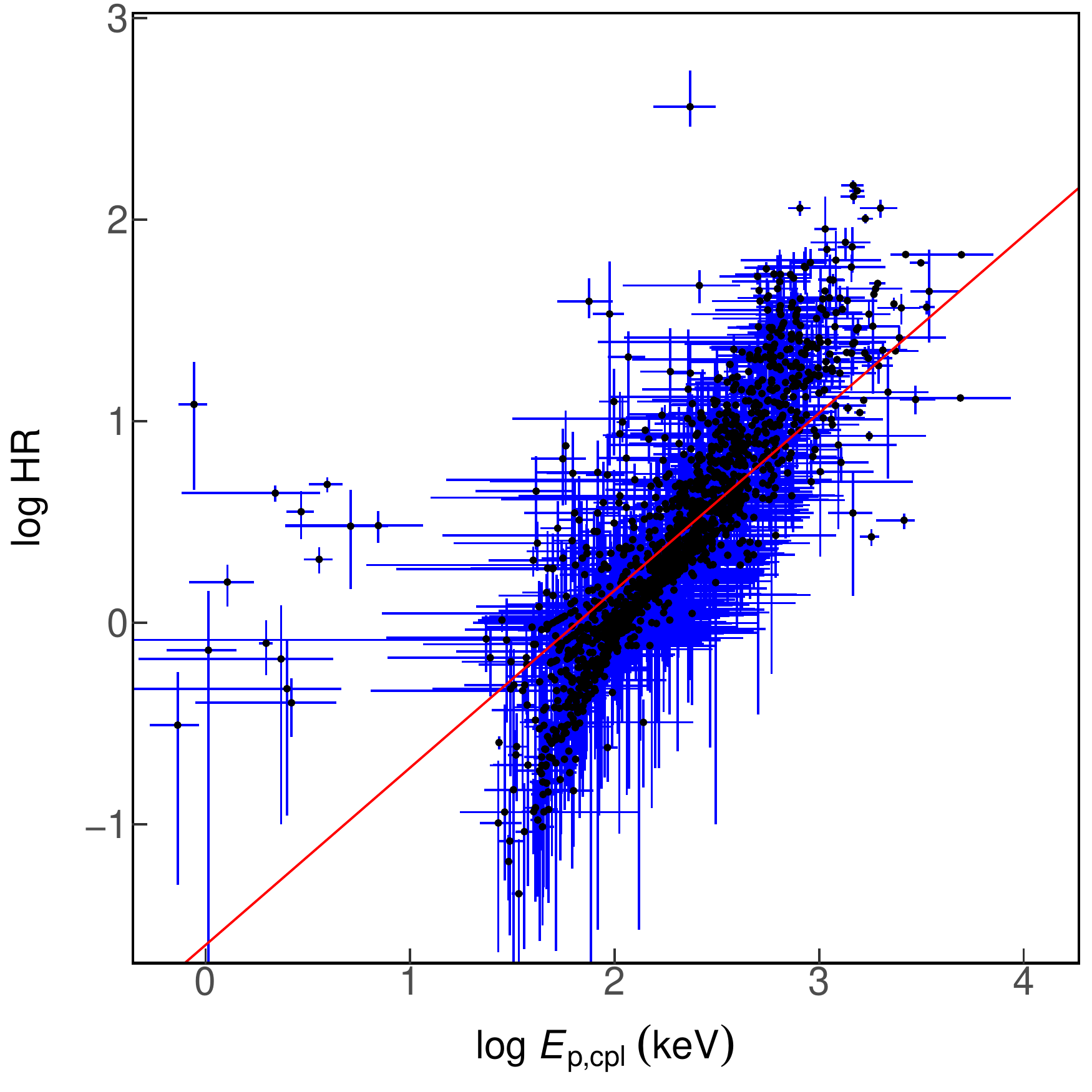}

\caption{
Scatter plot for $\log E_{\rm p,cpl}$ and $\log HR$.
The red line is our fit result. The formula of the red line is $\log E_{\rm p,cpl} = (0.54 \pm 0.013) \times \log HR + (2.1 \pm 0.0084)$. The description of every parameter is in Section \ref{sec:sample}. 
}
\label{fig:epplhr}
\end{figure*}

The correlation between $\log E_{\rm p,cpl}$ and $\log HR$ is:
\begin{equation} \label{eq:epplhr}
\log E_{\rm p,cpl} = (0.54 \pm 0.013) \times \log HR + (2.1 \pm 0.0084),
\end{equation}
where $E_{\rm p,cpl}$ is in unit of $\rm keV$. The adjusted $R^{2}$ is 0.65. The Pearson coefficient is $0.69 \pm 0.014$ with p-value $3.8 \times 10^{-211}$. The Spearman coefficient is $0.78 \pm 0.0094$ with p-value $6.5 \times 10^{-305}$. The Kendall $\tau$ coefficient is $0.61 \pm 0.0083$ with p-value $1.1 \times 10^{-267}$. The correlation ratio is $0.86 \pm 0.0023$. The cosine similarity is $0.67 \pm 0.008$. The scatter plot is in Figure \ref{fig:epplhr}. The GRB sample number is 1469.

\begin{figure*}[!b]
\centering
\includegraphics[width=0.45\textwidth]{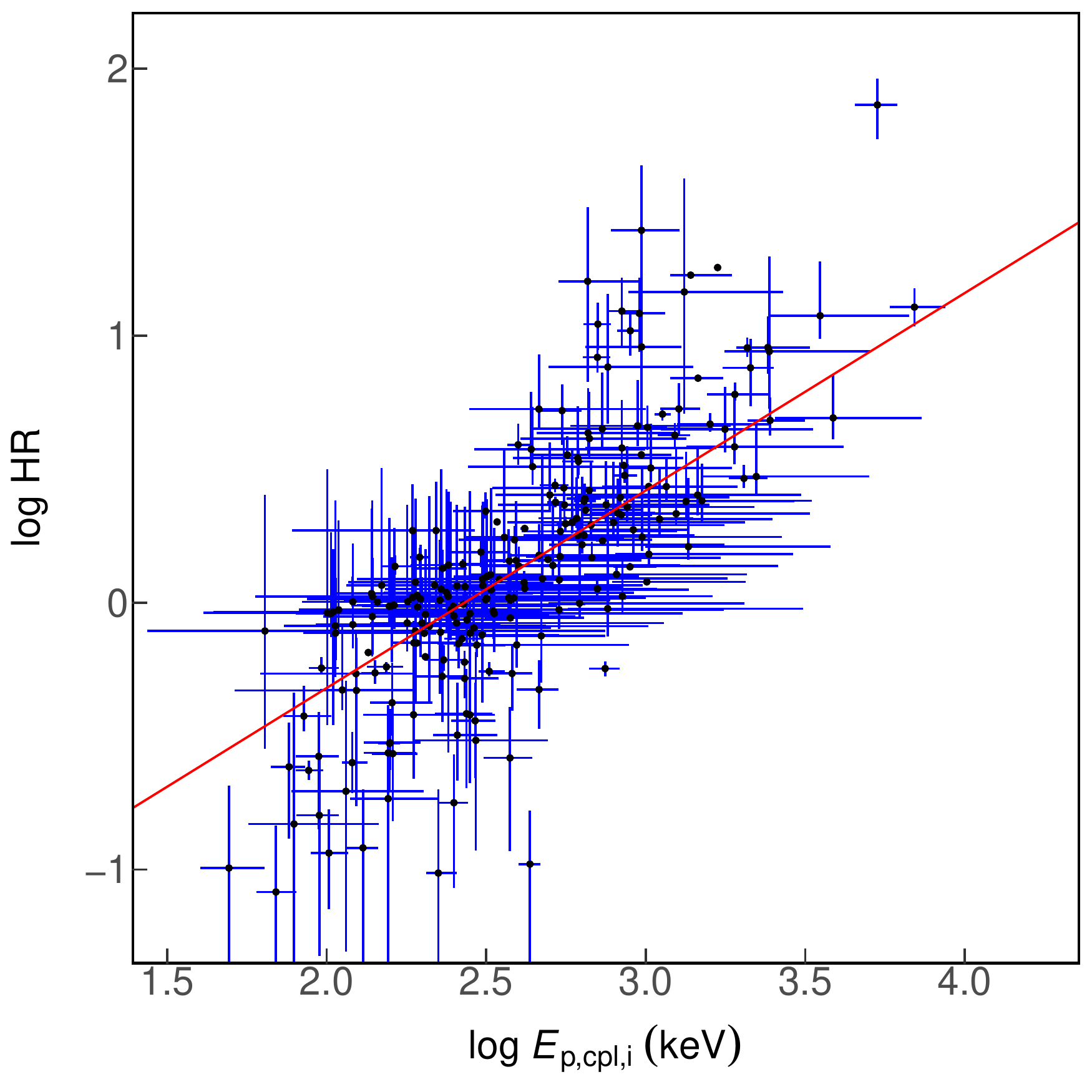}

\caption{
Scatter plot for $\log E_{\rm p,cpl,i}$ and $\log HR$.
The red line is our fit result. The formula of the red line is $\log E_{\rm p,cpl,i} = (0.55 \pm 0.036) \times \log HR + (2.6 \pm 0.018)$. The description of every parameter is in Section \ref{sec:sample}.
}
\label{fig:eplihr}
\end{figure*}

The correlation between $\log E_{\rm p,cpl,i}$ and $\log HR$ is:
\begin{equation} \label{eq:eplihr}
\log E_{\rm p,cpl,i} = (0.55 \pm 0.036) \times \log HR + (2.6 \pm 0.018),
\end{equation}
where $E_{\rm p,cpl,i}$ is in unit of $\rm keV$. The adjusted $R^{2}$ is 0.6. The Pearson coefficient is $0.63 \pm 0.03$ with p-value $4.3 \times 10^{-27}$. The Spearman coefficient is $0.68 \pm 0.028$ with p-value $1.9 \times 10^{-32}$. The Kendall $\tau$ coefficient is $0.49 \pm 0.023$ with p-value $2.2 \times 10^{-28}$. The correlation ratio is $0.93 \pm 0.0032$. The cosine similarity is $0.38 \pm 0.026$. The scatter plot is in Figure \ref{fig:eplihr}. The GRB sample number is 229.

\begin{figure*}[!b]
\centering
\includegraphics[width=0.45\textwidth]{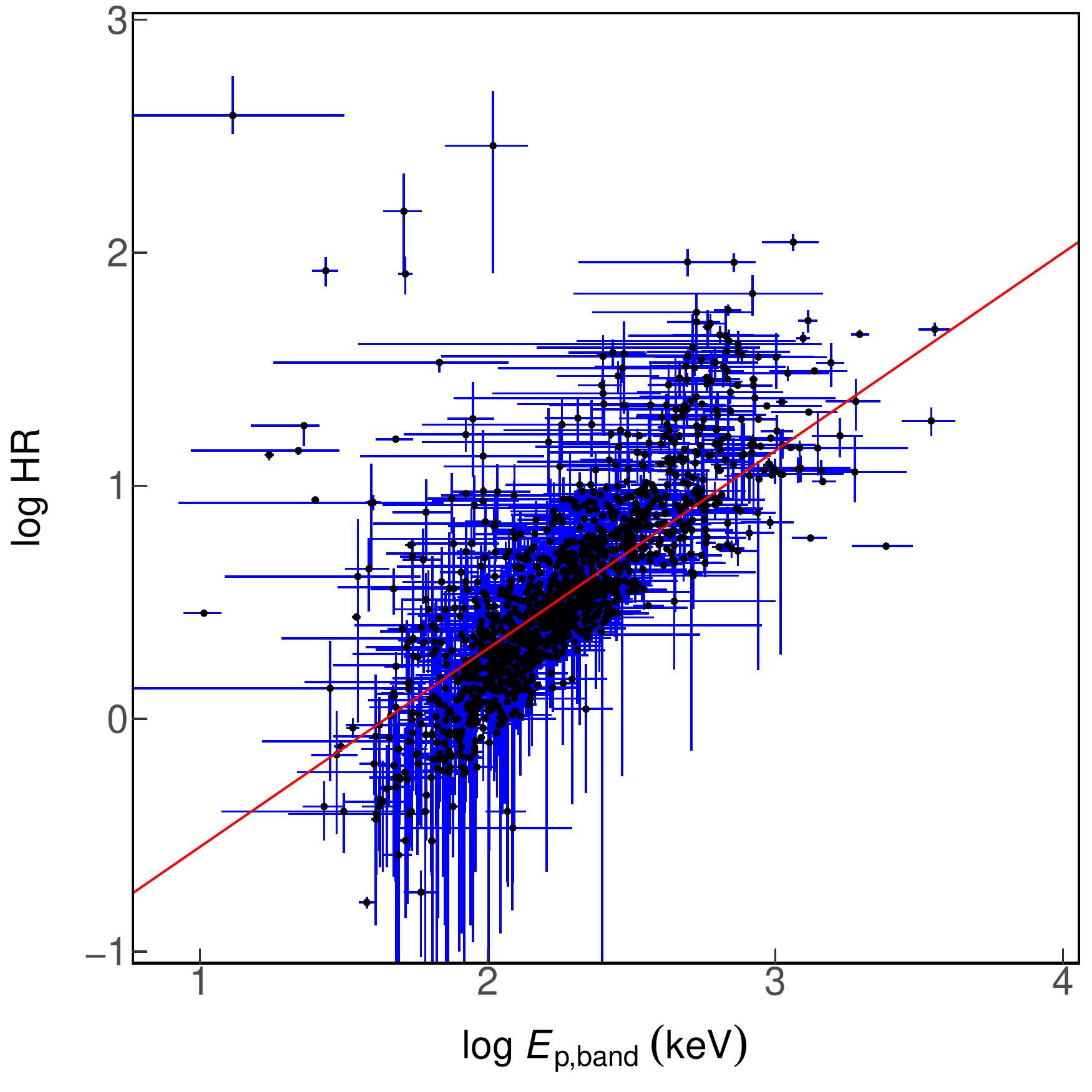}

\caption{
Scatter plot for $\log E_{\rm p,band}$ and $\log HR$.
The red line is our fit result. The formula of the red line is $\log E_{\rm p,band} = (0.47 \pm 0.017) \times \log HR + (2 \pm 0.01)$. The description of every parameter is in Section \ref{sec:sample}.}
\label{fig:epndhr}
\end{figure*}

The correlation between $\log E_{\rm p,band}$ and $\log HR$ is:
\begin{equation} \label{eq:epndhr}
\log E_{\rm p,band} = (0.47 \pm 0.017) \times \log HR + (2 \pm 0.01),
\end{equation}
where $E_{\rm p,band}$ is in unit of $\rm keV$. The adjusted $R^{2}$ is 0.51. The Pearson coefficient is $0.63 \pm 0.017$ with p-value $5.2 \times 10^{-149}$. The Spearman coefficient is $0.72 \pm 0.01$ with p-value $2.3 \times 10^{-214}$. The Kendall $\tau$ coefficient is $0.56 \pm 0.0081$ with p-value $5.9 \times 10^{-207}$. The correlation ratio is $0.9 \pm 0.0021$. The cosine similarity is $0.81 \pm 0.0079$. The scatter plot is in Figure \ref{fig:epndhr}. The GRB sample number is 1332.

\begin{figure*}[!b]
\centering
\includegraphics[width=0.45\textwidth]{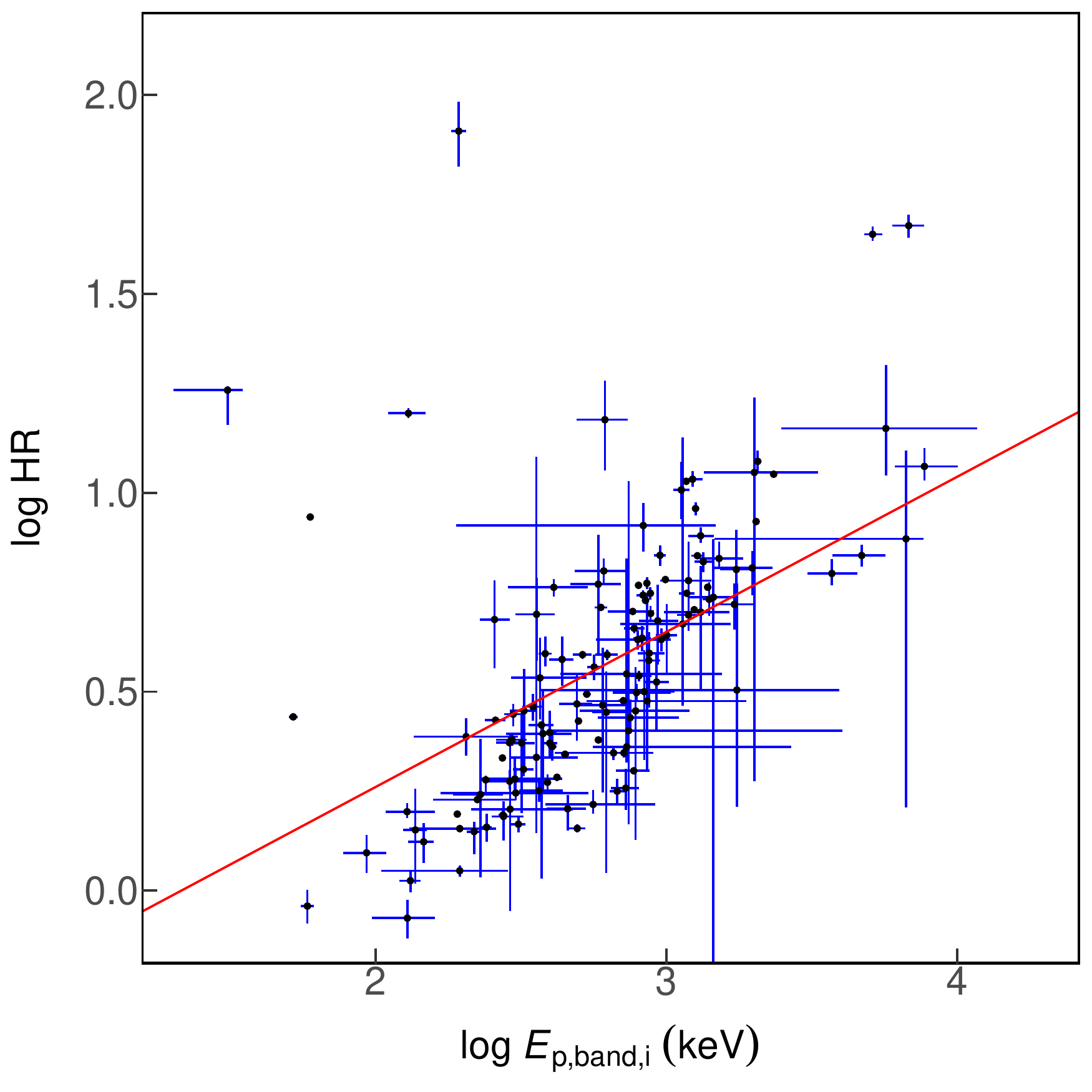}

\caption{
Scatter plot for $\log E_{\rm p,band,i}$ and $\log HR$.
The red line is our fit result. The formula of the red line is $\log E_{\rm p,band,i} = (0.58 \pm 0.066) \times \log HR + (2.4 \pm 0.041)$. The description of every parameter is in Section \ref{sec:sample}.
}
\label{fig:epdihr}
\end{figure*}

The correlation between $\log E_{\rm p,band,i}$ and $\log HR$ is:
\begin{equation} \label{eq:epdihr}
\log E_{\rm p,band,i} = (0.58 \pm 0.066) \times \log HR + (2.4 \pm 0.041),
\end{equation}
where $E_{\rm p,band,i}$ is in unit of $\rm keV$. The adjusted $R^{2}$ is 0.3. The Pearson coefficient is $0.48 \pm 0.045$ with p-value $4.6 \times 10^{-9}$. The Spearman coefficient is $0.6 \pm 0.035$ with p-value $1.4 \times 10^{-14}$. The Kendall $\tau$ coefficient is $0.47 \pm 0.028$ with p-value $4 \times 10^{-16}$. The correlation ratio is $0.94 \pm 0.0025$. The cosine similarity is $0.87 \pm 0.013$. The scatter plot is in Figure \ref{fig:epdihr}. The GRB sample number is 136.

\subsubsection{Correlations between parameters related with HR or peak energies}

\begin{figure*}[!b]
\centering
\includegraphics[width=0.45\textwidth]{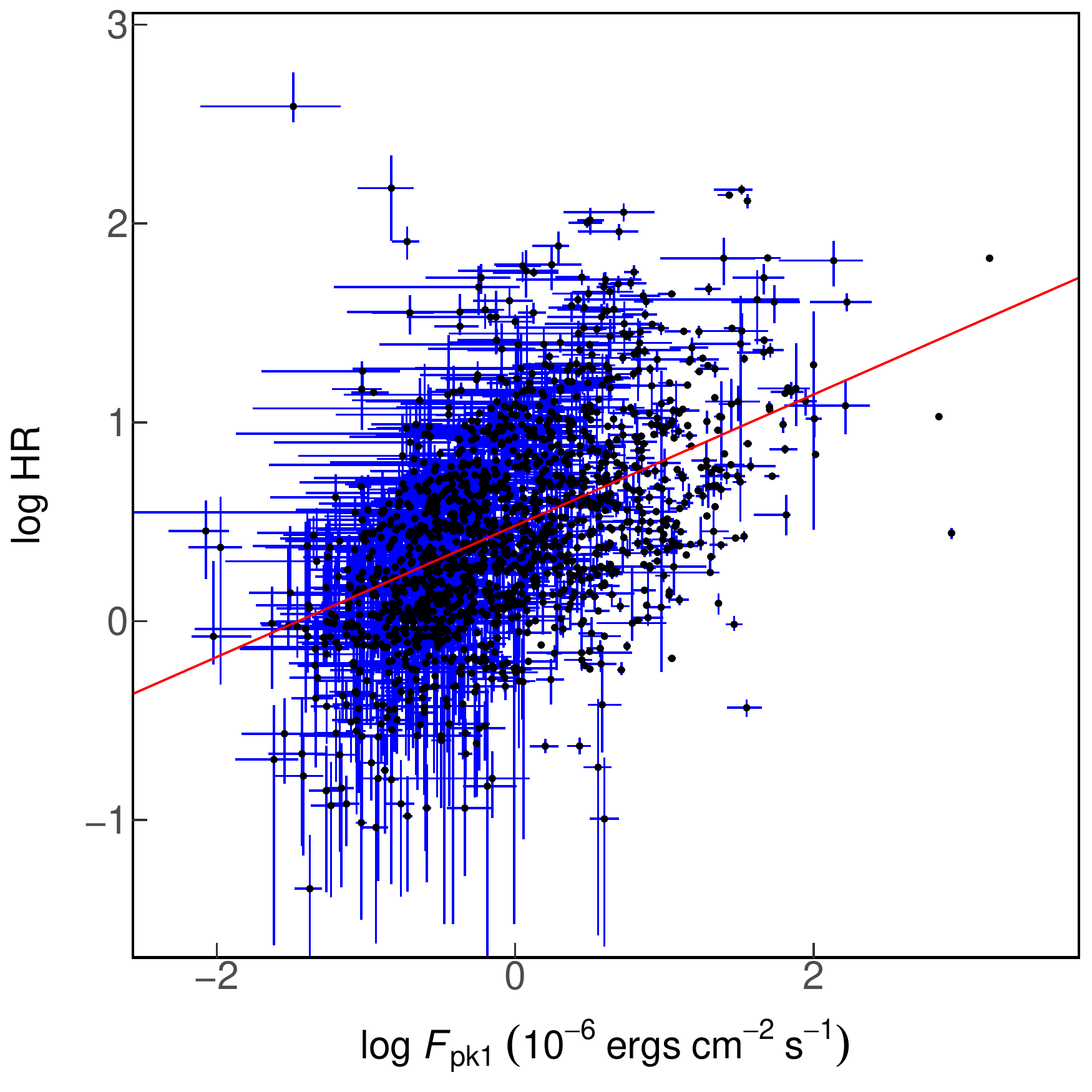}

\caption{
Scatter plot for $\log  F_{\rm pk1}$ and $\log HR$.
The red line is our fit result. The formula of the red line is $\log F_{\rm pk1} = (0.6 \pm 0.017) \times \log HR + (-0.34 \pm 0.0094)$. The description of every parameter is in Section \ref{sec:sample}.
}
\label{fig:fpk1hr}
\end{figure*}

The correlation between $\log  F_{\rm pk1}$ and $\log HR$ is:
\begin{equation} \label{eq:fpk1hr}
\log F_{\rm pk1} = (0.6 \pm 0.017) \times \log HR + (-0.34 \pm 0.0094),
\end{equation}
where $F_{\rm pk1}$ is peak energy flux of 1 $\rm s$ time bin in rest-frame 1-$10^{4}$ $\rm keV$ energy band and in unit of $\rm 10^{\rm -6} ~ ergs ~ cm^{\rm -2} ~ s^{\rm -1}$. The adjusted $R^{2}$ is 0.26. The Pearson coefficient is $0.45 \pm 0.011$ with p-value $3.9 \times 10^{-75}$. The Spearman coefficient is $0.47 \pm 0.01$ with p-value $7.9 \times 10^{-83}$. The Kendall $\tau$ coefficient is $0.33 \pm 0.0073$ with p-value $4.2 \times 10^{-79}$. The correlation ratio is $0.37 \pm 0.0044$. The cosine similarity is $0.29 \pm 0.011$. The scatter plot is in Figure \ref{fig:fpk1hr}. The GRB sample number is 1494. Though there is a correlation, one would say it is just a weak tendency.

In \citet{Mallozzi1995}, they found a correlation between mean peak energies and 256 $\rm ms$ peak photon flux $P_{pk2}$. With accumulated more data, we found $P_{pk2}$ and peak energy have no obvious correlation with adjusted $R_{2}$ smaller than 0.1, except $P_{\rm pk4}$ and $E_{\rm p,band}$.

\begin{figure*}[!b]
\centering
\includegraphics[width=0.45\textwidth]{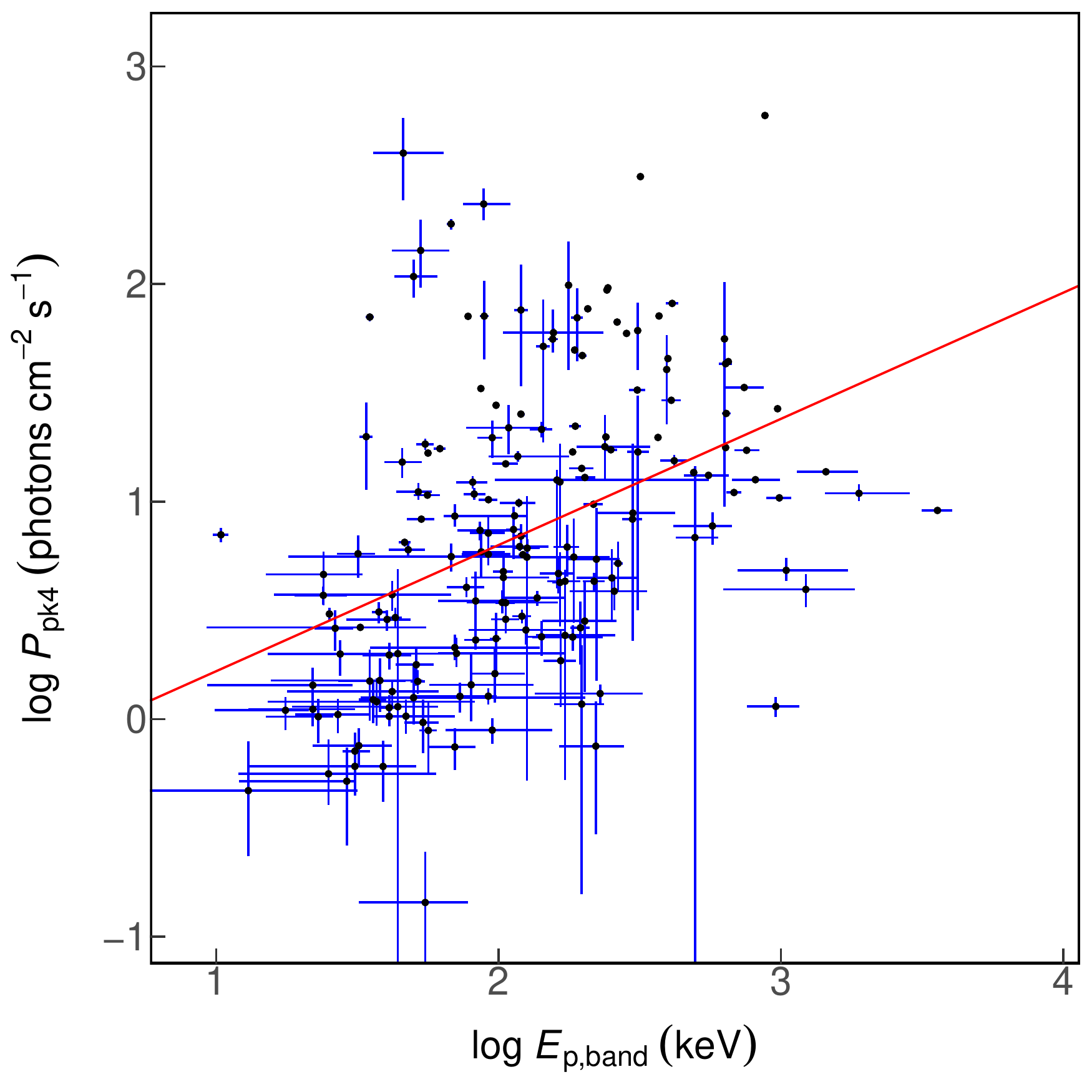}

\caption{
Scatter plot for $\log E_{\rm p,band}$ and $\log P_{\rm pk4}$.
The red line is our fit result. The formula of the red line is $\log E_{\rm p,band} = (0.28 \pm 0.03) \times \log P_{\rm pk4} + (1.8 \pm 0.033)$. The description of every parameter is in Section \ref{sec:sample}.
}
\label{fig:epndppk4}
\end{figure*}

The correlation between $\log E_{\rm p,band}$ and $\log P_{\rm pk4}$ is:
\begin{equation} \label{eq:epndppk4}
\log E_{\rm p,band} = (0.28 \pm 0.03) \times \log P_{\rm pk4} + (1.8 \pm 0.033),
\end{equation}
where $P_{\rm pk4}$ is peak photon flux of 1 $\rm s$ time bin in 10-1000 $\rm keV$ and in unit of $\rm photons ~ cm^{\rm -2} ~ s^{\rm -1}$. $E_{\rm p,band}$ is in unit of $\rm keV$. The adjusted $R^{2}$ is 0.19. The Pearson coefficient is $0.4 \pm 0.032$ with p-value $1.9 \times 10^{-8}$. The Spearman coefficient is $0.46 \pm 0.024$ with p-value $4.5 \times 10^{-11}$. The Kendall $\tau$ coefficient is $0.31 \pm 0.017$ with p-value $3.4 \times 10^{-10}$. The correlation ratio is $0.71 \pm 0.0087$. The cosine similarity is $0.81 \pm 0.017$. The scatter plot is in Figure \ref{fig:epndppk4}. The GRB sample number is 181. As one can see it from the figure, even this relation is not very not very tight. One would say there no much correlation between peak energies in spectra and peak photon fluxes in light curves.

Alternatively, we found a more tight correlation between peak energy flux and peak energy. Peak energy flux can be calculated from peak photon flux, so this correlation is natural.

\begin{figure*}[!b]
\centering
\includegraphics[width=0.45\textwidth]{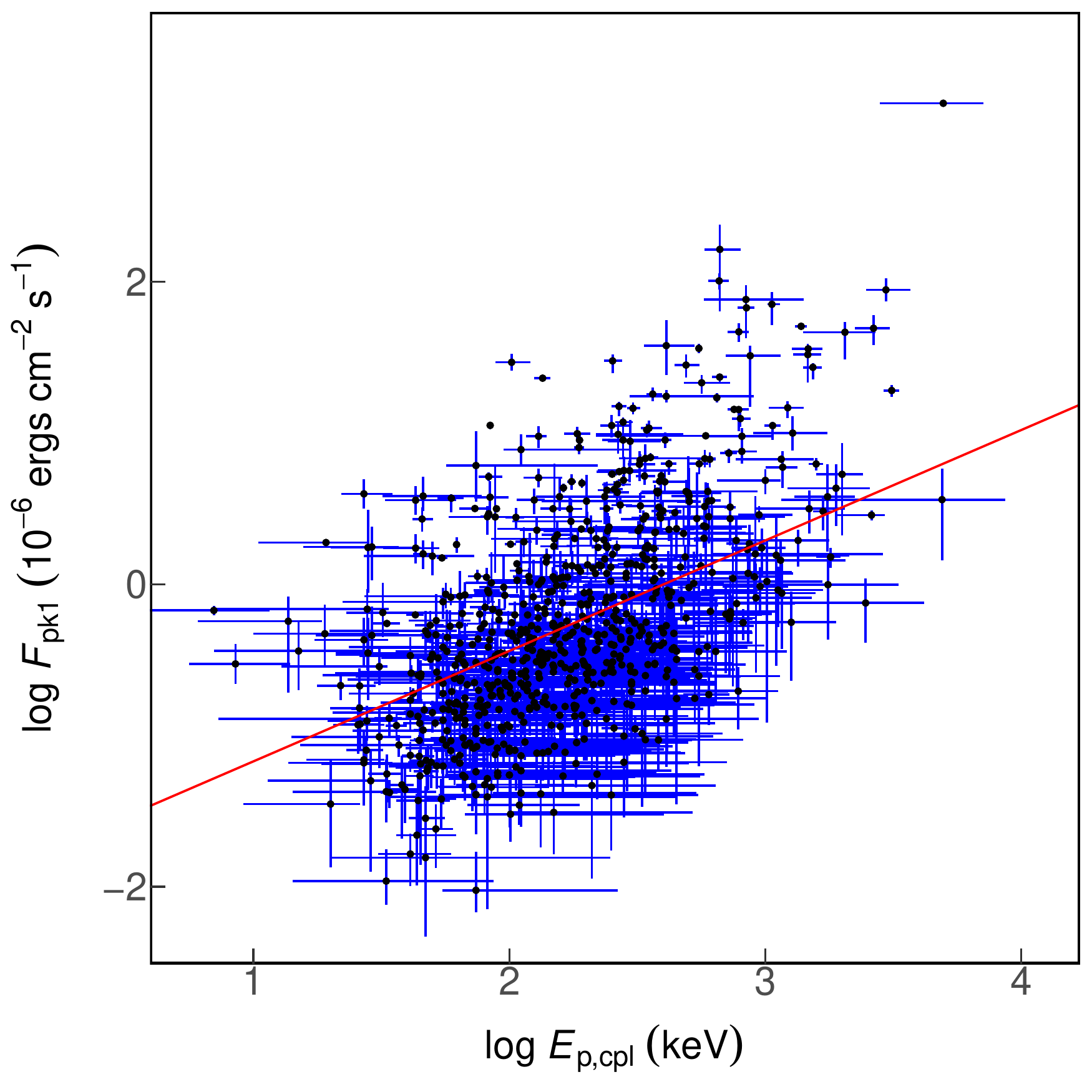}

\caption{
Scatter plot for $\log E_{\rm p,cpl}$ and $\log F_{\rm pk1}$.
The red line is our fit result. The formula of the red line is $\log E_{\rm p,cpl} = (0.31 \pm 0.012) \times \log F_{\rm pk1} + (2.3 \pm 0.0067)$. The description of every parameter is in Section \ref{sec:sample}.
}
\label{fig:epcplfpk1}
\end{figure*}

The correlation between $\log E_{\rm p,cpl}$ and $\log F_{\rm pk1}$ is:
\begin{equation} \label{eq:epcplfpk1}
\log E_{\rm p,cpl} = (0.31 \pm 0.012) \times \log F_{\rm pk1} + (2.3 \pm 0.0067),
\end{equation}
where $F_{\rm pk1}$ is peak energy flux of 1 $\rm s$ time bin in rest-frame 1-$10^{4}$ $\rm keV$ energy band and in unit of $\rm 10^{\rm -6} ~ ergs ~ cm^{\rm -2} ~ s^{\rm -1}$. $E_{\rm p,cpl}$ is in unit of $\rm keV$. The adjusted $R^{2}$ is 0.33. The Pearson coefficient is $0.48 \pm 0.018$ with p-value $1.7 \times 10^{-43}$. The Spearman coefficient is $0.47 \pm 0.018$ with p-value $1.6 \times 10^{-42}$. The Kendall $\tau$ coefficient is $0.33 \pm 0.013$ with p-value $3.4 \times 10^{-42}$. The correlation ratio is $0.9 \pm 0.0014$. The cosine similarity is $-0.22 \pm 0.0083$. The scatter plot is in Figure \ref{fig:epcplfpk1}. The GRB sample number is 749.

\begin{figure*}[!b]
\centering
\includegraphics[width=0.45\textwidth]{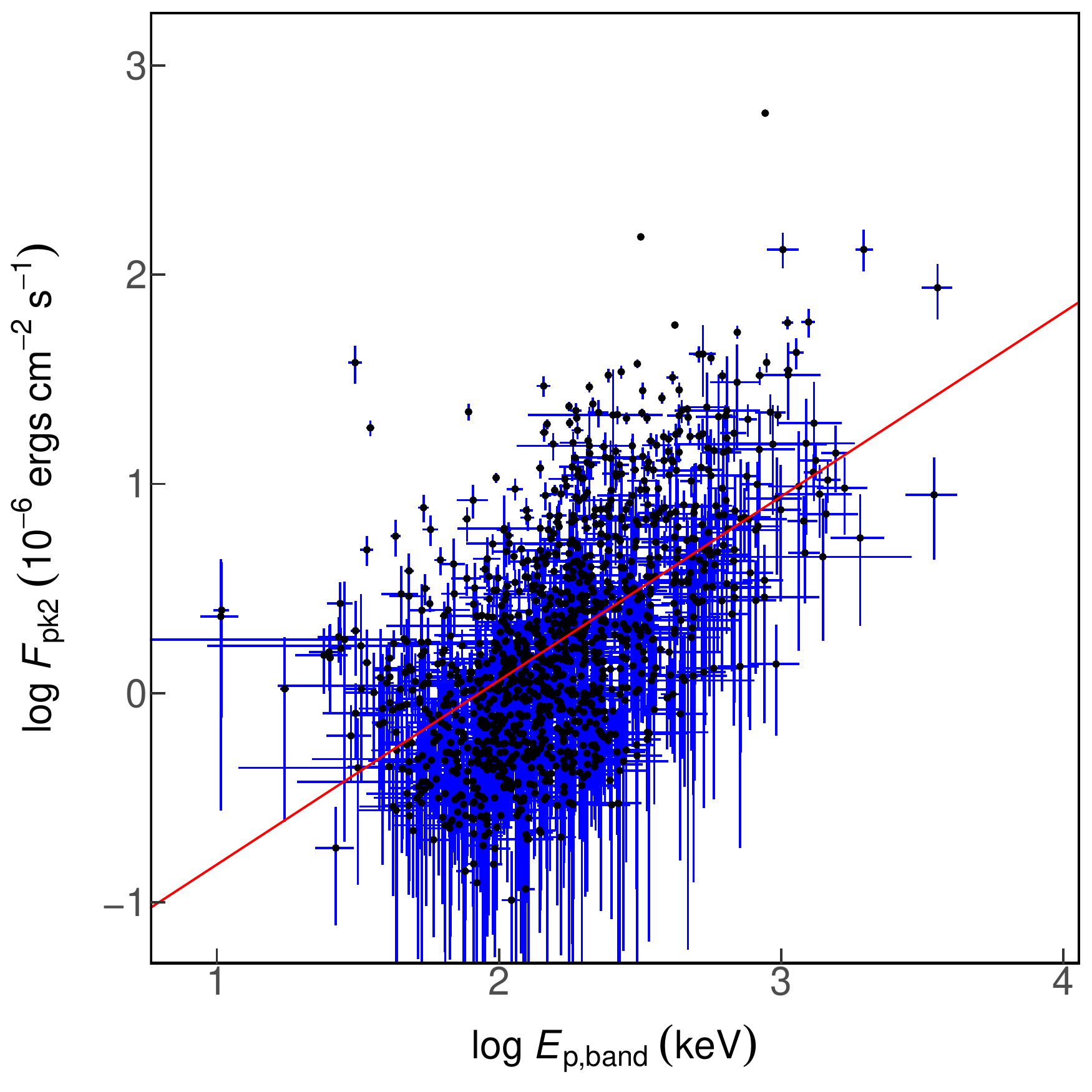}

\caption{
Scatter plot for $\log E_{\rm p,band}$ and $\log F_{\rm pk2}$.
The red line is our fit result. The formula of the red line is $\log E_{\rm p,band} = (0.28 \pm 0.012) \times \log F_{\rm pk2} + (2.2 \pm 0.0049)$. The description of every parameter is in Section \ref{sec:sample}.
}
\label{fig:epndfpk2}
\end{figure*}

The correlation between $\log E_{\rm p,band}$ and $\log F_{\rm pk2}$ is:
\begin{equation} \label{eq:epndfpk2}
\log E_{\rm p,band} = (0.28 \pm 0.012) \times \log F_{\rm pk2} + (2.2 \pm 0.0049),
\end{equation}
where $F_{\rm pk2}$ is peak energy flux of 64 $\rm ms$ time bin in rest-frame 1-$10^{4}$ $\rm keV$ energy band and in unit of $\rm 10^{\rm -6} ~ ergs ~ cm^{\rm -2} ~ s^{\rm -1}$. $E_{\rm p,band}$ is in unit of $\rm keV$. The adjusted $R^{2}$ is 0.33. The Pearson coefficient is $0.5 \pm 0.015$ with p-value $1.1 \times 10^{-73}$. The Spearman coefficient is $0.52 \pm 0.012$ with p-value $6.7 \times 10^{-81}$. The Kendall $\tau$ coefficient is $0.36 \pm 0.009$ with p-value $1.5 \times 10^{-75}$. The correlation ratio is $0.89 \pm 0.0024$. The cosine similarity is $0.4 \pm 0.015$. The scatter plot is in Figure \ref{fig:epndfpk2}. The GRB sample number is 1168. One can see from these two figures, the data are still widely distributed. What's more, as the flux is the energy of the photons multiplied by the number of the photons, the flux contains the information from the spectrum. Therefore, these two correlations does not reveal more information.

\subsection{Correlations about Lorentz factor $\Gamma_{0}$} \label{sec:gamma0}

\begin{figure*}[!b]
\centering
\includegraphics[width=0.45\textwidth]{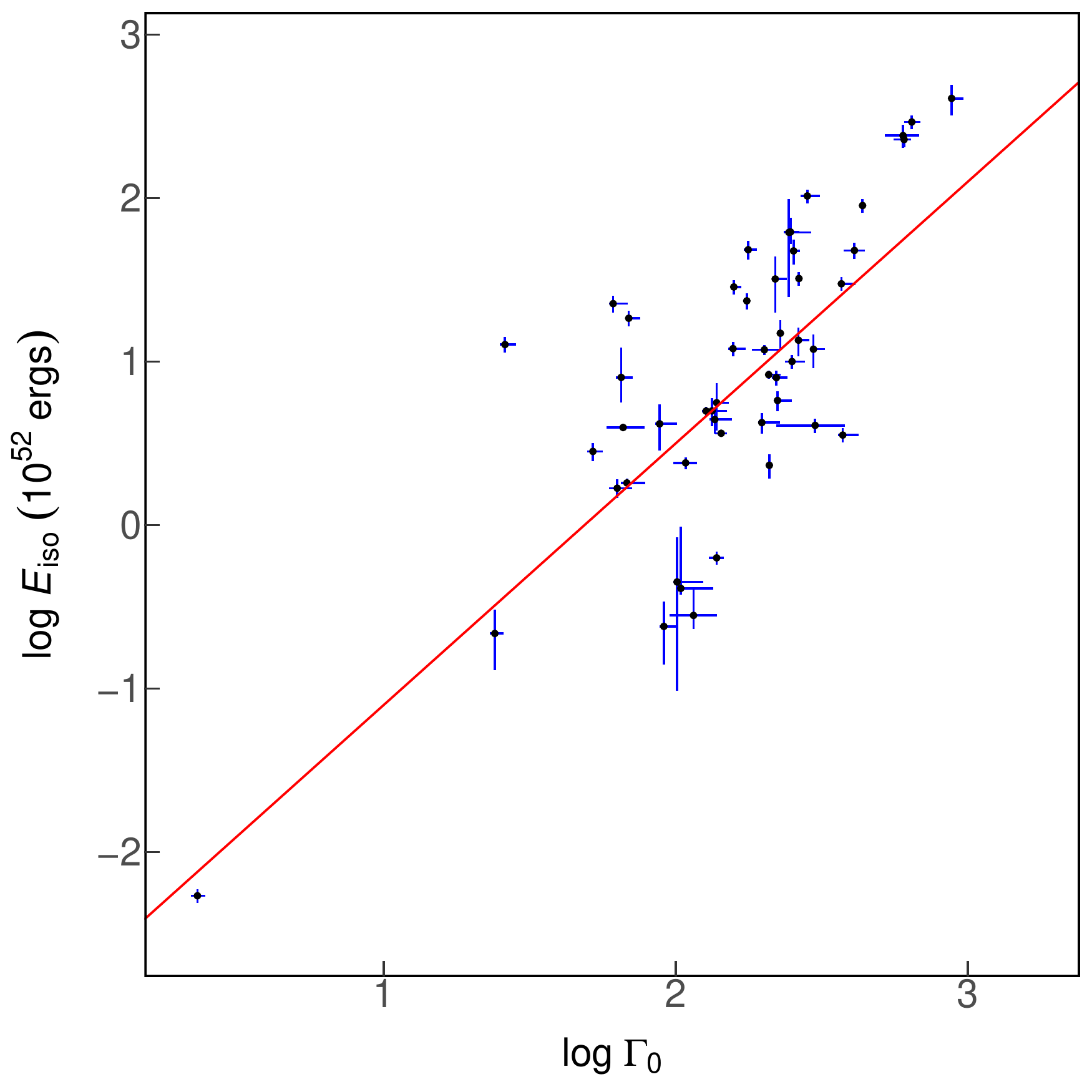}

\caption{
Scatter plot for $\log \Gamma_{0}$ and $\log E_{\rm iso}$.
The red line is our fit result. The formula of the red line is $\log \Gamma_{0} = (0.35 \pm 0.014) \times \log E_{\rm iso} + (1.9 \pm 0.018)$. The description of every parameter is in Section \ref{sec:sample}. The outlier is GRB 060218A.
}
\label{fig:gaa0eiso}
\end{figure*}

The correlation between $\log \Gamma_{0}$ and $\log E_{\rm iso}$ is:
\begin{equation} \label{eq:gaa0eiso}
\log \Gamma_{0} = (0.35 \pm 0.014) \times \log E_{\rm iso} + (1.9 \pm 0.018),
\end{equation}
where $E_{\rm iso}$ is in unit of $\rm 10^{\rm 52} ~ ergs$ and in rest-frame 1-$10^{4}$ $\rm keV$ energy band. The adjusted $R^{2}$ is 0.58. The Pearson coefficient is $0.75 \pm 0.016$ with p-value $1.9 \times 10^{-10}$. The Spearman coefficient is $0.68 \pm 0.02$ with p-value $4.4 \times 10^{-8}$. The Kendall $\tau$ coefficient is $0.52 \pm 0.019$ with p-value $9.5 \times 10^{-8}$. The correlation ratio is $0.67 \pm 0.0065$. The cosine similarity is $0.79 \pm 0.013$. The scatter plot is in Figure \ref{fig:gaa0eiso}. The GRB sample number is 51. Previous studies gave $ \log{\Gamma_{\rm 0}}=(0.269\pm0.002)\log{E_{\rm iso,52}} + (2.291\pm0.002)$ with 19 samples \citep{Liang2010} and $ \log{\Gamma_{\rm 0}}=(0.29\pm0.002)\log{E_{\rm iso,52}} + (1.96\pm0.002)$ with 38 samples \citep{Lv2012}, which are shown here for comparison. \footnote{Note that the notation $Q_x \equiv Q/10^x$ is used in the whole paper, and the unit is in cgs unit by default if they are not defined explicitly.} The positive correlation shows that the stronger bursts is likely to produce faster ejecta, which might reveals the acceleration mechanism of the central engine.

\begin{figure*}[!b]
\centering
\includegraphics[width=0.45\textwidth]{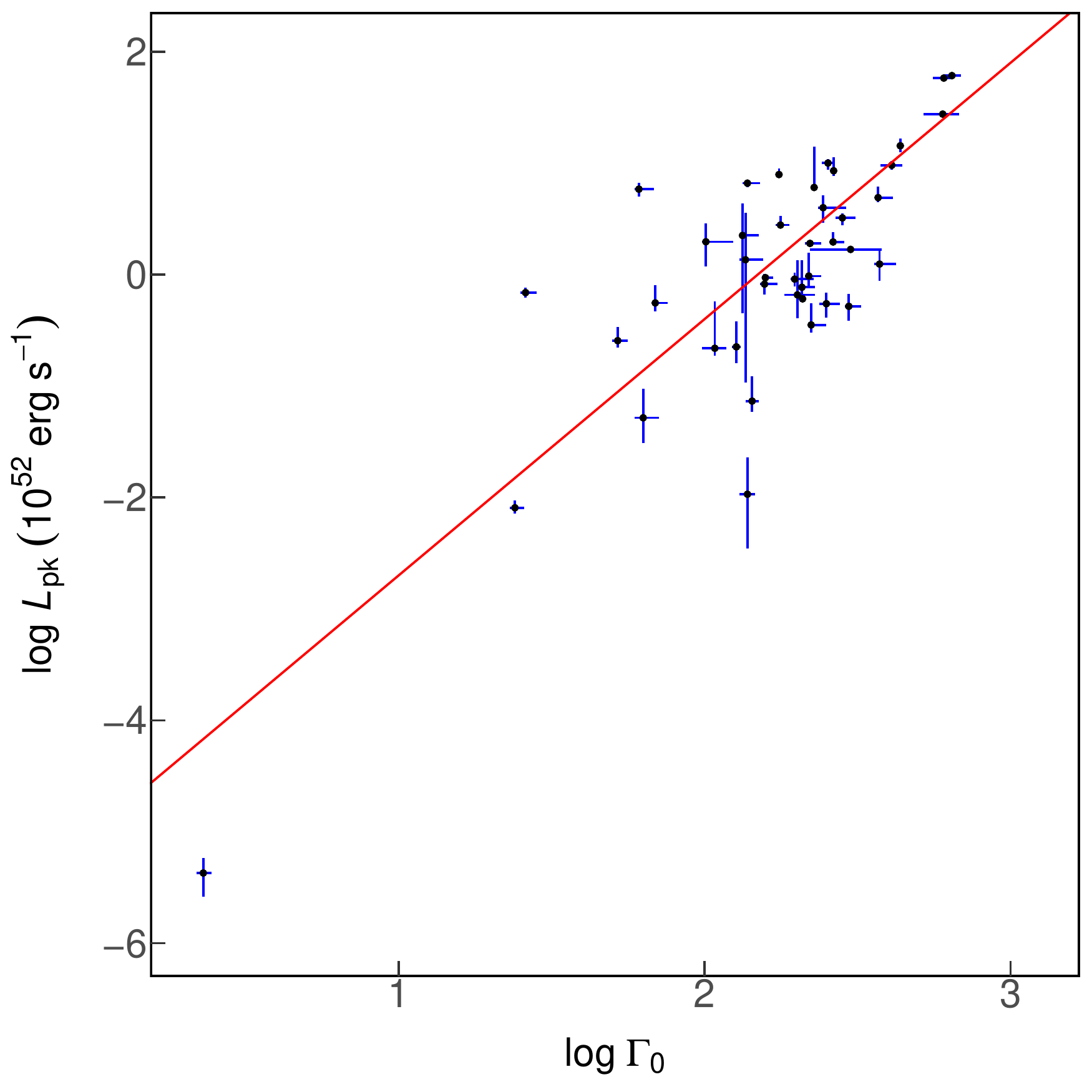}

\caption{
Scatter plot for $\log \Gamma_{0}$ and $\log L_{\rm pk}$.
The red line is our fit result. The formula of the red line is $\log \Gamma_{0} = (0.29 \pm 0.011) \times \log L_{\rm pk} + (2.2 \pm 0.01)$. The description of every parameter is in Section \ref{sec:sample}. The outlier is GRB 060218A with lowest $L_{\rm pk}$ and $\Gamma_{0}$
}
\label{fig:gaa0lpk}
\end{figure*}

The correlation between $\log \Gamma_{0}$ and $\log L_{\rm pk}$ is:
\begin{equation} \label{eq:gaa0lpk}
\log \Gamma_{0} = (0.29 \pm 0.011) \times \log L_{\rm pk} + (2.2 \pm 0.01),
\end{equation}
where $L_{\rm pk}$ is in unit of $\rm 10^{\rm 52} ~ erg ~ s^{\rm -1}$ and in 1-$10^{4}$ $\rm keV$ energy band. The adjusted $R^{2}$ is 0.66. The Pearson coefficient is $0.81 \pm 0.015$ with p-value $1.1 \times 10^{-10}$. The Spearman coefficient is $0.63 \pm 0.031$ with p-value $7 \times 10^{-6}$. The Kendall $\tau$ coefficient is $0.47 \pm 0.027$ with p-value $1 \times 10^{-5}$. The correlation ratio is $0.77 \pm 0.0061$. The cosine similarity is $0.16 \pm 0.026$. The scatter plot is in Figure \ref{fig:gaa0lpk}. The GRB sample number is 42. A previous study gave $ \log{\Gamma_{\rm 0}} = (0.28\pm0.002)\log{L_{\rm iso,52}} + (2.39\pm0.003)$ with 38 samples \citep{Lv2012}. This correlation is related to eq. (\ref{eq:gaa0eiso}).

\begin{figure*}[!b]
\centering
\includegraphics[width=0.45\textwidth]{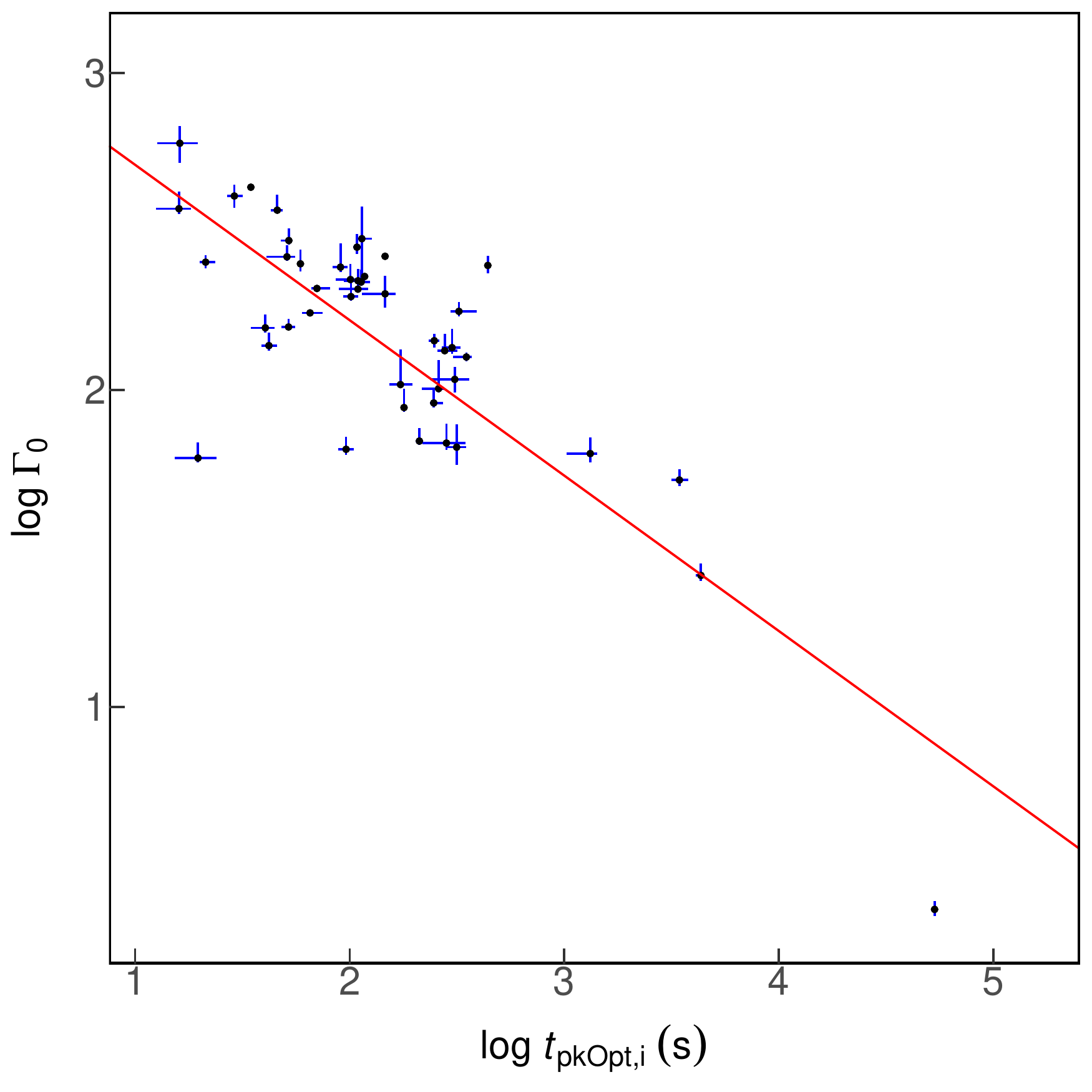}

\caption{
Scatter plot for $\log \Gamma_{0}$ and $\log t_{\rm pkOpt,i}$.
The red line is our fit result. The formula of the red line is $\log \Gamma_{0} = (-0.49 \pm 0.011) \times \log t_{\rm pkOpt,i} + (3.2 \pm 0.026)$. The description of every parameter is in Section \ref{sec:sample}. The outlier is GRB 060218A.
}
\label{fig:gaa0tpti}
\end{figure*}

The correlation between $\log \Gamma_{0}$ and $\log t_{\rm pkOpt,i}$ is:
\begin{equation} \label{eq:gaa0tpti}
\log \Gamma_{0} = (-0.49 \pm 0.011) \times \log t_{\rm pkOpt,i} + (3.2 \pm 0.026),
\end{equation}
where $t_{\rm pkOpt,i}$ is in unit of $\rm s$. The adjusted $R^{2}$ is 0.66. The Pearson coefficient is $-0.81 \pm 0.011$ with p-value $1.5 \times 10^{-11}$. The Spearman coefficient is $-0.64 \pm 0.019$ with p-value $2.6 \times 10^{-6}$. The Kendall $\tau$ coefficient is $-0.48 \pm 0.019$ with p-value $3.5 \times 10^{-6}$. The correlation ratio is $0.019 \pm 0.0085$. The cosine similarity is $0.9 \pm 0.0016$. The scatter plot is in Figure \ref{fig:gaa0tpti}. The GRB sample number is 45. A previous study gave $\log \Gamma_{0} = (-0.63 \pm 0.04) \times \log t_{\rm pkOpt,i} + (3.69 \pm 0.09)$ with 19 samples \citep{Liang2010} (Notice two of them are the peak times in X-rays). The anti-correlation mainly reveals how the GRB jets decelerate by the circum burst environment. Noticing the deceleration time is also related to the profile of the density of the environment, the correlation here may not be very tight.

\begin{figure*}[!b]
\centering
\includegraphics[width=0.45\textwidth]{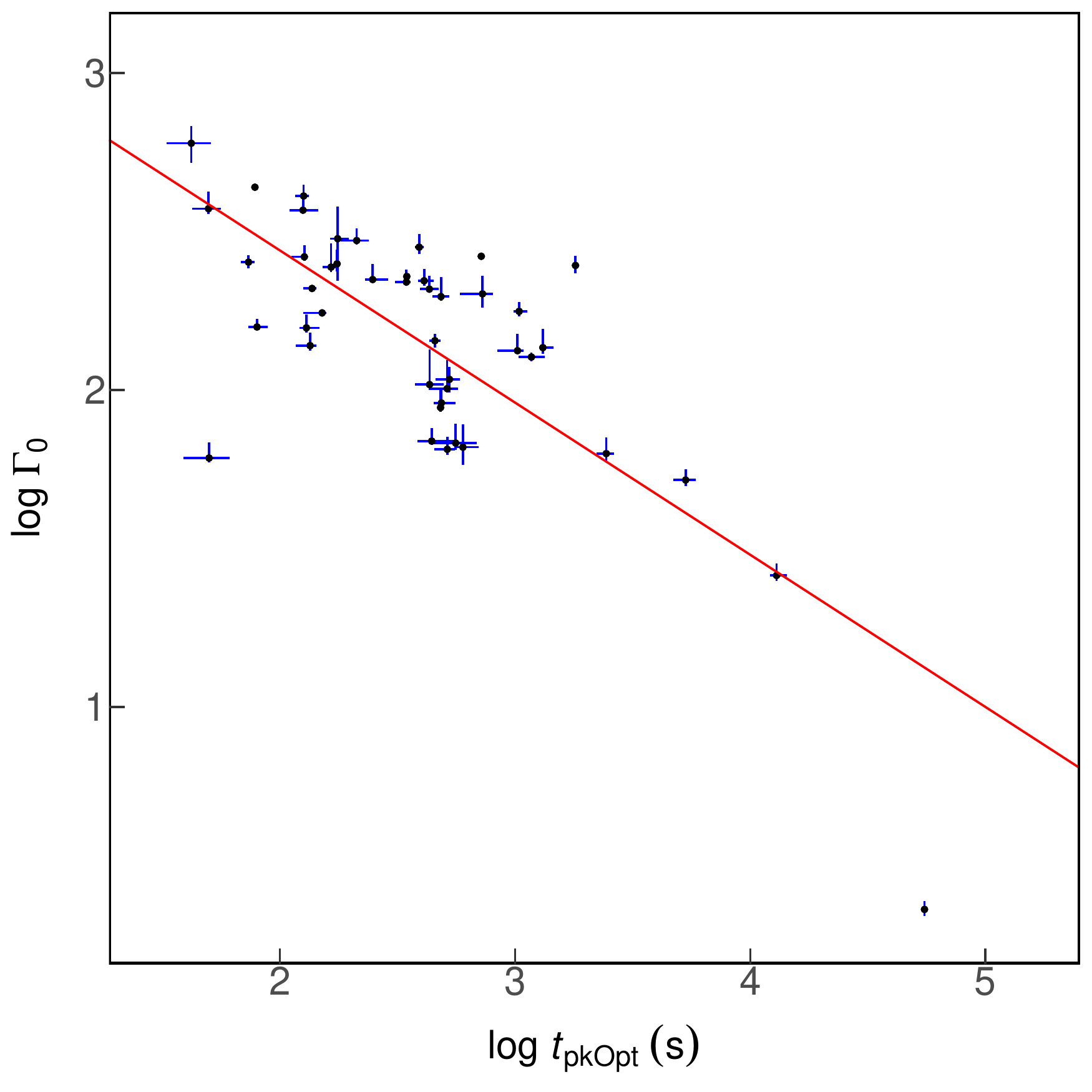}

\caption{
Scatter plot for $\log \Gamma_{0}$ and $\log t_{\rm pkOpt}$.
The red line is our fit result. The formula of the red line is $\log \Gamma_{0} = (-0.48 \pm 0.012) \times \log t_{\rm pkOpt} + (3.4 \pm 0.033)$. The description of every parameter is in Section \ref{sec:sample}. The outlier is GRB 060218A.
}
\label{fig:gaa0tppt}
\end{figure*}

The correlation between $\log \Gamma_{0}$ and $\log t_{\rm pkOpt}$ is:
\begin{equation} \label{eq:gaa0tppt}
\log \Gamma_{0} = (-0.48 \pm 0.012) \times \log t_{\rm pkOpt} + (3.4 \pm 0.033),
\end{equation}
where $t_{\rm pkOpt}$ is in unit of $\rm s$. The adjusted $R^{2}$ is 0.56. The Pearson coefficient is $-0.75 \pm 0.012$ with p-value $3.4 \times 10^{-9}$. The Spearman coefficient is $-0.59 \pm 0.022$ with p-value $2.4 \times 10^{-5}$. The Kendall $\tau$ coefficient is $-0.44 \pm 0.021$ with p-value $1.8 \times 10^{-5}$. The correlation ratio is $0.38 \pm 0.0084$. The cosine similarity is $0.93 \pm 0.0012$. The scatter plot is in Figure \ref{fig:gaa0tppt}. The GRB sample number is 45. This correlation is clearly worse than the previous one shown in equation \ref{eq:gaa0tpti}, which has the redshift correction. It implies the correlation between $\Gamma_{0}$ and $t_{\rm pkOpt,i}$ is intrinsic, and the time dilation due to redshift goes against the correlation.

\begin{figure*}[!b]
\centering
\includegraphics[width=0.45\textwidth]{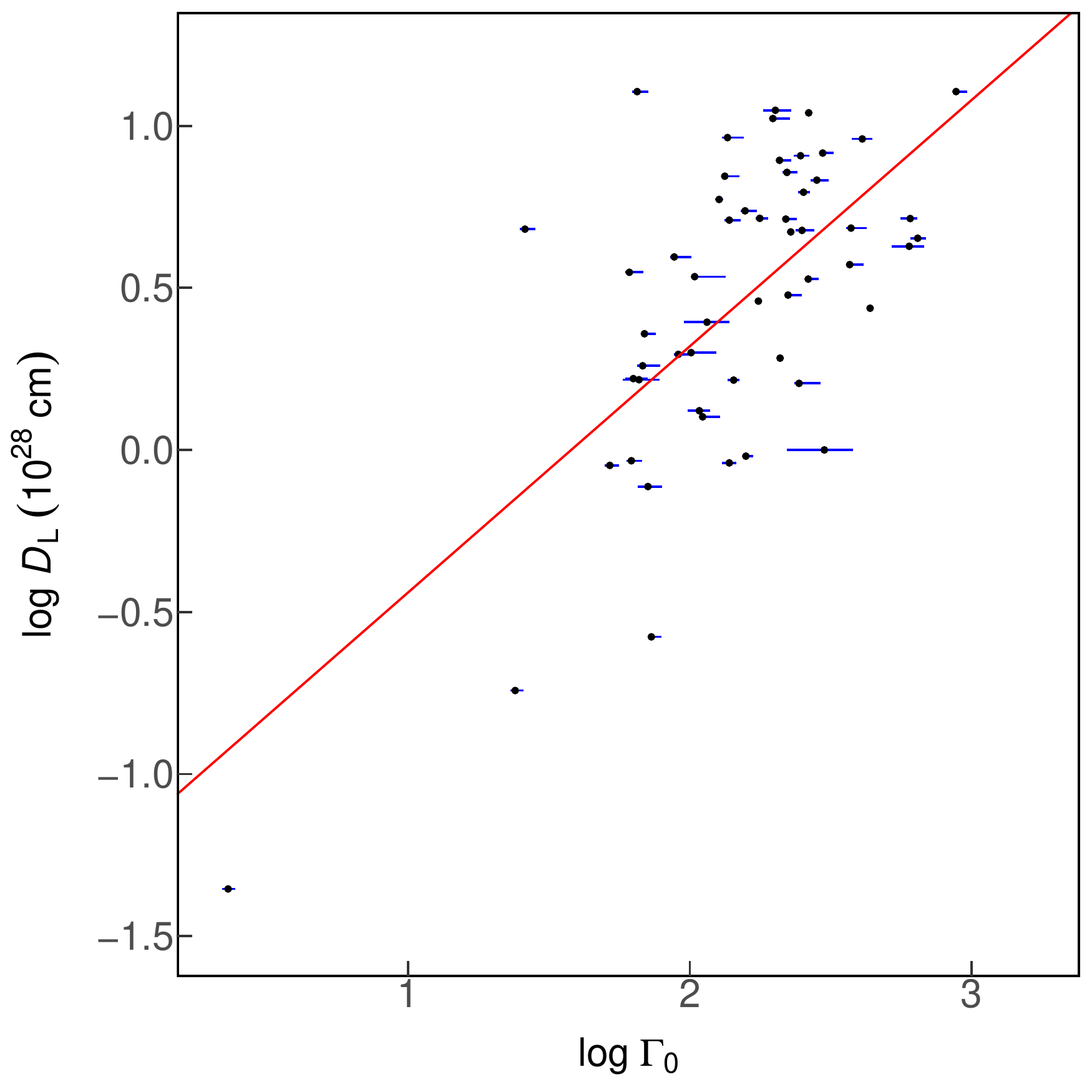}

\caption{
Scatter plot for $\log \Gamma_{0}$ and $\log D_{\rm L}$.
The red line is our fit result. The formula of the red line is $\log \Gamma_{0} = (0.58 \pm 0.0095) \times \log D_{\rm L} + (1.9 \pm 0.0074)$. The description of every parameter is in Section \ref{sec:sample}. The outlier is GRB 060218A.
}
\label{fig:gaa0dl}
\end{figure*}

The correlation between $\log \Gamma_{0}$ and $\log D_{\rm L}$ is:
\begin{equation} \label{eq:gaa0dl}
\log \Gamma_{0} = (0.58 \pm 0.0095) \times \log D_{\rm L} + (1.9 \pm 0.0074),
\end{equation}
where $D_{\rm L}$ is in unit of $\rm 10^{\rm 28} ~ cm$. The adjusted $R^{2}$ is 0.43. The Pearson coefficient is $0.67 \pm 0.01$ with p-value $2.6 \times 10^{-8}$. The Spearman coefficient is $0.5 \pm 0.019$ with p-value $9.6 \times 10^{-5}$. The Kendall $\tau$ coefficient is $0.35 \pm 0.014$ with p-value $1.5 \times 10^{-4}$. The correlation ratio is $0.89 \pm 0.00096$. The cosine similarity is $0.78 \pm 0.0014$. The scatter plot is in Figure \ref{fig:gaa0dl}. The GRB sample number is 55. The dependence on the distance indicates observational bias, that the low $\Gamma_{0}$ GRBs might not be able to be observed at longer distance. The other possibility is that the GRBs exploded earlier do have higher $\Gamma_{0}$.

\begin{figure*}[!b]
\centering
\includegraphics[width=0.45\textwidth]{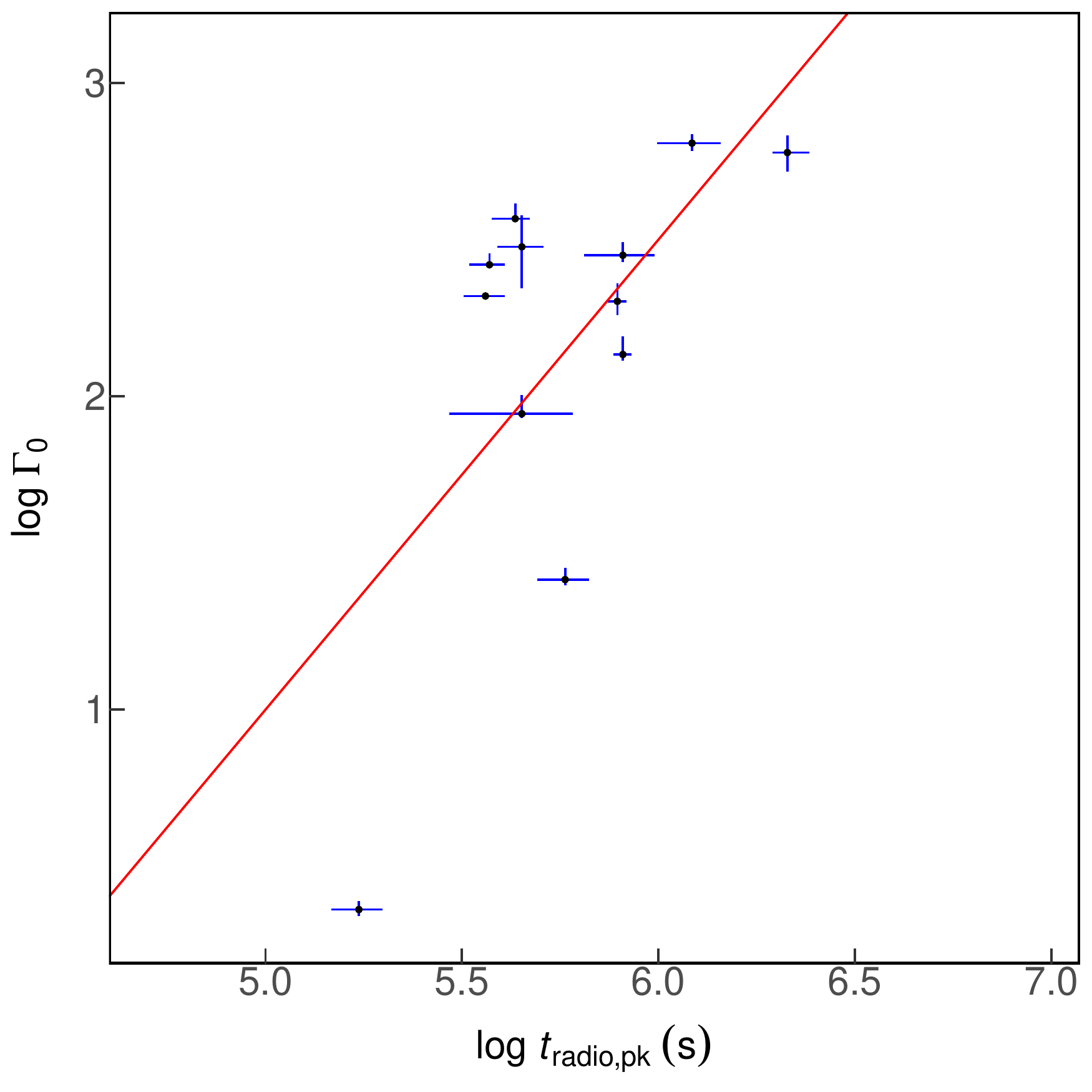}

\caption{
Scatter plot for $\log \Gamma_{0}$ and $\log t_{\rm radio,pk}$.
The red line is our fit result. The formula of the red line is $\log \Gamma_{0} = (1.5 \pm 0.14) \times \log t_{\rm radio,pk} + (-6.5 \pm 0.81)$. The description of every parameter is in Section \ref{sec:sample}.
}
\label{fig:gaa0trpk}
\end{figure*}

The correlation between $\log \Gamma_{0}$ and $\log t_{\rm radio,pk}$ is:
\begin{equation} \label{eq:gaa0trpk}
\log \Gamma_{0} = (1.5 \pm 0.14) \times \log t_{\rm radio,pk} + (-6.5 \pm 0.81),
\end{equation}
where $t_{\rm radio,pk}$ is in unit of $\rm s$. The adjusted $R^{2}$ is 0.39. The Pearson coefficient is $0.64 \pm 0.05$ with p-value $2.4 \times 10^{-2}$. The Spearman coefficient is $0.49 \pm 0.086$ with p-value 0.11. The Kendall $\tau$ coefficient is $0.37 \pm 0.081$ with p-value 0.1. The correlation ratio is $0.96 \pm 0.0013$. The cosine similarity is $0.97 \pm 0.0013$. The scatter plot is in Figure \ref{fig:gaa0trpk}. The GRB sample number is 12. Interestingly, not like eq. (\ref{eq:gaa0tpti}), $\Gamma_{0}$ here is positively correlated with $t_{\rm radio,pk}$. As $t_{\rm radio,pk}$ does not represent the deceleration time, higher $\Gamma_{0}$  may corresponds to higher kinetic energy, and the radio emission lasts longer.

\begin{figure*}[!b]
\centering
\includegraphics[width=0.45\textwidth]{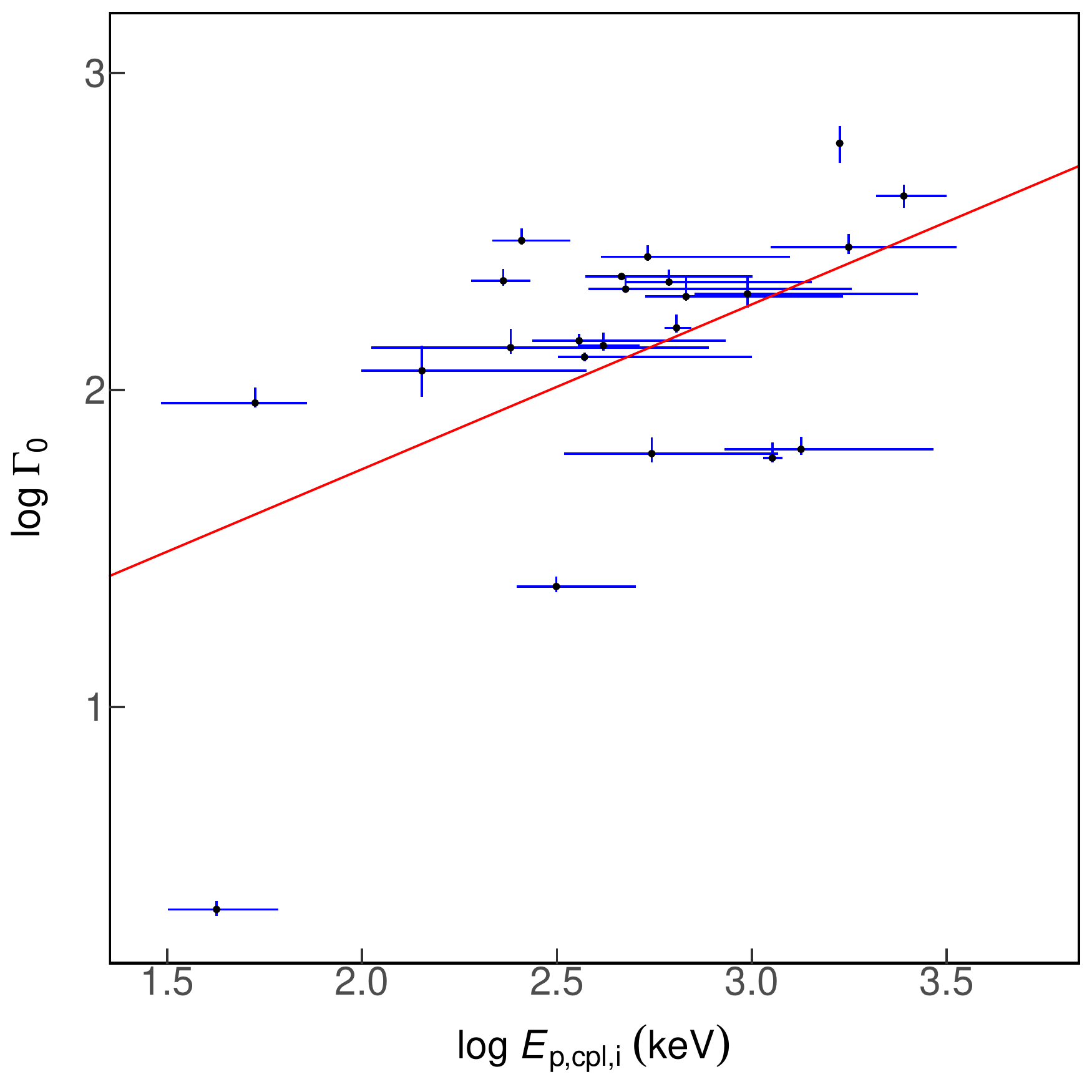}

\caption{
Scatter plot for $\log \Gamma_{0}$ and $\log E_{\rm p,cpl,i}$.
The red line is our fit result. The formula of the red line is $\log \Gamma_{0} = (0.52 \pm 0.084) \times \log E_{\rm p,cpl,i} + (0.71 \pm 0.22)$. The description of every parameter is in Section \ref{sec:sample}.
}
\label{fig:gaa0epli}
\end{figure*}

The correlation between $\log \Gamma_{0}$ and $\log E_{\rm p,cpl,i}$ is:
\begin{equation} \label{eq:gaa0epli}
\log \Gamma_{0} = (0.52 \pm 0.084) \times \log E_{\rm p,cpl,i} + (0.71 \pm 0.22),
\end{equation}
where $E_{\rm p,cpl,i}$ is in unit of $\rm keV$. The adjusted $R^{2}$ is 0.31. The Pearson coefficient is $0.53 \pm 0.061$ with p-value $9.9 \times 10^{-3}$. The Spearman coefficient is $0.38 \pm 0.087$ with p-value $7.7 \times 10^{-2}$. The Kendall $\tau$ coefficient is $0.27 \pm 0.07$ with p-value $6.8 \times 10^{-2}$. The correlation ratio is $0.52 \pm 0.031$. The cosine similarity is $0.98 \pm 0.0028$. The scatter plot is in Figure \ref{fig:gaa0epli}. The GRB sample number is 23. As $\Gamma_{0}$ is related to $E_{\rm iso}$ (as shown in eq. (\ref{eq:gaa0eiso})), and $E_{\rm iso}$ is related to $E_{\rm p}$ (as shown in eq. (\ref{eq:epndeiso})), it is not surprising that $\Gamma_{0}$ is related to $E_{\rm p}$.

\begin{figure*}[!b]
\centering
\includegraphics[width=0.45\textwidth]{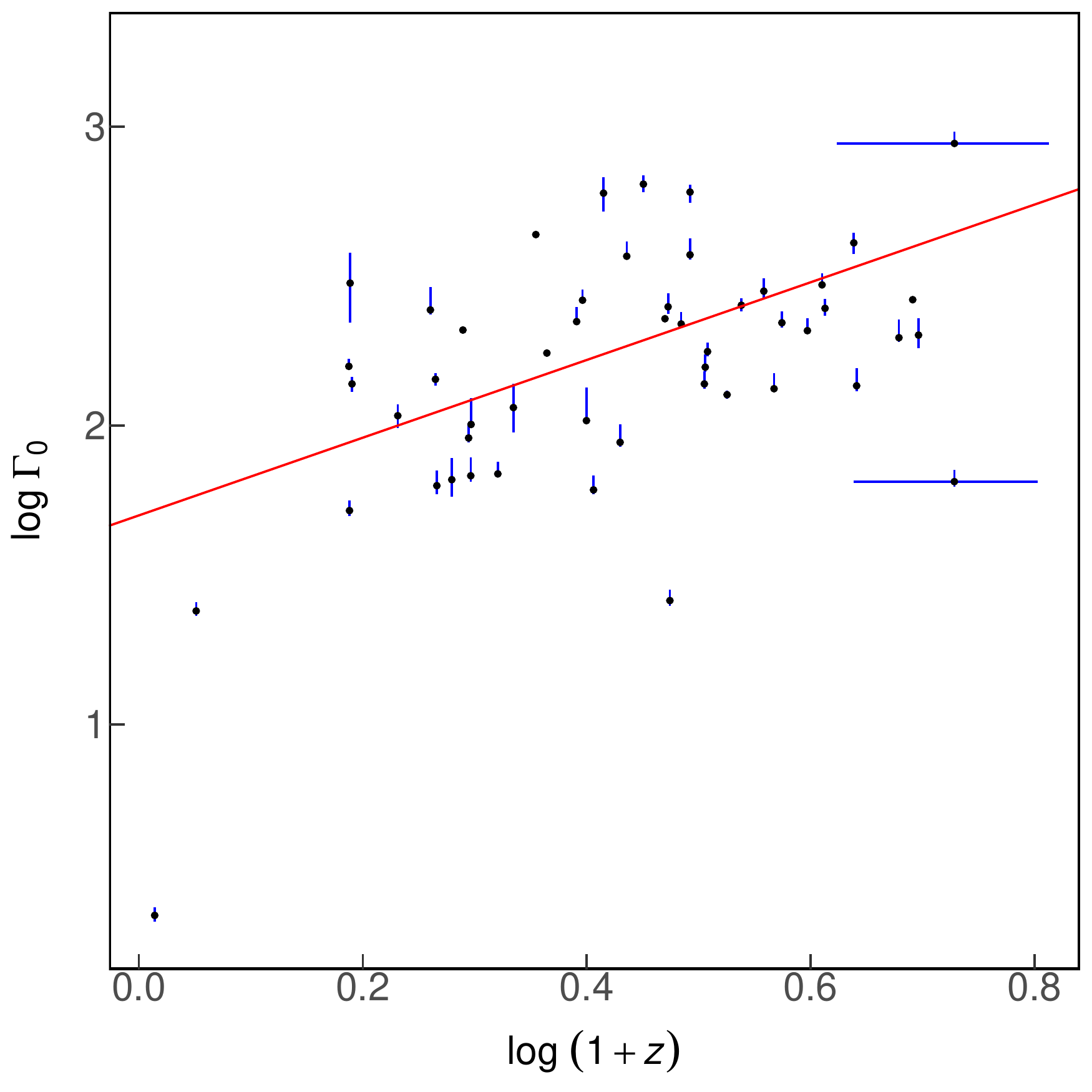}

\caption{
Scatter plot for $\log  \Gamma_{0}$ and $\log (1+z)$.
The red line is our fit result. The formula of the red line is $\log  \Gamma_{0} = (1.3 \pm 0.071) \times \log (1+z) + (1.7 \pm 0.03)$. The description of every parameter is in Section \ref{sec:sample}.
 The outlier is GRB 060218A.}
\label{fig:gaa0z}
\end{figure*}

The correlation between $\log  \Gamma_{0}$ and $\log (1+z)$ is:
\begin{equation} \label{eq:gaa0z}
\log  \Gamma_{0} = (1.3 \pm 0.071) \times \log (1+z) + (1.7 \pm 0.03),
\end{equation}
the adjusted $R^{2}$ is 0.25. The Pearson coefficient is $0.52 \pm 0.024$ with p-value $9.8 \times 10^{-5}$. The Spearman coefficient is $0.42 \pm 0.026$ with p-value $2.3 \times 10^{-3}$. The Kendall $\tau$ coefficient is $0.29 \pm 0.018$ with p-value $2.9 \times 10^{-3}$. The correlation ratio is $0.94 \pm 0.0011$. The cosine similarity is $0.95 \pm 0.0026$. The scatter plot is in Figure \ref{fig:gaa0z}. The GRB sample number is 51. This might be a selection effect, that further GRBs can be observed only if they are stronger and with faster jets. Otherwise, this may indicate the properties of GRBs evolve with cosmic time. This relation is intrinsically the same as shown in equation \ref{eq:gaa0dl}.

\begin{figure*}[!b]
\centering
\includegraphics[width=0.45\textwidth]{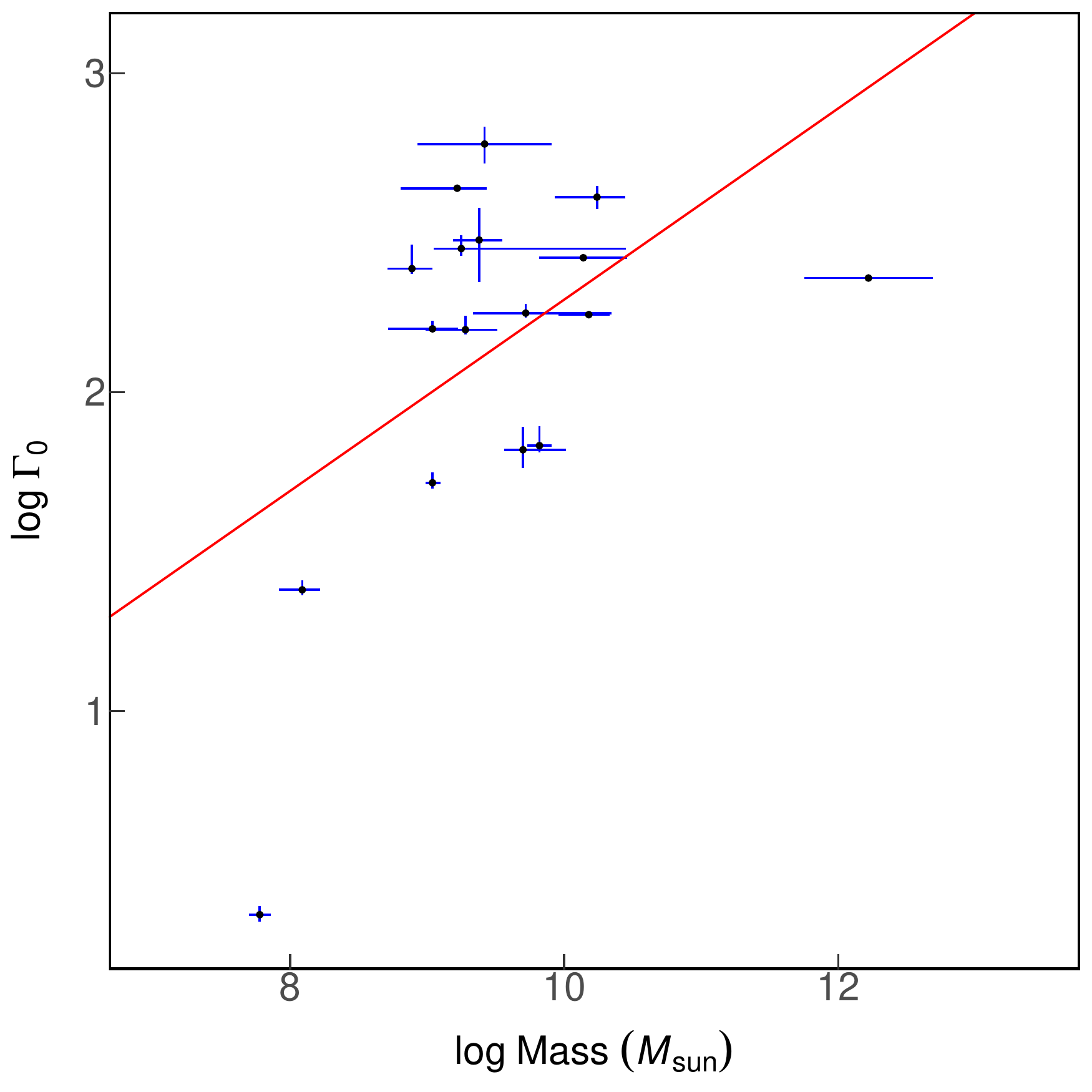}

\caption{
Scatter plot for $\log  \Gamma_{0}$ and $\log Mass$.
The red line is our fit result. The formula of the red line is $\log  \Gamma_{0} = (0.3 \pm 0.052) \times \log Mass + (-0.71 \pm 0.49)$. The description of every parameter is in Section \ref{sec:sample}. The outlier is GRB 060218A.
}
\label{fig:gaa0mass}
\end{figure*}

The correlation between $\log  \Gamma_{0}$ and $\log Mass$ is:
\begin{equation} \label{eq:gaa0mass}
\log  \Gamma_{0} = (0.3 \pm 0.052) \times \log Mass + (-0.71 \pm 0.49),
\end{equation}
where Mass is in unit of $M_{\bigodot}$. The adjusted $R^{2}$ is 0.25. The Pearson coefficient is $0.52 \pm 0.058$ with p-value $3.3 \times 10^{-2}$. The Spearman coefficient is $0.35 \pm 0.1$ with p-value 0.17. The Kendall $\tau$ coefficient is $0.26 \pm 0.078$ with p-value 0.14. The correlation ratio is $0.98 \pm 0.0029$. The cosine similarity is $0.98 \pm 0.0017$. The scatter plot is in Figure \ref{fig:gaa0mass}. The GRB sample number is 17. This weak correlation may imply that the stronger GRBs are harbored in more massive host galaxies.


\subsection{Some correlations about $t_{\rm pkX}$}

$t_{\rm pkX}$ is the peak time in the X-ray light curve. It is often taken as the deceleration time of the ejecta decelerated by the circum-burst medium. It mainly reveals the information of the initial Lorentz factor and the number density of the circum-burst medium.

\begin{figure*}[!b]
\centering
\includegraphics[width=0.45\textwidth]{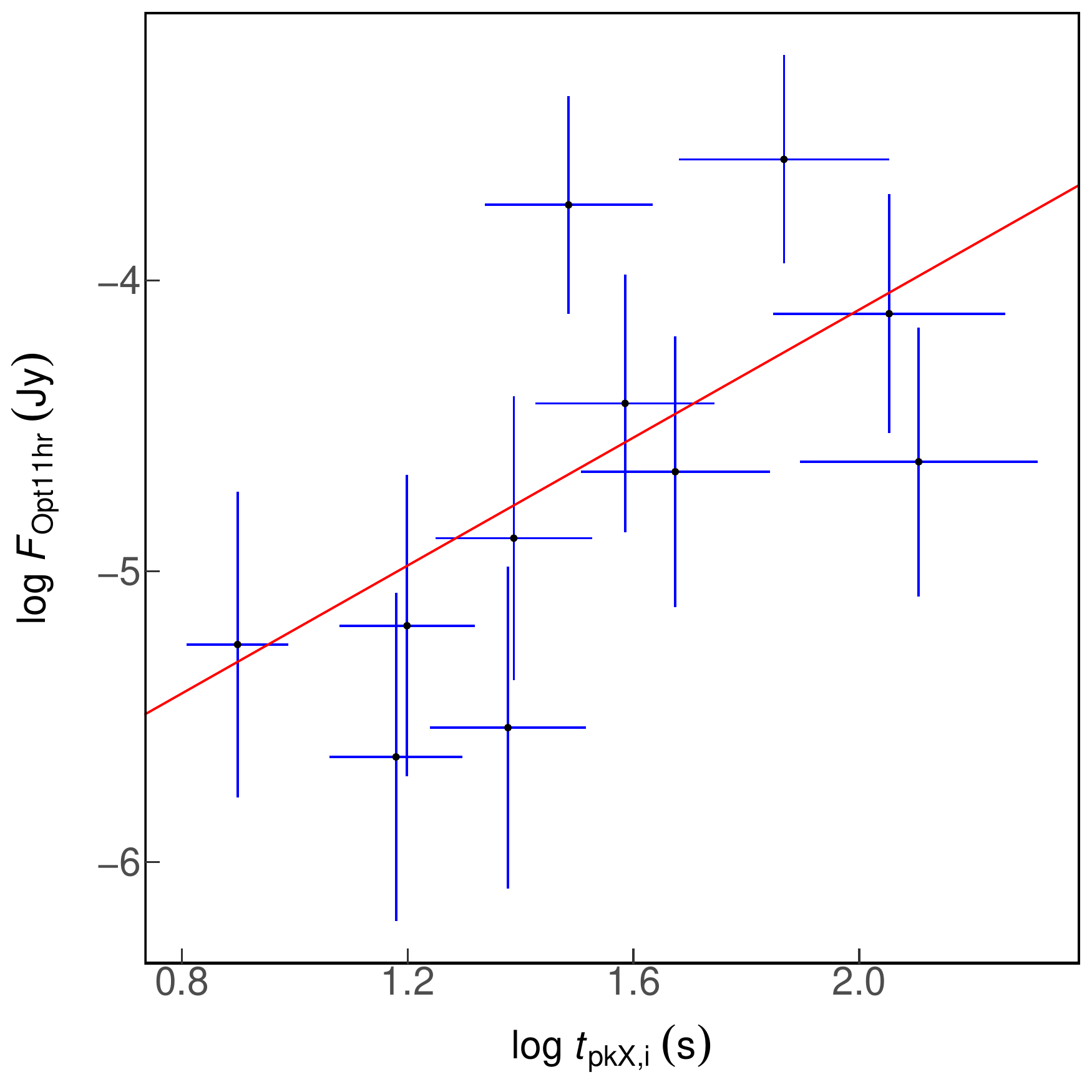}

\caption{
Scatter plot for $\log F_{\rm Opt11hr}$ and $\log t_{\rm pkX,i}$.
The red line is our fit result. The formula of the red line is $\log F_{\rm Opt11hr} = (1.1 \pm 0.44) \times \log t_{\rm pkX,i} + (-6.3 \pm 0.71)$. The description of every parameter is in Section \ref{sec:sample}.
}
\label{fig:fohrtpkxi}
\end{figure*}

The correlation between $\log F_{\rm Opt11hr}$ and $\log t_{\rm pkX,i}$ is:
\begin{equation} \label{eq:fohrtpkxi}
\log F_{\rm Opt11hr} = (1.1 \pm 0.44) \times \log t_{\rm pkX,i} + (-6.3 \pm 0.71),
\end{equation}
where $F_{\rm Opt11hr}$ is in unit of $\rm Jy$. $t_{\rm pkX,i}$ is in unit of $\rm s$. The adjusted $R^{2}$ is 0.36. The Pearson coefficient is $0.5 \pm 0.17$ with p-value 0.11. The Spearman coefficient is $0.54 \pm 0.18$ with p-value $8.9 \times 10^{-2}$. The Kendall $\tau$ coefficient is $0.39 \pm 0.15$ with p-value 0.1. The correlation ratio is $0.98 \pm 0.0049$. The cosine similarity is $-0.94 \pm 0.014$. The scatter plot is in Figure \ref{fig:fohrtpkxi}. The GRB sample number is 11. As the afterglow fades after the deceleration time, and the deceleration time is generally smaller than 11 hours, as shown in  Figure \ref{fig:fohrtpkxi}, the optical afterglow might be brighter if it is more closer to the deceleration time. Therefore, $F_{\rm Opt11hr}$ is roughly proportional to the $t_{\rm pkX,i}$.

\begin{figure*}[!b]
\centering
\includegraphics[width=0.45\textwidth]{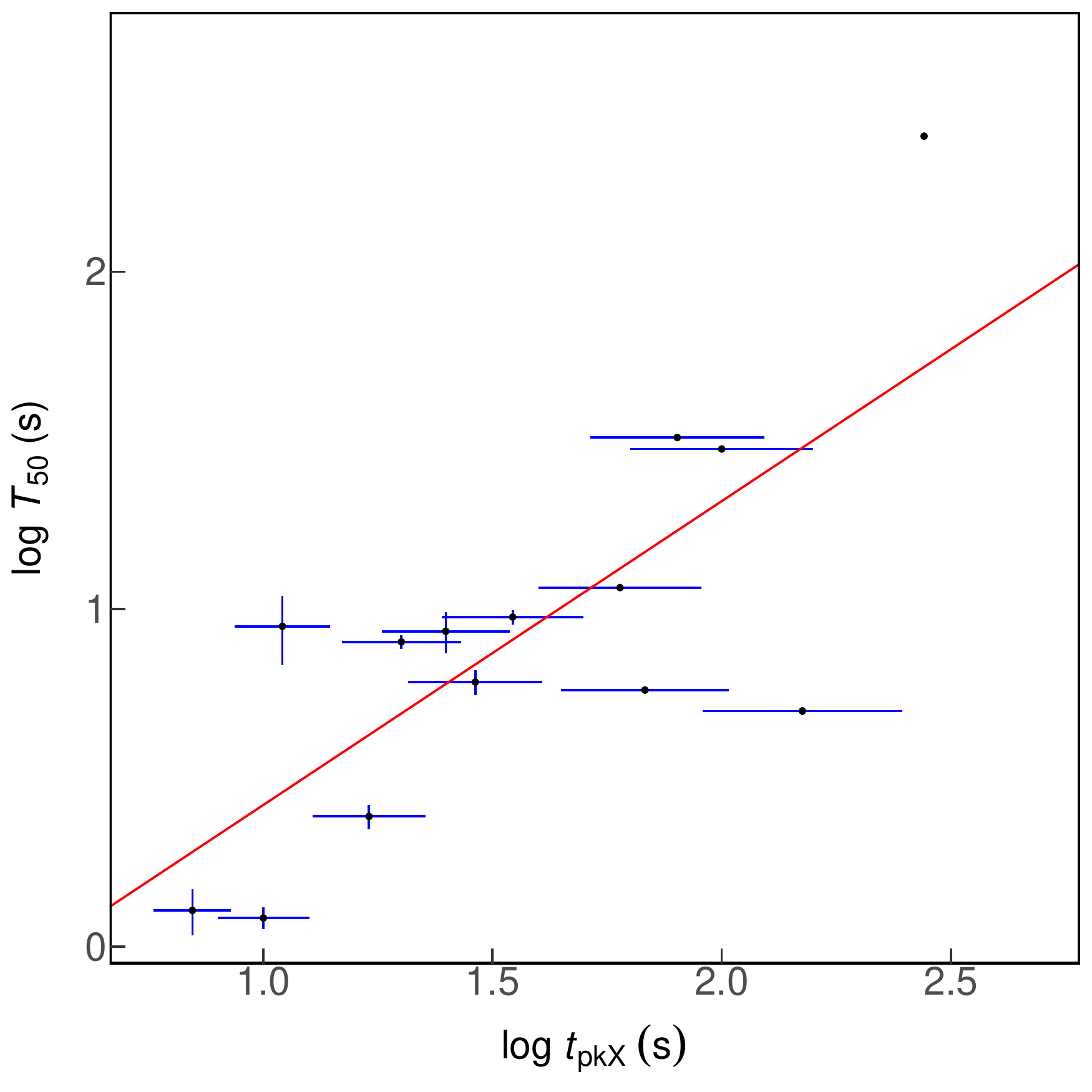}

\caption{
Scatter plot for $\log T_{\rm 50}$ and $\log t_{\rm pkX}$.
The red line is our fit result. The formula of the red line is $\log T_{\rm 50} = (0.9 \pm 0.12) \times \log t_{\rm pkX} + (-0.48 \pm 0.17)$. The description of every parameter is in Section \ref{sec:sample}.
}
\label{fig:t50tpkx}
\end{figure*}

The correlation between $\log T_{\rm 50}$ and $\log t_{\rm pkX}$ is:
\begin{equation} \label{eq:t50tpkx}
\log T_{\rm 50} = (0.9 \pm 0.12) \times \log t_{\rm pkX} + (-0.48 \pm 0.17),
\end{equation}
where $T_{\rm 50}$ is in unit of $\rm s$. $t_{\rm pkX}$ is in unit of $\rm s$. The adjusted $R^{2}$ is 0.57. The Pearson coefficient is $0.74 \pm 0.058$ with p-value $2.4 \times 10^{-3}$. The Spearman coefficient is $0.6 \pm 0.076$ with p-value $2.2 \times 10^{-2}$. The Kendall $\tau$ coefficient is $0.47 \pm 0.074$ with p-value $1.9 \times 10^{-2}$. The correlation ratio is $0.52 \pm 0.023$. The cosine similarity is $0.93 \pm 0.0085$. The scatter plot is in Figure \ref{fig:t50tpkx}. The GRB sample number is 14. The duration is positively correlated to the deceleration time, which may indicate that the luminosity of the GRBs are kept consistent, and longer time of duration sustains the ejecta decelerated at a later time. However, as can be seen from the figure, this relation far more than a tight correlation.

\subsection{Some correlations about $F_{\rm g}$}

The $\gamma-$ray fluence $F_{\rm g}$ is related to several other parameters of GRBs. However, as $F_{\rm g}$ is just an observational properties, which is highly affected by the distance comparing with the more intrinsic quantities like the luminosity and the total energy. Therefore, the correlations about $F_{\rm g}$ can not directly reveal the intrinsic properties. One would try to seek the correlations between the intrinsic properties. As the $F_{\rm g}$ is one of the easiest obtained quantities, we still list the correlations in the following. One may get clues from these correlations.

\begin{figure*}[!b]
\centering
\includegraphics[width=0.45\textwidth]{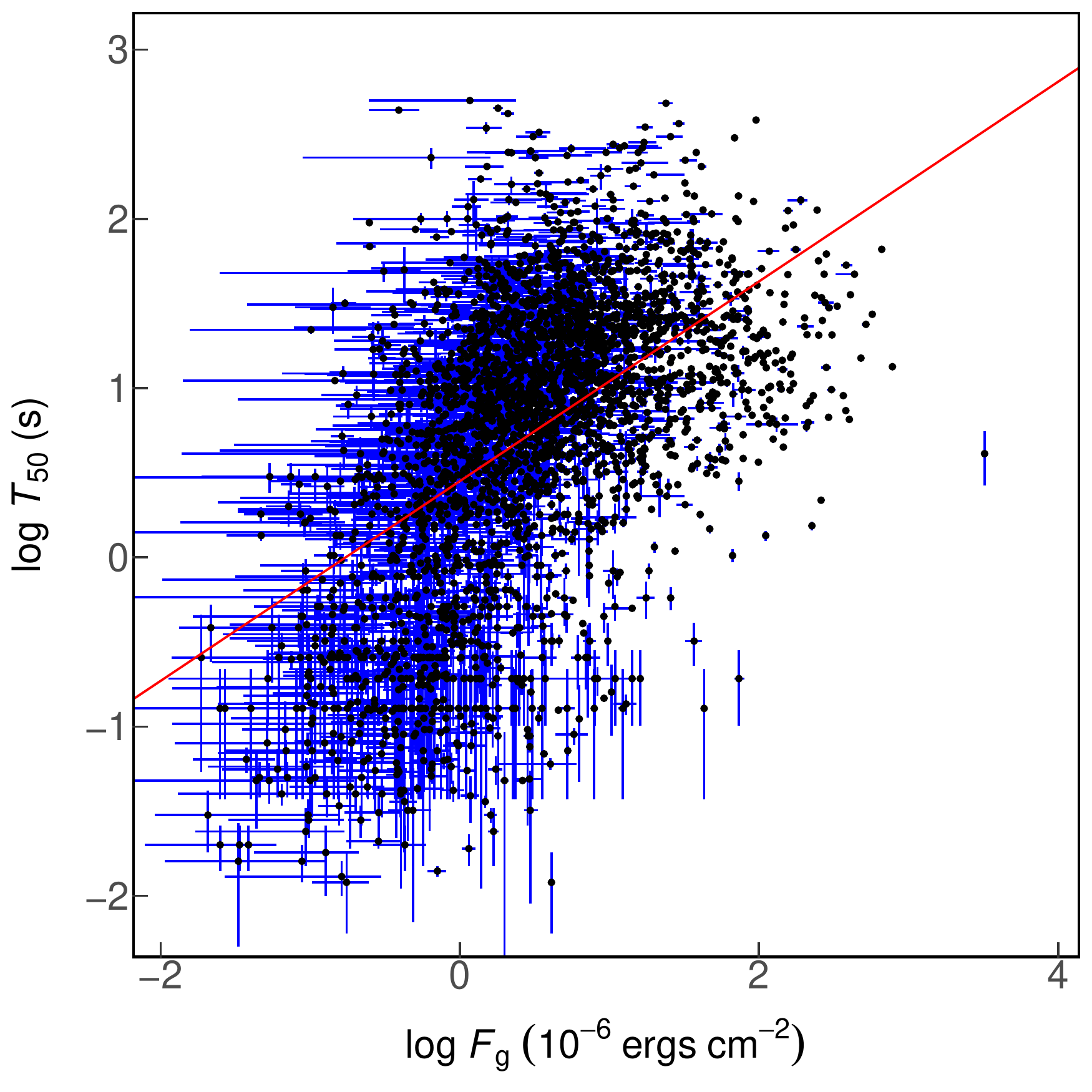}

\caption{
Scatter plot for $\log F_{\rm g}$ and $\log T_{\rm 50}$.
The red line is our fit result. The formula of the red line is $\log F_{\rm g} = (0.48 \pm 0.0051) \times \log T_{\rm 50} + (0.13 \pm 0.0057)$. The description of every parameter is in Section \ref{sec:sample}.
}
\label{fig:fgt50}
\end{figure*}

The correlation between $\log F_{\rm g}$ and $\log T_{\rm 50}$ is:
\begin{equation} \label{eq:fgt50}
\log F_{\rm g} = (0.48 \pm 0.0051) \times \log T_{\rm 50} + (0.13 \pm 0.0057),
\end{equation}
where $F_{\rm g}$ is in unit of $\rm 10^{\rm -6} ~ ergs ~ cm^{\rm -2}$ and in 20-2000 $\rm keV$ energy band. $T_{\rm 50}$ is in unit of $\rm s$. The adjusted $R^{2}$ is 0.3. The Pearson coefficient is $0.53 \pm 0.0044$ with p-value $6.2 \times 10^{-211}$. The Spearman coefficient is $0.53 \pm 0.0032$ with p-value $6.3 \times 10^{-212}$. The Kendall $\tau$ coefficient is $0.37 \pm 0.0023$ with p-value $5.7 \times 10^{-195}$. The correlation ratio is $0.15 \pm 0.0021$. The cosine similarity is $0.69 \pm 0.0039$. The scatter plot is in Figure \ref{fig:fgt50}. The GRB sample number is 2916. One can find two subgroups from the figure, which are defined as short GRBs and long GRBs.

\begin{figure*}[!b]
\centering
\includegraphics[width=0.45\textwidth]{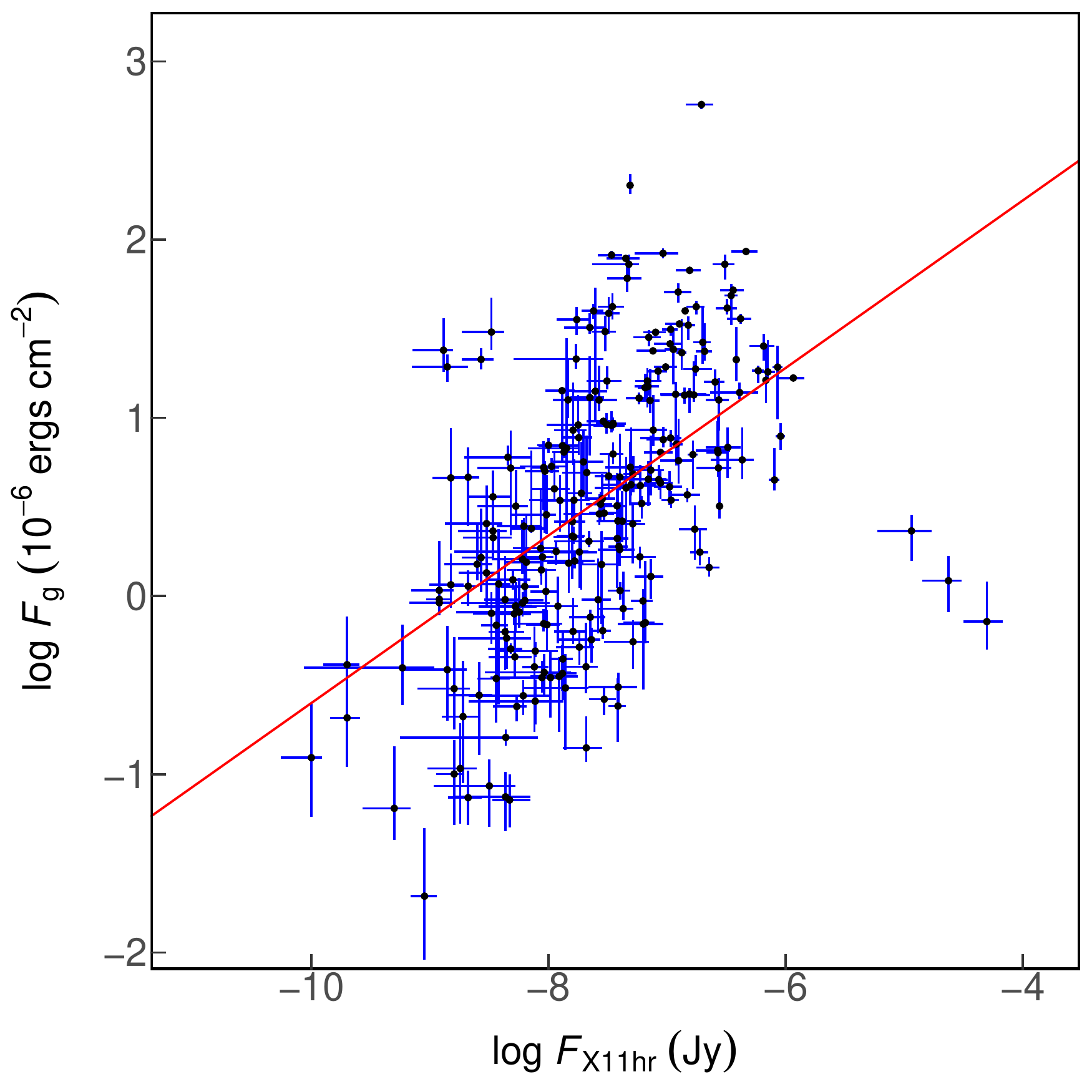}

\caption{
Scatter plot for $\log F_{\rm g}$ and $\log F_{\rm X11hr}$.
The red line is our fit result. The formula of the red line is $\log F_{\rm g} = (0.47 \pm 0.018) \times \log F_{\rm X11hr} + (4.1 \pm 0.14)$. The description of every parameter is in Section \ref{sec:sample}. GRB 060804, GRB 050916 and GRB 060719A are outliers with highest $F_{\rm X11hr}$.
}
\label{fig:fgfxhr}
\end{figure*}

The correlation between $\log F_{\rm g}$ and $\log F_{\rm X11hr}$ is:
\begin{equation} \label{eq:fgfxhr}
\log F_{\rm g} = (0.47 \pm 0.018) \times \log F_{\rm X11hr} + (4.1 \pm 0.14),
\end{equation}
where $F_{\rm g}$ is in unit of $\rm 10^{\rm -6} ~ ergs ~ cm^{\rm -2}$ and in 20-2000 $\rm keV$ energy band. $F_{\rm X11hr}$ is in unit of $\rm Jy$. The adjusted $R^{2}$ is 0.29. The Pearson coefficient is $0.52 \pm 0.017$ with p-value $6.7 \times 10^{-18}$. The Spearman coefficient is $0.57 \pm 0.016$ with p-value $2.1 \times 10^{-22}$. The Kendall $\tau$ coefficient is $0.4 \pm 0.013$ with p-value $2.2 \times 10^{-20}$. The correlation ratio is $0.98 \pm 0.00041$. The cosine similarity is $-0.51 \pm 0.012$. The scatter plot is in Figure \ref{fig:fgfxhr}. The GRB sample number is 239.

\begin{figure*}[!b]
\centering
\includegraphics[width=0.45\textwidth]{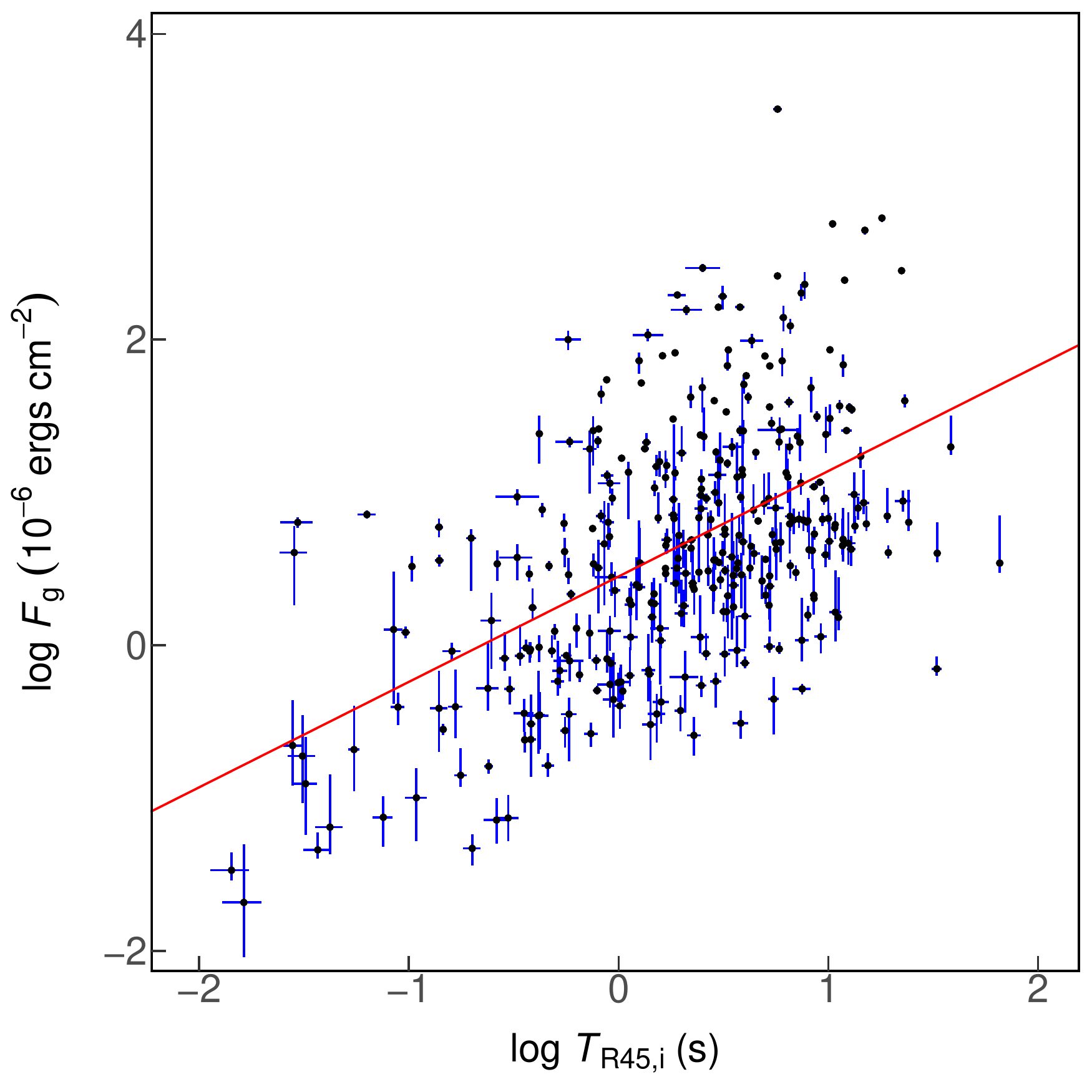}

\caption{
Scatter plot for $\log F_{\rm g}$ and $\log T_{\rm R45,i}$.
The red line is our fit result. The formula of the red line is $\log F_{\rm g} = (0.69 \pm 0.015) \times \log T_{\rm R45,i} + (0.45 \pm 0.0097)$. The description of every parameter is in Section \ref{sec:sample}.
}
\label{fig:fgtr45i}
\end{figure*}

The correlation between $\log F_{\rm g}$ and $\log T_{\rm R45,i}$ is:
\begin{equation} \label{eq:fgtr45i}
\log F_{\rm g} = (0.69 \pm 0.015) \times \log T_{\rm R45,i} + (0.45 \pm 0.0097),
\end{equation}
where $F_{\rm g}$ is in unit of $\rm 10^{\rm -6} ~ ergs ~ cm^{\rm -2}$ and in 20-2000 $\rm keV$ energy band. $T_{\rm R45,i}$ is in unit of $\rm s$. The adjusted $R^{2}$ is 0.28. The Pearson coefficient is $0.53 \pm 0.0089$ with p-value $1.3 \times 10^{-25}$. The Spearman coefficient is $0.48 \pm 0.0089$ with p-value $5 \times 10^{-21}$. The Kendall $\tau$ coefficient is $0.33 \pm 0.0068$ with p-value $5.6 \times 10^{-20}$. The correlation ratio is $0.24 \pm 0.0053$. The cosine similarity is $0.64 \pm 0.0064$. The scatter plot is in Figure \ref{fig:fgtr45i}. The GRB sample number is 337. Interestingly, the clustering effect in this figure is not obvious comparing with Figure \ref{fig:fgt50}.

In \citet{Lloyd2000}, they found a correlation between $F_{\rm g}$ and $E_{p,band}$, and also mentioned a correlation between $F_{\rm g}$ and $E_{p,band,i}$. They explained this correlation is due to an intrinsic relation between the burst rest-frame peak energy and the total radiated energy. They also found the internal shock model is consistent with their interpretation of the correlation, not external shock.

\begin{figure*}[!b]
\centering
\includegraphics[width=0.45\textwidth]{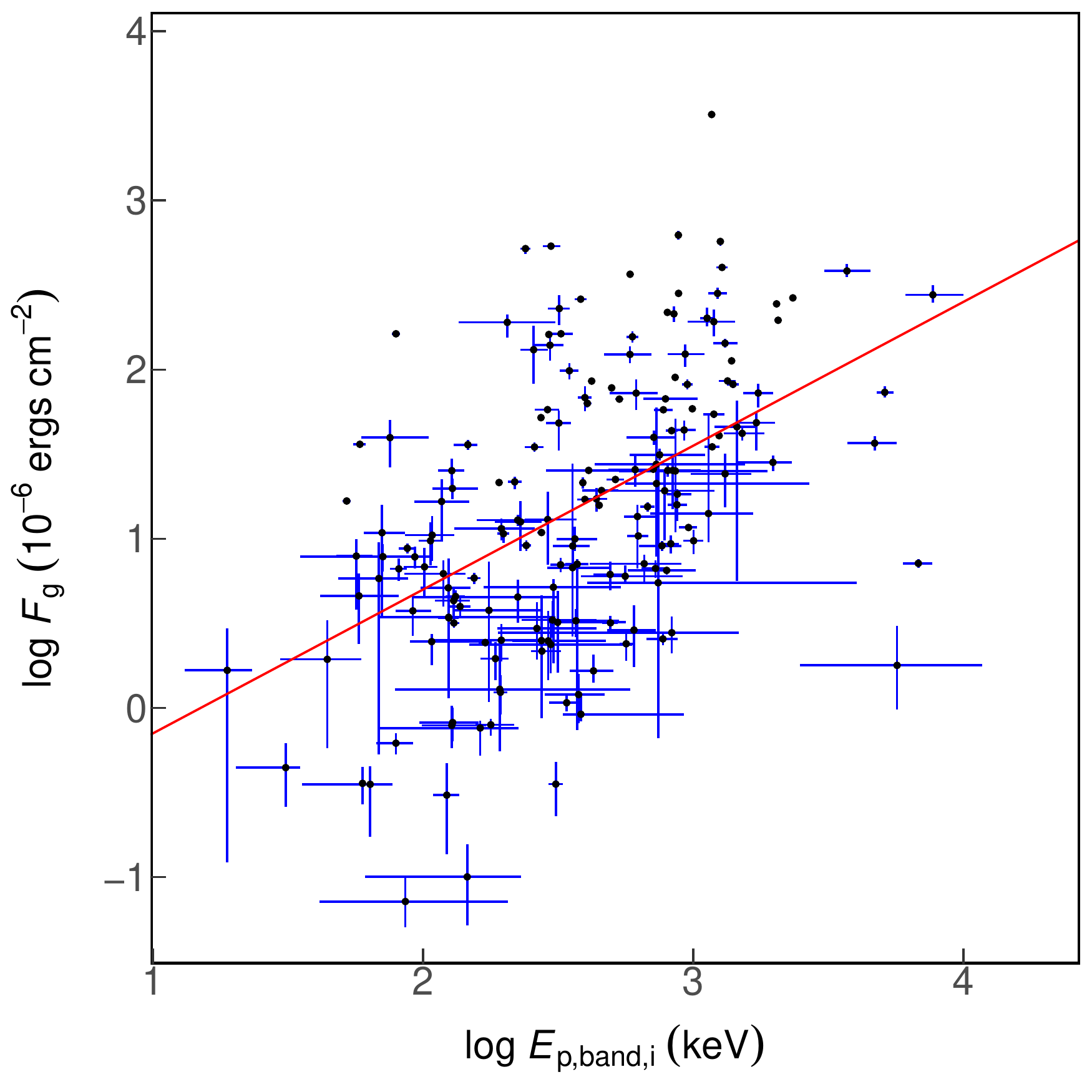}

\caption{
Scatter plot for $\log  F_{\rm g}$ and $\log E_{p,band,i}$.
The red line is our fit result. The formula of the red line is $\log  F_{\rm g} = (0.85 \pm 0.057) \times \log E_{p,band,i} + (-1 \pm 0.15)$. The description of every parameter is in Section \ref{sec:sample}.
}
\label{fig:fgepdi}
\end{figure*}

The correlation between $\log  F_{\rm g}$ and $\log E_{p,band,i}$ is:
\begin{equation} \label{eq:fgepdi}
\log  F_{\rm g} = (0.85 \pm 0.057) \times \log E_{p,band,i} + (-1 \pm 0.15),
\end{equation}
where $F_{\rm g}$ is in unit of $\rm 10^{\rm -6} ~ ergs ~ cm^{\rm -2}$ and in 20-2000 $\rm keV$ energy band. $E_{\rm p,band,i}$ is in unit of $\rm keV$. The adjusted $R^{2}$ is 0.26. The Pearson coefficient is $0.49 \pm 0.027$ with p-value $2.4 \times 10^{-12}$. The Spearman coefficient is $0.51 \pm 0.022$ with p-value $1.8 \times 10^{-13}$. The Kendall $\tau$ coefficient is $0.35 \pm 0.016$ with p-value $1.9 \times 10^{-12}$. The correlation ratio is $0.71 \pm 0.0047$. The cosine similarity is $0.85 \pm 0.0072$. The scatter plot is in Figure \ref{fig:fgepdi}. The GRB sample number is 179.

\begin{figure*}[!b]
\centering
\includegraphics[width=0.45\textwidth]{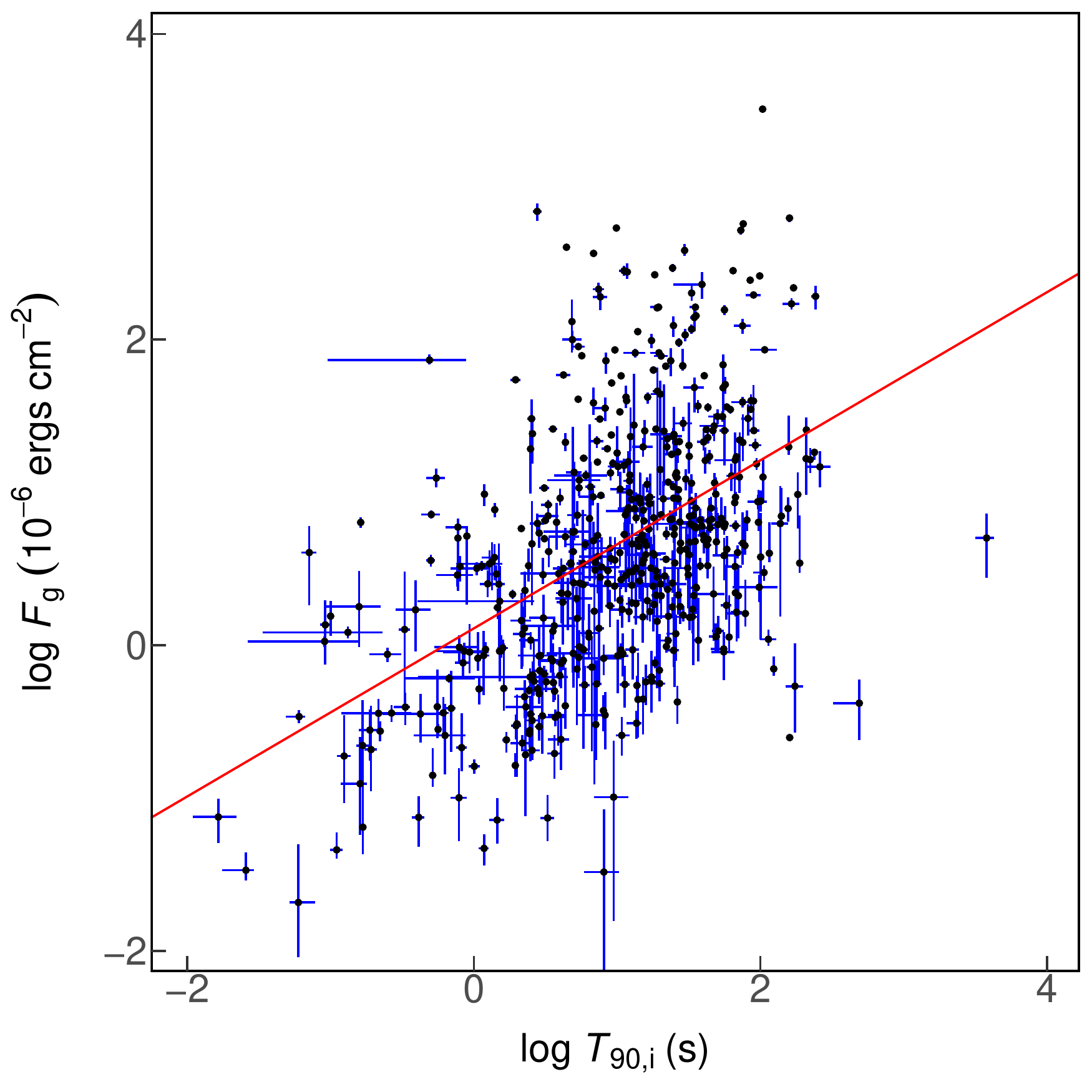}

\caption{
Scatter plot for $\log  F_{\rm g}$ and $\log T_{\rm 90,i}$.
The red line is our fit result. The formula of the red line is $\log  F_{\rm g} = (0.55 \pm 0.012) \times \log T_{\rm 90,i} + (0.11 \pm 0.016)$. The description of every parameter is in Section \ref{sec:sample}.
}
\label{fig:fgt90i}
\end{figure*}

The correlation between $\log  F_{\rm g}$ and $\log T_{\rm 90,i}$ is:
\begin{equation} \label{eq:fgt90i}
\log  F_{\rm g} = (0.55 \pm 0.012) \times \log T_{\rm 90,i} + (0.11 \pm 0.016),
\end{equation}
where $F_{\rm g}$ is in unit of $\rm 10^{\rm -6} ~ ergs ~ cm^{\rm -2}$ and in 20-2000 $\rm keV$ energy band. $T_{\rm 90,i}$ is in unit of $\rm s$. The adjusted $R^{2}$ is 0.24. The Pearson coefficient is $0.48 \pm 0.009$ with p-value $6.9 \times 10^{-32}$. The Spearman coefficient is $0.48 \pm 0.0081$ with p-value $3.2 \times 10^{-31}$. The Kendall $\tau$ coefficient is $0.33 \pm 0.006$ with p-value $2.5 \times 10^{-29}$. The correlation ratio is $0.21 \pm 0.0043$. The cosine similarity is $0.72 \pm 0.0063$. The scatter plot is in Figure \ref{fig:fgt90i}. The GRB sample number is 526.

\begin{figure*}[!b]
\centering
\includegraphics[width=0.45\textwidth]{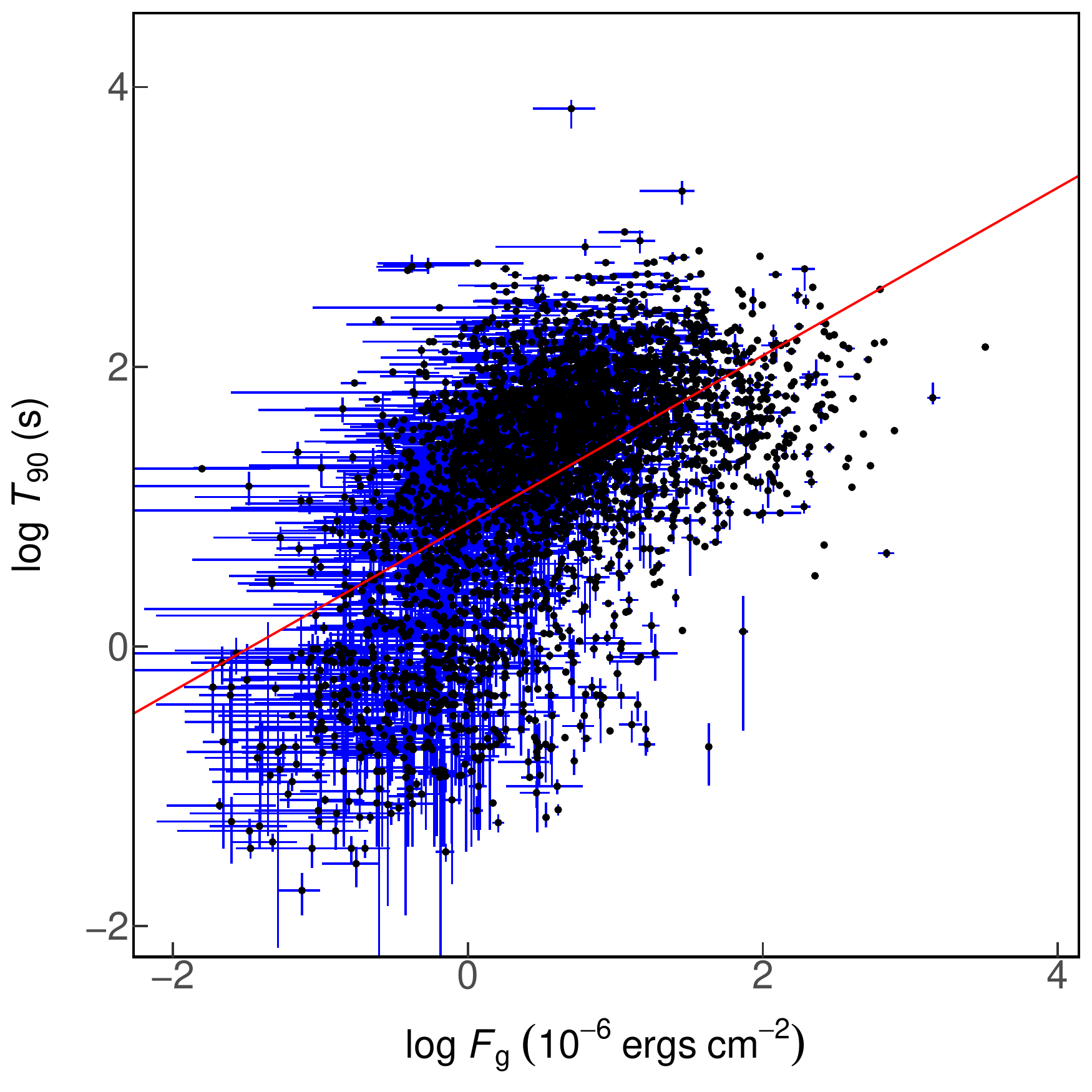}

\caption{
Scatter plot for $\log F_{\rm g}$ and $\log T_{\rm 90}$.
The red line is our fit result. The formula of the red line is $\log F_{\rm g} = (0.5 \pm 0.0049) \times \log T_{\rm 90} + (-0.083 \pm 0.0075)$. The description of every parameter is in Section \ref{sec:sample}.
}
\label{fig:fgt90}
\end{figure*}

The correlation between $\log F_{\rm g}$ and $\log T_{\rm 90}$ is:
\begin{equation} \label{eq:fgt90}
\log F_{\rm g} = (0.5 \pm 0.0049) \times \log T_{\rm 90} + (-0.083 \pm 0.0075),
\end{equation}
where $T_{\rm 90}$ is in unit of $\rm s$. $F_{\rm g}$ is in unit of $\rm 10^{\rm -6} ~ ergs ~ cm^{\rm -2}$ and in 20-2000 $\rm keV$ energy band. The adjusted $R^{2}$ is 0.32. The Pearson coefficient is $0.55 \pm 0.004$ with p-value $2.5 \times 10^{-276}$. The Spearman coefficient is $0.55 \pm 0.0029$ with p-value $8.2 \times 10^{-281}$. The Kendall $\tau$ coefficient is $0.38 \pm 0.0022$ with p-value $1.5 \times 10^{-257}$. The correlation ratio is $0.38 \pm 0.0018$. The cosine similarity is $0.72 \pm 0.0036$. The scatter plot is in Figure \ref{fig:fgt90}. The GRB sample number is 3532. Comparing with Figure \ref{fig:fgt90i}, the clustering effect is much more obvious in Figure \ref{fig:fgt90}. However, the rest frame duration should be more intrinsic. It could be a puzzle. Notice the sample number in Figure \ref{fig:fgt90i} is much fewer, which might cause the clustering effect not obvious.

\begin{figure*}[!b]
\centering
\includegraphics[width=0.45\textwidth]{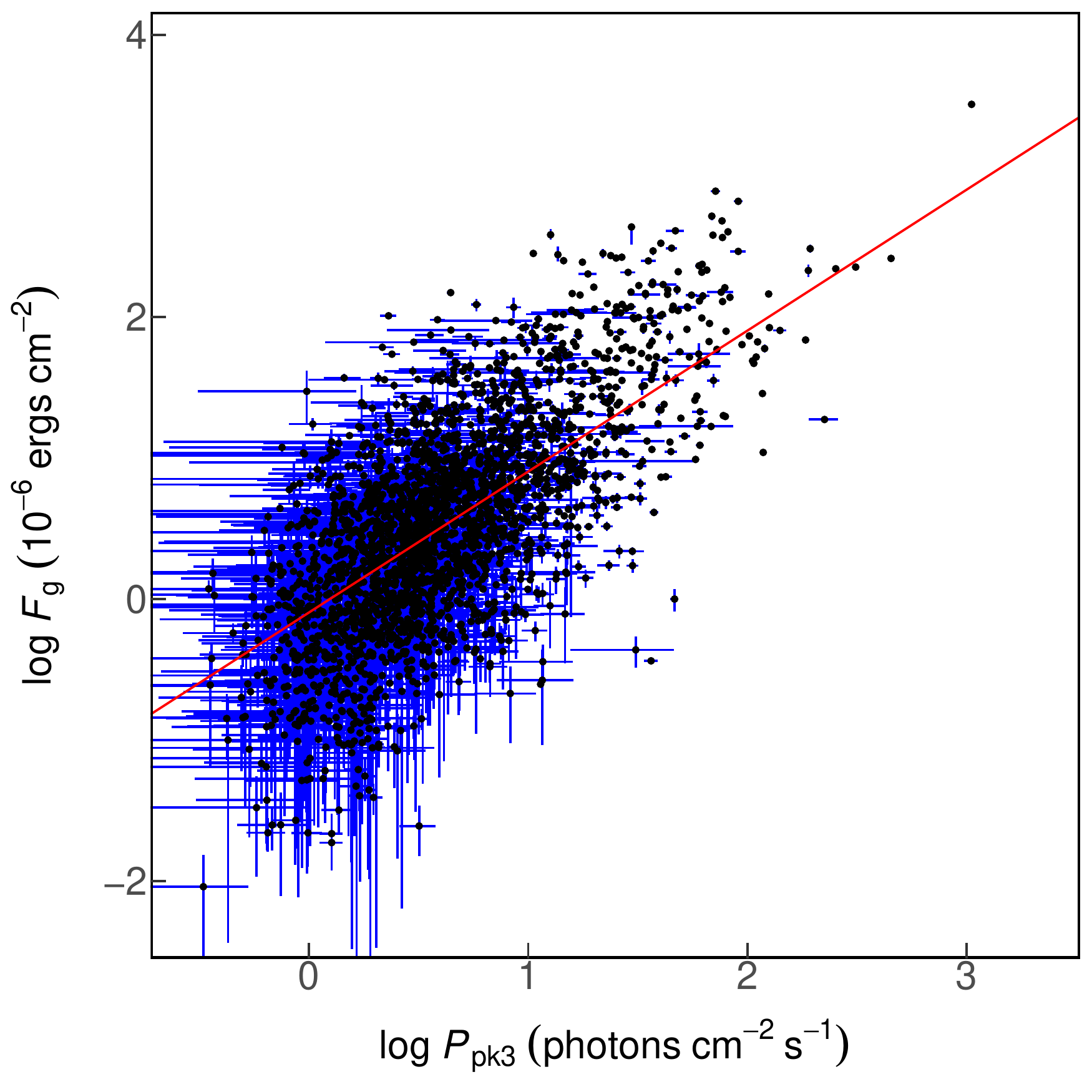}

\caption{
Scatter plot for $\log F_{\rm g}$ and $\log P_{\rm pk3}$.
The red line is our fit result. The formula of the red line is $\log F_{\rm g} = (1 \pm 0.02) \times \log P_{\rm pk3} + (-0.096 \pm 0.015)$. The description of every parameter is in Section \ref{sec:sample}.
}
\label{fig:fgppk3}
\end{figure*}

The correlation between $\log F_{\rm g}$ and $\log P_{\rm pk3}$ is:
\begin{equation} \label{eq:fgppk3}
\log F_{\rm g} = (1 \pm 0.02) \times \log P_{\rm pk3} + (-0.096 \pm 0.015),
\end{equation}
where $F_{\rm g}$ is in unit of $\rm 10^{\rm -6} ~ ergs ~ cm^{\rm -2}$ and in 20-2000 $\rm keV$ energy band. $P_{\rm pk3}$ is peak photon flux of 1024 $\rm ms$ time bin in 10-1000 $\rm keV$ and in unit of $\rm photons ~ cm^{\rm -2} ~ s^{\rm -1}$. The adjusted $R^{2}$ is 0.48. The Pearson coefficient is $0.65 \pm 0.0073$ with p-value $3.8 \times 10^{-304}$. The Spearman coefficient is $0.64 \pm 0.0049$ with p-value $3 \times 10^{-289}$. The Kendall $\tau$ coefficient is $0.46 \pm 0.004$ with p-value $7.7 \times 10^{-263}$. The correlation ratio is $0.064 \pm 0.0032$. The cosine similarity is $0.77 \pm 0.0052$. The scatter plot is in Figure \ref{fig:fgppk3}. The GRB sample number is 2531. It is quite natural that the fluence $F_{\rm g}$ is proportional to peak photon numbers $P_{\rm pk}$. There are similar relations which are not shown here.

\subsection{Some correlations about $F_{\rm X11hr}$}

\begin{figure*}[!b]
\centering
\includegraphics[width=0.45\textwidth]{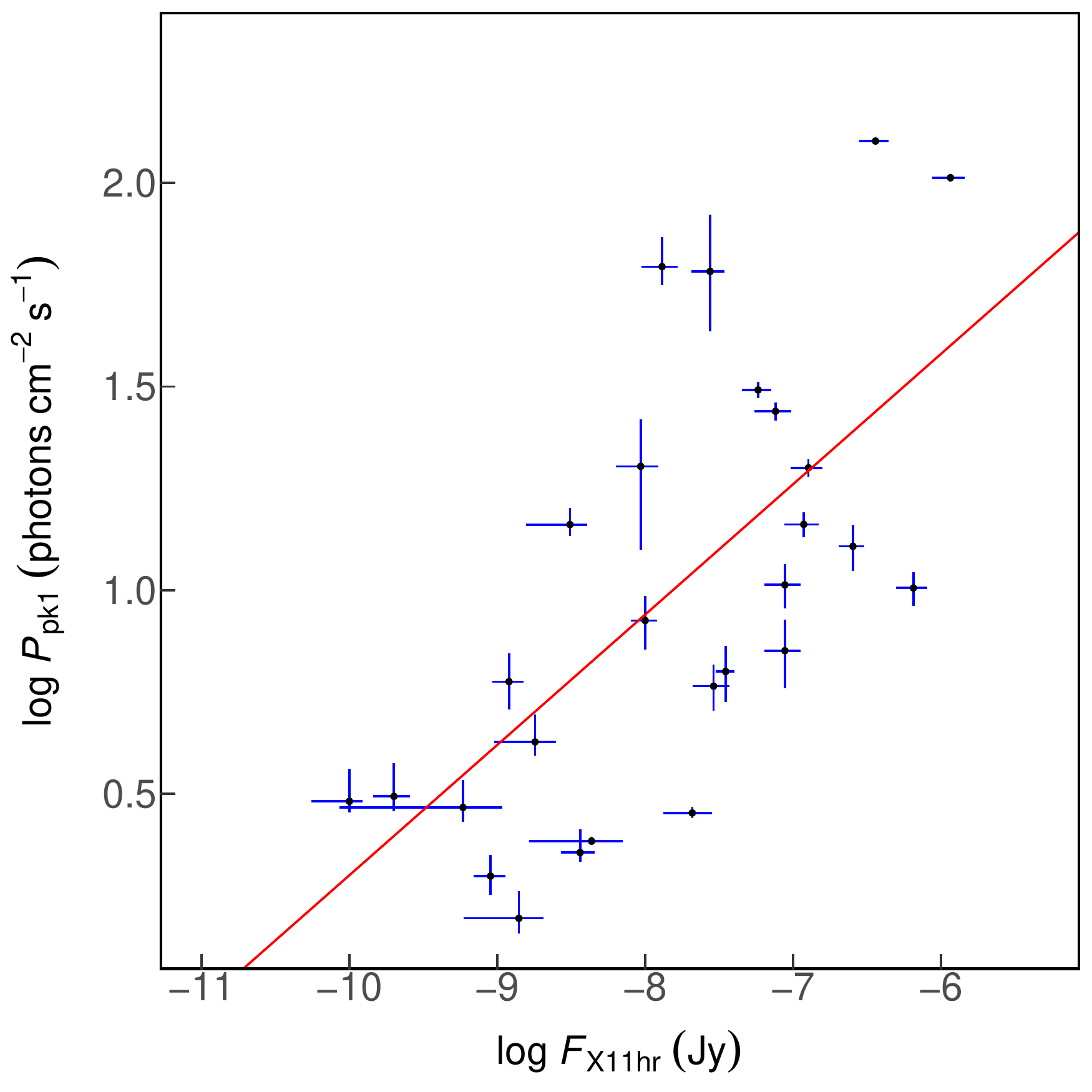}

\caption{
Scatter plot for $\log F_{\rm X11hr}$ and $\log P_{\rm pk1}$.
The red line is our fit result. The formula of the red line is $\log F_{\rm X11hr} = (1.4 \pm 0.086) \times \log P_{\rm pk1} + (-9.2 \pm 0.1)$. The description of every parameter is in Section \ref{sec:sample}.
}
\label{fig:fxhrppk1}
\end{figure*}

The correlation between $\log F_{\rm X11hr}$ and $\log P_{\rm pk1}$ is:
\begin{equation} \label{eq:fxhrppk1}
\log F_{\rm X11hr} = (1.4 \pm 0.086) \times \log P_{\rm pk1} + (-9.2 \pm 0.1),
\end{equation}
where $F_{\rm X11hr}$ is in unit of $\rm Jy$. $P_{\rm pk1}$ is peak photon flux of 64 $\rm ms$ time bin in 10-1000 $\rm keV$ and in unit of $\rm photons ~ cm^{\rm -2} ~ s^{\rm -1}$. The adjusted $R^{2}$ is 0.44. The Pearson coefficient is $0.66 \pm 0.026$ with p-value $1.8 \times 10^{-4}$. The Spearman coefficient is $0.68 \pm 0.03$ with p-value $1.1 \times 10^{-4}$. The Kendall $\tau$ coefficient is $0.47 \pm 0.027$ with p-value $5.8 \times 10^{-4}$. The correlation ratio is $0.98 \pm 0.0012$. The cosine similarity is $-0.83 \pm 0.0054$. The scatter plot is in Figure \ref{fig:fxhrppk1}. The GRB sample number is 27.

\begin{figure*}[!b]
\centering
\includegraphics[width=0.45\textwidth]{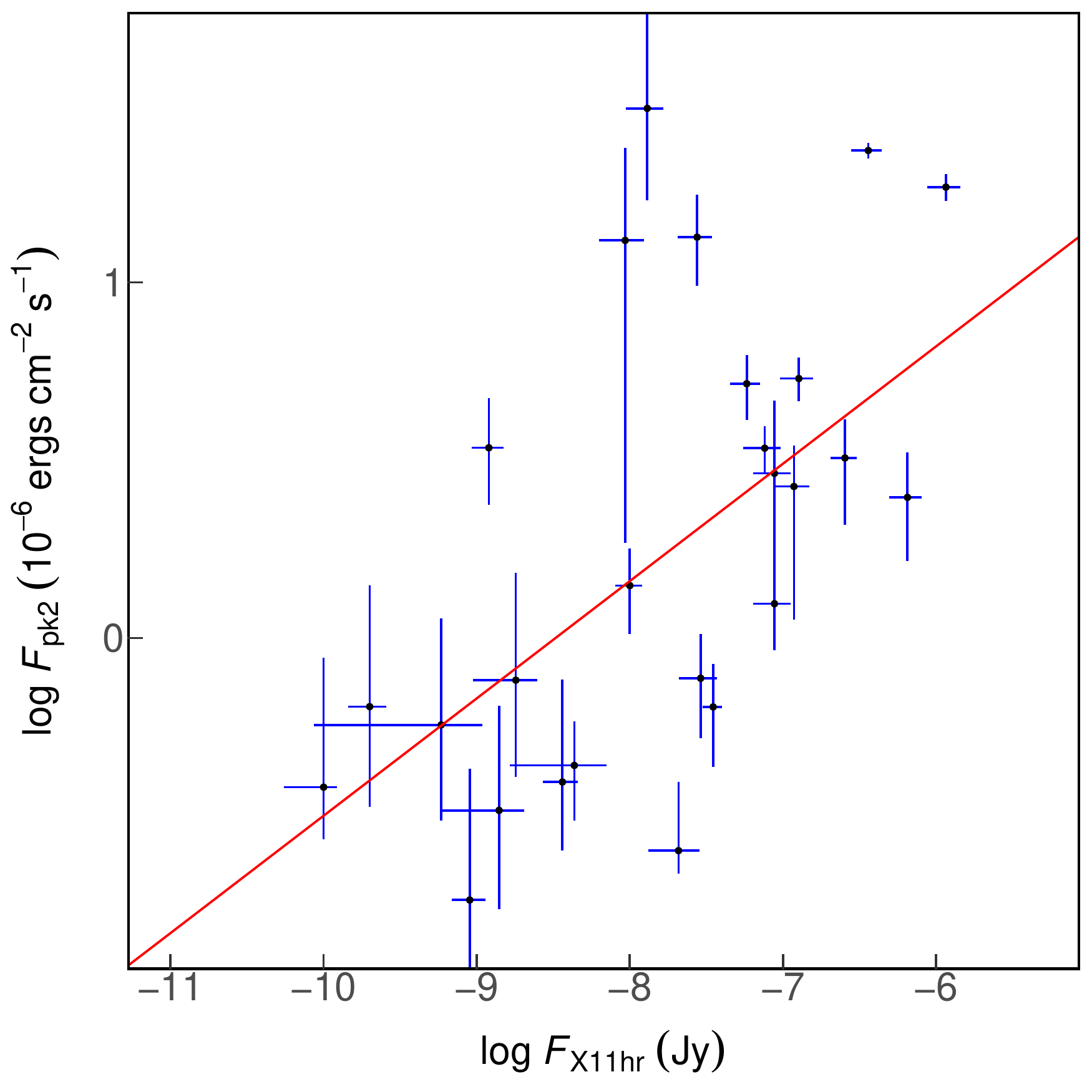}

\caption{
Scatter plot for $\log F_{\rm X11hr}$ and $\log F_{\rm pk2}$.
The red line is our fit result. The formula of the red line is $\log F_{\rm X11hr} = (0.93 \pm 0.13) \times \log F_{\rm pk2} + (-8.1 \pm 0.063)$. The description of every parameter is in Section \ref{sec:sample}.
}
\label{fig:fxhrfpk2}
\end{figure*}

The correlation between $\log F_{\rm X11hr}$ and $\log F_{\rm pk2}$ is:
\begin{equation} \label{eq:fxhrfpk2}
\log F_{\rm X11hr} = (0.93 \pm 0.13) \times \log F_{\rm pk2} + (-8.1 \pm 0.063),
\end{equation}
where $F_{\rm pk2}$ is peak energy flux of 64 $\rm ms$ time bin in rest-frame 1-$10^{4}$ $\rm keV$ energy band and in unit of $\rm 10^{\rm -6} ~ ergs ~ cm^{\rm -2} ~ s^{\rm -1}$. $F_{\rm X11hr}$ is in unit of $\rm Jy$. The adjusted $R^{2}$ is 0.33. The Pearson coefficient is $0.55 \pm 0.063$ with p-value $3.8 \times 10^{-3}$. The Spearman coefficient is $0.57 \pm 0.066$ with p-value $2.4 \times 10^{-3}$. The Kendall $\tau$ coefficient is $0.39 \pm 0.054$ with p-value $5.1 \times 10^{-3}$. The correlation ratio is $0.98 \pm 0.0016$. The cosine similarity is $-0.3 \pm 0.068$. The scatter plot is in Figure \ref{fig:fxhrfpk2}. The GRB sample number is 26.

\subsection{Some correlations about $F_{\rm Opt11hr}$}

$\log F_{\rm X11hr}$ and $\log F_{\rm Opt11hr}$ has a remarkable linear correlation, but not very strong. The hypothesis testing p-values of the whole formula, linear coefficient $a$, Pearson, Spearman and Kendall $\tau$ coefficients are all smaller than 0.05. The Pearson coefficient is $0.25 \pm 0.033$ with p-value $6.5 \times 10^{-5}$. The Spearman coefficient is $0.27 \pm 0.036$ with p-value $1.3 \times 10^{-5}$. The Kendall $\tau$ coefficient is $0.18 \pm 0.025$ with p-value $1.2 \times 10^{-5}$. The adjusted $R^{2}$ is 0.08. Because $\log F_{\rm X11hr}$ has correlations with $\log P_{\rm pk1}$ and $\log F_{\rm pk2}$, $\log F_{\rm Opt11hr}$ should have similar results.

\begin{figure*}[!b]
\centering
\includegraphics[width=0.45\textwidth]{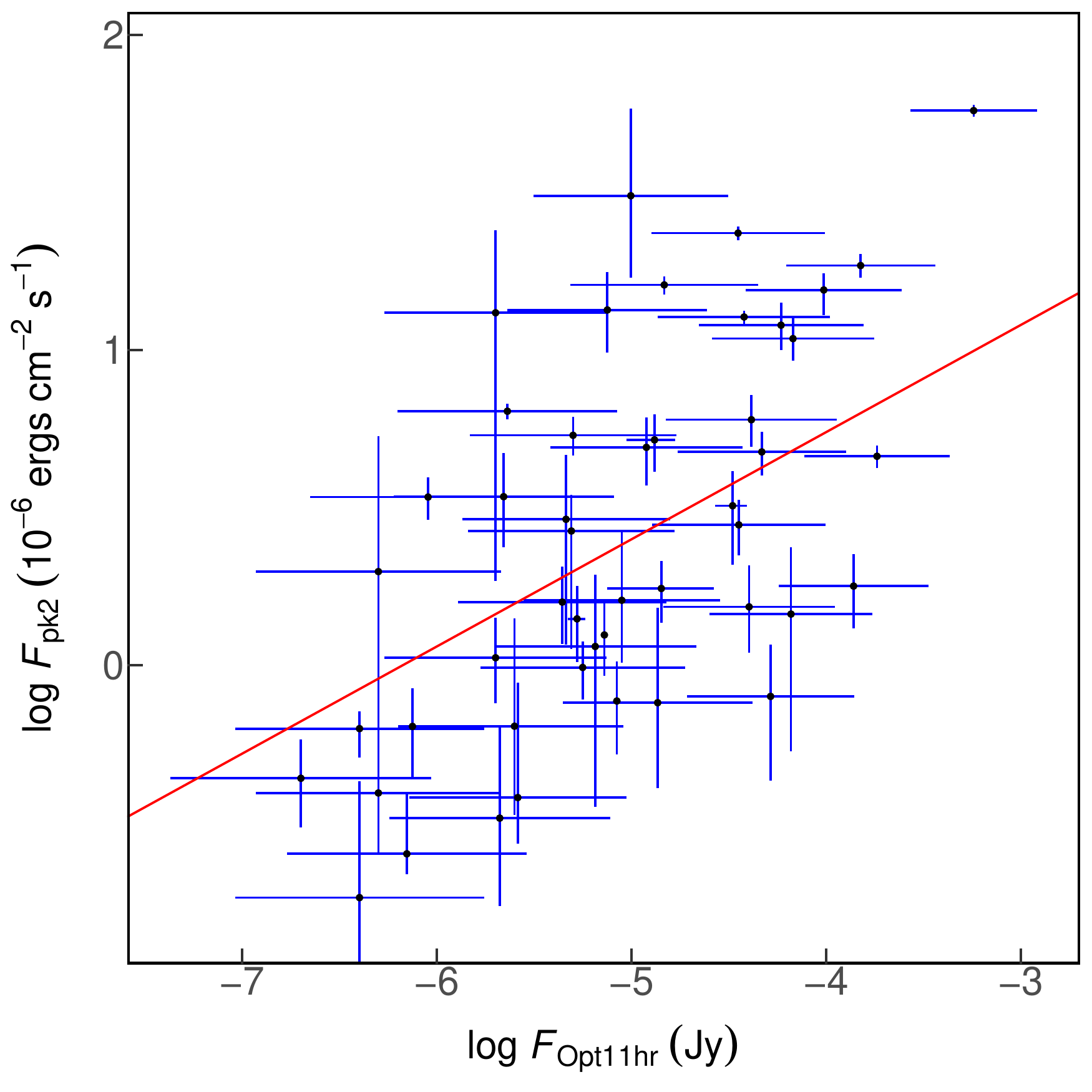}

\caption{
Scatter plot for $\log F_{\rm Opt11hr}$ and $\log F_{\rm pk2}$.
The red line is our fit result. The formula of the red line is $\log F_{\rm Opt11hr} = (0.76 \pm 0.14) \times \log F_{\rm pk2} + (-5.4 \pm 0.095)$. The description of every parameter is in Section \ref{sec:sample}.
}
\label{fig:fo11fpk2}
\end{figure*}

The correlation between $\log F_{\rm Opt11hr}$ and $\log F_{\rm pk2}$ is:
\begin{equation} \label{eq:fo11fpk2}
\log F_{\rm Opt11hr} = (0.76 \pm 0.14) \times \log F_{\rm pk2} + (-5.4 \pm 0.095),
\end{equation}
where $F_{\rm pk2}$ is peak energy flux of 64 $\rm ms$ time bin in rest-frame 1-$10^{4}$ $\rm keV$ energy band, and in unit of $\rm 10^{\rm -6} ~ ergs ~ cm^{\rm -2} ~ s^{\rm -1}$. $F_{\rm Opt11hr}$ is in unit of $\rm Jy$. The adjusted $R^{2}$ is 0.36. The Pearson coefficient is $0.51 \pm 0.075$ with p-value $2.5 \times 10^{-4}$. The Spearman coefficient is $0.49 \pm 0.077$ with p-value $4.5 \times 10^{-4}$. The Kendall $\tau$ coefficient is $0.35 \pm 0.058$ with p-value $5.3 \times 10^{-4}$. The correlation ratio is $0.96 \pm 0.0038$. The cosine similarity is $-0.46 \pm 0.043$. The scatter plot is in Figure \ref{fig:fo11fpk2}. The GRB sample number is 48.

\begin{figure*}[!b]
\centering
\includegraphics[width=0.45\textwidth]{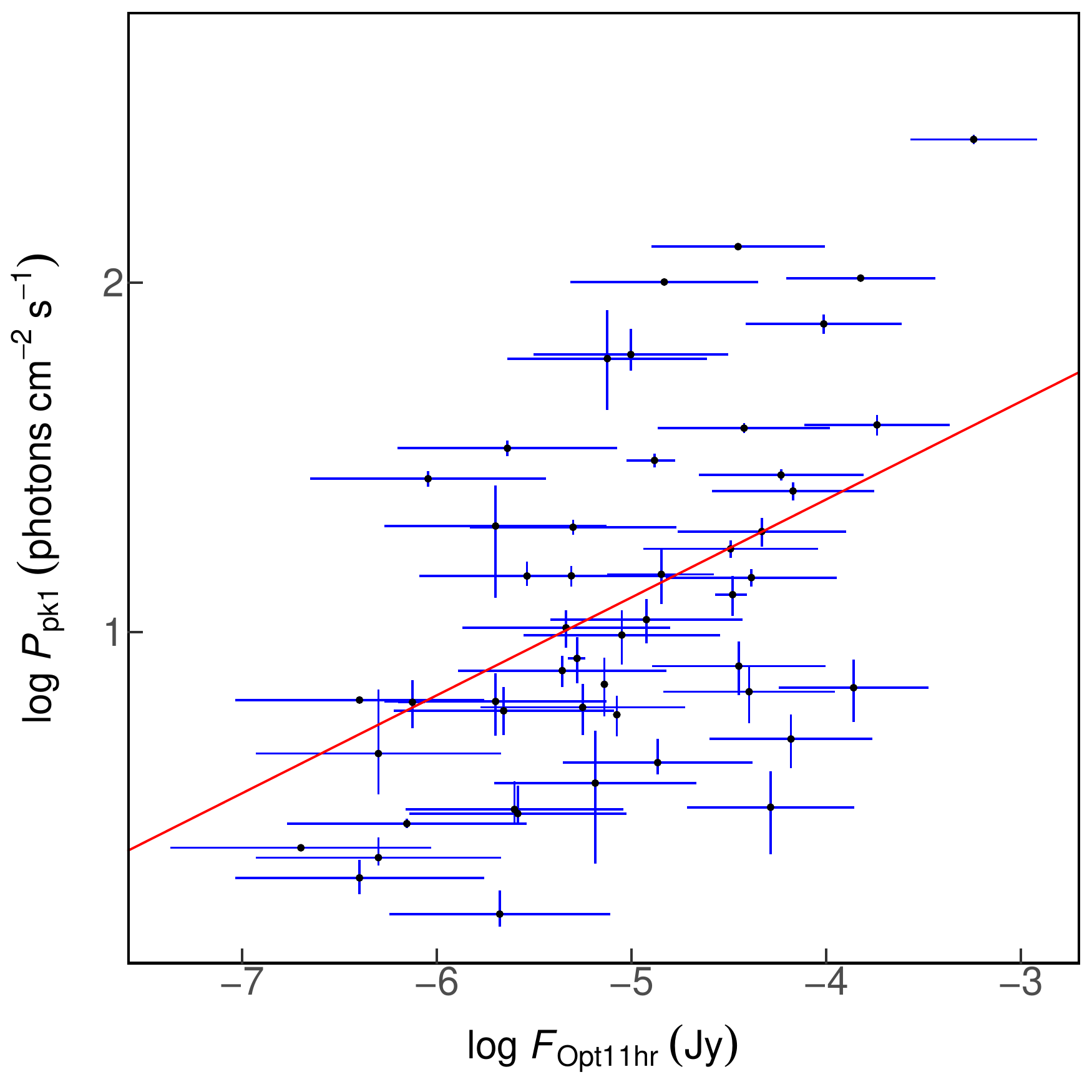}

\caption{
Scatter plot for $\log F_{\rm Opt11hr}$ and $\log P_{\rm pk1}$.
The red line is our fit result. The formula of the red line is $\log F_{\rm Opt11hr} = (0.88 \pm 0.14) \times \log P_{\rm pk1} + (-6 \pm 0.17)$. The description of every parameter is in Section \ref{sec:sample}.
}
\label{fig:fo11ppk1}
\end{figure*}

The correlation between $\log F_{\rm Opt11hr}$ and $\log P_{\rm pk1}$ is:
\begin{equation} \label{eq:fo11ppk1}
\log F_{\rm Opt11hr} = (0.88 \pm 0.14) \times \log P_{\rm pk1} + (-6 \pm 0.17),
\end{equation}
where $F_{\rm Opt11hr}$ is in unit of $\rm Jy$. $P_{\rm pk1}$ is peak photon flux of 64 $\rm ms$ time bin in 10-1000 $\rm keV$, and in unit of $\rm photons ~ cm^{\rm -2} ~ s^{\rm -1}$. The adjusted $R^{2}$ is 0.32. The Pearson coefficient is $0.49 \pm 0.065$ with p-value $2.8 \times 10^{-4}$. The Spearman coefficient is $0.46 \pm 0.07$ with p-value $8.2 \times 10^{-4}$. The Kendall $\tau$ coefficient is $0.33 \pm 0.053$ with p-value $8 \times 10^{-4}$. The correlation ratio is $0.97 \pm 0.0029$. The cosine similarity is $-0.85 \pm 0.008$. The scatter plot is in Figure \ref{fig:fo11ppk1}. The GRB sample number is 50.

\subsection{Some correlations about $F_{\rm radio,pk}$}

\begin{figure*}[!b]
\centering
\includegraphics[width=0.45\textwidth]{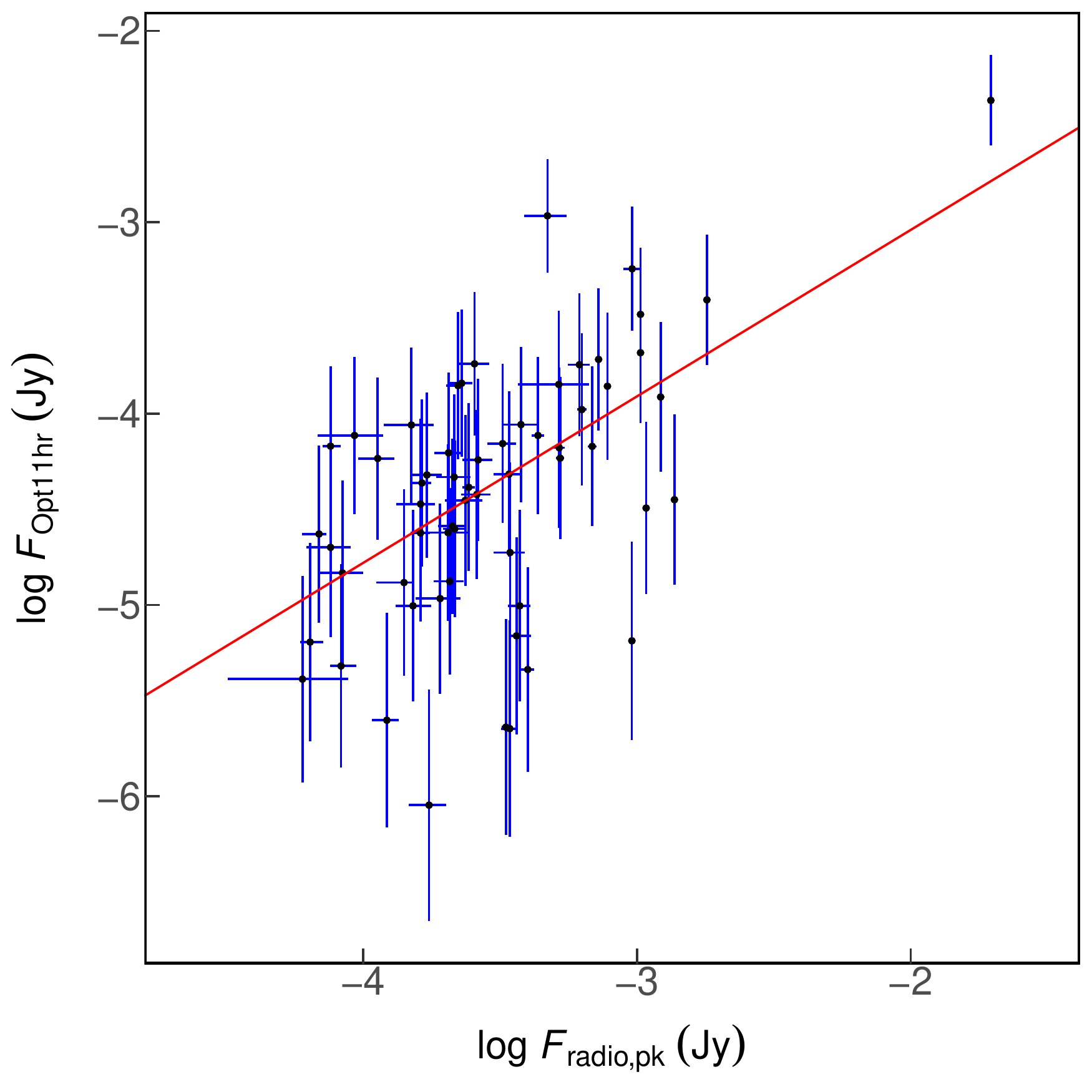}

\caption{
Scatter plot for $\log F_{\rm radio,pk}$ and $\log F_{\rm Opt11hr}$.
The red line is our fit result. The formula of the red line is $\log F_{\rm radio,pk} = (0.26 \pm 0.039) \times \log F_{\rm Opt11hr} + (-2.4 \pm 0.17)$. The description of every parameter is in Section \ref{sec:sample}. The outlier is GRB 030329A with highest $\log F_{\rm radio,pk}$ and $\log F_{\rm Opt11hr}$.
}
\label{fig:frapkfophr}
\end{figure*}

The correlation between $\log F_{\rm radio,pk}$ and $\log F_{\rm Opt11hr}$ is:
\begin{equation} \label{eq:frapkfophr}
\log F_{\rm radio,pk} = (0.26 \pm 0.039) \times \log F_{\rm Opt11hr} + (-2.4 \pm 0.17),
\end{equation}
where $F_{\rm Opt11hr}$ is in unit of $\rm Jy$. $F_{\rm radio,pk}$ is in unit of $\rm Jy$. The adjusted $R^{2}$ is 0.32. The Pearson coefficient is $0.47 \pm 0.059$ with p-value $1.3 \times 10^{-4}$. The Spearman coefficient is $0.41 \pm 0.071$ with p-value $1.3 \times 10^{-3}$. The Kendall $\tau$ coefficient is $0.28 \pm 0.052$ with p-value $1.4 \times 10^{-3}$. The correlation ratio is $0.56 \pm 0.028$. The cosine similarity is $0.99 \pm 0.002$. The scatter plot is in Figure \ref{fig:frapkfophr}. The GRB sample number is 60.

\begin{figure*}[!b]
\centering
\includegraphics[width=0.45\textwidth]{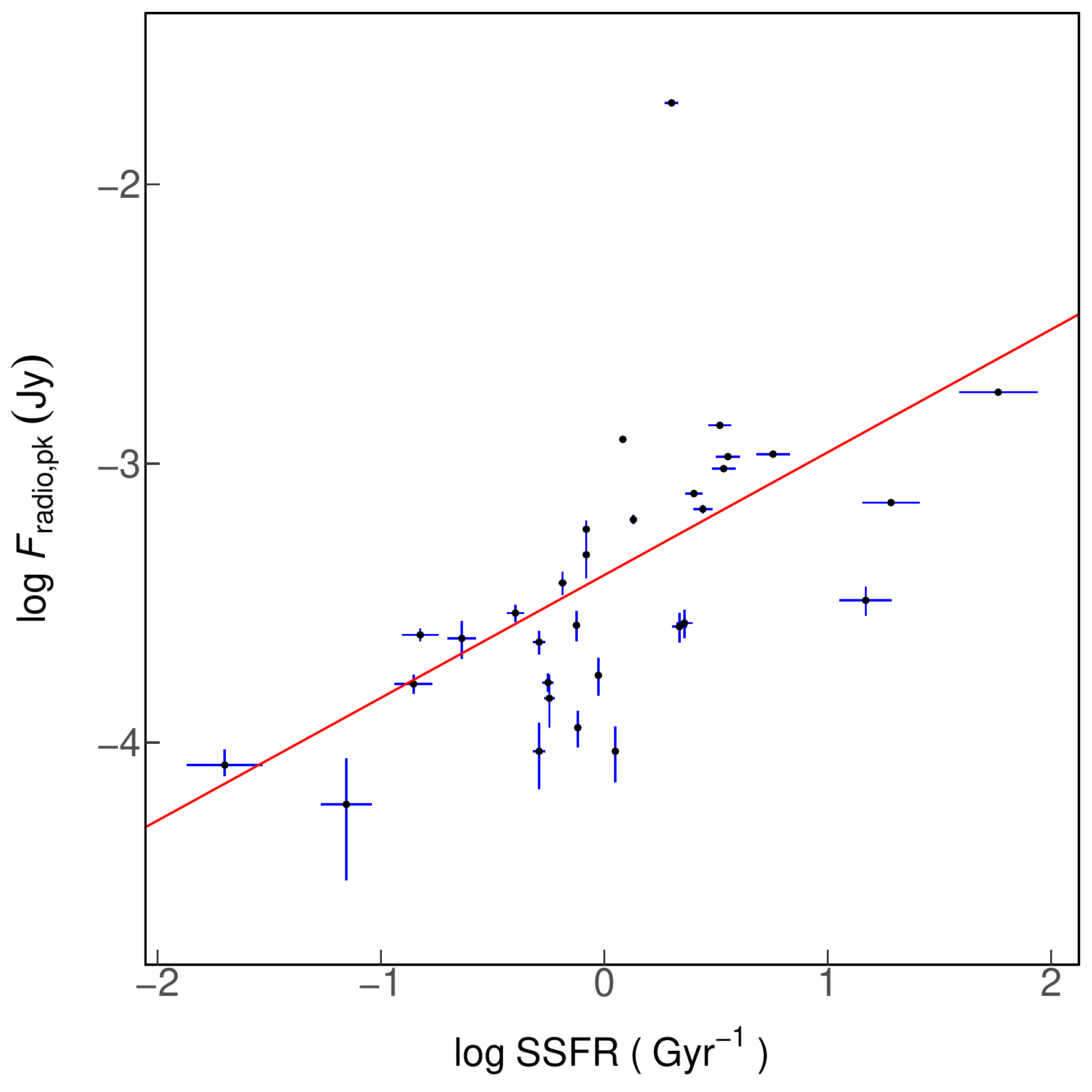}

\caption{
Scatter plot for $\log F_{\rm radio,pk}$ and $\log SSFR$.
The red line is our fit result. The formula of the red line is $\log F_{\rm radio,pk} = (0.44 \pm 0.027) \times \log SSFR + (-3.4 \pm 0.013)$. The description of every parameter is in Section \ref{sec:sample}. The outlier is GRB 030329A with highest $\log F_{\rm radio,pk}$.
}
\label{fig:frapkssfr}
\end{figure*}

The correlation between $\log F_{\rm radio,pk}$ and $\log SSFR$ is:
\begin{equation} \label{eq:frapkssfr}
\log F_{\rm radio,pk} = (0.44 \pm 0.027) \times \log SSFR + (-3.4 \pm 0.013),
\end{equation}
where $F_{\rm radio,pk}$ is in unit of $\rm Jy$. $\log SSFR$ is in unit of $\rm Gyr^{\rm -1}$. The adjusted $R^{2}$ is 0.34. The Pearson coefficient is $0.6 \pm 0.018$ with p-value $3.9 \times 10^{-4}$. The Spearman coefficient is $0.74 \pm 0.021$ with p-value $2.3 \times 10^{-6}$. The Kendall $\tau$ coefficient is $0.54 \pm 0.019$ with p-value $2 \times 10^{-5}$. The correlation ratio is $0.94 \pm 0.0024$. The cosine similarity is $0.023 \pm 0.018$. The scatter plot is in Figure \ref{fig:frapkssfr}. The GRB sample number is 31.

\subsection{Some correlations about $\alpha_{\rm spl}$}

\begin{figure*}[!b]
\centering
\includegraphics[width=0.45\textwidth]{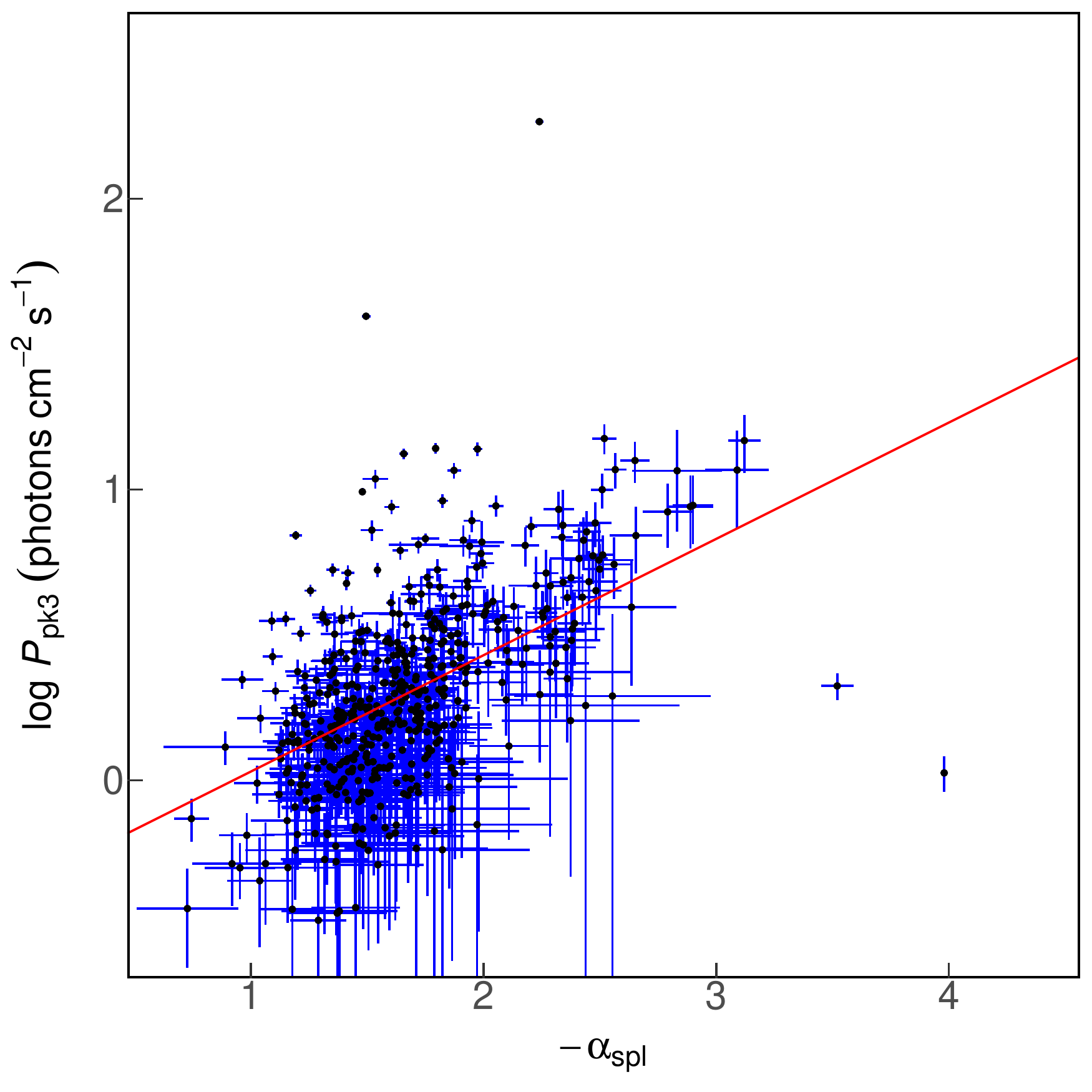}

\caption{
Scatter plot for $\log P_{\rm pk3}$ and $(-\alpha_{\rm spl})$.
The red line is our fit result. The formula of the red line is $\log P_{\rm pk3} = (0.4 \pm 0.023) \times (-\alpha_{\rm spl}) + (-0.37 \pm 0.036)$. The description of every parameter is in Section \ref{sec:sample}.
}
\label{fig:ppk3alspl}
\end{figure*}

The correlation between $\log P_{\rm pk3}$ and $(-\alpha_{\rm spl})$ is:
\begin{equation} \label{eq:ppk3alspl}
\log P_{\rm pk3} = (0.4 \pm 0.023) \times (-\alpha_{\rm spl}) + (-0.37 \pm 0.036),
\end{equation}
where $P_{\rm pk3}$ is peak photon flux of 1024 $\rm ms$ time bin in 10-1000 $\rm keV$, and in unit of $\rm photons ~ cm^{\rm -2} ~ s^{\rm -1}$. The adjusted $R^{2}$ is 0.29. The Pearson coefficient is $0.46 \pm 0.028$ with p-value $7.4 \times 10^{-27}$. The Spearman coefficient is $0.49 \pm 0.021$ with p-value $2.2 \times 10^{-31}$. The Kendall $\tau$ coefficient is $0.35 \pm 0.015$ with p-value $6.5 \times 10^{-31}$. The correlation ratio is $0.87 \pm 0.0027$. The cosine similarity is $0.7 \pm 0.018$. The scatter plot is in Figure \ref{fig:ppk3alspl}. The GRB sample number is 488. This might be a selection effect, that those GRBs with steep spectra (larger $(-\alpha_{\rm spl})$) can only be observed if they are strong, i.e., high peak photon flux.

\begin{figure*}[!b]
\centering
\includegraphics[width=0.45\textwidth]{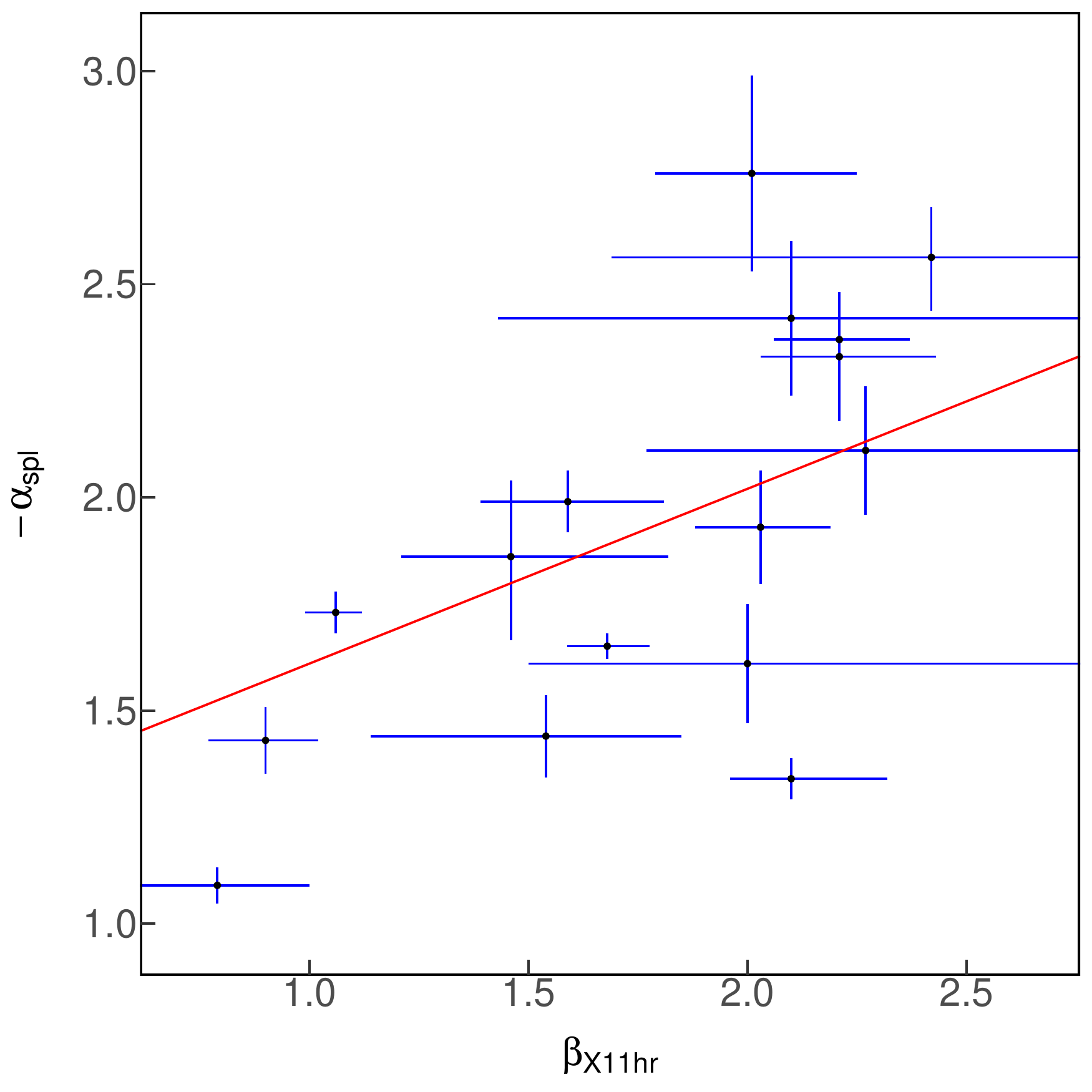}

\caption{
Scatter plot for $\beta_{\rm X11hr}$ and $(-\alpha_{\rm spl})$.
The red line is our fit result. The formula of the red line is $\beta_{\rm X11hr}= (0.72 \pm 0.24) \times (-\alpha_{\rm spl}) + (0.43 \pm 0.42)$. The description of every parameter is in Section \ref{sec:sample}.
}
\label{fig:behralspl}
\end{figure*}

The correlation between $\beta_{\rm X11hr}$ and $(-\alpha_{\rm spl})$ is:
\begin{equation} \label{eq:behralspl}
\beta_{\rm X11hr}= (0.72 \pm 0.24) \times (-\alpha_{\rm spl}) + (0.43 \pm 0.42),
\end{equation}
the adjusted $R^{2}$ is 0.43. The Pearson coefficient is $0.53 \pm 0.14$ with p-value $3.3 \times 10^{-2}$. The Spearman coefficient is $0.52 \pm 0.14$ with p-value $3.7 \times 10^{-2}$. The Kendall $\tau$ coefficient is $0.39 \pm 0.11$ with p-value $3.1 \times 10^{-2}$. The correlation ratio is $0.12 \pm 0.081$. The cosine similarity is $0.96 \pm 0.015$. The scatter plot is in Figure \ref{fig:behralspl}. The GRB sample number is 16. It tells those GRBs with steeper spectra in the prompt emission stage also have steeper spectra in the afterglow stage, even the main emtting energy bands are different. It is likely that the emission in these two stages share the same radiation mechanism, and with the cooling of the ejecta, the radiation bands shift from $\gamma-$rays to X-rays.

\begin{figure*}[!b]
\centering
\includegraphics[width=0.45\textwidth]{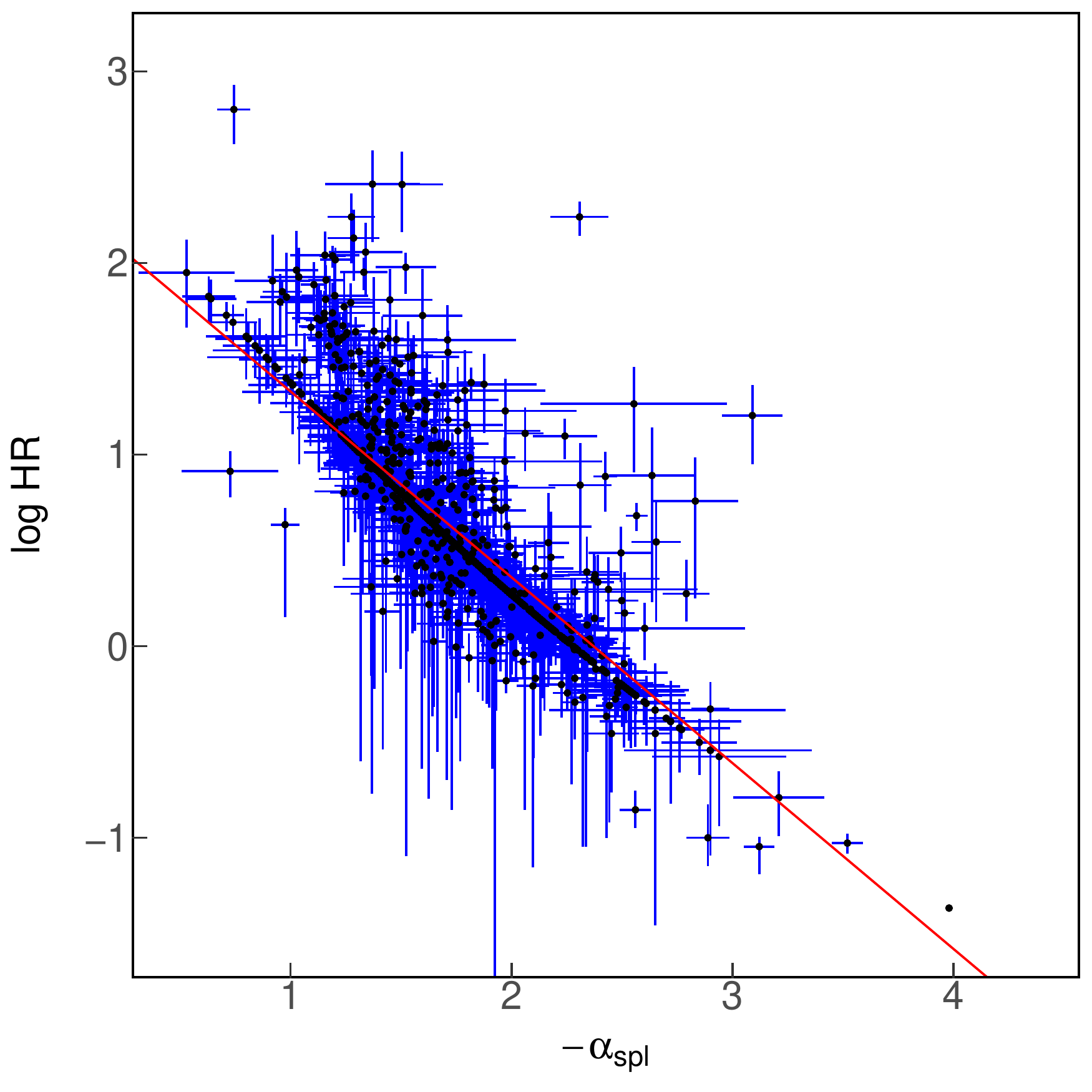}

\caption{
Scatter plot for $\log HR$ and $(-\alpha_{\rm spl})$.
The red line is our fit result. The formula of the red line is $\log HR = (-0.97 \pm 0.023) \times (-\alpha_{\rm spl}) + (2.3 \pm 0.039)$. The description of every parameter is in Section \ref{sec:sample}.
}
\label{fig:hralspl}
\end{figure*}

The correlation between $\log HR$ and $(-\alpha_{\rm spl})$ is:
\begin{equation} \label{eq:hralspl}
\log HR = (-0.97 \pm 0.023) \times (-\alpha_{\rm spl}) + (2.3 \pm 0.039),
\end{equation}
the adjusted $R^{2}$ is 0.69. The Pearson coefficient is $-0.73 \pm 0.016$ with p-value $1.6 \times 10^{-148}$. The Spearman coefficient is $-0.78 \pm 0.011$ with p-value $1.9 \times 10^{-176}$. The Kendall $\tau$ coefficient is $-0.6 \pm 0.01$ with p-value $8.4 \times 10^{-158}$. The correlation ratio is $0.71 \pm 0.0043$. The cosine similarity is $0.59 \pm 0.0093$. The scatter plot is in Figure \ref{fig:hralspl}. The GRB sample number is 874. One can see a double linear relation in the figure. This may indicate that the classification is important before the correlation analysis.

Because $\beta_{\rm X11hr}$ and $P_{\rm pk3}$ both correlate with $\alpha_{\rm spl}$, and $P_{\rm pk3}$ and $P_{\rm pk4}$ also have correlation, $\beta_{\rm X11hr}$ and $P_{\rm pk4}$ have correlation naturally.

\begin{figure*}[!b]
\centering
\includegraphics[width=0.45\textwidth]{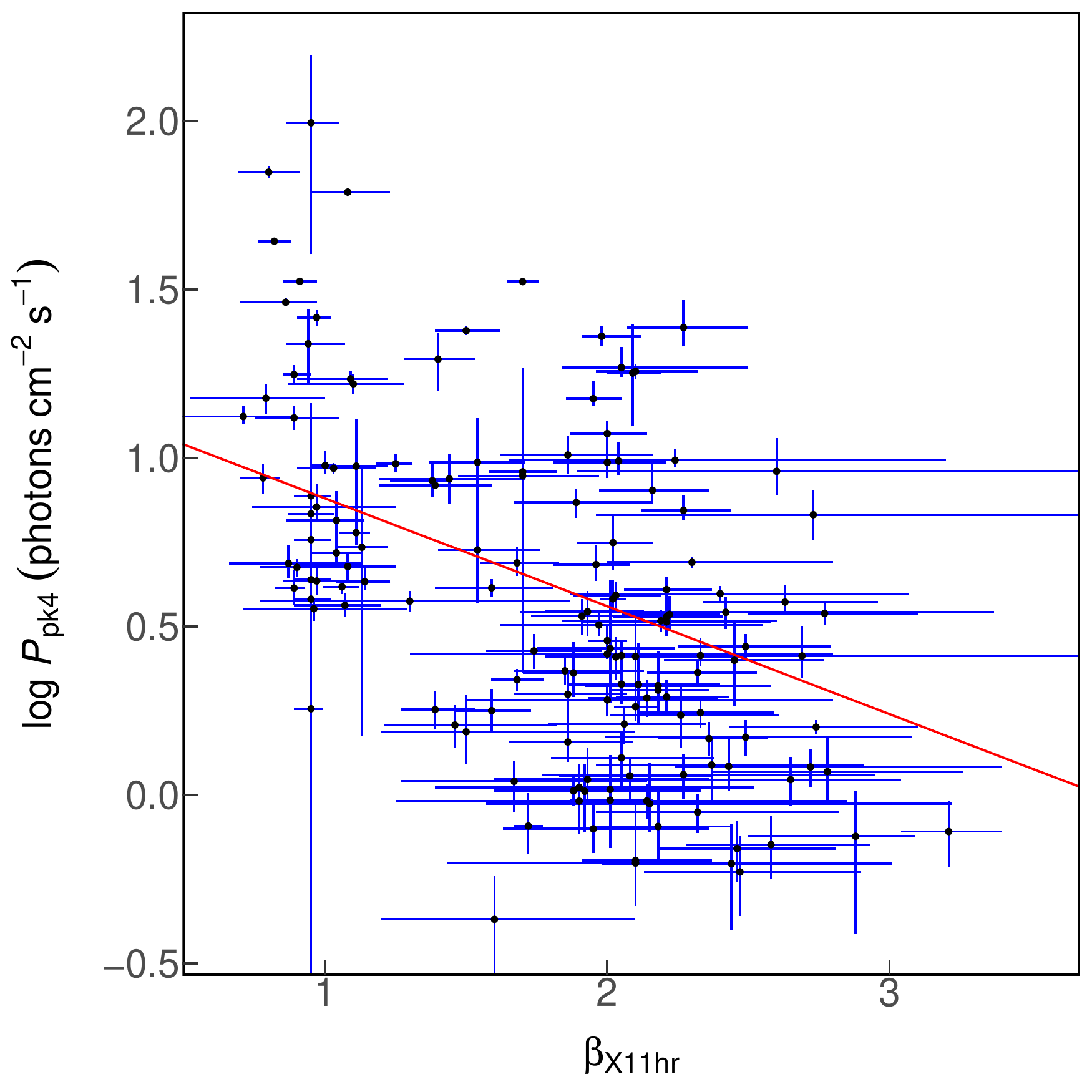}

\caption{
Scatter plot for $\log P_{\rm pk4}$ and $\beta_{\rm X11hr}$.
The red line is our fit result. The formula of the red line is $\log P_{\rm pk4} = (-0.32 \pm 0.047) \times \beta_{\rm X11hr} + (1.2 \pm 0.087)$. The description of every parameter is in Section \ref{sec:sample}.
}
\label{fig:ppk4behr}
\end{figure*}

The correlation between $\log P_{\rm pk4}$ and $\beta_{\rm X11hr}$ is:
\begin{equation} \label{eq:ppk4behr}
\log P_{\rm pk4} = (-0.32 \pm 0.047) \times \beta_{\rm X11hr} + (1.2 \pm 0.087),
\end{equation}
where $P_{\rm pk4}$ is peak photon flux of 1 $\rm s$ time bin in 10-1000 $\rm keV$, and in unit of $\rm photons ~ cm^{\rm -2} ~ s^{\rm -1}$. The adjusted $R^{2}$ is 0.29. The Pearson coefficient is $-0.44 \pm 0.058$ with p-value $1.5 \times 10^{-8}$. The Spearman coefficient is $-0.48 \pm 0.037$ with p-value $4.4 \times 10^{-10}$. The Kendall $\tau$ coefficient is $-0.33 \pm 0.028$ with p-value $2.9 \times 10^{-9}$. The correlation ratio is $0.72 \pm 0.012$. The cosine similarity is $0.6 \pm 0.018$. The scatter plot is in Figure \ref{fig:ppk4behr}. The GRB sample number is 151.

\subsection{Correlations about host galaxy offset}

\begin{figure*}[!b]
\centering
\includegraphics[width=0.45\textwidth]{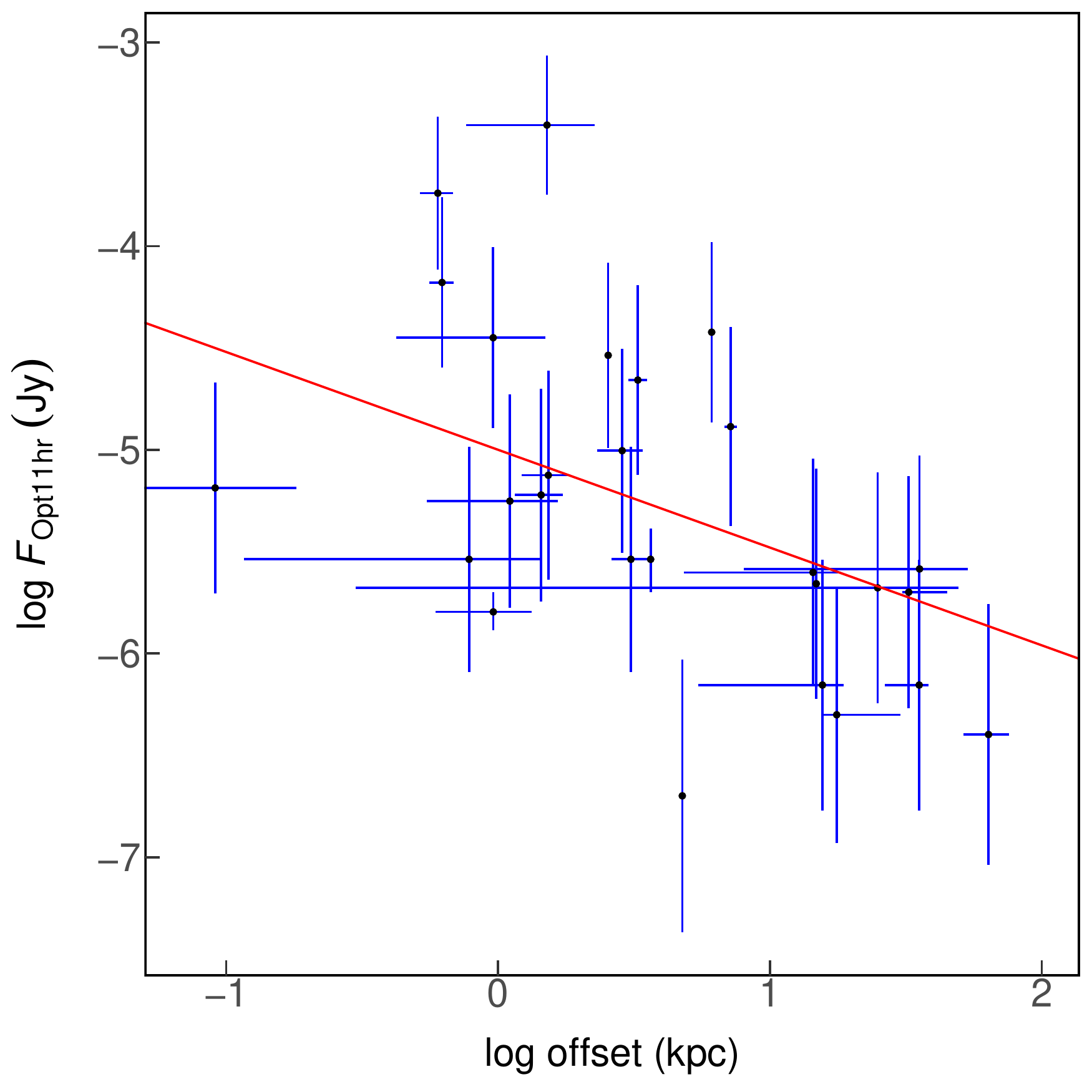}

\caption{
Scatter plot for $\log F_{\rm Opt11hr}$ and $\log offset$.
The red line is our fit result. The formula of the red line is $\log F_{\rm Opt11hr} = (-0.48 \pm 0.21) \times \log offset + (-5 \pm 0.16)$. The description of every parameter is in Section \ref{sec:sample}.
}
\label{fig:fohroffset}
\end{figure*}

The correlation between $\log F_{\rm Opt11hr}$ and $\log offset$ is:
\begin{equation} \label{eq:fohroffset}
\log F_{\rm Opt11hr} = (-0.48 \pm 0.21) \times \log offset + (-5 \pm 0.16),
\end{equation}
where $F_{\rm Opt11hr}$ is in unit of $\rm Jy$. Host galaxy offset is in unit of $\rm kpc$. The adjusted $R^{2}$ is 0.28. The Pearson coefficient is $-0.39 \pm 0.13$ with p-value $4.3 \times 10^{-2}$. The Spearman coefficient is $-0.46 \pm 0.12$ with p-value $1.6 \times 10^{-2}$. The Kendall $\tau$ coefficient is $-0.32 \pm 0.085$ with p-value $1.8 \times 10^{-2}$. The correlation ratio is $0.96 \pm 0.01$. The cosine similarity is $-0.6 \pm 0.12$. The scatter plot is in Figure \ref{fig:fohroffset}. The GRB sample number is 27.

\begin{figure*}[!b]
\centering
\includegraphics[width=0.45\textwidth]{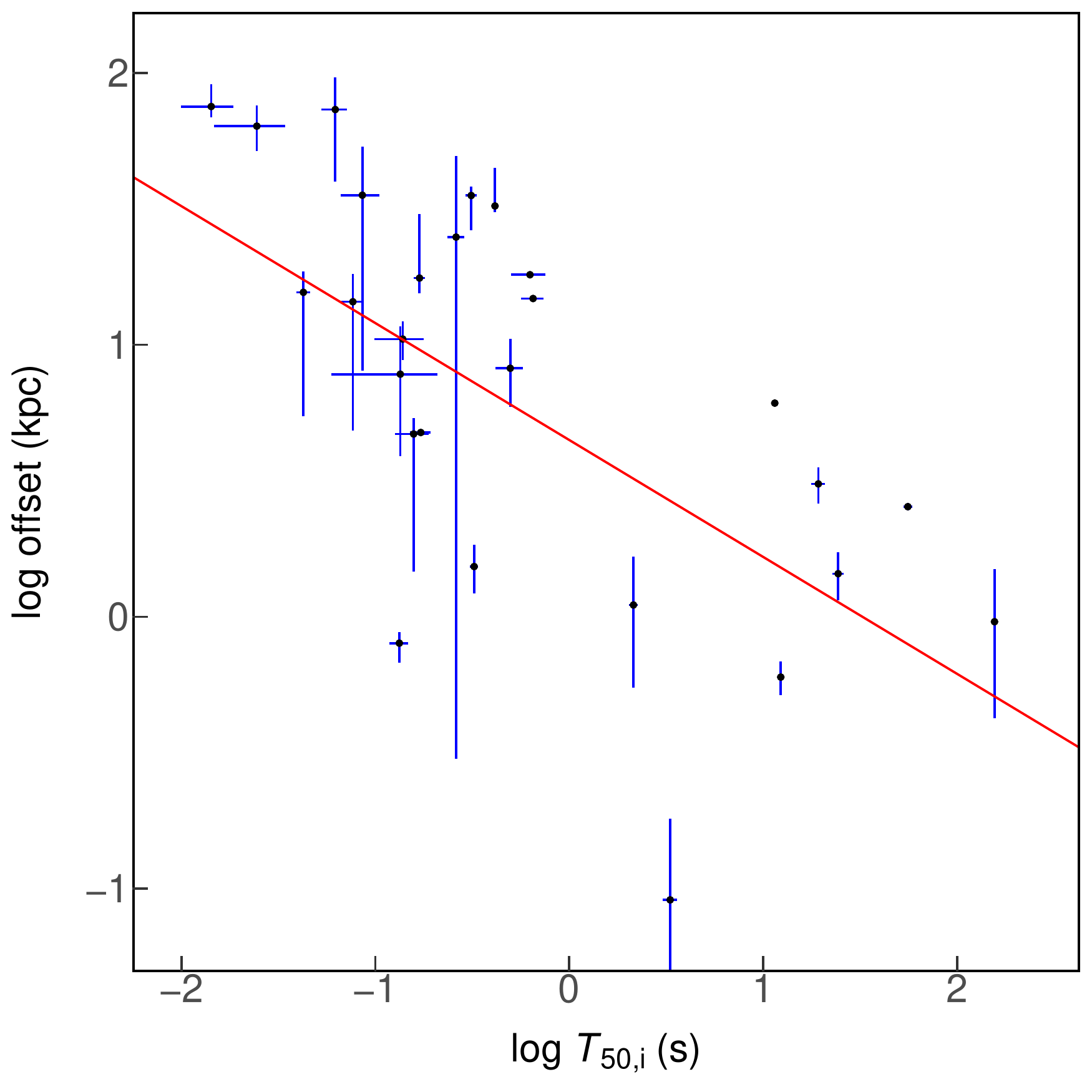}

\caption{
Scatter plot for $\log T_{\rm 50,i}$ and $\log offset$.
The red line is our fit result. The formula of the red line is $\log T_{\rm 50,i} = (-0.74 \pm 0.2) \times \log offset + (0.33 \pm 0.19)$. The description of every parameter is in Section \ref{sec:sample}.
}
\label{fig:t50ioffset}
\end{figure*}

The correlation between $\log T_{\rm 50,i}$ and $\log offset$ is:
\begin{equation} \label{eq:t50ioffset}
\log T_{\rm 50,i} = (-0.74 \pm 0.2) \times \log offset + (0.33 \pm 0.19),
\end{equation}
where host galaxy offset is in unit of $\rm kpc$. $T_{\rm 50,i}$ is in unit of $\rm s$. The adjusted $R^{2}$ is 0.4. The Pearson coefficient is $-0.56 \pm 0.076$ with p-value $2.6 \times 10^{-3}$. The Spearman coefficient is $-0.58 \pm 0.05$ with p-value $1.4 \times 10^{-3}$. The Kendall $\tau$ coefficient is $-0.42 \pm 0.042$ with p-value $2.2 \times 10^{-3}$. The correlation ratio is $0.46 \pm 0.05$. The cosine similarity is $-0.54 \pm 0.043$. The scatter plot is in Figure \ref{fig:t50ioffset}. The GRB sample number is 27. This is consistent with the classification of the long GRBs and the short GRBs. For short GRBs are thought occurring further from the center of the host galaxies, therefore, $T_{\rm 50,i}$  and offset is anti-correlated.

\begin{figure*}[!b]
\centering
\includegraphics[width=0.45\textwidth]{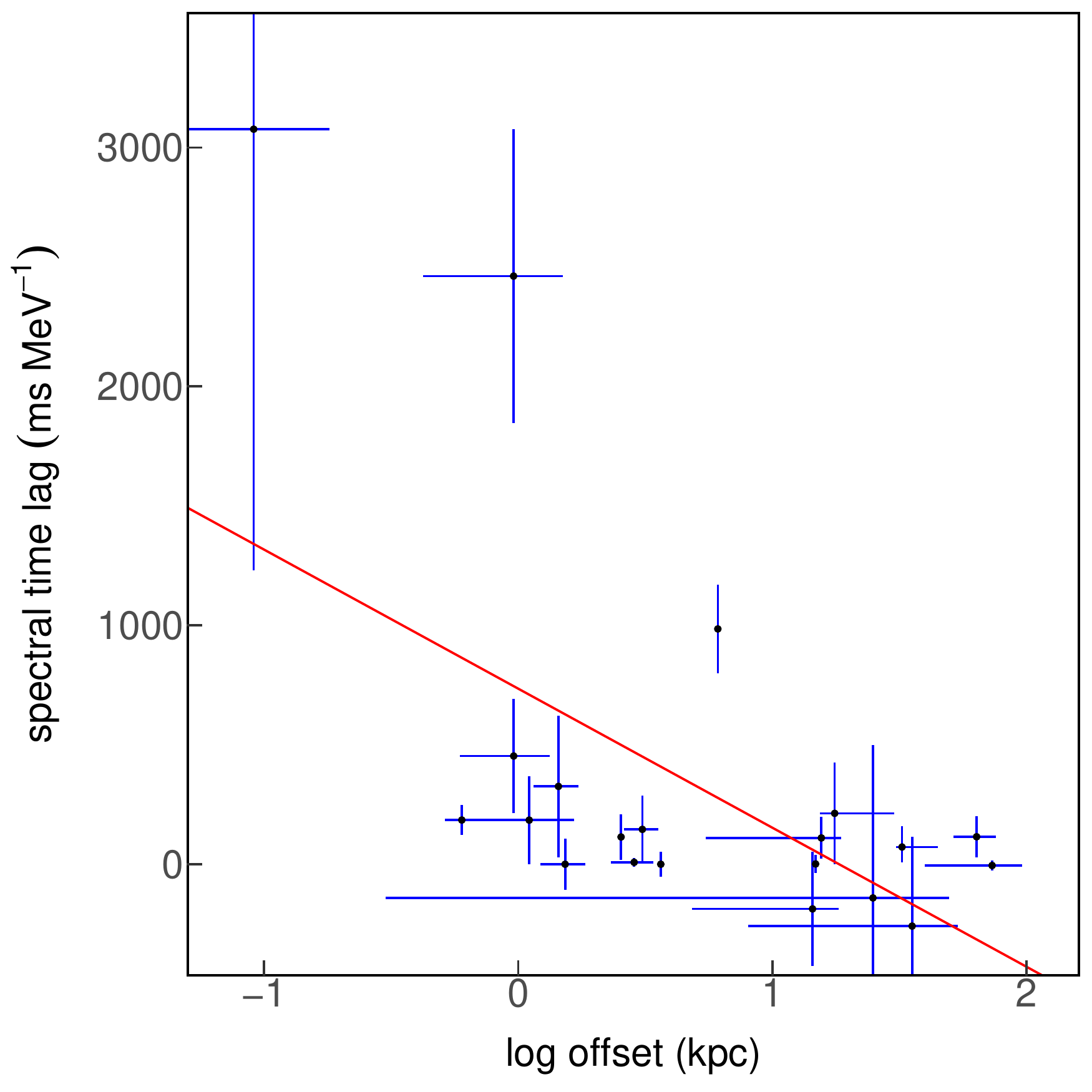}

\caption{
Scatter plot for spectral lag and $\log offset$.
The red line is our fit result. The formula of the red line is $spectral ~ lag = (-582 \pm 289) \times \log offset + (735 \pm 276)$. The description of every parameter is in Section \ref{sec:sample}.
}
\label{fig:lagoffset}
\end{figure*}

The correlation between spectral lag and $\log offset$ is:
\begin{equation} \label{eq:lagoffset}
spectral ~ lag = (-582 \pm 289) \times \log offset + (735 \pm 276),
\end{equation}
where Spectral time lag is in unit of $\rm ms ~ MeV^{\rm -1}$. Host galaxy offset is in unit of $\rm kpc$. The adjusted $R^{2}$ is 0.35. The Pearson coefficient is $-0.53 \pm 0.19$ with p-value $1.4 \times 10^{-2}$. The Spearman coefficient is $-0.41 \pm 0.15$ with p-value $6.5 \times 10^{-2}$. The Kendall $\tau$ coefficient is $-0.3 \pm 0.11$ with p-value $6.1 \times 10^{-2}$. The correlation ratio is $0.28 \pm 0.051$. The cosine similarity is $-0.17 \pm 0.18$. The scatter plot is in Figure \ref{fig:lagoffset}. The GRB sample number is 21.

\begin{figure*}[!b]
\centering
\includegraphics[width=0.45\textwidth]{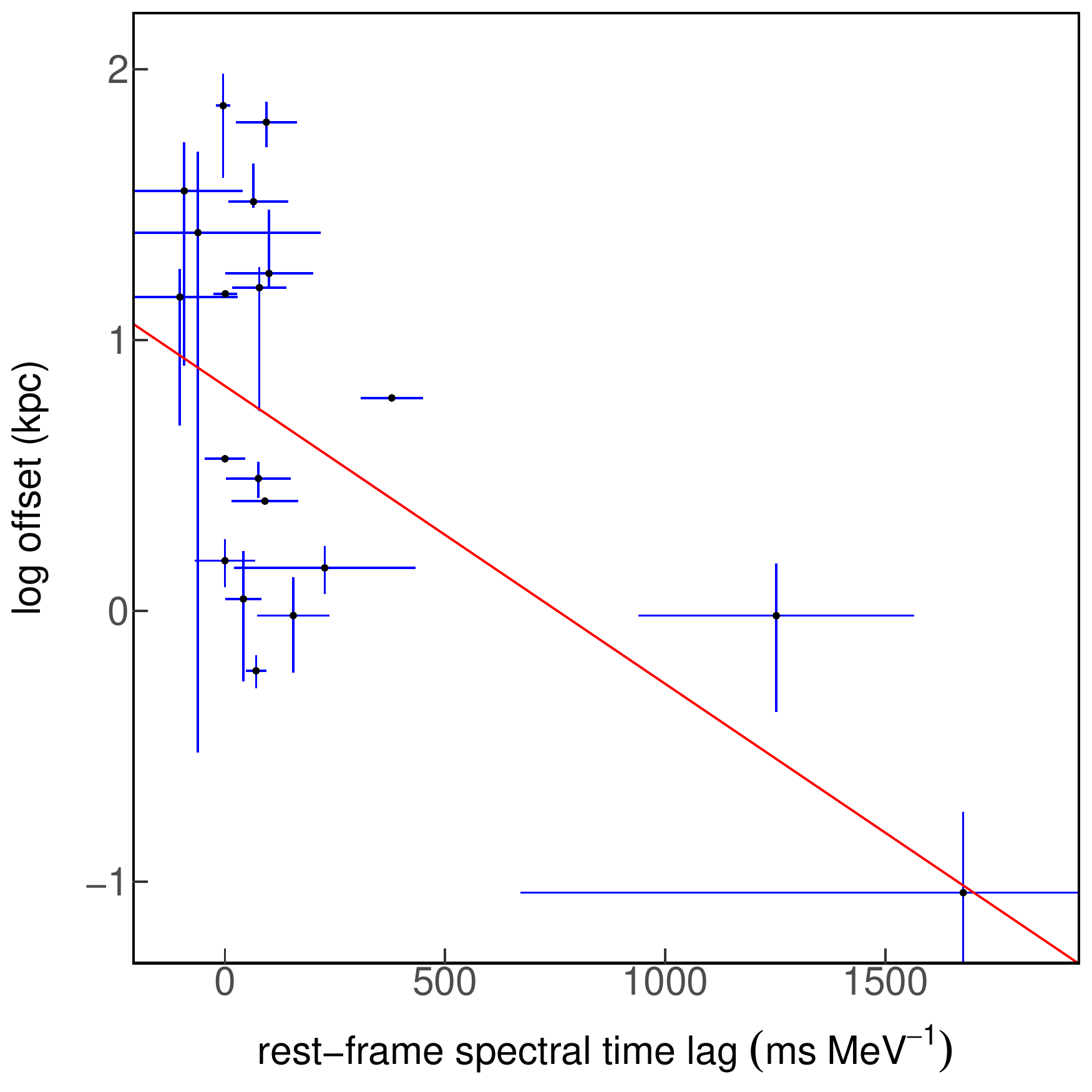}

\caption{
Scatter plot for rest-frame spectral lag and $\log offset$.
The red line is our fit result. The formula of the red line is $rest-frame ~ spectral ~ lag = (-306 \pm 154) \times \log offset + (396 \pm 151)$. The description of every parameter is in Section \ref{sec:sample}.
}
\label{fig:lagioffset}
\end{figure*}

The correlation between rest-frame spectral lag and $\log offset$ is:
\begin{equation} \label{eq:lagioffset}
rest-frame ~ spectral ~ lag = (-306 \pm 154) \times \log offset + (396 \pm 151),
\end{equation}
where rest-frame spectral lag is in unit of $\rm ms ~ MeV^{\rm -1}$. Host galaxy offset is in unit of $\rm kpc$. The adjusted $R^{2}$ is 0.35. The Pearson coefficient is $-0.53 \pm 0.19$ with p-value $1.6 \times 10^{-2}$. The Spearman coefficient is $-0.35 \pm 0.15$ with p-value 0.12. The Kendall $\tau$ coefficient is $-0.25 \pm 0.11$ with p-value 0.12. The correlation ratio is $0.29 \pm 0.05$. The cosine similarity is $-0.17 \pm 0.19$. The scatter plot is in Figure \ref{fig:lagioffset}. The GRB sample number is 20. This may also comes from the difference of long GRBs and short GRBs. Short GRBs are thought having smaller spectral lags. Noticing the large scattering, one may not get much information from it.

\subsection{Some correlations about $t_{\rm radio,pk}$}

\begin{figure*}[!b]
\centering
\includegraphics[width=0.45\textwidth]{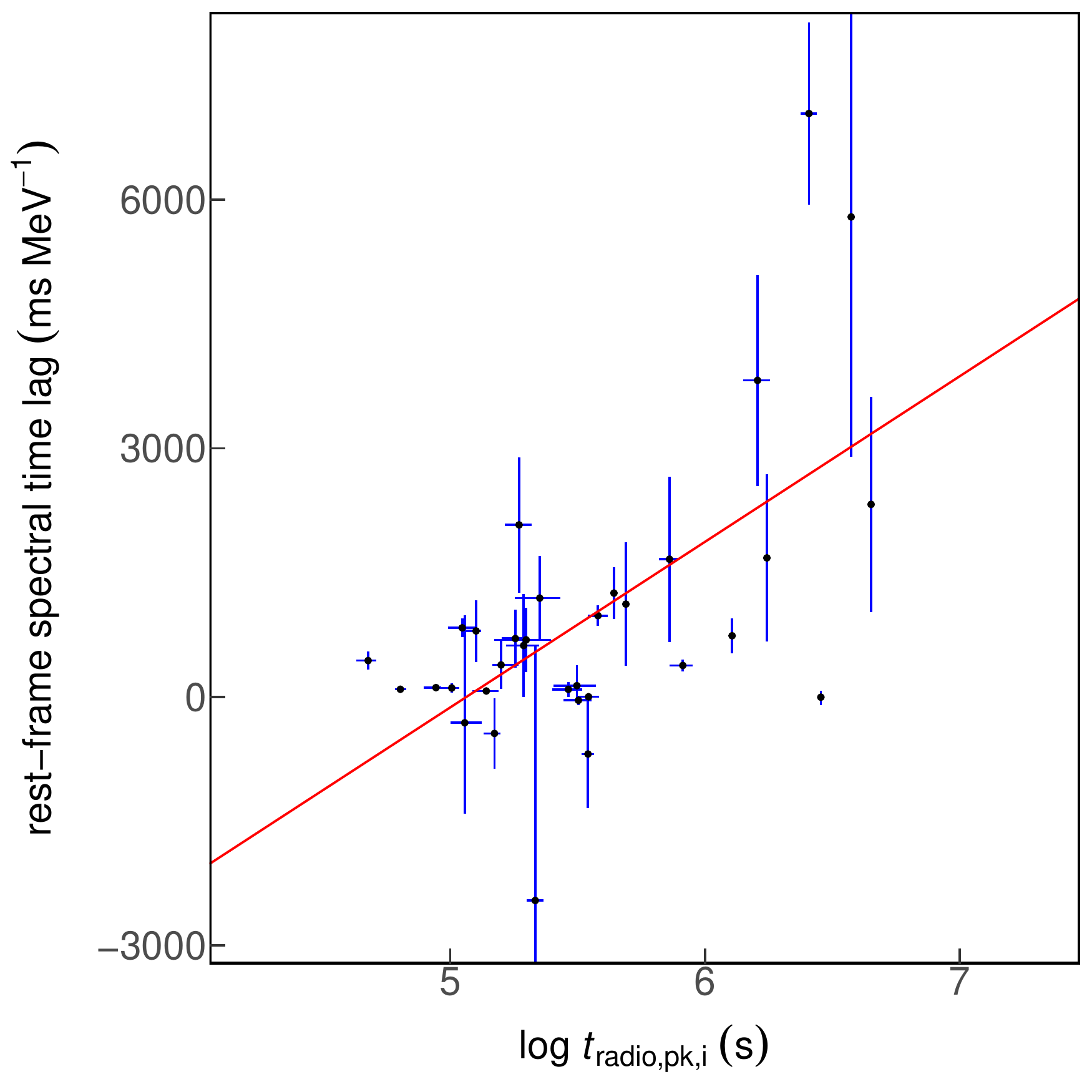}

\caption{
Scatter plot for rest-frame spectral lag and $\log t_{\rm radio,pk,i}$.
The red line is our fit result. The formula of the red line is $rest-frame ~ spectral ~ lag = (1998 \pm 431) \times \log t_{\rm radio,pk,i} + (-10118 \pm 2304)$. The description of every parameter is in Section \ref{sec:sample}.
}
\label{fig:lagitrapki}
\end{figure*}

The correlation between rest-frame spectral lag and $\log t_{\rm radio,pk,i}$ is:
\begin{equation} \label{eq:lagitrapki}
rest-frame ~ spectral ~ lag = (1998 \pm 431) \times \log t_{\rm radio,pk,i} + (-10118 \pm 2304),
\end{equation}
where rest-frame spectral lag is in unit of $\rm ms ~ MeV^{\rm -1}$. $t_{\rm radio,pk,i}$ is in unit of $\rm s$. The adjusted $R^{2}$ is 0.34. The Pearson coefficient is $0.53 \pm 0.072$ with p-value $1.3 \times 10^{-3}$. The Spearman coefficient is $0.38 \pm 0.076$ with p-value $2.7 \times 10^{-2}$. The Kendall $\tau$ coefficient is $0.27 \pm 0.056$ with p-value $2.6 \times 10^{-2}$. The correlation ratio is $0.33 \pm 0.061$. The cosine similarity is $0.48 \pm 0.075$. The scatter plot is in Figure \ref{fig:lagitrapki}. The GRB sample number is 33. This is an interesting correlations. The spectral lag is a quantity of the prompt emission, while the physical origin is still not clear. The peak time of the radio emission $t_{\rm radio,pk}$ is a quantity of the late afterglow. The spectral lag might be related to the radiation mechanism, while $t_{\rm radio,pk}$  is more likely related to the total energy and the environment. They are not likely to be related to each other. The reason of this correlation is not clear.

\begin{figure*}[!b]
\centering
\includegraphics[width=0.45\textwidth]{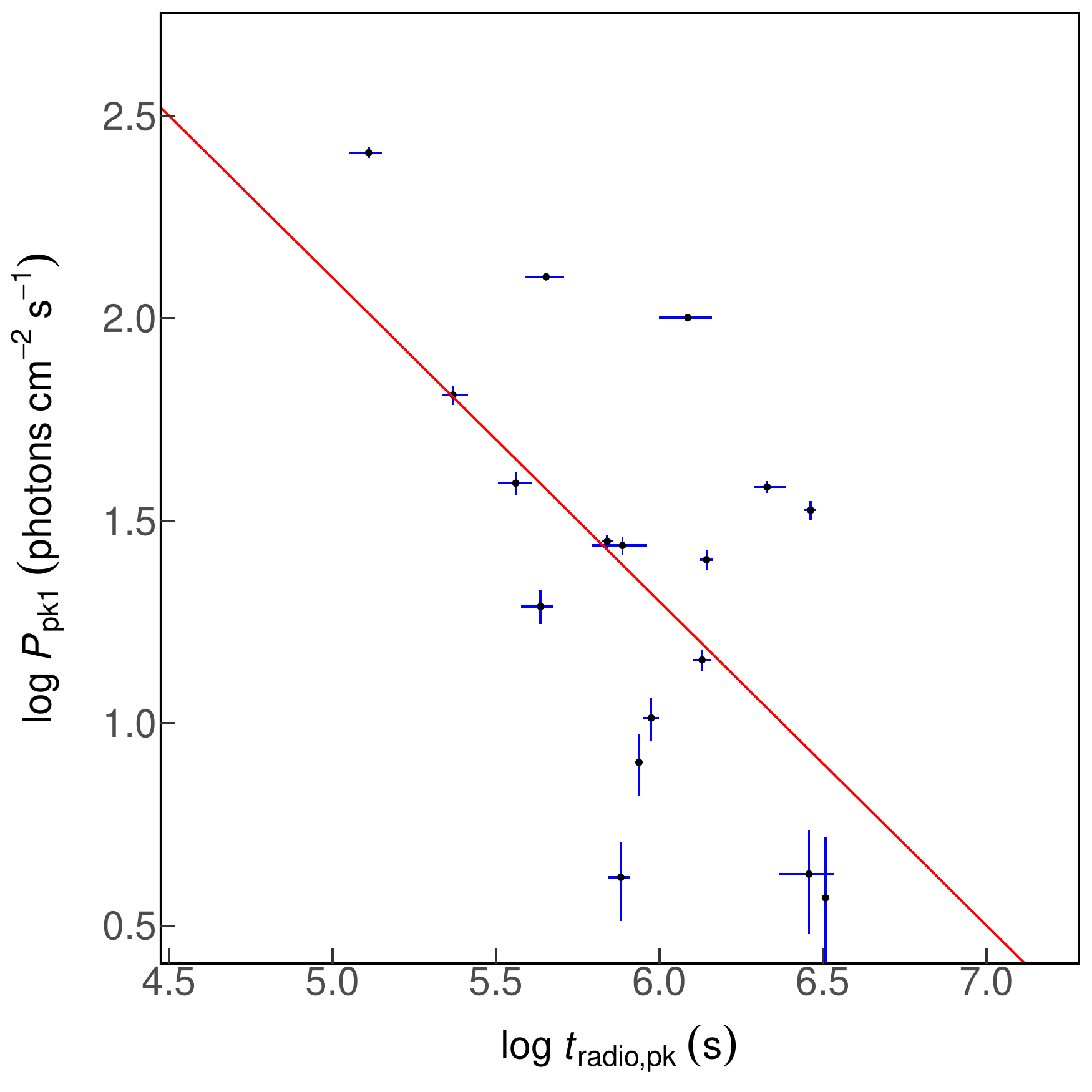}

\caption{
Scatter plot for $\log P_{\rm pk1}$ and $\log t_{\rm radio,pk}$.
The red line is our fit result. The formula of the red line is $\log P_{\rm pk1} = (-0.8 \pm 0.066) \times \log t_{\rm radio,pk} + (6.1 \pm 0.38)$. The description of every parameter is in Section \ref{sec:sample}.
}
\label{fig:ppk1trapk}
\end{figure*}

The correlation between $\log P_{\rm pk1}$ and $\log t_{\rm radio,pk}$ is:
\begin{equation} \label{eq:ppk1trapk}
\log P_{\rm pk1} = (-0.8 \pm 0.066) \times \log t_{\rm radio,pk} + (6.1 \pm 0.38),
\end{equation}
where $P_{\rm pk1}$ is peak photon flux of 64 $\rm ms$ time bin in 10-1000 $\rm keV$, and in unit of $\rm photons ~ cm^{\rm -2} ~ s^{\rm -1}$. $t_{\rm radio,pk}$ is in unit of $\rm s$. The adjusted $R^{2}$ is 0.31. The Pearson coefficient is $-0.59 \pm 0.032$ with p-value $1.3 \times 10^{-2}$. The Spearman coefficient is $-0.5 \pm 0.044$ with p-value $4.2 \times 10^{-2}$. The Kendall $\tau$ coefficient is $-0.36 \pm 0.035$ with p-value $4.8 \times 10^{-2}$. The correlation ratio is $0.98 \pm 0.0011$. The cosine similarity is $0.92 \pm 0.0077$. The scatter plot is in Figure \ref{fig:ppk1trapk}. The GRB sample number is 17.

\begin{figure*}[!b]
\centering
\includegraphics[width=0.45\textwidth]{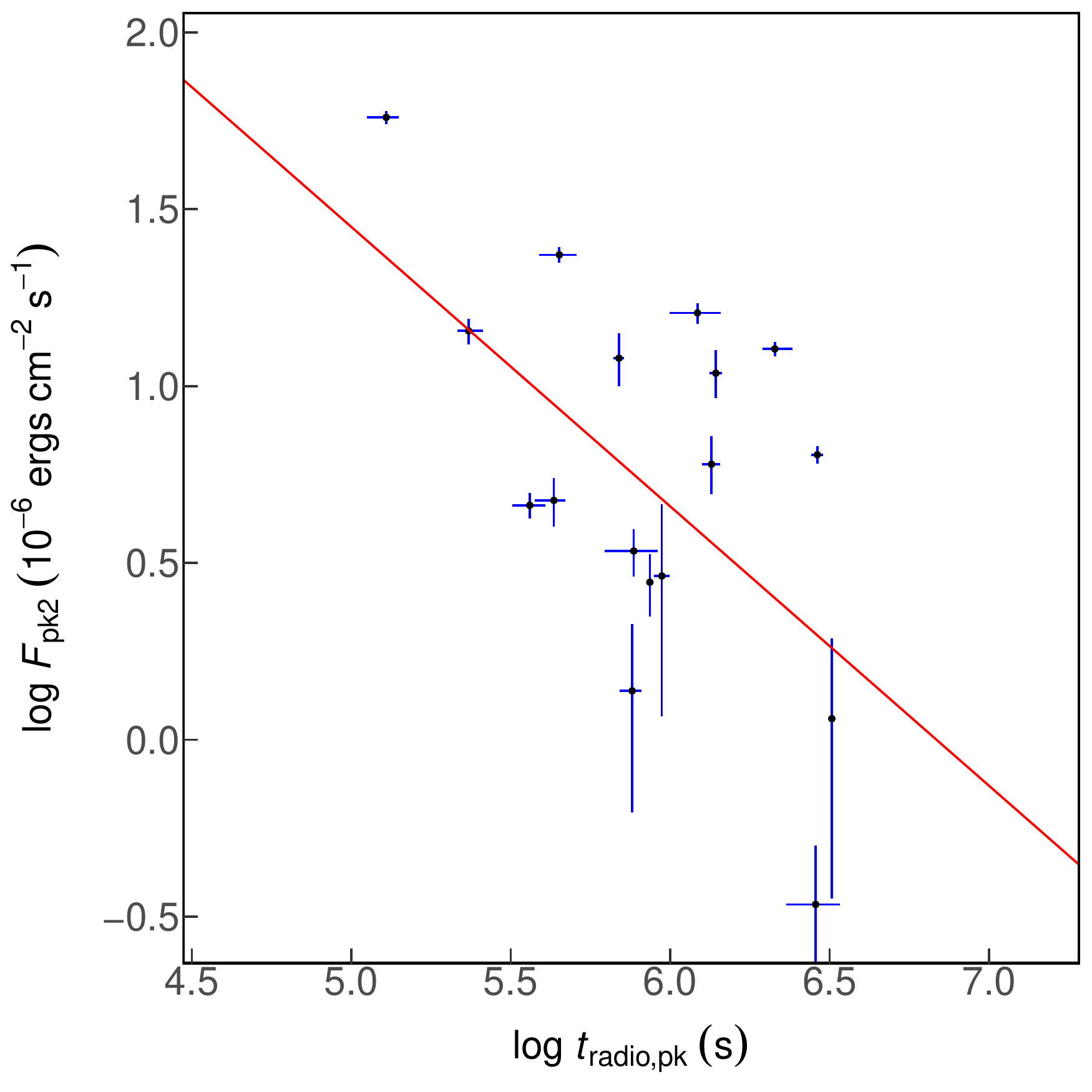}

\caption{
Scatter plot for $\log  F_{\rm pk2}$ and $\log t_{\rm radio,pk}$.
The red line is our fit result. The formula of the red line is $\log  F_{\rm pk2} = (-0.79 \pm 0.11) \times \log t_{\rm radio,pk} + (5.4 \pm 0.63)$. The description of every parameter is in Section \ref{sec:sample}.
}
\label{fig:fpk2trapk}
\end{figure*}

The correlation between $\log  F_{\rm pk2}$ and $\log t_{\rm radio,pk}$ is:
\begin{equation} \label{eq:fpk2trapk}
\log  F_{\rm pk2} = (-0.79 \pm 0.11) \times \log t_{\rm radio,pk} + (5.4 \pm 0.63),
\end{equation}
where $F_{\rm pk2}$ is peak energy flux of 64 $\rm ms$ time bin in rest-frame 1-$10^{4}$ $\rm keV$ energy band, and in unit of $\rm 10^{\rm -6} ~ ergs ~ cm^{\rm -2} ~ s^{\rm -1}$. $t_{\rm radio,pk}$ is in unit of $\rm s$. The adjusted $R^{2}$ is 0.27. The Pearson coefficient is $-0.54 \pm 0.049$ with p-value $2.5 \times 10^{-2}$. The Spearman coefficient is $-0.41 \pm 0.049$ with p-value 0.11. The Kendall $\tau$ coefficient is $-0.3 \pm 0.043$ with p-value $9.9 \times 10^{-2}$. The correlation ratio is $0.98 \pm 0.002$. The cosine similarity is $0.77 \pm 0.044$. The scatter plot is in Figure \ref{fig:fpk2trapk}. The GRB sample number is 17.

\subsection{Correlations about $N_{\rm H}$}

\begin{figure*}[!b]
\centering
\includegraphics[width=0.45\textwidth]{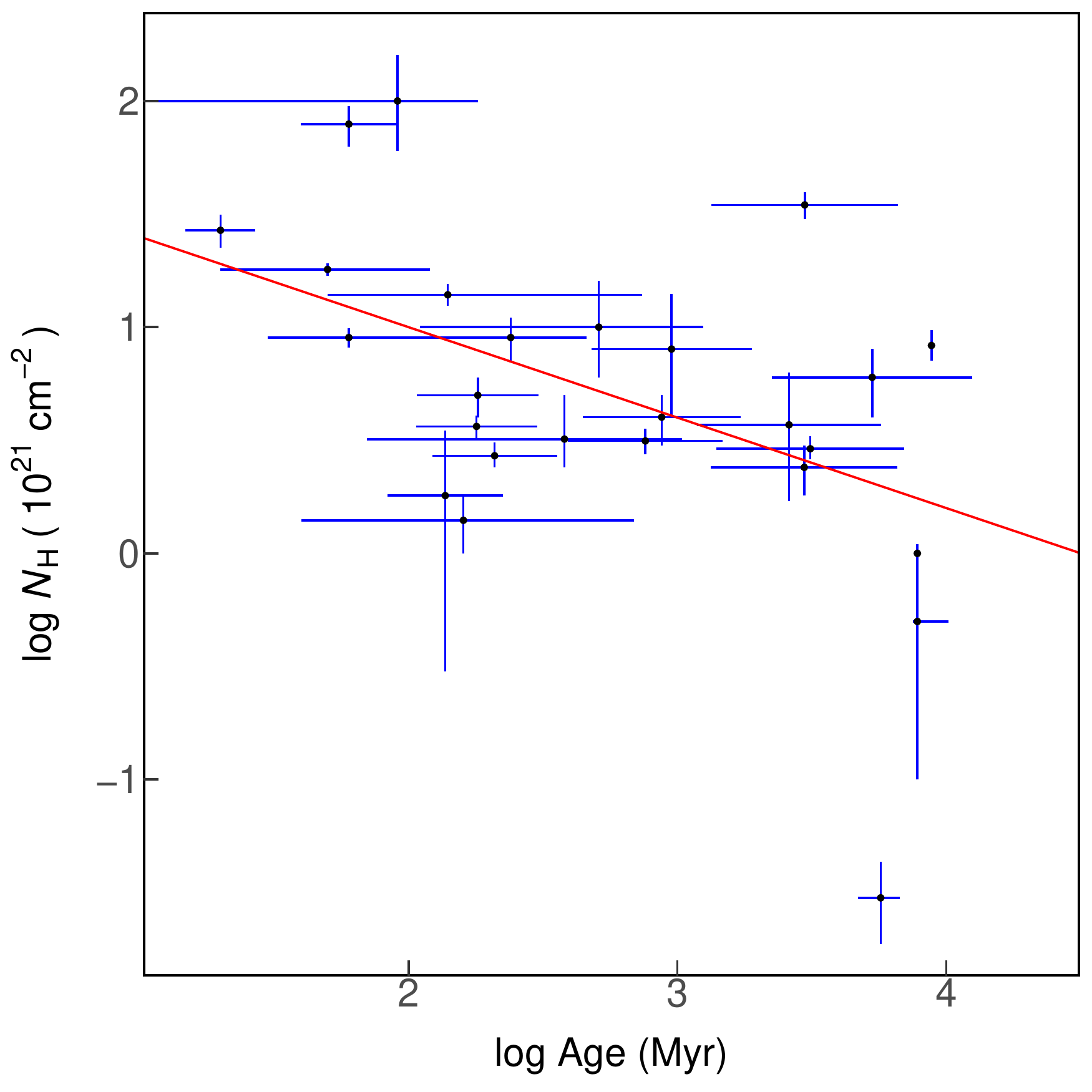}

\caption{
Scatter plot for $\log  N_{\rm H}$ and $\log Age$.
The red line is our fit result. The formula of the red line is $\log  N_{\rm H} = (-0.4 \pm 0.075) \times \log Age + (1.8 \pm 0.21)$. The description of every parameter is in Section \ref{sec:sample}.
}
\label{fig:nhage}
\end{figure*}

The correlation between $\log  N_{\rm H}$ and $\log Age$ is:
\begin{equation} \label{eq:nhage}
\log  N_{\rm H} = (-0.4 \pm 0.075) \times \log Age + (1.8 \pm 0.21),
\end{equation}
where $N_{\rm H}$ is in unit of $\rm 10^{\rm 21} ~ cm^{\rm -2}$. Age is in unit of $\rm Myr$. The adjusted $R^{2}$ is 0.25. The Pearson coefficient is $-0.49 \pm 0.077$ with p-value $1.2 \times 10^{-2}$. The Spearman coefficient is $-0.45 \pm 0.083$ with p-value $2.2 \times 10^{-2}$. The Kendall $\tau$ coefficient is $-0.32 \pm 0.067$ with p-value $2.1 \times 10^{-2}$. The correlation ratio is $0.78 \pm 0.034$. The cosine similarity is $0.53 \pm 0.055$. The scatter plot is in Figure \ref{fig:nhage}. The GRB sample number is 26. This may reveals the property of the evolution of the host galaxies, that more material has been formed into stars for the older galaxies.

\begin{figure*}[!b]
\centering
\includegraphics[width=0.45\textwidth]{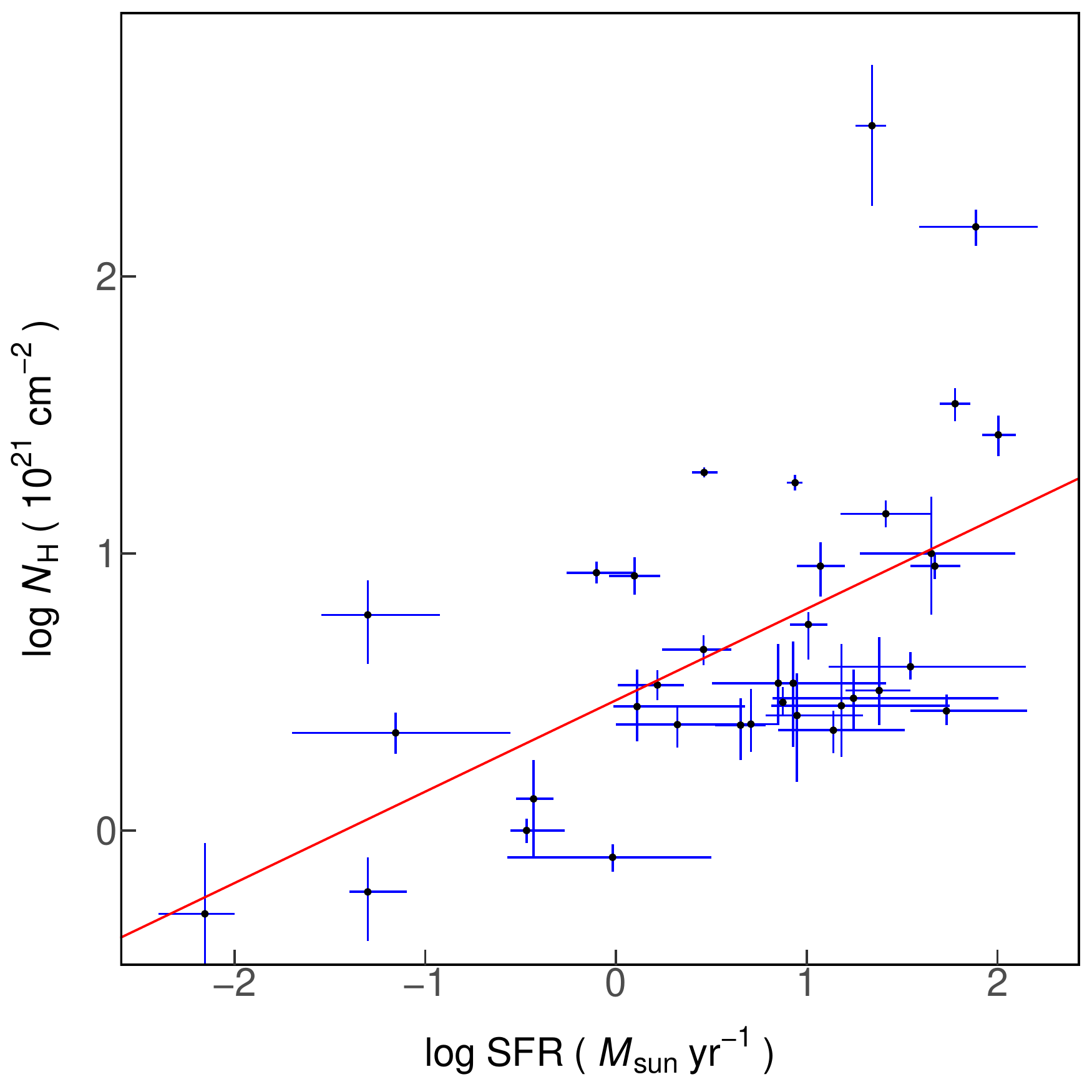}

\caption{
Scatter plot for $\log N_{\rm H}$ and $\log SFR$.
The red line is our fit result. The formula of the red line is $\log N_{\rm H} = (0.33 \pm 0.052) \times \log SFR + (0.47 \pm 0.049)$. The description of every parameter is in Section \ref{sec:sample}.
}
\label{fig:nhsfr}
\end{figure*}

The correlation between $\log N_{\rm H}$ and $\log SFR$ is:
\begin{equation} \label{eq:nhsfr}
\log N_{\rm H} = (0.33 \pm 0.052) \times \log SFR + (0.47 \pm 0.049),
\end{equation}
where $N_{\rm H}$ is in unit of $\rm 10^{\rm 21} ~ cm^{\rm -2}$. SFR is in unit of $\rm M_{\bigodot} ~ yr^{\rm -1}$. The adjusted $R^{2}$ is 0.31. The Pearson coefficient is $0.55 \pm 0.056$ with p-value $5.6 \times 10^{-4}$. The Spearman coefficient is $0.52 \pm 0.056$ with p-value $1.3 \times 10^{-3}$. The Kendall $\tau$ coefficient is $0.37 \pm 0.046$ with p-value $1.6 \times 10^{-3}$. The correlation ratio is $0.03 \pm 0.022$. The cosine similarity is $0.71 \pm 0.034$. The scatter plot is in Figure \ref{fig:nhsfr}. The GRB sample number is 36. It is clear that in a star forming galaxy, the material must be rich. Consequently, the column density of hydrogen is high.

\begin{figure*}[!b]
\centering
\includegraphics[width=0.45\textwidth]{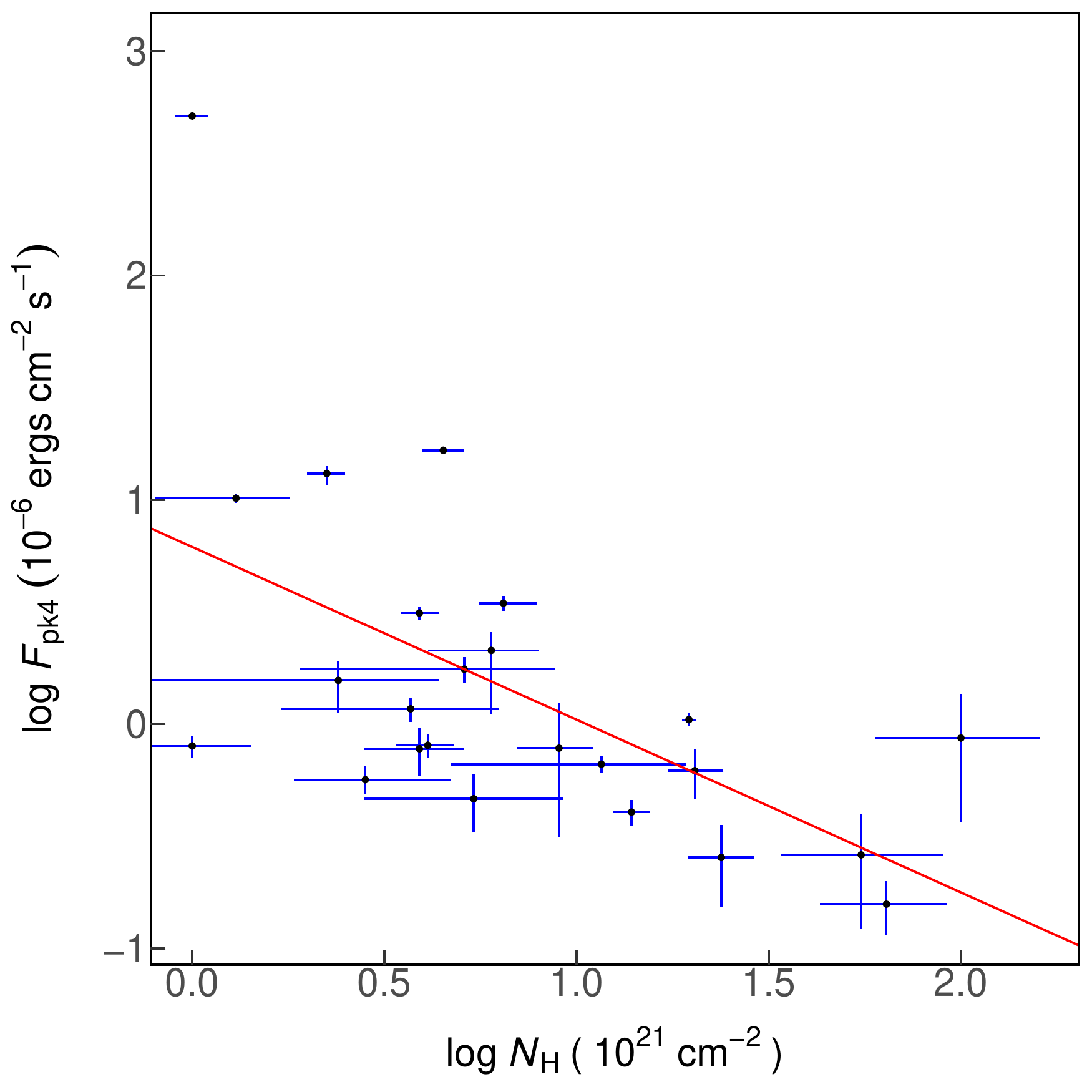}

\caption{
Scatter plot for $\log N_{\rm H}$ and $\log F_{\rm pk4}$.
The red line is our fit result. The formula of the red line is $\log N_{\rm H} = (-0.43 \pm 0.037) \times \log F_{\rm pk4} + (0.89 \pm 0.041)$. The description of every parameter is in Section \ref{sec:sample}.
}
\label{fig:nhfpk4}
\end{figure*}

The correlation between $\log N_{\rm H}$ and $\log F_{\rm pk4}$ is:
\begin{equation} \label{eq:nhfpk4}
\log N_{\rm H} = (-0.43 \pm 0.037) \times \log F_{\rm pk4} + (0.89 \pm 0.041),
\end{equation}
where $F_{\rm pk4}$ is peak energy flux of 1024 $\rm ms$ time bin in rest-frame 1-$10^{4}$ $\rm keV$ energy band, and in unit of $\rm 10^{\rm -6} ~ ergs ~ cm^{\rm -2} ~ s^{\rm -1}$. $N_{\rm H}$ is in unit of $\rm 10^{\rm 21} ~ cm^{\rm -2}$. The adjusted $R^{2}$ is 0.35. The Pearson coefficient is $-0.58 \pm 0.055$ with p-value $3.1 \times 10^{-3}$. The Spearman coefficient is $-0.58 \pm 0.078$ with p-value $3 \times 10^{-3}$. The Kendall $\tau$ coefficient is $-0.43 \pm 0.061$ with p-value $3.4 \times 10^{-3}$. The correlation ratio is $0.45 \pm 0.027$. The cosine similarity is $-0.16 \pm 0.05$. The scatter plot is in Figure \ref{fig:nhfpk4}. The GRB sample number is 24.

\begin{figure*}[!b]
\centering
\includegraphics[width=0.45\textwidth]{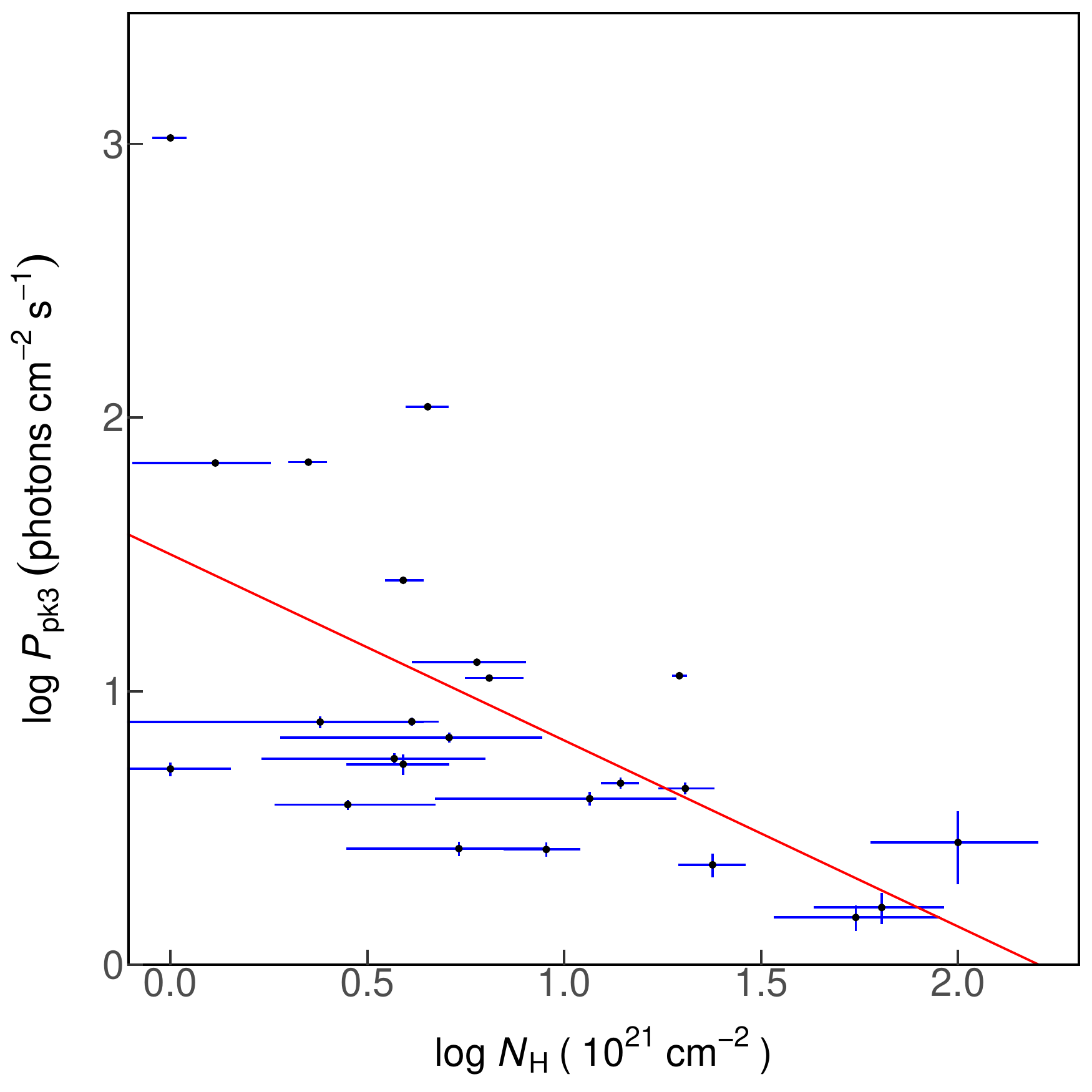}

\caption{
Scatter plot for $\log N_{\rm H}$ and $\log P_{\rm pk3}$.
The red line is our fit result. The formula of the red line is $\log N_{\rm H} = (-0.51 \pm 0.038) \times \log P_{\rm pk3} + (1.3 \pm 0.065)$. The description of every parameter is in Section \ref{sec:sample}.
}
\label{fig:nhppk3}
\end{figure*}

The correlation between $\log N_{\rm H}$ and $\log P_{\rm pk3}$ is:
\begin{equation} \label{eq:nhppk3}
\log N_{\rm H} = (-0.51 \pm 0.038) \times \log P_{\rm pk3} + (1.3 \pm 0.065),
\end{equation}
where $P_{\rm pk3}$ is peak photon flux of 1024 $\rm ms$ time bin in 10-1000 $\rm keV$, and in unit of $\rm photons ~ cm^{\rm -2} ~ s^{\rm -1}$. $N_{\rm H}$ is in unit of $\rm 10^{\rm 21} ~ cm^{\rm -2}$. The adjusted $R^{2}$ is 0.36. The Pearson coefficient is $-0.59 \pm 0.049$ with p-value $2.7 \times 10^{-3}$. The Spearman coefficient is $-0.61 \pm 0.058$ with p-value $1.5 \times 10^{-3}$. The Kendall $\tau$ coefficient is $-0.45 \pm 0.05$ with p-value $2.1 \times 10^{-3}$. The correlation ratio is $0.1 \pm 0.031$. The cosine similarity is $0.49 \pm 0.026$. The scatter plot is in Figure \ref{fig:nhppk3}. The GRB sample number is 24.

\subsection{Correlations about Age of the host galaxies}

\begin{figure*}[!b]
\centering
\includegraphics[width=0.45\textwidth]{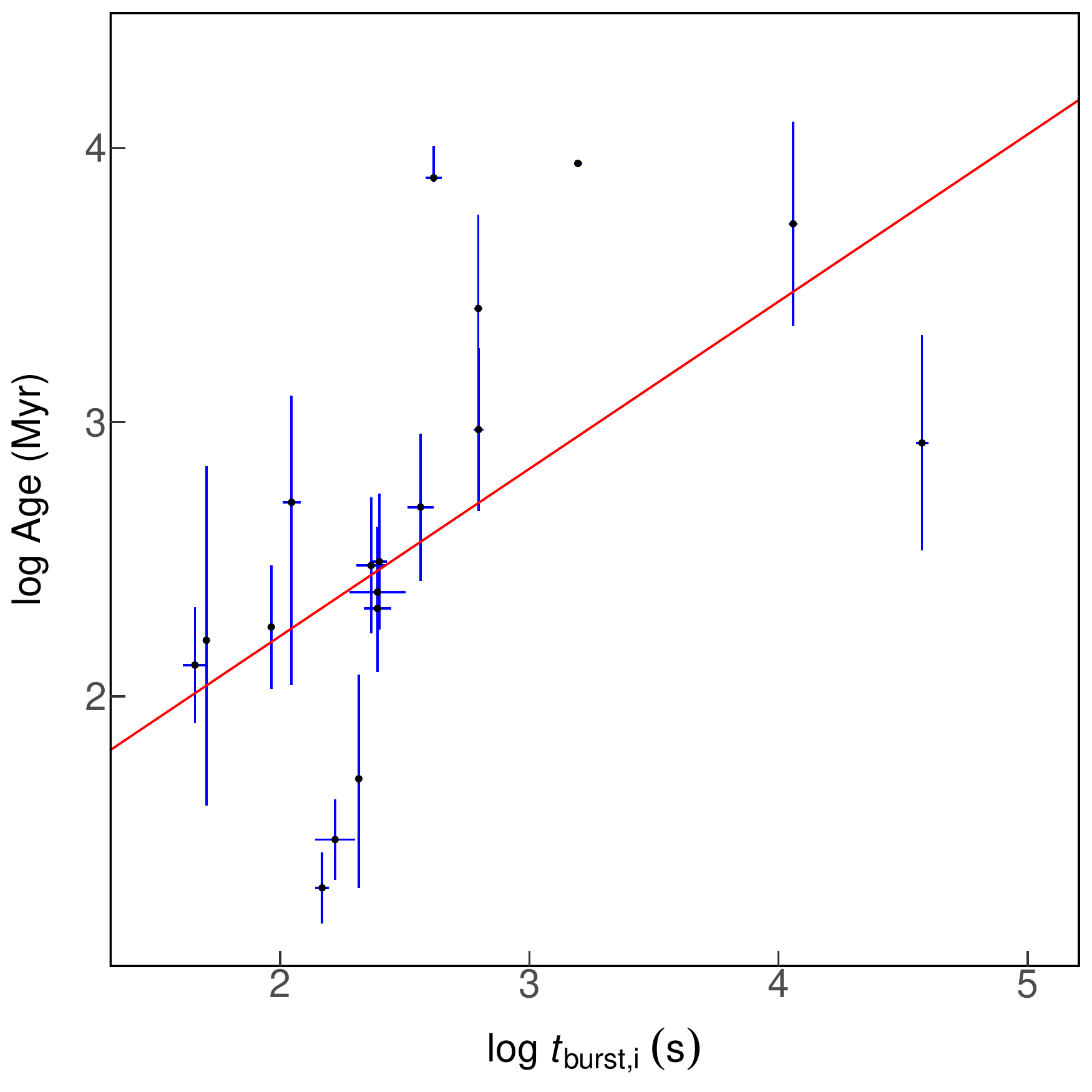}

\caption{
Scatter plot for $\log Age$ and $\log t_{\rm burst,i}$.
The red line is our fit result. The formula of the red line is $\log Age = (0.61 \pm 0.12) \times \log t_{\rm burst,i} + (1 \pm 0.33)$. The description of every parameter is in Section \ref{sec:sample}.
}
\label{fig:agetbti}
\end{figure*}

The correlation between $\log Age$ and $\log t_{\rm burst,i}$ is:
\begin{equation} \label{eq:agetbti}
\log Age = (0.61 \pm 0.12) \times \log t_{\rm burst,i} + (1 \pm 0.33),
\end{equation}
where Age is in unit of $\rm Myr$. $t_{\rm burst,i}$ is in unit of $\rm s$. The adjusted $R^{2}$ is 0.3. The Pearson coefficient is $0.54 \pm 0.093$ with p-value $2 \times 10^{-2}$. The Spearman coefficient is $0.69 \pm 0.094$ with p-value $1.6 \times 10^{-3}$. The Kendall $\tau$ coefficient is $0.51 \pm 0.08$ with p-value $3.5 \times 10^{-3}$. The correlation ratio is $0.044 \pm 0.032$. The cosine similarity is $0.96 \pm 0.0073$. The scatter plot is in Figure \ref{fig:agetbti}. The GRB sample number is 18. $t_{\rm burst,i}$ indicates the active time scale of the GRB central engine. It is unclear why the older host galaxies contain GRBs with longer activity.

\begin{figure*}[!b]
\centering
\includegraphics[width=0.45\textwidth]{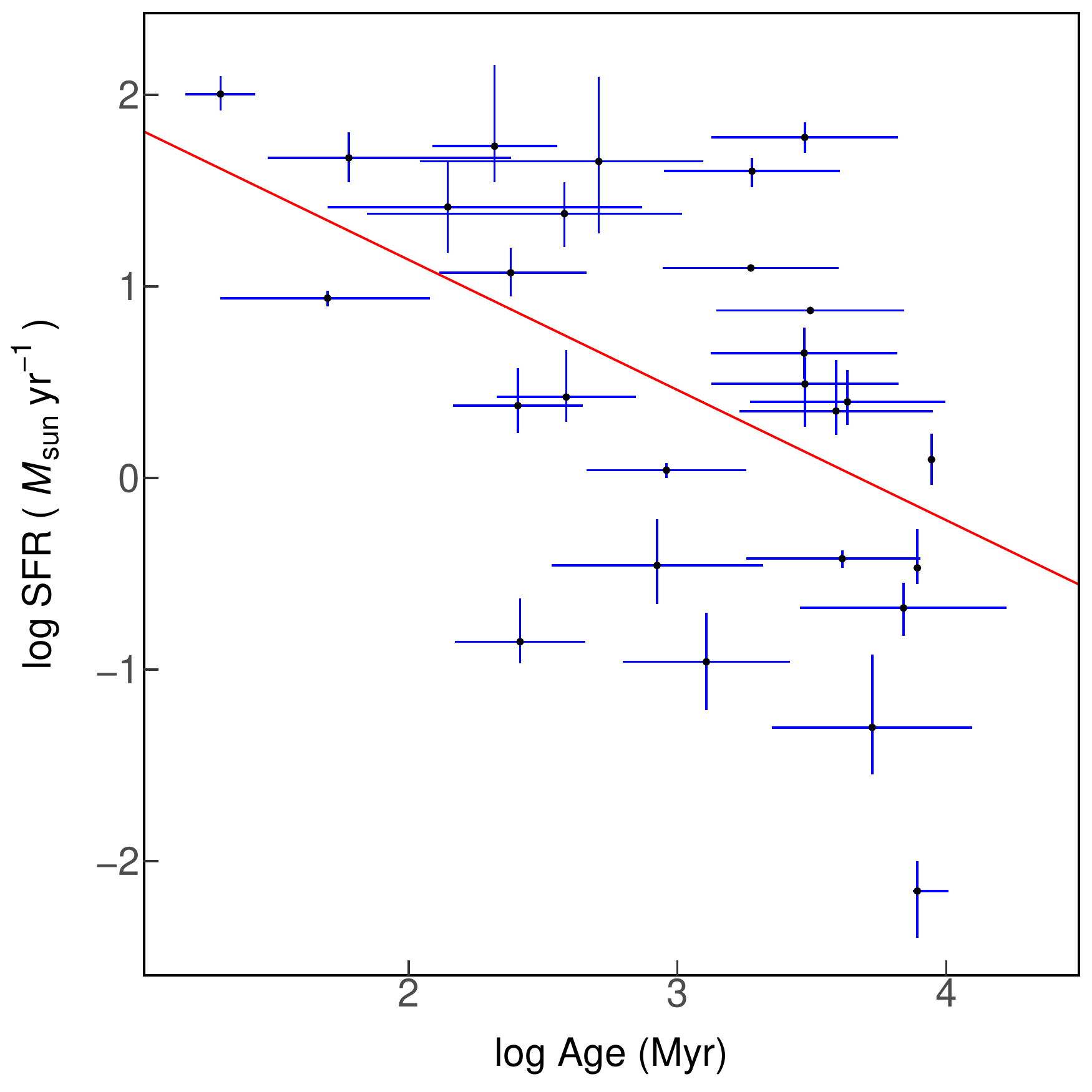}

\caption{
Scatter plot for $\log Age$ and $\log SFR$.
The red line is our fit result. The formula of the red line is $\log Age = (-0.38 \pm 0.059) \times \log SFR + (3.2 \pm 0.065)$. The description of every parameter is in Section \ref{sec:sample}.
}
\label{fig:agesfr}
\end{figure*}

The correlation between $\log Age$ and $\log SFR$ is:
\begin{equation} \label{eq:agesfr}
\log Age = (-0.38 \pm 0.059) \times \log SFR + (3.2 \pm 0.065),
\end{equation}
where SFR is in unit of $\rm M_{\bigodot} ~ yr^{\rm -1}$. Age is in unit of $\rm Myr$. The adjusted $R^{2}$ is 0.28. The Pearson coefficient is $-0.5 \pm 0.067$ with p-value $6.4 \times 10^{-3}$. The Spearman coefficient is $-0.49 \pm 0.079$ with p-value $7.6 \times 10^{-3}$. The Kendall $\tau$ coefficient is $-0.35 \pm 0.062$ with p-value $9.1 \times 10^{-3}$. The correlation ratio is $0.8 \pm 0.014$. The cosine similarity is $0.27 \pm 0.035$. The scatter plot is in Figure \ref{fig:agesfr}. The GRB sample number is 28. It is natural that star forming is more active in a younger galaxy. The correlation between the age and the SFR is an evidence.

\begin{figure*}[!b]
\centering
\includegraphics[width=0.45\textwidth]{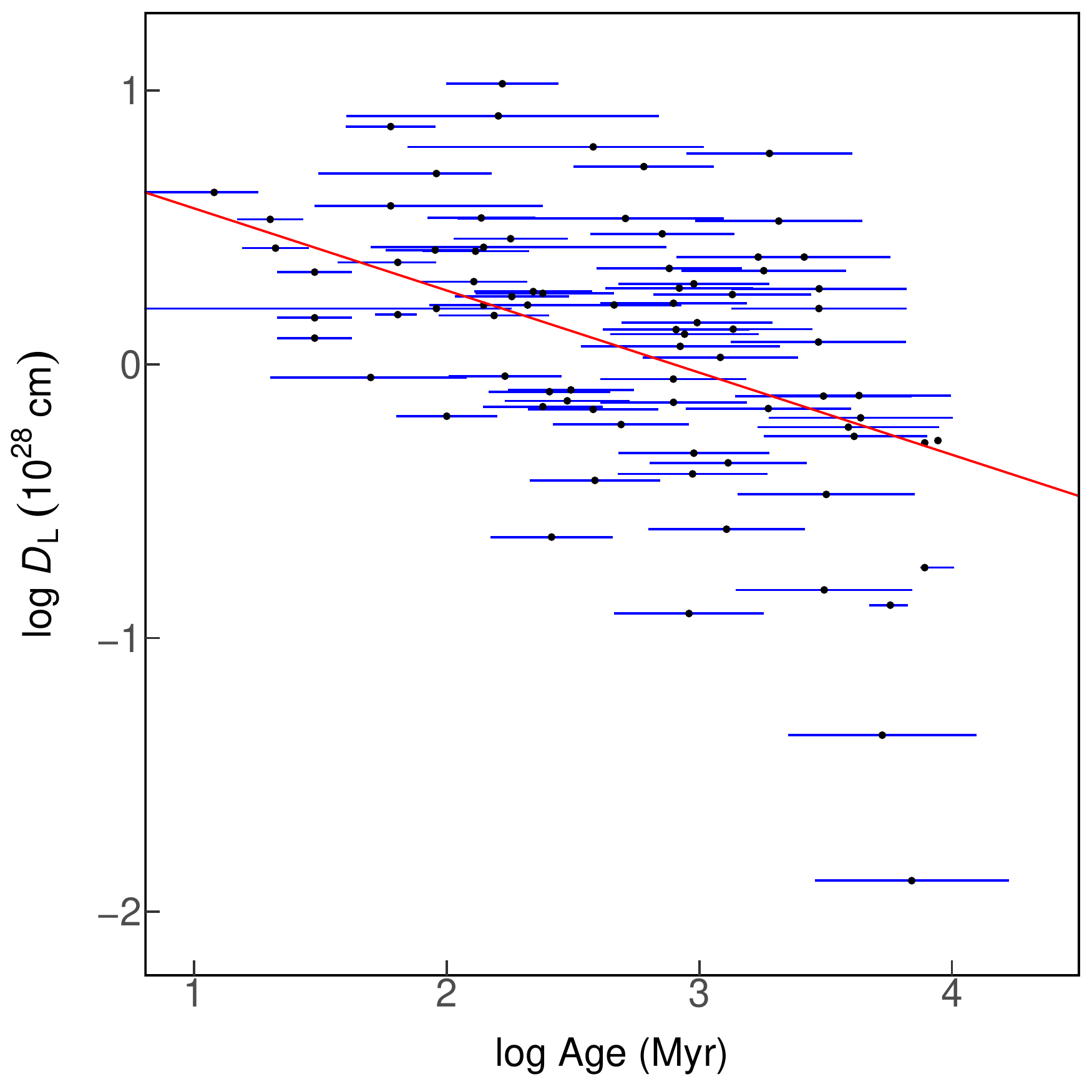}

\caption{
Scatter plot for $\log  Age$ and $\log D_{\rm L}$.
The red line is our fit result. The formula of the red line is $\log  Age = (-0.75 \pm 0.078) \times \log D_{\rm L} + (2.7 \pm 0.036)$. The description of every parameter is in Section \ref{sec:sample}.
}
\label{fig:agedl}
\end{figure*}

The correlation between $\log  Age$ and $\log D_{\rm L}$ is:
\begin{equation} \label{eq:agedl}
\log  Age = (-0.75 \pm 0.078) \times \log D_{\rm L} + (2.7 \pm 0.036),
\end{equation}
where $D_{\rm L}$ is in unit of $\rm 10^{\rm 28} ~ cm$. Age is in unit of $\rm Myr$. The adjusted $R^{2}$ is 0.26. The Pearson coefficient is $-0.3 \pm 0.081$ with p-value $6.9 \times 10^{-3}$. The Spearman coefficient is $-0.2 \pm 0.091$ with p-value $7.5 \times 10^{-2}$. The Kendall $\tau$ coefficient is $-0.14 \pm 0.064$ with p-value $7.3 \times 10^{-2}$. The correlation ratio is $0.99 \pm 0.002$. The cosine similarity is $-0.82 \pm 0.0058$. The scatter plot is in Figure \ref{fig:agedl}. The GRB sample number is 82.

\begin{figure*}[!b]
\centering
\includegraphics[width=0.45\textwidth]{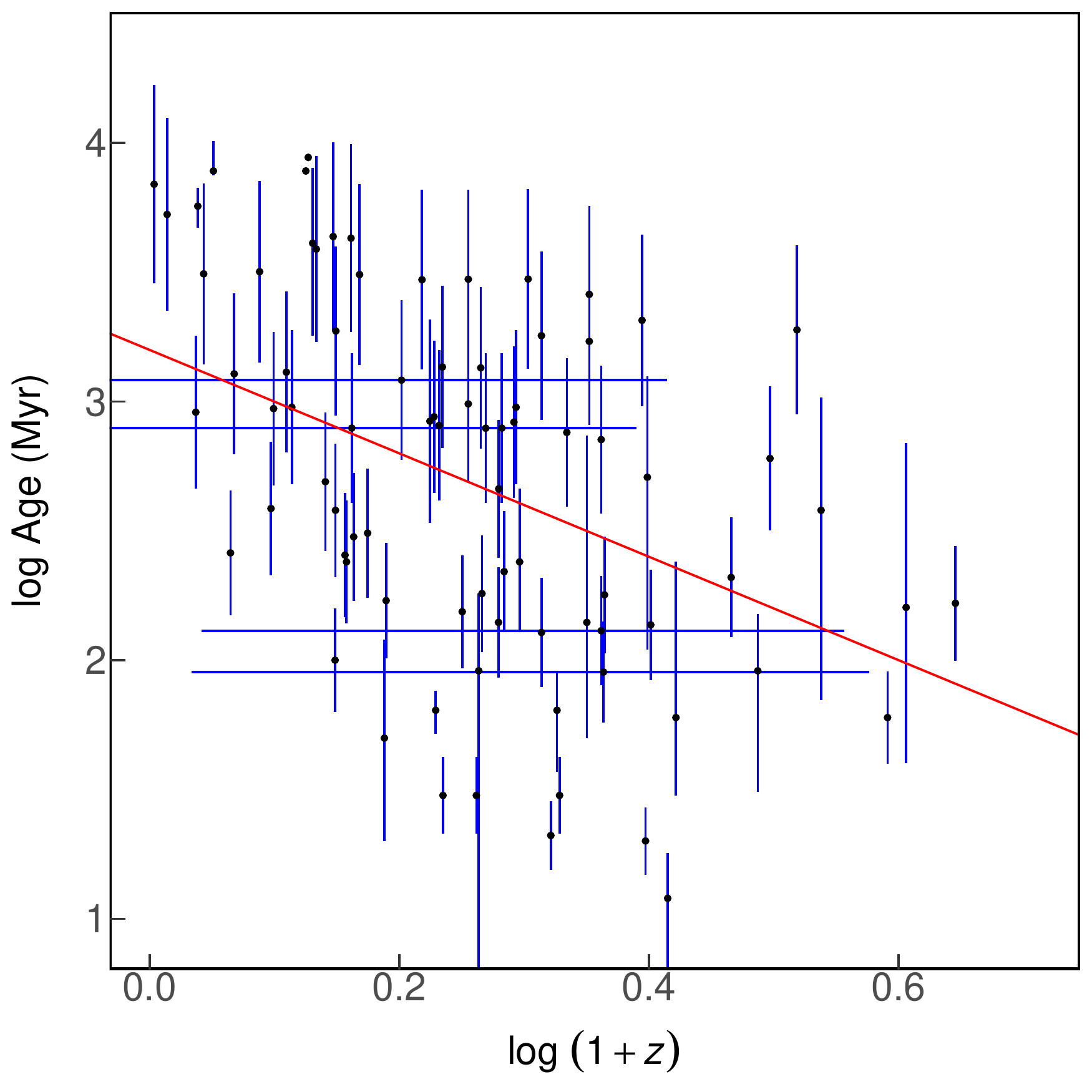}

\caption{
Scatter plot for $\log  Age$ and $\log (1+z)$.
The red line is our fit result. The formula of the red line is $\log  Age = (-2 \pm 0.38) \times \log (1+z) + (3.2 \pm 0.11)$. The description of every parameter is in Section \ref{sec:sample}.
}
\label{fig:agez}
\end{figure*}

The correlation between $\log  Age$ and $\log (1+z)$ is:
\begin{equation} \label{eq:agez}
\log  Age = (-2 \pm 0.38) \times \log (1+z) + (3.2 \pm 0.11),
\end{equation}
where Age is in unit of $\rm Myr$. The adjusted $R^{2}$ is 0.22. The Pearson coefficient is $-0.39 \pm 0.059$ with p-value $3 \times 10^{-4}$. The Spearman coefficient is $-0.43 \pm 0.049$ with p-value $5.5 \times 10^{-5}$. The Kendall $\tau$ coefficient is $-0.29 \pm 0.035$ with p-value $9.9 \times 10^{-5}$. The correlation ratio is $0.91 \pm 0.011$. The cosine similarity is $0.76 \pm 0.03$. The scatter plot is in Figure \ref{fig:agez}. The GRB sample number is 82. It is similar as shown in eq. (\ref{eq:agedl}). It is quite reasonable that the further galaxy is younger. However, the relation between age of the host galaxies and the redshift is not exactly because of the cosmological evolution. Considering the large scattering, it is hard to tell the special information of the GRB host galaxies from the general evolution of the universe.

Age also correlates with $E_{\rm p,cpl}$ and $E_{\rm p,cpl,i}$. We have shown the two correlations in section \ref{sec:gamma0}. 

\subsection{Some correlations about $L_{\rm radio,pk}$} \label{subsec:lradiopk}

We calculated radio luminosity $L_{\rm radio,pk}$ in rest-frame 8.46 $\rm GHz$. The formula is
$L_{\rm radio,pk}=4 \pi D_{\rm L}^{2} \times F_{\rm radio,pk} \times (1+z) \times 8.46 ~ \rm GHz$.
The unit of $L_{\rm radio,pk}$ is $\rm 10^{\rm 40} ~ erg ~ s^{\rm -1}$. We also found some interesting results.

\begin{figure*}[!b]
\centering
\includegraphics[width=0.45\textwidth]{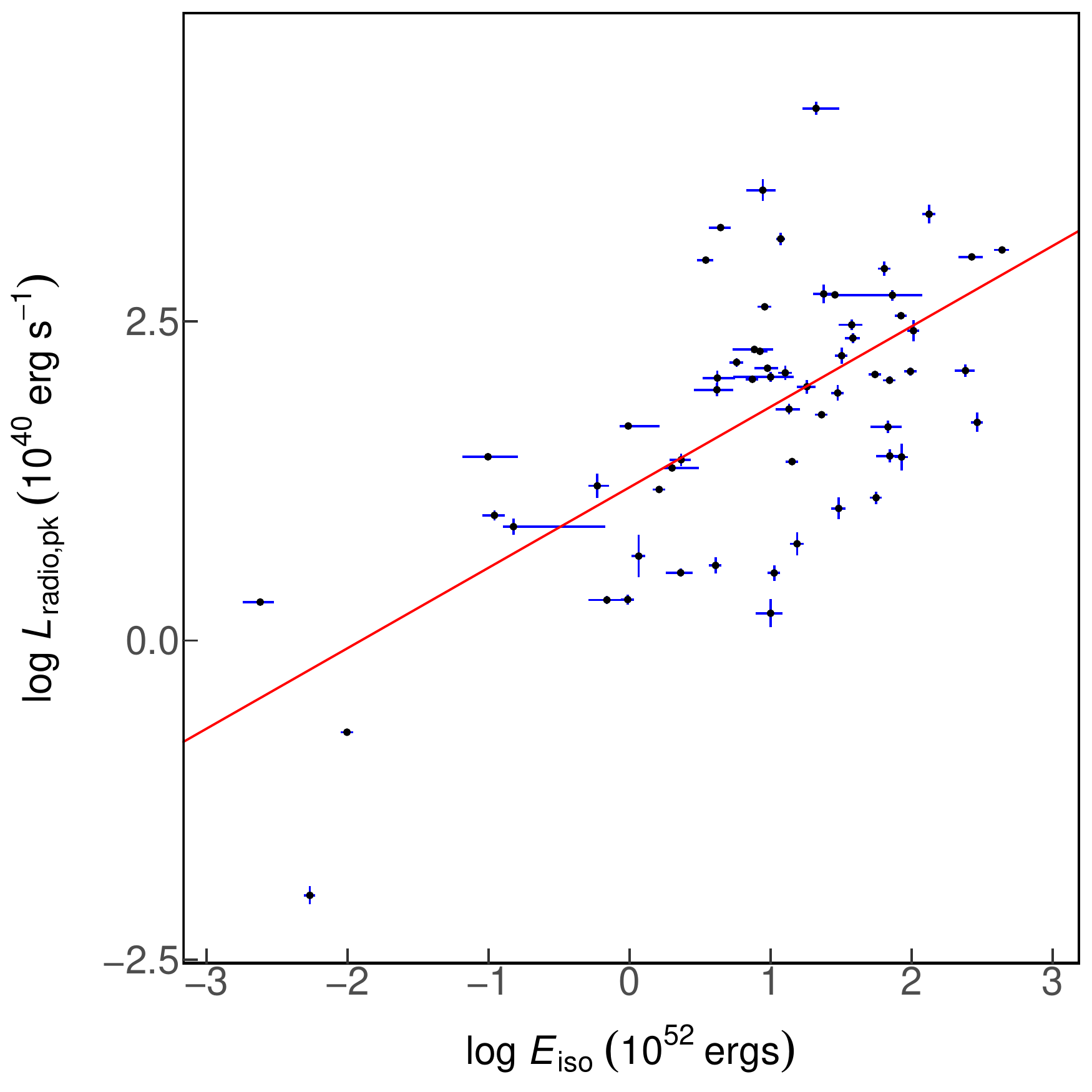}

\caption{
Scatter plot for $\log L_{\rm radio,pk}$ and $\log E_{\rm iso}$.
The red line is our fit result. The formula of the red line is $\log L_{\rm radio,pk} = (0.63 \pm 0.013) \times \log E_{\rm iso} + (1.2 \pm 0.019)$. The description of every parameter is in Section \ref{sec:sample}.
}
\label{fig:lradioeiso}
\end{figure*}

The correlation between $\log L_{\rm radio,pk}$ and $\log E_{\rm iso}$ is:
\begin{equation} \label{eq:lradioeiso}
\log L_{\rm radio,pk} = (0.63 \pm 0.013) \times \log E_{\rm iso} + (1.2 \pm 0.019),
\end{equation}
where $L_{\rm radio,pk}$ is in rest-frame 8.46 $\rm GHz$, the unit is $\rm 10^{\rm 40} ~ erg ~ s^{\rm -1}$. $E_{\rm iso}$ is in unit of $\rm 10^{\rm 52} ~ ergs$ and in rest-frame 1-$10^{4}$ $\rm keV$ energy band. The adjusted $R^{2}$ is 0.41. The Pearson coefficient is $0.64 \pm 0.0094$ with p-value $2.2 \times 10^{-8}$. The Spearman coefficient is $0.52 \pm 0.015$ with p-value $1.5 \times 10^{-5}$. The Kendall $\tau$ coefficient is $0.37 \pm 0.013$ with p-value $2.6 \times 10^{-5}$. The correlation ratio is $0.37 \pm 0.0057$. The cosine similarity is $0.81 \pm 0.0067$. The scatter plot is in Figure \ref{fig:lradioeiso}. The GRB sample number is 61. The peak time of the radio emission are often very late. The peak radio luminosity mainly represents the total kinetic energy and environment, while the isotropic equivalent $\gamma-$ray energy $E_{\rm iso}$  is also positively correlated to the total kinetic energy, though people like to assume they are proportional by an efficiency factor. Therefore, it is expectable about the correlation between $L_{\rm radio,pk}$ and $E_{\rm iso}$. As the data are largely scattered, the actually power law index between them is not clear. With cumulated data and classification, the indices might change.

\begin{figure*}[!b]
\centering
\includegraphics[width=0.45\textwidth]{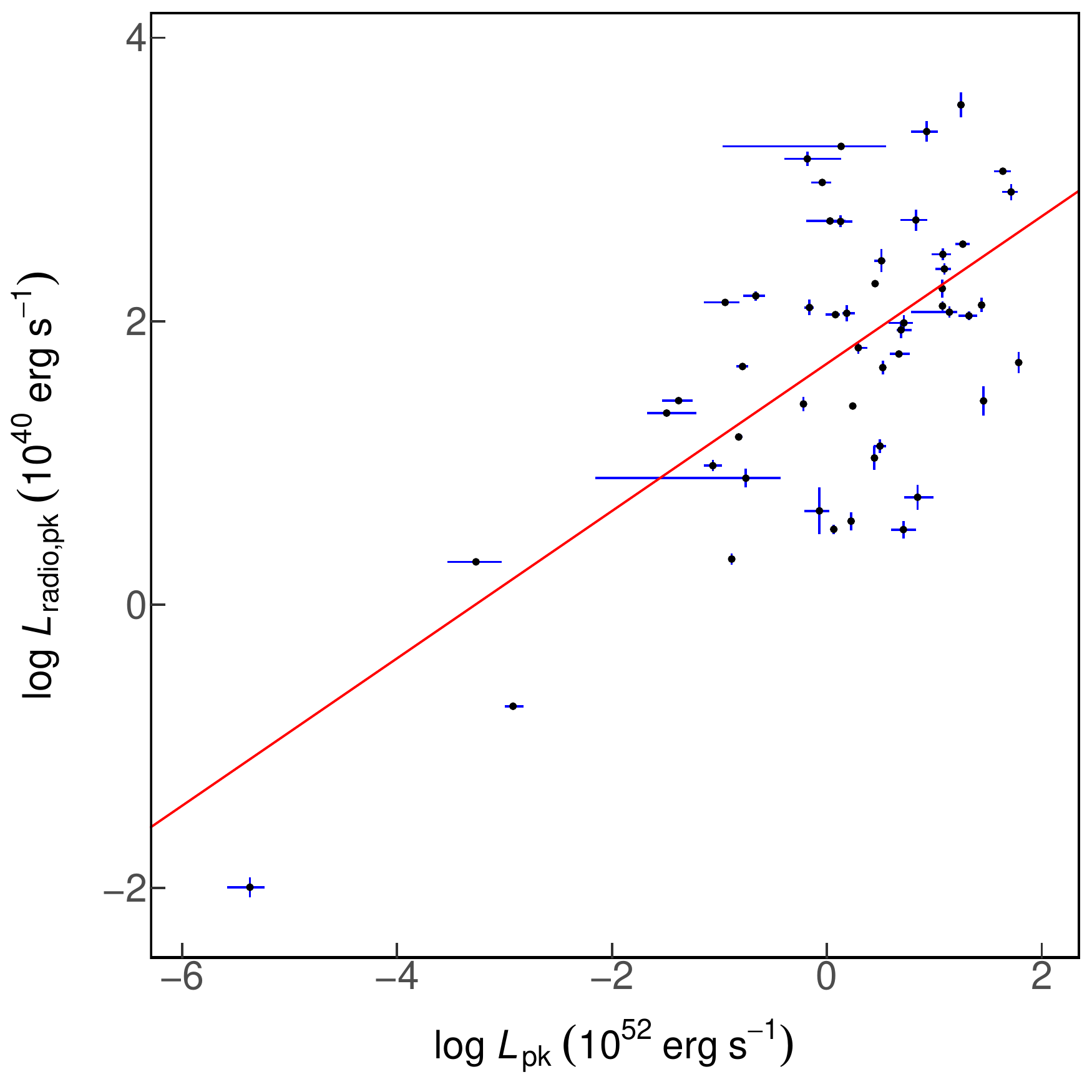}

\caption{
Scatter plot for $\log L_{\rm radio,pk}$ and $\log L_{\rm pk}$.
The red line is our fit result. The formula of the red line is $\log L_{\rm radio,pk} = (0.52 \pm 0.025) \times \log L_{\rm pk}+ (1.7 \pm 0.018)$. The description of every parameter is in Section \ref{sec:sample}.
}
\label{fig:lradiolpk}
\end{figure*}

The correlation between $\log L_{\rm radio,pk}$ and $\log L_{\rm pk}$ is:
\begin{equation} \label{eq:lradiolpk}
\log L_{\rm radio,pk} = (0.52 \pm 0.025) \times \log L_{\rm pk}+ (1.7 \pm 0.018),
\end{equation}
where $L_{\rm radio,pk}$ is in rest-frame 8.46 $\rm GHz$, the unit is $\rm 10^{\rm 40} ~ erg ~ s^{\rm -1}$. $L_{\rm pk}$ is in unit of $\rm 10^{\rm 52} ~ ergs ~ s^{\rm -1}$, and in 1-$10^{4}$ $\rm keV$ energy band. The adjusted $R^{2}$ is 0.45. The Pearson coefficient is $0.67 \pm 0.023$ with p-value $9.1 \times 10^{-8}$. The Spearman coefficient is $0.45 \pm 0.029$ with p-value $8.2 \times 10^{-4}$. The Kendall $\tau$ coefficient is $0.32 \pm 0.021$ with p-value $8.9 \times 10^{-4}$. The correlation ratio is $0.57 \pm 0.0055$. The cosine similarity is $0.4 \pm 0.027$. The scatter plot is in Figure \ref{fig:lradiolpk}. The GRB sample number is 51. This relation is similar to eq. (\ref{eq:lradioeiso}).

\begin{figure*}[!b]
\centering
\includegraphics[width=0.45\textwidth]{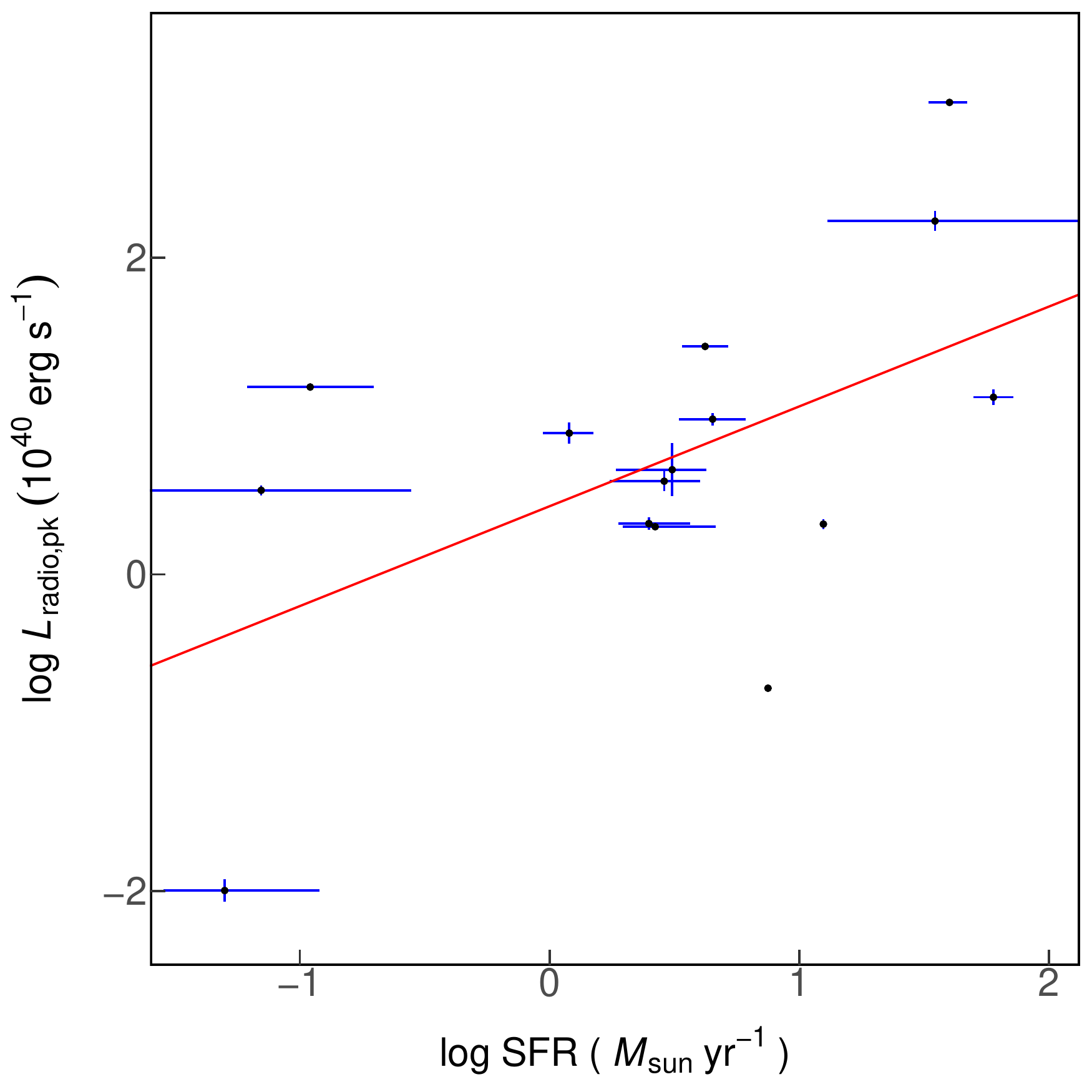}

\caption{
Scatter plot for $\log L_{\rm radio,pk}$ and $\log SFR$.
The red line is our fit result. The formula of the red line is $\log L_{\rm radio,pk} = (0.63 \pm 0.092) \times \log SFR+ (0.43 \pm 0.073)$. The description of every parameter is in Section \ref{sec:sample}.
}
\label{fig:lradiosfr}
\end{figure*}

The correlation between $\log L_{\rm radio,pk}$ and $\log SFR$ is:
\begin{equation} \label{eq:lradiosfr}
\log L_{\rm radio,pk} = (0.63 \pm 0.092) \times \log SFR+ (0.43 \pm 0.073),
\end{equation}
where $L_{\rm radio,pk}$ is in rest-frame 8.46 $\rm GHz$, the unit is $\rm 10^{\rm 40} ~ erg ~ s^{\rm -1}$. SFR is in unit of $\rm M_{\bigodot} ~ yr^{\rm -1}$. The adjusted $R^{2}$ is 0.25. The Pearson coefficient is $0.53 \pm 0.057$ with p-value $4 \times 10^{-2}$. The Spearman coefficient is $0.37 \pm 0.06$ with p-value 0.17. The Kendall $\tau$ coefficient is $0.25 \pm 0.052$ with p-value 0.18. The correlation ratio is $0.13 \pm 0.03$. The cosine similarity is $0.64 \pm 0.054$. The scatter plot is in Figure \ref{fig:lradiosfr}. The GRB sample number is 15. This relation may reveal the dependance of the radio afterglow and the environment of the GRBs. Higher star formation rate region implies higher number density of the environment.

\begin{figure*}[!b]
\centering
\includegraphics[width=0.45\textwidth]{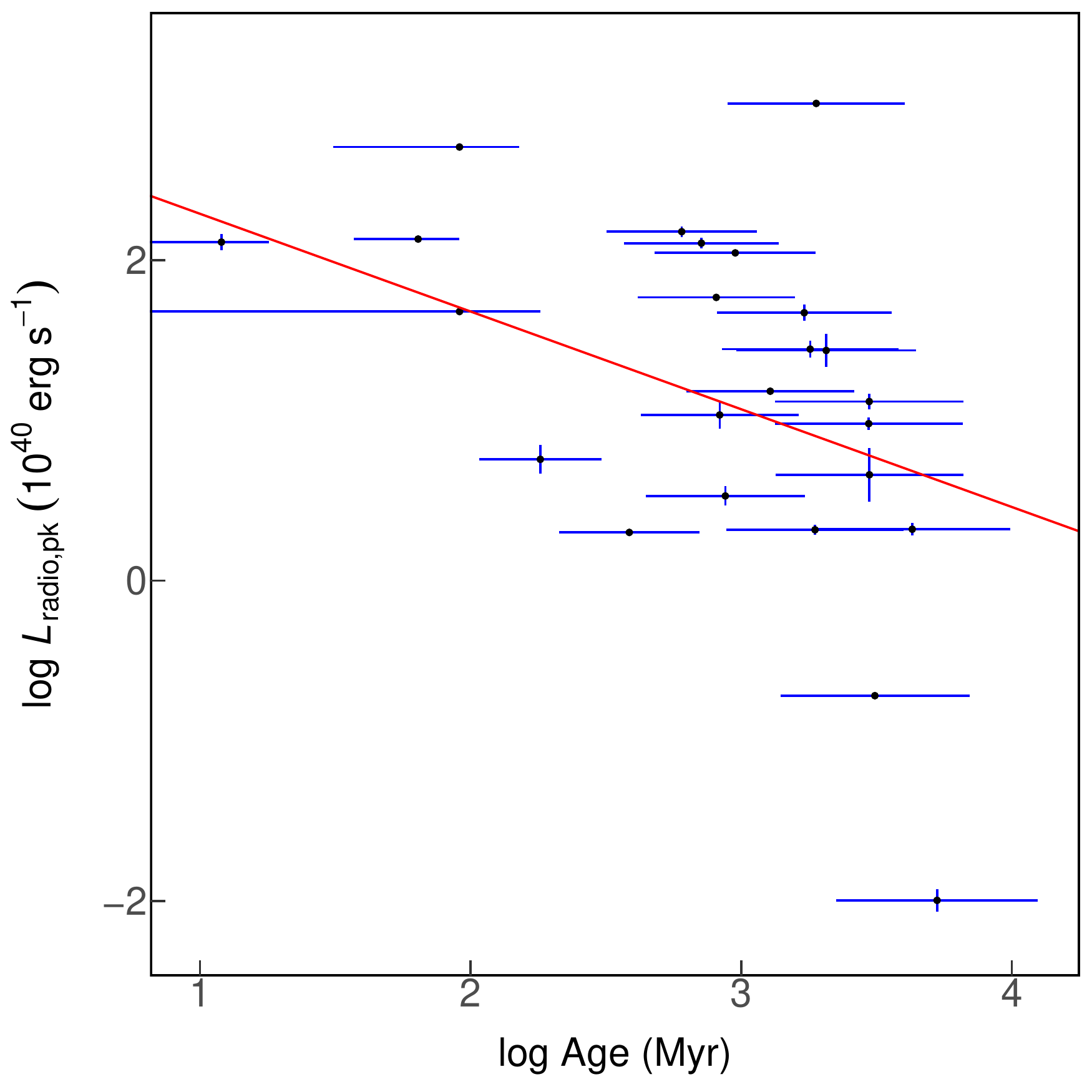}

\caption{
Scatter plot for $\log L_{\rm radio,pk}$ and $\log Age$.
The red line is our fit result. The formula of the red line is $\log L_{\rm radio,pk} = (-0.61 \pm 0.16) \times \log Age+ (2.9 \pm 0.48)$. The description of every parameter is in Section \ref{sec:sample}.
}
\label{fig:lradioage}
\end{figure*}

The correlation between $\log L_{\rm radio,pk}$ and $\log Age$ is:
\begin{equation} \label{eq:lradioage}
\log L_{\rm radio,pk} = (-0.61 \pm 0.16) \times \log Age+ (2.9 \pm 0.48),
\end{equation}
where $L_{\rm radio,pk}$ is in rest-frame 8.46 $\rm GHz$, the unit is $\rm 10^{\rm 40} ~ erg ~ s^{\rm -1}$. Age is in unit of $\rm Myr$. The adjusted $R^{2}$ is 0.21. The Pearson coefficient is $-0.43 \pm 0.085$ with p-value $3.4 \times 10^{-2}$. The Spearman coefficient is $-0.44 \pm 0.099$ with p-value $3.1 \times 10^{-2}$. The Kendall $\tau$ coefficient is $-0.31 \pm 0.077$ with p-value $3.3 \times 10^{-2}$. The correlation ratio is $0.66 \pm 0.038$. The cosine similarity is $0.64 \pm 0.024$. The scatter plot is in Figure \ref{fig:lradioage}. The GRB sample number is 24. As shown in eq. (\ref{eq:lradiosfr}), higher radio luminosity corresponds to more active star forming. And in eq. (\ref{eq:agesfr}), higher star formation rate corresponds to the younger host galaxy. These two relations derives to eq. (\ref{eq:lradioage}).

\begin{figure*}[!b]
\centering
\includegraphics[width=0.45\textwidth]{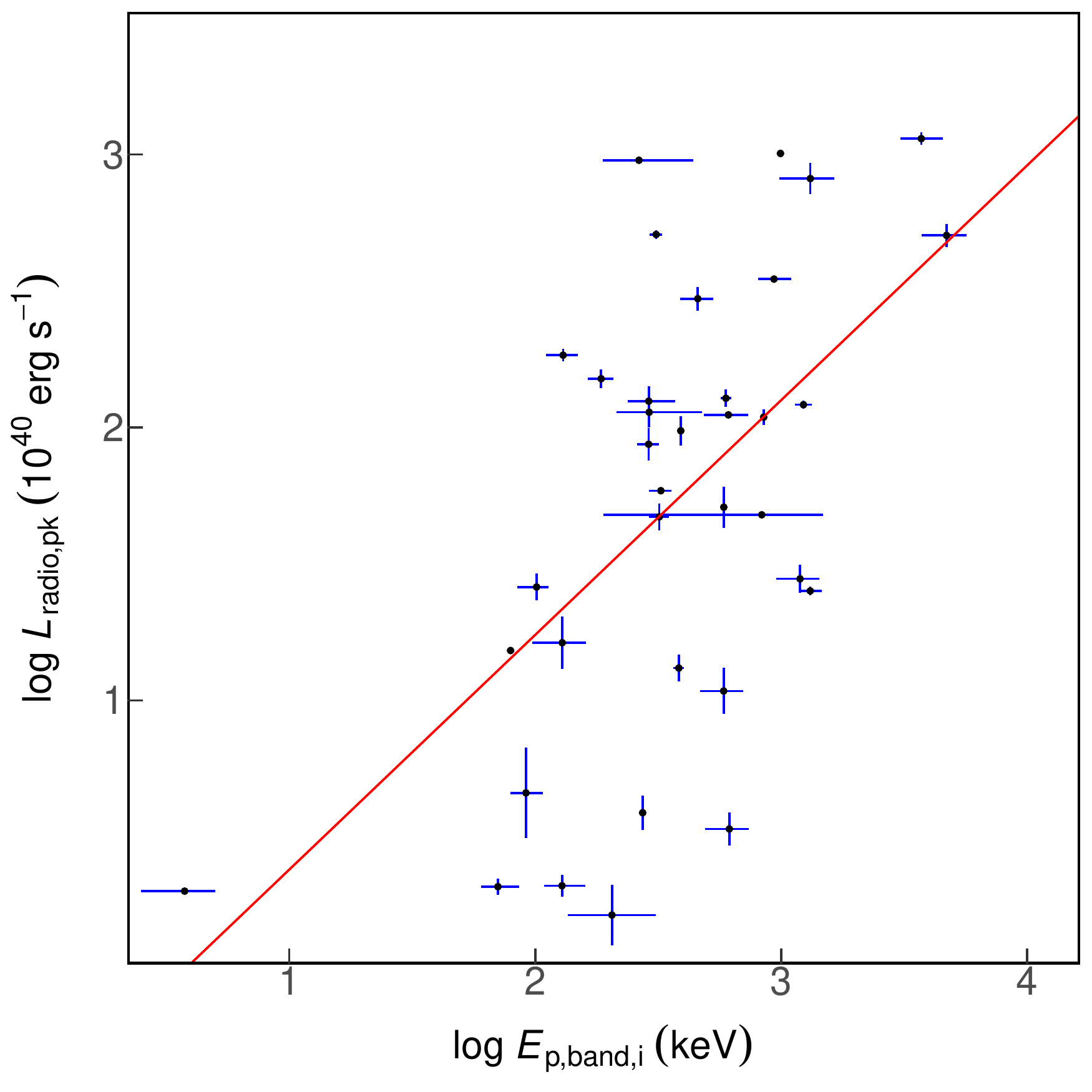}

\caption{
Scatter plot for $\log L_{\rm radio,pk}$ and $\log E_{\rm p,band,i}$.
The red line is our fit result. The formula of the red line is $\log L_{\rm radio,pk} = (0.86 \pm 0.054) \times \log E_{\rm p,band,i}+ (-0.48 \pm 0.14)$. The description of every parameter is in Section \ref{sec:sample}.
}
\label{fig:lradioepbandi}
\end{figure*}

The correlation between $\log L_{\rm radio,pk}$ and $\log E_{\rm p,band,i}$ is:
\begin{equation} \label{eq:lradioepbandi}
\log L_{\rm radio,pk} = (0.86 \pm 0.054) \times \log E_{\rm p,band,i}+ (-0.48 \pm 0.14),
\end{equation}
where $L_{\rm radio,pk}$ is in rest-frame 8.46 $\rm GHz$, the unit is $\rm 10^{\rm 40} ~ erg ~ s^{\rm -1}$. $E_{\rm p,band,i}$ is in unit of $\rm keV$. The adjusted $R^{2}$ is 0.32. The Pearson coefficient is $0.57 \pm 0.027$ with p-value $2.6 \times 10^{-4}$. The Spearman coefficient is $0.5 \pm 0.037$ with p-value $1.8 \times 10^{-3}$. The Kendall $\tau$ coefficient is $0.35 \pm 0.031$ with p-value $2.5 \times 10^{-3}$. The correlation ratio is $0.51 \pm 0.011$. The cosine similarity is $0.93 \pm 0.0032$. The scatter plot is in Figure \ref{fig:lradioepbandi}. The GRB sample number is 36. This relation might be from the Amati relation. As higher peak energy corresponds to higher total energy of the GRBs, it also derives to a brighter radio luminosity.

\begin{figure*}[!b]
\centering
\includegraphics[width=0.45\textwidth]{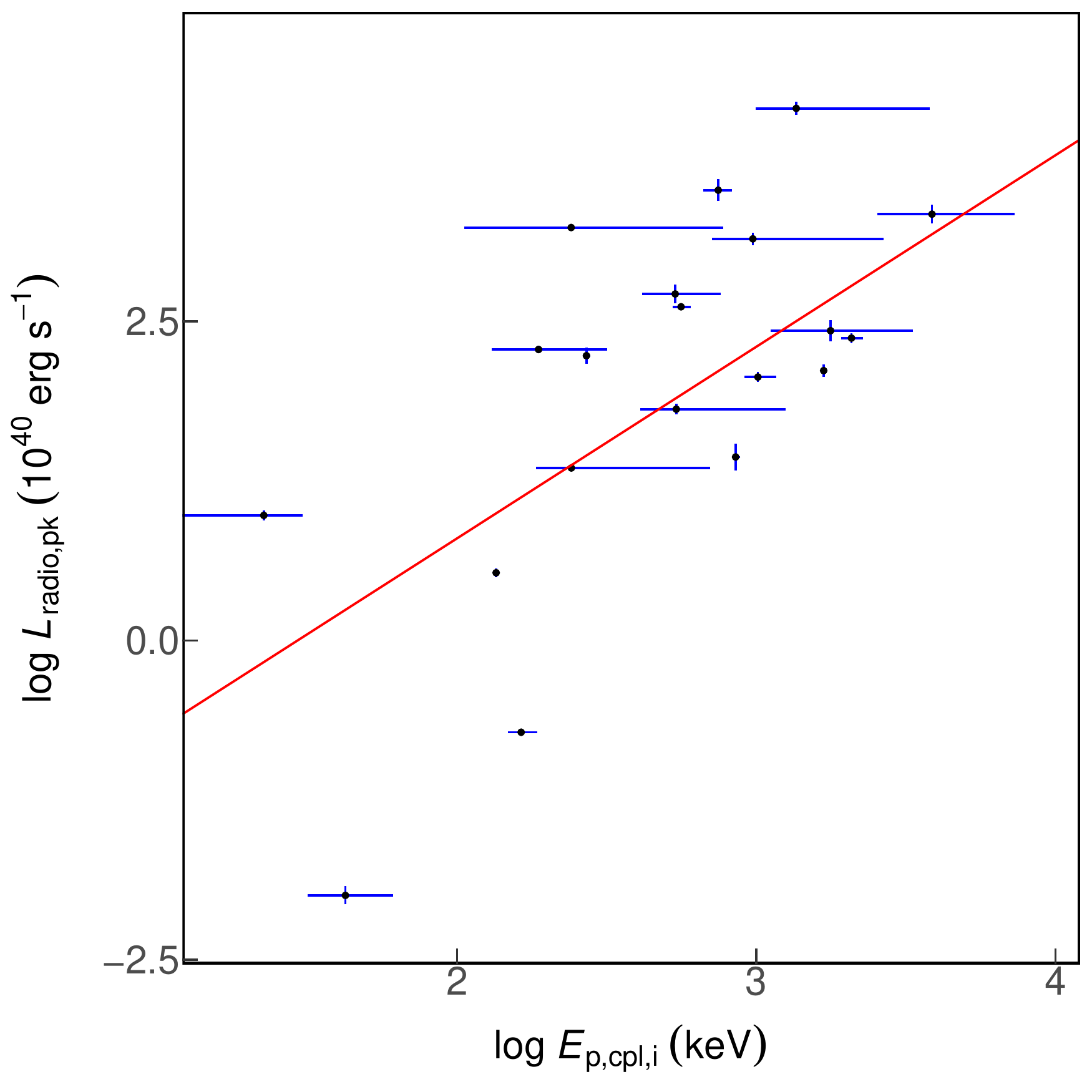}

\caption{
Scatter plot for $\log L_{\rm radio,pk}$ and $\log E_{\rm p,cpl,i}$.
The red line is our fit result. The formula of the red line is $\log L_{\rm radio,pk} = (1.5 \pm 0.2) \times \log E_{\rm p,cpl,i}+ (-2.2 \pm 0.55)$. The description of every parameter is in Section \ref{sec:sample}.
}
\label{fig:lradioepcpli}
\end{figure*}

The correlation between $\log L_{\rm radio,pk}$ and $\log E_{\rm p,cpl,i}$ is:
\begin{equation} \label{eq:lradioepcpli}
\log L_{\rm radio,pk} = (1.5 \pm 0.2) \times \log E_{\rm p,cpl,i}+ (-2.2 \pm 0.55),
\end{equation}
where $L_{\rm radio,pk}$ is in rest-frame 8.46 $\rm GHz$, the unit is $\rm 10^{\rm 40} ~ erg ~ s^{\rm -1}$. $E_{\rm p,cpl,i}$ is in unit of $\rm keV$. The adjusted $R^{2}$ is 0.42. The Pearson coefficient is $0.65 \pm 0.057$ with p-value $2.1 \times 10^{-3}$. The Spearman coefficient is $0.57 \pm 0.077$ with p-value $8.7 \times 10^{-3}$. The Kendall $\tau$ coefficient is $0.42 \pm 0.064$ with p-value $9.4 \times 10^{-3}$. The correlation ratio is $0.31 \pm 0.018$. The cosine similarity is $0.88 \pm 0.0075$. The scatter plot is in Figure \ref{fig:lradioepcpli}. The GRB sample number is 20. It is similar to eq. (\ref{eq:lradioepbandi}).

\begin{figure*}[!b]
\centering
\includegraphics[width=0.45\textwidth]{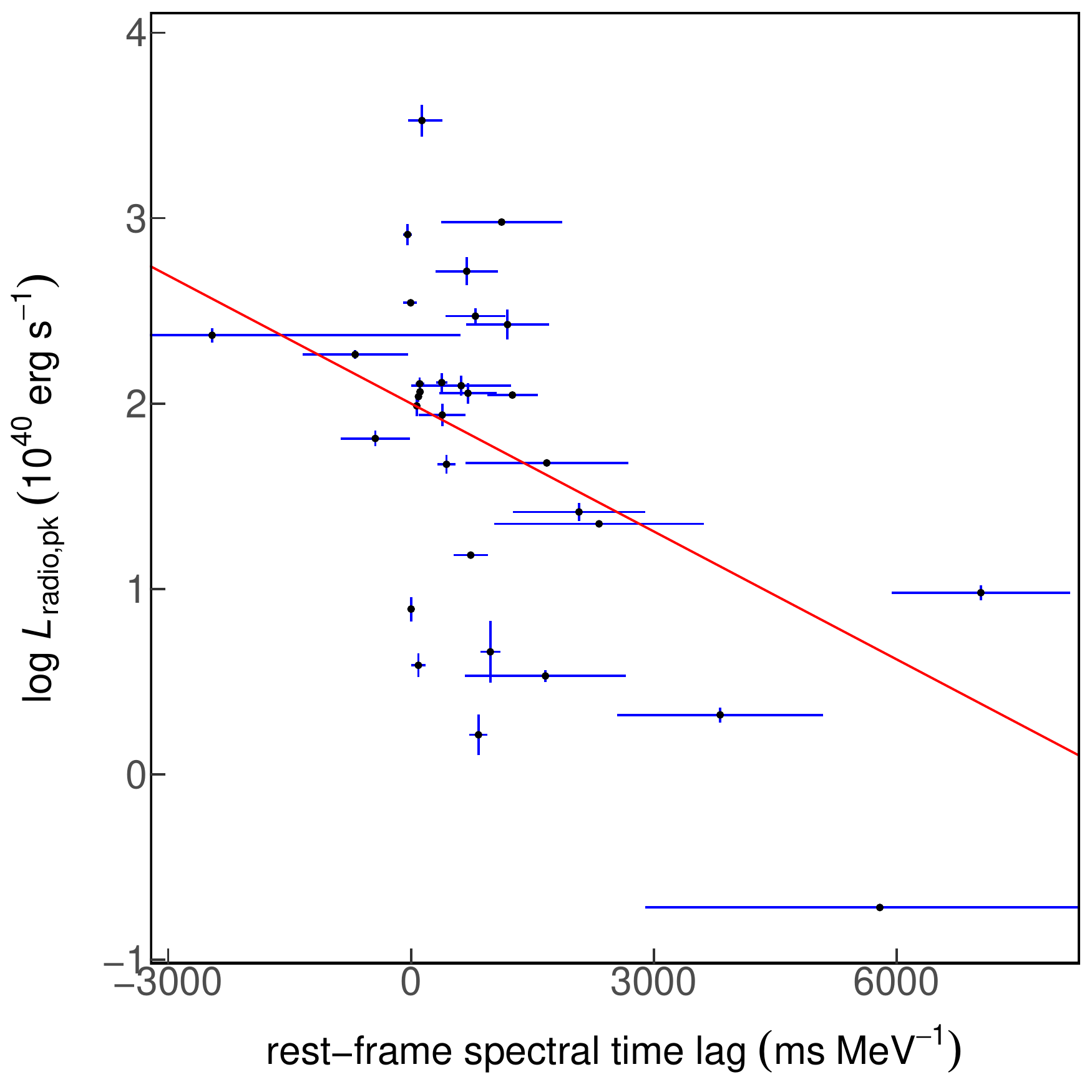}

\caption{
Scatter plot for $\log L_{\rm radio,pk}$ and rest-frame spectral lag.
The red line is our fit result. The formula of the red line is $\log L_{\rm radio,pk} = (-0.00023 \pm 0.000048) \times rest-frame ~ spectral ~ lag+ (2 \pm 0.063)$. The description of every parameter is in Section \ref{sec:sample}.
}
\label{fig:lradiolagi}
\end{figure*}

The correlation between $\log L_{\rm radio,pk}$ and rest-frame spectral lag is:
\begin{equation} \label{eq:lradiolagi}
\log L_{\rm radio,pk} = (-0.00023 \pm 0.000048) \times rest-frame ~ spectral ~ lag+ (2 \pm 0.063),
\end{equation}
where $L_{\rm radio,pk}$ is in rest-frame 8.46 $\rm GHz$, the unit is $\rm 10^{\rm 40} ~ erg ~ s^{\rm -1}$. Rest-frame spectral lag is in unit of $\rm ms ~ MeV^{\rm -1}$. The adjusted $R^{2}$ is 0.29. The Pearson coefficient is $-0.49 \pm 0.089$ with p-value $4.2 \times 10^{-3}$. The Spearman coefficient is $-0.4 \pm 0.081$ with p-value $2.2 \times 10^{-2}$. The Kendall $\tau$ coefficient is $-0.28 \pm 0.06$ with p-value $2.3 \times 10^{-2}$. The correlation ratio is $0.34 \pm 0.06$. The cosine similarity is $0.2 \pm 0.078$. The scatter plot is in Figure \ref{fig:lradiolagi}. The GRB sample number is 32. It is also quite puzzle why the radio luminosity if related to the spectral lag.

\section{Remarkable results of three parameters} \label{sec:threeresult}

In this section, we analyzed some good correlations between three parameters. Firstly the correlation should also have at least 10 GRBs. Adjusted $R^{2}$ should also be bigger than 0.2. The whole linear model, $a_{1}$ and $a_{2}$ have hypothesis testing p-values smaller than 0.05. For peak energy flux $F_{\rm pk}$ and peak photon flux $P_{\rm pk}$, both of them have four different time bins. If $F_{\rm pk}$ in all the four time bins correlates with another one same parameter, we just showed the best time bin result. It is the same for $P_{\rm pk}$. All the results are in machine readable tables. In the following, we will show all the remarkable results in order of increasing sample numbers.

The $\log \Gamma_{0}$-$(-\alpha_{\rm band})$-$\beta_{\rm X11hr}$ formula is:
\begin{equation} \label{eq:g0adbr}
\log \Gamma_{0} = (-0.39 \pm 0.09) \times (-\alpha_{\rm band}) + (-0.59 \pm 0.14) \times \beta_{\rm X11hr} + (3.4 \pm 0.21),
\end{equation}
the adjusted $R^{2}$ is 0.6552. The GRB sample number is 10.

The $\log SSFR$-$\log Mass$-$\log t_{\rm pkX,i}$ formula is:
\begin{equation} \label{eq:srmsti}
\log SSFR = (-0.47 \pm 0.14) \times \log Mass + (-0.76 \pm 0.18) \times \log t_{\rm pkX,i} + (5.5 \pm 1.3),
\end{equation}
where $\log SSFR$ is in unit of $\rm Gyr^{\rm -1}$. Mass is in unit of $M_{\bigodot}$. $t_{\rm pkX,i}$ is in unit of $\rm s$. The adjusted $R^{2}$ is 0.6759. The GRB sample number is 10.

The $\log t_{\rm pkX,i}$-$\log SSFR$-$\log T_{\rm 90,i}$ formula is:
\begin{equation} \label{eq:tisrti}
\log t_{\rm pkX,i} = (-0.31 \pm 0.13) \times \log SSFR + (0.84 \pm 0.18) \times \log T_{\rm 90,i} + (0.53 \pm 0.22),
\end{equation}
where $t_{\rm pkX,i}$ is in unit of $\rm s$. $\log SSFR$ is in unit of $\rm Gyr^{\rm -1}$. $T_{\rm 90,i}$ is in unit of $\rm s$. The adjusted $R^{2}$ is 0.7597. The GRB sample number is 10.

The $\log Age$-$\log  N_{\rm H}$-$\log t_{\rm burst,i}$ formula is:
\begin{equation} \label{eq:aenhti}
\log Age = (-0.87 \pm 0.25) \times \log  N_{\rm H} + (1 \pm 0.2) \times \log t_{\rm burst,i} + (0.78 \pm 0.59),
\end{equation}
where Age is in unit of $\rm Myr$. $N_{\rm H}$ is in unit of $\rm 10^{\rm 21} ~ cm^{\rm -2}$. $t_{\rm burst,i}$ is in unit of $\rm s$. The adjusted $R^{2}$ is 0.681. The GRB sample number is 10.

The $(-\beta_{\rm band})$-$\log offset$-$\log Age$ formula is:
\begin{equation}  \label{eq:bdotae}
(-\beta_{\rm band}) = (0.65 \pm 0.61) \times \log offset + (-0.42 \pm 0.49) \times \log Age + (3.5 \pm 1.3),
\end{equation}
where host galaxy offset is in unit of $\rm kpc$. Age is in unit of $\rm Myr$. The adjusted $R^{2}$ is 0.5508. The GRB sample number is 10.

The $\log E_{\rm iso}$-$\log t_{\rm pkOpt}$-$\log Age$ formula is:
\begin{equation} \label{eq:eottage}
\log E_{\rm iso} = (-1 \pm 0.05) \times \log t_{\rm pkOpt} + (-0.48 \pm 0.061) \times \log Age + (4.5 \pm 0.22),
\end{equation}
where $E_{\rm iso}$ is in unit of $\rm 10^{\rm 52} ~ ergs$ and in rest-frame 1-$10^{4}$ $\rm keV$ energy band. $t_{\rm pkOpt}$ is in unit of $\rm s$. Age is in unit of $\rm Myr$. The adjusted $R^{2}$ is 0.9079. The GRB sample number is 10.

The $\log \Gamma_{0}$-$\log L_{\rm radio,pk}$-$\log T_{\rm R45}$ formula is:
\begin{equation} \label{eq:g0lratr45}
\log \Gamma_{0} = (0.33 \pm 0.012) \times \log L_{\rm radio,pk} + (-0.96 \pm 0.086) \times \log T_{\rm R45} + (2.6 \pm 0.11),
\end{equation}
where $L_{\rm radio,pk}$ is in rest-frame 8.46 $\rm GHz$, the unit is $\rm 10^{\rm 40} ~ erg ~ s^{\rm -1}$. $T_{\rm R45}$ is in unit of $\rm s$. The adjusted $R^{2}$ is 0.6438. The GRB sample number is 10.

The $\beta_{\rm X11hr}$-$(-\alpha_{\rm spl})$-$\log Mass$ formula is:
\begin{equation} \label{eq:behrasplms}
\beta_{\rm X11hr} = (0.68 \pm 0.16) \times (-\alpha_{\rm spl}) + (0.24 \pm 0.068) \times \log Mass + (-1.9 \pm 0.76),
\end{equation}
where Mass is in unit of $M_{\bigodot}$. The adjusted $R^{2}$ is 0.5919. The GRB sample number is 11.

The $\log E_{\rm p,cpl}$-$\log HR$-$\log t_{\rm pkX}$ formula is:
\begin{equation} \label{eq:epcplhrtx}
\log E_{\rm p,cpl} = (0.59 \pm 0.1) \times \log HR + (0.15 \pm 0.11) \times \log t_{\rm pkX} + (1.8 \pm 0.19),
\end{equation}
where $E_{\rm p,cpl}$ is in unit of $\rm keV$. $t_{\rm pkX}$ is in unit of $\rm s$. The adjusted $R^{2}$ is 0.9356. The GRB sample number is 11.

The $\log E_{\rm iso}$-$\log t_{\rm burst,i}$-$\log t_{\rm radio,pk,i}$ formula is:
\begin{equation} \label{eq:esottitapki}
\log E_{\rm iso} = (-1.1 \pm 0.047) \times \log t_{\rm burst,i} + (-1.6 \pm 0.2) \times \log t_{\rm radio,pk,i} + (12 \pm 1.1),
\end{equation}
where $E_{\rm iso}$ is in unit of $\rm 10^{\rm 52} ~ ergs$ and in rest-frame 1-$10^{4}$ $\rm keV$ energy band. $t_{\rm burst,i}$ is in unit of $\rm s$. $t_{\rm radio,pk,i}$ is in unit of $\rm s$. The adjusted $R^{2}$ is 0.6305. The GRB sample number is 11.

The $\log F_{\rm radio,pk}$-$\log t_{\rm burst,i}$-$\log t_{\rm radio,pk,i}$ formula is:
\begin{equation} \label{eq:fratiirapki}
\log F_{\rm radio,pk} = (0.24 \pm 0.026) \times \log t_{\rm burst,i} + (0.66 \pm 0.055) \times \log t_{\rm radio,pk,i} + (-7.7 \pm 0.31),
\end{equation}
where $F_{\rm radio,pk}$ is in unit of $\rm Jy$. $t_{\rm burst,i}$ is in unit of $\rm s$. $t_{\rm radio,pk,i}$ is in unit of $\rm s$. The adjusted $R^{2}$ is 0.4834. The GRB sample number is 11.

The $\beta_{\rm X11hr}$-$\log P_{\rm pk2}$-$\log SSFR$ formula is:
\begin{equation} \label{eq:behrppk2ssfr}
\beta_{\rm X11hr} = (-0.57 \pm 0.13) \times \log P_{\rm pk2} + (-0.55 \pm 0.12) \times \log SSFR + (2.1 \pm 0.22),
\end{equation}
where $P_{\rm pk2}$ is peak photon flux of 256 $\rm ms$ time bin in 10-1000 $\rm keV$, and in unit of $\rm photons ~ cm^{\rm -2} ~ s^{\rm -1}$. $\log SSFR$ is in unit of $\rm Gyr^{\rm -1}$. The adjusted $R^{2}$ is 0.6811. The GRB sample number is 11.

The $\log \Gamma_{0}$-$\log t_{\rm radio,pk}$-$\log T_{\rm 50,i}$ formula is:
\begin{equation} \label{eq:g0tpkt50i}
\log \Gamma_{0} = (1.3 \pm 0.12) \times \log t_{\rm radio,pk} + (-0.74 \pm 0.068) \times \log T_{\rm 50,i} + (-4.5 \pm 0.73),
\end{equation}
where $t_{\rm radio,pk}$ is in unit of $\rm s$. $T_{\rm 50,i}$ is in unit of $\rm s$. The adjusted $R^{2}$ is 0.6372. The GRB sample number is 11.

The $\log Mass$-$\log T_{\rm 90}$-$\log t_{\rm pkX}$ formula is:
\begin{equation} \label{eq:mst90tpkx}
\log Mass = (-2.3 \pm 0.36) \times \log T_{\rm 90} + (1.7 \pm 0.37) \times \log t_{\rm pkX} + (9.7 \pm 0.88),
\end{equation}
where Mass is in unit of $M_{\bigodot}$. $T_{\rm 90}$ is in unit of $\rm s$. $t_{\rm pkX}$ is in unit of $\rm s$. The adjusted $R^{2}$ is 0.7206. The GRB sample number is 11.

The $\log L_{\rm pk}$-$\log L_{\rm radio,pk}$-$\log t_{\rm burst}$ formula is:
\begin{equation} \label{eq:lpklratbt}
\log L_{\rm pk} = (0.91 \pm 0.07) \times \log L_{\rm radio,pk} + (-0.98 \pm 0.24) \times \log t_{\rm burst} + (0.68 \pm 0.7),
\end{equation}
where $L_{\rm pk}$ is in unit of $\rm 10^{\rm 52} ~ ergs ~ s^{\rm -1}$, and in 1-$10^{4}$ $\rm keV$ energy band. $L_{\rm radio,pk}$ is in rest-frame 8.46 $\rm GHz$, the unit is $\rm 10^{\rm 40} ~ erg ~ s^{\rm -1}$. $t_{\rm burst}$ is in unit of $\rm s$. The adjusted $R^{2}$ is 0.8403. The GRB sample number is 11.

The rest-frame spectral lag-$\log L_{\rm radio,pk}$-$\log E_{\rm p,cpl}$ formula is:
\begin{equation} \label{eq:laglraepcpl}
rest-frame ~ spectral ~ lag = (-1263 \pm 580) \times \log L_{\rm radio,pk} + (-2816 \pm 951) \times \log E_{\rm p,cpl} + (10058 \pm 2111),
\end{equation}
where rest-frame spectral lag is in unit of $\rm ms ~ MeV^{\rm -1}$. $L_{\rm radio,pk}$ is in rest-frame 8.46 $\rm GHz$, the unit is $\rm 10^{\rm 40} ~ erg ~ s^{\rm -1}$. $E_{\rm p,cpl}$ is in unit of $\rm keV$. The adjusted $R^{2}$ is 0.7246. The GRB sample number is 11.

The $\log \Gamma_{0}$-$\log L_{\rm radio,pk}$-$\log t_{\rm pkOpt}$ formula is:
\begin{equation} \label{eq:g0lratpko}
\log \Gamma_{0} = (0.15 \pm 0.015) \times \log L_{\rm radio,pk} + (-0.57 \pm 0.023) \times \log t_{\rm pkOpt} + (3.4 \pm 0.093),
\end{equation}
where $L_{\rm radio,pk}$ is in rest-frame 8.46 $\rm GHz$, the unit is $\rm 10^{\rm 40} ~ erg ~ s^{\rm -1}$. $t_{\rm pkOpt}$ is in unit of $\rm s$. The adjusted $R^{2}$ is 0.9562. The GRB sample number is 11.

The $\log E_{\rm iso}$-$\log \Gamma_{0}$-$\log F_{\rm radio,pk}$ formula is:
\begin{equation} \label{eq:eog0frapk}
\log E_{\rm iso} = (1.2 \pm 0.094) \times \log \Gamma_{0} + (-2.2 \pm 0.35) \times \log F_{\rm radio,pk} + (-9.7 \pm 1.2),
\end{equation}
where $E_{\rm iso}$ is in unit of $\rm 10^{\rm 52} ~ ergs$ and in rest-frame 1-$10^{4}$ $\rm keV$ energy band. $F_{\rm radio,pk}$ is in unit of $\rm Jy$. The adjusted $R^{2}$ is 0.8299. The GRB sample number is 12.

The $\log F_{\rm g}$-$\log \Gamma_{0}$-$\log t_{\rm radio,pk,i}$ formula is:
\begin{equation} \label{eq:fgg0trapki}
\log F_{\rm g} = (0.64 \pm 0.08) \times \log \Gamma_{0} + (1.8 \pm 0.24) \times \log t_{\rm radio,pk,i} + (-9.9 \pm 1.2),
\end{equation}
where $F_{\rm g}$ is in unit of $\rm 10^{\rm -6} ~ ergs ~ cm^{\rm -2}$, and in 20-2000 $\rm keV$ energy band. $t_{\rm radio,pk,i}$ is in unit of $\rm s$. The adjusted $R^{2}$ is 0.7167. The GRB sample number is 12.

The Mag-$\log  F_{\rm Opt11hr}$-metallicity formula is:
\begin{equation} \label{eq:magfohrmety}
Mag = (1.1 \pm 0.64) \times \log  F_{\rm Opt11hr} + (-2.4 \pm 1.8) \times metallicity + (5.5 \pm 15),
\end{equation}
where Mag is in unit of magnitude. $F_{\rm Opt11hr}$ is in unit of $\rm Jy$. Metallicity is the value of $12+\log O/H$. The adjusted $R^{2}$ is 0.5847. The GRB sample number is 13.

The $\log L_{\rm pk}$-$\log offset$-$\log E_{\rm p,cpl,i}$ formula is:
\begin{equation} \label{eq:lpkofetepcpli}
\log L_{\rm pk} = (-0.94 \pm 0.26) \times \log offset + (1.1 \pm 0.33) \times \log E_{\rm p,cpl,i} + (-3.2 \pm 0.86),
\end{equation}
where $L_{\rm pk}$ is in unit of $\rm 10^{\rm 52} ~ erg ~ s^{\rm -1}$, and in 1-$10^{4}$ $\rm keV$ energy band. Host galaxy offset is in unit of $\rm kpc$. $E_{\rm p,cpl,i}$ is in unit of $\rm keV$. The adjusted $R^{2}$ is 0.4724. The GRB sample number is 13.

The $\log t_{\rm pkX}$-$\log T_{\rm 50}$-$\log HR$ formula is:
\begin{equation} \label{eq:tpkxt50hr}
\log t_{\rm pkX} = (0.59 \pm 0.057) \times \log T_{\rm 50} + (0.44 \pm 0.11) \times \log HR + (0.81 \pm 0.083),
\end{equation}
where $t_{\rm pkX}$ is in unit of $\rm s$. $T_{\rm 50}$ is in unit of $\rm s$. The adjusted $R^{2}$ is 0.7483. The GRB sample number is 13.

The $\log F_{\rm pk3}$-$\log T_{\rm 90}$-$\log F_{\rm X11hr}$ formula is:
\begin{equation} \label{eq:fpk3t90fx11}
\log F_{\rm pk3} = (-1 \pm 0.15) \times \log T_{\rm 90} + (0.54 \pm 0.063) \times \log F_{\rm X11hr} + (5.5 \pm 0.41),
\end{equation}
where $F_{\rm pk3}$ is peak energy flux of 256 $\rm ms$ time bin in rest-frame 1-$10^{4}$ $\rm keV$ energy band, and in unit of $\rm 10^{\rm -6} ~ ergs ~ cm^{\rm -2} ~ s^{\rm -1}$. $T_{\rm 90}$ is in unit of $\rm s$. $F_{\rm X11hr}$ is in unit of $\rm Jy$. The adjusted $R^{2}$ is 0.7421. The GRB sample number is 13.

The $\log P_{\rm pk2}$-$\log T_{\rm 90}$-$\log F_{\rm X11hr}$ formula is:
\begin{equation} \label{eq:ppk2t90fx11}
\log P_{\rm pk2} = (-0.97 \pm 0.12) \times \log T_{\rm 90} + (0.46 \pm 0.047) \times \log F_{\rm X11hr}  + (5.7 \pm 0.31),
\end{equation}
where $P_{\rm pk2}$ is peak photon flux of 256 $\rm ms$ time bin in 10-1000 $\rm keV$, and in unit of $\rm photons ~ cm^{\rm -2} ~ s^{\rm -1}$. $T_{\rm 90}$ is in unit of $\rm s$. $F_{\rm X11hr}$ is in unit of $\rm Jy$. The adjusted $R^{2}$ is 0.65. The GRB sample number is 13.

The $(-\alpha_{\rm spl})$-$variability_{1}$-$\log F_{\rm X11hr}$ formula is:
\begin{equation} \label{eq:alsplvar1fx11}
(-\alpha_{\rm spl}) = (0.48 \pm 0.76) \times variability_{1} + (-0.23 \pm 0.062) \times \log F_{\rm X11hr} + (0.44 \pm 0.48),
\end{equation}
where $F_{\rm X11hr}$ is in unit of $\rm Jy$. The adjusted $R^{2}$ is 0.4567. The GRB sample number is 13.

The $\log F_{\rm pk2}$-$(-\beta_{\rm band})$-$\log t_{\rm radio,pk}$ formula is:
\begin{equation} \label{eq:fpk2bendtrapk}
\log F_{\rm pk2} = (0.46 \pm 0.1) \times (-\beta_{\rm band}) + (-0.68 \pm 0.13) \times \log t_{\rm radio,pk} + (3.7 \pm 0.67),
\end{equation}
where $F_{\rm pk2}$ is peak energy flux of 64 $\rm ms$ time bin in rest-frame 1-$10^{4}$ $\rm keV$ energy band, and in unit of $\rm 10^{\rm -6} ~ ergs ~ cm^{\rm -2} ~ s^{\rm -1}$. $t_{\rm radio,pk}$ is in unit of $\rm s$. The adjusted $R^{2}$ is 0.499. The GRB sample number is 14.

The $(-\beta_{\rm band})$-$\log F_{\rm pk2}$-$\log F_{\rm radio,pk}$ formula is:
\begin{equation} \label{eq:bendfpk2frapk}
(-\beta_{\rm band}) = (0.64 \pm 0.16) \times \log F_{\rm pk2} + (-0.83 \pm 0.21) \times \log F_{\rm radio,pk} + (-1 \pm 0.7),
\end{equation}
where $F_{\rm pk2}$ is peak energy flux of 64 $\rm ms$ time bin in rest-frame 1-$10^{4}$ $\rm keV$ energy band, and in unit of $\rm 10^{\rm -6} ~ ergs ~ cm^{\rm -2} ~ s^{\rm -1}$. $F_{\rm radio,pk}$ is in unit of $\rm Jy$. The adjusted $R^{2}$ is 0.6076. The GRB sample number is 14.

The $\log HR$-$\log F_{\rm pk3}$-$\log Age$ formula is:
\begin{equation} \label{eq:hrfpk3age}
\log HR = (0.39 \pm 0.035) \times \log F_{\rm pk3} + (-0.24 \pm 0.065) \times \log Age + (1.2 \pm 0.19),
\end{equation}
where $F_{\rm pk3}$ is peak energy flux of 256 $\rm ms$ time bin in rest-frame 1-$10^{4}$ $\rm keV$ energy band, and in unit of $\rm 10^{\rm -6} ~ ergs ~ cm^{\rm -2} ~ s^{\rm -1}$. Age is in unit of $\rm Myr$. The adjusted $R^{2}$ is 0.5974. The GRB sample number is 14.

The $\log t_{\rm pkOpt}$-$\log L_{\rm pk}$-$\log SSFR$ formula is:
\begin{equation} \label{eq:tpkolpkssfr}
\log t_{\rm pkOpt} = (-0.47 \pm 0.014) \times \log L_{\rm pk} + (0.69 \pm 0.042) \times \log SSFR + (2.2 \pm 0.025),
\end{equation}
where $t_{\rm pkOpt}$ is in unit of $\rm s$. $L_{\rm pk}$ is in unit of $\rm 10^{\rm 52} ~ erg ~ s^{\rm -1}$, and in 1-$10^{4}$ $\rm keV$ energy band. $\log SSFR$ is in unit of $\rm Gyr^{\rm -1}$. The adjusted $R^{2}$ is 0.8097. The GRB sample number is 14.

The $\log t_{\rm pkOpt,i}$-Mag-$\log  N_{\rm H}$ formula is:
\begin{equation} \label{eq:topimagnh}
\log t_{\rm pkOpt,i} = (0.14 \pm 0.058) \times Mag + (0.74 \pm 0.19) \times \log  N_{\rm H} + (4.9 \pm 1.2),
\end{equation}
where $t_{\rm pkOpt,i}$ is in unit of $\rm s$. Mag is in unit of magnitude. $N_{\rm H}$ is in unit of $\rm 10^{\rm 21} ~ cm^{\rm -2}$. The adjusted $R^{2}$ is 0.4411. The GRB sample number is 14.

The $(-\beta_{\rm band})$-$\log P_{\rm pk1}$-$\log F_{\rm radio,pk}$ formula is:
\begin{equation} \label{eq:bendppk1frapk}
(-\beta_{\rm band}) = (0.59 \pm 0.13) \times \log P_{\rm pk1} + (-0.73 \pm 0.2) \times \log F_{\rm radio,pk} + (-0.99 \pm 0.61),
\end{equation}
where $P_{\rm pk1}$ is peak photon flux of 64 $\rm ms$ time bin in 10-1000 $\rm keV$, and in unit of $\rm photons ~ cm^{\rm -2} ~ s^{\rm -1}$. $F_{\rm radio,pk}$ is in unit of $\rm Jy$. The adjusted $R^{2}$ is 0.5843. The GRB sample number is 14.

The $\log HR$-$\log P_{\rm pk2}$-$\log Age$ formula is:
\begin{equation} \label{eq:hrppk2age}
\log HR = (0.38 \pm 0.039) \times \log P_{\rm pk2} + (-0.26 \pm 0.077) \times \log Age + (1 \pm 0.21),
\end{equation}
where $P_{\rm pk2}$ is peak photon flux of 256 $\rm ms$ time bin in 10-1000 $\rm keV$, and in unit of $\rm photons ~ cm^{\rm -2} ~ s^{\rm -1}$. Age is in unit of $\rm Myr$. The adjusted $R^{2}$ is 0.3174. The GRB sample number is 14.

The $\log L_{\rm pk}$-$\log L_{\rm radio,pk}$-$\log t_{\rm pkOpt}$ formula is:
\begin{equation} \label{eq:lpklratpko}
\log L_{\rm pk} = (0.62 \pm 0.068) \times \log L_{\rm radio,pk} + (-0.99 \pm 0.089) \times \log t_{\rm pkOpt} + (1.3 \pm 0.28),
\end{equation}
where $L_{\rm pk}$ is in unit of $\rm 10^{\rm 52} ~ ergs ~ s^{\rm -1}$, and in 1-$10^{4}$ $\rm keV$ energy band. $L_{\rm radio,pk}$ is in rest-frame 8.46 $\rm GHz$, the unit is $\rm 10^{\rm 40} ~ erg ~ s^{\rm -1}$. $t_{\rm pkOpt}$ is in unit of $\rm s$. The adjusted $R^{2}$ is 0.8128. The GRB sample number is 14.

The $\log Age$-$\log \theta_{\rm j}$-$\log  N_{\rm H}$ formula is:
\begin{equation} \label{eq:agethjnh}
\log Age = (0.94 \pm 0.3) \times \log \theta_{\rm j} + (-0.82 \pm 0.32) \times \log  N_{\rm H} + (4.5 \pm 0.37),
\end{equation}
where Age is in unit of $\rm Myr$. $\theta_{\rm j}$ is in unit of $\rm rad$. $N_{\rm H}$ is in unit of $\rm 10^{\rm 21} ~ cm^{\rm -2}$. The adjusted $R^{2}$ is 0.3373. The GRB sample number is 15.

The $\log F_{\rm g}$-$(-\alpha_{\rm spl})$-$\log  N_{\rm H}$ formula is:
\begin{equation} \label{eq:fgalsplnh}
\log F_{\rm g} = (-1.2 \pm 0.2) \times (-\alpha_{\rm spl}) + (0.39 \pm 0.089) \times \log  N_{\rm H} + (2.3 \pm 0.37),
\end{equation}
where $F_{\rm g}$ is in unit of $\rm 10^{\rm -6} ~ ergs ~ cm^{\rm -2}$, and in 20-2000 $\rm keV$ energy band. $N_{\rm H}$ is in unit of $\rm 10^{\rm 21} ~ cm^{\rm -2}$. The adjusted $R^{2}$ is 0.5037. The GRB sample number is 16.

The $\log F_{\rm pk2}$-$\log E_{\rm p,cpl}$-$\log F_{\rm X11hr}$ formula is:
\begin{equation} \label{eq:fpk2epcplfx11}
\log F_{\rm pk2} = (0.87 \pm 0.29) \times \log E_{\rm p,cpl} + (0.35 \pm 0.099) \times \log F_{\rm X11hr} + (0.81 \pm 0.7),
\end{equation}
where $F_{\rm pk2}$ is peak energy flux of 64 $\rm ms$ time bin in rest-frame 1-$10^{4}$ $\rm keV$ energy band, and in unit of $\rm 10^{\rm -6} ~ ergs ~ cm^{\rm -2} ~ s^{\rm -1}$. $E_{\rm p,cpl}$ is in unit of $\rm keV$. $F_{\rm X11hr}$ is in unit of $\rm Jy$. The adjusted $R^{2}$ is 0.6483. The GRB sample number is 16.

The $\log \Gamma_{0}$-$\log F_{\rm g}$-$\log Mass$ formula is:
\begin{equation} \label{eq:g0fgmass}
\log \Gamma_{0} = (0.57 \pm 0.035) \times \log F_{\rm g} + (0.25 \pm 0.038) \times \log Mass + (-1.1 \pm 0.35),
\end{equation}
where $F_{\rm g}$ is in unit of $\rm 10^{\rm -6} ~ ergs ~ cm^{\rm -2}$, and in 20-2000 $\rm keV$ energy band. Mass is in unit of $M_{\bigodot}$. The adjusted $R^{2}$ is 0.6784. The GRB sample number is 16.

The $(-\alpha_{\rm band})$-$variability_{3}$-$\log P_{\rm pk4}$ formula is:
\begin{equation} \label{eq:alndvar3ppk4}
(-\alpha_{\rm band}) = (-21 \pm 12) \times variability_{3} + (0.12 \pm 0.045) \times \log P_{\rm pk4} + (1.2 \pm 0.099),
\end{equation}
where $P_{\rm pk4}$ is peak photon flux of 1 $\rm s$ time bin in 10-1000 $\rm keV$, and in unit of $\rm photons ~ cm^{\rm -2} ~ s^{\rm -1}$. The adjusted $R^{2}$ is 0.4653. The GRB sample number is 16.

The $\log E_{\rm p,cpl,i}$-$\log L_{\rm radio,pk}$-$A_{\rm V}$ formula is:
\begin{equation} \label{eq:epcilraav}
\log E_{\rm p,cpl,i} = (0.28 \pm 0.036) \times \log L_{\rm radio,pk} + (-0.42 \pm 0.13) \times A_{\rm V} + (2.4 \pm 0.091),
\end{equation}
where $E_{\rm p,cpl,i}$ is in unit of $\rm keV$. $L_{\rm radio,pk}$ is in rest-frame 8.46 $\rm GHz$, the unit is $\rm 10^{\rm 40} ~ erg ~ s^{\rm -1}$. The adjusted $R^{2}$ is 0.5624. The GRB sample number is 16.

The $\log F_{\rm g}$-$(-\alpha_{\rm spl})$-$\log F_{\rm X11hr}$ formula is:
\begin{equation} \label{eq:fgalsplfx11}
\log F_{\rm g} = (-0.67 \pm 0.16) \times (-\alpha_{\rm spl}) + (0.43 \pm 0.078) \times \log F_{\rm X11hr} + (4.6 \pm 0.57),
\end{equation}
where $F_{\rm g}$ is in unit of $\rm 10^{\rm -6} ~ ergs ~ cm^{\rm -2}$, and in 20-2000 $\rm keV$ energy band. $F_{\rm X11hr}$ is in unit of $\rm Jy$. The adjusted $R^{2}$ is 0.4784. The GRB sample number is 17.

The $\log \Gamma_{0}$-$\log F_{\rm pk1}$-$\log Mass$ formula is:
\begin{equation} \label{eq:g0fpk1mass}
\log \Gamma_{0} = (0.37 \pm 0.028) \times \log F_{\rm pk1} + (0.23 \pm 0.045) \times \log Mass + (-0.22 \pm 0.43),
\end{equation}
where $F_{\rm pk1}$ is peak energy flux of 1 $\rm s$ time bin in rest-frame 1-$10^{4}$ $\rm keV$ energy band, and in unit of $\rm 10^{\rm -6} ~ ergs ~ cm^{\rm -2} ~ s^{\rm -1}$. Mass is in unit of $M_{\bigodot}$. The adjusted $R^{2}$ is 0.4798. The GRB sample number is 17.

The $\log (1+z)$-$\log F_{\rm pk2}$-$\log F_{\rm radio,pk}$ formula is:
\begin{equation} \label{eq:zfpk2frapk}
\log (1+z) = (-0.13 \pm 0.025) \times \log F_{\rm pk2} + (-0.29 \pm 0.027) \times \log F_{\rm radio,pk} + (-0.47 \pm 0.11),
\end{equation}
where $F_{\rm pk2}$ is peak energy flux of 64 $\rm ms$ time bin in rest-frame 1-$10^{4}$ $\rm keV$ energy band, and in unit of $\rm 10^{\rm -6} ~ ergs ~ cm^{\rm -2} ~ s^{\rm -1}$. $F_{\rm radio,pk}$ is in unit of $\rm Jy$. The adjusted $R^{2}$ is 0.6147. The GRB sample number is 17.

The $\log t_{\rm radio,pk}$-$\log F_{\rm radio,pk}$-$\log t_{\rm pkOpt,i}$ formula is:
\begin{equation} \label{eq:trapkfrapktpkoi}
\log t_{\rm radio,pk} = (-0.56 \pm 0.063) \times \log F_{\rm radio,pk} + (-0.2 \pm 0.021) \times \log t_{\rm pkOpt,i} + (4.2 \pm 0.22),
\end{equation}
where $t_{\rm radio,pk}$ is in unit of $\rm s$. $F_{\rm radio,pk}$ is in unit of $\rm Jy$. $t_{\rm pkOpt,i}$ is in unit of $\rm s$. The adjusted $R^{2}$ is 0.3468. The GRB sample number is 17.

The $\log  D_{\rm L}$-$\log \Gamma_{0}$-$\log Mass$ formula is:
\begin{equation} \label{eq:dlg0mass}
\log  D_{\rm L} = (0.73 \pm 0.05) \times \log \Gamma_{0} + (0.2 \pm 0.041) \times \log Mass + (-3.1 \pm 0.31),
\end{equation}
where $D_{\rm L}$ is in unit of $\rm 10^{\rm 28} ~ cm$. Mass is in unit of $M_{\bigodot}$. The adjusted $R^{2}$ is 0.7751. The GRB sample number is 17.

The spectral lag-$\log t_{\rm burst,i}$-$\log t_{\rm pkOpt,i}$ formula is:
\begin{equation} \label{eq:lagtbuitpkoi}
spectral ~ lag = (3801 \pm 940) \times \log t_{\rm burst,i} + (4028 \pm 1142) \times \log t_{\rm pkOpt,i} + (-13854 \pm 3488),
\end{equation}
where spectral time lag is in unit of $\rm ms ~ MeV^{\rm -1}$. $t_{\rm burst,i}$ is in unit of $\rm s$. $t_{\rm pkOpt,i}$ is in unit of $\rm s$. The adjusted $R^{2}$ is 0.4456. The GRB sample number is 17.

The $\log \Gamma_{0}$-$\log \theta_{\rm j}$-$\log E_{\rm p,cpl,i}$ formula is:
\begin{equation} \label{eq:g0thejepcpli}
\log \Gamma_{0} = (-0.64 \pm 0.11) \times \log \theta_{\rm j} + (0.63 \pm 0.13) \times \log E_{\rm p,cpl,i} + (-0.35 \pm 0.35),
\end{equation}
where $\theta_{\rm j}$ is in unit of $\rm rad$. $E_{\rm p,cpl,i}$ is in unit of $\rm keV$. The adjusted $R^{2}$ is 0.6444. The GRB sample number is 17.

The $\log L_{\rm radio,pk}$-$\log T_{\rm 90}$-$\log t_{\rm pkOpt,i}$ formula is:
\begin{equation} \label{eq:lrat90tpkoi}
\log L_{\rm radio,pk} = (1.6 \pm 0.11) \times \log T_{\rm 90} + (-1 \pm 0.033) \times \log t_{\rm pkOpt,i} + (1.3 \pm 0.2),
\end{equation}
where $L_{\rm radio,pk}$ is in rest-frame 8.46 $\rm GHz$, the unit is $\rm 10^{\rm 40} ~ erg ~ s^{\rm -1}$. $T_{\rm 90}$ is in unit of $\rm s$. $t_{\rm pkOpt,i}$ is in unit of $\rm s$. The adjusted $R^{2}$ is 0.5211. The GRB sample number is 17.

The $\log E_{\rm iso}$-$\log L_{\rm radio,pk}$-$\log t_{\rm pkOpt,i}$ formula is:
\begin{equation} \label{eq:eisolratpkoi}
\log E_{\rm iso} = (0.36 \pm 0.019) \times \log L_{\rm radio,pk} + (-0.51 \pm 0.03) \times \log t_{\rm pkOpt,i} + (1.3 \pm 0.097),
\end{equation}
where $E_{\rm iso}$ is in unit of $\rm 10^{\rm 52} ~ ergs$ and in rest-frame 1-$10^{4}$ $\rm keV$ energy band. $L_{\rm radio,pk}$ is in rest-frame 8.46 $\rm GHz$, the unit is $\rm 10^{\rm 40} ~ erg ~ s^{\rm -1}$. $t_{\rm pkOpt,i}$ is in unit of $\rm s$. The adjusted $R^{2}$ is 0.6016. The GRB sample number is 17.

The $\log T_{\rm R45}$-$\log F_{\rm X11hr}$-$\log t_{\rm radio,pk,i}$ formula is:
\begin{equation} \label{eq:tr45fx11trapki}
\log T_{\rm R45} = (-0.41 \pm 0.044) \times \log F_{\rm X11hr} + (-0.75 \pm 0.07) \times \log t_{\rm radio,pk,i} + (1.9 \pm 0.3),
\end{equation}
where $T_{\rm R45}$ is in unit of $\rm s$. $F_{\rm X11hr}$ is in unit of $\rm Jy$. $t_{\rm radio,pk,i}$ is in unit of $\rm s$. The adjusted $R^{2}$ is 0.3888. The GRB sample number is 18.

The $(-\beta_{\rm band})$-$\log P_{\rm pk1}$-$\log SSFR$ formula is:
\begin{equation} \label{eq:bendppk1ssfr}
(-\beta_{\rm band}) = (0.12 \pm 0.27) \times \log P_{\rm pk1} + (-0.75 \pm 0.57) \times \log SSFR + (2.4 \pm 0.63),
\end{equation}
where $P_{\rm pk1}$ is peak photon flux of 64 $\rm ms$ time bin in 10-1000 $\rm keV$, and in unit of $\rm photons ~ cm^{\rm -2} ~ s^{\rm -1}$. $\log SSFR$ is in unit of $\rm Gyr^{\rm -1}$. The adjusted $R^{2}$ is 0.4079. The GRB sample number is 18.

The $\log t_{\rm radio,pk,i}$-$\log T_{\rm 50}$-$\log SSFR$ formula is:
\begin{equation} \label{eq:trapkit50ssfr}
\log t_{\rm radio,pk,i} = (-0.44 \pm 0.018) \times \log T_{\rm 50} + (0.43 \pm 0.025) \times \log SSFR + (6.3 \pm 0.025),
\end{equation}
where $t_{\rm radio,pk,i}$ is in unit of $\rm s$. $T_{\rm 50}$ is in unit of $\rm s$. $\log SSFR$ is in unit of $\rm Gyr^{\rm -1}$. The adjusted $R^{2}$ is 0.6841. The GRB sample number is 18.

The $\log \theta_{\rm j}$-$variability_{3}$-$\log  N_{\rm H}$ formula is:
\begin{equation} \label{eq:thejvar3nh}
\log \theta_{\rm j} = (24 \pm 5.3) \times variability_{3} + (0.28 \pm 0.057) \times \log  N_{\rm H} + (-1.7 \pm 0.078),
\end{equation}
where $\theta_{\rm j}$ is in unit of $\rm rad$. $N_{\rm H}$ is in unit of $\rm 10^{\rm 21} ~ cm^{\rm -2}$. The adjusted $R^{2}$ is 0.4522. The GRB sample number is 18.

The metallicity-$(-\alpha_{\rm cpl})$-$\beta_{\rm X11hr}$ formula is:
\begin{equation} \label{eq:metalcplbex11}
metallicity = (-0.13 \pm 0.066) \times (-\alpha_{\rm cpl}) + (0.1 \pm 0.064) \times \beta_{\rm X11hr} + (8.6 \pm 0.13),
\end{equation}
where metallicity is the value of $12+\log O/H$. The adjusted $R^{2}$ is 0.3058. The GRB sample number is 19.

The $\log E_{\rm p,cpl,i}$-$\log HR$-$\log t_{\rm radio,pk,i}$ formula is:
\begin{equation} \label{eq:epcplihrtrapki}
\log E_{\rm p,cpl,i} = (0.64 \pm 0.099) \times \log HR + (-0.31 \pm 0.12) \times \log t_{\rm radio,pk,i} + (4.4 \pm 0.67),
\end{equation}
where $E_{\rm p,cpl,i}$ is in unit of $\rm keV$. $t_{\rm radio,pk,i}$ is in unit of $\rm s$. The adjusted $R^{2}$ is 0.6582. The GRB sample number is 19.

The $\log L_{\rm pk}$-$\log t_{\rm pkOpt}$-Mag formula is:
\begin{equation} \label{eq:lpktpkomag}
\log L_{\rm pk} = (-1.3 \pm 0.14) \times \log t_{\rm pkOpt} + (-0.22 \pm 0.06) \times Mag + (-1.1 \pm 1.3),
\end{equation}
where $L_{\rm pk}$ is in unit of $\rm 10^{\rm 52} ~ erg ~ s^{\rm -1}$, and in 1-$10^{4}$ $\rm keV$ energy band. $t_{\rm pkOpt}$ is in unit of $\rm s$. Mag is in unit of magnitude. The adjusted $R^{2}$ is 0.9072. The GRB sample number is 19.

The Mag-$\log (1+z)$-metallicity formula is:
\begin{equation} \label{eq:magzmet}
Mag = (-12 \pm 4.5) \times \log (1+z) + (-1.7 \pm 1.6) \times metallicity + (-2.5 \pm 13),
\end{equation}
where Mag is in unit of magnitude. Metallicity is the value of $12+\log O/H$. The adjusted $R^{2}$ is 0.7636. The GRB sample number is 19.

The $\log L_{\rm radio,pk}$-$\log F_{\rm Opt11hr}$-$\log E_{\rm p,cpl,i}$ formula is:
\begin{equation} \label{eq:lrafo11epci}
\log L_{\rm radio,pk} = (-0.75 \pm 0.19) \times \log F_{\rm Opt11hr} + (1.4 \pm 0.23) \times \log E_{\rm p,cpl,i} + (-5.1 \pm 0.95),
\end{equation}
where $L_{\rm radio,pk}$ is in rest-frame 8.46 $\rm GHz$, the unit is $\rm 10^{\rm 40} ~ erg ~ s^{\rm -1}$. $F_{\rm Opt11hr}$ is in unit of $\rm Jy$. $E_{\rm p,cpl,i}$ is in unit of $\rm keV$. The adjusted $R^{2}$ is 0.6282. The GRB sample number is 19.

The $\log F_{\rm pk3}$-$A_{\rm V}$-$\log SSFR$ formula is:
\begin{equation} \label{eq:fpk3avssfr}
\log F_{\rm pk3} = (-0.57 \pm 0.14) \times A_{\rm V} + (-0.48 \pm 0.076) \times \log SSFR + (1.2 \pm 0.13),
\end{equation}
where $F_{\rm pk3}$ is peak energy flux of 256 $\rm ms$ time bin in rest-frame 1-$10^{4}$ $\rm keV$ energy band, and in unit of $\rm 10^{\rm -6} ~ ergs ~ cm^{\rm -2} ~ s^{\rm -1}$. $\log SSFR$ is in unit of $\rm Gyr^{\rm -1}$. The adjusted $R^{2}$ is 0.363. The GRB sample number is 20.

The $\log  F_{\rm Opt11hr}$-$(-\alpha_{\rm cpl})$-$\log \Gamma_{0}$ formula is:
\begin{equation} \label{eq:fop11alcplg0}
\log  F_{\rm Opt11hr} = (0.34 \pm 0.24) \times (-\alpha_{\rm cpl}) + (-0.56 \pm 0.19) \times \log \Gamma_{0} + (-3.8 \pm 0.49),
\end{equation}
where $F_{\rm Opt11hr}$ is in unit of $\rm Jy$. The adjusted $R^{2}$ is 0.3106. The GRB sample number is 20.

The $\log \Gamma_{0}$-$\log  F_{\rm Opt11hr}$-$\log E_{\rm p,cpl,i}$ formula is:
\begin{equation} \label{eq:g0fop11epcpli}
\log \Gamma_{0} = (-0.23 \pm 0.079) \times \log  F_{\rm Opt11hr} + (0.61 \pm 0.098) \times \log E_{\rm p,cpl,i} + (-0.59 \pm 0.43),
\end{equation}
where $F_{\rm Opt11hr}$ is in unit of $\rm Jy$. $E_{\rm p,cpl,i}$ is in unit of $\rm keV$. The adjusted $R^{2}$ is 0.5677. The GRB sample number is 20.

The $\log  D_{\rm L}$-$\log F_{\rm radio,pk}$-$\log E_{\rm p,cpl,i}$ formula is:
\begin{equation} \label{eq:dlfrapkepcpli}
\log  D_{\rm L} = (-0.72 \pm 0.092) \times \log F_{\rm radio,pk} + (0.6 \pm 0.088) \times \log E_{\rm p,cpl,i} + (-3.7 \pm 0.37),
\end{equation}
where $D_{\rm L}$ is in unit of $\rm 10^{\rm 28} ~ cm$. $F_{\rm radio,pk}$ is in unit of $\rm Jy$. $E_{\rm p,cpl,i}$ is in unit of $\rm keV$. The adjusted $R^{2}$ is 0.5231. The GRB sample number is 20.

The $\log SFR$-$\log HR$-$(-\alpha_{\rm spl})$ formula is:
\begin{equation} \label{eq:sfrhralspl}
\log SFR = (-0.17 \pm 0.98) \times \log HR + (-0.59 \pm 1) \times (-\alpha_{\rm spl}) + (1.9 \pm 2.3),
\end{equation}
where SFR is in unit of $\rm M_{\bigodot} ~ yr^{\rm -1}$. The adjusted $R^{2}$ is 0.2678. The GRB sample number is 20.

The $\log SFR$-Mag-$\log E_{\rm p,cpl,i}$ formula is:
\begin{equation} \label{eq:sfrmagepcpli}
\log SFR = (-0.16 \pm 0.055) \times Mag + (0.78 \pm 0.4) \times \log E_{\rm p,cpl,i} + (-4.3 \pm 1.1),
\end{equation}
where SFR is in unit of $\rm M_{\bigodot} ~ yr^{\rm -1}$. Mag is in unit of magnitude. $E_{\rm p,cpl,i}$ is in unit of $\rm keV$. The adjusted $R^{2}$ is 0.8205. The GRB sample number is 20.

The $\log SFR$-$\log (1+z)$-$(-\alpha_{\rm spl})$ formula is:
\begin{equation} \label{eq:sfrzalspl}
\log SFR = (3.5 \pm 0.4) \times \log (1+z) + (-0.49 \pm 0.13) \times (-\alpha_{\rm spl}) + (0.49 \pm 0.23),
\end{equation}
where SFR is in unit of $\rm M_{\bigodot} ~ yr^{\rm -1}$. The adjusted $R^{2}$ is 0.4955. The GRB sample number is 20.

The $\log E_{\rm iso}$-$\log L_{\rm radio,pk}$-$\log E_{\rm p,cpl}$ formula is:
\begin{equation} \label{eq:eisolraepcpl}
\log E_{\rm iso} = (0.56 \pm 0.03) \times \log L_{\rm radio,pk} + (1 \pm 0.17) \times \log E_{\rm p,cpl} + (-2.5 \pm 0.35),
\end{equation}
where $E_{\rm iso}$ is in unit of $\rm 10^{\rm 52} ~ ergs$ and in rest-frame 1-$10^{4}$ $\rm keV$ energy band. $L_{\rm radio,pk}$ is in rest-frame 8.46 $\rm GHz$, the unit is $\rm 10^{\rm 40} ~ erg ~ s^{\rm -1}$. $E_{\rm p,cpl}$ is in unit of $\rm keV$. The adjusted $R^{2}$ is 0.7599. The GRB sample number is 20.

The $\log T_{\rm 90,i}$-$\log L_{\rm radio,pk}$-$(-\alpha_{\rm cpl})$ formula is:
\begin{equation} \label{eq:t90ilraacpl}
\log T_{\rm 90,i} = (-0.18 \pm 0.046) \times \log L_{\rm radio,pk} + (0.67 \pm 0.25) \times (-\alpha_{\rm cpl}) + (0.83 \pm 0.33),
\end{equation}
where $T_{\rm 90,i}$ is in unit of $\rm s$. $L_{\rm radio,pk}$ is in rest-frame 8.46 $\rm GHz$, the unit is $\rm 10^{\rm 40} ~ erg ~ s^{\rm -1}$. The adjusted $R^{2}$ is 0.3241. The GRB sample number is 20.

The $\log t_{\rm pkOpt}$-$\log  E_{\rm p,band}$-$\log t_{\rm burst}$ formula is:
\begin{equation} \label{eq:tpkoepndtbur}
\log t_{\rm pkOpt} = (-0.58 \pm 0.097) \times \log  E_{\rm p,band} + (0.5 \pm 0.027) \times \log t_{\rm burst} + (2.4 \pm 0.2),
\end{equation}
where $t_{\rm pkOpt}$ is in unit of $\rm s$. $E_{\rm p,band}$ is in unit of $\rm keV$. $t_{\rm burst}$ is in unit of $\rm s$. The adjusted $R^{2}$ is 0.2739. The GRB sample number is 21.

The $\log P_{\rm pk1}$-$\log F_{\rm g}$-metallicity formula is:
\begin{equation} \label{eq:ppk1fgmet}
\log P_{\rm pk1} = (0.44 \pm 0.027) \times \log F_{\rm g} + (0.87 \pm 0.24) \times metallicity + (-6.6 \pm 2.1),
\end{equation}
where $P_{\rm pk1}$ is peak photon flux of 64 $\rm ms$ time bin in 10-1000 $\rm keV$, and in unit of $\rm photons ~ cm^{\rm -2} ~ s^{\rm -1}$. $F_{\rm g}$ is in unit of $\rm 10^{\rm -6} ~ ergs ~ cm^{\rm -2}$, and in 20-2000 $\rm keV$ energy band. Metallicity is the value of $12+\log O/H$. The adjusted $R^{2}$ is 0.5787. The GRB sample number is 21.

The spectral lag-$\log (1+z)$-$\log F_{\rm pk3}$ formula is:
\begin{equation} \label{eq:lagzfpk3}
spectral ~ lag = (-28190 \pm 4027) \times \log (1+z) + (-8125 \pm 1343) \times \log F_{\rm pk3} + (17758 \pm 2296),
\end{equation}
where spectral time lag is in unit of $\rm ms ~ MeV^{\rm -1}$. $F_{\rm pk3}$ is peak energy flux of 256 $\rm ms$ time bin in rest-frame 1-$10^{4}$ $\rm keV$ energy band, and in unit of $\rm 10^{\rm -6} ~ ergs ~ cm^{\rm -2} ~ s^{\rm -1}$. The adjusted $R^{2}$ is 0.4887. The GRB sample number is 21.

The spectral lag-$\log (1+z)$-$\log P_{\rm pk2}$ formula is:
\begin{equation} \label{eq:zppk2lag}
spectral ~ lag = (-26436 \pm 3596) \times \log (1+z) + (-7866 \pm 1114) \times \log P_{\rm pk2} + (22841 \pm 2772),
\end{equation}
where spectral time lag is in unit of $\rm ms ~ MeV^{\rm -1}$. $P_{\rm pk2}$ is peak photon flux of 256 $\rm ms$ time bin in 10-1000 $\rm keV$, and in unit of $\rm photons ~ cm^{\rm -2} ~ s^{\rm -1}$. The adjusted $R^{2}$ is 0.3876. The GRB sample number is 21.

The $\log F_{\rm radio,pk}$-$(-\alpha_{\rm band})$-$\log Mass$ formula is:
\begin{equation} \label{eq:afrpkabandms}
\log F_{\rm radio,pk} = (0.58 \pm 0.16) \times (-\alpha_{\rm band}) + (-0.3 \pm 0.077) \times \log Mass + (-1.1 \pm 0.81),
\end{equation}
where $F_{\rm radio,pk}$ is in unit of $\rm Jy$. Mass is in unit of $M_{\bigodot}$. The adjusted $R^{2}$ is 0.434. The GRB sample number is 22.

The $\log F_{\rm pk2}$-$(-\alpha_{\rm cpl})$-$\log  F_{\rm Opt11hr}$ formula is:
\begin{equation} \label{eq:fpk2acplfo11}
\log F_{\rm pk2} = (-0.48 \pm 0.17) \times (-\alpha_{\rm cpl}) + (0.23 \pm 0.11) \times \log  F_{\rm Opt11hr} + (1.9 \pm 0.59),
\end{equation}
where $F_{\rm pk2}$ is peak energy flux of 64 $\rm ms$ time bin in rest-frame 1-$10^{4}$ $\rm keV$ energy band, and in unit of $\rm 10^{\rm -6} ~ ergs ~ cm^{\rm -2} ~ s^{\rm -1}$. $F_{\rm Opt11hr}$ is in unit of $\rm Jy$. The adjusted $R^{2}$ is 0.3327. The GRB sample number is 22.

The $\log SFR$-metallicity-$\log Age$ formula is:
\begin{equation} \label{eq:sfrmetage}
\log SFR = (1.3 \pm 0.38) \times metallicity + (-0.64 \pm 0.12) \times \log Age + (-9.1 \pm 3.3),
\end{equation}
where SFR is in unit of $\rm M_{\bigodot} ~ yr^{\rm -1}$. Metallicity is the value of $12+\log O/H$. Age is in unit of $\rm Myr$. The adjusted $R^{2}$ is 0.4076. The GRB sample number is 22.

The Mag-$\log  N_{\rm H}$-$\log E_{\rm p,cpl,i}$ formula is:
\begin{equation} \label{eq:magnhepli}
Mag = (-1.5 \pm 0.89) \times \log  N_{\rm H} + (-1.2 \pm 0.96) \times \log E_{\rm p,cpl,i} + (-17 \pm 2.5),
\end{equation}
where Mag is in unit of magnitude. $N_{\rm H}$ is in unit of $\rm 10^{\rm 21} ~ cm^{\rm -2}$. $E_{\rm p,cpl,i}$ is in unit of $\rm keV$. The adjusted $R^{2}$ is 0.2554. The GRB sample number is 22.

The $\log E_{\rm iso}$-$\log offset$-$\log E_{\rm p,cpl,i}$ formula is:
\begin{equation} \label{eq:eisoofetecpli}
\log E_{\rm iso} = (-0.8 \pm 0.21) \times \log offset + (1.3 \pm 0.3) \times \log E_{\rm p,cpl,i} + (-3.7 \pm 0.82),
\end{equation}
where $E_{\rm iso}$ is in unit of $\rm 10^{\rm 52} ~ ergs$ and in rest-frame 1-$10^{4}$ $\rm keV$ energy band. Host galaxy offset is in unit of $\rm kpc$. $E_{\rm p,cpl,i}$ is in unit of $\rm keV$. The adjusted $R^{2}$ is 0.5009. The GRB sample number is 22.

The Mag-$\log SFR$-$\log t_{\rm burst,i}$ formula is:
\begin{equation} \label{eq:magsfrtti}
Mag = (-1.5 \pm 0.54) \times \log SFR + (0.68 \pm 0.65) \times \log t_{\rm burst,i} + (-21 \pm 2),
\end{equation}
where Mag is in unit of magnitude. SFR is in unit of $\rm M_{\bigodot} ~ yr^{\rm -1}$. $t_{\rm burst,i}$ is in unit of $\rm s$. The adjusted $R^{2}$ is 0.8763. The GRB sample number is 22.

The $\log F_{\rm radio,pk}$-$\log (1+z)$-$variability_{1}$ formula is:
\begin{equation} \label{eq:frapkzvar1}
\log F_{\rm radio,pk} = (-0.74 \pm 0.13) \times \log (1+z) + (-0.75 \pm 0.87) \times variability_{1} + (-3.2 \pm 0.053),
\end{equation}
where $F_{\rm radio,pk}$ is in unit of $\rm Jy$. The adjusted $R^{2}$ is 0.286. The GRB sample number is 22.

The $\log T_{\rm R45,i}$-$\log L_{\rm radio,pk}$-$\log SSFR$ formula is:
\begin{equation} \label{eq:tr45ilrassfr}
\log T_{\rm R45,i} = (-0.19 \pm 0.011) \times \log L_{\rm radio,pk} + (0.3 \pm 0.026) \times \log SSFR + (0.61 \pm 0.016),
\end{equation}
where $T_{\rm R45,i}$ is in unit of $\rm s$. $L_{\rm radio,pk}$ is in rest-frame 8.46 $\rm GHz$, the unit is $\rm 10^{\rm 40} ~ erg ~ s^{\rm -1}$. $\log SSFR$ is in unit of $\rm Gyr^{\rm -1}$. The adjusted $R^{2}$ is 0.2284. The GRB sample number is 23.

The $\log F_{\rm X11hr}$-$\log F_{\rm pk2}$-$\beta_{\rm X11hr}$ formula is:
\begin{equation} \label{eq:fx11fpk2be11}
\log F_{\rm X11hr} = (0.65 \pm 0.13) \times \log F_{\rm pk2} + (-0.94 \pm 0.17) \times \beta_{\rm X11hr} + (-6.5 \pm 0.25),
\end{equation}
where $F_{\rm X11hr}$ is in unit of $\rm Jy$. $F_{\rm pk2}$ is peak energy flux of 64 $\rm ms$ time bin in rest-frame 1-$10^{4}$ $\rm keV$ energy band, and in unit of $\rm 10^{\rm -6} ~ ergs ~ cm^{\rm -2} ~ s^{\rm -1}$. The adjusted $R^{2}$ is 0.6052. The GRB sample number is 23.

The $\log E_{\rm iso}$-$\log \Gamma_{0}$-$\log E_{\rm p,cpl,i}$ formula is:
\begin{equation} \label{eq:eog0ecpli}
\log E_{\rm iso} = (1.3 \pm 0.11) \times \log \Gamma_{0} + (0.83 \pm 0.17) \times \log E_{\rm p,cpl,i} + (-4.3 \pm 0.3),
\end{equation}
where $E_{\rm iso}$ is in unit of $\rm 10^{\rm 52} ~ ergs$ and in rest-frame 1-$10^{4}$ $\rm keV$ energy band. $E_{\rm p,cpl,i}$ is in unit of $\rm keV$. The adjusted $R^{2}$ is 0.9021. The GRB sample number is 23.

The $\log E_{\rm p,band,i}$-$\log HR$-$\log SFR$ formula is:
\begin{equation} \label{eq:ebandihrsfr}
\log E_{\rm p,band,i} = (1.1 \pm 0.14) \times \log HR + (0.11 \pm 0.026) \times \log SFR + (2 \pm 0.081),
\end{equation}
where $E_{\rm p,band,i}$ is in unit of $\rm keV$. SFR is in unit of $\rm M_{\bigodot} ~ yr^{\rm -1}$. The adjusted $R^{2}$ is 0.6949. The GRB sample number is 23.

The Mag-$\log HR$-$\log E_{\rm p,band,i}$ formula is:
\begin{equation} \label{eq:maghrebandi}
Mag = (2.7 \pm 2.1) \times \log HR + (-2.4 \pm 1.4) \times \log E_{\rm p,band,i} + (-16 \pm 3.4),
\end{equation}
where Mag is in unit of magnitude. $E_{\rm p,band,i}$ is in unit of $\rm keV$. The adjusted $R^{2}$ is 0.3913. The GRB sample number is 23.

The $\log offset$-$\log T_{\rm 50}$-$\log P_{\rm pk4}$ formula is:
\begin{equation} \label{eq:ofett50ppk4}
\log offset = (-0.32 \pm 0.044) \times \log T_{\rm 50} + (-0.35 \pm 0.19) \times \log P_{\rm pk4} + (1.1 \pm 0.17),
\end{equation}
where host galaxy offset is in unit of $\rm kpc$. $T_{\rm 50}$ is in unit of $\rm s$. $P_{\rm pk4}$ is peak photon flux of 1 $\rm s$ time bin in 10-1000 $\rm keV$, and in unit of $\rm photons ~ cm^{\rm -2} ~ s^{\rm -1}$. The adjusted $R^{2}$ is 0.4552. The GRB sample number is 23.

The $\log T_{\rm 90}$-$variability_{2}$-$\log E_{\rm p,cpl,i}$ formula is:
\begin{equation} \label{eq:t90var2ecpli}
\log T_{\rm 90} = (3.5 \pm 0.93) \times variability_{2} + (0.36 \pm 0.13) \times \log E_{\rm p,cpl,i} + (0.18 \pm 0.38),
\end{equation}
where $T_{\rm 90}$ is in unit of $\rm s$. $E_{\rm p,cpl,i}$ is in unit of $\rm keV$. The adjusted $R^{2}$ is 0.3912. The GRB sample number is 23.

The $\log E_{\rm iso}$-$\log L_{\rm radio,pk}$-$variability_{3}$ formula is:
\begin{equation} \label{eq:eisolravar3}
\log E_{\rm iso} = (0.33 \pm 0.037) \times \log L_{\rm radio,pk} + (58 \pm 7.7) \times variability_{3} + (0.1 \pm 0.094),
\end{equation}
where $E_{\rm iso}$ is in unit of $\rm 10^{\rm 52} ~ ergs$ and in rest-frame 1-$10^{4}$ $\rm keV$ energy band. $L_{\rm radio,pk}$ is in rest-frame 8.46 $\rm GHz$, the unit is $\rm 10^{\rm 40} ~ erg ~ s^{\rm -1}$. The adjusted $R^{2}$ is 0.3608. The GRB sample number is 24.

The Mag-$(-\beta_{\rm band})$-$\log  F_{\rm Opt11hr}$ formula is:
\begin{equation} \label{eq:magbendfo11}
Mag = (0.48 \pm 0.85) \times (-\beta_{\rm band}) + (0.54 \pm 0.47) \times \log F_{\rm Opt11hr}  + (-20 \pm 3.4),
\end{equation}
where Mag is in unit of magnitude. $F_{\rm Opt11hr}$ is in unit of $\rm Jy$. The adjusted $R^{2}$ is 0.3541. The GRB sample number is 24.

The $\log P_{\rm pk1}$-$\log F_{\rm g}$-$\log  N_{\rm H}$ formula is:
\begin{equation} \label{eq:ppk1fgnh}
\log P_{\rm pk1} = (0.44 \pm 0.016) \times \log F_{\rm g} + (-0.28 \pm 0.043) \times \log  N_{\rm H} + (0.93 \pm 0.042),
\end{equation}
where $P_{\rm pk1}$ is peak photon flux of 64 $\rm ms$ time bin in 10-1000 $\rm keV$, and in unit of $\rm photons ~ cm^{\rm -2} ~ s^{\rm -1}$. $F_{\rm g}$ is in unit of $\rm 10^{\rm -6} ~ ergs ~ cm^{\rm -2}$, and in 20-2000 $\rm keV$ energy band. $N_{\rm H}$ is in unit of $\rm 10^{\rm 21} ~ cm^{\rm -2}$. The adjusted $R^{2}$ is 0.6225. The GRB sample number is 24.

The $\log t_{\rm pkOpt}$-$\log  N_{\rm H}$-$\log t_{\rm burst,i}$ formula is:
\begin{equation} \label{eq:toptnhtti}
\log t_{\rm pkOpt} = (0.52 \pm 0.08) \times \log  N_{\rm H} + (0.45 \pm 0.028) \times \log t_{\rm burst,i} + (1.1 \pm 0.092),
\end{equation}
where $t_{\rm pkOpt}$ is in unit of $\rm s$. $N_{\rm H}$ is in unit of $\rm 10^{\rm 21} ~ cm^{\rm -2}$. $t_{\rm burst,i}$ is in unit of $\rm s$. The adjusted $R^{2}$ is 0.3136. The GRB sample number is 24.

The $\log  F_{\rm Opt11hr}$-$\log SFR$-$\log t_{\rm burst,i}$ formula is:
\begin{equation} \label{eq:fo11sfrtti}
\log F_{\rm Opt11hr} = (-0.28 \pm 0.11) \times \log SFR + (0.56 \pm 0.14) \times \log t_{\rm burst,i} + (-6.1 \pm 0.41),
\end{equation}
where $F_{\rm Opt11hr}$ is in unit of $\rm Jy$. SFR is in unit of $\rm M_{\bigodot} ~ yr^{\rm -1}$. $t_{\rm burst,i}$ is in unit of $\rm s$. The adjusted $R^{2}$ is 0.4728. The GRB sample number is 24.

The $\log F_{\rm pk3}$-$\log (1+z)$-$\log  N_{\rm H}$ formula is:
\begin{equation} \label{eq:fpk3znh}
\log F_{\rm pk3} = (-1.5 \pm 0.24) \times \log (1+z) + (-0.5 \pm 0.12) \times \log  N_{\rm H} + (1.3 \pm 0.078),
\end{equation}
where $F_{\rm pk3}$ is peak energy flux of 256 $\rm ms$ time bin in rest-frame 1-$10^{4}$ $\rm keV$ energy band, and in unit of $\rm 10^{\rm -6} ~ ergs ~ cm^{\rm -2} ~ s^{\rm -1}$. $N_{\rm H}$ is in unit of $\rm 10^{\rm 21} ~ cm^{\rm -2}$. The adjusted $R^{2}$ is 0.4391. The GRB sample number is 24.

The $\log SFR$-$\log  F_{\rm Opt11hr}$-$\log  N_{\rm H}$ formula is:
\begin{equation} \label{eq:sfrfo11nh}
\log SFR = (-0.47 \pm 0.12) \times \log F_{\rm Opt11hr} + (1.2 \pm 0.22) \times \log  N_{\rm H} + (-2.3 \pm 0.55),
\end{equation}
where SFR is in unit of $\rm M_{\bigodot} ~ yr^{\rm -1}$. $F_{\rm Opt11hr}$ is in unit of $\rm Jy$. $N_{\rm H}$ is in unit of $\rm 10^{\rm 21} ~ cm^{\rm -2}$. The adjusted $R^{2}$ is 0.5387. The GRB sample number is 25.

The $A_{\rm V}$-$\log F_{\rm pk4}$-$(-\alpha_{\rm cpl})$ formula is:
\begin{equation} \label{eq:avfpk4acpl}
A_{\rm V} = (1 \pm 0.16) \times \log F_{\rm pk4} + (-0.72 \pm 0.2) \times (-\alpha_{\rm cpl}) + (1.7 \pm 0.23),
\end{equation}
where $F_{\rm pk4}$ is peak energy flux of 1024 $\rm ms$ time bin in rest-frame 1-$10^{4}$ $\rm keV$ energy band, and in unit of $\rm 10^{\rm -6} ~ ergs ~ cm^{\rm -2} ~ s^{\rm -1}$. The adjusted $R^{2}$ is 0.2497. The GRB sample number is 25.

The Mag-$\log L_{\rm pk}$-$(-\beta_{\rm band})$ formula is:
\begin{equation} \label{eq:maglpkbeband}
Mag = (-0.68 \pm 0.79) \times \log L_{\rm pk} + (0.5 \pm 0.77) \times (-\beta_{\rm band}) + (-22 \pm 1.8),
\end{equation}
where Mag is in unit of magnitude. $L_{\rm pk}$ is in unit of $\rm 10^{\rm 52} ~ erg ~ s^{\rm -1}$, and in 1-$10^{4}$ $\rm keV$ energy band. The adjusted $R^{2}$ is 0.227. The GRB sample number is 25.

The $\log  F_{\rm Opt11hr}$-$\log Mass$-$\log t_{\rm pkOpt,i}$ formula is:
\begin{equation} \label{eq:fo11masstoi}
\log F_{\rm Opt11hr} = (-0.27 \pm 0.092) \times \log Mass + (0.43 \pm 0.11) \times \log t_{\rm pkOpt,i} + (-2.9 \pm 0.97),
\end{equation}
where $F_{\rm Opt11hr}$ is in unit of $\rm Jy$. Mass is in unit of $M_{\bigodot}$. $t_{\rm pkOpt,i}$ is in unit of $\rm s$. The adjusted $R^{2}$ is 0.3158. The GRB sample number is 25.

The $A_{\rm V}$-$\log P_{\rm pk3}$-$(-\alpha_{\rm cpl})$ formula is:
\begin{equation} \label{eq:avppk3acpl}
A_{\rm V} = (1.2 \pm 0.19) \times \log P_{\rm pk3} + (-0.75 \pm 0.2) \times (-\alpha_{\rm cpl}) + (0.74 \pm 0.25),
\end{equation}
where $P_{\rm pk3}$ is peak photon flux of 1024 $\rm ms$ time bin in 10-1000 $\rm keV$, and in unit of $\rm photons ~ cm^{\rm -2} ~ s^{\rm -1}$. The adjusted $R^{2}$ is 0.2262. The GRB sample number is 25.

The $\log  N_{\rm H}$-$\log T_{\rm 90}$-$\log F_{\rm pk2}$ formula is:
\begin{equation} \label{eq:nht90fpk2}
\log  N_{\rm H} = (0.45 \pm 0.057) \times \log T_{\rm 90} + (-0.43 \pm 0.055) \times \log F_{\rm pk2} + (0.23 \pm 0.099),
\end{equation}
where $N_{\rm H}$ is in unit of $\rm 10^{\rm 21} ~ cm^{\rm -2}$. $T_{\rm 90}$ is in unit of $\rm s$. $F_{\rm pk2}$ is peak energy flux of 64 $\rm ms$ time bin in rest-frame 1-$10^{4}$ $\rm keV$ energy band, and in unit of $\rm 10^{\rm -6} ~ ergs ~ cm^{\rm -2} ~ s^{\rm -1}$. The adjusted $R^{2}$ is 0.4444. The GRB sample number is 25.

The rest-frame spectral lag-$variability_{3}$-$\log \theta_{\rm j}$ formula is:
\begin{equation} \label{eq:lagivar3thej}
rest-frame ~ spectral ~ lag = (-54909 \pm 24240) \times variability_{3} + (1590 \pm 590) \times \log \theta_{\rm j} + (3315 \pm 906),
\end{equation}
where rest-frame spectral lag is in unit of $\rm ms ~ MeV^{\rm -1}$. $\theta_{\rm j}$ is in unit of $\rm rad$. The adjusted $R^{2}$ is 0.4137. The GRB sample number is 25.

The $\log SFR$-$A_{\rm V}$-$\log Age$ formula is:
\begin{equation} \label{eq:sfravage}
\log SFR = (0.43 \pm 0.16) \times A_{\rm V} + (-0.77 \pm 0.11) \times \log Age + (2.5 \pm 0.33),
\end{equation}
where SFR is in unit of $\rm M_{\bigodot} ~ yr^{\rm -1}$. Age is in unit of $\rm Myr$. The adjusted $R^{2}$ is 0.4286. The GRB sample number is 26.

The $\log  N_{\rm H}$-$\log Age$-$\log T_{\rm 90,i}$ formula is:
\begin{equation} \label{eq:nhaget90i}
\log  N_{\rm H} = (-0.41 \pm 0.071) \times \log Age + (0.41 \pm 0.054) \times \log T_{\rm 90,i} + (1.2 \pm 0.22),
\end{equation}
where $N_{\rm H}$ is in unit of $\rm 10^{\rm 21} ~ cm^{\rm -2}$. Age is in unit of $\rm Myr$. $T_{\rm 90,i}$ is in unit of $\rm s$. The adjusted $R^{2}$ is 0.4491. The GRB sample number is 26.

The $\log Age$-$\log E_{\rm iso}$-$\log  N_{\rm H}$ formula is:
\begin{equation} \label{eq:ageeisonh}
\log Age = (-0.21 \pm 0.062) \times \log E_{\rm iso} + (-0.51 \pm 0.16) \times \log  N_{\rm H} + (3.1 \pm 0.094),
\end{equation}
where Age is in unit of $\rm Myr$. $E_{\rm iso}$ is in unit of $\rm 10^{\rm 52} ~ ergs$ and in rest-frame 1-$10^{4}$ $\rm keV$ energy band. $N_{\rm H}$ is in unit of $\rm 10^{\rm 21} ~ cm^{\rm -2}$. The adjusted $R^{2}$ is 0.3666. The GRB sample number is 26.

The $\log Age$-$\log (1+z)$-$\log  N_{\rm H}$ formula is:
\begin{equation} \label{eq:ageznh}
\log Age = (-2.7 \pm 0.58) \times \log (1+z) + (-0.41 \pm 0.17) \times \log  N_{\rm H} + (3.8 \pm 0.11),
\end{equation}
where Age is in unit of $\rm Myr$. $N_{\rm H}$ is in unit of $\rm 10^{\rm 21} ~ cm^{\rm -2}$. The adjusted $R^{2}$ is 0.5434. The GRB sample number is 26.

The $\log T_{\rm 90,i}$-$\beta_{\rm X11hr}$-$\log offset$ formula is:
\begin{equation} \label{eq:t90ibex11ofet}
\log T_{\rm 90,i} = (-0.6 \pm 0.17) \times \beta_{\rm X11hr} + (-0.55 \pm 0.16) \times \log offset + (1.6 \pm 0.31),
\end{equation}
where $T_{\rm 90,i}$ is in unit of $\rm s$. Host galaxy offset is in unit of $\rm kpc$. The adjusted $R^{2}$ is 0.4588. The GRB sample number is 27.

The $\log T_{\rm 50}$-$\log F_{\rm g}$-$\log offset$ formula is:
\begin{equation} \label{eq:t50fgofet}
\log T_{\rm 50} = (0.56 \pm 0.067) \times \log F_{\rm g} + (-0.49 \pm 0.15) \times \log offset + (0.29 \pm 0.14),
\end{equation}
where $T_{\rm 50}$ is in unit of $\rm s$. $F_{\rm g}$ is in unit of $\rm 10^{\rm -6} ~ ergs ~ cm^{\rm -2}$, and in 20-2000 $\rm keV$ energy band. Host galaxy offset is in unit of $\rm kpc$. The adjusted $R^{2}$ is 0.5799. The GRB sample number is 27.

The $\log  F_{\rm Opt11hr}$-$\log F_{\rm pk2}$-$A_{\rm V}$ formula is:
\begin{equation} \label{eq:fo11fpk2av}
\log F_{\rm Opt11hr} = (0.63 \pm 0.2) \times \log F_{\rm pk2} + (-0.4 \pm 0.2) \times A_{\rm V} + (-4.9 \pm 0.2),
\end{equation}
where $F_{\rm Opt11hr}$ is in unit of $\rm Jy$. $F_{\rm pk2}$ is peak energy flux of 64 $\rm ms$ time bin in rest-frame 1-$10^{4}$ $\rm keV$ energy band, and in unit of $\rm 10^{\rm -6} ~ ergs ~ cm^{\rm -2} ~ s^{\rm -1}$. The adjusted $R^{2}$ is 0.406. The GRB sample number is 27.

The $\log F_{\rm g}$-$\log F_{\rm X11hr}$-$\log Age$ formula is:
\begin{equation} \label{eq:fgfx11age}
\log F_{\rm g} = (0.72 \pm 0.073) \times \log F_{\rm X11hr} + (-0.43 \pm 0.098) \times \log Age + (7.1 \pm 0.57),
\end{equation}
where $F_{\rm g}$ is in unit of $\rm 10^{\rm -6} ~ ergs ~ cm^{\rm -2}$, and in 20-2000 $\rm keV$ energy band. $F_{\rm X11hr}$ is in unit of $\rm Jy$. Age is in unit of $\rm Myr$. The adjusted $R^{2}$ is 0.6765. The GRB sample number is 27.

The Mag-$\log L_{\rm pk}$-$\log SFR$ formula is:
\begin{equation} \label{eq:maglpksfr}
Mag = (-0.54 \pm 0.38) \times \log L_{\rm pk} + (-1.5 \pm 0.56) \times \log SFR + (-20 \pm 0.71),
\end{equation}
where Mag is in unit of magnitude. $L_{\rm pk}$ is in unit of $\rm 10^{\rm 52} ~ erg ~ s^{\rm -1}$, and in 1-$10^{4}$ $\rm keV$ energy band. SFR is in unit of $\rm M_{\bigodot} ~ yr^{\rm -1}$. The adjusted $R^{2}$ is 0.837. The GRB sample number is 27.

The $\log  F_{\rm Opt11hr}$-metallicity-$\log Age$ formula is:
\begin{equation} \label{eq:fo11metage}
\log F_{\rm Opt11hr} = (-1.3 \pm 0.42) \times metallicity + (0.55 \pm 0.17) \times \log Age + (4.9 \pm 3.7),
\end{equation}
where $F_{\rm Opt11hr}$ is in unit of $\rm Jy$. Metallicity is the value of $12+\log O/H$. Age is in unit of $\rm Myr$. The adjusted $R^{2}$ is 0.471. The GRB sample number is 27.

The $\log Mass$-$\log offset$-$\log Age$ formula is:
\begin{equation} \label{eq:massofetage}
\log Mass = (0.42 \pm 0.15) \times \log offset + (0.69 \pm 0.15) \times \log Age + (7.9 \pm 0.37),
\end{equation}
where Mass is in unit of $M_{\bigodot}$. Host galaxy offset is in unit of $\rm kpc$. Age is in unit of $\rm Myr$. The adjusted $R^{2}$ is 0.6363. The GRB sample number is 27.

The $\log SFR$-$\log Age$-$\log Mass$ formula is:
\begin{equation} \label{eq:sfragemass}
\log SFR = (-0.49 \pm 0.092) \times \log Age + (0.8 \pm 0.092) \times \log Mass + (-5.6 \pm 0.98),
\end{equation}
where SFR is in unit of $\rm M_{\bigodot} ~ yr^{\rm -1}$. Age is in unit of $\rm Myr$. Mass is in unit of $M_{\bigodot}$. The adjusted $R^{2}$ is 0.6711. The GRB sample number is 28.

The $\log T_{\rm 90}$-$\beta_{\rm X11hr}$-$\log offset$ formula is:
\begin{equation} \label{eq:t90bex11ofet}
\log T_{\rm 90} = (-0.62 \pm 0.18) \times \beta_{\rm X11hr} + (-0.62 \pm 0.17) \times \log offset + (2 \pm 0.32),
\end{equation}
where $T_{\rm 90}$ is in unit of $\rm s$. Host galaxy offset is in unit of $\rm kpc$. The adjusted $R^{2}$ is 0.5046. The GRB sample number is 28.

The $\log Mass$-$\log  E_{\rm p,band}$-$\log SFR$ formula is:
\begin{equation} \label{eq:massebandsfr}
\log Mass = (0.45 \pm 0.1) \times \log  E_{\rm p,band} + (0.61 \pm 0.064) \times \log SFR + (8.2 \pm 0.21),
\end{equation}
where Mass is in unit of $M_{\bigodot}$. $E_{\rm p,band}$ is in unit of $\rm keV$. SFR is in unit of $\rm M_{\bigodot} ~ yr^{\rm -1}$. The adjusted $R^{2}$ is 0.5525. The GRB sample number is 28.

The $\log  F_{\rm Opt11hr}$-$\log F_{\rm pk2}$-$\beta_{\rm X11hr}$ formula is:
\begin{equation} \label{eq:fo11fpk2bex11}
\log F_{\rm Opt11hr} = (0.67 \pm 0.18) \times \log F_{\rm pk2} + (-0.41 \pm 0.21) \times \beta_{\rm X11hr} + (-5 \pm 0.3),
\end{equation}
where $F_{\rm Opt11hr}$ is in unit of $\rm Jy$. $F_{\rm pk2}$ is peak energy flux of 64 $\rm ms$ time bin in rest-frame 1-$10^{4}$ $\rm keV$ energy band, and in unit of $\rm 10^{\rm -6} ~ ergs ~ cm^{\rm -2} ~ s^{\rm -1}$. The adjusted $R^{2}$ is 0.4429. The GRB sample number is 28.

The $\log F_{\rm g}$-$\log \Gamma_{0}$-$\log F_{\rm X11hr}$ formula is:
\begin{equation} \label{eq:fgg0fx11}
\log F_{\rm g} = (0.51 \pm 0.05) \times \log \Gamma_{0} + (0.37 \pm 0.056) \times \log F_{\rm X11hr} + (2.5 \pm 0.43),
\end{equation}
where $F_{\rm g}$ is in unit of $\rm 10^{\rm -6} ~ ergs ~ cm^{\rm -2}$, and in 20-2000 $\rm keV$ energy band. $F_{\rm X11hr}$ is in unit of $\rm Jy$. The adjusted $R^{2}$ is 0.2285. The GRB sample number is 28.

The $\log Mass$-$\log P_{\rm pk1}$-$\log Age$ formula is:
\begin{equation} \label{eq:massppk1age}
\log Mass = (-0.55 \pm 0.1) \times \log P_{\rm pk1} + (0.52 \pm 0.096) \times \log Age + (9.2 \pm 0.25),
\end{equation}
where Mass is in unit of $M_{\bigodot}$. $P_{\rm pk1}$ is peak photon flux of 64 $\rm ms$ time bin in 10-1000 $\rm keV$, and in unit of $\rm photons ~ cm^{\rm -2} ~ s^{\rm -1}$. Age is in unit of $\rm Myr$. The adjusted $R^{2}$ is 0.3665. The GRB sample number is 28.

The $\log HR$-$(-\alpha_{\rm band})$-$\log F_{\rm radio,pk}$ formula is:
\begin{equation} \label{eq:hrabandfrapk}
\log HR = (-0.51 \pm 0.087) \times (-\alpha_{\rm band}) + (0.29 \pm 0.07) \times \log F_{\rm radio,pk} + (2.1 \pm 0.28),
\end{equation}
where $F_{\rm radio,pk}$ is in unit of $\rm Jy$. The adjusted $R^{2}$ is 0.4247. The GRB sample number is 29.

The $\log Mass$-$(-\beta_{\rm band})$-metallicity formula is:
\begin{equation} \label{eq:massbbandmet}
\log Mass = (-0.24 \pm 0.18) \times (-\beta_{\rm band}) + (0.89 \pm 0.26) \times metallicity + (2.3 \pm 2.3),
\end{equation}
where Mass is in unit of $M_{\bigodot}$. Metallicity is the value of $12+\log O/H$. The adjusted $R^{2}$ is 0.2659. The GRB sample number is 29.

The spectral lag-$\log F_{\rm pk2}$-$\log T_{\rm R45,i}$ formula is:
\begin{equation} \label{eq:fpk2tr45ilag}
spectral ~ lag = (-4955 \pm 1139) \times \log F_{\rm pk2} + (4402 \pm 716) \times \log T_{\rm R45,i} + (4946 \pm 784),
\end{equation}
where spectral time lag is in unit of $\rm ms ~ MeV^{\rm -1}$. $F_{\rm pk2}$ is peak energy flux of 64 $\rm ms$ time bin in rest-frame 1-$10^{4}$ $\rm keV$ energy band, and in unit of $\rm 10^{\rm -6} ~ ergs ~ cm^{\rm -2} ~ s^{\rm -1}$. $T_{\rm R45,i}$ is in unit of $\rm s$. The adjusted $R^{2}$ is 0.2428. The GRB sample number is 29.

The $\log  F_{\rm Opt11hr}$-$\log F_{\rm pk3}$-$\log T_{\rm 50,i}$ formula is:
\begin{equation} \label{eq:fo11fpk3t50i}
\log F_{\rm Opt11hr} = (0.69 \pm 0.14) \times \log F_{\rm pk3} + (0.49 \pm 0.15) \times \log T_{\rm 50,i} + (-5.4 \pm 0.15),
\end{equation}
where $F_{\rm Opt11hr}$ is in unit of $\rm Jy$. $F_{\rm pk3}$ is peak energy flux of 256 $\rm ms$ time bin in rest-frame 1-$10^{4}$ $\rm keV$ energy band, and in unit of $\rm 10^{\rm -6} ~ ergs ~ cm^{\rm -2} ~ s^{\rm -1}$. $T_{\rm 50,i}$ is in unit of $\rm s$. The adjusted $R^{2}$ is 0.3968. The GRB sample number is 29.

The $\log Mass$-metallicity-$\log E_{\rm p,band,i}$ formula is:
\begin{equation} \label{eq:massmetebandi}
\log Mass = (0.82 \pm 0.22) \times metallicity + (0.55 \pm 0.1) \times \log E_{\rm p,band,i} + (1.1 \pm 1.9),
\end{equation}
where Mass is in unit of $M_{\bigodot}$. Metallicity is the value of $12+\log O/H$. $E_{\rm p,band,i}$ is in unit of $\rm keV$. The adjusted $R^{2}$ is 0.2796. The GRB sample number is 29.

The $\log  F_{\rm Opt11hr}$-$\log P_{\rm pk2}$-$\log T_{\rm 50,i}$ formula is:
\begin{equation} \label{eq:fo11ppk2t50i}
\log F_{\rm Opt11hr} = (0.61 \pm 0.15) \times \log P_{\rm pk2} + (0.5 \pm 0.15) \times \log T_{\rm 50,i} + (-5.8 \pm 0.23),
\end{equation}
where $F_{\rm Opt11hr}$ is in unit of $\rm Jy$. $P_{\rm pk2}$ is peak photon flux of 256 $\rm ms$ time bin in 10-1000 $\rm keV$, and in unit of $\rm photons ~ cm^{\rm -2} ~ s^{\rm -1}$. $T_{\rm 50,i}$ is in unit of $\rm s$. The adjusted $R^{2}$ is 0.2895. The GRB sample number is 29.

The Mag-$\log  D_{\rm L}$-$(-\beta_{\rm band})$ formula is:
\begin{equation} \label{eq:magdlbeband}
Mag = (-1.4 \pm 1.1) \times \log  D_{\rm L} + (0.38 \pm 0.69) \times (-\beta_{\rm band}) + (-21 \pm 1.9),
\end{equation}
where Mag is in unit of magnitude. $D_{\rm L}$ is in unit of $\rm 10^{\rm 28} ~ cm$. The adjusted $R^{2}$ is 0.2703. The GRB sample number is 30.

The $\log t_{\rm pkOpt,i}$-$\log E_{\rm p,cpl}$-$\log  F_{\rm Opt11hr}$ formula is:
\begin{equation} \label{eq:totiecplfo11}
\log t_{\rm pkOpt,i} = (-0.66 \pm 0.15) \times \log E_{\rm p,cpl} + (0.32 \pm 0.1) \times \log F_{\rm Opt11hr} + (4.9 \pm 0.52),
\end{equation}
where $t_{\rm pkOpt,i}$ is in unit of $\rm s$. $E_{\rm p,cpl}$ is in unit of $\rm keV$. $F_{\rm Opt11hr}$ is in unit of $\rm Jy$. The adjusted $R^{2}$ is 0.3993. The GRB sample number is 30.

The Mag-$\log T_{\rm 90}$-$(-\beta_{\rm band})$ formula is:
\begin{equation} \label{eq:t90bbandmag}
Mag = (-1.3 \pm 0.95) \times \log T_{\rm 90} + (0.48 \pm 0.67) \times (-\beta_{\rm band}) + (-20 \pm 2.4),
\end{equation}
where Mag is in unit of magnitude. $T_{\rm 90}$ is in unit of $\rm s$. The adjusted $R^{2}$ is 0.3291. The GRB sample number is 30.

The $\log \Gamma_{0}$-$\log \theta_{\rm j}$-$\log T_{\rm 50,i}$ formula is:
\begin{equation} \label{eq:g0thejt50i}
\log \Gamma_{0} = (-0.45 \pm 0.077) \times \log \theta_{\rm j} + (-0.35 \pm 0.031) \times \log T_{\rm 50,i} + (2 \pm 0.12),
\end{equation}
where $\theta_{\rm j}$ is in unit of $\rm rad$. $T_{\rm 50,i}$ is in unit of $\rm s$. The adjusted $R^{2}$ is 0.2545. The GRB sample number is 30.

The $\log P_{\rm pk3}$-$\log (1+z)$-$\log t_{\rm pkOpt}$ formula is:
\begin{equation} \label{eq:ppk3ztpko}
\log P_{\rm pk3} = (-2.6 \pm 0.047) \times \log (1+z) + (-0.45 \pm 0.017) \times \log t_{\rm pkOpt} + (3.1 \pm 0.039),
\end{equation}
where $P_{\rm pk3}$ is peak photon flux of 1024 $\rm ms$ time bin in 10-1000 $\rm keV$, and in unit of $\rm photons ~ cm^{\rm -2} ~ s^{\rm -1}$. $t_{\rm pkOpt}$ is in unit of $\rm s$. The adjusted $R^{2}$ is 0.5827. The GRB sample number is 30.

The Mag-$\log (1+z)$-$(-\beta_{\rm band})$ formula is:
\begin{equation} \label{eq:magzbband}
Mag = (-3.4 \pm 2.3) \times \log (1+z) + (0.36 \pm 0.69) \times (-\beta_{\rm band}) + (-20 \pm 2.1),
\end{equation}
where Mag is in unit of magnitude. The adjusted $R^{2}$ is 0.3315. The GRB sample number is 30.

The $\log L_{\rm pk}$-$\log L_{\rm radio,pk}$-$\log  E_{\rm p,band}$ formula is:
\begin{equation} \label{eq:lpklraepba}
\log L_{\rm pk} = (0.36 \pm 0.048) \times \log L_{\rm radio,pk} + (1.2 \pm 0.13) \times \log  E_{\rm p,band} + (-2.9 \pm 0.25),
\end{equation}
where $L_{\rm pk}$ is in unit of $\rm 10^{\rm 52} ~ ergs ~ s^{\rm -1}$, and in 1-$10^{4}$ $\rm keV$ energy band. $L_{\rm radio,pk}$ is in rest-frame 8.46 $\rm GHz$, the unit is $\rm 10^{\rm 40} ~ erg ~ s^{\rm -1}$. $E_{\rm p,band}$ is in unit of $\rm keV$. The adjusted $R^{2}$ is 0.5213. The GRB sample number is 30.

The $\log Mass$-$\log L_{\rm radio,pk}$-$\log SSFR$ formula is:
\begin{equation} \label{eq:masslrassfr}
\log Mass = (0.36 \pm 0.07) \times \log L_{\rm radio,pk} + (-0.52 \pm 0.18) \times \log SSFR + (9 \pm 0.086),
\end{equation}
where Mass is in unit of $M_{\bigodot}$. $L_{\rm radio,pk}$ is in rest-frame 8.46 $\rm GHz$, the unit is $\rm 10^{\rm 40} ~ erg ~ s^{\rm -1}$. $\log SSFR$ is in unit of $\rm Gyr^{\rm -1}$. The adjusted $R^{2}$ is 0.3256. The GRB sample number is 30.

The $\log (1+z)$-$\log \Gamma_{0}$-$\log  N_{\rm H}$ formula is:
\begin{equation} \label{eq:zg0nh}
\log (1+z) = (0.25 \pm 0.016) \times \log \Gamma_{0} + (0.13 \pm 0.021) \times \log  N_{\rm H} + (-0.2 \pm 0.036),
\end{equation}
where $N_{\rm H}$ is in unit of $\rm 10^{\rm 21} ~ cm^{\rm -2}$. The adjusted $R^{2}$ is 0.481. The GRB sample number is 31.

The Mag-$\log SSFR$-$\log Mass$ formula is:
\begin{equation} \label{eq:magssfrmass}
Mag = (-0.95 \pm 0.76) \times \log SSFR + (-1.9 \pm 0.47) \times \log Mass + (-2.2 \pm 4.5),
\end{equation}
where Mag is in unit of magnitude. $\log SSFR$ is in unit of $\rm Gyr^{\rm -1}$. Mass is in unit of $M_{\bigodot}$. The adjusted $R^{2}$ is 0.8329. The GRB sample number is 31.

The $\log t_{\rm pkOpt,i}$-$\log  F_{\rm Opt11hr}$-$\log  N_{\rm H}$ formula is:
\begin{equation} \label{eq:totifo11nh}
\log t_{\rm pkOpt,i} = (0.51 \pm 0.095) \times \log F_{\rm Opt11hr} + (0.38 \pm 0.11) \times \log  N_{\rm H} + (4.2 \pm 0.43),
\end{equation}
where $t_{\rm pkOpt,i}$ is in unit of $\rm s$. $F_{\rm Opt11hr}$ is in unit of $\rm Jy$. $N_{\rm H}$ is in unit of $\rm 10^{\rm 21} ~ cm^{\rm -2}$. The adjusted $R^{2}$ is 0.4855. The GRB sample number is 32.

The $\log SFR$-$\log  N_{\rm H}$-$\log Mass$ formula is:
\begin{equation} \label{eq:sfrnhmass}
\log SFR = (0.58 \pm 0.16) \times \log  N_{\rm H} + (0.66 \pm 0.11) \times \log Mass + (-5.9 \pm 0.98),
\end{equation}
where SFR is in unit of $\rm M_{\bigodot} ~ yr^{\rm -1}$. $N_{\rm H}$ is in unit of $\rm 10^{\rm 21} ~ cm^{\rm -2}$. Mass is in unit of $M_{\bigodot}$. The adjusted $R^{2}$ is 0.6262. The GRB sample number is 32.

The $\log L_{\rm pk}$-$\log P_{\rm pk4}$-$\log F_{\rm radio,pk}$ formula is:
\begin{equation} \label{eq:lpkppk4frapk}
\log L_{\rm pk} = (0.8 \pm 0.089) \times \log P_{\rm pk4} + (-1.3 \pm 0.097) \times \log F_{\rm radio,pk} + (-5.4 \pm 0.38),
\end{equation}
where $L_{\rm pk}$ is in unit of $\rm 10^{\rm 52} ~ erg ~ s^{\rm -1}$, and in 1-$10^{4}$ $\rm keV$ energy band. $P_{\rm pk4}$ is peak photon flux of 1 $\rm s$ time bin in 10-1000 $\rm keV$, and in unit of $\rm photons ~ cm^{\rm -2} ~ s^{\rm -1}$. $F_{\rm radio,pk}$ is in unit of $\rm Jy$. The adjusted $R^{2}$ is 0.2764. The GRB sample number is 32.

The $(-\alpha_{\rm band})$-$variability_{3}$-$A_{\rm V}$ formula is:
\begin{equation} \label{eq:abandvar3av}
(-\alpha_{\rm band}) = (-15 \pm 6.8) \times variability_{3} + (-0.13 \pm 0.068) \times A_{\rm V} + (1.3 \pm 0.097),
\end{equation}
the adjusted $R^{2}$ is 0.2283. The GRB sample number is 32.

The rest-frame spectral lag-$\log L_{\rm radio,pk}$-$\log t_{\rm radio,pk}$ formula is:
\begin{equation} \label{eq:lagilratra}
rest-frame ~ spectral ~ lag = (-1029 \pm 290) \times \log L_{\rm radio,pk} + (1663 \pm 355) \times \log t_{\rm radio,pk} + (-7091 \pm 1663),
\end{equation}
where rest-frame spectral lag is in unit of $\rm ms ~ MeV^{\rm -1}$. $L_{\rm radio,pk}$ is in rest-frame 8.46 $\rm GHz$, the unit is $\rm 10^{\rm 40} ~ erg ~ s^{\rm -1}$. $t_{\rm radio,pk}$ is in unit of $\rm s$. The adjusted $R^{2}$ is 0.4842. The GRB sample number is 32.

The $\log L_{\rm radio,pk}$-$A_{\rm V}$-$\log E_{\rm p,band,i}$ formula is:
\begin{equation} \label{eq:lraavepbai}
\log L_{\rm radio,pk} = (-0.31 \pm 0.083) \times A_{\rm V} + (0.72 \pm 0.056) \times \log E_{\rm p,band,i} + (0.11 \pm 0.17),
\end{equation}
where $L_{\rm radio,pk}$ is in rest-frame 8.46 $\rm GHz$, the unit is $\rm 10^{\rm 40} ~ erg ~ s^{\rm -1}$. $E_{\rm p,band,i}$ is in unit of $\rm keV$. The adjusted $R^{2}$ is 0.4928. The GRB sample number is 32.

The $\log HR$-$\log E_{\rm p,cpl}$-$\log Age$ formula is:
\begin{equation} \label{eq:hrecplage}
\log HR = (0.79 \pm 0.18) \times \log E_{\rm p,cpl} + (-0.15 \pm 0.07) \times \log Age + (-1 \pm 0.54),
\end{equation}
where $E_{\rm p,cpl}$ is in unit of $\rm keV$. Age is in unit of $\rm Myr$. The adjusted $R^{2}$ is 0.9192. The GRB sample number is 33.

The $\log  F_{\rm Opt11hr}$-$\log HR$-$\log P_{\rm pk2}$ formula is:
\begin{equation} \label{eq:fo11hrppk2}
\log F_{\rm Opt11hr} = (0.65 \pm 0.21) \times \log HR + (0.66 \pm 0.15) \times \log P_{\rm pk2} + (-5.9 \pm 0.2),
\end{equation}
where $F_{\rm Opt11hr}$ is in unit of $\rm Jy$. $P_{\rm pk2}$ is peak photon flux of 256 $\rm ms$ time bin in 10-1000 $\rm keV$, and in unit of $\rm photons ~ cm^{\rm -2} ~ s^{\rm -1}$. The adjusted $R^{2}$ is 0.2795. The GRB sample number is 33.

The Mag-$\log SFR$-$\log Mass$ formula is:
\begin{equation} \label{eq:magsfrmass}
Mag = (-0.85 \pm 0.66) \times \log SFR + (-1.3 \pm 0.73) \times \log Mass + (-7.4 \pm 6.5),
\end{equation}
where Mag is in unit of magnitude. SFR is in unit of $\rm M_{\bigodot} ~ yr^{\rm -1}$. Mass is in unit of $M_{\bigodot}$. The adjusted $R^{2}$ is 0.8705. The GRB sample number is 33.

The $\log L_{\rm radio,pk}$-$A_{\rm V}$-$\log Mass$ formula is:
\begin{equation} \label{eq:lraavmass}
\log L_{\rm radio,pk} = (-0.24 \pm 0.066) \times A_{\rm V} + (0.5 \pm 0.088) \times \log Mass + (-3.1 \pm 0.82),
\end{equation}
where $L_{\rm radio,pk}$ is in rest-frame 8.46 $\rm GHz$, the unit is $\rm 10^{\rm 40} ~ erg ~ s^{\rm -1}$. Mass is in unit of $M_{\bigodot}$. The adjusted $R^{2}$ is 0.2234. The GRB sample number is 34.

The $\log L_{\rm pk}$-$\beta_{\rm X11hr}$-$\log t_{\rm pkOpt,i}$ formula is:
\begin{equation} \label{eq:lpkbx11toti}
\log L_{\rm pk} = (-0.39 \pm 0.1) \times \beta_{\rm X11hr} + (-0.46 \pm 0.05) \times \log t_{\rm pkOpt,i} + (1.7 \pm 0.13),
\end{equation}
where $L_{\rm pk}$ is in unit of $\rm 10^{\rm 52} ~ erg ~ s^{\rm -1}$, and in 1-$10^{4}$ $\rm keV$ energy band. $t_{\rm pkOpt,i}$ is in unit of $\rm s$. The adjusted $R^{2}$ is 0.2613. The GRB sample number is 34.

The $\log Age$-$\log E_{\rm iso}$-$(-\alpha_{\rm cpl})$ formula is:
\begin{equation} \label{eq:ageeisoacpl}
\log Age = (-0.27 \pm 0.047) \times \log E_{\rm iso} + (0.48 \pm 0.18) \times (-\alpha_{\rm cpl}) + (2.1 \pm 0.19),
\end{equation}
where Age is in unit of $\rm Myr$. $E_{\rm iso}$ is in unit of $\rm 10^{\rm 52} ~ ergs$ and in rest-frame 1-$10^{4}$ $\rm keV$ energy band. The adjusted $R^{2}$ is 0.4063. The GRB sample number is 34.

The $\log P_{\rm pk1}$-$\log L_{\rm pk}$-$\log  F_{\rm Opt11hr}$ formula is:
\begin{equation} \label{eq:ppk1lpkfo11}
\log P_{\rm pk1} = (0.22 \pm 0.028) \times \log L_{\rm pk} + (0.21 \pm 0.049) \times \log F_{\rm Opt11hr} + (2.2 \pm 0.24),
\end{equation}
where $P_{\rm pk1}$ is peak photon flux of 64 $\rm ms$ time bin in 10-1000 $\rm keV$, and in unit of $\rm photons ~ cm^{\rm -2} ~ s^{\rm -1}$. $L_{\rm pk}$ is in unit of $\rm 10^{\rm 52} ~ erg ~ s^{\rm -1}$, and in 1-$10^{4}$ $\rm keV$ energy band. $F_{\rm Opt11hr}$ is in unit of $\rm Jy$. The adjusted $R^{2}$ is 0.4753. The GRB sample number is 34.

The $\log F_{\rm g}$-$variability_{3}$-$\log T_{\rm 50,i}$ formula is:
\begin{equation} \label{eq:fgvar3t50i}
\log F_{\rm g} = (59 \pm 13) \times variability_{3} + (0.45 \pm 0.053) \times \log T_{\rm 50,i} + (0.16 \pm 0.11),
\end{equation}
where $F_{\rm g}$ is in unit of $\rm 10^{\rm -6} ~ ergs ~ cm^{\rm -2}$, and in 20-2000 $\rm keV$ energy band. $T_{\rm 50,i}$ is in unit of $\rm s$. The adjusted $R^{2}$ is 0.388. The GRB sample number is 34.

The $\log Age$-$\log  D_{\rm L}$-$\log E_{\rm p,cpl}$ formula is:
\begin{equation} \label{eq:agedlepcpl}
\log Age = (-0.83 \pm 0.14) \times \log  D_{\rm L} + (-0.5 \pm 0.18) \times \log E_{\rm p,cpl} + (3.9 \pm 0.43),
\end{equation}
where Age is in unit of $\rm Myr$. $D_{\rm L}$ is in unit of $\rm 10^{\rm 28} ~ cm$. $E_{\rm p,cpl}$ is in unit of $\rm keV$. The adjusted $R^{2}$ is 0.4641. The GRB sample number is 35.

The rest-frame spectral lag-$\log  D_{\rm L}$-$\log SFR$ formula is:
\begin{equation} \label{eq:lagidlsfr}
rest-frame ~ spectral ~ lag = (-9353 \pm 1291) \times \log  D_{\rm L} + (2618 \pm 725) \times \log SFR + (2406 \pm 525),
\end{equation}
where rest-frame spectral lag is in unit of $\rm ms ~ MeV^{\rm -1}$. $D_{\rm L}$ is in unit of $\rm 10^{\rm 28} ~ cm$. SFR is in unit of $\rm M_{\bigodot} ~ yr^{\rm -1}$. The adjusted $R^{2}$ is 0.384. The GRB sample number is 35.

The $\log SFR$-$\log E_{\rm iso}$-$\log  N_{\rm H}$ formula is:
\begin{equation} \label{eq:sfreisonh}
\log SFR = (0.35 \pm 0.052) \times \log E_{\rm iso} + (0.83 \pm 0.094) \times \log  N_{\rm H} + (-0.033 \pm 0.088),
\end{equation}
where SFR is in unit of $\rm M_{\bigodot} ~ yr^{\rm -1}$. $E_{\rm iso}$ is in unit of $\rm 10^{\rm 52} ~ ergs$ and in rest-frame 1-$10^{4}$ $\rm keV$ energy band. $N_{\rm H}$ is in unit of $\rm 10^{\rm 21} ~ cm^{\rm -2}$. The adjusted $R^{2}$ is 0.4156. The GRB sample number is 35.

The Mag-$\log  F_{\rm Opt11hr}$-$\log E_{\rm p,cpl,i}$ formula is:
\begin{equation} \label{eq:magfo11ecpli}
Mag = (0.89 \pm 0.41) \times \log F_{\rm Opt11hr} + (-1.2 \pm 0.83) \times \log E_{\rm p,cpl,i} + (-13 \pm 2.7),
\end{equation}
where Mag is in unit of magnitude. $F_{\rm Opt11hr}$ is in unit of $\rm Jy$. $E_{\rm p,cpl,i}$ is in unit of $\rm keV$. The adjusted $R^{2}$ is 0.4763. The GRB sample number is 35.

The $(-\alpha_{\rm cpl})$-$\log T_{\rm 90}$-$\log Age$ formula is:
\begin{equation} \label{eq:acplt90age}
(-\alpha_{\rm cpl}) = (0.15 \pm 0.045) \times \log T_{\rm 90} + (0.2 \pm 0.083) \times \log Age + (0.35 \pm 0.2),
\end{equation}
where $T_{\rm 90}$ is in unit of $\rm s$. Age is in unit of $\rm Myr$. The adjusted $R^{2}$ is 0.2569. The GRB sample number is 35.

The Mag-$\log \theta_{\rm j}$-$\log Mass$ formula is:
\begin{equation} \label{eq:magthejmass}
Mag = (0.61 \pm 0.76) \times \log \theta_{\rm j} + (-1.8 \pm 0.5) \times \log Mass + (-2.7 \pm 4.7),
\end{equation}
where Mag is in unit of magnitude. $\theta_{\rm j}$ is in unit of $\rm rad$. Mass is in unit of $M_{\bigodot}$. The adjusted $R^{2}$ is 0.878. The GRB sample number is 35.

The $\log Age$-$\log (1+z)$-$\log E_{\rm p,cpl}$ formula is:
\begin{equation} \label{eq:agezecpl}
\log Age = (-2 \pm 0.63) \times \log (1+z) + (-0.6 \pm 0.19) \times \log E_{\rm p,cpl} + (4.6 \pm 0.44),
\end{equation}
where Age is in unit of $\rm Myr$. $E_{\rm p,cpl}$ is in unit of $\rm keV$. The adjusted $R^{2}$ is 0.4777. The GRB sample number is 35.

The $\log Age$-$\log (1+z)$-$(-\alpha_{\rm cpl})$ formula is:
\begin{equation} \label{eq:agezacpl}
\log Age = (-2.2 \pm 0.64) \times \log (1+z) + (0.45 \pm 0.2) \times (-\alpha_{\rm cpl}) + (2.8 \pm 0.28),
\end{equation}
where Age is in unit of $\rm Myr$. The adjusted $R^{2}$ is 0.4567. The GRB sample number is 35.

The $\log SFR$-$\log  D_{\rm L}$-$\log  N_{\rm H}$ formula is:
\begin{equation} \label{eq:sfrdlnh}
\log SFR = (1.3 \pm 0.12) \times \log  D_{\rm L} + (0.7 \pm 0.083) \times \log  N_{\rm H} + (-0.061 \pm 0.081),
\end{equation}
where SFR is in unit of $\rm M_{\bigodot} ~ yr^{\rm -1}$. $D_{\rm L}$ is in unit of $\rm 10^{\rm 28} ~ cm$. $N_{\rm H}$ is in unit of $\rm 10^{\rm 21} ~ cm^{\rm -2}$. The adjusted $R^{2}$ is 0.6648. The GRB sample number is 36.

The $\log \Gamma_{0}$-$\log F_{\rm g}$-$\log  F_{\rm Opt11hr}$ formula is:
\begin{equation} \label{eq:g0fgfo11}
\log \Gamma_{0} = (0.26 \pm 0.028) \times \log F_{\rm g} + (-0.21 \pm 0.06) \times \log F_{\rm Opt11hr} + (0.94 \pm 0.27),
\end{equation}
where $F_{\rm g}$ is in unit of $\rm 10^{\rm -6} ~ ergs ~ cm^{\rm -2}$, and in 20-2000 $\rm keV$ energy band. $F_{\rm Opt11hr}$ is in unit of $\rm Jy$. The adjusted $R^{2}$ is 0.3764. The GRB sample number is 36.

The $\log \Gamma_{0}$-$\log F_{\rm pk1}$-$\log  F_{\rm Opt11hr}$ formula is:
\begin{equation} \label{eq:g0fpk1fo11}
\log \Gamma_{0} = (0.2 \pm 0.027) \times \log F_{\rm pk1} + (-0.2 \pm 0.063) \times \log F_{\rm Opt11hr} + (1.3 \pm 0.29),
\end{equation}
where $F_{\rm pk1}$ is peak energy flux of 1 $\rm s$ time bin in rest-frame 1-$10^{4}$ $\rm keV$ energy band, and in unit of $\rm 10^{\rm -6} ~ ergs ~ cm^{\rm -2} ~ s^{\rm -1}$. $F_{\rm Opt11hr}$ is in unit of $\rm Jy$. The adjusted $R^{2}$ is 0.3141. The GRB sample number is 36.

The $\log  E_{\rm p,band}$-$variability_{3}$-$\log T_{\rm 90,i}$ formula is:
\begin{equation} \label{eq:ebandvar3t90i}
\log  E_{\rm p,band} = (23 \pm 5) \times variability_{3} + (0.2 \pm 0.046) \times \log T_{\rm 90,i} + (1.7 \pm 0.085),
\end{equation}
where $E_{\rm p,band}$ is in unit of $\rm keV$. $T_{\rm 90,i}$ is in unit of $\rm s$. The adjusted $R^{2}$ is 0.287. The GRB sample number is 36.

The $\log SFR$-$\log (1+z)$-$\log  N_{\rm H}$ formula is:
\begin{equation} \label{eq:sfrznh}
\log SFR = (4.1 \pm 0.42) \times \log (1+z) + (0.73 \pm 0.082) \times \log  N_{\rm H} + (-1.1 \pm 0.12),
\end{equation}
where SFR is in unit of $\rm M_{\bigodot} ~ yr^{\rm -1}$. $N_{\rm H}$ is in unit of $\rm 10^{\rm 21} ~ cm^{\rm -2}$. The adjusted $R^{2}$ is 0.6871. The GRB sample number is 36.

The $\log E_{\rm p,band,i}$-$\log L_{\rm radio,pk}$-$\log T_{\rm 90}$ formula is:
\begin{equation} \label{eq:epbilrat90}
\log E_{\rm p,band,i} = (0.32 \pm 0.023) \times \log L_{\rm radio,pk} + (0.37 \pm 0.036) \times \log T_{\rm 90} + (1.4 \pm 0.072),
\end{equation}
where $E_{\rm p,band,i}$ is in unit of $\rm keV$. $L_{\rm radio,pk}$ is in rest-frame 8.46 $\rm GHz$, the unit is $\rm 10^{\rm 40} ~ erg ~ s^{\rm -1}$. $T_{\rm 90}$ is in unit of $\rm s$. The adjusted $R^{2}$ is 0.4479. The GRB sample number is 36.

The $\log E_{\rm iso}$-$\log L_{\rm radio,pk}$-$\log  E_{\rm p,band}$ formula is:
\begin{equation} \label{eq:eisolraepba}
\log E_{\rm iso} = (0.43 \pm 0.039) \times \log L_{\rm radio,pk} + (1.3 \pm 0.11) \times \log  E_{\rm p,band} + (-2.5 \pm 0.21),
\end{equation}
where $E_{\rm iso}$ is in unit of $\rm 10^{\rm 52} ~ ergs$ and in rest-frame 1-$10^{4}$ $\rm keV$ energy band. $L_{\rm radio,pk}$ is in rest-frame 8.46 $\rm GHz$, the unit is $\rm 10^{\rm 40} ~ erg ~ s^{\rm -1}$. $E_{\rm p,band}$ is in unit of $\rm keV$. The adjusted $R^{2}$ is 0.6844. The GRB sample number is 36.

The Mag-$A_{\rm V}$-$\log E_{\rm p,cpl,i}$ formula is:
\begin{equation} \label{eq:avecplimag}
Mag = (-0.59 \pm 0.45) \times A_{\rm V} + (-1.1 \pm 0.81) \times \log E_{\rm p,cpl,i} + (-18 \pm 2.2),
\end{equation}
where Mag is in unit of magnitude. $E_{\rm p,cpl,i}$ is in unit of $\rm keV$. The adjusted $R^{2}$ is 0.2498. The GRB sample number is 37.

The $\log F_{\rm g}$-$\log  E_{\rm p,band}$-$\log t_{\rm radio,pk}$ formula is:
\begin{equation} \label{eq:fgebandtrapk}
\log F_{\rm g} = (1.3 \pm 0.12) \times \log  E_{\rm p,band} + (-0.54 \pm 0.091) \times \log t_{\rm radio,pk} + (1.7 \pm 0.67),
\end{equation}
where $F_{\rm g}$ is in unit of $\rm 10^{\rm -6} ~ ergs ~ cm^{\rm -2}$, and in 20-2000 $\rm keV$ energy band. $E_{\rm p,band}$ is in unit of $\rm keV$. $t_{\rm radio,pk}$ is in unit of $\rm s$. The adjusted $R^{2}$ is 0.404. The GRB sample number is 37.

The $\log  F_{\rm Opt11hr}$-$\log F_{\rm radio,pk}$-$\log T_{\rm R45,i}$ formula is:
\begin{equation} \label{eq:fo11frapktr45i}
\log F_{\rm Opt11hr} = (0.84 \pm 0.14) \times \log F_{\rm radio,pk} + (0.48 \pm 0.16) \times \log T_{\rm R45,i} + (-1.6 \pm 0.51),
\end{equation}
where $F_{\rm Opt11hr}$ is in unit of $\rm Jy$. $F_{\rm radio,pk}$ is in unit of $\rm Jy$. $T_{\rm R45,i}$ is in unit of $\rm s$. The adjusted $R^{2}$ is 0.4266. The GRB sample number is 37.

The $\log F_{\rm radio,pk}$-$\log HR$-$\log T_{\rm 50,i}$ formula is:
\begin{equation} \label{eq:frapkhrt50i}
\log F_{\rm radio,pk} = (0.34 \pm 0.05) \times \log HR + (0.18 \pm 0.023) \times \log T_{\rm 50,i} + (-3.9 \pm 0.029),
\end{equation}
where $F_{\rm radio,pk}$ is in unit of $\rm Jy$. $T_{\rm 50,i}$ is in unit of $\rm s$. The adjusted $R^{2}$ is 0.255. The GRB sample number is 37.

The $\log L_{\rm pk}$-$variability_{1}$-$(-\alpha_{\rm band})$ formula is:
\begin{equation} \label{eq:lpkvar1aband}
\log L_{\rm pk} = (4.3 \pm 2.3) \times variability_{1} + (-0.95 \pm 0.51) \times (-\alpha_{\rm band}) + (1.1 \pm 0.54),
\end{equation}
where $L_{\rm pk}$ is in unit of $\rm 10^{\rm 52} ~ erg ~ s^{\rm -1}$, and in 1-$10^{4}$ $\rm keV$ energy band. The adjusted $R^{2}$ is 0.2743. The GRB sample number is 37.

The $\log F_{\rm pk3}$-$\log E_{\rm p,cpl}$-$\log t_{\rm burst}$ formula is:
\begin{equation} \label{eq:fpk3ecpltt}
\log F_{\rm pk3} = (0.35 \pm 0.12) \times \log E_{\rm p,cpl} + (-0.25 \pm 0.035) \times \log t_{\rm burst} + (-0.23 \pm 0.27),
\end{equation}
where $F_{\rm pk3}$ is peak energy flux of 256 $\rm ms$ time bin in rest-frame 1-$10^{4}$ $\rm keV$ energy band, and in unit of $\rm 10^{\rm -6} ~ ergs ~ cm^{\rm -2} ~ s^{\rm -1}$. $E_{\rm p,cpl}$ is in unit of $\rm keV$. $t_{\rm burst}$ is in unit of $\rm s$. The adjusted $R^{2}$ is 0.2004. The GRB sample number is 38.

The $\log T_{\rm 90}$-$\log F_{\rm g}$-$\log offset$ formula is:
\begin{equation} \label{eq:t90fgofet}
\log T_{\rm 90} = (0.65 \pm 0.039) \times \log F_{\rm g} + (-0.39 \pm 0.094) \times \log offset + (0.47 \pm 0.082),
\end{equation}
where $T_{\rm 90}$ is in unit of $\rm s$. $F_{\rm g}$ is in unit of $\rm 10^{\rm -6} ~ ergs ~ cm^{\rm -2}$, and in 20-2000 $\rm keV$ energy band. Host galaxy offset is in unit of $\rm kpc$. The adjusted $R^{2}$ is 0.6172. The GRB sample number is 38.

The $\log (1+z)$-$\log P_{\rm pk4}$-$\log F_{\rm radio,pk}$ formula is:
\begin{equation} \label{eq:zppk4frapk}
\log (1+z) = (-0.11 \pm 0.019) \times \log P_{\rm pk4} + (-0.24 \pm 0.031) \times \log F_{\rm radio,pk} + (-0.33 \pm 0.12),
\end{equation}
where $P_{\rm pk4}$ is peak photon flux of 1 $\rm s$ time bin in 10-1000 $\rm keV$, and in unit of $\rm photons ~ cm^{\rm -2} ~ s^{\rm -1}$. $F_{\rm radio,pk}$ is in unit of $\rm Jy$. The adjusted $R^{2}$ is 0.3862. The GRB sample number is 38.

The $\log P_{\rm pk1}$-$variability_{1}$-$(-\alpha_{\rm cpl})$ formula is:
\begin{equation} \label{eq:ppk1var1acpl}
\log P_{\rm pk1} = (-1.7 \pm 0.91) \times variability_{1} + (0.57 \pm 0.074) \times (-\alpha_{\rm cpl}) + (0.54 \pm 0.09),
\end{equation}
where $P_{\rm pk1}$ is peak photon flux of 64 $\rm ms$ time bin in 10-1000 $\rm keV$, and in unit of $\rm photons ~ cm^{\rm -2} ~ s^{\rm -1}$. The adjusted $R^{2}$ is 0.4813. The GRB sample number is 38.

The $\log L_{\rm pk}$-$variability_{2}$-$(-\alpha_{\rm band})$ formula is:
\begin{equation} \label{eq:lpkvar2aband}
\log L_{\rm pk} = (4.9 \pm 1.3) \times variability_{2} + (-0.59 \pm 0.38) \times (-\alpha_{\rm band}) + (0.19 \pm 0.48),
\end{equation}
where $L_{\rm pk}$ is in unit of $\rm 10^{\rm 52} ~ erg ~ s^{\rm -1}$, and in 1-$10^{4}$ $\rm keV$ energy band. The adjusted $R^{2}$ is 0.2358. The GRB sample number is 38.

The $\log L_{\rm pk}$-$variability_{3}$-$\log \theta_{\rm j}$ formula is:
\begin{equation} \label{eq:lpkvar3thej}
\log L_{\rm pk} = (47 \pm 6.9) \times variability_{3} + (-0.97 \pm 0.12) \times \log \theta_{\rm j} + (-1.3 \pm 0.17),
\end{equation}
where $L_{\rm pk}$ is in unit of $\rm 10^{\rm 52} ~ erg ~ s^{\rm -1}$, and in 1-$10^{4}$ $\rm keV$ energy band. $\theta_{\rm j}$ is in unit of $\rm rad$. The adjusted $R^{2}$ is 0.3097. The GRB sample number is 38.

The $\log L_{\rm radio,pk}$-$\log T_{\rm 50}$-$\log L_{\rm pk}$ formula is:
\begin{equation} \label{eq:lrat50lpk}
\log L_{\rm radio,pk} = (0.44 \pm 0.038) \times \log T_{\rm 50} + (0.61 \pm 0.029) \times \log L_{\rm pk} + (1.2 \pm 0.052),
\end{equation}
where $L_{\rm radio,pk}$ is in rest-frame 8.46 $\rm GHz$, the unit is $\rm 10^{\rm 40} ~ erg ~ s^{\rm -1}$. $T_{\rm 50}$ is in unit of $\rm s$. $L_{\rm pk}$ is in unit of $\rm 10^{\rm 52} ~ ergs ~ s^{\rm -1}$, and in 1-$10^{4}$ $\rm keV$ energy band. The adjusted $R^{2}$ is 0.6609. The GRB sample number is 38.

The $\log t_{\rm pkOpt}$-$\log HR$-$\log \Gamma_{0}$ formula is:
\begin{equation} \label{eq:tothrg0}
\log t_{\rm pkOpt} = (-0.28 \pm 0.045) \times \log HR + (-1.2 \pm 0.047) \times \log \Gamma_{0} + (5.3 \pm 0.11),
\end{equation}
where $t_{\rm pkOpt}$ is in unit of $\rm s$. The adjusted $R^{2}$ is 0.4806. The GRB sample number is 39.

The $\log P_{\rm pk4}$-$\log t_{\rm burst}$-$\log t_{\rm pkOpt}$ formula is:
\begin{equation} \label{eq:ppk4tttpko}
\log P_{\rm pk4} = (-0.21 \pm 0.017) \times \log t_{\rm burst} + (-0.34 \pm 0.035) \times \log t_{\rm pkOpt} + (2.1 \pm 0.09),
\end{equation}
where $P_{\rm pk4}$ is peak photon flux of 1 $\rm s$ time bin in 10-1000 $\rm keV$, and in unit of $\rm photons ~ cm^{\rm -2} ~ s^{\rm -1}$. $t_{\rm burst}$ is in unit of $\rm s$. $t_{\rm pkOpt}$ is in unit of $\rm s$. The adjusted $R^{2}$ is 0.3097. The GRB sample number is 39.

The $\log F_{\rm X11hr}$-$variability_{2}$-$\log  F_{\rm Opt11hr}$ formula is:
\begin{equation} \label{eq:fx11var2fo11}
\log F_{\rm X11hr} = (2.2 \pm 0.58) \times variability_{2} + (0.3 \pm 0.097) \times \log F_{\rm Opt11hr} + (-6.3 \pm 0.51),
\end{equation}
where $F_{\rm X11hr}$ is in unit of $\rm Jy$. $F_{\rm Opt11hr}$ is in unit of $\rm Jy$. The adjusted $R^{2}$ is 0.3125. The GRB sample number is 39.

The $\log F_{\rm g}$-$\beta_{\rm X11hr}$-$\log Age$ formula is:
\begin{equation} \label{eq:fgbe11age}
\log F_{\rm g} = (-0.54 \pm 0.16) \times \beta_{\rm X11hr} + (-0.53 \pm 0.12) \times \log Age + (2.7 \pm 0.34),
\end{equation}
where $F_{\rm g}$ is in unit of $\rm 10^{\rm -6} ~ ergs ~ cm^{\rm -2}$, and in 20-2000 $\rm keV$ energy band. Age is in unit of $\rm Myr$. The adjusted $R^{2}$ is 0.4175. The GRB sample number is 40.

The $\log  F_{\rm Opt11hr}$-$\log F_{\rm pk2}$-$\log T_{\rm 90,i}$ formula is:
\begin{equation} \label{eq:fopfpk2t90i}
\log F_{\rm Opt11hr} = (0.65 \pm 0.14) \times \log F_{\rm pk2} + (0.43 \pm 0.11) \times \log T_{\rm 90,i} + (-5.6 \pm 0.13),
\end{equation}
where $F_{\rm Opt11hr}$ is in unit of $\rm Jy$. $F_{\rm pk2}$ is peak energy flux of 64 $\rm ms$ time bin in rest-frame 1-$10^{4}$ $\rm keV$ energy band, and in unit of $\rm 10^{\rm -6} ~ ergs ~ cm^{\rm -2} ~ s^{\rm -1}$. $T_{\rm 90,i}$ is in unit of $\rm s$. The adjusted $R^{2}$ is 0.5628. The GRB sample number is 40.

The $\log Age$-$\log HR$-$\beta_{\rm X11hr}$ formula is:
\begin{equation} \label{eq:agehrbe11}
\log Age = (-0.63 \pm 0.14) \times \log HR + (0.25 \pm 0.11) \times \beta_{\rm X11hr} + (2.4 \pm 0.2),
\end{equation}
where Age is in unit of $\rm Myr$. The adjusted $R^{2}$ is 0.231. The GRB sample number is 40.

The $\log  D_{\rm L}$-$\log \theta_{\rm j}$-$\log Age$ formula is:
\begin{equation} \label{eq:dlthejage}
\log  D_{\rm L} = (-0.46 \pm 0.074) \times \log \theta_{\rm j} + (-0.31 \pm 0.052) \times \log Age + (0.44 \pm 0.21),
\end{equation}
where $D_{\rm L}$ is in unit of $\rm 10^{\rm 28} ~ cm$. $\theta_{\rm j}$ is in unit of $\rm rad$. Age is in unit of $\rm Myr$. The adjusted $R^{2}$ is 0.3565. The GRB sample number is 40.

The $\log  F_{\rm Opt11hr}$-$\log \theta_{\rm j}$-$\log t_{\rm pkOpt}$ formula is:
\begin{equation} \label{eq:fo11thejtopt}
\log F_{\rm Opt11hr} = (0.45 \pm 0.19) \times \log \theta_{\rm j} + (0.32 \pm 0.1) \times \log t_{\rm pkOpt}  + (-4.8 \pm 0.37),
\end{equation}
where $F_{\rm Opt11hr}$ is in unit of $\rm Jy$. $\theta_{\rm j}$ is in unit of $\rm rad$. $t_{\rm pkOpt}$ is in unit of $\rm s$. The adjusted $R^{2}$ is 0.2278. The GRB sample number is 40.

The $\log E_{\rm p,cpl,i}$-$\log F_{\rm g}$-$\log SFR$ formula is:
\begin{equation} \label{eq:ecplifgsfr}
\log E_{\rm p,cpl,i} = (0.39 \pm 0.043) \times \log F_{\rm g} + (0.12 \pm 0.032) \times \log SFR + (2.1 \pm 0.052),
\end{equation}
where $E_{\rm p,cpl,i}$ is in unit of $\rm keV$. $F_{\rm g}$ is in unit of $\rm 10^{\rm -6} ~ ergs ~ cm^{\rm -2}$, and in 20-2000 $\rm keV$ energy band. SFR is in unit of $\rm M_{\bigodot} ~ yr^{\rm -1}$. The adjusted $R^{2}$ is 0.4204. The GRB sample number is 41.

The $\log F_{\rm g}$-$\log  F_{\rm Opt11hr}$-metallicity formula is:
\begin{equation} \label{eq:fgfo11met}
\log F_{\rm g} = (0.27 \pm 0.066) \times \log F_{\rm Opt11hr} + (-0.64 \pm 0.25) \times metallicity + (7.4 \pm 1.9),
\end{equation}
where $F_{\rm g}$ is in unit of $\rm 10^{\rm -6} ~ ergs ~ cm^{\rm -2}$, and in 20-2000 $\rm keV$ energy band. $F_{\rm Opt11hr}$ is in unit of $\rm Jy$. Metallicity is the value of $12+\log O/H$. The adjusted $R^{2}$ is 0.242. The GRB sample number is 41.

The $\log \Gamma_{0}$-$\log P_{\rm pk4}$-$A_{\rm V}$ formula is:
\begin{equation} \label{eq:g0ppk4av}
\log \Gamma_{0} = (0.29 \pm 0.03) \times \log P_{\rm pk4} + (-0.26 \pm 0.13) \times A_{\rm V} + (2.1 \pm 0.042),
\end{equation}
where $P_{\rm pk4}$ is peak photon flux of 1 $\rm s$ time bin in 10-1000 $\rm keV$, and in unit of $\rm photons ~ cm^{\rm -2} ~ s^{\rm -1}$. The adjusted $R^{2}$ is 0.2218. The GRB sample number is 41.

The $\log t_{\rm pkOpt,i}$-$\log T_{\rm R45}$-$\log E_{\rm p,cpl}$ formula is:
\begin{equation} \label{eq:totitr45ecpl}
\log t_{\rm pkOpt,i} = (0.67 \pm 0.053) \times \log T_{\rm R45} + (-0.61 \pm 0.12) \times \log E_{\rm p,cpl} + (2.7 \pm 0.25),
\end{equation}
where $t_{\rm pkOpt,i}$ is in unit of $\rm s$. $T_{\rm R45}$ is in unit of $\rm s$. $E_{\rm p,cpl}$ is in unit of $\rm keV$. The adjusted $R^{2}$ is 0.3051. The GRB sample number is 41.

The $\log SFR$-$A_{\rm V}$-$\log t_{\rm burst,i}$ formula is:
\begin{equation} \label{eq:sfravtti}
\log SFR = (0.26 \pm 0.079) \times A_{\rm V} + (-0.46 \pm 0.077) \times \log t_{\rm burst,i} + (1.7 \pm 0.25),
\end{equation}
where SFR is in unit of $\rm M_{\bigodot} ~ yr^{\rm -1}$. $t_{\rm burst,i}$ is in unit of $\rm s$. The adjusted $R^{2}$ is 0.2794. The GRB sample number is 42.

The $\log L_{\rm pk}$-$\log F_{\rm g}$-$\log \Gamma_{0}$ formula is:
\begin{equation} \label{eq:lpkfgg0}
\log L_{\rm pk} = (0.34 \pm 0.057) \times \log F_{\rm g} + (2 \pm 0.07) \times \log \Gamma_{0} + (-4.8 \pm 0.14),
\end{equation}
where $L_{\rm pk}$ is in unit of $\rm 10^{\rm 52} ~ erg ~ s^{\rm -1}$, and in 1-$10^{4}$ $\rm keV$ energy band. $F_{\rm g}$ is in unit of $\rm 10^{\rm -6} ~ ergs ~ cm^{\rm -2}$, and in 20-2000 $\rm keV$ energy band. The adjusted $R^{2}$ is 0.6858. The GRB sample number is 42.

The $\log E_{\rm p,cpl,i}$-$\log F_{\rm pk1}$-$\log SFR$ formula is:
\begin{equation} \label{eq:ecplifpk1sfr}
\log E_{\rm p,cpl,i} = (0.25 \pm 0.036) \times \log F_{\rm pk1} + (0.13 \pm 0.032) \times \log SFR + (2.4 \pm 0.038),
\end{equation}
where $E_{\rm p,cpl,i}$ is in unit of $\rm keV$. $F_{\rm pk1}$ is peak energy flux of 1 $\rm s$ time bin in rest-frame 1-$10^{4}$ $\rm keV$ energy band, and in unit of $\rm 10^{\rm -6} ~ ergs ~ cm^{\rm -2} ~ s^{\rm -1}$. SFR is in unit of $\rm M_{\bigodot} ~ yr^{\rm -1}$. The adjusted $R^{2}$ is 0.3181. The GRB sample number is 42.

The $\log SFR$-$\log L_{\rm pk}$-$\log  F_{\rm Opt11hr}$ formula is:
\begin{equation} \label{eq:sfrlpkfo11}
\log SFR = (0.3 \pm 0.053) \times \log L_{\rm pk} + (-0.33 \pm 0.08) \times \log F_{\rm Opt11hr} + (-0.88 \pm 0.4),
\end{equation}
where SFR is in unit of $\rm M_{\bigodot} ~ yr^{\rm -1}$. $L_{\rm pk}$ is in unit of $\rm 10^{\rm 52} ~ erg ~ s^{\rm -1}$, and in 1-$10^{4}$ $\rm keV$ energy band. $F_{\rm Opt11hr}$ is in unit of $\rm Jy$. The adjusted $R^{2}$ is 0.3027. The GRB sample number is 42.

The $(-\beta_{\rm band})$-$\log F_{\rm g}$-$\log  N_{\rm H}$ formula is:
\begin{equation} \label{eq:bbabdfgnh}
(-\beta_{\rm band}) = (0.21 \pm 0.087) \times \log F_{\rm g} + (-0.22 \pm 0.15) \times \log  N_{\rm H} + (2.2 \pm 0.2),
\end{equation}
where $F_{\rm g}$ is in unit of $\rm 10^{\rm -6} ~ ergs ~ cm^{\rm -2}$, and in 20-2000 $\rm keV$ energy band. $N_{\rm H}$ is in unit of $\rm 10^{\rm 21} ~ cm^{\rm -2}$. The adjusted $R^{2}$ is 0.2912. The GRB sample number is 43.

The $\log  F_{\rm Opt11hr}$-$\log F_{\rm X11hr}$-$\log t_{\rm pkOpt,i}$ formula is:
\begin{equation} \label{eq:fo11fx11ttoi}
\log F_{\rm Opt11hr} = (0.25 \pm 0.1) \times \log F_{\rm X11hr} + (0.49 \pm 0.096) \times \log t_{\rm pkOpt,i} + (-3.9 \pm 0.77),
\end{equation}
where $F_{\rm Opt11hr}$ is in unit of $\rm Jy$. $F_{\rm X11hr}$ is in unit of $\rm Jy$. $t_{\rm pkOpt,i}$ is in unit of $\rm s$. The adjusted $R^{2}$ is 0.3049. The GRB sample number is 43.

The Mag-$\log E_{\rm iso}$-$\log  N_{\rm H}$ formula is:
\begin{equation} \label{eq:mageisonh}
Mag = (-1.1 \pm 0.34) \times \log E_{\rm iso} + (-1.5 \pm 0.61) \times \log  N_{\rm H} + (-19 \pm 0.6),
\end{equation}
where Mag is in unit of magnitude. $E_{\rm iso}$ is in unit of $\rm 10^{\rm 52} ~ ergs$ and in rest-frame 1-$10^{4}$ $\rm keV$ energy band. $N_{\rm H}$ is in unit of $\rm 10^{\rm 21} ~ cm^{\rm -2}$. The adjusted $R^{2}$ is 0.4373. The GRB sample number is 44.

The Mag-$\log  F_{\rm Opt11hr}$-$\log t_{\rm burst,i}$ formula is:
\begin{equation} \label{eq:magfo11tti}
Mag = (0.39 \pm 0.39) \times \log F_{\rm Opt11hr} + (1.4 \pm 0.57) \times \log t_{\rm burst,i} + (-22 \pm 2.9),
\end{equation}
where Mag is in unit of magnitude. $F_{\rm Opt11hr}$ is in unit of $\rm Jy$. $t_{\rm burst,i}$ is in unit of $\rm s$. The adjusted $R^{2}$ is 0.4612. The GRB sample number is 44.

The Mag-$\log Mass$-$\log E_{\rm p,cpl,i}$ formula is:
\begin{equation} \label{eq:magmassecpli}
Mag = (-1.9 \pm 0.45) \times \log Mass + (-0.37 \pm 0.76) \times \log E_{\rm p,cpl,i} + (-2.1 \pm 4.4),
\end{equation}
where Mag is in unit of magnitude. Mass is in unit of $M_{\bigodot}$. $E_{\rm p,cpl,i}$ is in unit of $\rm keV$. The adjusted $R^{2}$ is 0.8675. The GRB sample number is 44.

The $\log E_{\rm p,band,i}$-$\log \theta_{\rm j}$-$\log Mass$ formula is:
\begin{equation} \label{eq:ebandithejmass}
\log E_{\rm p,band,i} = (-0.55 \pm 0.088) \times \log \theta_{\rm j} + (0.19 \pm 0.045) \times \log Mass + (0.06 \pm 0.45),
\end{equation}
where $E_{\rm p,band,i}$ is in unit of $\rm keV$. $\theta_{\rm j}$ is in unit of $\rm rad$. Mass is in unit of $M_{\bigodot}$. The adjusted $R^{2}$ is 0.2478. The GRB sample number is 44.

The $\log \Gamma_{0}$-$\log (1+z)$-$\log P_{\rm pk4}$ formula is:
\begin{equation} \label{eq:g0zppk4}
\log \Gamma_{0} = (1.7 \pm 0.086) \times \log (1+z) + (0.4 \pm 0.022) \times \log P_{\rm pk4} + (1.2 \pm 0.04),
\end{equation}
where $P_{\rm pk4}$ is peak photon flux of 1 $\rm s$ time bin in 10-1000 $\rm keV$, and in unit of $\rm photons ~ cm^{\rm -2} ~ s^{\rm -1}$. The adjusted $R^{2}$ is 0.6095. The GRB sample number is 44.

The Mag-$\log (1+z)$-$\log  N_{\rm H}$ formula is:
\begin{equation} \label{eq:magznh}
Mag = (-5.3 \pm 1.9) \times \log (1+z) + (-1.2 \pm 0.6) \times \log  N_{\rm H} + (-18 \pm 0.9),
\end{equation}
where Mag is in unit of magnitude. $N_{\rm H}$ is in unit of $\rm 10^{\rm 21} ~ cm^{\rm -2}$. The adjusted $R^{2}$ is 0.4667. The GRB sample number is 44.

The $\log E_{\rm iso}$-$\log E_{\rm p,cpl,i}$-$\log t_{\rm pkOpt,i}$ formula is:
\begin{equation} \label{eq:eisoecplittoi}
\log E_{\rm iso} = (0.9 \pm 0.11) \times \log E_{\rm p,cpl,i} + (-0.44 \pm 0.056) \times \log t_{\rm pkOpt,i} + (-0.66 \pm 0.37),
\end{equation}
where $E_{\rm iso}$ is in unit of $\rm 10^{\rm 52} ~ ergs$ and in rest-frame 1-$10^{4}$ $\rm keV$ energy band. $E_{\rm p,cpl,i}$ is in unit of $\rm keV$. $t_{\rm pkOpt,i}$ is in unit of $\rm s$. The adjusted $R^{2}$ is 0.6507. The GRB sample number is 45.

The $\log t_{\rm pkOpt}$-$\log P_{\rm pk4}$-$\log E_{\rm p,cpl}$ formula is:
\begin{equation} \label{eq:totppk4ecpl}
\log t_{\rm pkOpt} = (-0.63 \pm 0.046) \times \log P_{\rm pk4} + (-0.43 \pm 0.11) \times \log E_{\rm p,cpl} + (3.8 \pm 0.24),
\end{equation}
where $t_{\rm pkOpt}$ is in unit of $\rm s$. $P_{\rm pk4}$ is peak photon flux of 1 $\rm s$ time bin in 10-1000 $\rm keV$, and in unit of $\rm photons ~ cm^{\rm -2} ~ s^{\rm -1}$. $E_{\rm p,cpl}$ is in unit of $\rm keV$. The adjusted $R^{2}$ is 0.3112. The GRB sample number is 45.

The $\log  D_{\rm L}$-$\log T_{\rm 50,i}$-$\log t_{\rm radio,pk,i}$ formula is:
\begin{equation} \label{eq:dlt50itrapki}
\log  D_{\rm L} = (-0.27 \pm 0.0097) \times \log T_{\rm 50,i} + (-0.51 \pm 0.02) \times \log t_{\rm radio,pk,i} + (3.5 \pm 0.11),
\end{equation}
where $D_{\rm L}$ is in unit of $\rm 10^{\rm 28} ~ cm$. $T_{\rm 50,i}$ is in unit of $\rm s$. $t_{\rm radio,pk,i}$ is in unit of $\rm s$. The adjusted $R^{2}$ is 0.205. The GRB sample number is 45.

The $\log t_{\rm pkOpt,i}$-$\log T_{\rm 90}$-$\log \Gamma_{0}$ formula is:
\begin{equation} \label{eq:totit90g0}
\log t_{\rm pkOpt,i} = (0.29 \pm 0.026) \times \log T_{\rm 90} + (-1.4 \pm 0.028) \times \log \Gamma_{0} + (4.6 \pm 0.067),
\end{equation}
where $t_{\rm pkOpt,i}$ is in unit of $\rm s$. $T_{\rm 90}$ is in unit of $\rm s$. The adjusted $R^{2}$ is 0.7005. The GRB sample number is 45.

The $\log SFR$-$\log E_{\rm iso}$-$\log  F_{\rm Opt11hr}$ formula is:
\begin{equation} \label{eq:sfreisofo11}
\log SFR = (0.38 \pm 0.049) \times \log E_{\rm iso} + (-0.34 \pm 0.073) \times \log F_{\rm Opt11hr} + (-1.1 \pm 0.35),
\end{equation}
where SFR is in unit of $\rm M_{\bigodot} ~ yr^{\rm -1}$. $E_{\rm iso}$ is in unit of $\rm 10^{\rm 52} ~ ergs$ and in rest-frame 1-$10^{4}$ $\rm keV$ energy band. $F_{\rm Opt11hr}$ is in unit of $\rm Jy$. The adjusted $R^{2}$ is 0.3405. The GRB sample number is 46.

The $\log Mass$-metallicity-$\log SFR$ formula is:
\begin{equation} \label{eq:massmetsfr}
\log Mass = (0.8 \pm 0.18) \times metallicity + (0.55 \pm 0.054) \times \log SFR + (2.2 \pm 1.6),
\end{equation}
where Mass is in unit of $M_{\bigodot}$. Metallicity is the value of $12+\log O/H$. SFR is in unit of $\rm M_{\bigodot} ~ yr^{\rm -1}$. The adjusted $R^{2}$ is 0.6178. The GRB sample number is 46.

The $\log E_{\rm iso}$-$\log \theta_{\rm j}$-$\log t_{\rm radio,pk,i}$ formula is:
\begin{equation} \label{eq:eisothejtrapki}
\log E_{\rm iso} = (-0.98 \pm 0.18) \times \log \theta_{\rm j} + (-0.98 \pm 0.071) \times \log t_{\rm radio,pk,i} + (5.3 \pm 0.53),
\end{equation}
where $E_{\rm iso}$ is in unit of $\rm 10^{\rm 52} ~ ergs$ and in rest-frame 1-$10^{4}$ $\rm keV$ energy band. $\theta_{\rm j}$ is in unit of $\rm rad$. $t_{\rm radio,pk,i}$ is in unit of $\rm s$. The adjusted $R^{2}$ is 0.2374. The GRB sample number is 46.

The $\log E_{\rm iso}$-$\log L_{\rm radio,pk}$-$\log \theta_{\rm j}$ formula is:
\begin{equation} \label{eq:eisolrathej}
\log E_{\rm iso} = (0.68 \pm 0.022) \times \log L_{\rm radio,pk} + (-0.78 \pm 0.12) \times \log \theta_{\rm j} + (-1.2 \pm 0.12),
\end{equation}
where $E_{\rm iso}$ is in unit of $\rm 10^{\rm 52} ~ ergs$ and in rest-frame 1-$10^{4}$ $\rm keV$ energy band. $L_{\rm radio,pk}$ is in rest-frame 8.46 $\rm GHz$, the unit is $\rm 10^{\rm 40} ~ erg ~ s^{\rm -1}$. $\theta_{\rm j}$ is in unit of $\rm rad$. The adjusted $R^{2}$ is 0.5004. The GRB sample number is 46.

The $\log  F_{\rm Opt11hr}$-$\log Age$-$\log T_{\rm R45,i}$ formula is:
\begin{equation} \label{eq:fo11agetr45i}
\log F_{\rm Opt11hr} = (0.3 \pm 0.13) \times \log Age + (0.73 \pm 0.11) \times \log T_{\rm R45,i} + (-5.8 \pm 0.34),
\end{equation}
where $F_{\rm Opt11hr}$ is in unit of $\rm Jy$. Age is in unit of $\rm Myr$. $T_{\rm R45,i}$ is in unit of $\rm s$. The adjusted $R^{2}$ is 0.3872. The GRB sample number is 47.

The $\log P_{\rm pk4}$-$\log  E_{\rm p,band}$-$\log F_{\rm X11hr}$ formula is:
\begin{equation} \label{eq:ppk4ebandfx11}
\log P_{\rm pk4} = (0.49 \pm 0.093) \times \log  E_{\rm p,band} + (0.42 \pm 0.043) \times \log F_{\rm X11hr} + (2.8 \pm 0.28),
\end{equation}
where $P_{\rm pk4}$ is peak photon flux of 1 $\rm s$ time bin in 10-1000 $\rm keV$, and in unit of $\rm photons ~ cm^{\rm -2} ~ s^{\rm -1}$. $E_{\rm p,band}$ is in unit of $\rm keV$. $F_{\rm X11hr}$ is in unit of $\rm Jy$. The adjusted $R^{2}$ is 0.6621. The GRB sample number is 47.

The $\log  D_{\rm L}$-$\log HR$-$\log \Gamma_{0}$ formula is:
\begin{equation} \label{eq:dlhrga0}
\log  D_{\rm L} = (-0.24 \pm 0.031) \times \log HR + (0.65 \pm 0.026) \times \log \Gamma_{0} + (-0.8 \pm 0.053),
\end{equation}
where $D_{\rm L}$ is in unit of $\rm 10^{\rm 28} ~ cm$. The adjusted $R^{2}$ is 0.3027. The GRB sample number is 47.

The $\log F_{\rm g}$-$\log P_{\rm pk1}$-$\log  F_{\rm Opt11hr}$ formula is:
\begin{equation} \label{eq:fgppk1fo11}
\log F_{\rm g} = (1.1 \pm 0.11) \times \log P_{\rm pk1} + (0.35 \pm 0.084) \times \log F_{\rm Opt11hr} + (1.5 \pm 0.52),
\end{equation}
where $F_{\rm g}$ is in unit of $\rm 10^{\rm -6} ~ ergs ~ cm^{\rm -2}$, and in 20-2000 $\rm keV$ energy band. $P_{\rm pk1}$ is peak photon flux of 64 $\rm ms$ time bin in 10-1000 $\rm keV$, and in unit of $\rm photons ~ cm^{\rm -2} ~ s^{\rm -1}$. $F_{\rm Opt11hr}$ is in unit of $\rm Jy$. The adjusted $R^{2}$ is 0.6153. The GRB sample number is 48.

The $\log HR$-$variability_{3}$-$(-\beta_{\rm band})$ formula is:
\begin{equation} \label{eq:hrvar3bband}
\log HR = (7.9 \pm 3.2) \times variability_{3} + (-0.18 \pm 0.099) \times (-\beta_{\rm band}) + (0.81 \pm 0.24),
\end{equation}
The adjusted $R^{2}$ is 0.251. The GRB sample number is 48.

The $\log L_{\rm radio,pk}$-$\log L_{\rm pk}$-$A_{\rm V}$ formula is:
\begin{equation} \label{eq:lralpkav}
\log L_{\rm radio,pk} = (0.5 \pm 0.026) \times \log L_{\rm pk} + (-0.35 \pm 0.076) \times A_{\rm V} + (1.9 \pm 0.053),
\end{equation}
where $L_{\rm radio,pk}$ is in rest-frame 8.46 $\rm GHz$, the unit is $\rm 10^{\rm 40} ~ erg ~ s^{\rm -1}$. $L_{\rm pk}$ is in unit of $\rm 10^{\rm 52} ~ ergs ~ s^{\rm -1}$, and in 1-$10^{4}$ $\rm keV$ energy band. The adjusted $R^{2}$ is 0.49. The GRB sample number is 48.

The $\log L_{\rm pk}$-$\log F_{\rm g}$-$\log F_{\rm radio,pk}$ formula is:
\begin{equation} \label{eq:lpkfgfrapk}
\log L_{\rm pk} = (0.82 \pm 0.054) \times \log F_{\rm g} + (-0.85 \pm 0.06) \times \log F_{\rm radio,pk} + (-3.8 \pm 0.22),
\end{equation}
where $L_{\rm pk}$ is in unit of $\rm 10^{\rm 52} ~ erg ~ s^{\rm -1}$, and in 1-$10^{4}$ $\rm keV$ energy band. $F_{\rm g}$ is in unit of $\rm 10^{\rm -6} ~ ergs ~ cm^{\rm -2}$, and in 20-2000 $\rm keV$ energy band. $F_{\rm radio,pk}$ is in unit of $\rm Jy$. The adjusted $R^{2}$ is 0.4055. The GRB sample number is 49.

The $\log L_{\rm pk}$-$\log F_{\rm g}$-$\log t_{\rm radio,pk,i}$ formula is:
\begin{equation} \label{eq:lpkfgtrapki}
\log L_{\rm pk} = (0.74 \pm 0.055) \times \log F_{\rm g} + (-0.66 \pm 0.074) \times \log t_{\rm radio,pk,i} + (2.9 \pm 0.46),
\end{equation}
where $L_{\rm pk}$ is in unit of $\rm 10^{\rm 52} ~ erg ~ s^{\rm -1}$, and in 1-$10^{4}$ $\rm keV$ energy band. $F_{\rm g}$ is in unit of $\rm 10^{\rm -6} ~ ergs ~ cm^{\rm -2}$, and in 20-2000 $\rm keV$ energy band. $t_{\rm radio,pk,i}$ is in unit of $\rm s$. The adjusted $R^{2}$ is 0.353. The GRB sample number is 49.

The $\log  F_{\rm Opt11hr}$-$\log  N_{\rm H}$-$\log Mass$ formula is:
\begin{equation} \label{eq:fo11nhmass}
\log F_{\rm Opt11hr} = (-0.43 \pm 0.16) \times \log  N_{\rm H} + (-0.52 \pm 0.11) \times \log Mass + (0.074 \pm 1),
\end{equation}
where $F_{\rm Opt11hr}$ is in unit of $\rm Jy$. $N_{\rm H}$ is in unit of $\rm 10^{\rm 21} ~ cm^{\rm -2}$. Mass is in unit of $M_{\bigodot}$. The adjusted $R^{2}$ is 0.3336. The GRB sample number is 49.

The $\log F_{\rm g}$-$variability_{3}$-$\log T_{\rm 90,i}$ formula is:
\begin{equation} \label{eq:fgvar3t90i}
\log F_{\rm g} = (43 \pm 6.6) \times variability_{3} + (0.78 \pm 0.039) \times \log T_{\rm 90,i} + (-0.12 \pm 0.071),
\end{equation}
where $F_{\rm g}$ is in unit of $\rm 10^{\rm -6} ~ ergs ~ cm^{\rm -2}$, and in 20-2000 $\rm keV$ energy band. $T_{\rm 90,i}$ is in unit of $\rm s$. The adjusted $R^{2}$ is 0.4482. The GRB sample number is 49.

The $\log F_{\rm g}$-$variability_{3}$-$\log T_{\rm R45,i}$ formula is:
\begin{equation} \label{eq:fgvar3tr45i}
\log F_{\rm g} = (47 \pm 6.9) \times variability_{3} + (0.78 \pm 0.054) \times \log T_{\rm R45,i} + (0.51 \pm 0.06),
\end{equation}
where $F_{\rm g}$ is in unit of $\rm 10^{\rm -6} ~ ergs ~ cm^{\rm -2}$, and in 20-2000 $\rm keV$ energy band. $T_{\rm R45,i}$ is in unit of $\rm s$. The adjusted $R^{2}$ is 0.278. The GRB sample number is 49.

The $\log F_{\rm g}$-$\log (1+z)$-$variability_{3}$ formula is:
\begin{equation} \label{eq:fgzvar3}
\log F_{\rm g} = (-1.9 \pm 0.1) \times \log (1+z) + (35 \pm 6.9) \times variability_{3} + (1.7 \pm 0.086),
\end{equation}
where $F_{\rm g}$ is in unit of $\rm 10^{\rm -6} ~ ergs ~ cm^{\rm -2}$, and in 20-2000 $\rm keV$ energy band. The adjusted $R^{2}$ is 0.3132. The GRB sample number is 49.

The $\log F_{\rm g}$-$\log  D_{\rm L}$-$variability_{3}$ formula is:
\begin{equation} \label{eq:fgdlvar3}
\log F_{\rm g} = (-0.86 \pm 0.042) \times \log  D_{\rm L} + (37 \pm 6.9) \times variability_{3} + (1.3 \pm 0.069),
\end{equation}
where $F_{\rm g}$ is in unit of $\rm 10^{\rm -6} ~ ergs ~ cm^{\rm -2}$, and in 20-2000 $\rm keV$ energy band. $D_{\rm L}$ is in unit of $\rm 10^{\rm 28} ~ cm$. The adjusted $R^{2}$ is 0.2954. The GRB sample number is 50.

The $\log HR$-$\log E_{\rm iso}$-$\log F_{\rm radio,pk}$ formula is:
\begin{equation} \label{eq:hreisofrapk}
\log HR = (0.24 \pm 0.018) \times \log E_{\rm iso} + (0.33 \pm 0.049) \times \log F_{\rm radio,pk} + (1.4 \pm 0.17),
\end{equation}
where $E_{\rm iso}$ is in unit of $\rm 10^{\rm 52} ~ ergs$ and in rest-frame 1-$10^{4}$ $\rm keV$ energy band. $F_{\rm radio,pk}$ is in unit of $\rm Jy$. The adjusted $R^{2}$ is 0.2991. The GRB sample number is 50.

The $\log Age$-$\log  F_{\rm Opt11hr}$-$\log SSFR$ formula is:
\begin{equation} \label{eq:agefopssfr}
\log Age = (0.29 \pm 0.063) \times \log F_{\rm Opt11hr} + (-0.44 \pm 0.069) \times \log SSFR + (4.1 \pm 0.31),
\end{equation}
where Age is in unit of $\rm Myr$. $F_{\rm Opt11hr}$ is in unit of $\rm Jy$. $\log SSFR$ is in unit of $\rm Gyr^{\rm -1}$. The adjusted $R^{2}$ is 0.3771. The GRB sample number is 50.

The $\log E_{\rm p,band,i}$-$\log HR$-$\log \theta_{\rm j}$ formula is:
\begin{equation} \label{eq:ebandihrthej}
\log E_{\rm p,band,i} = (0.61 \pm 0.12) \times \log HR + (-0.33 \pm 0.087) \times \log \theta_{\rm j} + (2 \pm 0.11),
\end{equation}
where $E_{\rm p,band,i}$ is in unit of $\rm keV$. $\theta_{\rm j}$ is in unit of $\rm rad$. The adjusted $R^{2}$ is 0.4157. The GRB sample number is 51.

The $\log SFR$-$\log T_{\rm 90}$-$\log  F_{\rm Opt11hr}$ formula is:
\begin{equation} \label{eq:sfrt90fo11}
\log SFR = (0.5 \pm 0.068) \times \log T_{\rm 90} + (-0.4 \pm 0.072) \times \log F_{\rm Opt11hr} + (-2.1 \pm 0.36),
\end{equation}
where SFR is in unit of $\rm M_{\bigodot} ~ yr^{\rm -1}$. $T_{\rm 90}$ is in unit of $\rm s$. $F_{\rm Opt11hr}$ is in unit of $\rm Jy$. The adjusted $R^{2}$ is 0.3356. The GRB sample number is 51.

The $\log L_{\rm radio,pk}$-$\log E_{\rm iso}$-$\log F_{\rm Opt11hr}$ formula is:
\begin{equation} \label{eq:lraeisofo11}
\log L_{\rm radio,pk} = (0.67 \pm 0.023) \times \log E_{\rm iso} + (-0.23 \pm 0.074) \times \log F_{\rm Opt11hr} + (0.18 \pm 0.31),
\end{equation}
where $L_{\rm radio,pk}$ is in rest-frame 8.46 $\rm GHz$, the unit is $\rm 10^{\rm 40} ~ erg ~ s^{\rm -1}$. $E_{\rm iso}$ is in unit of $\rm 10^{\rm 52} ~ ergs$ and in rest-frame 1-$10^{4}$ $\rm keV$ energy band. $F_{\rm Opt11hr}$ is in unit of $\rm Jy$. The adjusted $R^{2}$ is 0.4846. The GRB sample number is 52.

The $\log  F_{\rm Opt11hr}$-$\log Age$-$\log Mass$ formula is:
\begin{equation} \label{eq:fo11agemass}
\log F_{\rm Opt11hr} = (0.42 \pm 0.11) \times \log Age + (-0.45 \pm 0.098) \times \log Mass + (-1.8 \pm 1),
\end{equation}
where $F_{\rm Opt11hr}$ is in unit of $\rm Jy$. Age is in unit of $\rm Myr$. Mass is in unit of $M_{\bigodot}$. The adjusted $R^{2}$ is 0.3177. The GRB sample number is 53.

The $\beta_{\rm X11hr}$-$\log  E_{\rm p,band}$-$\log F_{\rm X11hr}$ formula is:
\begin{equation} \label{eq:be11ebandfx11}
\beta_{\rm X11hr} = (-0.39 \pm 0.12) \times \log  E_{\rm p,band} + (-0.37 \pm 0.091) \times \log F_{\rm X11hr} + (-0.27 \pm 0.81),
\end{equation}
where $E_{\rm p,band}$ is in unit of $\rm keV$. $F_{\rm X11hr}$ is in unit of $\rm Jy$. The adjusted $R^{2}$ is 0.3324. The GRB sample number is 53.

The $\log E_{\rm p,band,i}$-$\log HR$-$\log Mass$ formula is:
\begin{equation} \label{eq:ebandihrmass}
\log E_{\rm p,band,i} = (0.83 \pm 0.1) \times \log HR + (0.12 \pm 0.033) \times \log Mass + (1.1 \pm 0.32),
\end{equation}
where $E_{\rm p,band,i}$ is in unit of $\rm keV$. Mass is in unit of $M_{\bigodot}$. The adjusted $R^{2}$ is 0.6666. The GRB sample number is 53.

The $\log Age$-$\log HR$-$\log SSFR$ formula is:
\begin{equation} \label{eq:agehrssfr}
\log Age = (-0.41 \pm 0.13) \times \log HR + (-0.45 \pm 0.061) \times \log SSFR + (2.8 \pm 0.083),
\end{equation}
where Age is in unit of $\rm Myr$. $\log SSFR$ is in unit of $\rm Gyr^{\rm -1}$. The adjusted $R^{2}$ is 0.2396. The GRB sample number is 53.

The Mag-$\log L_{\rm pk}$-$\log P_{\rm pk4}$ formula is:
\begin{equation} \label{eq:lpkppk4mag}
Mag = (-1.1 \pm 0.28) \times \log L_{\rm pk} + (0.7 \pm 0.54) \times \log P_{\rm pk4} + (-21 \pm 0.48),
\end{equation}
where Mag is in unit of magnitude. $L_{\rm pk}$ is in unit of $\rm 10^{\rm 52} ~ erg ~ s^{\rm -1}$, and in 1-$10^{4}$ $\rm keV$ energy band. $P_{\rm pk4}$ is peak photon flux of 1 $\rm s$ time bin in 10-1000 $\rm keV$, and in unit of $\rm photons ~ cm^{\rm -2} ~ s^{\rm -1}$. The adjusted $R^{2}$ is 0.3453. The GRB sample number is 53.

The $\log Mass$-$\log L_{\rm pk}$-metallicity formula is:
\begin{equation} \label{eq:masslpkmet}
\log Mass = (0.17 \pm 0.038) \times \log L_{\rm pk} + (1.2 \pm 0.2) \times metallicity + (-0.7 \pm 1.7),
\end{equation}
where Mass is in unit of $M_{\bigodot}$. $L_{\rm pk}$ is in unit of $\rm 10^{\rm 52} ~ erg ~ s^{\rm -1}$, and in 1-$10^{4}$ $\rm keV$ energy band. Metallicity is the value of $12+\log O/H$. The adjusted $R^{2}$ is 0.3348. The GRB sample number is 53.

The $\log  F_{\rm Opt11hr}$-$\log Mass$-$\log t_{\rm burst,i}$ formula is:
\begin{equation} \label{eq:fo11masstti}
\log F_{\rm Opt11hr} = (-0.29 \pm 0.11) \times \log Mass + (0.63 \pm 0.12) \times \log t_{\rm burst,i} + (-3.8 \pm 1.2),
\end{equation}
where $F_{\rm Opt11hr}$ is in unit of $\rm Jy$. Mass is in unit of $M_{\bigodot}$. $t_{\rm burst,i}$ is in unit of $\rm s$. The adjusted $R^{2}$ is 0.2911. The GRB sample number is 53.

The $\log Age$-$\log (1+z)$-$\log  F_{\rm Opt11hr}$ formula is:
\begin{equation} \label{eq:agezfo11}
\log Age = (-1.7 \pm 0.34) \times \log (1+z) + (0.17 \pm 0.062) \times \log F_{\rm Opt11hr} + (3.9 \pm 0.3),
\end{equation}
where Age is in unit of $\rm Myr$. $F_{\rm Opt11hr}$ is in unit of $\rm Jy$. The adjusted $R^{2}$ is 0.2579. The GRB sample number is 53.

The $\log L_{\rm radio,pk}$-$\log E_{\rm iso}$-$A_{\rm V}$ formula is:
\begin{equation} \label{eq:lraeisoav}
\log L_{\rm radio,pk} = (0.59 \pm 0.015) \times \log E_{\rm iso} + (-0.27 \pm 0.055) \times A_{\rm V} + (1.3 \pm 0.041),
\end{equation}
where $L_{\rm radio,pk}$ is in rest-frame 8.46 $\rm GHz$, the unit is $\rm 10^{\rm 40} ~ erg ~ s^{\rm -1}$. $E_{\rm iso}$ is in unit of $\rm 10^{\rm 52} ~ ergs$ and in rest-frame 1-$10^{4}$ $\rm keV$ energy band. The adjusted $R^{2}$ is 0.5084. The GRB sample number is 53.

The $\log Mass$-$\beta_{\rm X11hr}$-Mag formula is:
\begin{equation} \label{eq:massbe11mag}
\log Mass = (0.12 \pm 0.081) \times \beta_{\rm X11hr} + (-0.11 \pm 0.029) \times Mag + (7 \pm 0.62),
\end{equation}
where Mass is in unit of $M_{\bigodot}$. Mag is in unit of magnitude. The adjusted $R^{2}$ is 0.8391. The GRB sample number is 54.

The $\log F_{\rm g}$-$\beta_{\rm X11hr}$-$\log SSFR$ formula is:
\begin{equation} \label{eq:fgbex11ssfr}
\log F_{\rm g} = (-0.49 \pm 0.11) \times \beta_{\rm X11hr} + (0.4 \pm 0.051) \times \log SSFR + (1.5 \pm 0.17),
\end{equation}
where $F_{\rm g}$ is in unit of $\rm 10^{\rm -6} ~ ergs ~ cm^{\rm -2}$, and in 20-2000 $\rm keV$ energy band. $\log SSFR$ is in unit of $\rm Gyr^{\rm -1}$. The adjusted $R^{2}$ is 0.3069. The GRB sample number is 54.

The $\log T_{\rm 90,i}$-$\log  E_{\rm p,band}$-$\log F_{\rm X11hr}$ formula is:
\begin{equation} \label{eq:t90iebandfx11}
\log T_{\rm 90,i} = (0.34 \pm 0.054) \times \log  E_{\rm p,band} + (0.21 \pm 0.026) \times \log F_{\rm X11hr} + (1.9 \pm 0.26),
\end{equation}
where $T_{\rm 90,i}$ is in unit of $\rm s$. $E_{\rm p,band}$ is in unit of $\rm keV$. $F_{\rm X11hr}$ is in unit of $\rm Jy$. The adjusted $R^{2}$ is 0.2015. The GRB sample number is 54.

The $\log F_{\rm radio,pk}$-$\log  F_{\rm Opt11hr}$-$\log t_{\rm radio,pk,i}$ formula is:
\begin{equation} \label{eq:frapkfop11hrtrapki}
\log F_{\rm radio,pk} = (0.27 \pm 0.04) \times \log F_{\rm Opt11hr} + (0.25 \pm 0.04) \times \log t_{\rm radio,pk,i} + (-3.7 \pm 0.31),
\end{equation}
where $F_{\rm radio,pk}$ is in unit of $\rm Jy$. $F_{\rm Opt11hr}$ is in unit of $\rm Jy$. $t_{\rm radio,pk,i}$ is in unit of $\rm s$. The adjusted $R^{2}$ is 0.4246. The GRB sample number is 54.

The $\log  F_{\rm Opt11hr}$-$\log F_{\rm radio,pk}$-$\log T_{\rm 90,i}$ formula is:
\begin{equation} \label{eq:fopfrapkt90i}
\log F_{\rm Opt11hr} = (0.89 \pm 0.12) \times \log F_{\rm radio,pk} + (0.32 \pm 0.11) \times \log T_{\rm 90,i} + (-1.6 \pm 0.46),
\end{equation}
where $F_{\rm Opt11hr}$ is in unit of $\rm Jy$. $F_{\rm radio,pk}$ is in unit of $\rm Jy$. $T_{\rm 90,i}$ is in unit of $\rm s$. The adjusted $R^{2}$ is 0.4369. The GRB sample number is 54.

The $\log  D_{\rm L}$-$\log F_{\rm radio,pk}$-$A_{\rm V}$ formula is:
\begin{equation} \label{eq:dlfrapkav}
\log  D_{\rm L} = (-0.46 \pm 0.025) \times \log F_{\rm radio,pk} + (-0.12 \pm 0.028) \times A_{\rm V} + (-1.2 \pm 0.084),
\end{equation}
where $D_{\rm L}$ is in unit of $\rm 10^{\rm 28} ~ cm$. $F_{\rm radio,pk}$ is in unit of $\rm Jy$. The adjusted $R^{2}$ is 0.237. The GRB sample number is 54.

The $\log Mass$-$\log T_{\rm 50}$-$\log SFR$ formula is:
\begin{equation} \label{eq:masst50sfr}
\log Mass = (0.34 \pm 0.077) \times \log T_{\rm 50} + (0.62 \pm 0.047) \times \log SFR + (8.7 \pm 0.11),
\end{equation}
where Mass is in unit of $M_{\bigodot}$. $T_{\rm 50}$ is in unit of $\rm s$. SFR is in unit of $\rm M_{\bigodot} ~ yr^{\rm -1}$. The adjusted $R^{2}$ is 0.6319. The GRB sample number is 54.

The $\log  F_{\rm Opt11hr}$-$\log (1+z)$-$\log F_{\rm radio,pk}$ formula is:
\begin{equation} \label{eq:fo11zfrapk}
\log F_{\rm Opt11hr} = (-0.86 \pm 0.37) \times \log (1+z) + (0.79 \pm 0.15) \times \log F_{\rm radio,pk} + (-1.3 \pm 0.45),
\end{equation}
where $F_{\rm Opt11hr}$ is in unit of $\rm Jy$. $F_{\rm radio,pk}$ is in unit of $\rm Jy$. The adjusted $R^{2}$ is 0.4305. The GRB sample number is 54.

The $\log  t_{\rm pkOpt}$-$\log F_{\rm g}$-$\log F_{\rm Opt11hr}$ formula is:
\begin{equation} \label{eq:topfgfo11}
\log t_{\rm pkOpt} = (-0.34 \pm 0.03) \times \log F_{\rm g} + (0.27 \pm 0.061) \times \log F_{\rm Opt11hr} + (4 \pm 0.28),
\end{equation}
where $F_{\rm Opt11hr}$ is in unit of $\rm Jy$. $t_{\rm pkOpt}$ is in unit of $\rm s$. $F_{\rm g}$ is in unit of $\rm 10^{\rm -6} ~ ergs ~ cm^{\rm -2}$, and in 20-2000 $\rm keV$ energy band. The adjusted $R^{2}$ is 0.309. The GRB sample number is 55.

The $\log SFR$-metallicity-$A_{\rm V}$ formula is:
\begin{equation} \label{eq:sfrmetav}
\log SFR = (1.1 \pm 0.24) \times metallicity + (0.24 \pm 0.07) \times A_{\rm V} + (-8.9 \pm 2),
\end{equation}
where SFR is in unit of $\rm M_{\bigodot} ~ yr^{\rm -1}$. Metallicity is the value of $12+\log O/H$. The adjusted $R^{2}$ is 0.265. The GRB sample number is 55.

The Mag-$\log Mass$-$\log t_{\rm burst,i}$ formula is:
\begin{equation} \label{eq:magmasstti}
Mag = (-1.7 \pm 0.46) \times \log Mass + (0.43 \pm 0.44) \times \log t_{\rm burst,i} + (-5.4 \pm 4.9),
\end{equation}
where Mag is in unit of magnitude. Mass is in unit of $M_{\bigodot}$. $t_{\rm burst,i}$ is in unit of $\rm s$. The adjusted $R^{2}$ is 0.8658. The GRB sample number is 57.

The Mag-$\log T_{\rm 90}$-$\log t_{\rm burst,i}$ formula is:
\begin{equation} \label{eq:magt90tti}
Mag = (-1.1 \pm 0.81) \times \log T_{\rm 90} + (1.1 \pm 0.4) \times \log t_{\rm burst,i} + (-21 \pm 1.8),
\end{equation}
where Mag is in unit of magnitude. $T_{\rm 90}$ is in unit of $\rm s$. $t_{\rm burst,i}$ is in unit of $\rm s$. The adjusted $R^{2}$ is 0.2815. The GRB sample number is 57.

The $\log HR$-$\log \theta_{\rm j}$-$\log E_{\rm p,cpl,i}$ formula is:
\begin{equation} \label{eq:hrthejecpli}
\log HR = (0.18 \pm 0.064) \times \log \theta_{\rm j} + (0.73 \pm 0.098) \times \log E_{\rm p,cpl,i} + (-1.6 \pm 0.26),
\end{equation}
where $\theta_{\rm j}$ is in unit of $\rm rad$. $E_{\rm p,cpl,i}$ is in unit of $\rm keV$. The adjusted $R^{2}$ is 0.605. The GRB sample number is 57.

The $\log L_{\rm pk}$-$\log \theta_{\rm j}$-$\log t_{\rm burst,i}$ formula is:
\begin{equation} \label{eq:lpkthejtti}
\log L_{\rm pk} = (-0.75 \pm 0.12) \times \log \theta_{\rm j} + (-0.43 \pm 0.047) \times \log t_{\rm burst,i} + (0.15 \pm 0.21),
\end{equation}
where $L_{\rm pk}$ is in unit of $\rm 10^{\rm 52} ~ erg ~ s^{\rm -1}$, and in 1-$10^{4}$ $\rm keV$ energy band. $\theta_{\rm j}$ is in unit of $\rm rad$. $t_{\rm burst,i}$ is in unit of $\rm s$. The adjusted $R^{2}$ is 0.2276. The GRB sample number is 57.

The $\log F_{\rm g}$-$(-\beta_{\rm band})$-$\beta_{\rm X11hr}$ formula is:
\begin{equation} \label{eq:fgbbandbx11}
\log F_{\rm g} = (0.22 \pm 0.13) \times (-\beta_{\rm band}) + (-0.69 \pm 0.1) \times \beta_{\rm X11hr} + (1.5 \pm 0.4),
\end{equation}
where $F_{\rm g}$ is in unit of $\rm 10^{\rm -6} ~ ergs ~ cm^{\rm -2}$, and in 20-2000 $\rm keV$ energy band. The adjusted $R^{2}$ is 0.4589. The GRB sample number is 58.

The $\log SFR$-$\log E_{\rm iso}$-metallicity formula is:
\begin{equation} \label{eq:sfreisomet}
\log SFR = (0.28 \pm 0.028) \times \log E_{\rm iso} + (0.97 \pm 0.2) \times metallicity + (-7.8 \pm 1.7),
\end{equation}
where SFR is in unit of $\rm M_{\bigodot} ~ yr^{\rm -1}$. $E_{\rm iso}$ is in unit of $\rm 10^{\rm 52} ~ ergs$ and in rest-frame 1-$10^{4}$ $\rm keV$ energy band. Metallicity is the value of $12+\log O/H$. The adjusted $R^{2}$ is 0.3087. The GRB sample number is 58.

The $\log F_{\rm g}$-$\log  E_{\rm p,band}$-$\beta_{\rm X11hr}$ formula is:
\begin{equation} \label{eq:fgebandbx11}
\log F_{\rm g} = (0.84 \pm 0.089) \times \log  E_{\rm p,band} + (-0.49 \pm 0.082) \times \beta_{\rm X11hr} + (-0.1 \pm 0.28),
\end{equation}
where $F_{\rm g}$ is in unit of $\rm 10^{\rm -6} ~ ergs ~ cm^{\rm -2}$, and in 20-2000 $\rm keV$ energy band. $E_{\rm p,band}$ is in unit of $\rm keV$. The adjusted $R^{2}$ is 0.6254. The GRB sample number is 58.

The $\log  E_{\rm p,band}$-$variability_{3}$-$\log F_{\rm g}$ formula is:
\begin{equation} \label{eq:ebandvar3fg}
\log  E_{\rm p,band} = (11 \pm 2.1) \times variability_{3} + (0.22 \pm 0.023) \times \log F_{\rm g} + (1.8 \pm 0.051),
\end{equation}
where $E_{\rm p,band}$ is in unit of $\rm keV$. $F_{\rm g}$ is in unit of $\rm 10^{\rm -6} ~ ergs ~ cm^{\rm -2}$, and in 20-2000 $\rm keV$ energy band. The adjusted $R^{2}$ is 0.3281. The GRB sample number is 58.

The $\log SFR$-$\log  D_{\rm L}$-metallicity formula is:
\begin{equation} \label{eq:sfrdlmet}
\log SFR = (1.2 \pm 0.066) \times \log  D_{\rm L} + (0.89 \pm 0.17) \times metallicity + (-7.1 \pm 1.5),
\end{equation}
where SFR is in unit of $\rm M_{\bigodot} ~ yr^{\rm -1}$. $D_{\rm L}$ is in unit of $\rm 10^{\rm 28} ~ cm$. Metallicity is the value of $12+\log O/H$. The adjusted $R^{2}$ is 0.584. The GRB sample number is 59.

The $\log F_{\rm pk1}$-$\log  E_{\rm p,band}$-$\beta_{\rm X11hr}$ formula is:
\begin{equation} \label{eq:fpk1ebandbx11}
\log F_{\rm pk1} = (0.98 \pm 0.091) \times \log  E_{\rm p,band} + (-0.33 \pm 0.07) \times \beta_{\rm X11hr} + (-1.5 \pm 0.25),
\end{equation}
where $F_{\rm pk1}$ is peak energy flux of 1 $\rm s$ time bin in rest-frame 1-$10^{4}$ $\rm keV$ energy band, and in unit of $\rm 10^{\rm -6} ~ ergs ~ cm^{\rm -2} ~ s^{\rm -1}$. $E_{\rm p,band}$ is in unit of $\rm keV$. The adjusted $R^{2}$ is 0.5634. The GRB sample number is 59.

The $\log L_{\rm pk}$-Mag-$\log T_{\rm 50,i}$ formula is:
\begin{equation} \label{eq:lpkmagt50i}
\log L_{\rm pk} = (-0.13 \pm 0.037) \times Mag + (-0.6 \pm 0.086) \times \log T_{\rm 50,i} + (-2.1 \pm 0.79),
\end{equation}
where $L_{\rm pk}$ is in unit of $\rm 10^{\rm 52} ~ erg ~ s^{\rm -1}$, and in 1-$10^{4}$ $\rm keV$ energy band. Mag is in unit of magnitude. $T_{\rm 50,i}$ is in unit of $\rm s$. The adjusted $R^{2}$ is 0.3217. The GRB sample number is 59.

The $\log SFR$-$\log (1+z)$-metallicity formula is:
\begin{equation} \label{eq:sfrzmet}
\log SFR = (4.6 \pm 0.26) \times \log (1+z) + (0.79 \pm 0.17) \times metallicity + (-7.3 \pm 1.4),
\end{equation}
where SFR is in unit of $\rm M_{\bigodot} ~ yr^{\rm -1}$. Metallicity is the value of $12+\log O/H$. The adjusted $R^{2}$ is 0.6034. The GRB sample number is 59.

The $\log F_{\rm g}$-$\log L_{\rm pk}$-metallicity formula is:
\begin{equation} \label{eq:fglpkmet}
\log F_{\rm g} = (0.35 \pm 0.018) \times \log L_{\rm pk} + (-0.61 \pm 0.18) \times metallicity + (6.3 \pm 1.5),
\end{equation}
where $F_{\rm g}$ is in unit of $\rm 10^{\rm -6} ~ ergs ~ cm^{\rm -2}$, and in 20-2000 $\rm keV$ energy band. $L_{\rm pk}$ is in unit of $\rm 10^{\rm 52} ~ erg ~ s^{\rm -1}$, and in 1-$10^{4}$ $\rm keV$ energy band. Metallicity is the value of $12+\log O/H$. The adjusted $R^{2}$ is 0.2819. The GRB sample number is 61.

The $\log E_{\rm iso}$-$\log L_{\rm radio,pk}$-$\log t_{\rm radio,pk}$ formula is:
\begin{equation} \label{eq:eisolratra}
\log E_{\rm iso} = (0.71 \pm 0.015) \times \log L_{\rm radio,pk} + (-0.76 \pm 0.052) \times \log t_{\rm radio,pk} + (4.2 \pm 0.31),
\end{equation}
where $E_{\rm iso}$ is in unit of $\rm 10^{\rm 52} ~ ergs$ and in rest-frame 1-$10^{4}$ $\rm keV$ energy band. $L_{\rm radio,pk}$ is in rest-frame 8.46 $\rm GHz$, the unit is $\rm 10^{\rm 40} ~ erg ~ s^{\rm -1}$. $t_{\rm radio,pk}$ is in unit of $\rm s$. The adjusted $R^{2}$ is 0.4832. The GRB sample number is 61.

The $\log L_{\rm radio,pk}$-$\log E_{\rm iso}$-$\log T_{\rm 90,i}$ formula is:
\begin{equation} \label{eq:lraeisot90i}
\log L_{\rm radio,pk} = (0.66 \pm 0.013) \times \log E_{\rm iso} + (-0.44 \pm 0.029) \times \log T_{\rm 90,i} + (1.7 \pm 0.039),
\end{equation}
where $L_{\rm radio,pk}$ is in rest-frame 8.46 $\rm GHz$, the unit is $\rm 10^{\rm 40} ~ erg ~ s^{\rm -1}$. $E_{\rm iso}$ is in unit of $\rm 10^{\rm 52} ~ ergs$ and in rest-frame 1-$10^{4}$ $\rm keV$ energy band. $T_{\rm 90,i}$ is in unit of $\rm s$. The adjusted $R^{2}$ is 0.4534. The GRB sample number is 61.

The $\log Mass$-$A_{\rm V}$-$\log E_{\rm p,cpl,i}$ formula is:
\begin{equation} \label{eq:massavecpli}
\log Mass = (0.32 \pm 0.093) \times A_{\rm V} + (0.48 \pm 0.13) \times \log E_{\rm p,cpl,i} + (8.2 \pm 0.36),
\end{equation}
where Mass is in unit of $M_{\bigodot}$. $E_{\rm p,cpl,i}$ is in unit of $\rm keV$. The adjusted $R^{2}$ is 0.2425. The GRB sample number is 62.

The Mag-$\log E_{\rm iso}$-$A_{\rm V}$ formula is:
\begin{equation} \label{eq:mageisoav}
Mag = (-0.84 \pm 0.31) \times \log E_{\rm iso} + (-0.49 \pm 0.36) \times A_{\rm V} + (-20 \pm 0.46),
\end{equation}
where Mag is in unit of magnitude. $E_{\rm iso}$ is in unit of $\rm 10^{\rm 52} ~ ergs$ and in rest-frame 1-$10^{4}$ $\rm keV$ energy band. The adjusted $R^{2}$ is 0.291. The GRB sample number is 62.

The $\log  D_{\rm L}$-$\log F_{\rm radio,pk}$-$\log t_{\rm radio,pk,i}$ formula is:
\begin{equation} \label{eq:dlfrapktrapki}
\log  D_{\rm L} = (-0.41 \pm 0.026) \times \log F_{\rm radio,pk} + (-0.29 \pm 0.019) \times \log t_{\rm radio,pk,i} + (0.58 \pm 0.17),
\end{equation}
where $D_{\rm L}$ is in unit of $\rm 10^{\rm 28} ~ cm$. $F_{\rm radio,pk}$ is in unit of $\rm Jy$. $t_{\rm radio,pk,i}$ is in unit of $\rm s$. The adjusted $R^{2}$ is 0.242. The GRB sample number is 62.

The $\log L_{\rm pk}$-Mag-$\log T_{\rm 90,i}$ formula is:
\begin{equation} \label{eq:lpkmagt90i}
\log L_{\rm pk} = (-0.13 \pm 0.036) \times Mag + (-0.66 \pm 0.094) \times \log T_{\rm 90,i} + (-1.8 \pm 0.74),
\end{equation}
where $L_{\rm pk}$ is in unit of $\rm 10^{\rm 52} ~ erg ~ s^{\rm -1}$, and in 1-$10^{4}$ $\rm keV$ energy band. Mag is in unit of magnitude. $T_{\rm 90,i}$ is in unit of $\rm s$. The adjusted $R^{2}$ is 0.3448. The GRB sample number is 62.

The $\log SFR$-$\log E_{\rm iso}$-$\log Mass$ formula is:
\begin{equation} \label{eq:sfreisomass}
\log SFR = (0.19 \pm 0.035) \times \log E_{\rm iso} + (0.64 \pm 0.064) \times \log Mass + (-5.5 \pm 0.6),
\end{equation}
where SFR is in unit of $\rm M_{\bigodot} ~ yr^{\rm -1}$. $E_{\rm iso}$ is in unit of $\rm 10^{\rm 52} ~ ergs$ and in rest-frame 1-$10^{4}$ $\rm keV$ energy band. Mass is in unit of $M_{\bigodot}$. The adjusted $R^{2}$ is 0.5535. The GRB sample number is 63.

The $\log Mass$-$\log E_{\rm iso}$-metallicity formula is:
\begin{equation} \label{eq:masseisomet}
\log Mass = (0.19 \pm 0.033) \times \log E_{\rm iso} + (1.2 \pm 0.17) \times metallicity + (-0.96 \pm 1.5),
\end{equation}
where Mass is in unit of $M_{\bigodot}$. $E_{\rm iso}$ is in unit of $\rm 10^{\rm 52} ~ ergs$ and in rest-frame 1-$10^{4}$ $\rm keV$ energy band. Metallicity is the value of $12+\log O/H$. The adjusted $R^{2}$ is 0.3948. The GRB sample number is 63.

The $\log SSFR$-$\log F_{\rm g}$-$\log Age$ formula is:
\begin{equation} \label{eq:ssfrfgage}
\log SSFR = (0.26 \pm 0.022) \times \log F_{\rm g} + (-0.41 \pm 0.054) \times \log Age + (0.88 \pm 0.15),
\end{equation}
where $\log SSFR$ is in unit of $\rm Gyr^{\rm -1}$. $F_{\rm g}$ is in unit of $\rm 10^{\rm -6} ~ ergs ~ cm^{\rm -2}$, and in 20-2000 $\rm keV$ energy band. Age is in unit of $\rm Myr$. The adjusted $R^{2}$ is 0.2893. The GRB sample number is 63.

The $\log \theta_{\rm j}$-$\log T_{\rm 90}$-$\log t_{\rm burst,i}$ formula is:
\begin{equation} \label{eq:thejt90tti}
\log \theta_{\rm j} = (-0.36 \pm 0.022) \times \log T_{\rm 90} + (0.14 \pm 0.019) \times \log t_{\rm burst,i} + (-0.82 \pm 0.06),
\end{equation}
where $\theta_{\rm j}$ is in unit of $\rm rad$. $T_{\rm 90}$ is in unit of $\rm s$. $t_{\rm burst,i}$ is in unit of $\rm s$. The adjusted $R^{2}$ is 0.2592. The GRB sample number is 63.

The $\log  D_{\rm L}$-$\log \theta_{\rm j}$-$\log E_{\rm p,cpl,i}$ formula is:
\begin{equation} \label{eq:dlthejecpli}
\log  D_{\rm L} = (-0.47 \pm 0.048) \times \log \theta_{\rm j} + (0.33 \pm 0.057) \times \log E_{\rm p,cpl,i} + (-0.88 \pm 0.14),
\end{equation}
where $D_{\rm L}$ is in unit of $\rm 10^{\rm 28} ~ cm$. $\theta_{\rm j}$ is in unit of $\rm rad$. $E_{\rm p,cpl,i}$ is in unit of $\rm keV$. The adjusted $R^{2}$ is 0.299. The GRB sample number is 63.

The $\log  D_{\rm L}$-$\log \theta_{\rm j}$-$\log t_{\rm burst,i}$ formula is:
\begin{equation} \label{eq:dlthejtti}
\log  D_{\rm L} = (-0.38 \pm 0.044) \times \log \theta_{\rm j} + (-0.16 \pm 0.011) \times \log t_{\rm burst,i} + (0.54 \pm 0.073),
\end{equation}
where $D_{\rm L}$ is in unit of $\rm 10^{\rm 28} ~ cm$. $\theta_{\rm j}$ is in unit of $\rm rad$. $t_{\rm burst,i}$ is in unit of $\rm s$. The adjusted $R^{2}$ is 0.2289. The GRB sample number is 63.

The $\log Mass$-$\log  D_{\rm L}$-metallicity formula is:
\begin{equation} \label{eq:massdlmet}
\log Mass = (0.61 \pm 0.09) \times \log  D_{\rm L} + (1.1 \pm 0.17) \times metallicity + (-0.21 \pm 1.4),
\end{equation}
where Mass is in unit of $M_{\bigodot}$. $D_{\rm L}$ is in unit of $\rm 10^{\rm 28} ~ cm$. Metallicity is the value of $12+\log O/H$. The adjusted $R^{2}$ is 0.4144. The GRB sample number is 64.

The Mag-$\log  F_{\rm Opt11hr}$-$\log Mass$ formula is:
\begin{equation} \label{eq:magfo11mass}
Mag = (0.29 \pm 0.32) \times \log F_{\rm Opt11hr} + (-1.7 \pm 0.43) \times \log Mass + (-3.2 \pm 3.8),
\end{equation}
where Mag is in unit of magnitude. $F_{\rm Opt11hr}$ is in unit of $\rm Jy$. Mass is in unit of $M_{\bigodot}$. The adjusted $R^{2}$ is 0.8621. The GRB sample number is 64.

The $\log (1+z)$-$\log \theta_{\rm j}$-$\log  N_{\rm H}$ formula is:
\begin{equation} \label{eq:zthejnh}
\log (1+z) = (-0.21 \pm 0.021) \times \log \theta_{\rm j} + (0.14 \pm 0.014) \times \log  N_{\rm H} + (0.066 \pm 0.026),
\end{equation}
where $\theta_{\rm j}$ is in unit of $\rm rad$. $N_{\rm H}$ is in unit of $\rm 10^{\rm 21} ~ cm^{\rm -2}$. The adjusted $R^{2}$ is 0.3328. The GRB sample number is 64.

The $\log  D_{\rm L}$-$\log \theta_{\rm j}$-$\log  N_{\rm H}$ formula is:
\begin{equation} \label{eq:dlthejnh}
\log  D_{\rm L} = (-0.58 \pm 0.052) \times \log \theta_{\rm j} + (0.32 \pm 0.035) \times \log  N_{\rm H} + (-0.45 \pm 0.064),
\end{equation}
where $D_{\rm L}$ is in unit of $\rm 10^{\rm 28} ~ cm$. $\theta_{\rm j}$ is in unit of $\rm rad$. $N_{\rm H}$ is in unit of $\rm 10^{\rm 21} ~ cm^{\rm -2}$. The adjusted $R^{2}$ is 0.3219. The GRB sample number is 64.

The $\log Mass$-$\log (1+z)$-metallicity formula is:
\begin{equation} \label{eq:masszmet}
\log Mass = (2.3 \pm 0.46) \times \log (1+z) + (1.1 \pm 0.16) \times metallicity + (-0.52 \pm 1.4),
\end{equation}
where Mass is in unit of $M_{\bigodot}$. Metallicity is the value of $12+\log O/H$. The adjusted $R^{2}$ is 0.4282. The GRB sample number is 64.

The Mag-$\log (1+z)$-$\log  F_{\rm Opt11hr}$ formula is:
\begin{equation} \label{eq:zfoptmag}
Mag = (-3.3 \pm 1.6) \times \log (1+z) + (0.54 \pm 0.31) \times \log F_{\rm Opt11hr} + (-17 \pm 1.5),
\end{equation}
where Mag is in unit of magnitude. $F_{\rm Opt11hr}$ is in unit of $\rm Jy$. The adjusted $R^{2}$ is 0.3878. The GRB sample number is 64.

The $\log F_{\rm g}$-$(-\beta_{\rm band})$-$\log F_{\rm X11hr}$ formula is:
\begin{equation} \label{eq:fgbbandfx11}
\log F_{\rm g} = (0.11 \pm 0.15) \times (-\beta_{\rm band}) + (0.64 \pm 0.046) \times \log F_{\rm X11hr} + (5.2 \pm 0.59),
\end{equation}
where $F_{\rm g}$ is in unit of $\rm 10^{\rm -6} ~ ergs ~ cm^{\rm -2}$, and in 20-2000 $\rm keV$ energy band. $F_{\rm X11hr}$ is in unit of $\rm Jy$. The adjusted $R^{2}$ is 0.4304. The GRB sample number is 65.

The $\log F_{\rm g}$-$\log  E_{\rm p,band}$-$\log F_{\rm X11hr}$ formula is:
\begin{equation} \label{eq:fgebandfx11}
\log F_{\rm g} = (1.1 \pm 0.079) \times \log  E_{\rm p,band} + (0.47 \pm 0.043) \times \log F_{\rm X11hr} + (2 \pm 0.39),
\end{equation}
where $F_{\rm g}$ is in unit of $\rm 10^{\rm -6} ~ ergs ~ cm^{\rm -2}$, and in 20-2000 $\rm keV$ energy band. $E_{\rm p,band}$ is in unit of $\rm keV$. $F_{\rm X11hr}$ is in unit of $\rm Jy$. The adjusted $R^{2}$ is 0.8306. The GRB sample number is 65.

The $\log F_{\rm pk1}$-$\log  E_{\rm p,band}$-$\log F_{\rm X11hr}$ formula is:
\begin{equation} \label{eq:fpk1ebandfx11}
\log F_{\rm pk1} = (1.1 \pm 0.088) \times \log  E_{\rm p,band} + (0.37 \pm 0.045) \times \log F_{\rm X11hr} + (0.37 \pm 0.38),
\end{equation}
where $F_{\rm pk1}$ is peak energy flux of 1 $\rm s$ time bin in rest-frame 1-$10^{4}$ $\rm keV$ energy band, and in unit of $\rm 10^{\rm -6} ~ ergs ~ cm^{\rm -2} ~ s^{\rm -1}$. $E_{\rm p,band}$ is in unit of $\rm keV$. $F_{\rm X11hr}$ is in unit of $\rm Jy$. The adjusted $R^{2}$ is 0.7179. The GRB sample number is 66.

The $\log E_{\rm p,band,i}$-$\log F_{\rm g}$-$\log \theta_{\rm j}$ formula is:
\begin{equation} \label{eq:ebandifgthej}
\log E_{\rm p,band,i} = (0.22 \pm 0.019) \times \log F_{\rm g} + (-0.29 \pm 0.066) \times \log \theta_{\rm j} + (2 \pm 0.085),
\end{equation}
where $E_{\rm p,band,i}$ is in unit of $\rm keV$. $F_{\rm g}$ is in unit of $\rm 10^{\rm -6} ~ ergs ~ cm^{\rm -2}$, and in 20-2000 $\rm keV$ energy band. $\theta_{\rm j}$ is in unit of $\rm rad$. The adjusted $R^{2}$ is 0.223. The GRB sample number is 66.

The $\log P_{\rm pk4}$-$\log  E_{\rm p,band}$-$\log  F_{\rm Opt11hr}$ formula is:
\begin{equation} \label{eq:ppk4ebandfo11}
\log P_{\rm pk4} = (0.58 \pm 0.1) \times \log  E_{\rm p,band} + (0.27 \pm 0.055) \times \log F_{\rm Opt11hr} + (0.83 \pm 0.37),
\end{equation}
where $P_{\rm pk4}$ is peak photon flux of 1 $\rm s$ time bin in 10-1000 $\rm keV$, and in unit of $\rm photons ~ cm^{\rm -2} ~ s^{\rm -1}$. $E_{\rm p,band}$ is in unit of $\rm keV$. $F_{\rm Opt11hr}$ is in unit of $\rm Jy$. The adjusted $R^{2}$ is 0.4235. The GRB sample number is 67.

The $\log Age$-$\log (1+z)$-$\log HR$ formula is:
\begin{equation} \label{eq:agezhr}
\log Age = (-2.1 \pm 0.45) \times \log (1+z) + (-0.36 \pm 0.12) \times \log HR + (3.4 \pm 0.13),
\end{equation}
where Age is in unit of $\rm Myr$. The adjusted $R^{2}$ is 0.3026. The GRB sample number is 67.

The $\log SSFR$-$\log Age$-$\log Mass$ formula is:
\begin{equation} \label{eq:ssfragemass}
\log SSFR = (-0.39 \pm 0.059) \times \log Age + (-0.29 \pm 0.061) \times \log Mass + (3.8 \pm 0.61),
\end{equation}
where $\log SSFR$ is in unit of $\rm Gyr^{\rm -1}$. Age is in unit of $\rm Myr$. Mass is in unit of $M_{\bigodot}$. The adjusted $R^{2}$ is 0.2948. The GRB sample number is 68.

The $\log SSFR$-$\log E_{\rm iso}$-$\log Age$ formula is:
\begin{equation} \label{eq:ssfreisoage}
\log SSFR = (0.15 \pm 0.013) \times \log E_{\rm iso} + (-0.33 \pm 0.051) \times \log Age + (0.88 \pm 0.14),
\end{equation}
where $\log SSFR$ is in unit of $\rm Gyr^{\rm -1}$. $E_{\rm iso}$ is in unit of $\rm 10^{\rm 52} ~ ergs$ and in rest-frame 1-$10^{4}$ $\rm keV$ energy band. Age is in unit of $\rm Myr$. The adjusted $R^{2}$ is 0.2254. The GRB sample number is 68.

The $\log E_{\rm p,band,i}$-$\log F_{\rm g}$-$\log Mass$ formula is:
\begin{equation} \label{eq:ebandifgmass}
\log E_{\rm p,band,i} = (0.19 \pm 0.023) \times \log F_{\rm g} + (0.14 \pm 0.028) \times \log Mass + (0.95 \pm 0.28),
\end{equation}
where $E_{\rm p,band,i}$ is in unit of $\rm keV$. $F_{\rm g}$ is in unit of $\rm 10^{\rm -6} ~ ergs ~ cm^{\rm -2}$, and in 20-2000 $\rm keV$ energy band. Mass is in unit of $M_{\bigodot}$. The adjusted $R^{2}$ is 0.2071. The GRB sample number is 68.

The $\log SFR$-$\log  D_{\rm L}$-$\log Mass$ formula is:
\begin{equation} \label{eq:sfrdlmass}
\log SFR = (0.76 \pm 0.076) \times \log  D_{\rm L} + (0.53 \pm 0.056) \times \log Mass + (-4.5 \pm 0.52),
\end{equation}
where SFR is in unit of $\rm M_{\bigodot} ~ yr^{\rm -1}$. $D_{\rm L}$ is in unit of $\rm 10^{\rm 28} ~ cm$. Mass is in unit of $M_{\bigodot}$. The adjusted $R^{2}$ is 0.6629. The GRB sample number is 69.

The $\log T_{\rm 90}$-$\log \theta_{\rm j}$-$\log E_{\rm p,band,i}$ formula is:
\begin{equation} \label{eq:t90thejebandi}
\log T_{\rm 90} = (-0.43 \pm 0.057) \times \log \theta_{\rm j} + (0.34 \pm 0.036) \times \log E_{\rm p,band,i} + (0.21 \pm 0.082),
\end{equation}
where $T_{\rm 90}$ is in unit of $\rm s$. $\theta_{\rm j}$ is in unit of $\rm rad$. $E_{\rm p,band,i}$ is in unit of $\rm keV$. The adjusted $R^{2}$ is 0.2107. The GRB sample number is 69.

The $\log SFR$-$\log (1+z)$-$\log Mass$ formula is:
\begin{equation} \label{eq:sfrzmass}
\log SFR = (2.9 \pm 0.28) \times \log (1+z) + (0.46 \pm 0.058) \times \log Mass + (-4.6 \pm 0.5),
\end{equation}
where SFR is in unit of $\rm M_{\bigodot} ~ yr^{\rm -1}$. Mass is in unit of $M_{\bigodot}$. The adjusted $R^{2}$ is 0.687. The GRB sample number is 69.

The Mag-$\log (1+z)$-$A_{\rm V}$ formula is:
\begin{equation} \label{eq:magzav}
Mag = (-4.2 \pm 1.6) \times \log (1+z) + (-0.52 \pm 0.32) \times A_{\rm V} + (-19 \pm 0.79),
\end{equation}
where Mag is in unit of magnitude. The adjusted $R^{2}$ is 0.408. The GRB sample number is 69.

The $\log t_{\rm burst,i}$-$\log L_{\rm pk}$-$\log  F_{\rm Opt11hr}$ formula is:
\begin{equation} \label{eq:ttilpkfo11}
\log t_{\rm burst,i} = (-0.18 \pm 0.021) \times \log L_{\rm pk} + (0.21 \pm 0.047) \times \log F_{\rm Opt11hr} + (3.4 \pm 0.24),
\end{equation}
where $t_{\rm burst,i}$ is in unit of $\rm s$. $L_{\rm pk}$ is in unit of $\rm 10^{\rm 52} ~ erg ~ s^{\rm -1}$, and in 1-$10^{4}$ $\rm keV$ energy band. $F_{\rm Opt11hr}$ is in unit of $\rm Jy$. The adjusted $R^{2}$ is 0.2132. The GRB sample number is 70.

The $\log F_{\rm X11hr}$-$\log F_{\rm g}$-$\log \theta_{\rm j}$ formula is:
\begin{equation} \label{eq:fx11fgthej}
\log F_{\rm X11hr} = (0.43 \pm 0.039) \times \log F_{\rm g} + (0.51 \pm 0.061) \times \log \theta_{\rm j} + (-6.9 \pm 0.089),
\end{equation}
where $F_{\rm X11hr}$ is in unit of $\rm Jy$. $F_{\rm g}$ is in unit of $\rm 10^{\rm -6} ~ ergs ~ cm^{\rm -2}$, and in 20-2000 $\rm keV$ energy band. $\theta_{\rm j}$ is in unit of $\rm rad$. The adjusted $R^{2}$ is 0.3077. The GRB sample number is 71.

The $\log Mass$-$\log P_{\rm pk4}$-Mag formula is:
\begin{equation} \label{eq:massppk4mag}
\log Mass = (-0.18 \pm 0.09) \times \log P_{\rm pk4} + (-0.14 \pm 0.023) \times Mag + (6.8 \pm 0.5),
\end{equation}
where Mass is in unit of $M_{\bigodot}$. $P_{\rm pk4}$ is peak photon flux of 1 $\rm s$ time bin in 10-1000 $\rm keV$, and in unit of $\rm photons ~ cm^{\rm -2} ~ s^{\rm -1}$. Mag is in unit of magnitude. The adjusted $R^{2}$ is 0.8775. The GRB sample number is 72.

The $\log F_{\rm pk2}$-$(-\beta_{\rm band})$-$\log E_{\rm p,band,i}$ formula is:
\begin{equation} \label{eq:fpk2bbandebandi}
\log F_{\rm pk2} = (0.46 \pm 0.12) \times (-\beta_{\rm band}) + (0.38 \pm 0.062) \times \log E_{\rm p,band,i} + (-1.5 \pm 0.27),
\end{equation}
where $F_{\rm pk2}$ is peak energy flux of 64 $\rm ms$ time bin in rest-frame 1-$10^{4}$ $\rm keV$ energy band, and in unit of $\rm 10^{\rm -6} ~ ergs ~ cm^{\rm -2} ~ s^{\rm -1}$. $E_{\rm p,band,i}$ is in unit of $\rm keV$. The adjusted $R^{2}$ is 0.2559. The GRB sample number is 73.

The $\log Mass$-$\log HR$-Mag formula is:
\begin{equation} \label{eq:masshrmag}
\log Mass = (-0.33 \pm 0.14) \times \log HR + (-0.12 \pm 0.025) \times Mag + (7.1 \pm 0.54),
\end{equation}
where Mass is in unit of $M_{\bigodot}$. Mag is in unit of magnitude. The adjusted $R^{2}$ is 0.8604. The GRB sample number is 73.

The $\log E_{\rm iso}$-$\log P_{\rm pk1}$-$\log E_{\rm p,band,i}$ formula is:
\begin{equation} \label{eq:eisoppk1ebandi}
\log E_{\rm iso} = (0.42 \pm 0.051) \times \log P_{\rm pk1} + (0.92 \pm 0.065) \times \log E_{\rm p,band,i}   + (-1.9 \pm 0.16),
\end{equation}
where $E_{\rm iso}$ is in unit of $\rm 10^{\rm 52} ~ ergs$ and in rest-frame 1-$10^{4}$ $\rm keV$ energy band. $P_{\rm pk1}$ is peak photon flux of 64 $\rm ms$ time bin in 10-1000 $\rm keV$, and in unit of $\rm photons ~ cm^{\rm -2} ~ s^{\rm -1}$. $E_{\rm p,band,i}$ is in unit of $\rm keV$. The adjusted $R^{2}$ is 0.2987. The GRB sample number is 73.

The $\log F_{\rm pk2}$-$\log (1+z)$-$(-\beta_{\rm band})$ formula is:
\begin{equation} \label{eq:fpk2zbband}
\log F_{\rm pk2} = (-0.93 \pm 0.14) \times \log (1+z) + (0.5 \pm 0.13) \times (-\beta_{\rm band}) + (-0.18 \pm 0.31),
\end{equation}
where $F_{\rm pk2}$ is peak energy flux of 64 $\rm ms$ time bin in rest-frame 1-$10^{4}$ $\rm keV$ energy band, and in unit of $\rm 10^{\rm -6} ~ ergs ~ cm^{\rm -2} ~ s^{\rm -1}$. The adjusted $R^{2}$ is 0.2389. The GRB sample number is 73.

The $\log Mass$-$(-\beta_{\rm band})$-$\log E_{\rm p,band,i}$ formula is:
\begin{equation} \label{eq:massbbandebandi}
\log Mass = (-0.16 \pm 0.12) \times (-\beta_{\rm band}) + (0.65 \pm 0.097) \times \log E_{\rm p,band,i} + (8.2 \pm 0.36),
\end{equation}
where Mass is in unit of $M_{\bigodot}$. $E_{\rm p,band,i}$ is in unit of $\rm keV$. The adjusted $R^{2}$ is 0.2063. The GRB sample number is 74.

The $\log E_{\rm p,band,i}$-$\log  D_{\rm L}$-$\log Mass$ formula is:
\begin{equation} \label{eq:ebandidlmass}
\log E_{\rm p,band,i} = (0.35 \pm 0.039) \times \log  D_{\rm L} + (0.13 \pm 0.03) \times \log Mass + (1.1 \pm 0.28),
\end{equation}
where $E_{\rm p,band,i}$ is in unit of $\rm keV$. $D_{\rm L}$ is in unit of $\rm 10^{\rm 28} ~ cm$. Mass is in unit of $M_{\bigodot}$. The adjusted $R^{2}$ is 0.2109. The GRB sample number is 74.

The $\log T_{\rm 50,i}$-$(-\alpha_{\rm cpl})$-$\log t_{\rm burst,i}$ formula is:
\begin{equation} \label{eq:t50iacpltti}
\log T_{\rm 50,i} = (0.22 \pm 0.072) \times (-\alpha_{\rm cpl}) + (0.33 \pm 0.02) \times \log t_{\rm burst,i} + (-0.16 \pm 0.11),
\end{equation}
where $T_{\rm 50,i}$ is in unit of $\rm s$. $t_{\rm burst,i}$ is in unit of $\rm s$. The adjusted $R^{2}$ is 0.2056. The GRB sample number is 75.

The $\log F_{\rm pk1}$-$\log HR$-$\log SFR$ formula is:
\begin{equation} \label{eq:fpk1hrsfr}
\log F_{\rm pk1} = (0.93 \pm 0.11) \times \log HR + (-0.17 \pm 0.035) \times \log SFR + (-0.059 \pm 0.041),
\end{equation}
where $F_{\rm pk1}$ is peak energy flux of 1 $\rm s$ time bin in rest-frame 1-$10^{4}$ $\rm keV$ energy band, and in unit of $\rm 10^{\rm -6} ~ ergs ~ cm^{\rm -2} ~ s^{\rm -1}$. SFR is in unit of $\rm M_{\bigodot} ~ yr^{\rm -1}$. The adjusted $R^{2}$ is 0.2425. The GRB sample number is 76.

The $\log L_{\rm pk}$-$\log \theta_{\rm j}$-$\log Mass$ formula is:
\begin{equation} \label{eq:lpkthejmass}
\log L_{\rm pk} = (-1.5 \pm 0.1) \times \log \theta_{\rm j} + (0.32 \pm 0.06) \times \log Mass + (-4.9 \pm 0.56),
\end{equation}
where $L_{\rm pk}$ is in unit of $\rm 10^{\rm 52} ~ erg ~ s^{\rm -1}$, and in 1-$10^{4}$ $\rm keV$ energy band. $\theta_{\rm j}$ is in unit of $\rm rad$. Mass is in unit of $M_{\bigodot}$. The adjusted $R^{2}$ is 0.2767. The GRB sample number is 76.

The $\log \theta_{\rm j}$-$\log (1+z)$-$\log F_{\rm X11hr}$ formula is:
\begin{equation} \label{eq:thejzfx11}
\log \theta_{\rm j} = (-0.84 \pm 0.073) \times \log (1+z) + (0.13 \pm 0.021) \times \log F_{\rm X11hr} + (0.073 \pm 0.14),
\end{equation}
where $\theta_{\rm j}$ is in unit of $\rm rad$. $F_{\rm X11hr}$ is in unit of $\rm Jy$. The adjusted $R^{2}$ is 0.2238. The GRB sample number is 76.

The $\log  D_{\rm L}$-$\log F_{\rm g}$-$\log Age$ formula is:
\begin{equation} \label{eq:dlfgage}
\log  D_{\rm L} = (0.12 \pm 0.01) \times \log F_{\rm g} + (-0.29 \pm 0.036) \times \log Age + (0.77 \pm 0.1),
\end{equation}
where $D_{\rm L}$ is in unit of $\rm 10^{\rm 28} ~ cm$. $F_{\rm g}$ is in unit of $\rm 10^{\rm -6} ~ ergs ~ cm^{\rm -2}$, and in 20-2000 $\rm keV$ energy band. Age is in unit of $\rm Myr$. The adjusted $R^{2}$ is 0.3209. The GRB sample number is 77.

The $\log F_{\rm g}$-$\log P_{\rm pk1}$-$A_{\rm V}$ formula is:
\begin{equation} \label{eq:fgppk1av}
\log F_{\rm g} = (1.1 \pm 0.04) \times \log P_{\rm pk1} + (-0.2 \pm 0.08) \times A_{\rm V} + (-0.08 \pm 0.083),
\end{equation}
where $F_{\rm g}$ is in unit of $\rm 10^{\rm -6} ~ ergs ~ cm^{\rm -2}$, and in 20-2000 $\rm keV$ energy band. $P_{\rm pk1}$ is peak photon flux of 64 $\rm ms$ time bin in 10-1000 $\rm keV$, and in unit of $\rm photons ~ cm^{\rm -2} ~ s^{\rm -1}$. The adjusted $R^{2}$ is 0.4258. The GRB sample number is 77.

The $\log P_{\rm pk3}$-$\log (1+z)$-$(-\beta_{\rm band})$ formula is:
\begin{equation} \label{eq:ppk3zbband}
\log P_{\rm pk3} = (-0.82 \pm 0.067) \times \log (1+z) + (0.33 \pm 0.14) \times (-\beta_{\rm band}) + (0.64 \pm 0.33),
\end{equation}
where $P_{\rm pk3}$ is peak photon flux of 1024 $\rm ms$ time bin in 10-1000 $\rm keV$, and in unit of $\rm photons ~ cm^{\rm -2} ~ s^{\rm -1}$. The adjusted $R^{2}$ is 0.2232. The GRB sample number is 77.

The $\log T_{\rm R45,i}$-$\log F_{\rm g}$-$\log t_{\rm pkOpt}$ formula is:
\begin{equation} \label{eq:tr45ifgtopt}
\log T_{\rm R45,i} = (0.35 \pm 0.013) \times \log F_{\rm g} + (0.38 \pm 0.014) \times \log t_{\rm pkOpt} + (-0.76 \pm 0.043),
\end{equation}
where $T_{\rm R45,i}$ is in unit of $\rm s$. $F_{\rm g}$ is in unit of $\rm 10^{\rm -6} ~ ergs ~ cm^{\rm -2}$, and in 20-2000 $\rm keV$ energy band. $t_{\rm pkOpt}$ is in unit of $\rm s$. The adjusted $R^{2}$ is 0.3476. The GRB sample number is 78.

The $\log P_{\rm pk4}$-$\log F_{\rm g}$-$\log SFR$ formula is:
\begin{equation} \label{eq:ppk4fgsfr}
\log P_{\rm pk4} = (0.45 \pm 0.022) \times \log F_{\rm g} + (-0.14 \pm 0.021) \times \log SFR + (0.45 \pm 0.028),
\end{equation}
where $P_{\rm pk4}$ is peak photon flux of 1 $\rm s$ time bin in 10-1000 $\rm keV$, and in unit of $\rm photons ~ cm^{\rm -2} ~ s^{\rm -1}$. $F_{\rm g}$ is in unit of $\rm 10^{\rm -6} ~ ergs ~ cm^{\rm -2}$, and in 20-2000 $\rm keV$ energy band. SFR is in unit of $\rm M_{\bigodot} ~ yr^{\rm -1}$. The adjusted $R^{2}$ is 0.3386. The GRB sample number is 79.

The $\log SFR$-$\log E_{\rm iso}$-$A_{\rm V}$ formula is:
\begin{equation} \label{eq:sfreisoav}
\log SFR = (0.32 \pm 0.028) \times \log E_{\rm iso} + (0.29 \pm 0.073) \times A_{\rm V} + (0.25 \pm 0.069),
\end{equation}
where SFR is in unit of $\rm M_{\bigodot} ~ yr^{\rm -1}$. $E_{\rm iso}$ is in unit of $\rm 10^{\rm 52} ~ ergs$ and in rest-frame 1-$10^{4}$ $\rm keV$ energy band. The adjusted $R^{2}$ is 0.3195. The GRB sample number is 81.

The $\log P_{\rm pk4}$-$\log L_{\rm pk}$-$(-\beta_{\rm band})$ formula is:
\begin{equation} \label{eq:ppk4lpkbband}
\log P_{\rm pk4} = (0.28 \pm 0.04) \times \log L_{\rm pk} + (0.18 \pm 0.14) \times (-\beta_{\rm band}) + (0.39 \pm 0.32),
\end{equation}
where $P_{\rm pk4}$ is peak photon flux of 1 $\rm s$ time bin in 10-1000 $\rm keV$, and in unit of $\rm photons ~ cm^{\rm -2} ~ s^{\rm -1}$. $L_{\rm pk}$ is in unit of $\rm 10^{\rm 52} ~ erg ~ s^{\rm -1}$, and in 1-$10^{4}$ $\rm keV$ energy band. The adjusted $R^{2}$ is 0.2003. The GRB sample number is 81.

The $\log  D_{\rm L}$-$\log \theta_{\rm j}$-$\log Mass$ formula is:
\begin{equation} \label{eq:dlthejmass}
\log  D_{\rm L} = (-0.54 \pm 0.034) \times \log \theta_{\rm j} + (0.23 \pm 0.028) \times \log Mass + (-2.4 \pm 0.26),
\end{equation}
where $D_{\rm L}$ is in unit of $\rm 10^{\rm 28} ~ cm$. $\theta_{\rm j}$ is in unit of $\rm rad$. Mass is in unit of $M_{\bigodot}$. The adjusted $R^{2}$ is 0.3388. The GRB sample number is 81.

The $\log  F_{\rm Opt11hr}$-$\beta_{\rm X11hr}$-$\log Mass$ formula is:
\begin{equation} \label{eq:fo11bx11mass}
\log F_{\rm Opt11hr} = (-0.3 \pm 0.098) \times \beta_{\rm X11hr} + (-0.34 \pm 0.074) \times \log Mass + (-1.5 \pm 0.7),
\end{equation}
where $F_{\rm Opt11hr}$ is in unit of $\rm Jy$. Mass is in unit of $M_{\bigodot}$. The adjusted $R^{2}$ is 0.252. The GRB sample number is 82.

The $\log Mass$-$\log T_{\rm 90}$-Mag formula is:
\begin{equation} \label{eq:masst90mag}
\log Mass = (0.49 \pm 0.094) \times \log T_{\rm 90} + (-0.12 \pm 0.022) \times Mag + (6.2 \pm 0.45),
\end{equation}
where Mass is in unit of $M_{\bigodot}$. $T_{\rm 90}$ is in unit of $\rm s$. Mag is in unit of magnitude. The adjusted $R^{2}$ is 0.8661. The GRB sample number is 82.

The $\log t_{\rm burst,i}$-$\log P_{\rm pk4}$-$\log  F_{\rm Opt11hr}$ formula is:
\begin{equation} \label{eq:ttippk4fo11}
\log t_{\rm burst,i} = (-0.29 \pm 0.055) \times \log P_{\rm pk4} + (0.27 \pm 0.043) \times \log F_{\rm Opt11hr} + (3.8 \pm 0.22),
\end{equation}
where $t_{\rm burst,i}$ is in unit of $\rm s$. $P_{\rm pk4}$ is peak photon flux of 1 $\rm s$ time bin in 10-1000 $\rm keV$, and in unit of $\rm photons ~ cm^{\rm -2} ~ s^{\rm -1}$. $F_{\rm Opt11hr}$ is in unit of $\rm Jy$. The adjusted $R^{2}$ is 0.2451. The GRB sample number is 83.

The $\log P_{\rm pk4}$-$\log  D_{\rm L}$-$\log t_{\rm pkOpt}$ formula is:
\begin{equation} \label{eq:ppk4dltopt}
\log P_{\rm pk4} = (-0.6 \pm 0.031) \times \log  D_{\rm L} + (-0.48 \pm 0.021) \times \log t_{\rm pkOpt} + (2.2 \pm 0.052),
\end{equation}
where $P_{\rm pk4}$ is peak photon flux of 1 $\rm s$ time bin in 10-1000 $\rm keV$, and in unit of $\rm photons ~ cm^{\rm -2} ~ s^{\rm -1}$. $D_{\rm L}$ is in unit of $\rm 10^{\rm 28} ~ cm$. $t_{\rm pkOpt}$ is in unit of $\rm s$. The adjusted $R^{2}$ is 0.38. The GRB sample number is 84.

The $\log  F_{\rm Opt11hr}$-$\log F_{\rm X11hr}$-$\log Mass$ formula is:
\begin{equation} \label{eq:fo11fx11mass}
\log F_{\rm Opt11hr} = (0.22 \pm 0.079) \times \log F_{\rm X11hr} + (-0.38 \pm 0.069) \times \log Mass + (0.0033 \pm 0.84),
\end{equation}
where $F_{\rm Opt11hr}$ is in unit of $\rm Jy$. $F_{\rm X11hr}$ is in unit of $\rm Jy$. Mass is in unit of $M_{\bigodot}$. The adjusted $R^{2}$ is 0.2405. The GRB sample number is 88.

The $\log SFR$-$\log  D_{\rm L}$-$A_{\rm V}$ formula is:
\begin{equation} \label{eq:sfrdlav}
\log SFR = (1.1 \pm 0.055) \times \log  D_{\rm L} + (0.24 \pm 0.061) \times A_{\rm V} + (0.24 \pm 0.057),
\end{equation}
where SFR is in unit of $\rm M_{\bigodot} ~ yr^{\rm -1}$. $D_{\rm L}$ is in unit of $\rm 10^{\rm 28} ~ cm$. The adjusted $R^{2}$ is 0.4998. The GRB sample number is 90.

The $\log SFR$-$\log (1+z)$-$A_{\rm V}$ formula is:
\begin{equation} \label{eq:sfrzav}
\log SFR = (3.8 \pm 0.2) \times \log (1+z) + (0.2 \pm 0.056) \times A_{\rm V} + (-0.66 \pm 0.066),
\end{equation}
where SFR is in unit of $\rm M_{\bigodot} ~ yr^{\rm -1}$. The adjusted $R^{2}$ is 0.5415. The GRB sample number is 90.

The $\log HR$-$\log T_{\rm 50}$-$\log E_{\rm p,band,i}$ formula is:
\begin{equation} \label{eq:hrt50ebandi}
\log HR = (-0.11 \pm 0.02) \times \log T_{\rm 50} + (0.33 \pm 0.032) \times \log E_{\rm p,band,i} + (-0.2 \pm 0.097),
\end{equation}
where $T_{\rm 50}$ is in unit of $\rm s$. $E_{\rm p,band,i}$ is in unit of $\rm keV$. The adjusted $R^{2}$ is 0.243. The GRB sample number is 96.

The $\log F_{\rm g}$-$\beta_{\rm X11hr}$-$\log Mass$ formula is:
\begin{equation} \label{eq:fgbx11mass}
\log F_{\rm g} = (-0.53 \pm 0.077) \times \beta_{\rm X11hr} + (-0.19 \pm 0.049) \times \log Mass + (3.4 \pm 0.43),
\end{equation}
where $F_{\rm g}$ is in unit of $\rm 10^{\rm -6} ~ ergs ~ cm^{\rm -2}$, and in 20-2000 $\rm keV$ energy band. Mass is in unit of $M_{\bigodot}$. The adjusted $R^{2}$ is 0.2965. The GRB sample number is 97.

The $\log E_{\rm iso}$-$\log F_{\rm X11hr}$-$\log E_{\rm p,cpl,i}$ formula is:
\begin{equation} \label{eq:eisofx11ecpli}
\log E_{\rm iso} = (0.55 \pm 0.05) \times \log F_{\rm X11hr} + (0.9 \pm 0.12) \times \log E_{\rm p,cpl,i} + (2.1 \pm 0.52),
\end{equation}
where $E_{\rm iso}$ is in unit of $\rm 10^{\rm 52} ~ ergs$ and in rest-frame 1-$10^{4}$ $\rm keV$ energy band. $F_{\rm X11hr}$ is in unit of $\rm Jy$. $E_{\rm p,cpl,i}$ is in unit of $\rm keV$. The adjusted $R^{2}$ is 0.386. The GRB sample number is 97.

The $\log P_{\rm pk4}$-$\log F_{\rm g}$-$\log  N_{\rm H}$ formula is:
\begin{equation} \label{eq:ppk4fgnh}
\log P_{\rm pk4} = (0.47 \pm 0.019) \times \log F_{\rm g} + (-0.18 \pm 0.037) \times \log  N_{\rm H} + (0.39 \pm 0.033),
\end{equation}
where $P_{\rm pk4}$ is peak photon flux of 1 $\rm s$ time bin in 10-1000 $\rm keV$, and in unit of $\rm photons ~ cm^{\rm -2} ~ s^{\rm -1}$. $F_{\rm g}$ is in unit of $\rm 10^{\rm -6} ~ ergs ~ cm^{\rm -2}$, and in 20-2000 $\rm keV$ energy band. $N_{\rm H}$ is in unit of $\rm 10^{\rm 21} ~ cm^{\rm -2}$. The adjusted $R^{2}$ is 0.4773. The GRB sample number is 104.

The $\log Mass$-$\log  F_{\rm Opt11hr}$-$A_{\rm V}$ formula is:
\begin{equation} \label{eq:massfo11av}
\log Mass = (-0.31 \pm 0.049) \times \log F_{\rm Opt11hr} + (0.14 \pm 0.047) \times A_{\rm V} + (7.9 \pm 0.25),
\end{equation}
where Mass is in unit of $M_{\bigodot}$. $F_{\rm Opt11hr}$ is in unit of $\rm Jy$. The adjusted $R^{2}$ is 0.2312. The GRB sample number is 104.

The $\log F_{\rm g}$-$\log L_{\rm pk}$-$\log  N_{\rm H}$ formula is:
\begin{equation} \label{eq:fglpknh}
\log F_{\rm g} = (0.38 \pm 0.017) \times \log L_{\rm pk} + (-0.26 \pm 0.05) \times \log  N_{\rm H} + (1.1 \pm 0.04),
\end{equation}
where $F_{\rm g}$ is in unit of $\rm 10^{\rm -6} ~ ergs ~ cm^{\rm -2}$, and in 20-2000 $\rm keV$ energy band. $L_{\rm pk}$ is in unit of $\rm 10^{\rm 52} ~ erg ~ s^{\rm -1}$, and in 1-$10^{4}$ $\rm keV$ energy band. $N_{\rm H}$ is in unit of $\rm 10^{\rm 21} ~ cm^{\rm -2}$. The adjusted $R^{2}$ is 0.2658. The GRB sample number is 104.

The $\log  F_{\rm Opt11hr}$-$\log Mass$-$\log T_{\rm 50,i}$ formula is:
\begin{equation} \label{eq:fo11masst50i}
\log F_{\rm Opt11hr} = (-0.37 \pm 0.063) \times \log Mass + (0.42 \pm 0.077) \times \log T_{\rm 50,i} + (-1.9 \pm 0.61),
\end{equation}
where $F_{\rm Opt11hr}$ is in unit of $\rm Jy$. Mass is in unit of $M_{\bigodot}$. $T_{\rm 50,i}$ is in unit of $\rm s$. The adjusted $R^{2}$ is 0.2729. The GRB sample number is 105.

The $\log F_{\rm pk1}$-$(-\alpha_{\rm cpl})$-$\beta_{\rm X11hr}$ formula is:
\begin{equation} \label{eq:fpk1acplbx11}
\log F_{\rm pk1} = (-0.37 \pm 0.11) \times (-\alpha_{\rm cpl}) + (-0.35 \pm 0.069) \times \beta_{\rm X11hr} + (0.89 \pm 0.16),
\end{equation}
where $F_{\rm pk1}$ is peak energy flux of 1 $\rm s$ time bin in rest-frame 1-$10^{4}$ $\rm keV$ energy band, and in unit of $\rm 10^{\rm -6} ~ ergs ~ cm^{\rm -2} ~ s^{\rm -1}$. The adjusted $R^{2}$ is 0.2512. The GRB sample number is 108.

The $\log F_{\rm g}$-$\log  E_{\rm p,band}$-$\log  F_{\rm Opt11hr}$ formula is:
\begin{equation} \label{eq:fgebandfo11}
\log F_{\rm g} = (1.3 \pm 0.085) \times \log  E_{\rm p,band} + (0.21 \pm 0.041) \times \log F_{\rm Opt11hr} + (-0.71 \pm 0.29),
\end{equation}
where $F_{\rm g}$ is in unit of $\rm 10^{\rm -6} ~ ergs ~ cm^{\rm -2}$, and in 20-2000 $\rm keV$ energy band. $E_{\rm p,band}$ is in unit of $\rm keV$. $F_{\rm Opt11hr}$ is in unit of $\rm Jy$. The adjusted $R^{2}$ is 0.6036. The GRB sample number is 108.

The $\log F_{\rm pk1}$-$\log E_{\rm p,cpl}$-$\beta_{\rm X11hr}$ formula is:
\begin{equation} \label{eq:fpk1ecplbx11}
\log F_{\rm pk1} = (0.83 \pm 0.1) \times \log E_{\rm p,cpl} + (-0.36 \pm 0.065) \times \beta_{\rm X11hr} + (-1.3 \pm 0.26),
\end{equation}
where $F_{\rm pk1}$ is peak energy flux of 1 $\rm s$ time bin in rest-frame 1-$10^{4}$ $\rm keV$ energy band, and in unit of $\rm 10^{\rm -6} ~ ergs ~ cm^{\rm -2} ~ s^{\rm -1}$. $E_{\rm p,cpl}$ is in unit of $\rm keV$. The adjusted $R^{2}$ is 0.4823. The GRB sample number is 108.

The $\log Mass$-$\log L_{\rm pk}$-$A_{\rm V}$ formula is:
\begin{equation} \label{eq:masslpkav}
\log Mass = (0.24 \pm 0.031) \times \log L_{\rm pk} + (0.3 \pm 0.068) \times A_{\rm V} + (9.3 \pm 0.065),
\end{equation}
where Mass is in unit of $M_{\bigodot}$. $L_{\rm pk}$ is in unit of $\rm 10^{\rm 52} ~ erg ~ s^{\rm -1}$, and in 1-$10^{4}$ $\rm keV$ energy band. The adjusted $R^{2}$ is 0.208. The GRB sample number is 110.

The $\log P_{\rm pk3}$-$\log (1+z)$-$\log F_{\rm g}$ formula is:
\begin{equation} \label{eq:ppk3zfg}
\log P_{\rm pk3} = (-0.73 \pm 0.057) \times \log (1+z) + (0.47 \pm 0.011) \times \log F_{\rm g} + (0.74 \pm 0.033),
\end{equation}
where $P_{\rm pk3}$ is peak photon flux of 1024 $\rm ms$ time bin in 10-1000 $\rm keV$, and in unit of $\rm photons ~ cm^{\rm -2} ~ s^{\rm -1}$. $F_{\rm g}$ is in unit of $\rm 10^{\rm -6} ~ ergs ~ cm^{\rm -2}$, and in 20-2000 $\rm keV$ energy band. The adjusted $R^{2}$ is 0.6368. The GRB sample number is 112.

The $\log  F_{\rm Opt11hr}$-$\log Mass$-$\log T_{\rm R45,i}$ formula is:
\begin{equation} \label{eq:fo11masstr45i}
\log F_{\rm Opt11hr} = (-0.38 \pm 0.062) \times \log Mass + (0.42 \pm 0.083) \times \log T_{\rm R45,i} + (-1.6 \pm 0.59),
\end{equation}
where $F_{\rm Opt11hr}$ is in unit of $\rm Jy$. Mass is in unit of $M_{\bigodot}$. $T_{\rm R45,i}$ is in unit of $\rm s$. The adjusted $R^{2}$ is 0.2692. The GRB sample number is 115.

The $\log F_{\rm g}$-$(-\beta_{\rm band})$-$\log T_{\rm R45,i}$ formula is:
\begin{equation} \label{eq:fgbbandtr45i}
\log F_{\rm g} = (0.19 \pm 0.11) \times (-\beta_{\rm band}) + (0.89 \pm 0.027) \times \log T_{\rm R45,i} + (0.33 \pm 0.25),
\end{equation}
where $F_{\rm g}$ is in unit of $\rm 10^{\rm -6} ~ ergs ~ cm^{\rm -2}$, and in 20-2000 $\rm keV$ energy band. $T_{\rm R45,i}$ is in unit of $\rm s$. The adjusted $R^{2}$ is 0.3425. The GRB sample number is 117.

The $\log F_{\rm g}$-$\log  E_{\rm p,band}$-$\log T_{\rm R45,i}$ formula is:
\begin{equation} \label{eq:fgebandtr45i}
\log F_{\rm g} = (1.1 \pm 0.056) \times \log  E_{\rm p,band} + (0.69 \pm 0.035) \times \log T_{\rm R45,i} + (-1.4 \pm 0.11),
\end{equation}
where $F_{\rm g}$ is in unit of $\rm 10^{\rm -6} ~ ergs ~ cm^{\rm -2}$, and in 20-2000 $\rm keV$ energy band. $E_{\rm p,band}$ is in unit of $\rm keV$. $T_{\rm R45,i}$ is in unit of $\rm s$. The adjusted $R^{2}$ is 0.6001. The GRB sample number is 117.

The $\log  E_{\rm p,band}$-$\log T_{\rm R45}$-$\log P_{\rm pk4}$ formula is:
\begin{equation} \label{eq:ebandtr45ppk4}
\log  E_{\rm p,band} = (0.19 \pm 0.028) \times \log T_{\rm R45} + (0.31 \pm 0.043) \times \log P_{\rm pk4} + (1.6 \pm 0.047),
\end{equation}
where $E_{\rm p,band}$ is in unit of $\rm keV$. $T_{\rm R45}$ is in unit of $\rm s$. $P_{\rm pk4}$ is peak photon flux of 1 $\rm s$ time bin in 10-1000 $\rm keV$, and in unit of $\rm photons ~ cm^{\rm -2} ~ s^{\rm -1}$. The adjusted $R^{2}$ is 0.2789. The GRB sample number is 118.

The $\log  F_{\rm Opt11hr}$-$\log F_{\rm g}$-$\log Mass$ formula is:
\begin{equation} \label{eq:fo11fgmass}
\log F_{\rm Opt11hr} = (0.33 \pm 0.055) \times \log F_{\rm g} + (-0.34 \pm 0.063) \times \log Mass + (-2 \pm 0.6),
\end{equation}
where $F_{\rm Opt11hr}$ is in unit of $\rm Jy$. $F_{\rm g}$ is in unit of $\rm 10^{\rm -6} ~ ergs ~ cm^{\rm -2}$, and in 20-2000 $\rm keV$ energy band. Mass is in unit of $M_{\bigodot}$. The adjusted $R^{2}$ is 0.2536. The GRB sample number is 120.

The $\log E_{\rm iso}$-$\log F_{\rm X11hr}$-$\beta_{\rm X11hr}$ formula is:
\begin{equation} \label{eq:eisofx11bx11}
\log E_{\rm iso} = (0.55 \pm 0.045) \times \log F_{\rm X11hr} + (-0.3 \pm 0.083) \times \beta_{\rm X11hr} + (5 \pm 0.29),
\end{equation}
where $E_{\rm iso}$ is in unit of $\rm 10^{\rm 52} ~ ergs$ and in rest-frame 1-$10^{4}$ $\rm keV$ energy band. $F_{\rm X11hr}$ is in unit of $\rm Jy$. The adjusted $R^{2}$ is 0.2423. The GRB sample number is 121.

The $\beta_{\rm X11hr}$-$\log P_{\rm pk4}$-$\log T_{\rm R45,i}$ formula is:
\begin{equation} \label{eq:bx11ppk4tr45i}
\beta_{\rm X11hr} = (-0.61 \pm 0.095) \times \log P_{\rm pk4} + (-0.16 \pm 0.067) \times \log T_{\rm R45,i} + (2.1 \pm 0.076),
\end{equation}
where $P_{\rm pk4}$ is peak photon flux of 1 $\rm s$ time bin in 10-1000 $\rm keV$, and in unit of $\rm photons ~ cm^{\rm -2} ~ s^{\rm -1}$. $T_{\rm R45,i}$ is in unit of $\rm s$. The adjusted $R^{2}$ is 0.3002. The GRB sample number is 123.

The $\log T_{\rm 90}$-$\log E_{\rm iso}$-$\log  N_{\rm H}$ formula is:
\begin{equation} \label{eq:t90eisonh}
\log T_{\rm 90} = (0.31 \pm 0.012) \times \log E_{\rm iso} + (0.17 \pm 0.032) \times \log  N_{\rm H} + (1.3 \pm 0.024),
\end{equation}
where $T_{\rm 90}$ is in unit of $\rm s$. $E_{\rm iso}$ is in unit of $\rm 10^{\rm 52} ~ ergs$ and in rest-frame 1-$10^{4}$ $\rm keV$ energy band. $N_{\rm H}$ is in unit of $\rm 10^{\rm 21} ~ cm^{\rm -2}$. The adjusted $R^{2}$ is 0.2526. The GRB sample number is 125.

The $\log Mass$-$\log  D_{\rm L}$-$\log  F_{\rm Opt11hr}$ formula is:
\begin{equation} \label{eq:massdlfo11}
\log Mass = (0.48 \pm 0.066) \times \log  D_{\rm L} + (-0.22 \pm 0.049) \times \log F_{\rm Opt11hr} + (8.3 \pm 0.24),
\end{equation}
where Mass is in unit of $M_{\bigodot}$. $D_{\rm L}$ is in unit of $\rm 10^{\rm 28} ~ cm$. $F_{\rm Opt11hr}$ is in unit of $\rm Jy$. The adjusted $R^{2}$ is 0.2064. The GRB sample number is 126.

The $\log  D_{\rm L}$-$\log \theta_{\rm j}$-$A_{\rm V}$ formula is:
\begin{equation} \label{eq:dlavthej}
\log  D_{\rm L} = (-0.44 \pm 0.036) \times \log \theta_{\rm j} + (-0.13 \pm 0.029) \times A_{\rm V} + (0.034 \pm 0.049),
\end{equation}
where $D_{\rm L}$ is in unit of $\rm 10^{\rm 28} ~ cm$. $\theta_{\rm j}$ is in unit of $\rm rad$. The adjusted $R^{2}$ is 0.2105. The GRB sample number is 126.

The $\log Mass$-$\log (1+z)$-$\log  F_{\rm Opt11hr}$ formula is:
\begin{equation} \label{eq:masszfo11}
\log Mass = (1.2 \pm 0.21) \times \log (1+z) + (-0.22 \pm 0.048) \times \log F_{\rm Opt11hr} + (8 \pm 0.23),
\end{equation}
where Mass is in unit of $M_{\bigodot}$. $F_{\rm Opt11hr}$ is in unit of $\rm Jy$. The adjusted $R^{2}$ is 0.2156. The GRB sample number is 126.

The $\log F_{\rm g}$-$\log L_{\rm pk}$-$(-\beta_{\rm band})$ formula is:
\begin{equation} \label{eq:fglpkbband}
\log F_{\rm g} = (0.39 \pm 0.02) \times \log L_{\rm pk} + (0.19 \pm 0.1) \times (-\beta_{\rm band}) + (0.67 \pm 0.24),
\end{equation}
where $F_{\rm g}$ is in unit of $\rm 10^{\rm -6} ~ ergs ~ cm^{\rm -2}$, and in 20-2000 $\rm keV$ energy band. $L_{\rm pk}$ is in unit of $\rm 10^{\rm 52} ~ erg ~ s^{\rm -1}$, and in 1-$10^{4}$ $\rm keV$ energy band. The adjusted $R^{2}$ is 0.2296. The GRB sample number is 127.

The $\log E_{\rm p,band,i}$-$\log T_{\rm 90}$-$\log L_{\rm pk}$ formula is:
\begin{equation} \label{eq:ebandit90lpk}
\log E_{\rm p,band,i} = (0.24 \pm 0.023) \times \log T_{\rm 90} + (0.29 \pm 0.012) \times  \log L_{\rm pk}+ (2.1 \pm 0.04),
\end{equation}
where $E_{\rm p,band,i}$ is in unit of $\rm keV$. $T_{\rm 90}$ is in unit of $\rm s$. $L_{\rm pk}$ is in unit of $\rm 10^{\rm 52} ~ erg ~ s^{\rm -1}$, and in 1-$10^{4}$ $\rm keV$ energy band. The adjusted $R^{2}$ is 0.4998. The GRB sample number is 127.

The $\log F_{\rm X11hr}$-$\beta_{\rm X11hr}$-$\log T_{\rm 50,i}$ formula is:
\begin{equation} \label{eq:fx11bx11t50i}
\log F_{\rm X11hr} = (-0.38 \pm 0.067) \times \beta_{\rm X11hr} + (0.36 \pm 0.038) \times \log T_{\rm 50,i} + (-7 \pm 0.11),
\end{equation}
where $F_{\rm X11hr}$ is in unit of $\rm Jy$. $T_{\rm 50,i}$ is in unit of $\rm s$. The adjusted $R^{2}$ is 0.2084. The GRB sample number is 129.

The $\log P_{\rm pk4}$-$\log  D_{\rm L}$-$\beta_{\rm X11hr}$ formula is:
\begin{equation} \label{eq:ppk4dlbx11}
\log P_{\rm pk4} = (-0.26 \pm 0.035) \times \log  D_{\rm L} + (-0.32 \pm 0.049) \times \beta_{\rm X11hr} + (1.2 \pm 0.082),
\end{equation}
where $P_{\rm pk4}$ is peak photon flux of 1 $\rm s$ time bin in 10-1000 $\rm keV$, and in unit of $\rm photons ~ cm^{\rm -2} ~ s^{\rm -1}$. $D_{\rm L}$ is in unit of $\rm 10^{\rm 28} ~ cm$. The adjusted $R^{2}$ is 0.3358. The GRB sample number is 130.

The $\log T_{\rm 50,i}$-$\log E_{\rm iso}$-$\log Mass$ formula is:
\begin{equation} \label{eq:t50ieisomass}
\log T_{\rm 50,i} = (0.29 \pm 0.0082) \times \log E_{\rm iso} + (-0.15 \pm 0.022) \times \log Mass + (2.1 \pm 0.21),
\end{equation}
where $T_{\rm 50,i}$ is in unit of $\rm s$. $E_{\rm iso}$ is in unit of $\rm 10^{\rm 52} ~ ergs$ and in rest-frame 1-$10^{4}$ $\rm keV$ energy band. Mass is in unit of $M_{\bigodot}$. The adjusted $R^{2}$ is 0.2199. The GRB sample number is 130.

The $\log F_{\rm g}$-$\log L_{\rm pk}$-$\log Mass$ formula is:
\begin{equation} \label{eq:fglpkmass}
\log F_{\rm g} = (0.44 \pm 0.017) \times \log L_{\rm pk} + (-0.26 \pm 0.035) \times \log Mass + (3.5 \pm 0.33),
\end{equation}
where $F_{\rm g}$ is in unit of $\rm 10^{\rm -6} ~ ergs ~ cm^{\rm -2}$, and in 20-2000 $\rm keV$ energy band. $L_{\rm pk}$ is in unit of $\rm 10^{\rm 52} ~ erg ~ s^{\rm -1}$, and in 1-$10^{4}$ $\rm keV$ energy band. Mass is in unit of $M_{\bigodot}$. The adjusted $R^{2}$ is 0.3385. The GRB sample number is 130.

The $\log P_{\rm pk4}$-$\log (1+z)$-$\beta_{\rm X11hr}$ formula is:
\begin{equation} \label{eq:ppk4zbx11}
\log P_{\rm pk4} = (-0.63 \pm 0.11) \times \log (1+z) + (-0.31 \pm 0.048) \times \beta_{\rm X11hr} + (1.4 \pm 0.084),
\end{equation}
where $P_{\rm pk4}$ is peak photon flux of 1 $\rm s$ time bin in 10-1000 $\rm keV$, and in unit of $\rm photons ~ cm^{\rm -2} ~ s^{\rm -1}$. The adjusted $R^{2}$ is 0.3504. The GRB sample number is 130.

The $\log T_{\rm 90}$-$variability_{1}$-$\log E_{\rm iso}$ formula is:
\begin{equation} \label{eq:t90var1eiso}
\log T_{\rm 90} = (0.42 \pm 0.22) \times variability_{1} + (0.22 \pm 0.0096) \times \log E_{\rm iso} + (1.4 \pm 0.017),
\end{equation}
where $T_{\rm 90}$ is in unit of $\rm s$. $E_{\rm iso}$ is in unit of $\rm 10^{\rm 52} ~ ergs$ and in rest-frame 1-$10^{4}$ $\rm keV$ energy band. The adjusted $R^{2}$ is 0.2188. The GRB sample number is 132.

The $\log F_{\rm X11hr}$-$\beta_{\rm X11hr}$-$\log T_{\rm 90,i}$ formula is:
\begin{equation} \label{eq:fx11bx11t90i}
\log F_{\rm X11hr} = (-0.33 \pm 0.064) \times \beta_{\rm X11hr} + (0.41 \pm 0.035) \times \log T_{\rm 90,i} + (-7.3 \pm 0.11),
\end{equation}
where $F_{\rm X11hr}$ is in unit of $\rm Jy$. $T_{\rm 90,i}$ is in unit of $\rm s$. The adjusted $R^{2}$ is 0.2234. The GRB sample number is 133.

The $(-\alpha_{\rm band})$-$\log HR$-$\log T_{\rm 90,i}$ formula is:
\begin{equation} \label{eq:abandhrt90i}
(-\alpha_{\rm band}) = (-0.41 \pm 0.064) \times \log HR + (0.11 \pm 0.02) \times \log T_{\rm 90,i} + (1.1 \pm 0.044),
\end{equation}
where $T_{\rm 90,i}$ is in unit of $\rm s$. The adjusted $R^{2}$ is 0.2783. The GRB sample number is 135.

The $\log HR$-$(-\alpha_{\rm band})$-$\log E_{\rm p,band,i}$ formula is:
\begin{equation} \label{eq:hrabandebandi}
\log HR = (-0.33 \pm 0.072) \times (-\alpha_{\rm band}) + (0.34 \pm 0.04) \times \log E_{\rm p,band,i} + (-0.033 \pm 0.14),
\end{equation}
where $E_{\rm p,band,i}$ is in unit of $\rm keV$. The adjusted $R^{2}$ is 0.442. The GRB sample number is 136.

The $\log L_{\rm pk}$-$\log P_{\rm pk4}$-$\log E_{\rm p,cpl,i}$ formula is:
\begin{equation} \label{eq:lpkppk4ecpli}
\log L_{\rm pk} = (0.87 \pm 0.054) \times \log P_{\rm pk4} + (0.63 \pm 0.097) \times \log E_{\rm p,cpl,i} + (-2.3 \pm 0.25),
\end{equation}
where $L_{\rm pk}$ is in unit of $\rm 10^{\rm 52} ~ erg ~ s^{\rm -1}$, and in 1-$10^{4}$ $\rm keV$ energy band. $P_{\rm pk4}$ is peak photon flux of 1 $\rm s$ time bin in 10-1000 $\rm keV$, and in unit of $\rm photons ~ cm^{\rm -2} ~ s^{\rm -1}$. $E_{\rm p,cpl,i}$ is in unit of $\rm keV$. The adjusted $R^{2}$ is 0.2905. The GRB sample number is 136.

The $\log F_{\rm g}$-$\log L_{\rm pk}$-$\log E_{\rm p,cpl,i}$ formula is:
\begin{equation} \label{eq:fglpkecpli}
\log F_{\rm g} = (0.16 \pm 0.022) \times \log L_{\rm pk} + (0.45 \pm 0.076) \times \log E_{\rm p,cpl,i} + (-0.63 \pm 0.2),
\end{equation}
where $F_{\rm g}$ is in unit of $\rm 10^{\rm -6} ~ ergs ~ cm^{\rm -2}$, and in 20-2000 $\rm keV$ energy band. $L_{\rm pk}$ is in unit of $\rm 10^{\rm 52} ~ erg ~ s^{\rm -1}$, and in 1-$10^{4}$ $\rm keV$ energy band. $E_{\rm p,cpl,i}$ is in unit of $\rm keV$. The adjusted $R^{2}$ is 0.2238. The GRB sample number is 139.

The $\log L_{\rm pk}$-$\log F_{\rm g}$-$\log t_{\rm burst,i}$ formula is:
\begin{equation} \label{eq:lpkfgtti}
\log L_{\rm pk} = (0.57 \pm 0.025) \times \log F_{\rm g} + (-0.46 \pm 0.027) \times \log t_{\rm burst,i} + (0.58 \pm 0.062),
\end{equation}
where $L_{\rm pk}$ is in unit of $\rm 10^{\rm 52} ~ erg ~ s^{\rm -1}$, and in 1-$10^{4}$ $\rm keV$ energy band. $F_{\rm g}$ is in unit of $\rm 10^{\rm -6} ~ ergs ~ cm^{\rm -2}$, and in 20-2000 $\rm keV$ energy band. $t_{\rm burst,i}$ is in unit of $\rm s$. The adjusted $R^{2}$ is 0.2877. The GRB sample number is 140.

The $\log E_{\rm iso}$-$\log T_{\rm R45}$-$\log t_{\rm burst,i}$ formula is:
\begin{equation} \label{eq:eisotr45tti}
\log E_{\rm iso} = (0.86 \pm 0.024) \times \log T_{\rm R45} + (-0.37 \pm 0.022) \times \log t_{\rm burst,i} + (0.66 \pm 0.05),
\end{equation}
where $E_{\rm iso}$ is in unit of $\rm 10^{\rm 52} ~ ergs$ and in rest-frame 1-$10^{4}$ $\rm keV$ energy band. $T_{\rm R45}$ is in unit of $\rm s$. $t_{\rm burst,i}$ is in unit of $\rm s$. The adjusted $R^{2}$ is 0.2667. The GRB sample number is 140.

The $\log Mass$-$\log (1+z)$-$A_{\rm V}$ formula is:
\begin{equation} \label{eq:masszav}
\log Mass = (1.8 \pm 0.23) \times \log (1+z) + (0.19 \pm 0.045) \times A_{\rm V} + (8.8 \pm 0.099),
\end{equation}
where Mass is in unit of $M_{\bigodot}$. The adjusted $R^{2}$ is 0.2586. The GRB sample number is 146.

The $\log E_{\rm iso}$-$\log T_{\rm 90}$-$\beta_{\rm X11hr}$ formula is:
\begin{equation} \label{eq:eisot90bx11}
\log E_{\rm iso} = (0.82 \pm 0.025) \times \log T_{\rm 90} + (-0.39 \pm 0.052) \times \beta_{\rm X11hr} + (-0.028 \pm 0.098),
\end{equation}
where $E_{\rm iso}$ is in unit of $\rm 10^{\rm 52} ~ ergs$ and in rest-frame 1-$10^{4}$ $\rm keV$ energy band. $T_{\rm 90}$ is in unit of $\rm s$. The adjusted $R^{2}$ is 0.516. The GRB sample number is 149.

The $\log P_{\rm pk4}$-$\log E_{\rm p,cpl}$-$\log F_{\rm X11hr}$ formula is:
\begin{equation} \label{eq:ppk4ecplfx11}
\log P_{\rm pk4} = (0.18 \pm 0.05) \times \log E_{\rm p,cpl} + (0.22 \pm 0.013) \times \log F_{\rm X11hr} + (1.7 \pm 0.14),
\end{equation}
where $P_{\rm pk4}$ is peak photon flux of 1 $\rm s$ time bin in 10-1000 $\rm keV$, and in unit of $\rm photons ~ cm^{\rm -2} ~ s^{\rm -1}$. $E_{\rm p,cpl}$ is in unit of $\rm keV$. $F_{\rm X11hr}$ is in unit of $\rm Jy$. The adjusted $R^{2}$ is 0.2295. The GRB sample number is 152.

The $\log F_{\rm X11hr}$-$\log E_{\rm iso}$-$\log  F_{\rm Opt11hr}$ formula is:
\begin{equation} \label{eq:fx11eisofo11}
\log F_{\rm X11hr} = (0.28 \pm 0.016) \times \log E_{\rm iso} + (0.19 \pm 0.045) \times \log F_{\rm Opt11hr} + (-6.7 \pm 0.23),
\end{equation}
where $F_{\rm X11hr}$ is in unit of $\rm Jy$. $E_{\rm iso}$ is in unit of $\rm 10^{\rm 52} ~ ergs$ and in rest-frame 1-$10^{4}$ $\rm keV$ energy band. $F_{\rm Opt11hr}$ is in unit of $\rm Jy$. The adjusted $R^{2}$ is 0.2229. The GRB sample number is 153.

The $\log E_{\rm iso}$-$\log T_{\rm 50}$-$\log F_{\rm X11hr}$ formula is:
\begin{equation} \label{eq:eisot50fx11}
\log E_{\rm iso} = (0.63 \pm 0.021) \times \log T_{\rm 50} + (0.45 \pm 0.028) \times \log F_{\rm X11hr} + (3.1 \pm 0.22),
\end{equation}
where $E_{\rm iso}$ is in unit of $\rm 10^{\rm 52} ~ ergs$ and in rest-frame 1-$10^{4}$ $\rm keV$ energy band. $T_{\rm 50}$ is in unit of $\rm s$. $F_{\rm X11hr}$ is in unit of $\rm Jy$. The adjusted $R^{2}$ is 0.3408. The GRB sample number is 153.

The $\log F_{\rm g}$-$\log (1+z)$-$\beta_{\rm X11hr}$ formula is:
\begin{equation} \label{eq:fgzbx11}
\log F_{\rm g} = (0.71 \pm 0.15) \times \log (1+z) + (-0.56 \pm 0.064) \times \beta_{\rm X11hr} + (1.3 \pm 0.11),
\end{equation}
where $F_{\rm g}$ is in unit of $\rm 10^{\rm -6} ~ ergs ~ cm^{\rm -2}$, and in 20-2000 $\rm keV$ energy band. The adjusted $R^{2}$ is 0.2953. The GRB sample number is 153.

The $\log F_{\rm pk1}$-$\log HR$-$\log Mass$ formula is:
\begin{equation} \label{eq:fpk1hrmass}
\log F_{\rm pk1} = (1 \pm 0.07) \times \log HR + (-0.15 \pm 0.03) \times \log Mass + (1.1 \pm 0.29),
\end{equation}
where $F_{\rm pk1}$ is peak energy flux of 1 $\rm s$ time bin in rest-frame 1-$10^{4}$ $\rm keV$ energy band, and in unit of $\rm 10^{\rm -6} ~ ergs ~ cm^{\rm -2} ~ s^{\rm -1}$. Mass is in unit of $M_{\bigodot}$. The adjusted $R^{2}$ is 0.3515. The GRB sample number is 157.

The $\log E_{\rm iso}$-$\log F_{\rm pk1}$-$\log E_{\rm p,band,i}$ formula is:
\begin{equation} \label{eq:eisofpk1ebandi}
\log E_{\rm iso} = (0.26 \pm 0.025) \times \log F_{\rm pk1} + (1.2 \pm 0.048) \times \log E_{\rm p,band,i} + (-2.2 \pm 0.12),
\end{equation}
where $E_{\rm iso}$ is in unit of $\rm 10^{\rm 52} ~ ergs$ and in rest-frame 1-$10^{4}$ $\rm keV$ energy band. $F_{\rm pk1}$ is peak energy flux of 1 $\rm s$ time bin in rest-frame 1-$10^{4}$ $\rm keV$ energy band, and in unit of $\rm 10^{\rm -6} ~ ergs ~ cm^{\rm -2} ~ s^{\rm -1}$. $E_{\rm p,band,i}$ is in unit of $\rm keV$. The adjusted $R^{2}$ is 0.564. The GRB sample number is 158.

The $\log E_{\rm iso}$-$\log T_{\rm 90}$-$\log F_{\rm X11hr}$ formula is:
\begin{equation} \label{eq:eisot90fx11}
\log E_{\rm iso} = (0.77 \pm 0.022) \times \log T_{\rm 90} + (0.36 \pm 0.028) \times \log F_{\rm X11hr} + (2 \pm 0.23),
\end{equation}
where $E_{\rm iso}$ is in unit of $\rm 10^{\rm 52} ~ ergs$ and in rest-frame 1-$10^{4}$ $\rm keV$ energy band. $T_{\rm 90}$ is in unit of $\rm s$. $F_{\rm X11hr}$ is in unit of $\rm Jy$. The adjusted $R^{2}$ is 0.3803. The GRB sample number is 159.

The $\log F_{\rm g}$-$\log L_{\rm pk}$-$\log  F_{\rm Opt11hr}$ formula is:
\begin{equation} \label{eq:fglpkfo11}
\log F_{\rm g} = (0.42 \pm 0.018) \times \log L_{\rm pk} + (0.28 \pm 0.034) \times \log F_{\rm Opt11hr} + (2.2 \pm 0.16),
\end{equation}
where $F_{\rm g}$ is in unit of $\rm 10^{\rm -6} ~ ergs ~ cm^{\rm -2}$, and in 20-2000 $\rm keV$ energy band. $L_{\rm pk}$ is in unit of $\rm 10^{\rm 52} ~ erg ~ s^{\rm -1}$, and in 1-$10^{4}$ $\rm keV$ energy band. $F_{\rm Opt11hr}$ is in unit of $\rm Jy$. The adjusted $R^{2}$ is 0.3953. The GRB sample number is 171.

The $\log T_{\rm 90}$-$\log E_{\rm iso}$-$\log Mass$ formula is:
\begin{equation} \label{eq:t90eisomass}
\log T_{\rm 90} = (0.41 \pm 0.0082) \times \log E_{\rm iso} + (-0.17 \pm 0.021) \times \log Mass + (2.9 \pm 0.2),
\end{equation}
where $T_{\rm 90}$ is in unit of $\rm s$. $E_{\rm iso}$ is in unit of $\rm 10^{\rm 52} ~ ergs$ and in rest-frame 1-$10^{4}$ $\rm keV$ energy band. Mass is in unit of $M_{\bigodot}$. The adjusted $R^{2}$ is 0.3543. The GRB sample number is 172.

The $\log P_{\rm pk4}$-$\log F_{\rm g}$-$\log t_{\rm burst,i}$ formula is:
\begin{equation} \label{eq:ppk4fgtti}
\log P_{\rm pk4} = (0.43 \pm 0.021) \times \log F_{\rm g} + (-0.12 \pm 0.02) \times \log t_{\rm burst,i} + (0.45 \pm 0.043),
\end{equation}
where $P_{\rm pk4}$ is peak photon flux of 1 $\rm s$ time bin in 10-1000 $\rm keV$, and in unit of $\rm photons ~ cm^{\rm -2} ~ s^{\rm -1}$. $F_{\rm g}$ is in unit of $\rm 10^{\rm -6} ~ ergs ~ cm^{\rm -2}$, and in 20-2000 $\rm keV$ energy band. $t_{\rm burst,i}$ is in unit of $\rm s$. The adjusted $R^{2}$ is 0.3963. The GRB sample number is 172.

The $\log P_{\rm pk4}$-$\log HR$-$\log F_{\rm X11hr}$ formula is:
\begin{equation} \label{eq:ppk4hrfx11}
\log P_{\rm pk4} = (0.28 \pm 0.047) \times \log HR + (0.23 \pm 0.016) \times \log F_{\rm X11hr} + (2.2 \pm 0.12),
\end{equation}
where $P_{\rm pk4}$ is peak photon flux of 1 $\rm s$ time bin in 10-1000 $\rm keV$, and in unit of $\rm photons ~ cm^{\rm -2} ~ s^{\rm -1}$. $F_{\rm X11hr}$ is in unit of $\rm Jy$. The adjusted $R^{2}$ is 0.2558. The GRB sample number is 182.

The $\log F_{\rm g}$-$\log T_{\rm 50,i}$-$\log E_{\rm p,cpl,i}$ formula is:
\begin{equation} \label{eq:fgt50iecpli}
\log F_{\rm g} = (0.48 \pm 0.026) \times \log T_{\rm 50,i} + (0.57 \pm 0.056) \times \log E_{\rm p,cpl,i} + (-1.3 \pm 0.14),
\end{equation}
where $F_{\rm g}$ is in unit of $\rm 10^{\rm -6} ~ ergs ~ cm^{\rm -2}$, and in 20-2000 $\rm keV$ energy band. $T_{\rm 50,i}$ is in unit of $\rm s$. $E_{\rm p,cpl,i}$ is in unit of $\rm keV$. The adjusted $R^{2}$ is 0.4764. The GRB sample number is 185.

The $\log P_{\rm pk4}$-$\log T_{\rm 90}$-$\log F_{\rm X11hr}$ formula is:
\begin{equation} \label{eq:ppk4t90fx11}
\log P_{\rm pk4} = (-0.13 \pm 0.013) \times \log T_{\rm 90} + (0.29 \pm 0.015) \times \log F_{\rm X11hr} + (2.8 \pm 0.12),
\end{equation}
where $P_{\rm pk4}$ is peak photon flux of 1 $\rm s$ time bin in 10-1000 $\rm keV$, and in unit of $\rm photons ~ cm^{\rm -2} ~ s^{\rm -1}$. $T_{\rm 90}$ is in unit of $\rm s$. $F_{\rm X11hr}$ is in unit of $\rm Jy$. The adjusted $R^{2}$ is 0.2226. The GRB sample number is 216.

The $\log F_{\rm g}$-$\log L_{\rm pk}$-$\log T_{\rm 50,i}$ formula is:
\begin{equation} \label{eq:fglpkt50i}
\log F_{\rm g} = (0.39 \pm 0.013) \times \log L_{\rm pk} + (0.55 \pm 0.022) \times \log T_{\rm 50,i} + (0.38 \pm 0.019),
\end{equation}
where $F_{\rm g}$ is in unit of $\rm 10^{\rm -6} ~ ergs ~ cm^{\rm -2}$, and in 20-2000 $\rm keV$ energy band. $L_{\rm pk}$ is in unit of $\rm 10^{\rm 52} ~ erg ~ s^{\rm -1}$, and in 1-$10^{4}$ $\rm keV$ energy band. $T_{\rm 50,i}$ is in unit of $\rm s$. The adjusted $R^{2}$ is 0.4039. The GRB sample number is 218.

The $\log HR$-$variability_{1}$-$(-\beta_{\rm band})$ formula is:
\begin{equation} \label{eq:hrvar1bband}
\log HR = (-0.54 \pm 0.49) \times variability_{1} + (-0.15 \pm 0.054) \times (-\beta_{\rm band}) + (1 \pm 0.14),
\end{equation}
the adjusted $R^{2}$ is 0.2164. The GRB sample number is 219.

The $\log L_{\rm pk}$-$\log F_{\rm g}$-$A_{\rm V}$ formula is:
\begin{equation} \label{eq:lpkfgav}
\log L_{\rm pk} = (0.51 \pm 0.021) \times \log F_{\rm g} + (-0.21 \pm 0.043) \times A_{\rm V} + (-0.35 \pm 0.037),
\end{equation}
where $L_{\rm pk}$ is in unit of $\rm 10^{\rm 52} ~ erg ~ s^{\rm -1}$, and in 1-$10^{4}$ $\rm keV$ energy band. $F_{\rm g}$ is in unit of $\rm 10^{\rm -6} ~ ergs ~ cm^{\rm -2}$, and in 20-2000 $\rm keV$ energy band. The adjusted $R^{2}$ is 0.2064. The GRB sample number is 220.

The $\log F_{\rm g}$-$variability_{1}$-$\log P_{\rm pk1}$ formula is:
\begin{equation} \label{eq:fgvar1ppk1}
\log F_{\rm g} = (-1.9 \pm 0.58) \times variability_{1} + (0.83 \pm 0.038) \times \log P_{\rm pk1} + (0.41 \pm 0.055),
\end{equation}
where $F_{\rm g}$ is in unit of $\rm 10^{\rm -6} ~ ergs ~ cm^{\rm -2}$, and in 20-2000 $\rm keV$ energy band. $P_{\rm pk1}$ is peak photon flux of 64 $\rm ms$ time bin in 10-1000 $\rm keV$, and in unit of $\rm photons ~ cm^{\rm -2} ~ s^{\rm -1}$. The adjusted $R^{2}$ is 0.4694. The GRB sample number is 227.

The $\log E_{\rm iso}$-$(-\alpha_{\rm cpl})$-$\log E_{\rm p,cpl,i}$ formula is:
\begin{equation} \label{eq:eisoacplecpli}
\log E_{\rm iso} = (0.3 \pm 0.071) \times (-\alpha_{\rm cpl}) + (0.85 \pm 0.08) \times \log E_{\rm p,cpl,i} + (-2.2 \pm 0.23),
\end{equation}
where $E_{\rm iso}$ is in unit of $\rm 10^{\rm 52} ~ ergs$ and in rest-frame 1-$10^{4}$ $\rm keV$ energy band. $E_{\rm p,cpl,i}$ is in unit of $\rm keV$. The adjusted $R^{2}$ is 0.2692. The GRB sample number is 228.

The $\log F_{\rm X11hr}$-$\log F_{\rm g}$-$\log  F_{\rm Opt11hr}$ formula is:
\begin{equation} \label{eq:fx11fgfo11}
\log F_{\rm X11hr} = (0.52 \pm 0.023) \times \log F_{\rm g} + (0.16 \pm 0.026) \times \log F_{\rm Opt11hr} + (-7.1 \pm 0.13),
\end{equation}
where $F_{\rm X11hr}$ is in unit of $\rm Jy$. $F_{\rm g}$ is in unit of $\rm 10^{\rm -6} ~ ergs ~ cm^{\rm -2}$, and in 20-2000 $\rm keV$ energy band. $F_{\rm Opt11hr}$ is in unit of $\rm Jy$. The adjusted $R^{2}$ is 0.3134. The GRB sample number is 232.

The $\log F_{\rm g}$-$\log  D_{\rm L}$-$\log E_{\rm p,cpl}$ formula is:
\begin{equation} \label{eq:fgdlecpl}
\log F_{\rm g} = (0.27 \pm 0.029) \times \log  D_{\rm L} + (0.63 \pm 0.059) \times \log E_{\rm p,cpl} + (-1.1 \pm 0.13),
\end{equation}
where $F_{\rm g}$ is in unit of $\rm 10^{\rm -6} ~ ergs ~ cm^{\rm -2}$, and in 20-2000 $\rm keV$ energy band. $D_{\rm L}$ is in unit of $\rm 10^{\rm 28} ~ cm$. $E_{\rm p,cpl}$ is in unit of $\rm keV$. The adjusted $R^{2}$ is 0.2516. The GRB sample number is 248.

The $\log  D_{\rm L}$-$\log T_{\rm 90}$-$\log E_{\rm p,cpl,i}$ formula is:
\begin{equation} \label{eq:dlt90ecpli}
\log  D_{\rm L} = (0.2 \pm 0.0065) \times \log T_{\rm 90} + (0.29 \pm 0.028) \times \log E_{\rm p,cpl,i} + (-0.52 \pm 0.071),
\end{equation}
where $D_{\rm L}$ is in unit of $\rm 10^{\rm 28} ~ cm$. $T_{\rm 90}$ is in unit of $\rm s$. $E_{\rm p,cpl,i}$ is in unit of $\rm keV$. The adjusted $R^{2}$ is 0.2192. The GRB sample number is 257.

The $\log F_{\rm g}$-$\log L_{\rm pk}$-$\log T_{\rm 90,i}$ formula is:
\begin{equation} \label{eq:fglpkt90i}
\log F_{\rm g} = (0.37 \pm 0.011) \times \log L_{\rm pk} + (0.63 \pm 0.017) \times \log T_{\rm 90,i} + (0.1 \pm 0.022),
\end{equation}
where $F_{\rm g}$ is in unit of $\rm 10^{\rm -6} ~ ergs ~ cm^{\rm -2}$, and in 20-2000 $\rm keV$ energy band. $L_{\rm pk}$ is in unit of $\rm 10^{\rm 52} ~ erg ~ s^{\rm -1}$, and in 1-$10^{4}$ $\rm keV$ energy band. $T_{\rm 90,i}$ is in unit of $\rm s$. The adjusted $R^{2}$ is 0.4429. The GRB sample number is 312.

The $\log F_{\rm g}$-$\log T_{\rm R45}$-$\log E_{\rm p,cpl}$ formula is:
\begin{equation} \label{eq:fgtr45ecpl}
\log F_{\rm g} = (0.59 \pm 0.019) \times \log T_{\rm R45} + (0.55 \pm 0.045) \times \log E_{\rm p,cpl} + (-1.2 \pm 0.098),
\end{equation}
where $F_{\rm g}$ is in unit of $\rm 10^{\rm -6} ~ ergs ~ cm^{\rm -2}$, and in 20-2000 $\rm keV$ energy band. $T_{\rm R45}$ is in unit of $\rm s$. $E_{\rm p,cpl}$ is in unit of $\rm keV$. The adjusted $R^{2}$ is 0.5627. The GRB sample number is 362.

The $\log P_{\rm pk4}$-$\log  D_{\rm L}$-$\log F_{\rm g}$ formula is:
\begin{equation} \label{eq:ppk4dlfg}
\log P_{\rm pk4} = (-0.25 \pm 0.013) \times \log  D_{\rm L} + (0.39 \pm 0.013) \times \log F_{\rm g} + (0.45 \pm 0.011),
\end{equation}
where $P_{\rm pk4}$ is peak photon flux of 1 $\rm s$ time bin in 10-1000 $\rm keV$, and in unit of $\rm photons ~ cm^{\rm -2} ~ s^{\rm -1}$. $D_{\rm L}$ is in unit of $\rm 10^{\rm 28} ~ cm$. $F_{\rm g}$ is in unit of $\rm 10^{\rm -6} ~ ergs ~ cm^{\rm -2}$, and in 20-2000 $\rm keV$ energy band. The adjusted $R^{2}$ is 0.3421. The GRB sample number is 419.

The $\log HR$-$\log F_{\rm pk2}$-$(-\alpha_{\rm spl})$ formula is:
\begin{equation} \label{eq:hrfpk2aspl}
\log HR = (0.17 \pm 0.036) \times \log F_{\rm pk2} + (-0.9 \pm 0.039) \times (-\alpha_{\rm spl}) + (2.2 \pm 0.075),
\end{equation}
where $F_{\rm pk2}$ is peak energy flux of 64 $\rm ms$ time bin in rest-frame 1-$10^{4}$ $\rm keV$ energy band, and in unit of $\rm 10^{\rm -6} ~ ergs ~ cm^{\rm -2} ~ s^{\rm -1}$. The adjusted $R^{2}$ is 0.5718. The GRB sample number is 490.

The $\log HR$-$\log T_{\rm 50}$-$(-\alpha_{\rm spl})$ formula is:
\begin{equation} \label{eq:hrt50aspl}
\log HR = (-0.11 \pm 0.014) \times \log T_{\rm 50} + (-0.87 \pm 0.035) \times (-\alpha_{\rm spl}) + (2.2 \pm 0.056),
\end{equation}
where $T_{\rm 50}$ is in unit of $\rm s$. The adjusted $R^{2}$ is 0.6123. The GRB sample number is 497.

The $\log HR$-$\log F_{\rm pk1}$-$\log E_{\rm p,cpl}$ formula is:
\begin{equation} \label{eq:hrfpk1ecpl}
\log HR = (0.15 \pm 0.02) \times \log F_{\rm pk1} + (0.92 \pm 0.045) \times \log E_{\rm p,cpl} + (-1.8 \pm 0.11),
\end{equation}
where $F_{\rm pk1}$ is peak energy flux of 1 $\rm s$ time bin in rest-frame 1-$10^{4}$ $\rm keV$ energy band, and in unit of $\rm 10^{\rm -6} ~ ergs ~ cm^{\rm -2} ~ s^{\rm -1}$. $E_{\rm p,cpl}$ is in unit of $\rm keV$. The adjusted $R^{2}$ is 0.8214. The GRB sample number is 713.

The $\log P_{\rm pk3}$-$\log F_{\rm g}$-$(-\alpha_{\rm cpl})$ formula is:
\begin{equation} \label{eq:ppk3fgacpl}
\log P_{\rm pk3} = (0.38 \pm 0.015) \times \log F_{\rm g} + (0.15 \pm 0.016) \times (-\alpha_{\rm cpl}) + (0.21 \pm 0.015),
\end{equation}
where $P_{\rm pk3}$ is peak photon flux of 1024 $\rm ms$ time bin in 10-1000 $\rm keV$, and in unit of $\rm photons ~ cm^{\rm -2} ~ s^{\rm -1}$. $F_{\rm g}$ is in unit of $\rm 10^{\rm -6} ~ ergs ~ cm^{\rm -2}$, and in 20-2000 $\rm keV$ energy band. The adjusted $R^{2}$ is 0.4674. The GRB sample number is 913.

The $\log HR$-$\log F_{\rm pk4}$-$(-\alpha_{\rm cpl})$ formula is:
\begin{equation} \label{eq:hrfpk4acpl}
\log HR = (0.42 \pm 0.024) \times \log F_{\rm pk4} + (-0.31 \pm 0.025) \times (-\alpha_{\rm cpl}) + (0.96 \pm 0.026),
\end{equation}
where $F_{\rm pk4}$ is peak energy flux of 1024 $\rm ms$ time bin in rest-frame 1-$10^{4}$ $\rm keV$ energy band, and in unit of $\rm 10^{\rm -6} ~ ergs ~ cm^{\rm -2} ~ s^{\rm -1}$. The adjusted $R^{2}$ is 0.3125. The GRB sample number is 937.

The $\log HR$-$\log P_{\rm pk1}$-$(-\beta_{\rm band})$ formula is:
\begin{equation} \label{eq:hrppk1bband}
\log HR = (0.15 \pm 0.018) \times \log P_{\rm pk1} + (-0.18 \pm 0.034) \times (-\beta_{\rm band}) + (0.83 \pm 0.085),
\end{equation}
where $P_{\rm pk1}$ is peak photon flux of 64 $\rm ms$ time bin in 10-1000 $\rm keV$, and in unit of $\rm photons ~ cm^{\rm -2} ~ s^{\rm -1}$. The adjusted $R^{2}$ is 0.2471. The GRB sample number is 1092.

The $\log F_{\rm g}$-$\log P_{\rm pk1}$-$\log  E_{\rm p,band}$ formula is:
\begin{equation} \label{eq:fgppk1eband}
\log F_{\rm g} = (0.58 \pm 0.027) \times \log P_{\rm pk1} + (0.63 \pm 0.025) \times \log  E_{\rm p,band} + (-1.1 \pm 0.053),
\end{equation}
where $F_{\rm g}$ is in unit of $\rm 10^{\rm -6} ~ ergs ~ cm^{\rm -2}$, and in 20-2000 $\rm keV$ energy band. $P_{\rm pk1}$ is peak photon flux of 64 $\rm ms$ time bin in 10-1000 $\rm keV$, and in unit of $\rm photons ~ cm^{\rm -2} ~ s^{\rm -1}$. $E_{\rm p,band}$ is in unit of $\rm keV$. The adjusted $R^{2}$ is 0.4116. The GRB sample number is 1121.

The $\log  E_{\rm p,band}$-$\log HR$-$\log F_{\rm pk2}$ formula is:
\begin{equation} \label{eq:ebandhrfpk2}
\log  E_{\rm p,band} = (0.42 \pm 0.019) \times \log HR + (0.14 \pm 0.011) \times \log F_{\rm pk2} + (2 \pm 0.0099),
\end{equation}
where $E_{\rm p,band}$ is in unit of $\rm keV$. $F_{\rm pk2}$ is peak energy flux of 64 $\rm ms$ time bin in rest-frame 1-$10^{4}$ $\rm keV$ energy band, and in unit of $\rm 10^{\rm -6} ~ ergs ~ cm^{\rm -2} ~ s^{\rm -1}$. The adjusted $R^{2}$ is 0.6307. The GRB sample number is 1130.

The $\log T_{\rm 50}$-$\log HR$-$(-\alpha_{\rm cpl})$ formula is:
\begin{equation} \label{eq:t50hracpl}
\log T_{\rm 50} = (-0.44 \pm 0.022) \times \log HR + (0.35 \pm 0.032) \times (-\alpha_{\rm cpl}) + (0.53 \pm 0.04),
\end{equation}
where $T_{\rm 50}$ is in unit of $\rm s$. The adjusted $R^{2}$ is 0.2224. The GRB sample number is 1140.

The $\log HR$-$\log T_{\rm 50}$-$(-\beta_{\rm band})$ formula is:
\begin{equation} \label{eq:hrt50band}
\log HR = (-0.14 \pm 0.0073) \times \log T_{\rm 50} + (-0.11 \pm 0.024) \times (-\beta_{\rm band}) + (0.97 \pm 0.06),
\end{equation}
where $T_{\rm 50}$ is in unit of $\rm s$. The adjusted $R^{2}$ is 0.201. The GRB sample number is 1183.

The $\log F_{\rm pk4}$-$(-\alpha_{\rm band})$-$\log  E_{\rm p,band}$ formula is:
\begin{equation} \label{eq:fpk4abandeband}
\log F_{\rm pk4} = (0.11 \pm 0.025) \times (-\alpha_{\rm band}) + (0.81 \pm 0.025) \times \log  E_{\rm p,band} + (-1.8 \pm 0.058),
\end{equation}
where $F_{\rm pk4}$ is peak energy flux of 1024 $\rm ms$ time bin in rest-frame 1-$10^{4}$ $\rm keV$ energy band, and in unit of $\rm 10^{\rm -6} ~ ergs ~ cm^{\rm -2} ~ s^{\rm -1}$. $E_{\rm p,band}$ is in unit of $\rm keV$. The adjusted $R^{2}$ is 0.3073. The GRB sample number is 1192.

The $\log T_{\rm 90}$-$\log F_{\rm g}$-$(-\alpha_{\rm cpl})$ formula is:
\begin{equation} \label{eq:t90fgacpl}
\log T_{\rm 90} = (0.56 \pm 0.013) \times \log F_{\rm g} + (0.32 \pm 0.03) \times (-\alpha_{\rm cpl}) + (0.6 \pm 0.032),
\end{equation}
where $T_{\rm 90}$ is in unit of $\rm s$. $F_{\rm g}$ is in unit of $\rm 10^{\rm -6} ~ ergs ~ cm^{\rm -2}$, and in 20-2000 $\rm keV$ energy band. The adjusted $R^{2}$ is 0.3263. The GRB sample number is 1347.

\section{Discussions} \label{sec:discuss}
There are also many other interesting quantities which are not listed in this data sample, mainly because the quantities are limited to a few GRBs, or they are not quite well defined or not widely accepted. People may want to use the data shown here as an reservoir, and add any other data interested to perform the statistical studies.
With Fourier transformation of the prompt light curves, the slope in the frequency domain. Properties of the precursors are not included neither, though they are thought no much different from the prompt emission themselves \citep{2008ApJ...685L..19B,2009A&A...505..569B}.
Because of the universal behavior of the afterglow light curves \citep{2006ApJ...642..354Z}, there are quite a few parameters including the temporal decay index, the ending time, and spectral index in each phase (they are in X-ray, and mainly contains only one segment, i.e., no breaks in the spectra), while the phases include the steep decay phase, the plateau phase, the normal decay phase and the jet break phase. There are also parameters for the X-ray flares in the afterglow, such as the number of the flares, typical duration, rising temporal index, decay temporal index, spectrum index and luminosities.
Beside the spectral indices, there are also plenty of detailed spectral lines from the afterglows and from the host galaxies, which can be used as characteristic quantities.
One can use those data and the data listed in this paper for a combined analysis.
On the other hand, the combination of the parameters can also be taken as independent parameters, such as the average luminosity $L_{\rm iso} \equiv \frac{(1+z) E_{\rm iso}}{T_{90}}$, and the spectral index difference of the Band spectrum $\alpha-\beta$. For the detailed study on special quantities, one can dig more data from the original data sets. For example, $P_{\rm pk}$ and $F_{\rm pk}$ have four time bins mainly because different authors are interested in different time bins. It is not proper to simply take them equal, and four different values are gathered in this work. In most cases, for each GRB, only a few time bins are available. One could produce the $P_{\rm pk}$ for 1024 ms bin (for example) if the original light curves are available. Therefore, digging into the original data from the satellites will provide much more extra data than them shown in this work. On the other hand, the light curves in different bands are also interesting, of which the corresponding quantities are not shown here. One could expand the parameters by digging into the raw data.

One should be careful of using these data because of the variety sources of the data. With different instruments, the energy band is different and the sensitivity is different. For example, the $T_{90}$ is energy dependent, therefore, $T_{90}$ from different instruments should not be taken as the same parameter if the energy band for different instruments differs too much. A proper way to handle this selection effect is to convert the light curves in the same energy band, which needs the spectrum models. However, in many cases, the spectrum model for each GRB is not the same, and for some GRBs, it is possible that two or more models are the same goodness for the spectrum fitting. Therefore, it is model dependent. For the complexity and the model dependent problem, we only include the direct available data, while leaving these converting works in the future. The same problem also apply to $T_{50}$, spectral lag, HR, fluence $F$, photon flux $P$, luminosity $L$ and isotropic equivalent energy $E_{\rm iso}$. To avoid these problems, we have converted them into the same band, either in a mostly used energy band, or a full band (like $1-10,000$keV), by employing the spectral fitting parameters. However, these conversions are model depended as for different bursts, the spectral fitting model might be different, and the real model might not be the best fitting model.
Even the data can be converted into the values in the same band, there is still cosmological effects. The same band is not the same band in the rest frame. Because of the cosmological redshift, in order to compare in the same band, one should convert the quantities in the same band in the rest frame.
However, even in the same band in the rest frame, the cosmological evolution effect is not removable. It is still unclear on how the GRBs evolves with the universe evolution.
On the other hand, with plenty of the data at different redshift, it is possible to study the evolution.

To avoid the problems arisen by the selection effects, we may mainly concentrate on the physical quantities (like luminosities, energies) rather than the observational values (like fluence, photon flux), and we may use the quantities in the rest frame rather than in the observer-frame. However, they both rely on the redshift detection, which is hard and consequently the number of the sample is much smaller. As a balance, one may use the data as many as possible, while keep in mind about the reliability of the GRBs without redshift obtained. On the other hand, the relations between observational quantities may reveal which is the selection effect. If one finds the properties only shown in the brighter GRBs (more photons observed), those may probably be not the intrinsic connections.

Even with the selection effect considered, one may see from the figures shown in the figure sets, the correlations are not tight. Though they are all consistent with the previously found correlations, such as the Amati relation \citep{Amati2002}, Ghirlanda relation \citep{Ghirlanda2004}, Yonetoku relation \citep{Yonetoku2004} relation, Liang-Zhang relation \citep{2005ApJ...633..611L} etc.,  either they can be grouped into several sub-groups (e.g., Amati relation is different for LGRBs and SGRBs), or the selected sample are still not tight enough (e.g., the standard candle relation is not good enough for precise cosmography \citep{2005ApJ...633..603X,Wang2011}). The probable reason might be the intrinsic variety of the GRBs. From the morphological view, they can be divided into different groups based on different properties. Based on the duration, there are  LGRBs, SGRBs, and some of them may be classified into intermediate GRBs, or ultra-long GRBs \citep{Levan2014,Boer2015,Greiner2015}. The physical mechanism responsible for ultra-long GRBs are so far unclear. \citet{Greiner2015} provided a striking clue for ultra-long GRB 111209A, which is driven by spin-down radiation from a highly magnetized millisecond pulsar, known as a magnetar. \citet{Gompertz2017} made efforts to place constraints on the magnetar model. Based on spectrum (or hardness ratio), there are soft GRBs, hard GRBs, and some are very soft and being classified into X-ray rich GRBs or even X-ray flashes. Based on the luminosity, there are high luminosity GRBs and low luminosity GRBs.
Based on the connection with other phenomena, there are SN Ic connected GRBs, no SN connected GRBs into a very dim flux, kilonova connected GRBs and strong GeV connected GRBs.
From the physical origin view, there are also different subgroups. Based on the progenitors, they can be massive stars, BH-NS binaries, NS-NS binaries, and even BH-WD binaries and NS-WD binaries. Based on the central engine, they can be BZ mechanism dominated, NDAF dominated and magentar dominated. Based on the radiation mechanism, they can be synchrotron radiation dominated, inverse Compton scattering dominated or photosphere emission dominated.

Therefore, the direct analysis may not reveal the underline pattern, mainly because of the selection effect and the clustering effect. For the selection effect, one should try to figure out each factor, if possible. For the clustering effect, one should try to find the subgroups from the full sample. A widely known clustering is the long-soft GRBs and the short-hard GRBs. However, it is very likely the GRBs should be classified into more subgroups. From the point of view of the central engine, there might be BH accretion disk system powered or neutron star powered. From the view of the progenitors, there might be massive star collapsar, BH-WD merger, BH-star merger for LGRBs, while BH-NS merger, NS-NS merger for SGRBs. The other possibilities, such as NS-star merger and NS-WD merger, might also be hidden in the sample. From the view of the radiation mechanism, there might be photosphere emission, synchrotron radiation, inverse Compton scattering (depending on the origins of the seed photons and the electrons/protons, much more subgroups could be needed) etc. From point of the view of the ejecta, it could be higher relativistic or mildly relativistic. From the view of the environment, the number density could be uniformly distributed or wind like distribution, and it could be a dense environment or a thin one. For the optical counterpart, it could be supernova connected, kilonova connected, no supernova connected, or even a dim optical counterpart. From the view of the host galaxy, there might be a spiral galaxy, a field galaxy, or other types. More detailed or other classification criteria can be proposed.
The different origins or mechanisms represent in the properties of the data, and should be revealed from the detailed and comprehensive study on the data.

The clustering and the correlation analysis actually affect each other. With a proper classification, such as if the NS-WD merger GRBs are gathered by some properties, the correlation between two or three parameters might be much more tighter, and the correlation might be more useful for a standard candle relation, or for a redshift indicator etc. On the other hand, the correlation can be used as indicator for clustering. For example, the $E_{\rm p}-E_{\rm iso}$ relation is often used as one of the indicator for a certain GRB being a long or short GRB. The GRB lies on the Amati relation is more likely a collapsar originated GRB, while the outlier is more likely a merger originated GRB. Therefore, with more detailed clustering, the correlation inside a certain group of GRBs is more reliable and reveals deeper of the nature, which need more and more data accumulation. Once the pure subset has been found, it could be used as a standard candle, and could be an ideal tool for the cosmology into high redshift region.

For the clustering, an example is the classification of the hard SGRBs and the soft LGRBs. There are also other independent classifications, like the ultra-long GRBs, low luminosity GRBs. These are classified by one or two parameters of the GRBs.
Machine learning is a  set of promising methods, such as the k-means, SVC,  Principal Component Analysis \citep{ZhouZH2016}. They are able to in a much higher dimension, and in high nonlinear manners. Similar as has been done in other areas of astrophysics \citep[e.g.][]{2016A&A...595A..82E}.
However, the machine learning methods are often like black boxes. One is easy to get the result of the clustering, but is difficult to find the criteria of the clustering. One of the main aims to figure out the black box, which reveals the intrinsic properties, similar as the clustering process shown in \citet{2009ApJ...703.1696Z}.
Except for clustering, machine learning can also for parameter predicting. For example, to predict the redshift, similar as being done to the SDSS galaxies \citep{2015MNRAS.452.4183H}.
As the lack of the massive data as used in general usage of machine learning, deep learning may also be a promising approach.

High energy radiation in GeV band has been detected in quite a few \textit{Fermi} GRBs. From the binned light curves, they are likely from the afterglow \citep{Ghisellini2010}. However, there are many sources may produce GeV photons, such as, the synchrotron and the synchrotron self inverse Compton scattering (SSC) emission of the long-lasting forward external shock can contribute; the GRB central engines are continually active and the SSC emission of the continued  internal shocks may produce GeV photons; in a relativistic reverse shock formed, the prompt optical/X-ray/$\gamma$-ray photons are inversely Compton scattered by the accelerated electrons in the forward shock region; and the external inverse Compton scattering (EIC) in the late-afterglow phase caused by X-ray flares may also give rise to GeV emission \citep{2009MNRAS.396.1163Z}.
It is still not clear about which part is dominated in the high energy radiation. With the date base of the information in other aspects, the high energy radiation may be used to compare with the data base, and it may reveal some connections. One may use these connections to figure out the origin of the high energy GeV emission, while the high energy data may come from all of the related instruments including \textit{Fermi}/LAT \citep{Ackermann2013}, HAWC \citep{2012APh....35..641A}, H.E.S.S. \citep{2009A&A...495..505A}, MAGIC \citep{2007ApJ...667..358A} and Wukong (DAMPE) \citep{2017APh....95....6C}.

Gravitational wave burst is accompanied with the SGRBs, as they are believed from double compact object (NS-NS or NS-BH). It has been identified the NS-NS origin since the detection of GW170817/GRB 170817A \citep{2017PhRvL.119p1101A,2017ApJ...848L..12A}. However, the observed GRB 170817A was a weak SGRB, which is much different from the normal SGRBs. As the distance of GRB 170817A is much closer than the other SGRBs, the birth rate is consequently much higher than the others, if we consider they belong to two different subgroups. It is possible that NS-NS mergers may produce weak SGRBs, whereas BH-NS mergers may produce strong SGRBs, which are the mostly observed normal SGRBs. There are other gravitational wave bursts have been observed, such as GW150914 \citep{2016PhRvL.116f1102A}. They have confirmed the presence of a BH that is tens of solar masses (for GW150914, it was the merger of $36^{+5}_{-4} {\rm M}_{\odot}$ BH and $29^{+4}_{-4} {\rm M}_{\odot}$ BH \citep{2016PhRvL.116f1102A}). Therefore, comparing with NS-NS merger, which is the merger of several $ {\rm M}_{\odot}$ object and several $ {\rm M}_{\odot}$ object, the BH-NS merger is the merger of tens of $ {\rm M}_{\odot}$ object and several $ {\rm M}_{\odot}$ object. The BH-NS merger might be naturally stronger and consequently produces the strong SGRBs.
If the hypothesis is correct, these two different objects should have different properties on other aspects. In the hundreds of the observed SGRBs, these two origins should both been included. Considering the operation of the LIGO was just a few months, an NS-NS GRB (GRB 170817A) was observed. It indicates there are much more this kind of GRBs have been observed and been archived when LIGO was not in operation. With the comprehensive table of different aspects of all the GRBs, one may try to figure out the subgroups in the SGRBs. The NS-NS and NS-BH origin may be distinguished, and maybe more subgroups of other origins can be uncovered.
Until now, the BH-NS merger has not been confirmed by the gravitational wave detector. With the future operation with longer time, and with the enhanced instruments, the BH-NS merger should be identified.

\section{Conclusion} \label{sec:conclusion}
In this work, we collected a large number of data for 6289 gamma-ray bursts (GRBs)  from different literature, GCN (Gamma-ray Coordinates Network), website data bases and the calculations. The data include four parts: basic information, prompt emission, afterglow and host galaxy (in total 46 items for each GRB). With this complete table, we performed a comprehensive statistical study. When we did statistics, we also changed some parameters from observer-frame to rest-frame, we use label ``i" to remark. For example, the duration of 5\% to 95\% $\gamma$-ray fluence ($T_{\rm 90}$) is in observer-frame, $T_{\rm 90,i}$ is $T_{\rm 90}$ in rest-frame.This work includes the following 6 items: (1) We imputed the missing errors with multiple imputation \citep{Rubin1987,Rubin1996} by chained equations (MICE); (2) We calculated a small part of peak energy flux ($F_{\rm pk}$) in rest-frame 1-$10^{\rm 4}$ $\rm keV$ energy band, peak photon flux ($P_{\rm pk}$) in observer-frame 10-1000 $\rm keV$ energy band, fluence ($F_{\rm g}$) in observer-frame 20-2000 $\rm keV$ energy band, hardness ratio (HR) between observer-frame 100-2000 and 20-100 $\rm keV$ energy band, isotropic $\gamma$-ray energy ($E_{\rm iso}$)  in rest-frame $1-10^{\rm 4}$ $\rm keV$ energy band and peak luminosity ($L_{\rm pk}$)  in rest-frame 1-$10^{\rm 4}$ $\rm keV$ energy band; (3) We got all the histograms for every parameter in observer-frame and some parameters in rest-frame; (4) We got all the scatter plots between arbitrary two parameters with at least 5 GRBs; (5) We calculated the linear correlation coefficients and nonlinear correlation ratio between arbitrary two parameters with at least 5 GRBs, and we excluded some correlations with  trivial  correlations, like $T_{\rm 90}$ and $T_{\rm 90,i}$. We do not need to put such correlations in the results. We also considered all the errors using Monte Carlo (MC) method; (6) We did the linear regression between arbitrary two parameters and three parameters with at least 5 GRBs, we also excluded some trivial  results. We used MC method to include all the errors. Then we analyzed some interesting results. Because there are many data and results, we made figure sets and several machine readable tables, which can be found in the electric version.  Just a small part of the results are shown in this paper as examples. With this complete catalog, we can find more important relations, and we can reveal the GRBs' intrinsic properties.

We discussed the defects of this comprehensive sample, which are mainly from the uncertainty and inconsistent of different instruments, and the selection effect. To reveal more physical principles, one should try to classify the GRBs into more precise subgroups based on their physical origin, and the classification itself is a process to reveal the intrinsic properties of the GRBs. With the detailed classifications, the correlations inside each group may be more tighter and more physical. The  correlations are then can be used to study the radiation mechanism as well as the high energy radiation, to be indicators as standard candle or pseudo redshift, and to study the gravitational waves of compact binary mergers.

\section*{Acknowledgments}
We thank the anonymous referee for the critically reading the manuscript and suggesting substantial improvements.
YCZ thanks the helpful discussion with Tsvi Piran, Bing Zhang, Zigao Dai, Kwong Sang Cheng, Daming Wei, Yongfeng Huang, Xiangyu Wang, Xuefeng Wu, Yizhong Fan, Enwei Liang, Fayin Wang, Yunwei Yu, Shuangxi Yi, Shiyong Liu, Haijun Tian, Gaochao Liu, Jun Liu, Deyi Ma, Sheng Cui, Reetanjali Moharana, Dingxiong Wang, Weihua Lei, Qingwen Wu, Jumpei Takata, Yan Wang, Biping Gong, Wei Xie, Wei Chen, Chao Yang, Lixiong Gan, Jiuzhou Wang, Wenbo Ma, Jun Tian and Shuaibing Ma. YCZ also thanks Jing Lv, Jingwen Xing, Yanhui Han, Max Oberndorfer, Chujun Yi, Zhengfu Xiong, Hualei Wang, Shaoping Huang, Xiaohao Cui and Yuan Xue for their statistical studies on GRBs. Their works were mainly for their bachelor's degree theses or part of their researches. Those were prototypes or trials in different aspects of this comprehensive statistical work, since it was conceived in 2010 and was started in 2012. The data were collected manually, were stored in Google spreadsheet, and were checked by eyes and by codes. The codes in this work were compiled in R and Python. This work is supported by the National Basic Research Program of China (973 Program, Grant No. 2014CB845800), by the National Natural Science Foundation of China (Grants Nos. U1738132, 11773010, 11601267 and U1231101) and by the Humanity and Social Science Foundation of MOE of China (20171304).

\bibliographystyle{aasjournal}
\nocite{*}
\bibliography{bigtable}

\setlength{\voffset}{-25mm}

\clearpage

\maxdeadcycles=1000

\clearpage

\clearpage

\clearpage

\clearpage

\clearpage

\clearpage

\clearpage

\clearpage

\clearpage

\clearpage

\clearpage

\clearpage

\clearpage

\clearpage

\clearpage

\clearpage

\clearpage

\clearpage

\clearpage

\clearpage

\clearpage

\clearpage

\clearpage

\clearpage

\clearpage

\clearpage

\clearpage

\clearpage

\clearpage

\clearpage

\clearpage

\clearpage

\clearpage

\clearpage

\clearpage

\clearpage

\clearpage

\clearpage

\clearpage

\clearpage

\clearpage
\setlength{\voffset}{-25mm}

\clearpage
\begin{figure*}

\includegraphics[width=0.45\textwidth]{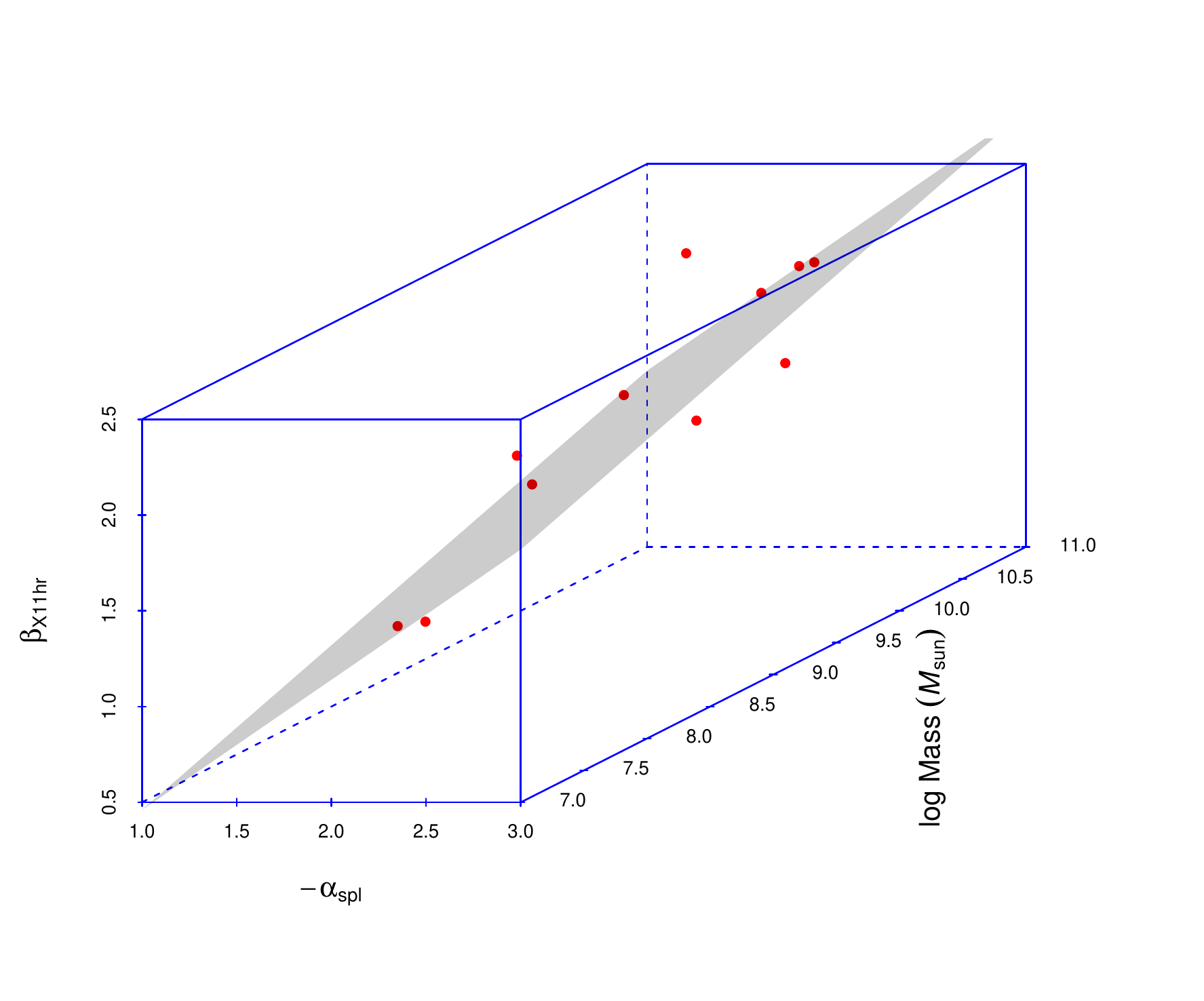}
\includegraphics[width=0.45\textwidth]{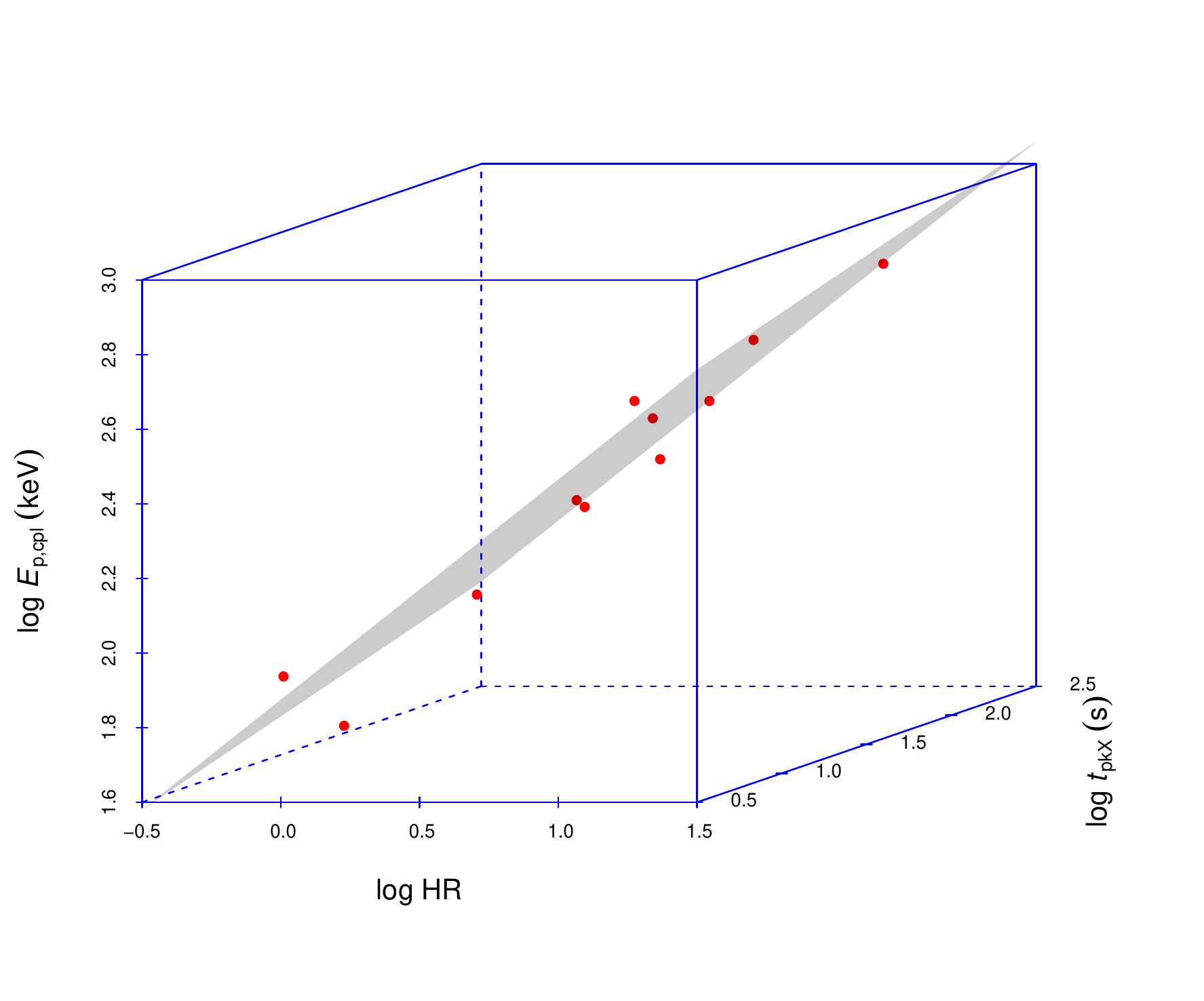}

\includegraphics[width=0.45\textwidth]{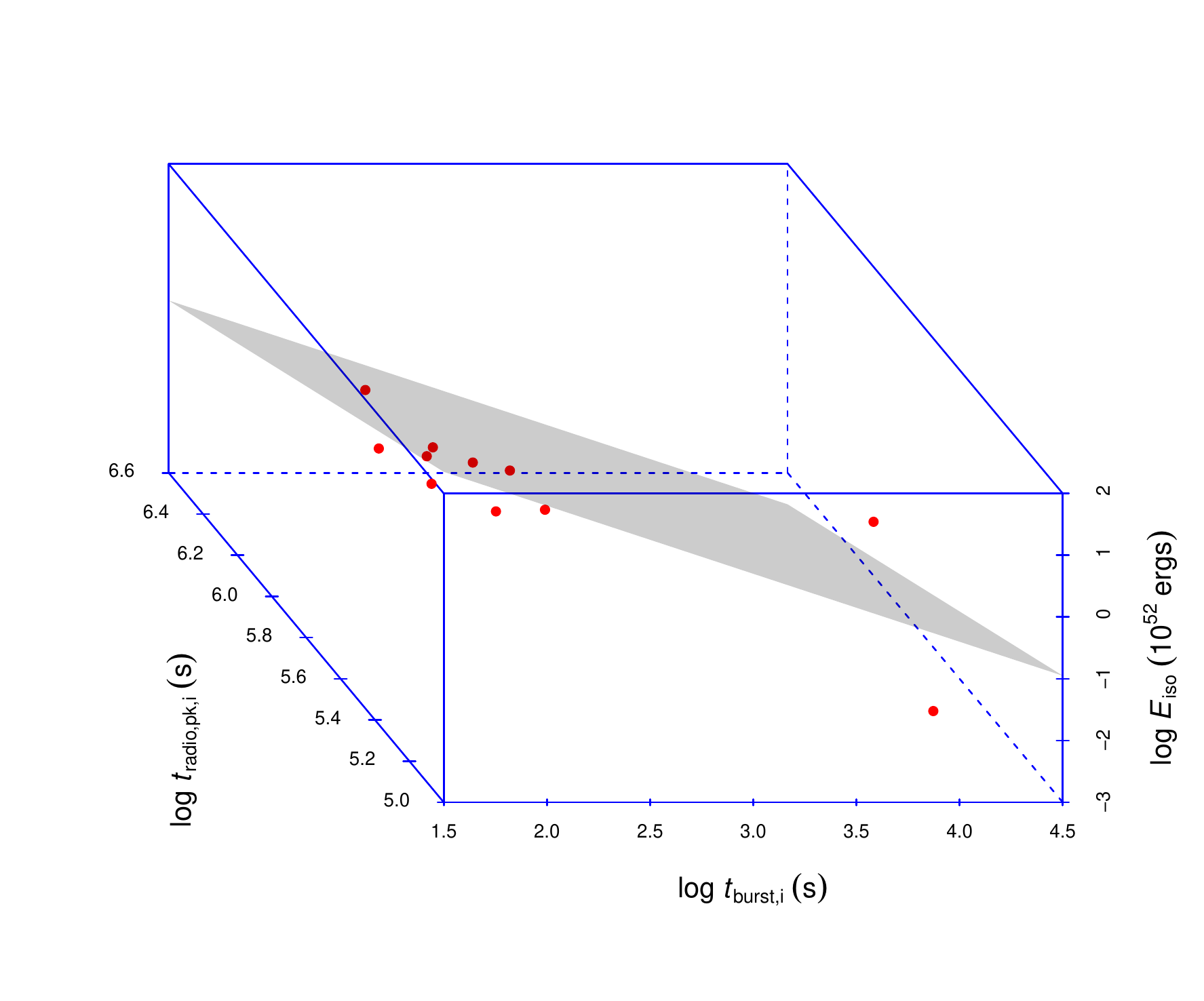}
\includegraphics[width=0.45\textwidth]{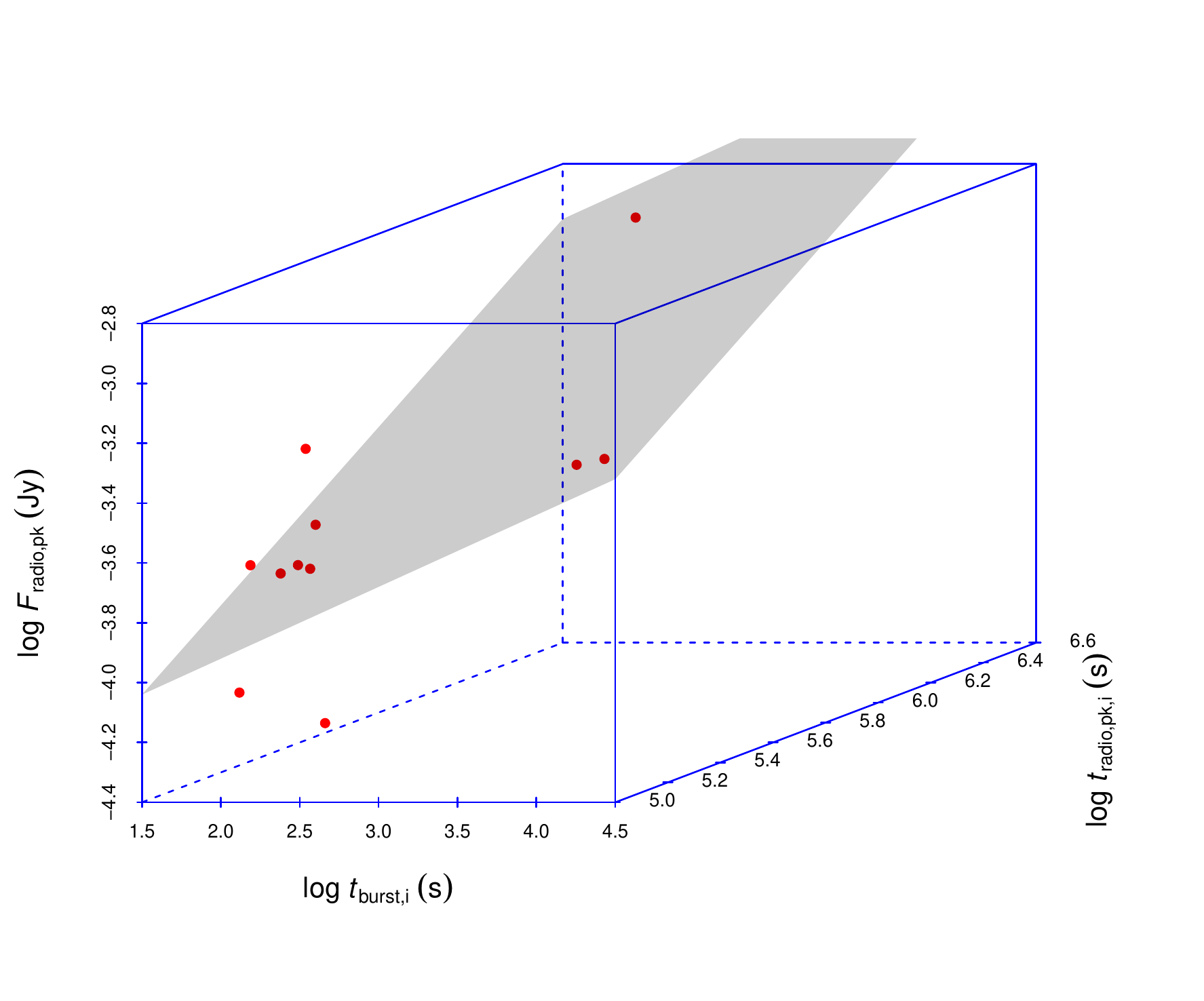}

\includegraphics[width=0.45\textwidth]{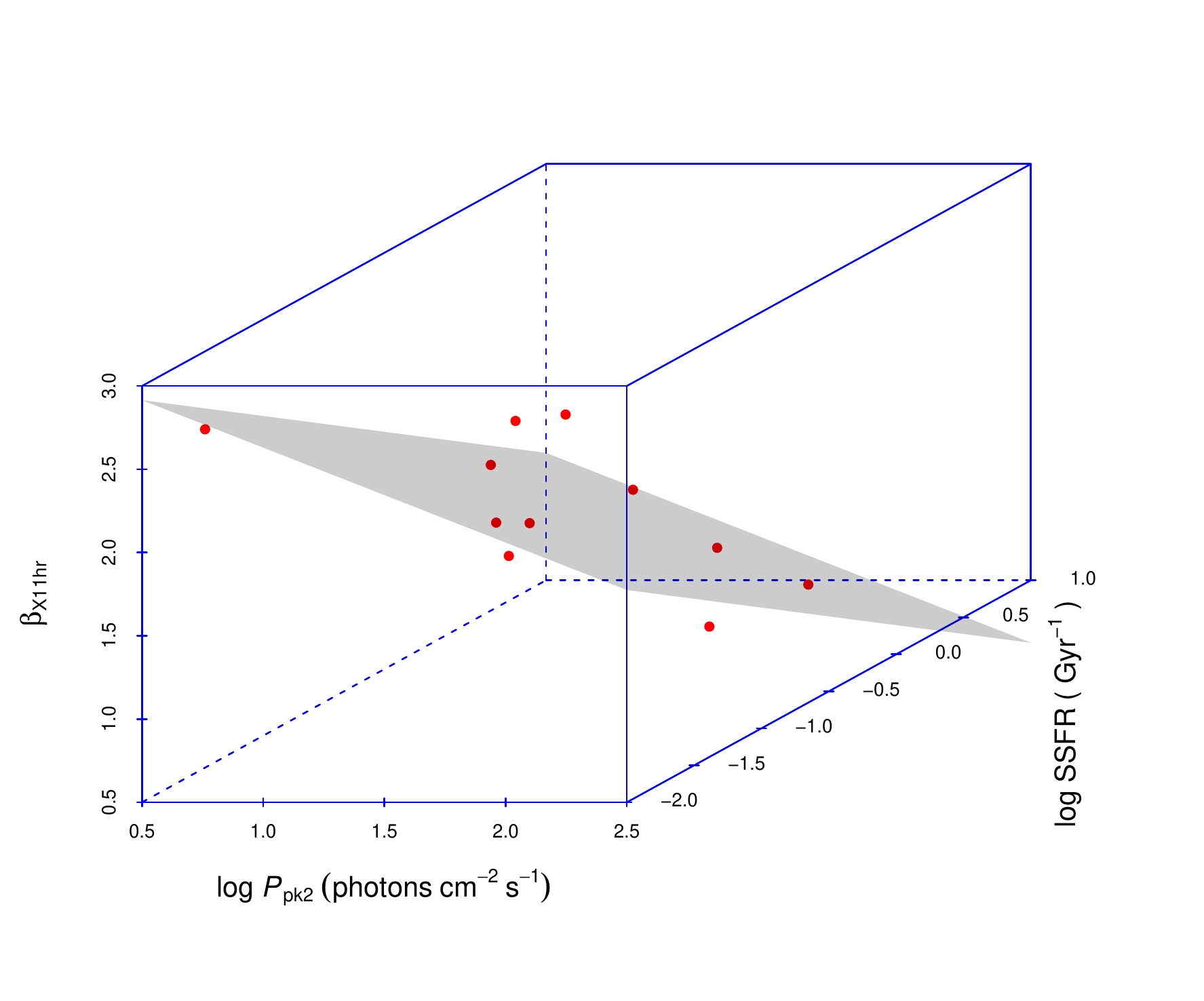}
\includegraphics[width=0.45\textwidth]{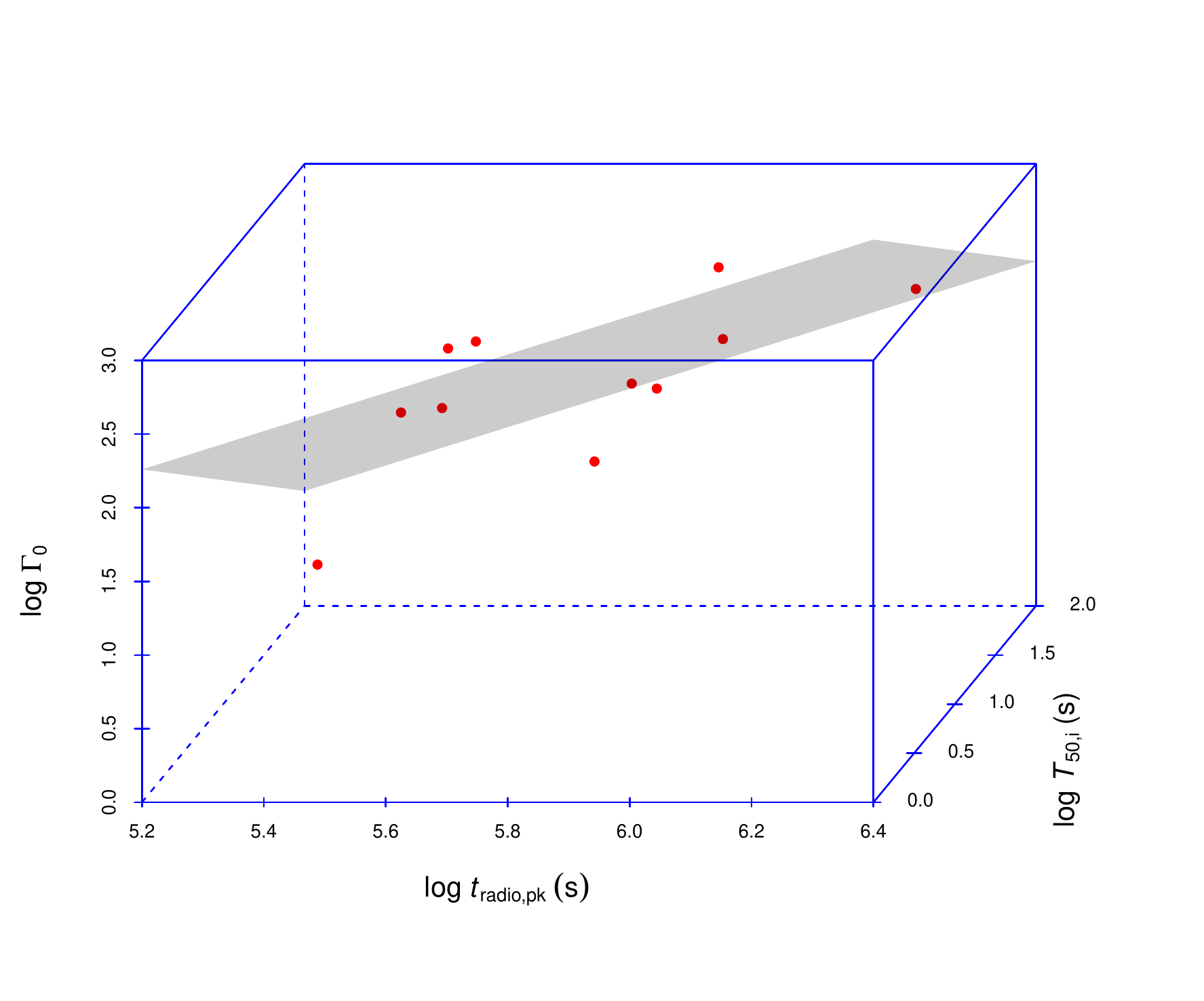}

\center{Fig. \ref{fig:three}---Continued}
\end{figure*}


\clearpage
\begin{figure*}

\includegraphics[width=0.45\textwidth]{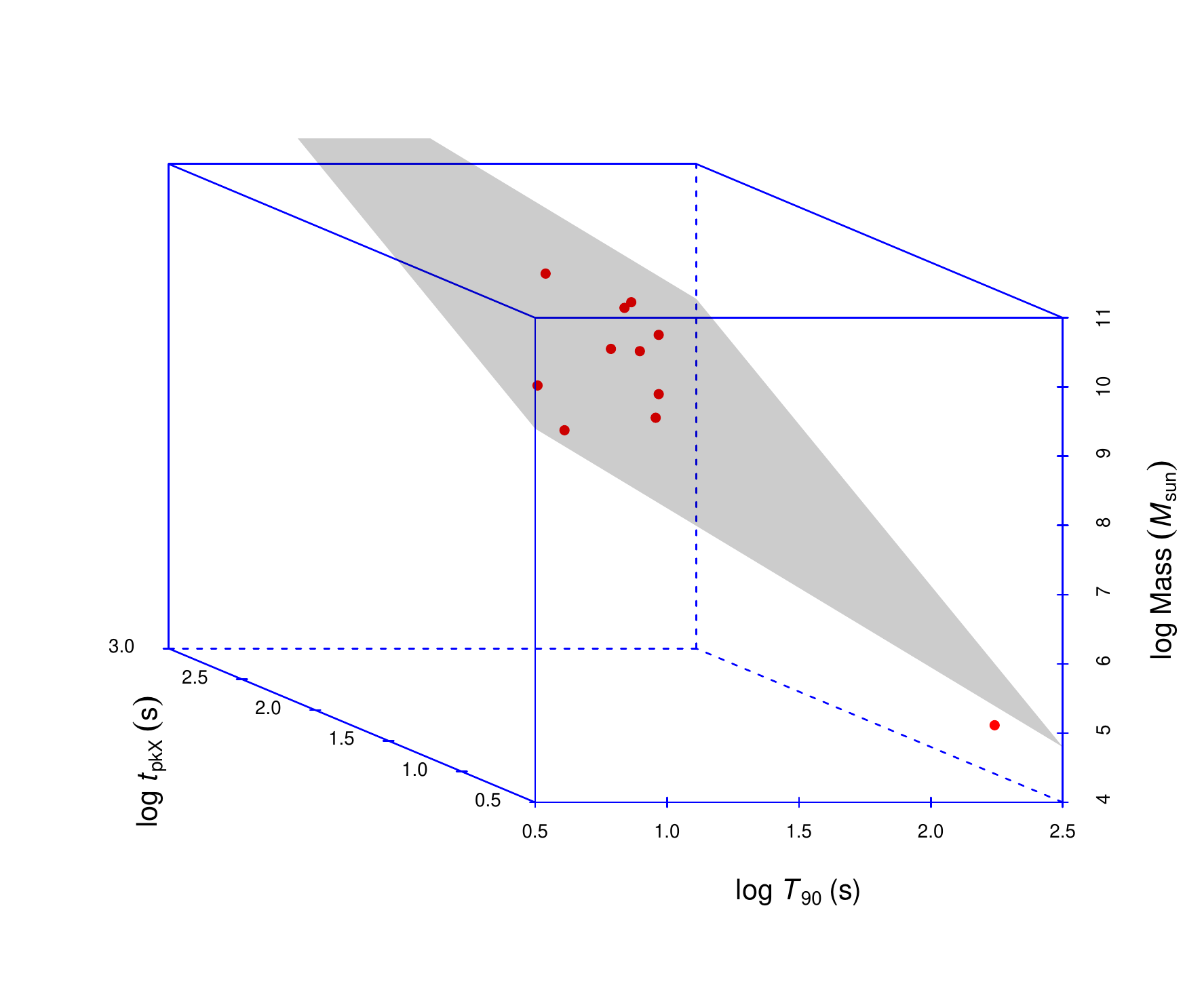}
\includegraphics[width=0.45\textwidth]{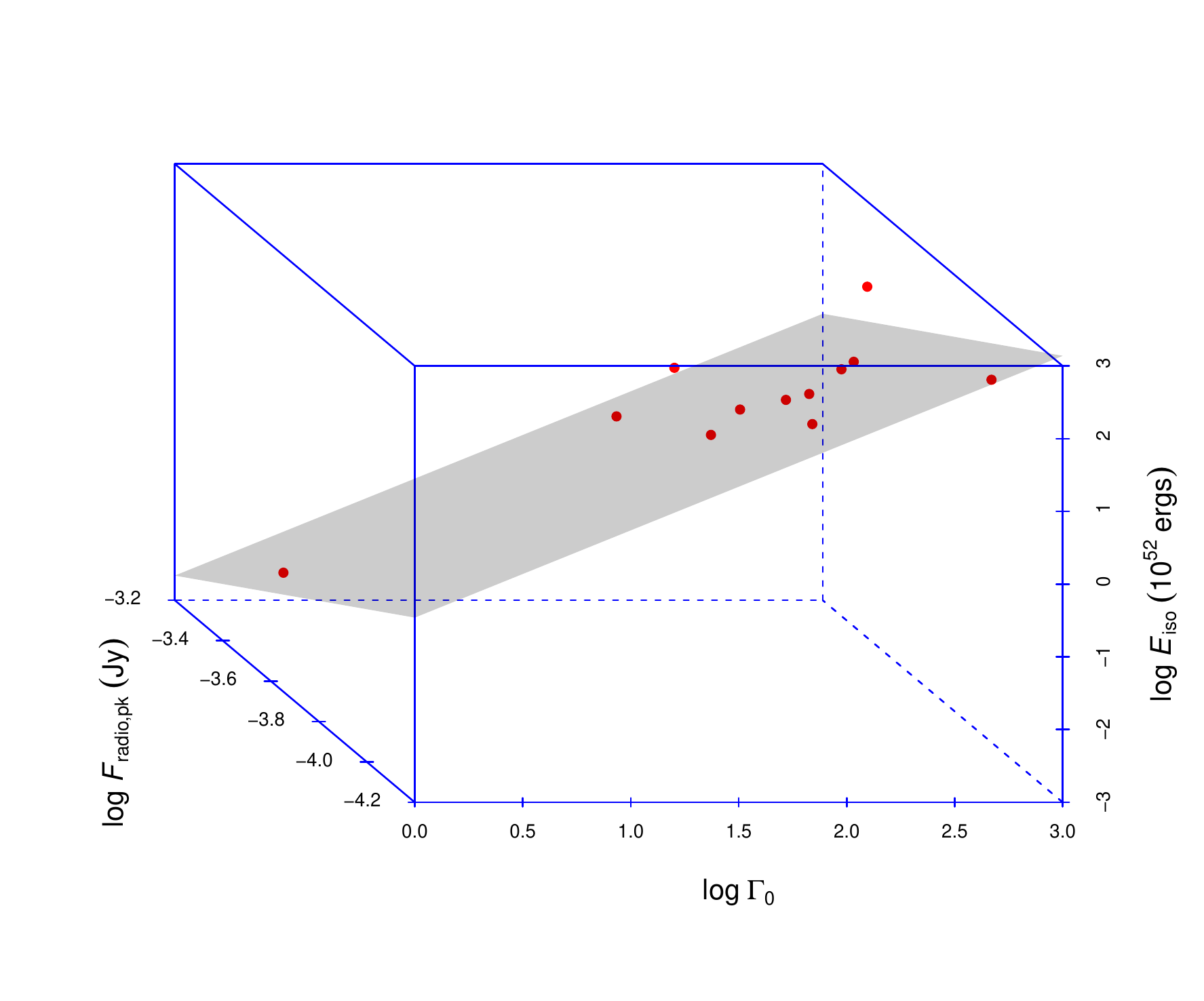}

\includegraphics[width=0.45\textwidth]{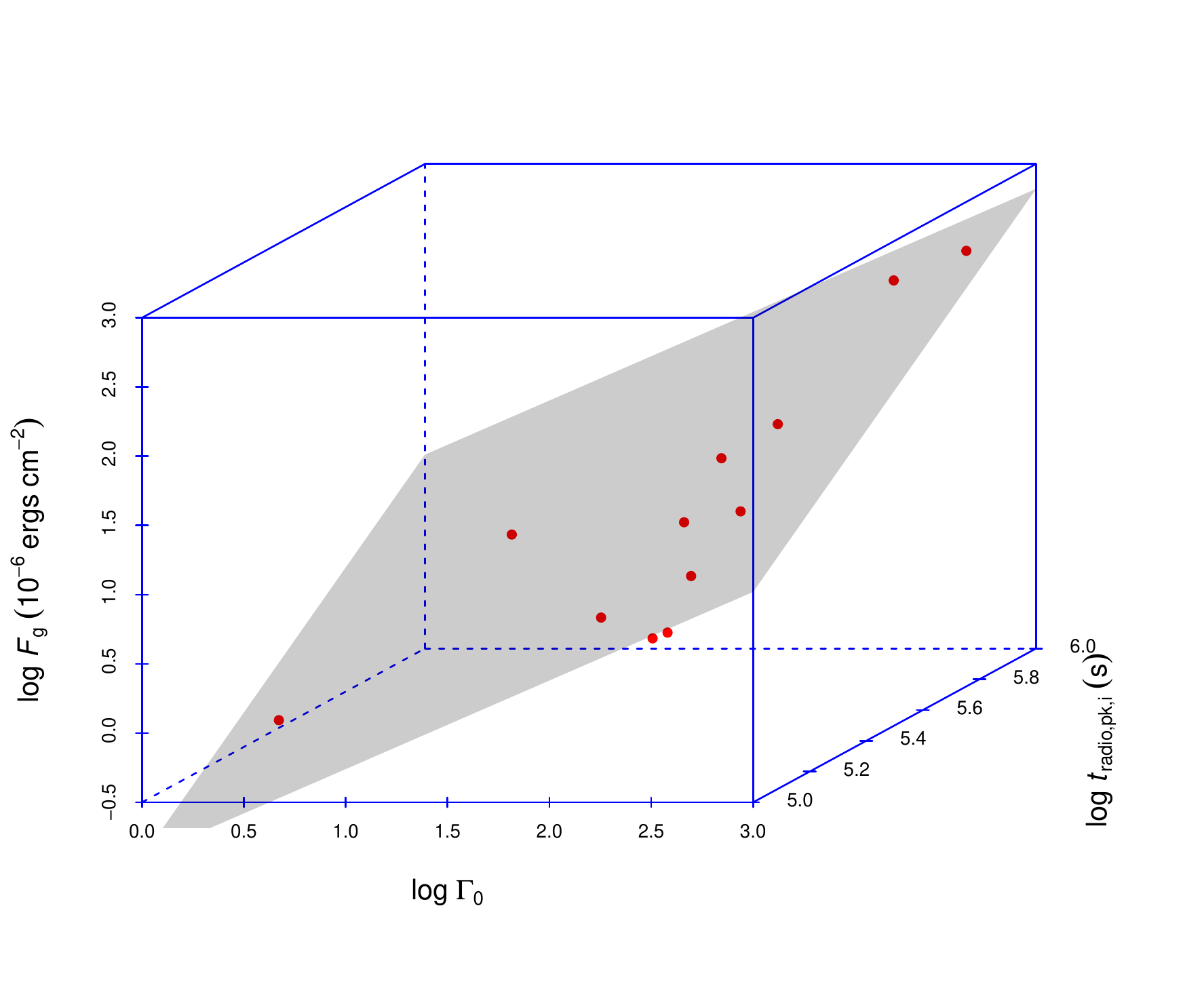}
\includegraphics[width=0.45\textwidth]{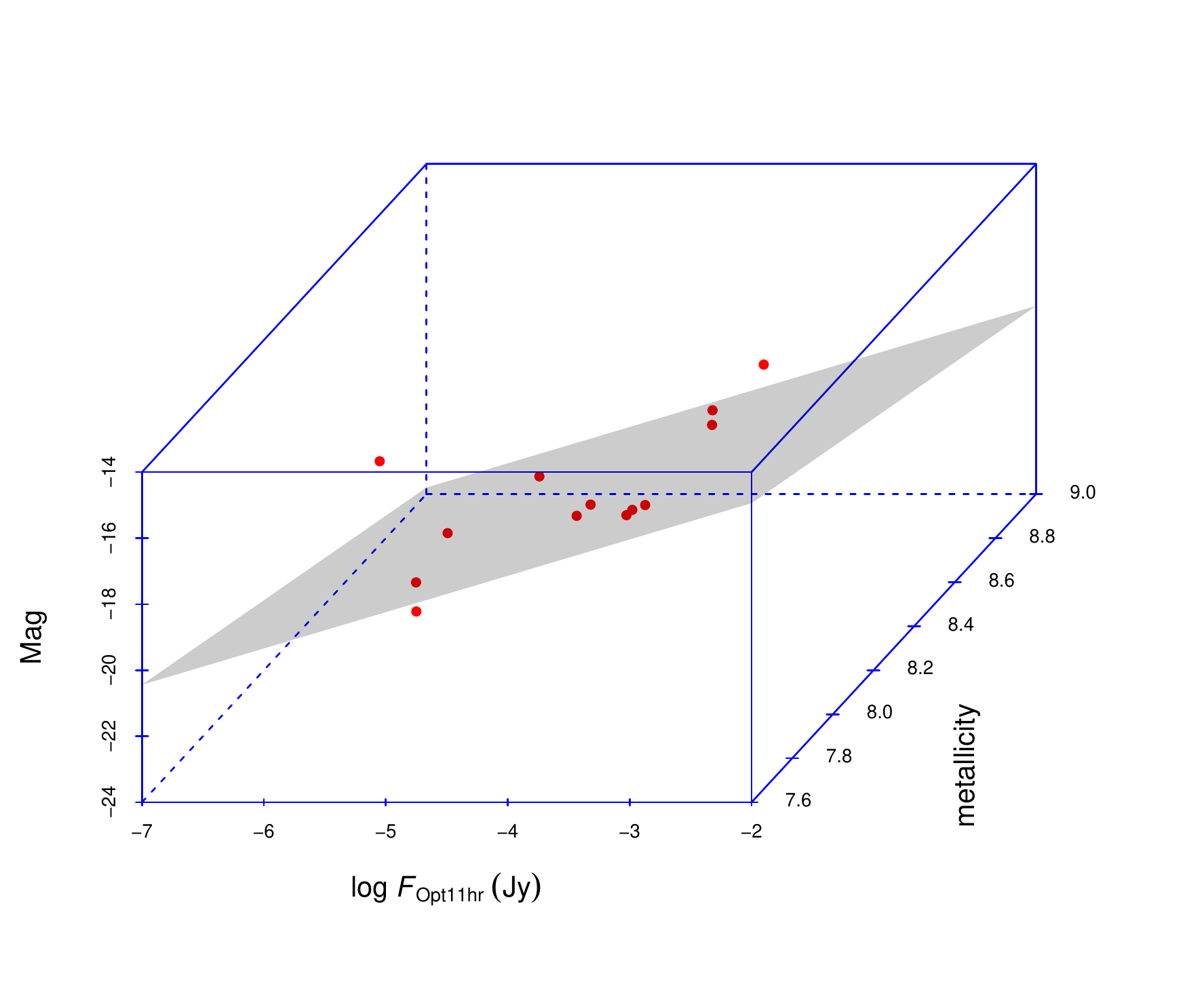}

\includegraphics[width=0.45\textwidth]{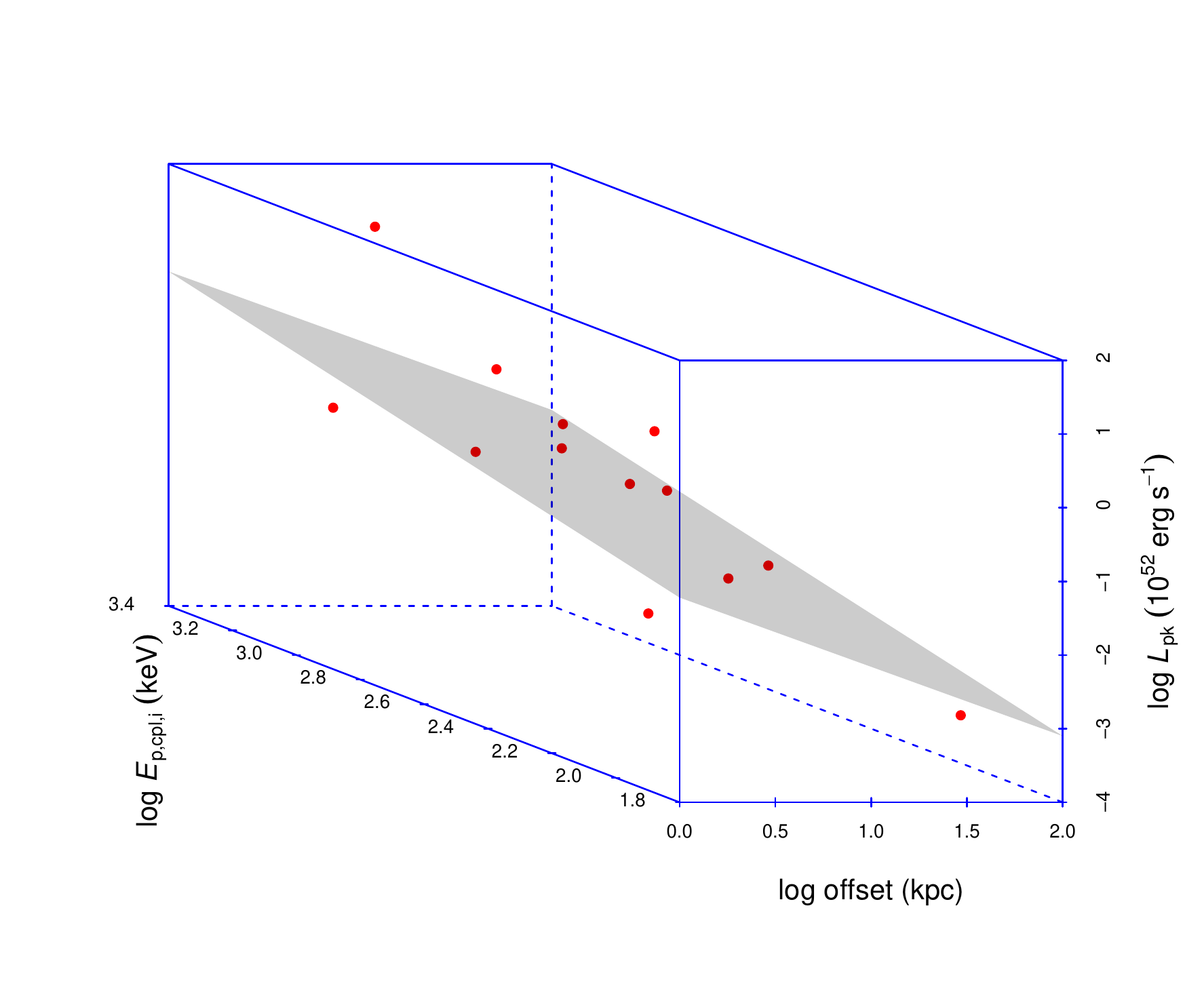}
\includegraphics[width=0.45\textwidth]{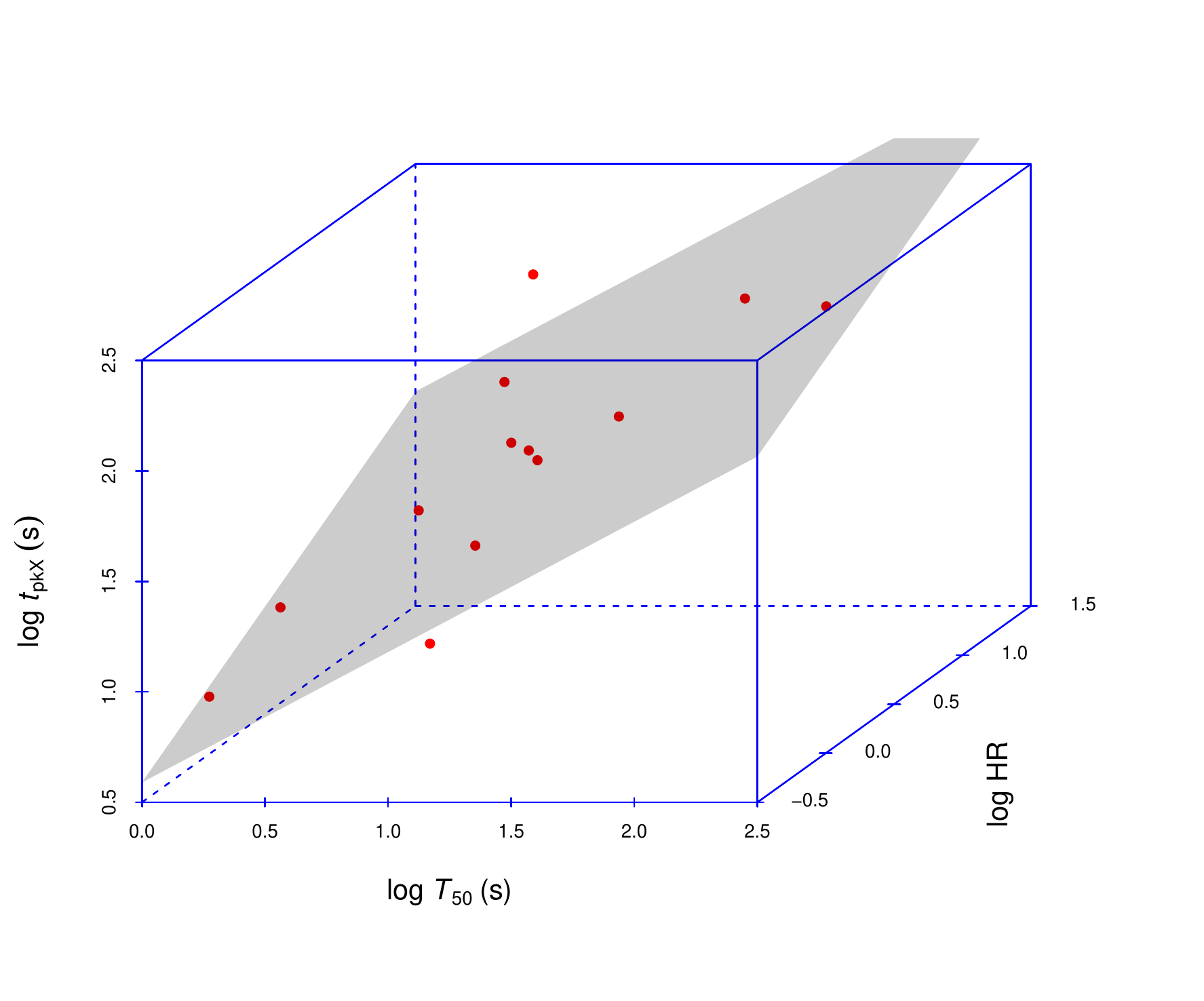}

\center{Fig. \ref{fig:three}---Continued}
\end{figure*}


\clearpage
\begin{figure*}

\includegraphics[width=0.45\textwidth]{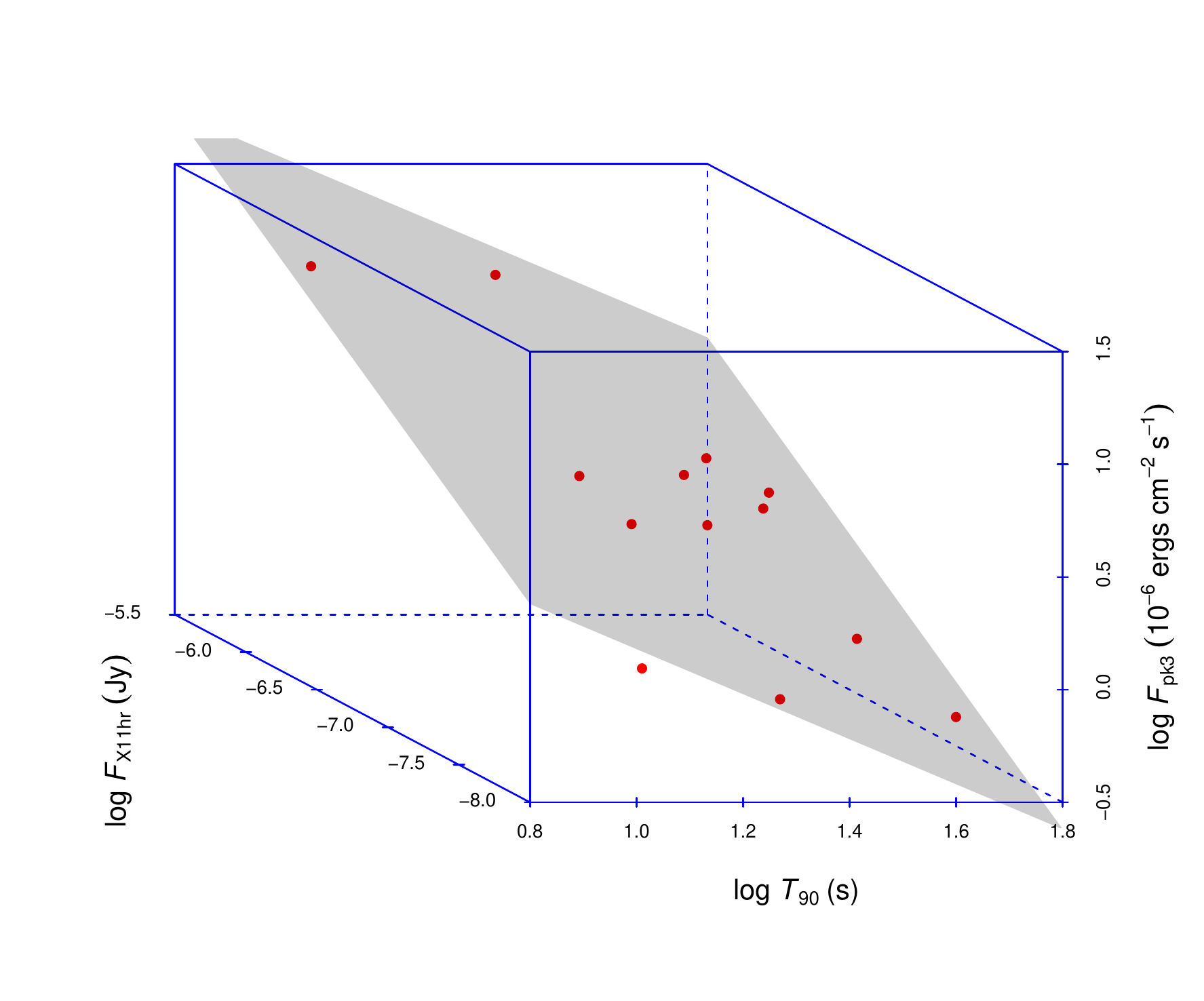}
\includegraphics[width=0.45\textwidth]{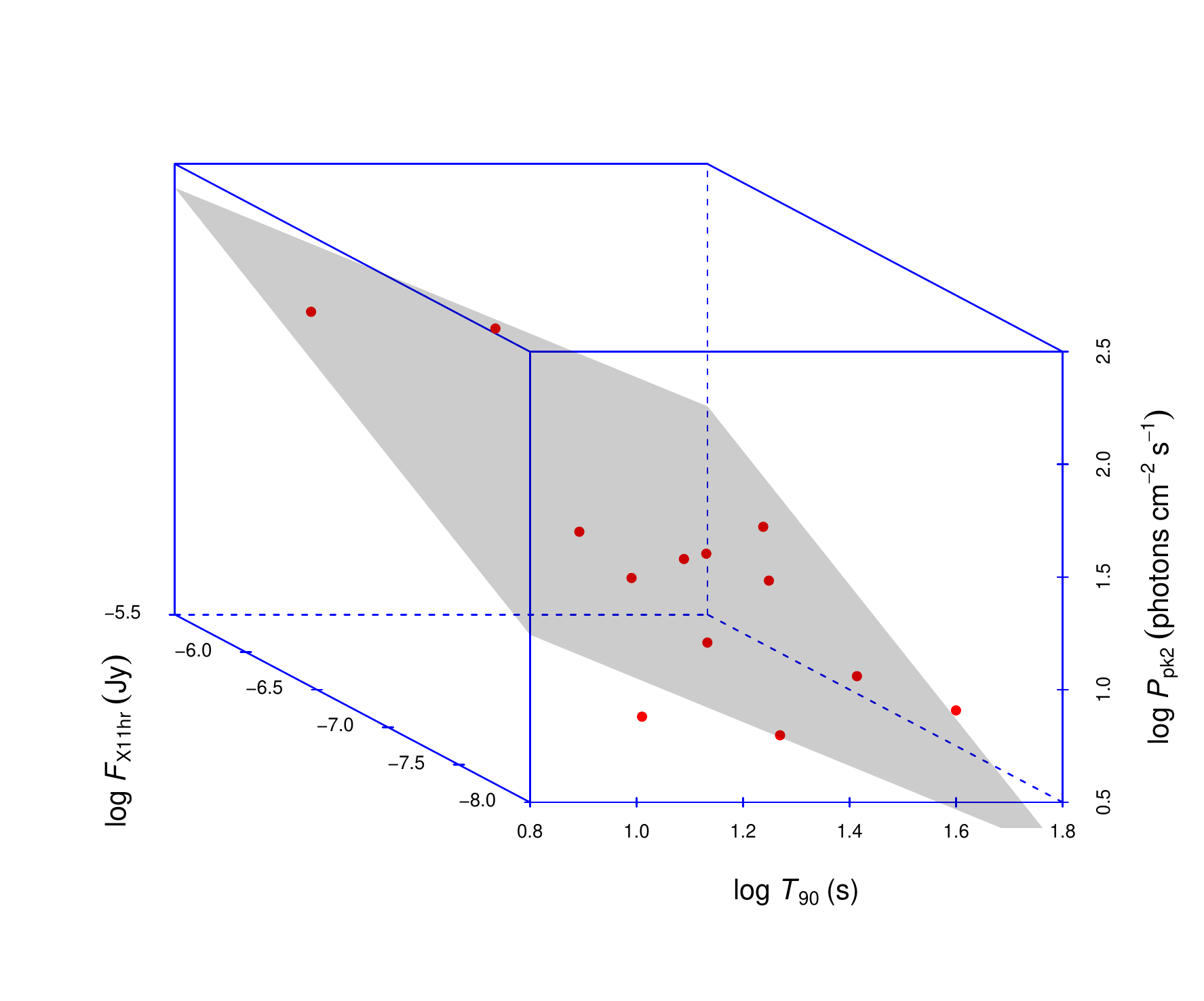}

\includegraphics[width=0.45\textwidth]{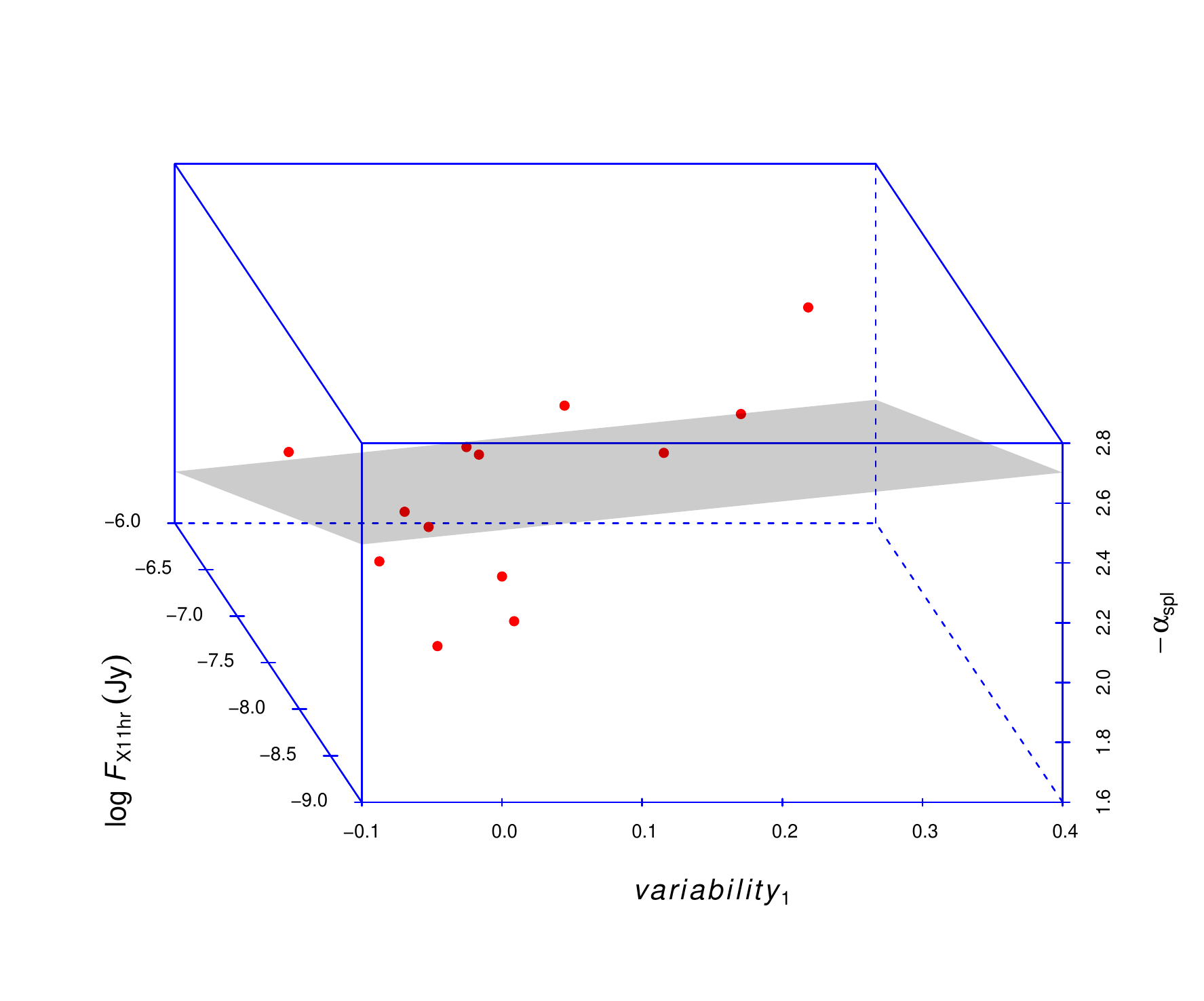}
\includegraphics[width=0.45\textwidth]{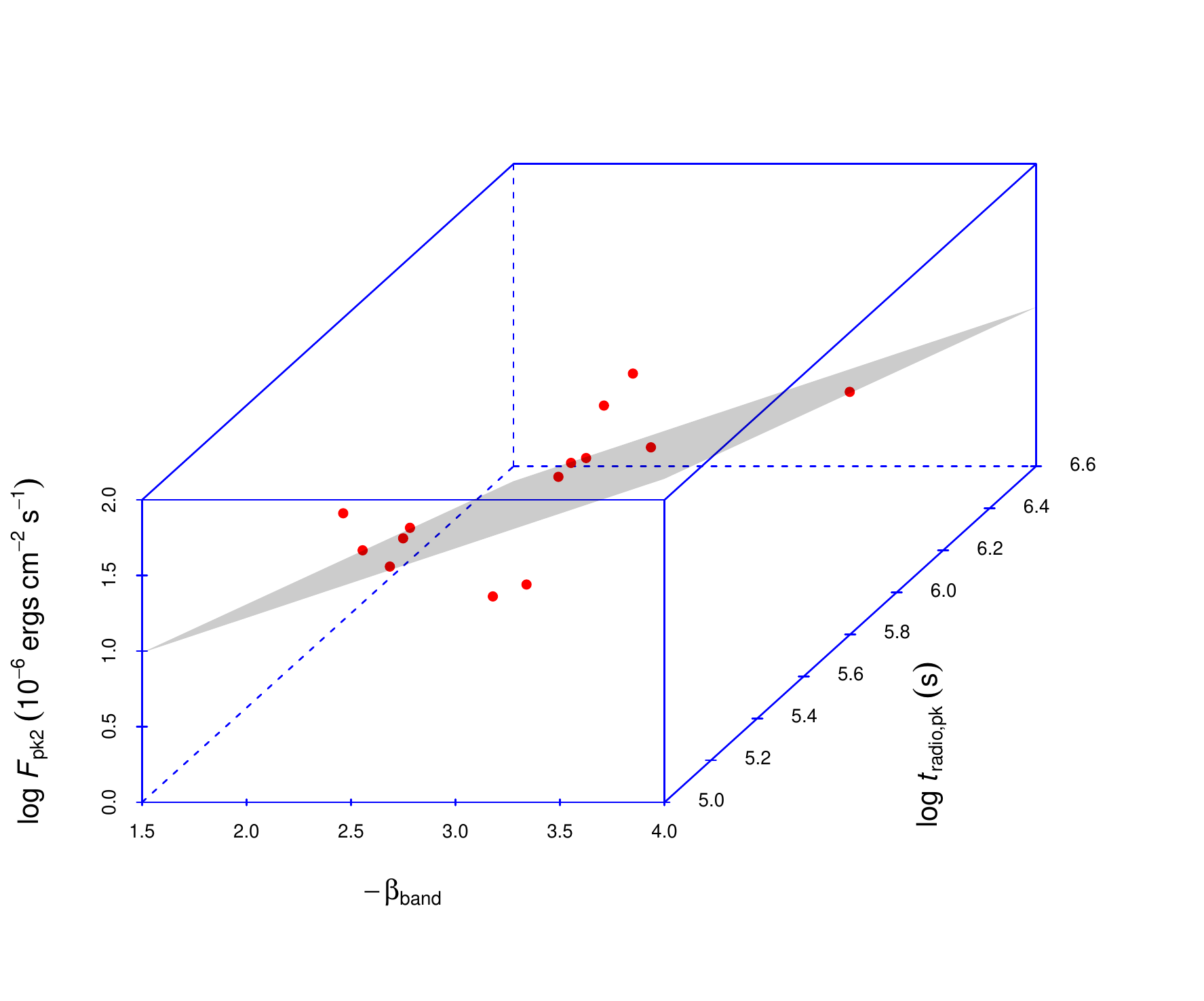}

\includegraphics[width=0.45\textwidth]{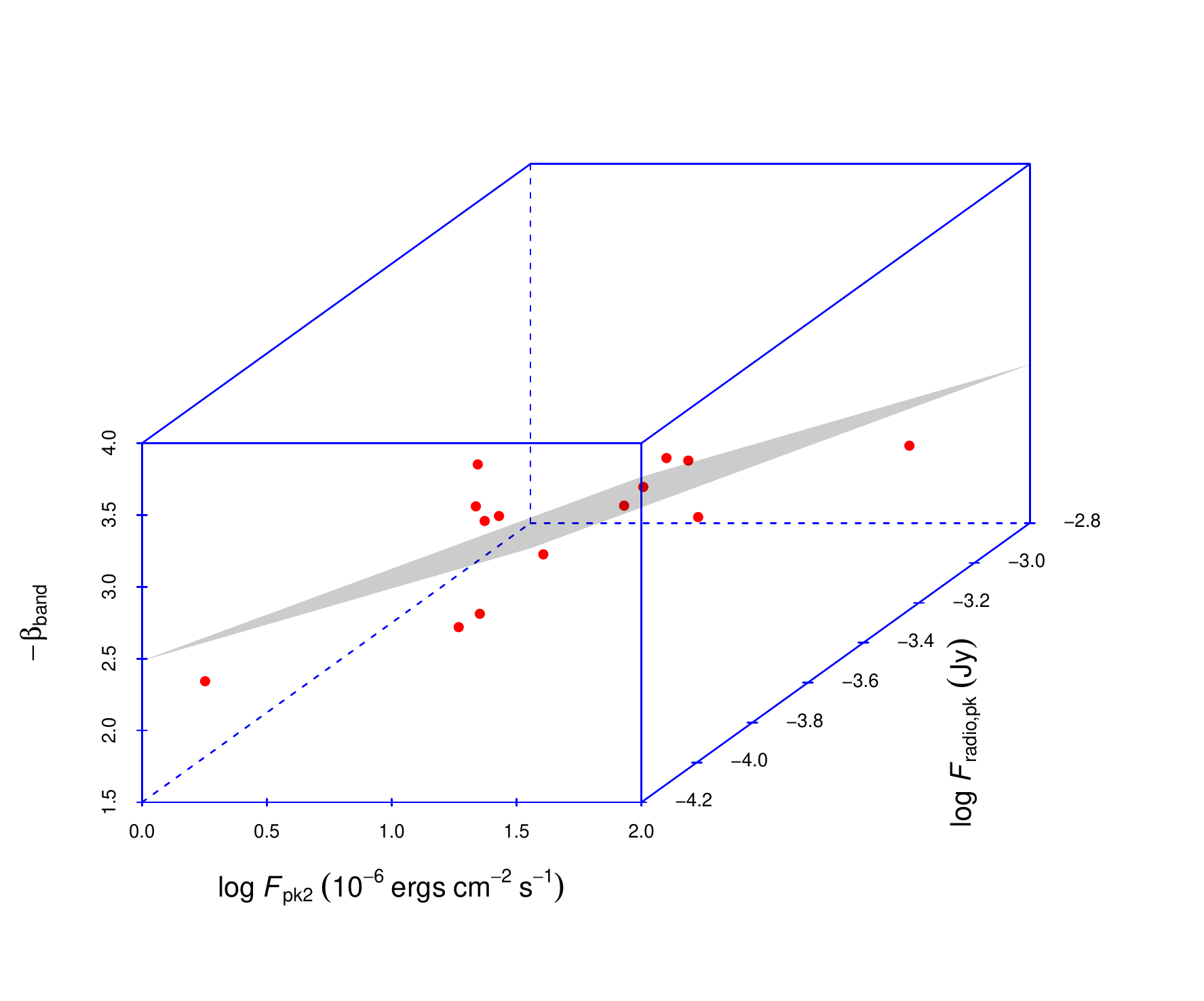}
\includegraphics[width=0.45\textwidth]{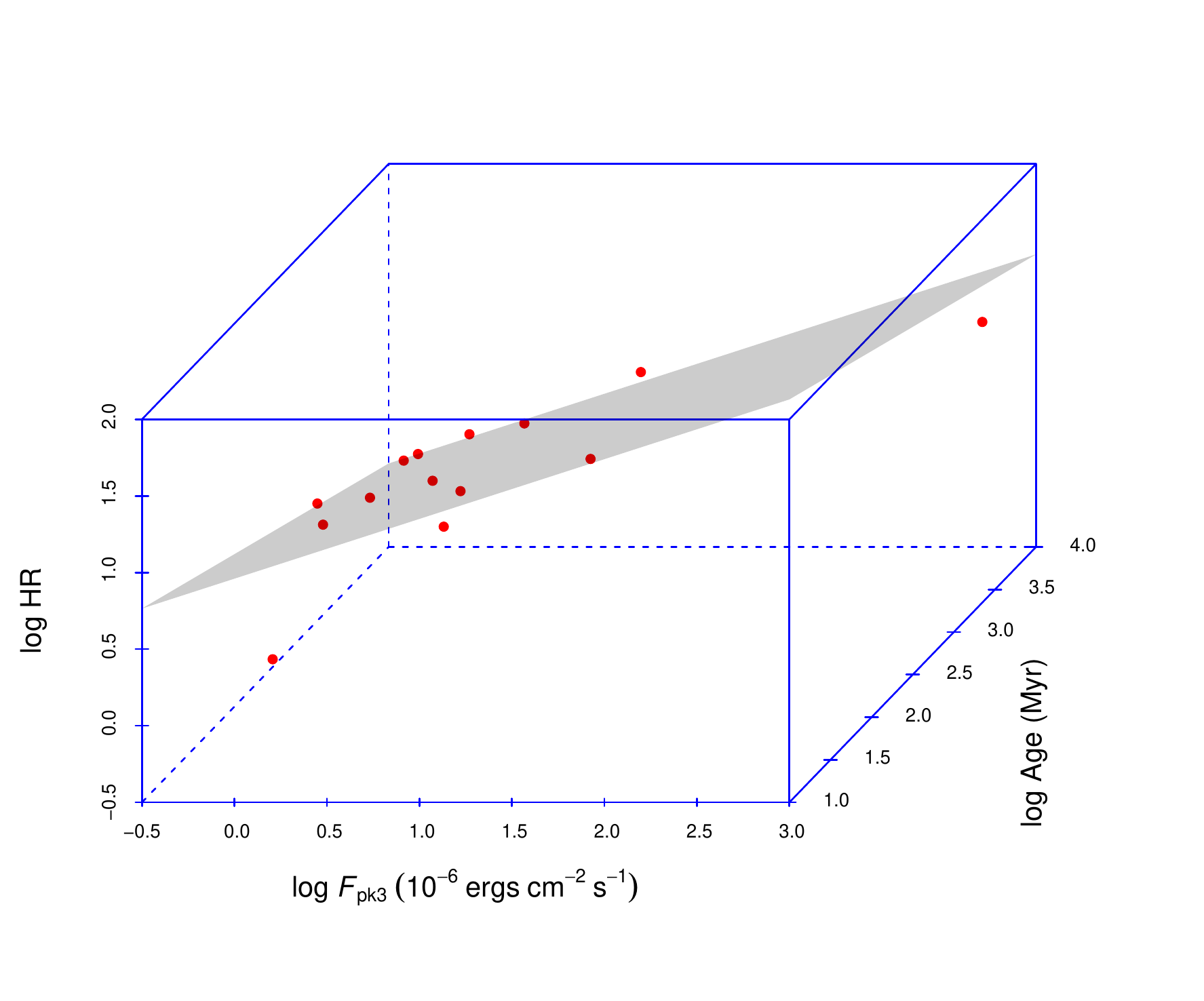}

\center{Fig. \ref{fig:three}---Continued}
\end{figure*}


\clearpage
\begin{figure*}

\includegraphics[width=0.45\textwidth]{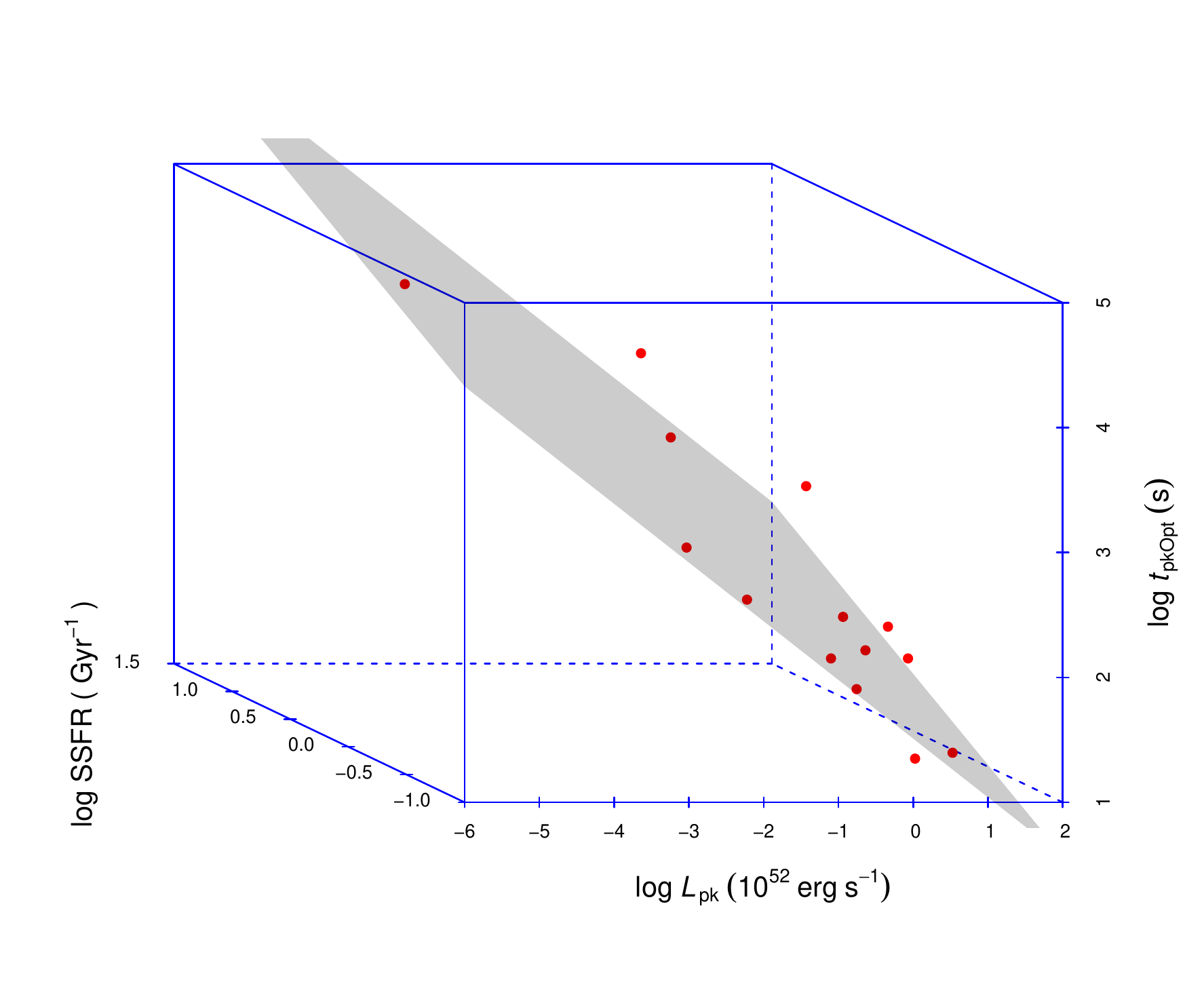}
\includegraphics[width=0.45\textwidth]{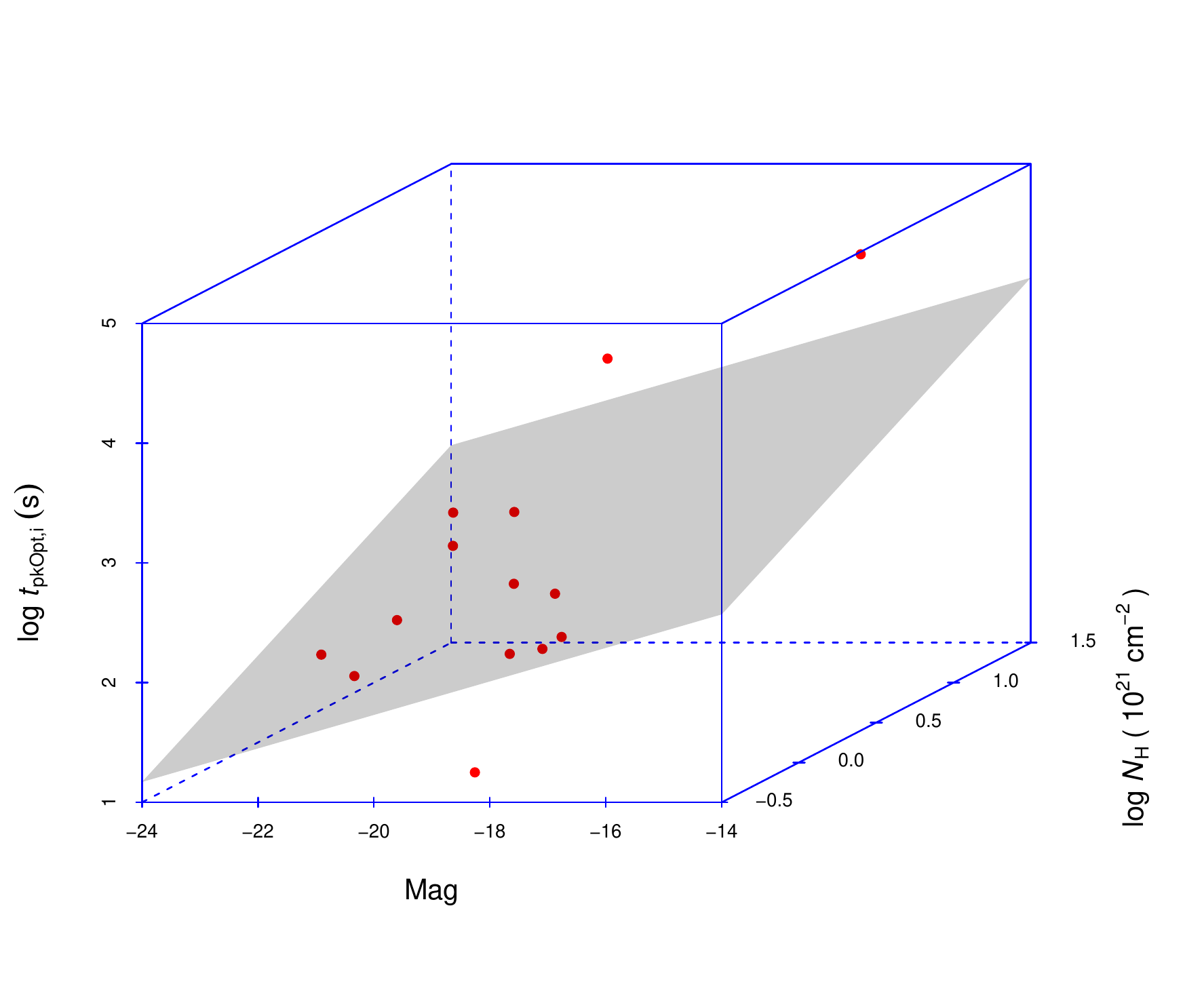}

\includegraphics[width=0.45\textwidth]{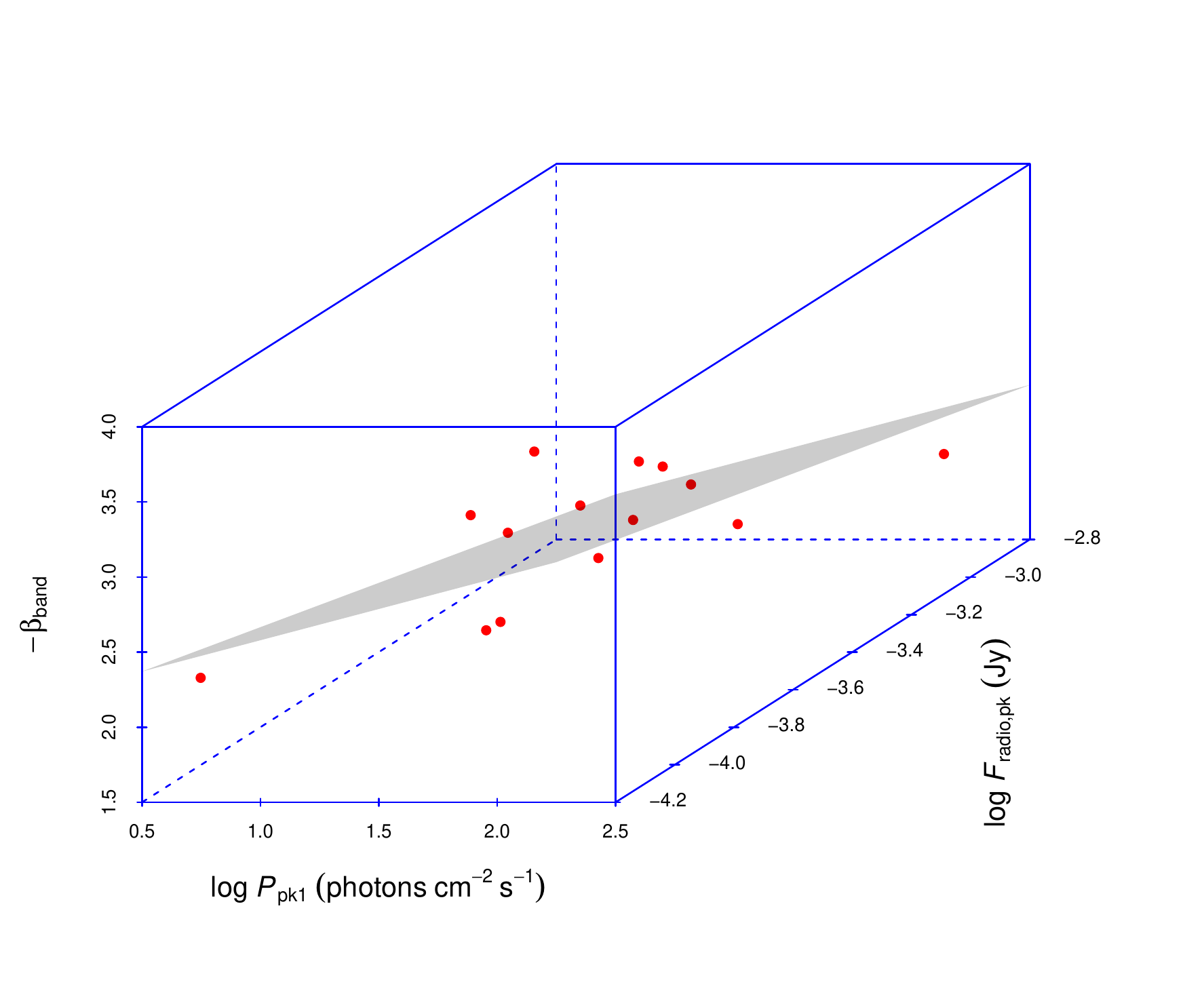}
\includegraphics[width=0.45\textwidth]{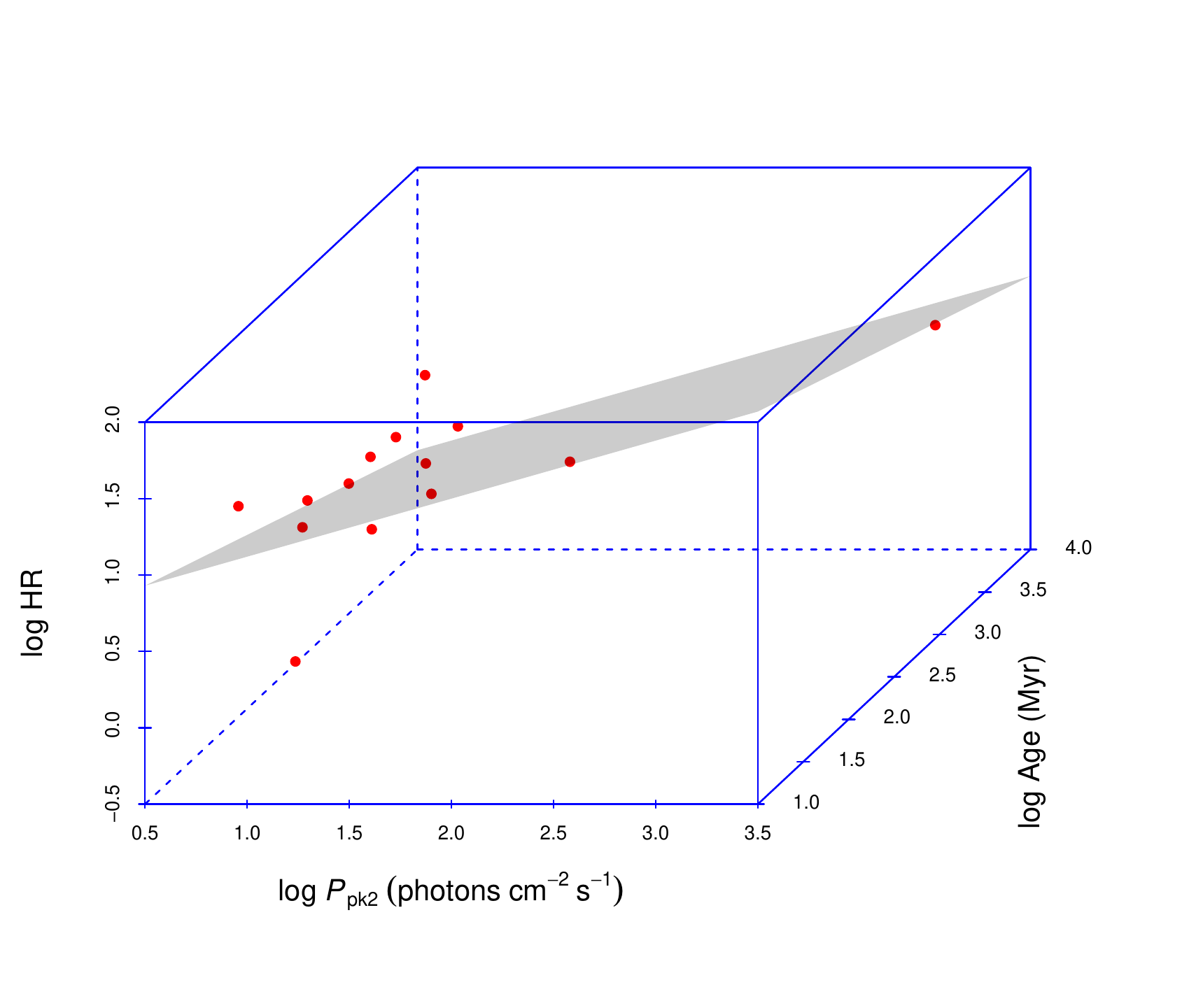}

\includegraphics[width=0.45\textwidth]{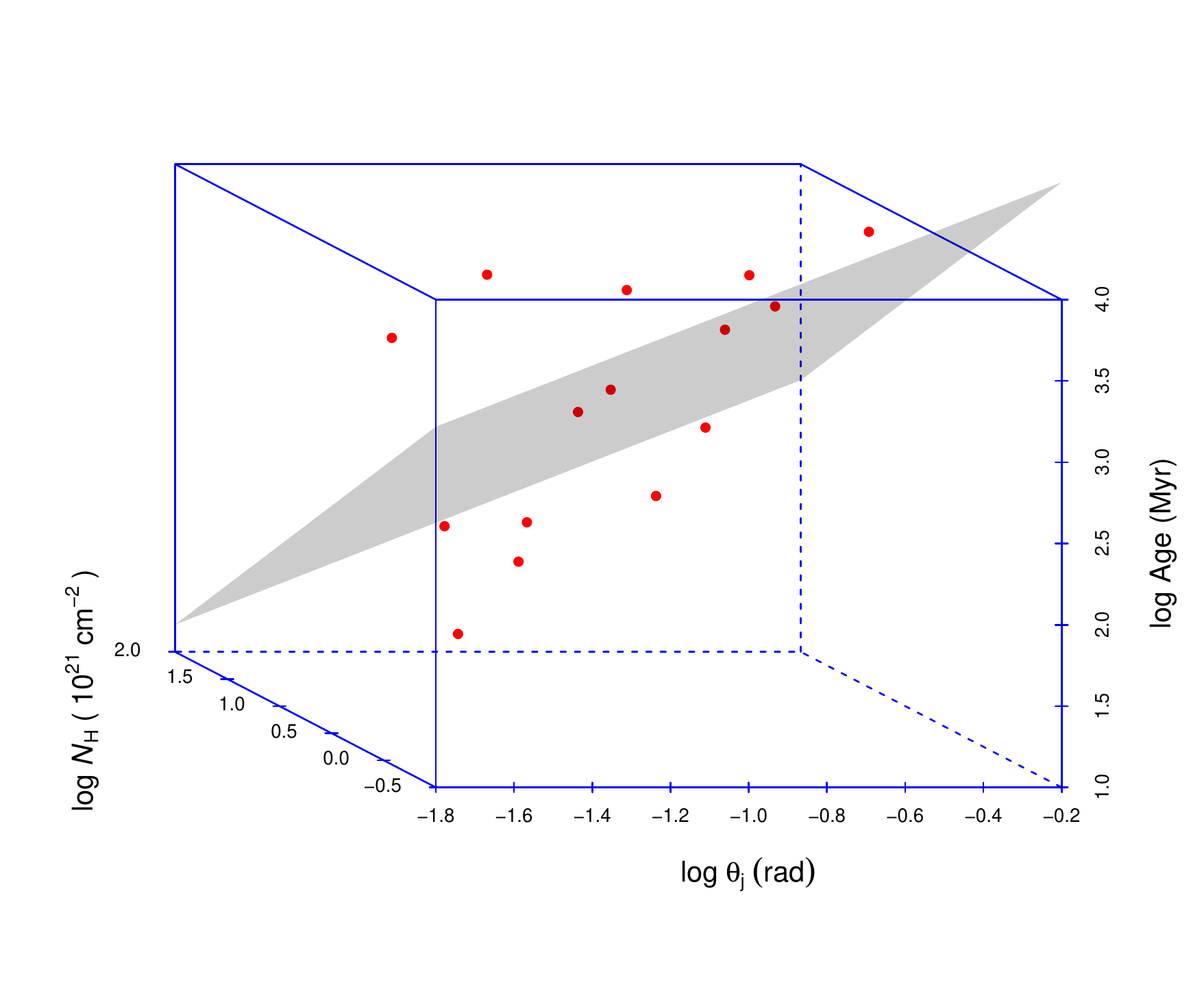}
\includegraphics[width=0.45\textwidth]{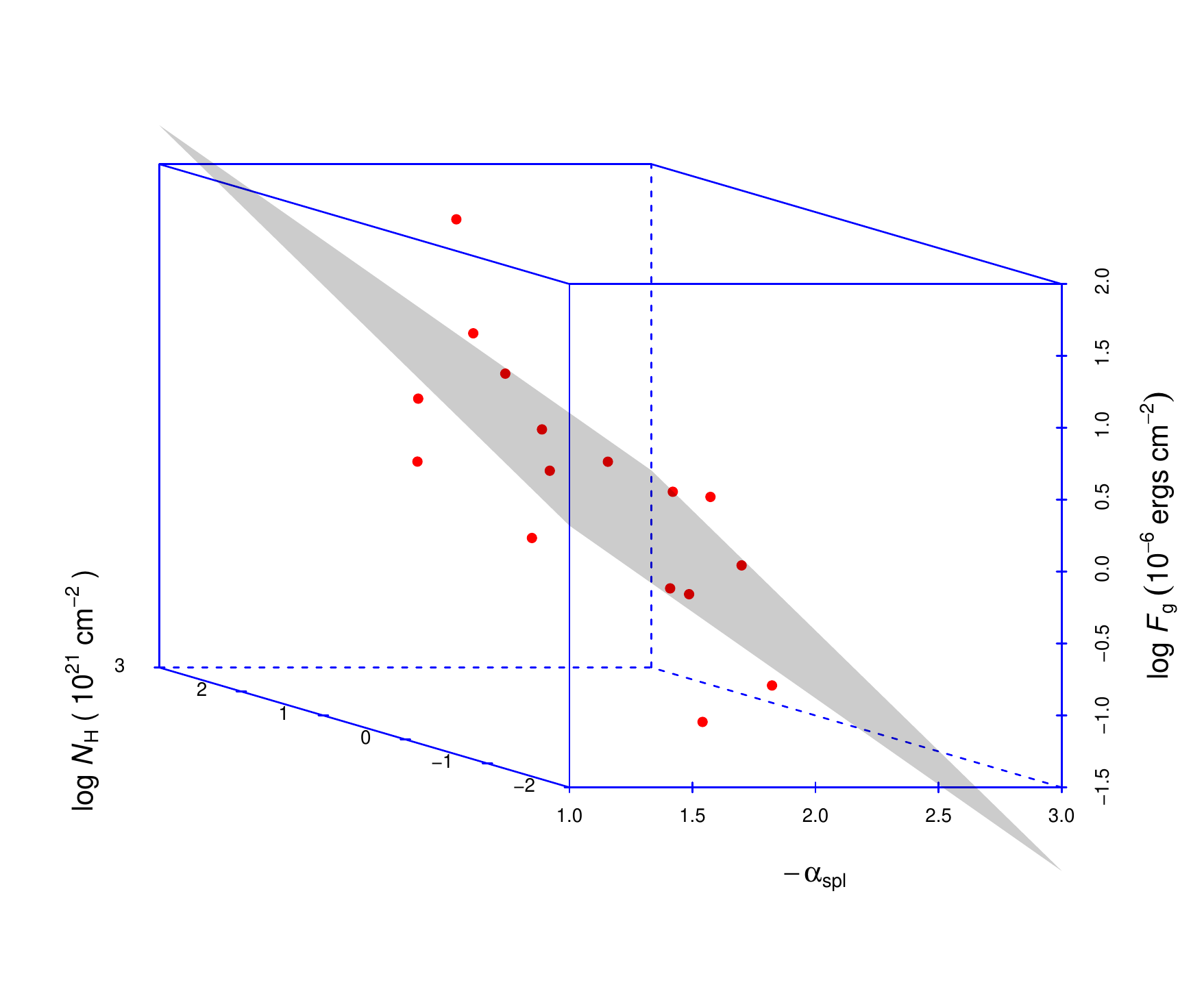}

\center{Fig. \ref{fig:three}---Continued}
\end{figure*}


\clearpage
\begin{figure*}

\includegraphics[width=0.45\textwidth]{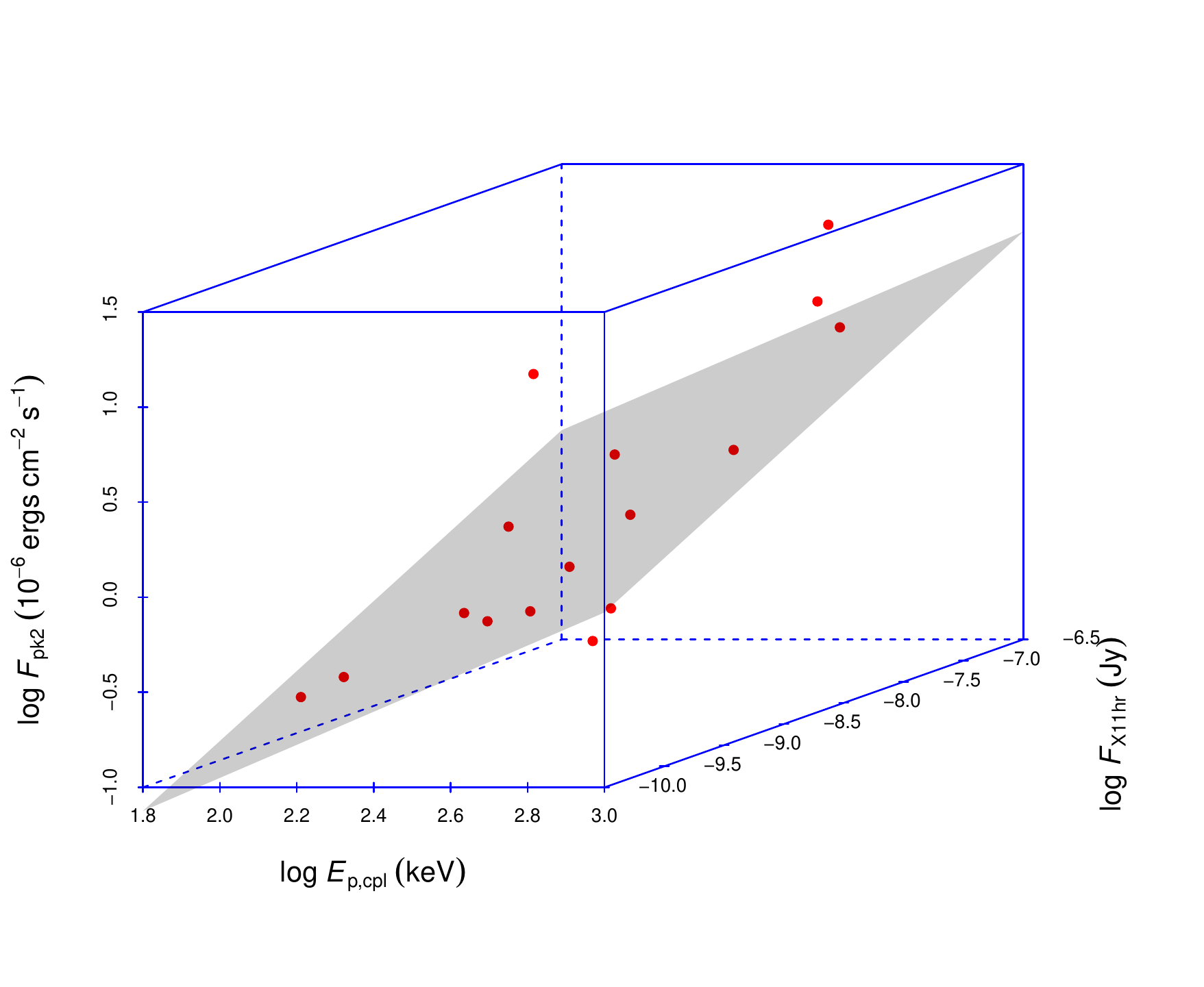}
\includegraphics[width=0.45\textwidth]{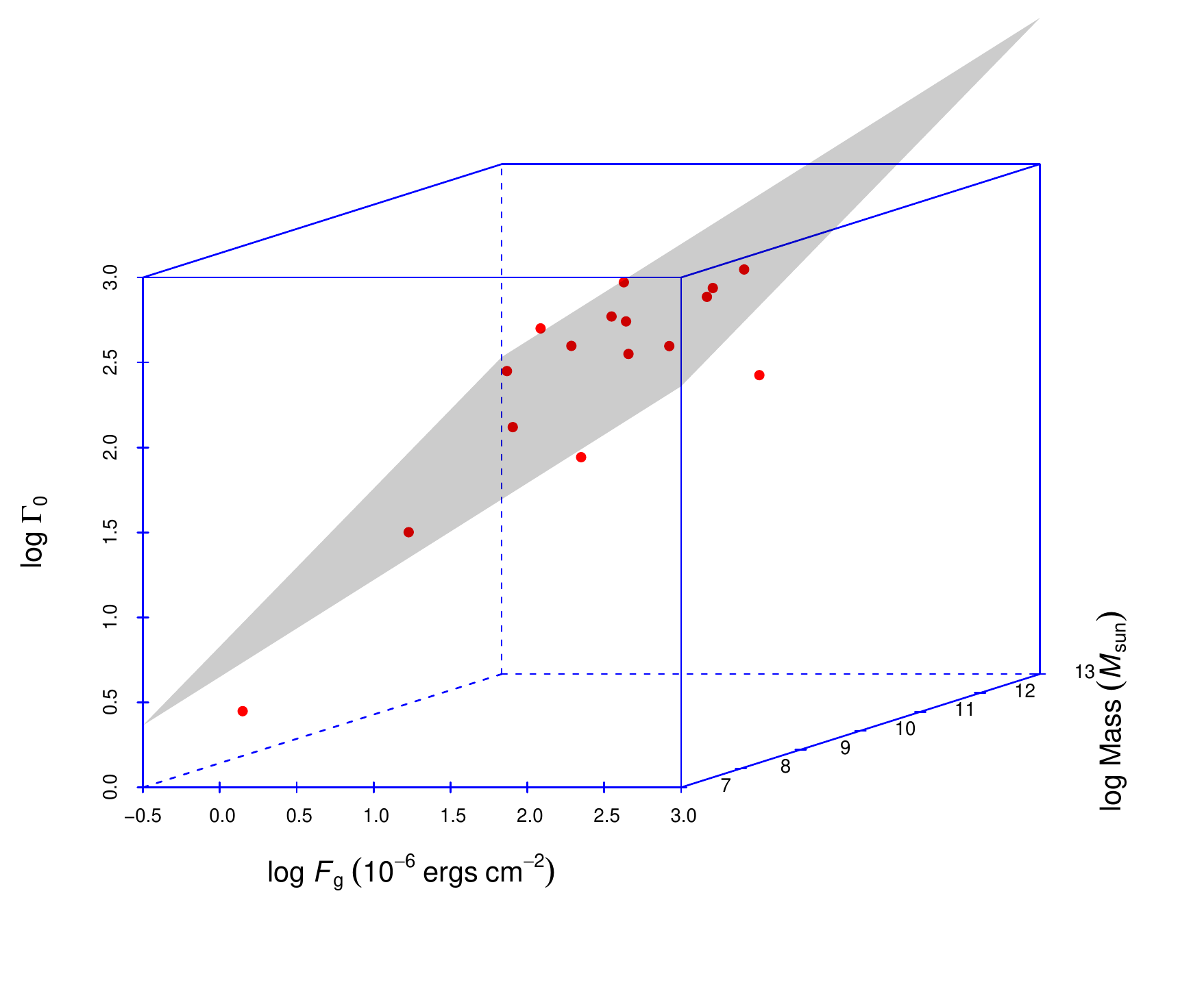}

\includegraphics[width=0.45\textwidth]{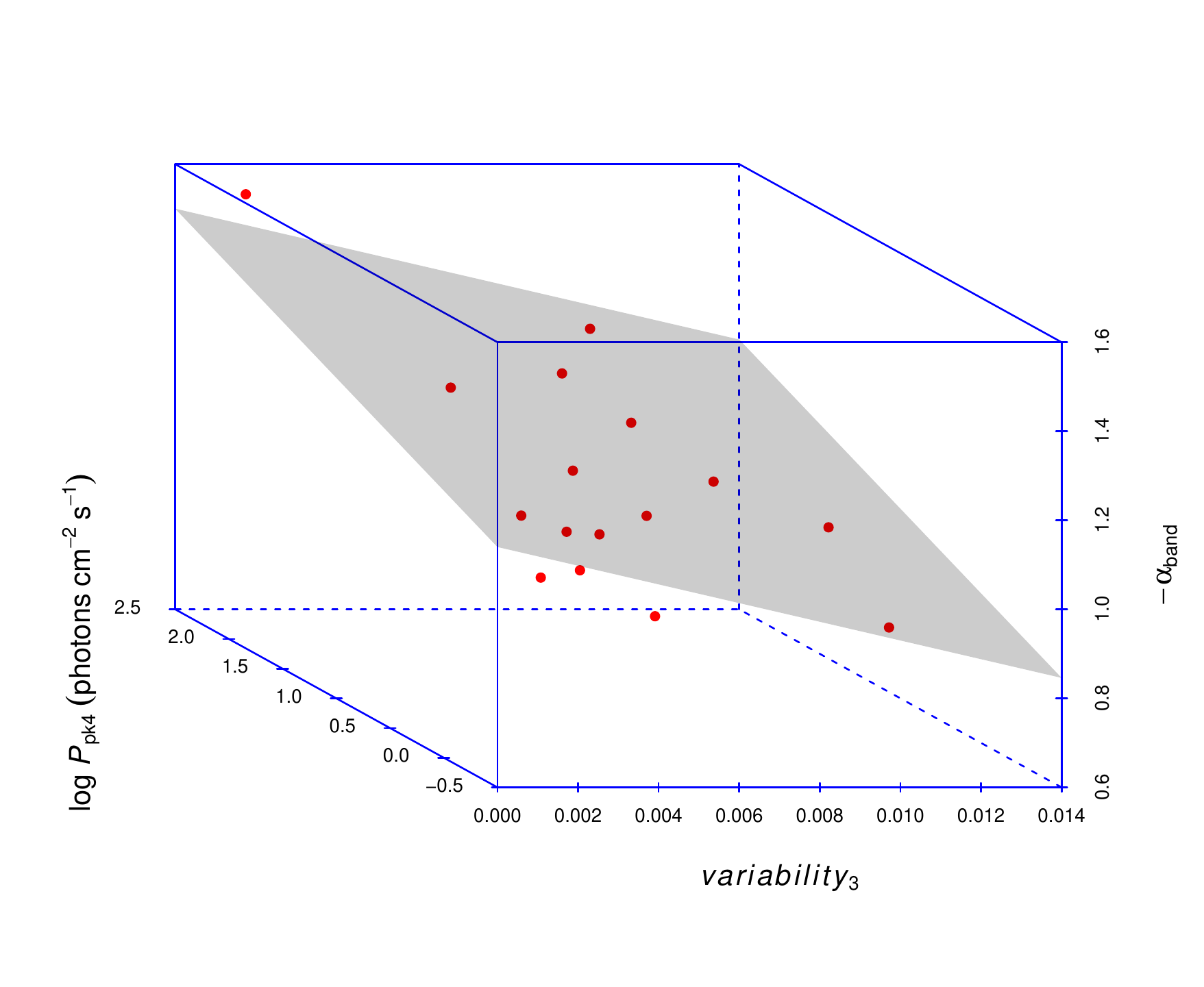}
\includegraphics[width=0.45\textwidth]{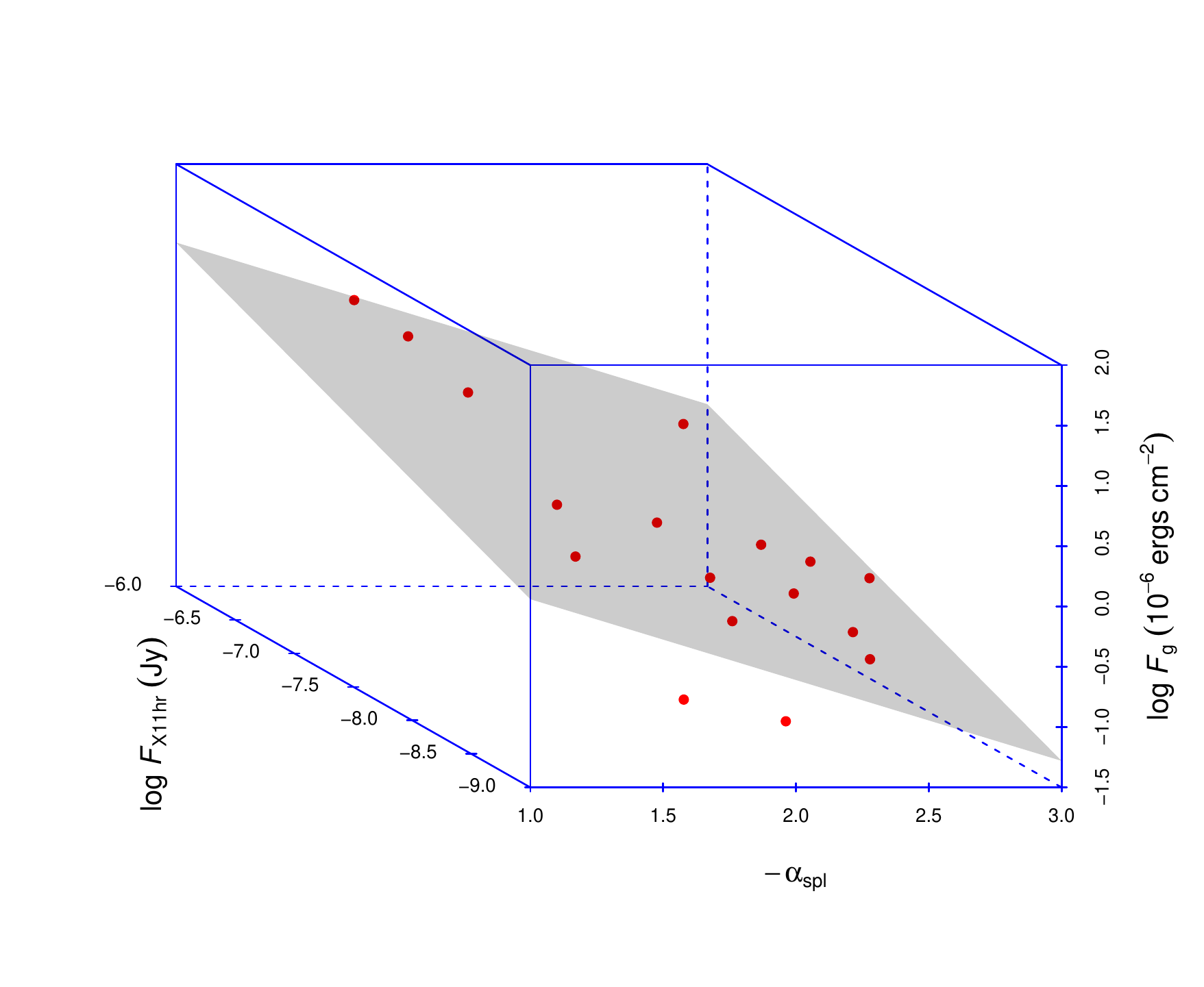}

\includegraphics[width=0.45\textwidth]{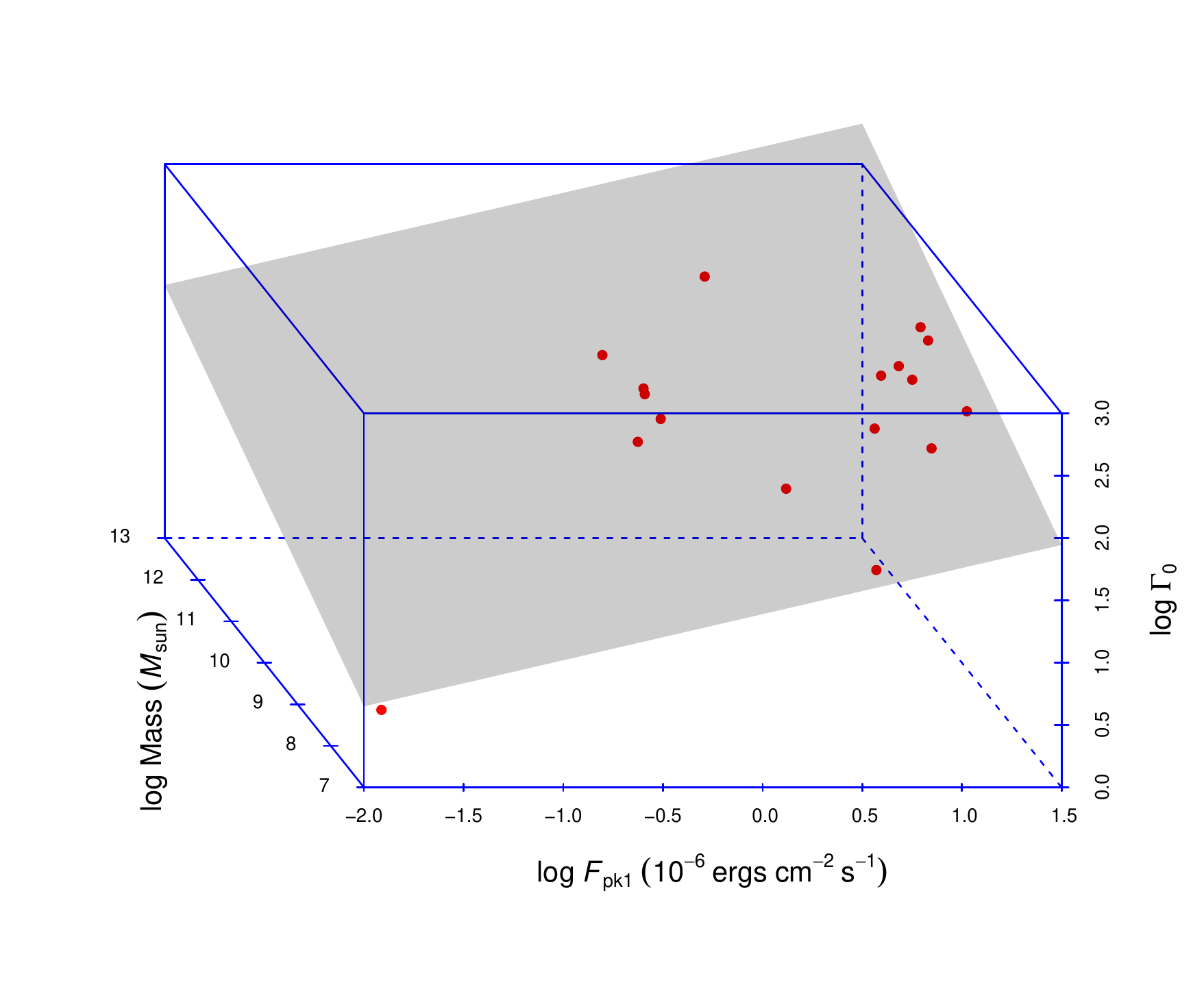}
\includegraphics[width=0.45\textwidth]{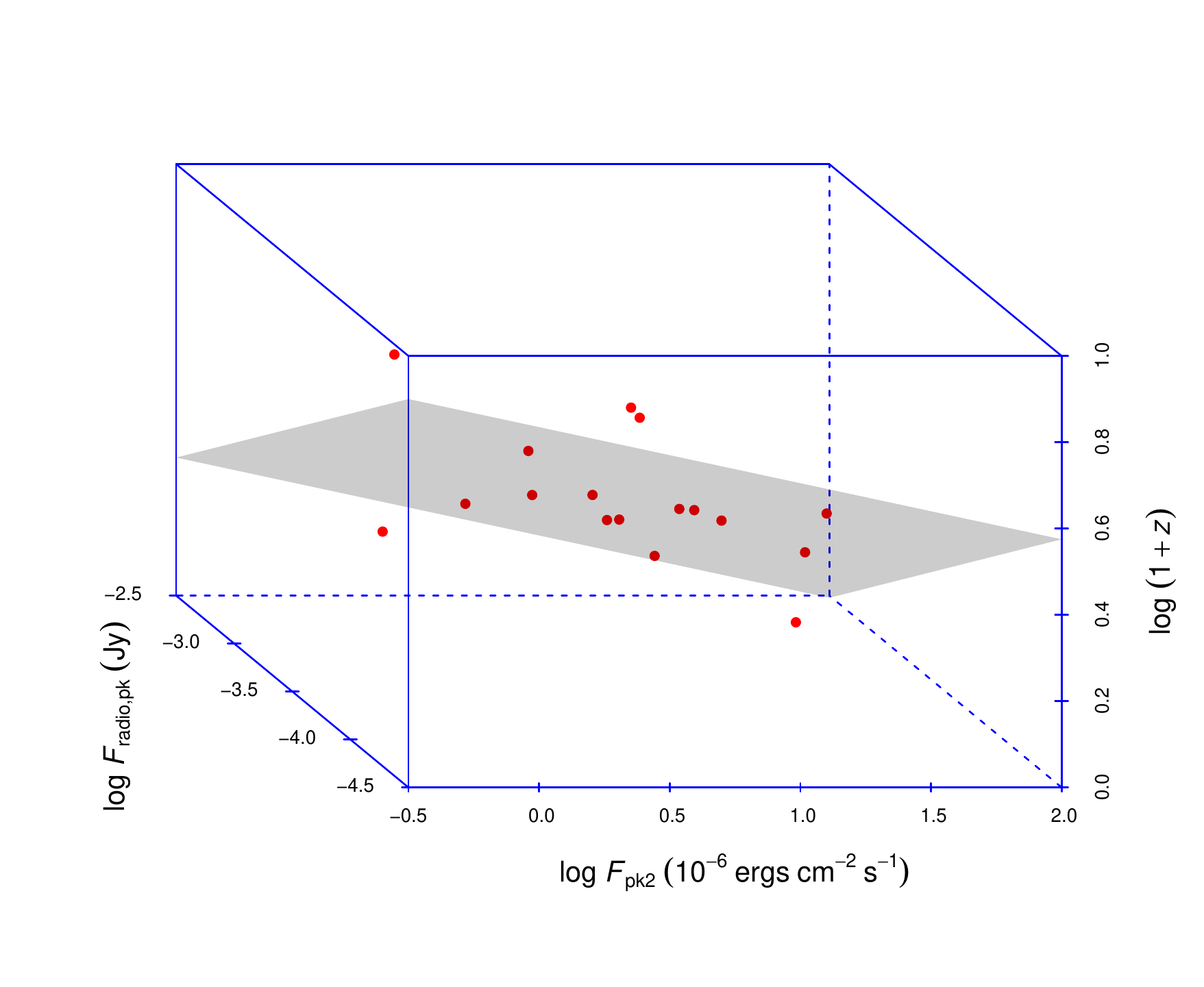}

\center{Fig. \ref{fig:three}---Continued}
\end{figure*}


\clearpage
\begin{figure*}

\includegraphics[width=0.45\textwidth]{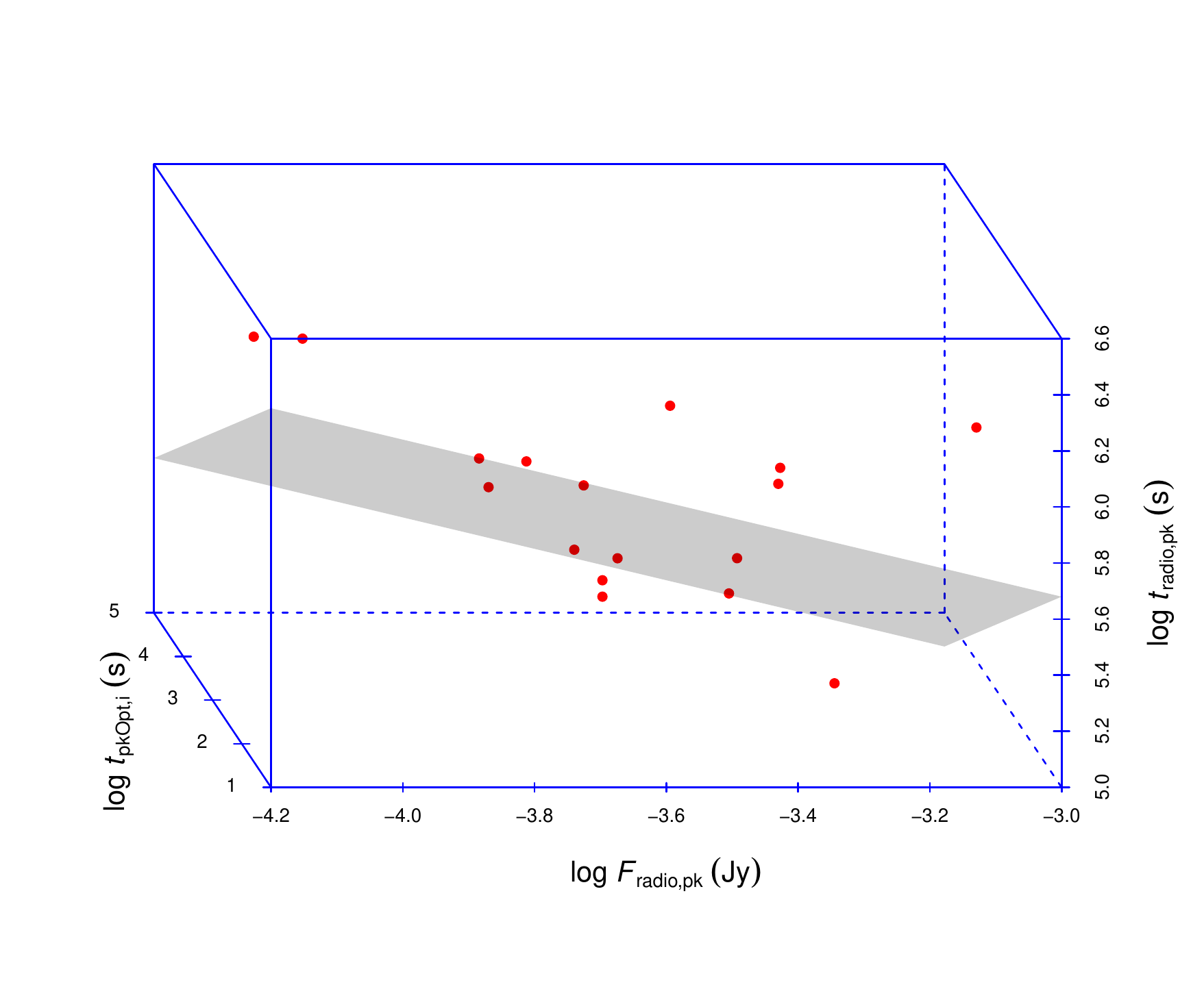}
\includegraphics[width=0.45\textwidth]{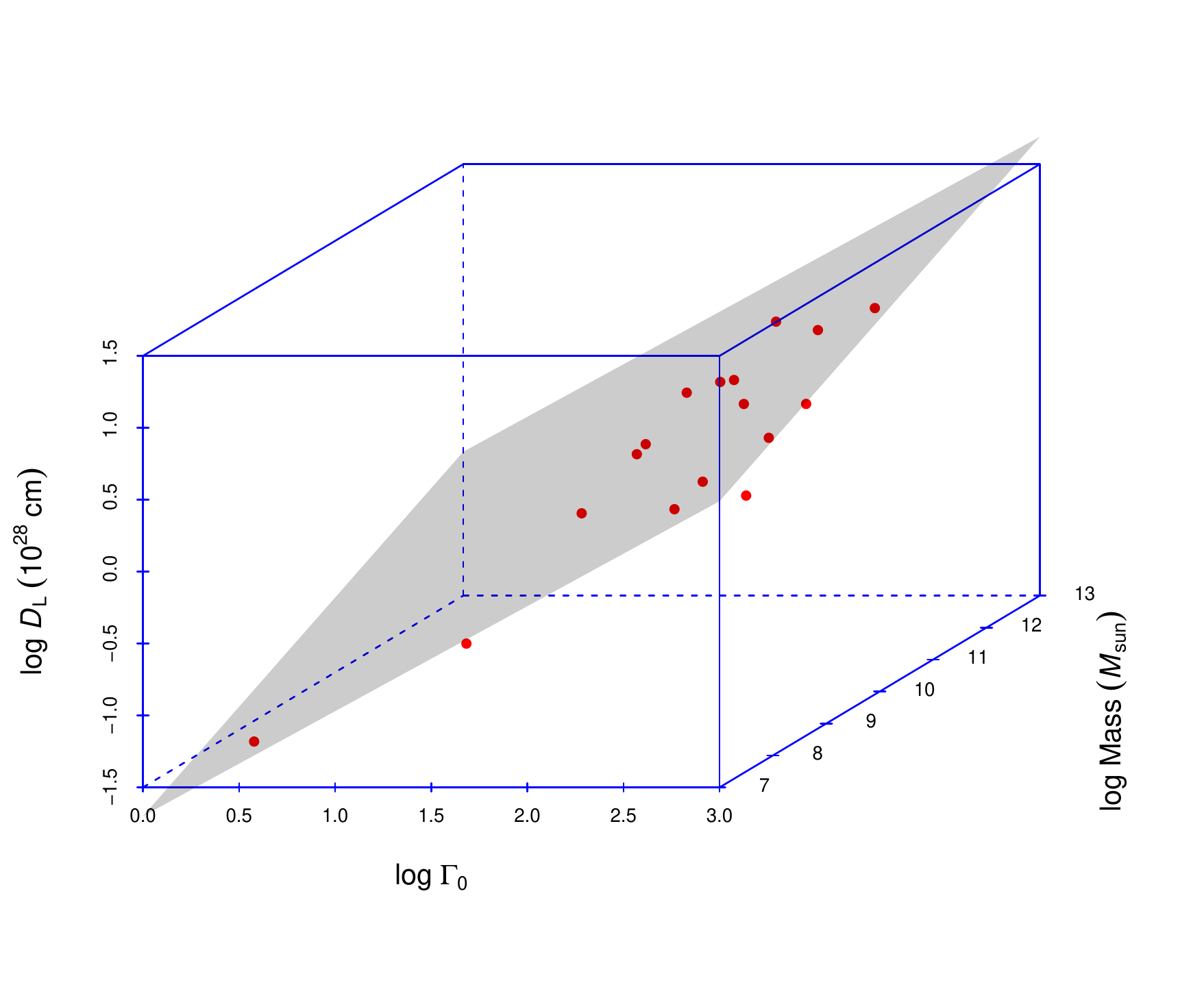}

\includegraphics[width=0.45\textwidth]{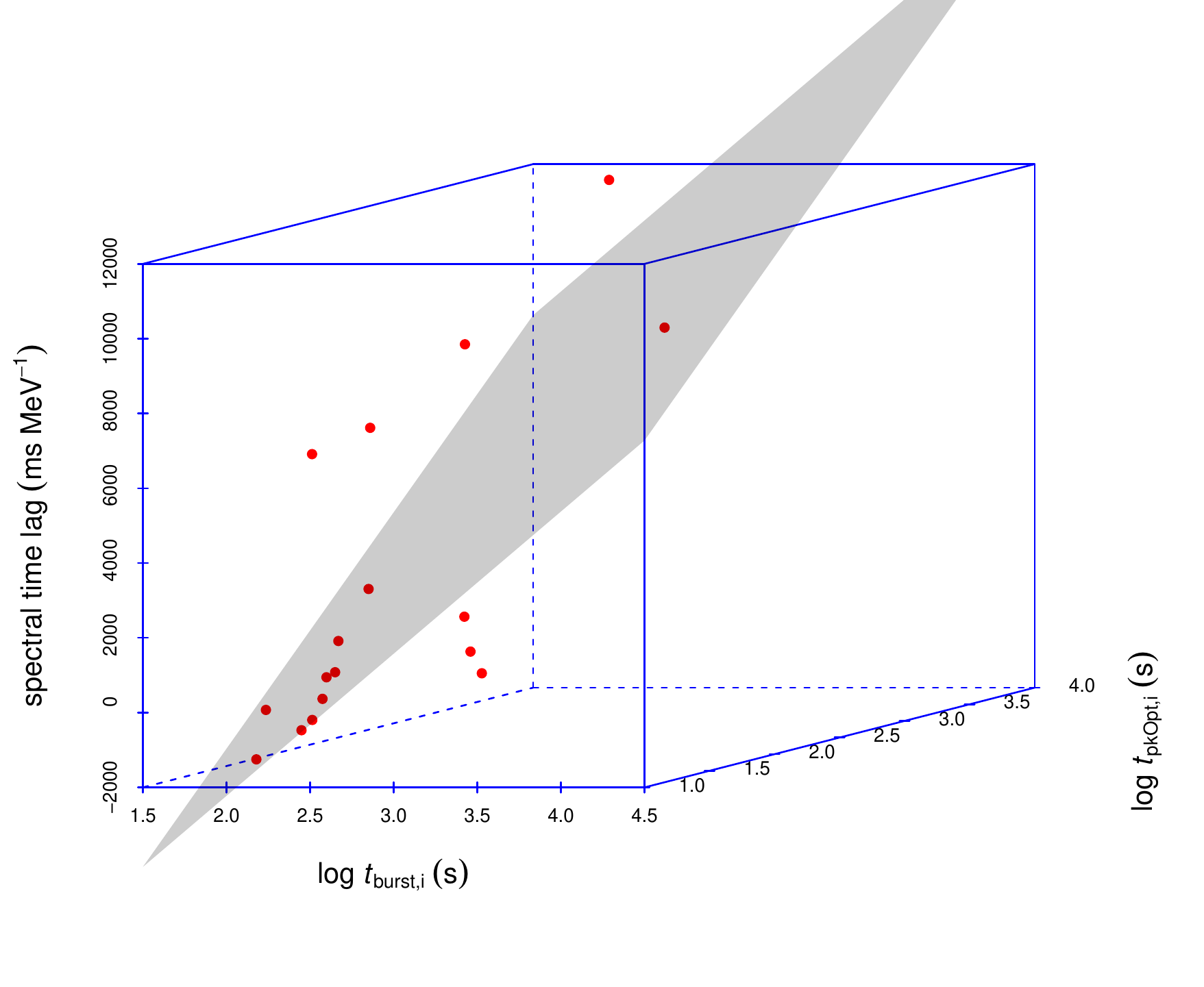}
\includegraphics[width=0.45\textwidth]{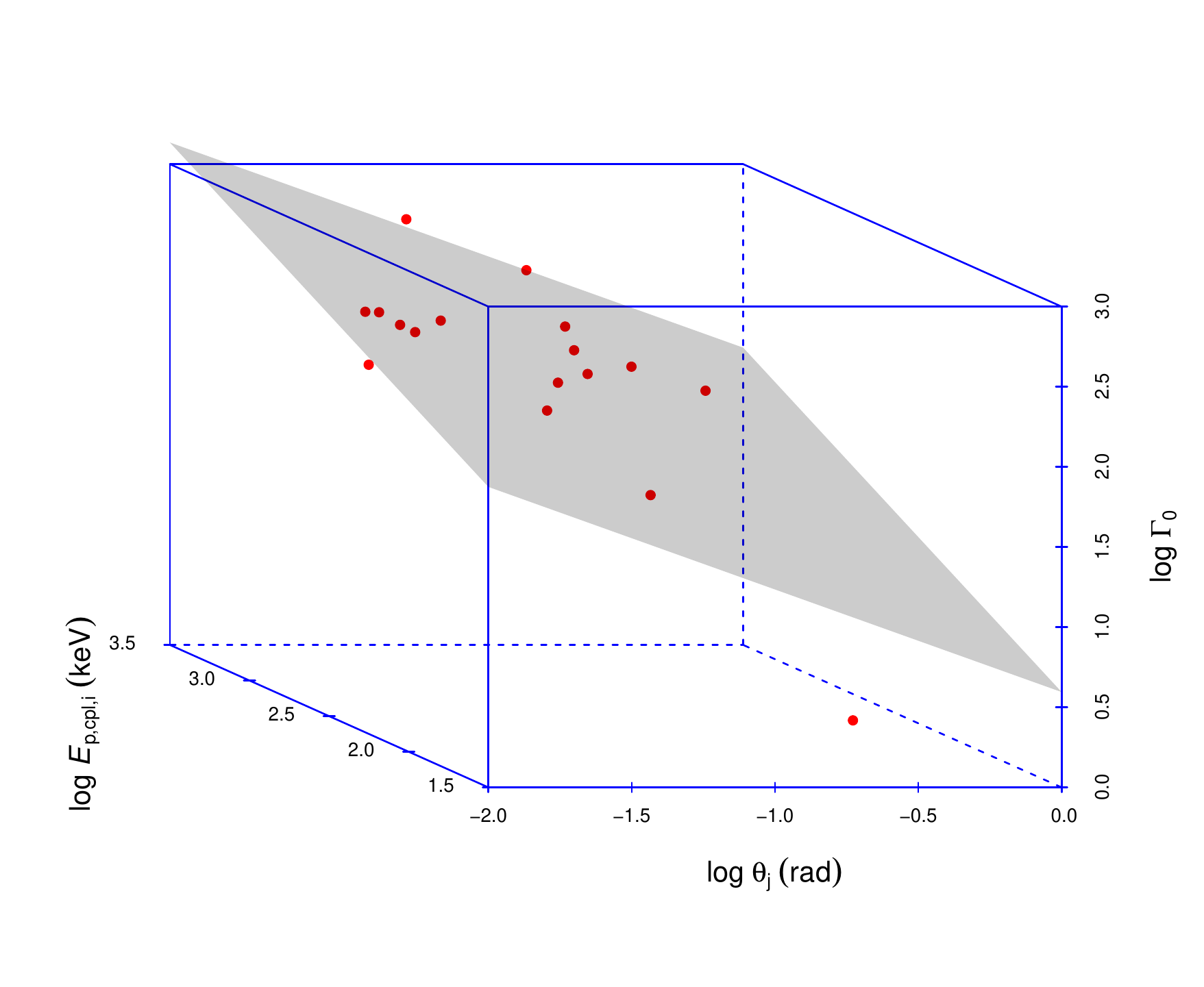}

\includegraphics[width=0.45\textwidth]{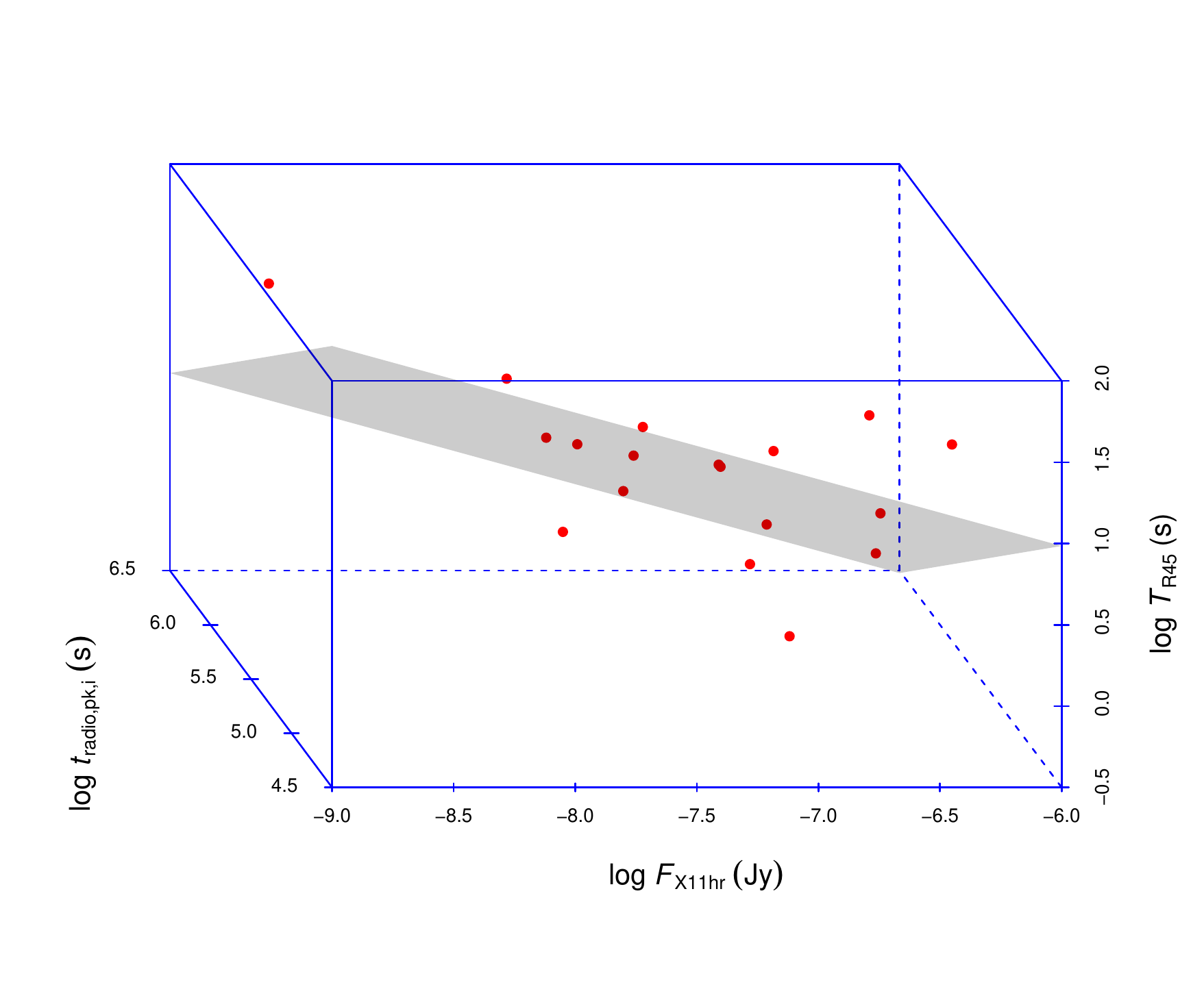}
\includegraphics[width=0.45\textwidth]{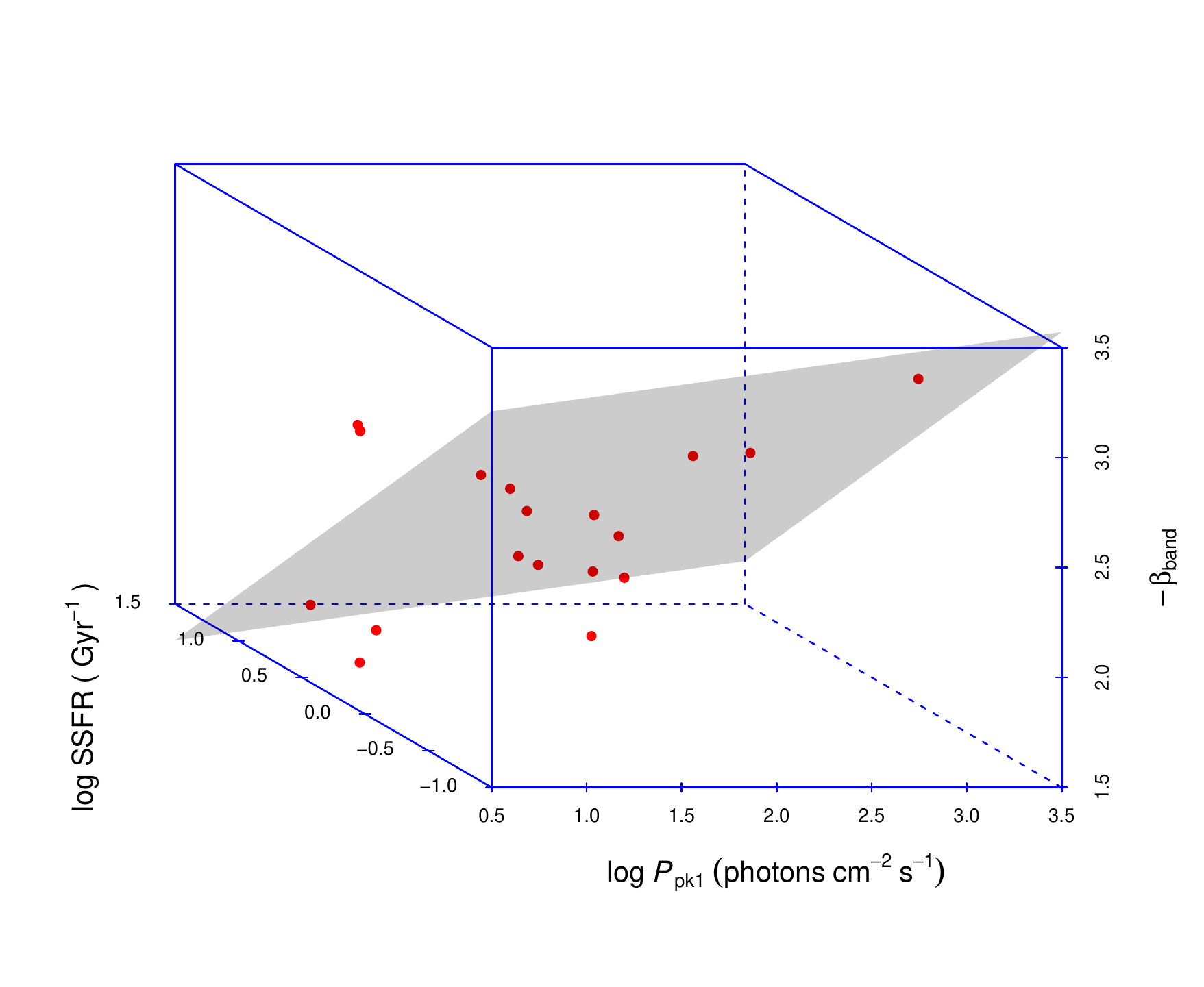}

\center{Fig. \ref{fig:three}---Continued}
\end{figure*}


\clearpage
\begin{figure*}

\includegraphics[width=0.45\textwidth]{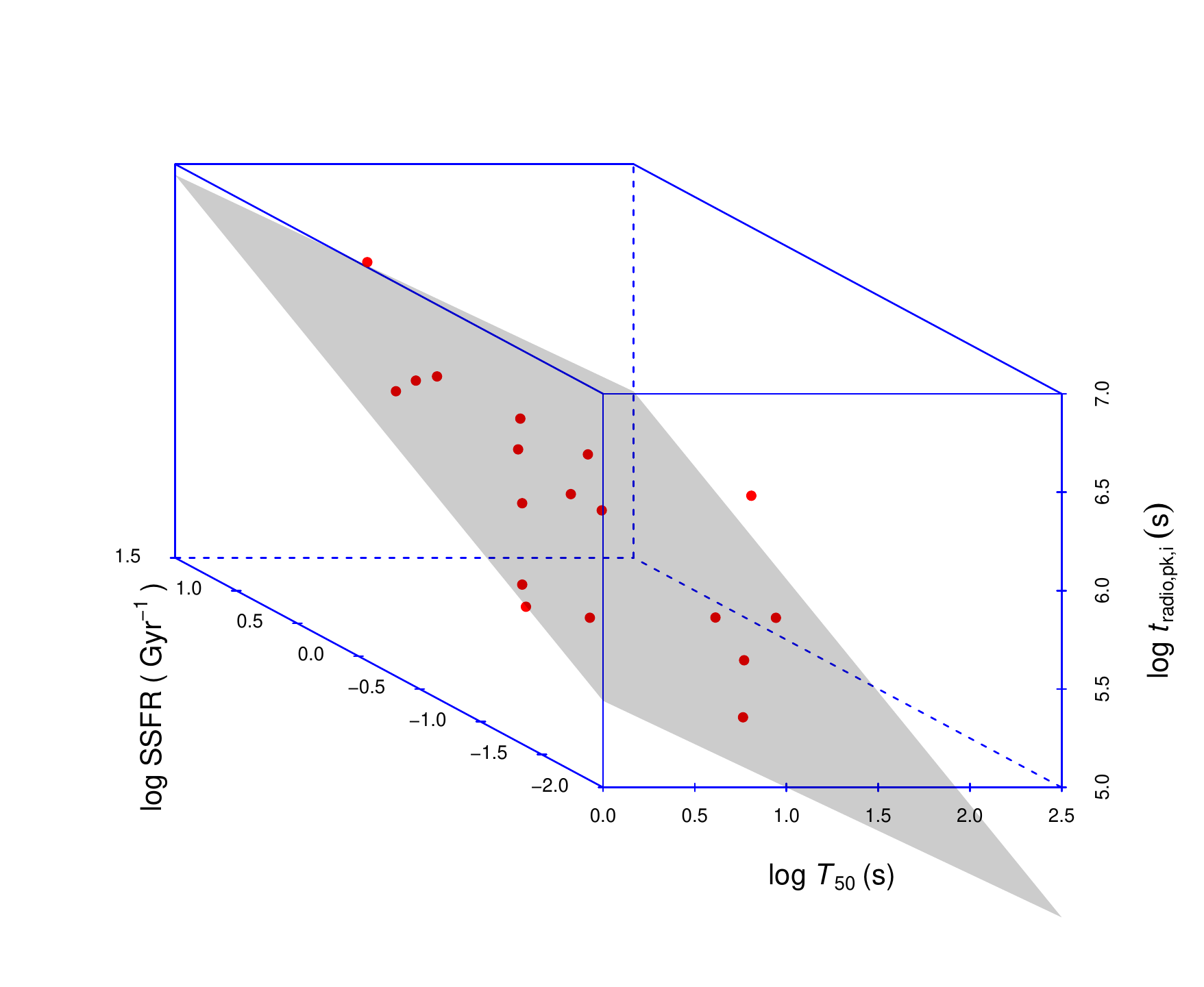}
\includegraphics[width=0.45\textwidth]{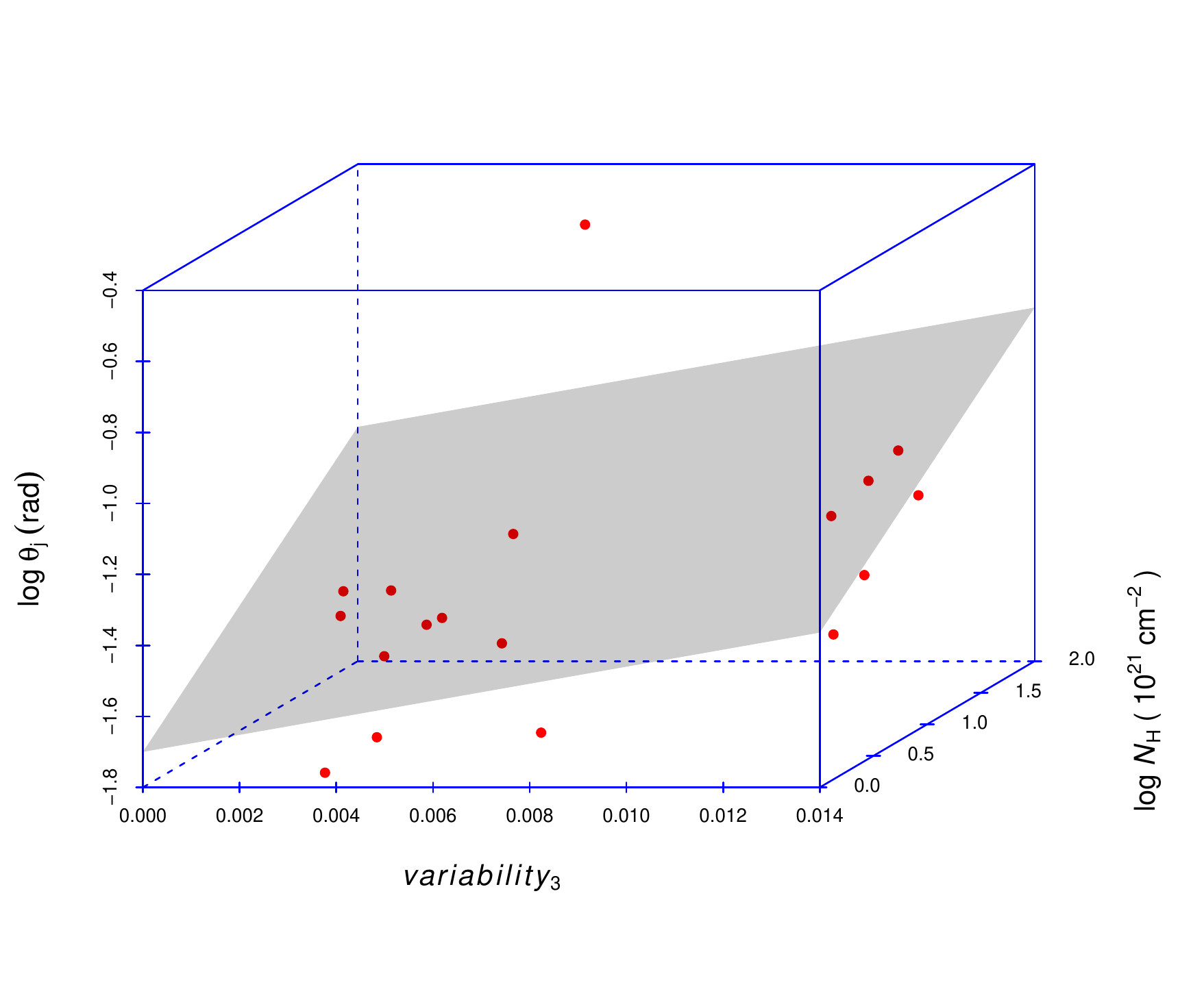}

\includegraphics[width=0.45\textwidth]{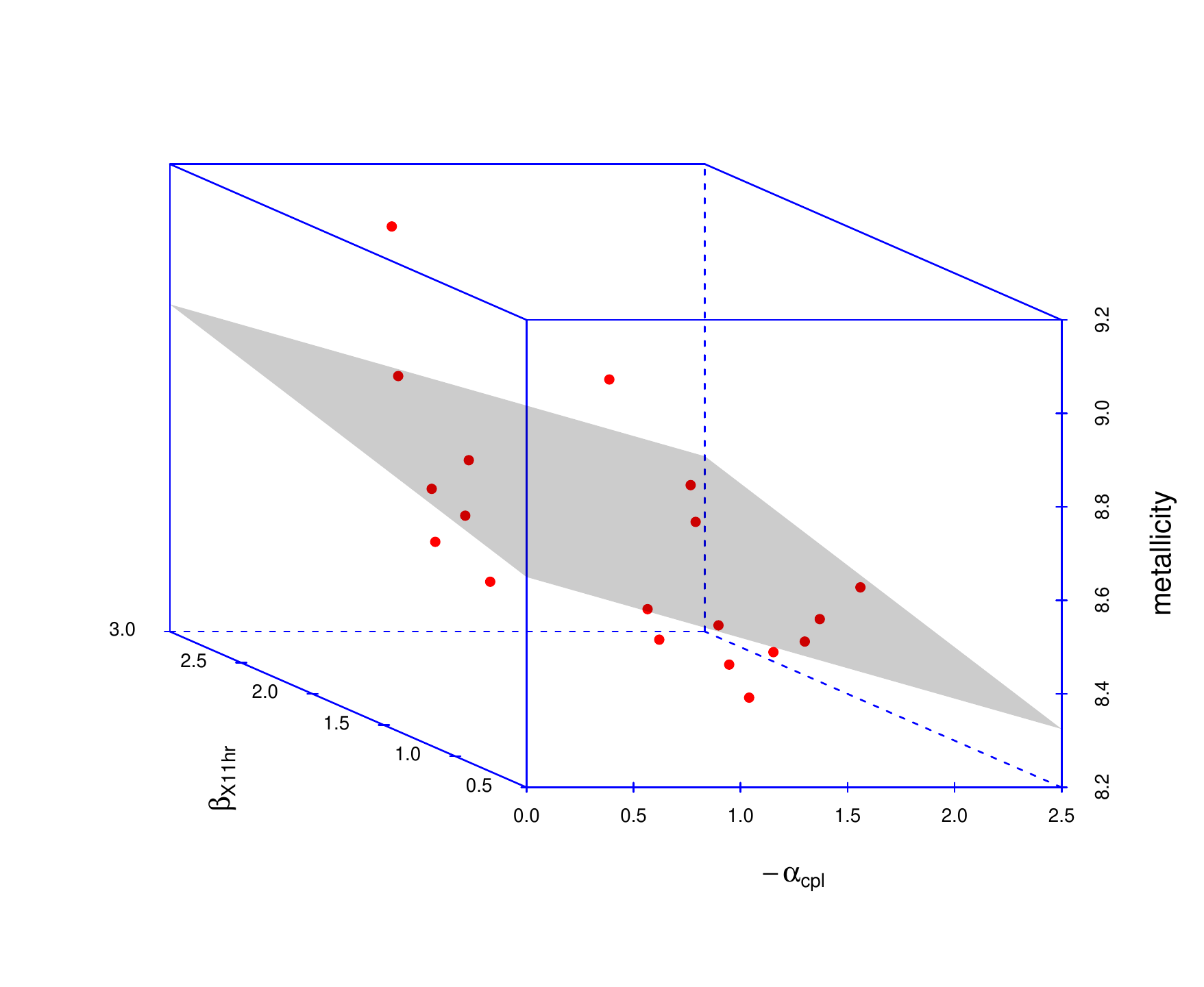}
\includegraphics[width=0.45\textwidth]{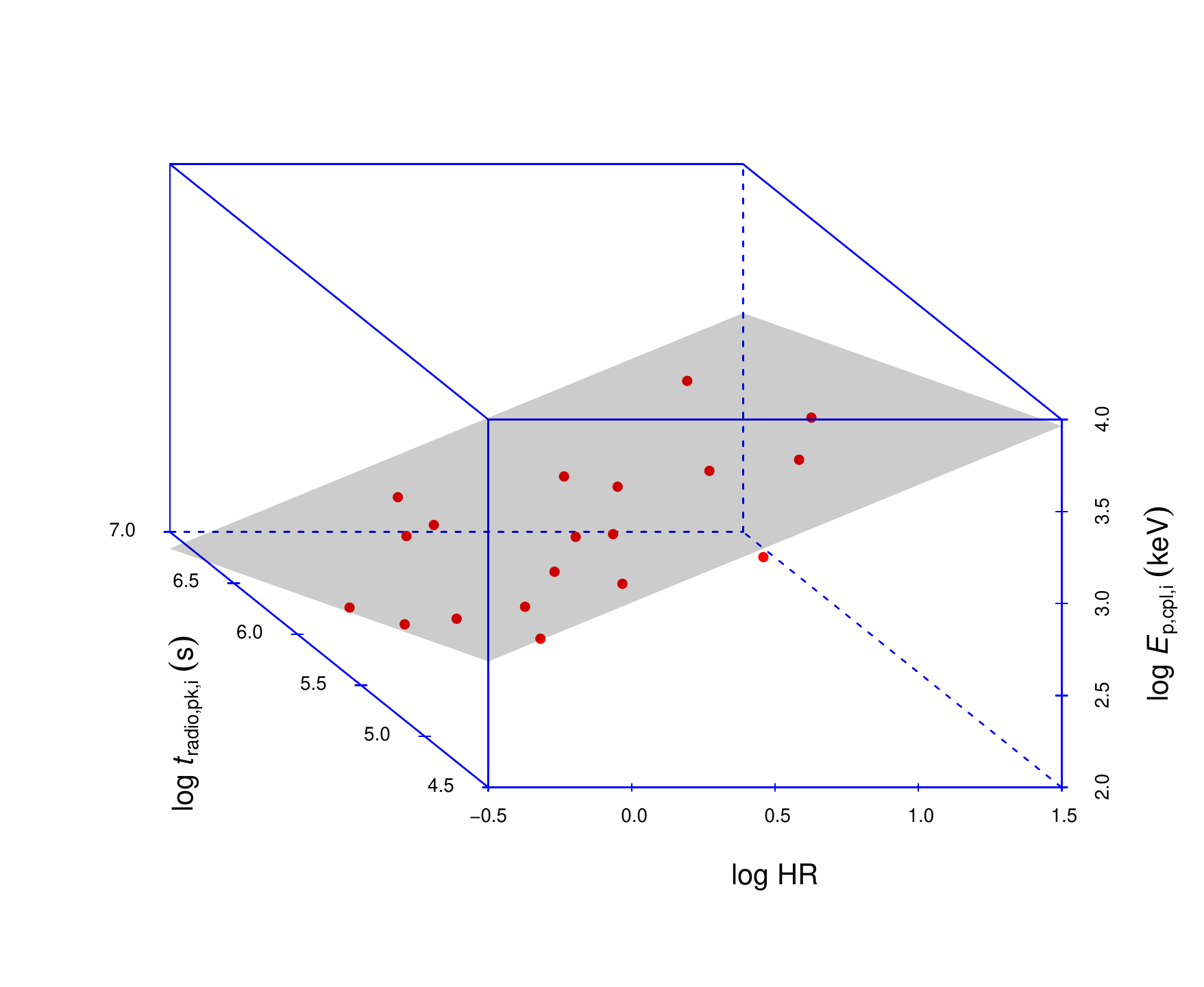}

\includegraphics[width=0.45\textwidth]{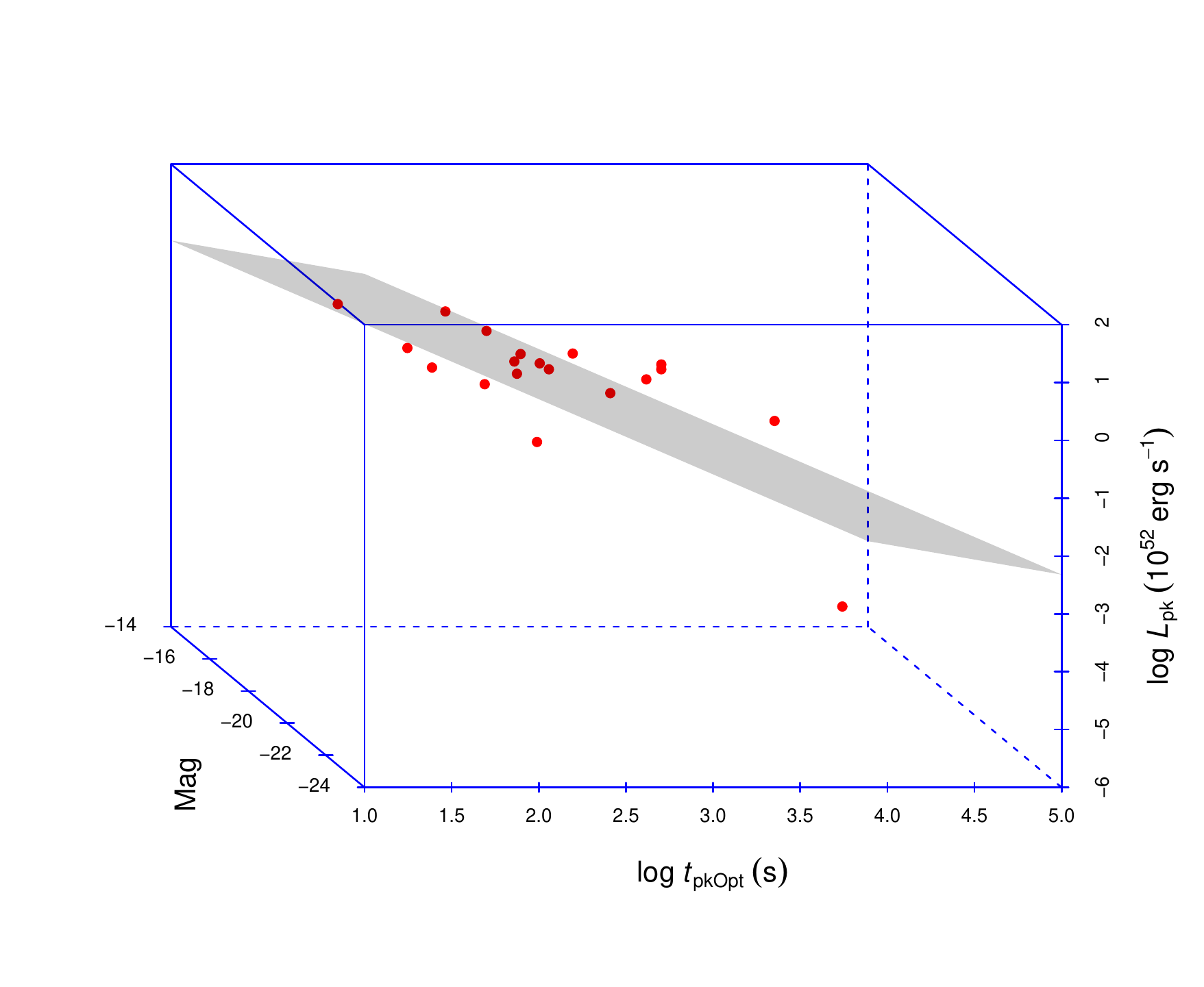}
\includegraphics[width=0.45\textwidth]{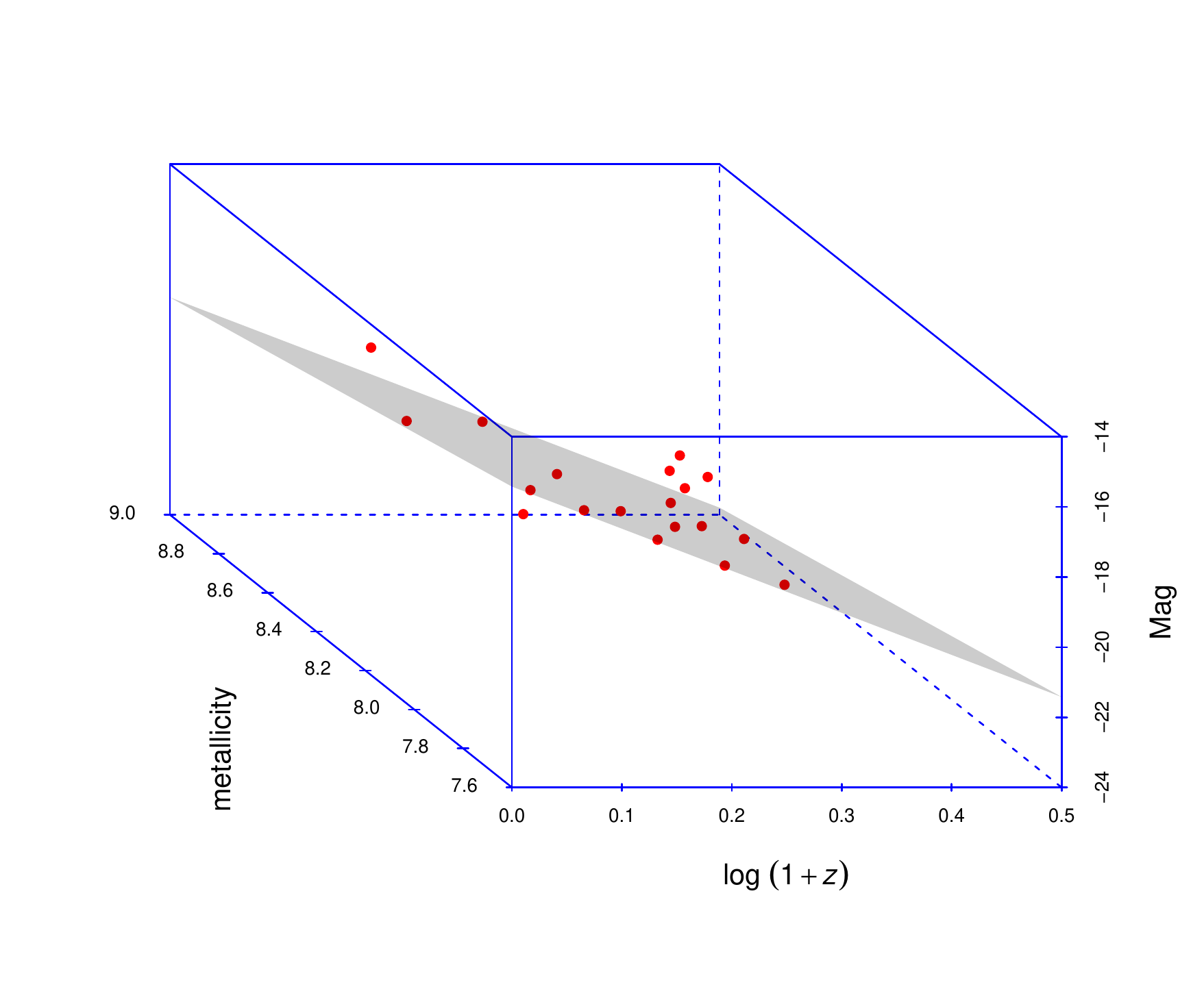}

\center{Fig. \ref{fig:three}---Continued}
\end{figure*}


\clearpage
\begin{figure*}

\includegraphics[width=0.45\textwidth]{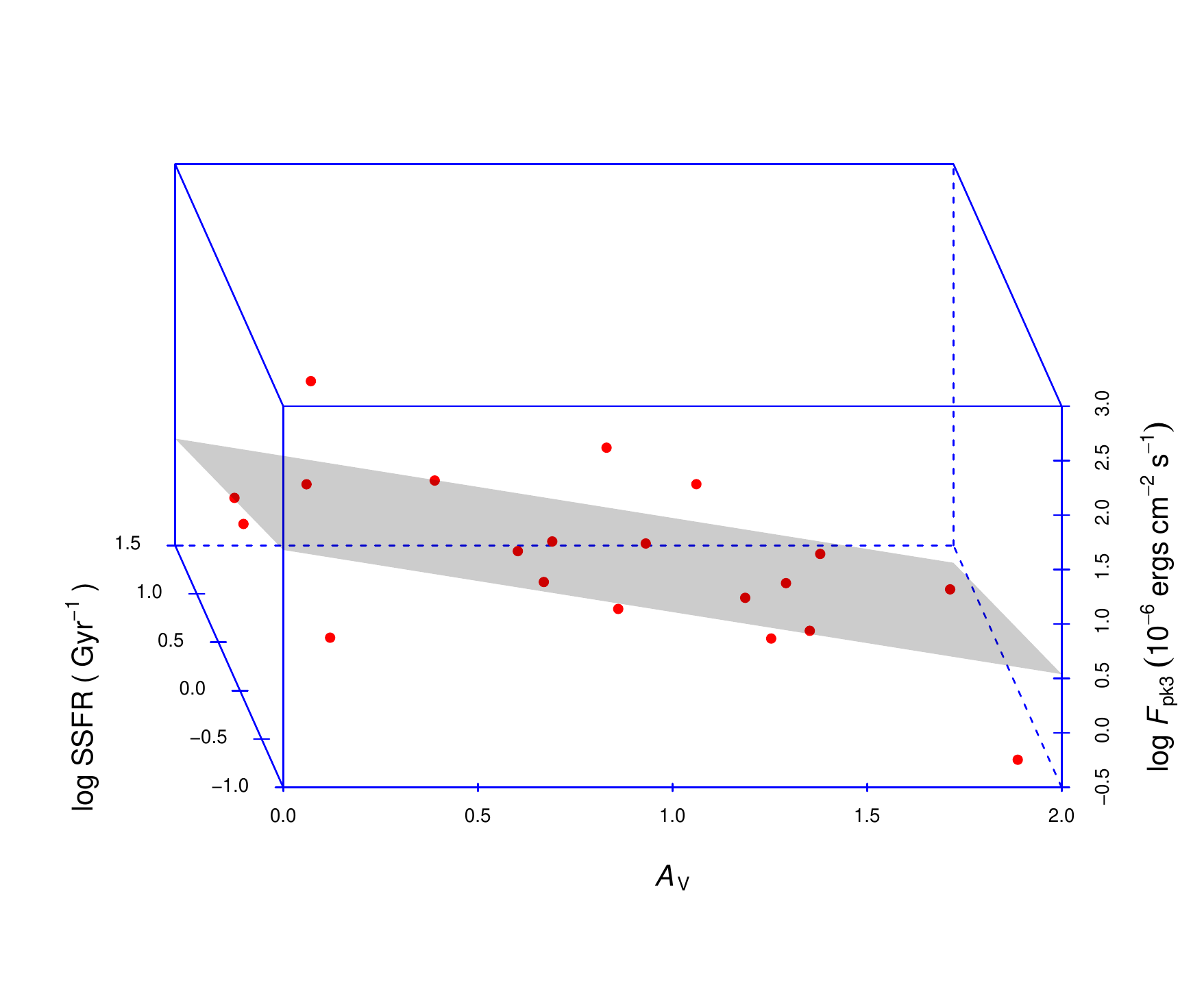}
\includegraphics[width=0.45\textwidth]{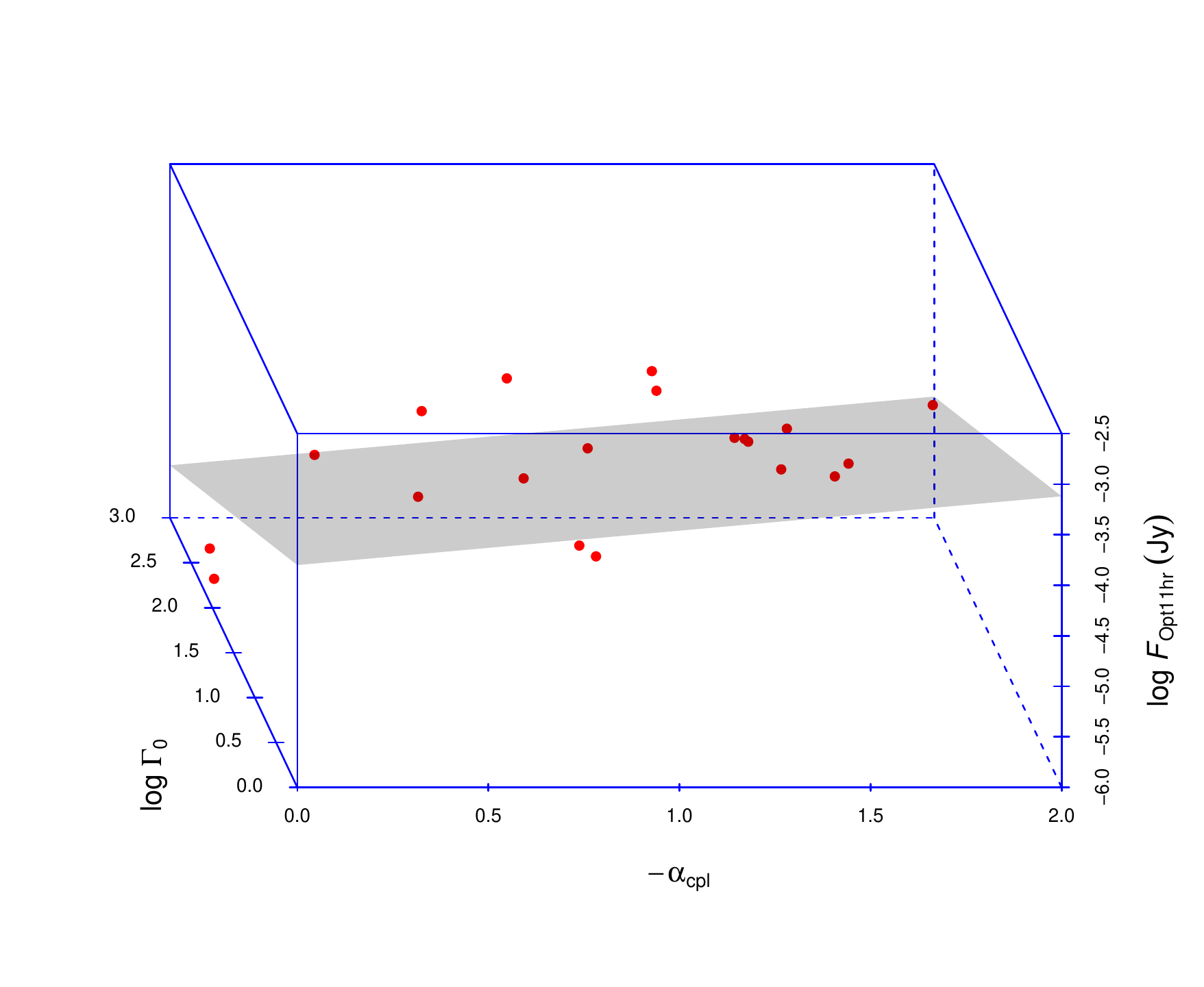}

\includegraphics[width=0.45\textwidth]{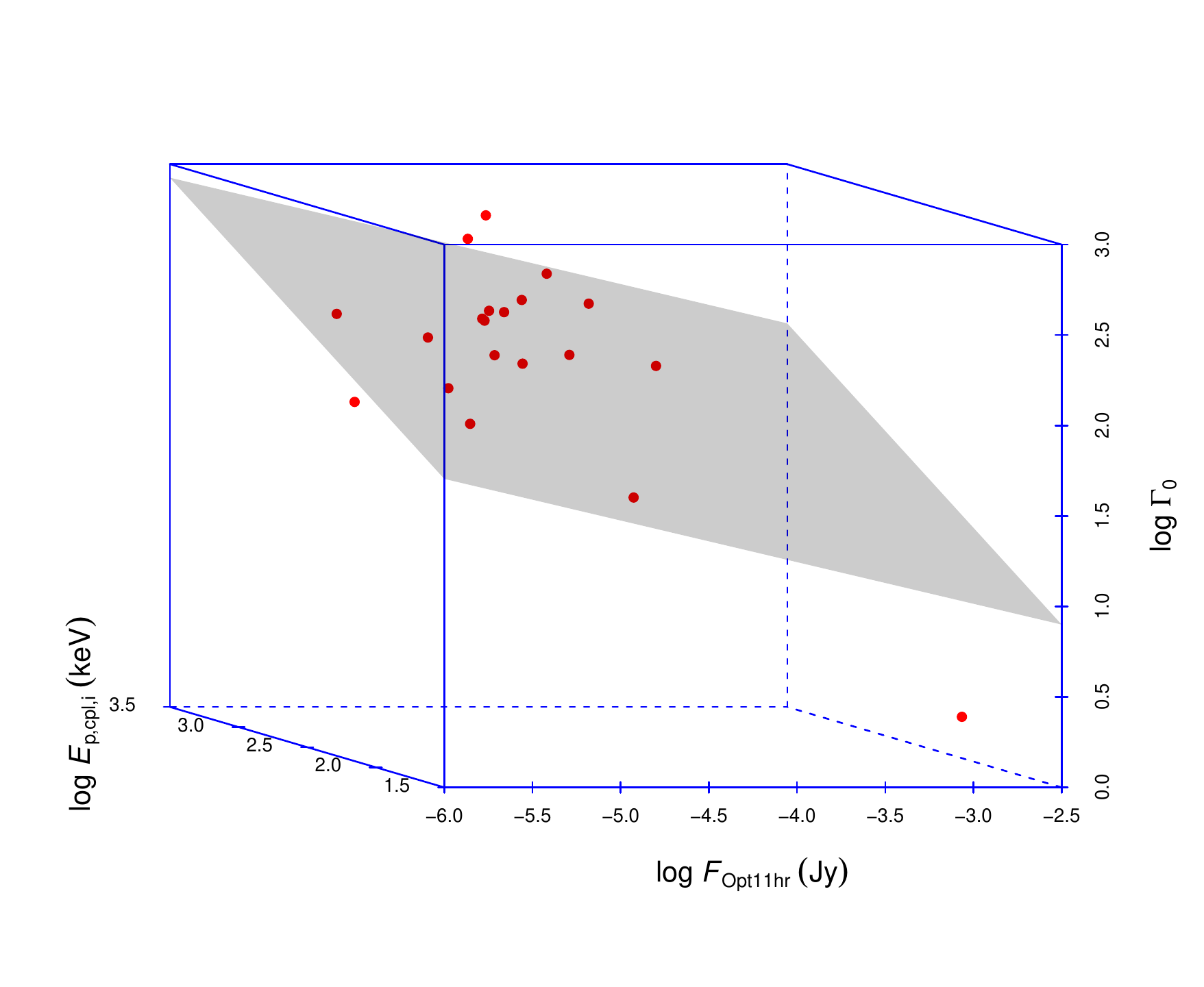}
\includegraphics[width=0.45\textwidth]{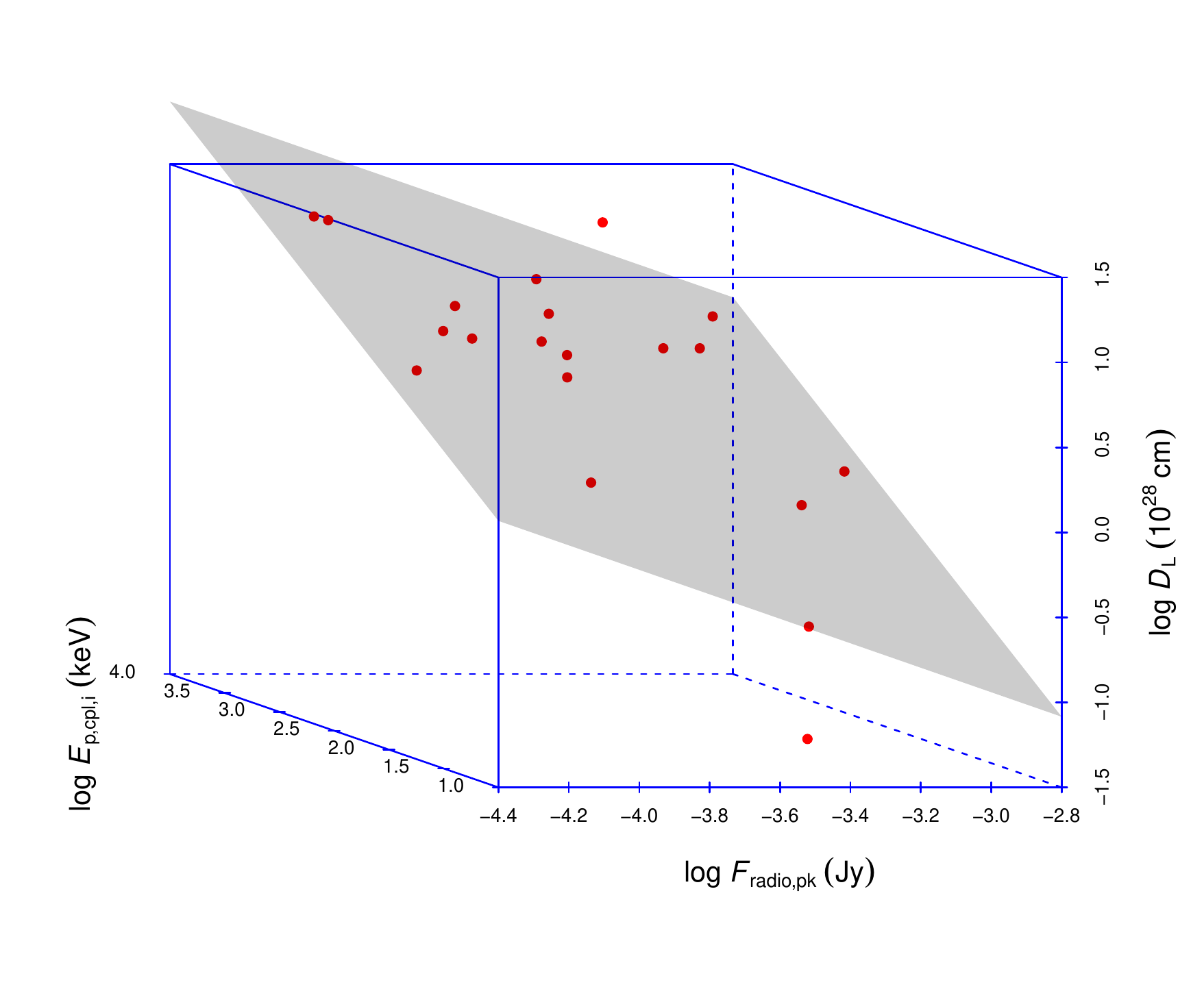}

\includegraphics[width=0.45\textwidth]{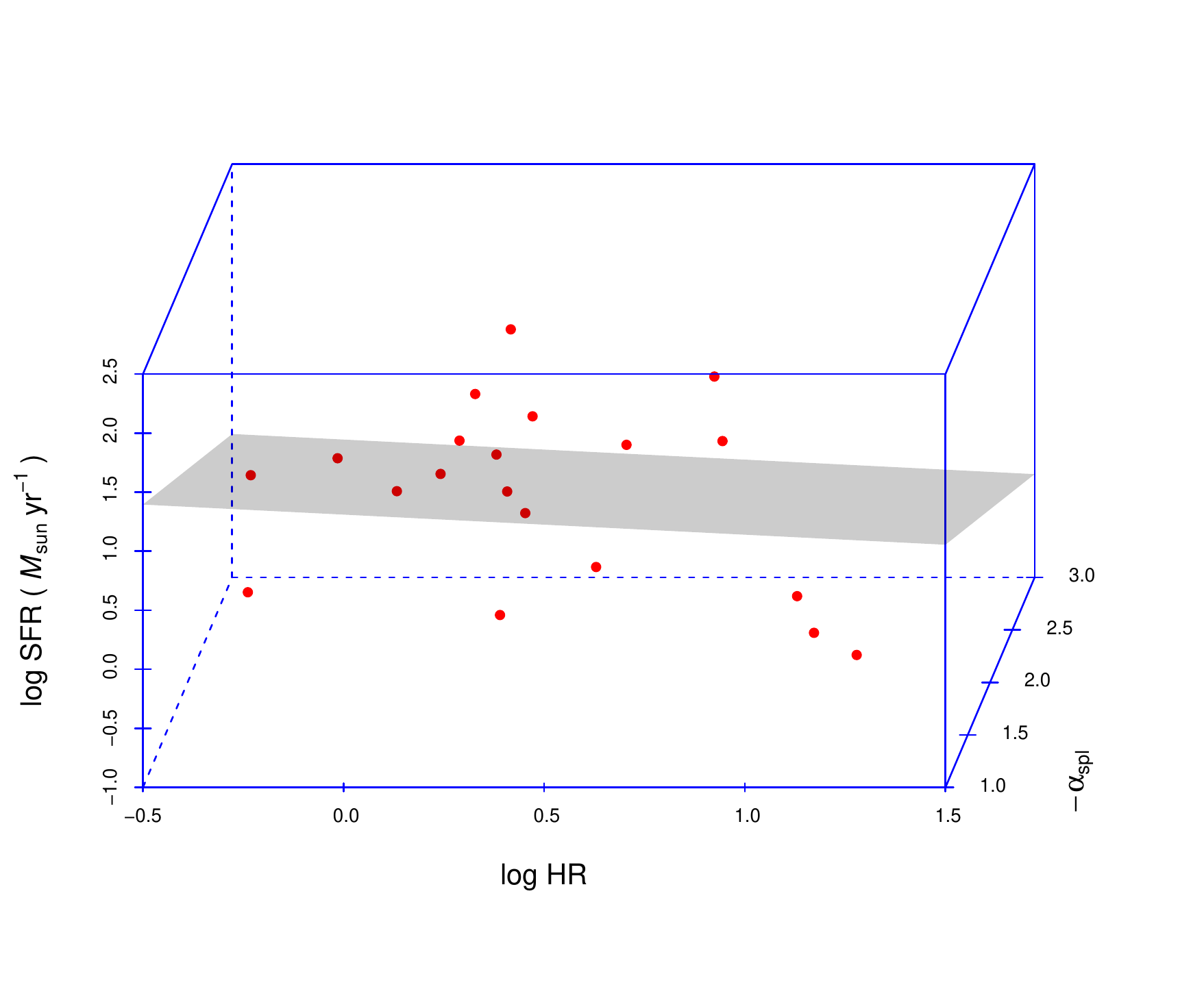}
\includegraphics[width=0.45\textwidth]{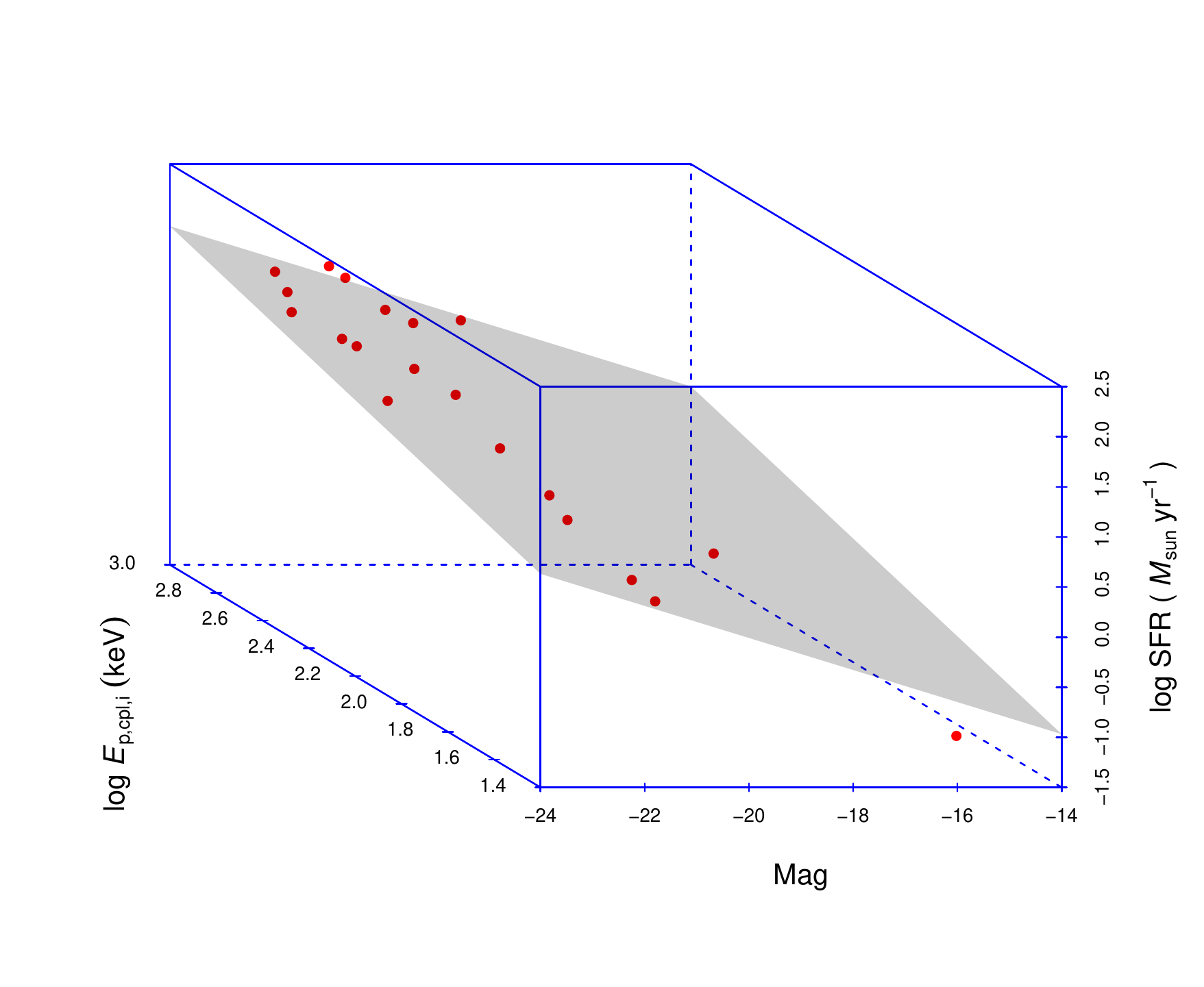}

\center{Fig. \ref{fig:three}---Continued}
\end{figure*}


\clearpage
\begin{figure*}

\includegraphics[width=0.45\textwidth]{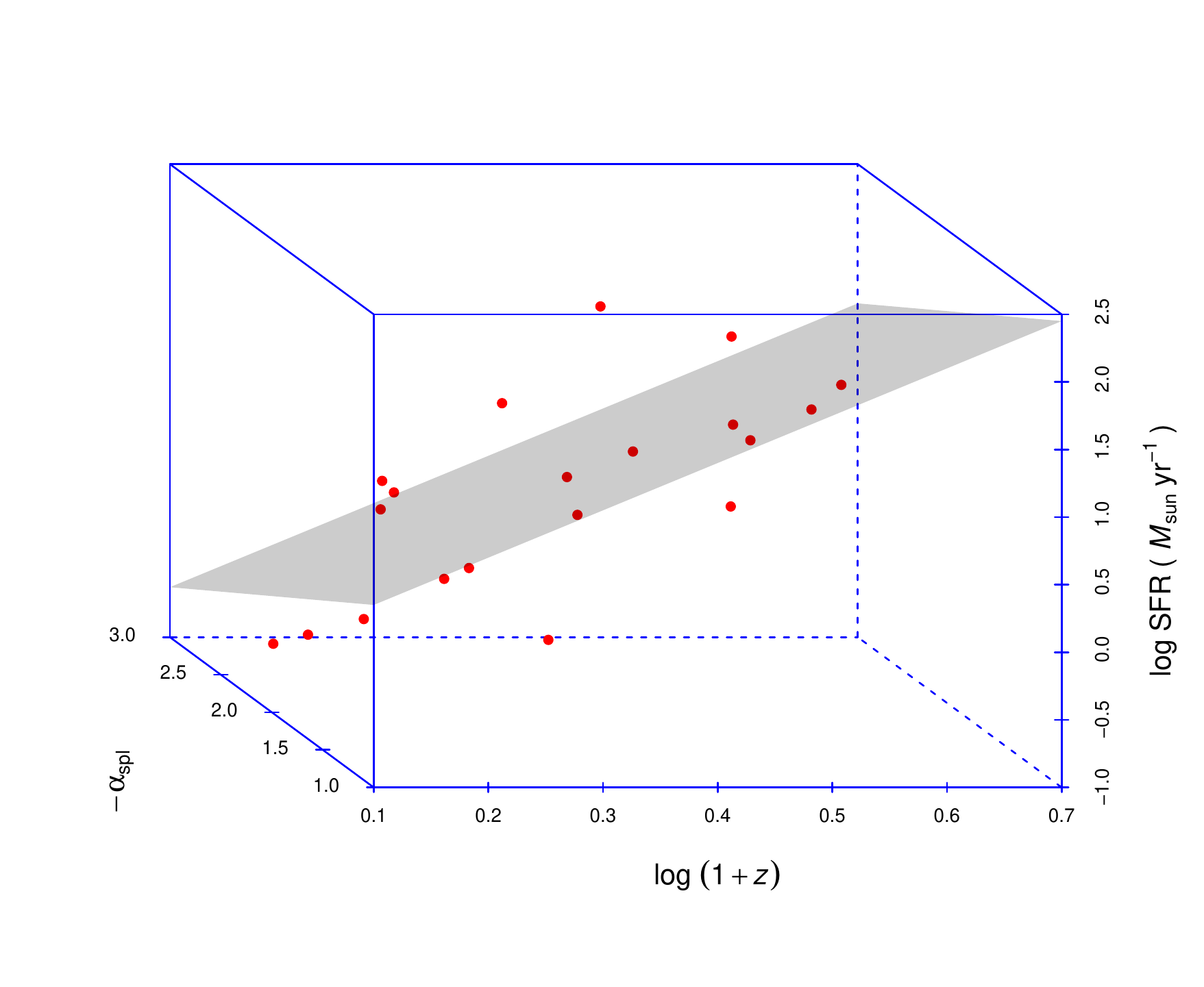}
\includegraphics[width=0.45\textwidth]{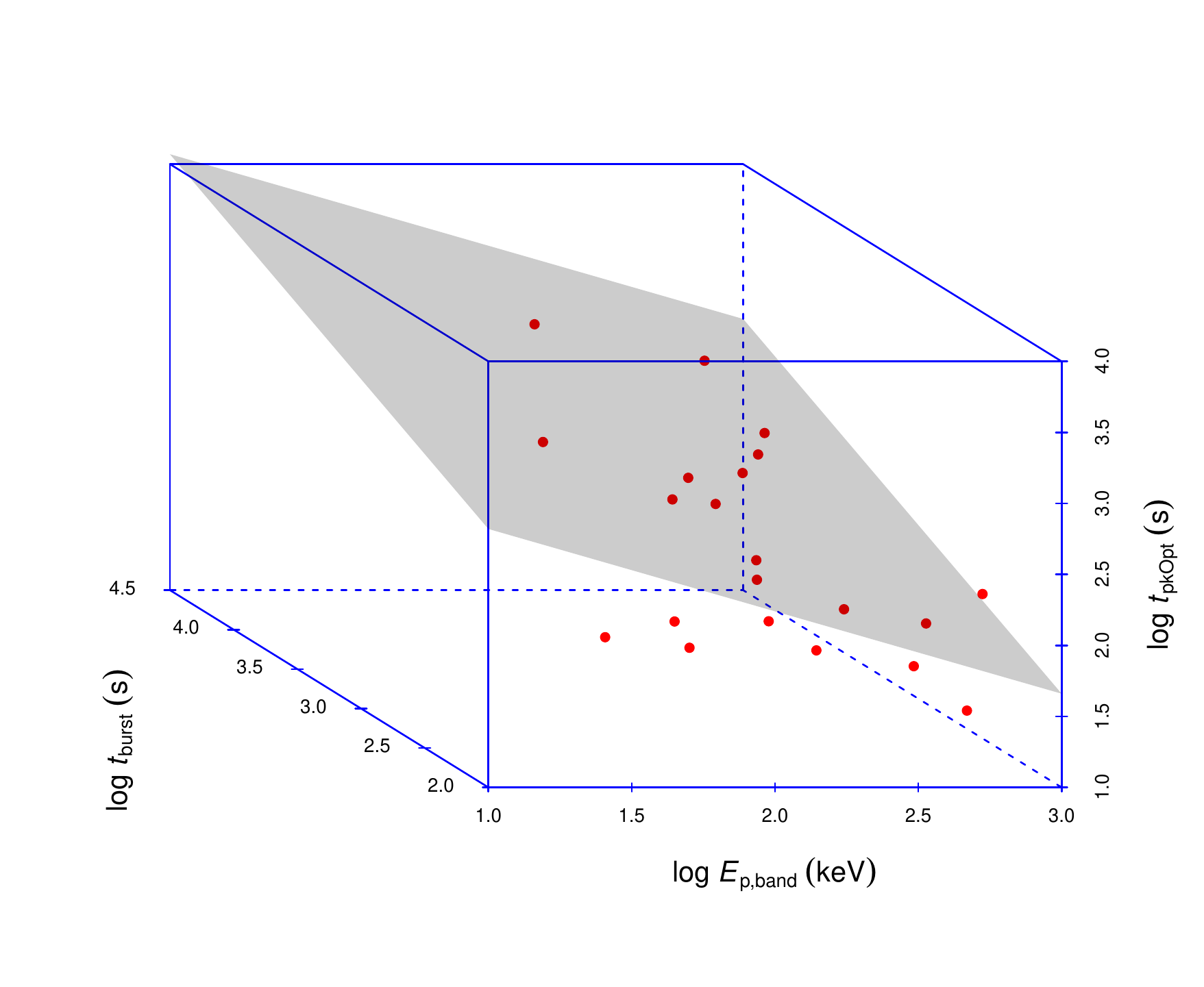}

\includegraphics[width=0.45\textwidth]{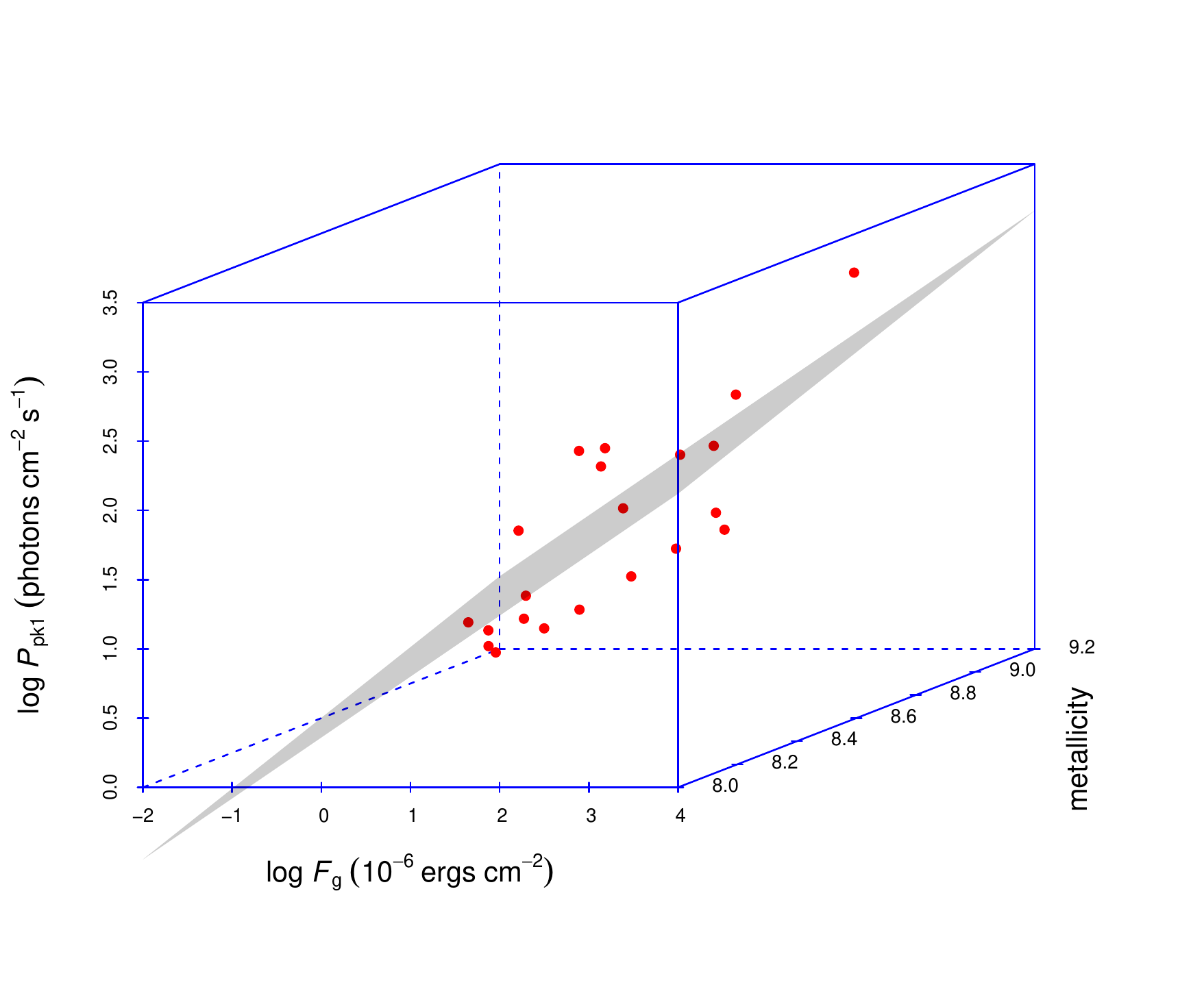}
\includegraphics[width=0.45\textwidth]{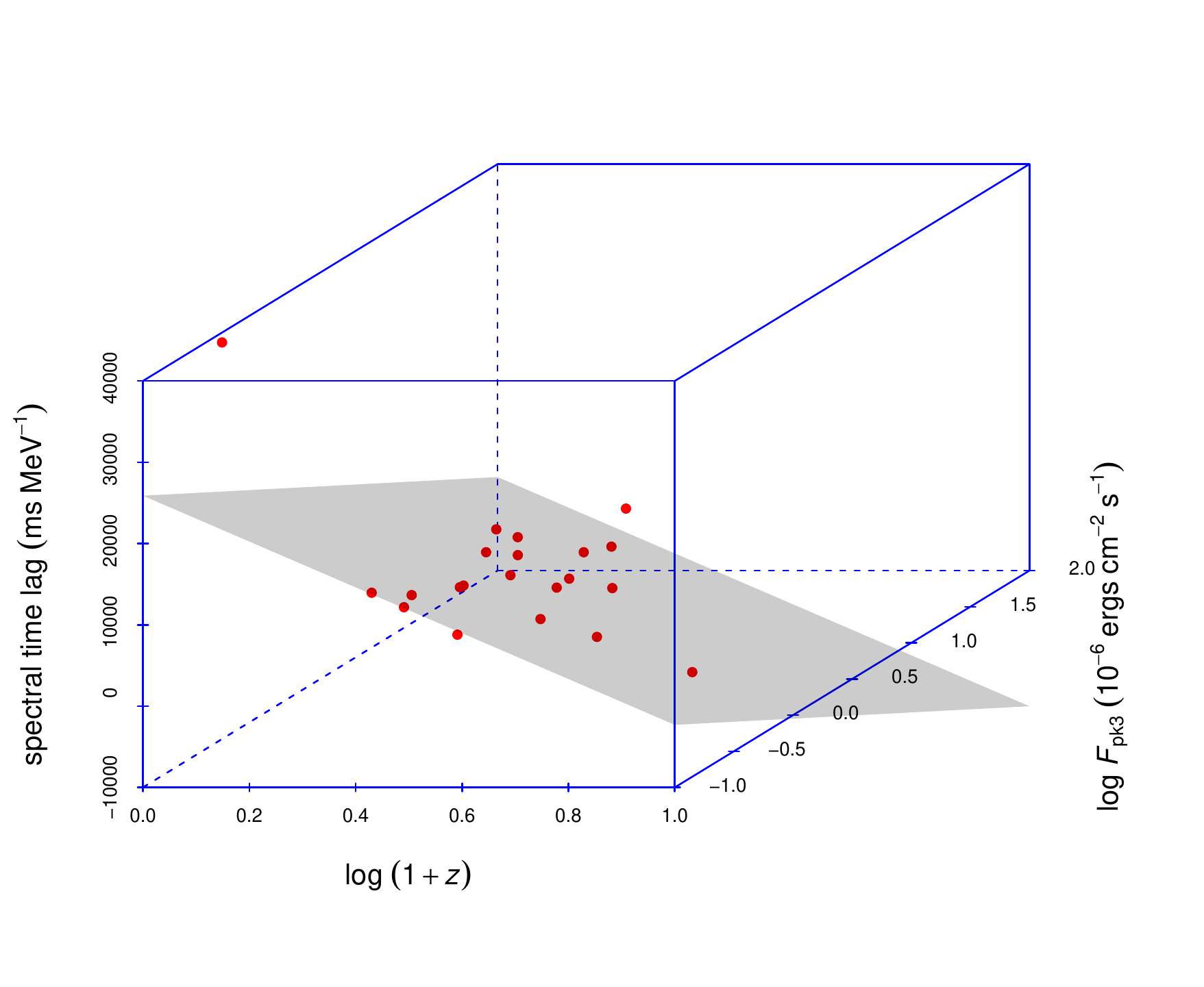}

\includegraphics[width=0.45\textwidth]{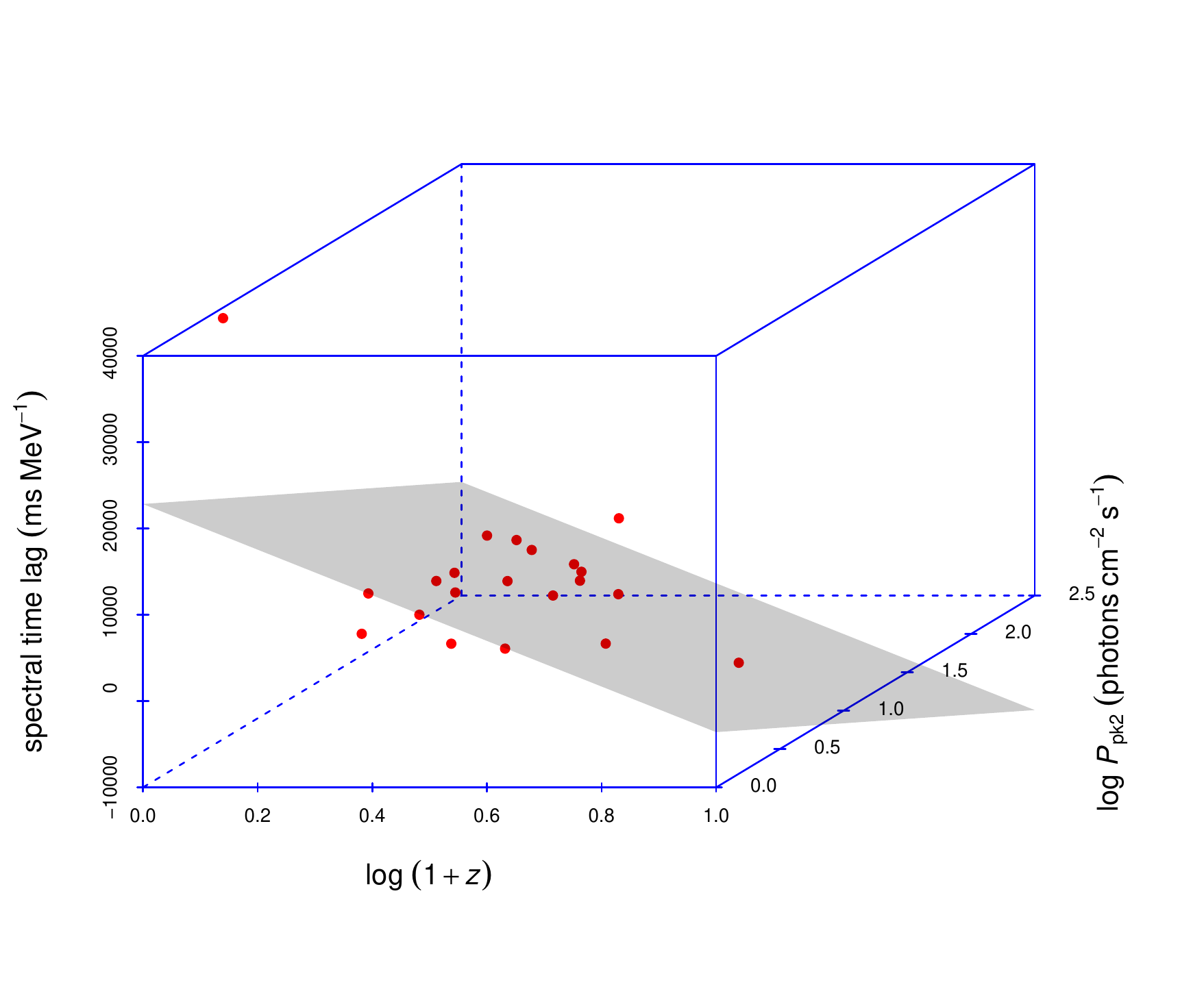}
\includegraphics[width=0.45\textwidth]{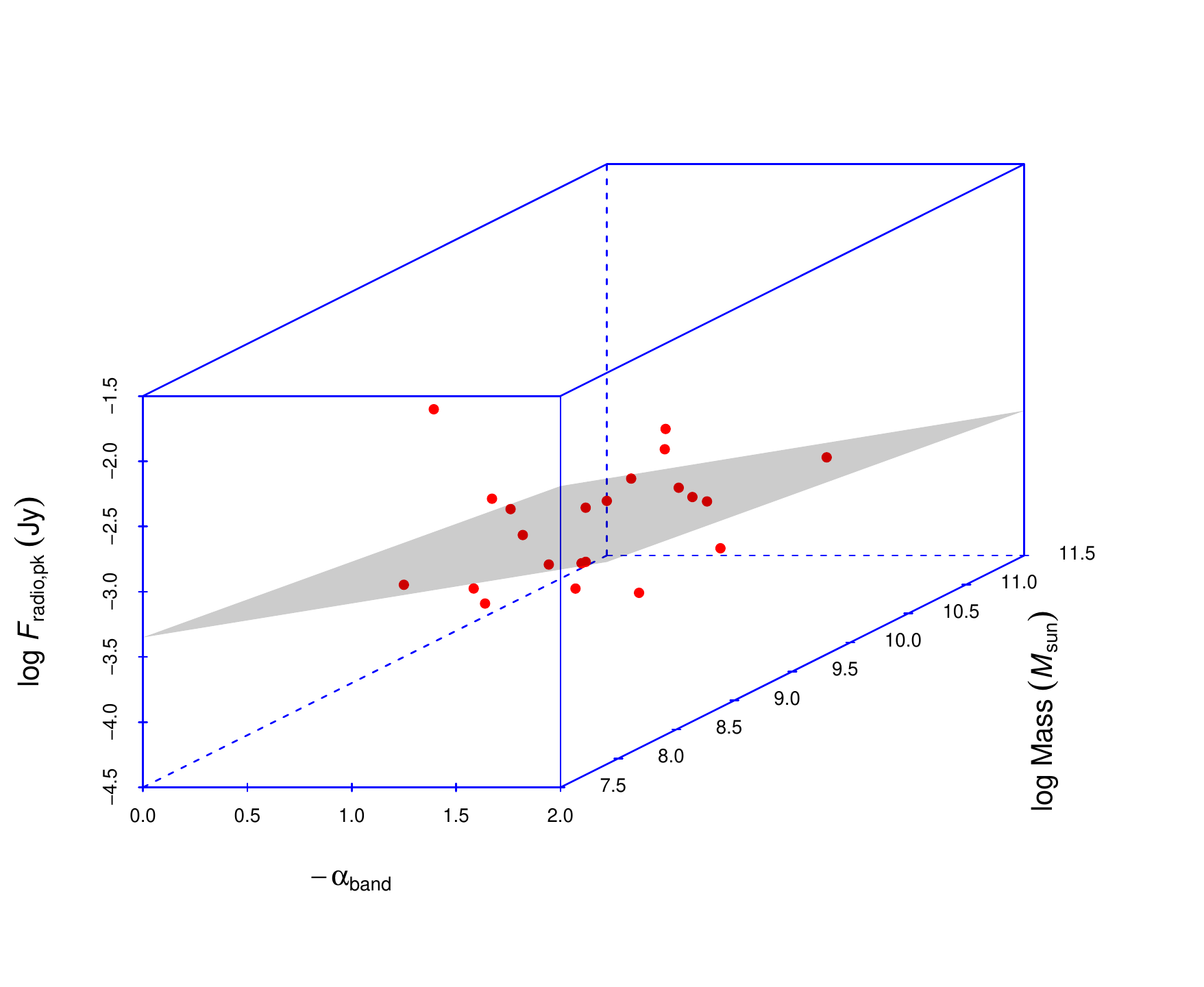}

\center{Fig. \ref{fig:three}---Continued}
\end{figure*}


\clearpage
\begin{figure*}

\includegraphics[width=0.45\textwidth]{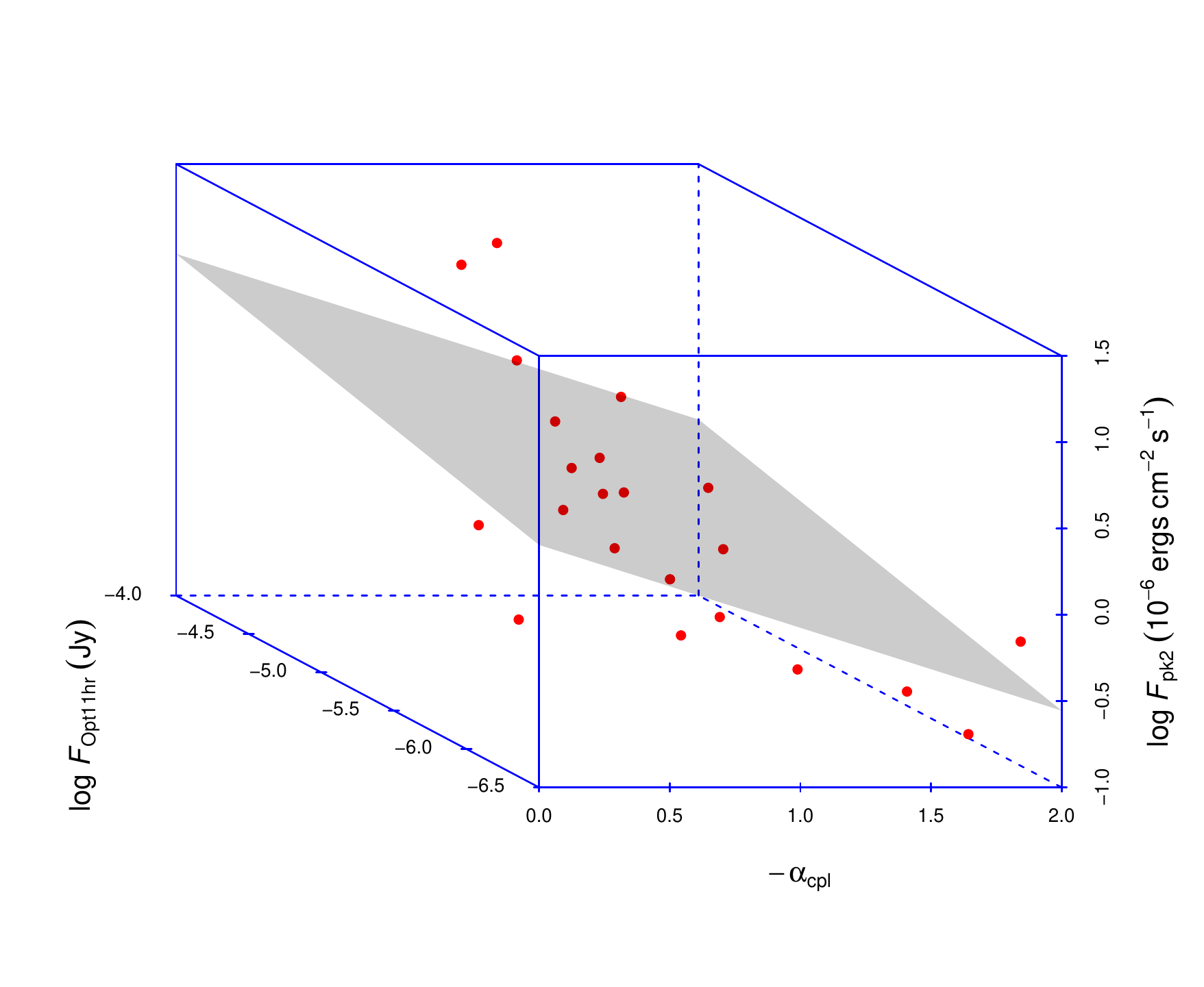}
\includegraphics[width=0.45\textwidth]{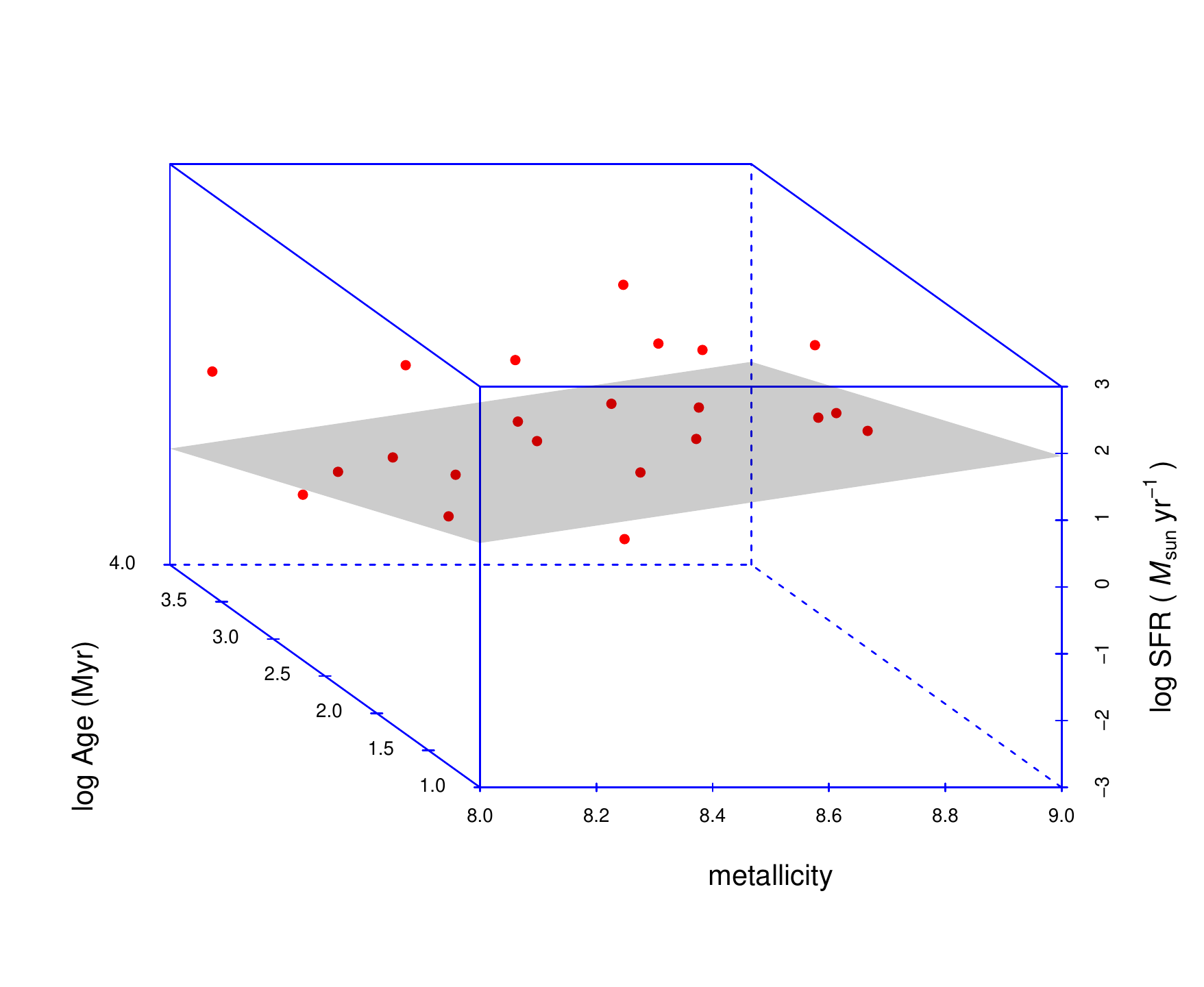}

\includegraphics[width=0.45\textwidth]{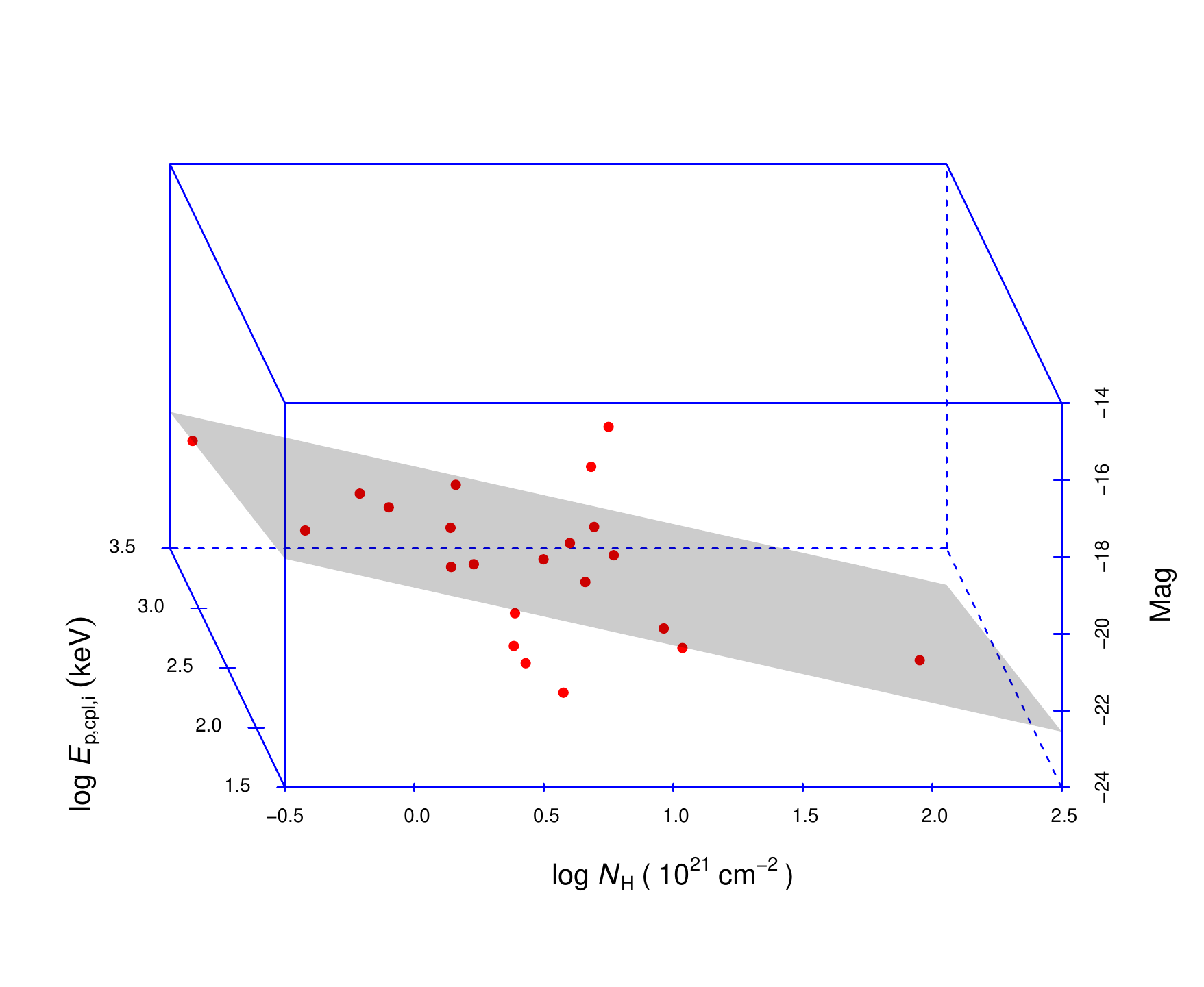}
\includegraphics[width=0.45\textwidth]{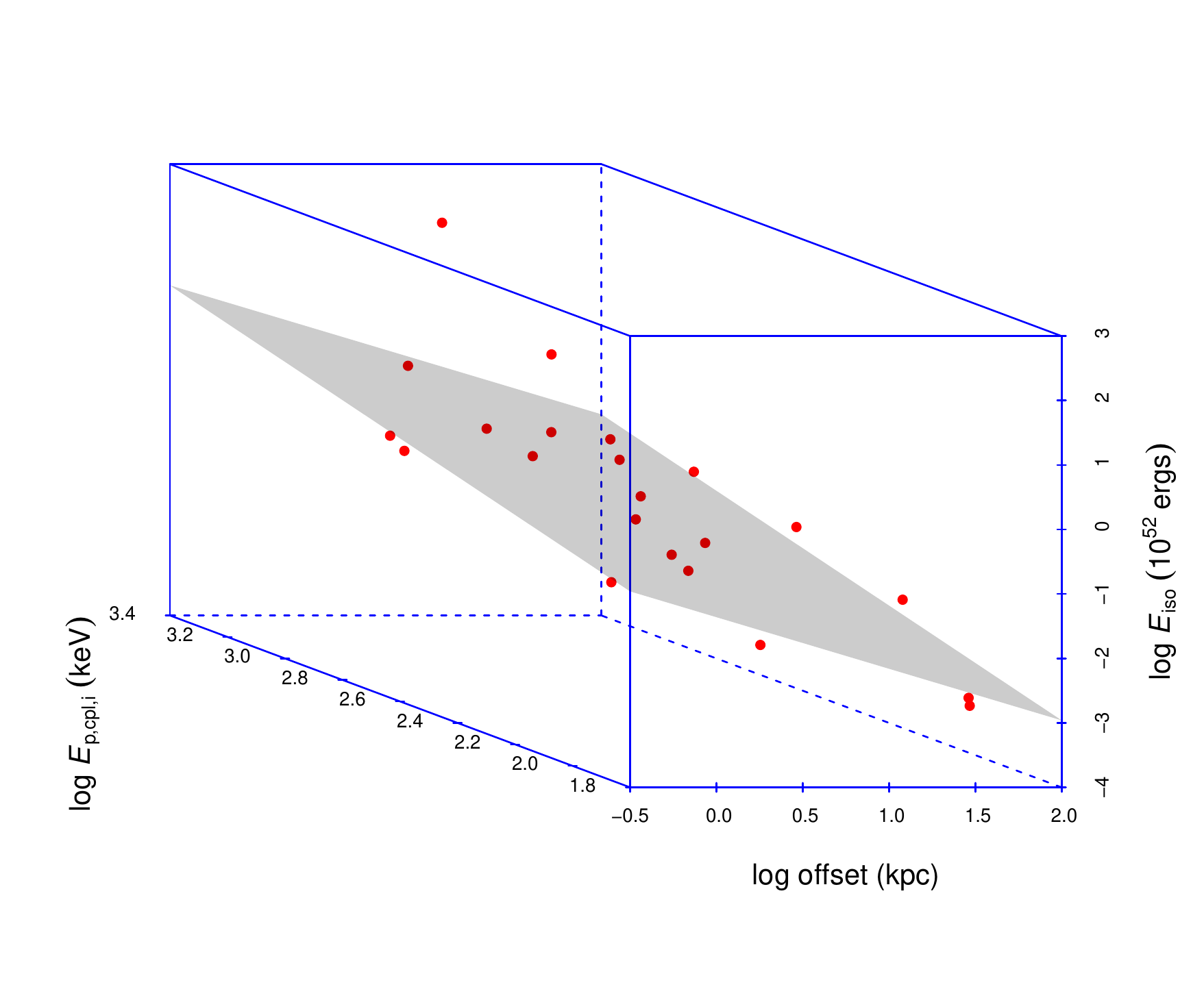}

\includegraphics[width=0.45\textwidth]{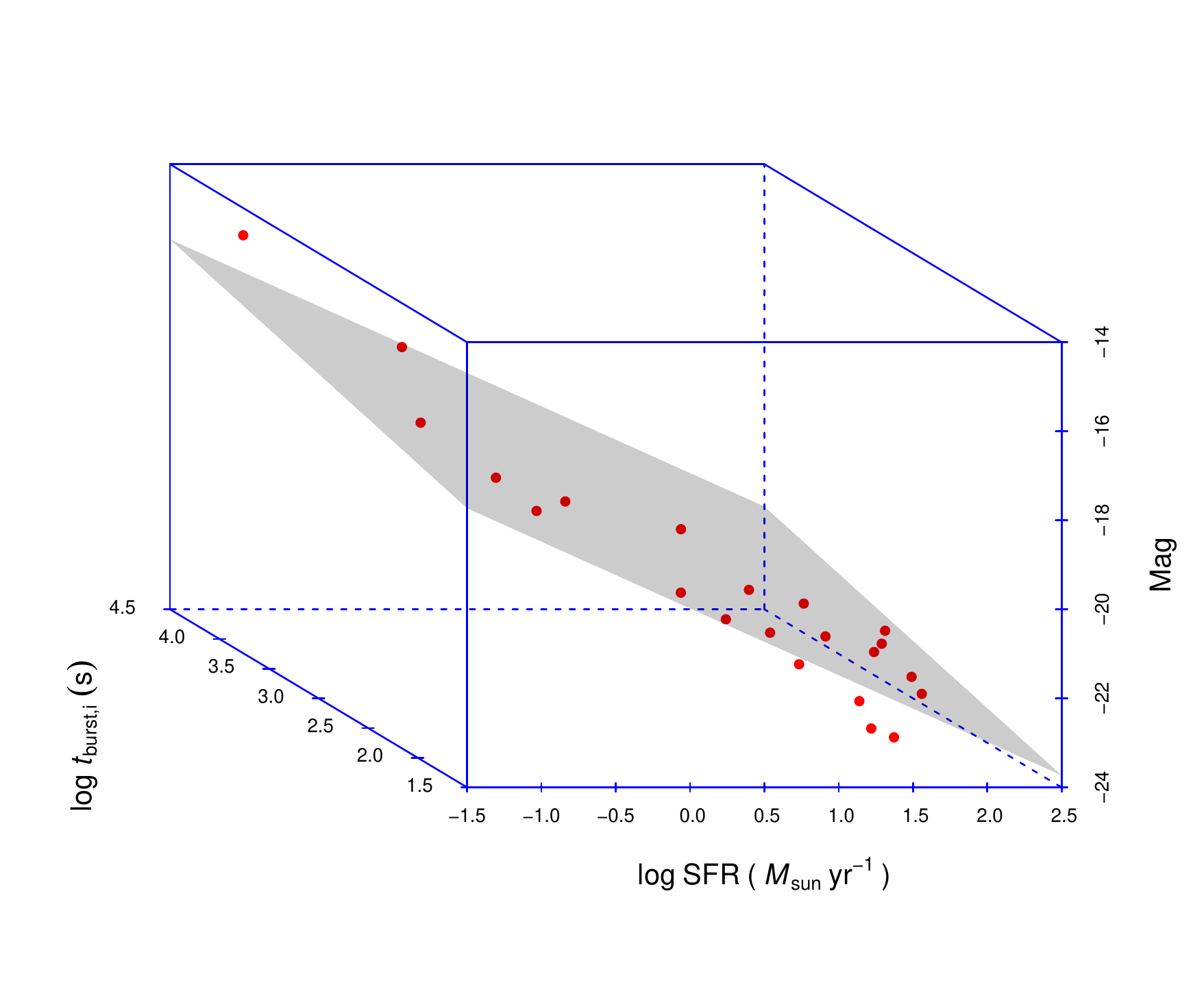}
\includegraphics[width=0.45\textwidth]{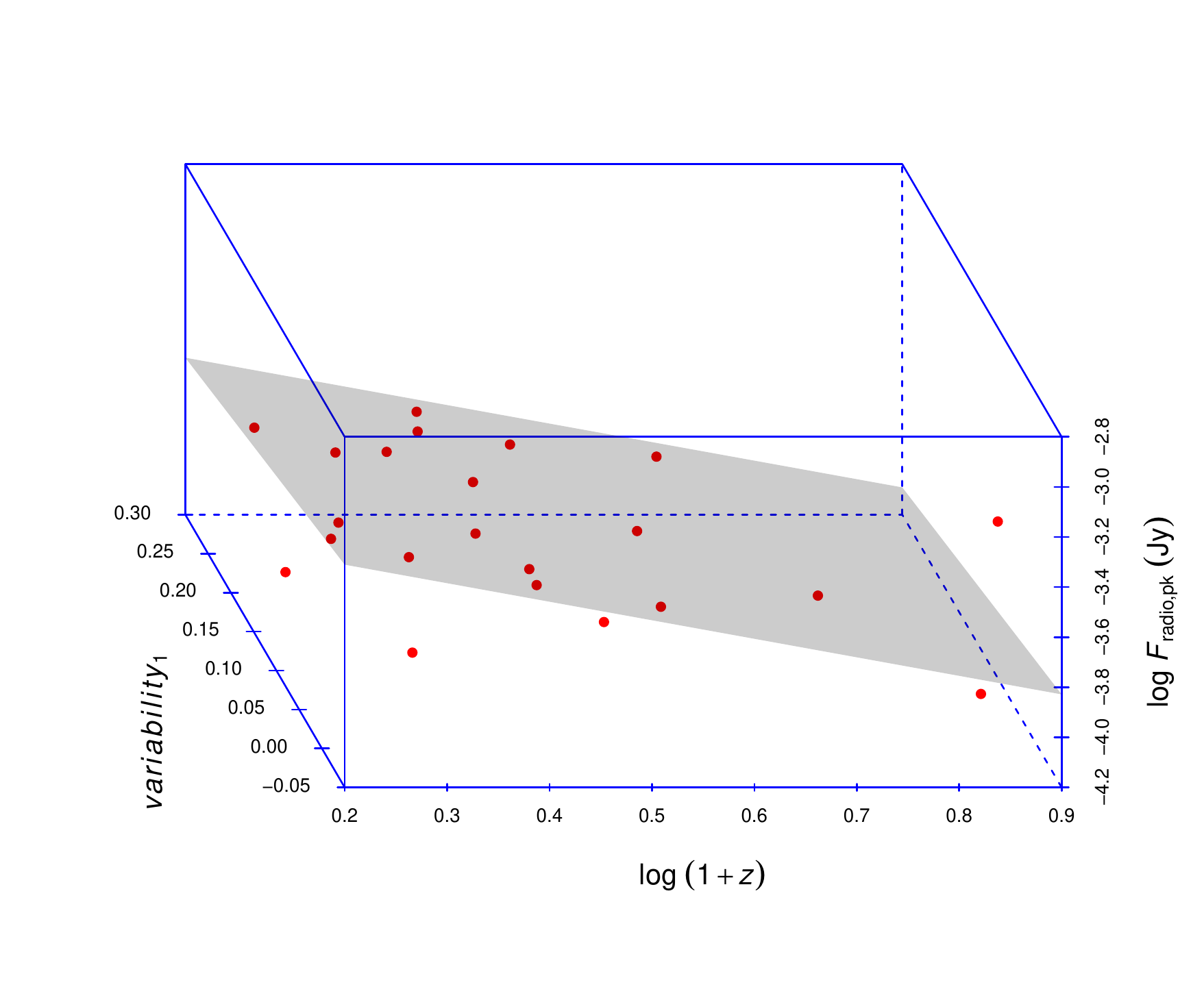}

\center{Fig. \ref{fig:three}---Continued}
\end{figure*}


\clearpage
\begin{figure*}

\includegraphics[width=0.45\textwidth]{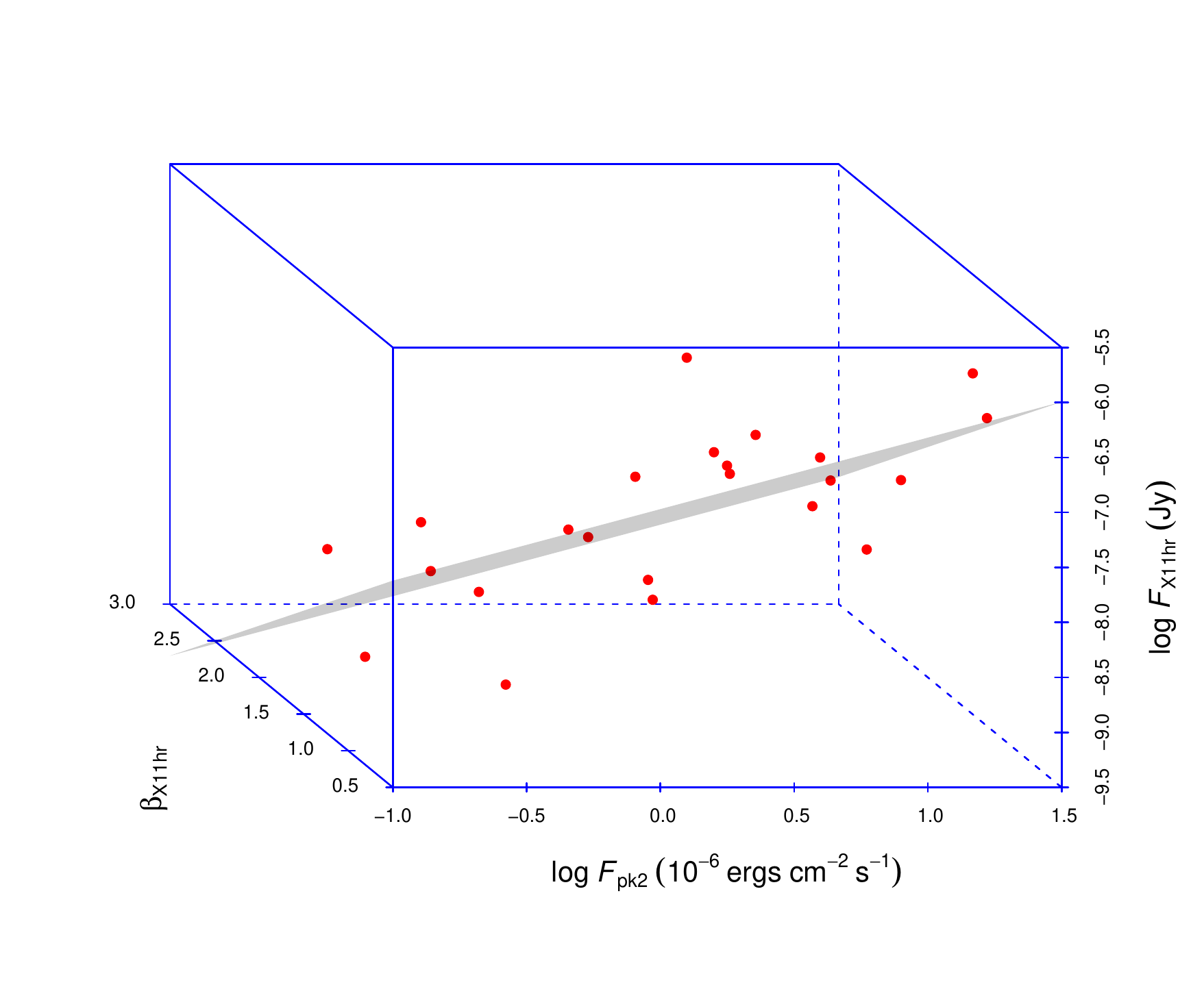}
\includegraphics[width=0.45\textwidth]{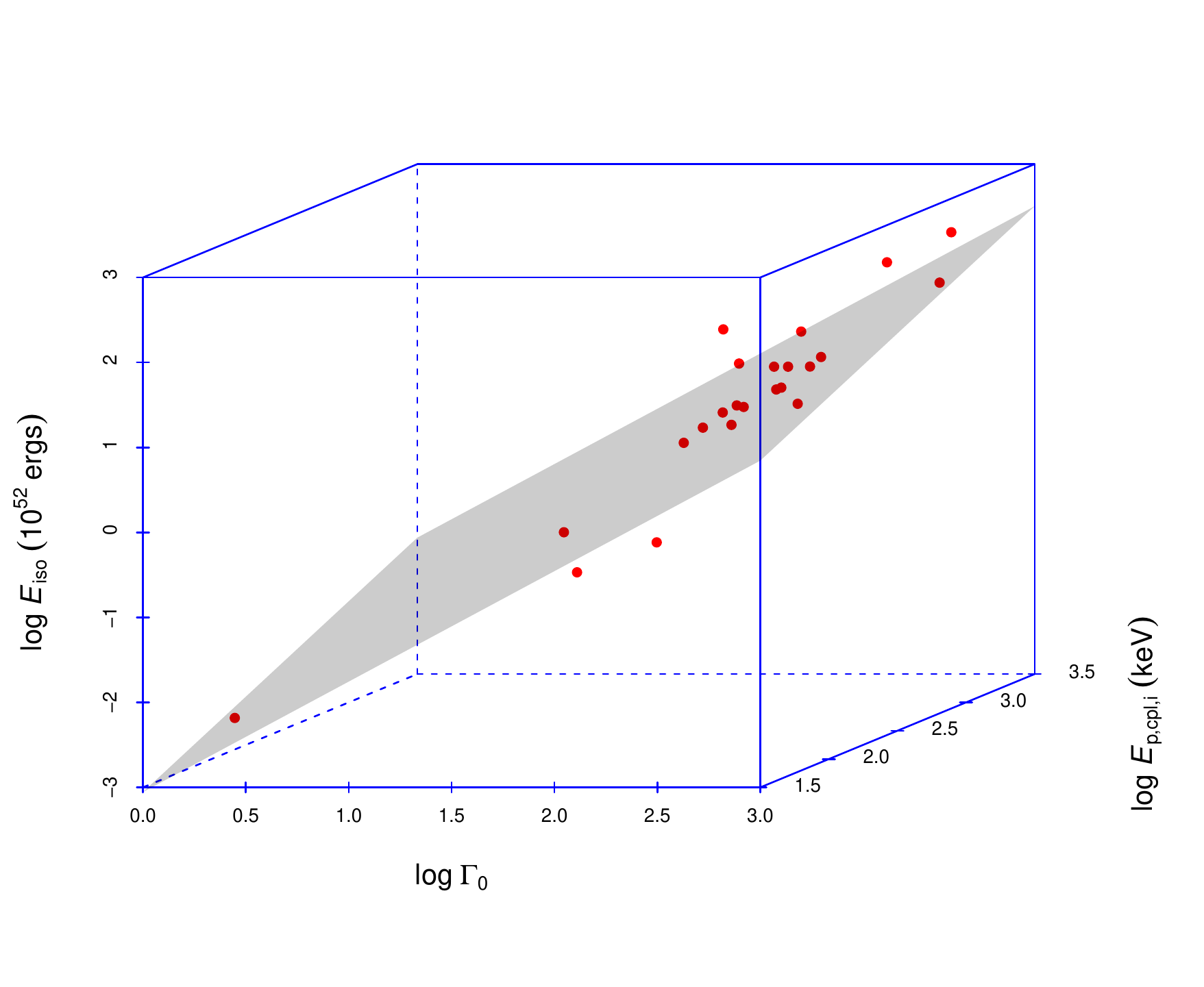}

\includegraphics[width=0.45\textwidth]{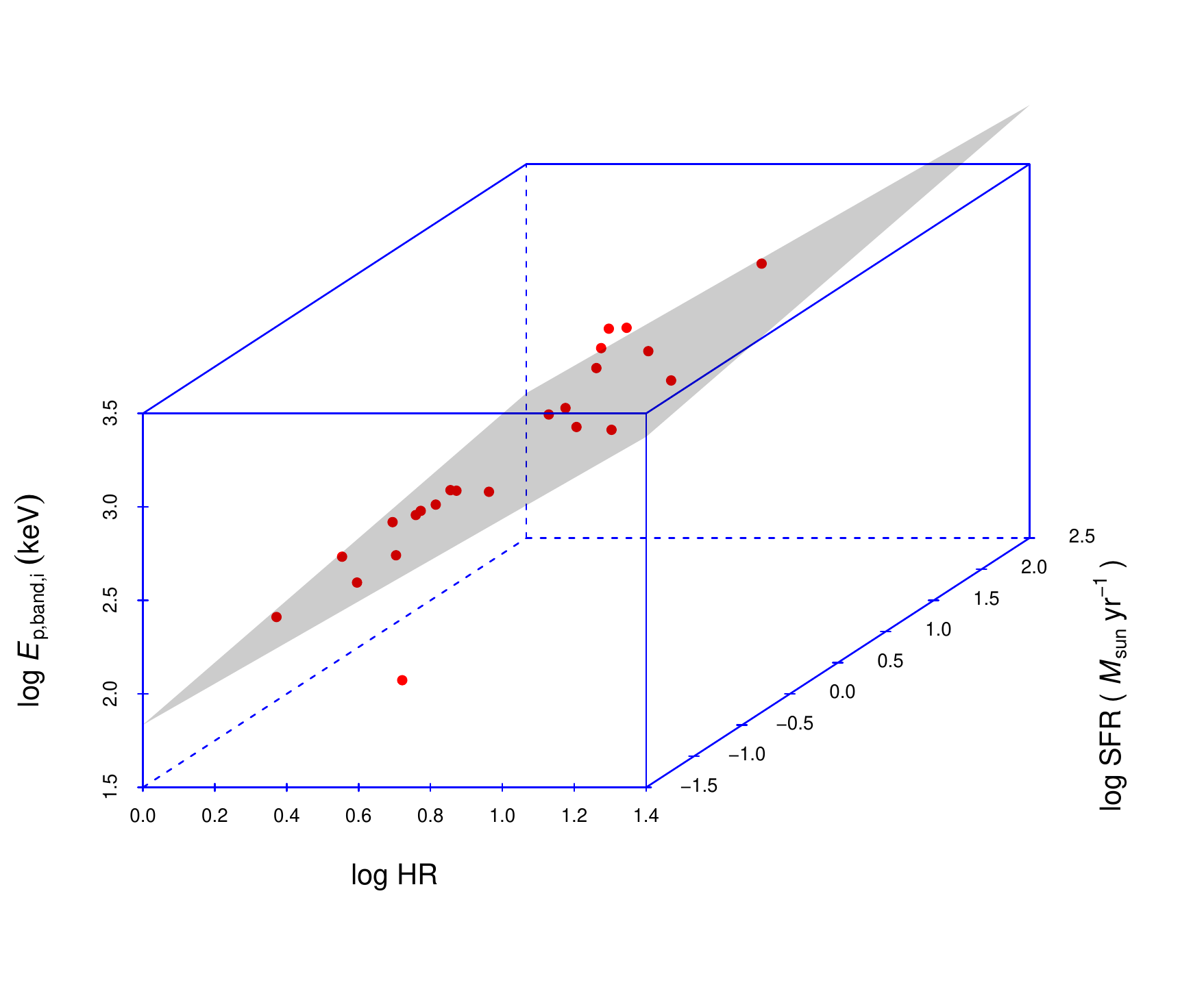}
\includegraphics[width=0.45\textwidth]{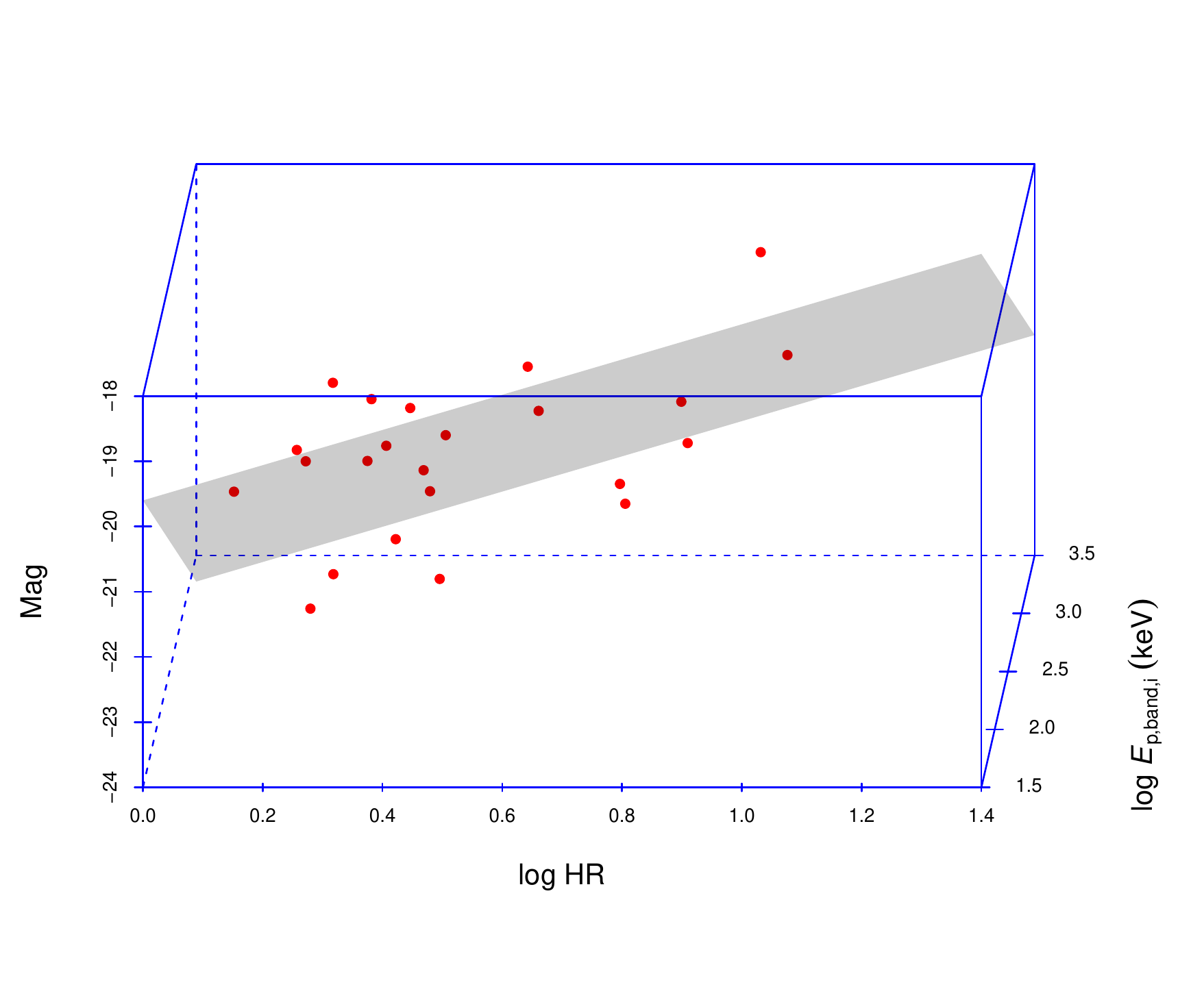}

\includegraphics[width=0.45\textwidth]{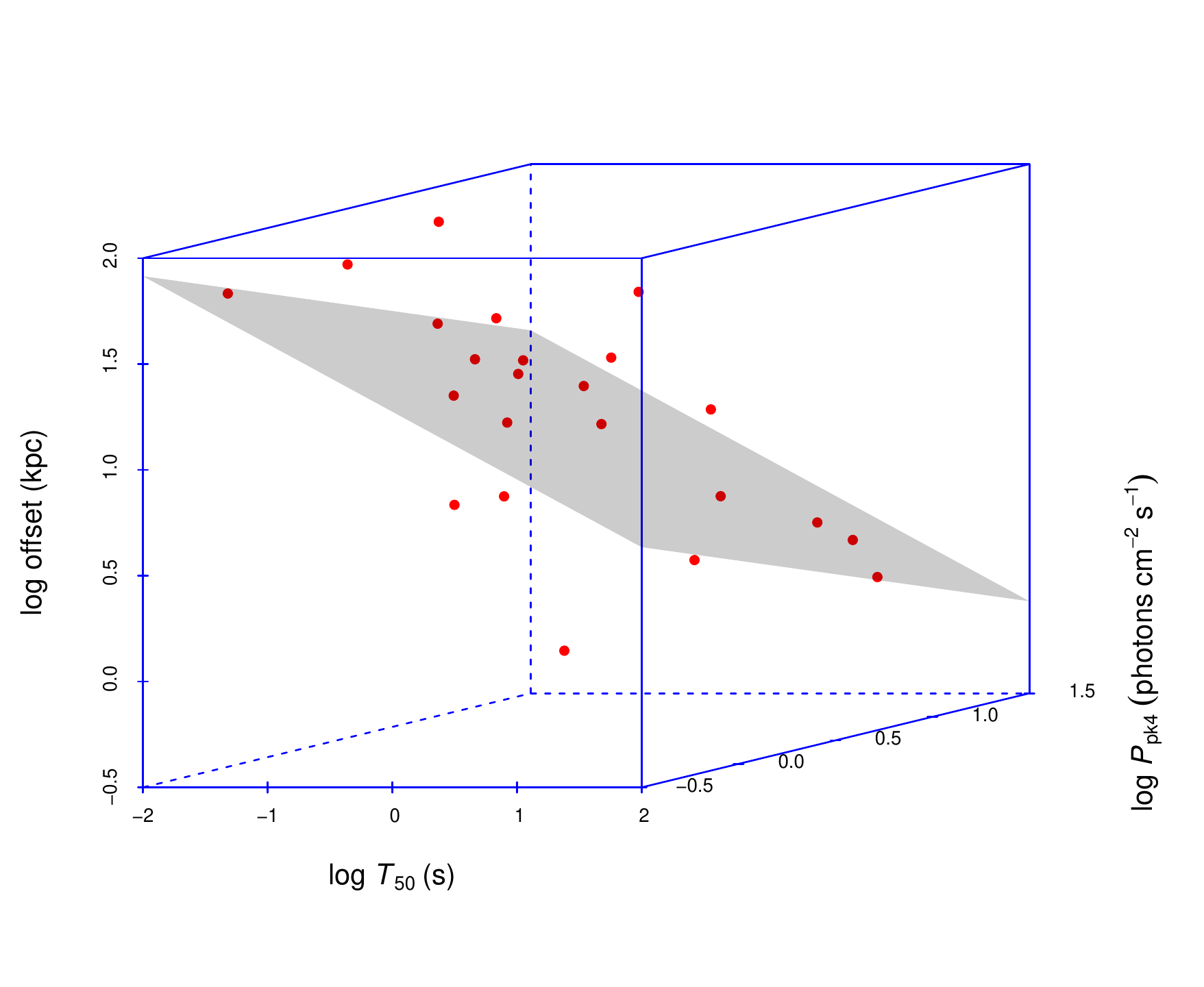}
\includegraphics[width=0.45\textwidth]{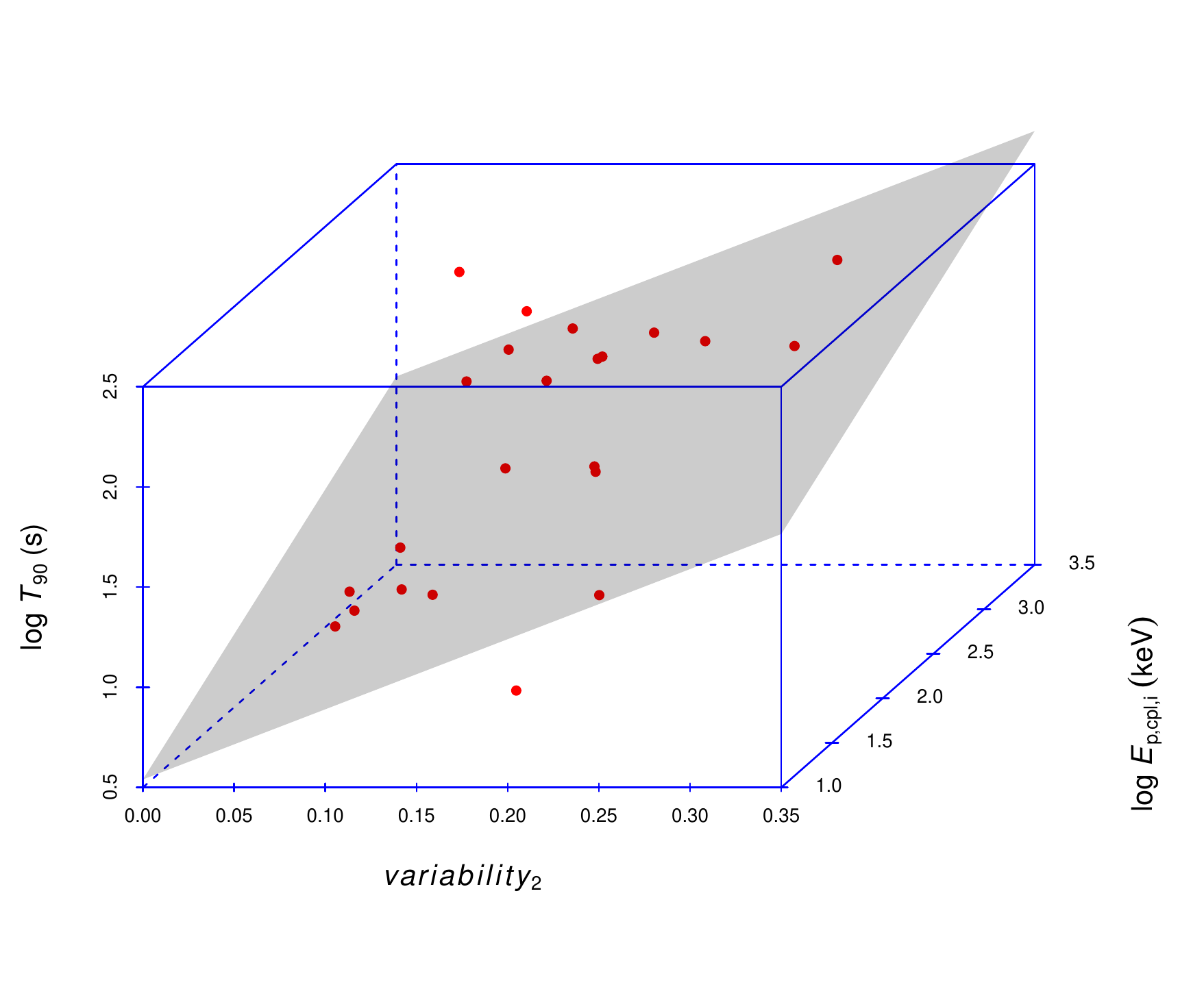}

\center{Fig. \ref{fig:three}---Continued}
\end{figure*}


\clearpage
\begin{figure*}

\includegraphics[width=0.45\textwidth]{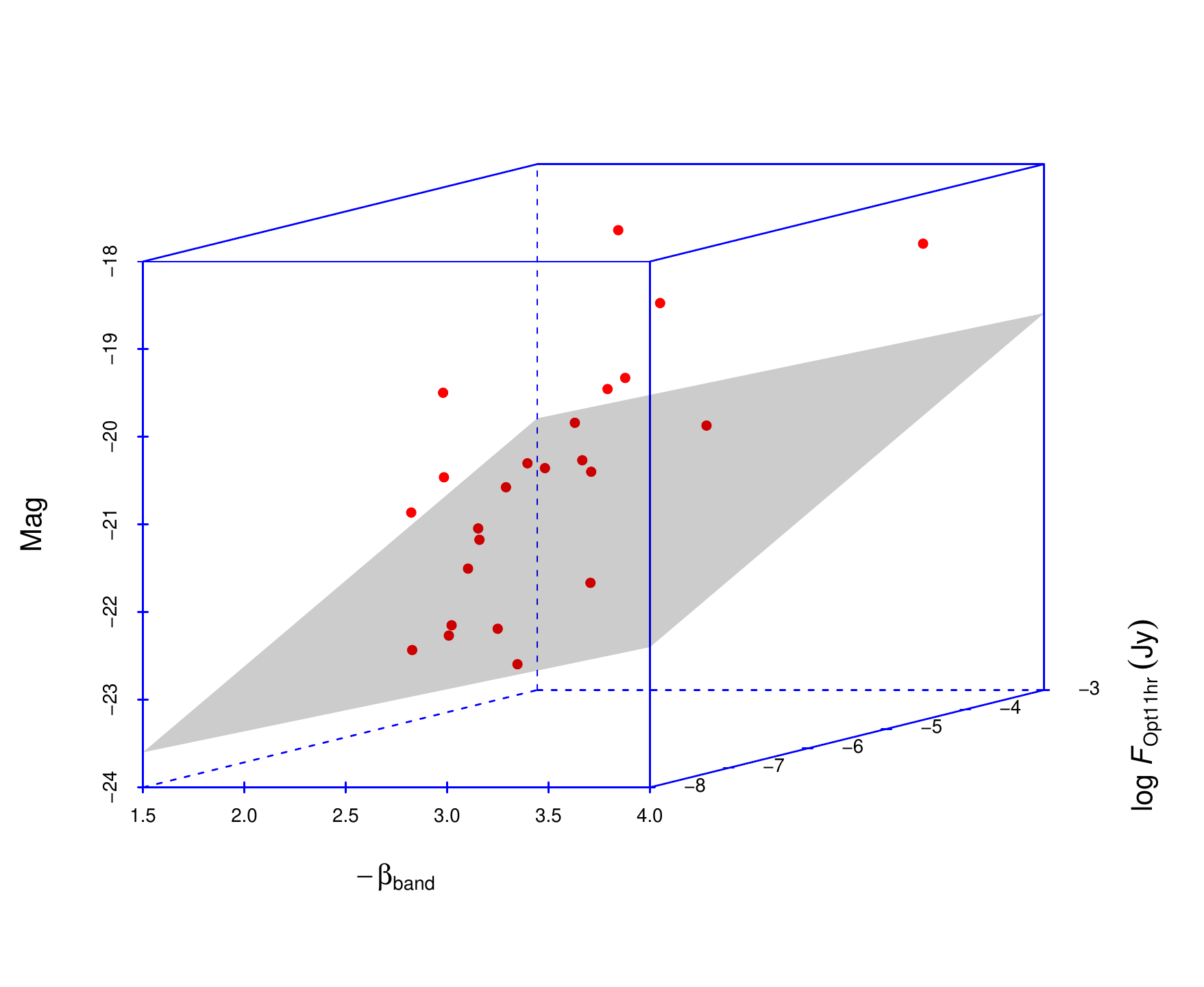}
\includegraphics[width=0.45\textwidth]{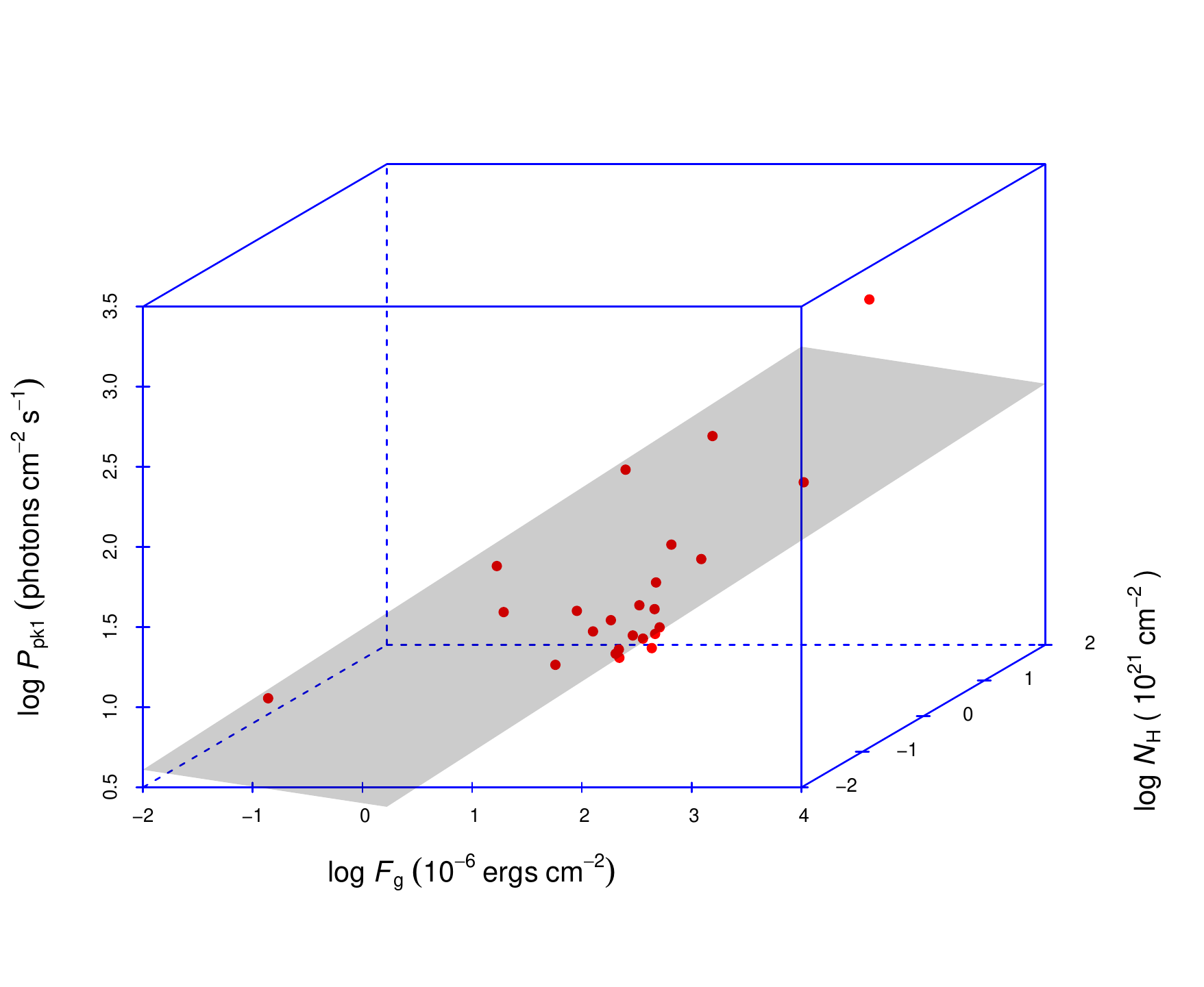}

\includegraphics[width=0.45\textwidth]{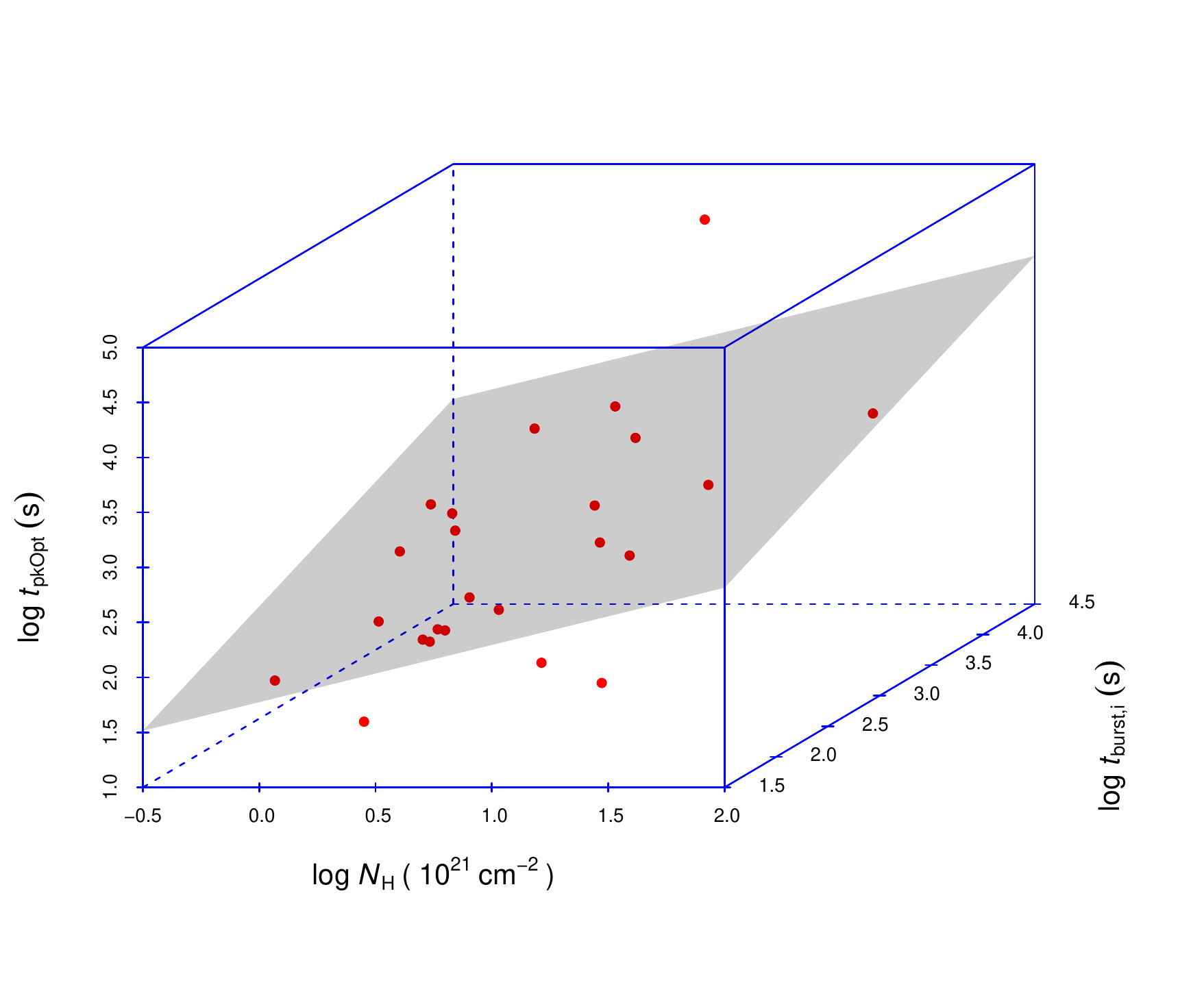}
\includegraphics[width=0.45\textwidth]{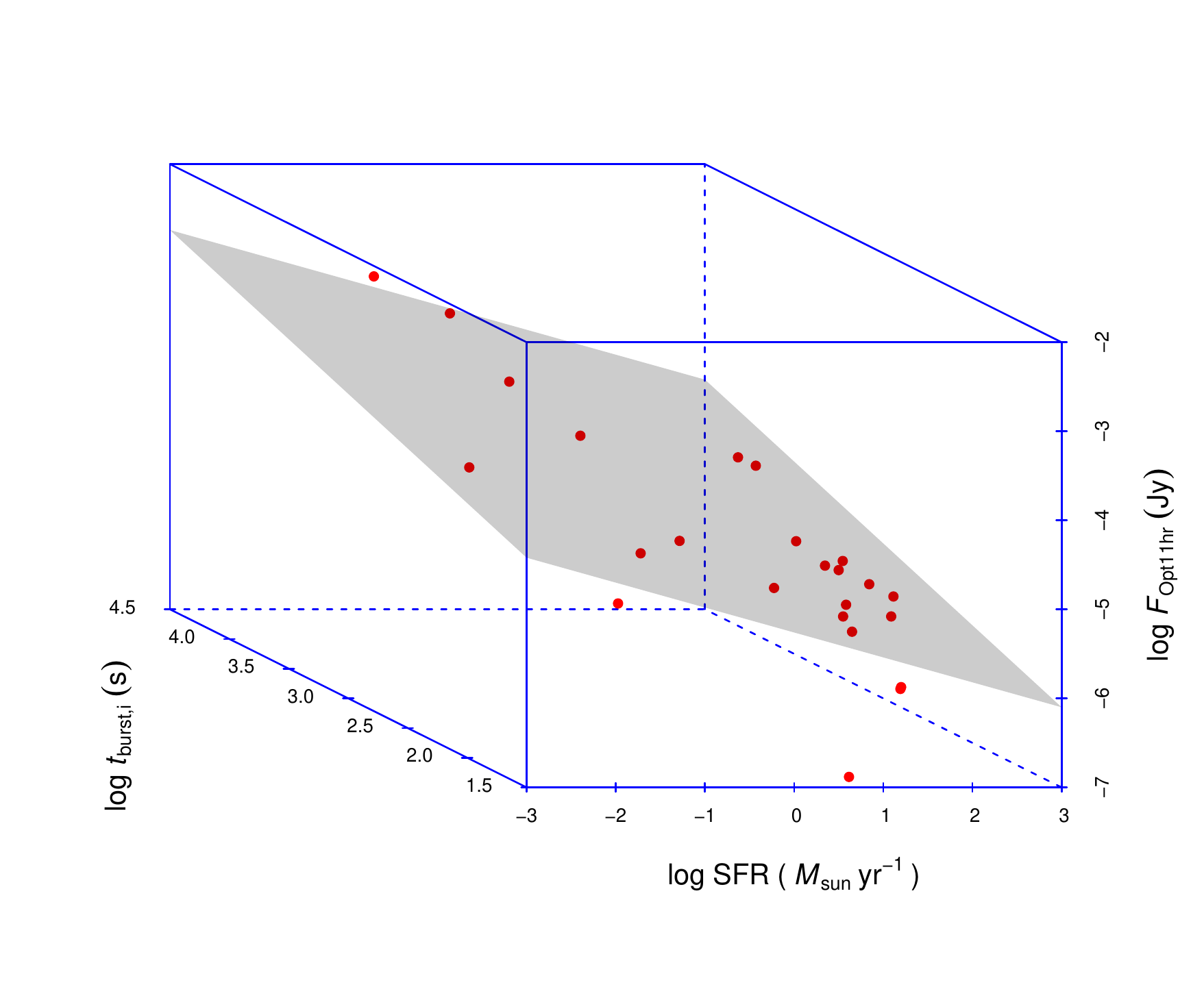}

\includegraphics[width=0.45\textwidth]{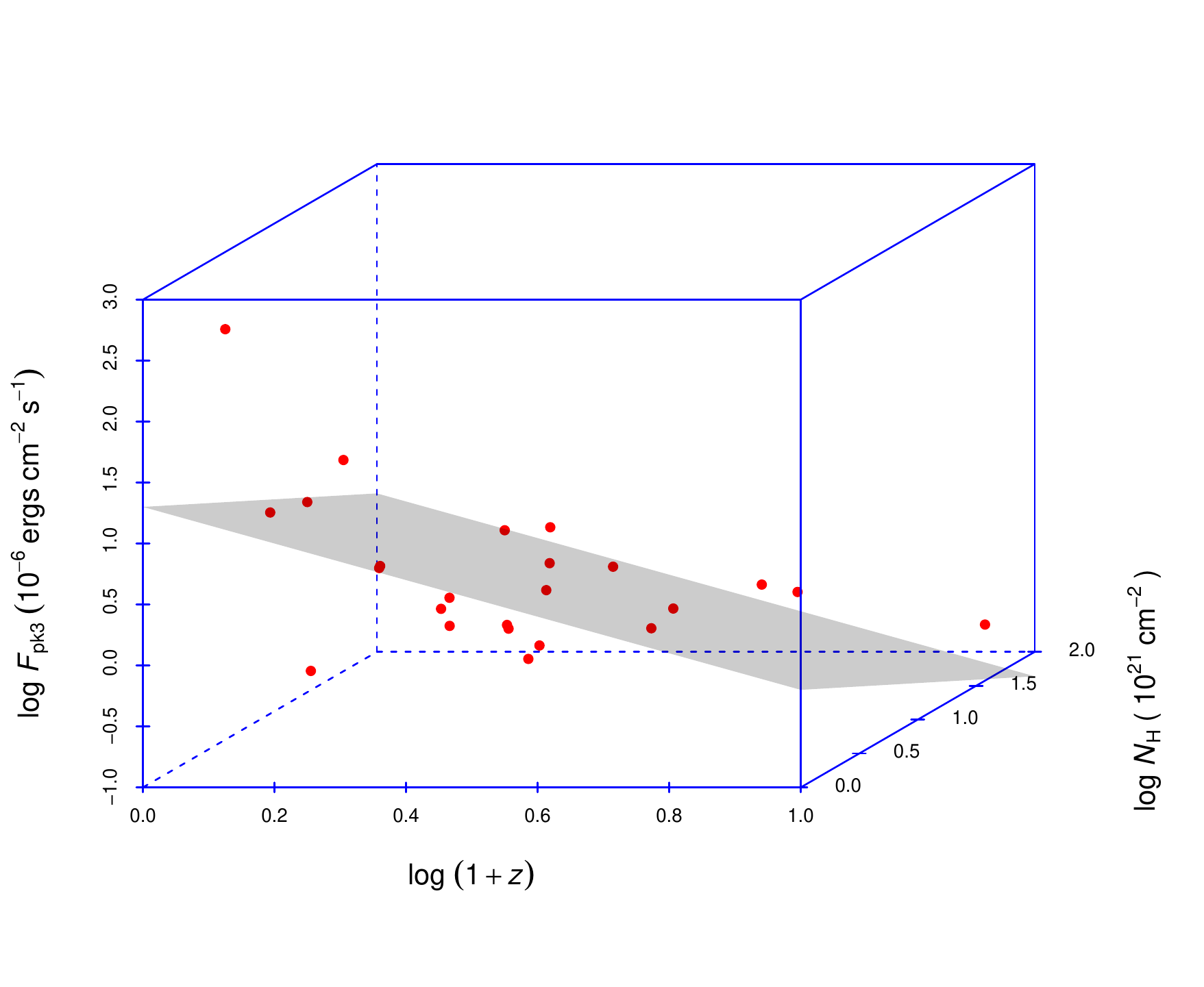}
\includegraphics[width=0.45\textwidth]{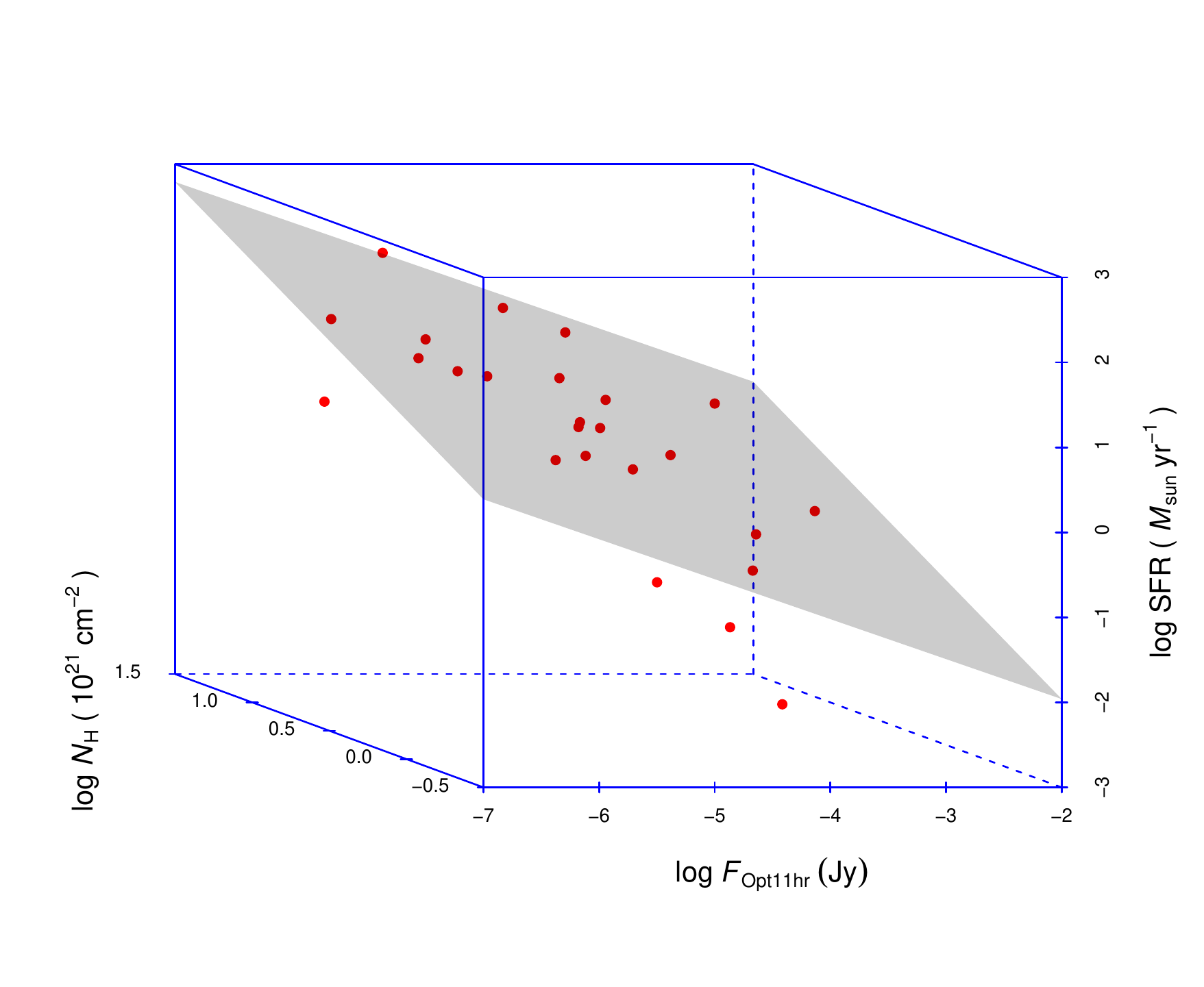}

\center{Fig. \ref{fig:three}---Continued}
\end{figure*}


\clearpage
\begin{figure*}

\includegraphics[width=0.45\textwidth]{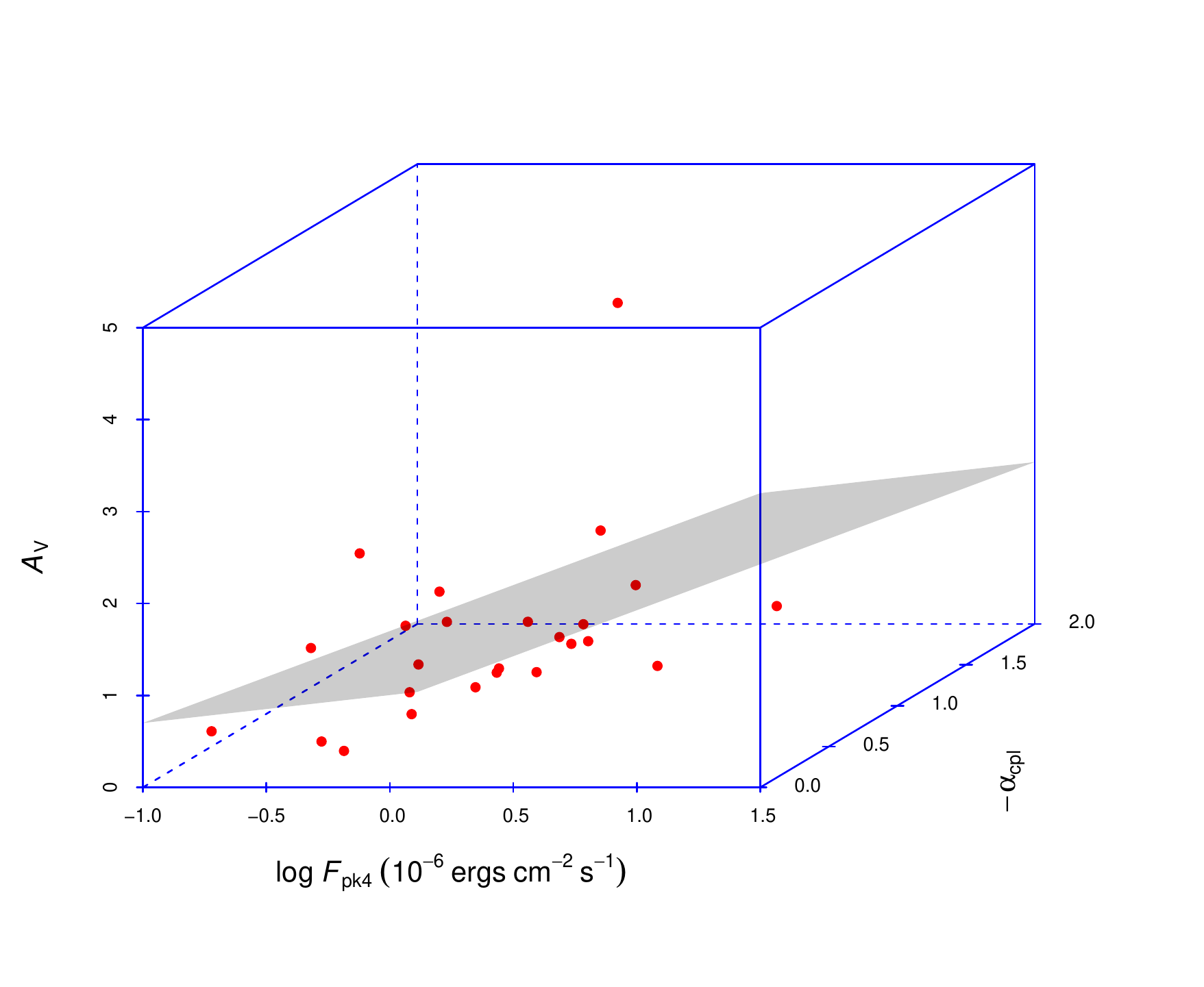}
\includegraphics[width=0.45\textwidth]{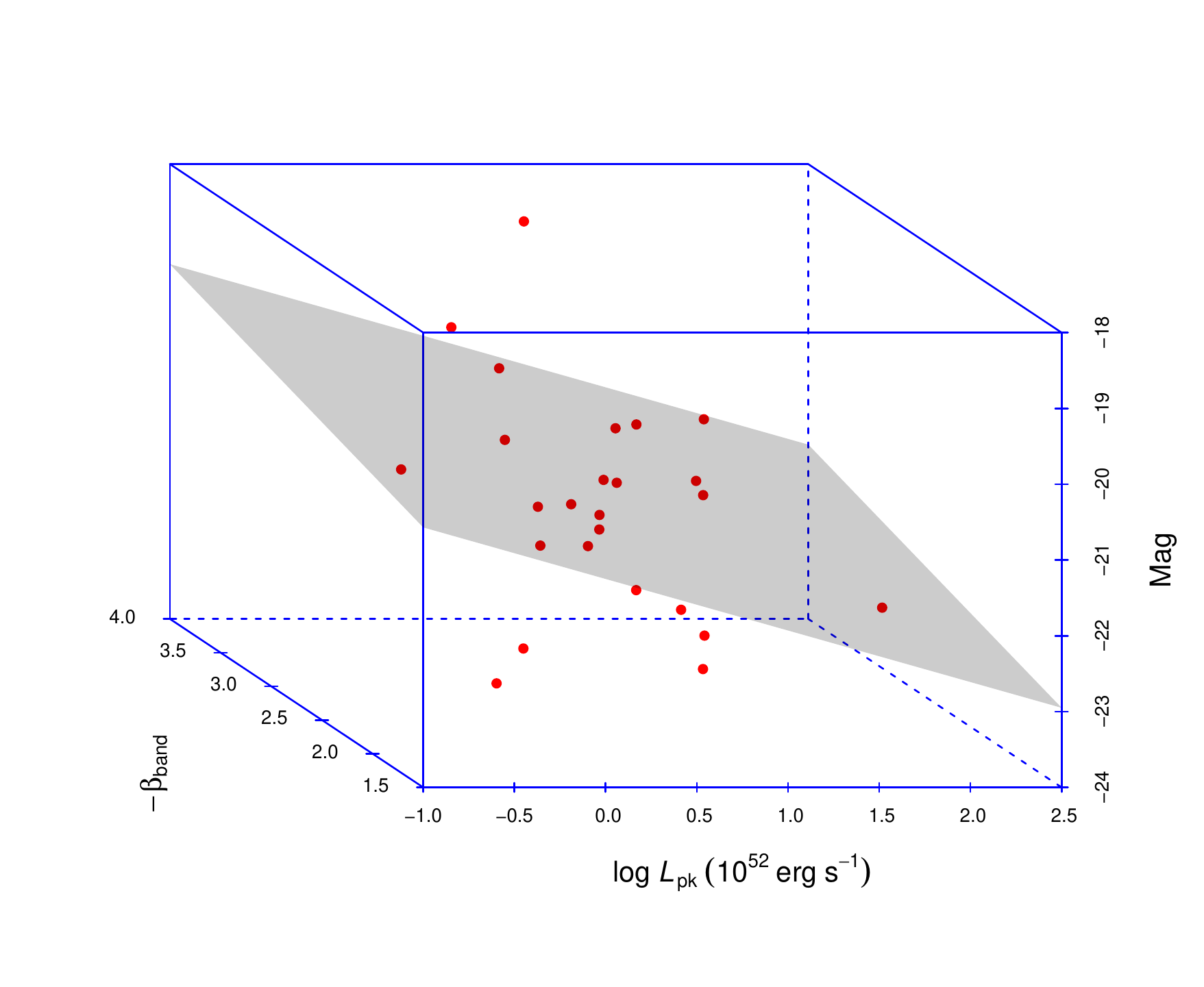}

\includegraphics[width=0.45\textwidth]{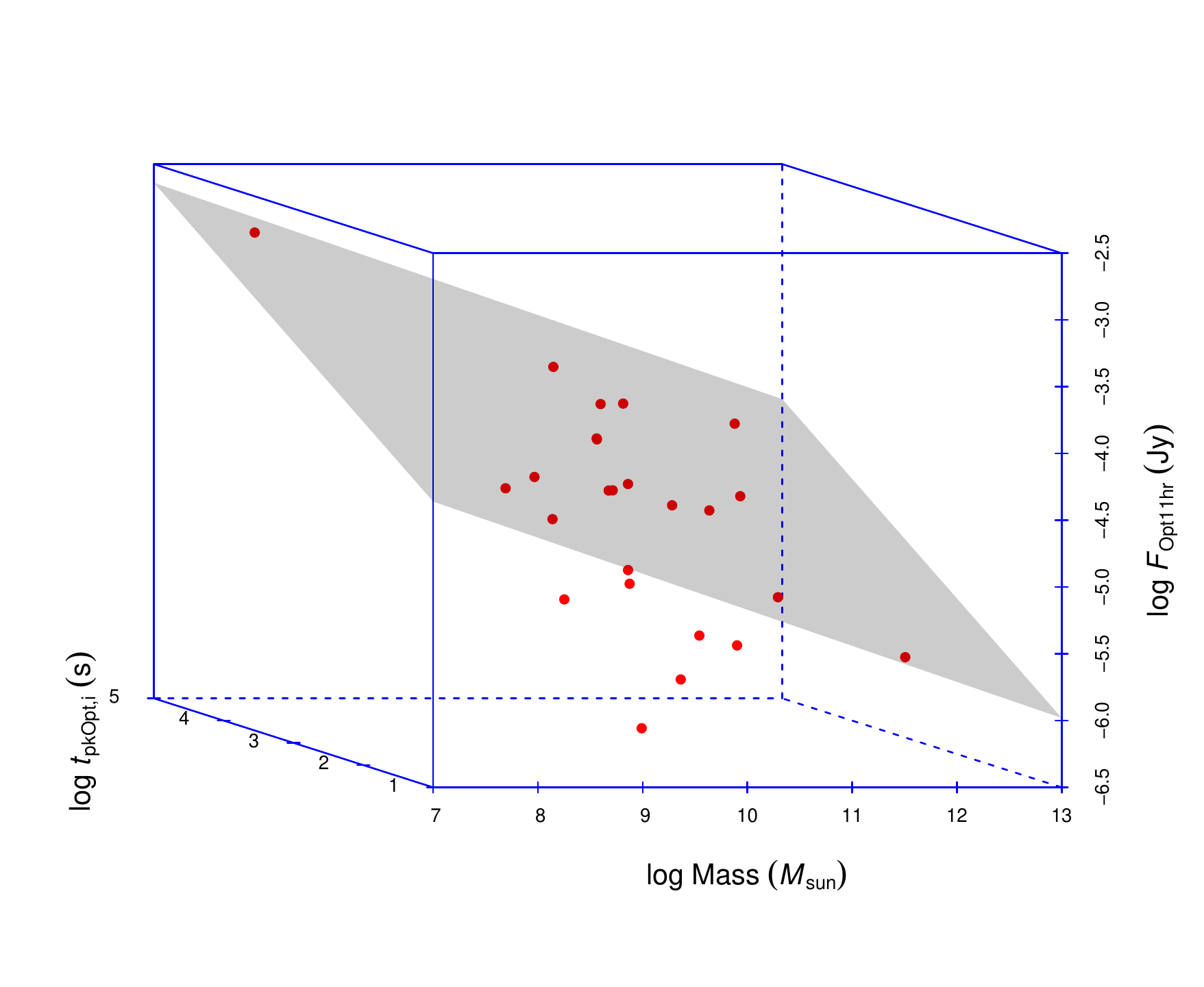}
\includegraphics[width=0.45\textwidth]{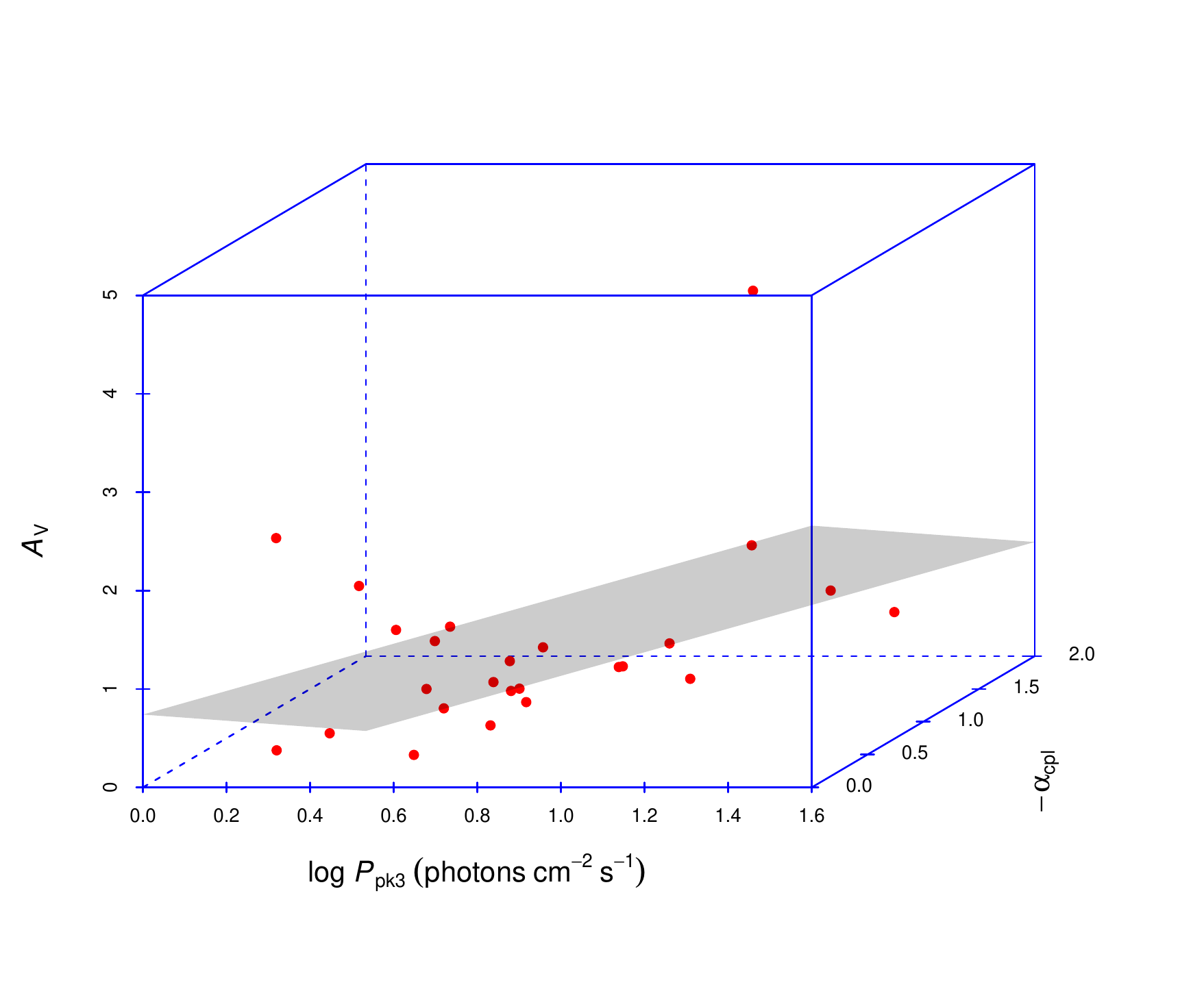}

\includegraphics[width=0.45\textwidth]{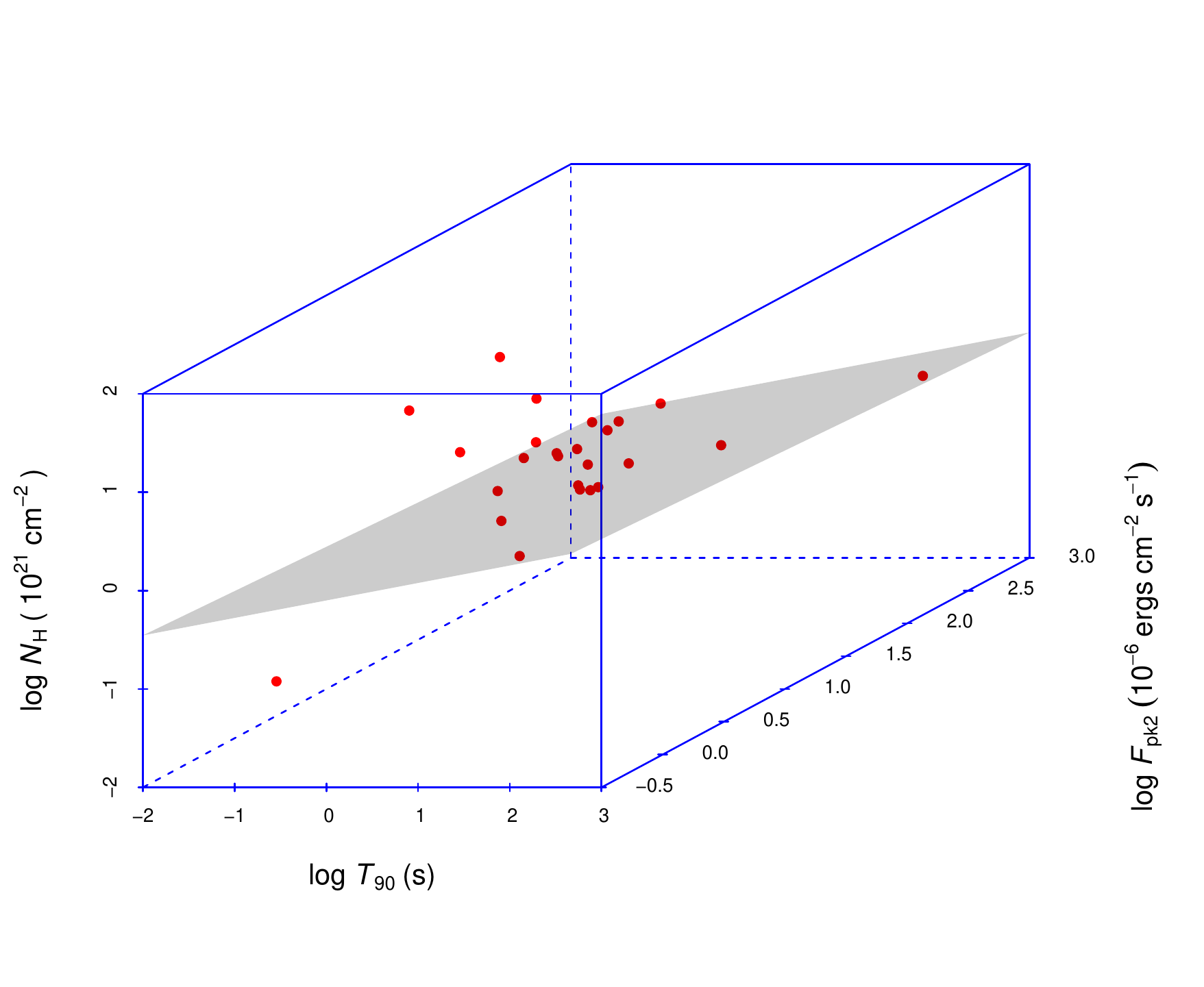}
\includegraphics[width=0.45\textwidth]{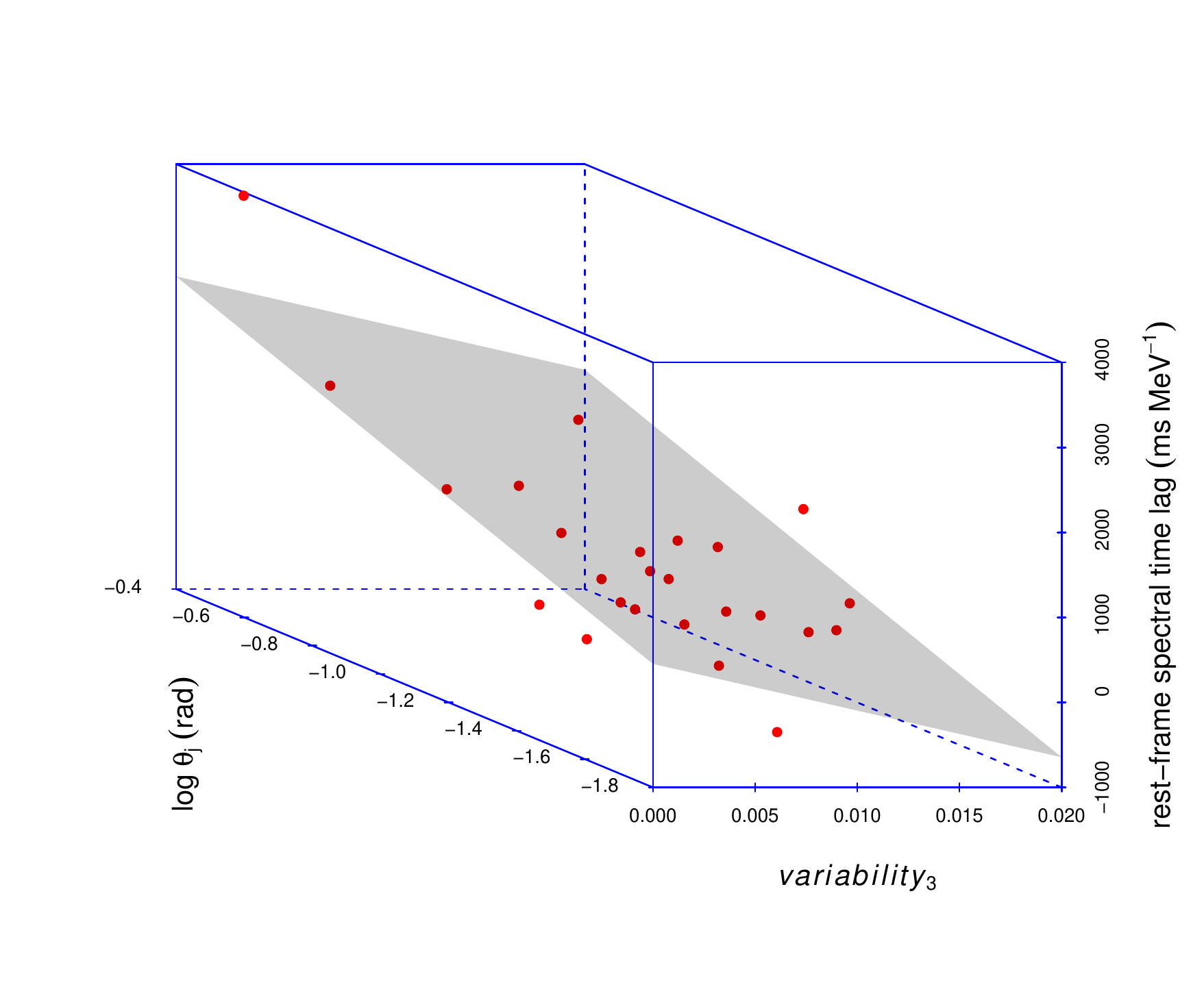}

\center{Fig. \ref{fig:three}---Continued}
\end{figure*}


\clearpage
\begin{figure*}

\includegraphics[width=0.45\textwidth]{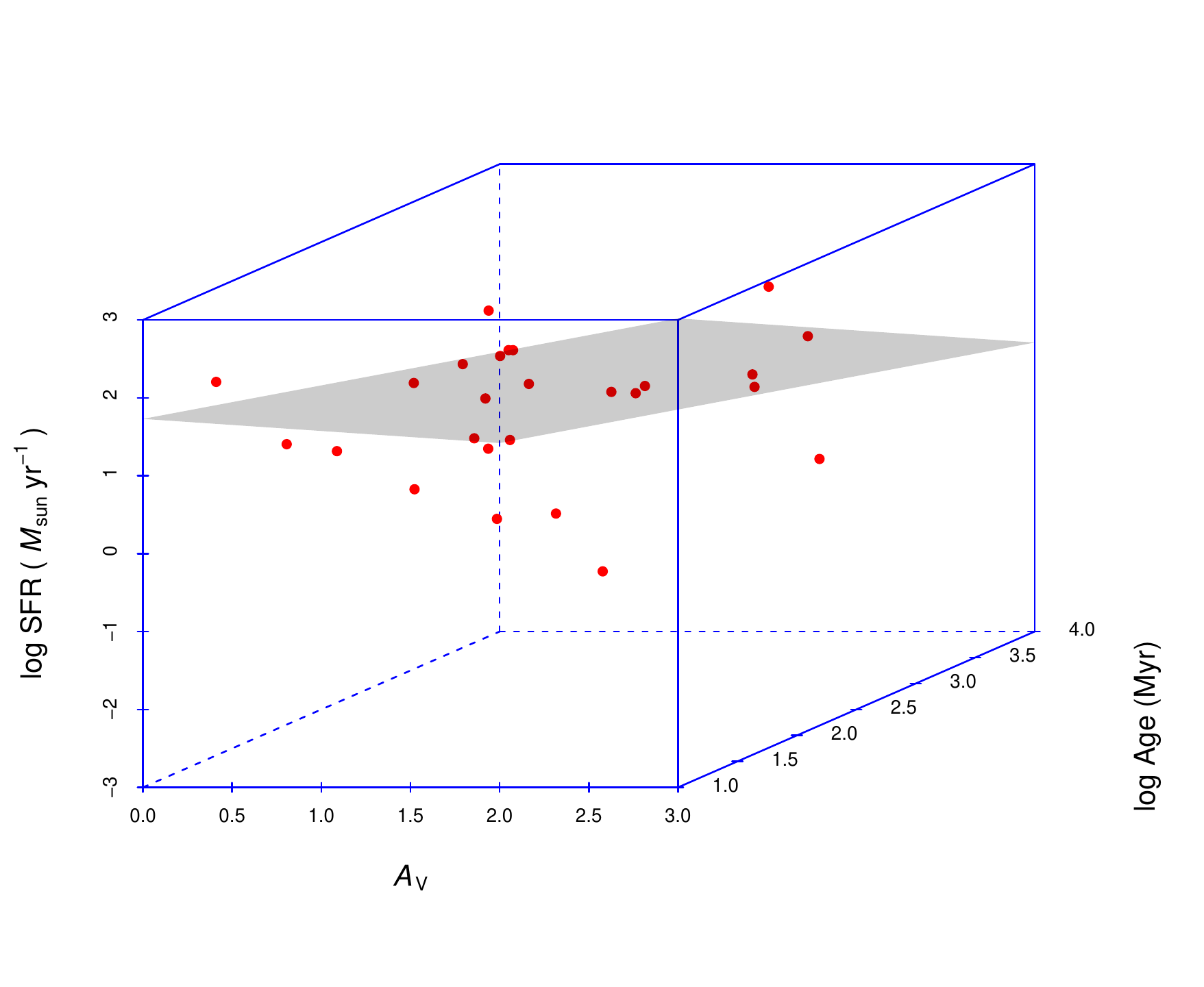}
\includegraphics[width=0.45\textwidth]{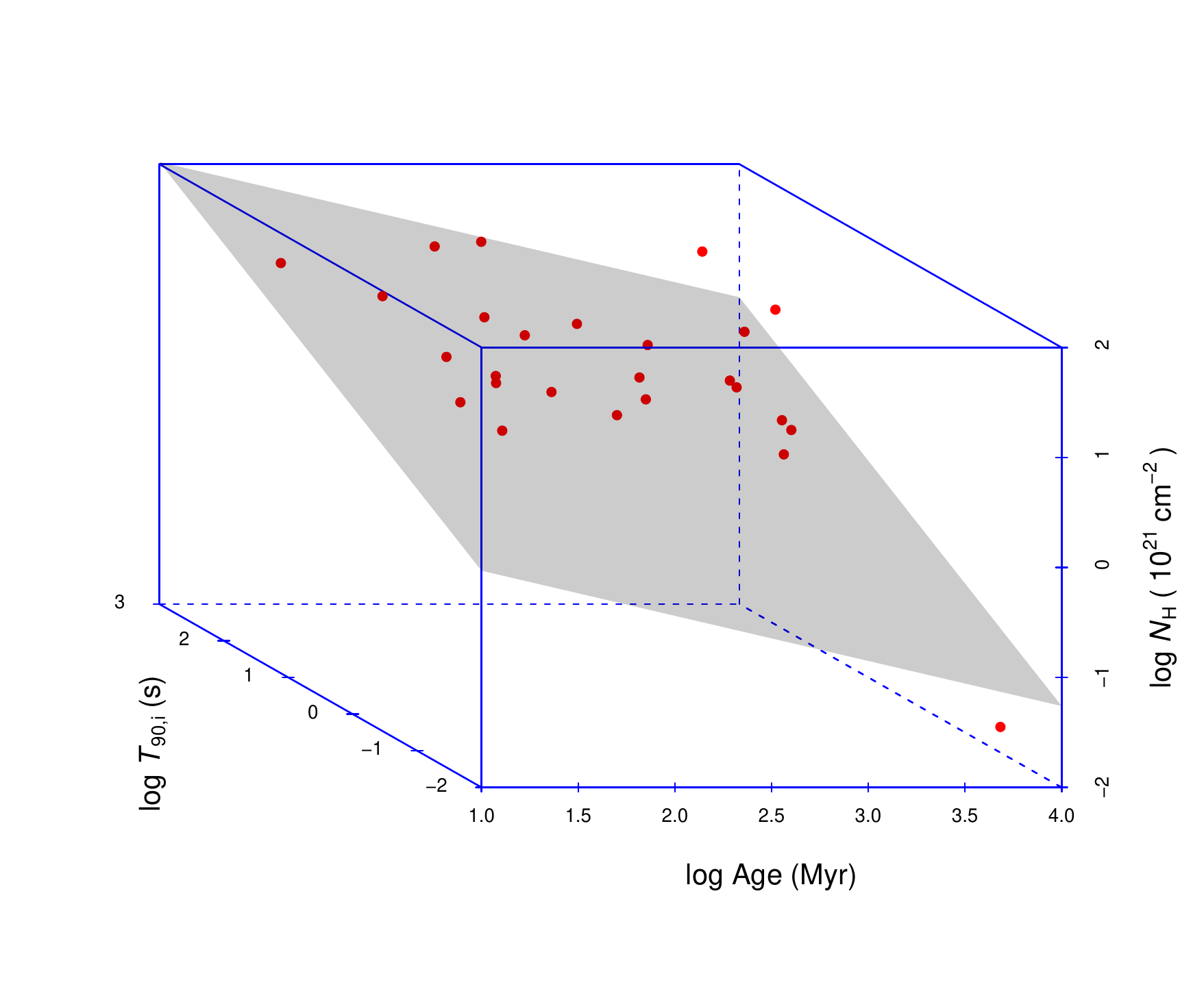}

\includegraphics[width=0.45\textwidth]{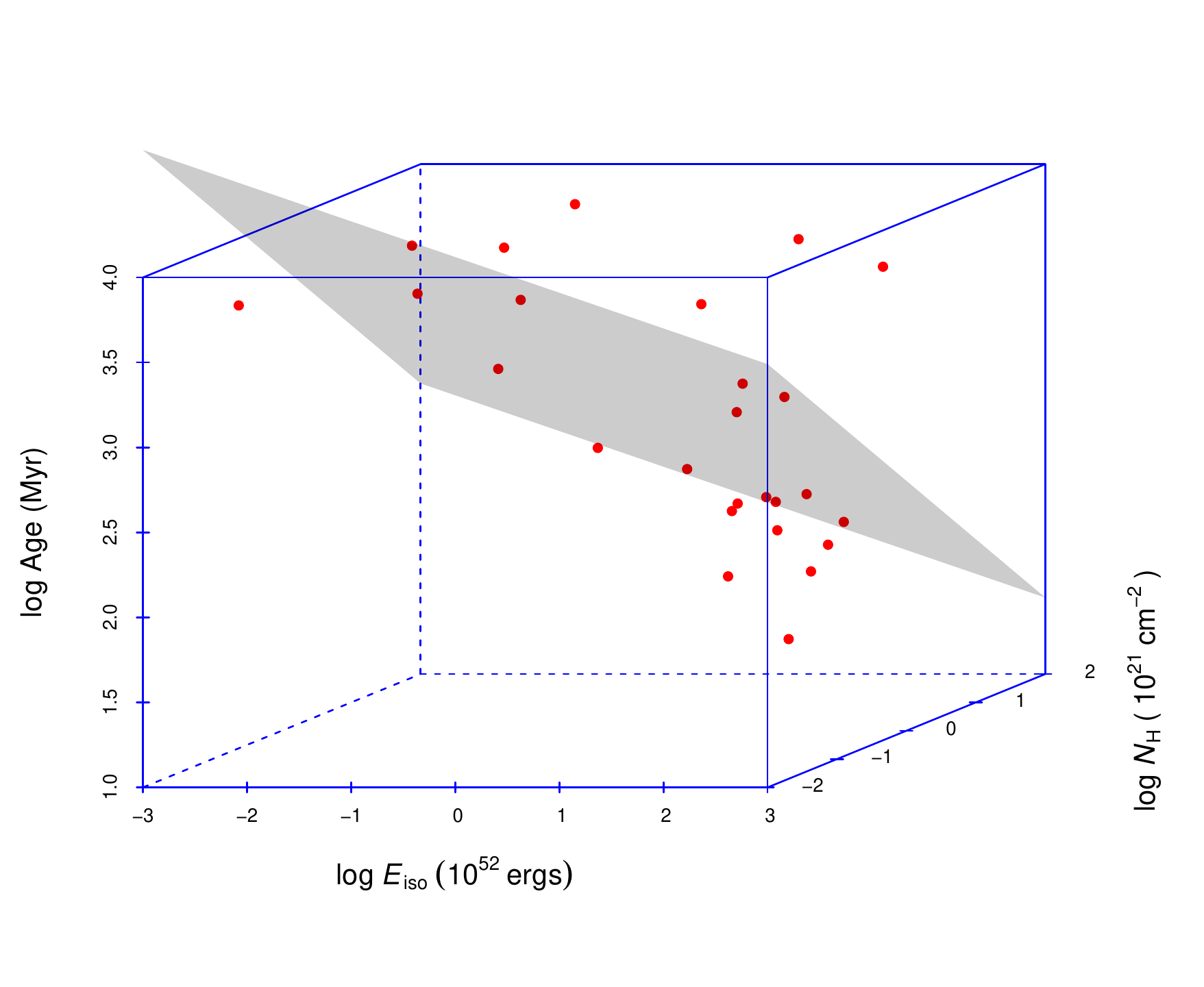}
\includegraphics[width=0.45\textwidth]{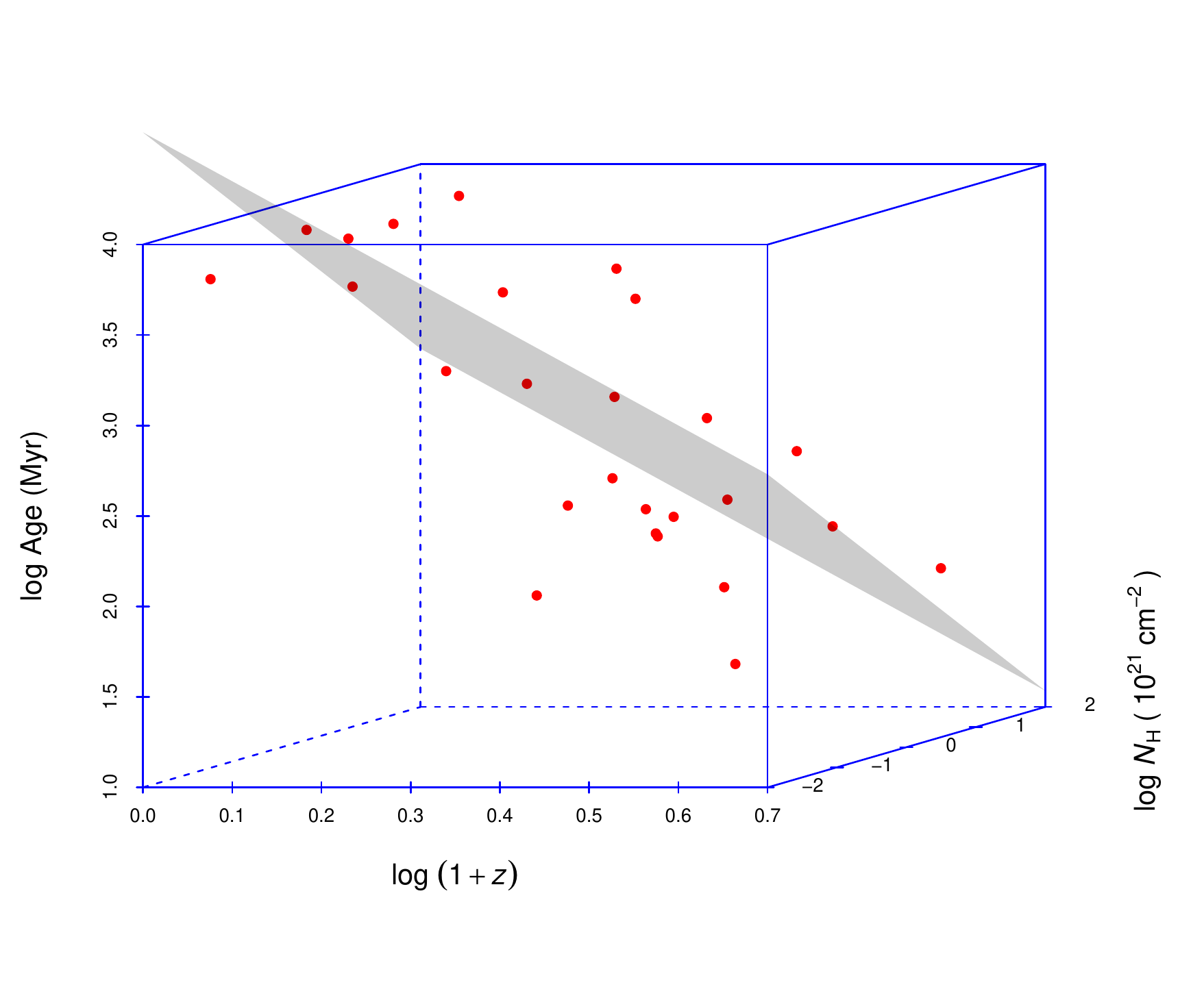}

\includegraphics[width=0.45\textwidth]{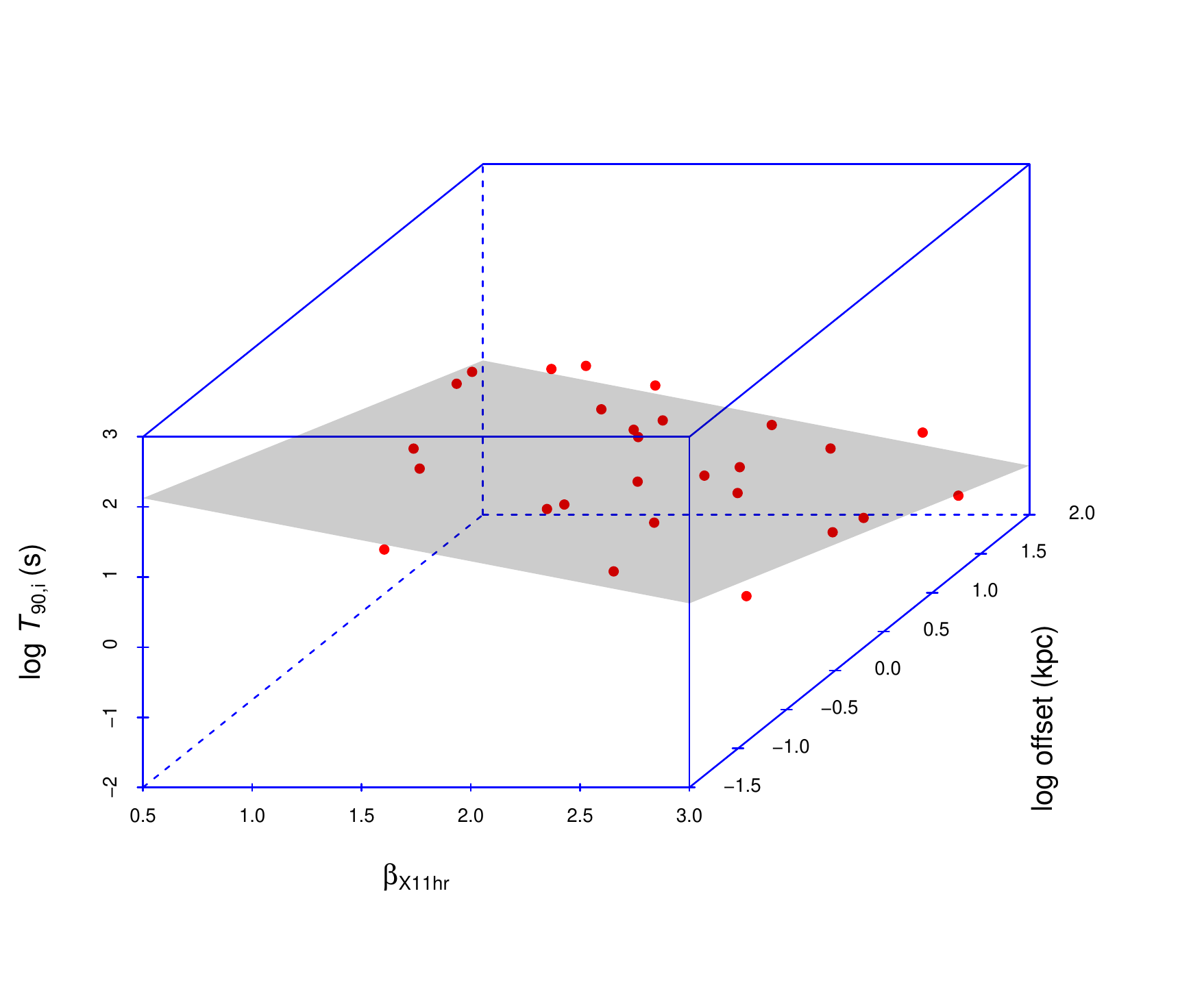}
\includegraphics[width=0.45\textwidth]{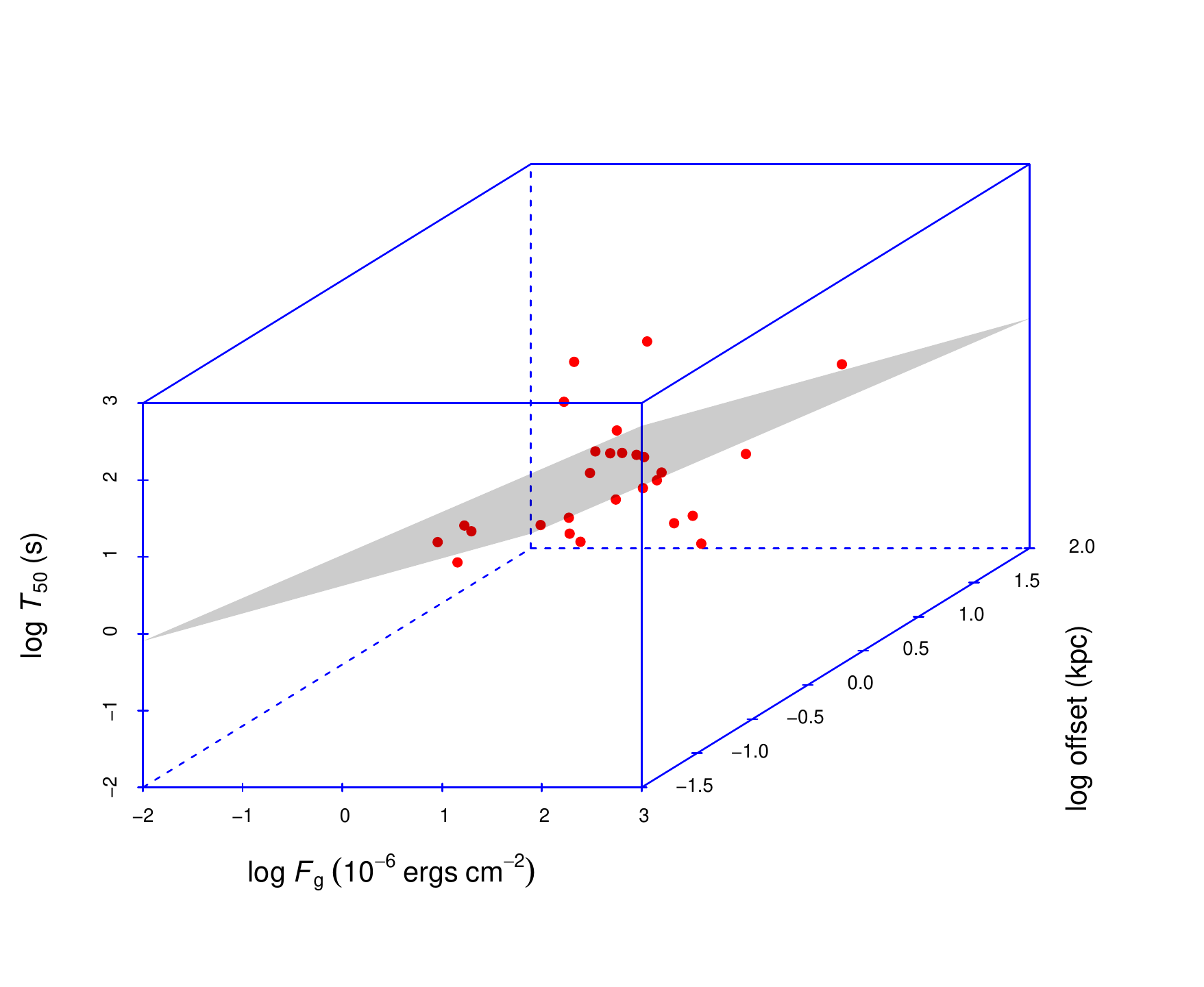}

\center{Fig. \ref{fig:three}---Continued}
\end{figure*}


\clearpage
\begin{figure*}

\includegraphics[width=0.45\textwidth]{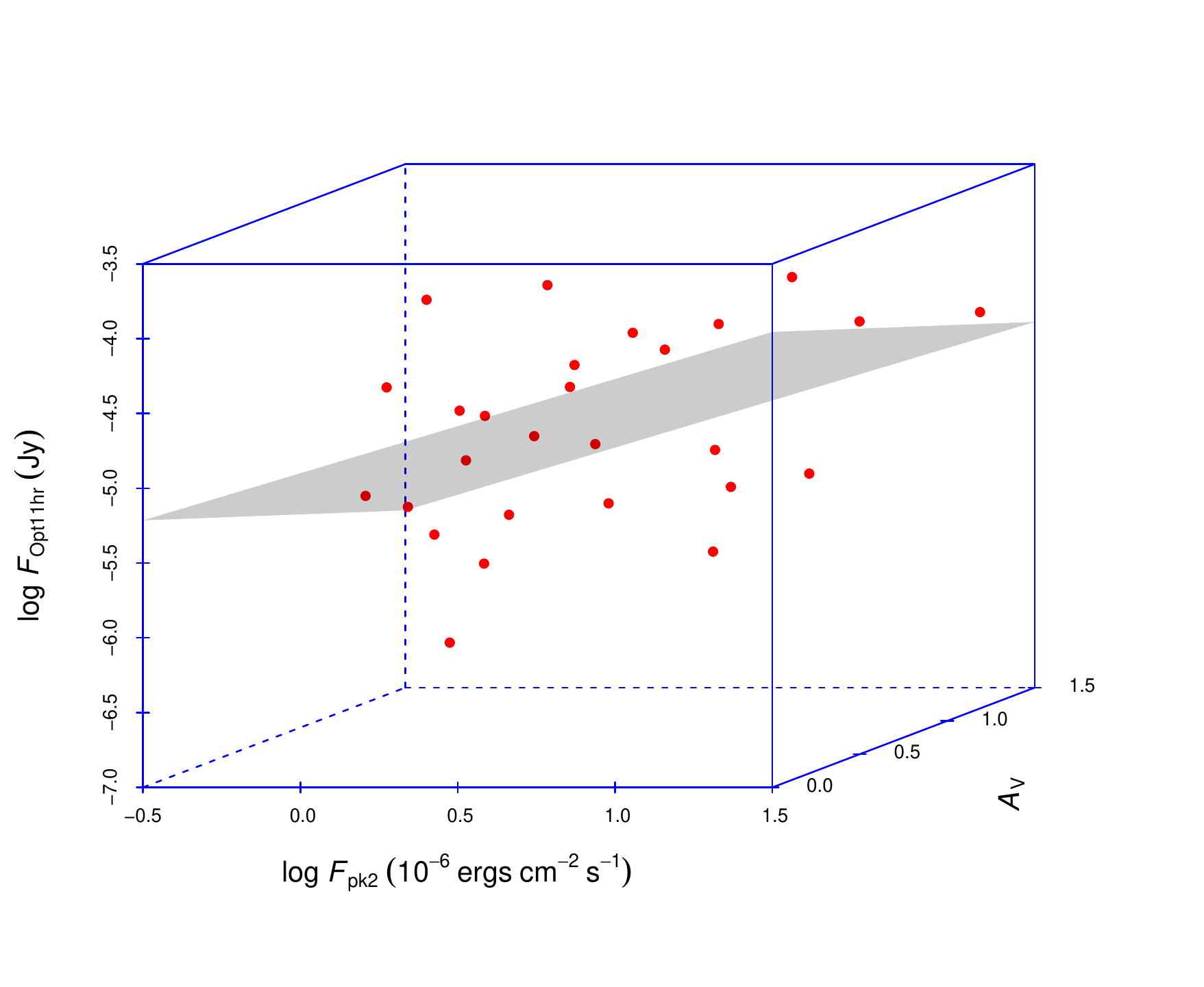}
\includegraphics[width=0.45\textwidth]{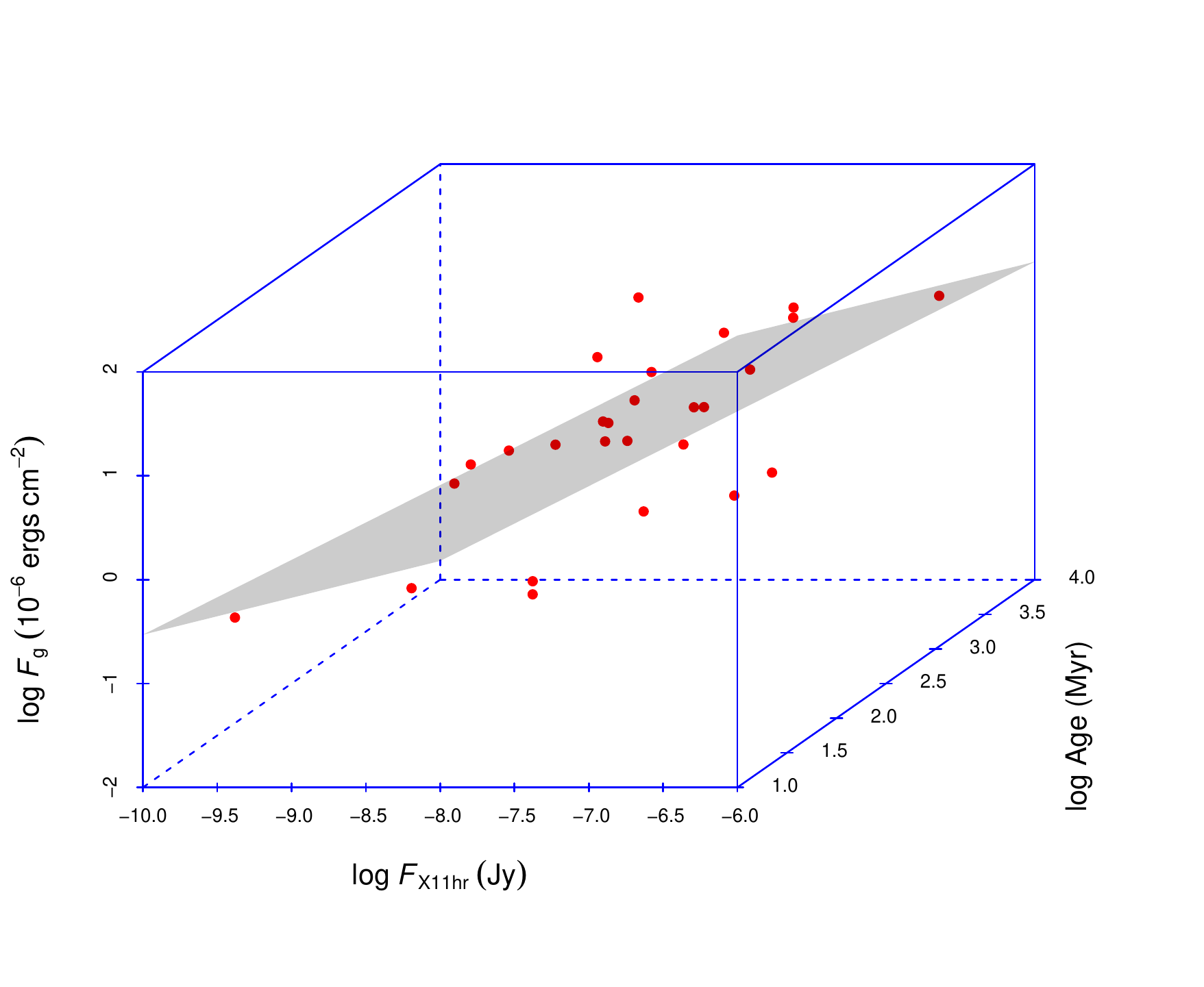}

\includegraphics[width=0.45\textwidth]{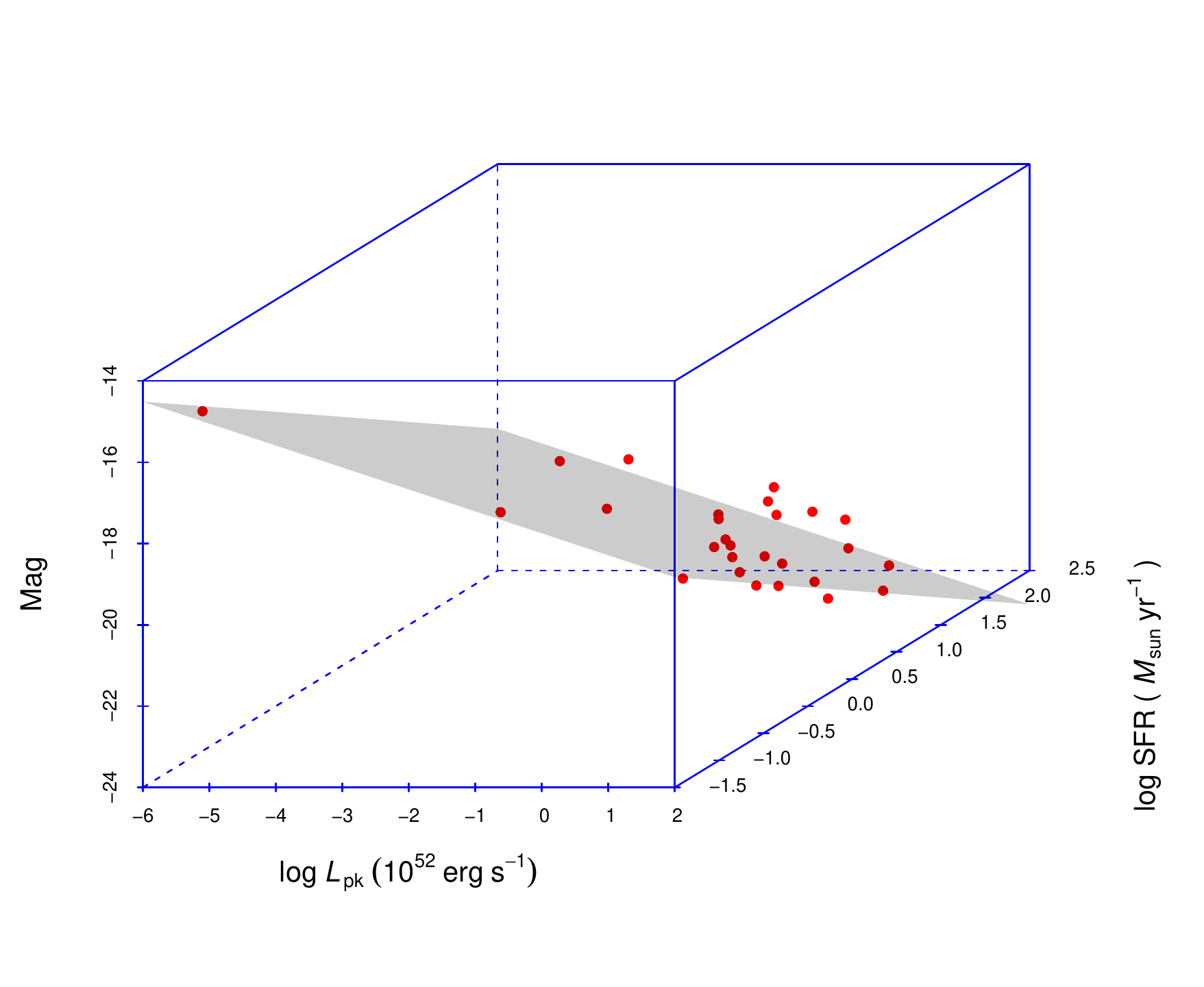}
\includegraphics[width=0.45\textwidth]{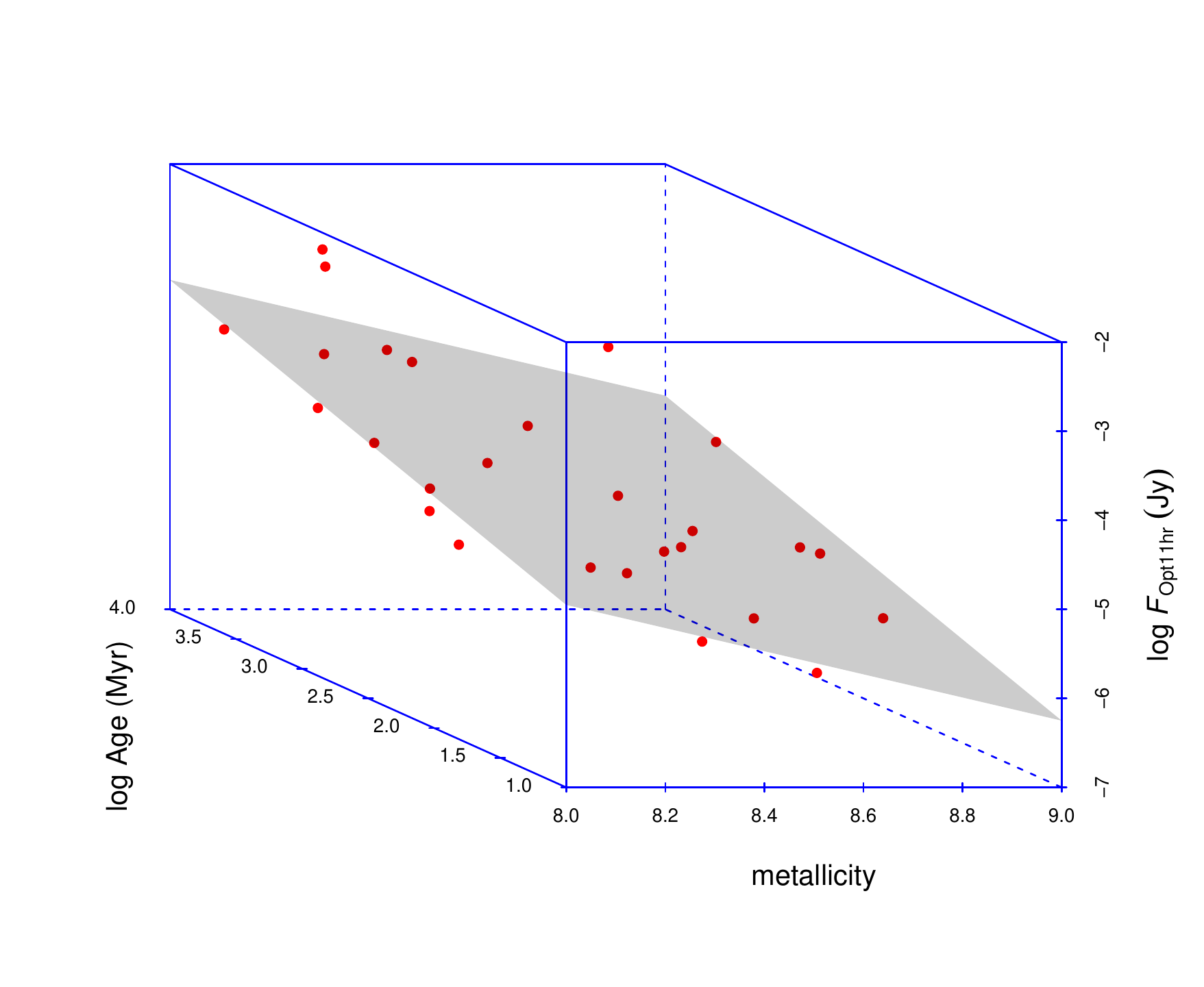}

\includegraphics[width=0.45\textwidth]{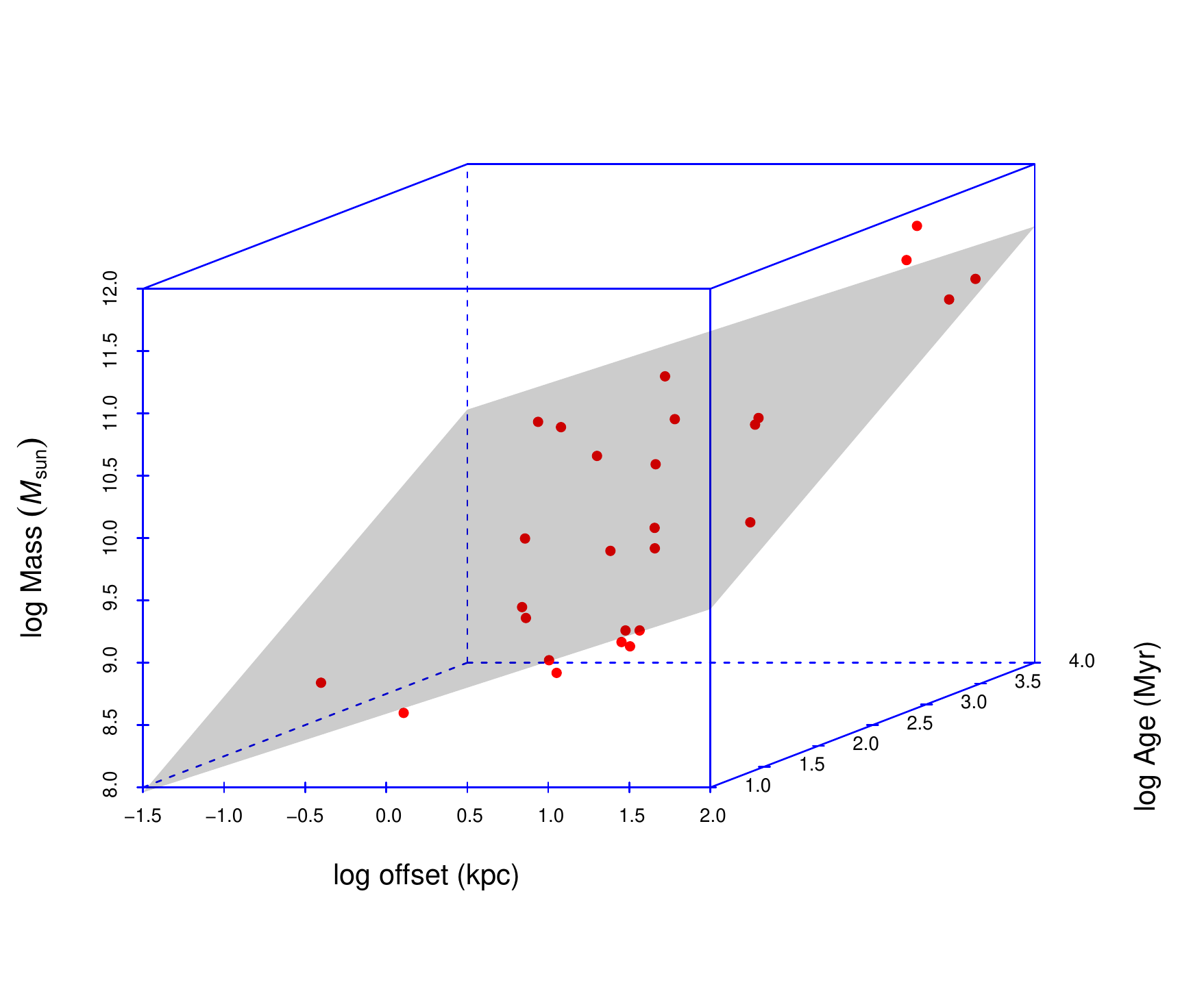}
\includegraphics[width=0.45\textwidth]{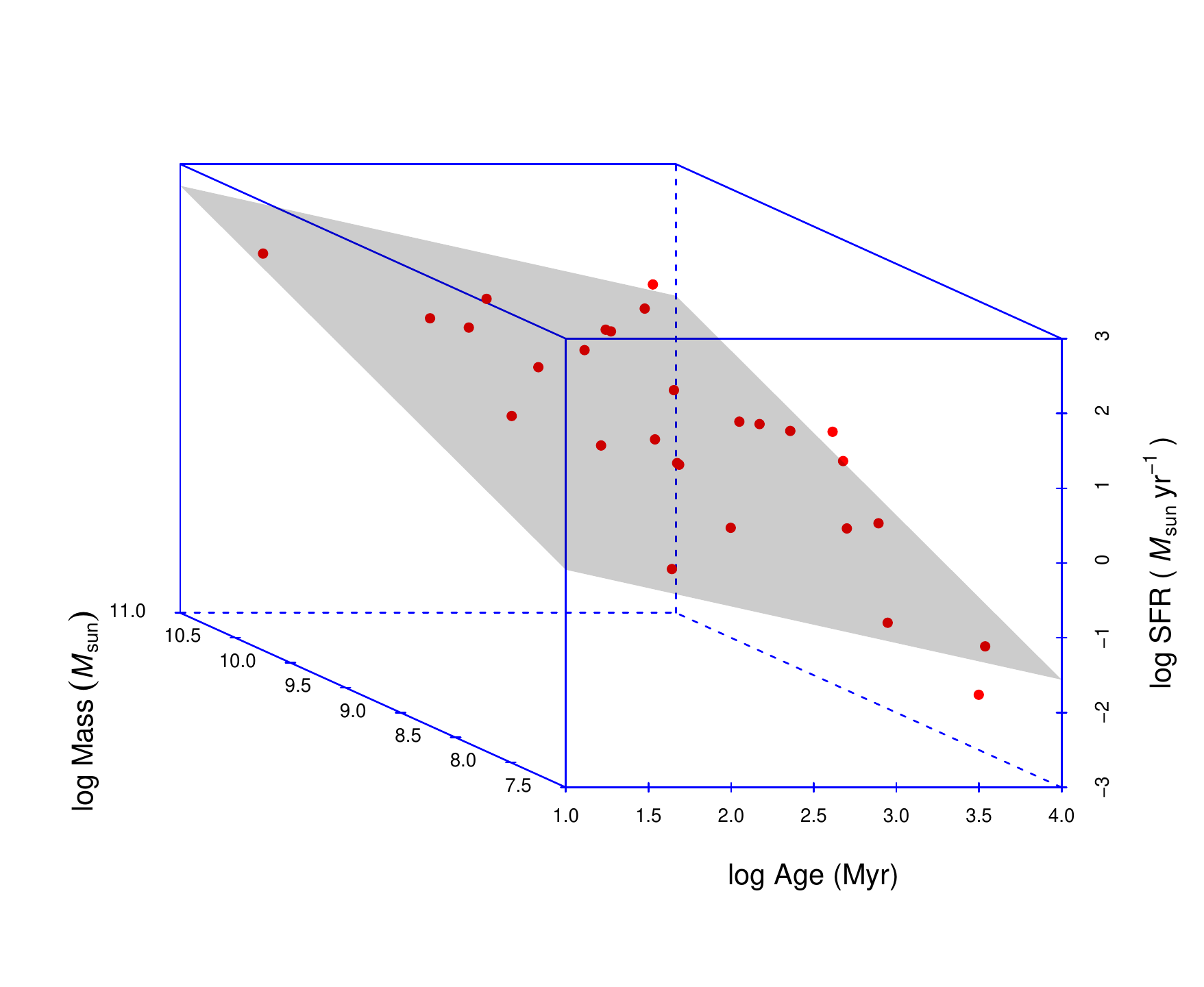}

\center{Fig. \ref{fig:three}---Continued}
\end{figure*}


\clearpage
\begin{figure*}

\includegraphics[width=0.45\textwidth]{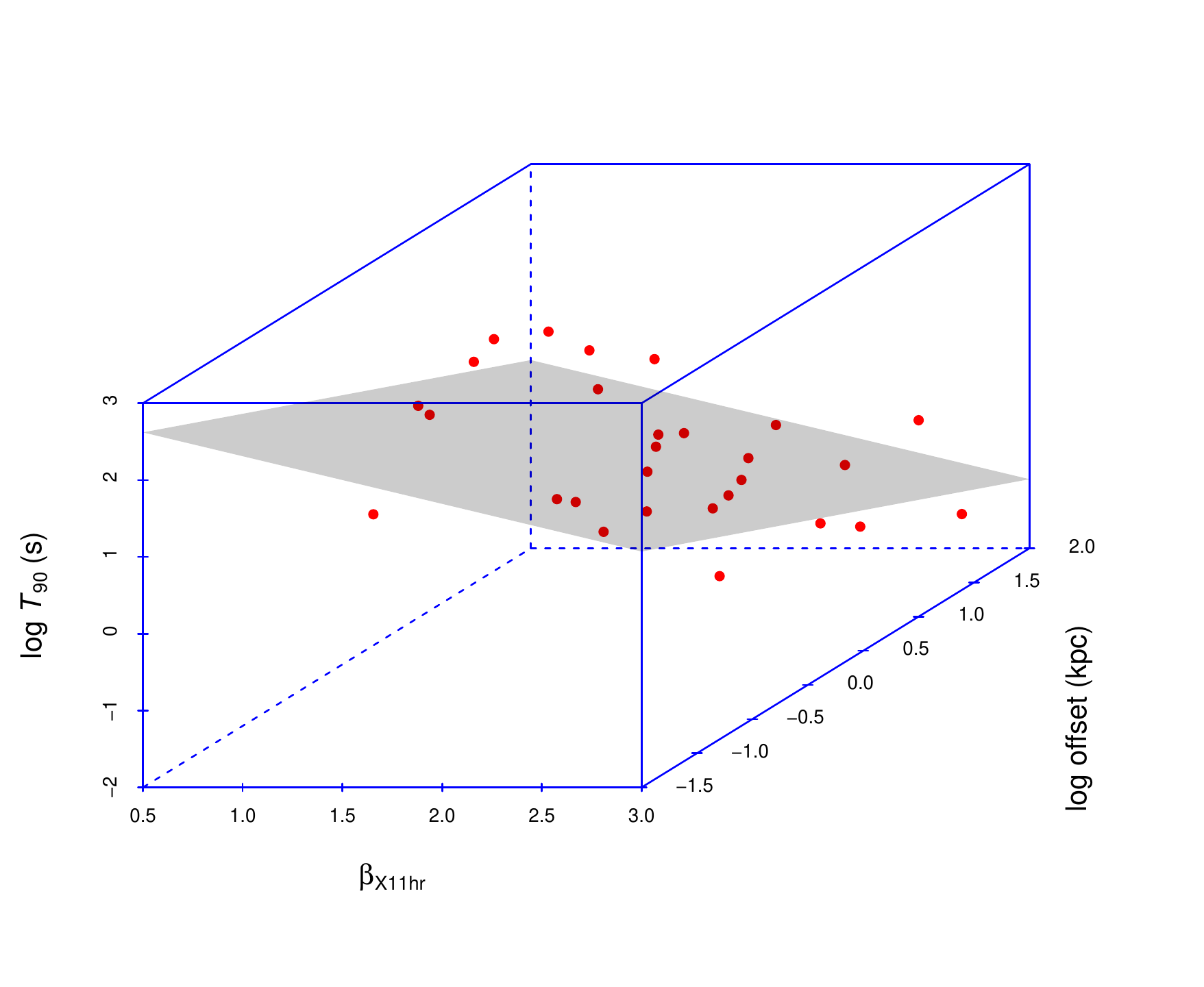}
\includegraphics[width=0.45\textwidth]{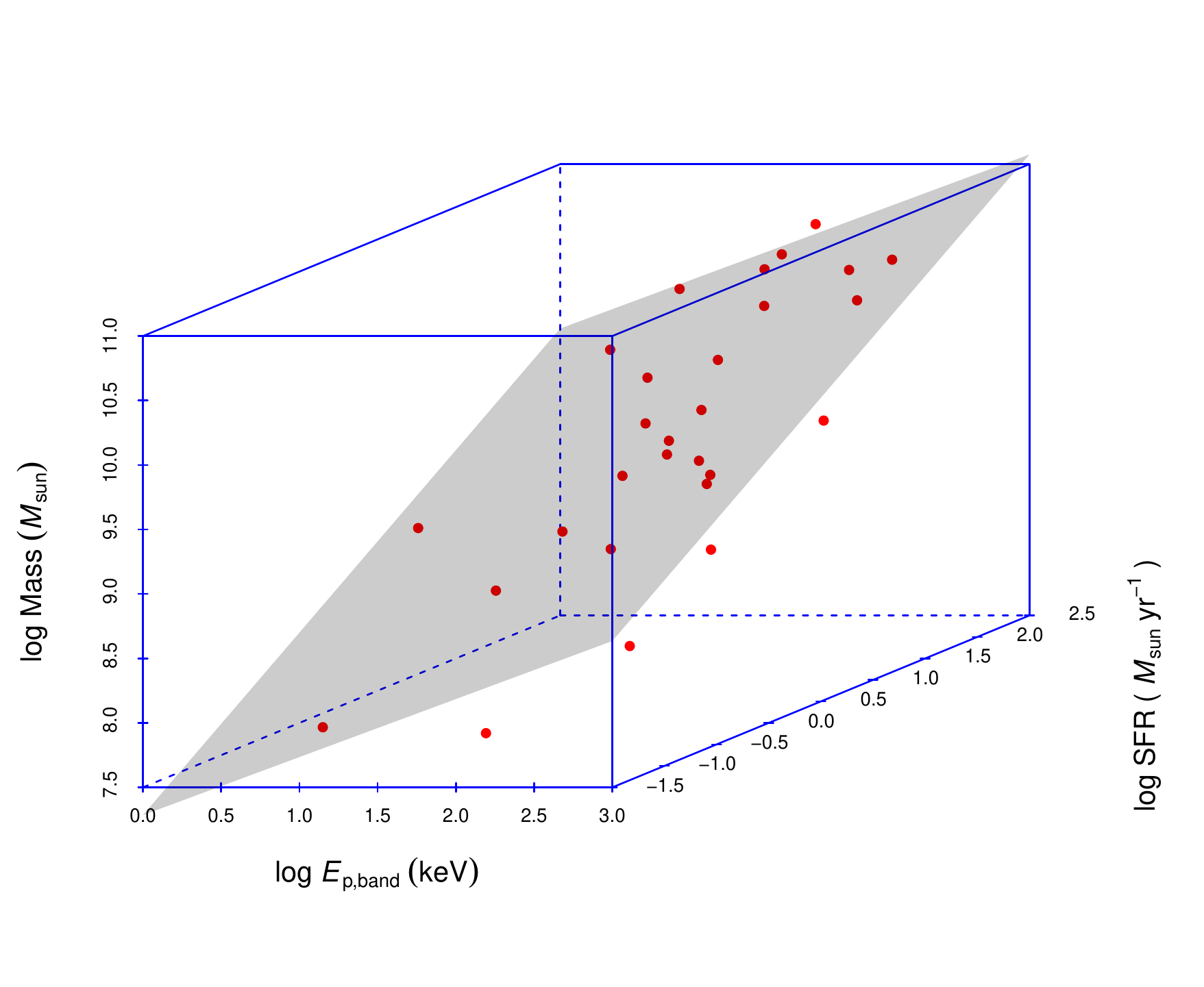}

\includegraphics[width=0.45\textwidth]{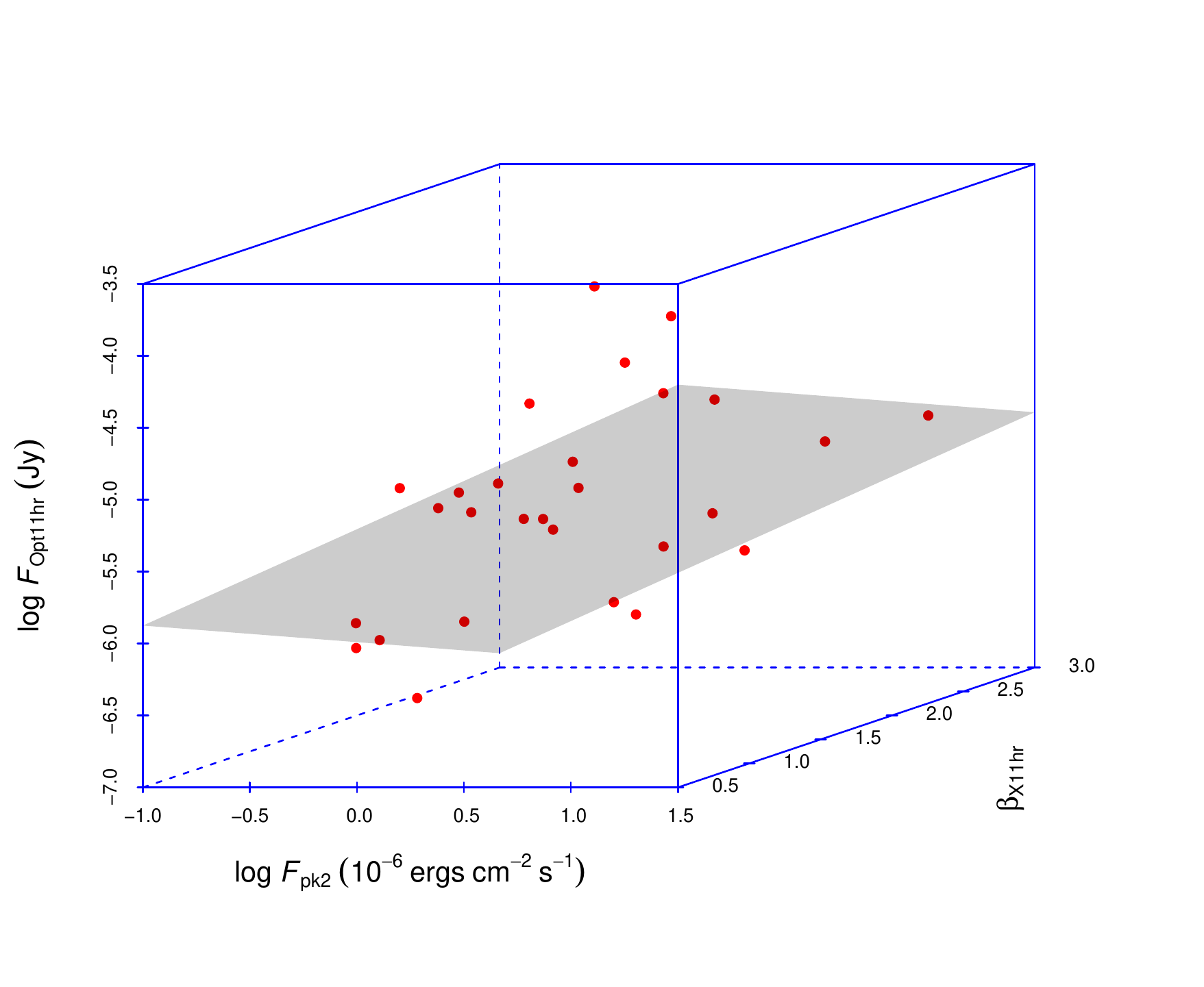}
\includegraphics[width=0.45\textwidth]{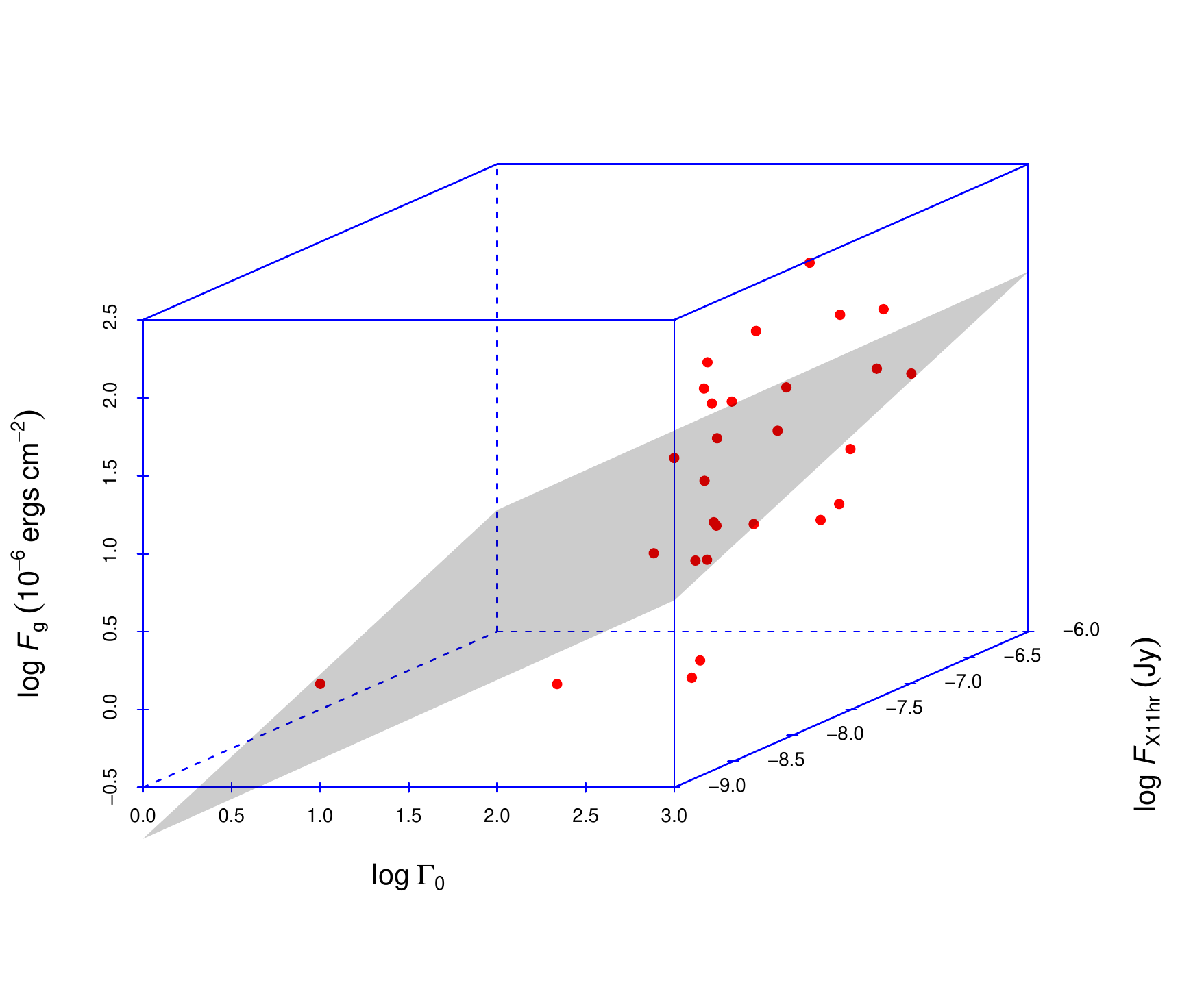}

\includegraphics[width=0.45\textwidth]{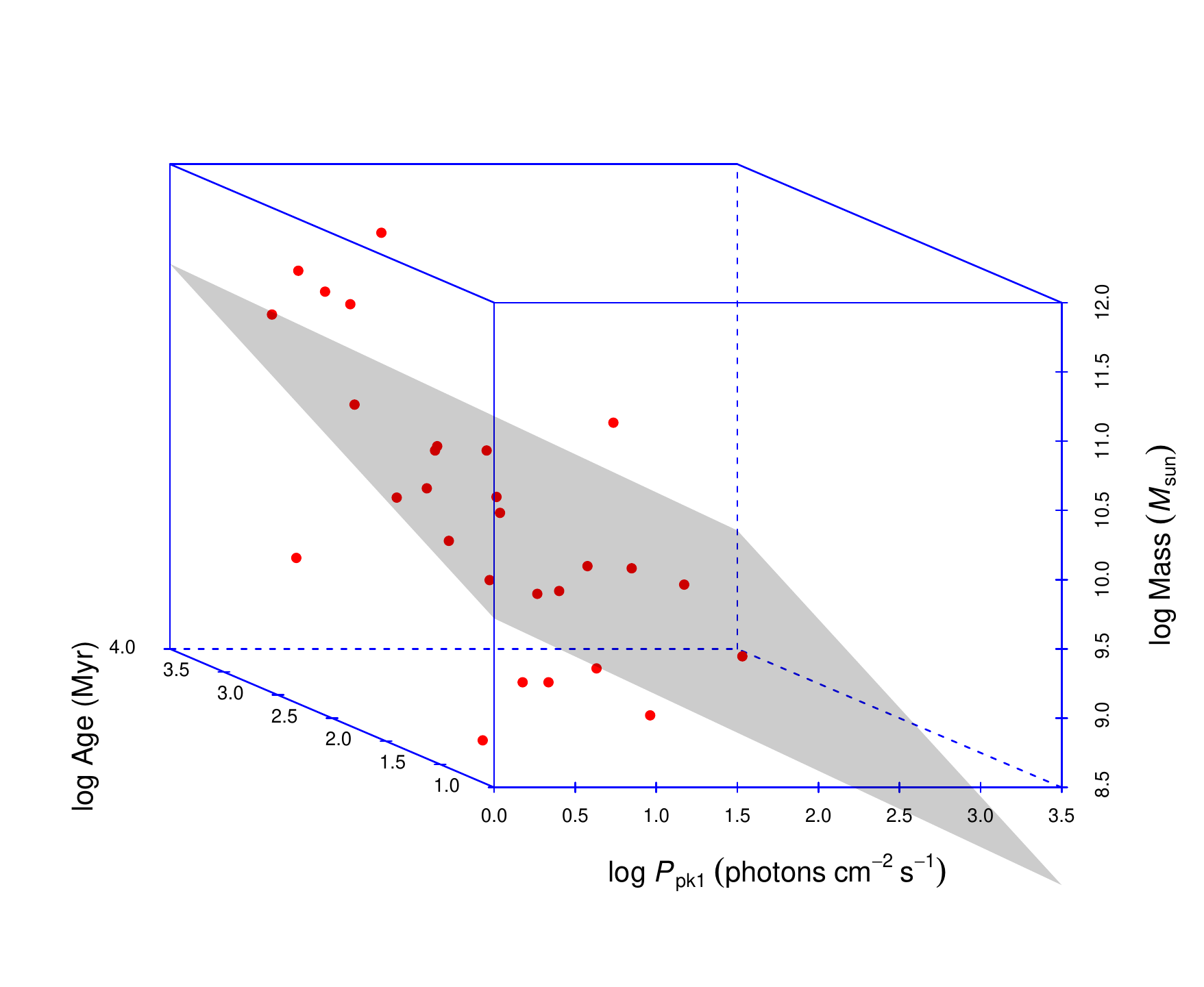}
\includegraphics[width=0.45\textwidth]{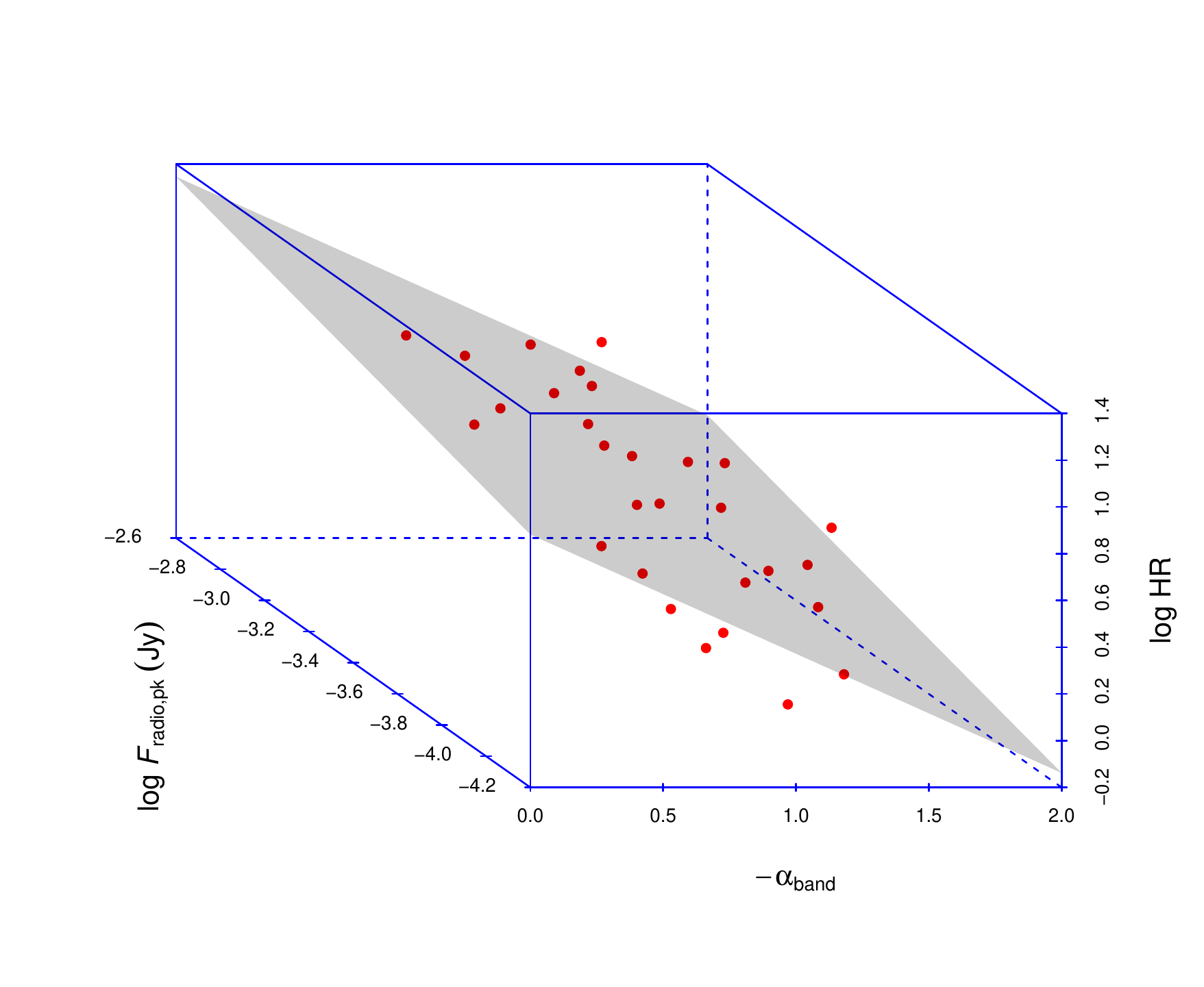}

\center{Fig. \ref{fig:three}---Continued}
\end{figure*}


\clearpage
\begin{figure*}

\includegraphics[width=0.45\textwidth]{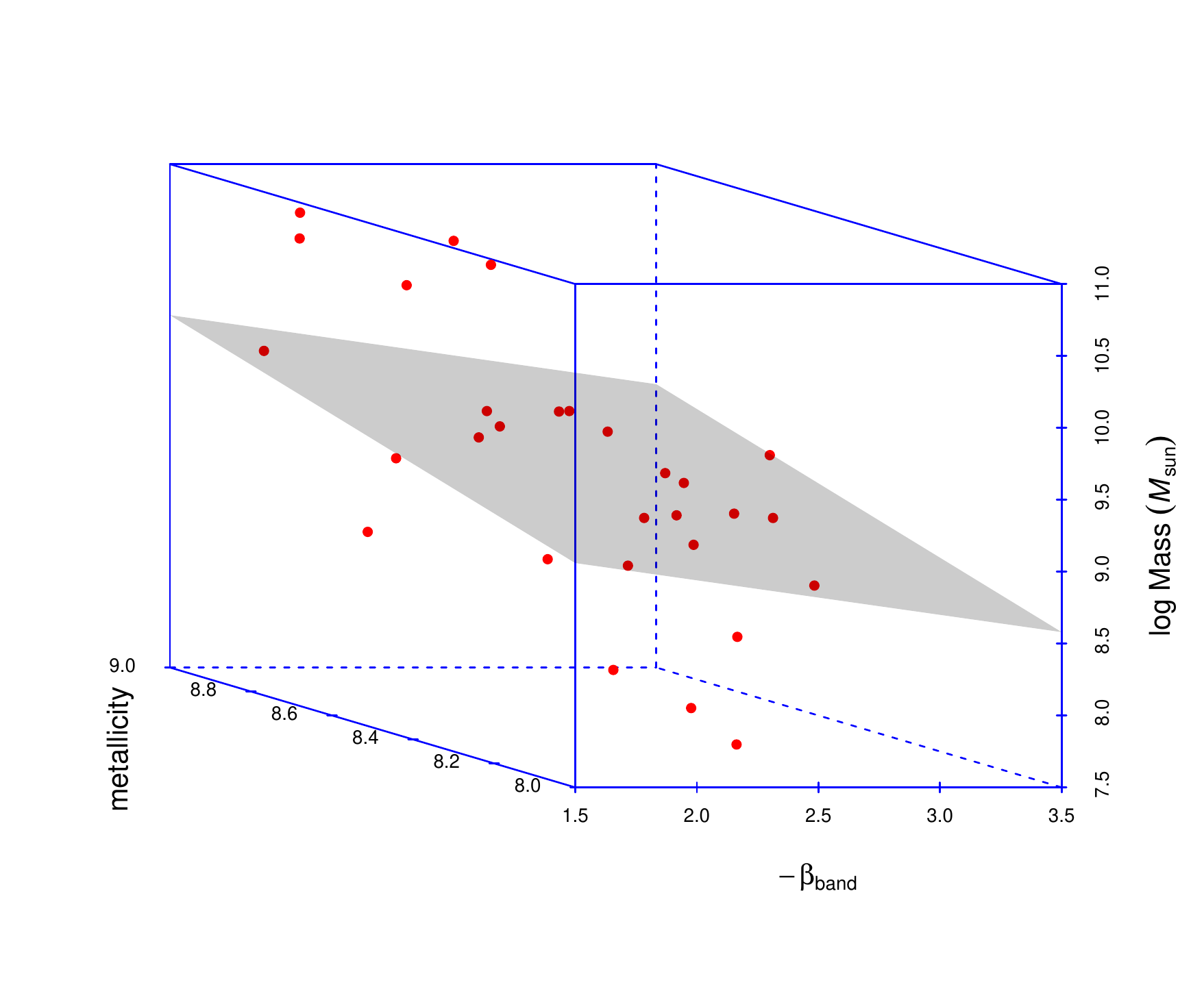}
\includegraphics[width=0.45\textwidth]{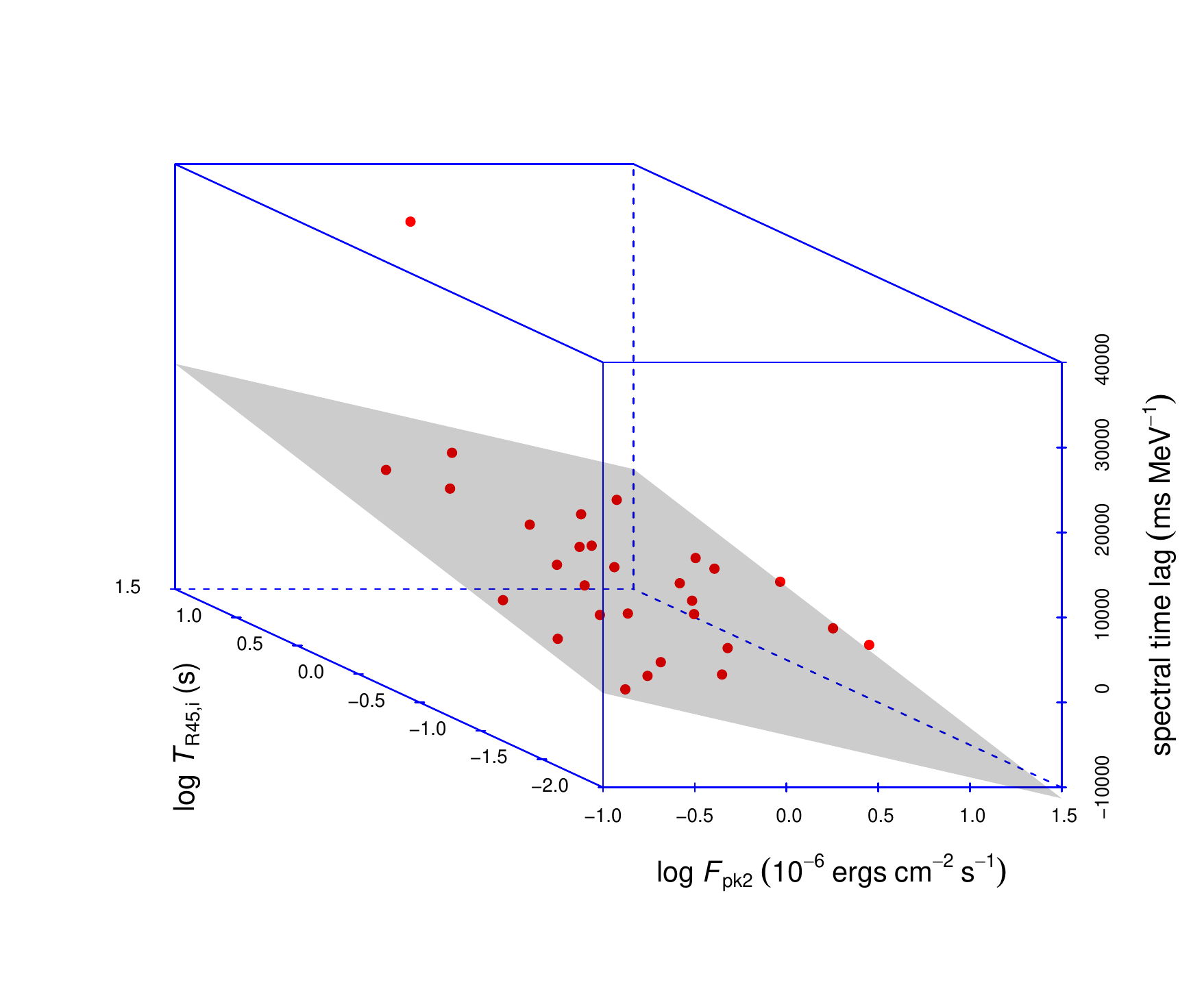}

\includegraphics[width=0.45\textwidth]{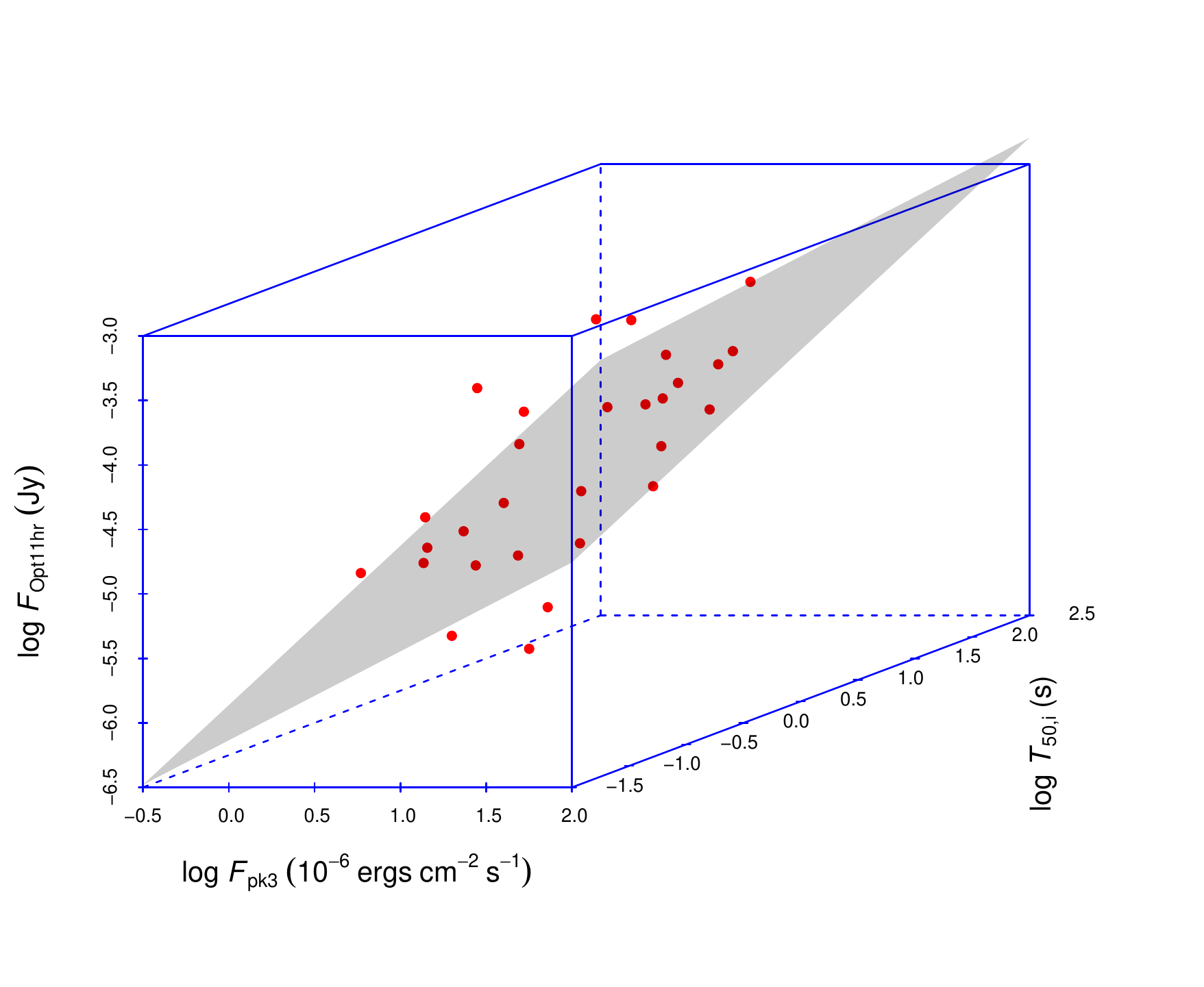}
\includegraphics[width=0.45\textwidth]{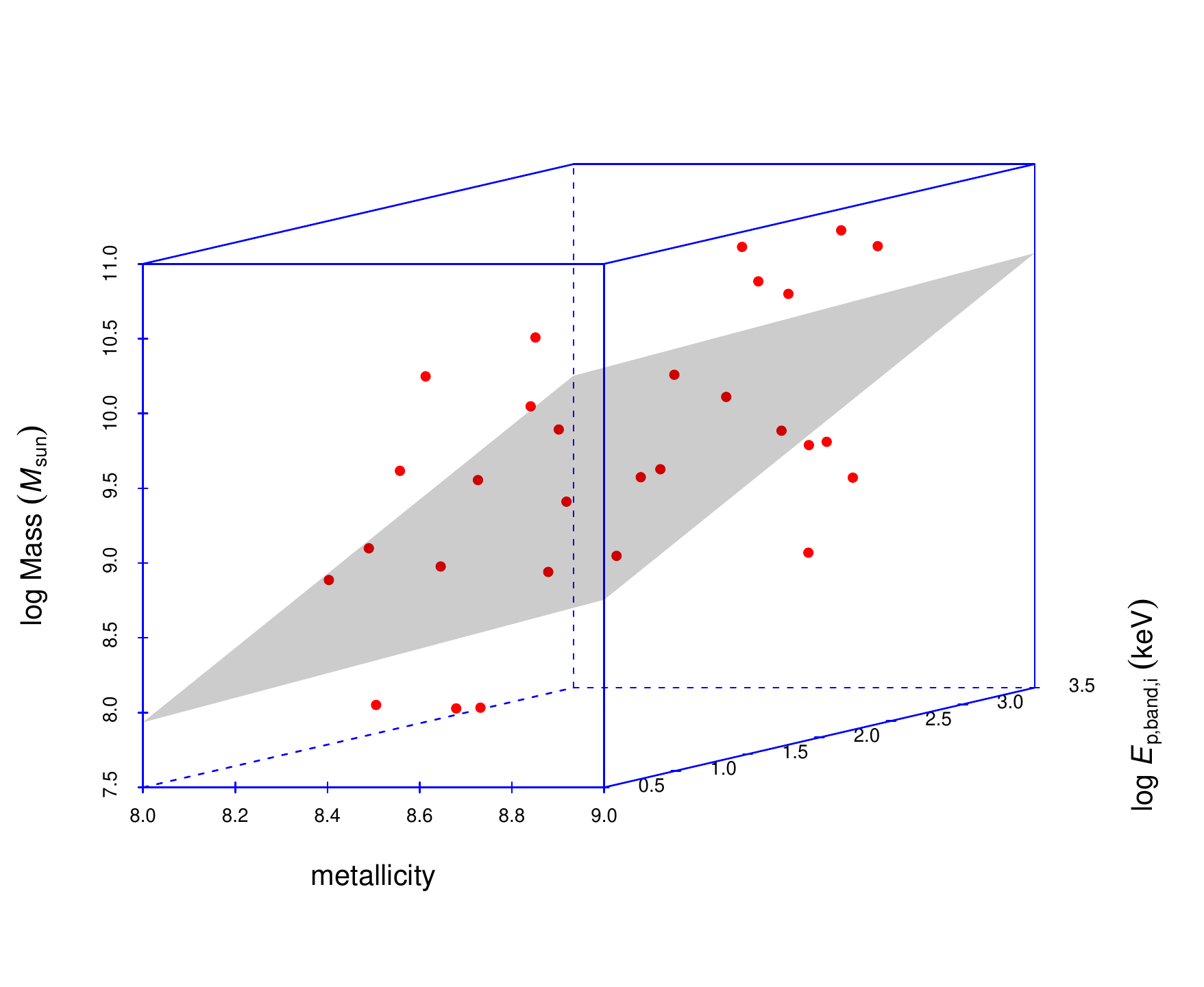}

\includegraphics[width=0.45\textwidth]{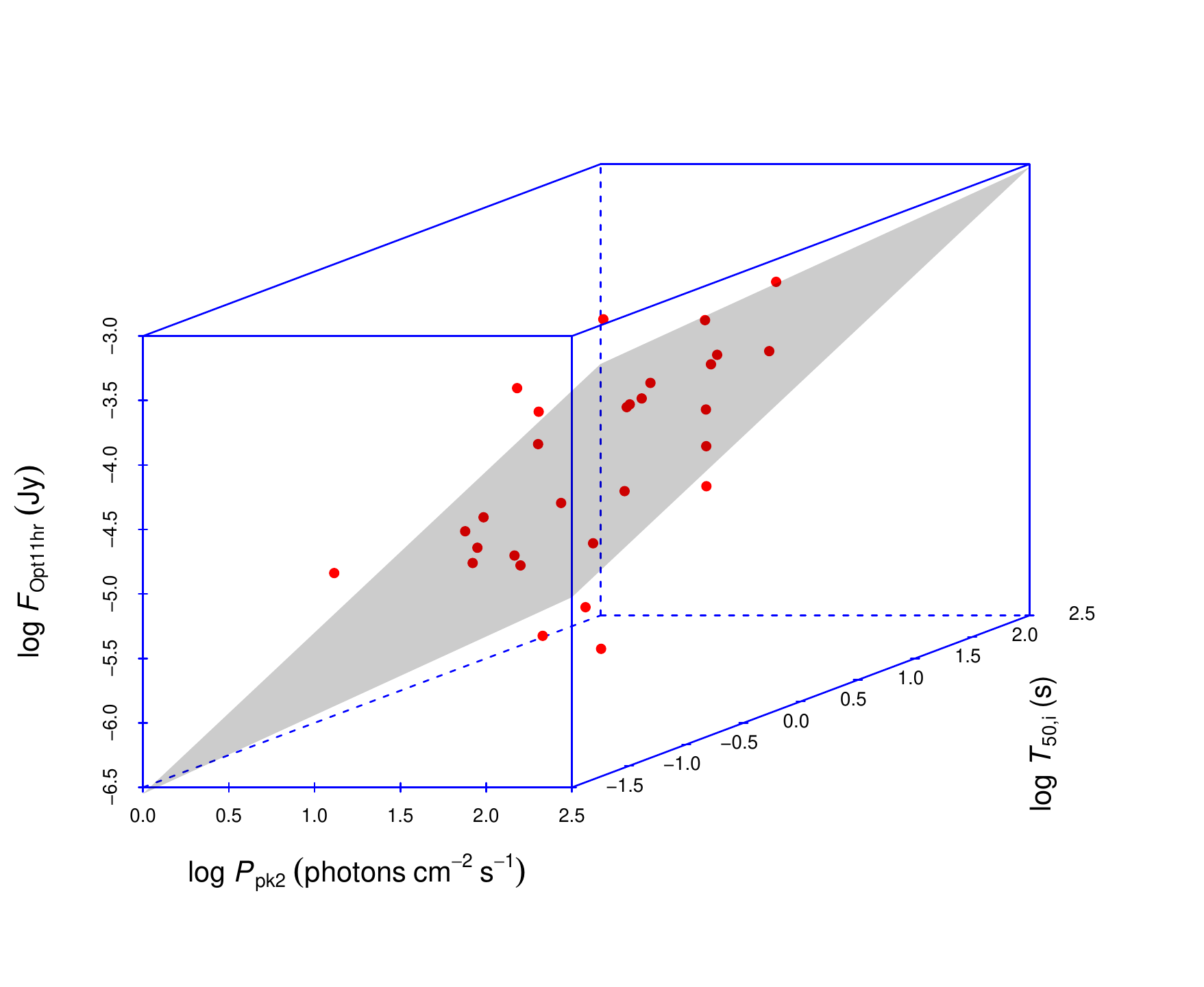}
\includegraphics[width=0.45\textwidth]{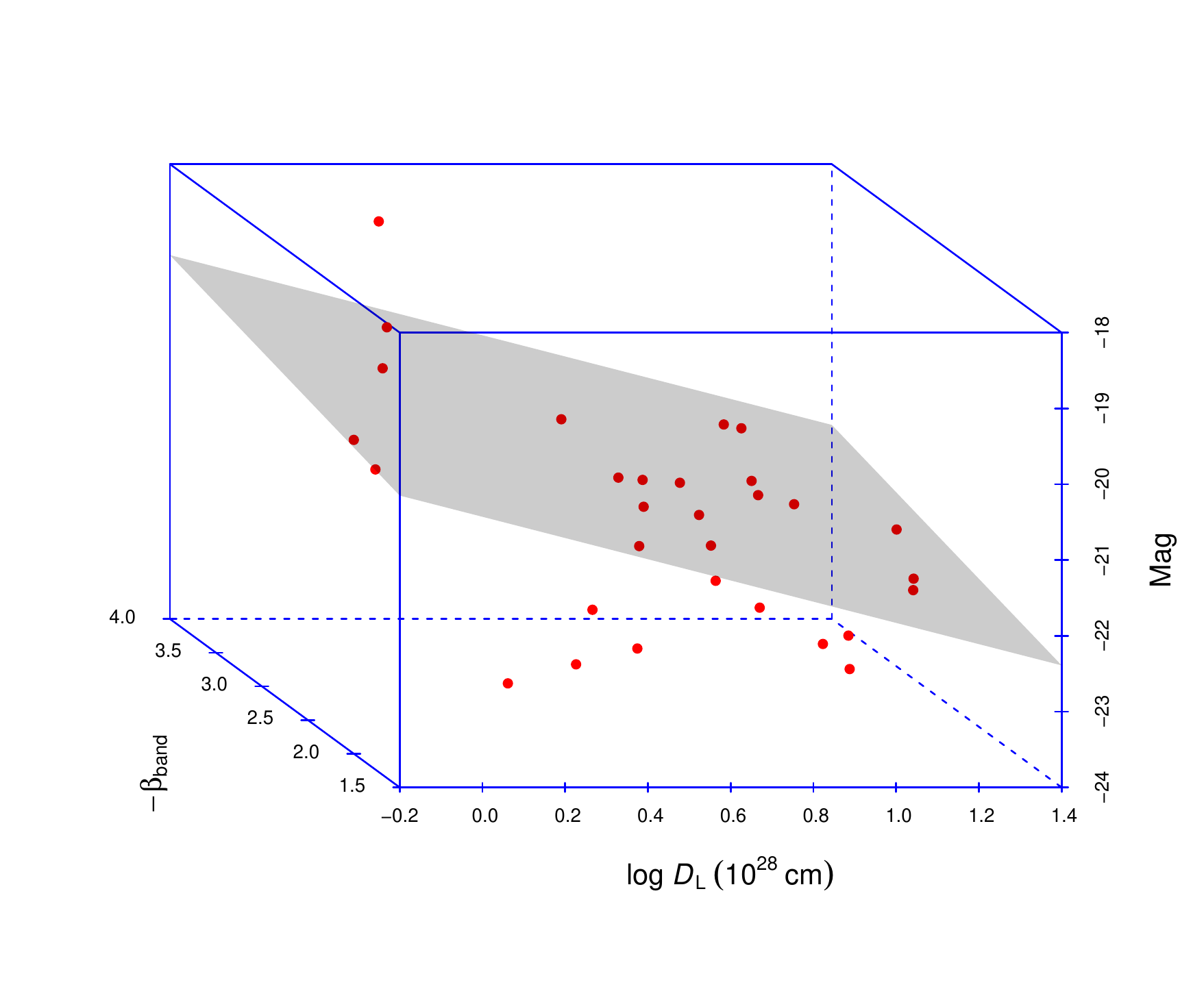}

\center{Fig. \ref{fig:three}---Continued}
\end{figure*}


\clearpage
\begin{figure*}

\includegraphics[width=0.45\textwidth]{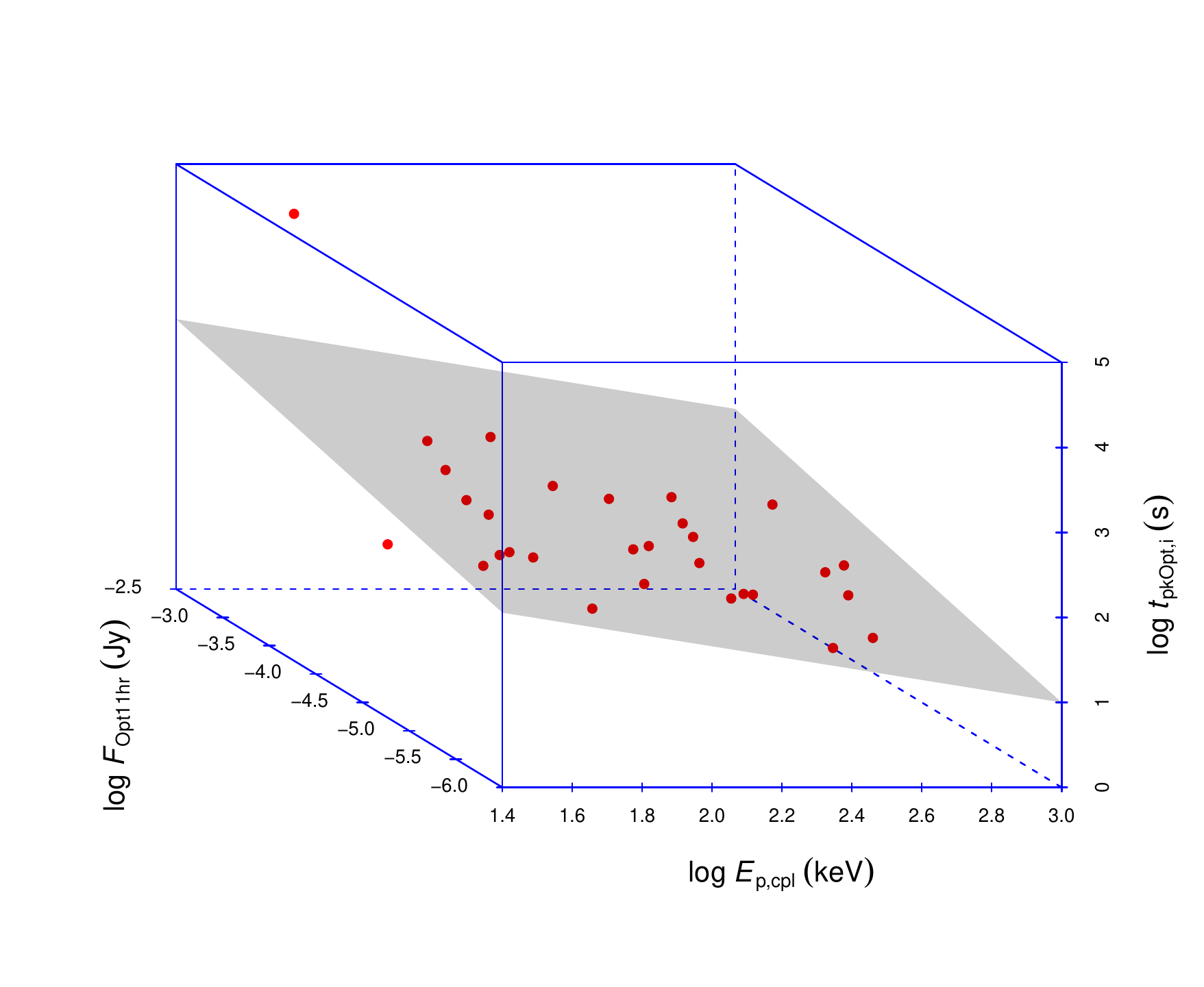}
\includegraphics[width=0.45\textwidth]{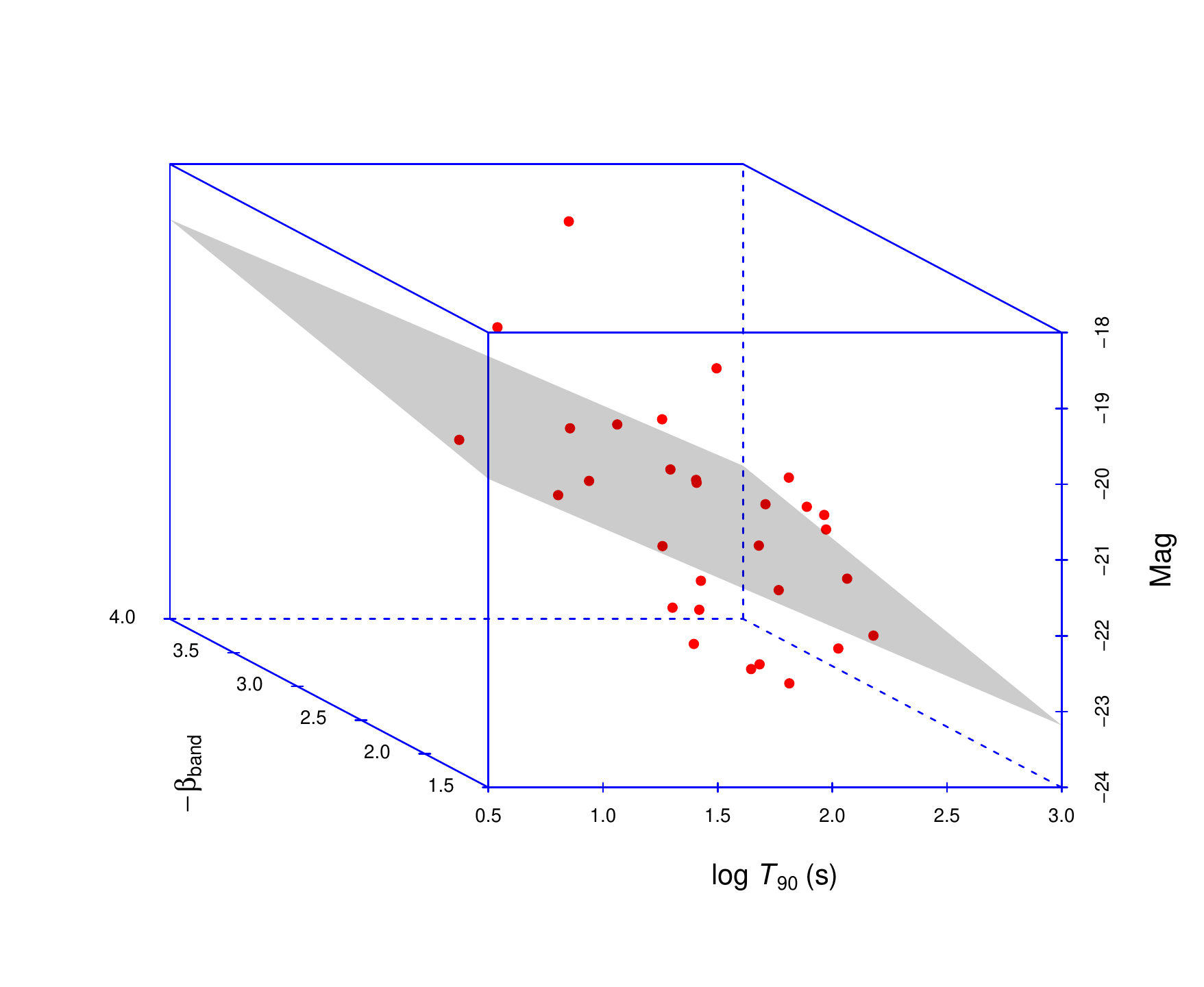}

\includegraphics[width=0.45\textwidth]{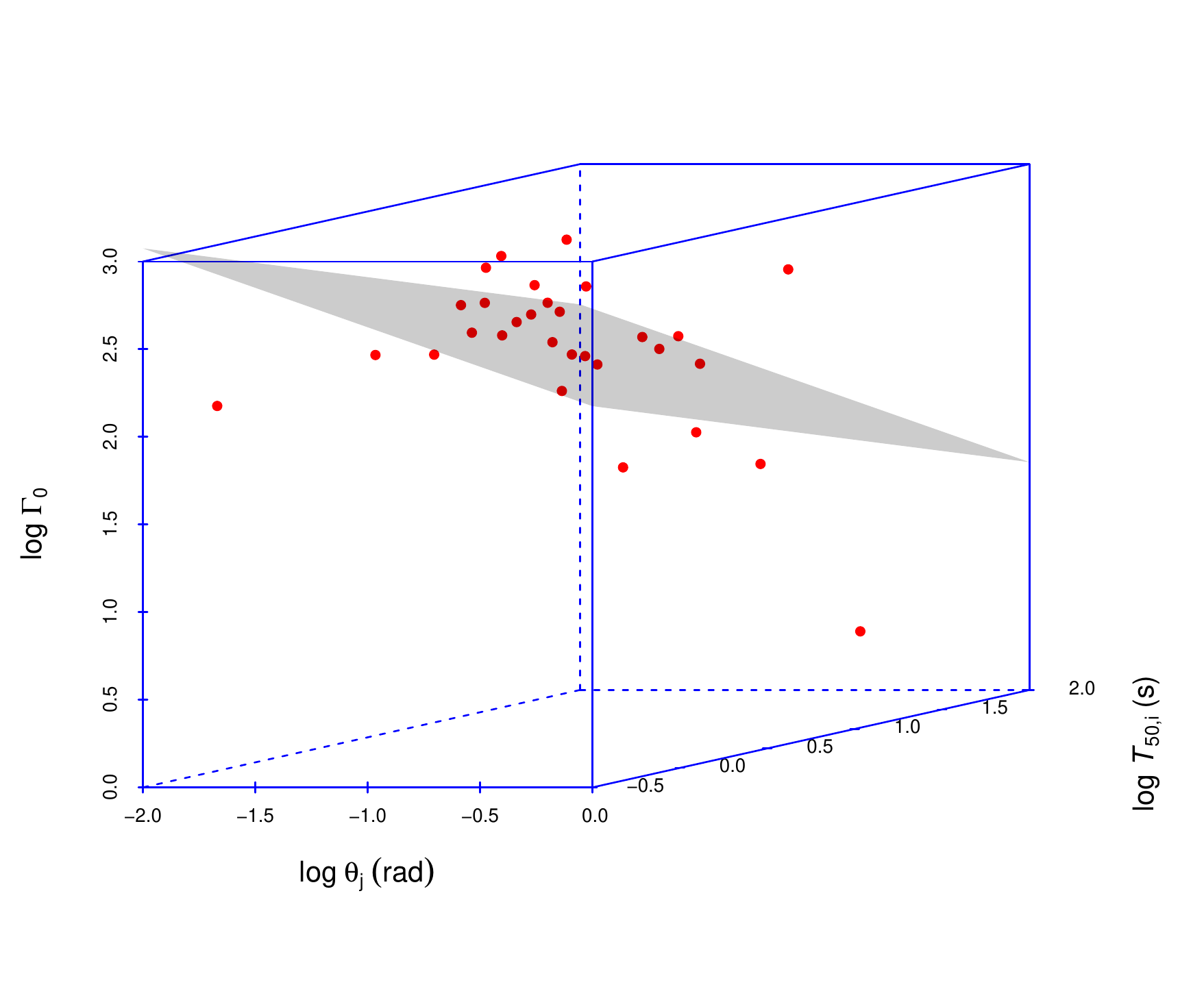}
\includegraphics[width=0.45\textwidth]{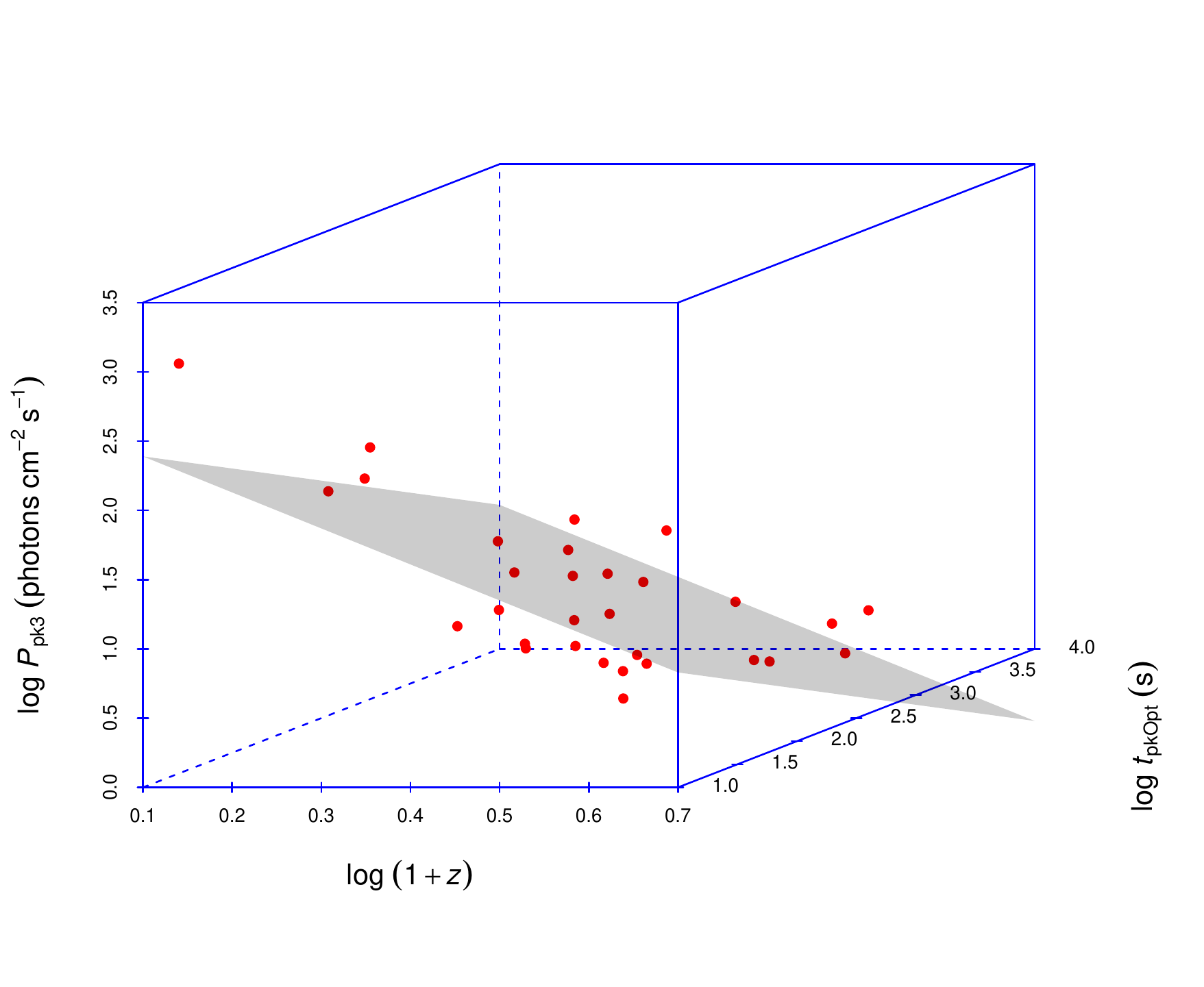}

\includegraphics[width=0.45\textwidth]{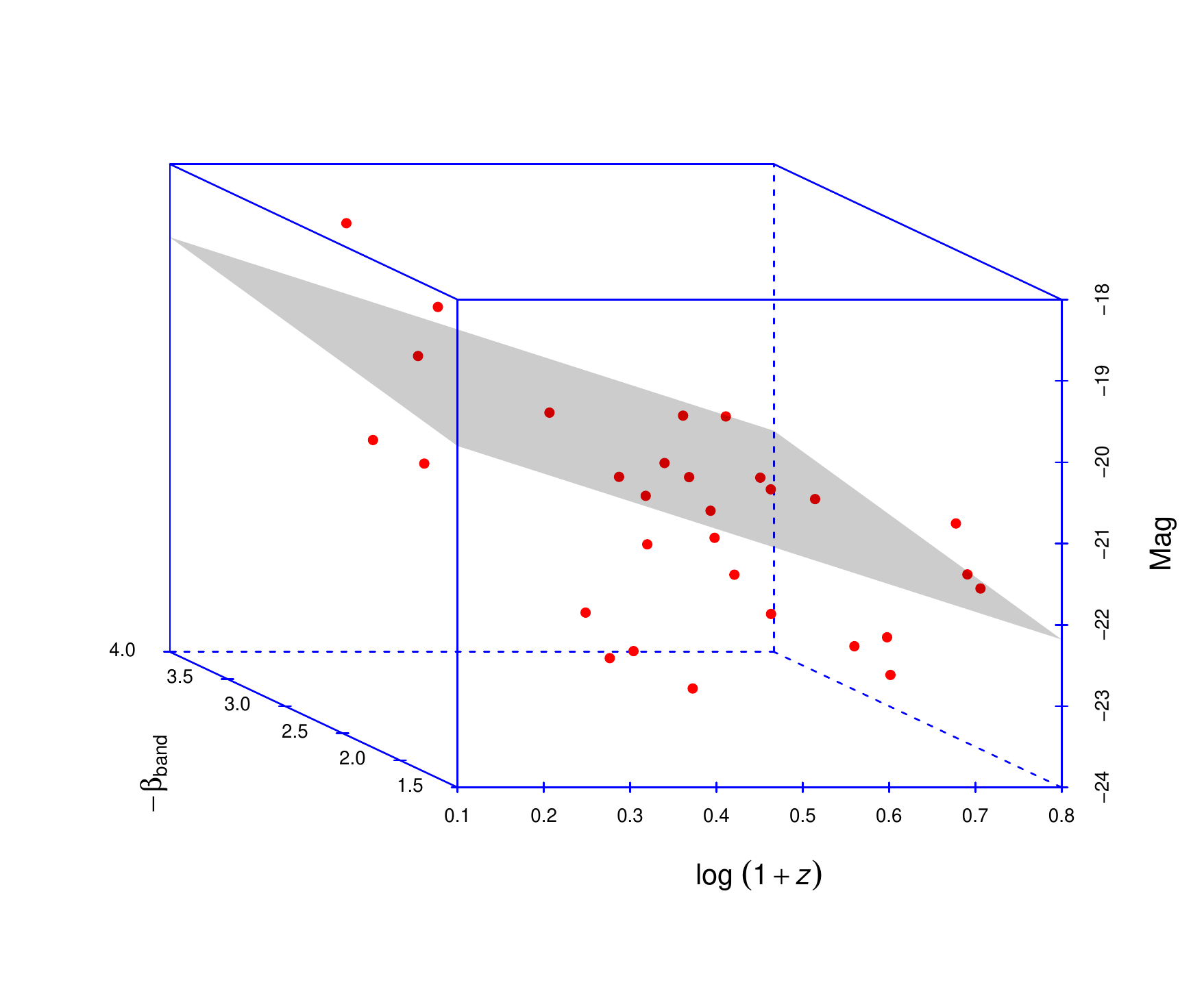}
\includegraphics[width=0.45\textwidth]{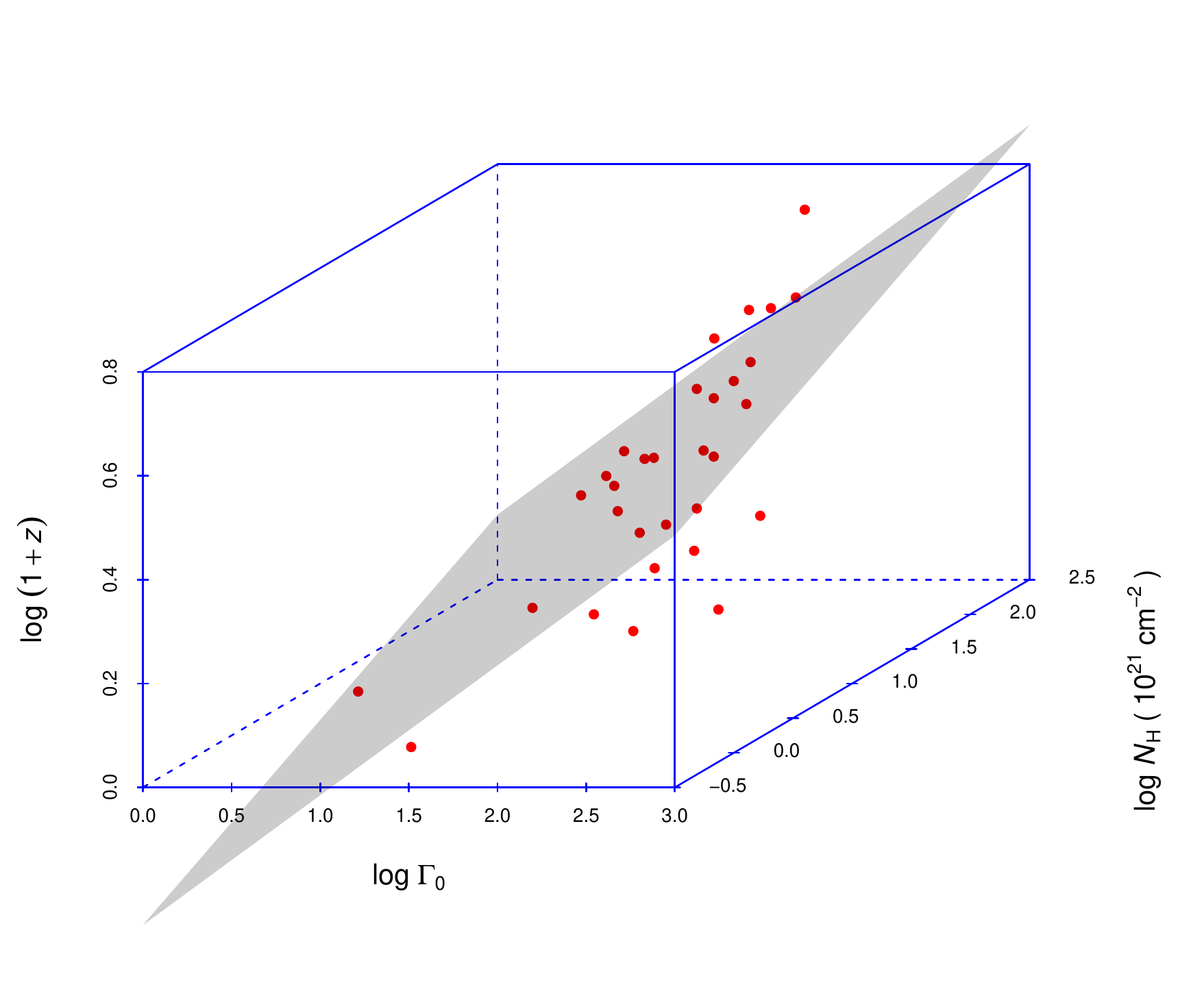}

\center{Fig. \ref{fig:three}---Continued}
\end{figure*}


\clearpage
\begin{figure*}

\includegraphics[width=0.45\textwidth]{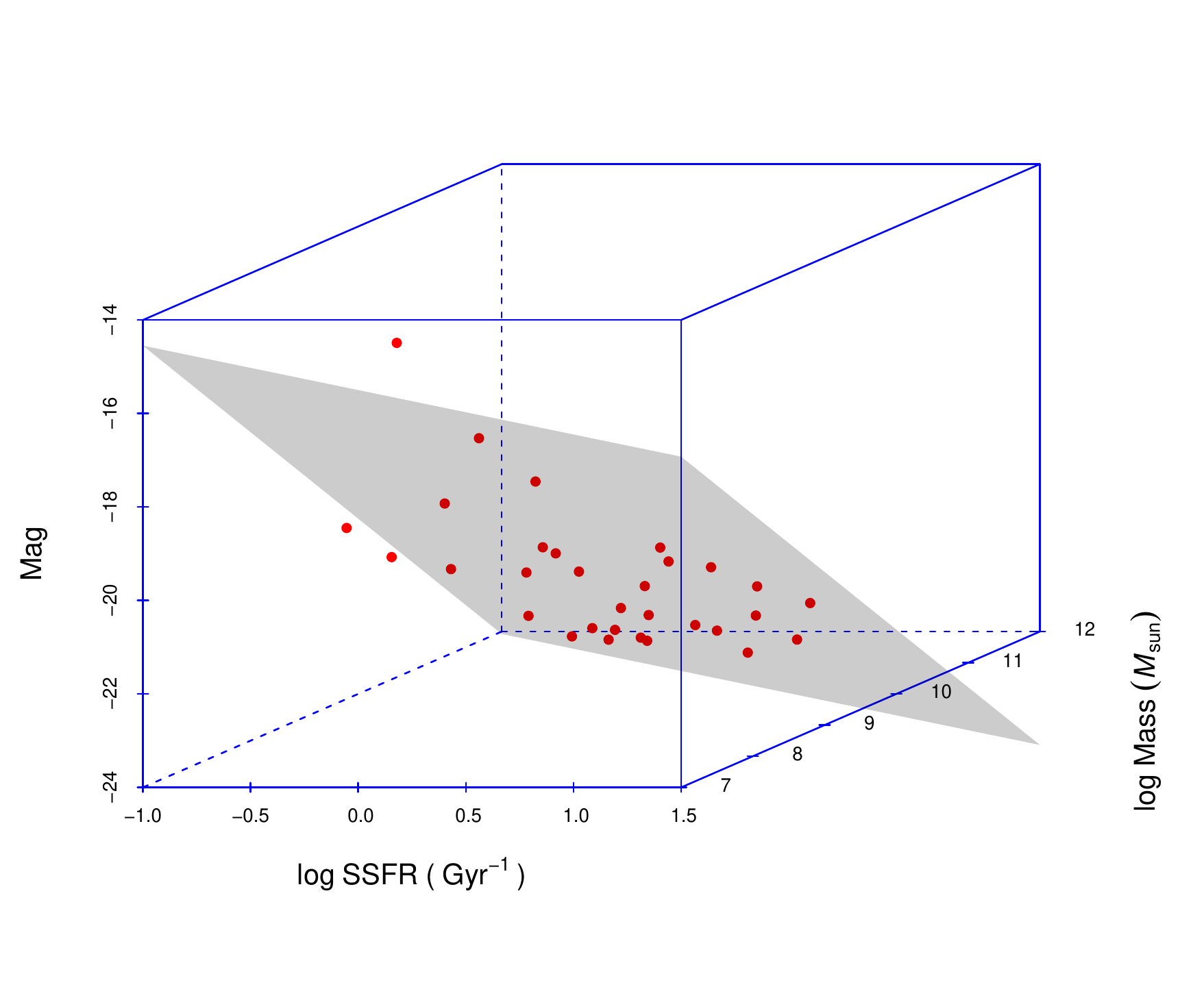}
\includegraphics[width=0.45\textwidth]{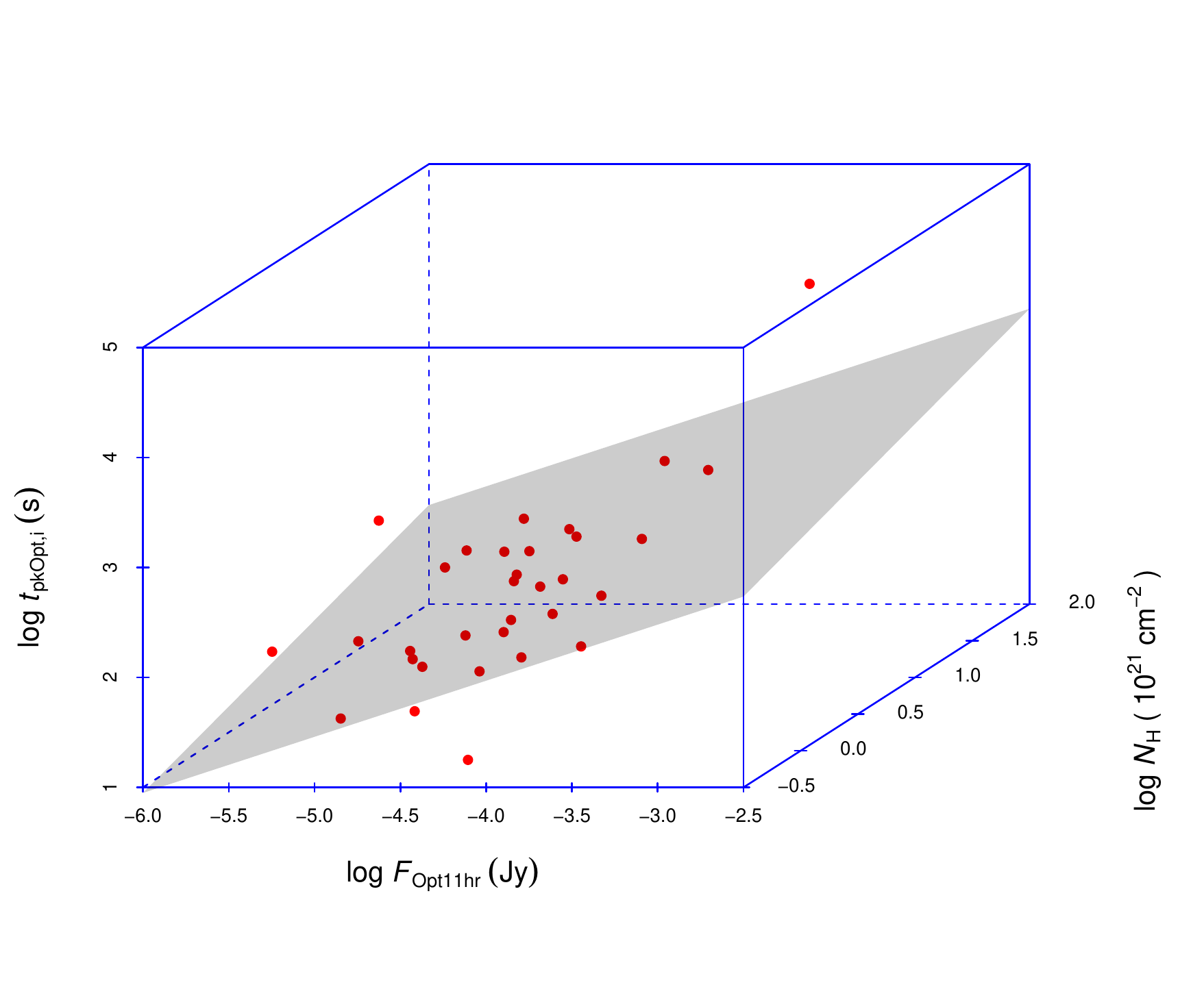}

\includegraphics[width=0.45\textwidth]{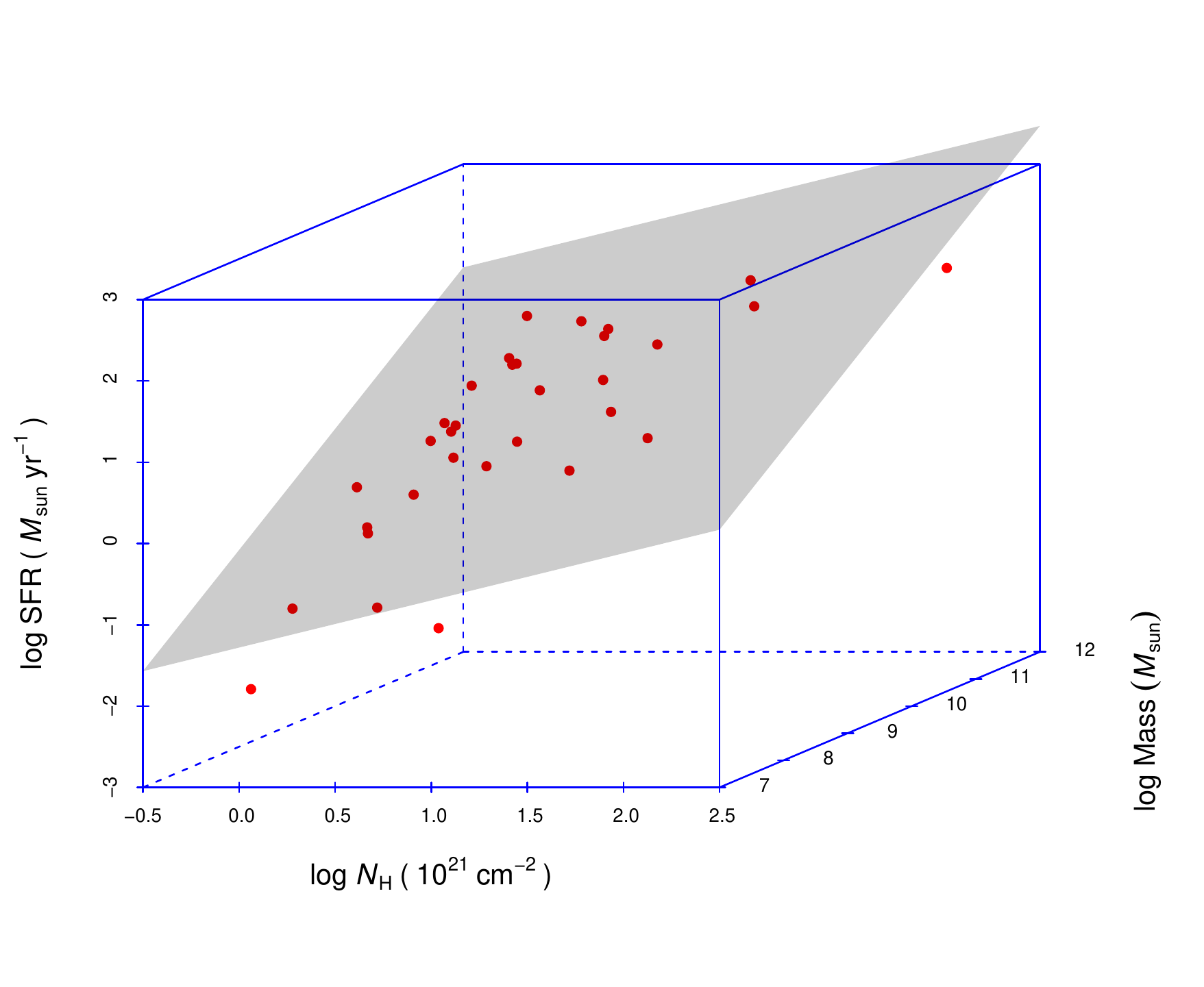}
\includegraphics[width=0.45\textwidth]{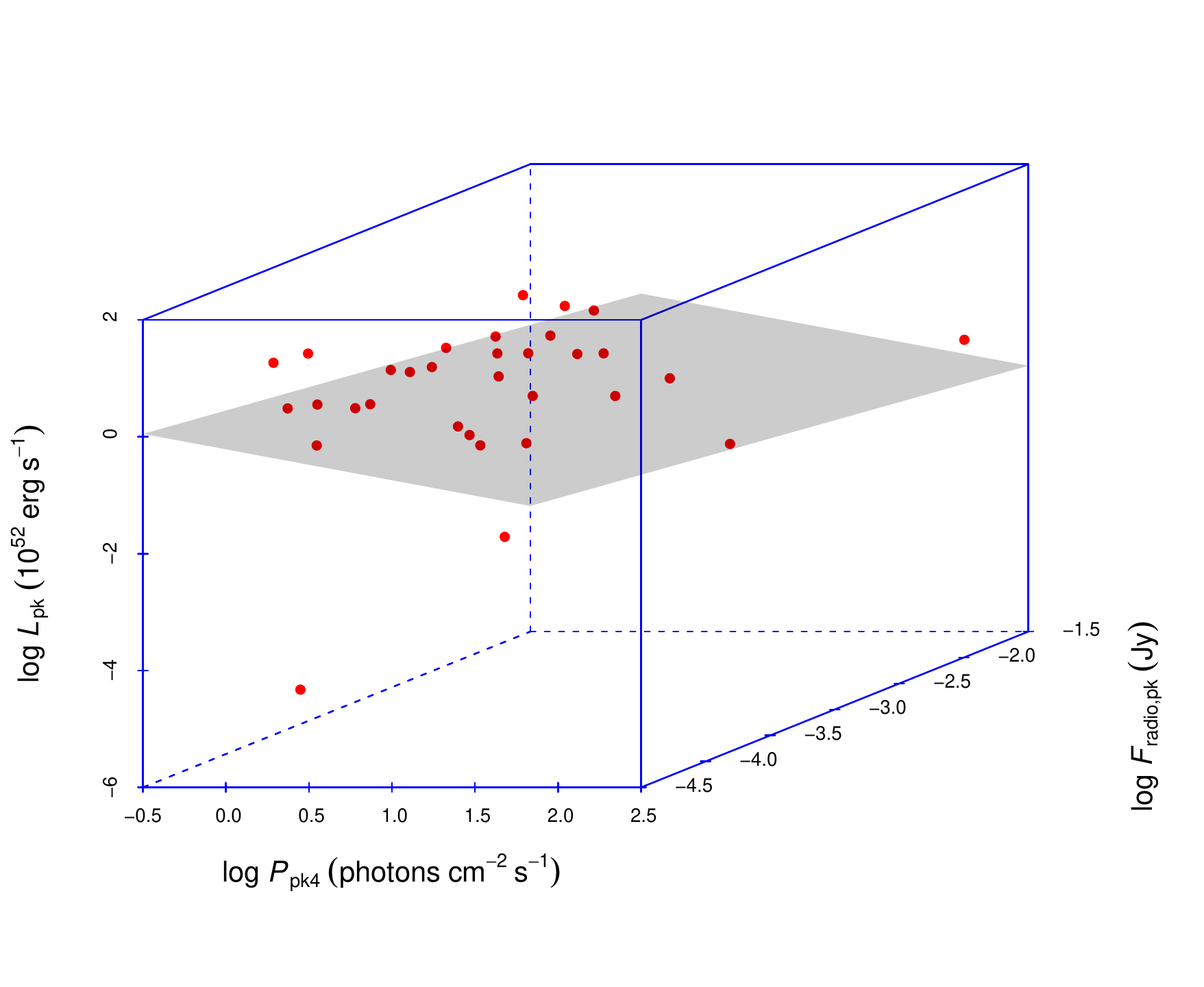}

\includegraphics[width=0.45\textwidth]{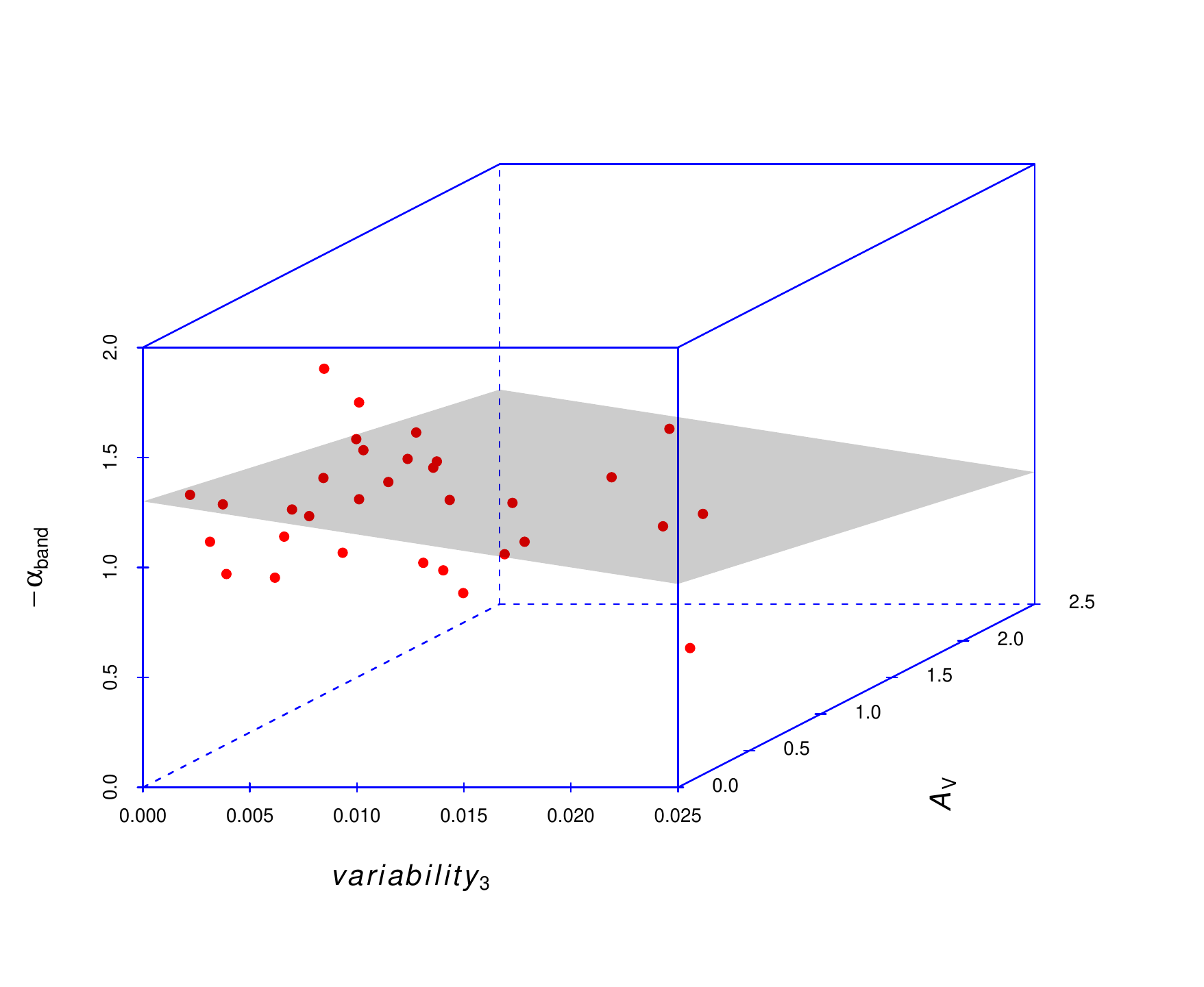}
\includegraphics[width=0.45\textwidth]{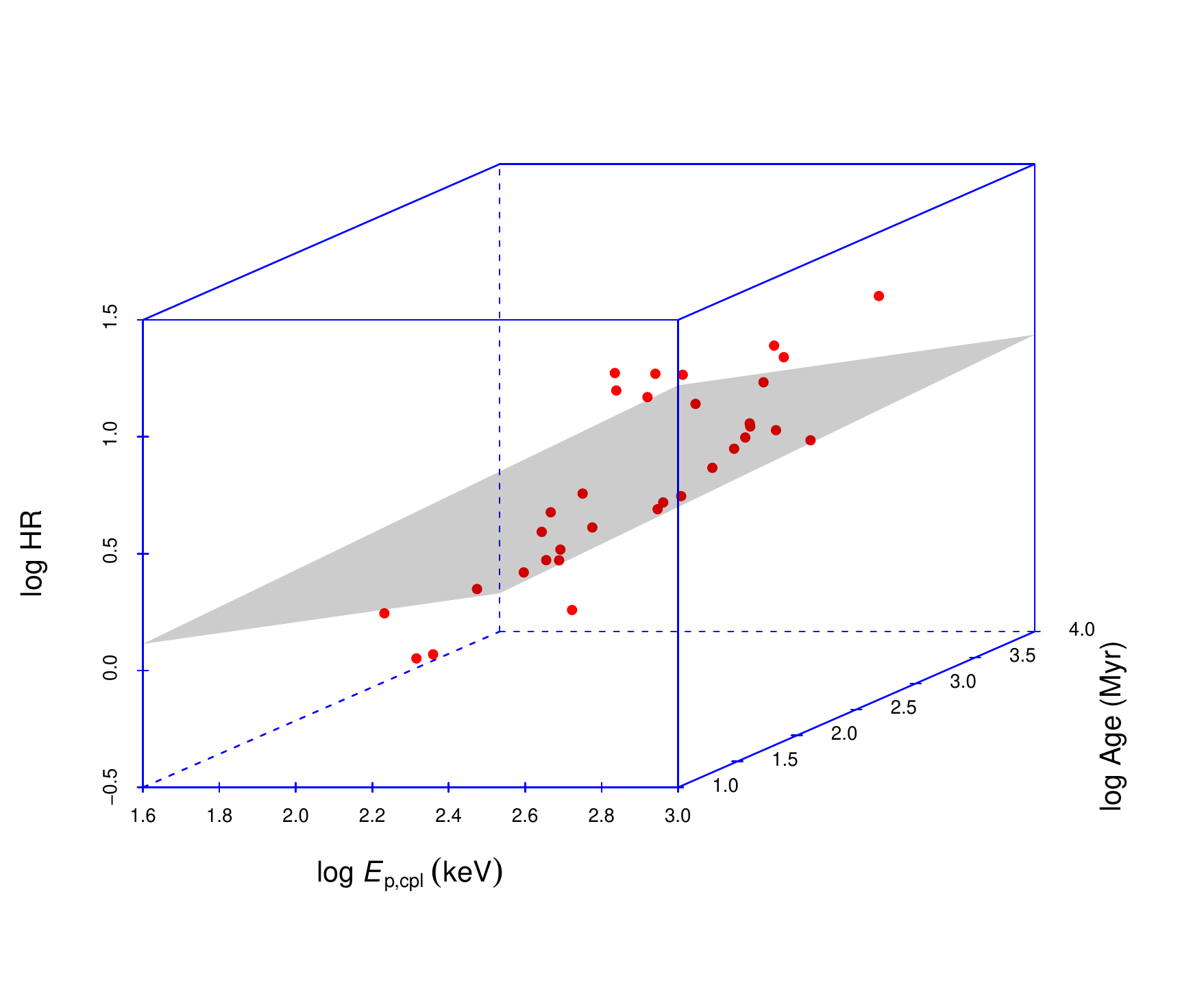}

\center{Fig. \ref{fig:three}---Continued}
\end{figure*}


\clearpage
\begin{figure*}

\includegraphics[width=0.45\textwidth]{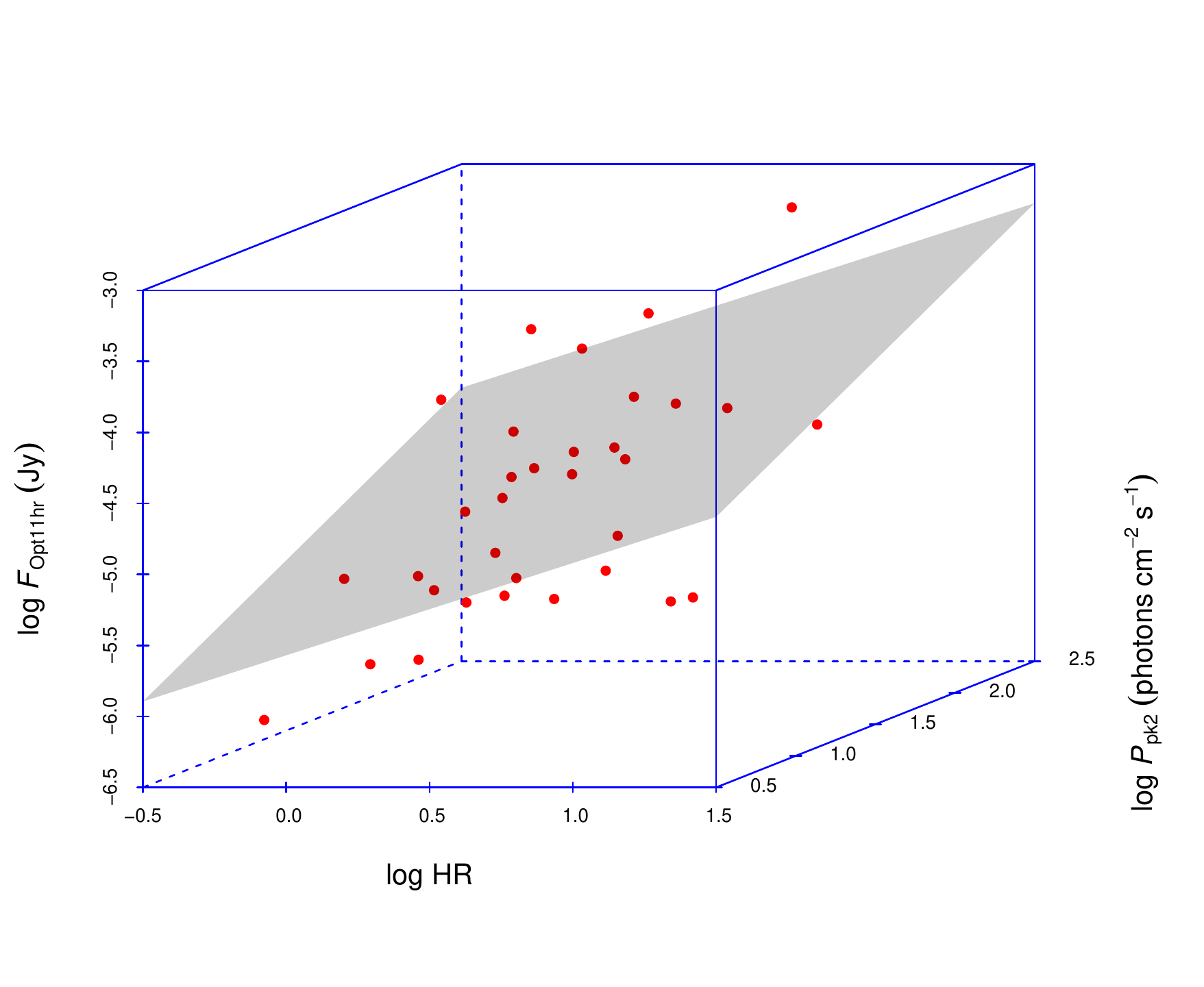}
\includegraphics[width=0.45\textwidth]{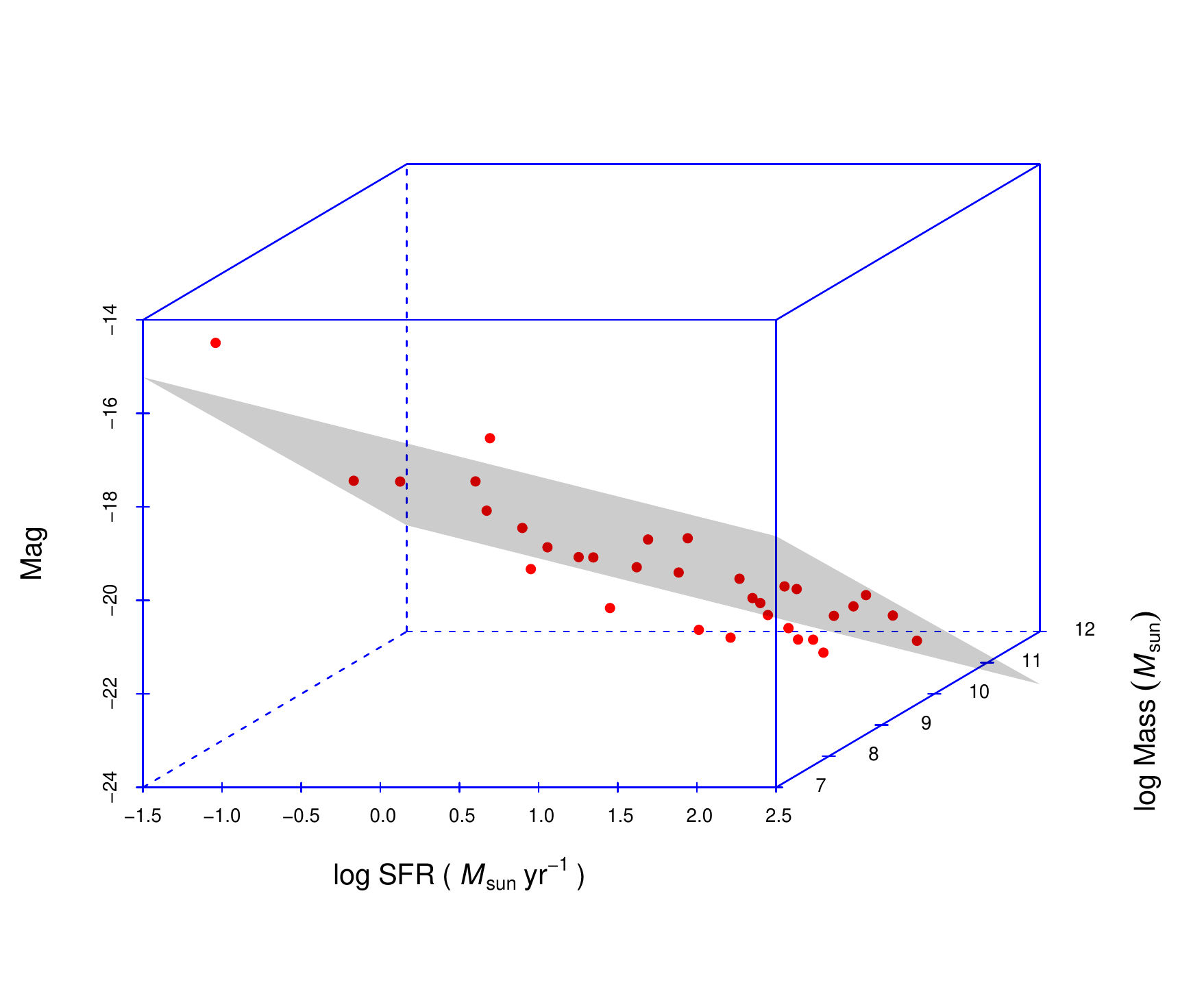}

\includegraphics[width=0.45\textwidth]{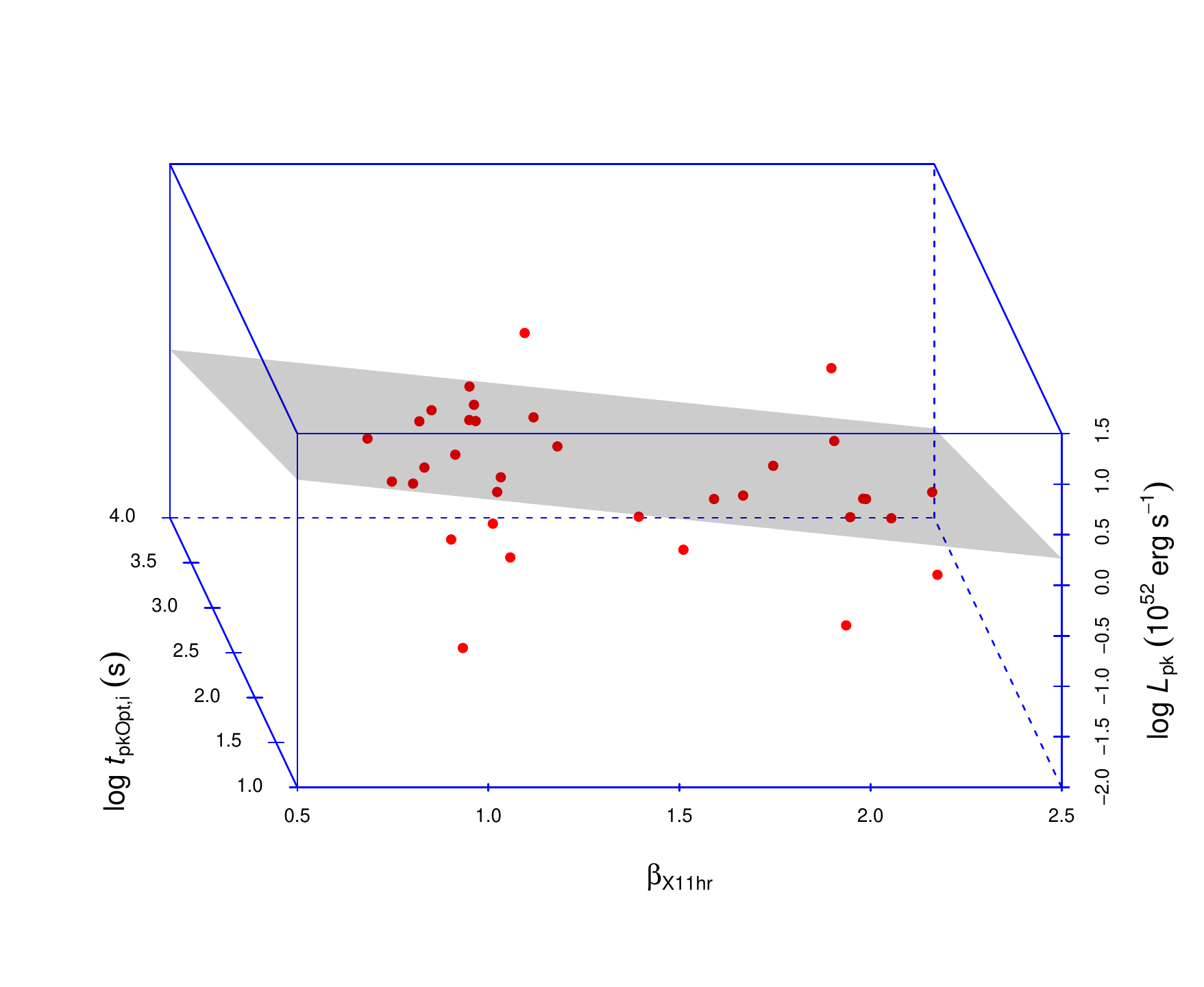}
\includegraphics[width=0.45\textwidth]{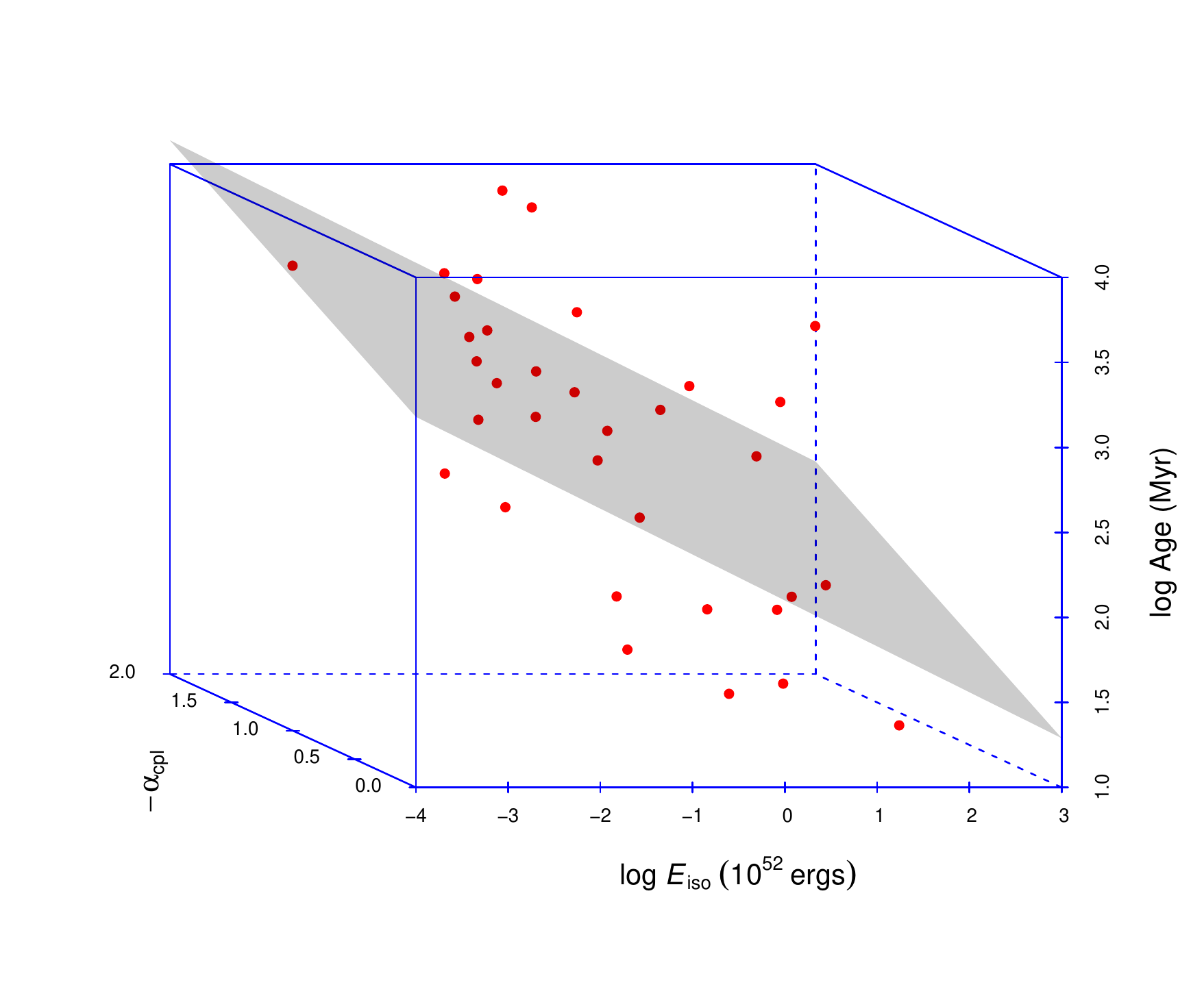}

\includegraphics[width=0.45\textwidth]{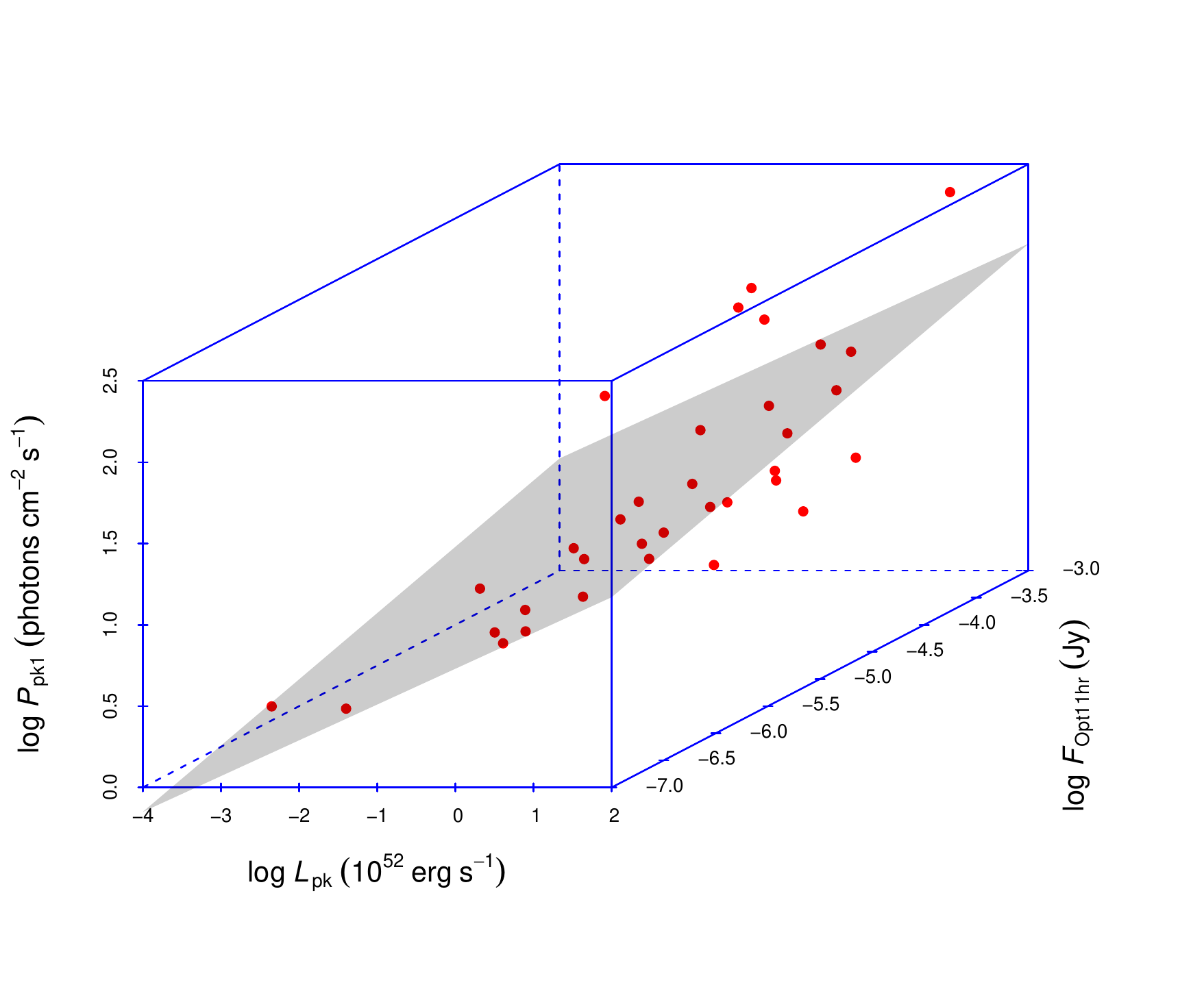}
\includegraphics[width=0.45\textwidth]{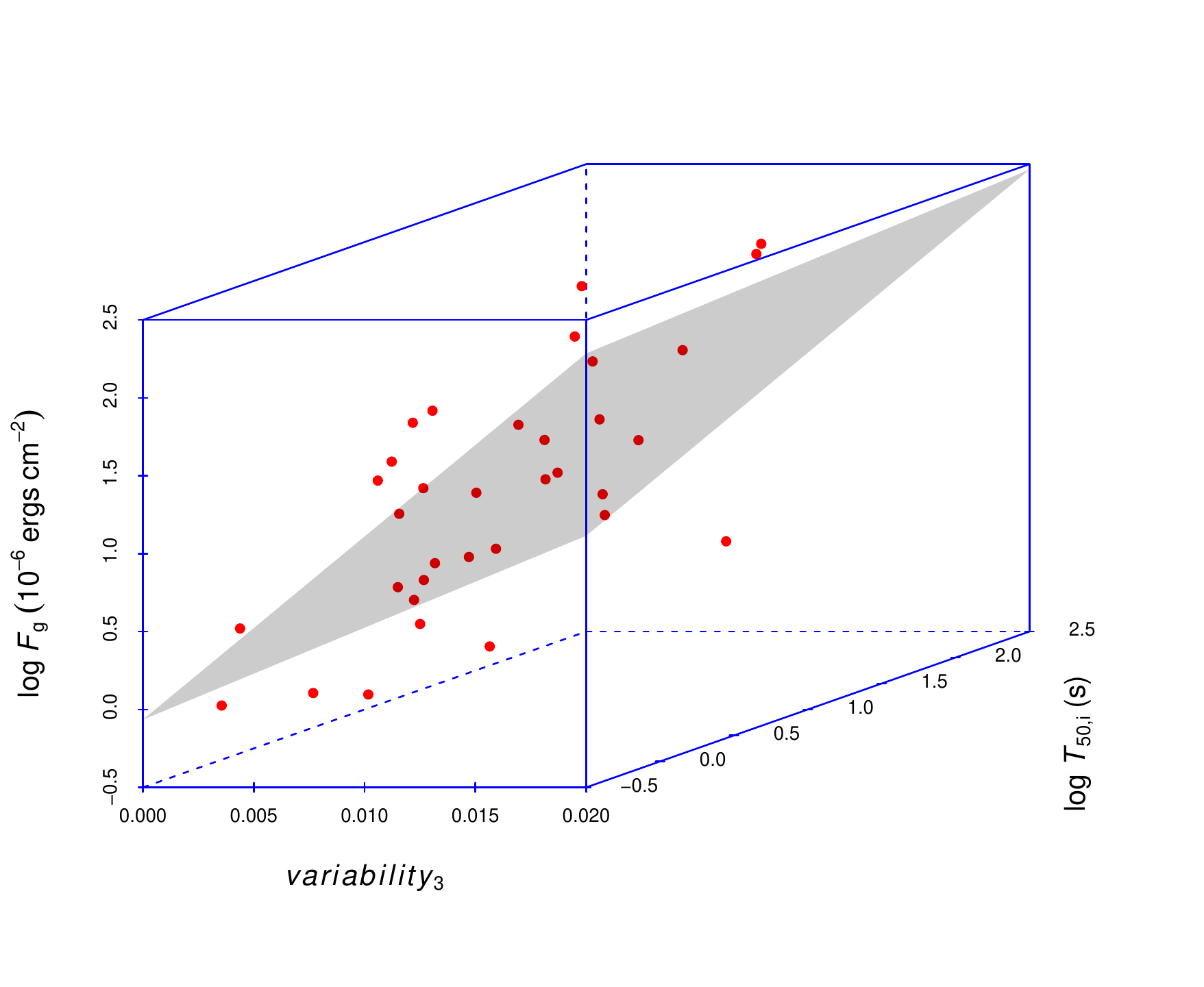}

\center{Fig. \ref{fig:three}---Continued}
\end{figure*}


\clearpage
\begin{figure*}

\includegraphics[width=0.45\textwidth]{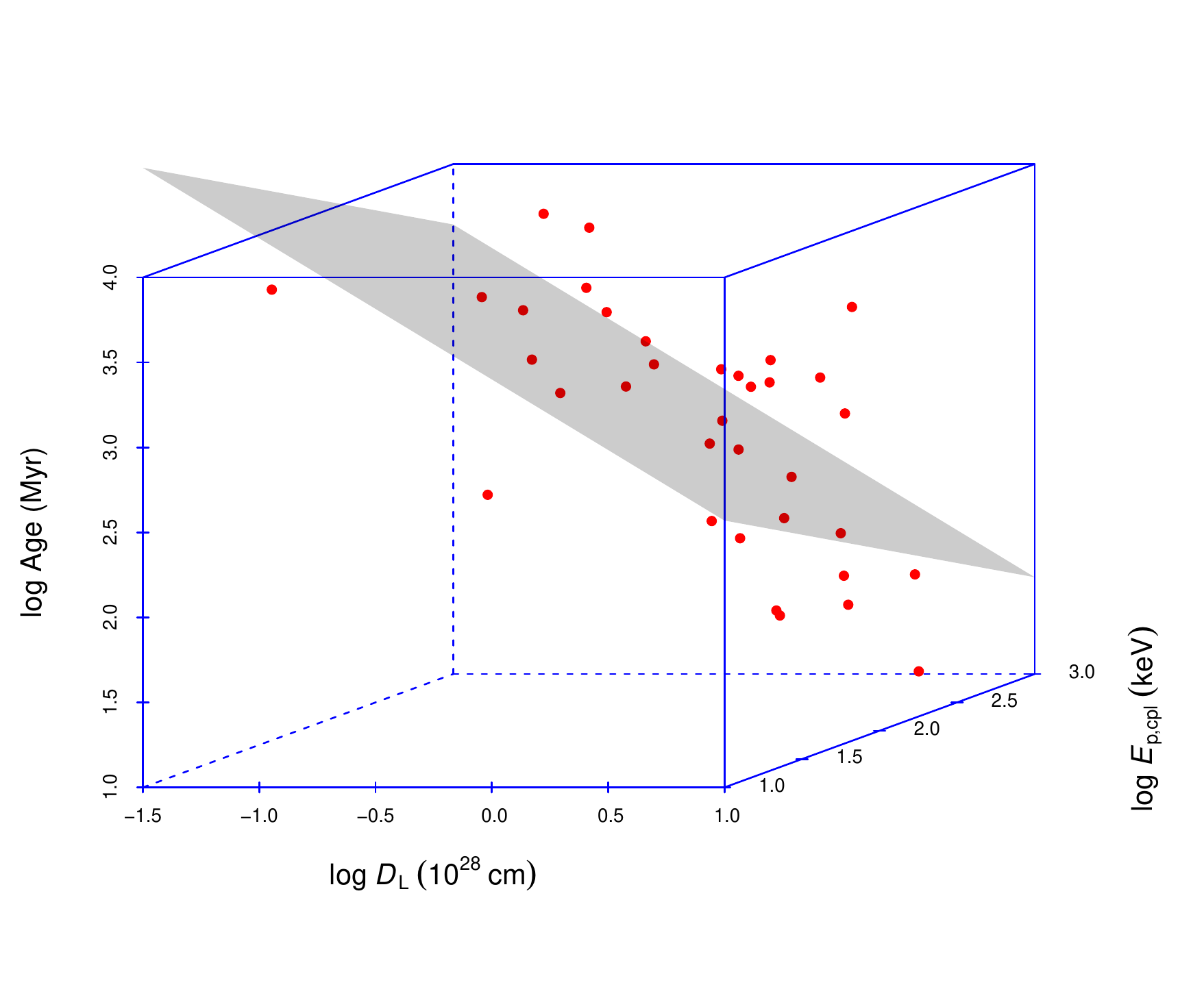}
\includegraphics[width=0.45\textwidth]{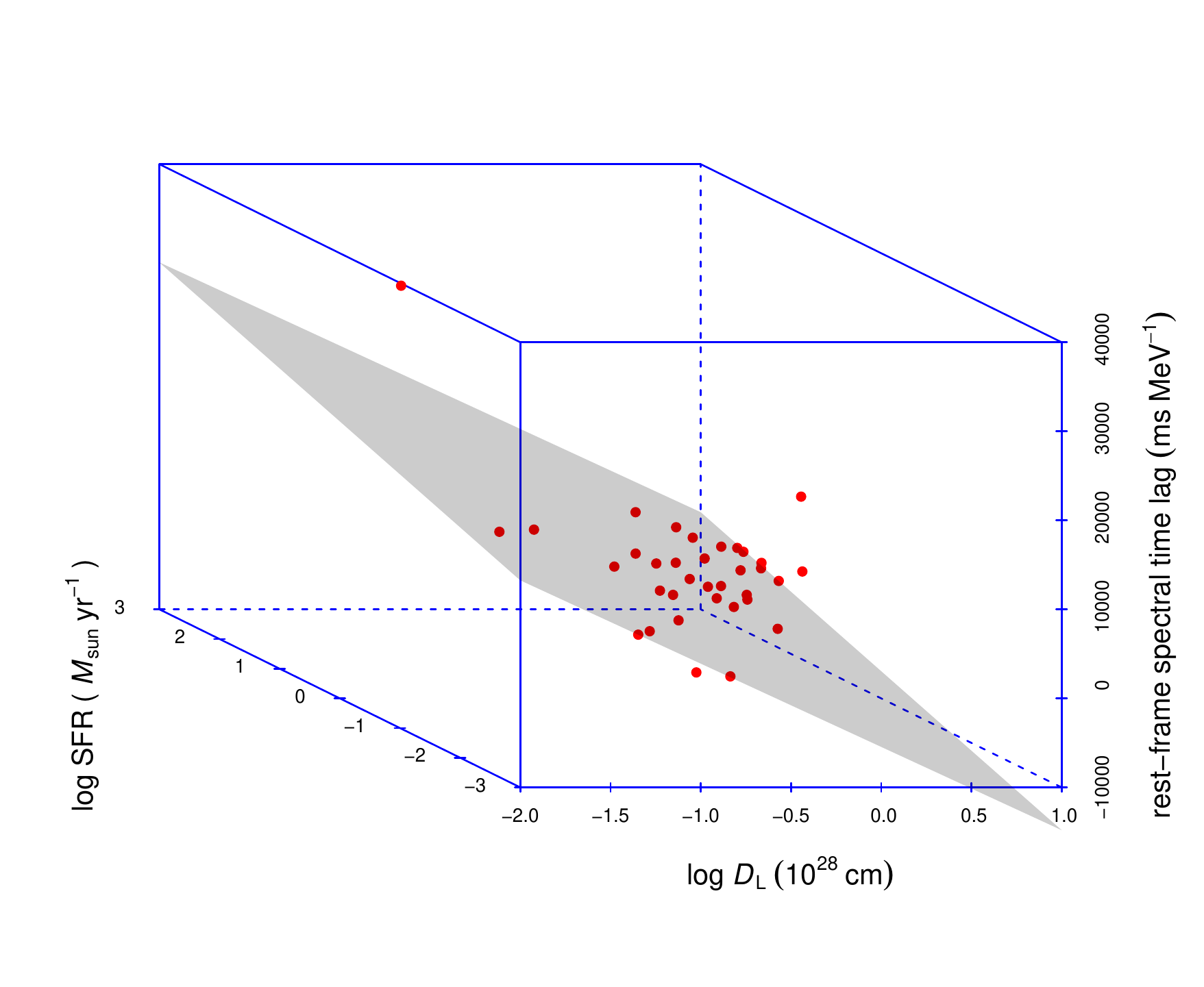}

\includegraphics[width=0.45\textwidth]{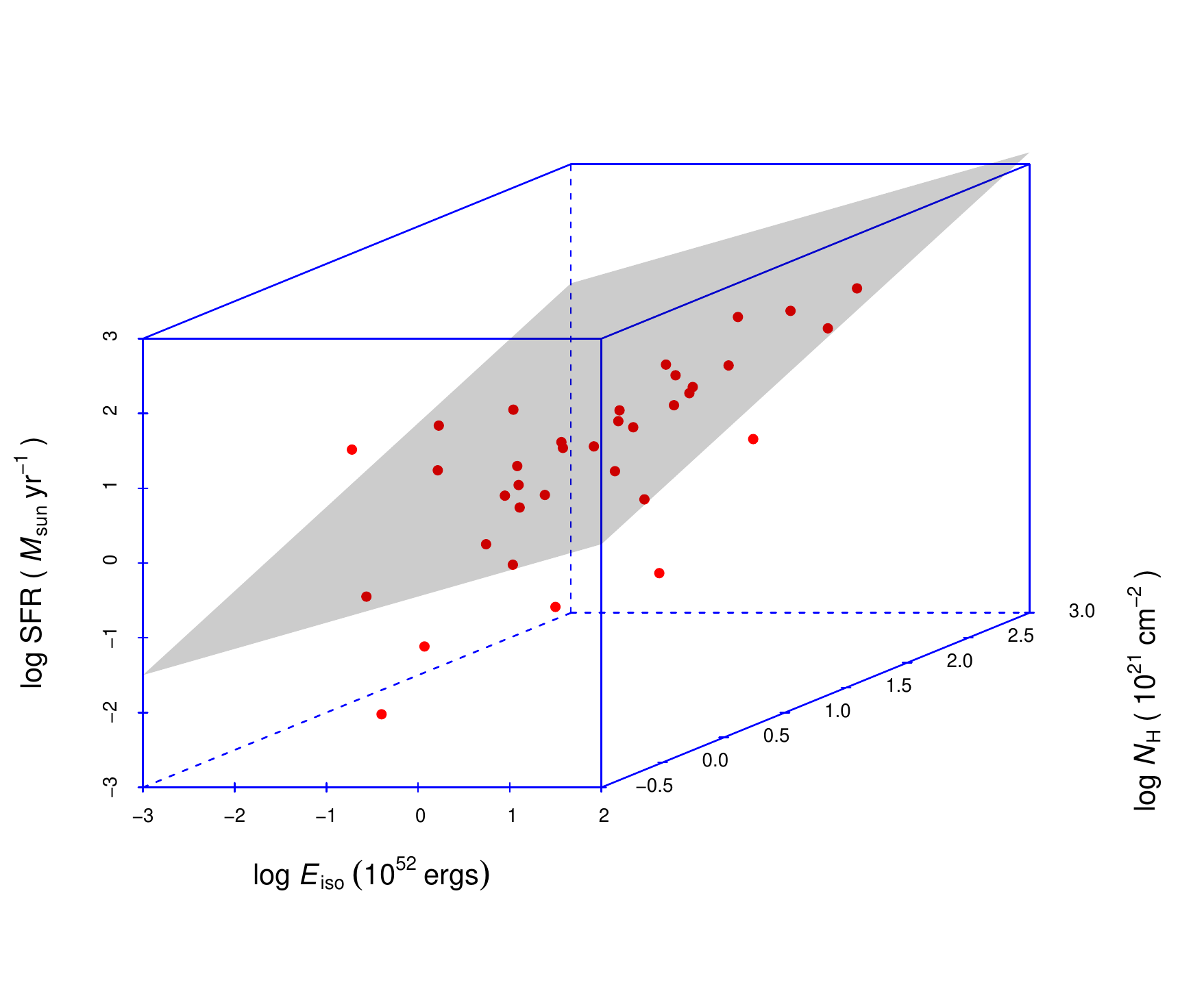}
\includegraphics[width=0.45\textwidth]{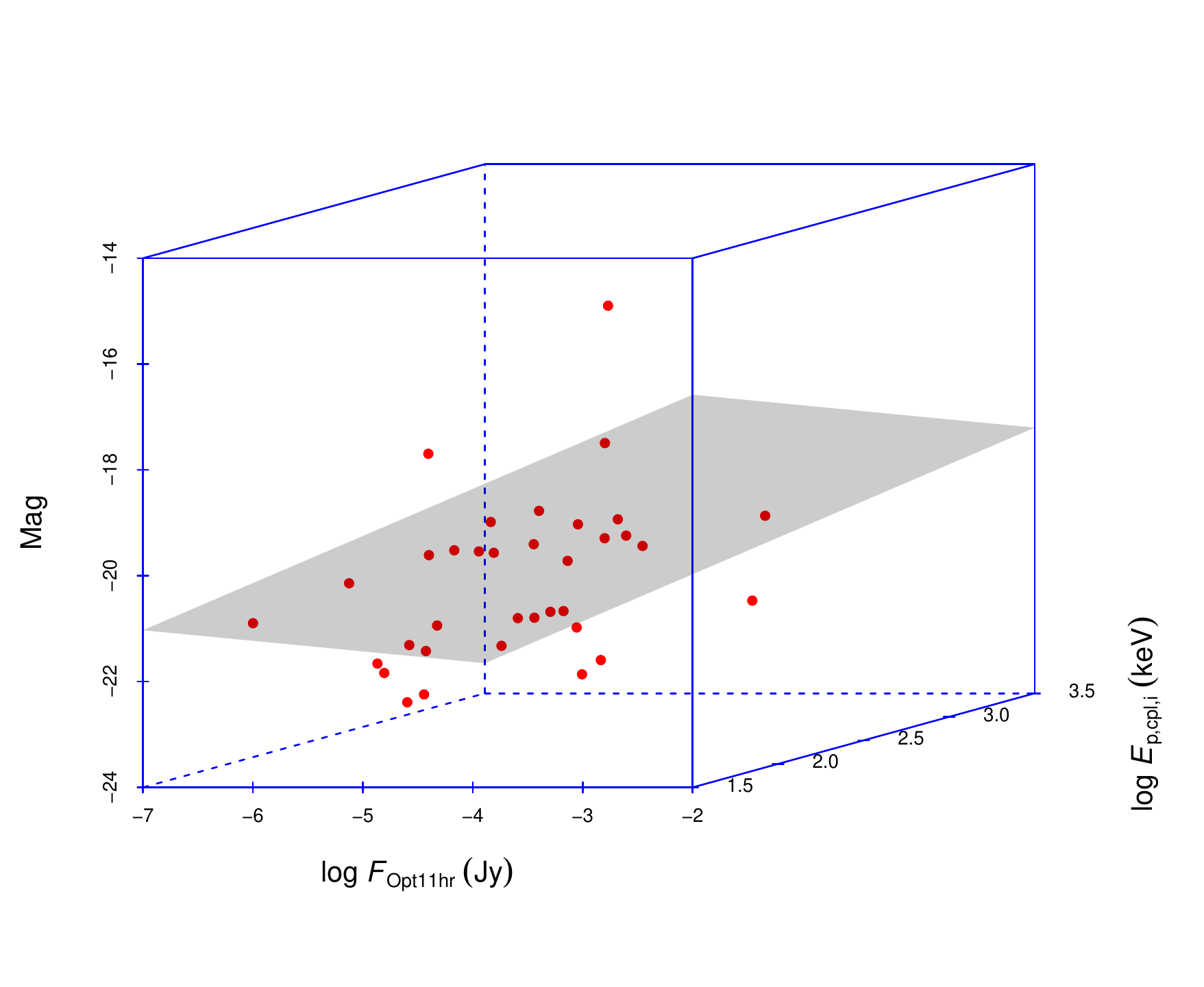}

\includegraphics[width=0.45\textwidth]{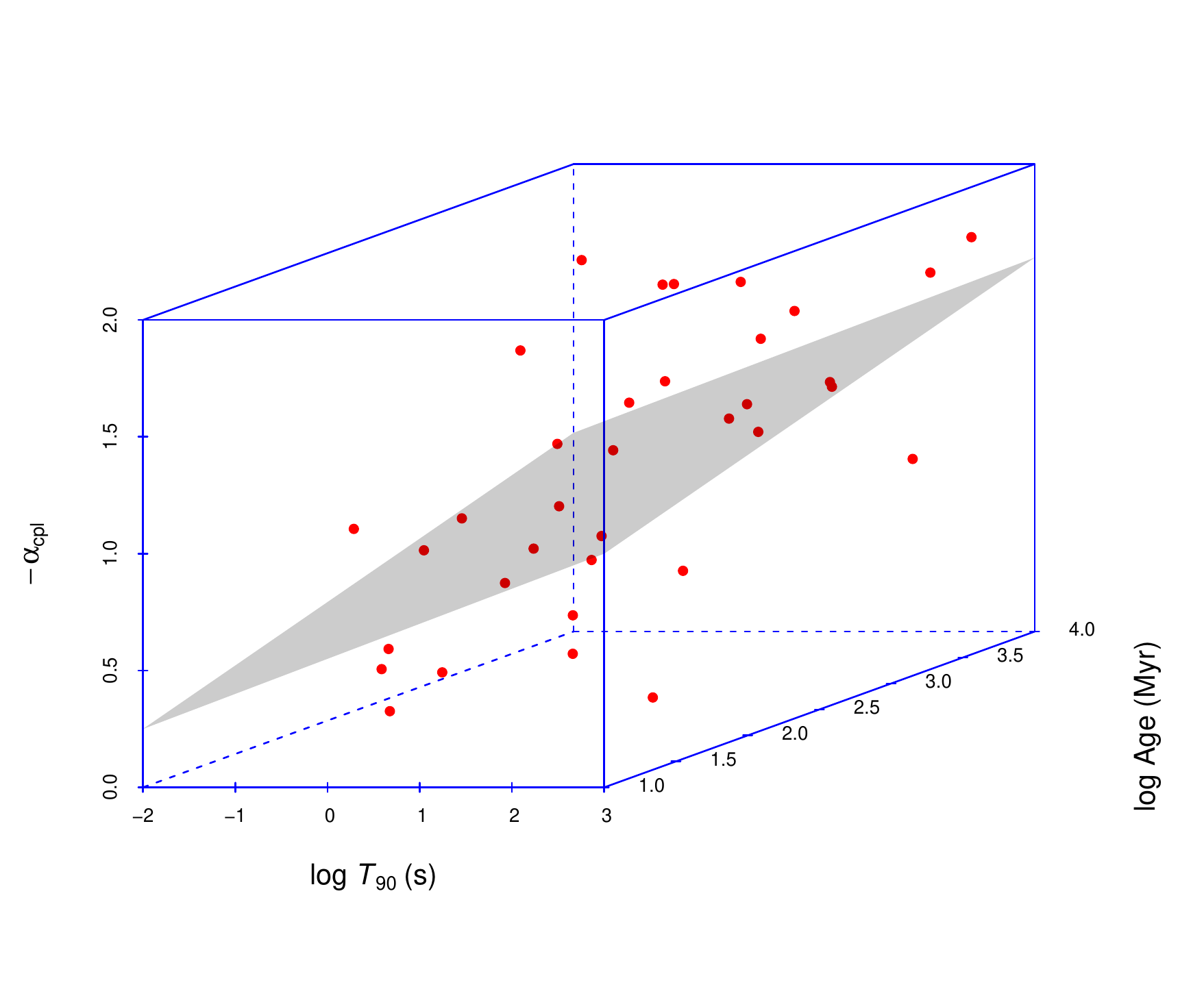}
\includegraphics[width=0.45\textwidth]{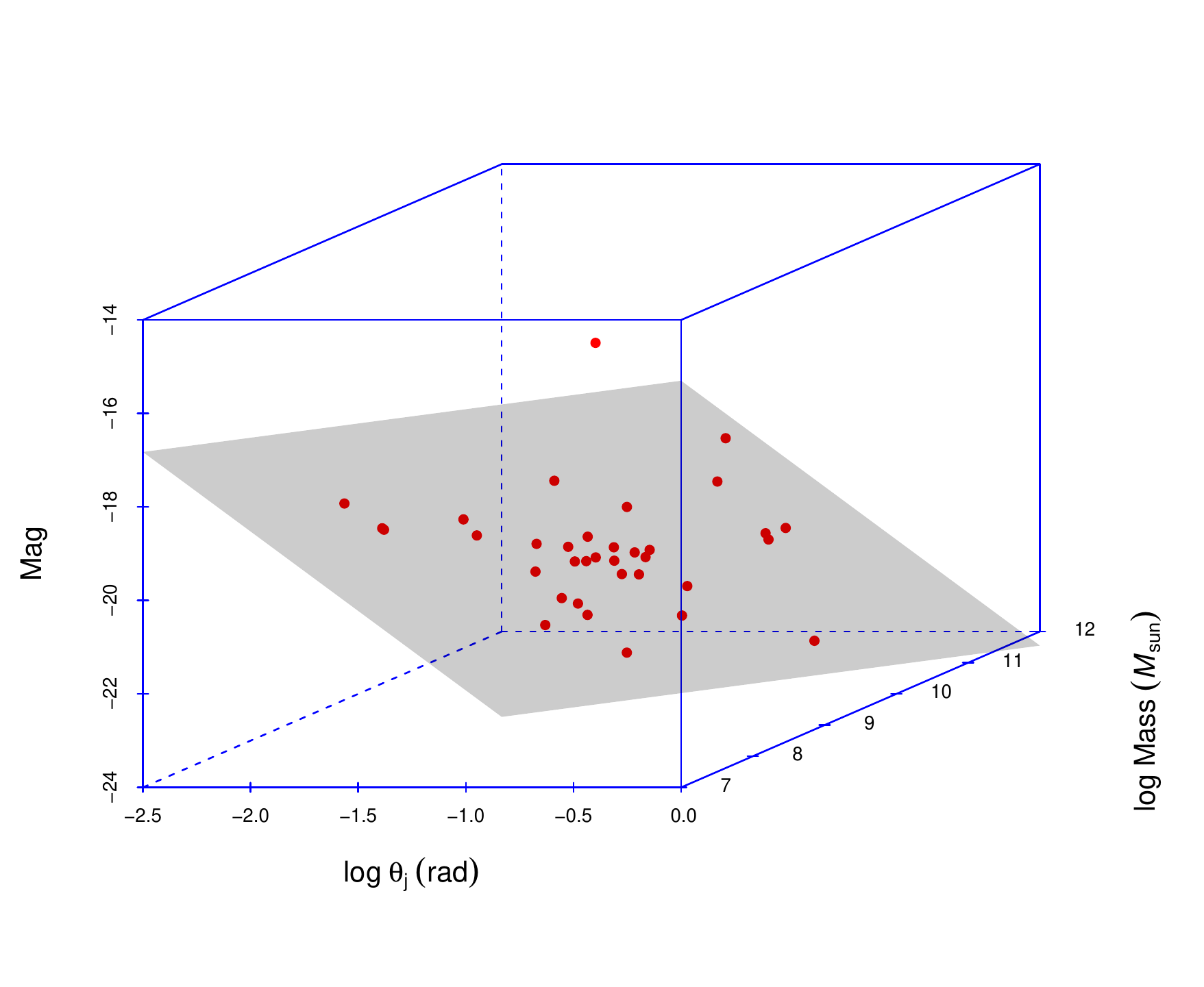}

\center{Fig. \ref{fig:three}---Continued}
\end{figure*}


\clearpage
\begin{figure*}

\includegraphics[width=0.45\textwidth]{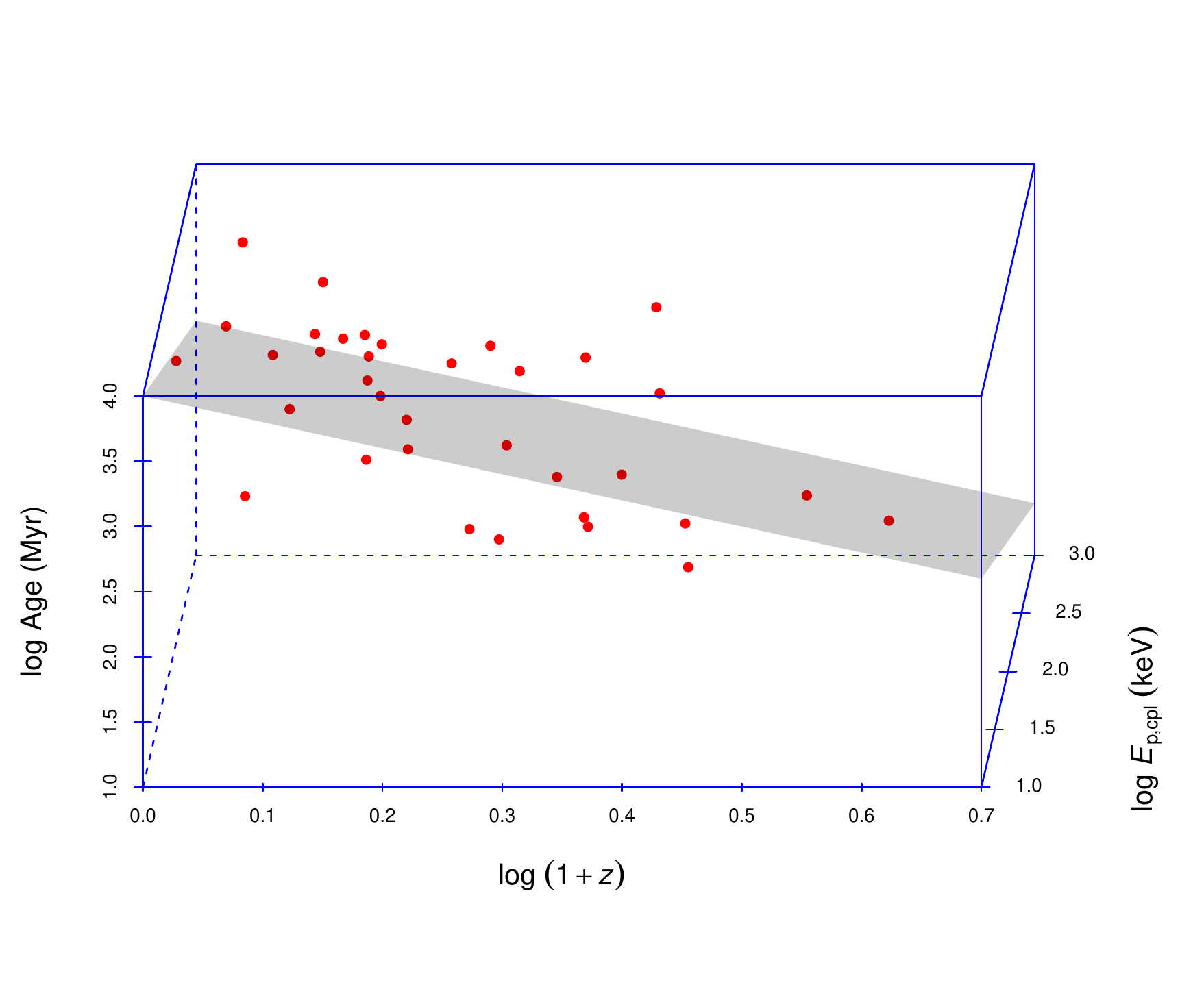}
\includegraphics[width=0.45\textwidth]{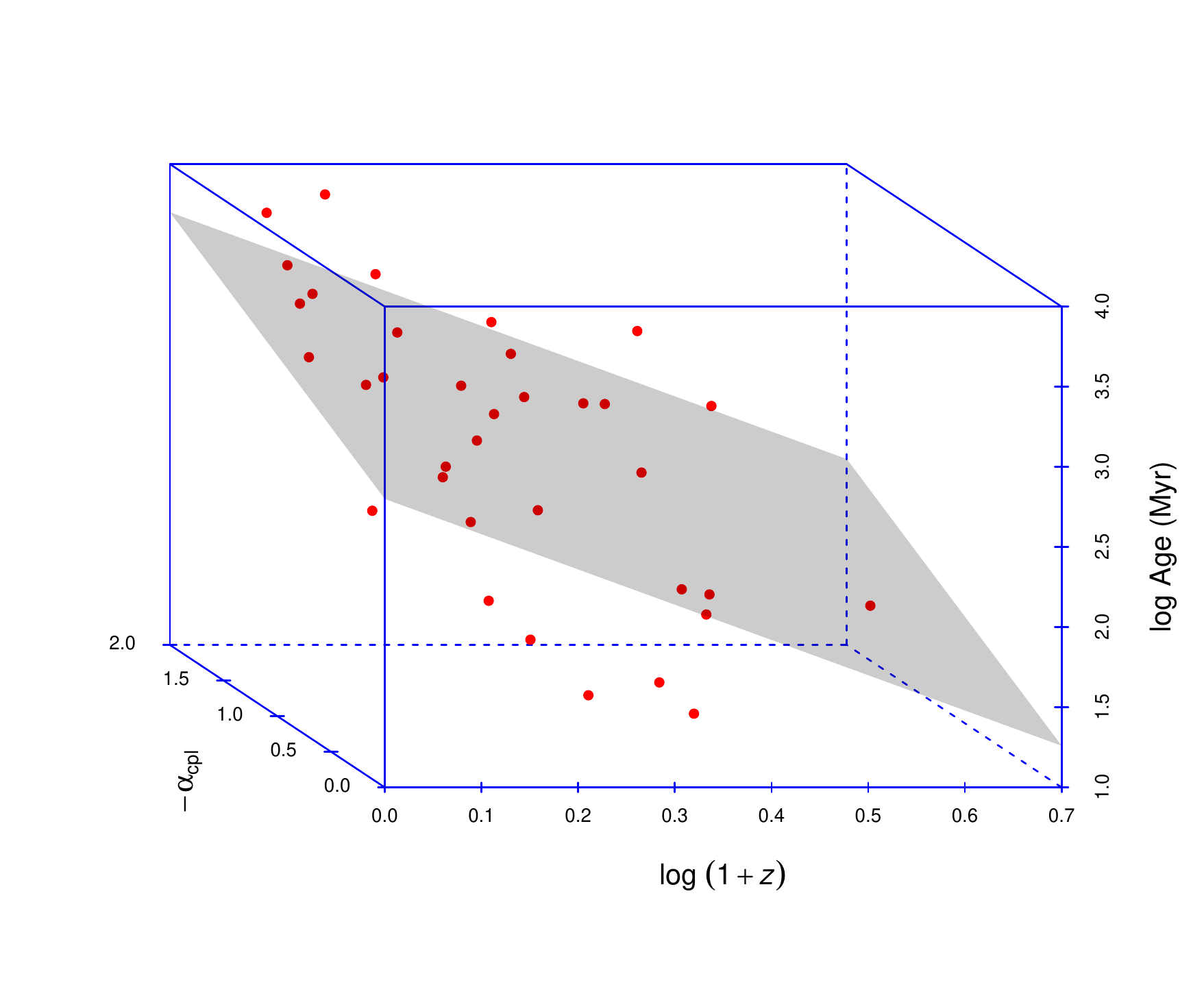}

\includegraphics[width=0.45\textwidth]{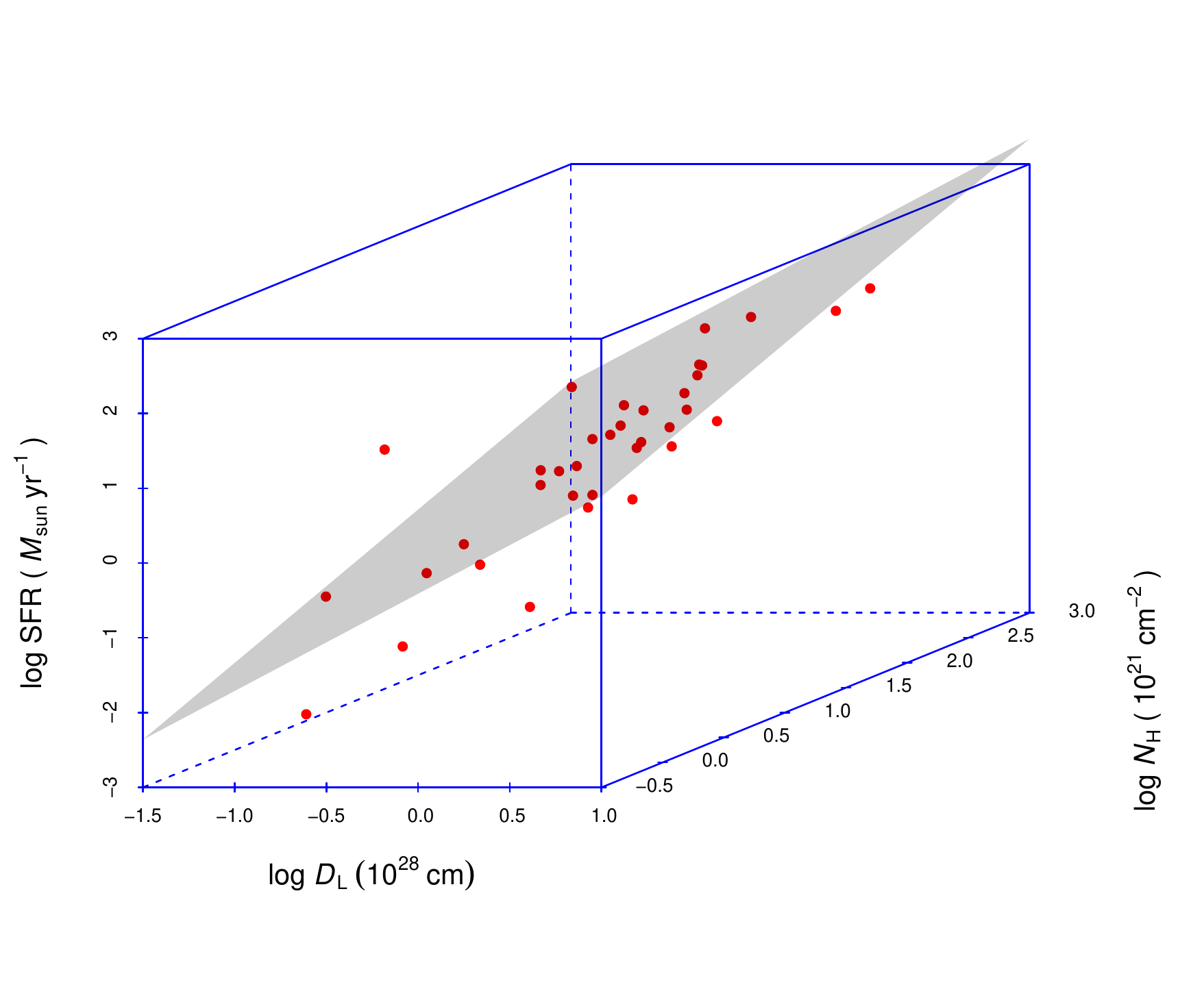}
\includegraphics[width=0.45\textwidth]{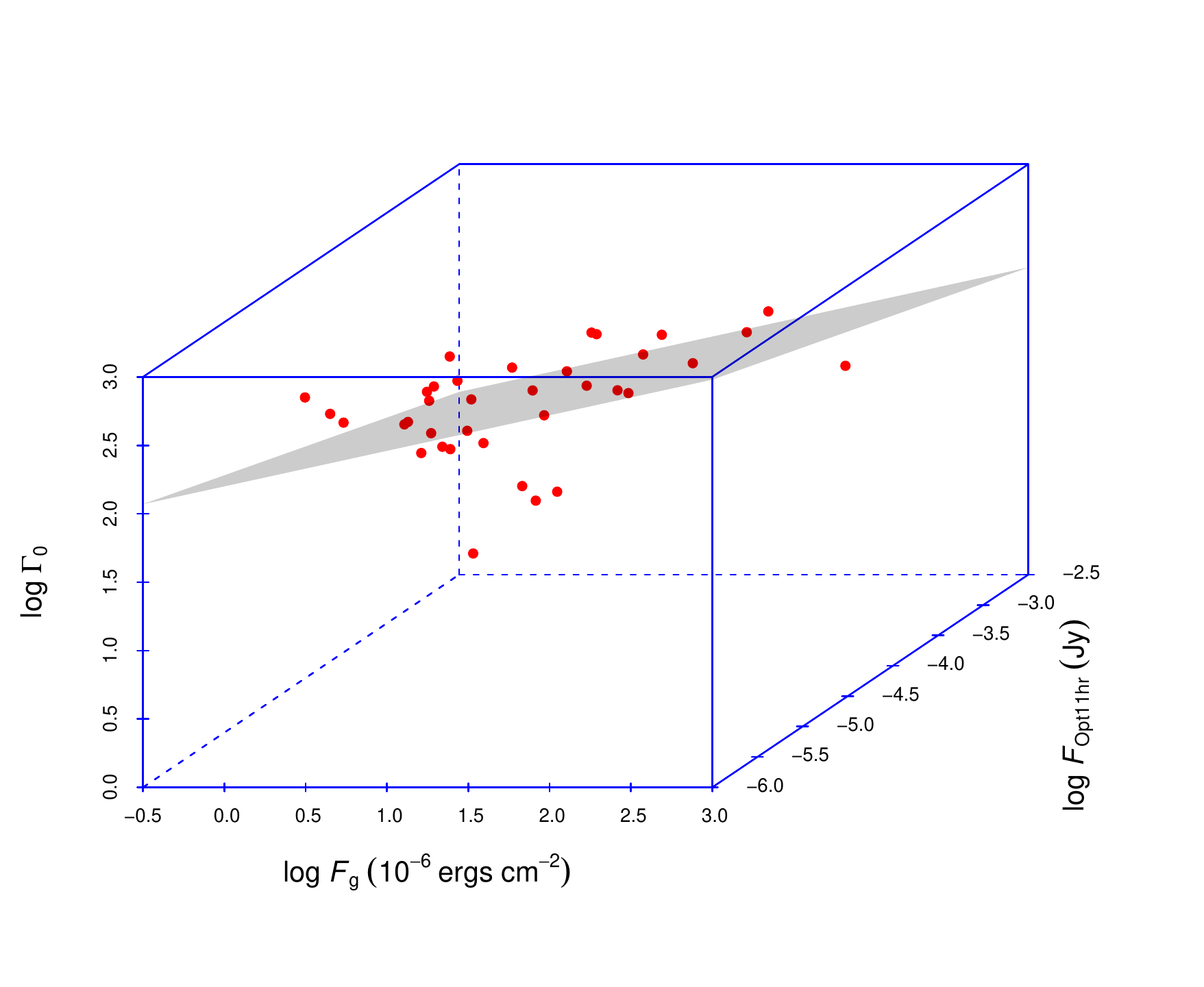}

\includegraphics[width=0.45\textwidth]{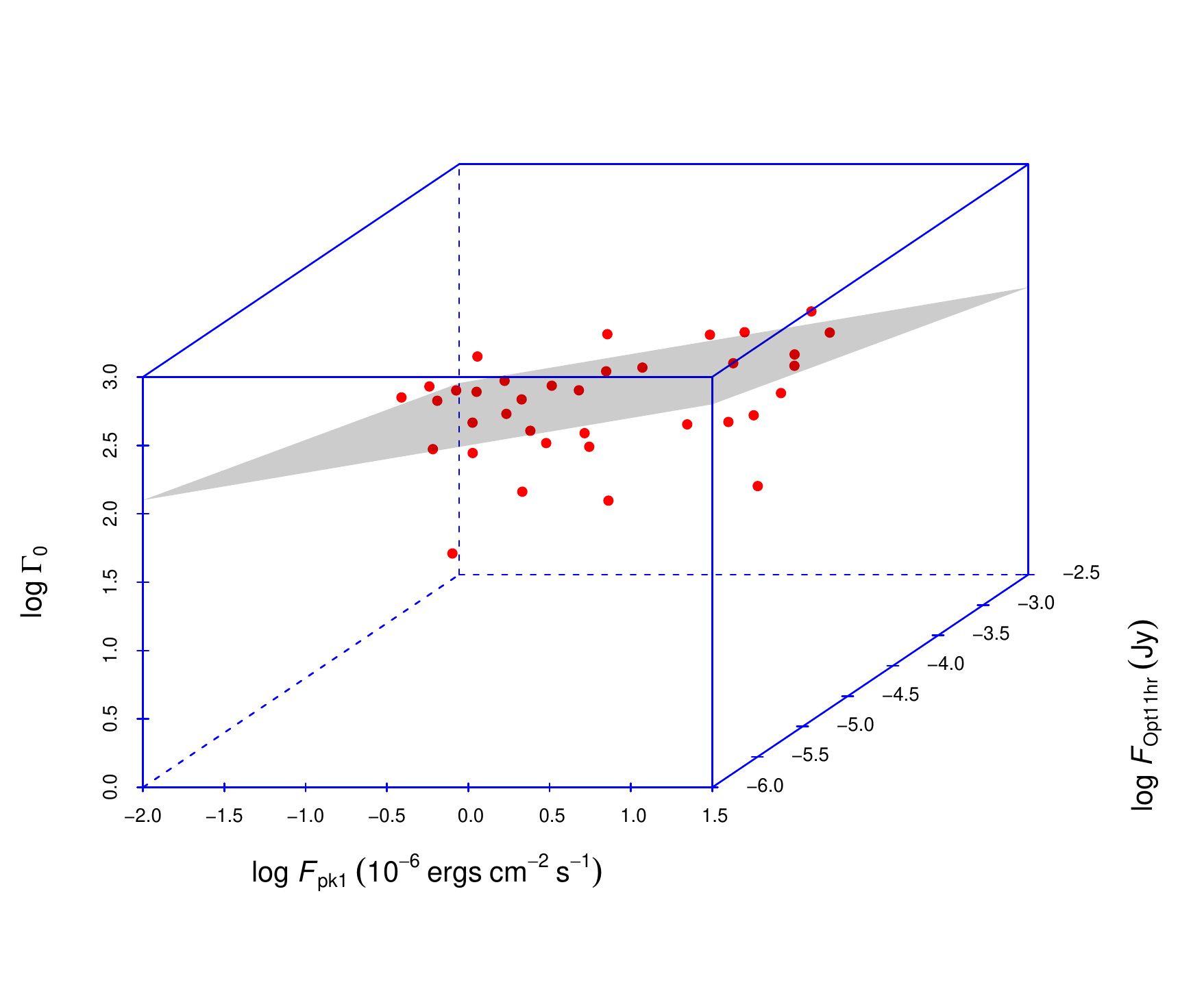}
\includegraphics[width=0.45\textwidth]{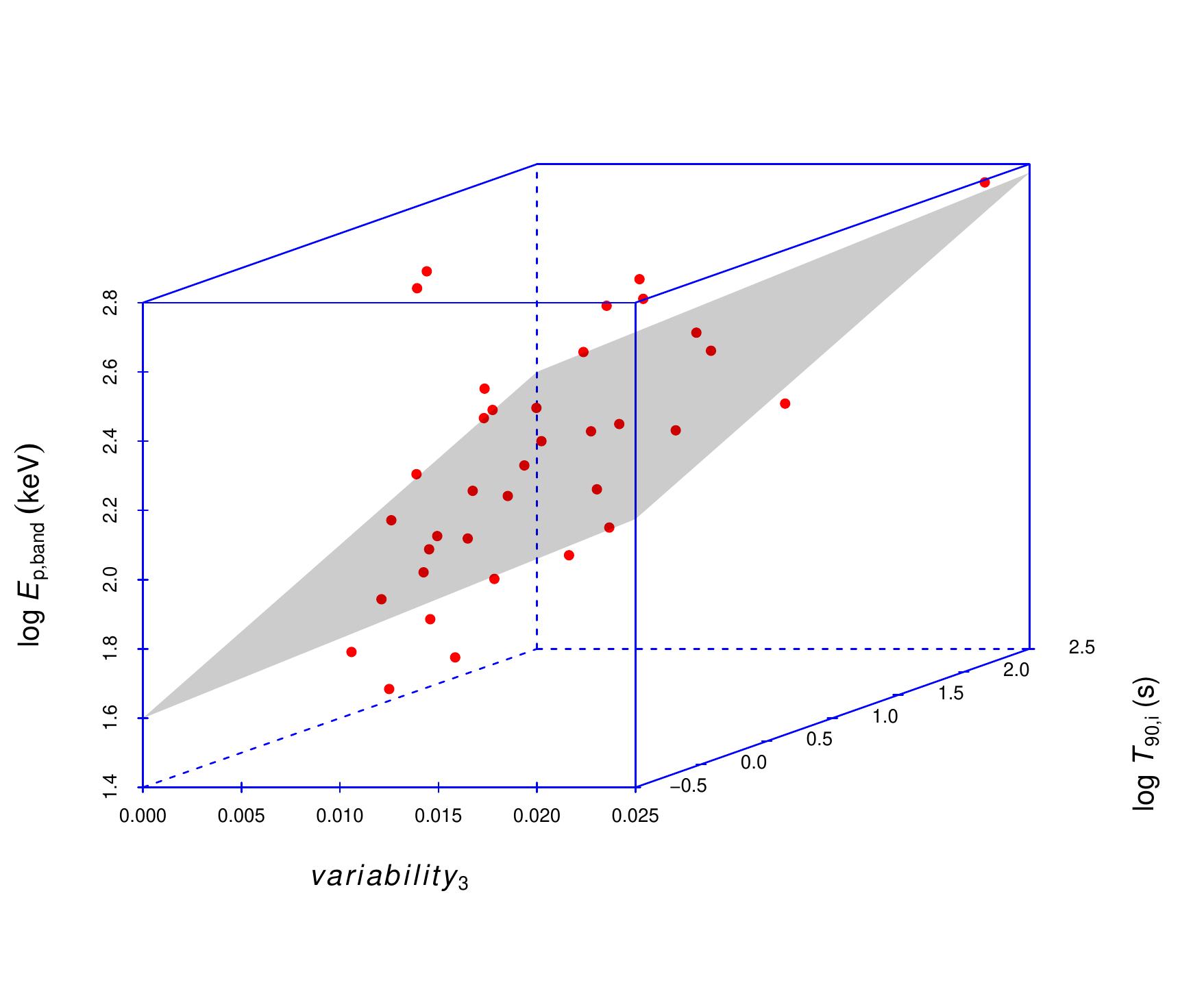}

\center{Fig. \ref{fig:three}---Continued}
\end{figure*}


\clearpage
\begin{figure*}

\includegraphics[width=0.45\textwidth]{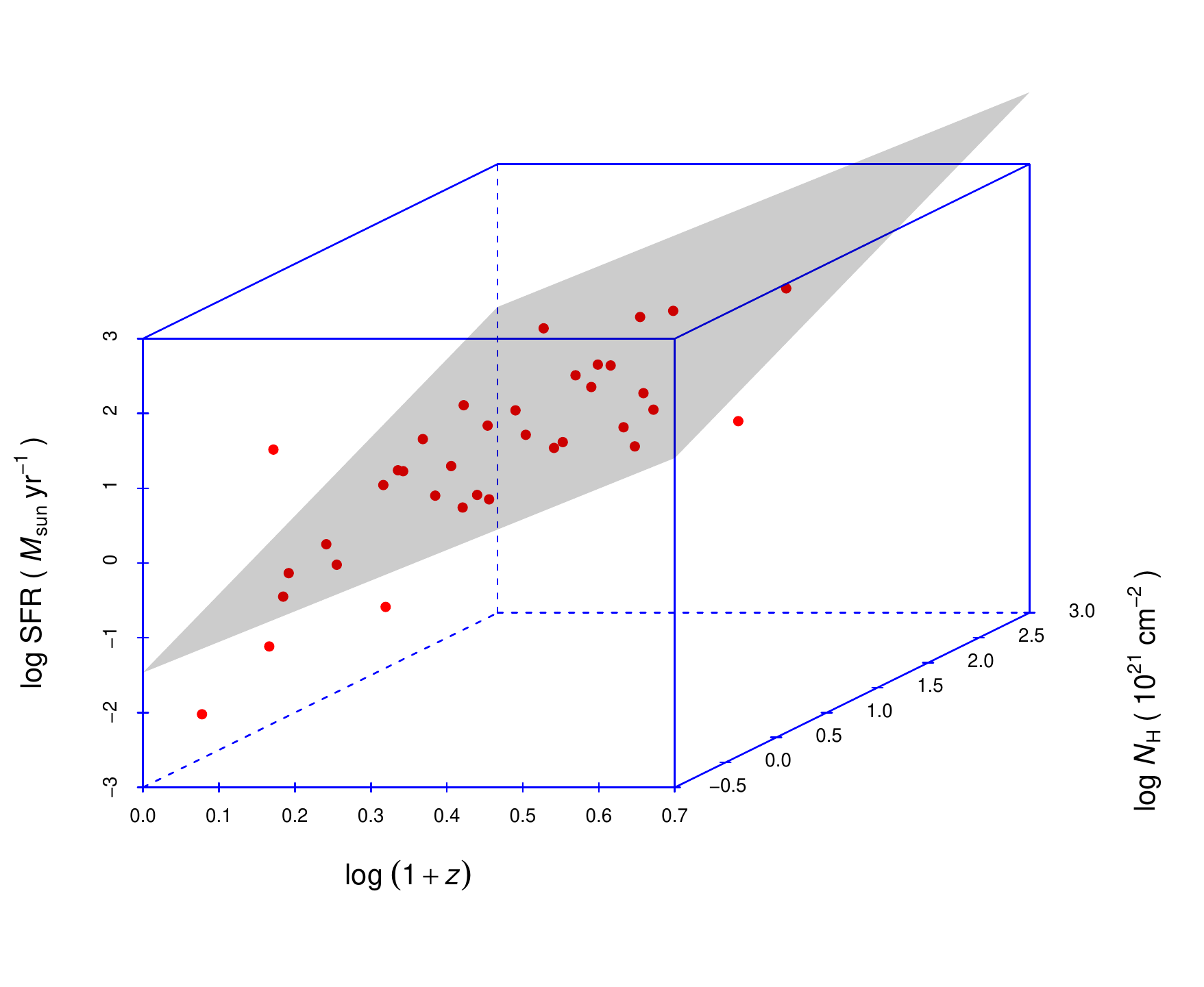}
\includegraphics[width=0.45\textwidth]{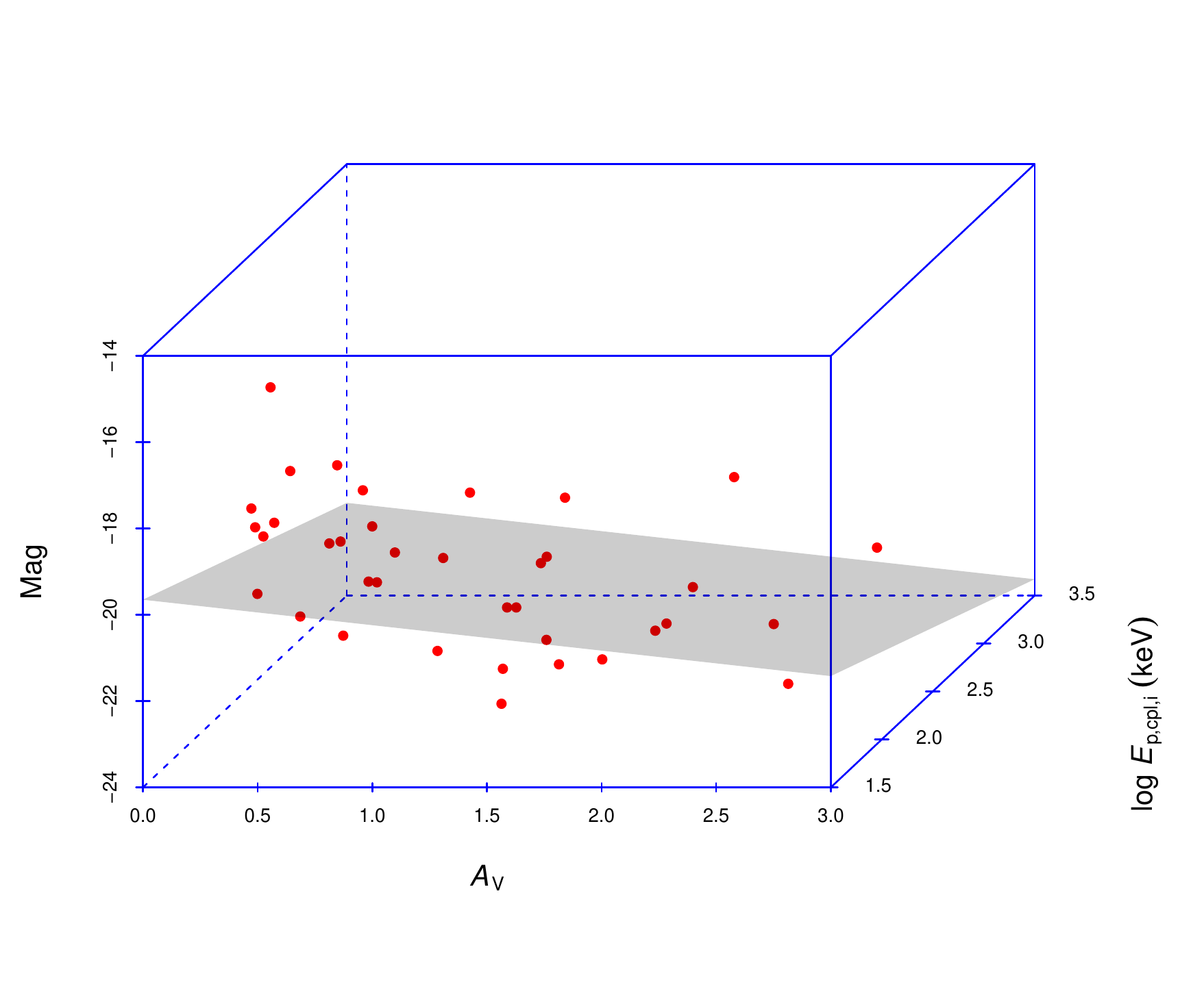}

\includegraphics[width=0.45\textwidth]{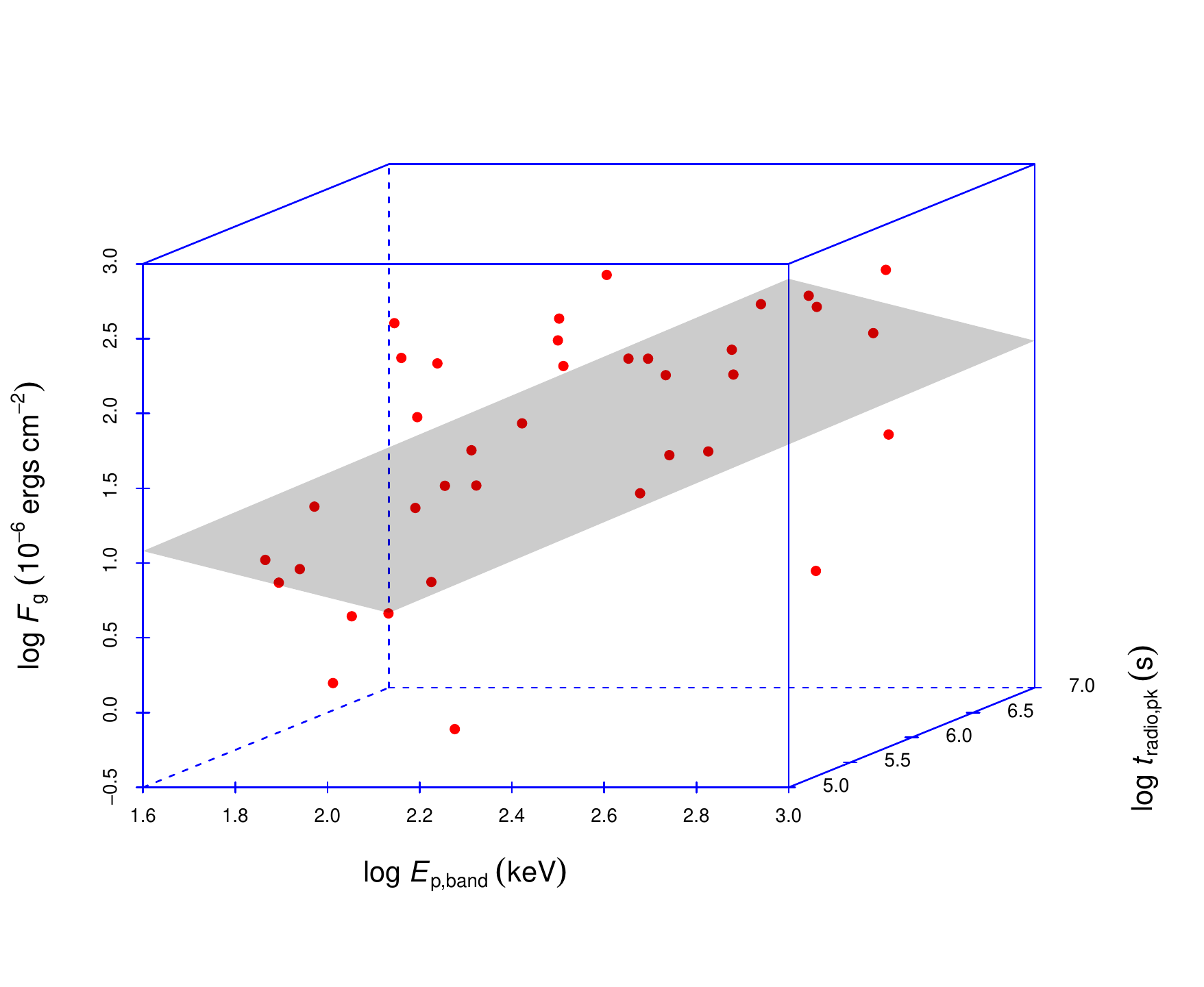}
\includegraphics[width=0.45\textwidth]{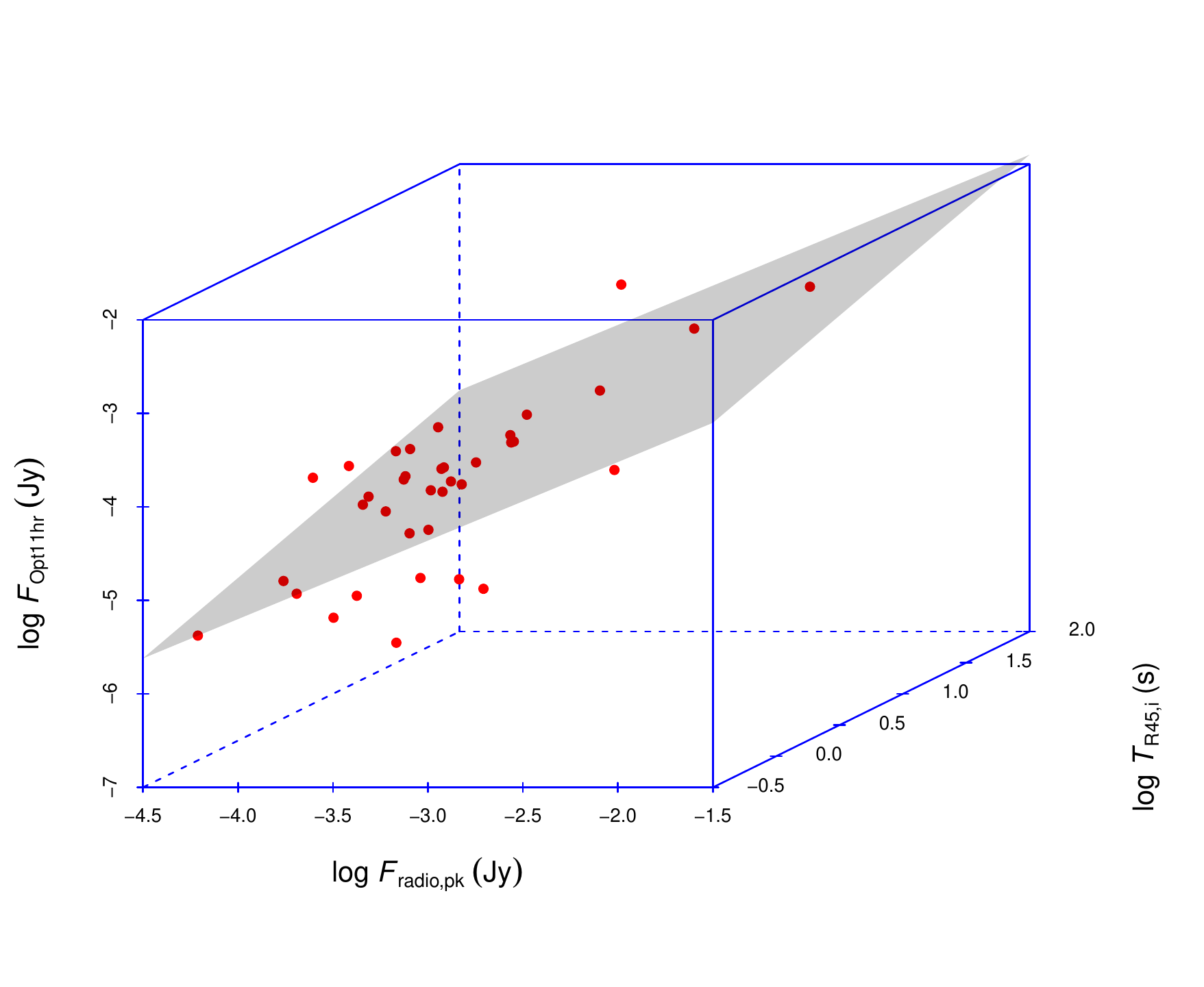}

\includegraphics[width=0.45\textwidth]{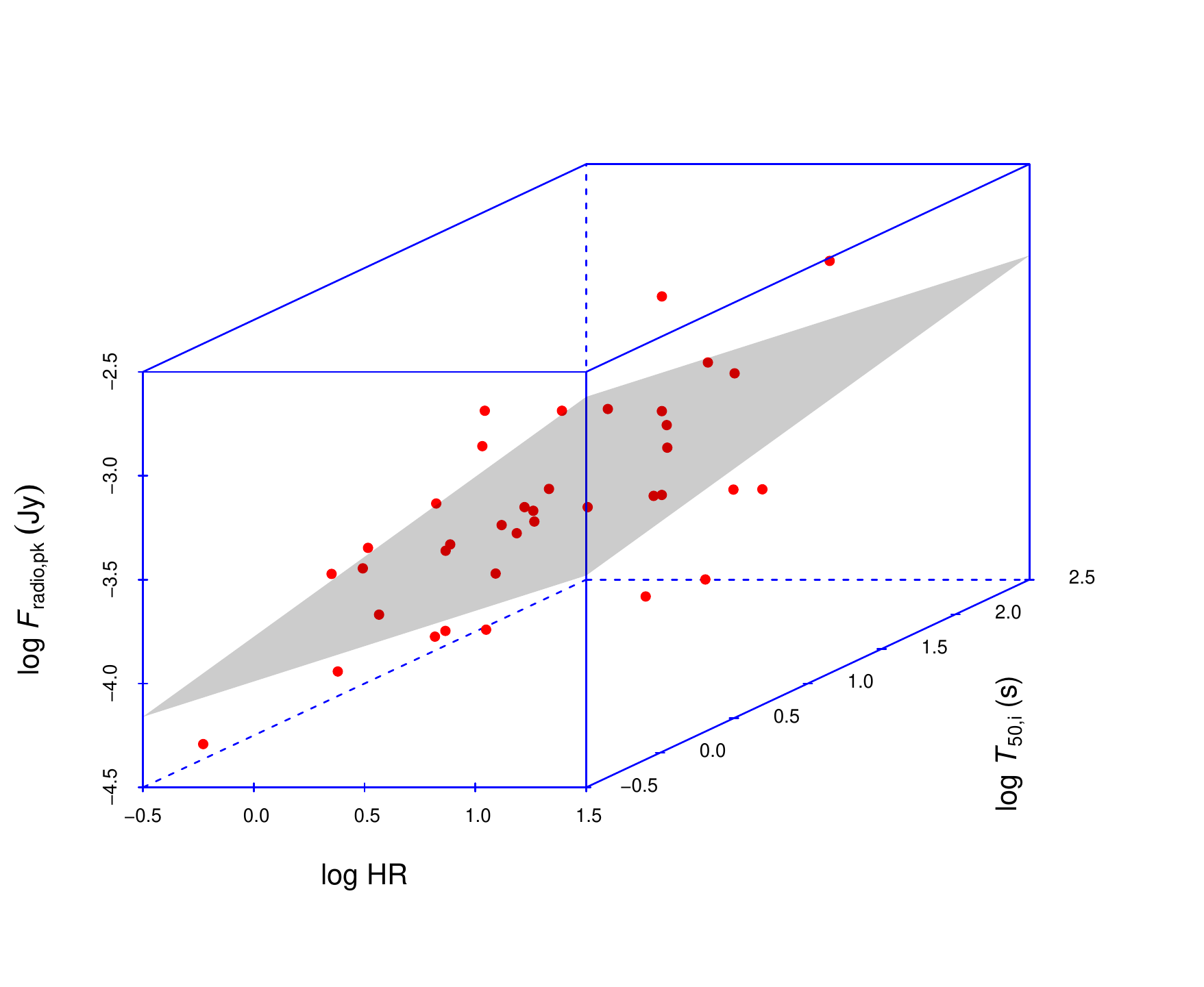}
\includegraphics[width=0.45\textwidth]{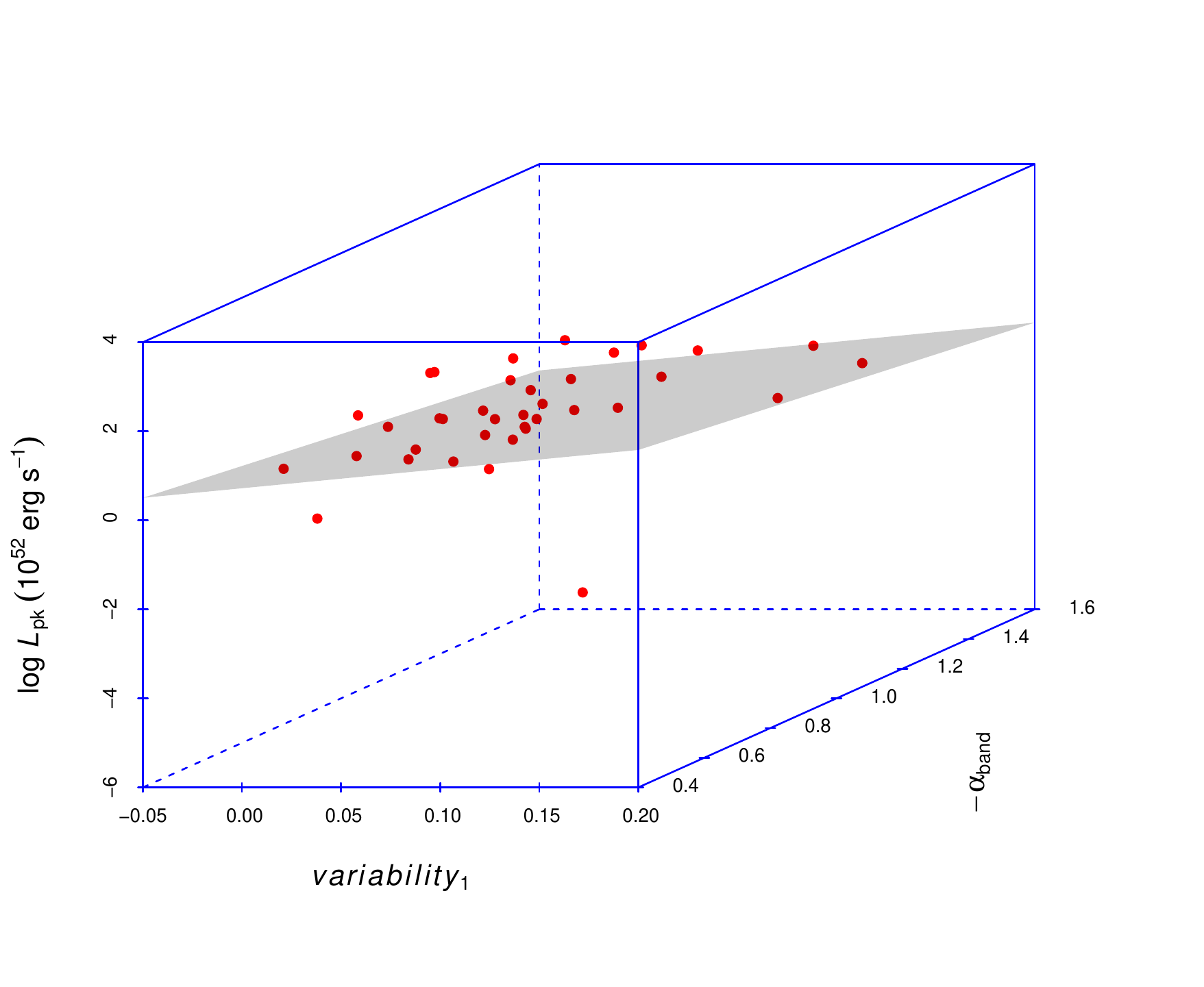}

\center{Fig. \ref{fig:three}---Continued}
\end{figure*}


\clearpage
\begin{figure*}

\includegraphics[width=0.45\textwidth]{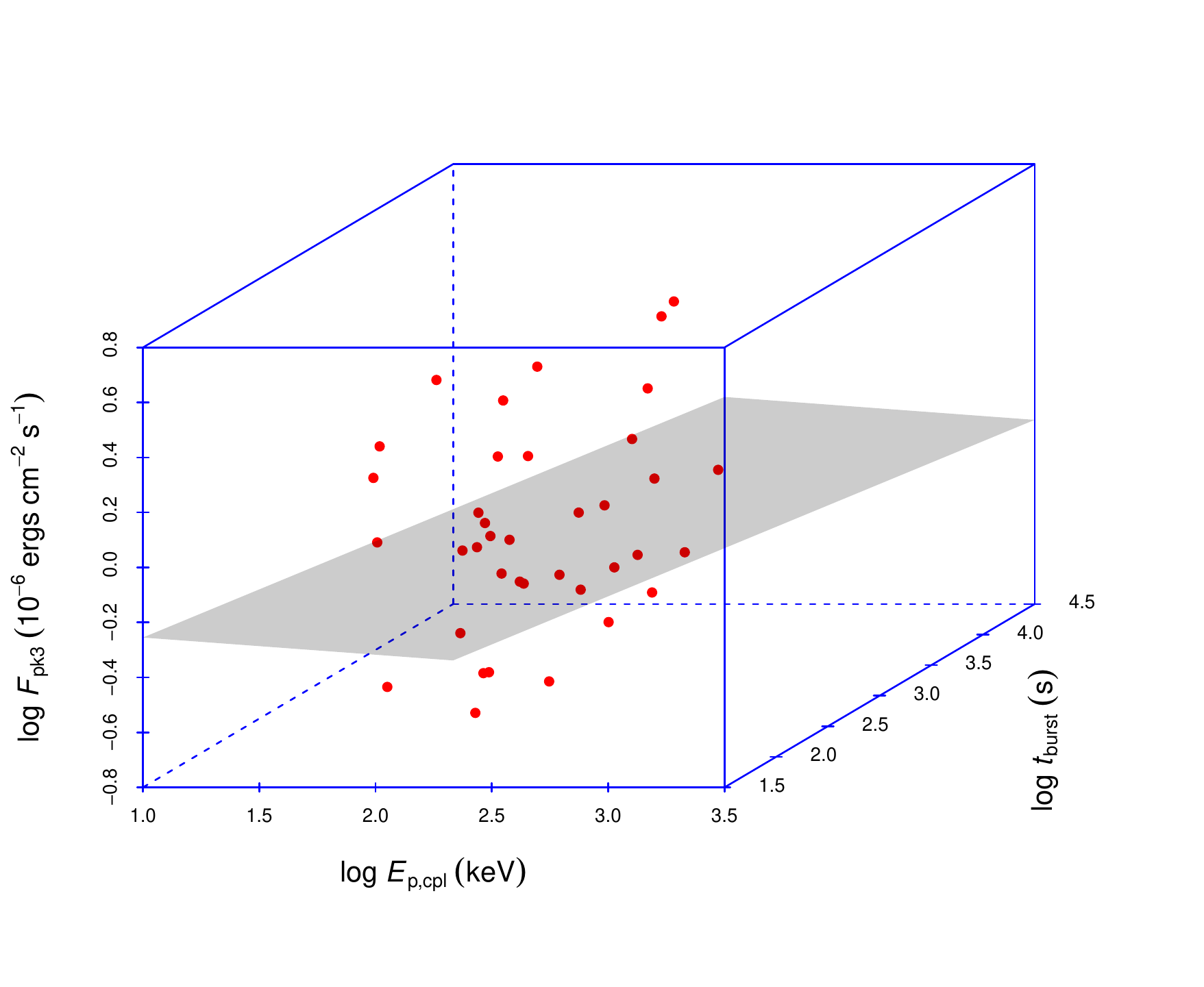}
\includegraphics[width=0.45\textwidth]{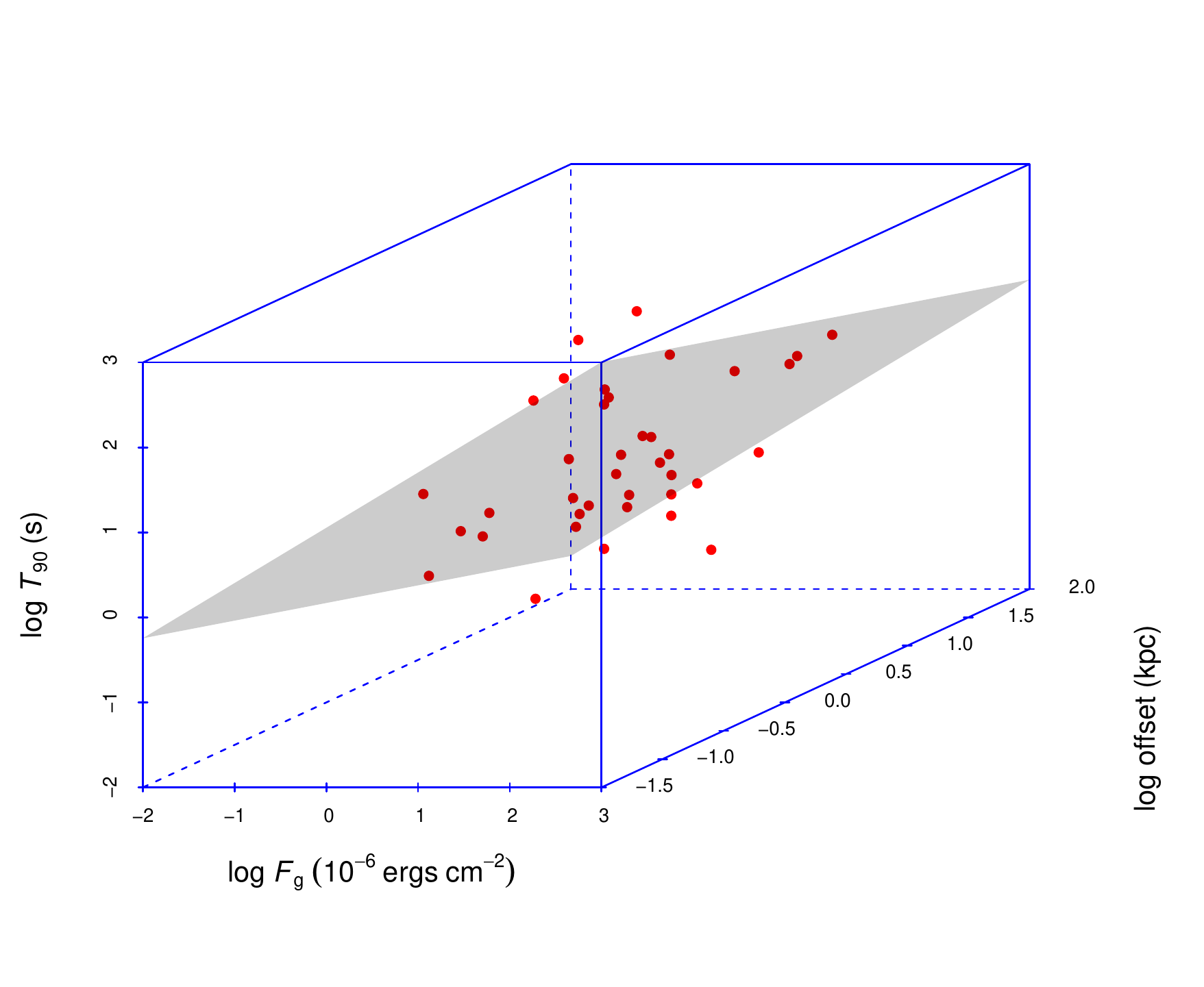}

\includegraphics[width=0.45\textwidth]{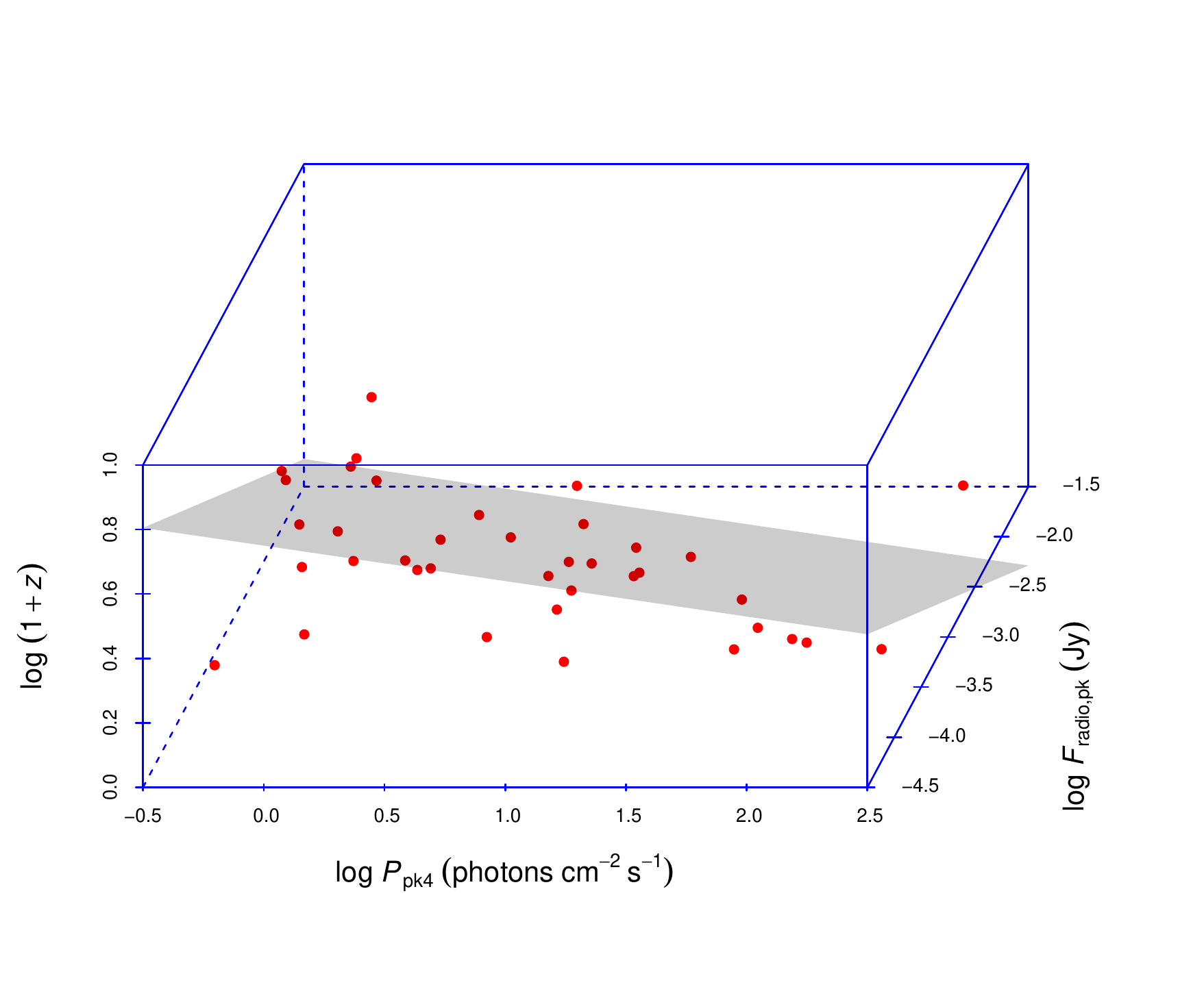}
\includegraphics[width=0.45\textwidth]{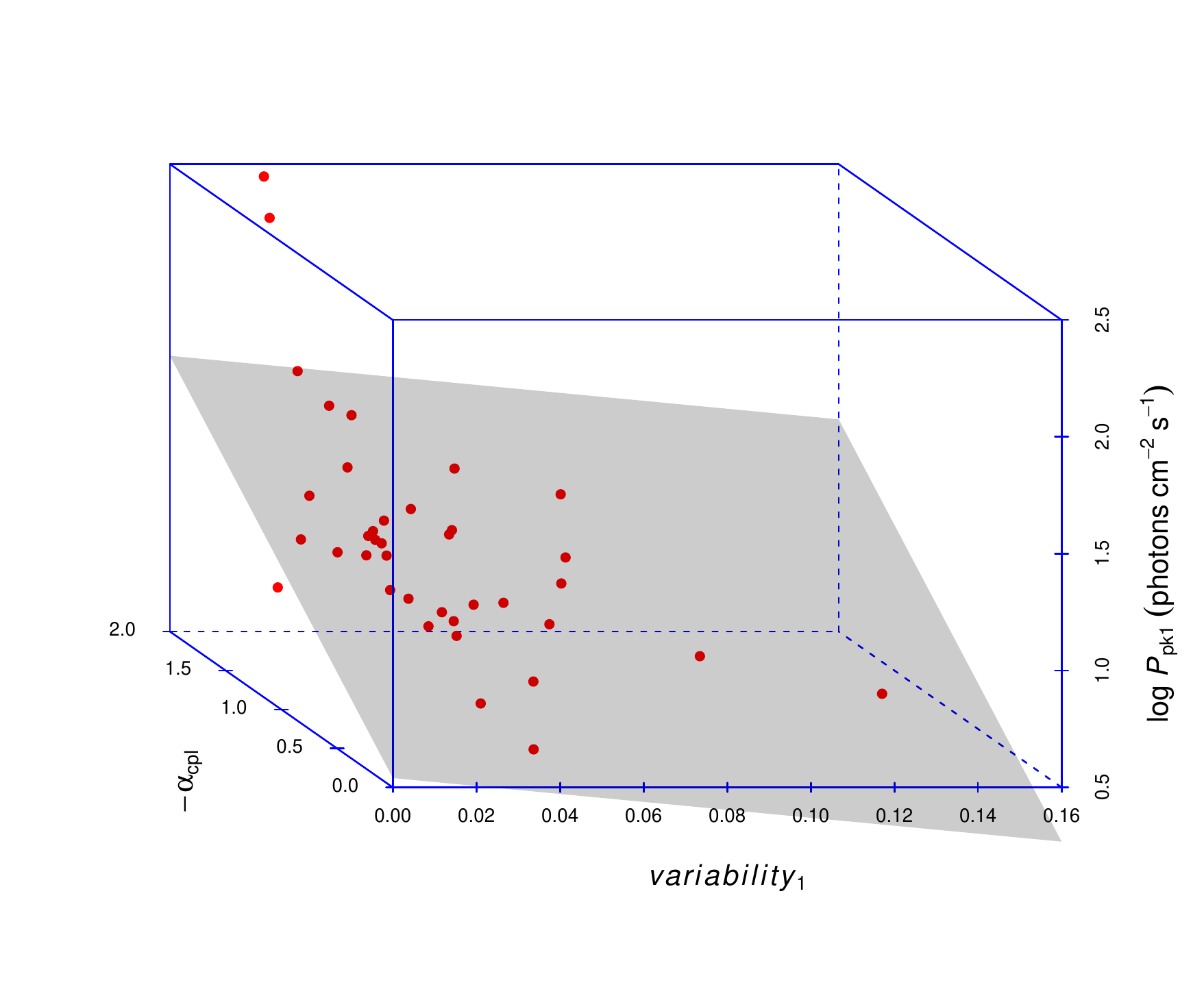}

\includegraphics[width=0.45\textwidth]{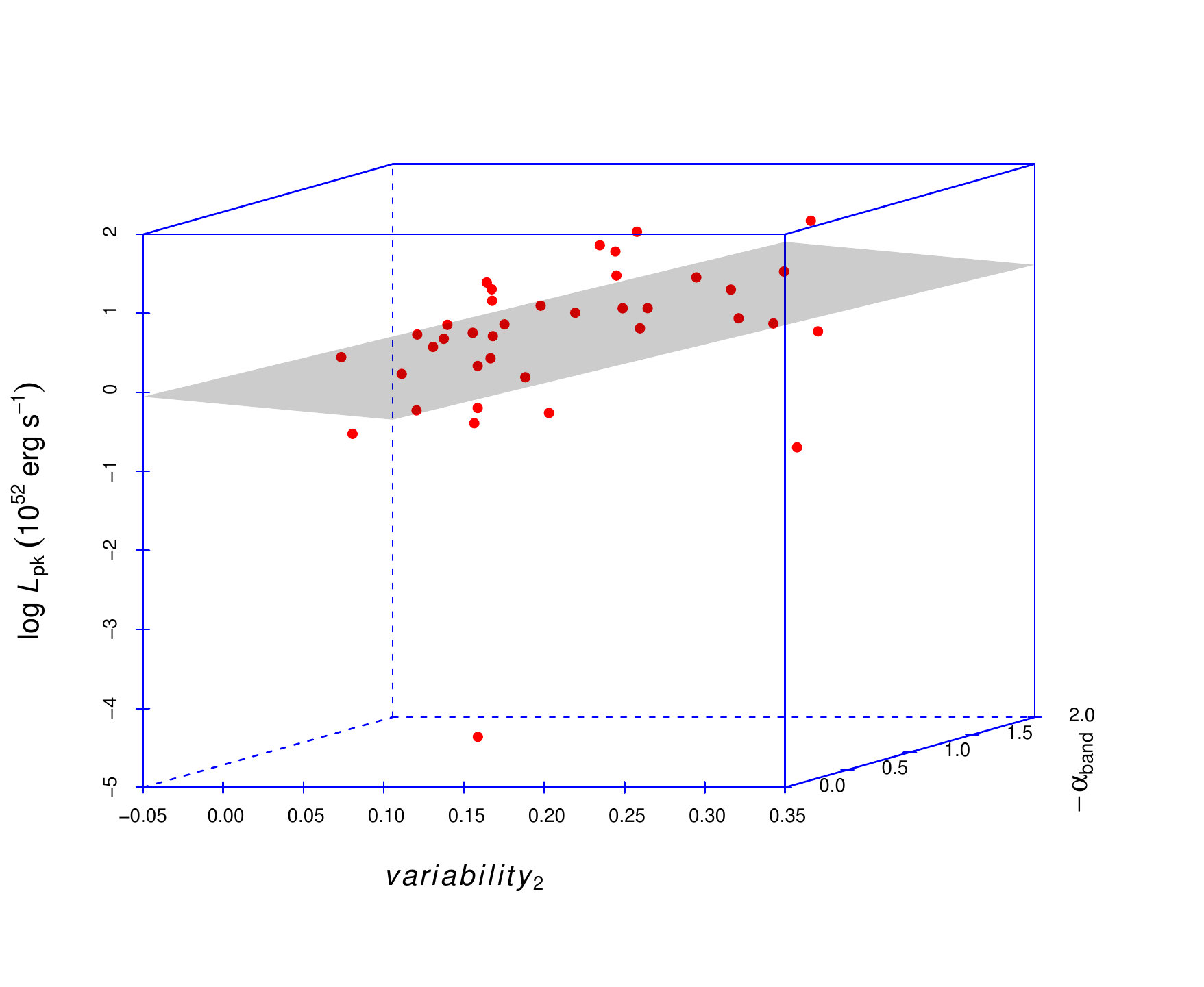}
\includegraphics[width=0.45\textwidth]{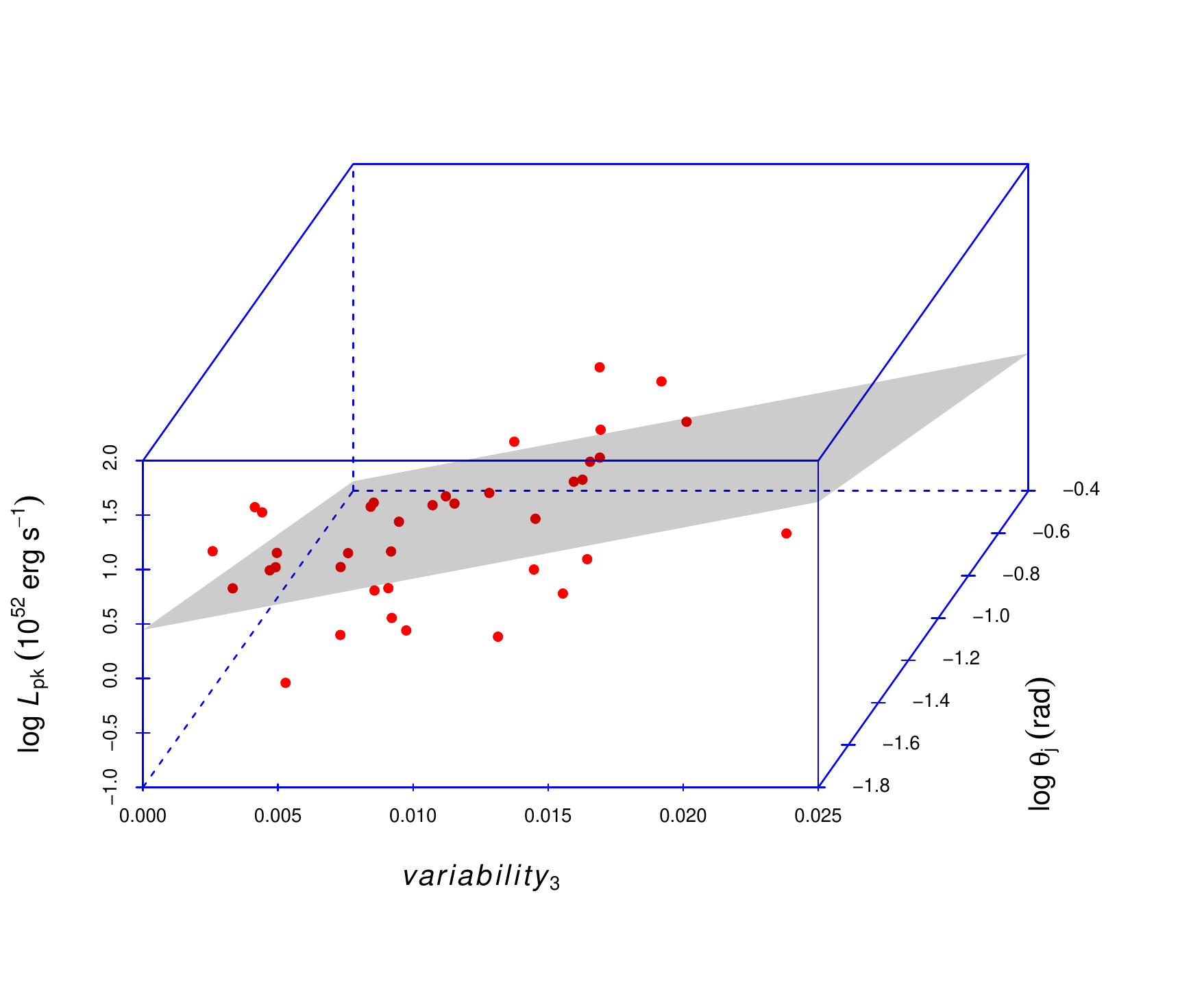}

\center{Fig. \ref{fig:three}---Continued}
\end{figure*}


\clearpage
\begin{figure*}

\includegraphics[width=0.45\textwidth]{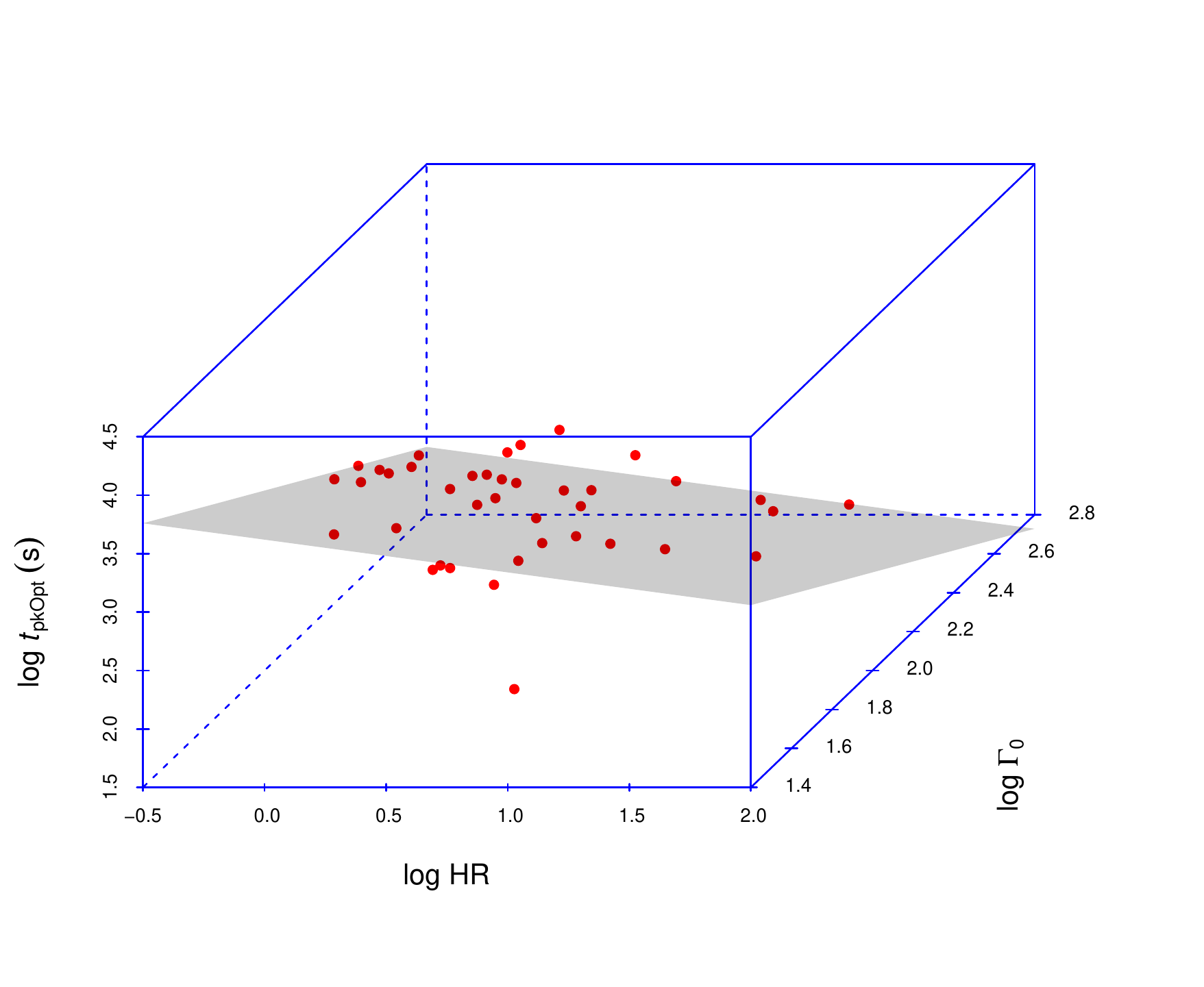}
\includegraphics[width=0.45\textwidth]{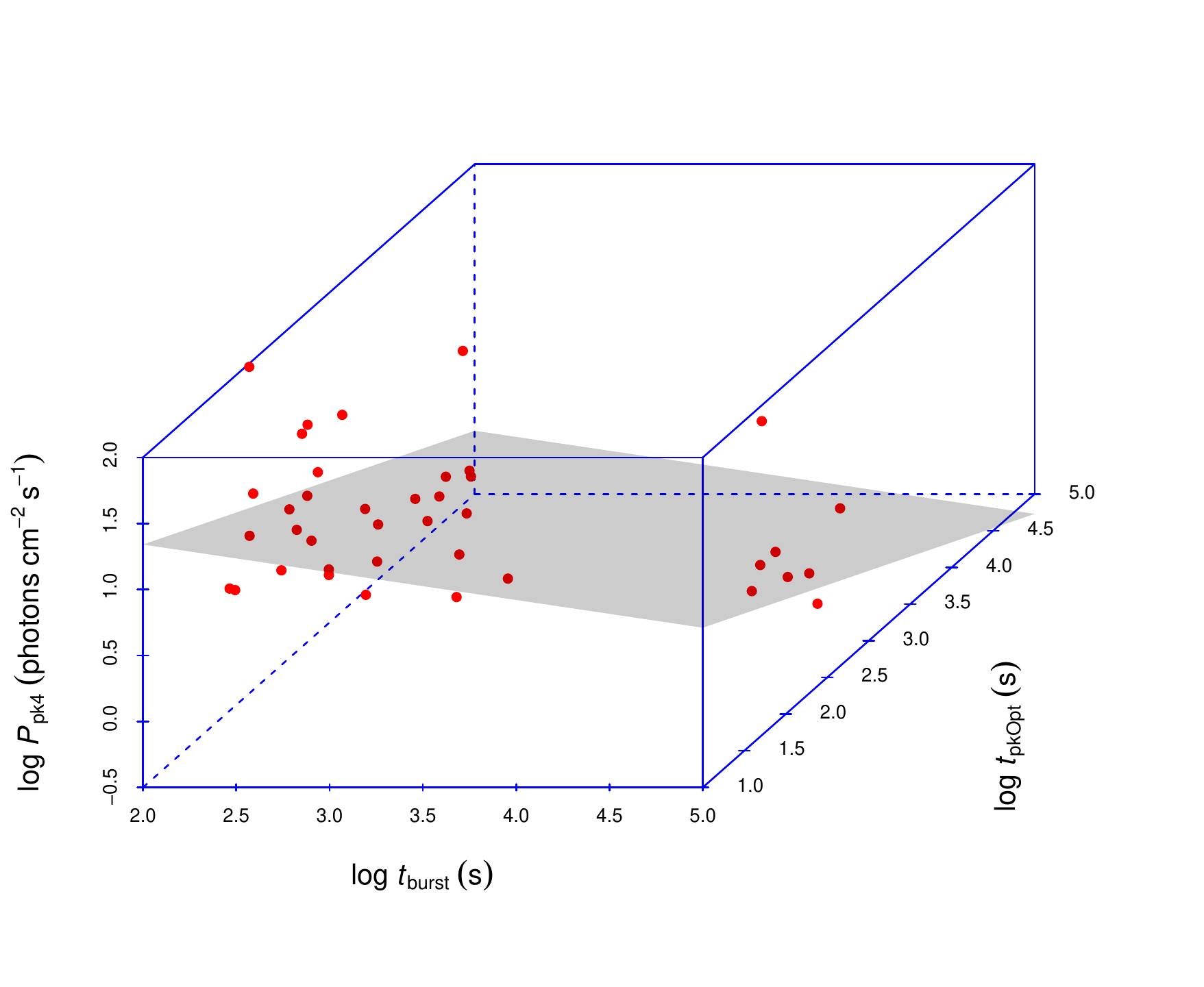}

\includegraphics[width=0.45\textwidth]{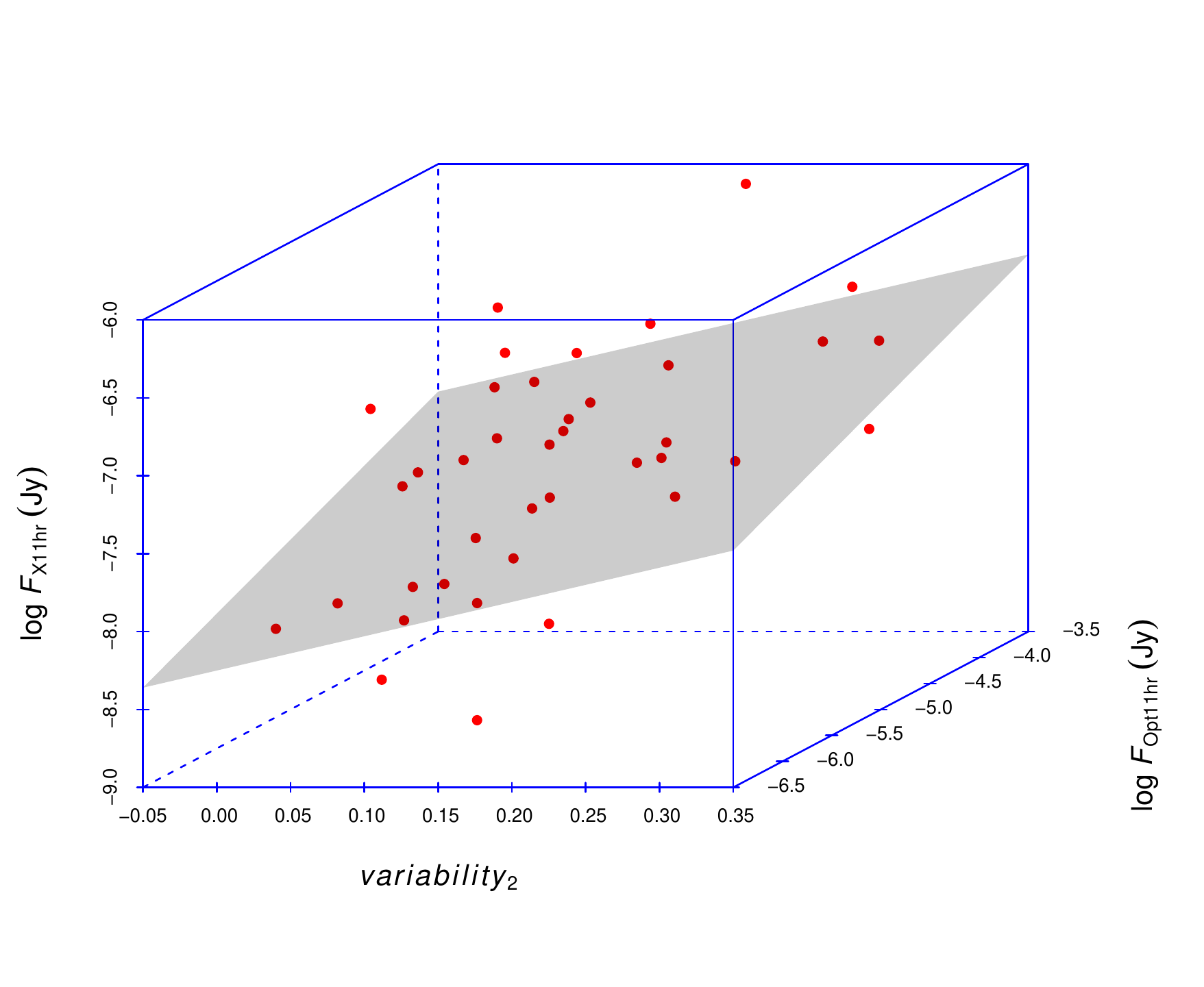}
\includegraphics[width=0.45\textwidth]{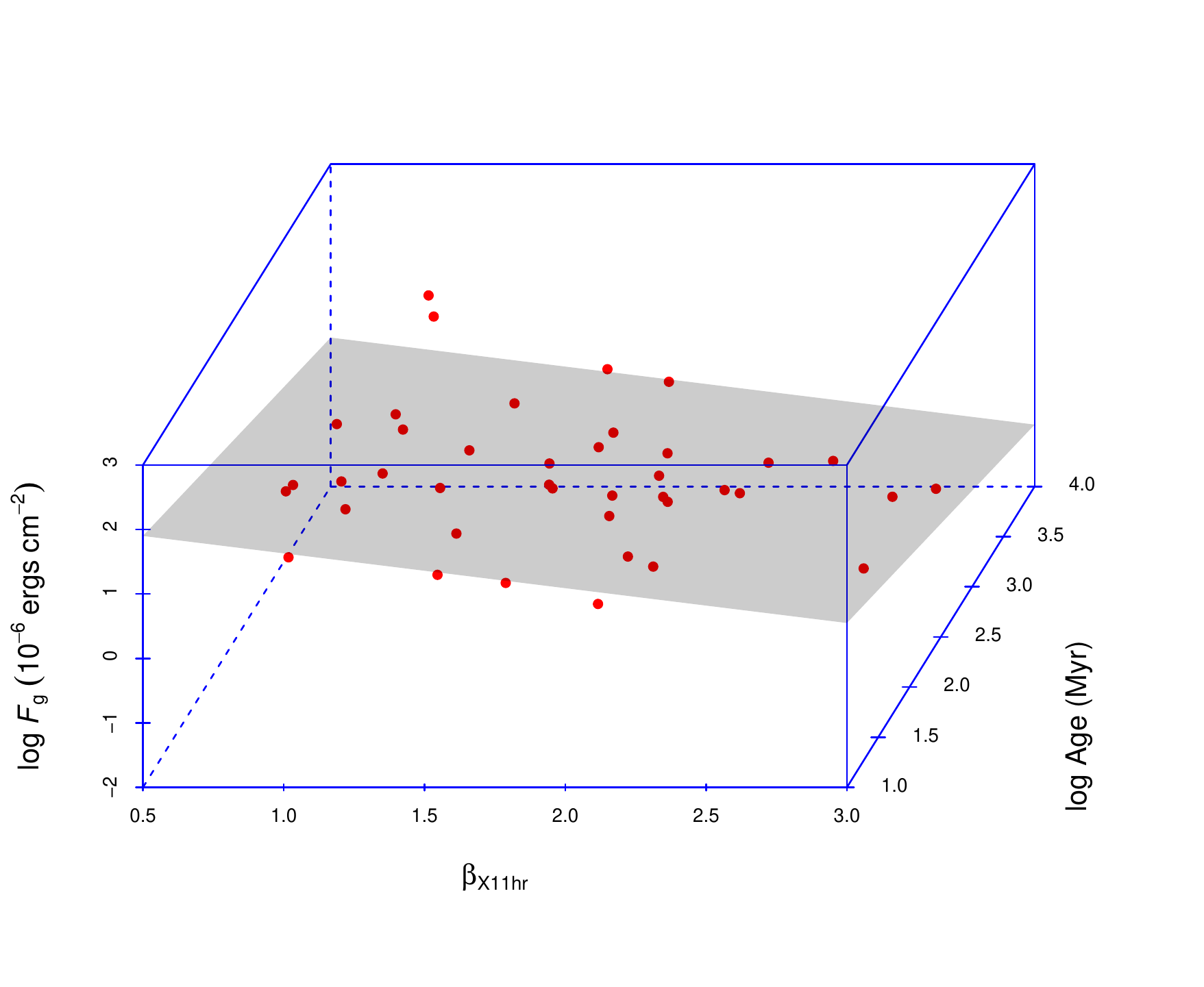}

\includegraphics[width=0.45\textwidth]{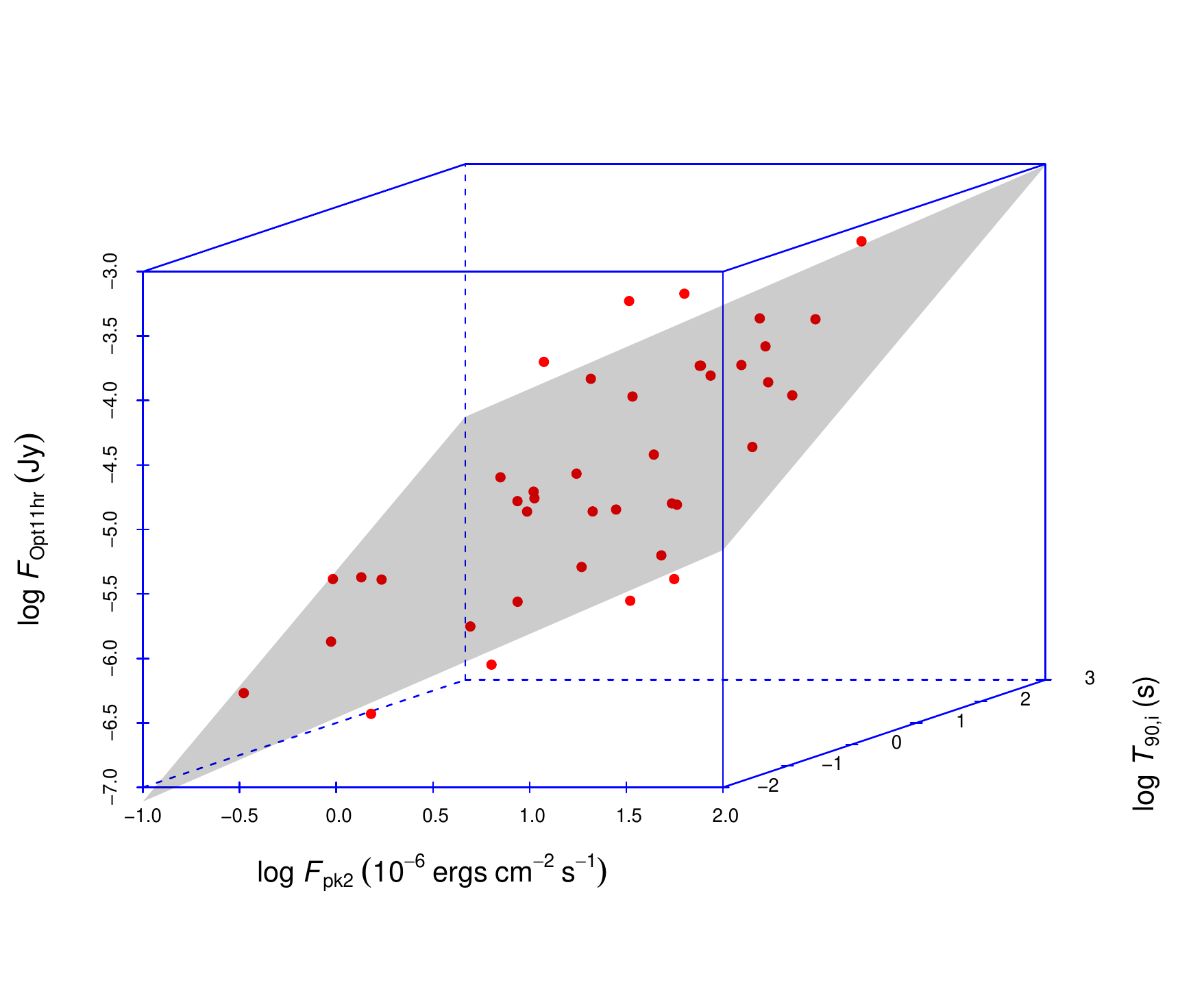}
\includegraphics[width=0.45\textwidth]{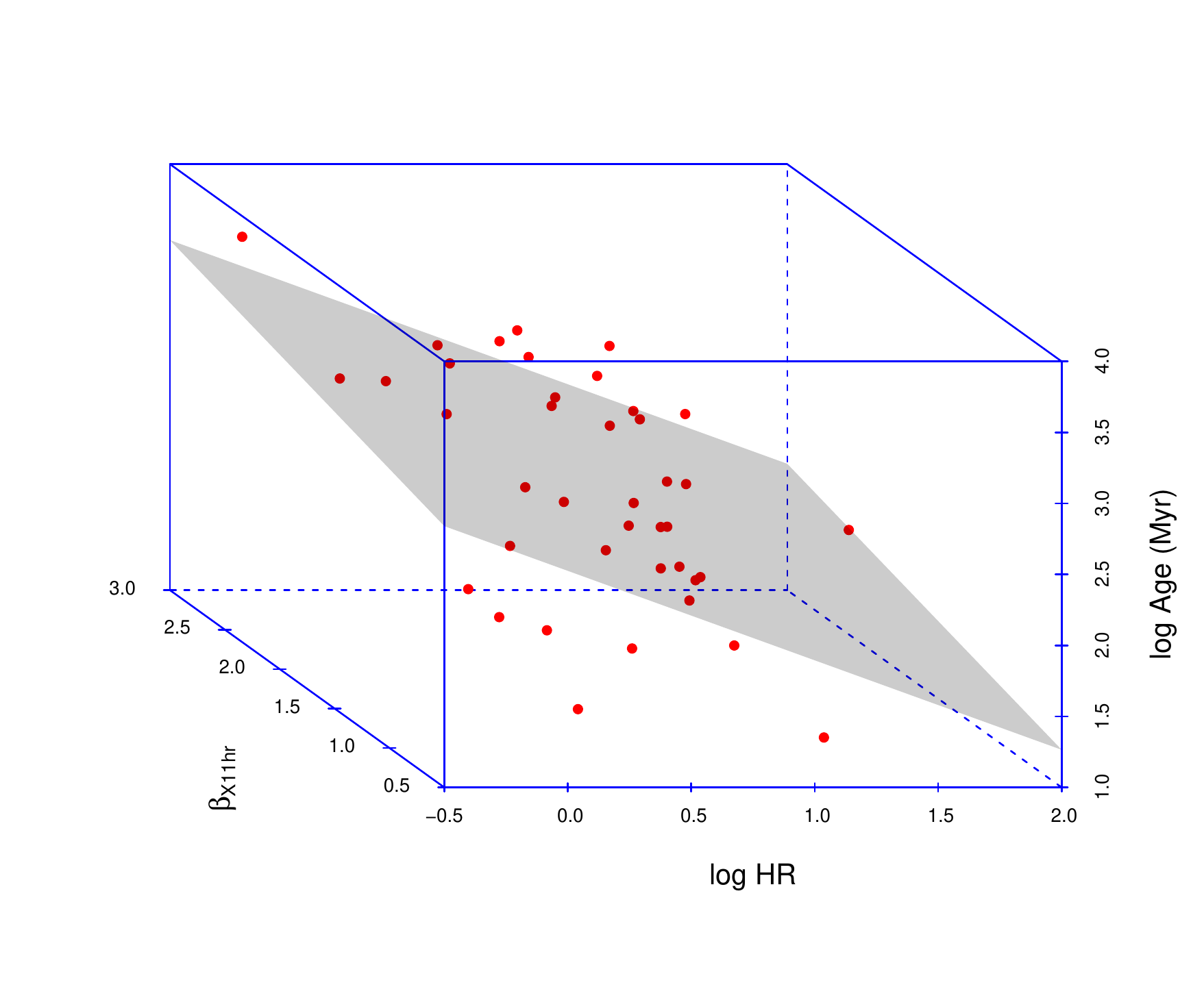}

\center{Fig. \ref{fig:three}---Continued}
\end{figure*}


\clearpage
\begin{figure*}

\includegraphics[width=0.45\textwidth]{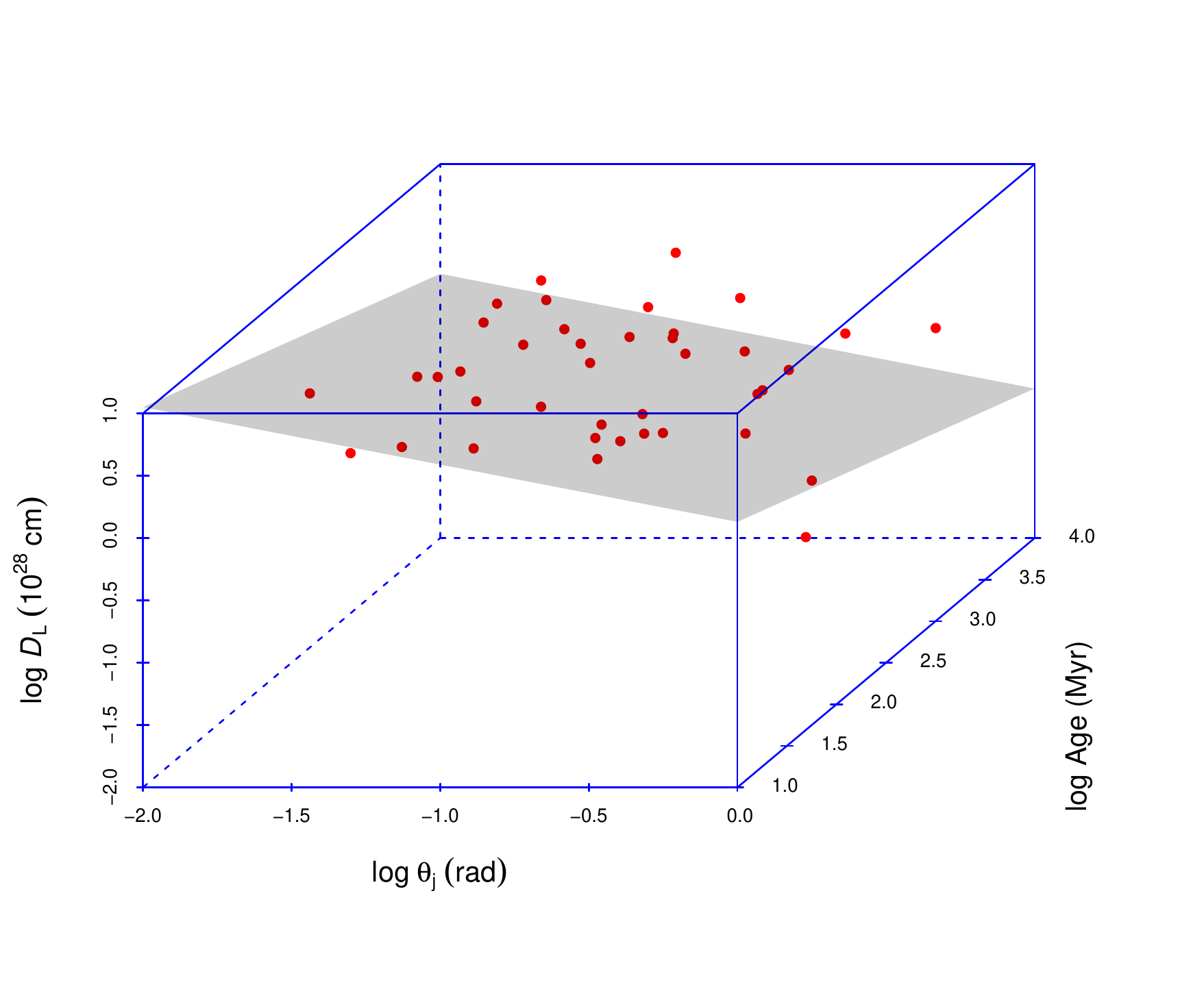}
\includegraphics[width=0.45\textwidth]{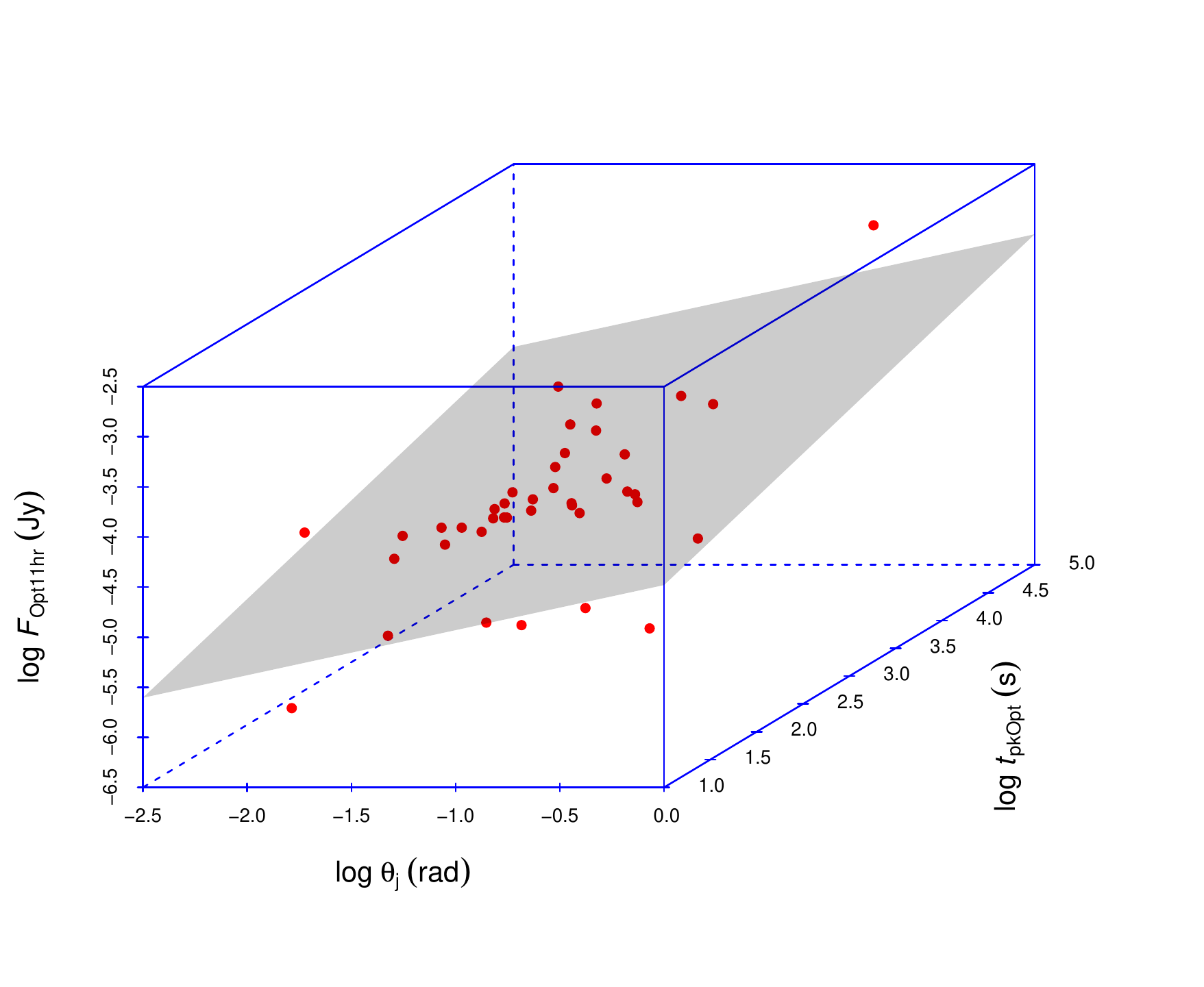}

\includegraphics[width=0.45\textwidth]{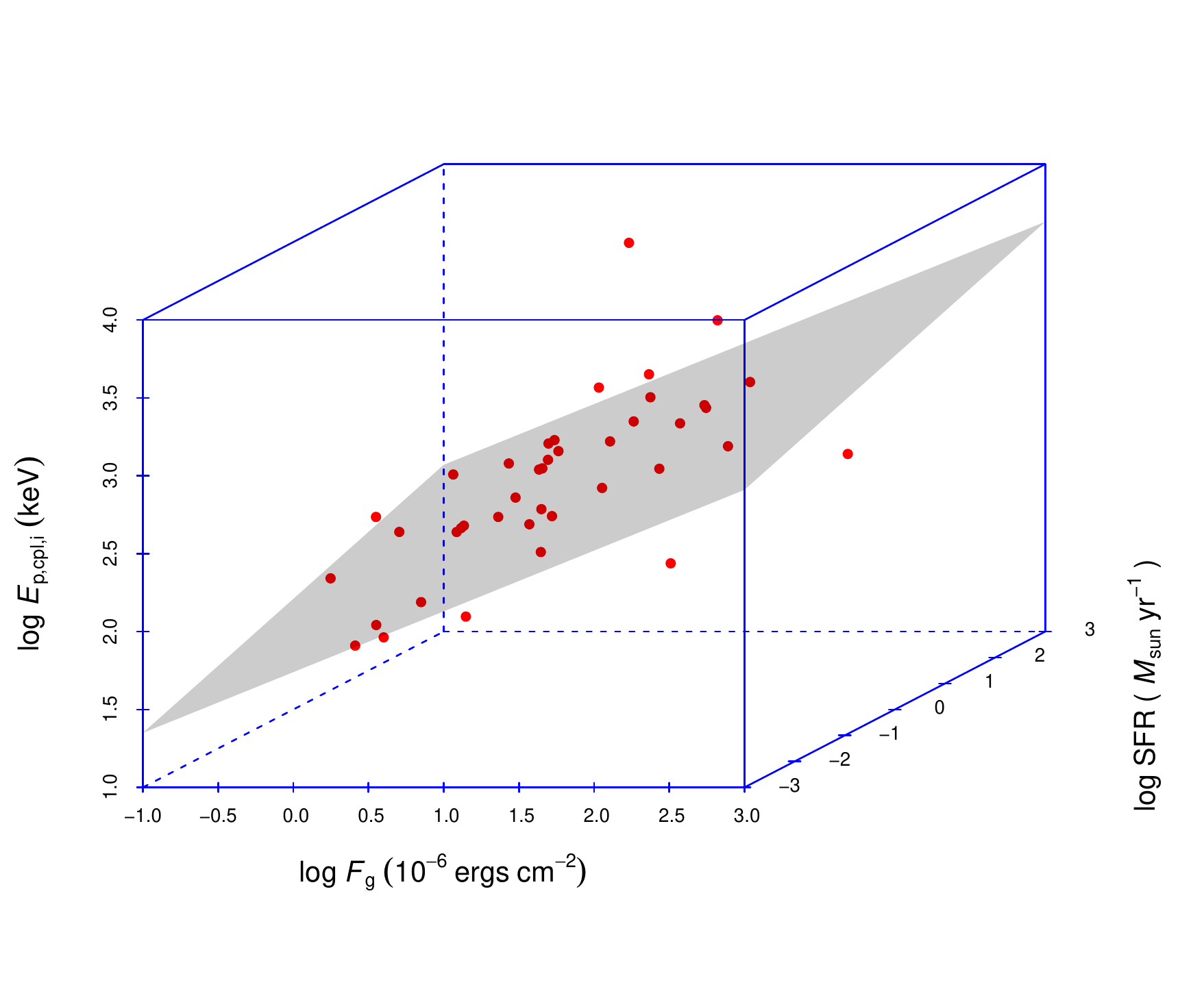}
\includegraphics[width=0.45\textwidth]{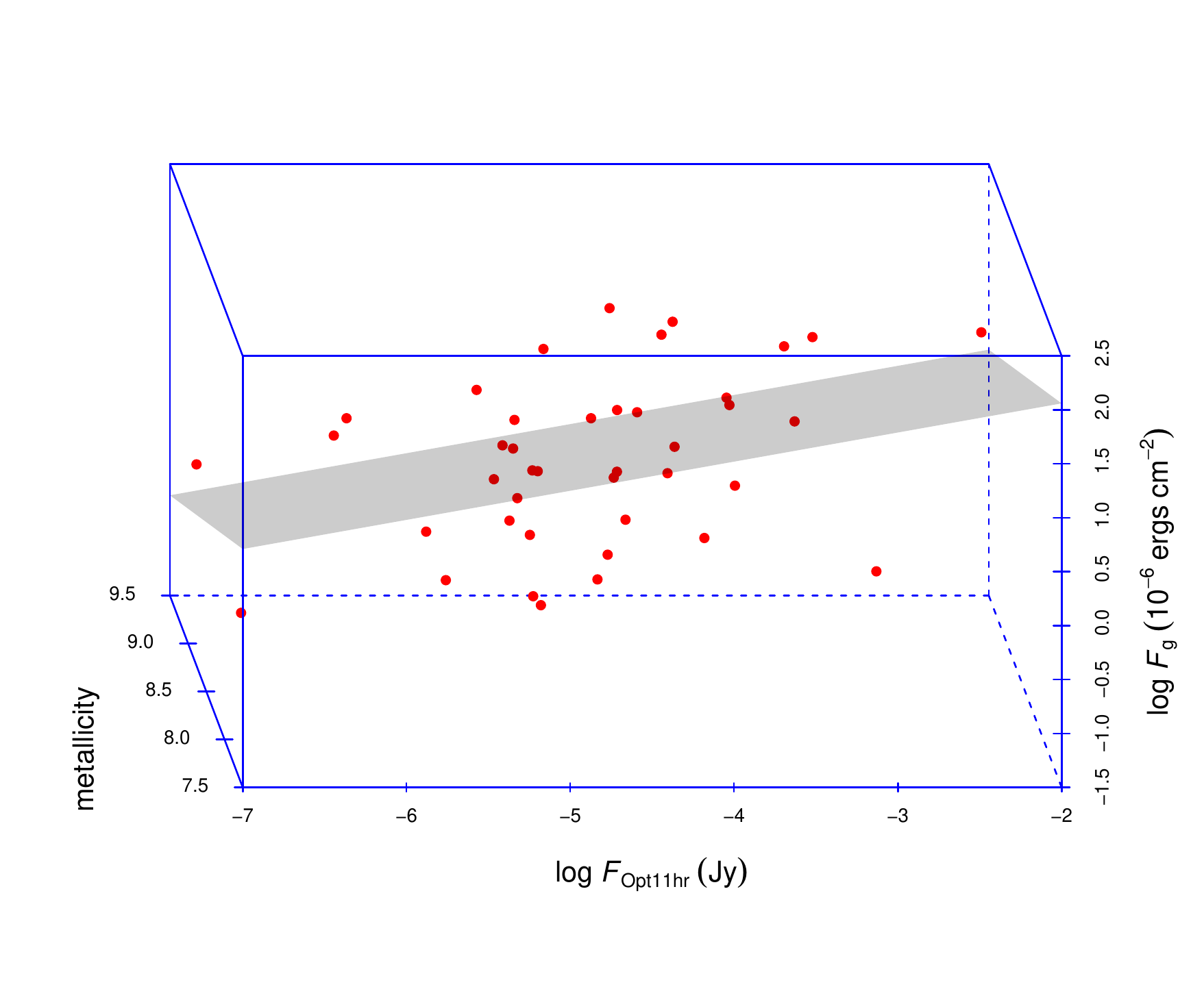}

\includegraphics[width=0.45\textwidth]{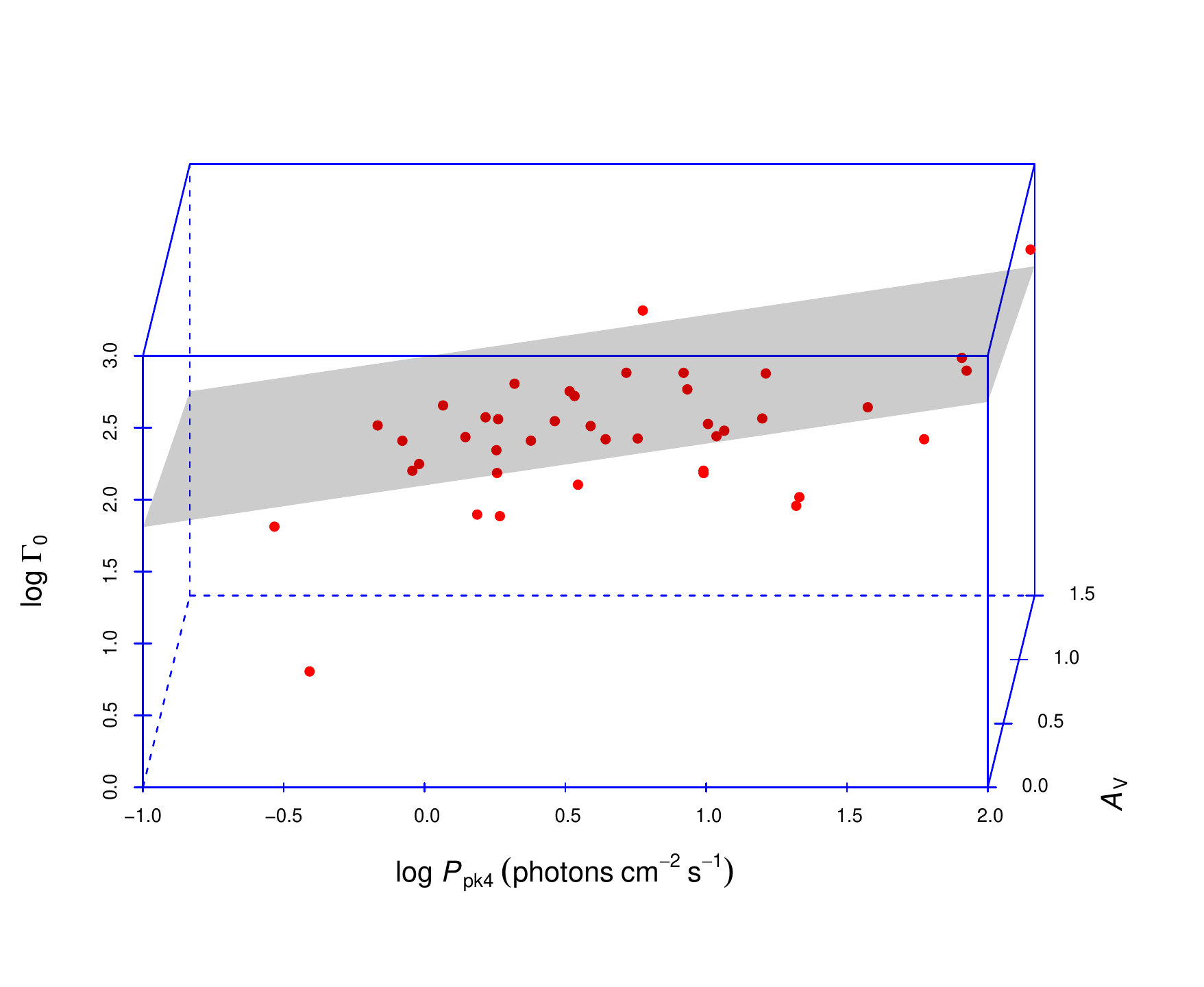}
\includegraphics[width=0.45\textwidth]{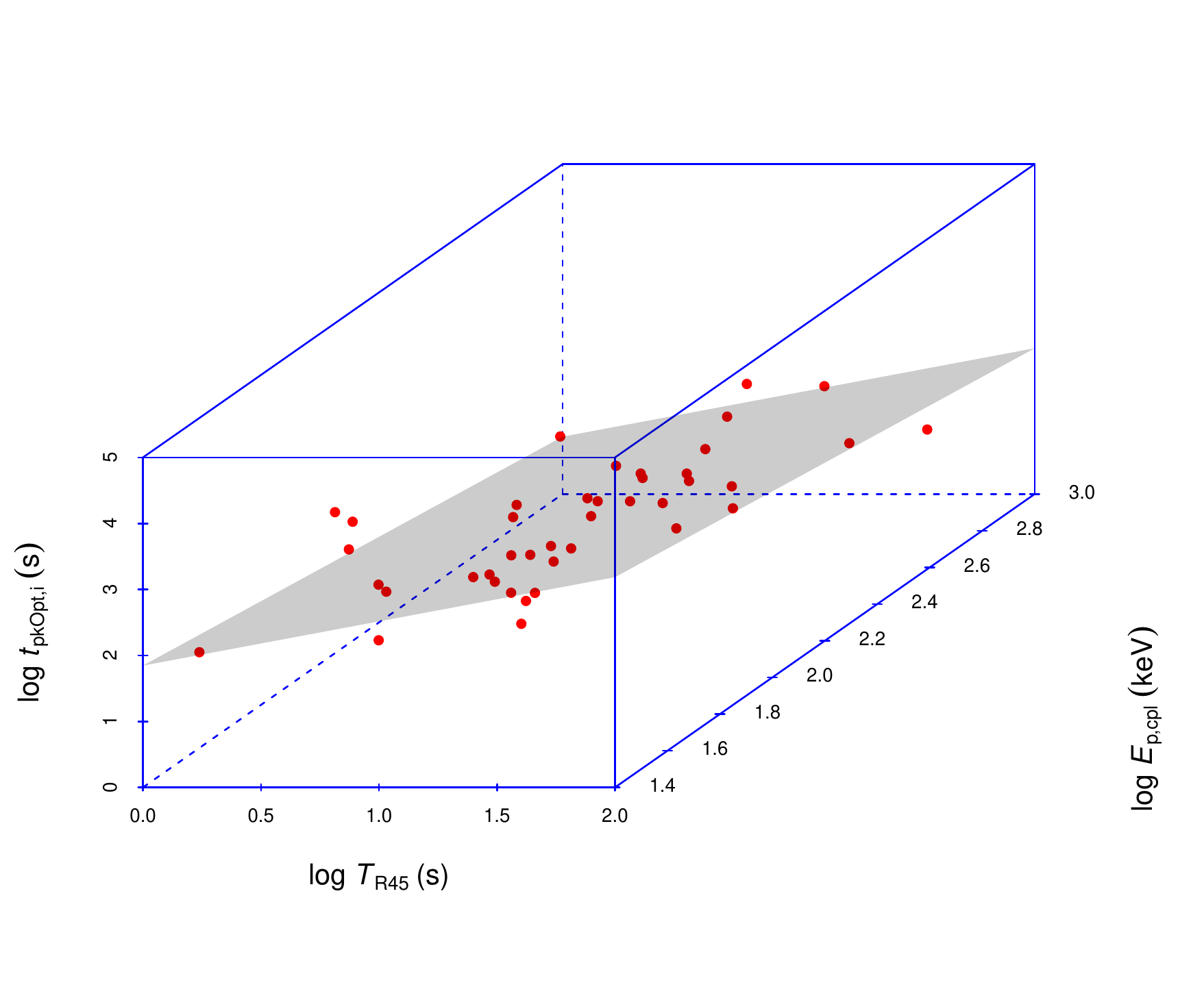}

\center{Fig. \ref{fig:three}---Continued}
\end{figure*}


\clearpage
\begin{figure*}

\includegraphics[width=0.45\textwidth]{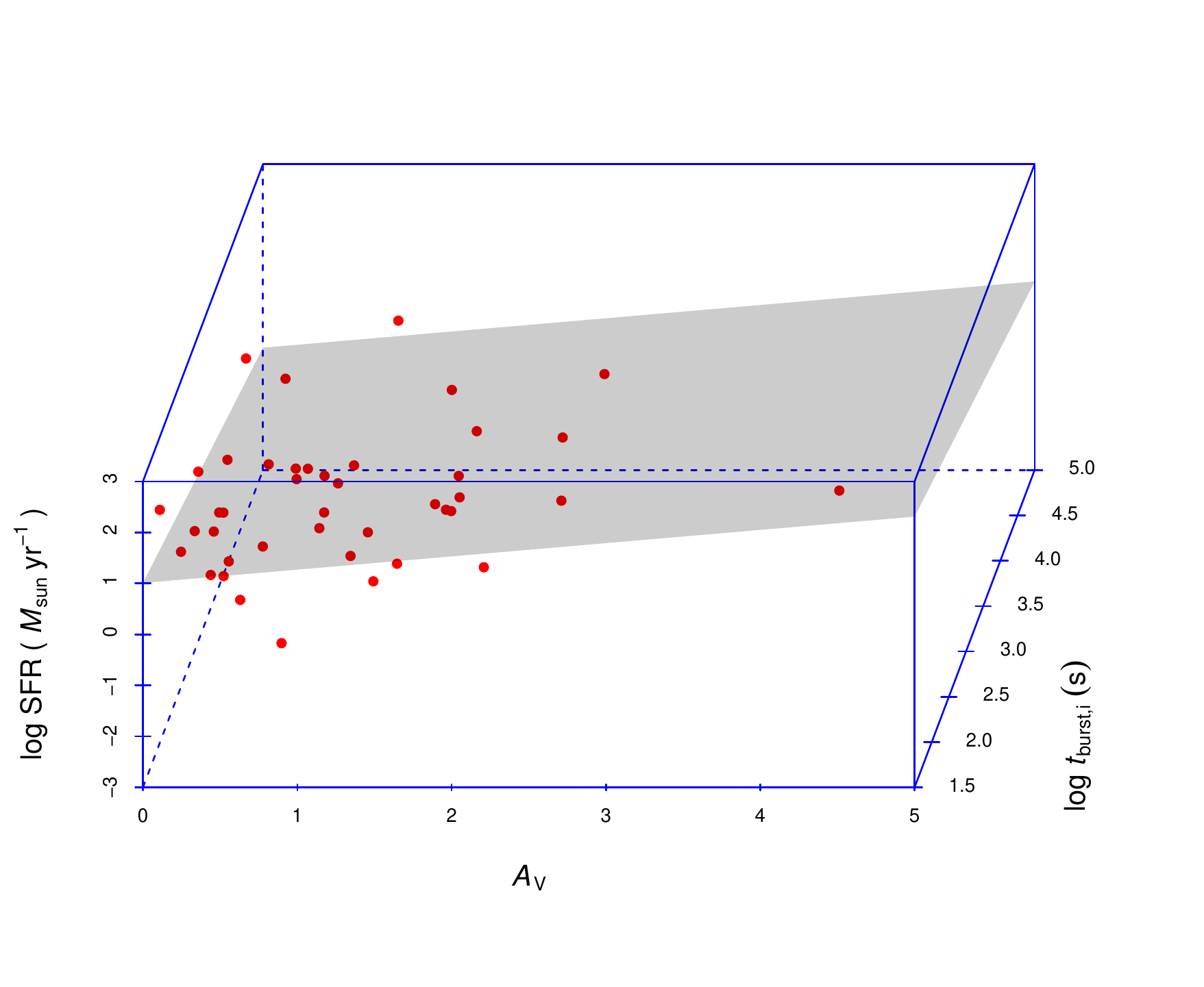}
\includegraphics[width=0.45\textwidth]{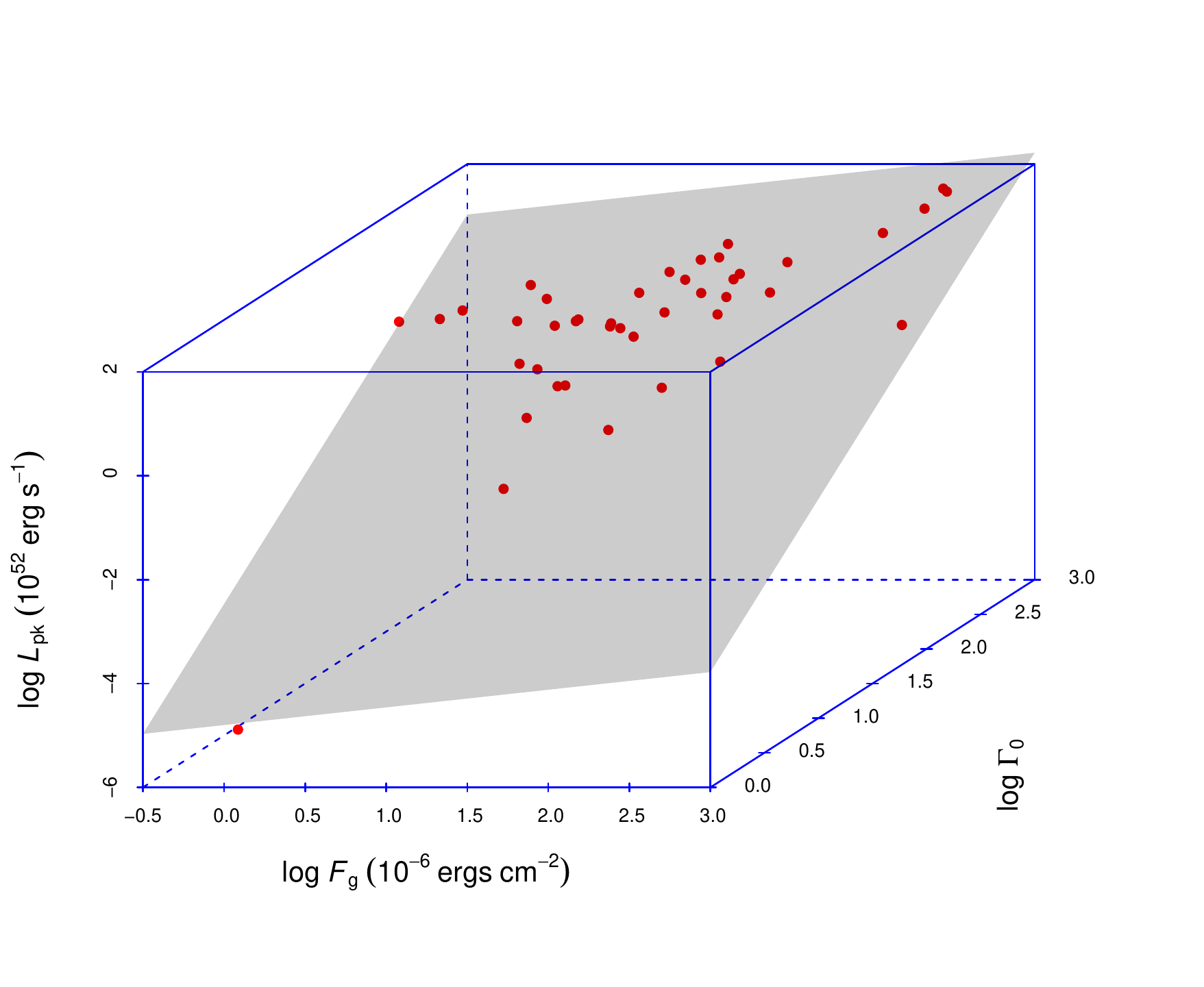}

\includegraphics[width=0.45\textwidth]{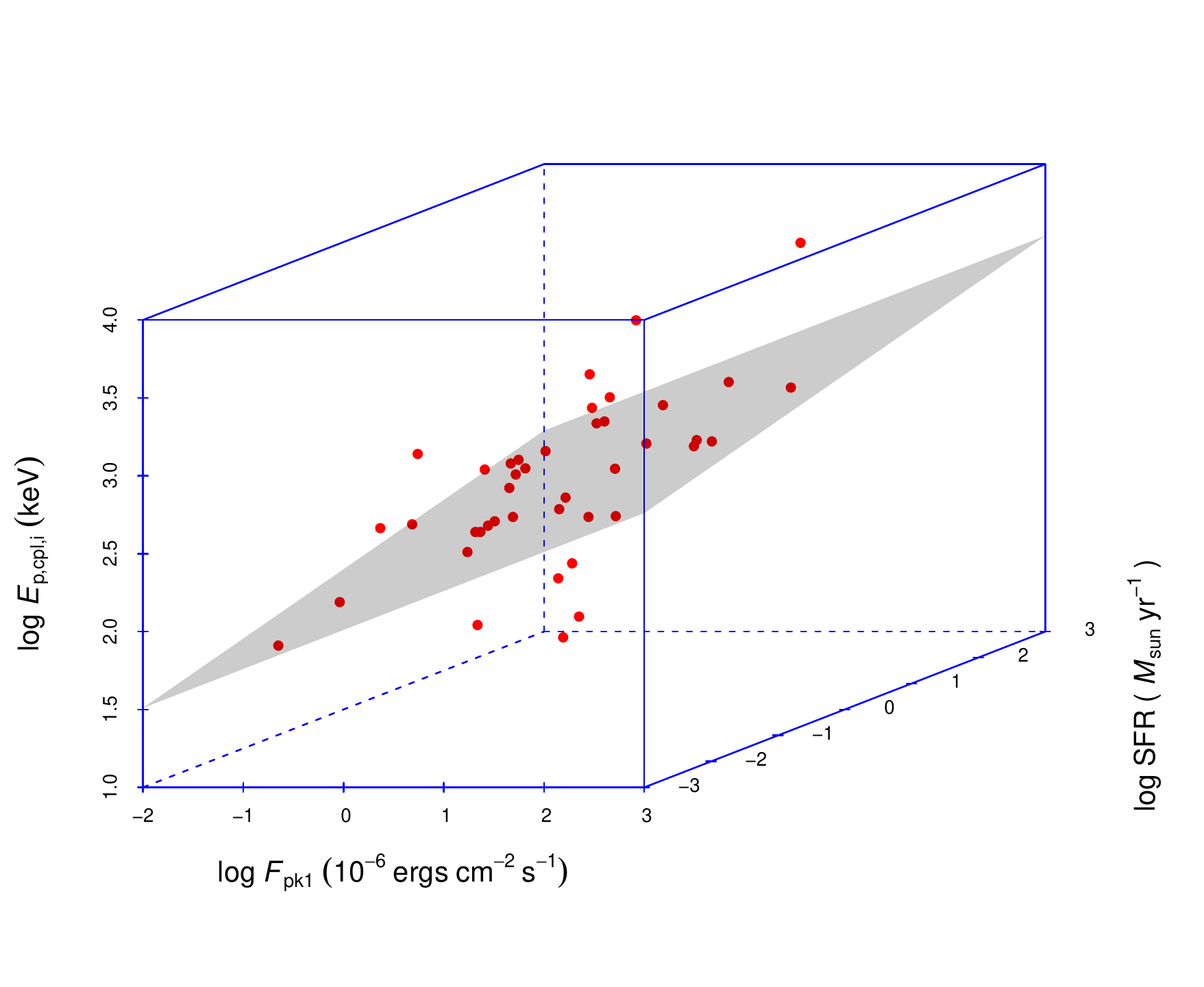}
\includegraphics[width=0.45\textwidth]{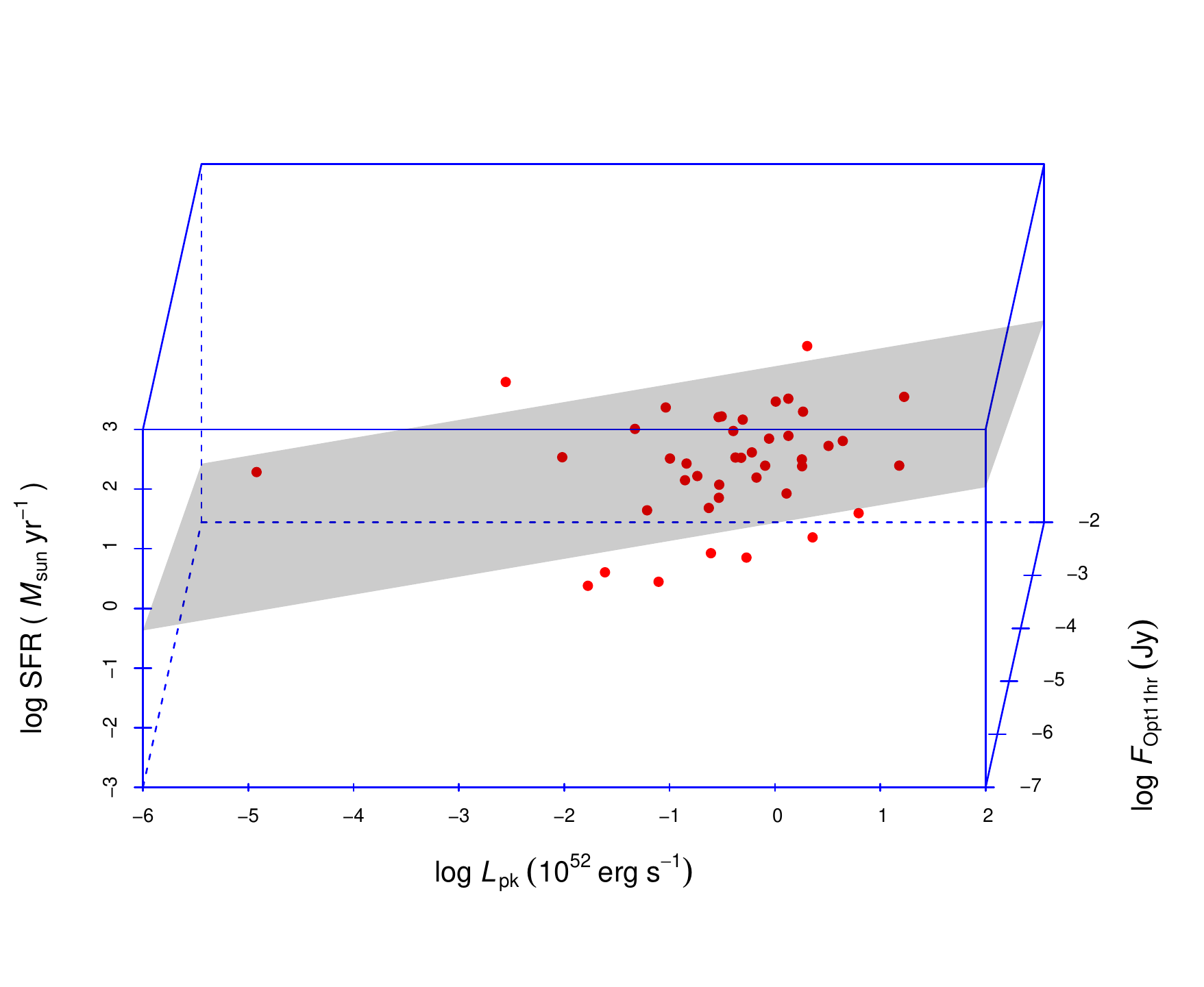}

\includegraphics[width=0.45\textwidth]{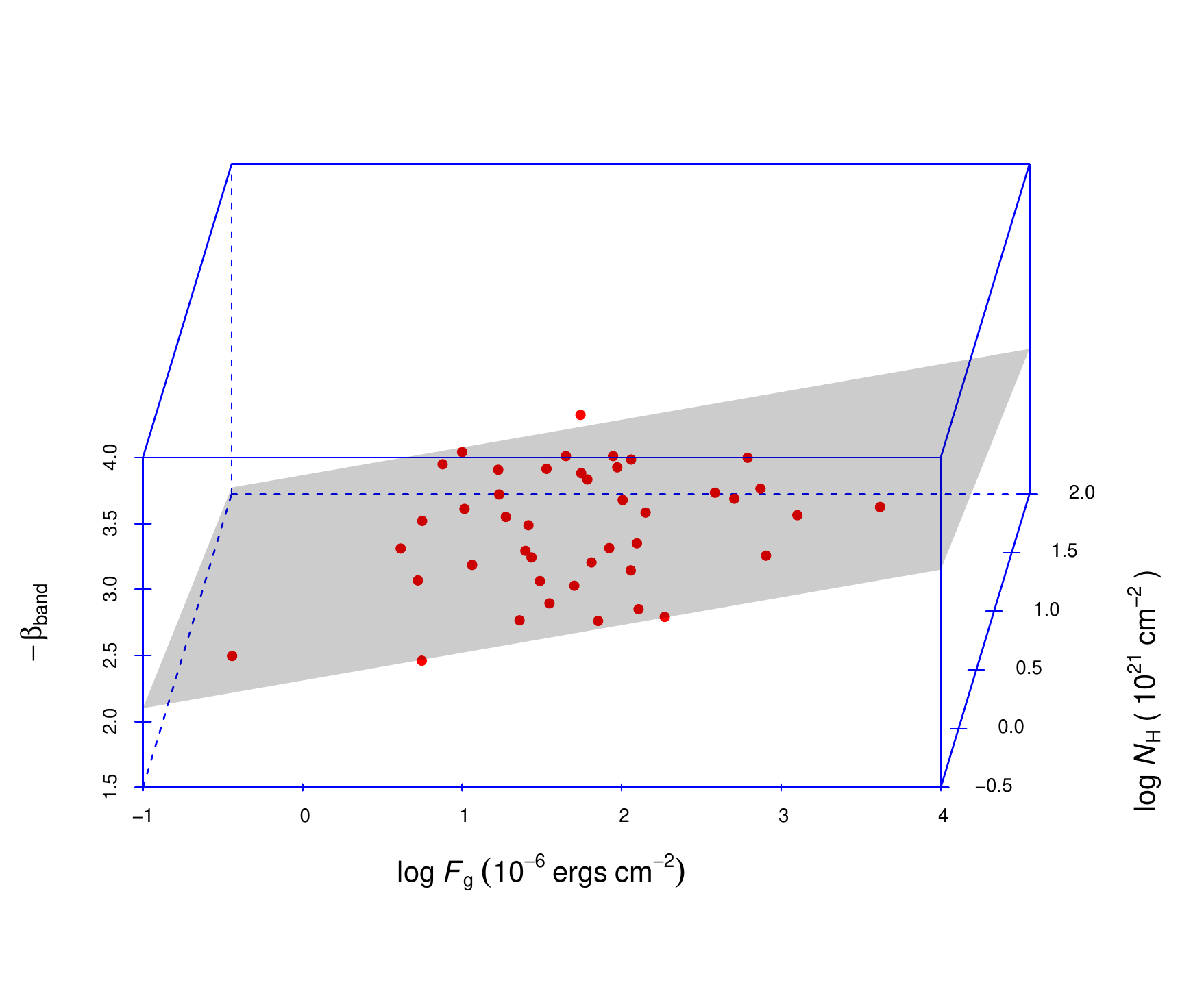}
\includegraphics[width=0.45\textwidth]{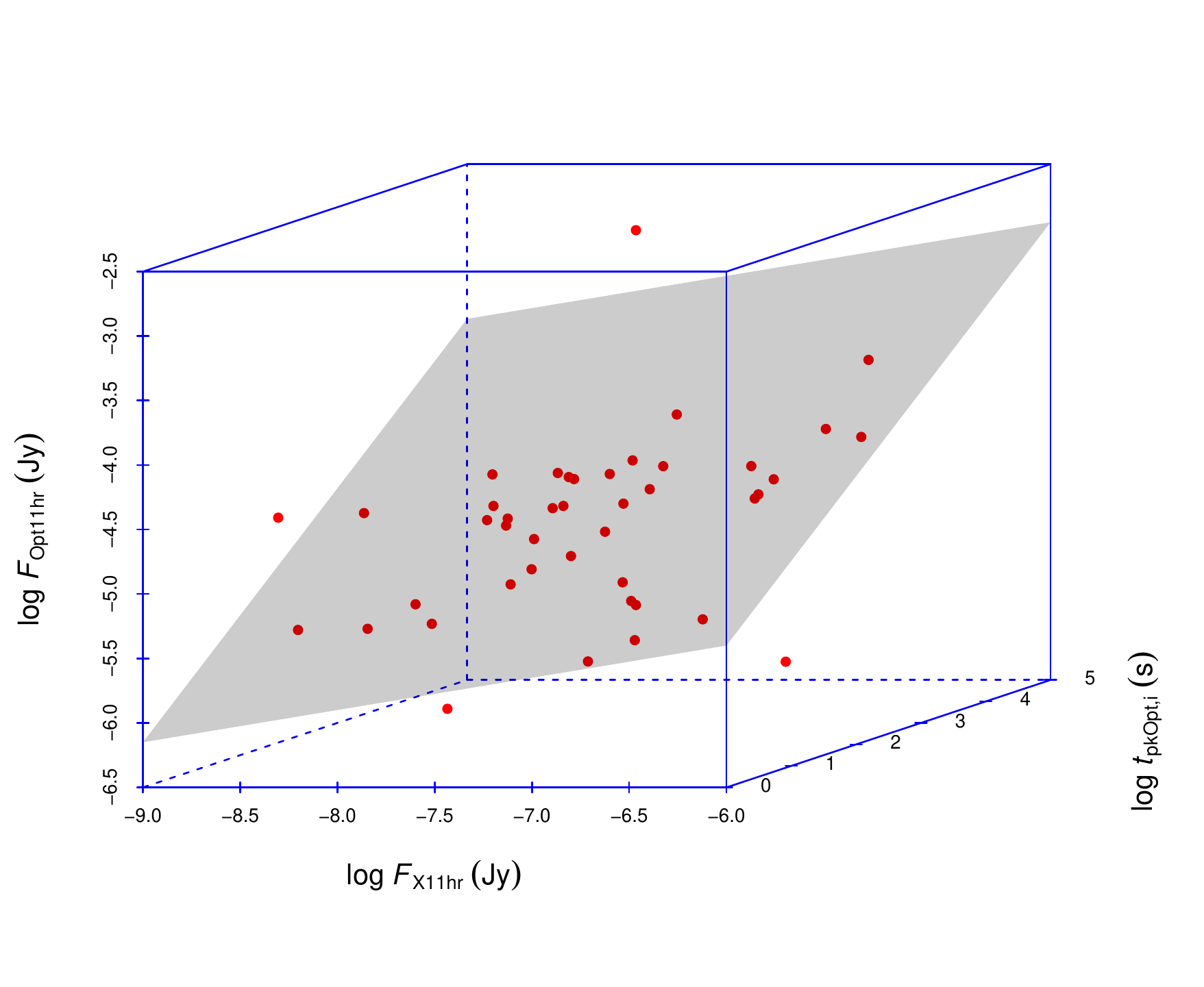}

\center{Fig. \ref{fig:three}---Continued}
\end{figure*}


\clearpage
\begin{figure*}

\includegraphics[width=0.45\textwidth]{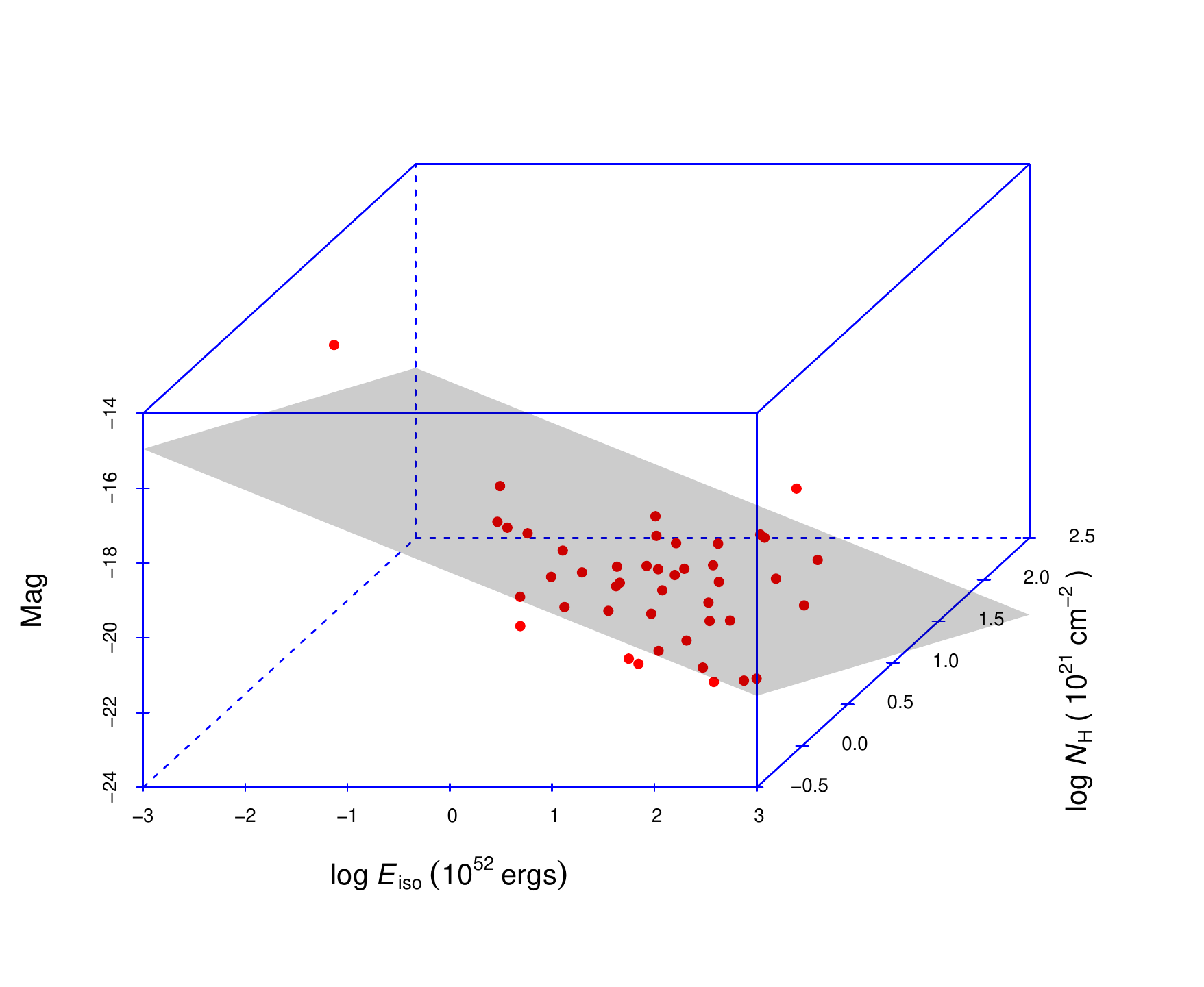}
\includegraphics[width=0.45\textwidth]{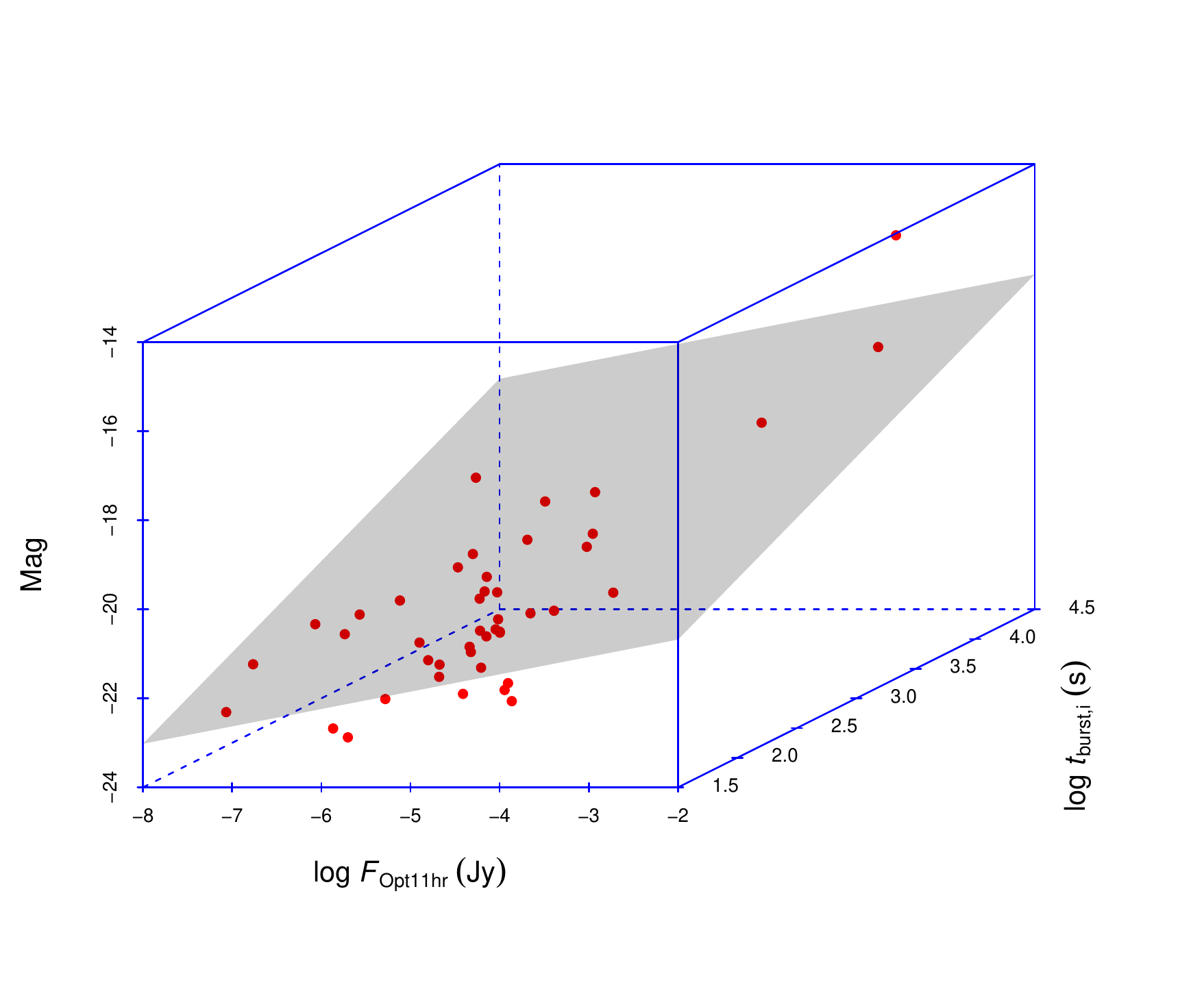}

\includegraphics[width=0.45\textwidth]{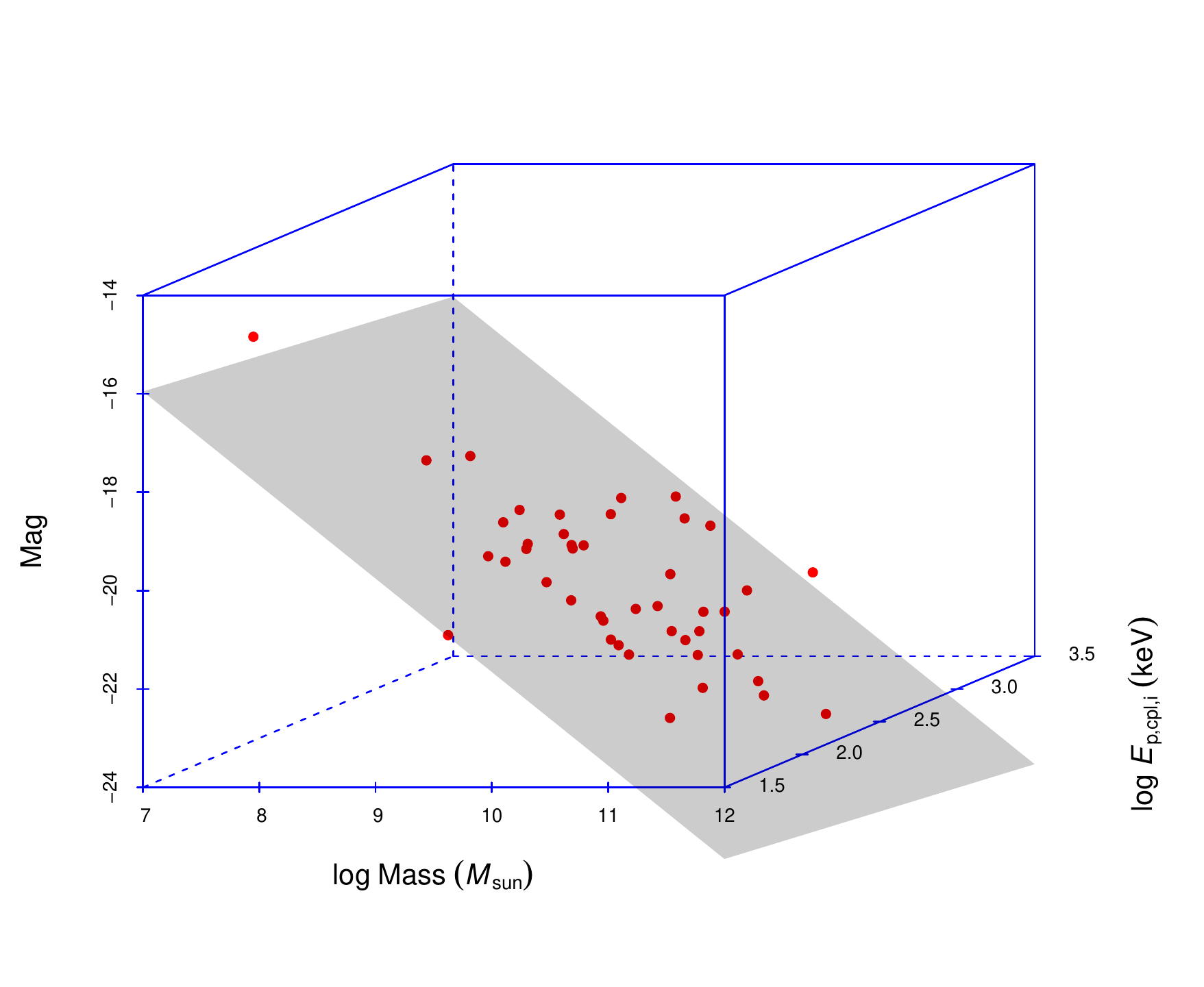}
\includegraphics[width=0.45\textwidth]{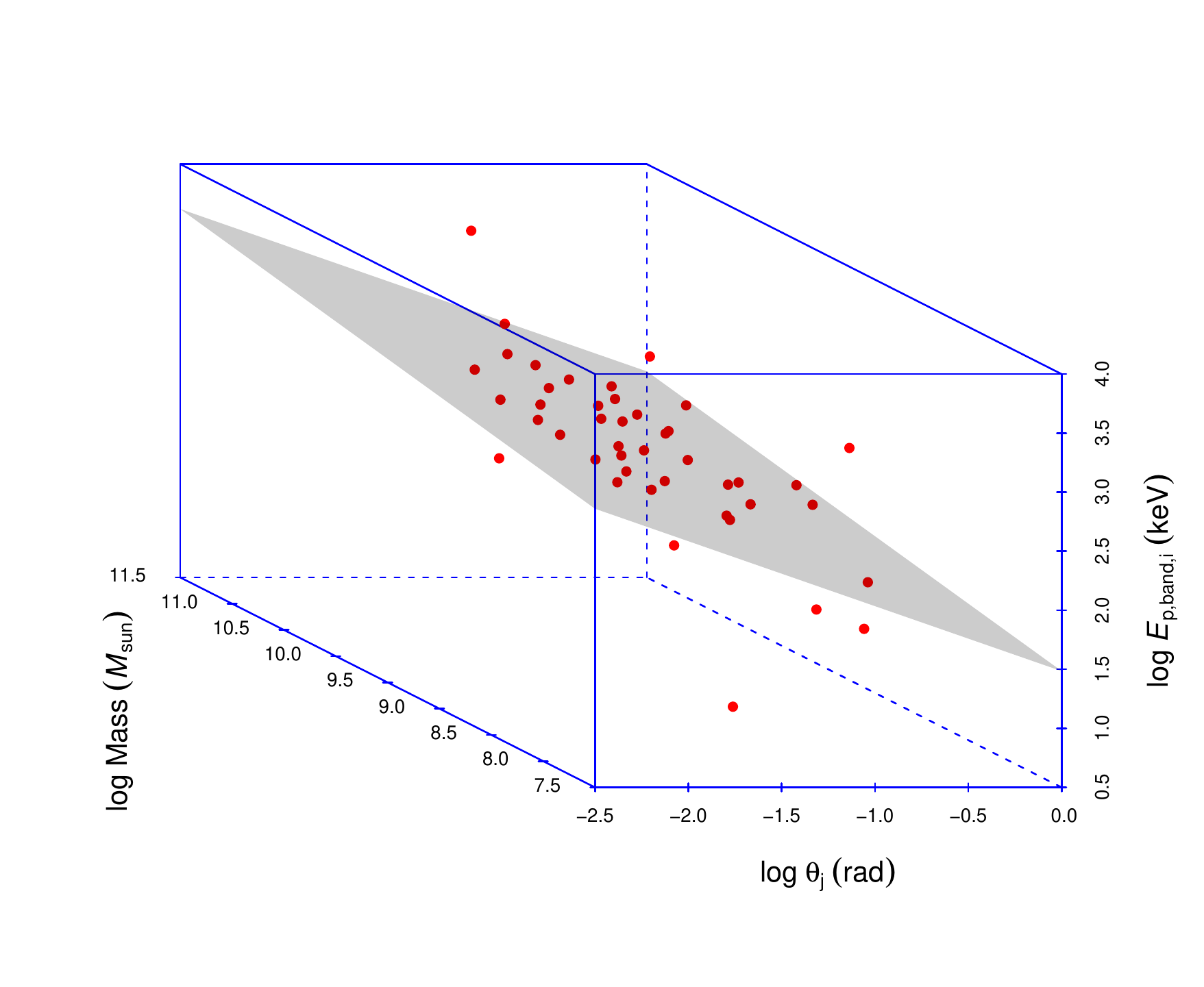}

\includegraphics[width=0.45\textwidth]{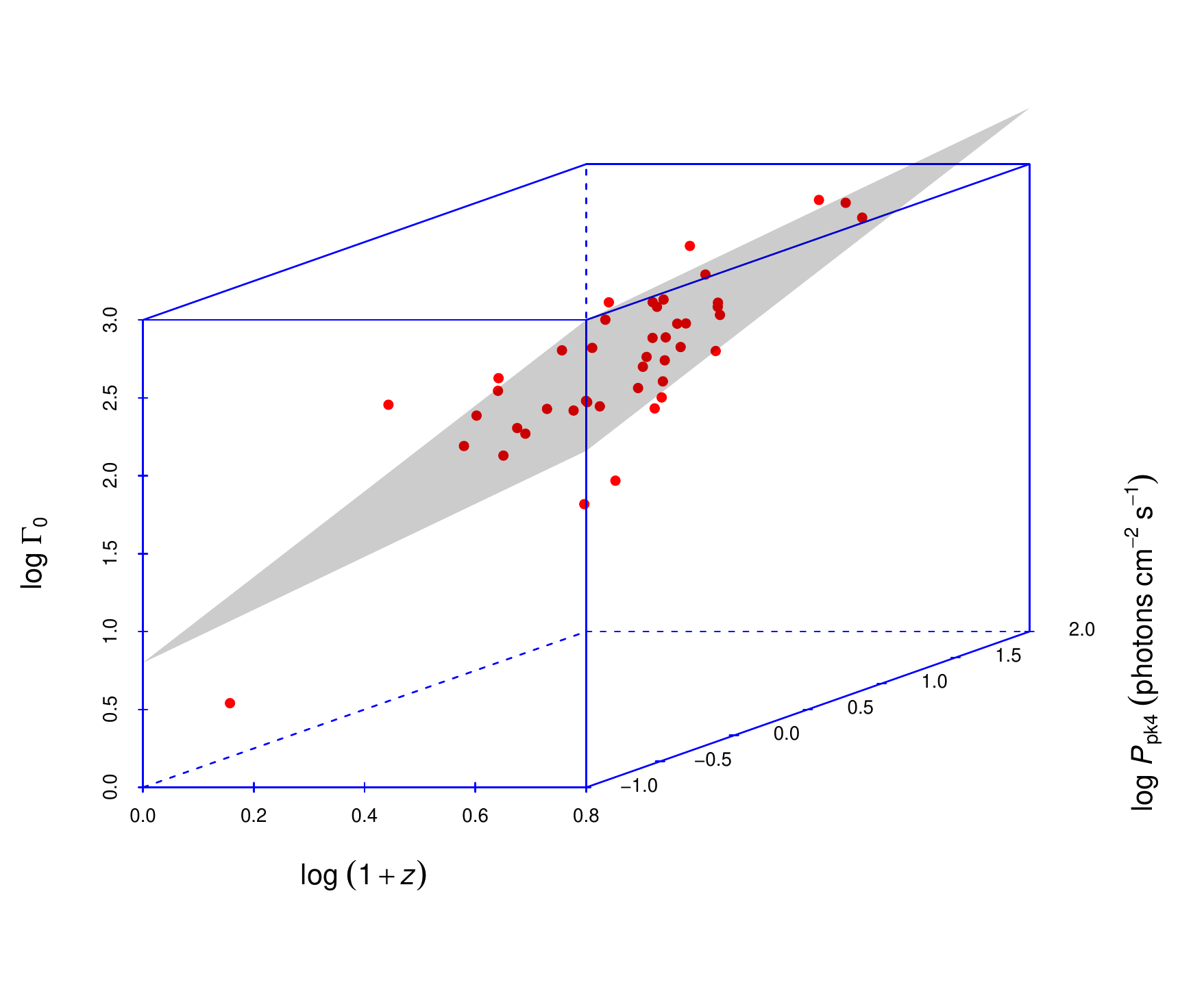}
\includegraphics[width=0.45\textwidth]{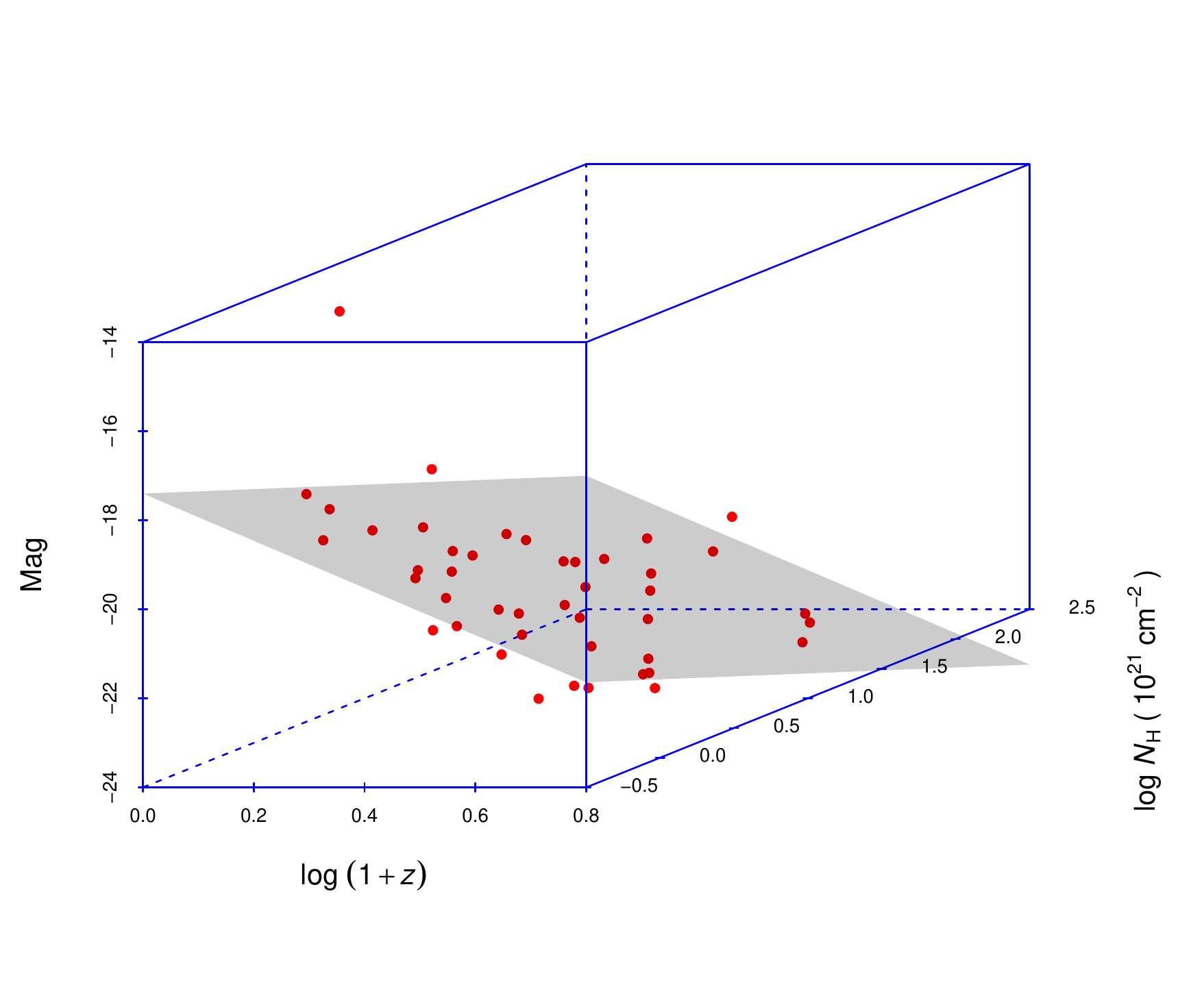}

\center{Fig. \ref{fig:three}---Continued}
\end{figure*}


\clearpage
\begin{figure*}

\includegraphics[width=0.45\textwidth]{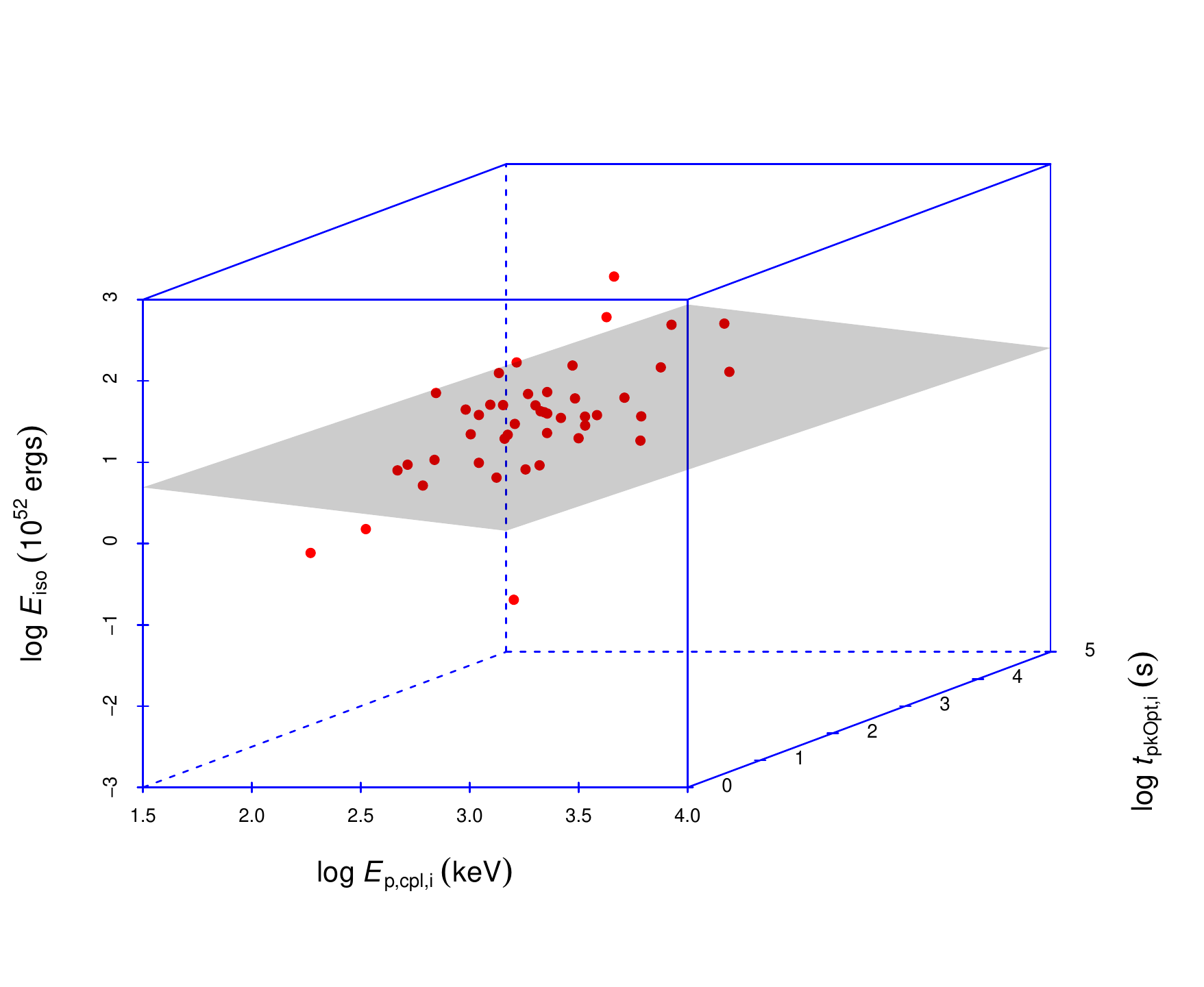}
\includegraphics[width=0.45\textwidth]{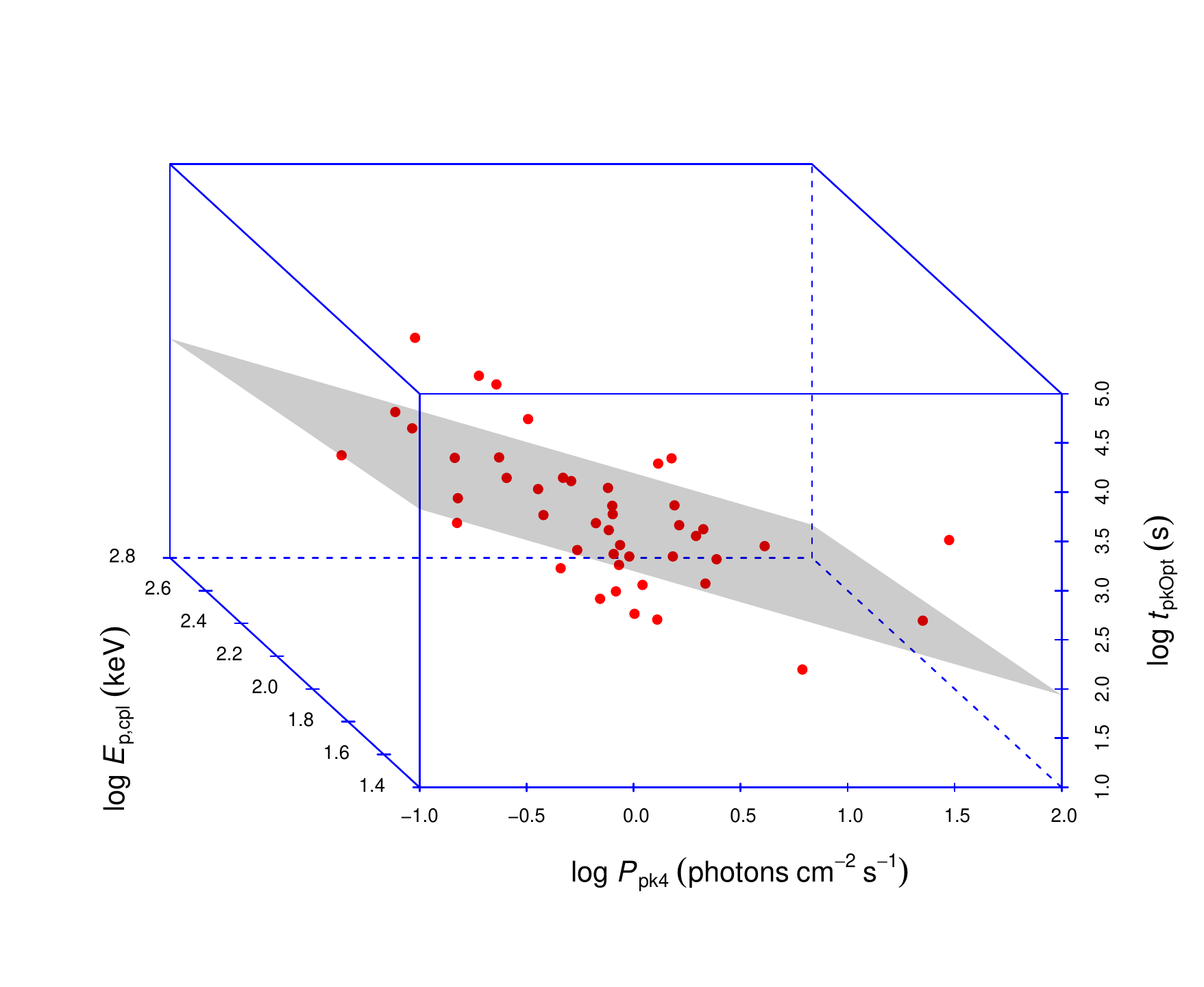}

\includegraphics[width=0.45\textwidth]{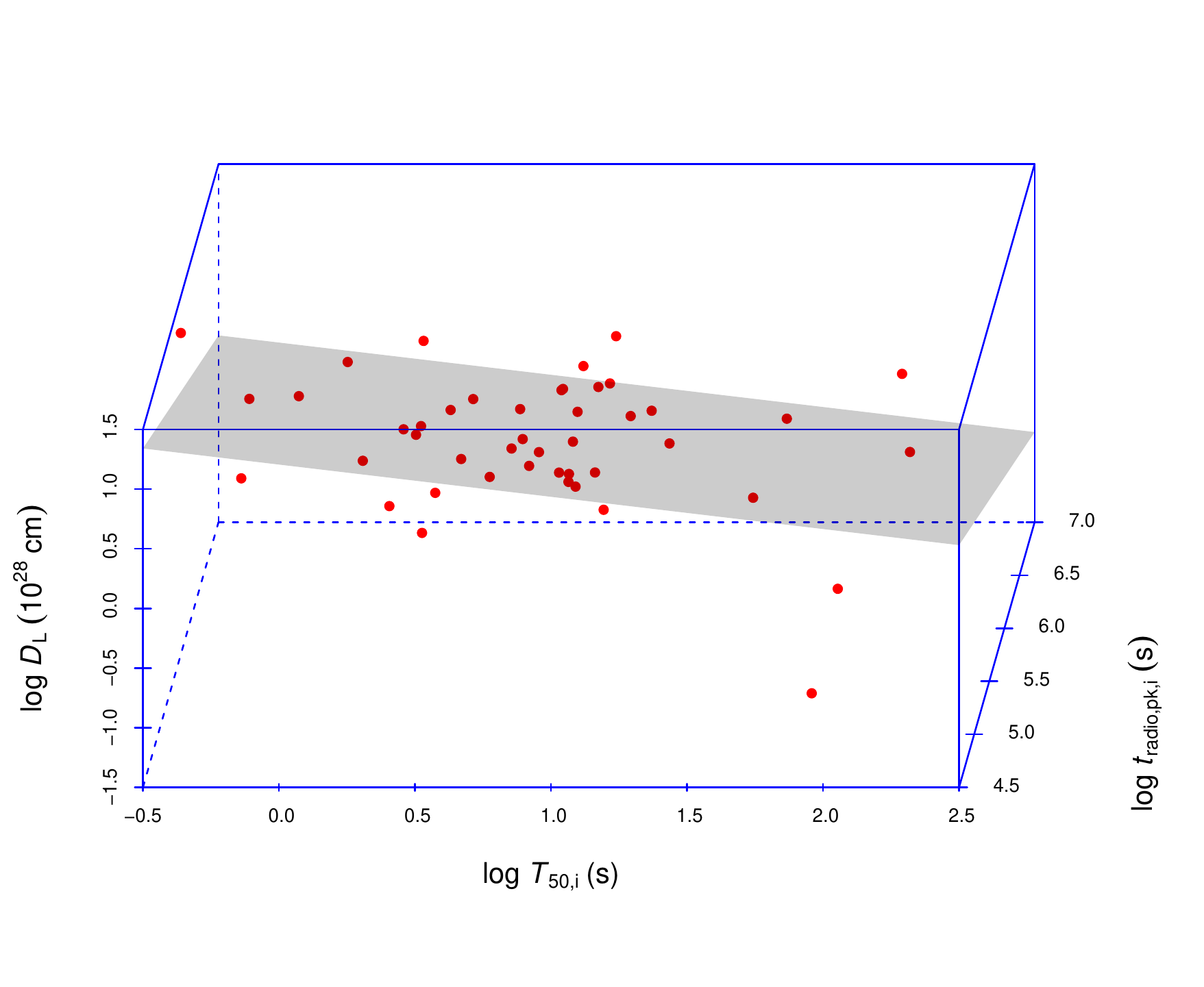}
\includegraphics[width=0.45\textwidth]{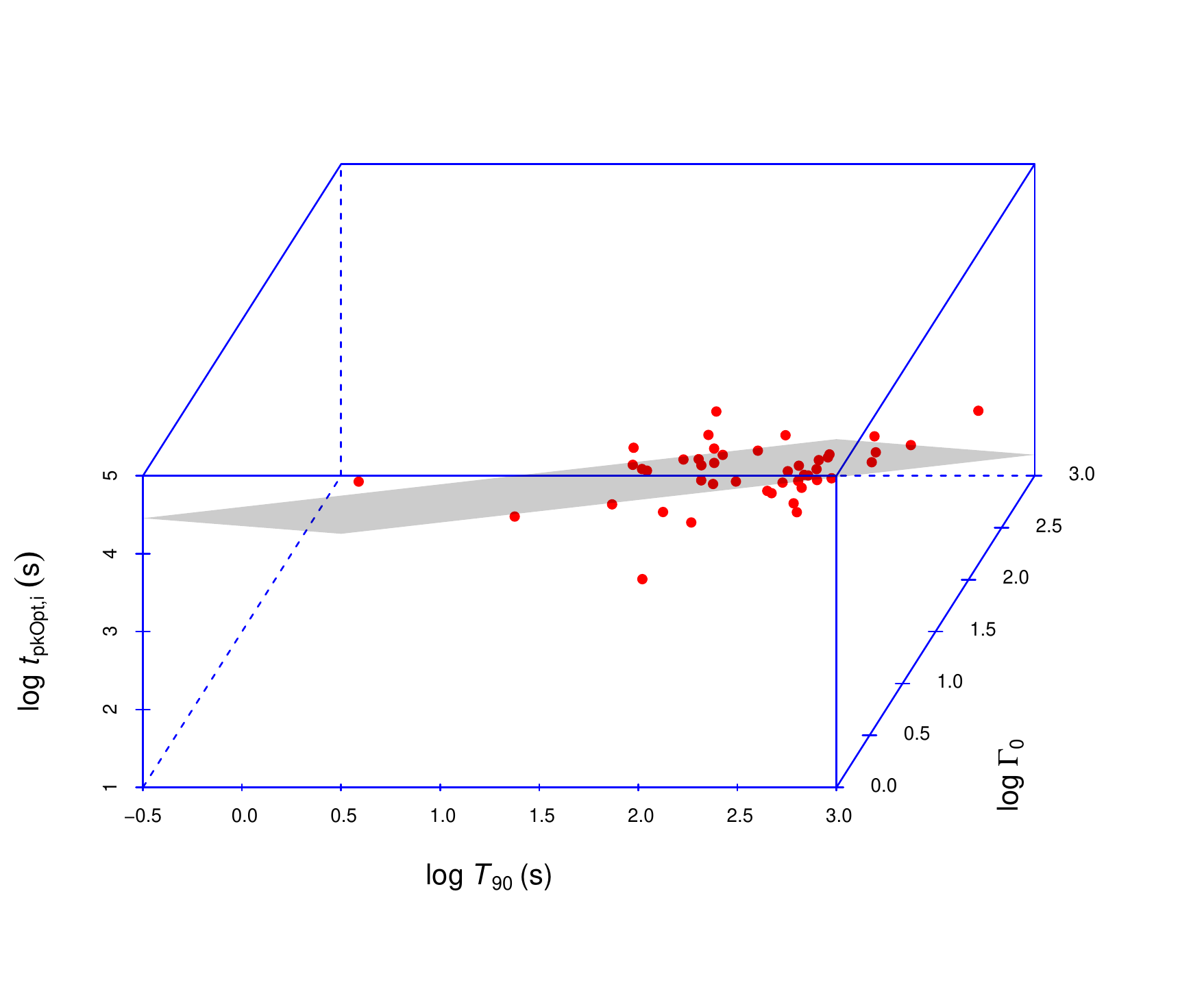}

\includegraphics[width=0.45\textwidth]{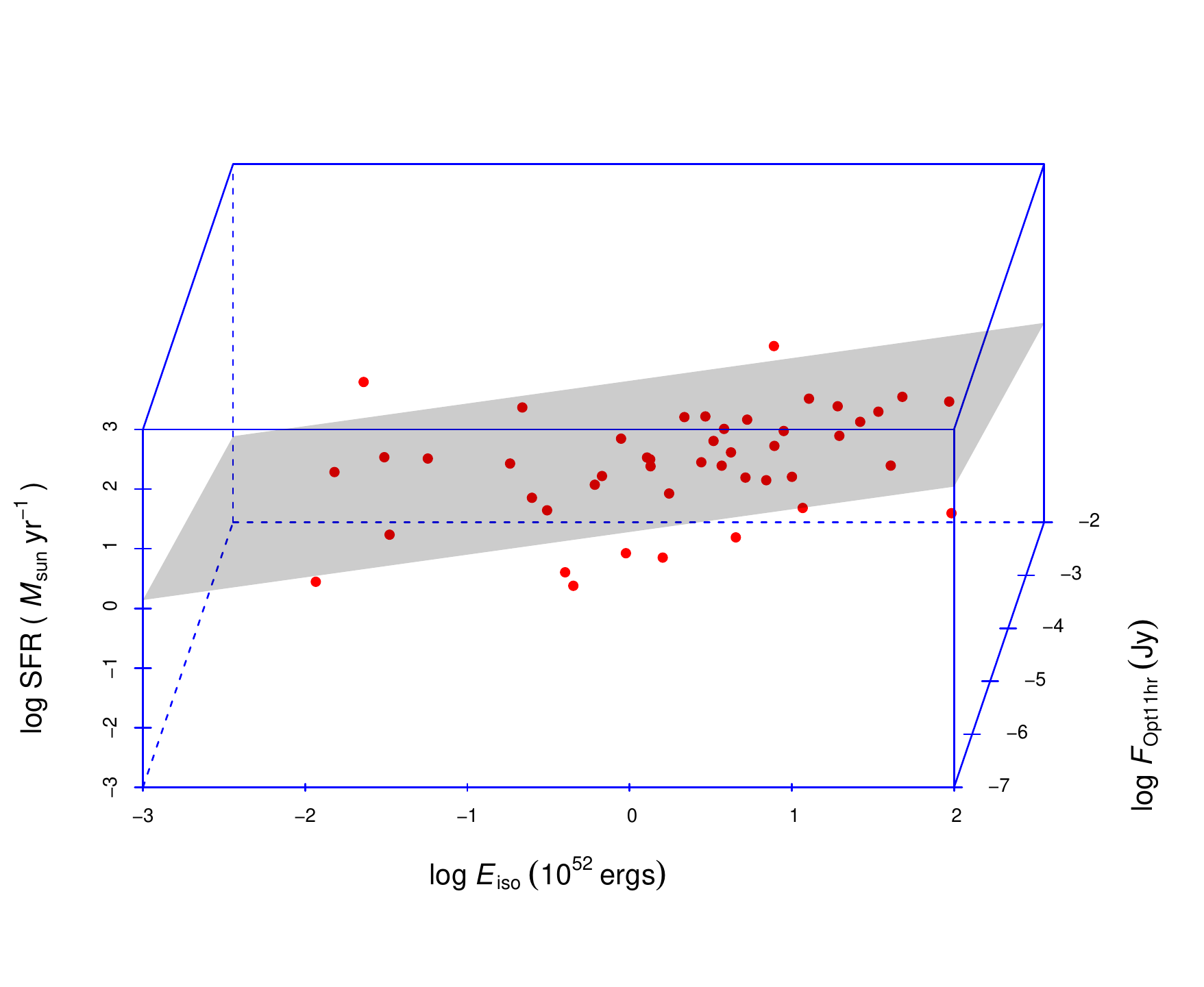}
\includegraphics[width=0.45\textwidth]{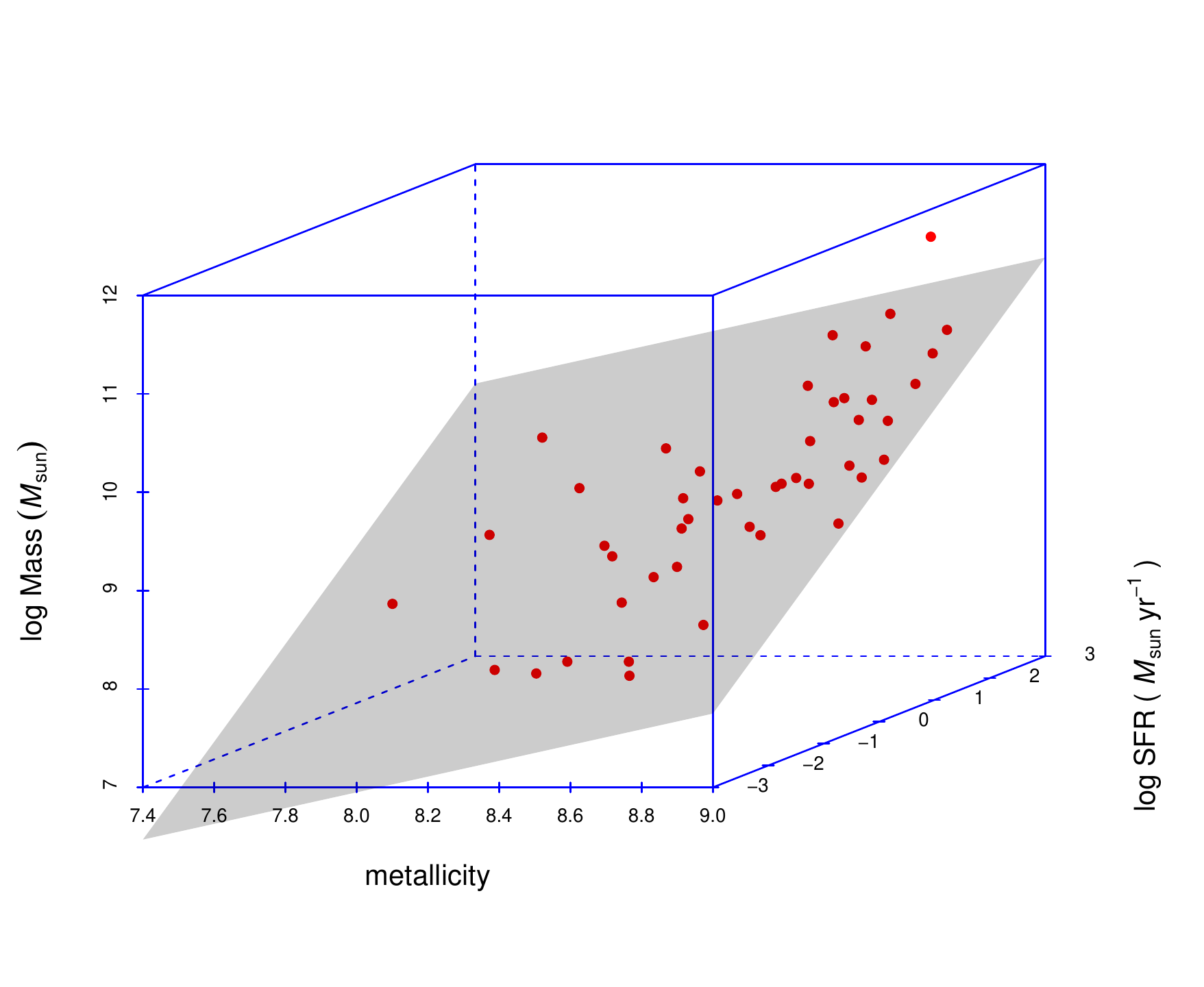}

\center{Fig. \ref{fig:three}---Continued}
\end{figure*}


\clearpage
\begin{figure*}

\includegraphics[width=0.45\textwidth]{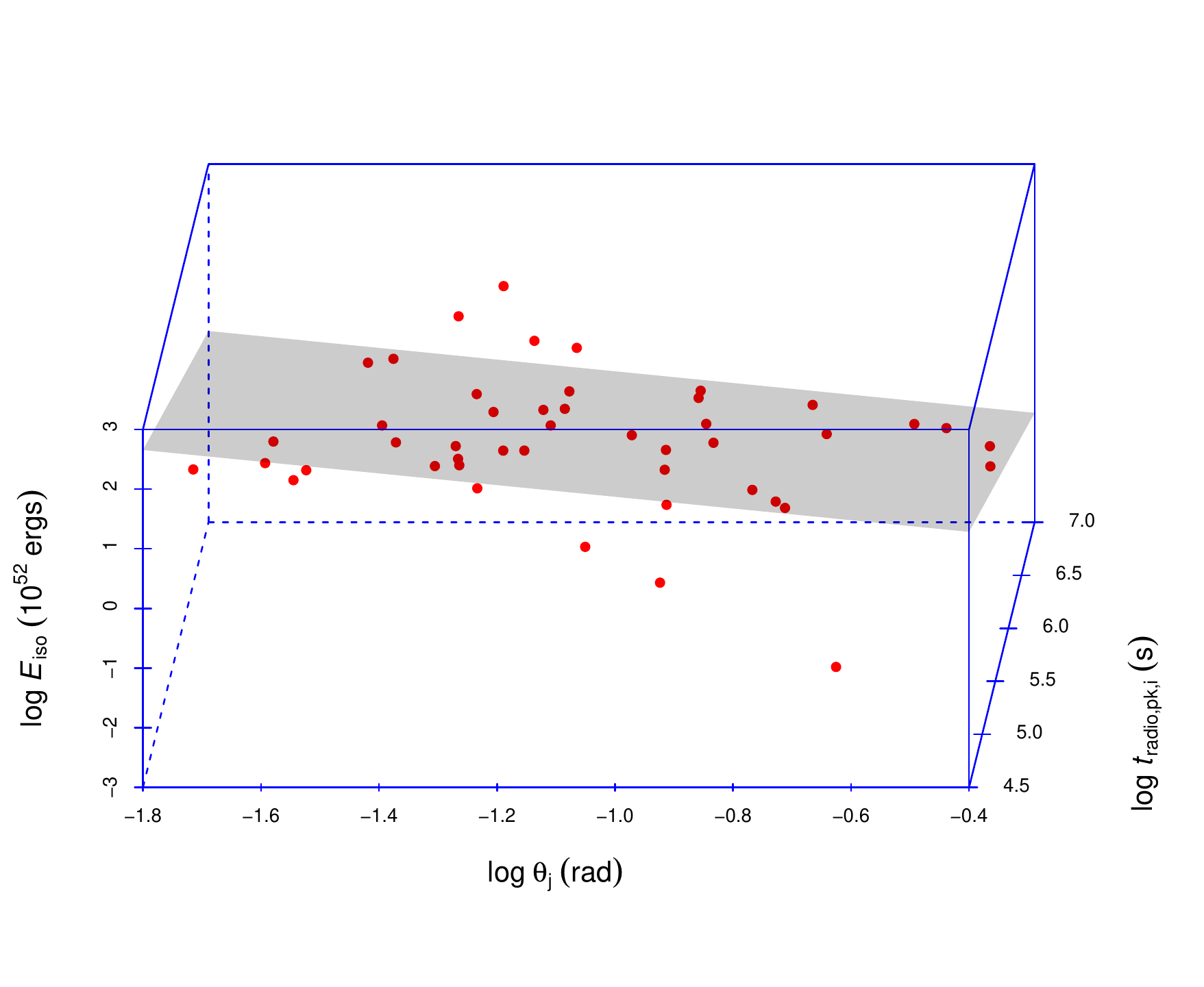}
\includegraphics[width=0.45\textwidth]{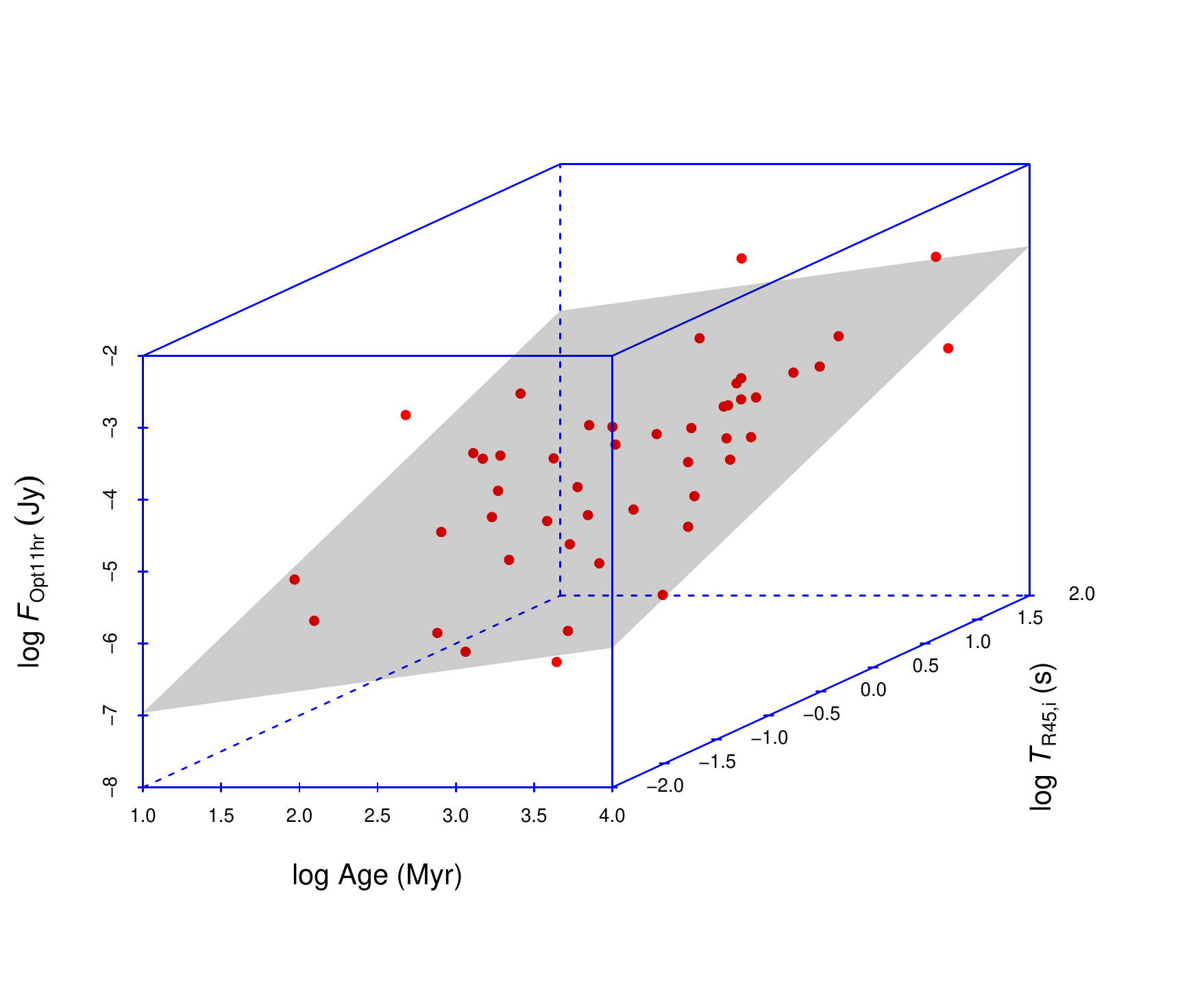}

\includegraphics[width=0.45\textwidth]{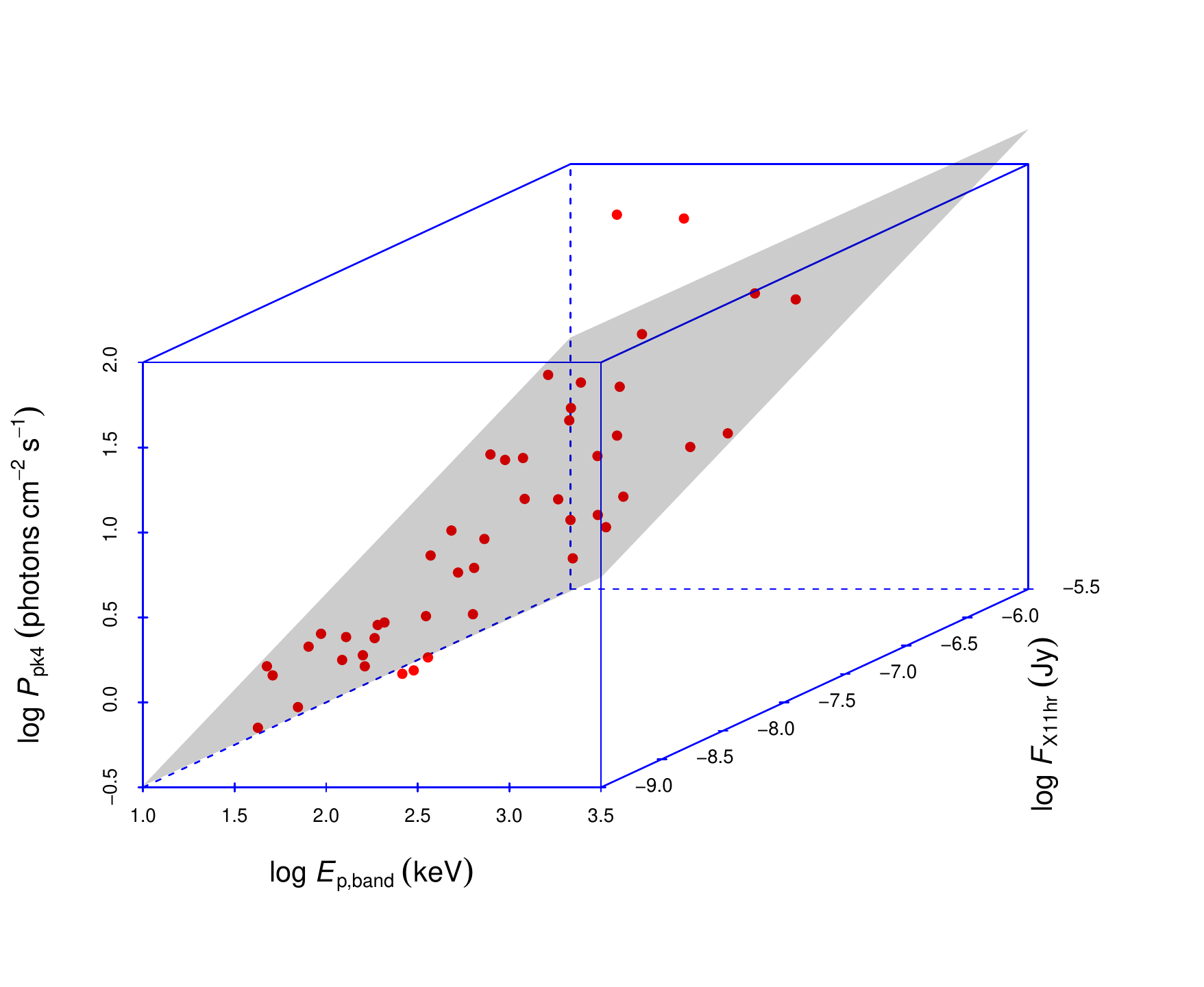}
\includegraphics[width=0.45\textwidth]{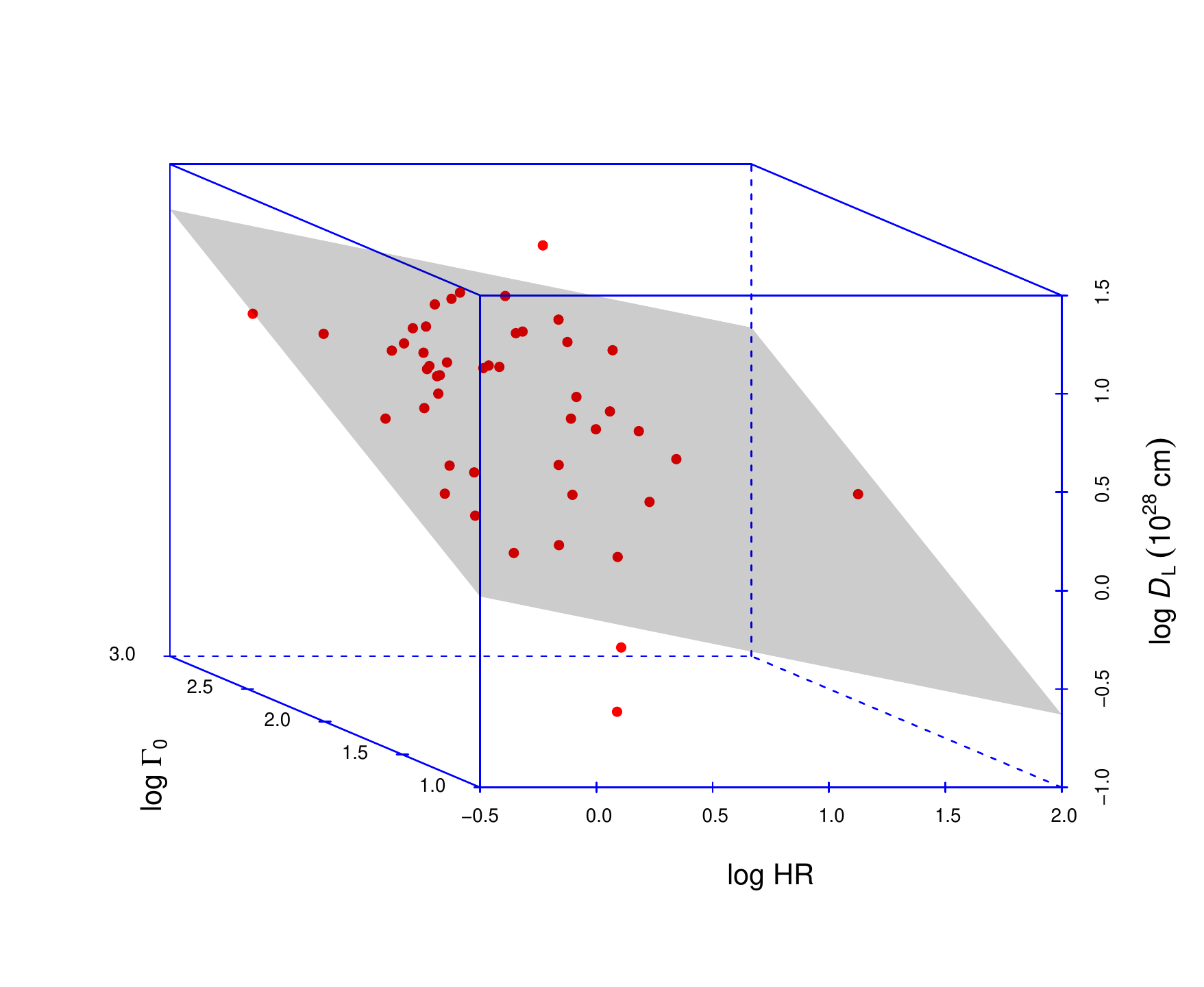}

\includegraphics[width=0.45\textwidth]{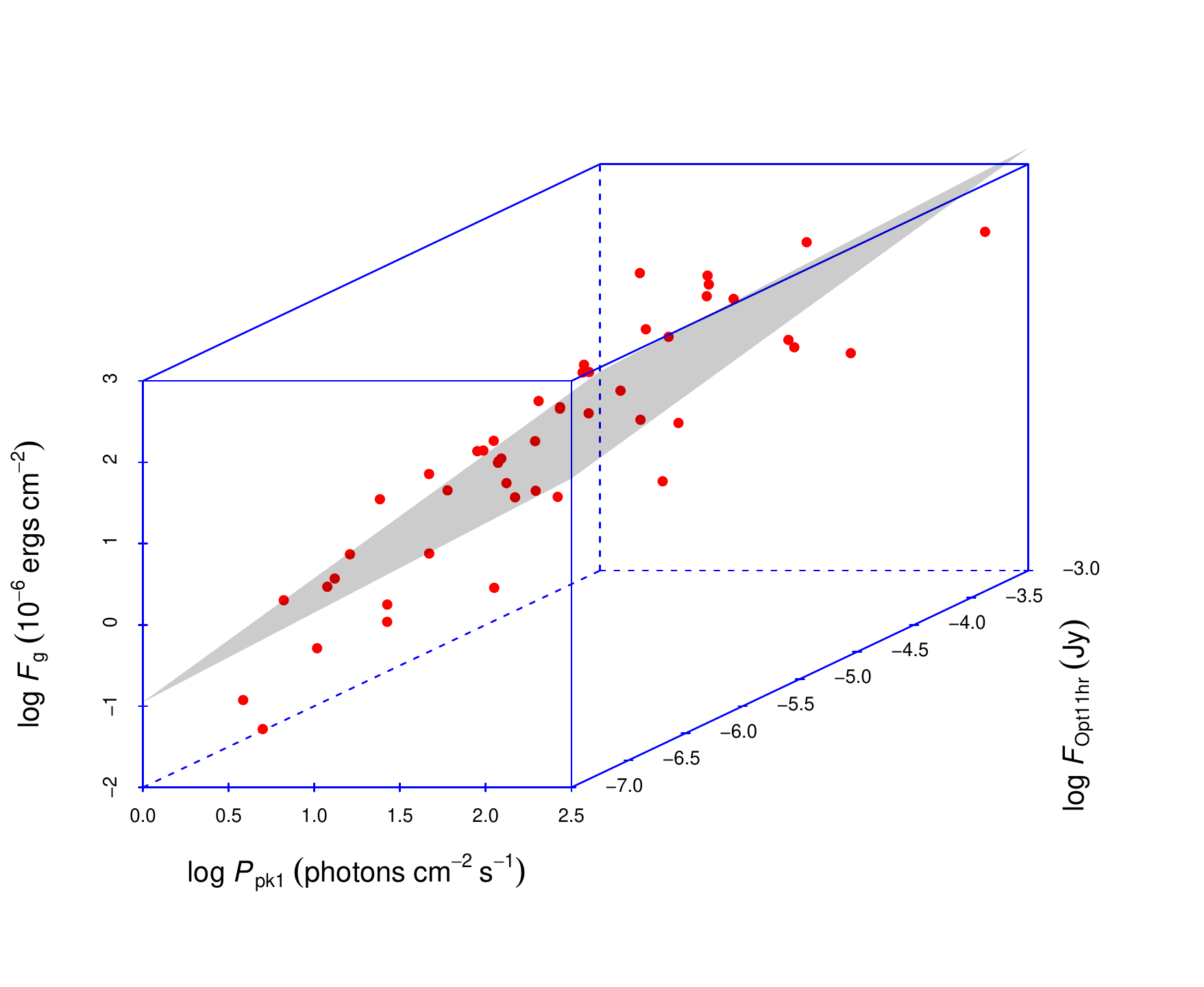}
\includegraphics[width=0.45\textwidth]{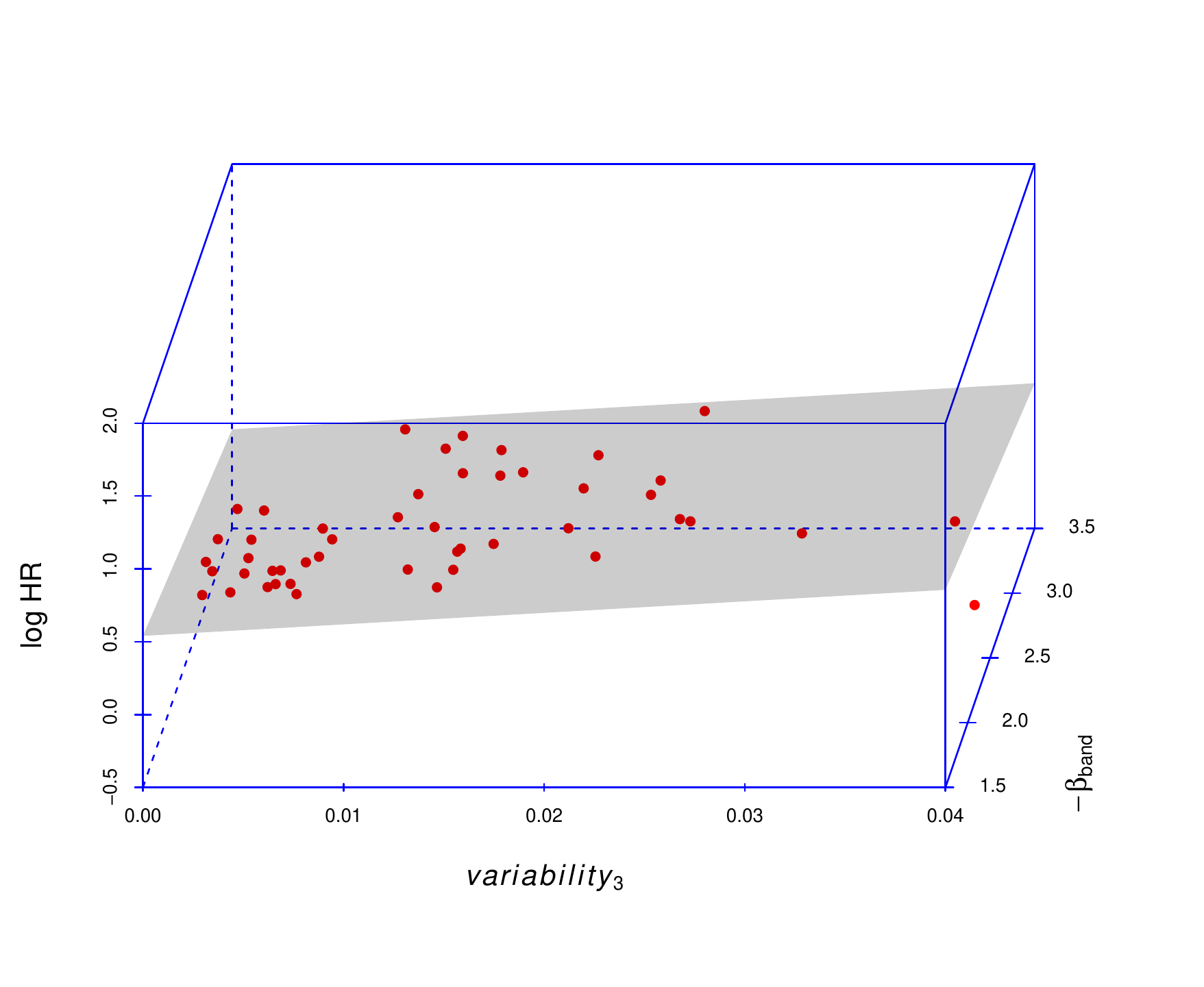}

\center{Fig. \ref{fig:three}---Continued}
\end{figure*}


\clearpage
\begin{figure*}

\includegraphics[width=0.45\textwidth]{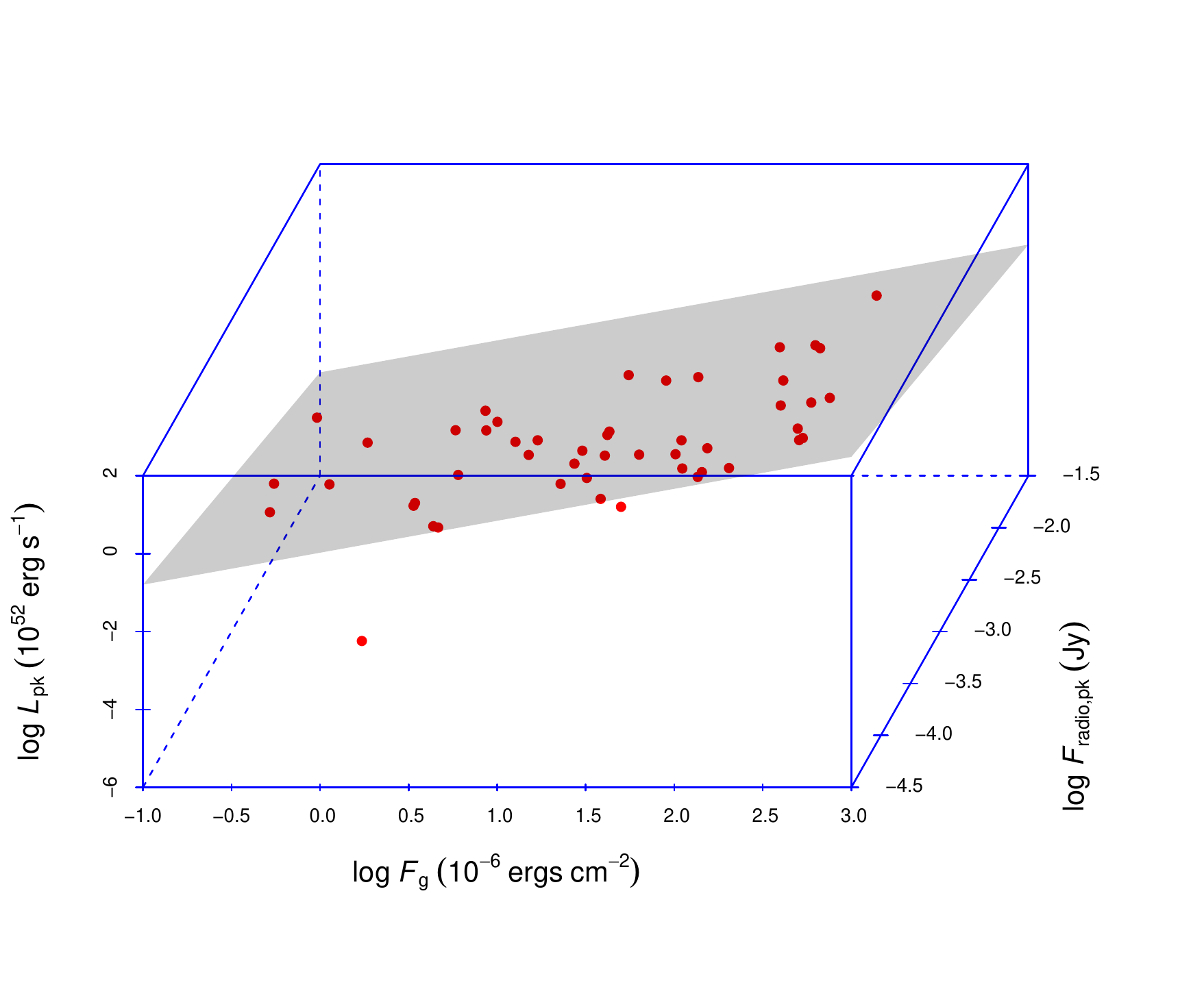}
\includegraphics[width=0.45\textwidth]{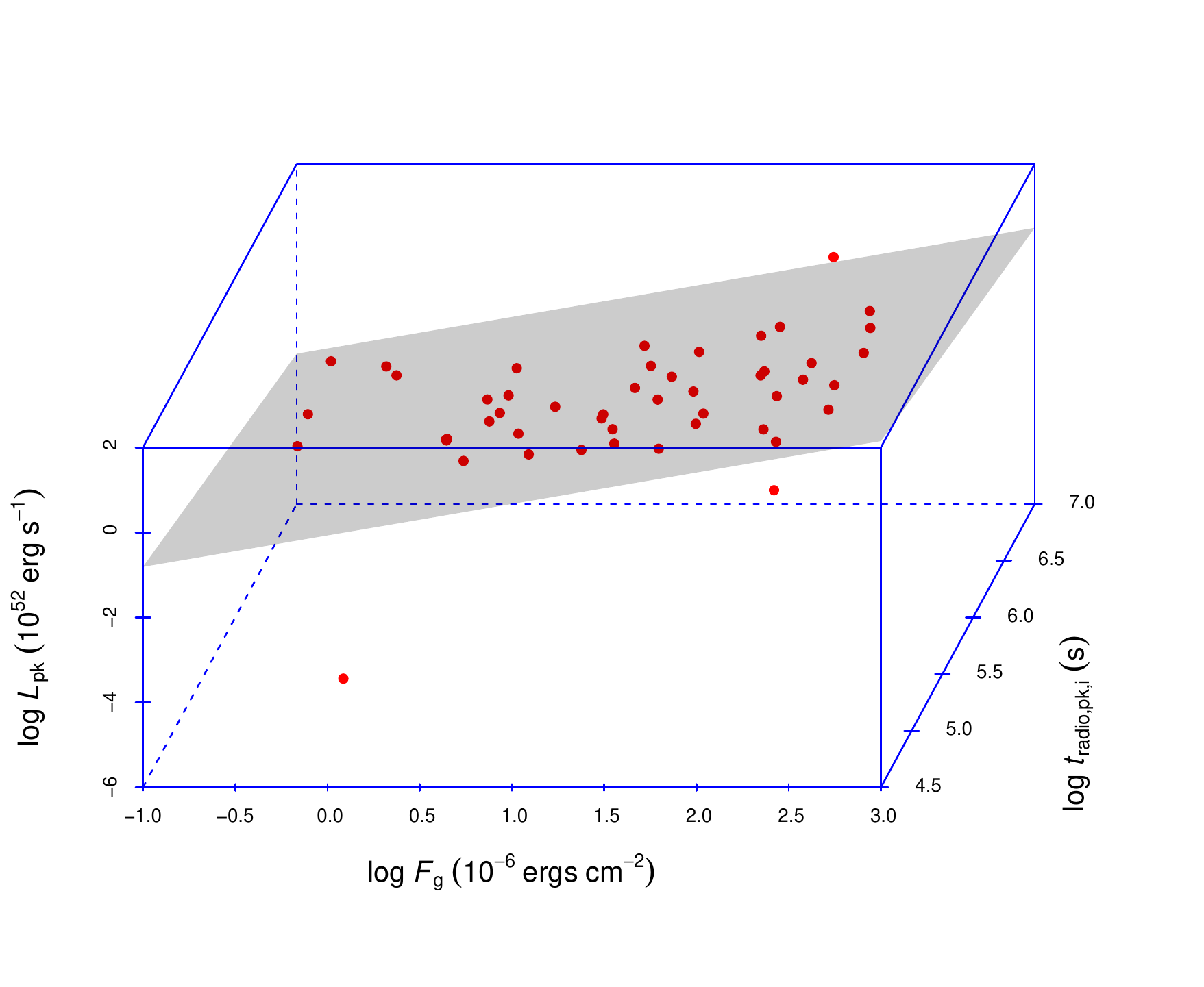}

\includegraphics[width=0.45\textwidth]{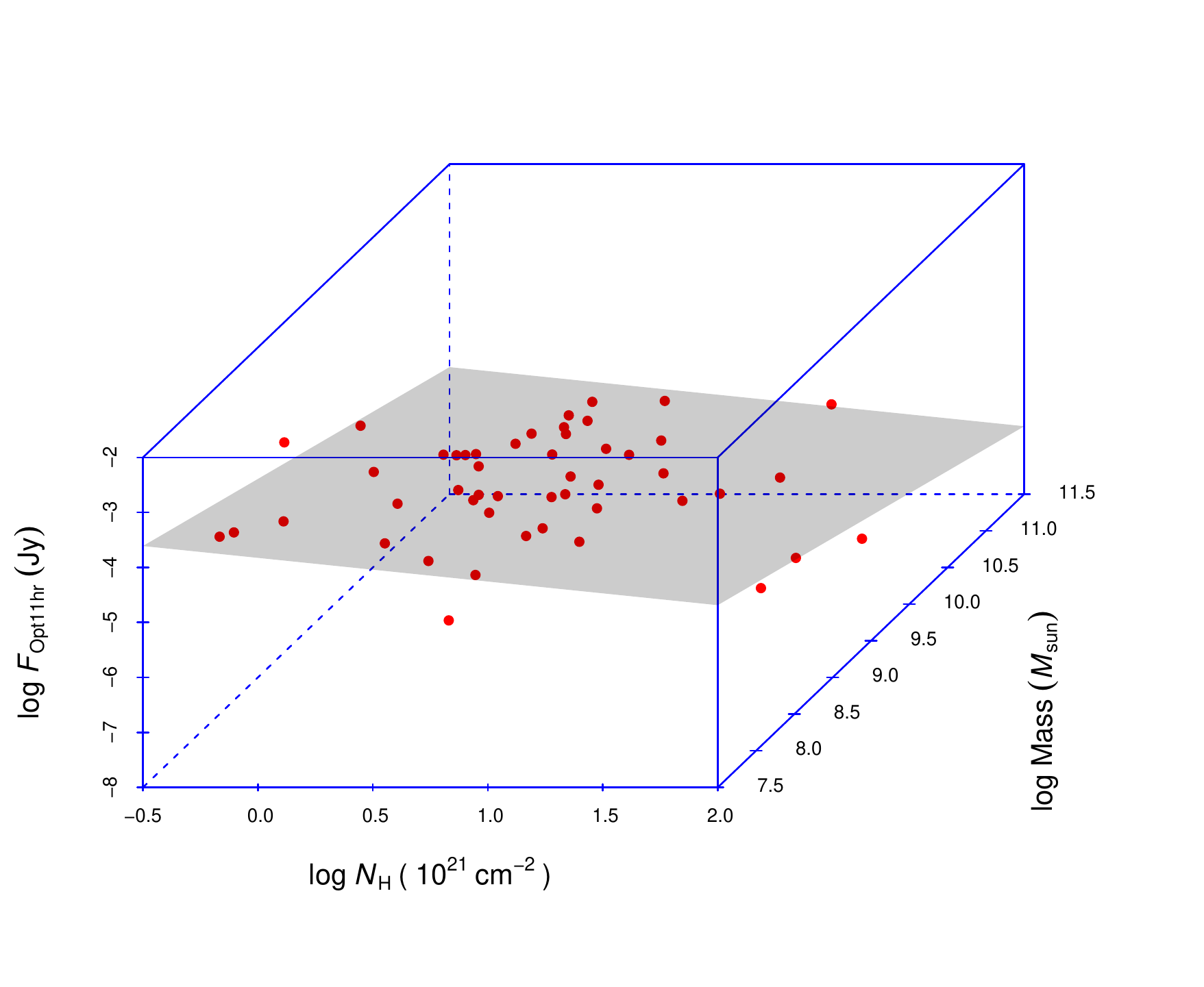}
\includegraphics[width=0.45\textwidth]{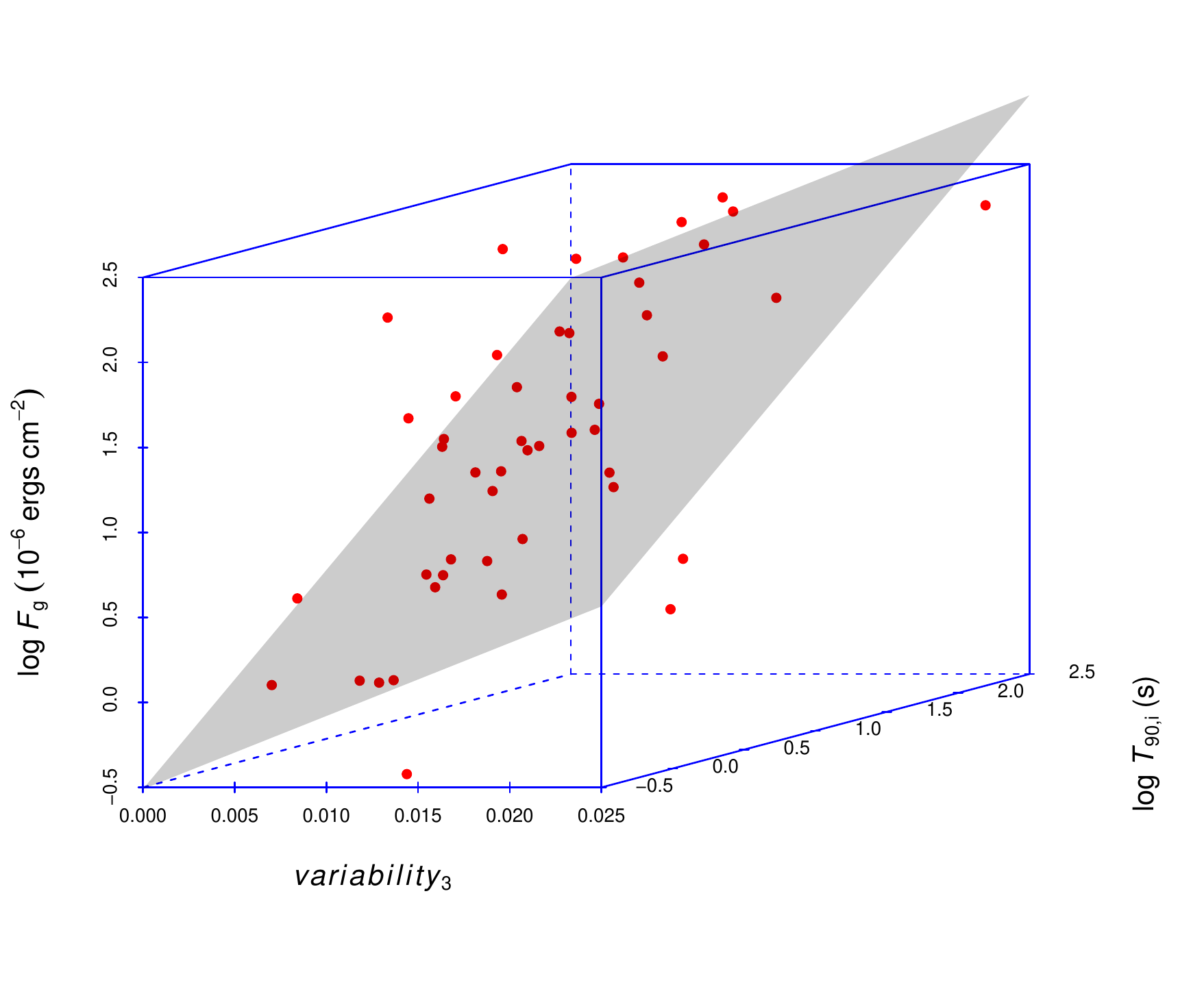}

\includegraphics[width=0.45\textwidth]{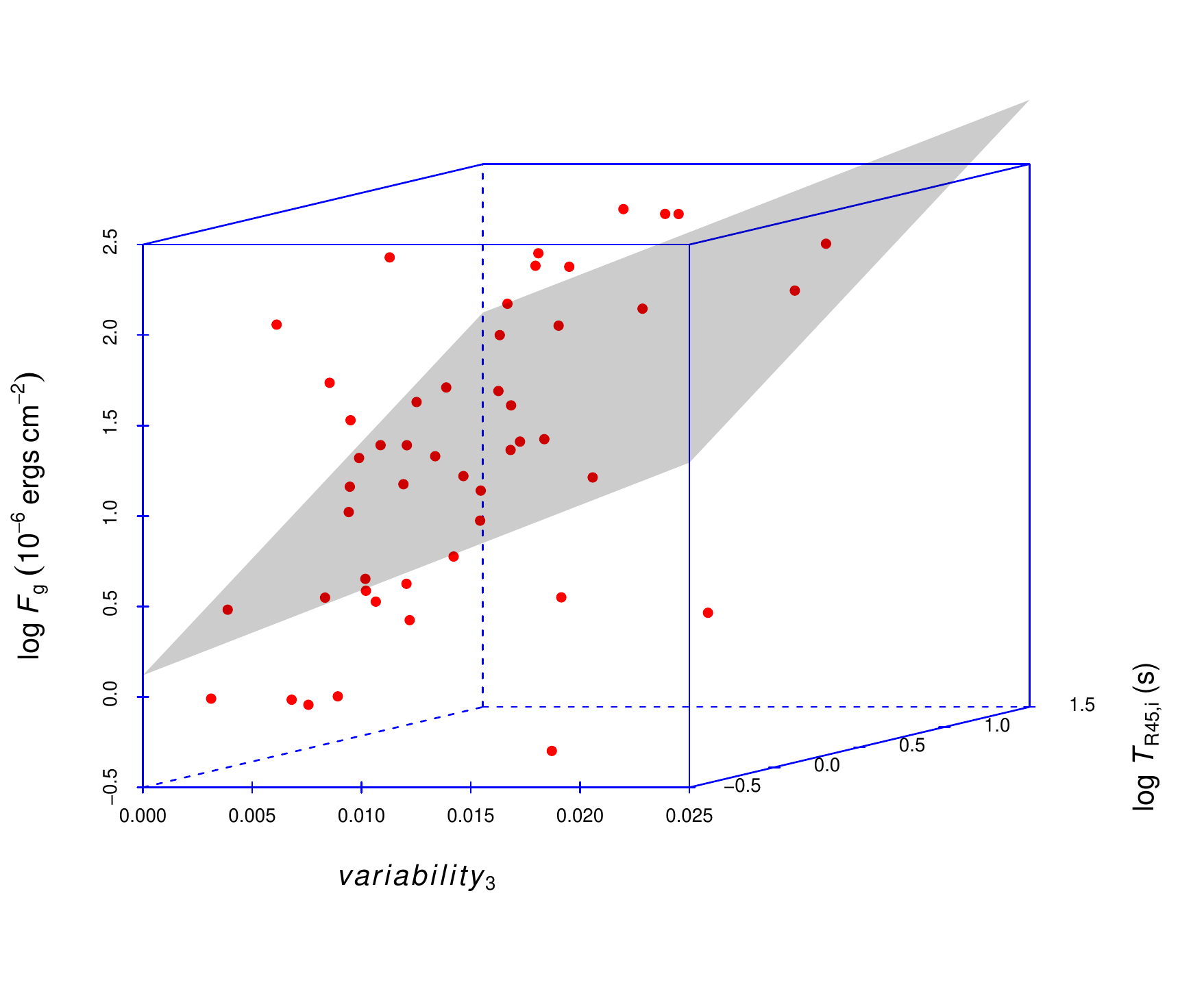}
\includegraphics[width=0.45\textwidth]{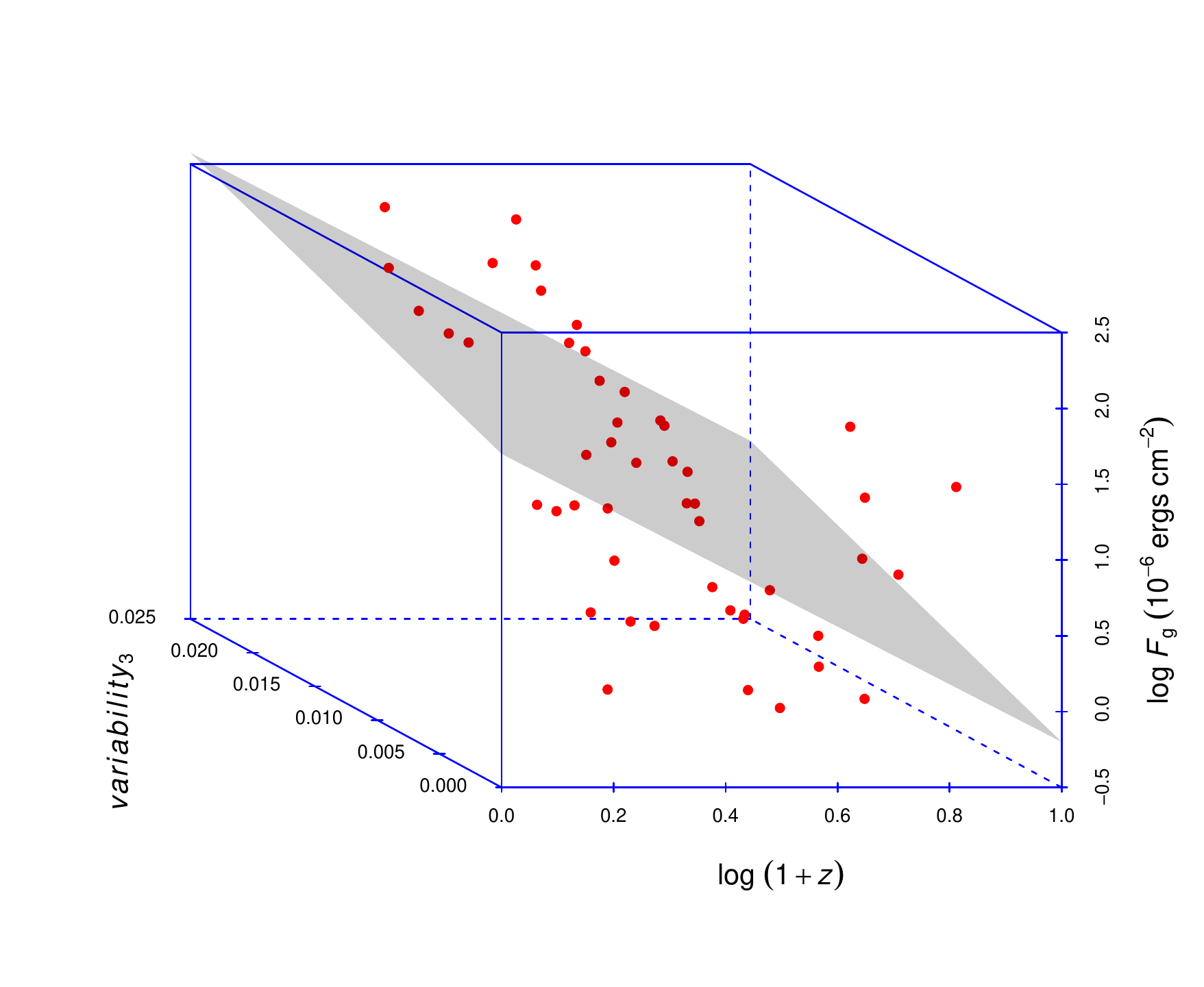}

\center{Fig. \ref{fig:three}---Continued}
\end{figure*}


\clearpage
\begin{figure*}

\includegraphics[width=0.45\textwidth]{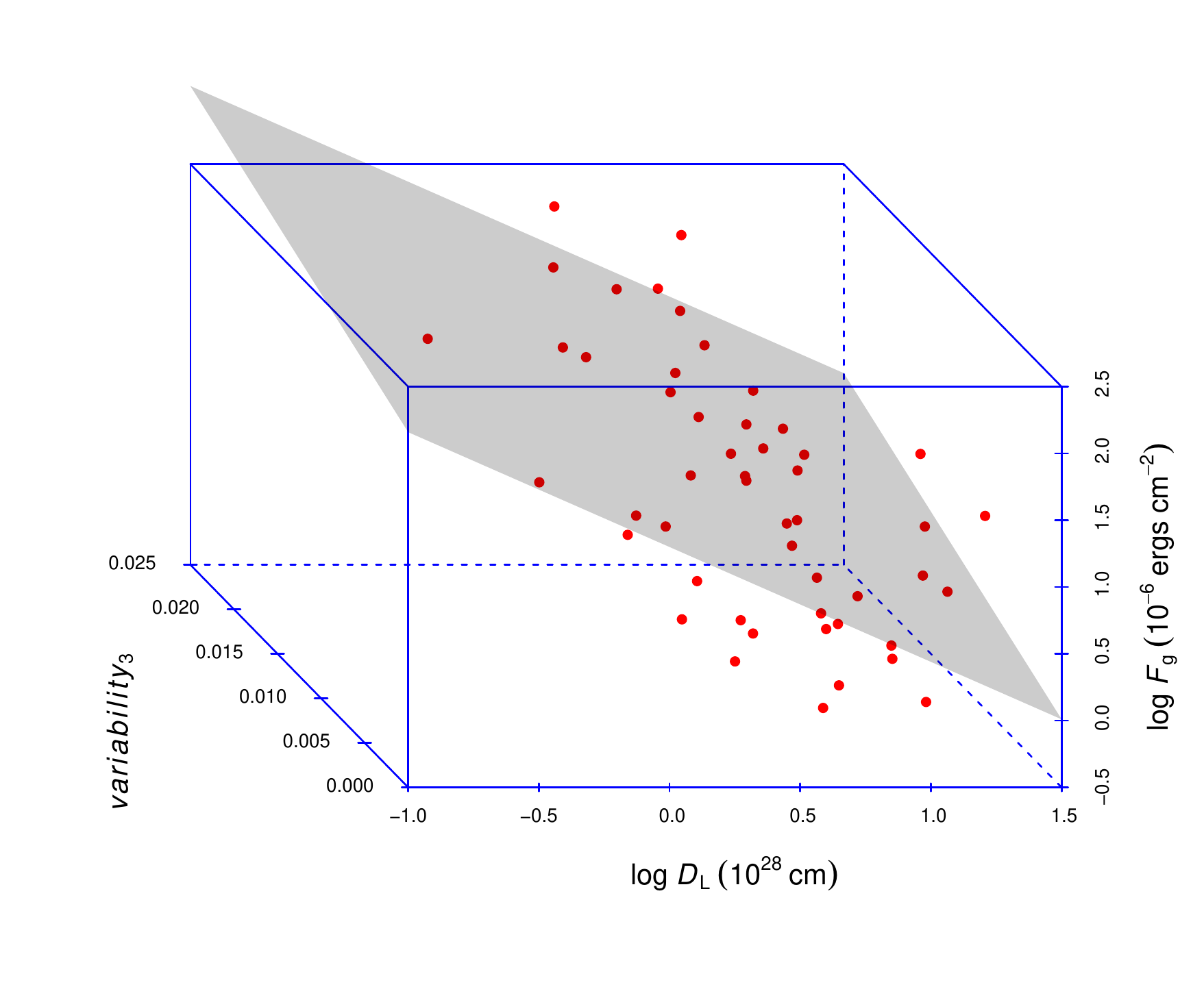}
\includegraphics[width=0.45\textwidth]{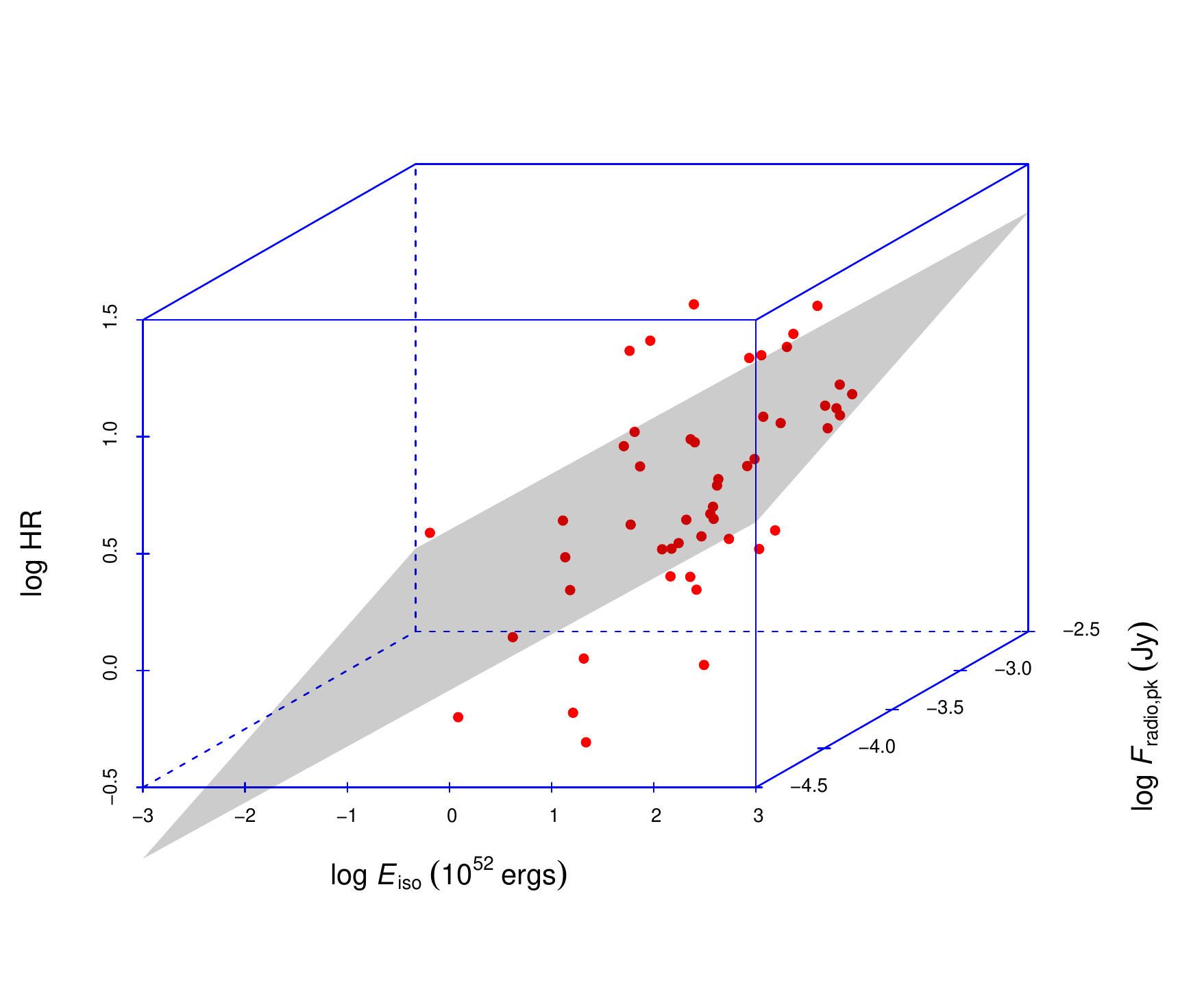}

\includegraphics[width=0.45\textwidth]{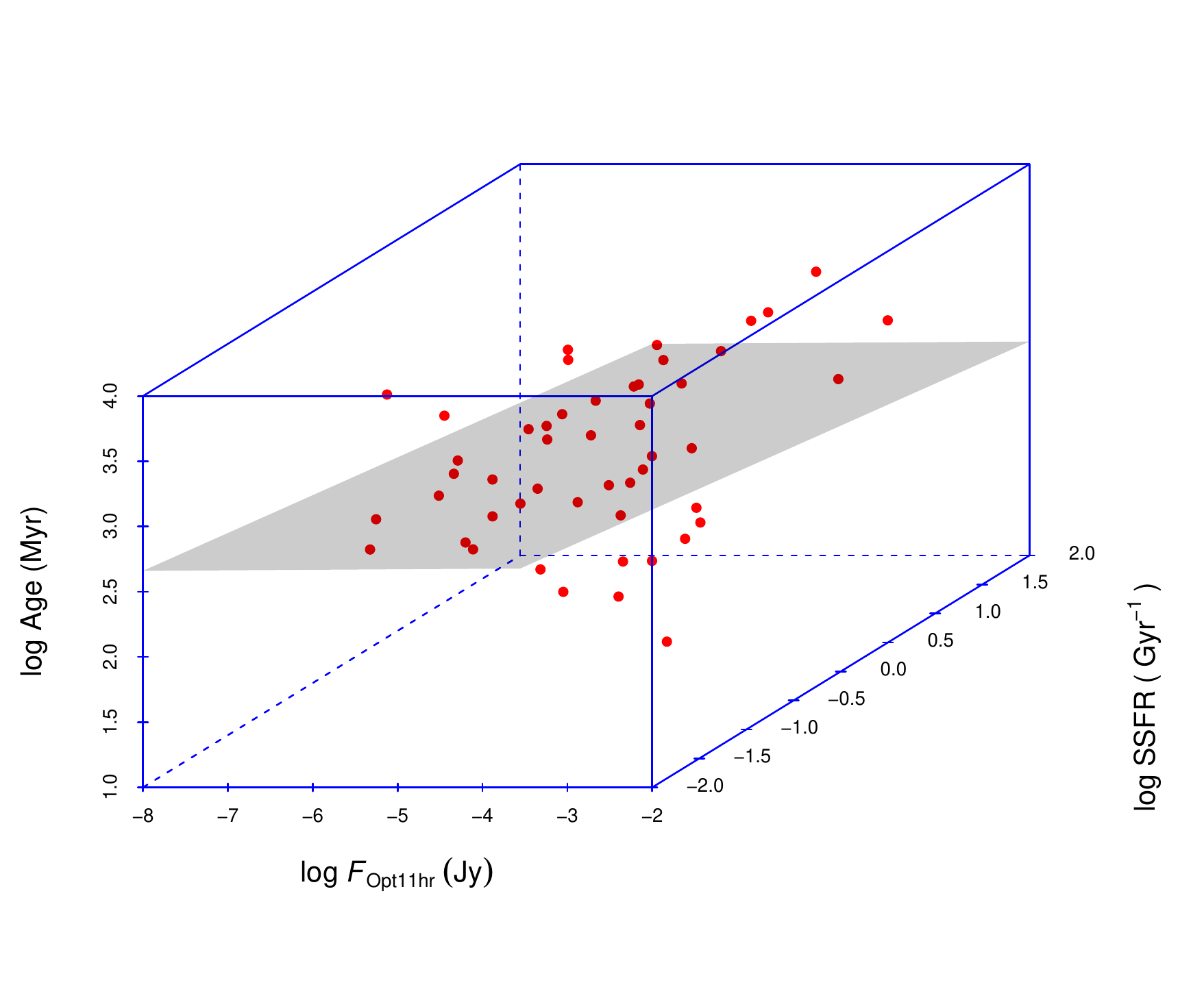}
\includegraphics[width=0.45\textwidth]{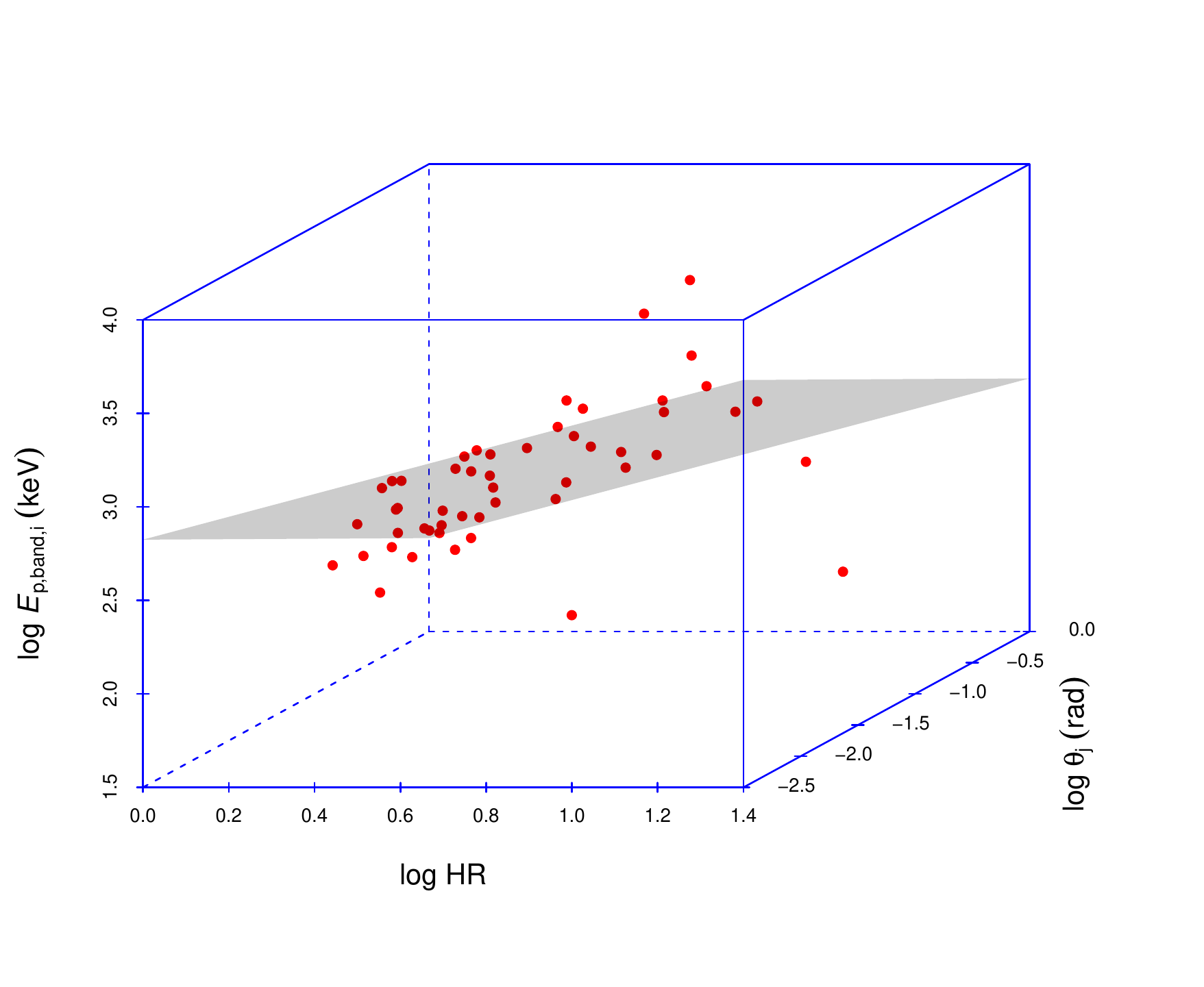}

\includegraphics[width=0.45\textwidth]{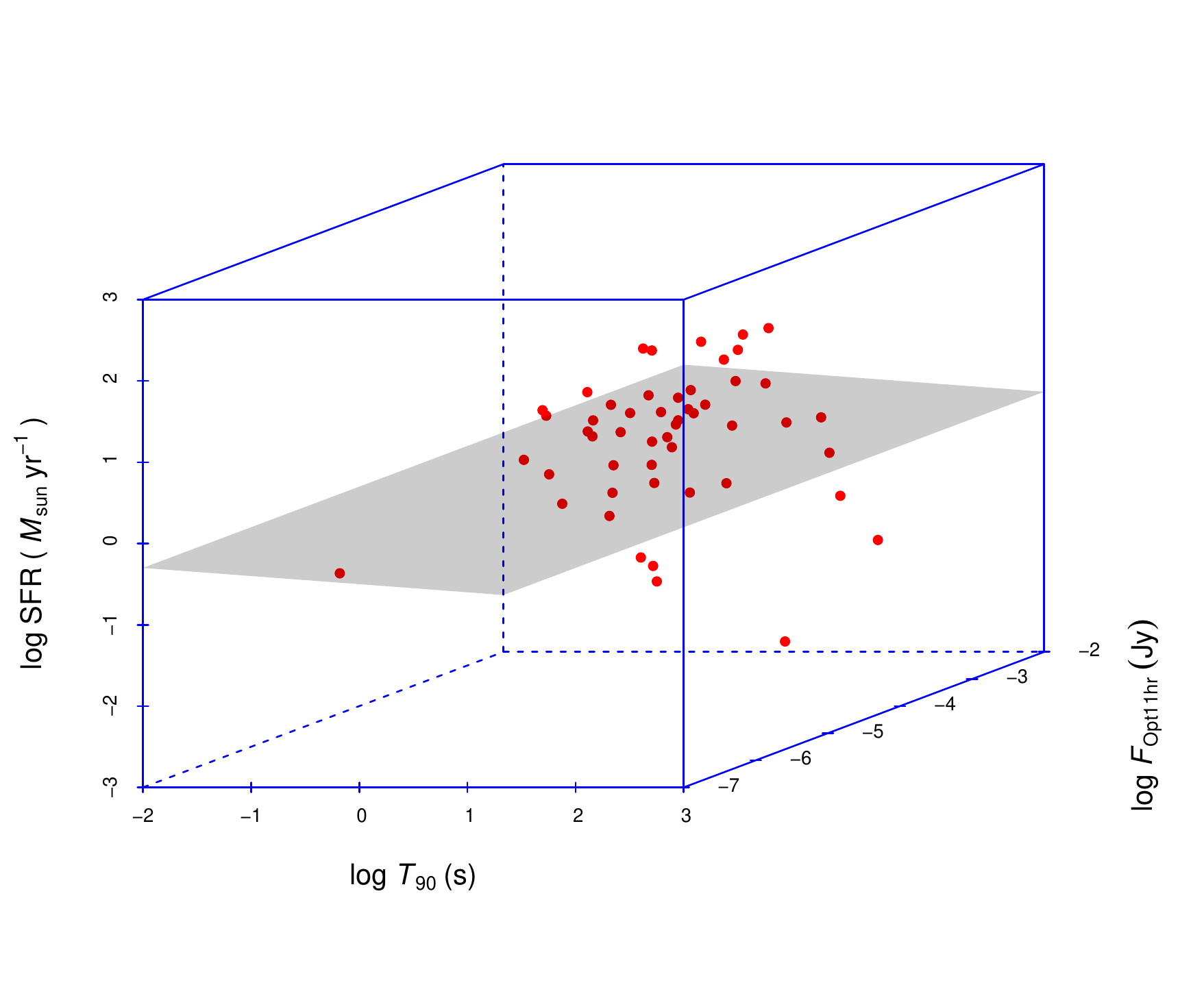}
\includegraphics[width=0.45\textwidth]{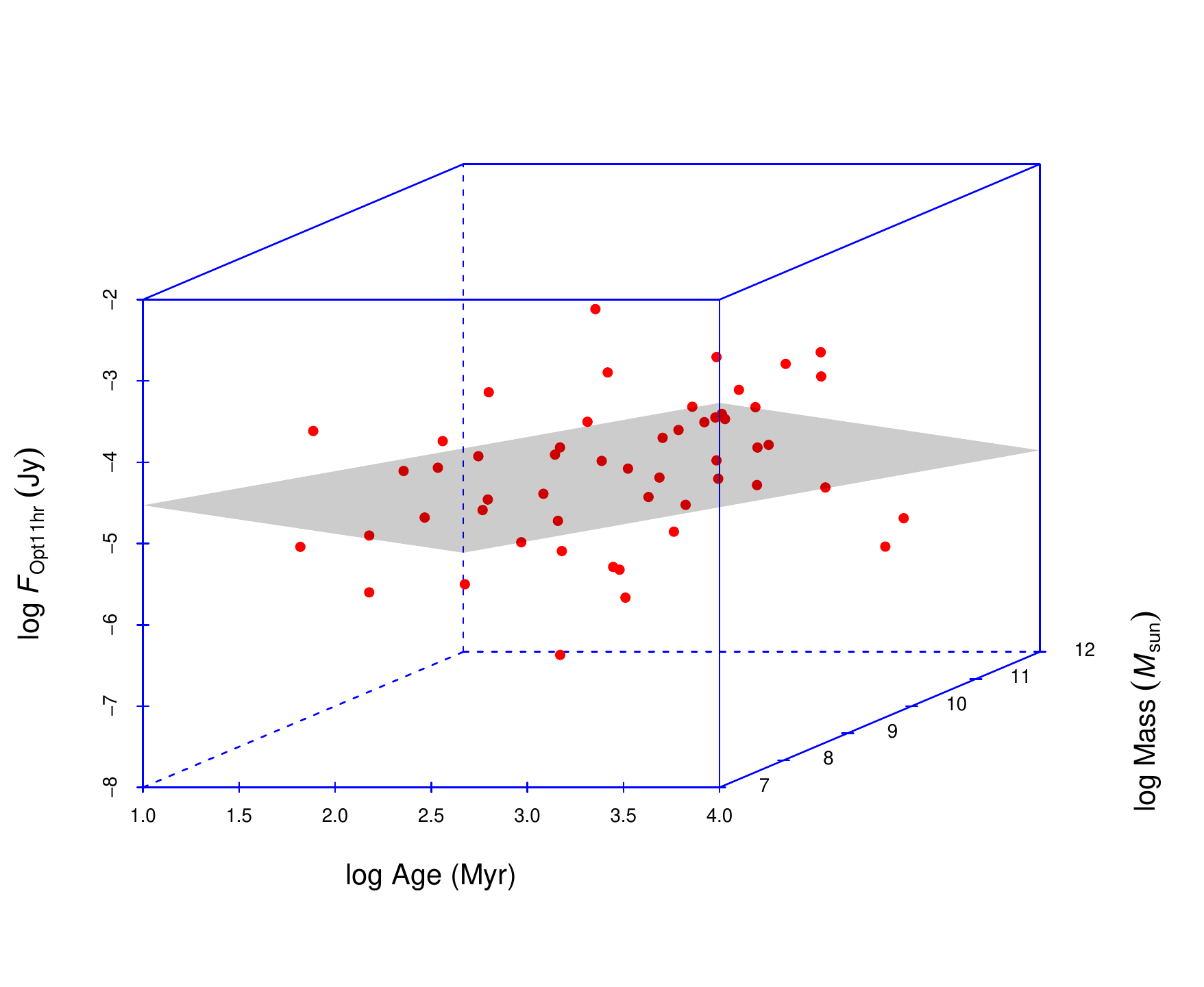}

\center{Fig. \ref{fig:three}---Continued}
\end{figure*}


\clearpage
\begin{figure*}

\includegraphics[width=0.45\textwidth]{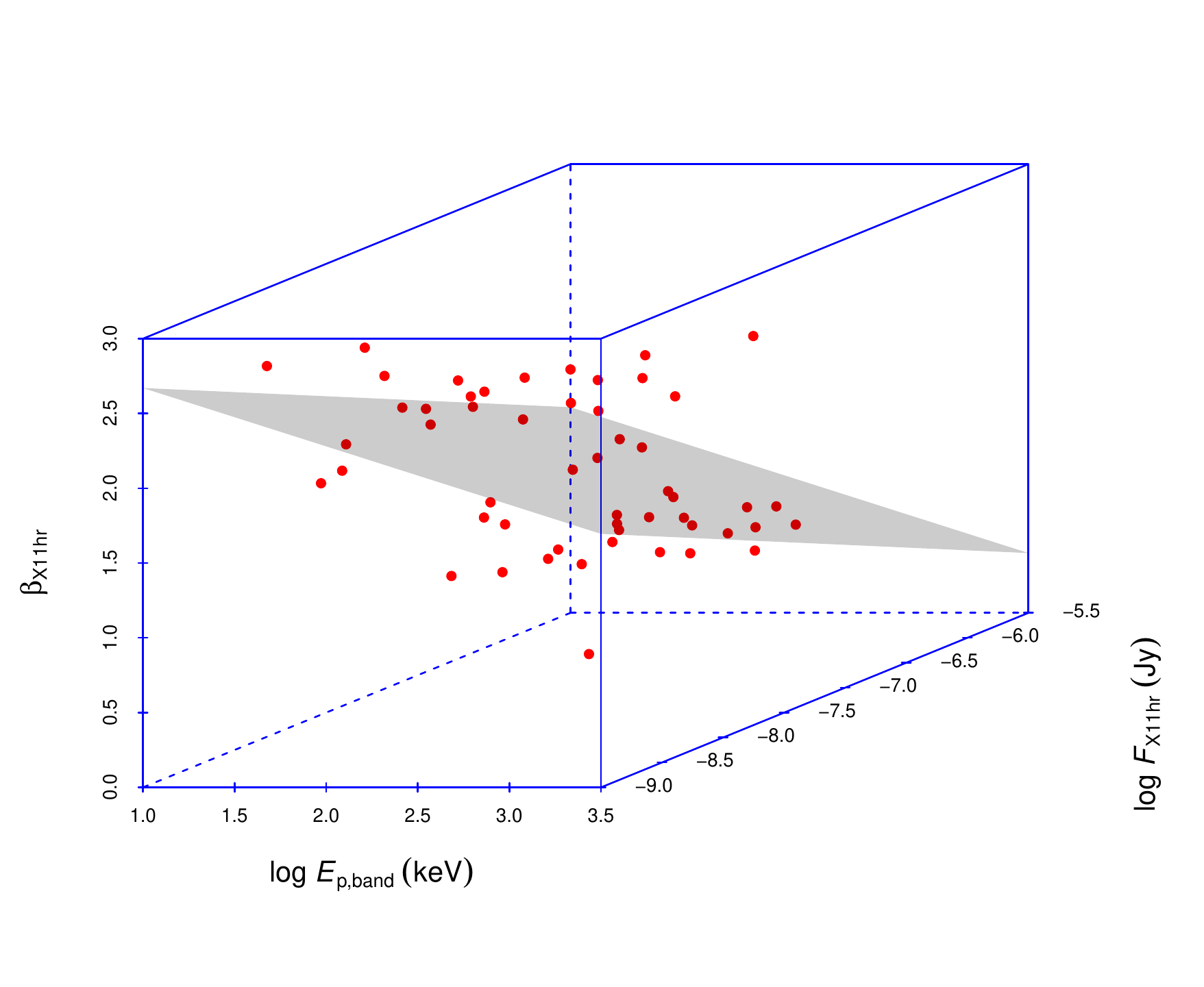}
\includegraphics[width=0.45\textwidth]{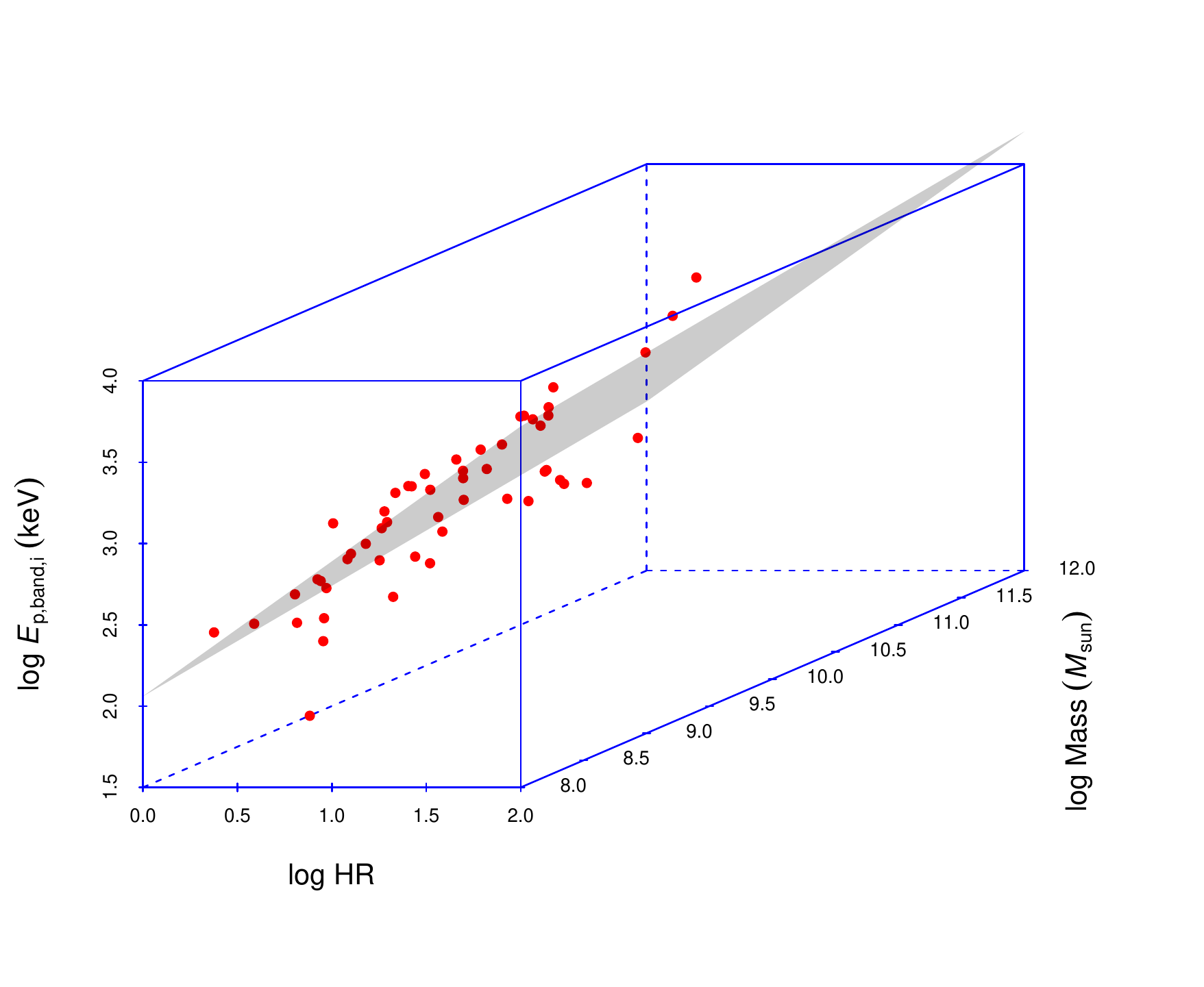}

\includegraphics[width=0.45\textwidth]{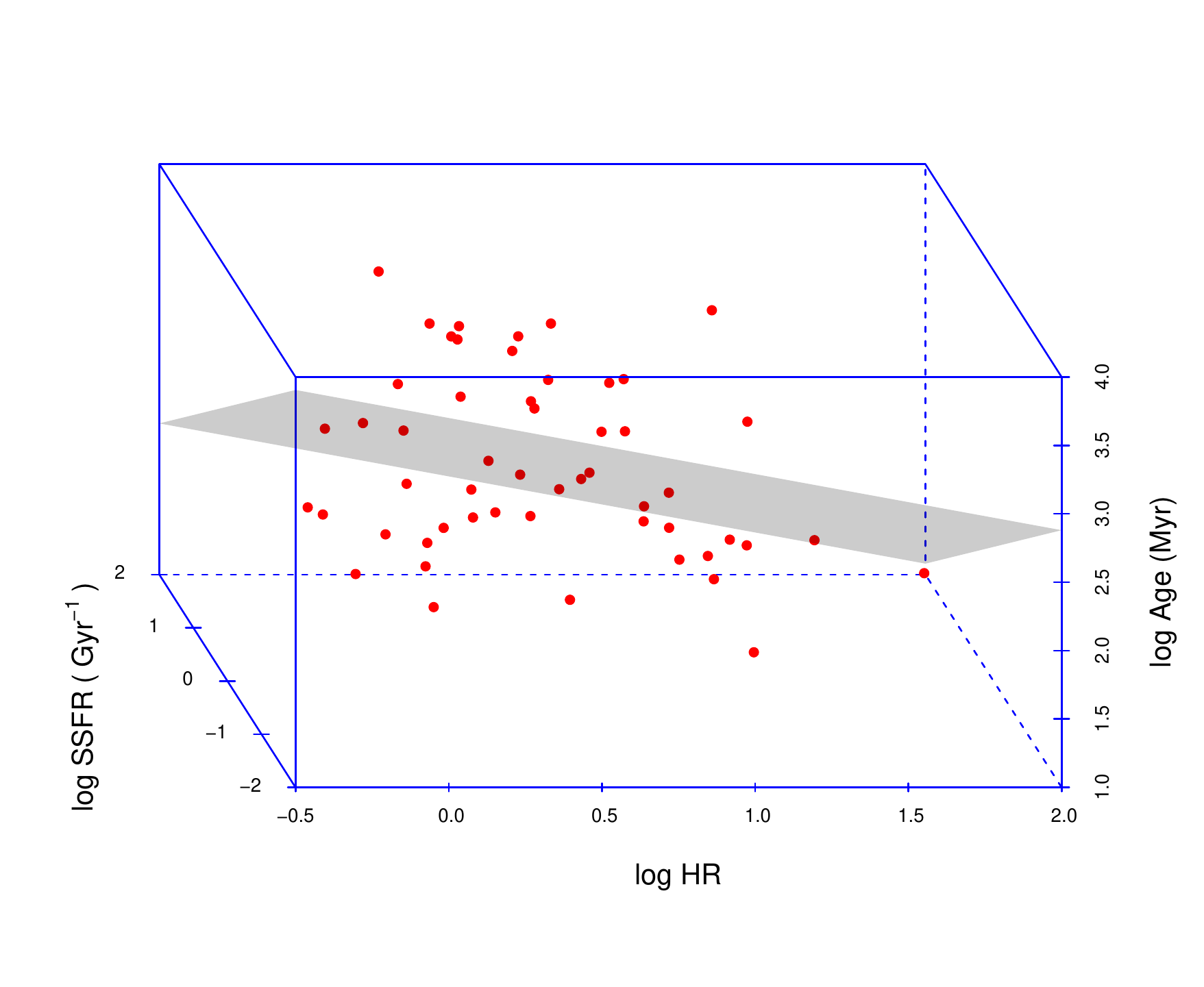}
\includegraphics[width=0.45\textwidth]{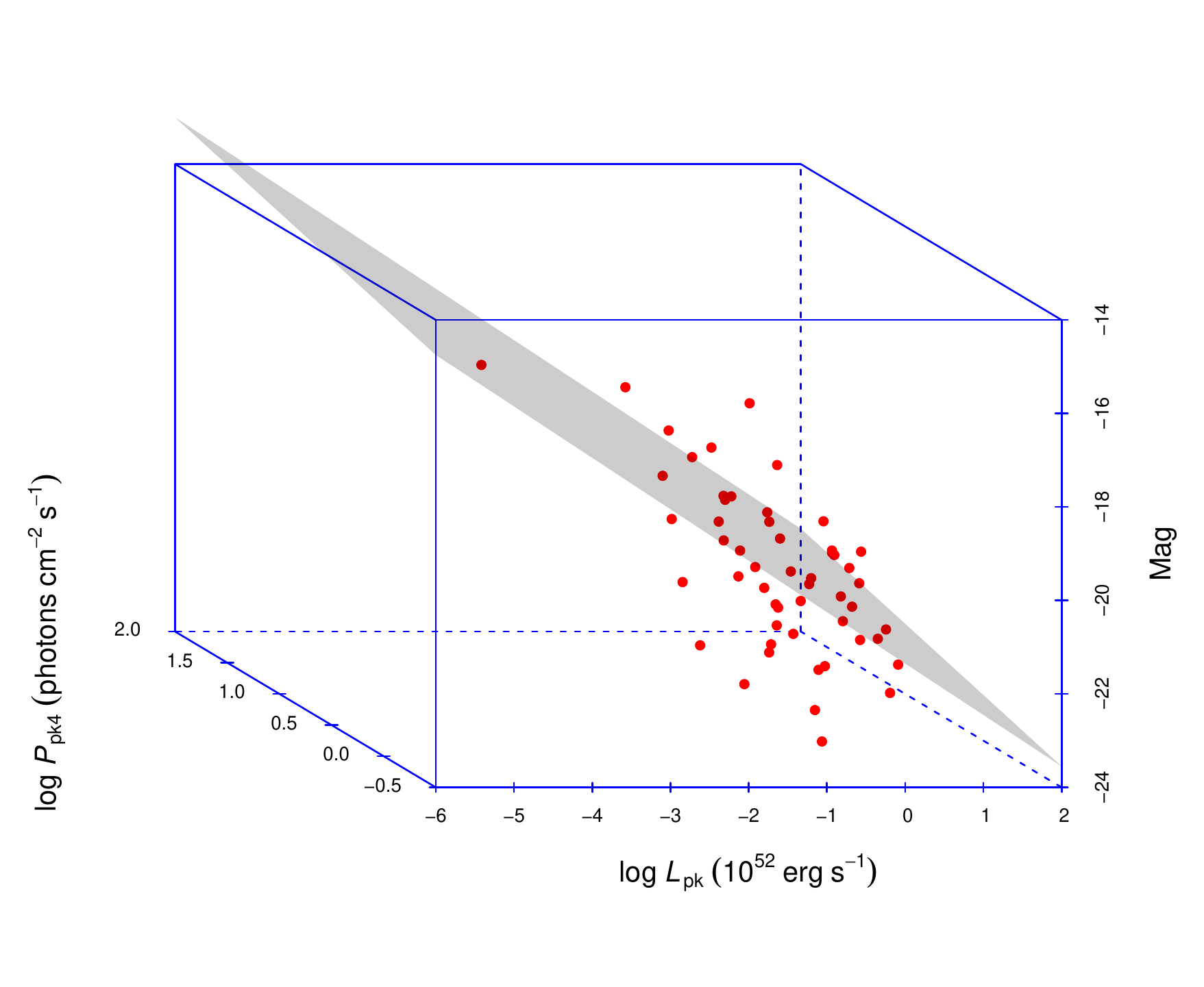}

\includegraphics[width=0.45\textwidth]{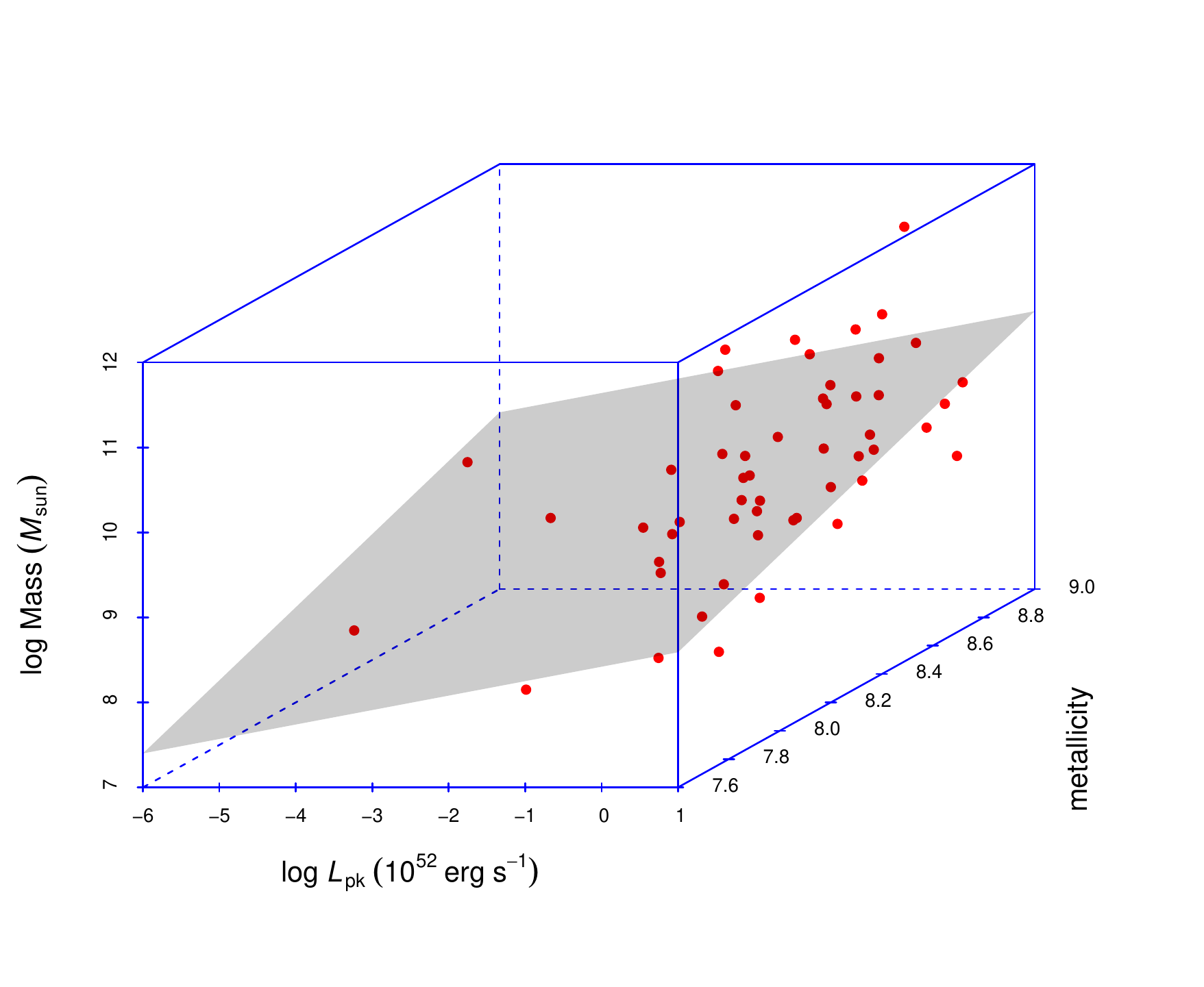}
\includegraphics[width=0.45\textwidth]{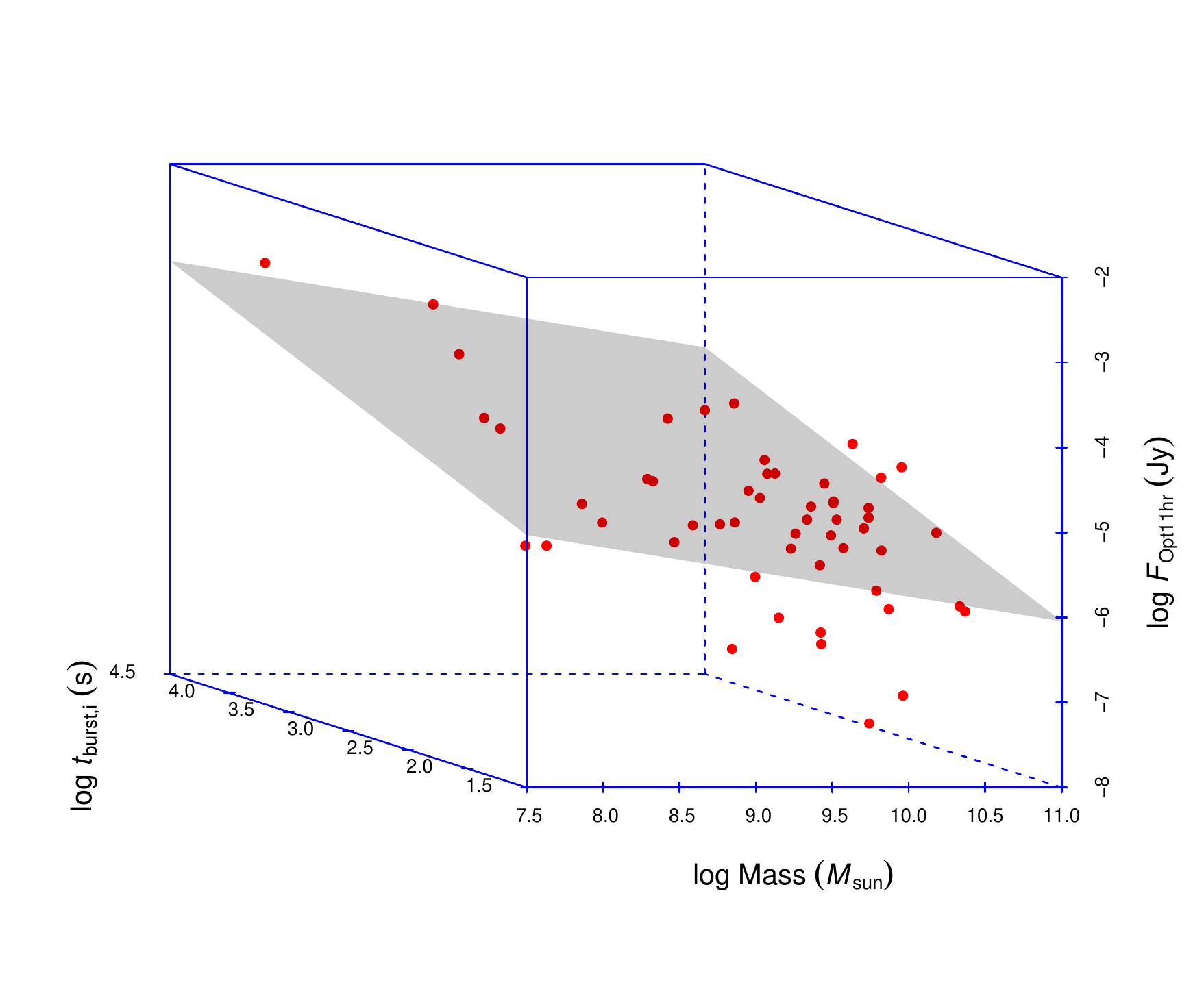}

\center{Fig. \ref{fig:three}---Continued}
\end{figure*}


\clearpage
\begin{figure*}

\includegraphics[width=0.45\textwidth]{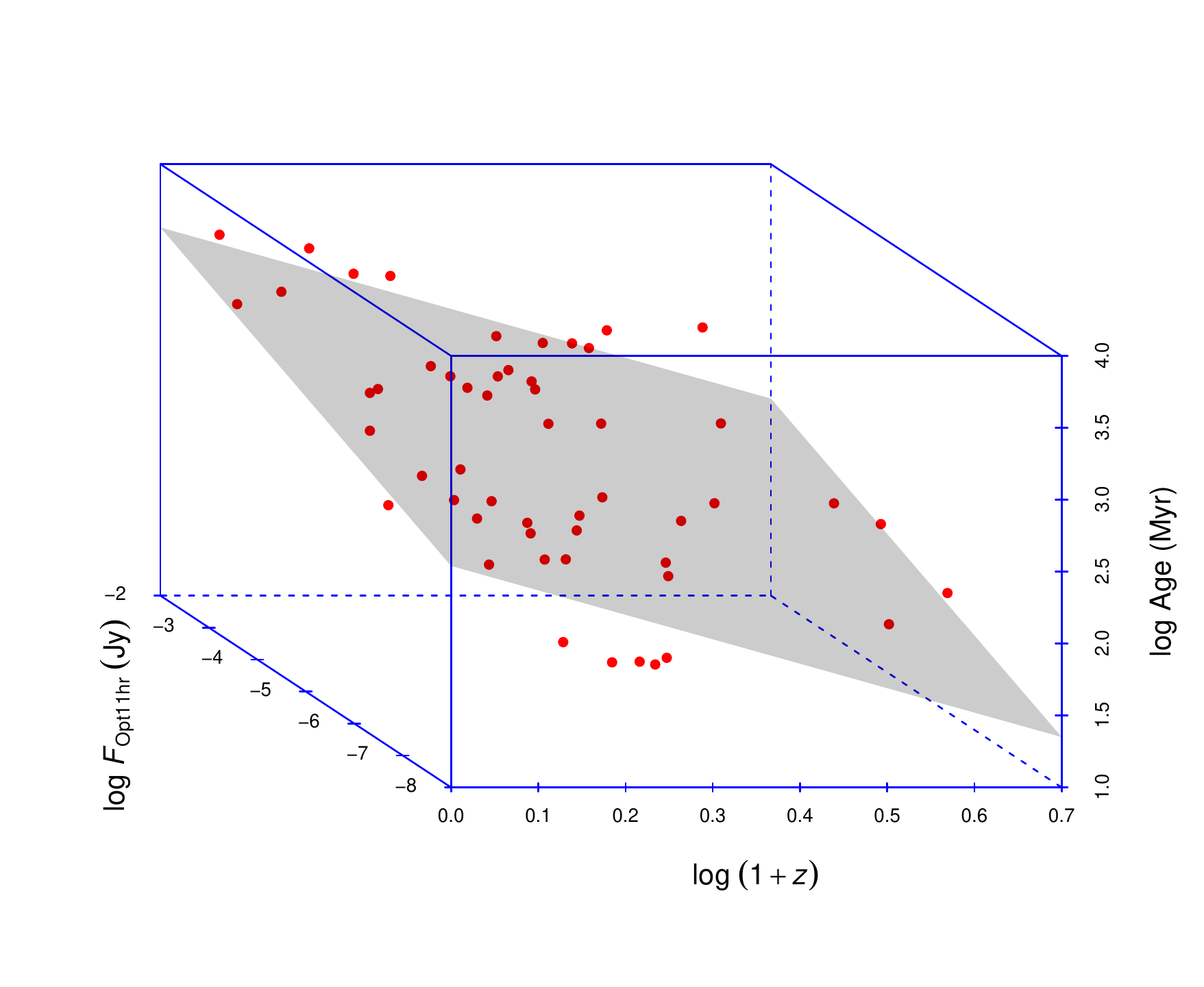}
\includegraphics[width=0.45\textwidth]{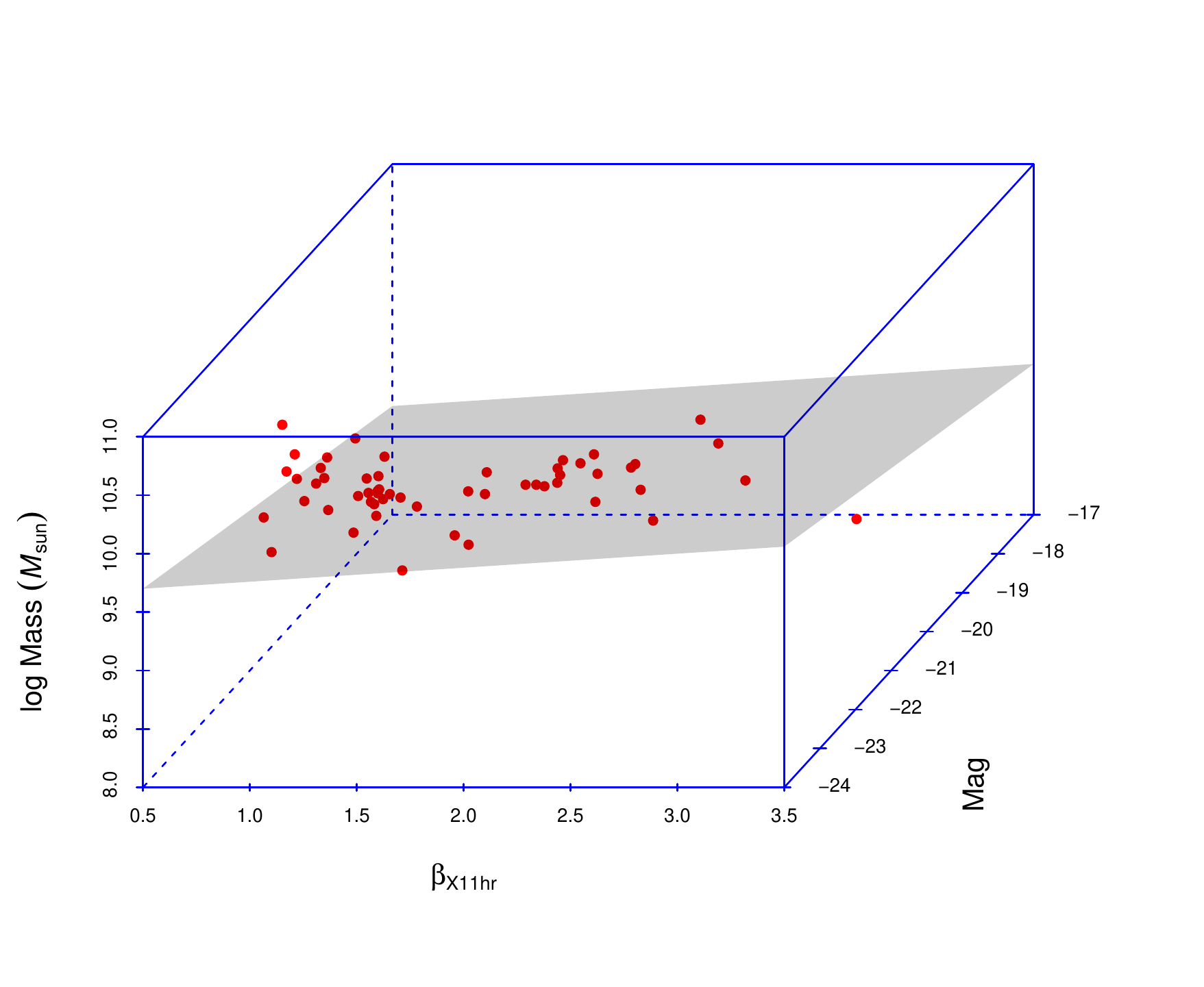}

\includegraphics[width=0.45\textwidth]{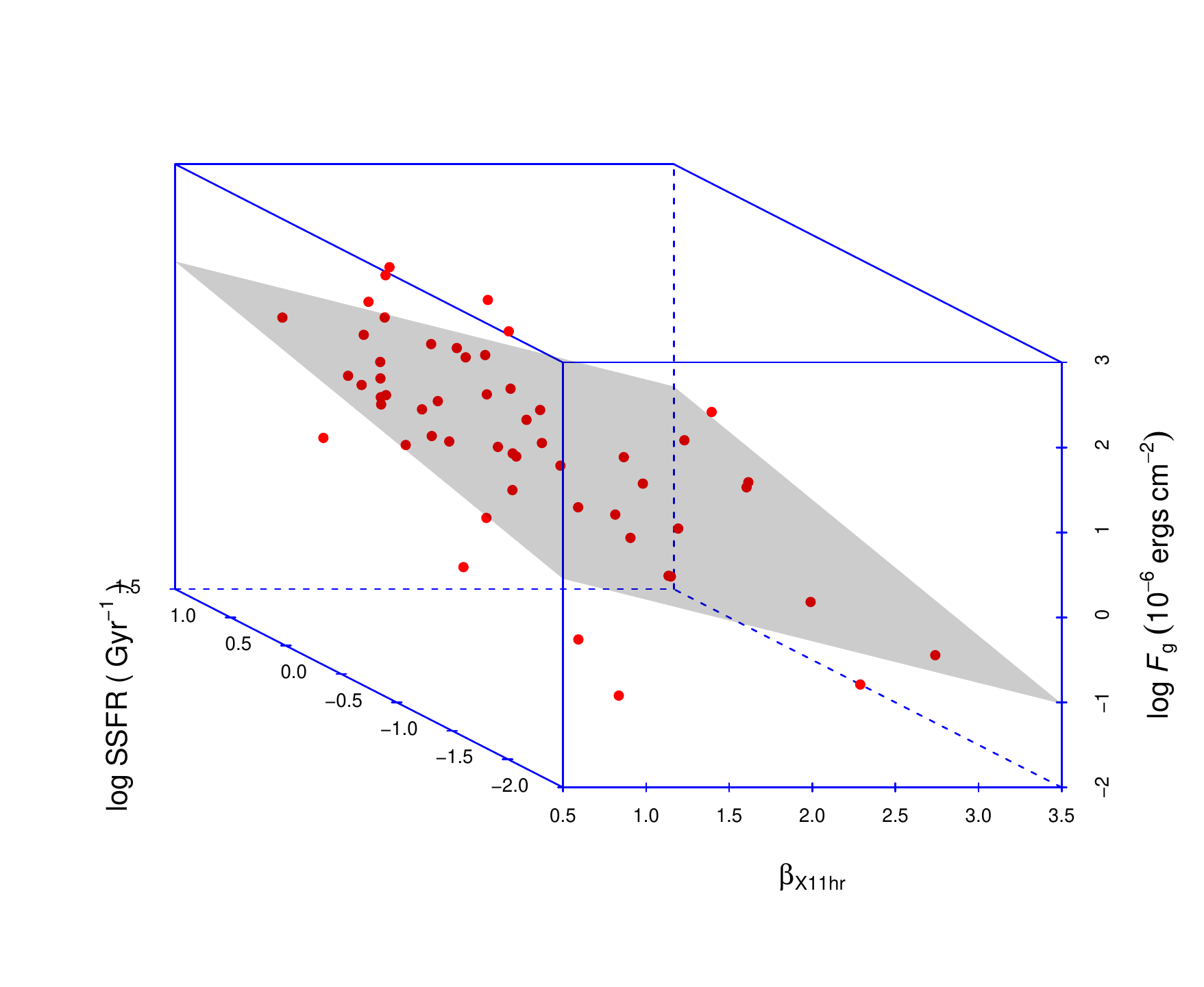}
\includegraphics[width=0.45\textwidth]{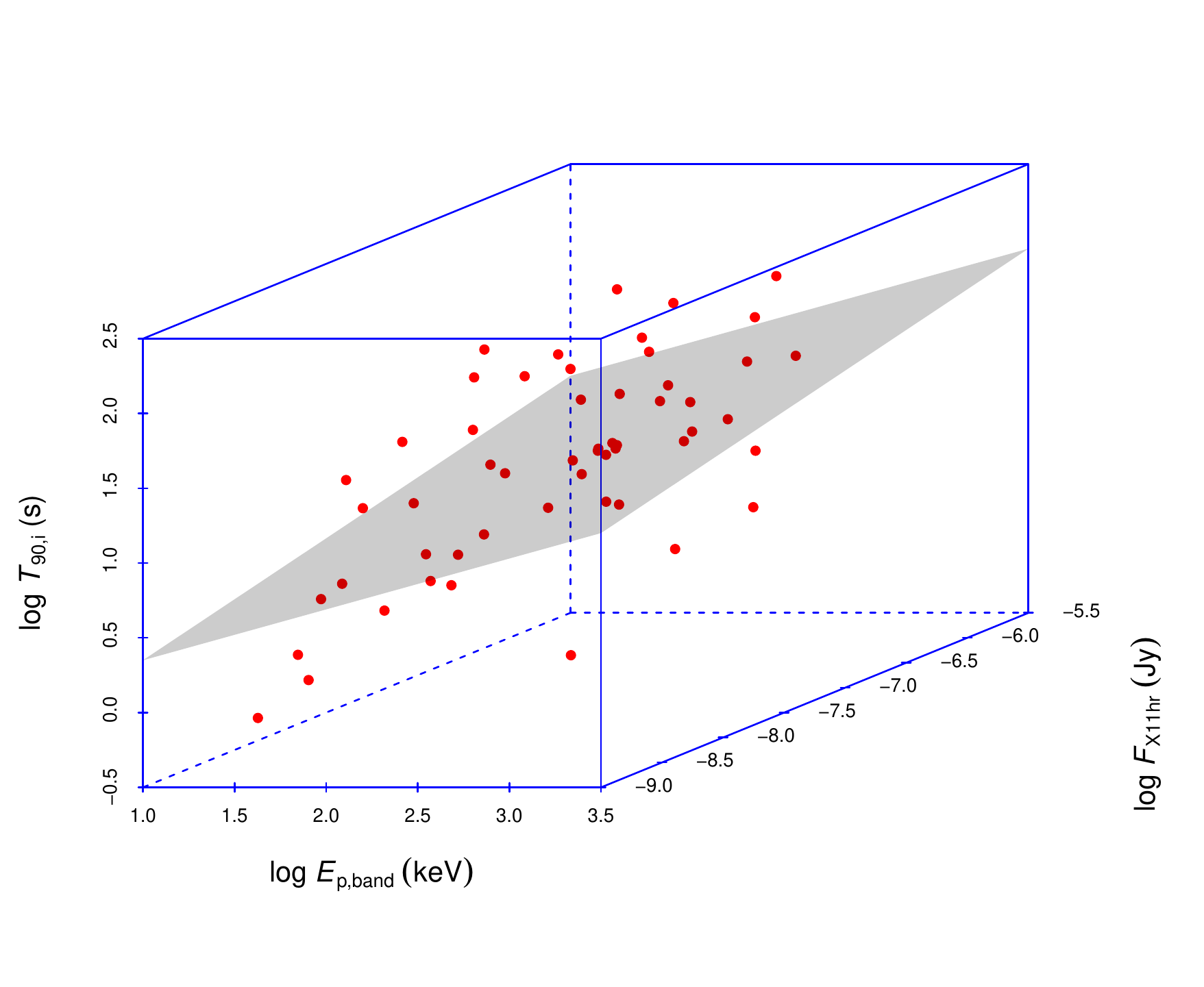}

\includegraphics[width=0.45\textwidth]{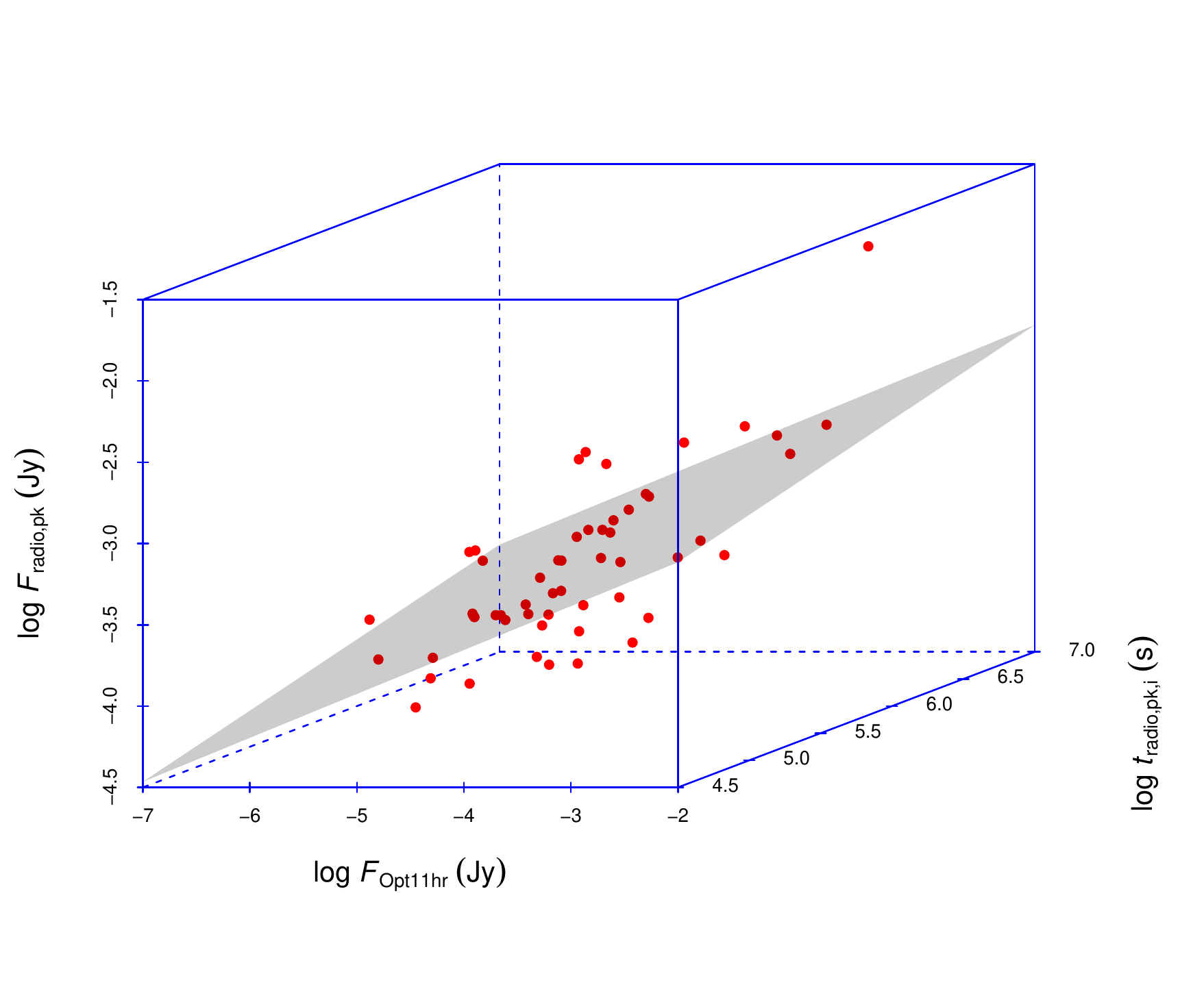}
\includegraphics[width=0.45\textwidth]{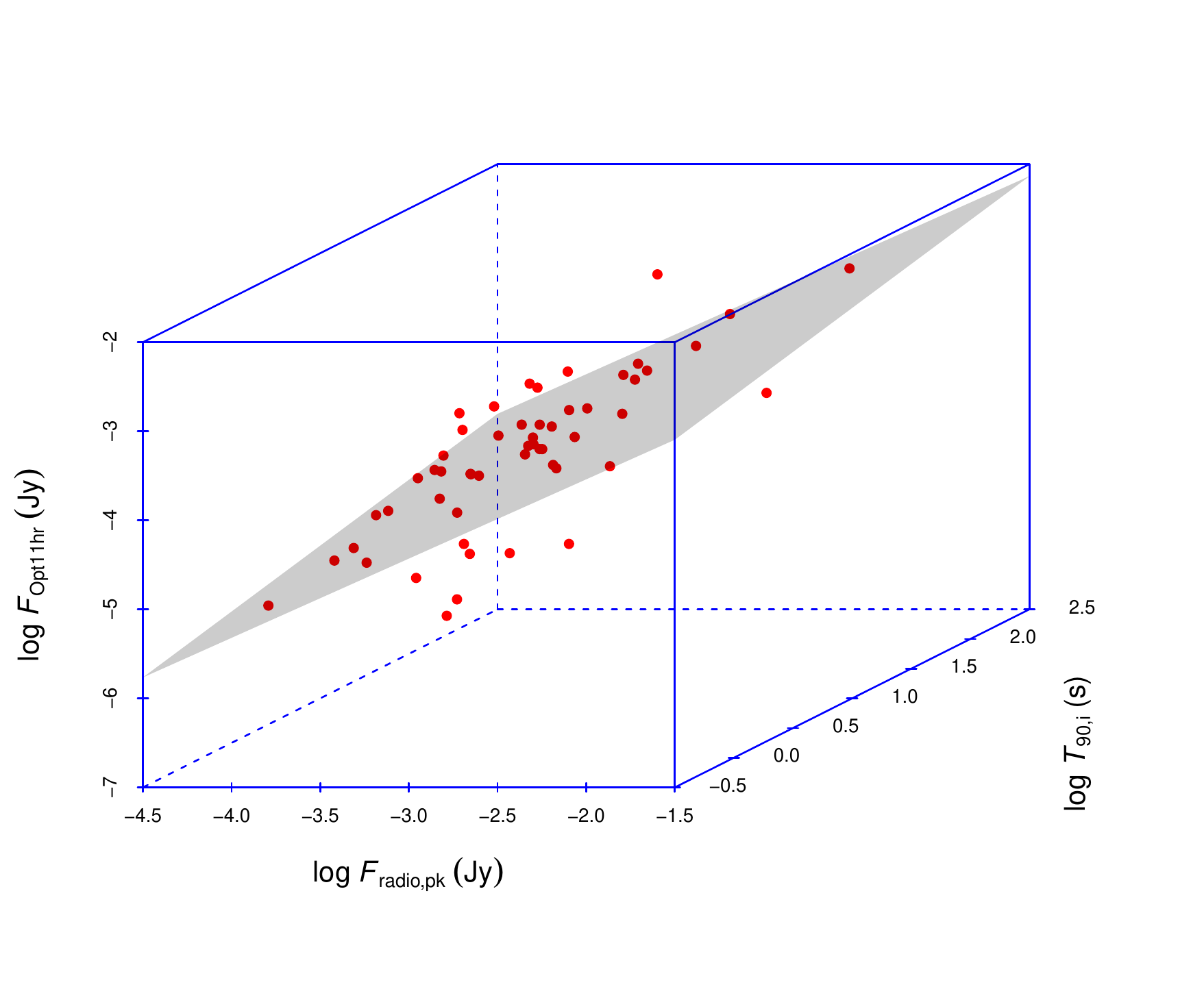}

\center{Fig. \ref{fig:three}---Continued}
\end{figure*}


\clearpage
\begin{figure*}

\includegraphics[width=0.45\textwidth]{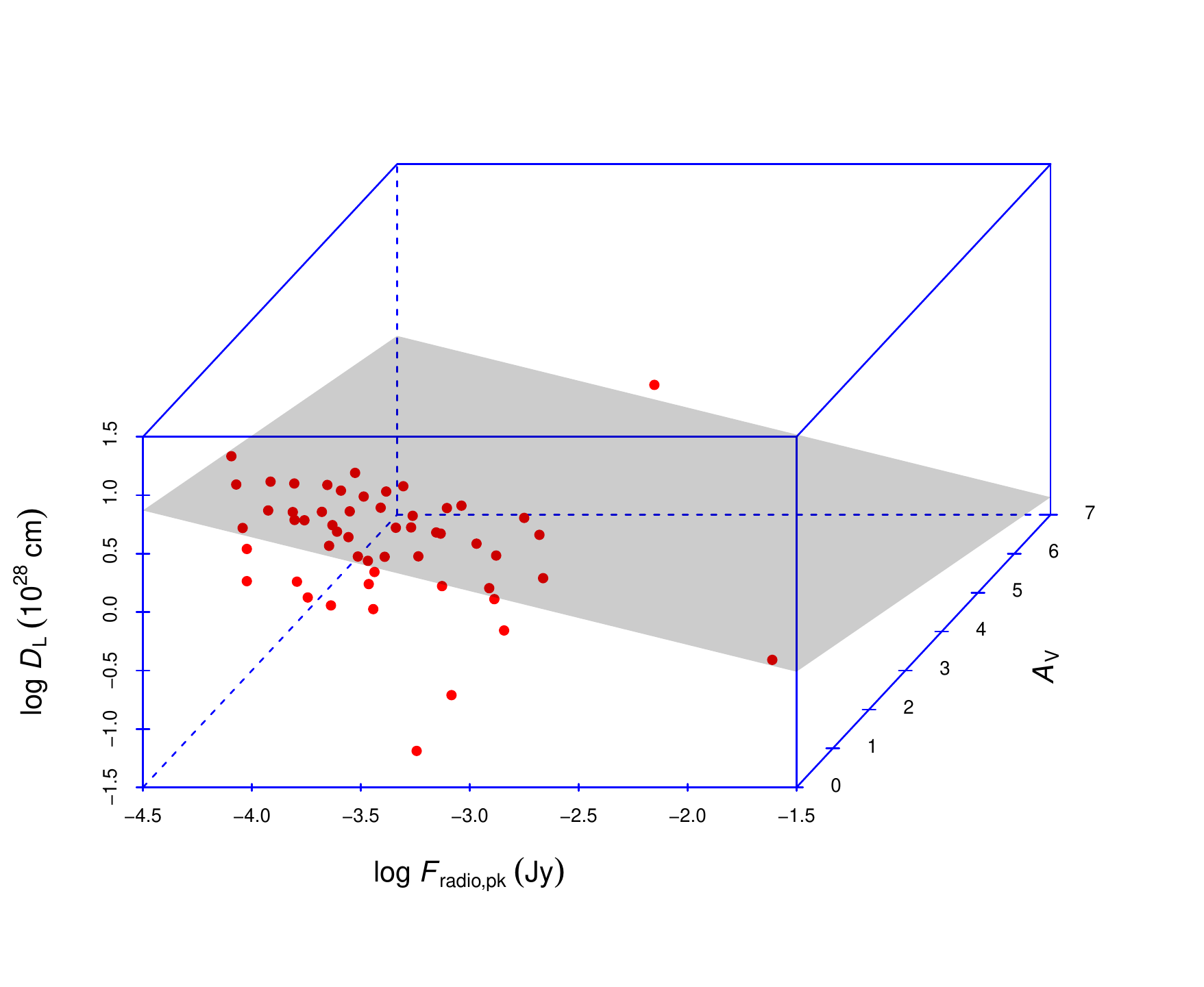}
\includegraphics[width=0.45\textwidth]{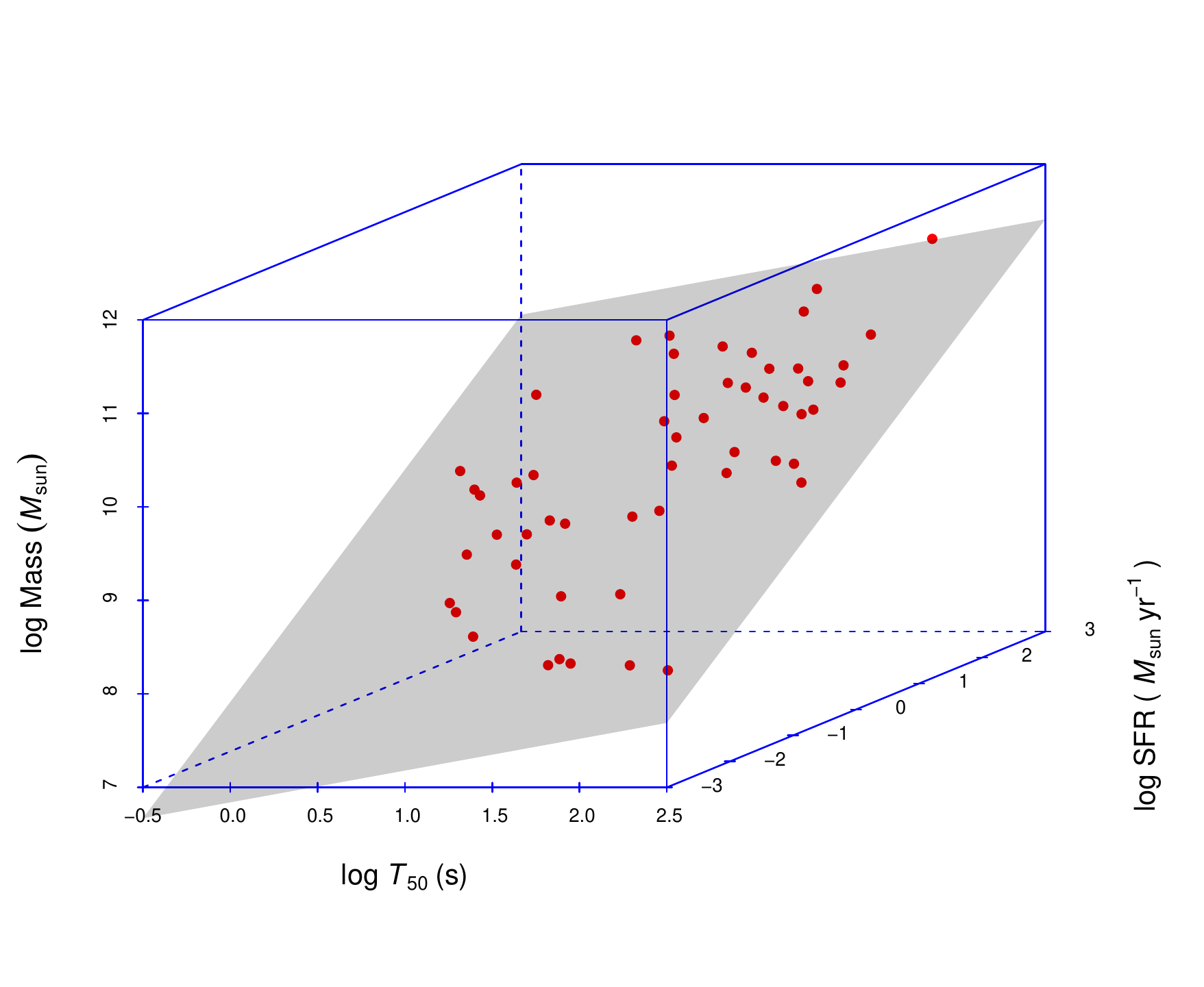}

\includegraphics[width=0.45\textwidth]{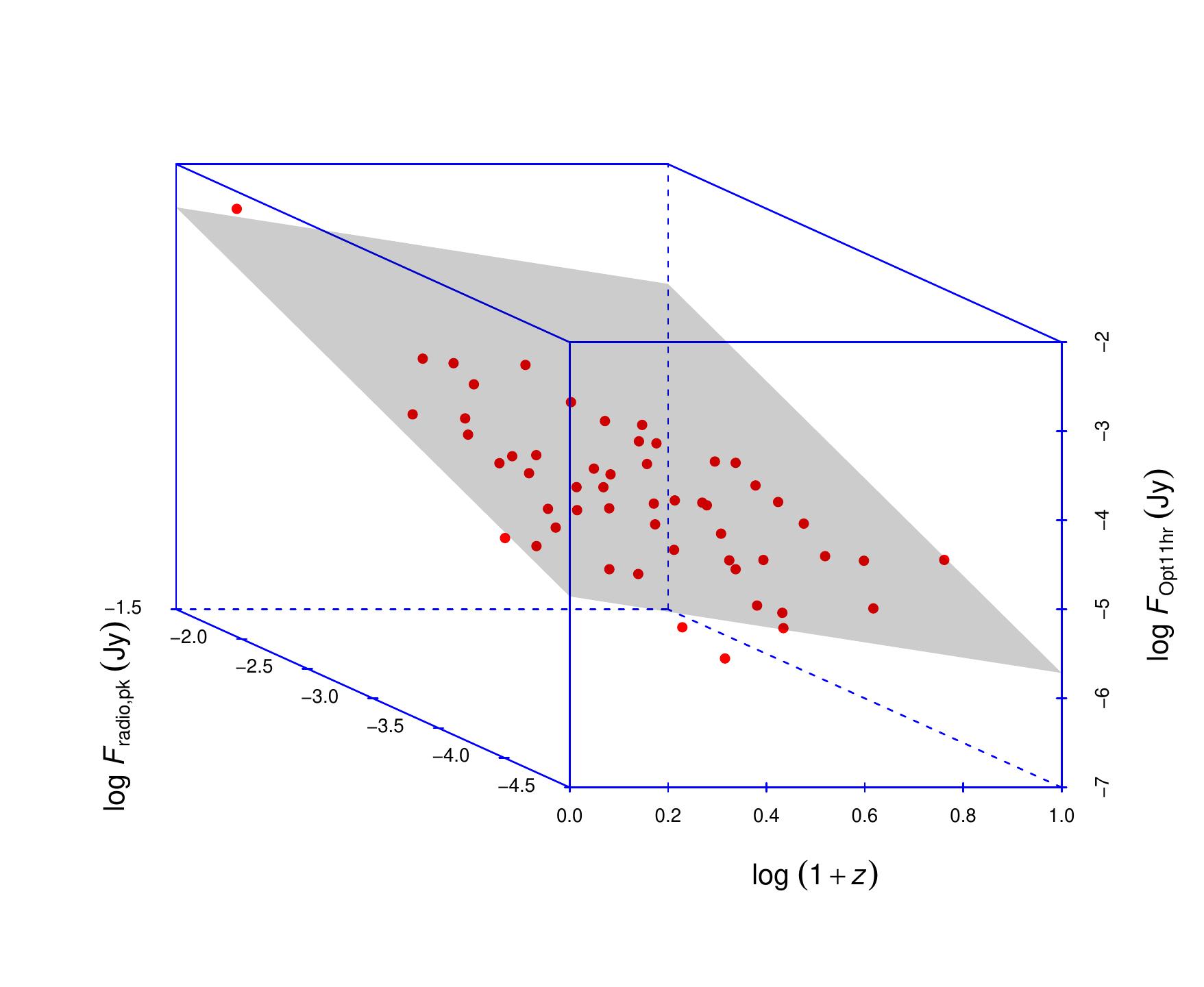}
\includegraphics[width=0.45\textwidth]{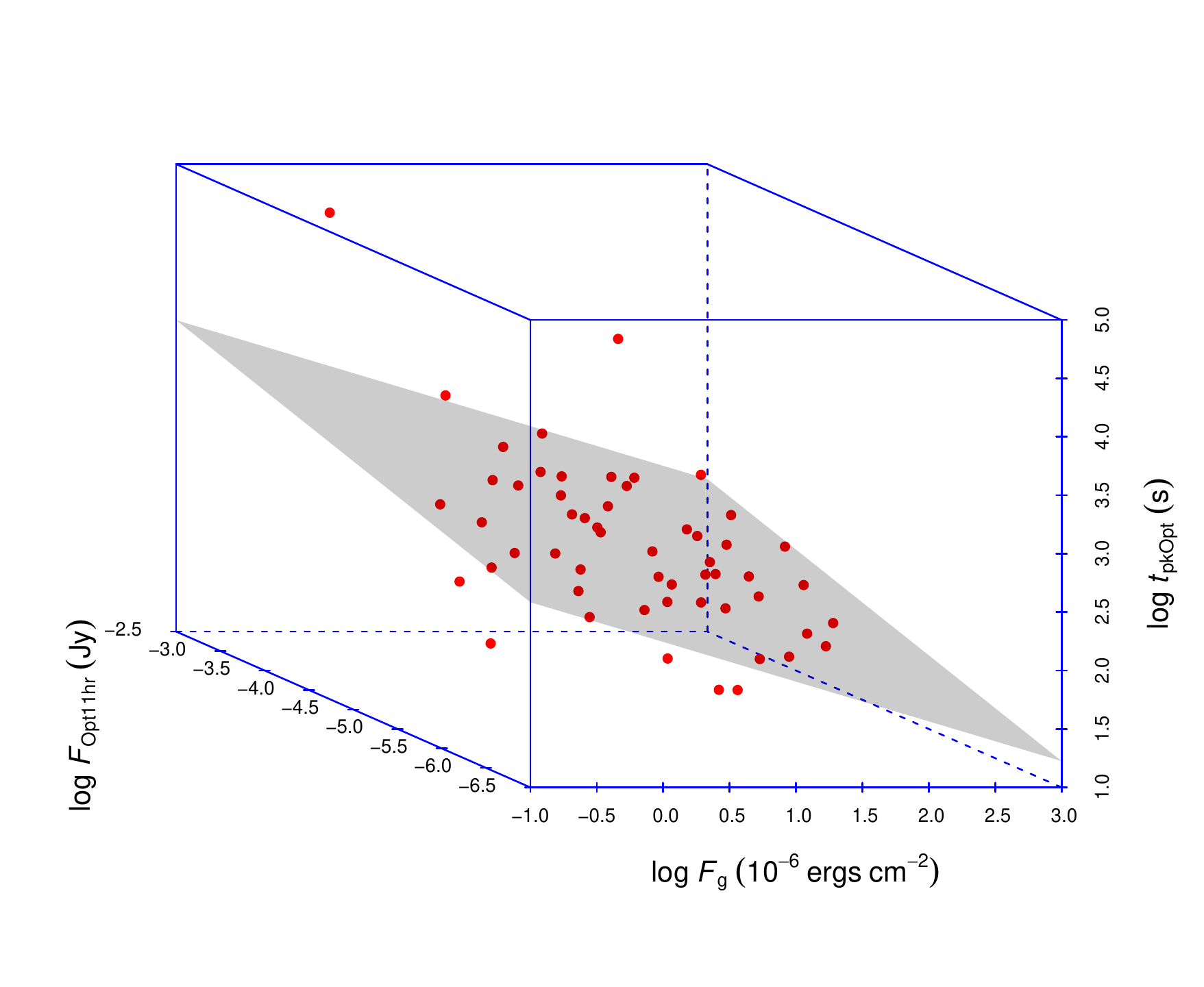}

\includegraphics[width=0.45\textwidth]{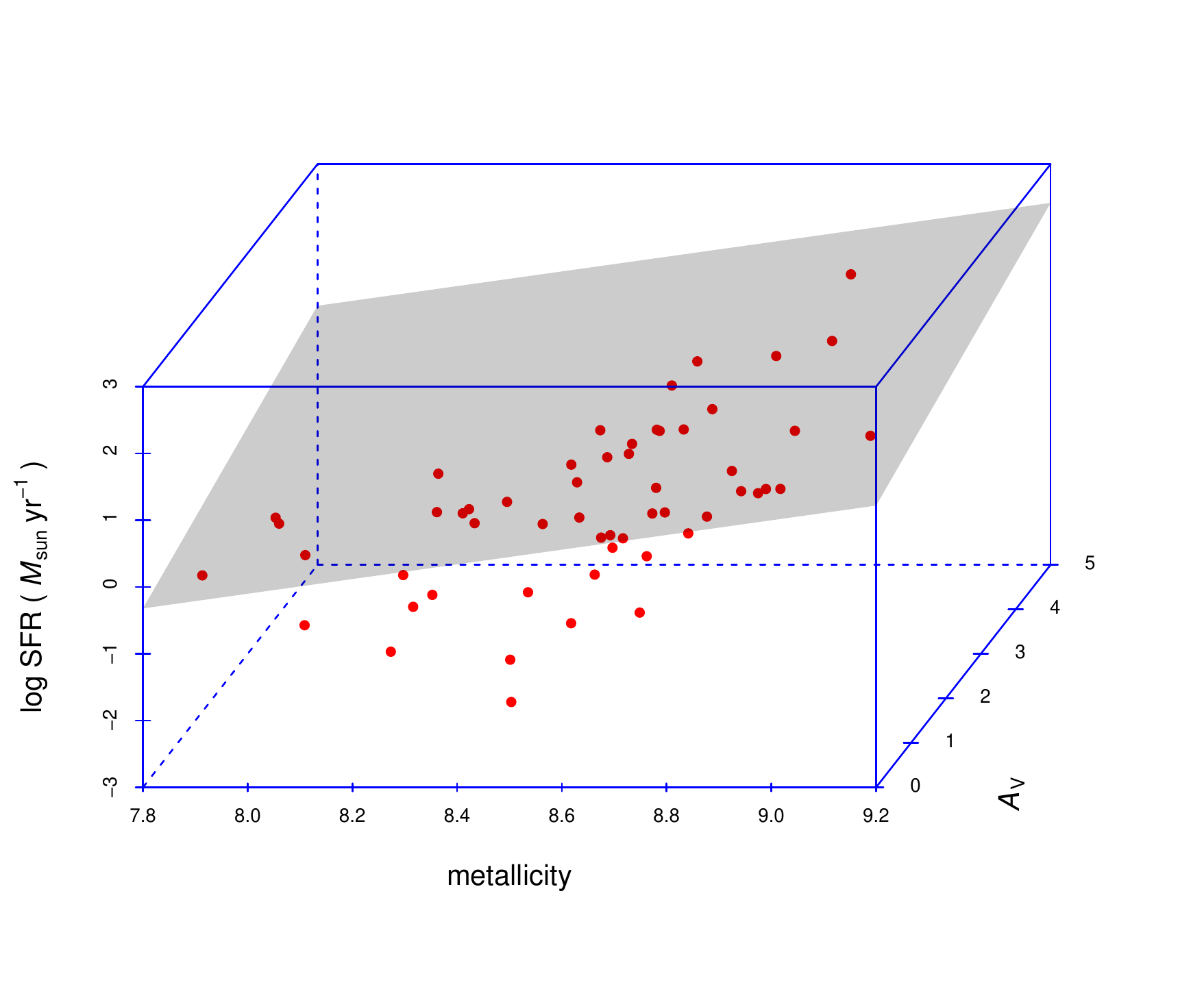}
\includegraphics[width=0.45\textwidth]{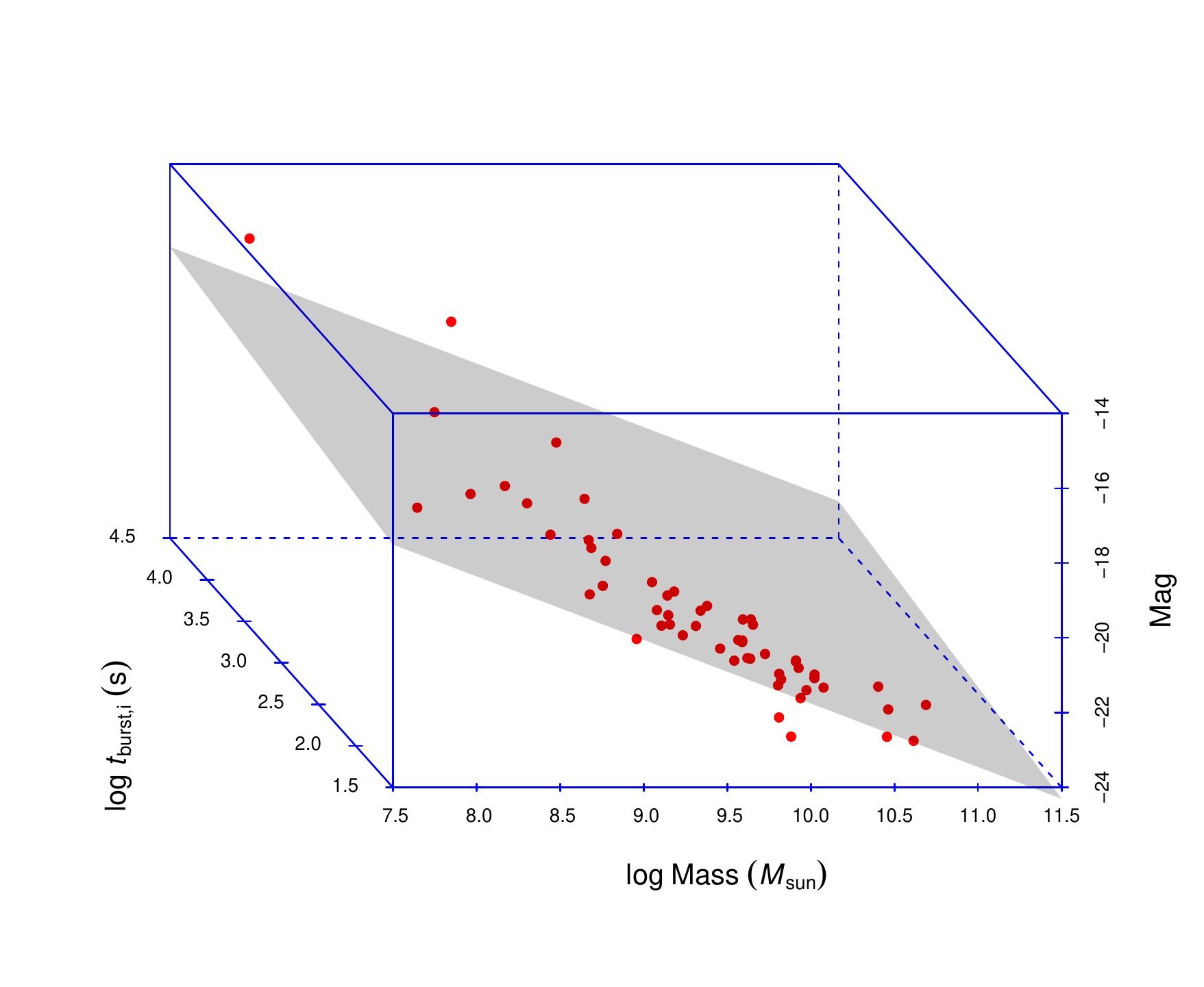}

\center{Fig. \ref{fig:three}---Continued}
\end{figure*}


\clearpage
\begin{figure*}

\includegraphics[width=0.45\textwidth]{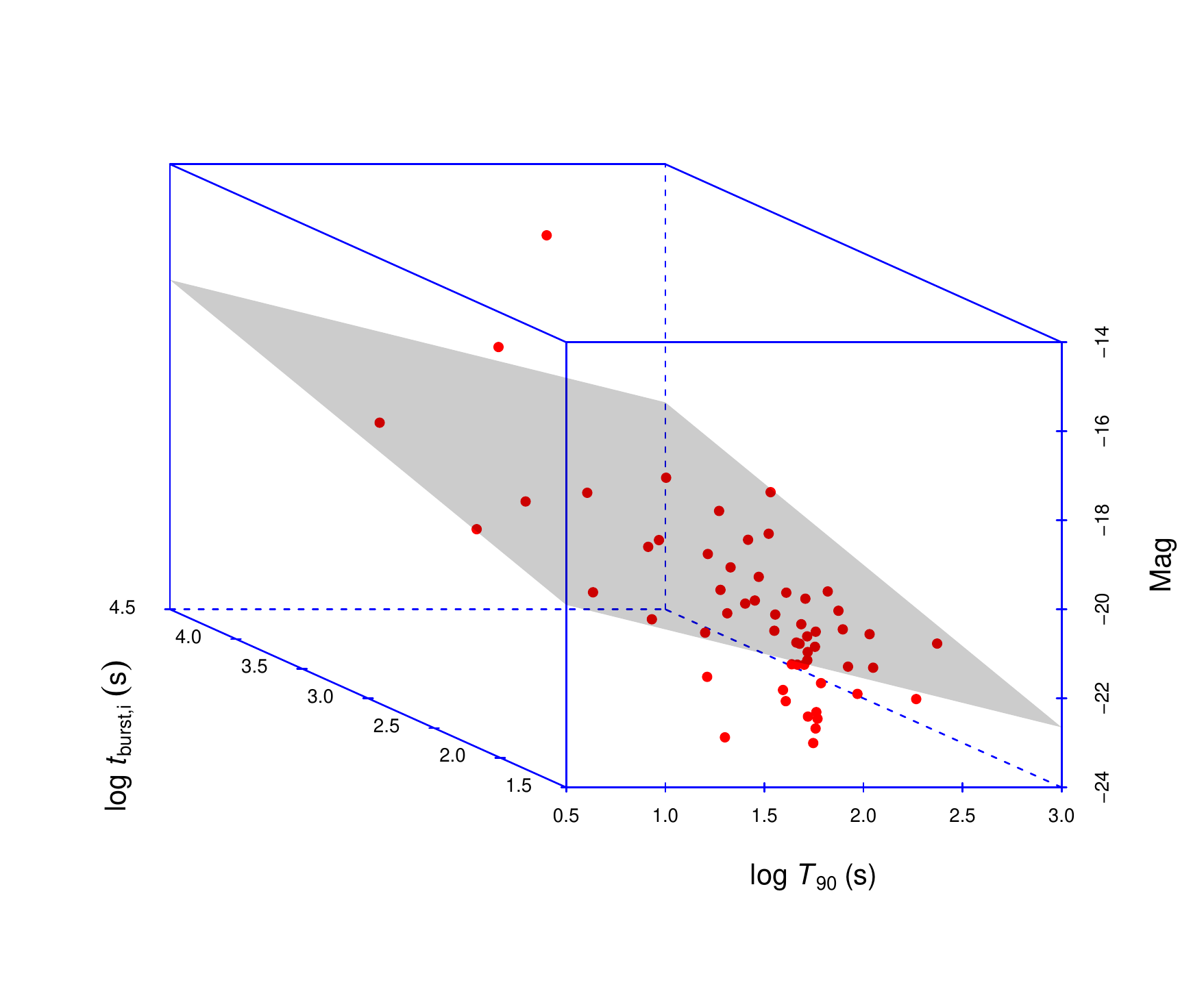}
\includegraphics[width=0.45\textwidth]{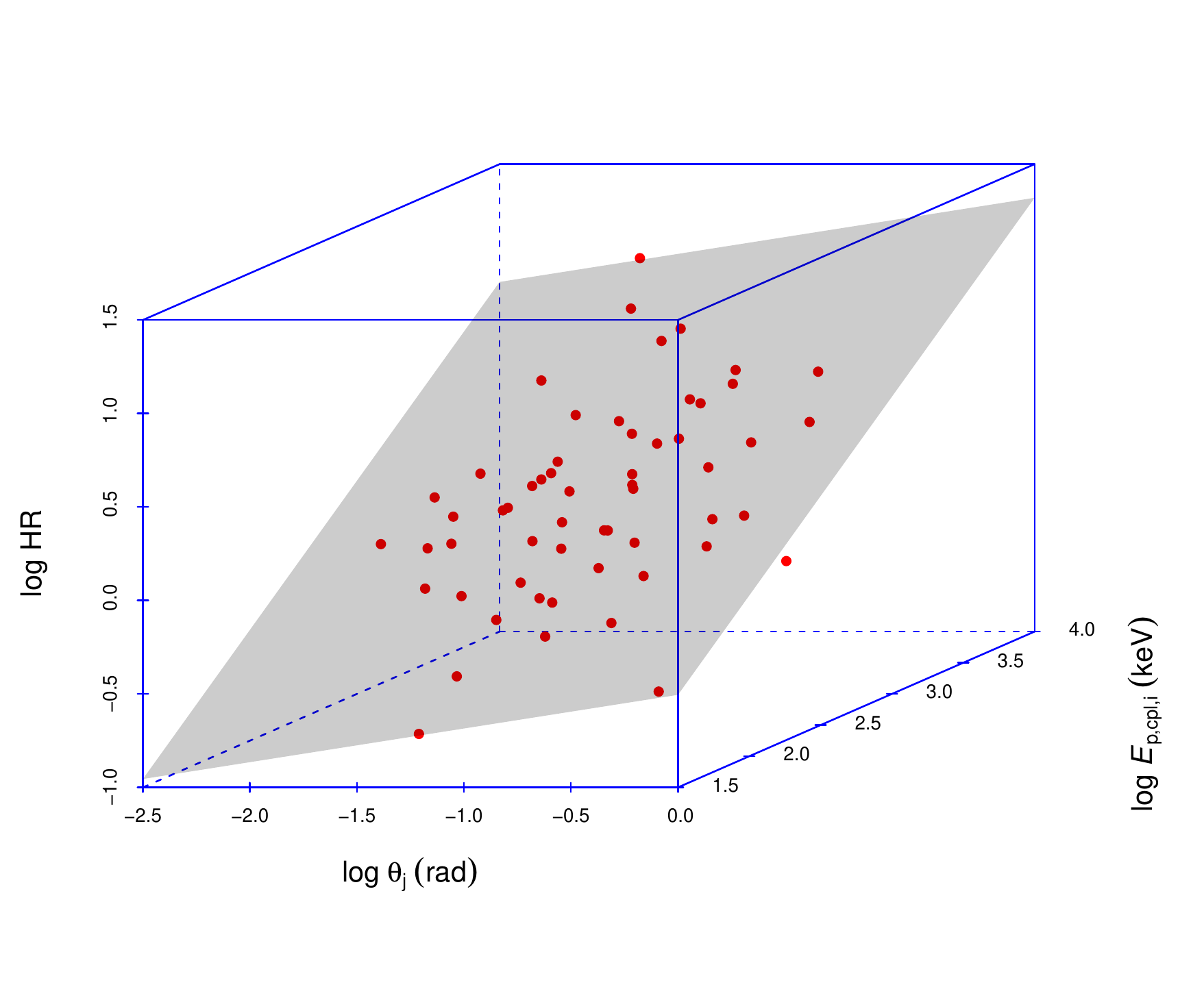}

\includegraphics[width=0.45\textwidth]{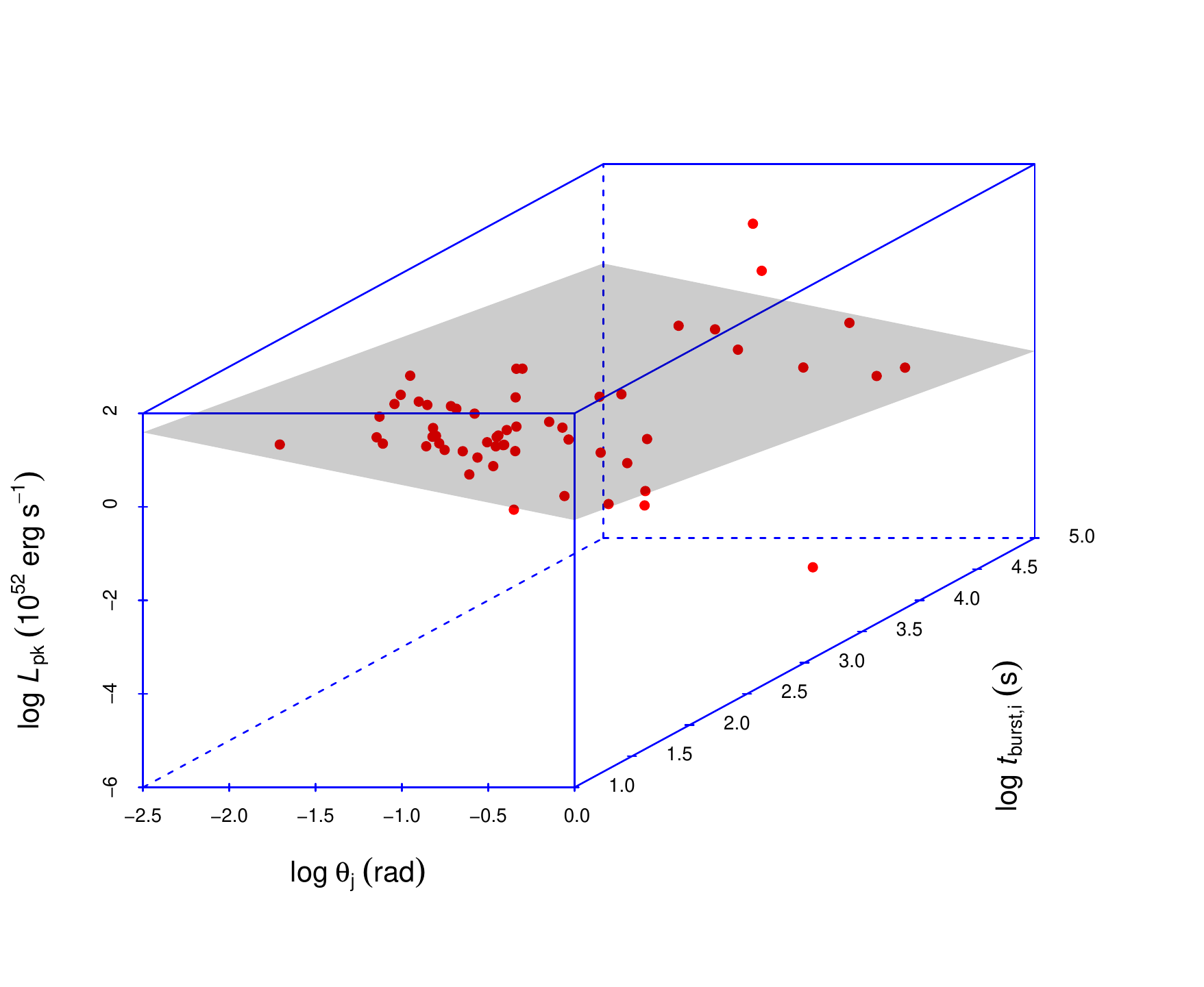}
\includegraphics[width=0.45\textwidth]{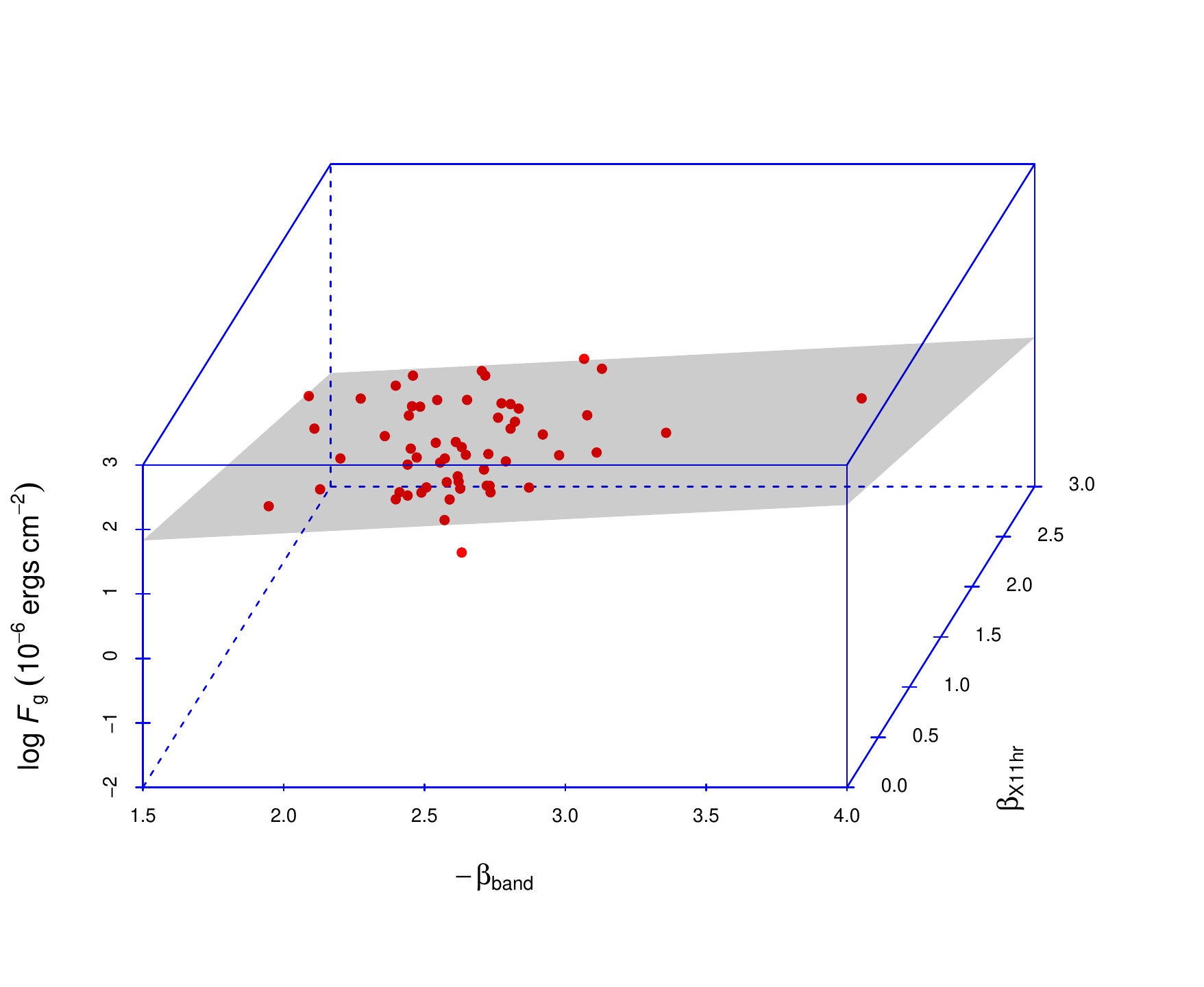}

\includegraphics[width=0.45\textwidth]{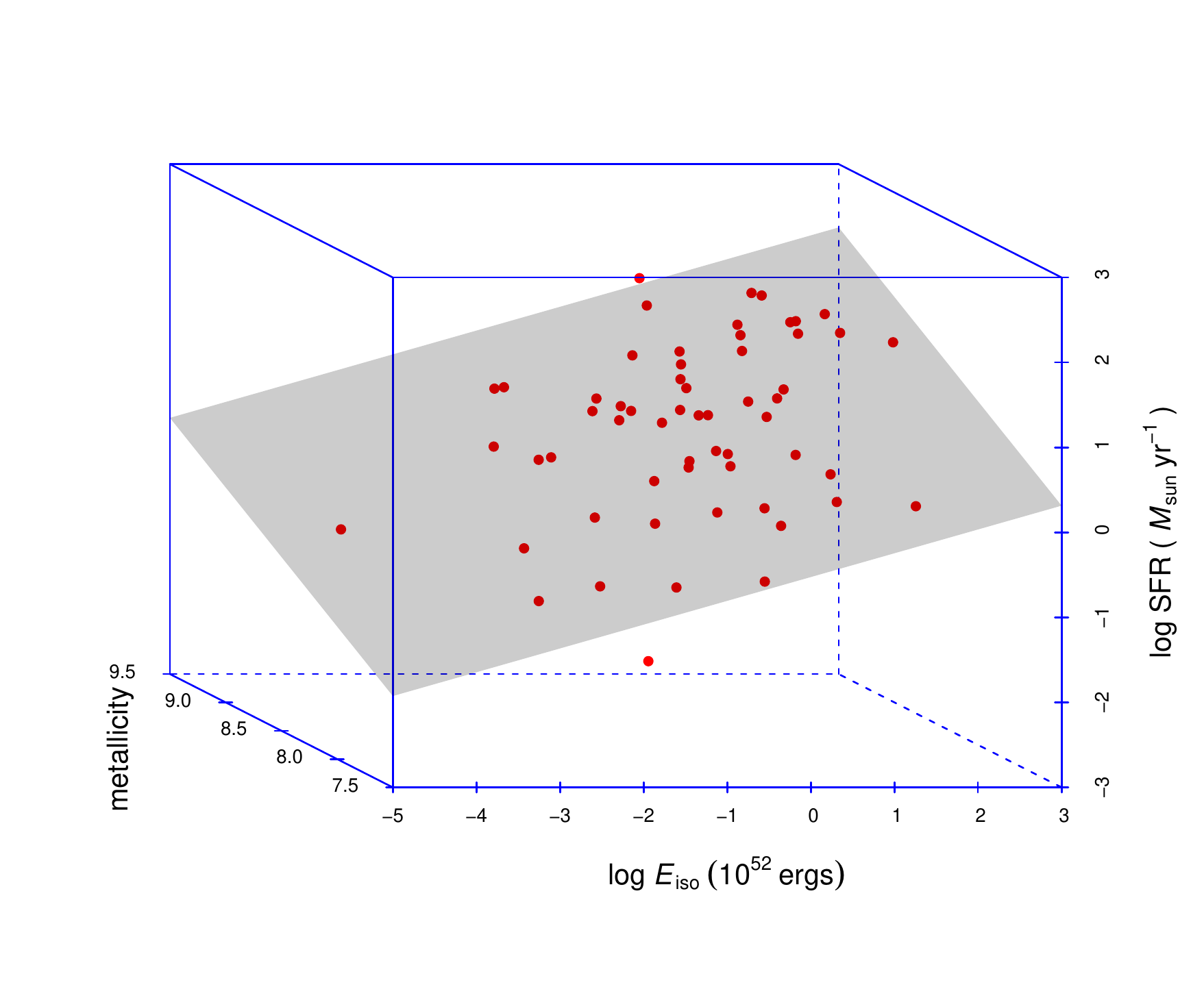}
\includegraphics[width=0.45\textwidth]{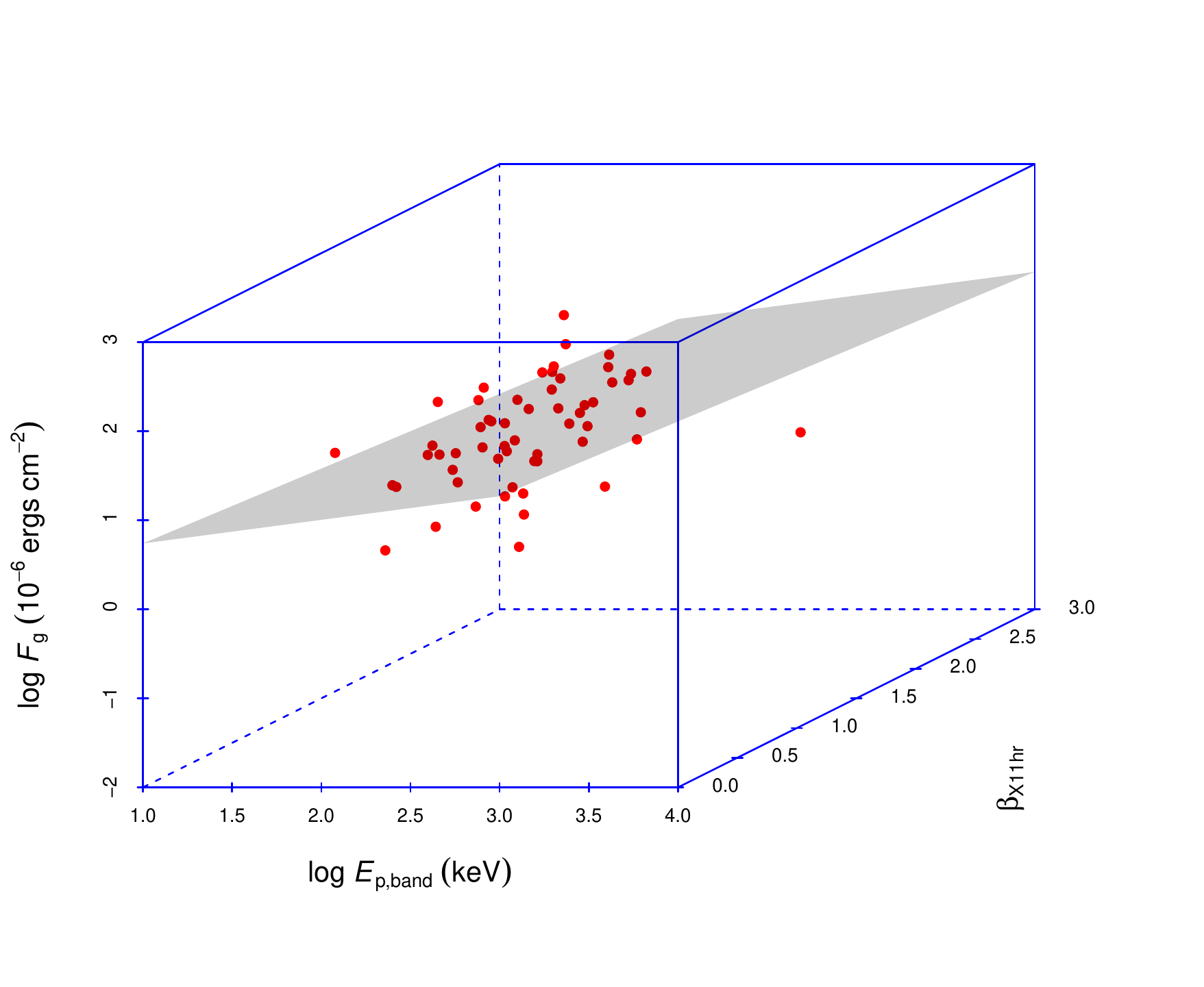}

\center{Fig. \ref{fig:three}---Continued}
\end{figure*}


\clearpage
\begin{figure*}

\includegraphics[width=0.45\textwidth]{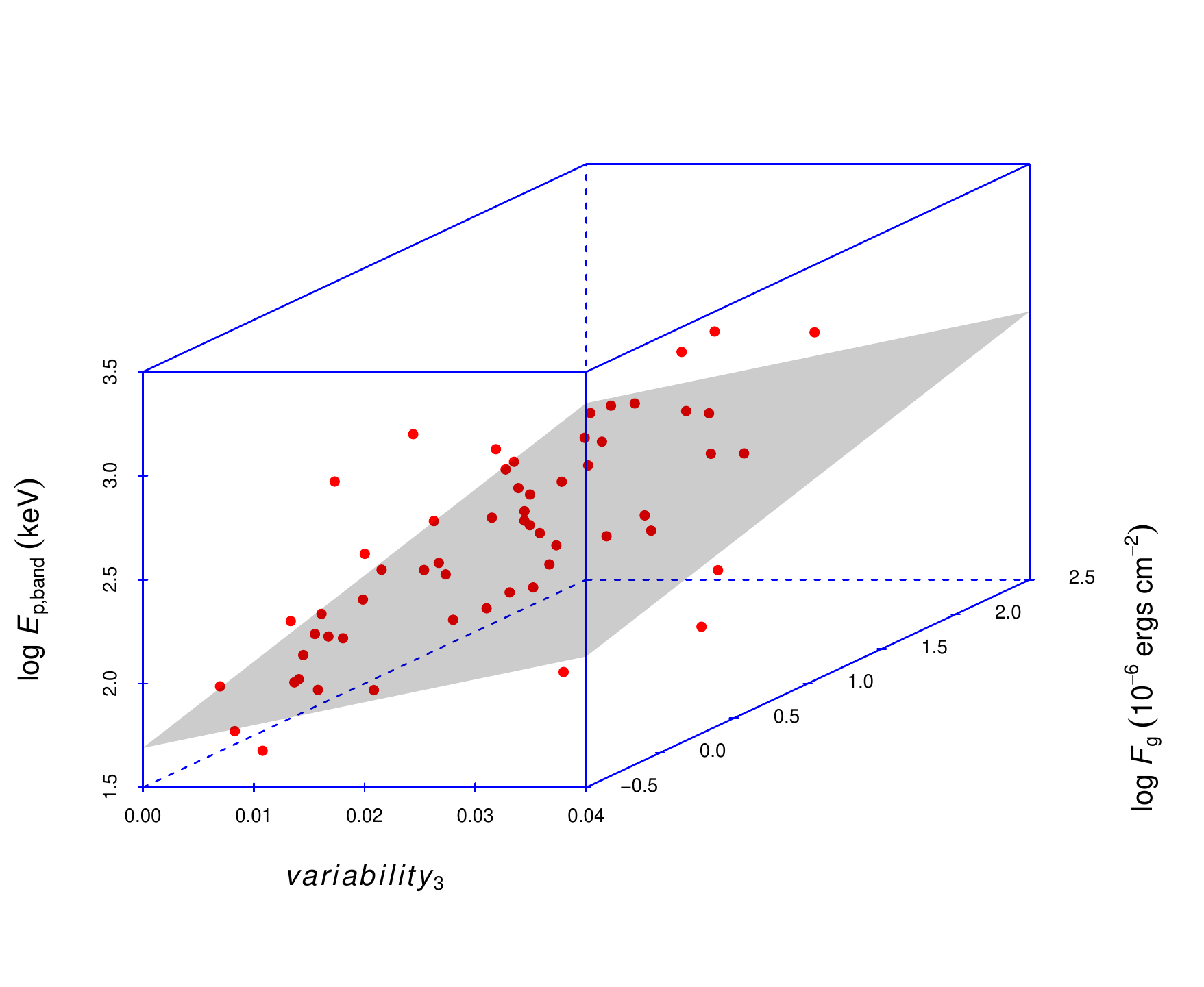}
\includegraphics[width=0.45\textwidth]{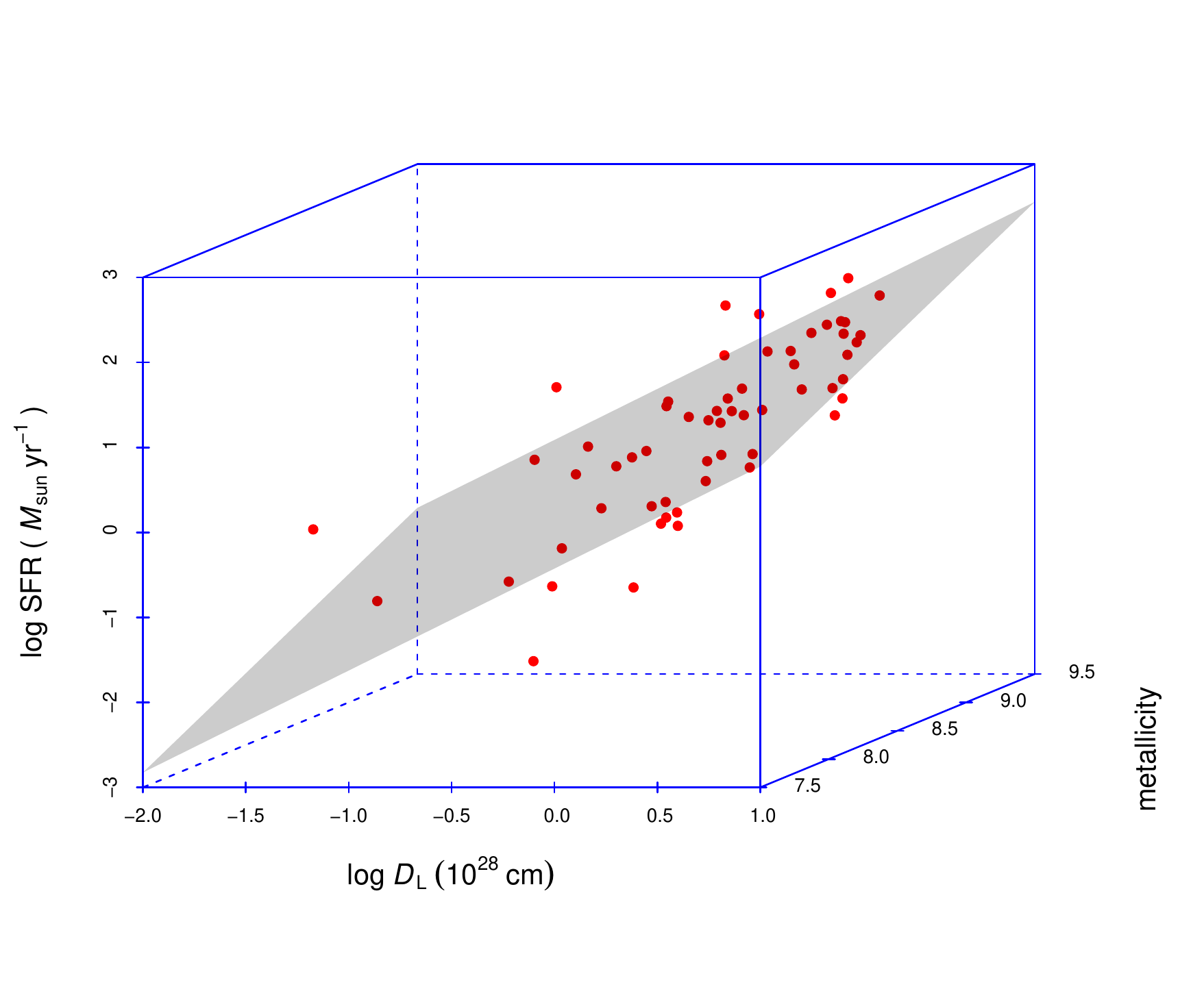}

\includegraphics[width=0.45\textwidth]{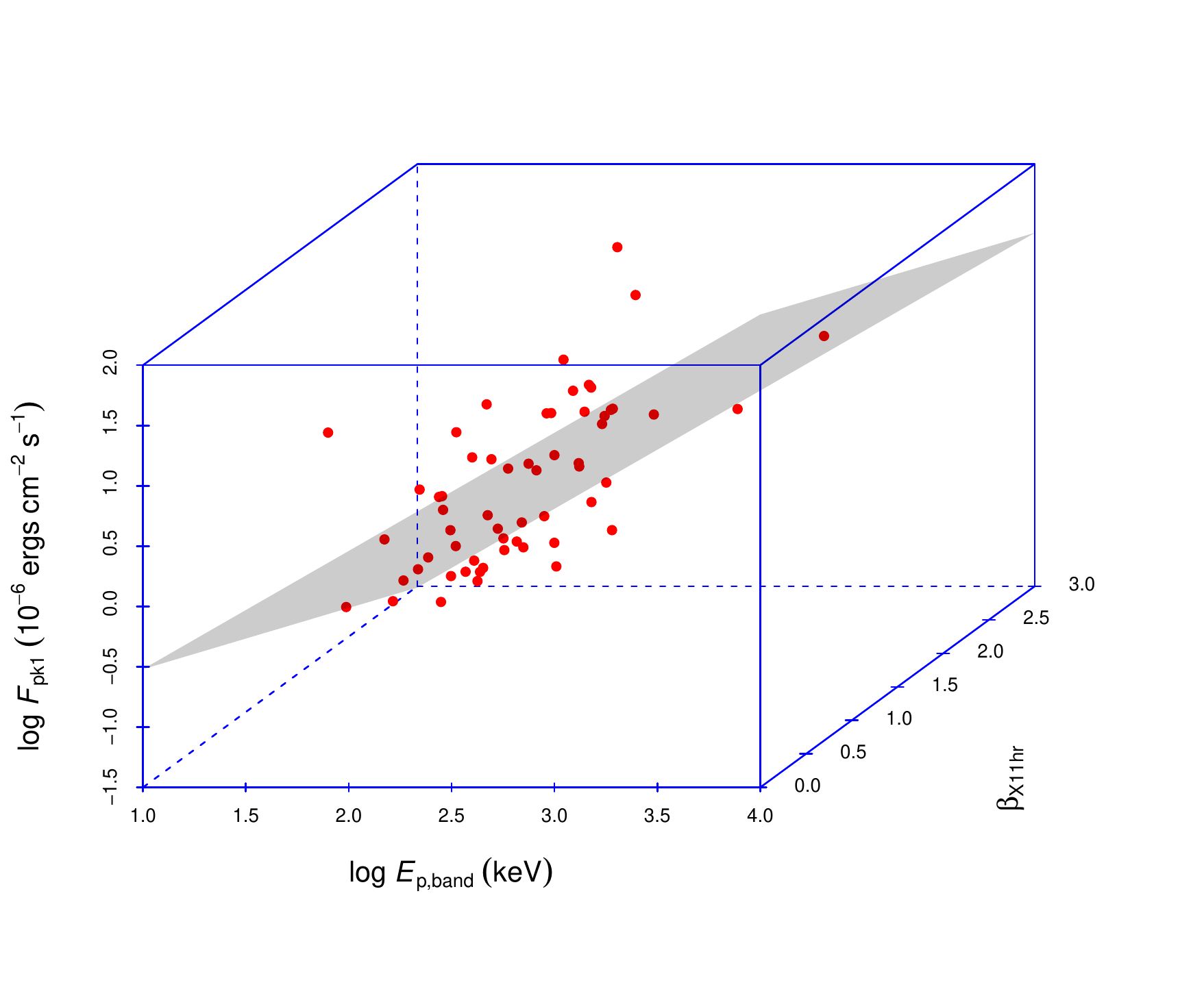}
\includegraphics[width=0.45\textwidth]{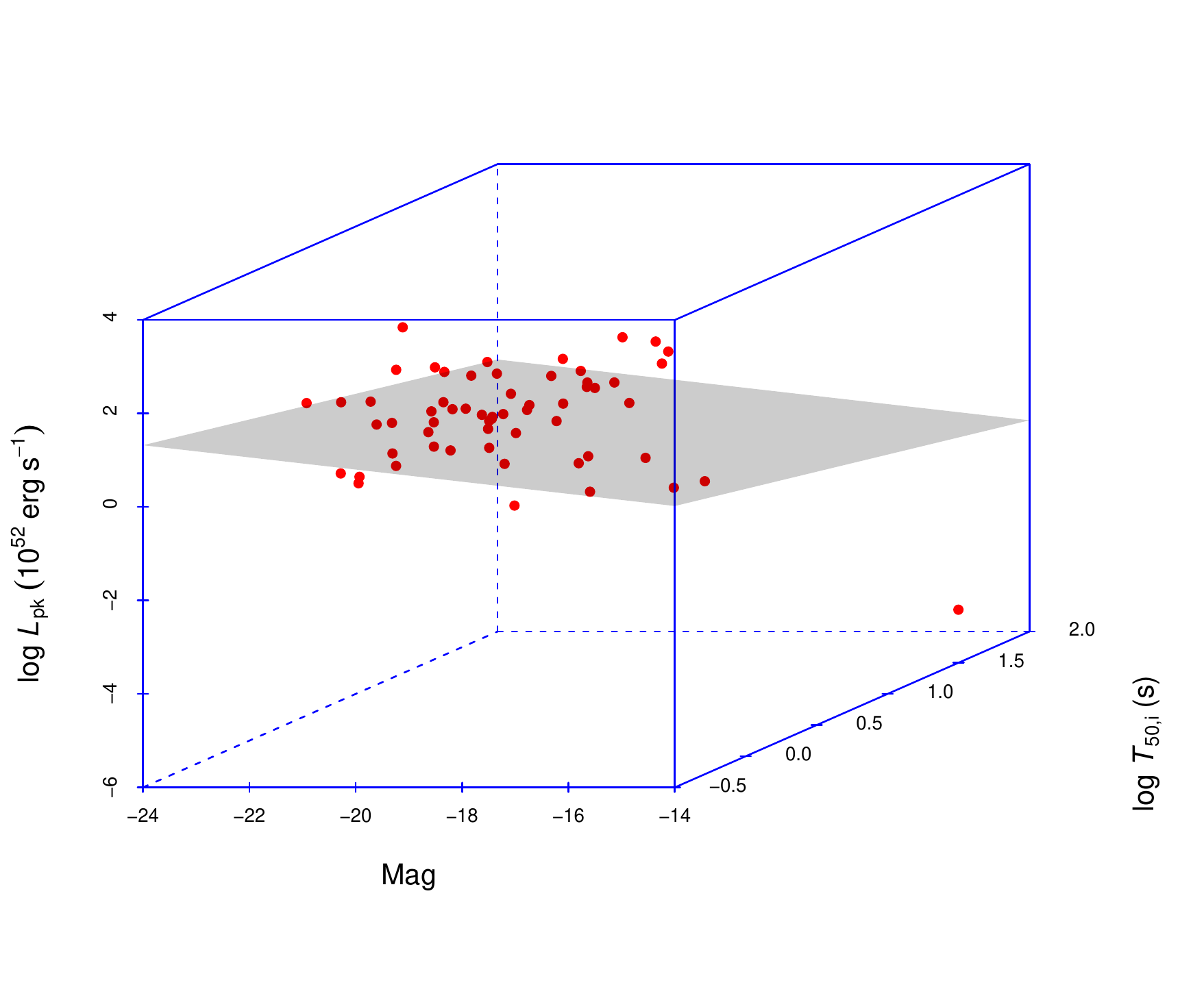}

\includegraphics[width=0.45\textwidth]{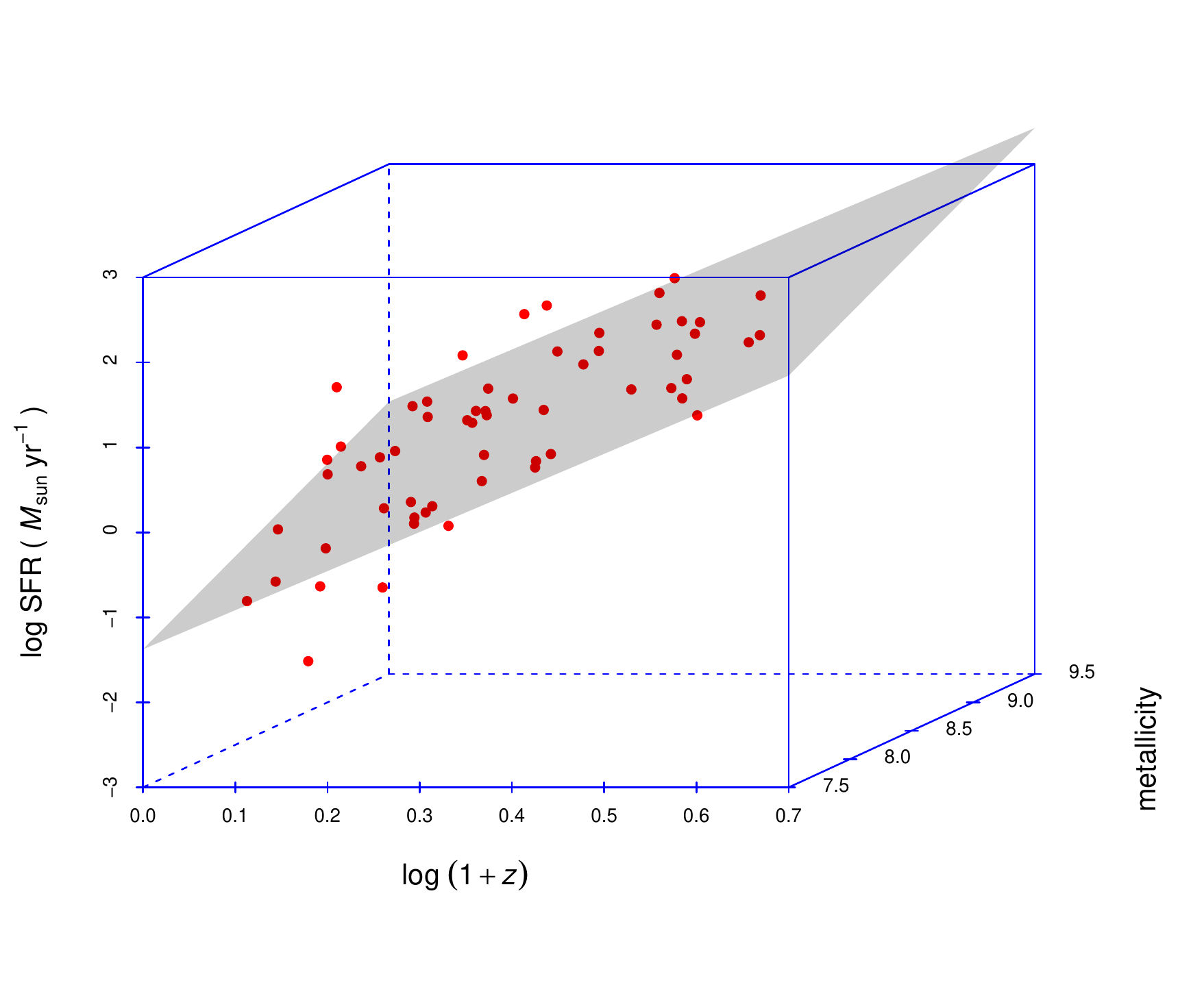}
\includegraphics[width=0.45\textwidth]{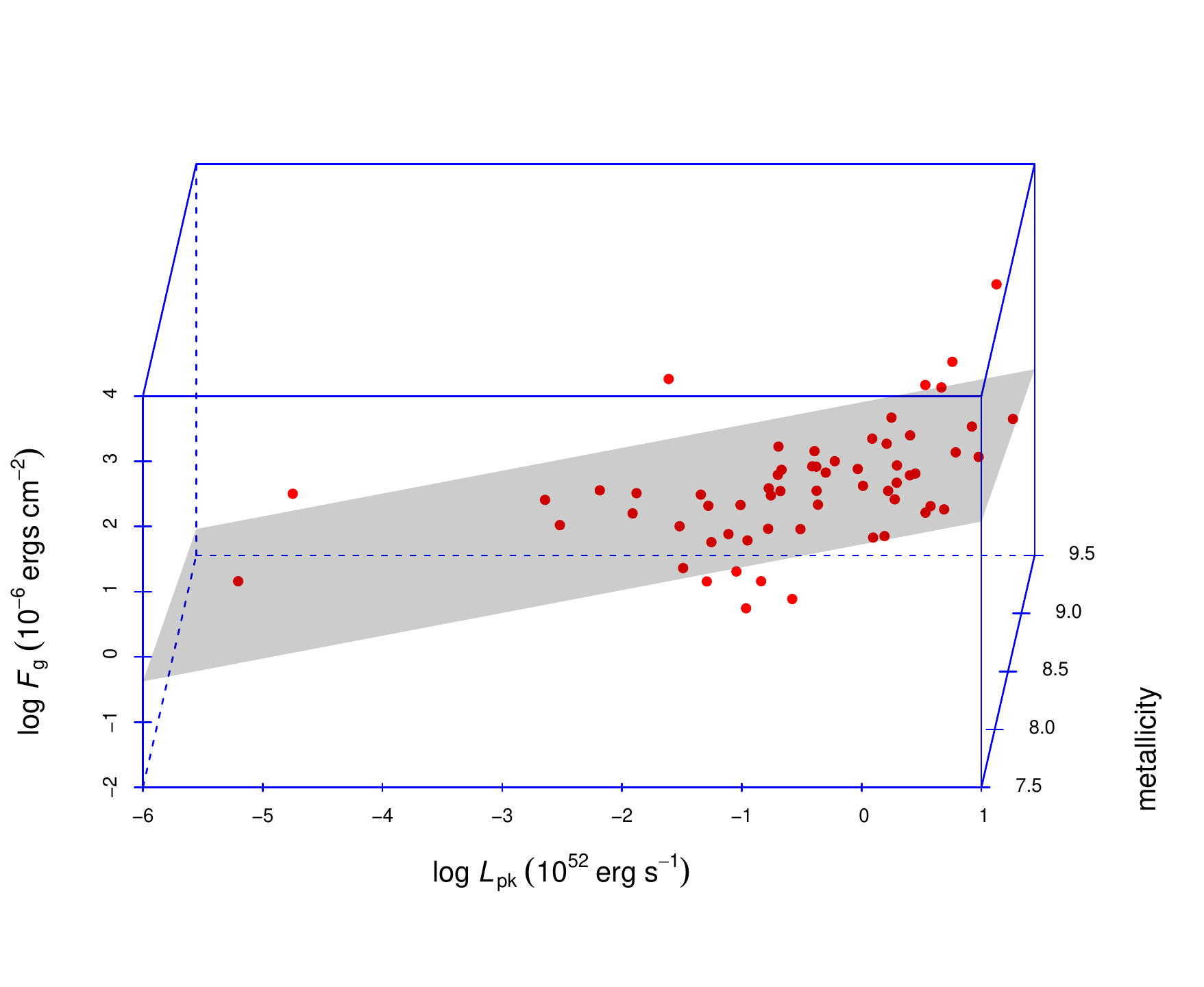}

\center{Fig. \ref{fig:three}---Continued}
\end{figure*}


\clearpage
\begin{figure*}

\includegraphics[width=0.45\textwidth]{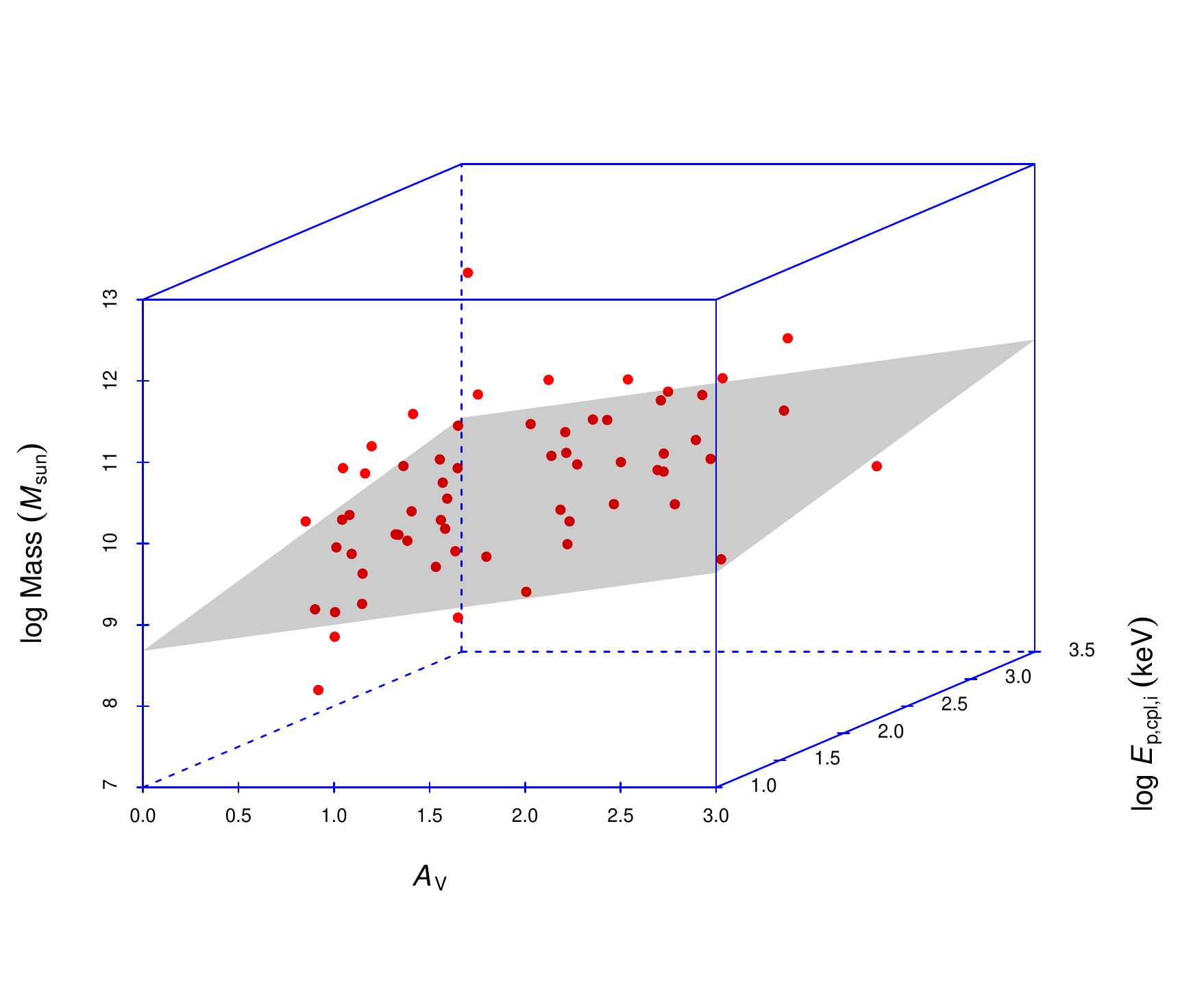}
\includegraphics[width=0.45\textwidth]{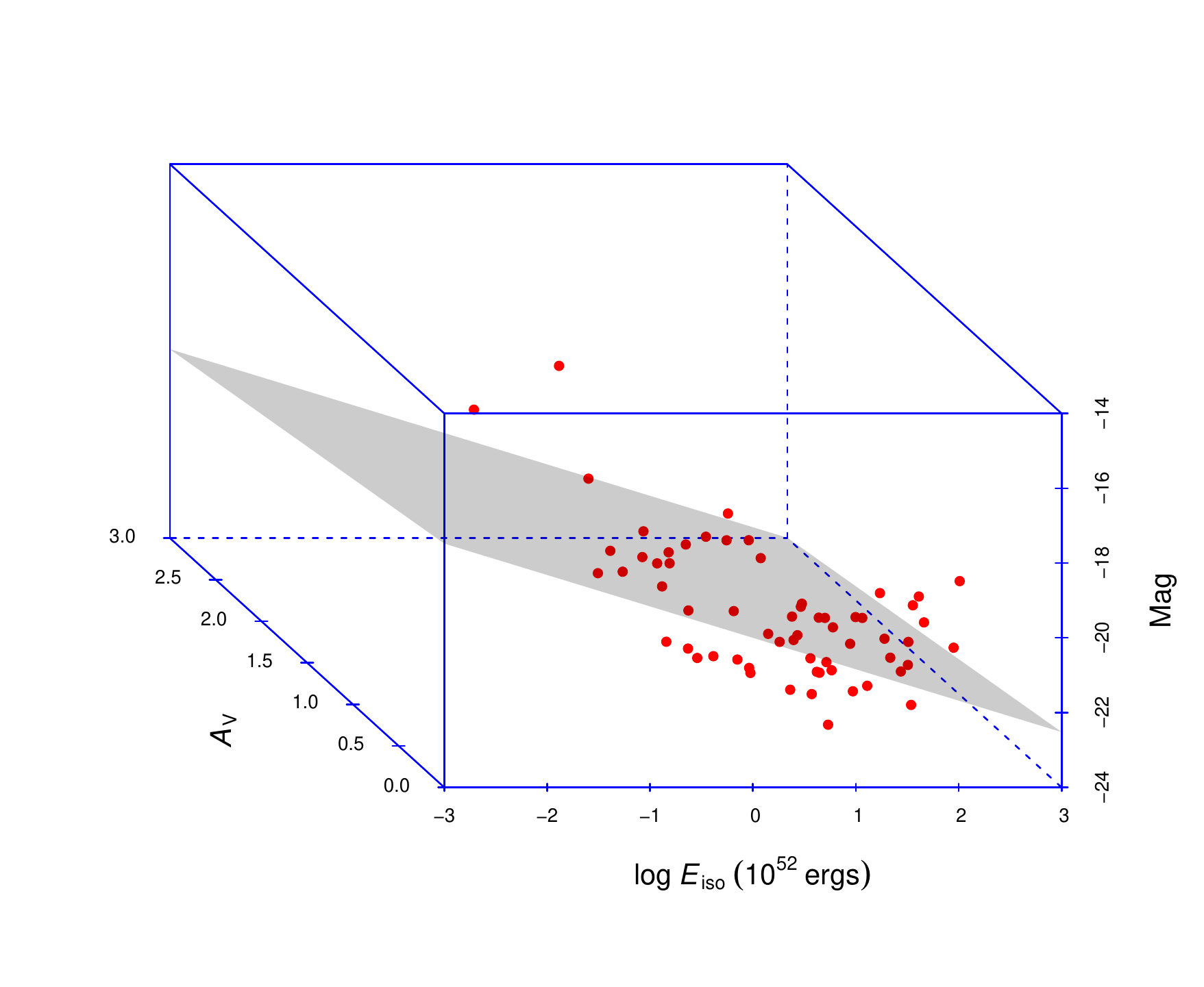}

\includegraphics[width=0.45\textwidth]{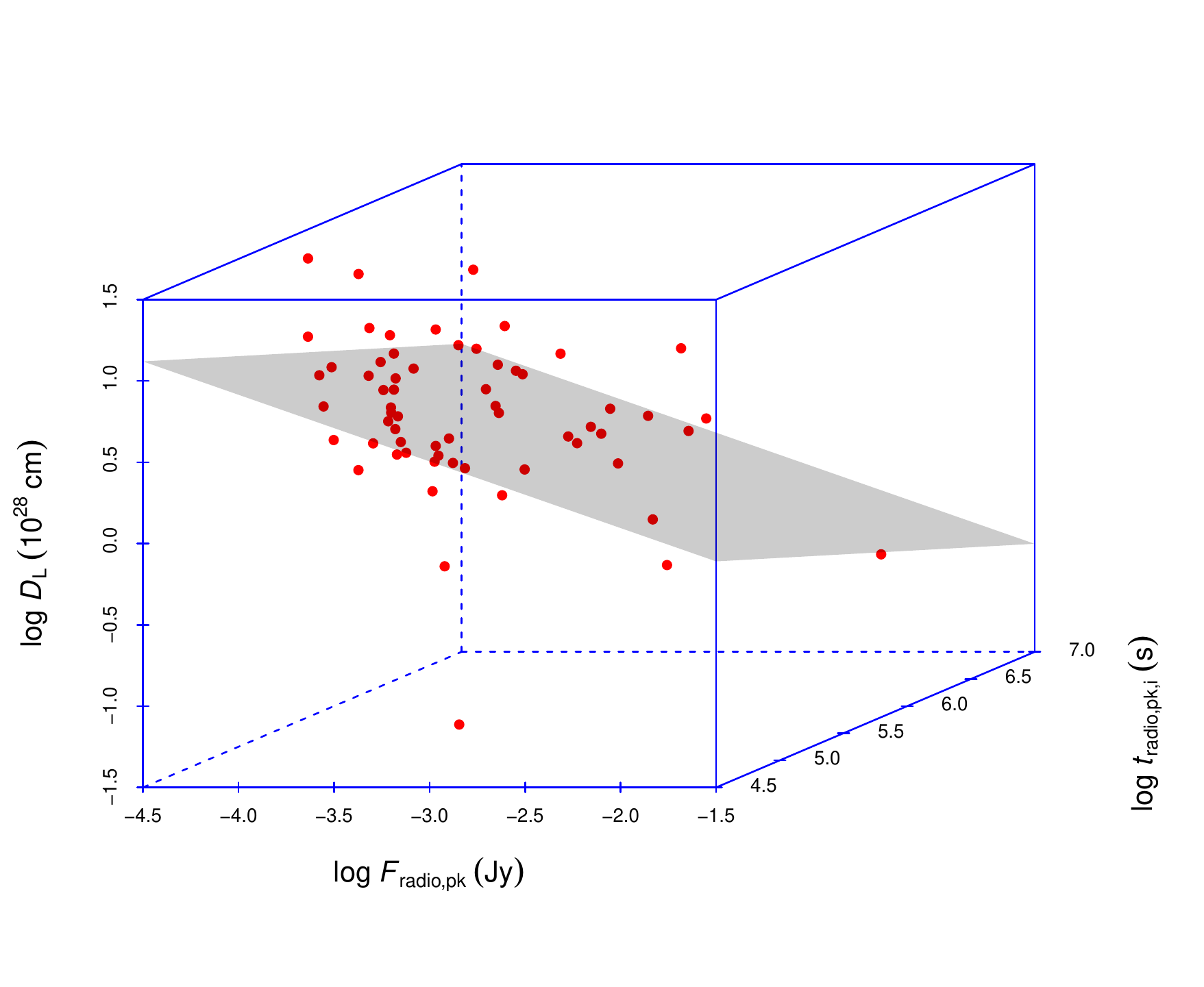}
\includegraphics[width=0.45\textwidth]{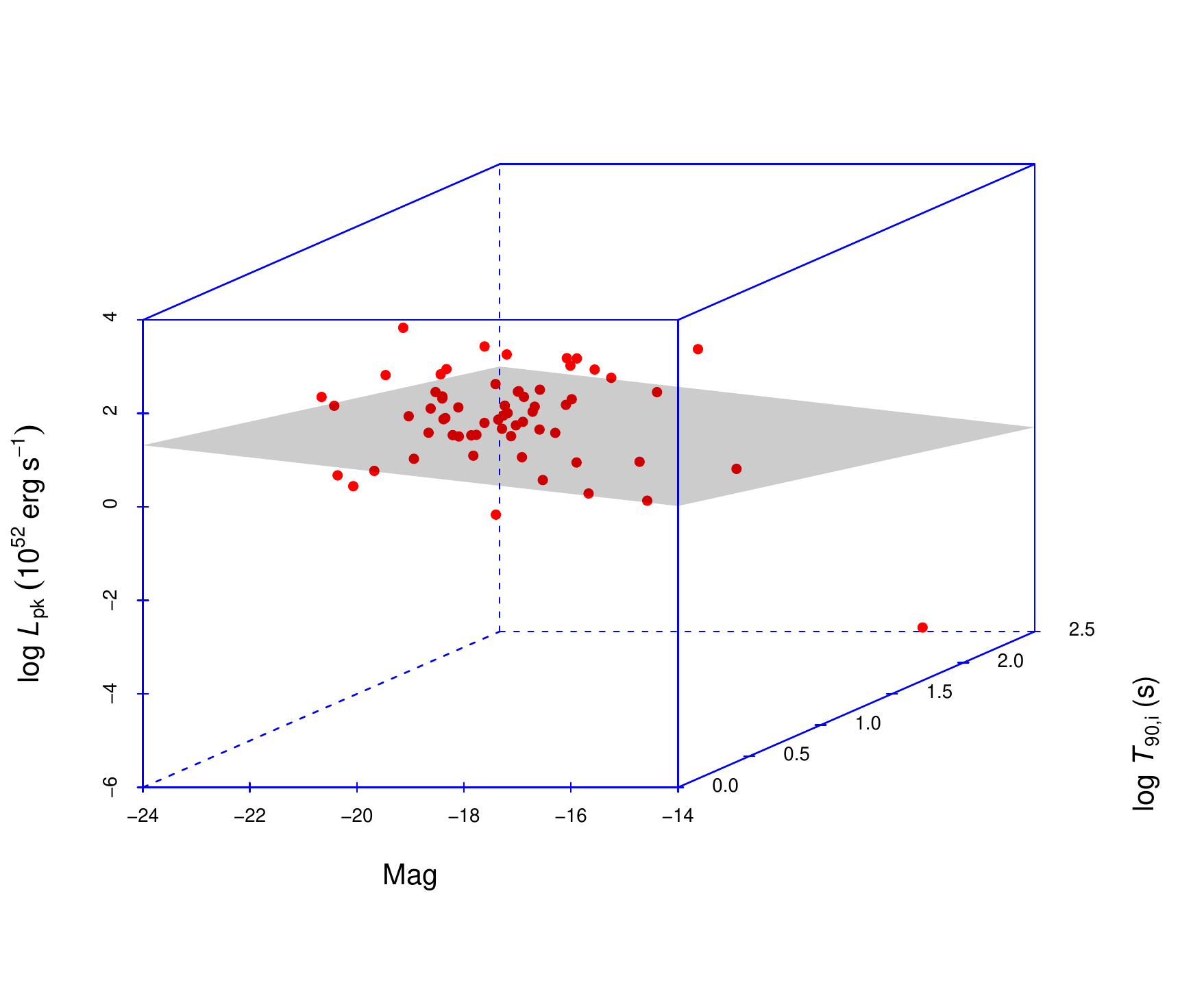}

\includegraphics[width=0.45\textwidth]{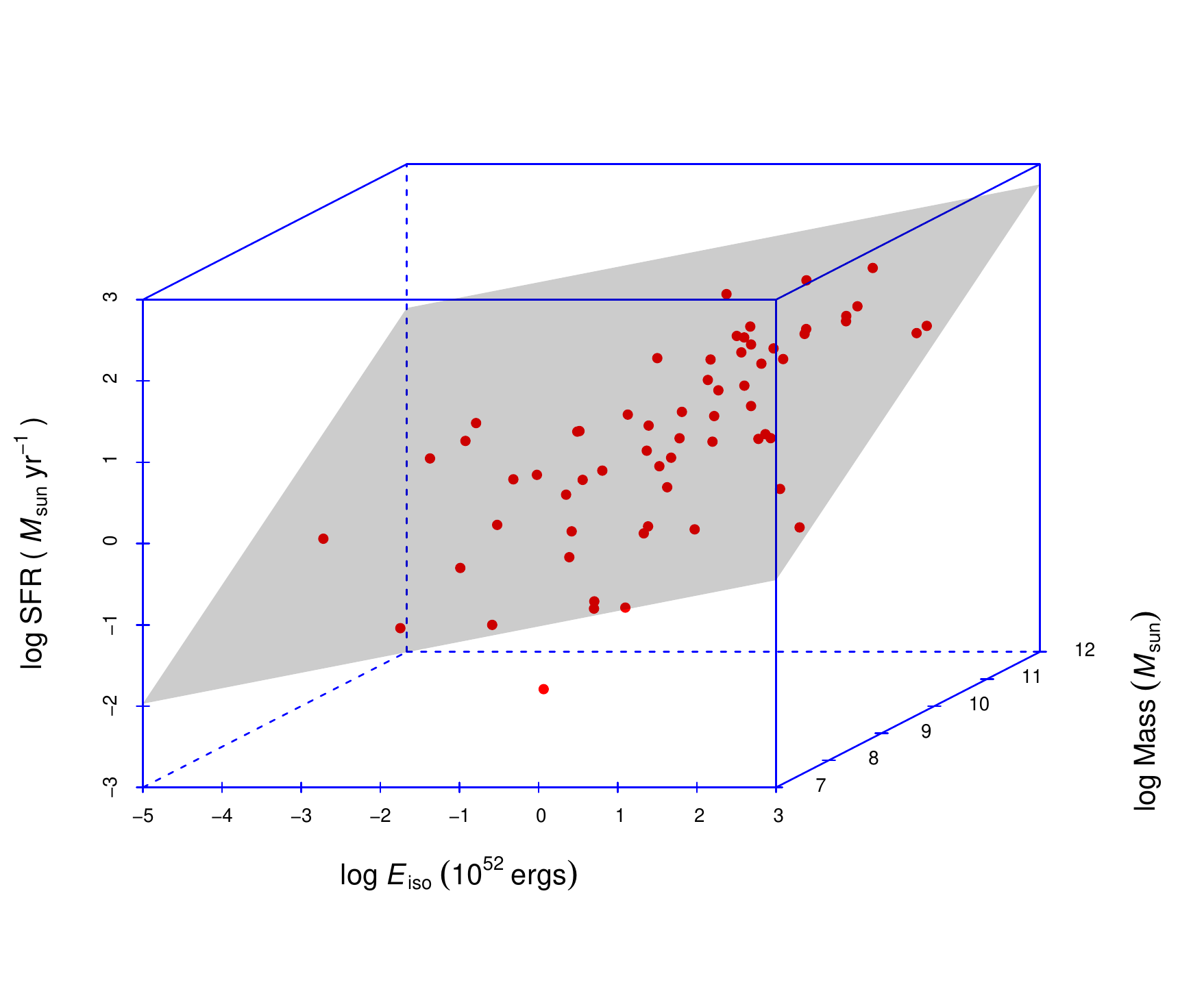}
\includegraphics[width=0.45\textwidth]{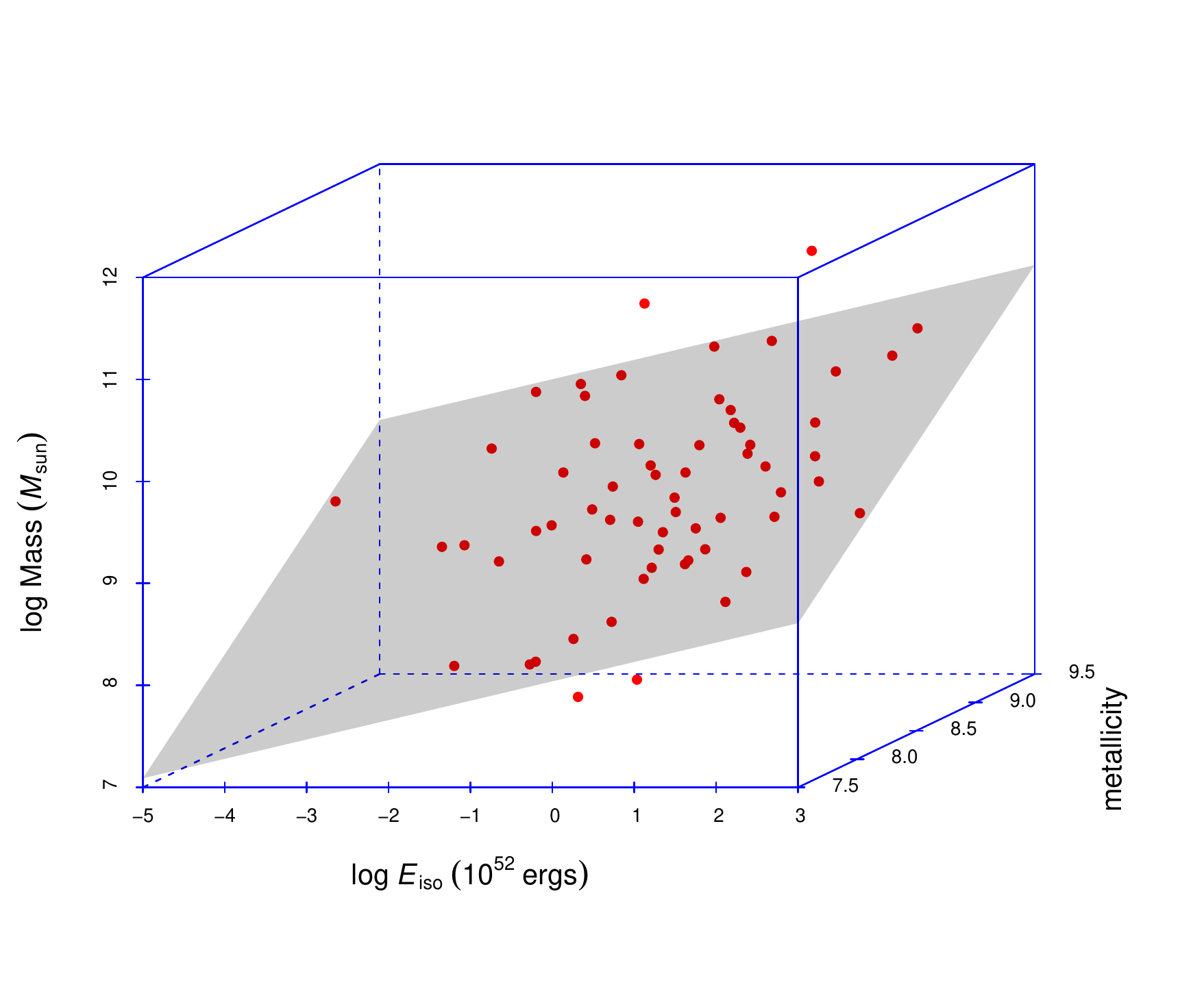}

\center{Fig. \ref{fig:three}---Continued}
\end{figure*}


\clearpage
\begin{figure*}

\includegraphics[width=0.45\textwidth]{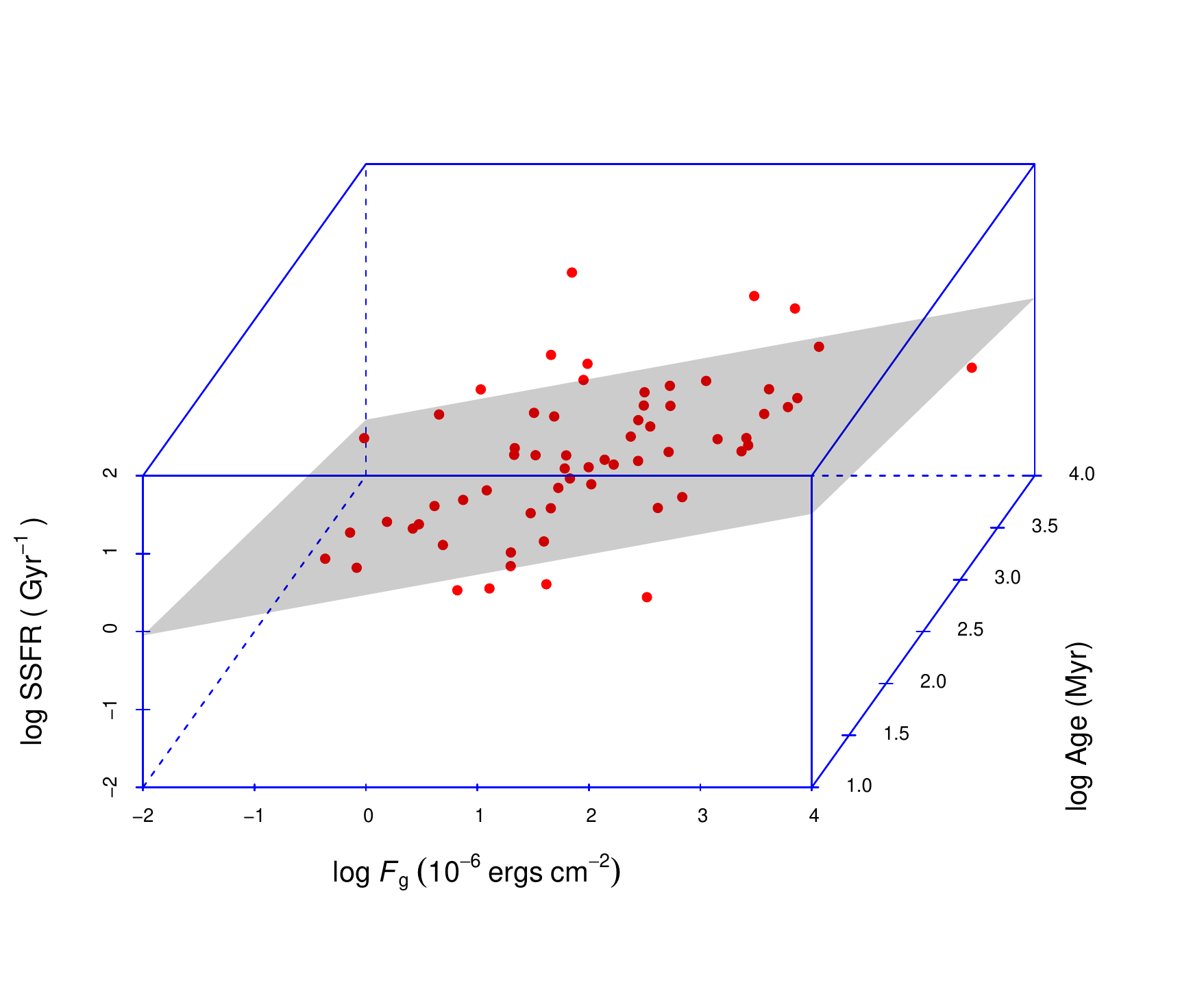}
\includegraphics[width=0.45\textwidth]{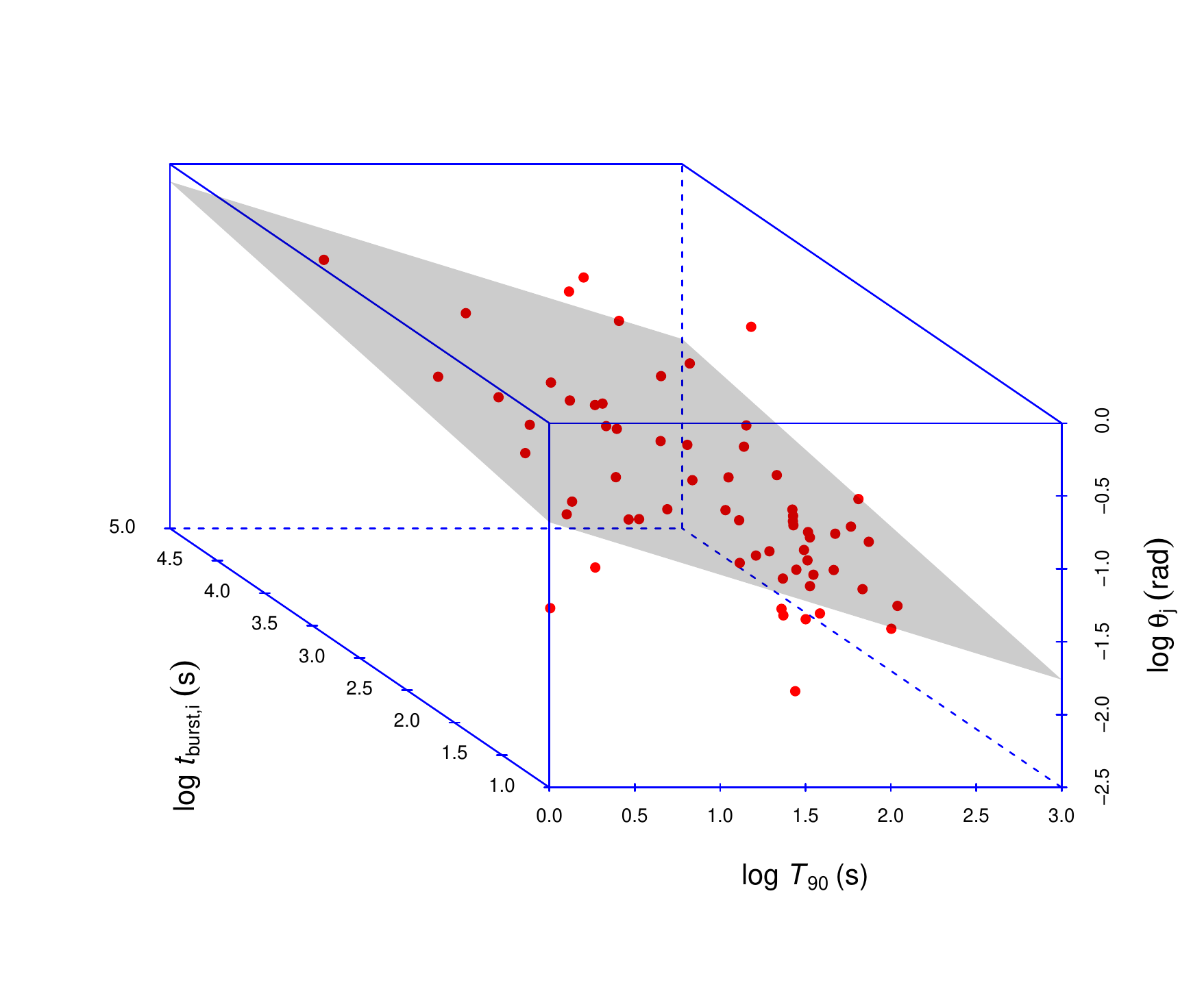}

\includegraphics[width=0.45\textwidth]{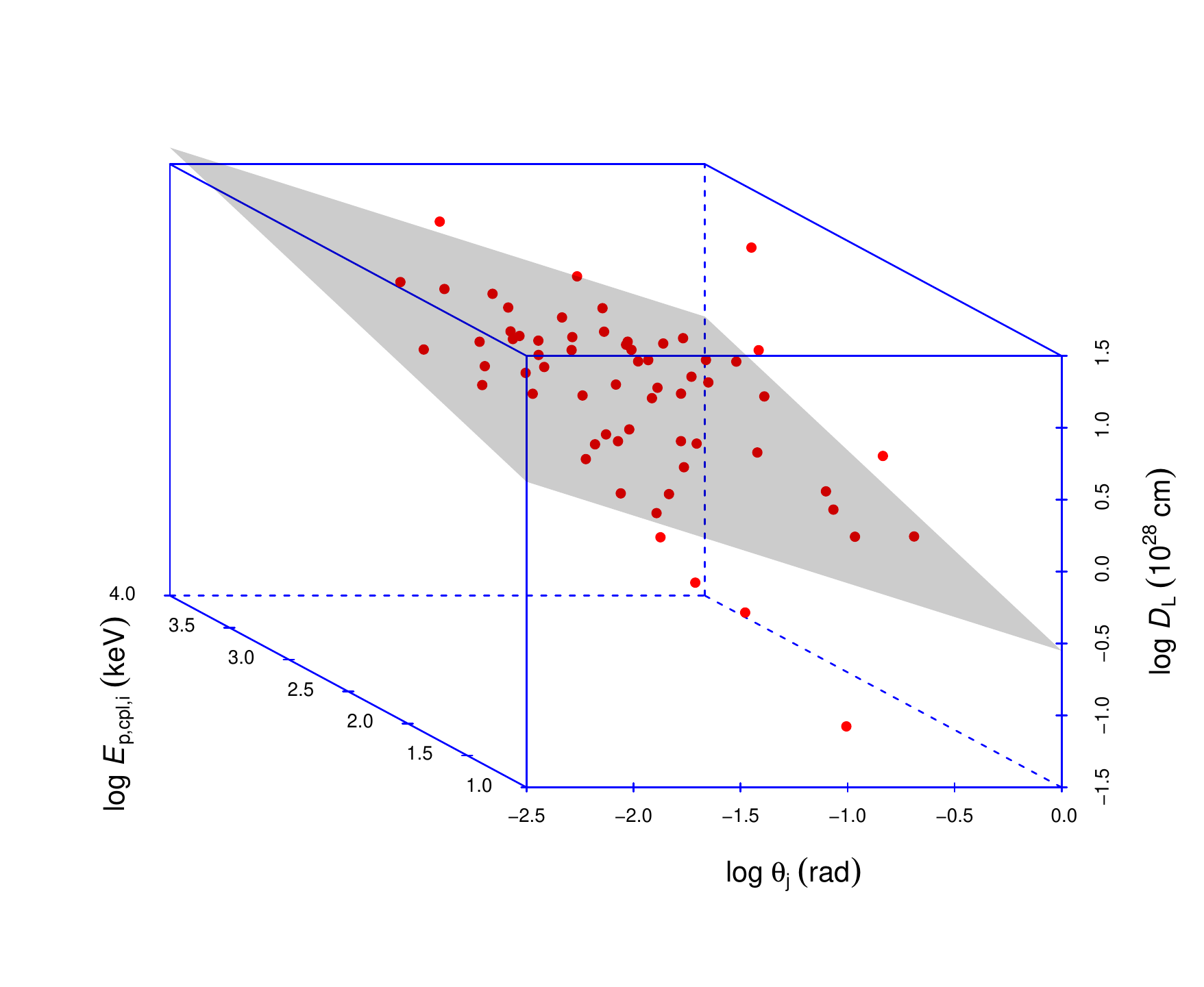}
\includegraphics[width=0.45\textwidth]{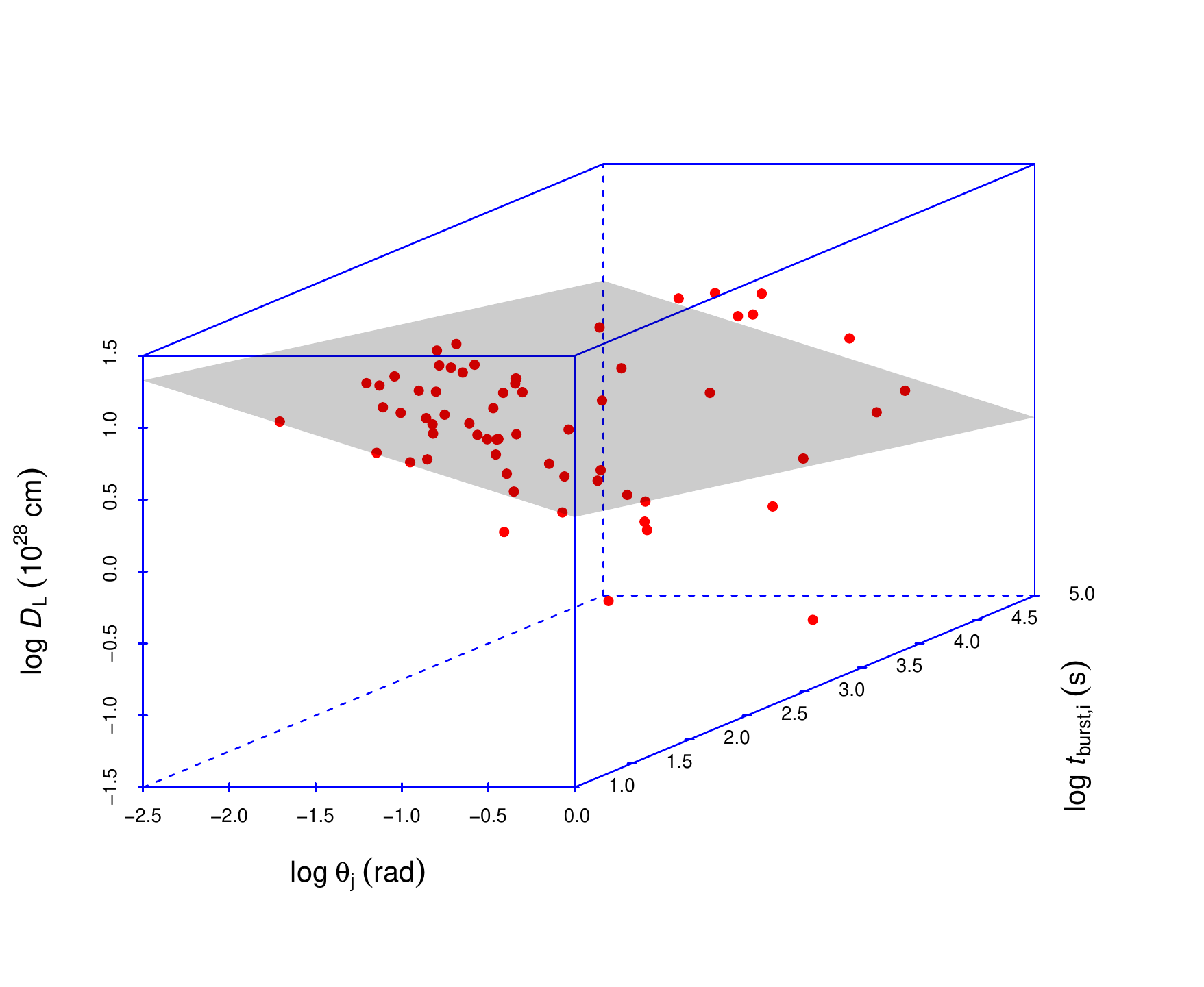}

\includegraphics[width=0.45\textwidth]{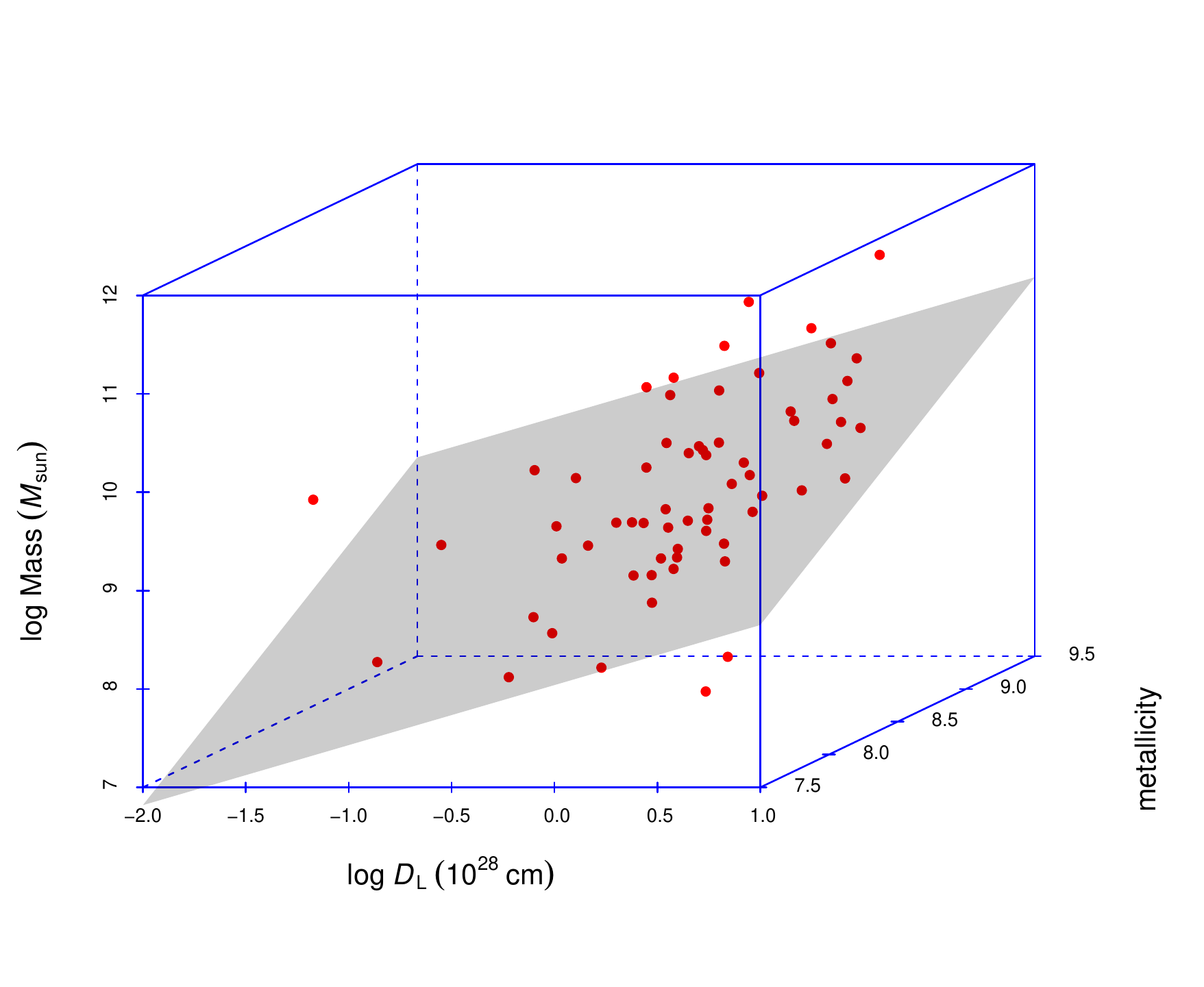}
\includegraphics[width=0.45\textwidth]{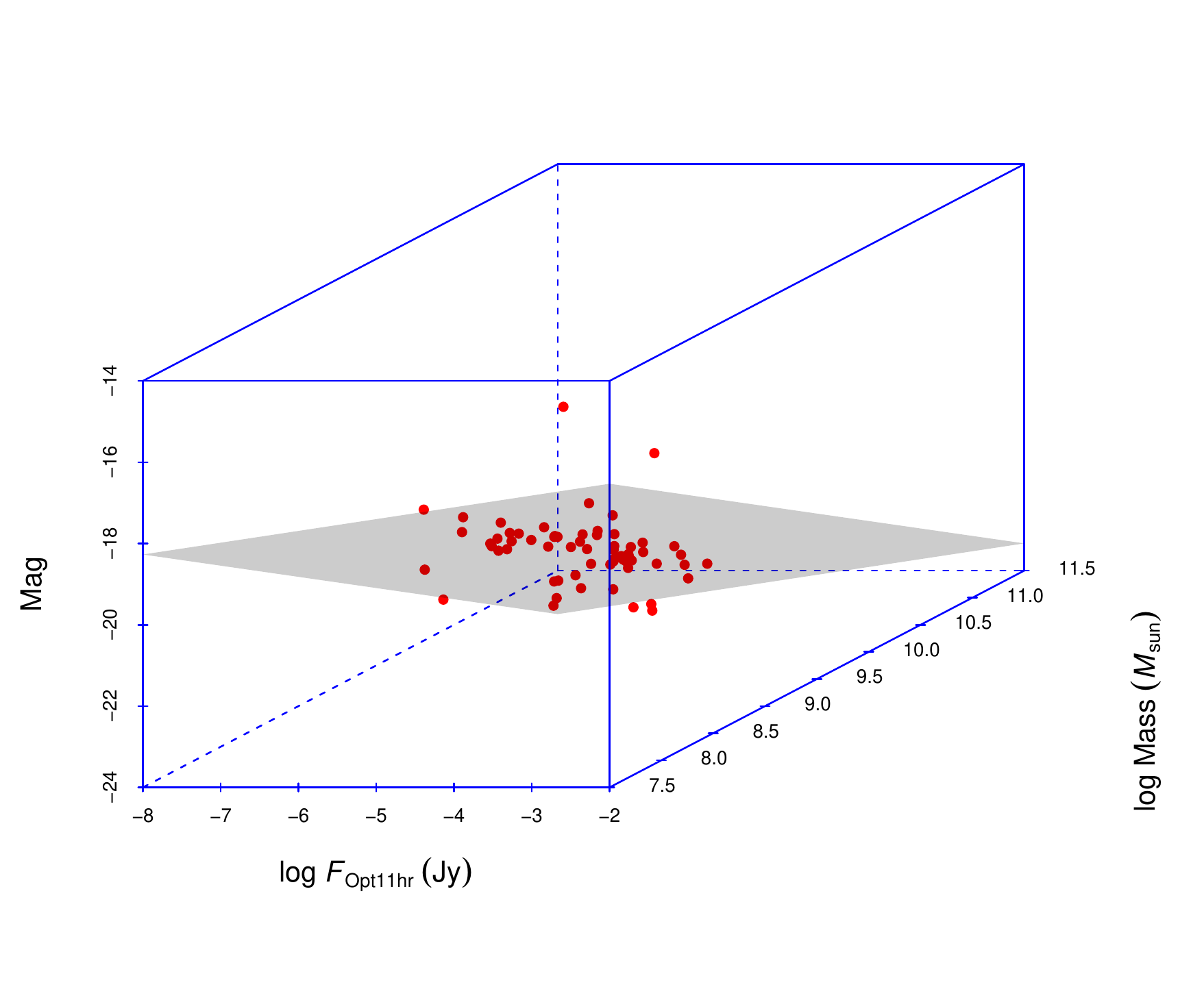}

\center{Fig. \ref{fig:three}---Continued}
\end{figure*}


\clearpage
\begin{figure*}

\includegraphics[width=0.45\textwidth]{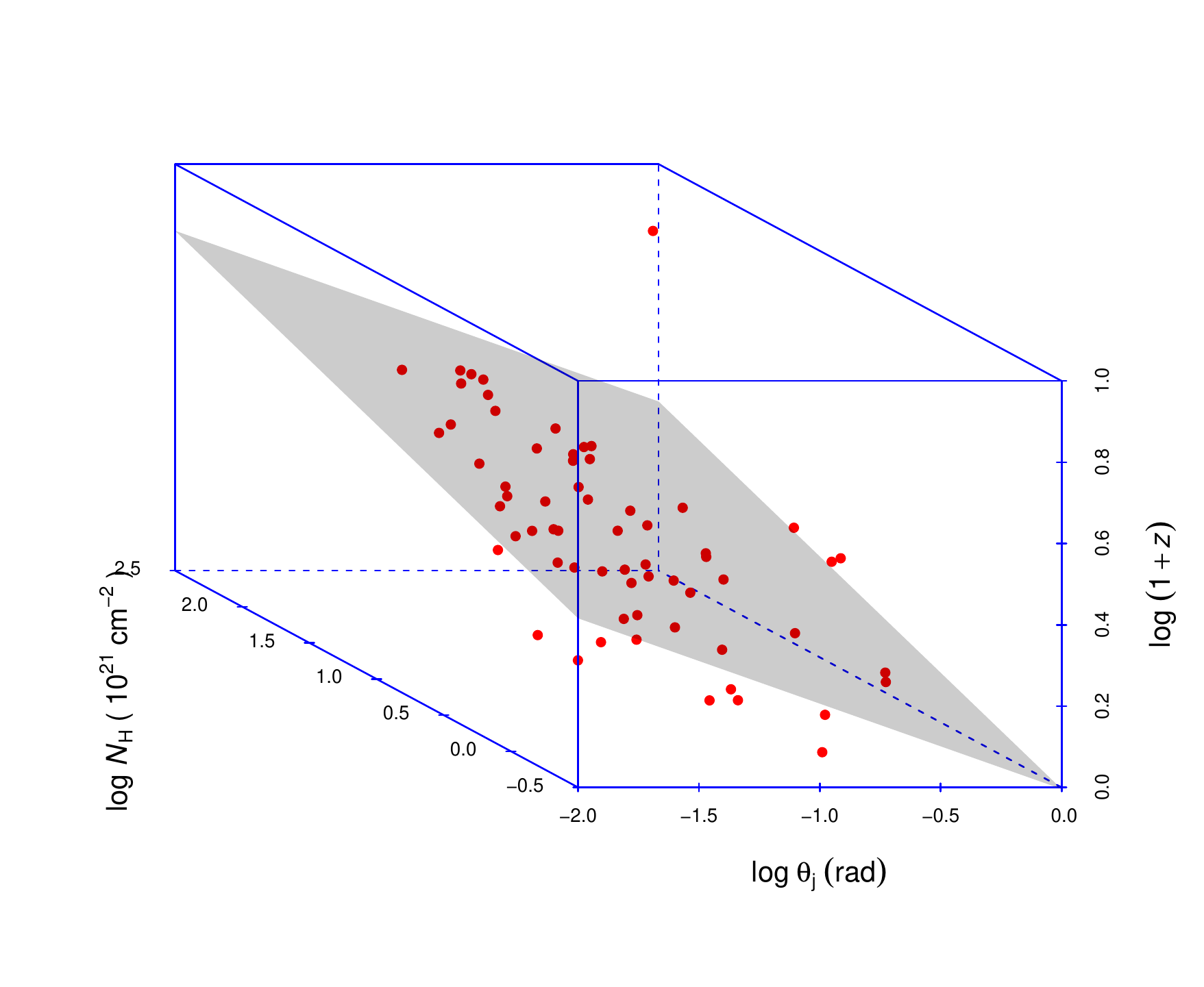}
\includegraphics[width=0.45\textwidth]{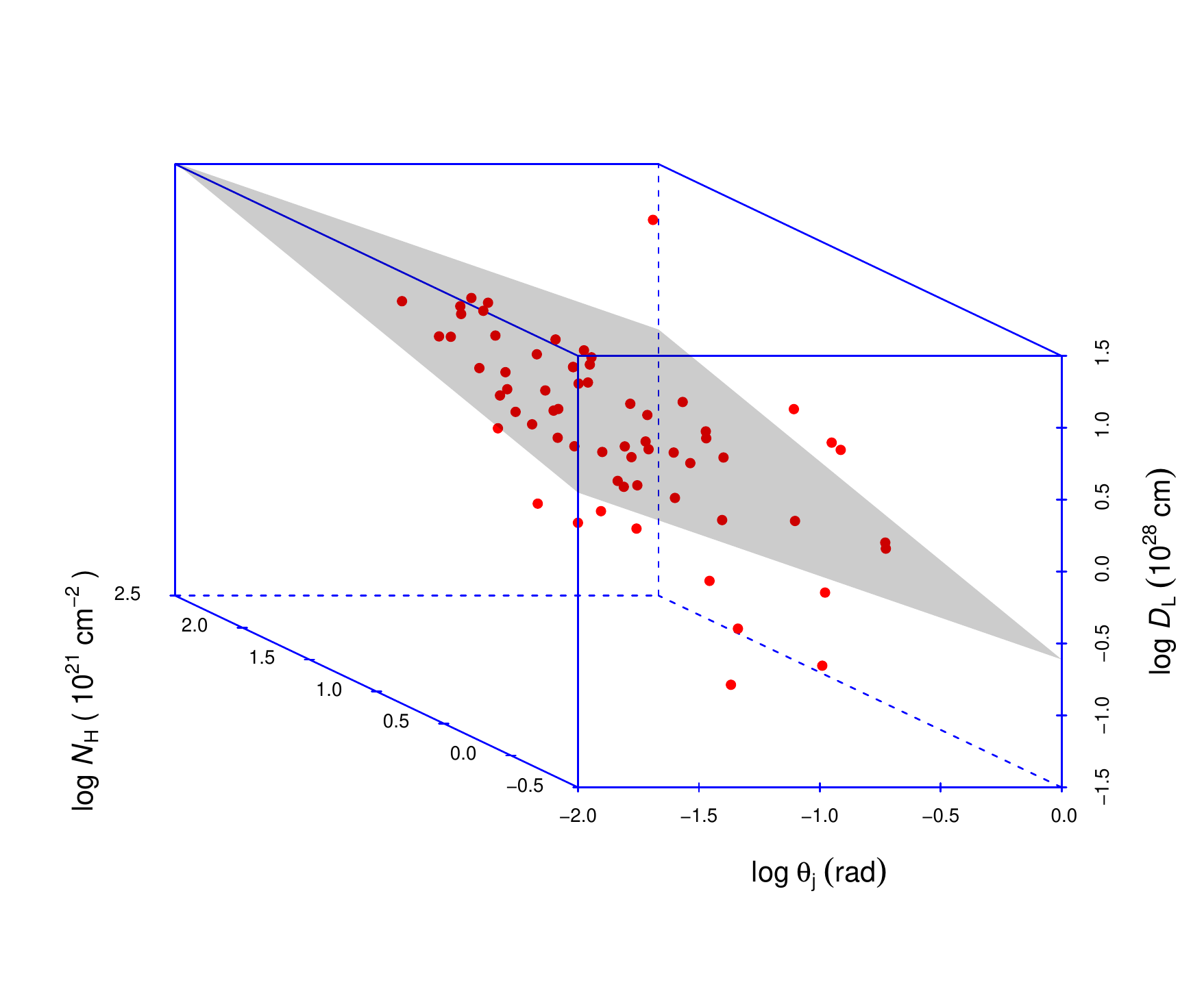}

\includegraphics[width=0.45\textwidth]{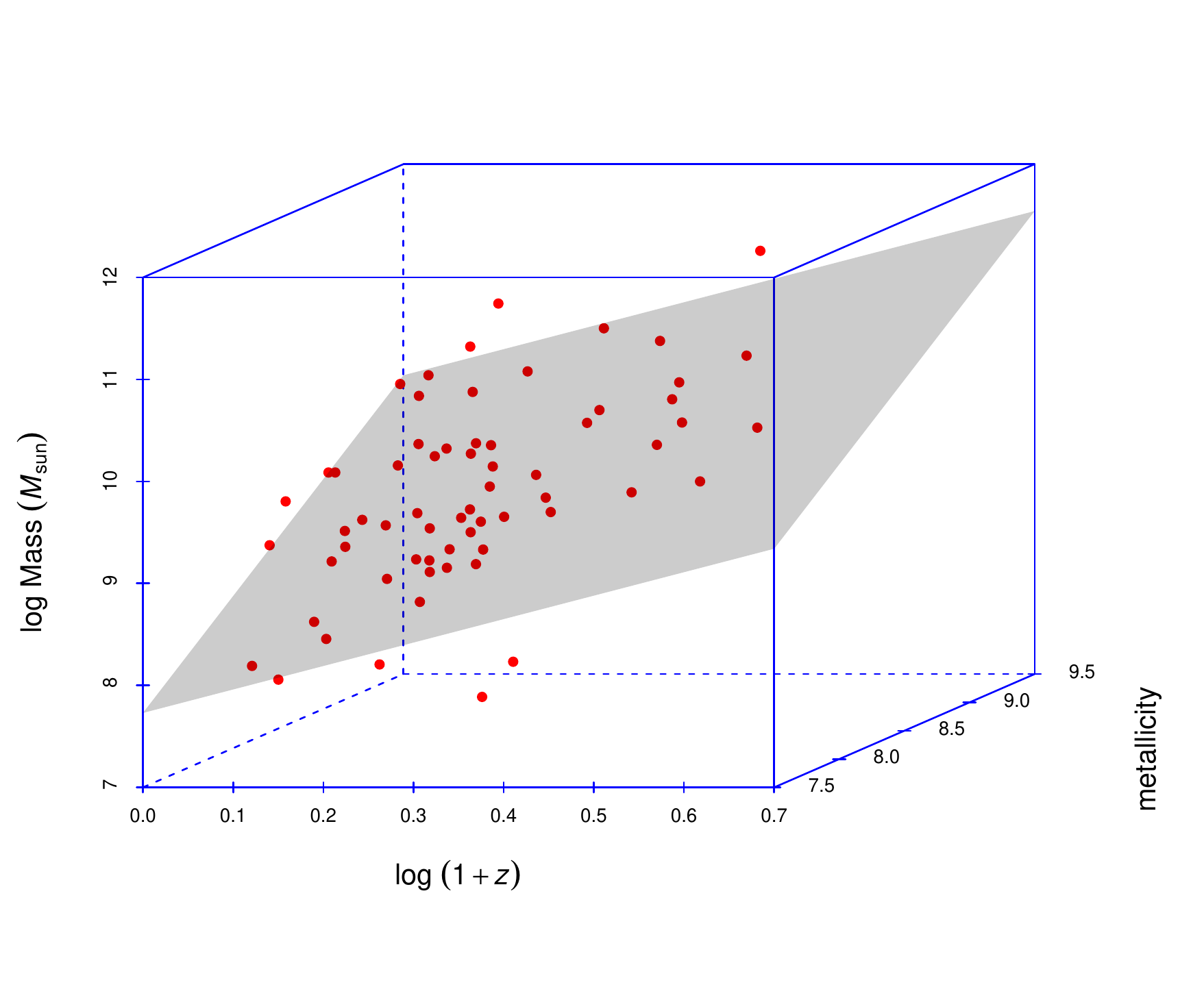}
\includegraphics[width=0.45\textwidth]{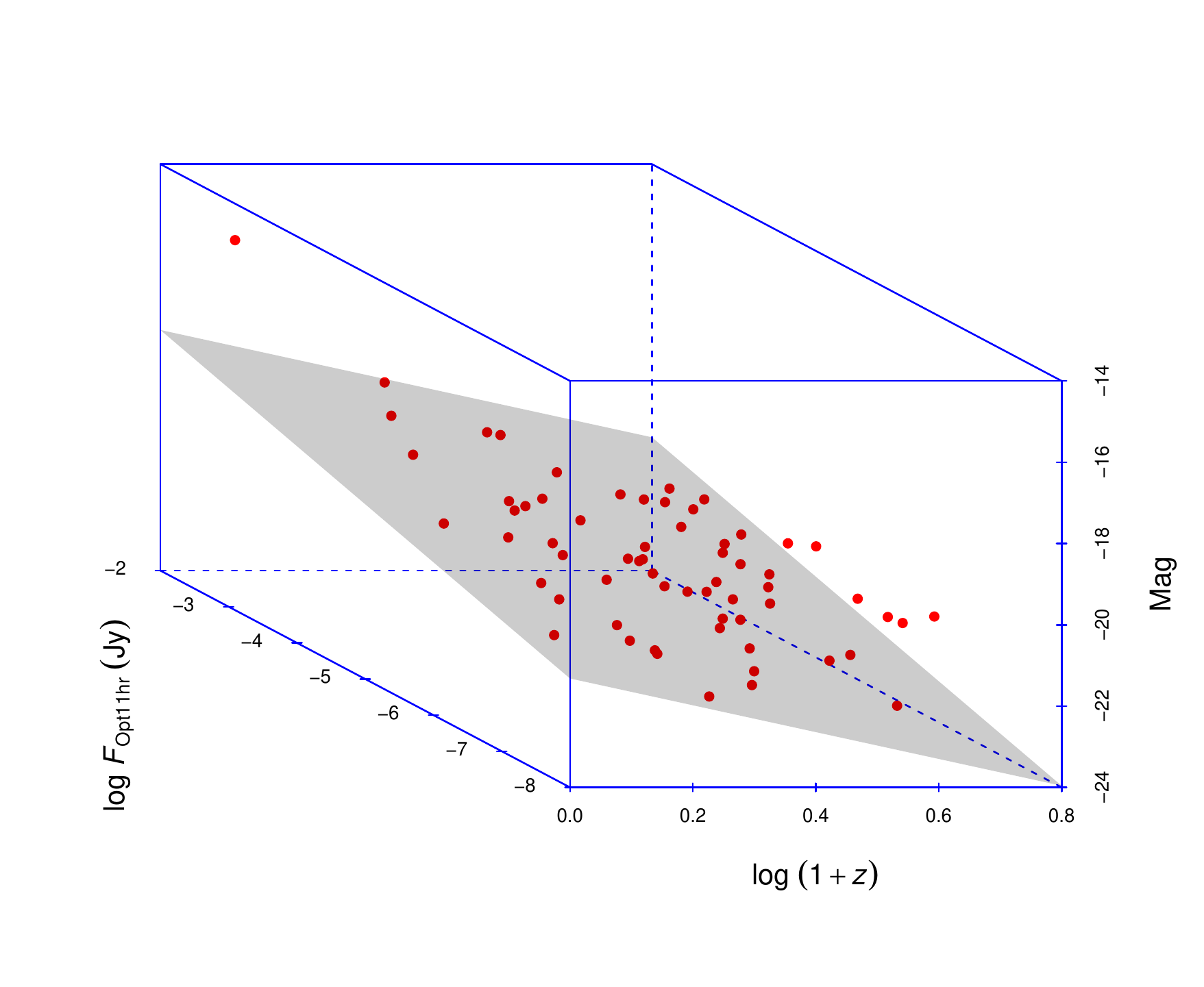}

\includegraphics[width=0.45\textwidth]{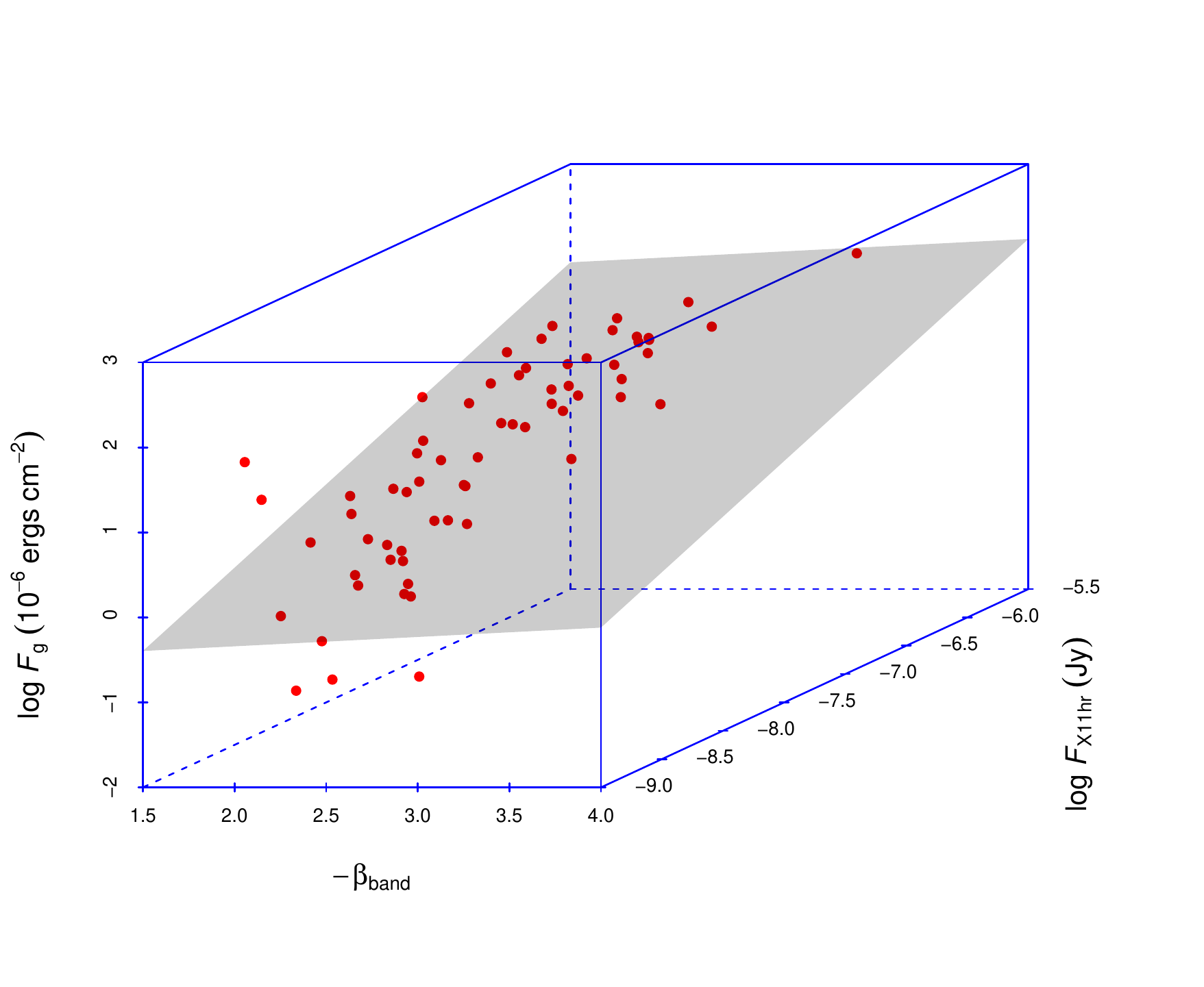}
\includegraphics[width=0.45\textwidth]{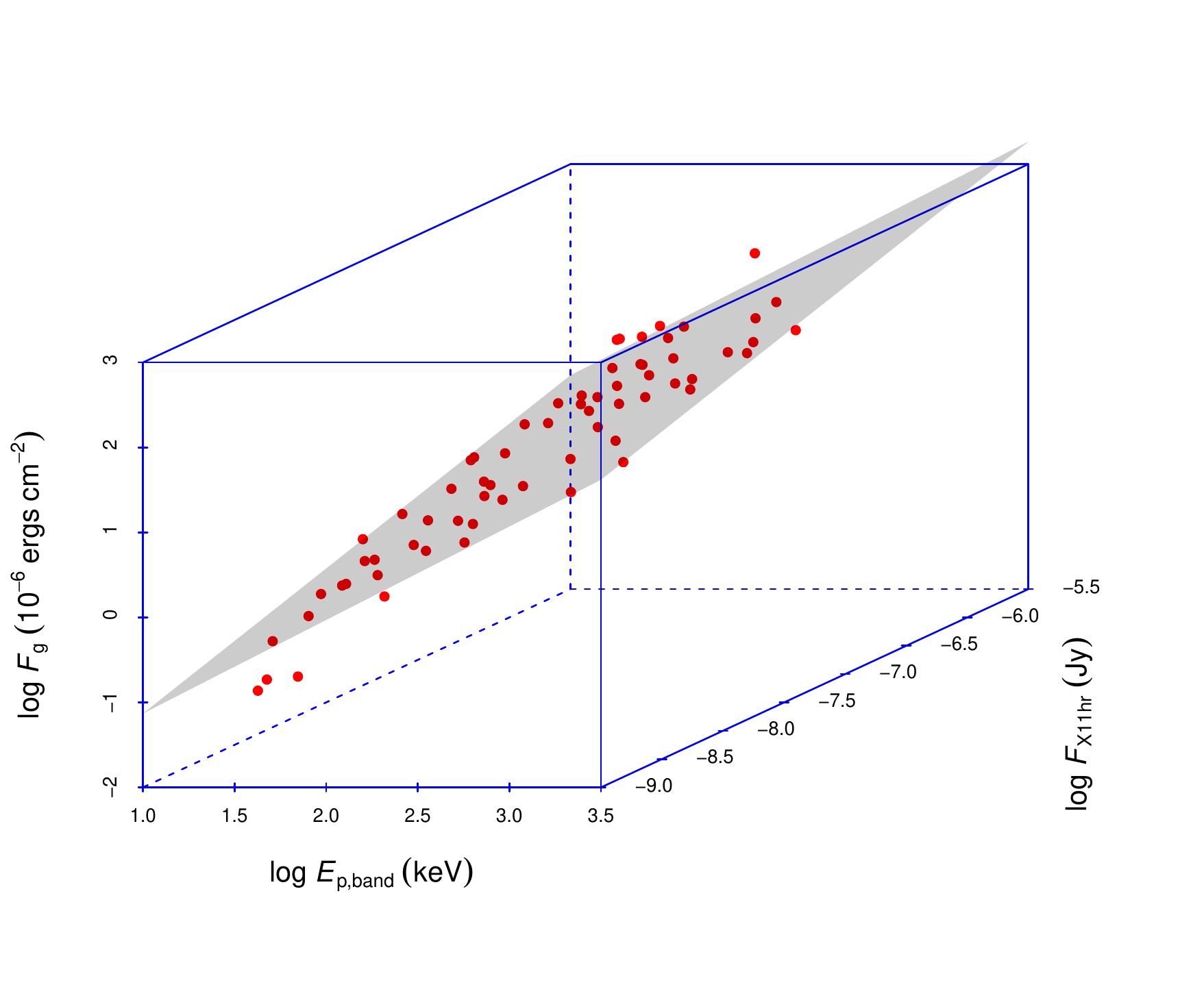}

\center{Fig. \ref{fig:three}---Continued}
\end{figure*}


\clearpage
\begin{figure*}

\includegraphics[width=0.45\textwidth]{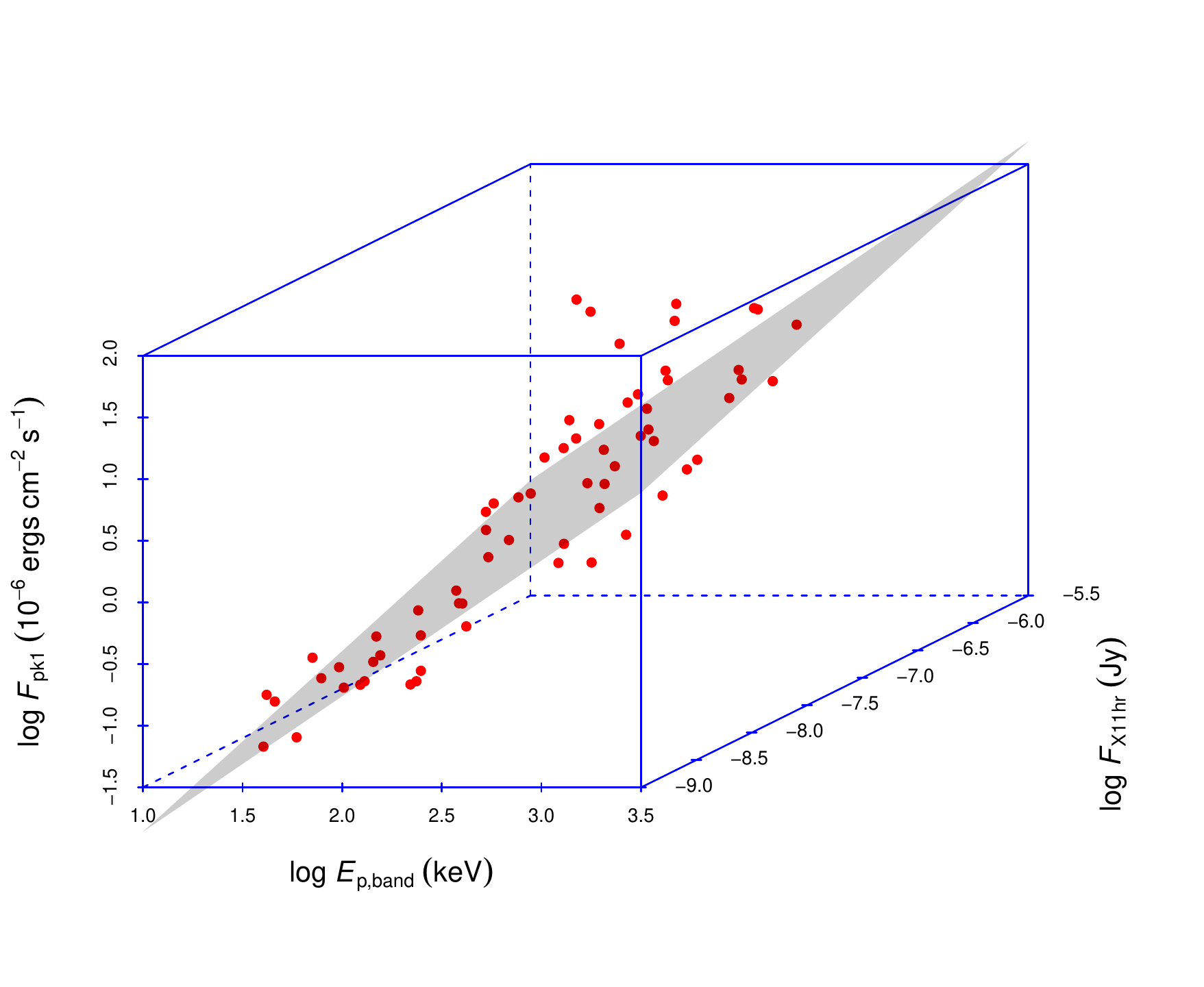}
\includegraphics[width=0.45\textwidth]{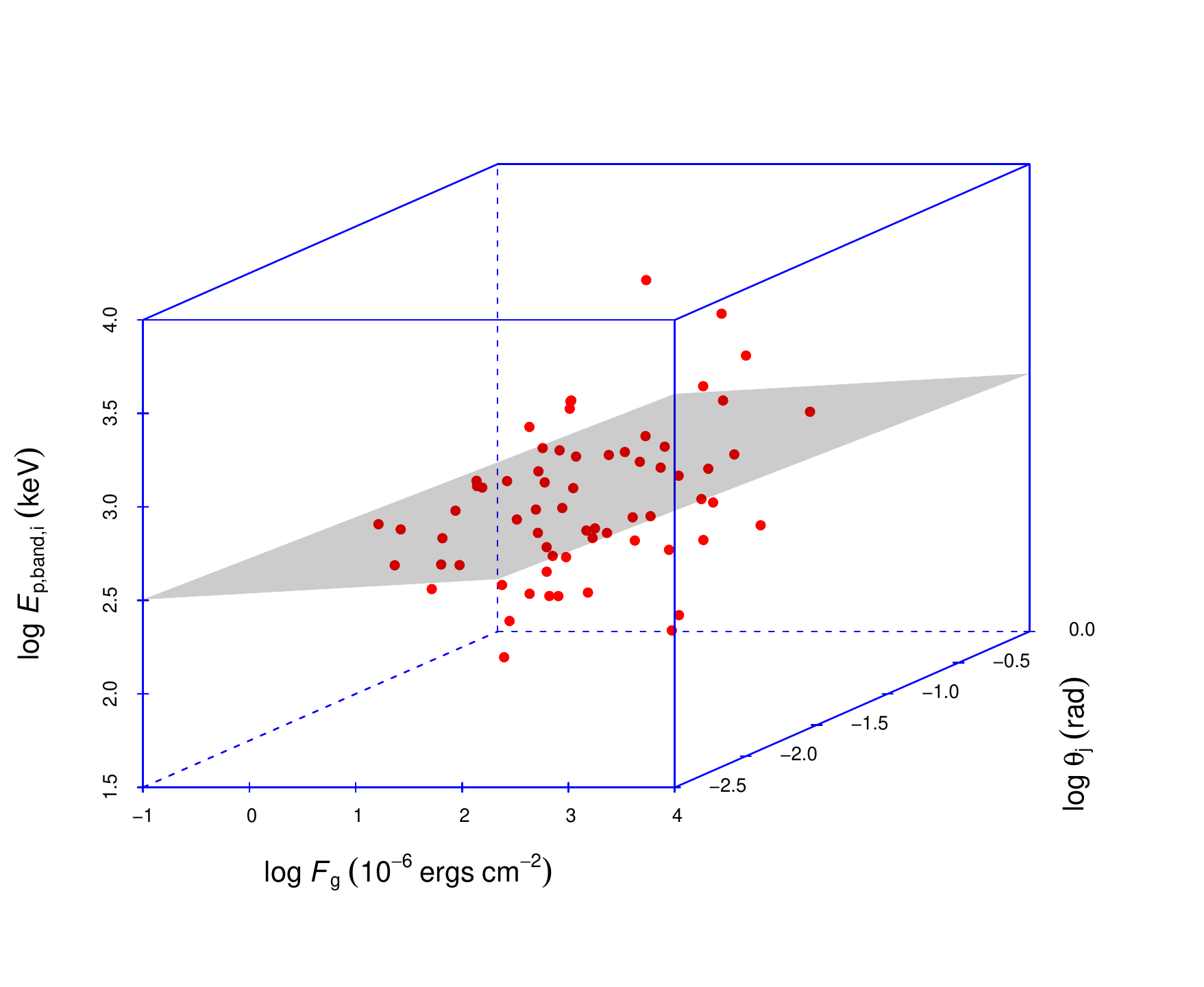}

\includegraphics[width=0.45\textwidth]{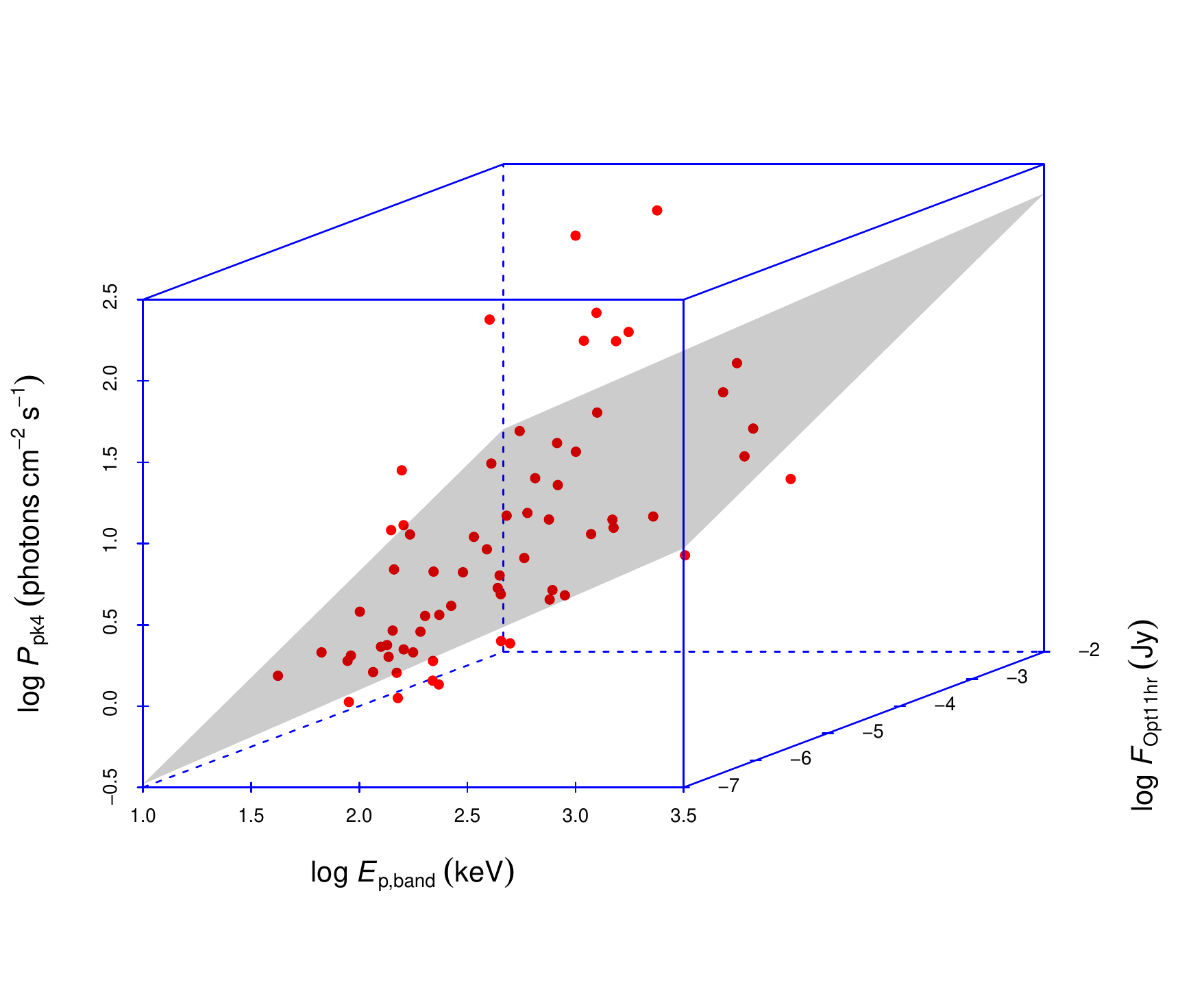}
\includegraphics[width=0.45\textwidth]{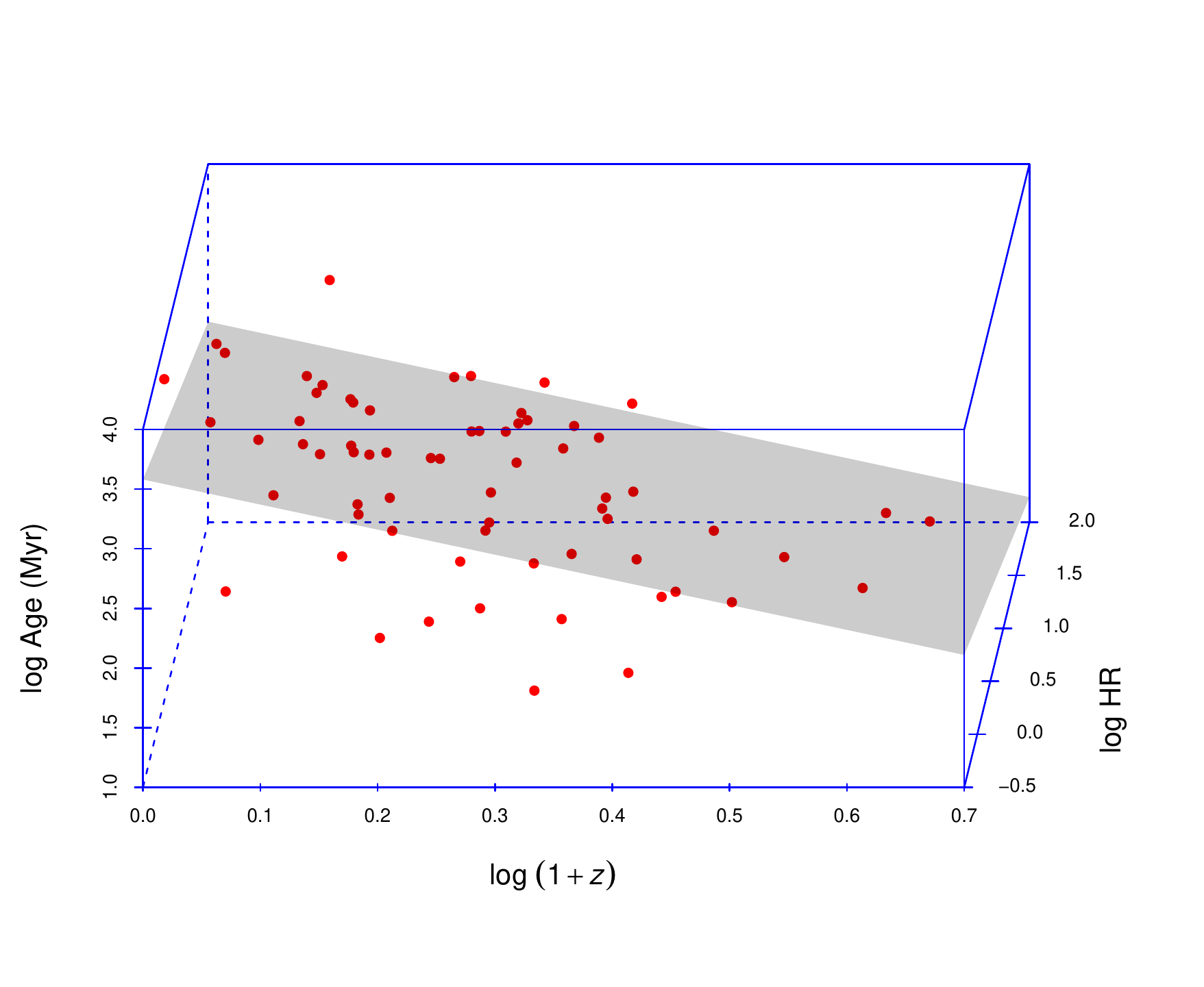}

\includegraphics[width=0.45\textwidth]{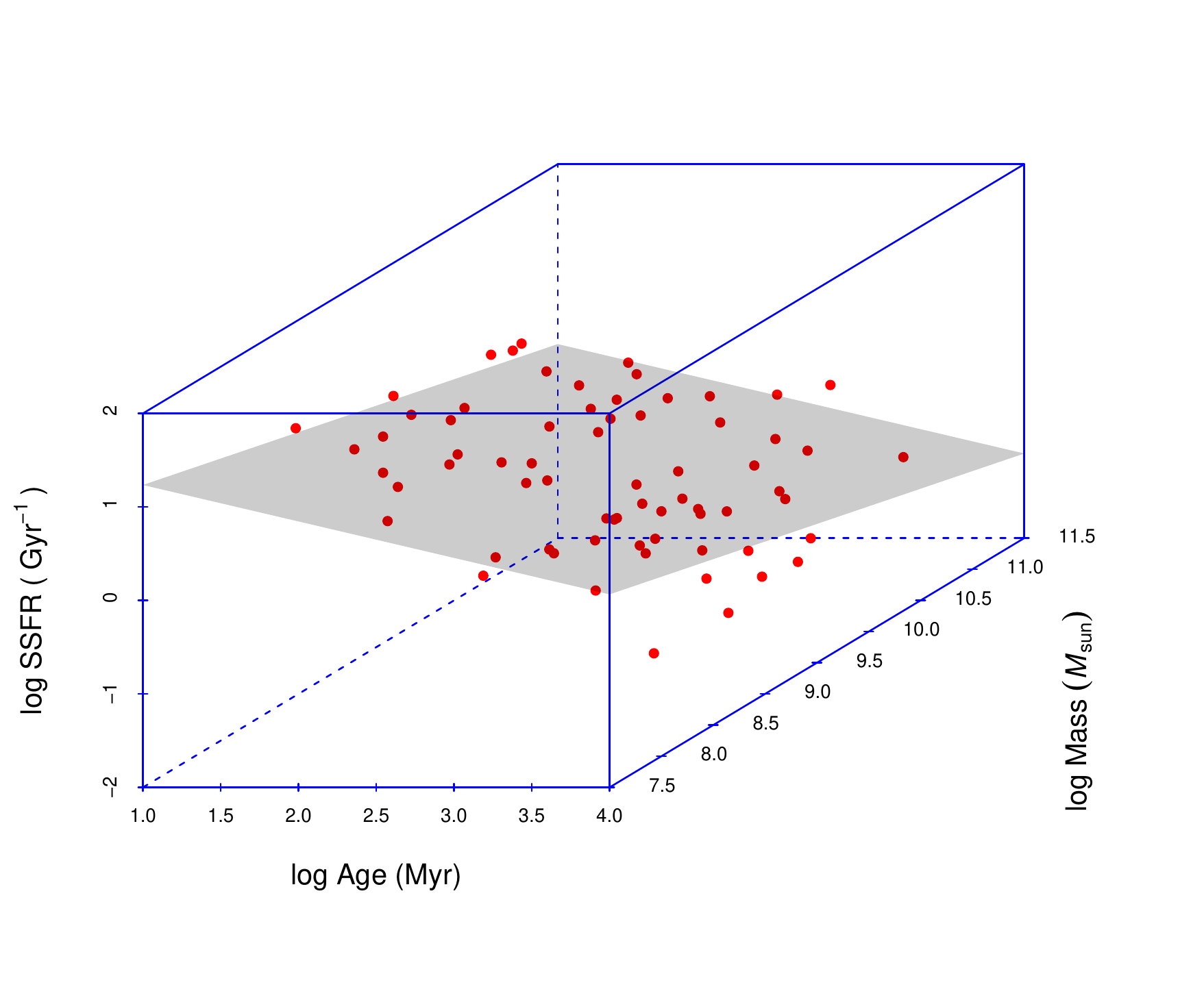}
\includegraphics[width=0.45\textwidth]{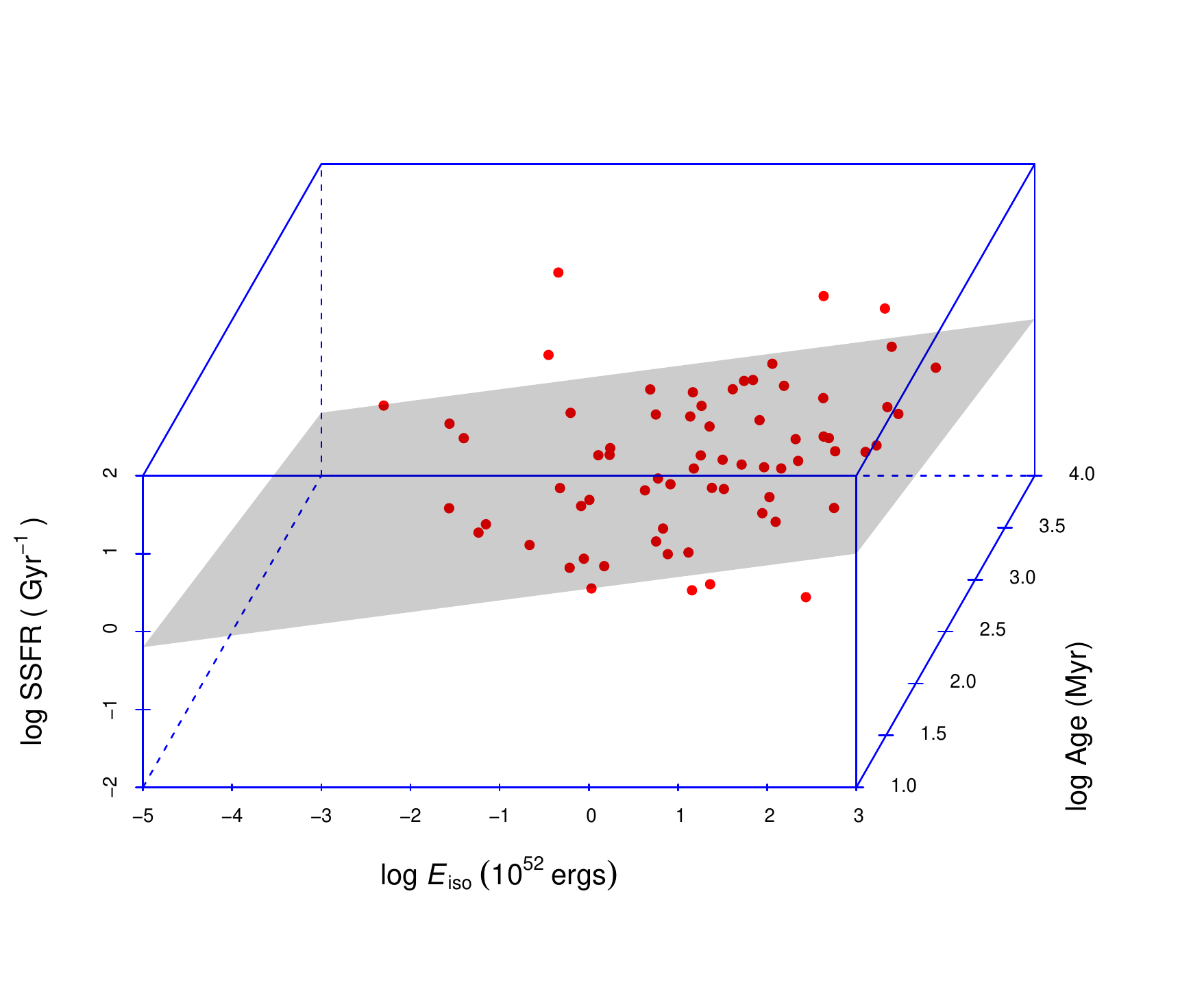}

\center{Fig. \ref{fig:three}---Continued}
\end{figure*}


\clearpage
\begin{figure*}

\includegraphics[width=0.45\textwidth]{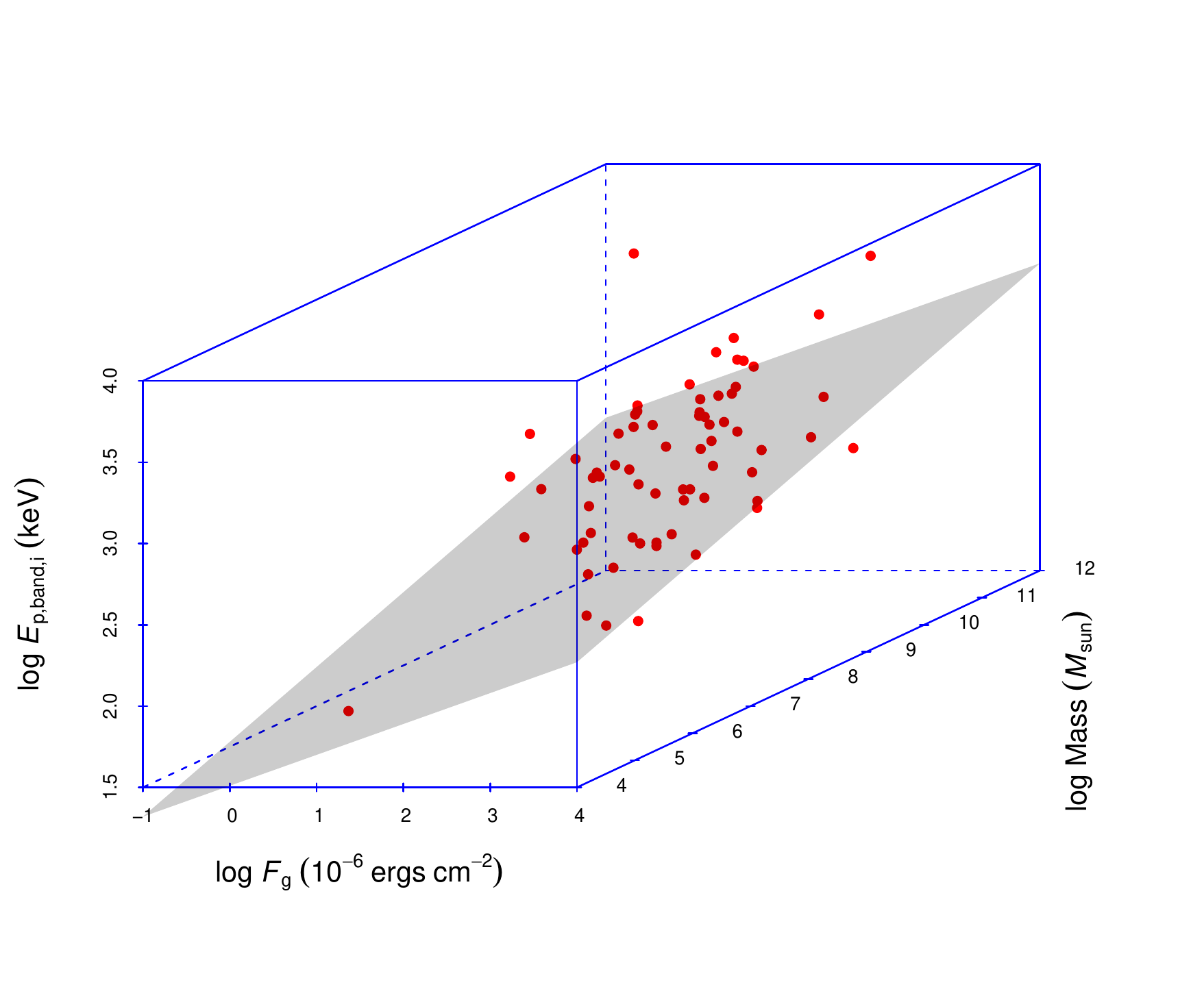}
\includegraphics[width=0.45\textwidth]{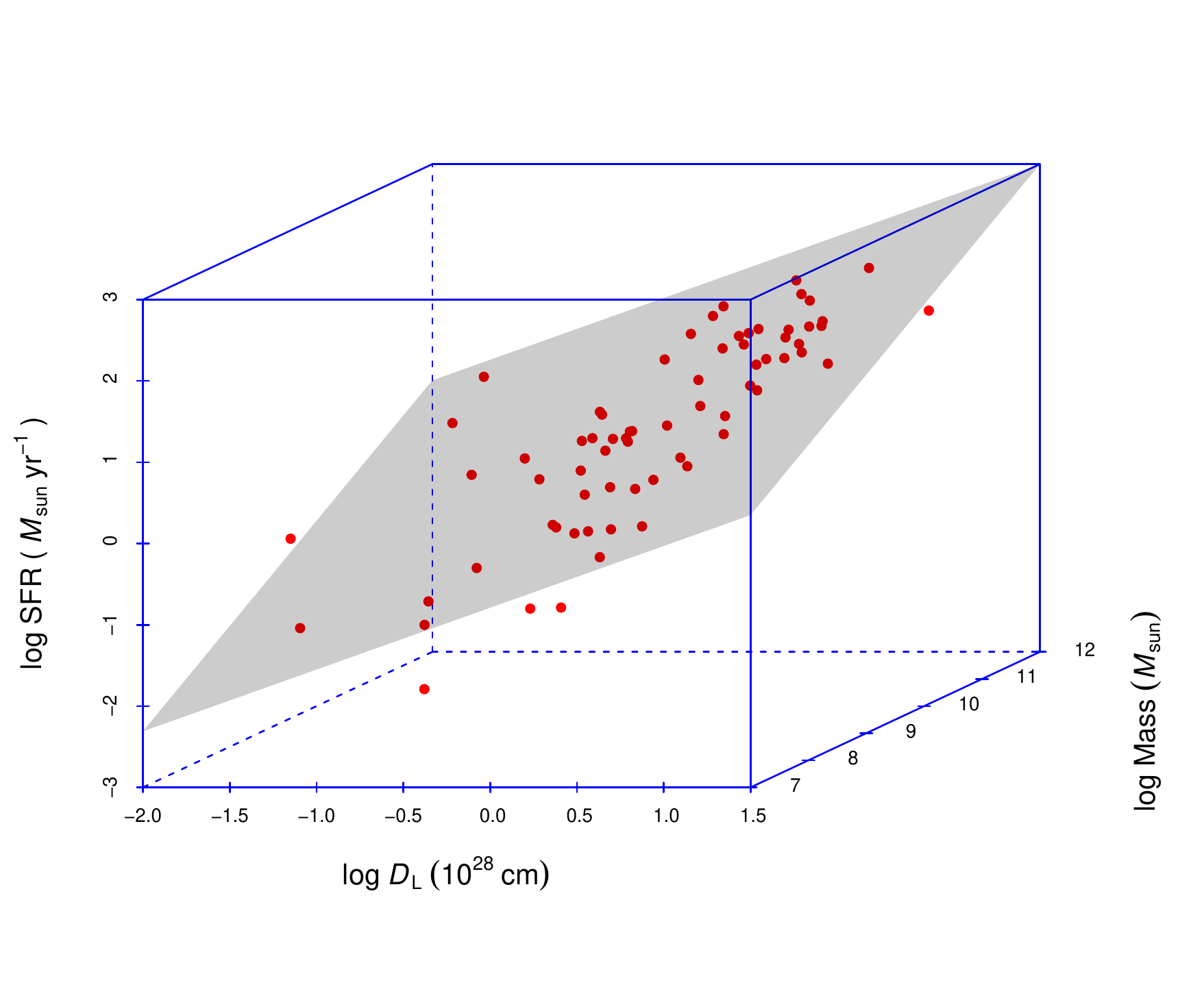}

\includegraphics[width=0.45\textwidth]{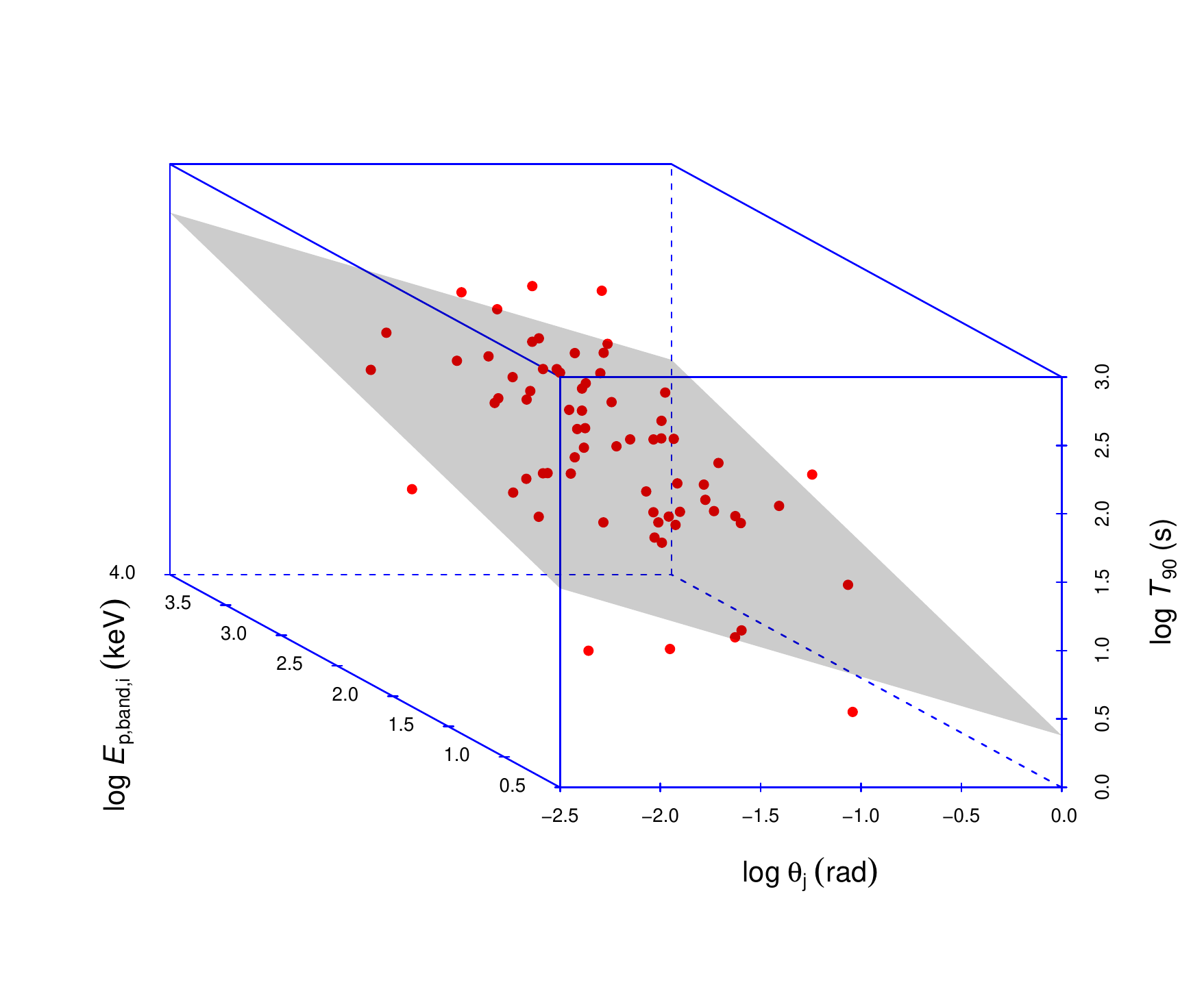}
\includegraphics[width=0.45\textwidth]{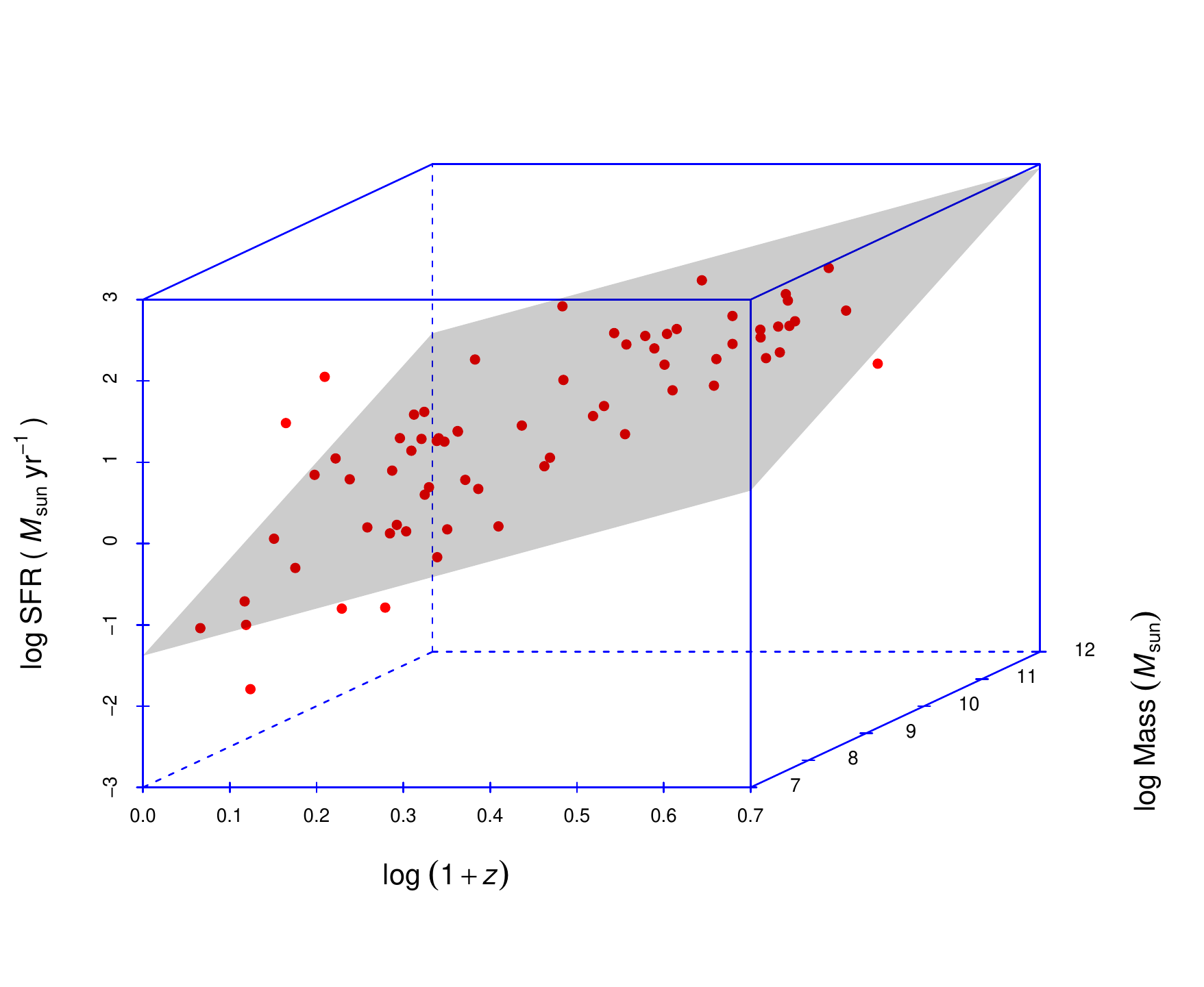}

\includegraphics[width=0.45\textwidth]{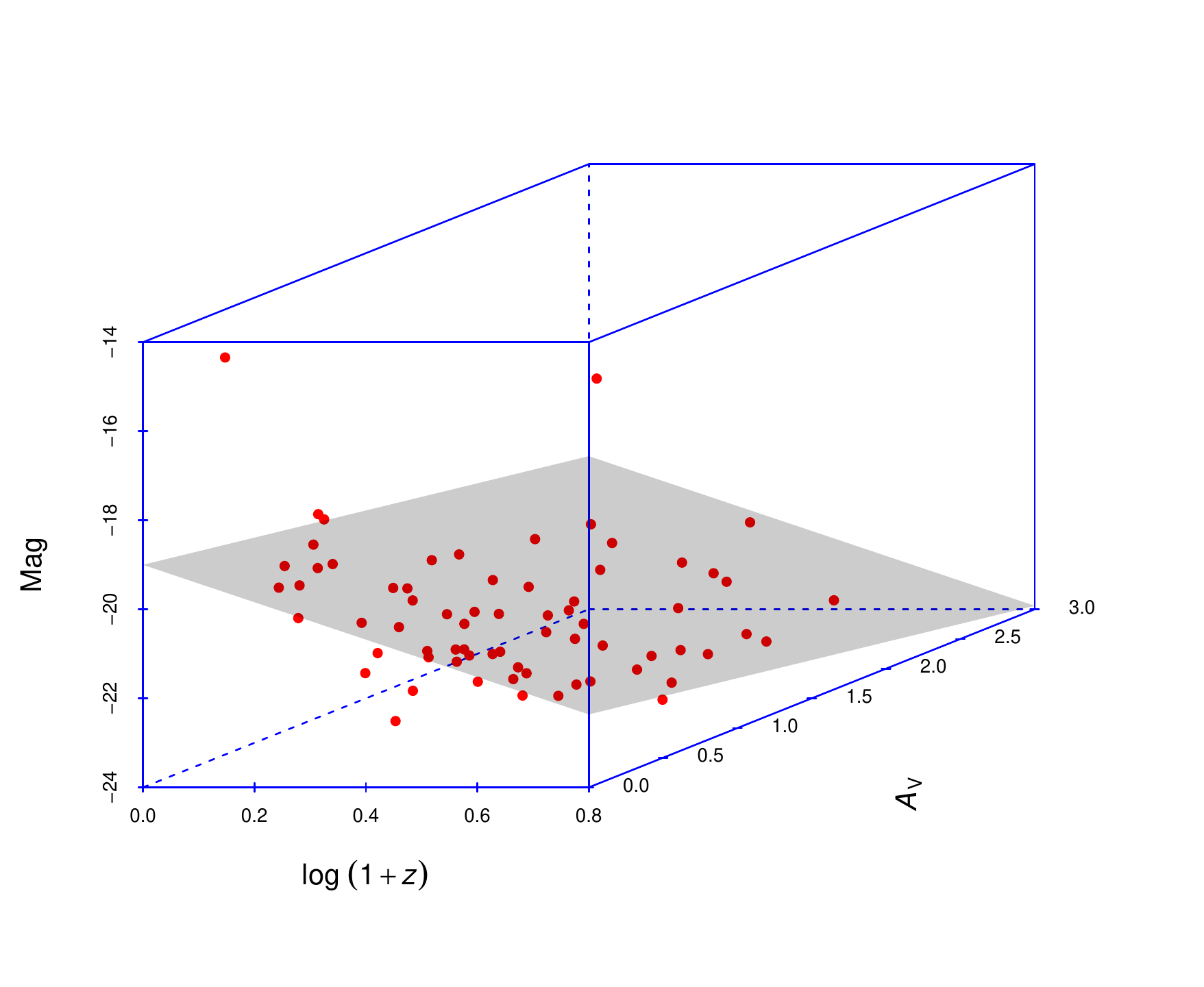}
\includegraphics[width=0.45\textwidth]{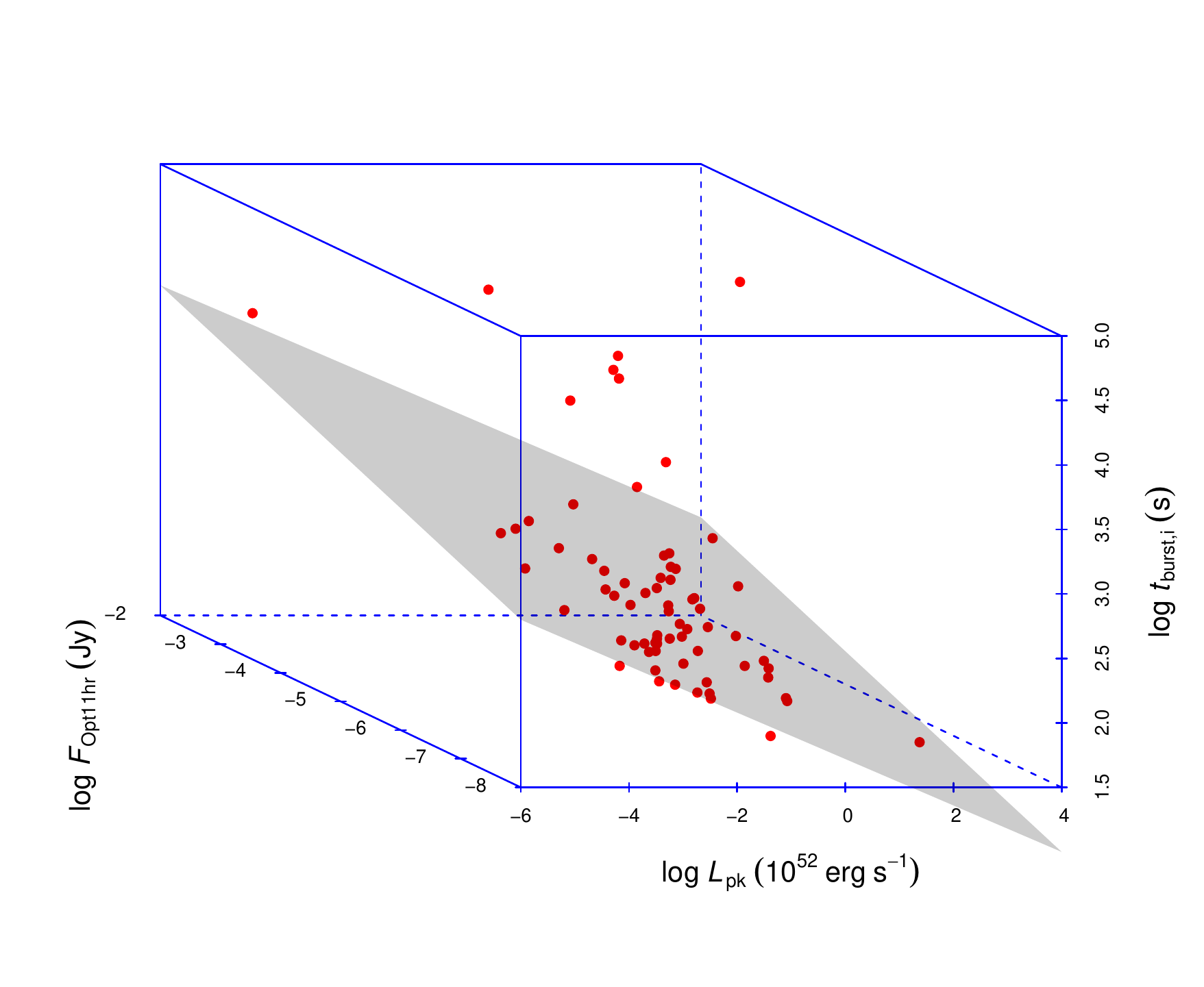}

\center{Fig. \ref{fig:three}---Continued}
\end{figure*}


\clearpage
\begin{figure*}

\includegraphics[width=0.45\textwidth]{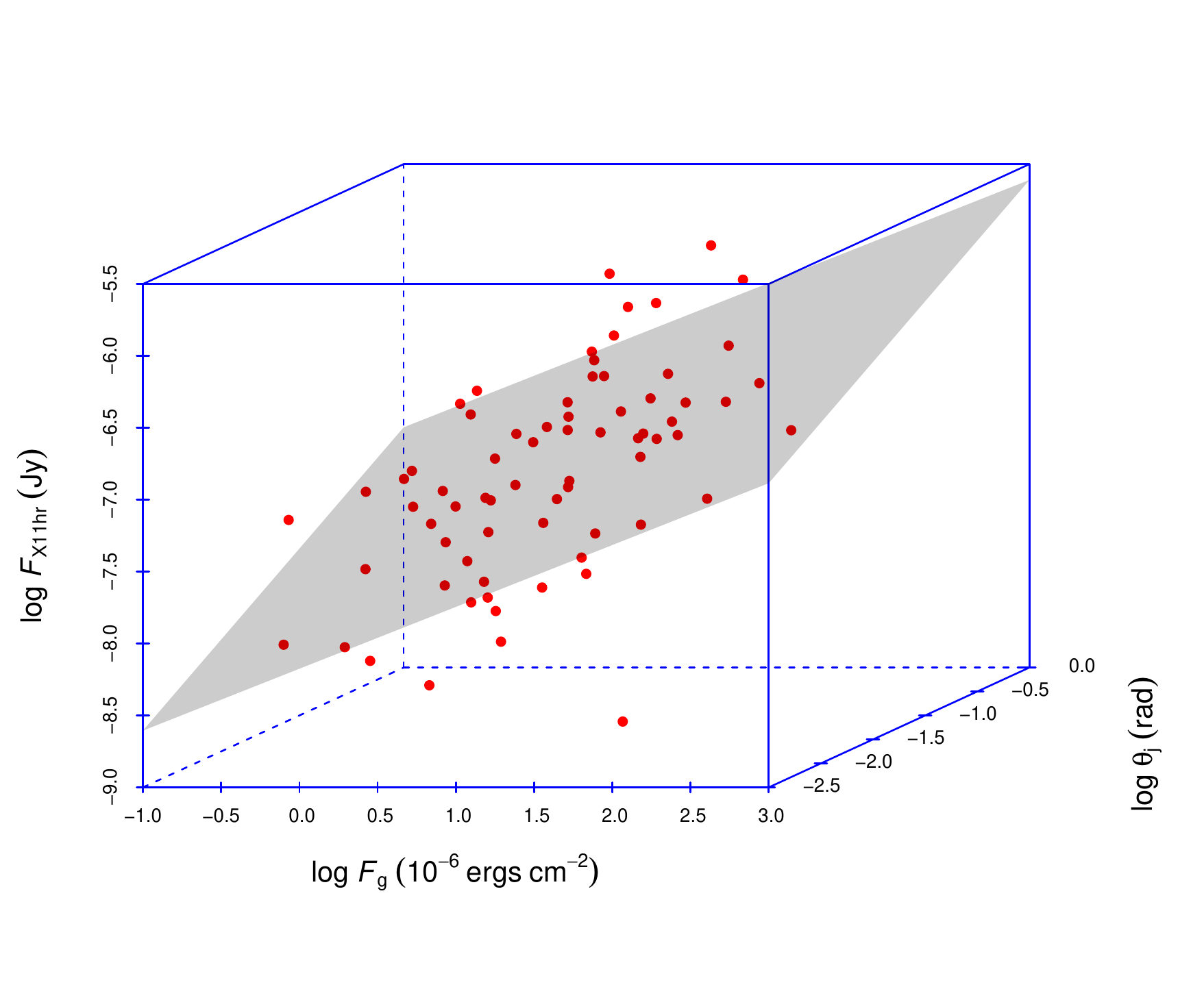}
\includegraphics[width=0.45\textwidth]{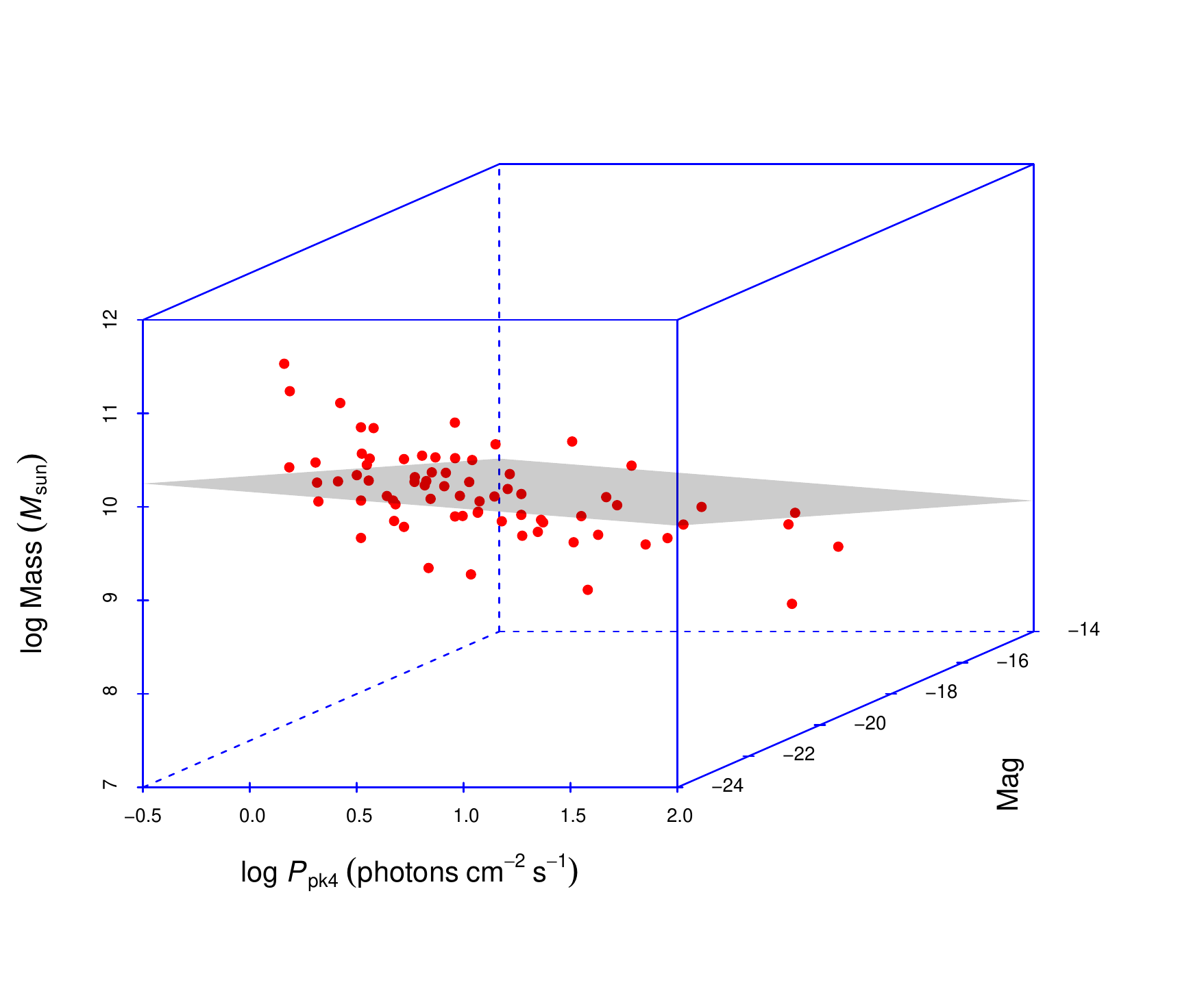}

\includegraphics[width=0.45\textwidth]{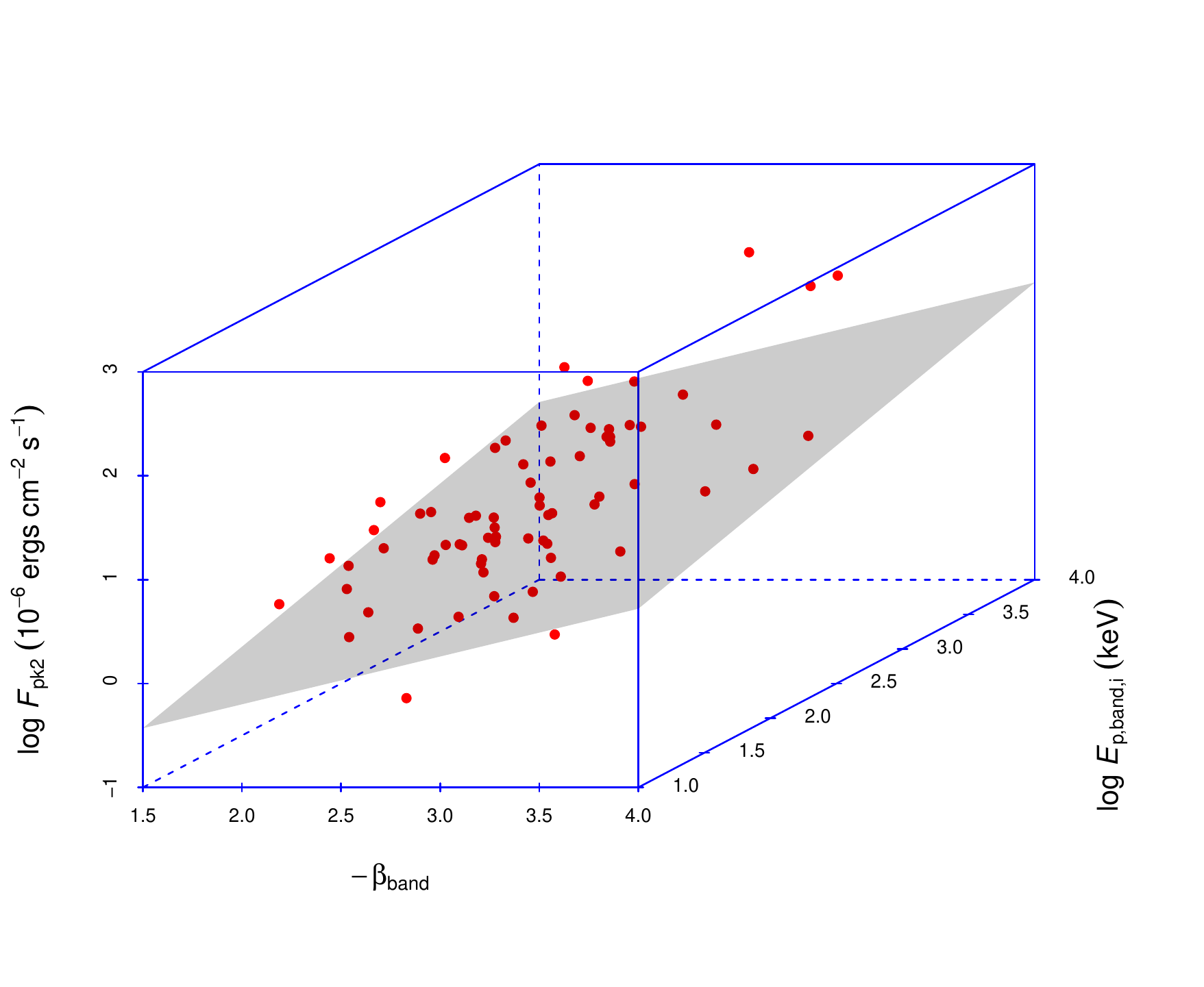}
\includegraphics[width=0.45\textwidth]{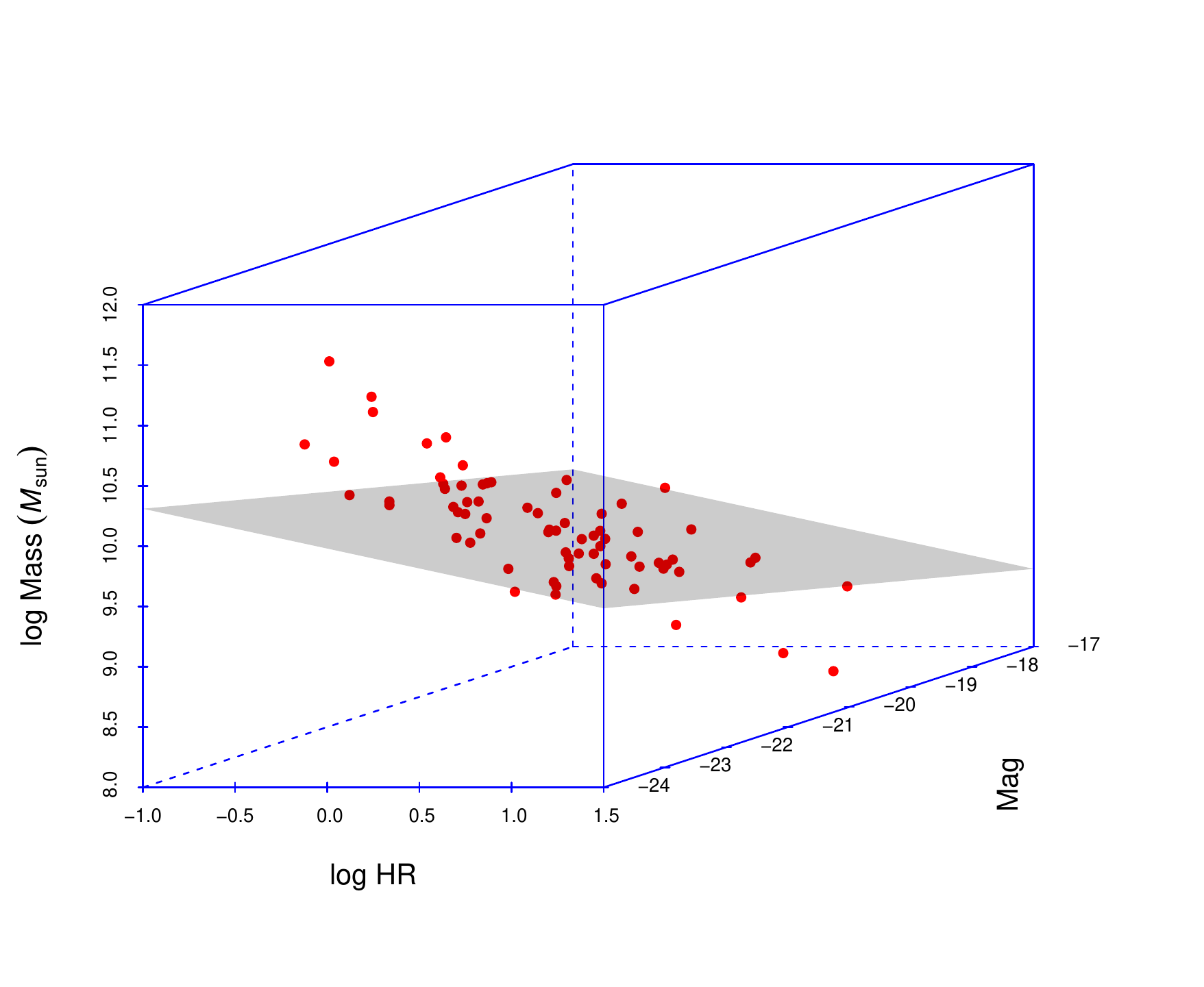}

\includegraphics[width=0.45\textwidth]{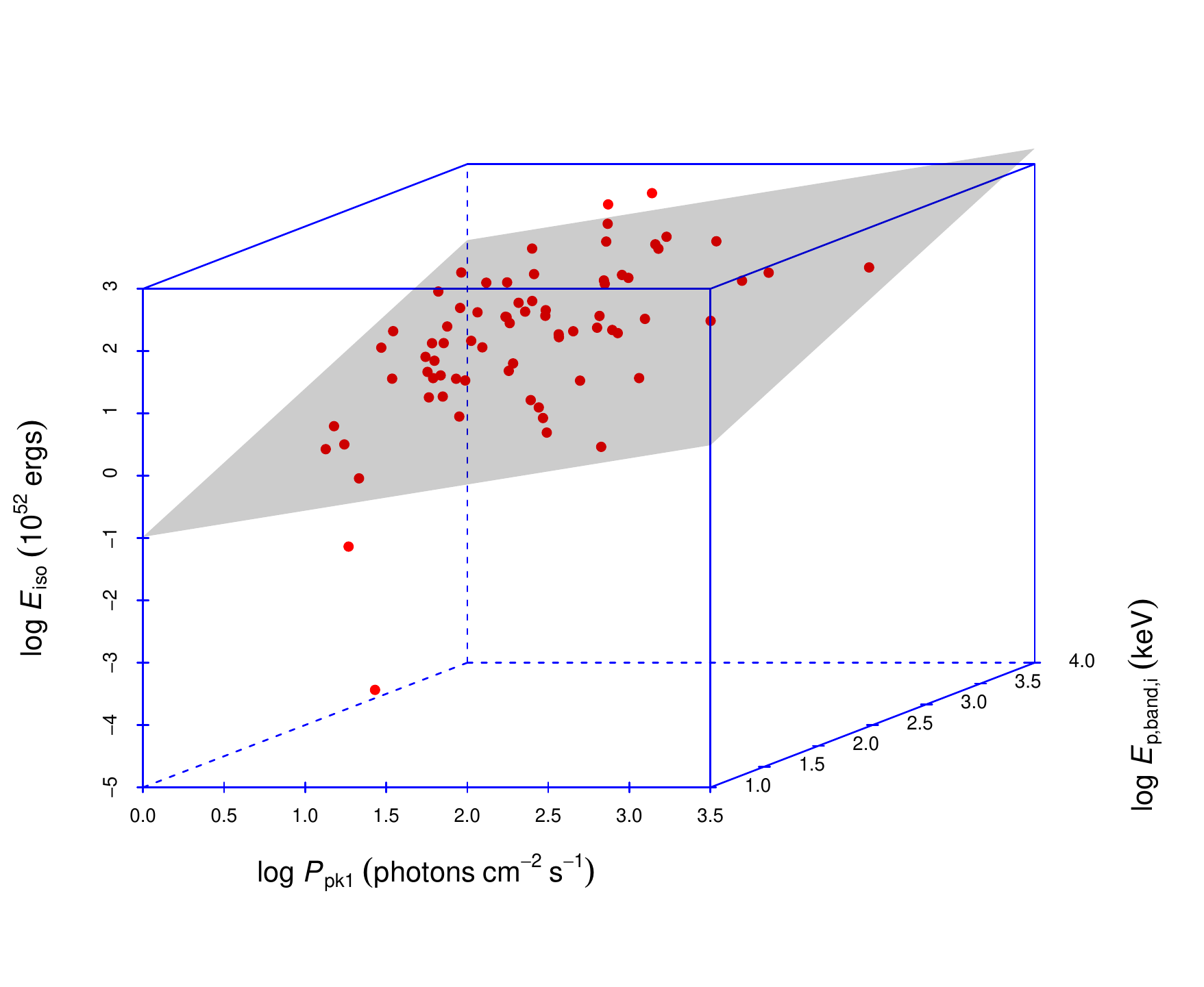}
\includegraphics[width=0.45\textwidth]{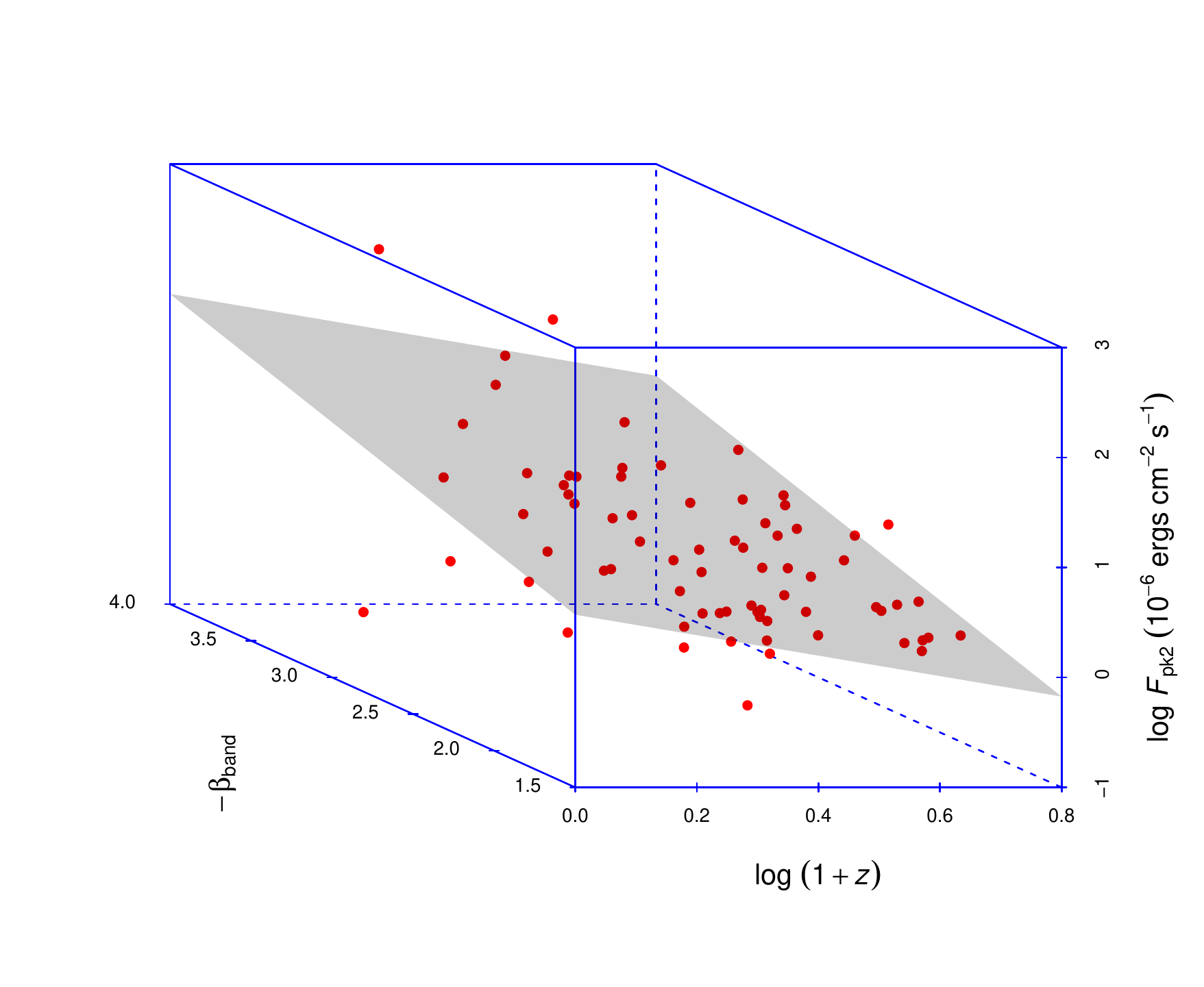}

\center{Fig. \ref{fig:three}---Continued}
\end{figure*}


\clearpage
\begin{figure*}

\includegraphics[width=0.45\textwidth]{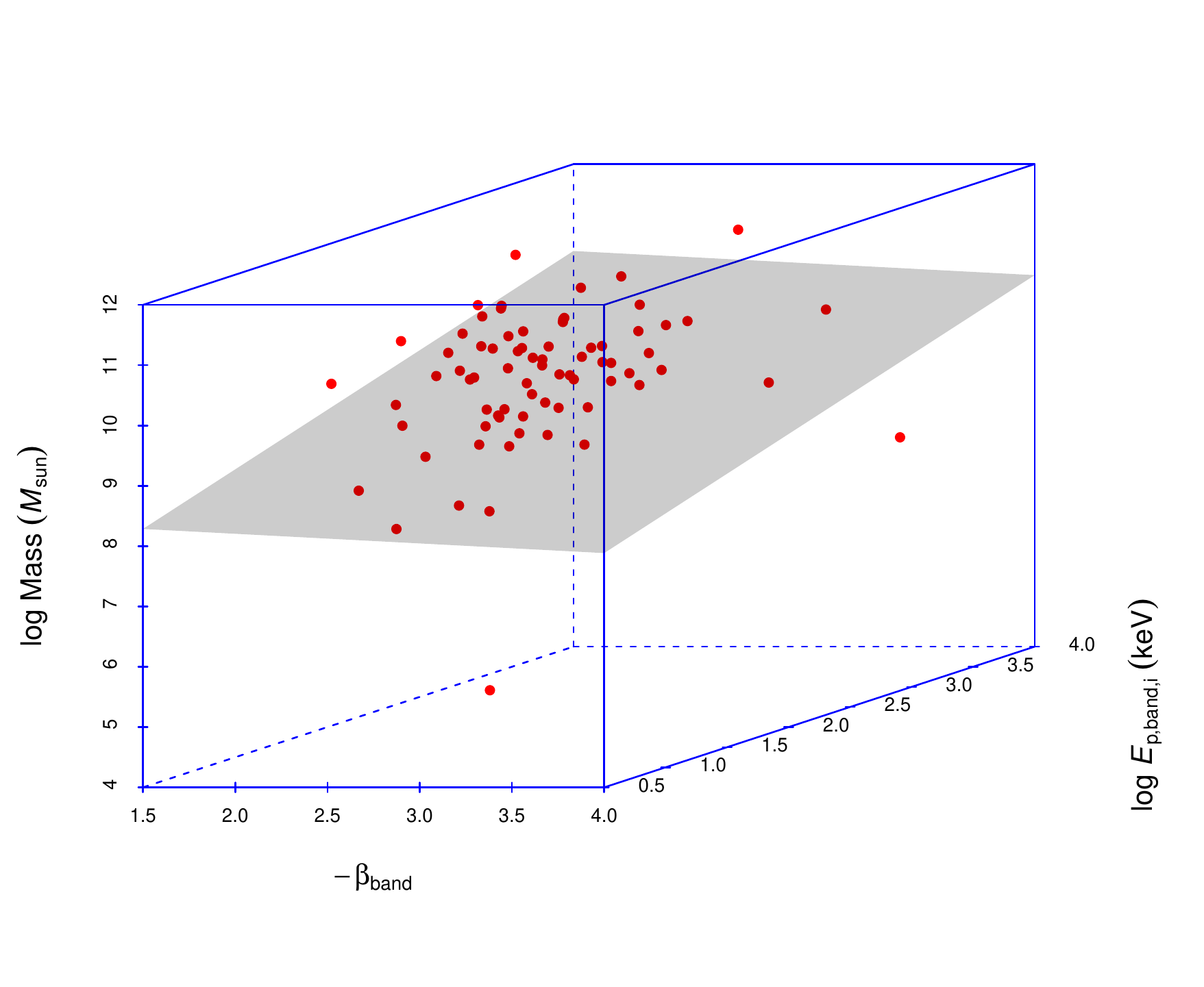}
\includegraphics[width=0.45\textwidth]{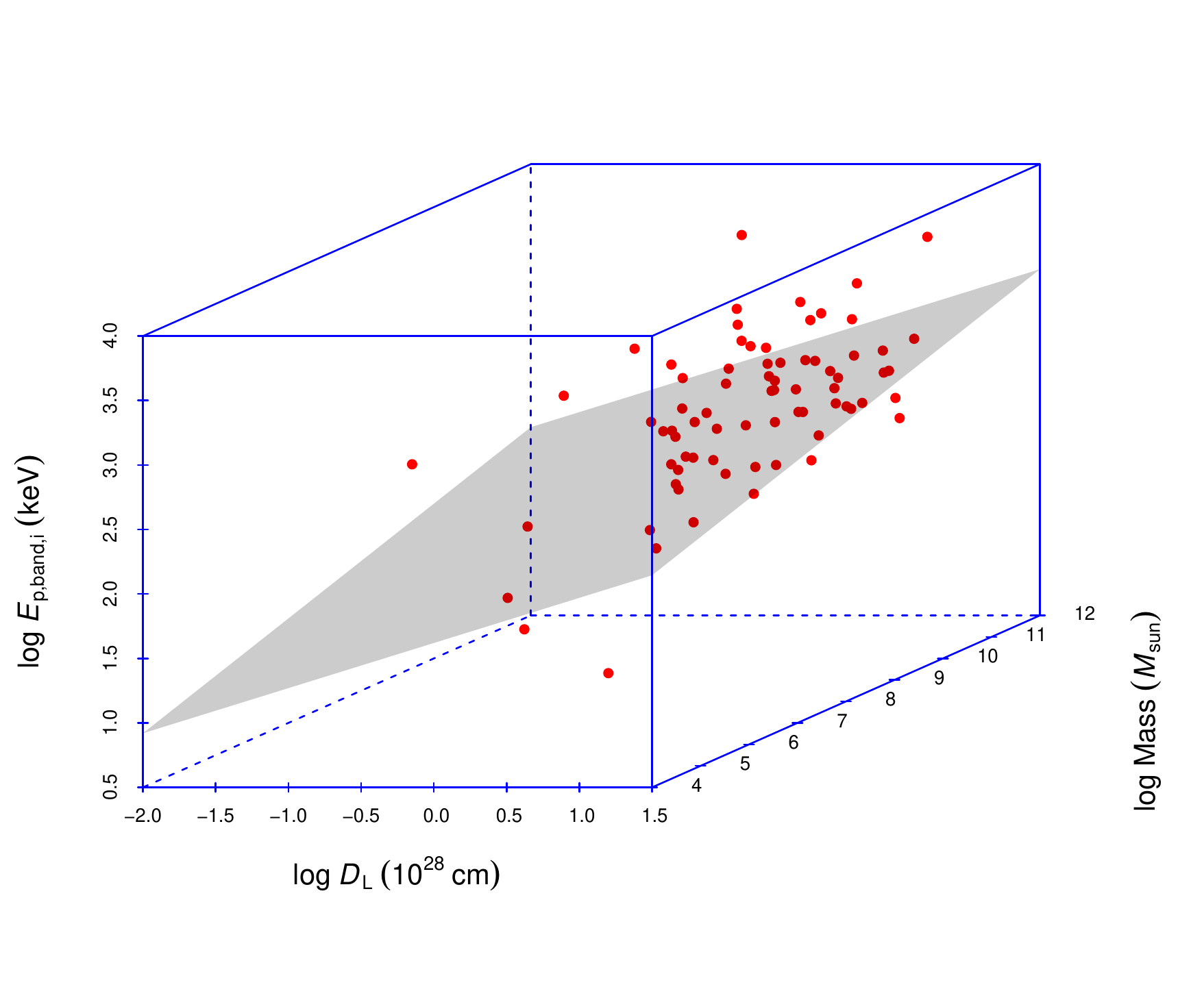}

\includegraphics[width=0.45\textwidth]{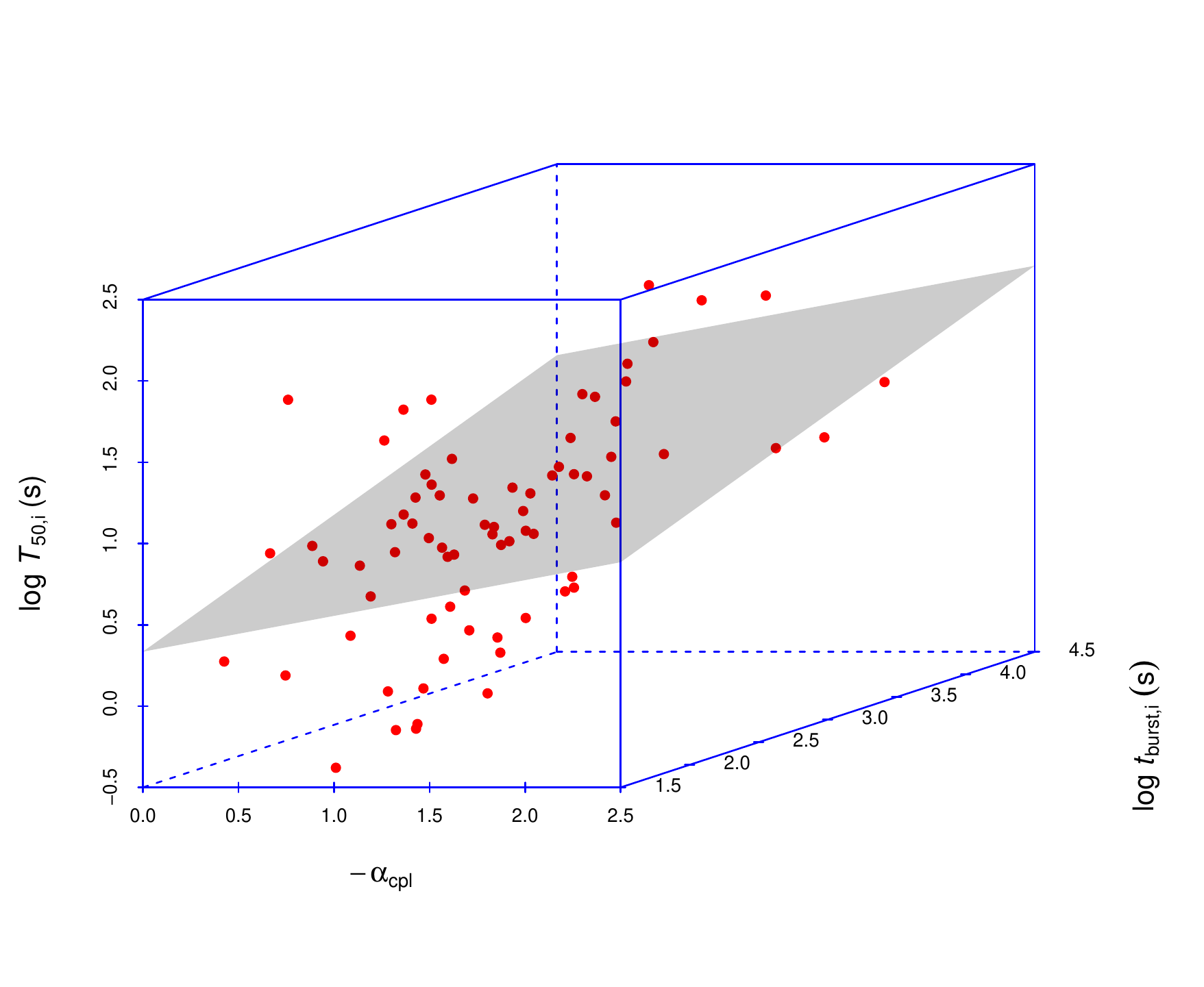}
\includegraphics[width=0.45\textwidth]{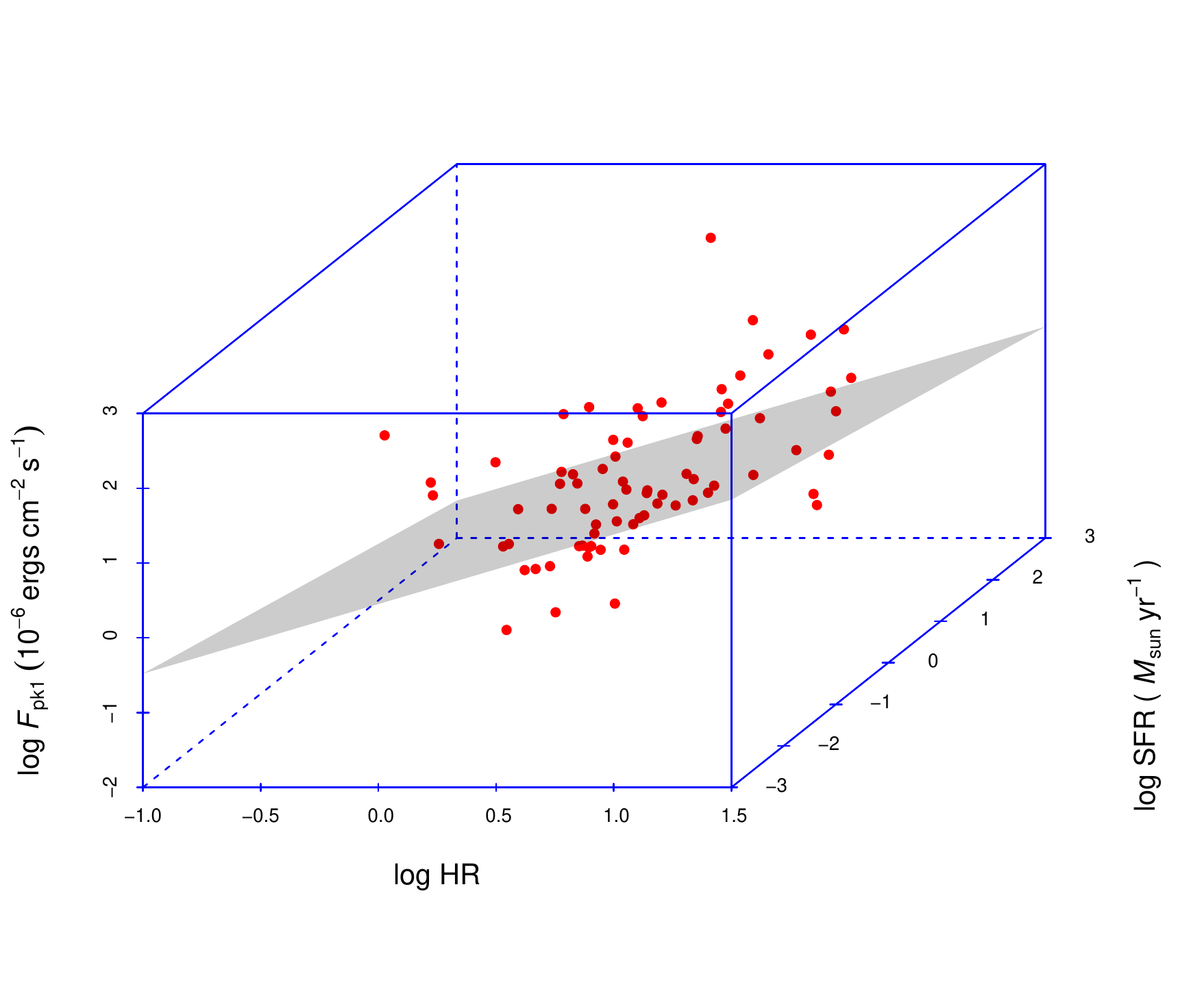}

\includegraphics[width=0.45\textwidth]{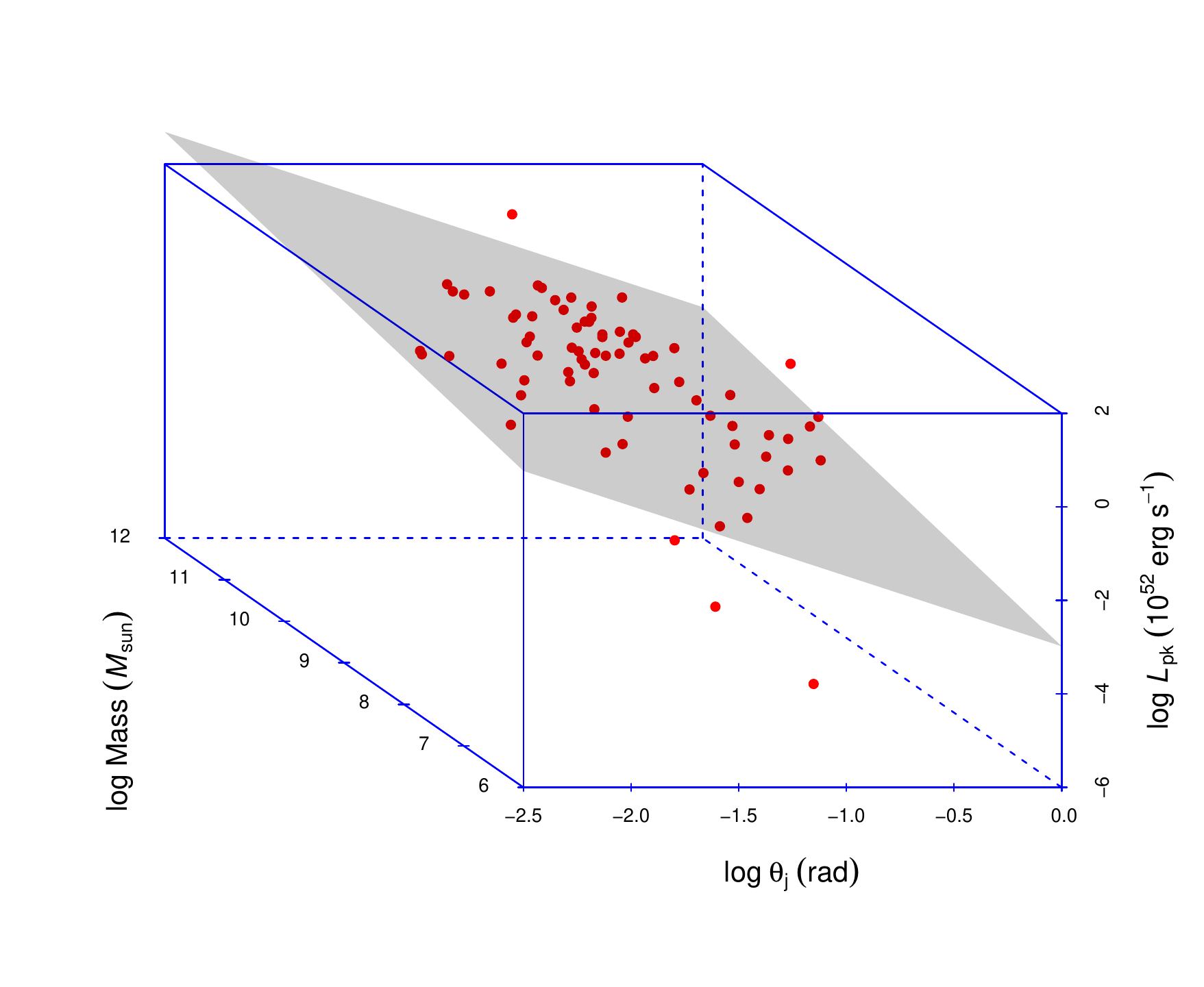}
\includegraphics[width=0.45\textwidth]{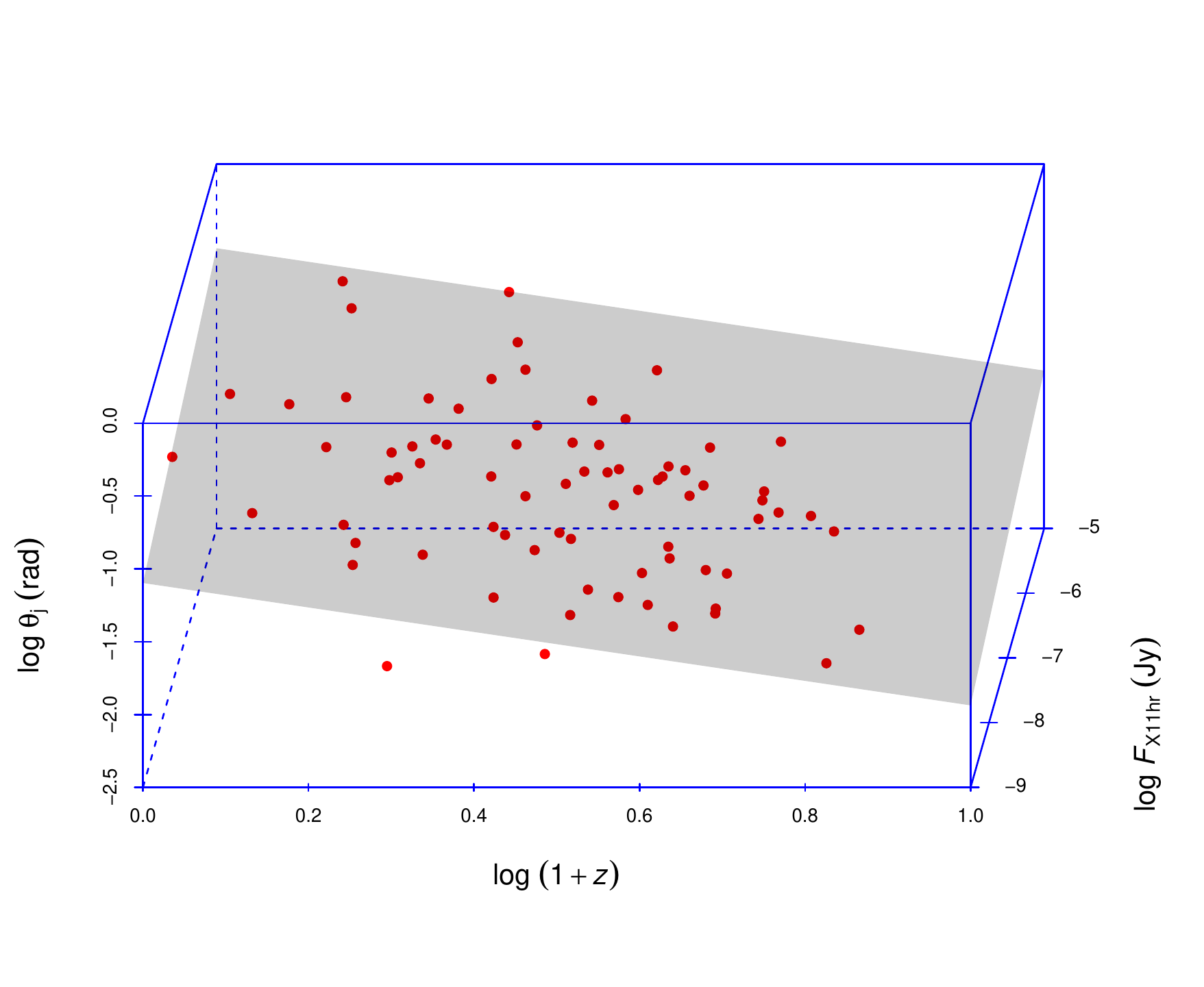}

\center{Fig. \ref{fig:three}---Continued}
\end{figure*}


\clearpage
\begin{figure*}

\includegraphics[width=0.45\textwidth]{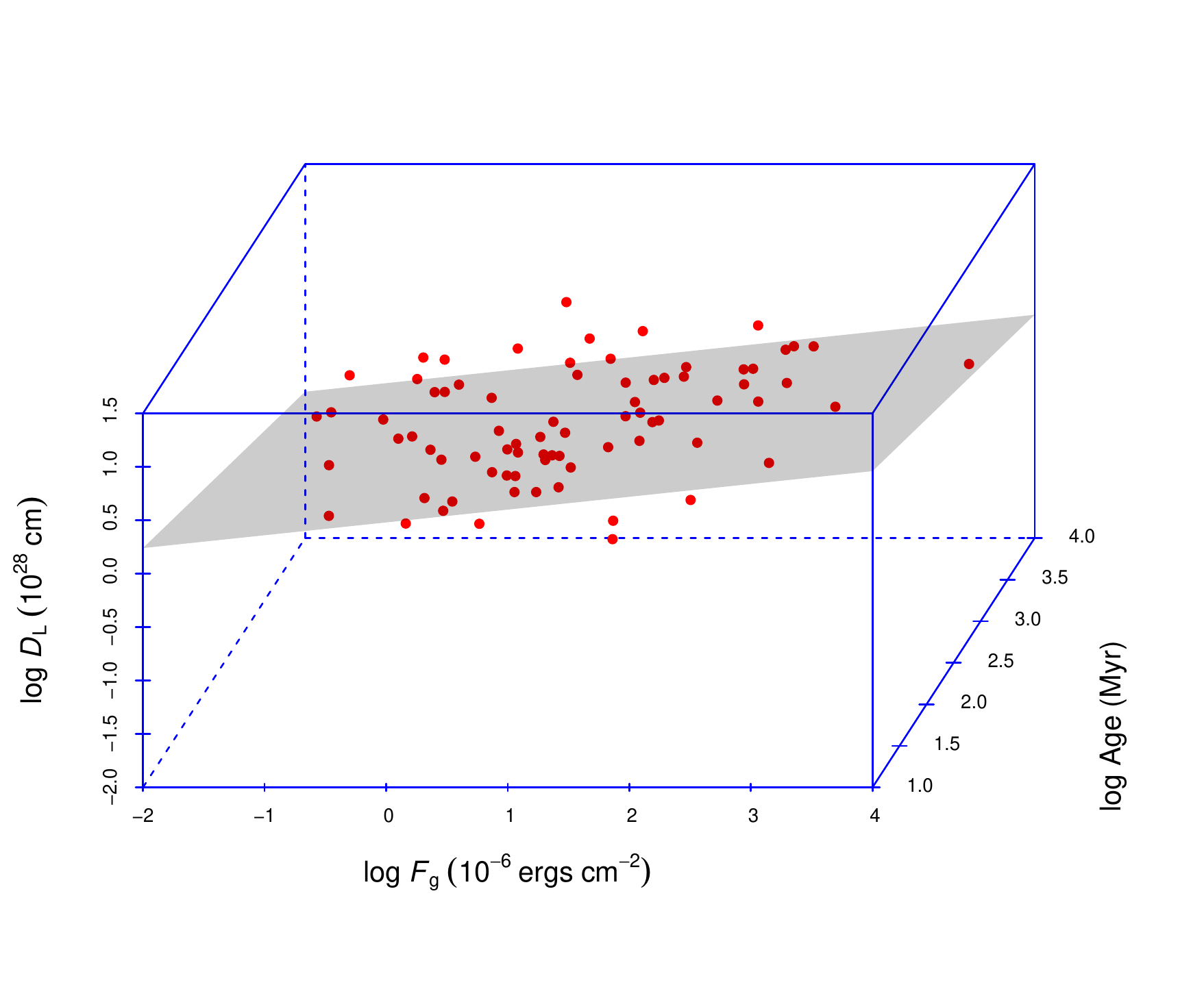}
\includegraphics[width=0.45\textwidth]{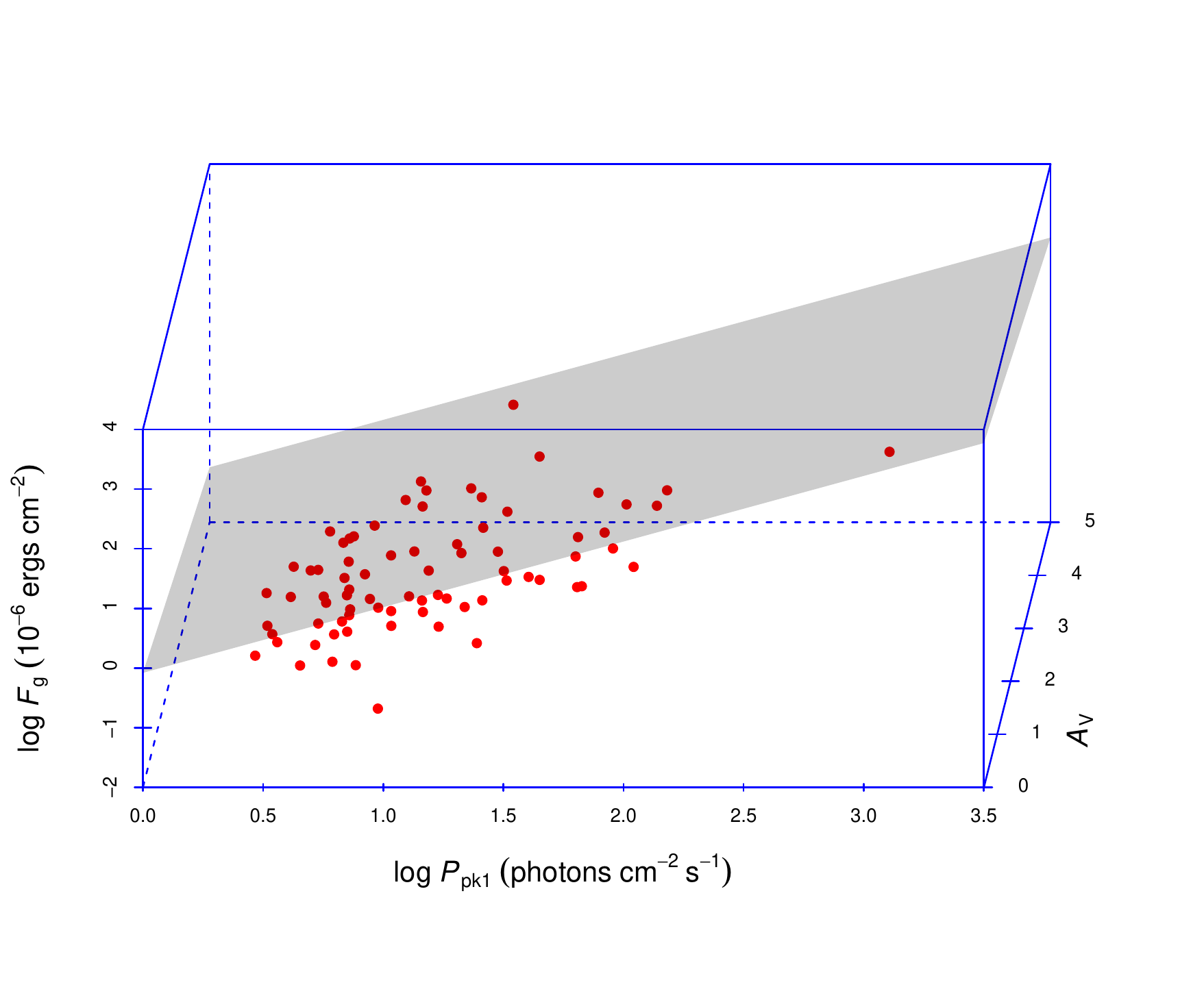}

\includegraphics[width=0.45\textwidth]{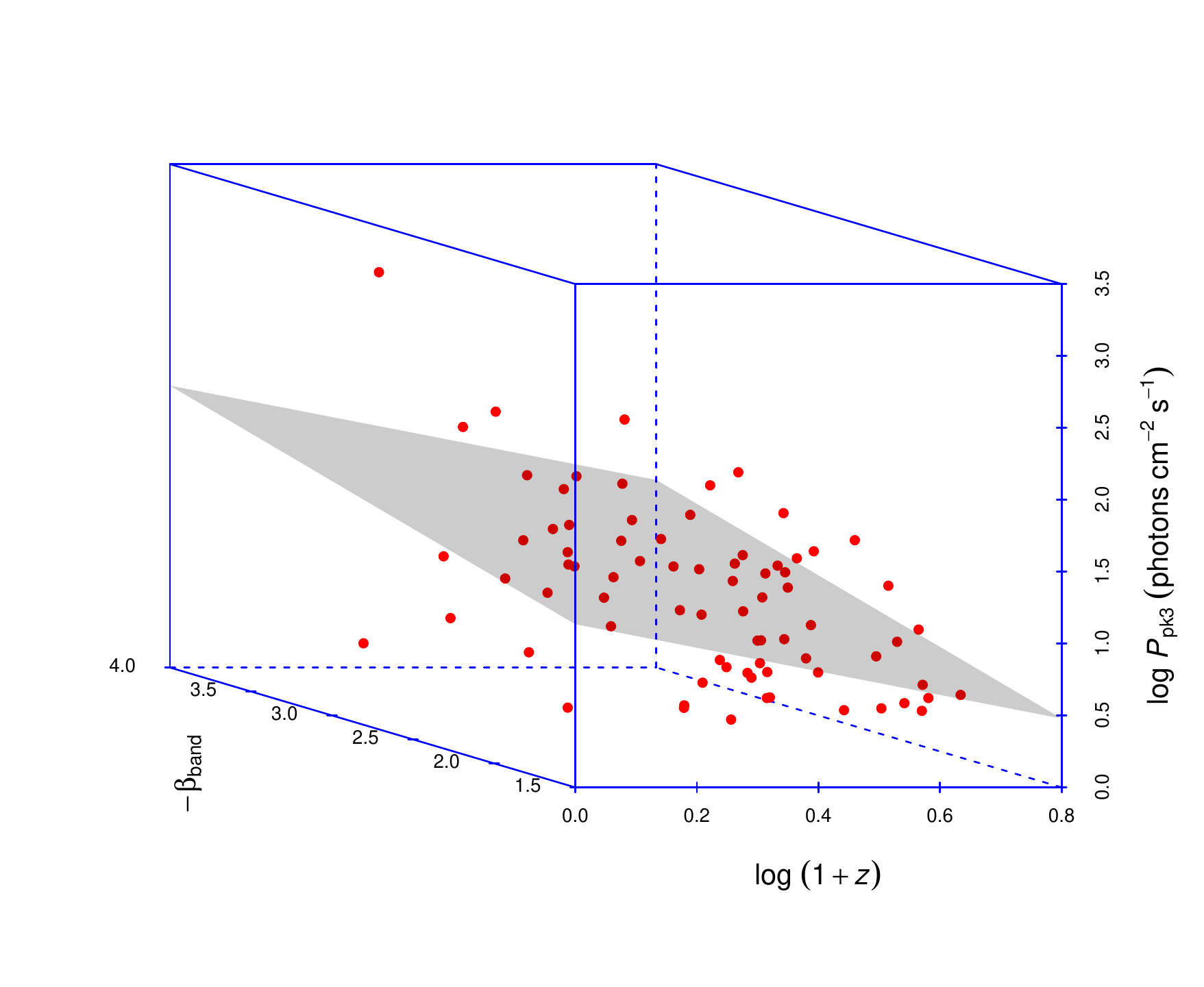}
\includegraphics[width=0.45\textwidth]{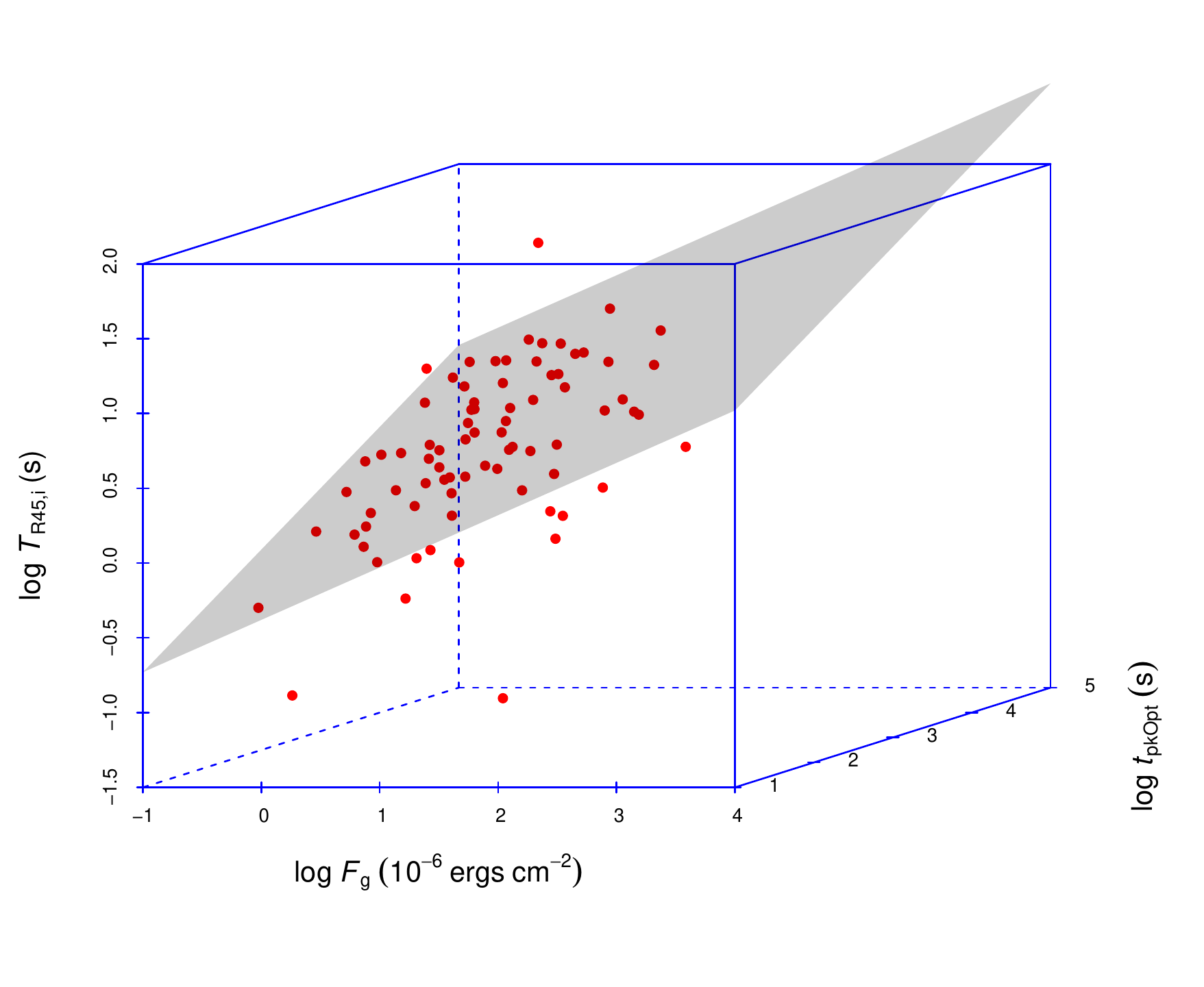}

\includegraphics[width=0.45\textwidth]{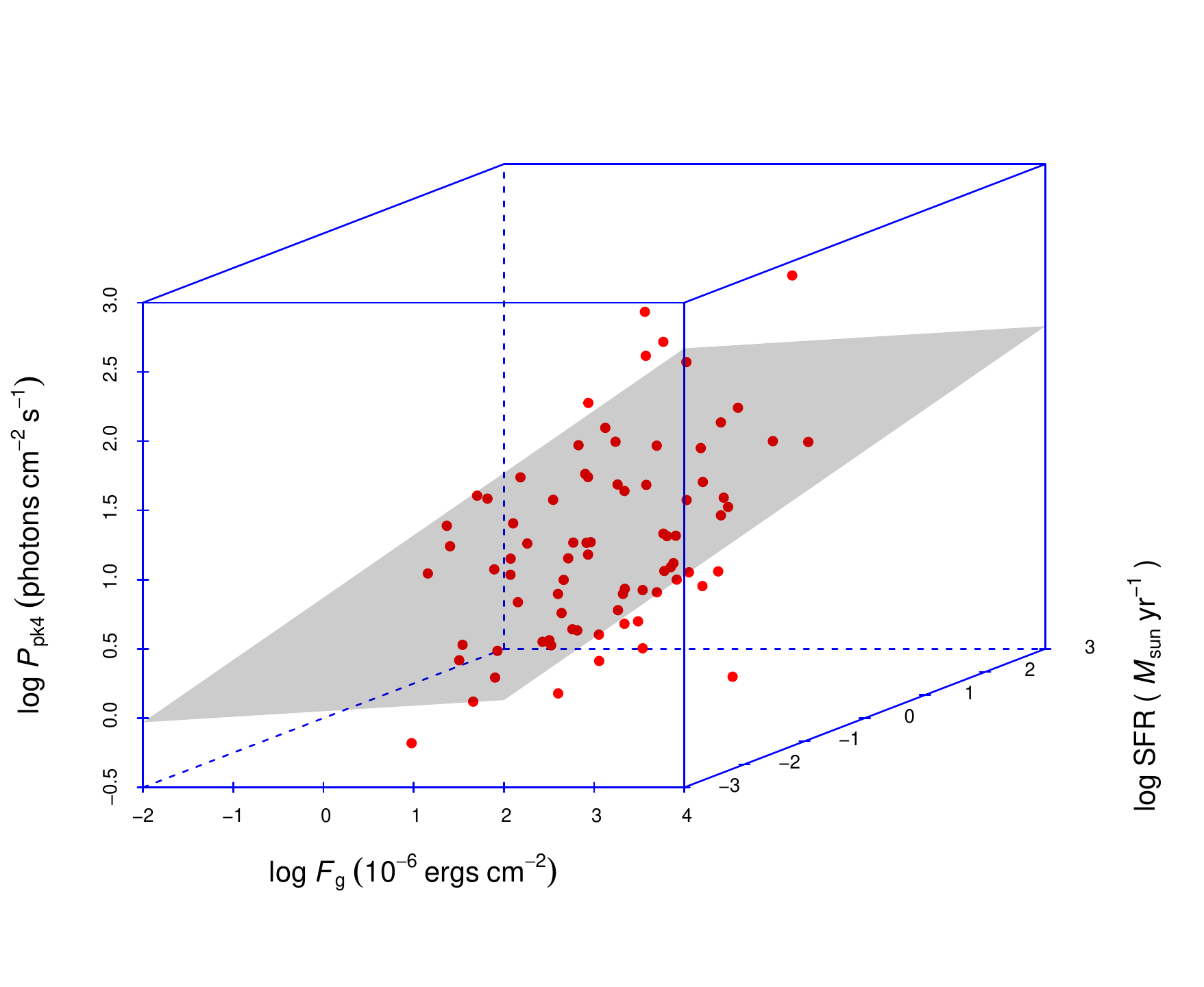}
\includegraphics[width=0.45\textwidth]{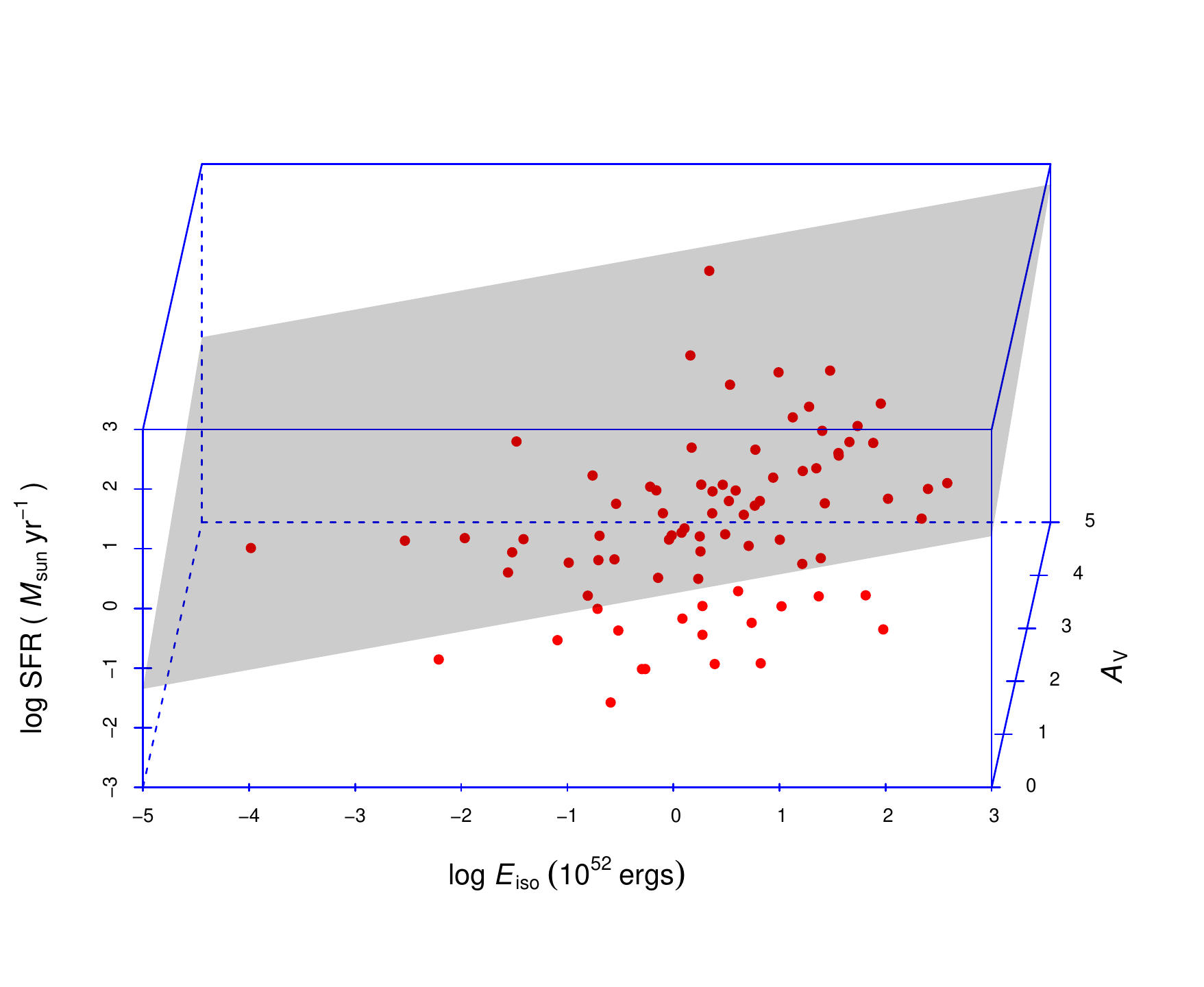}

\center{Fig. \ref{fig:three}---Continued}
\end{figure*}


\clearpage
\begin{figure*}

\includegraphics[width=0.45\textwidth]{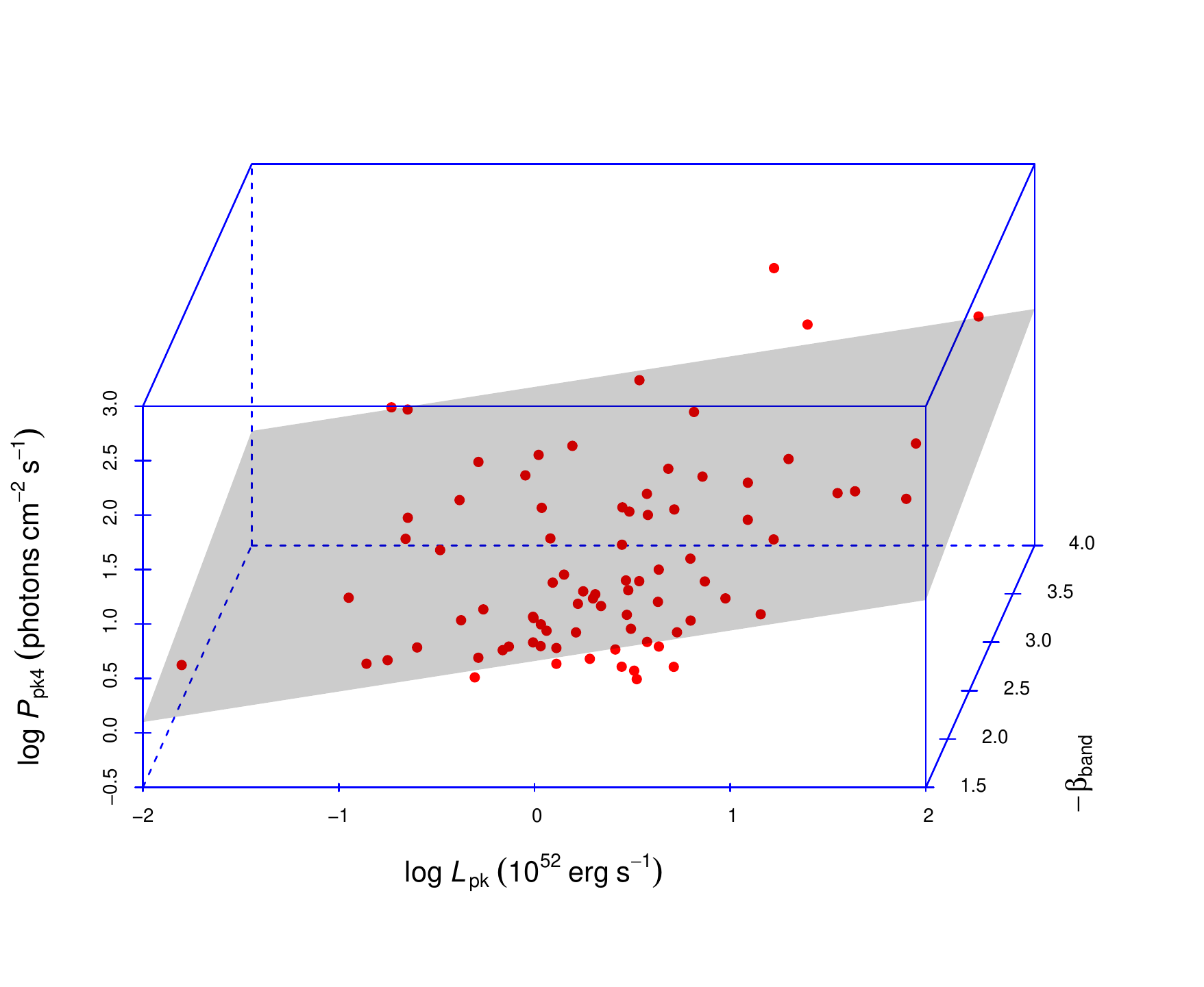}
\includegraphics[width=0.45\textwidth]{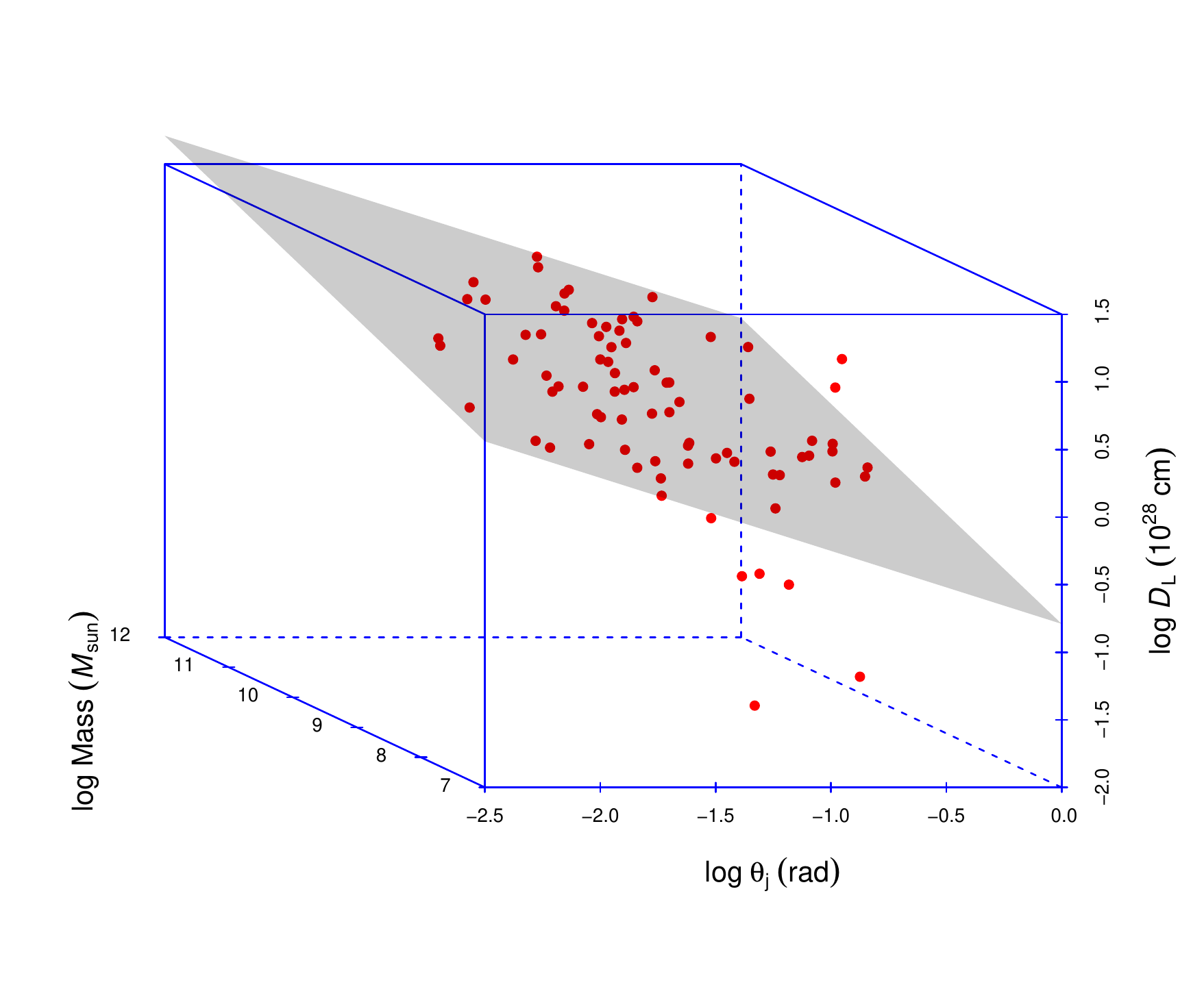}

\includegraphics[width=0.45\textwidth]{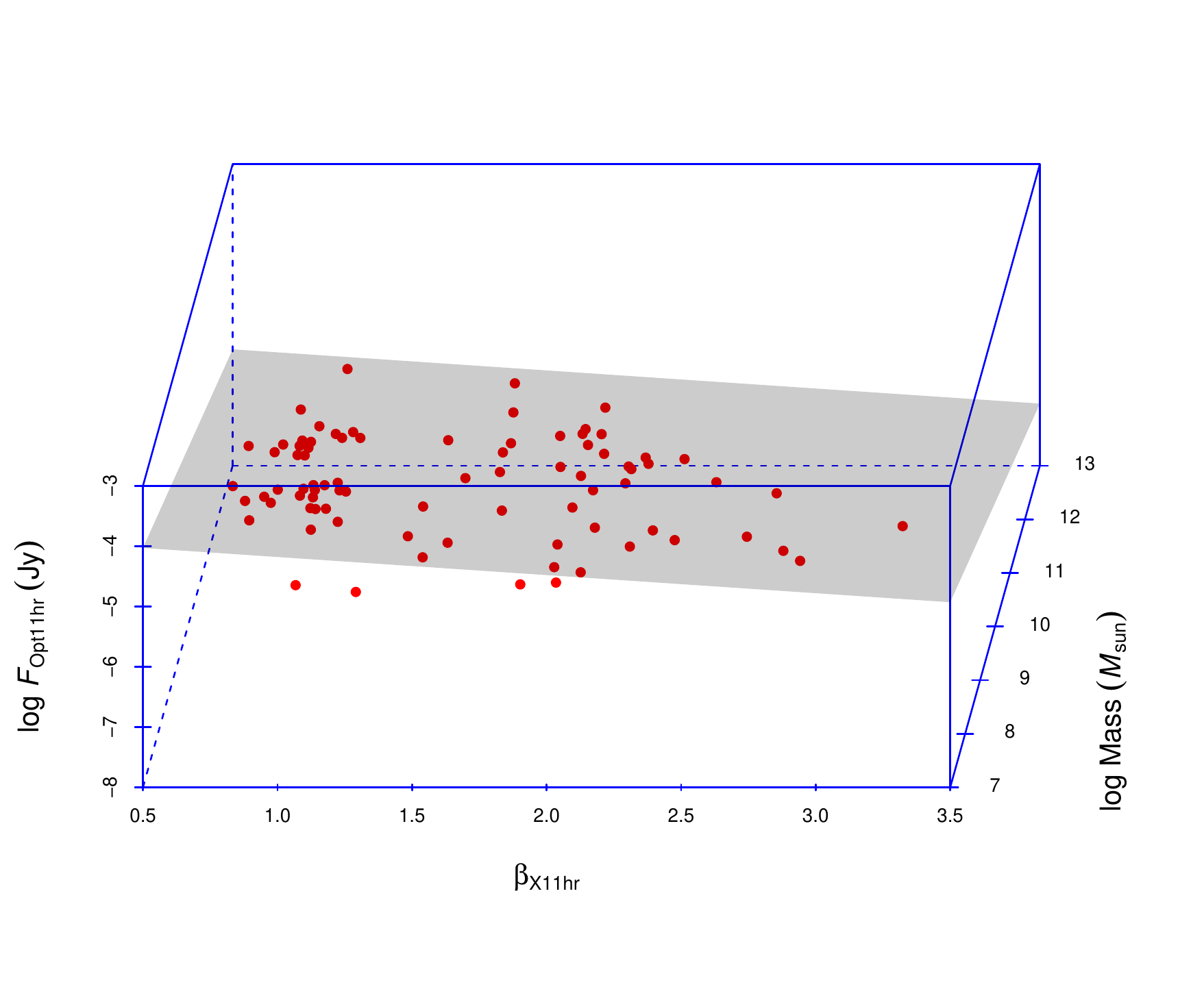}
\includegraphics[width=0.45\textwidth]{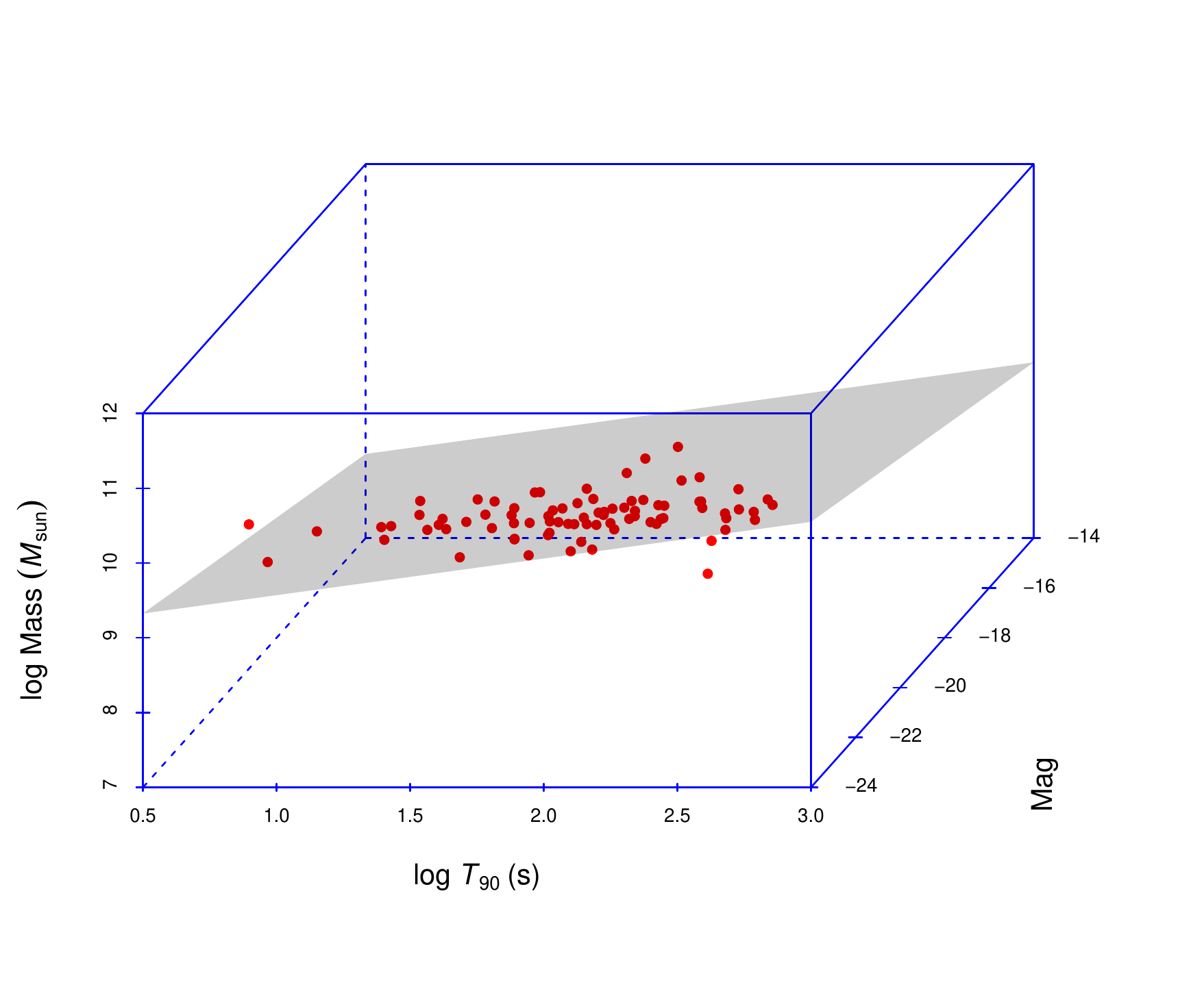}

\includegraphics[width=0.45\textwidth]{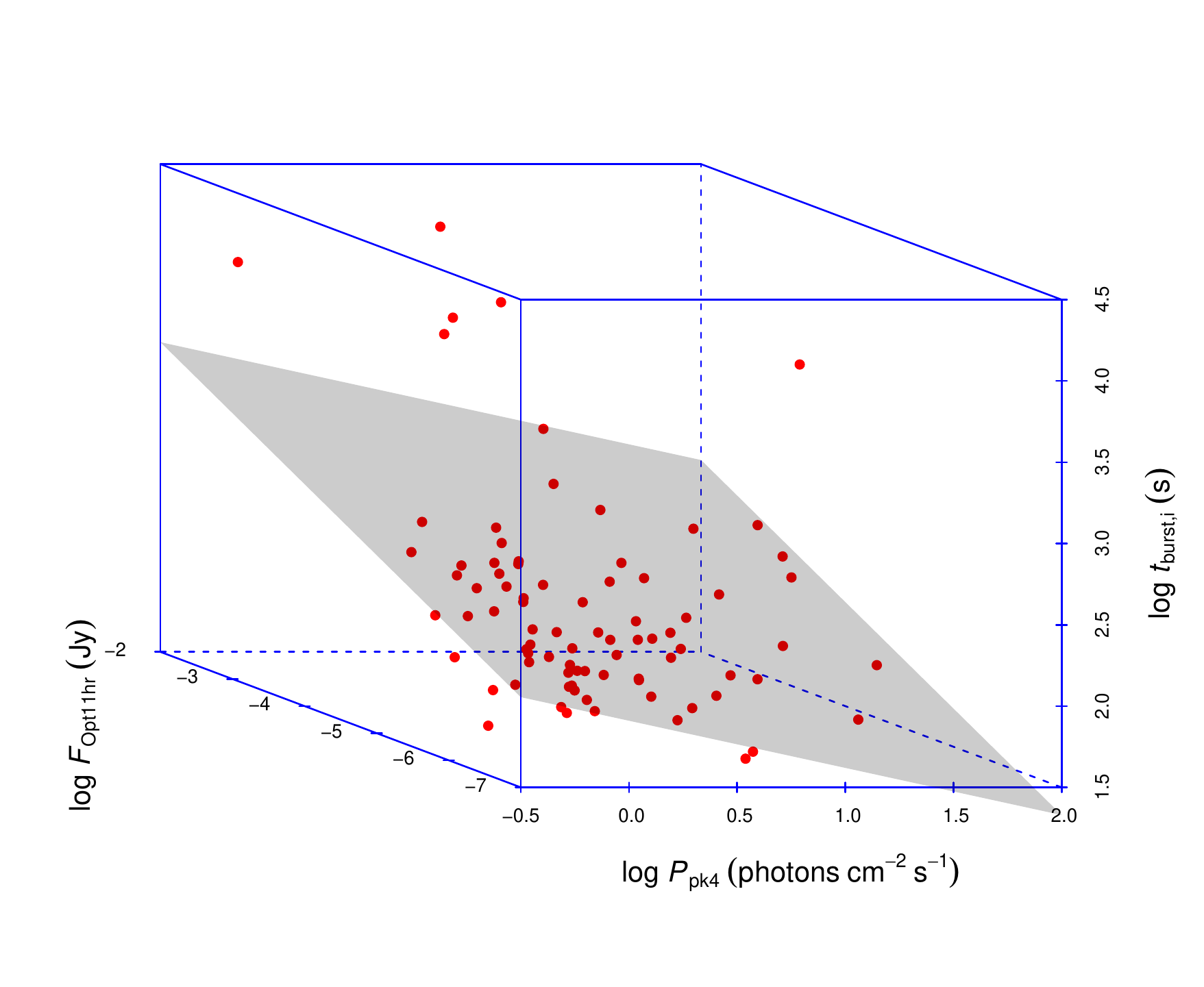}
\includegraphics[width=0.45\textwidth]{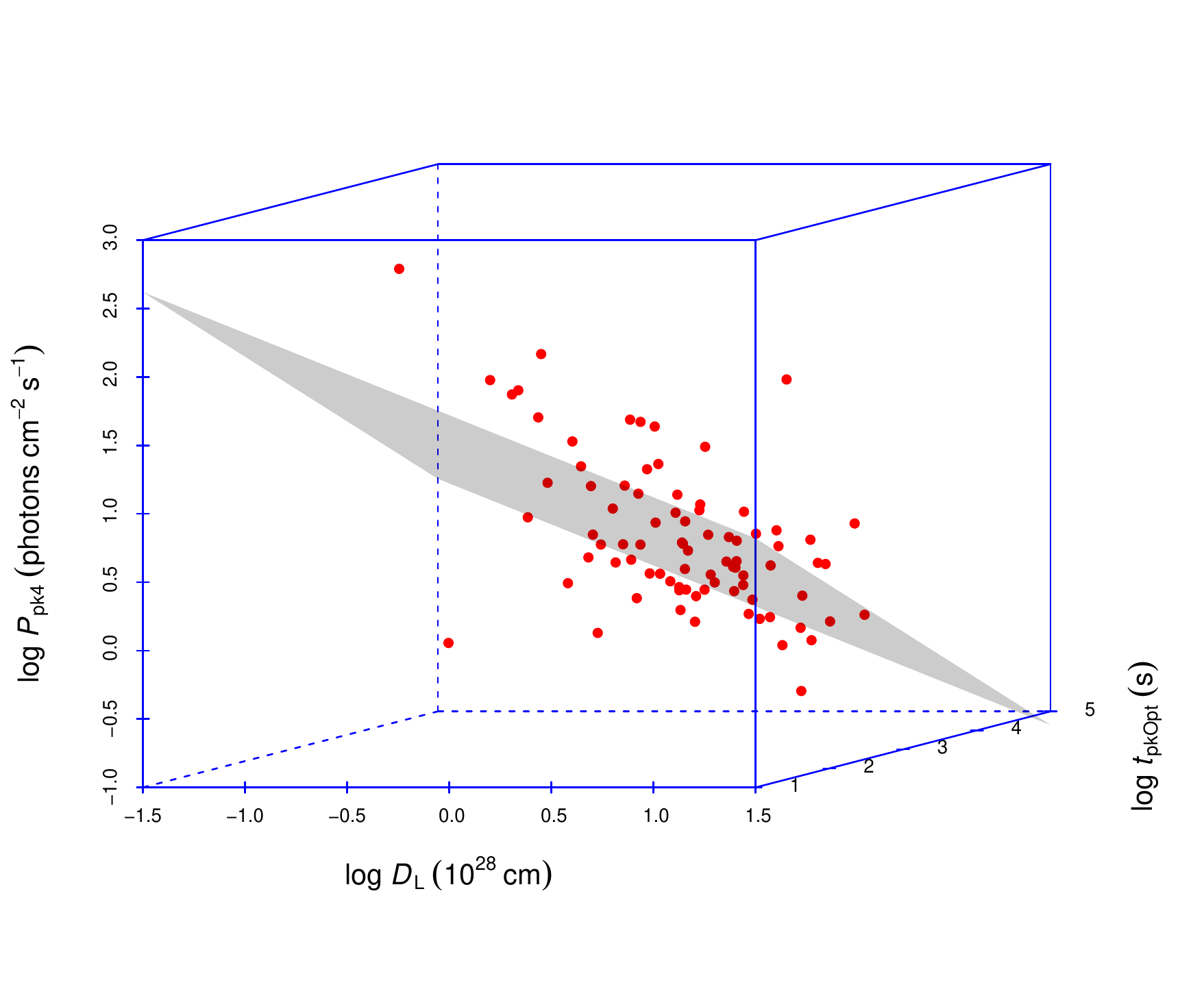}

\center{Fig. \ref{fig:three}---Continued}
\end{figure*}


\clearpage
\begin{figure*}

\includegraphics[width=0.45\textwidth]{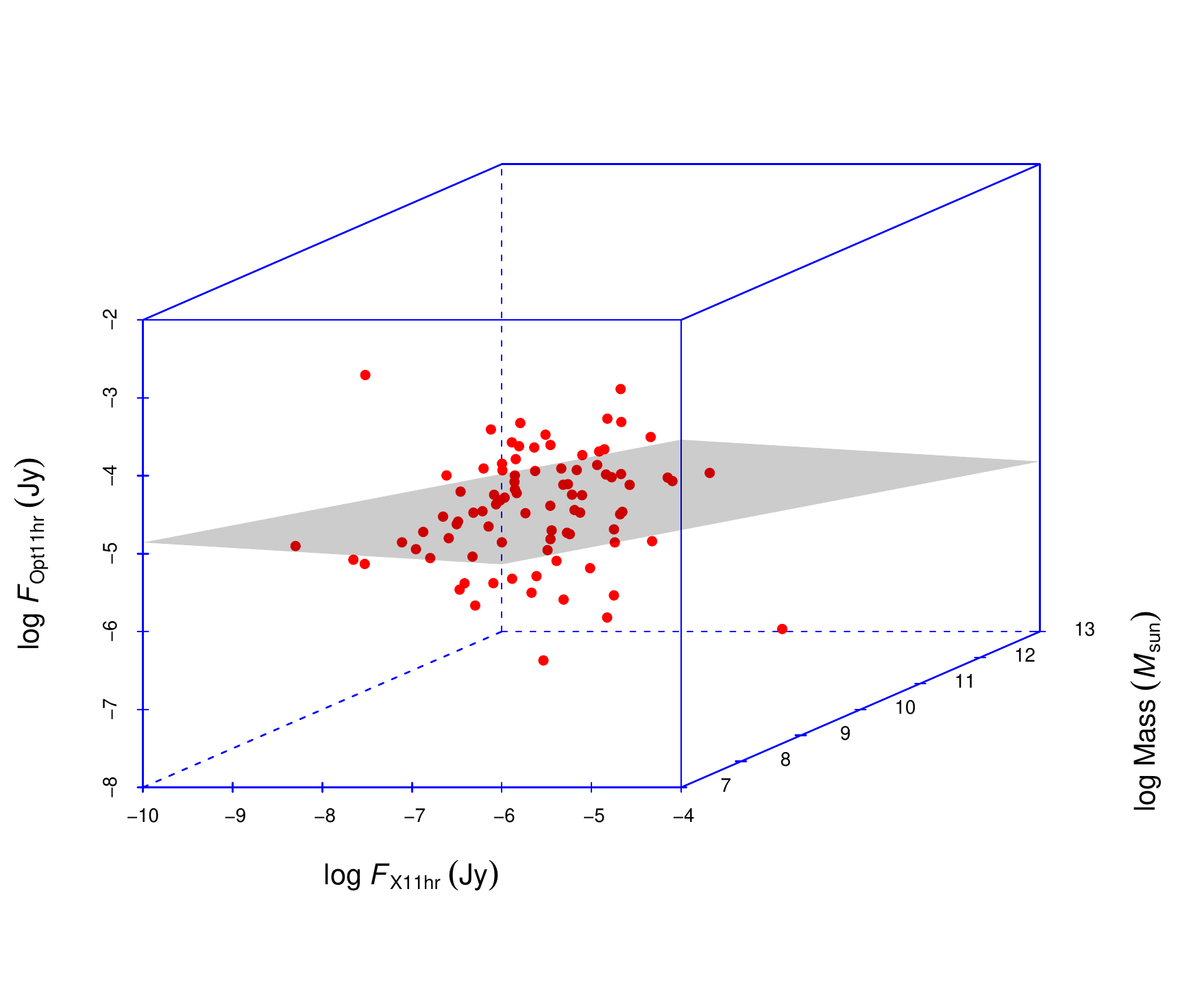}
\includegraphics[width=0.45\textwidth]{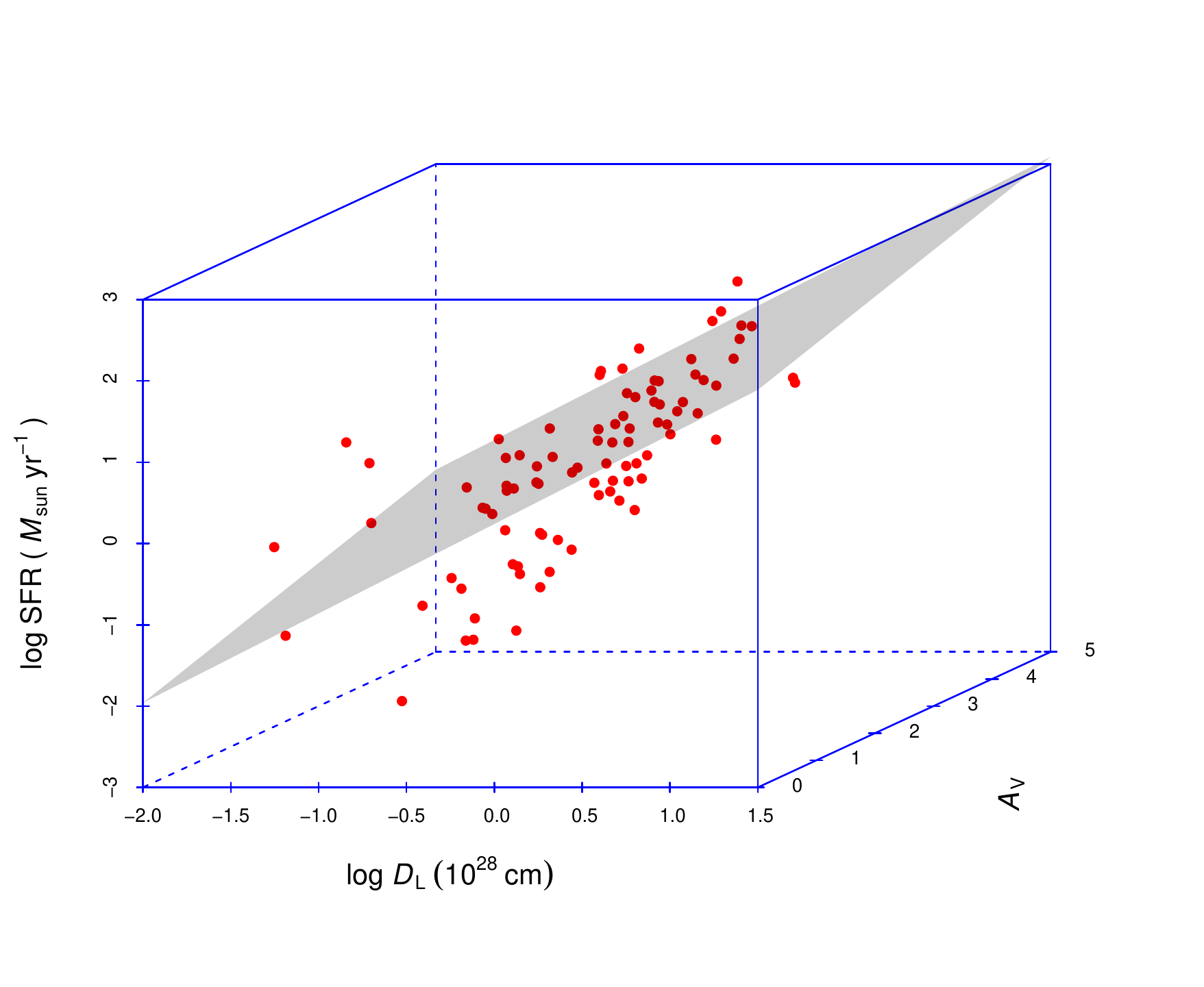}

\includegraphics[width=0.45\textwidth]{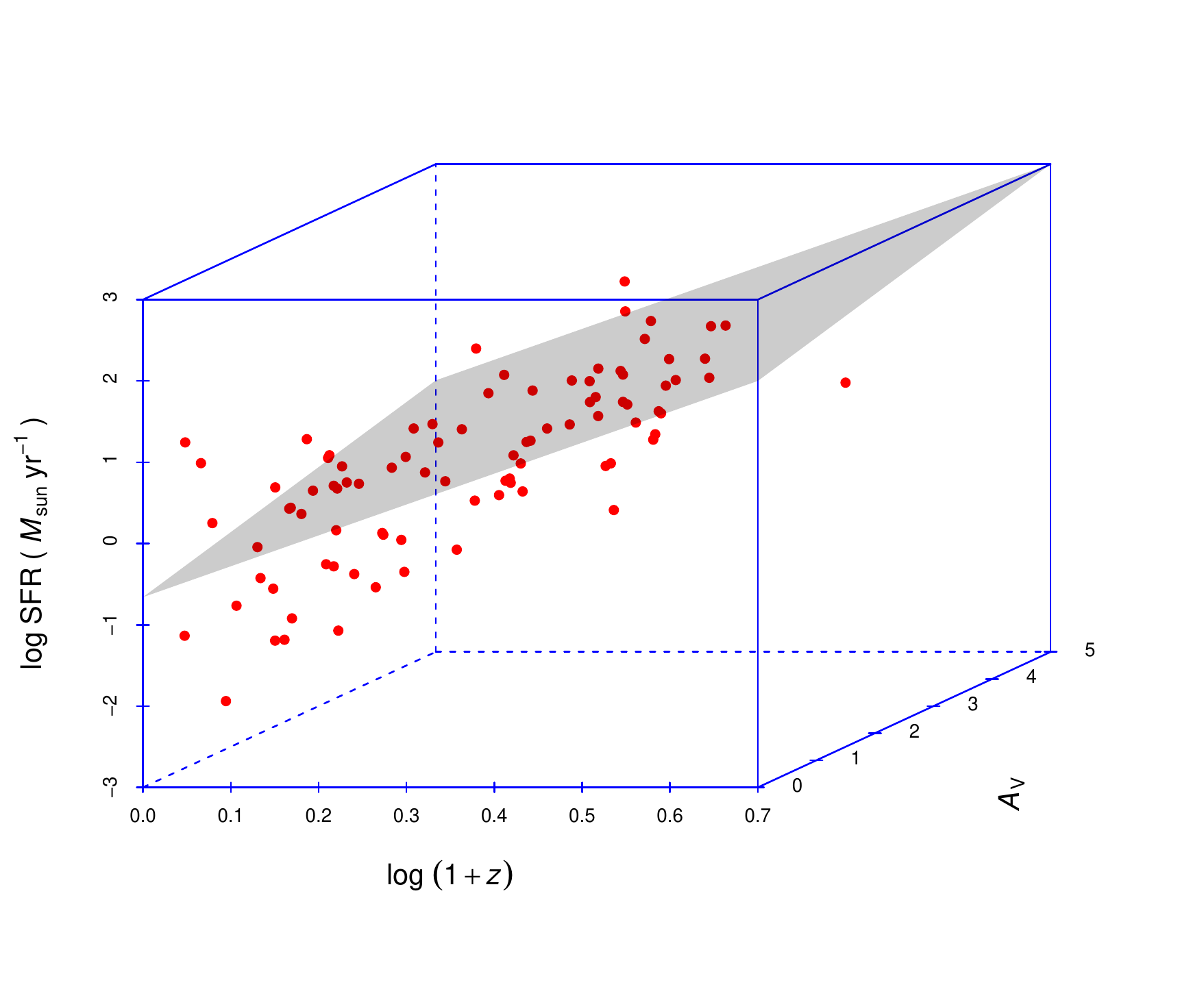}
\includegraphics[width=0.45\textwidth]{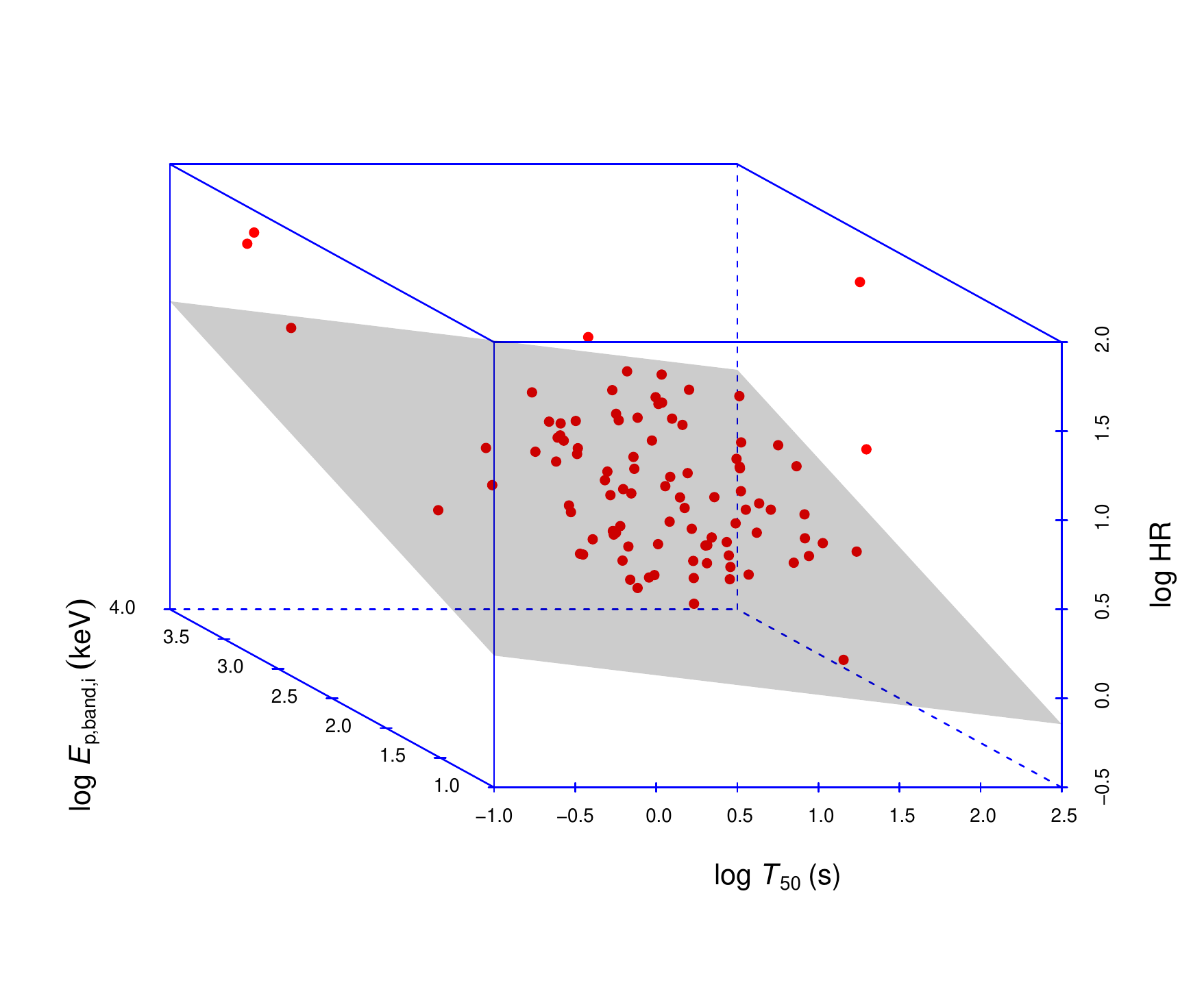}

\includegraphics[width=0.45\textwidth]{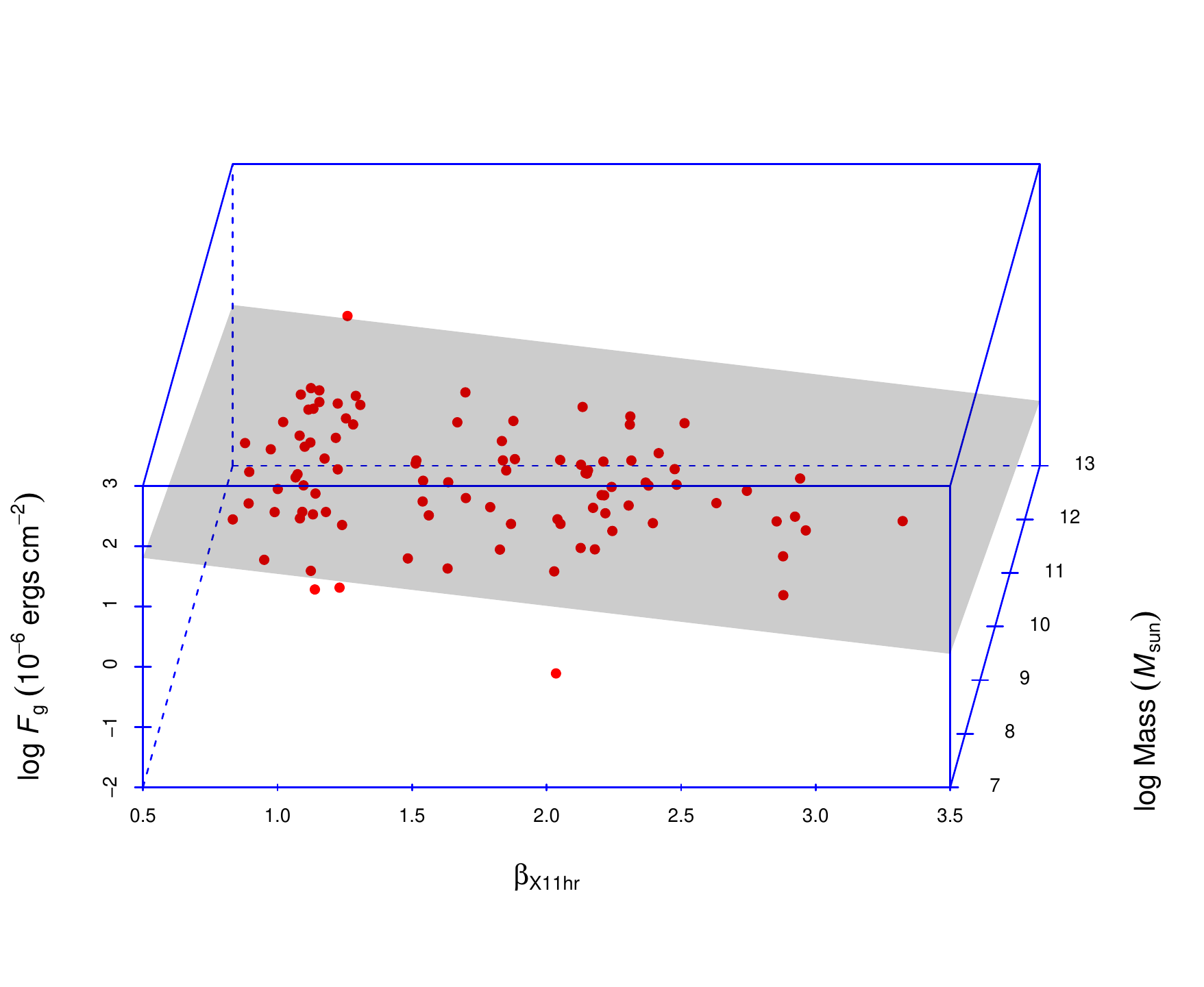}
\includegraphics[width=0.45\textwidth]{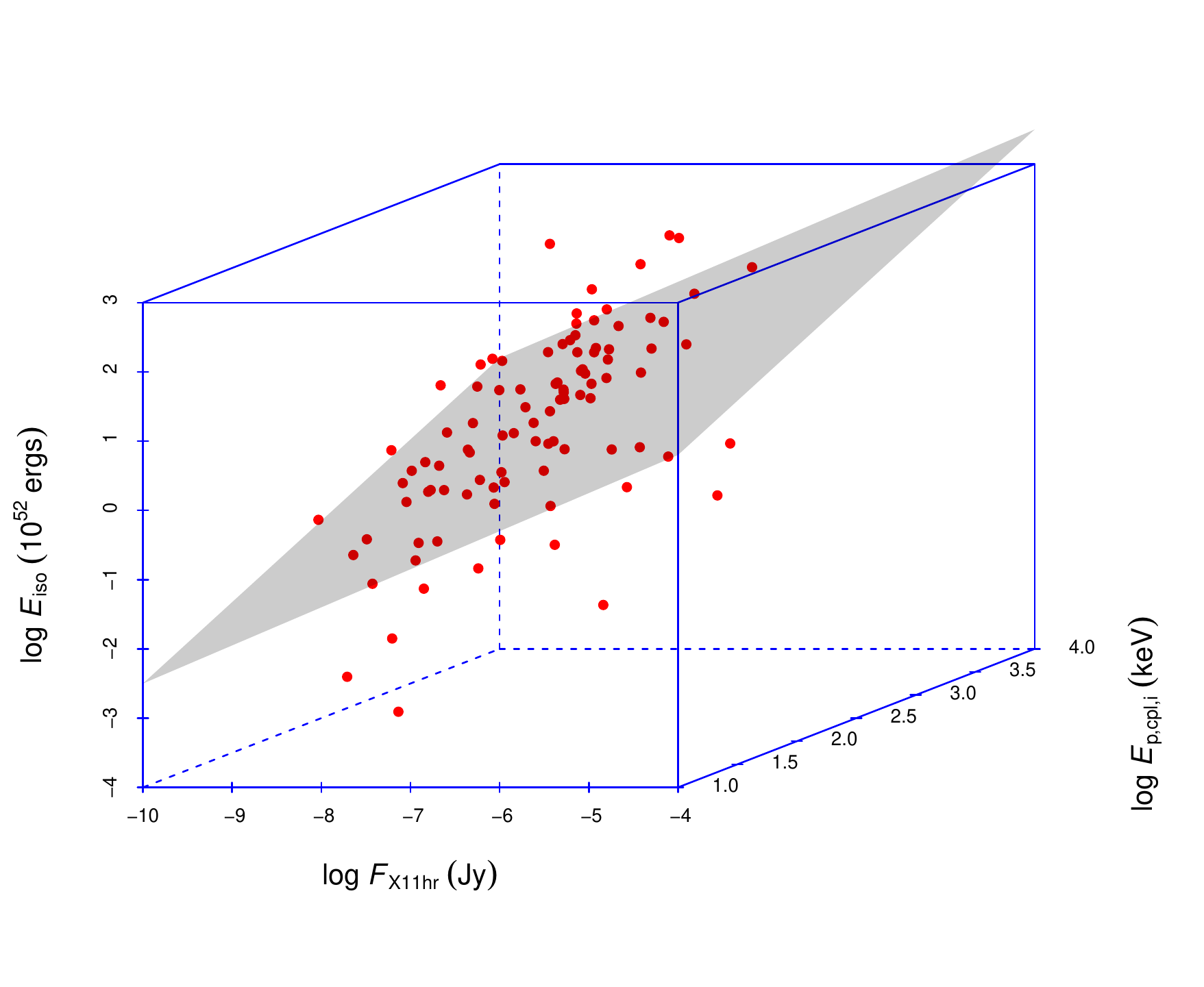}

\center{Fig. \ref{fig:three}---Continued}
\end{figure*}


\clearpage
\begin{figure*}

\includegraphics[width=0.45\textwidth]{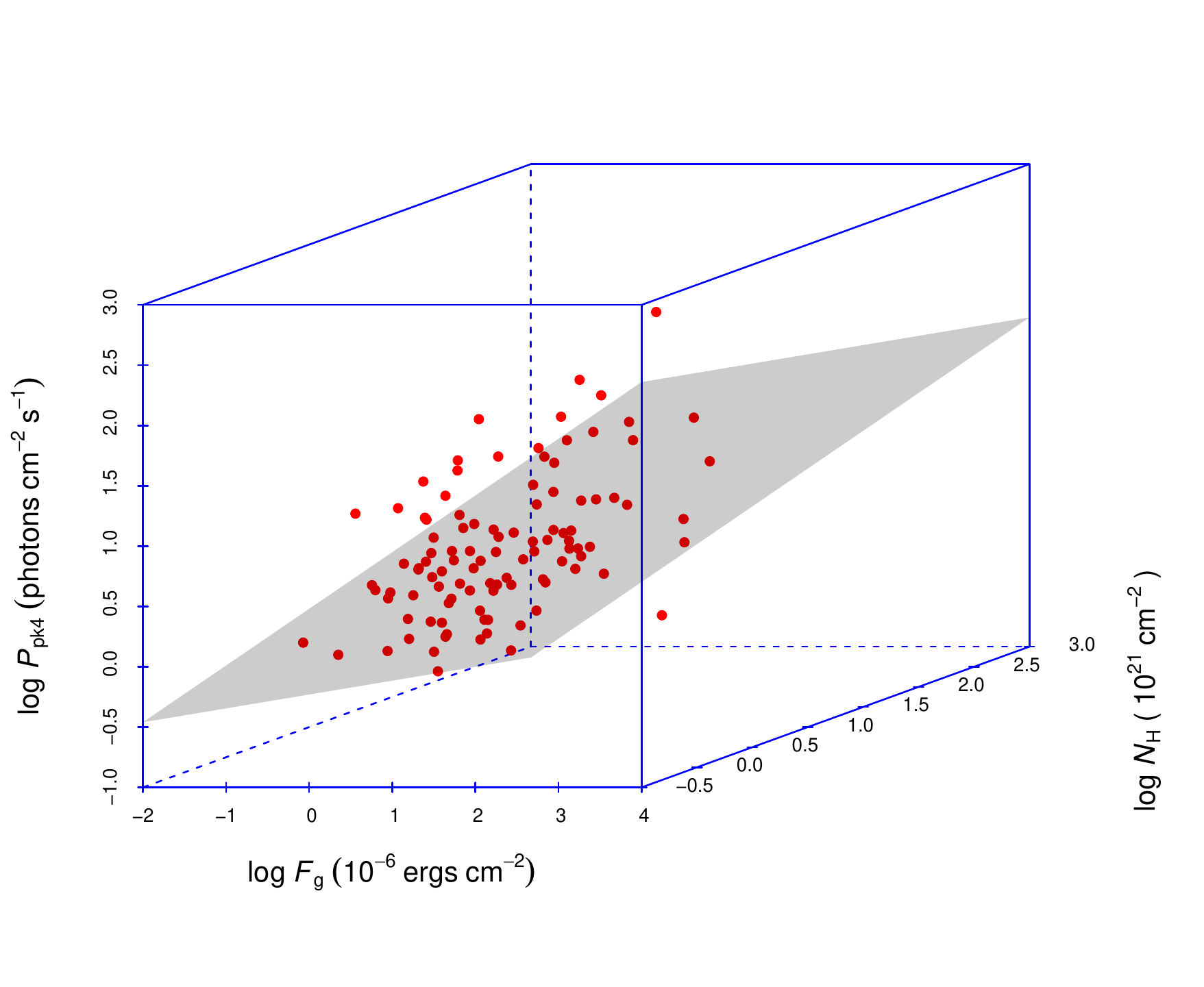}
\includegraphics[width=0.45\textwidth]{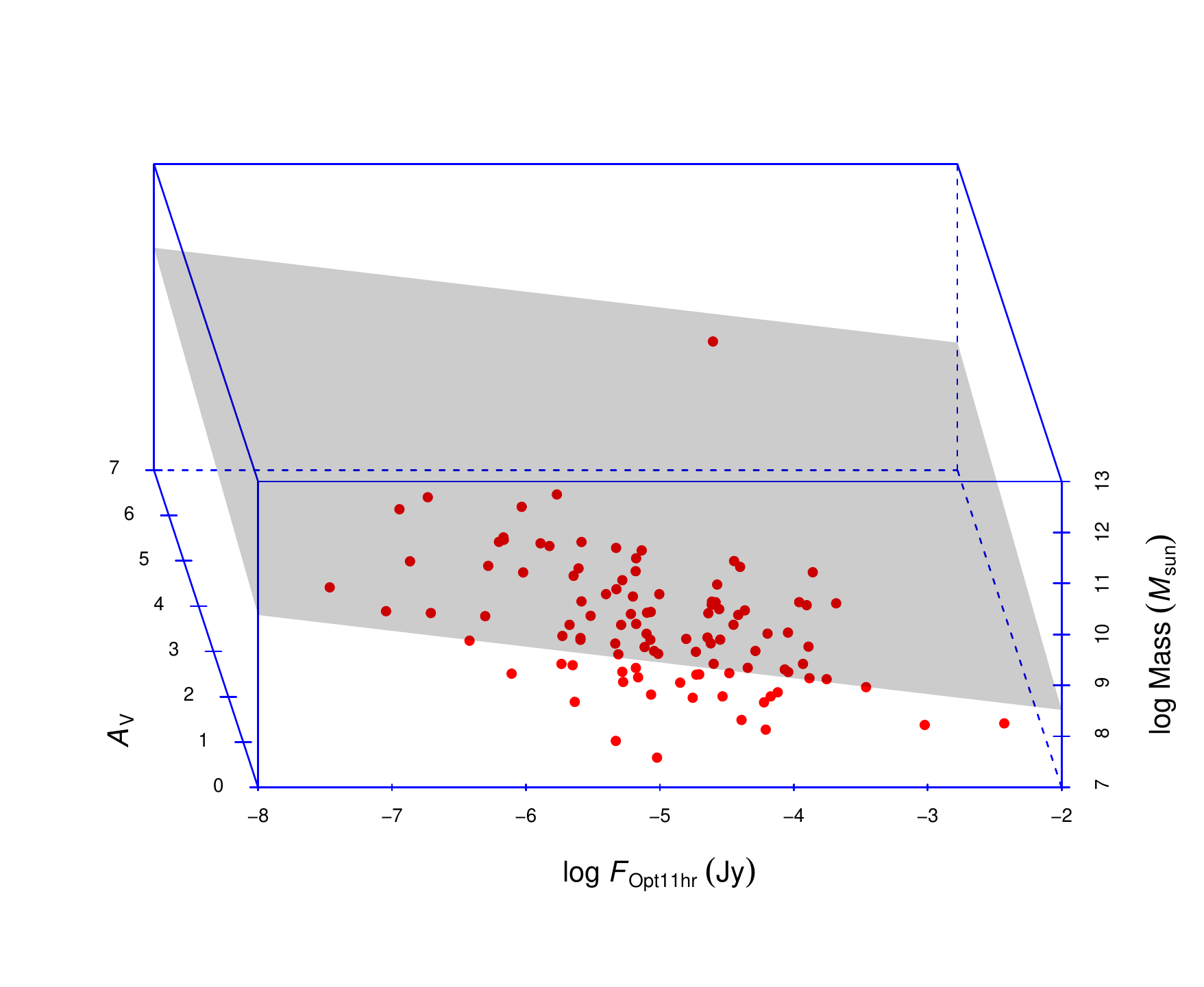}

\includegraphics[width=0.45\textwidth]{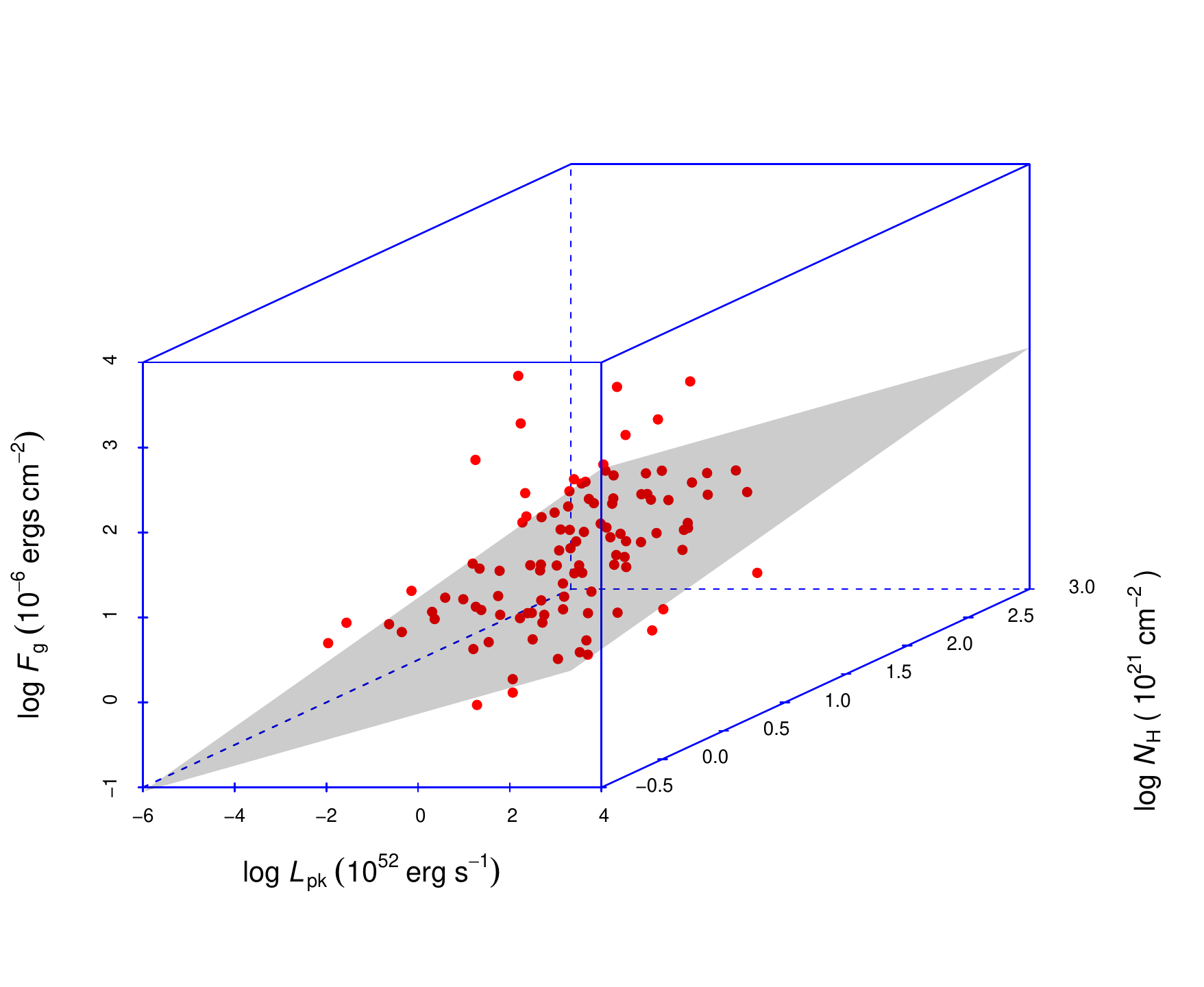}
\includegraphics[width=0.45\textwidth]{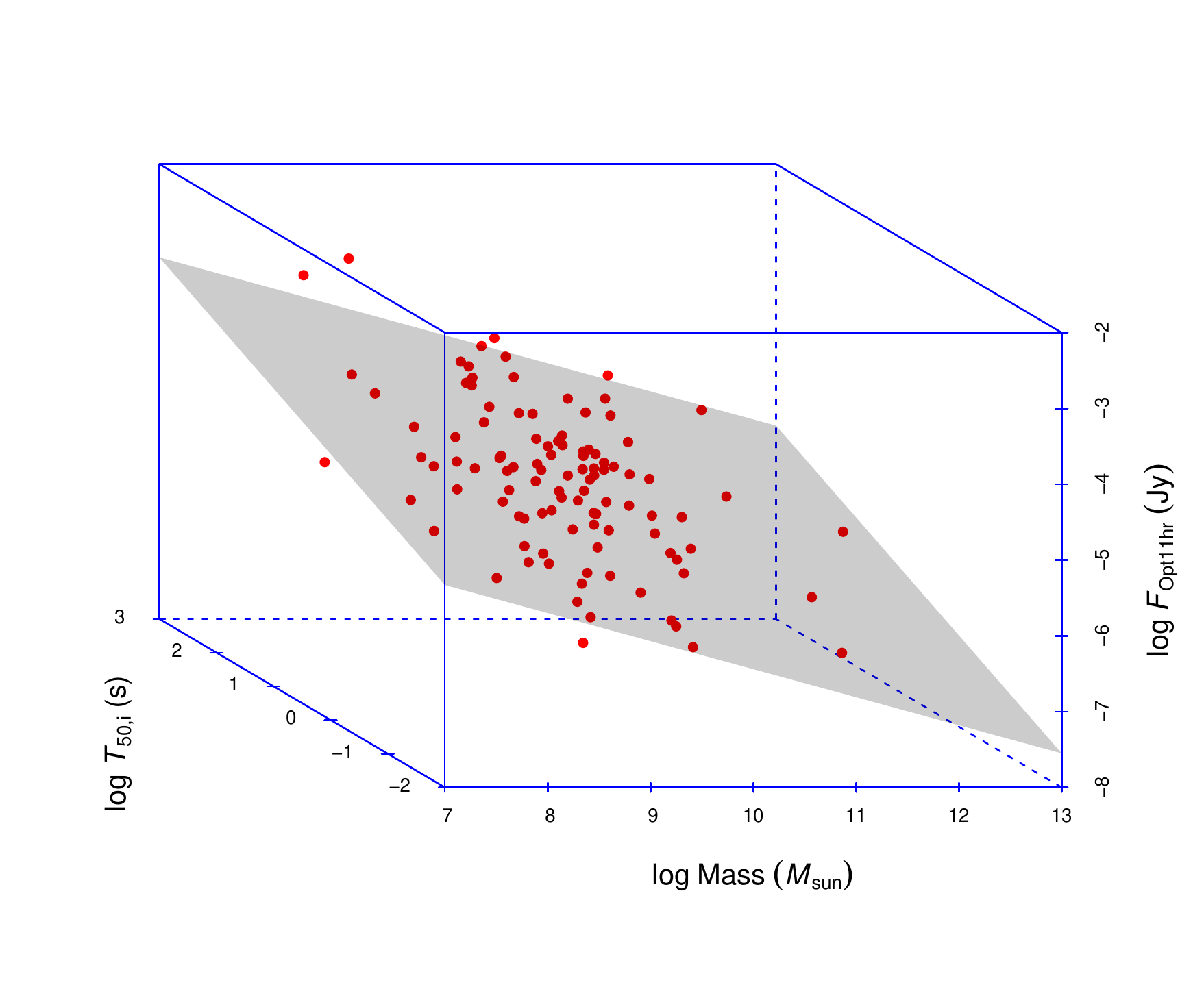}

\includegraphics[width=0.45\textwidth]{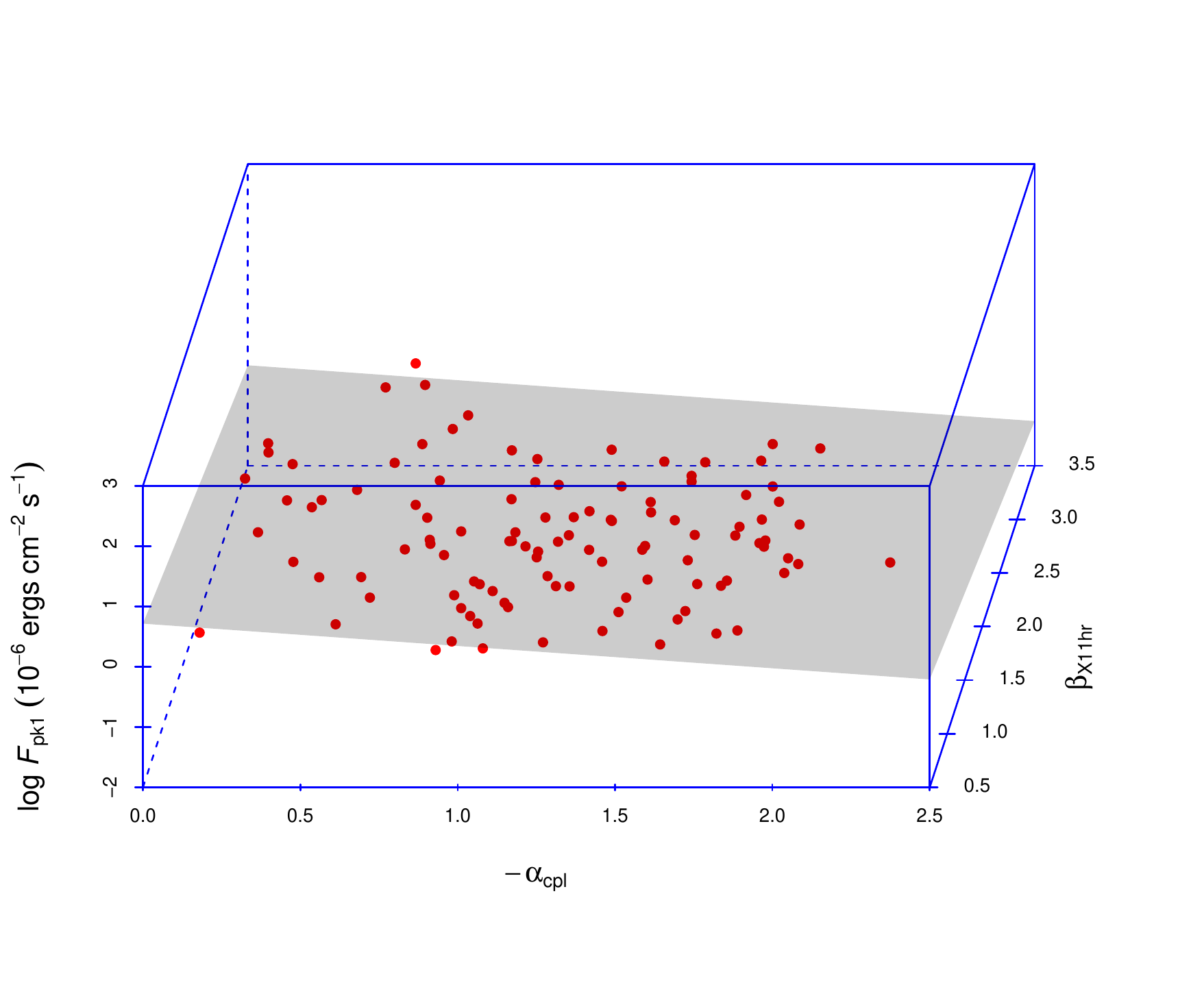}
\includegraphics[width=0.45\textwidth]{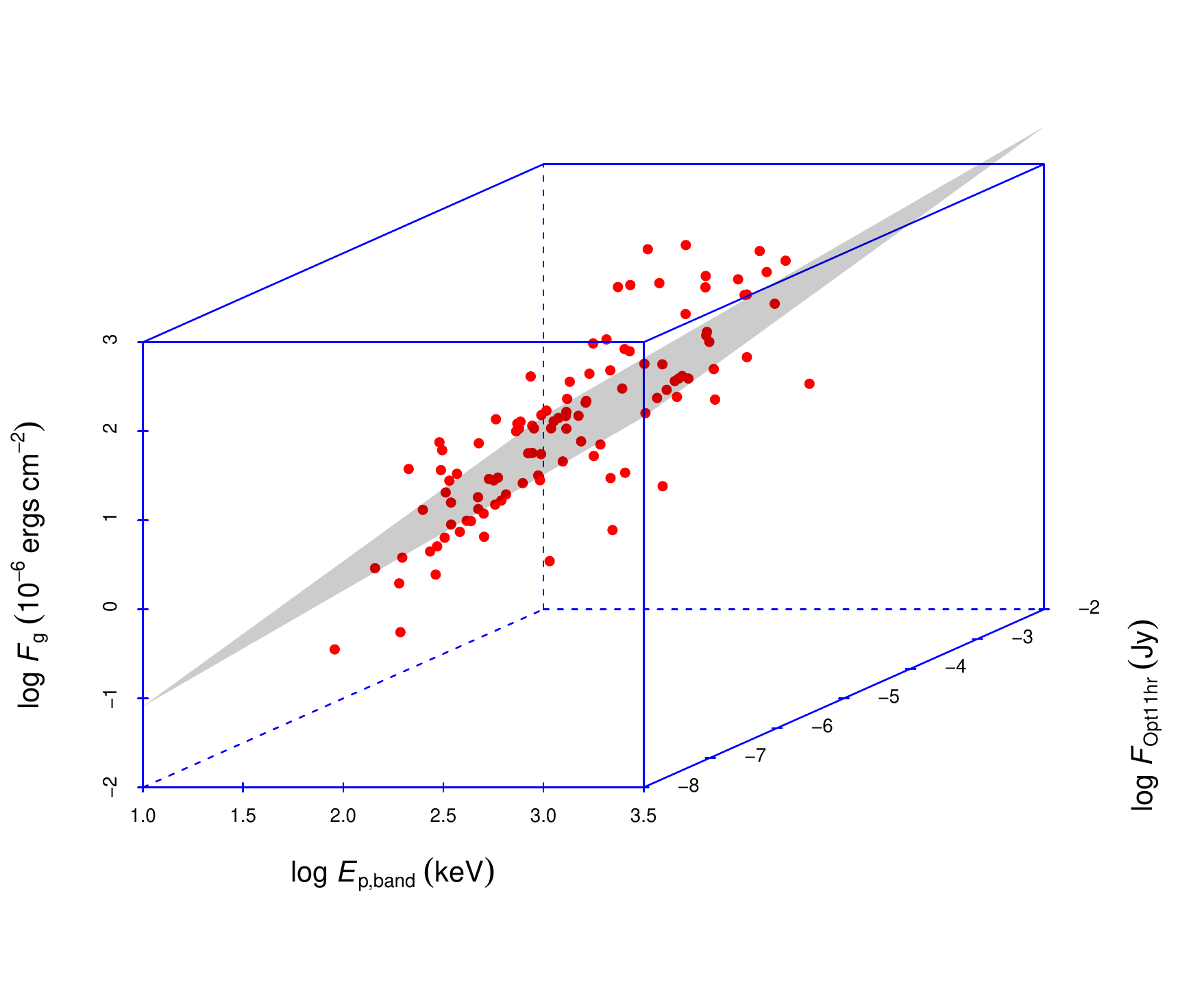}

\center{Fig. \ref{fig:three}---Continued}
\end{figure*}


\clearpage
\begin{figure*}

\includegraphics[width=0.45\textwidth]{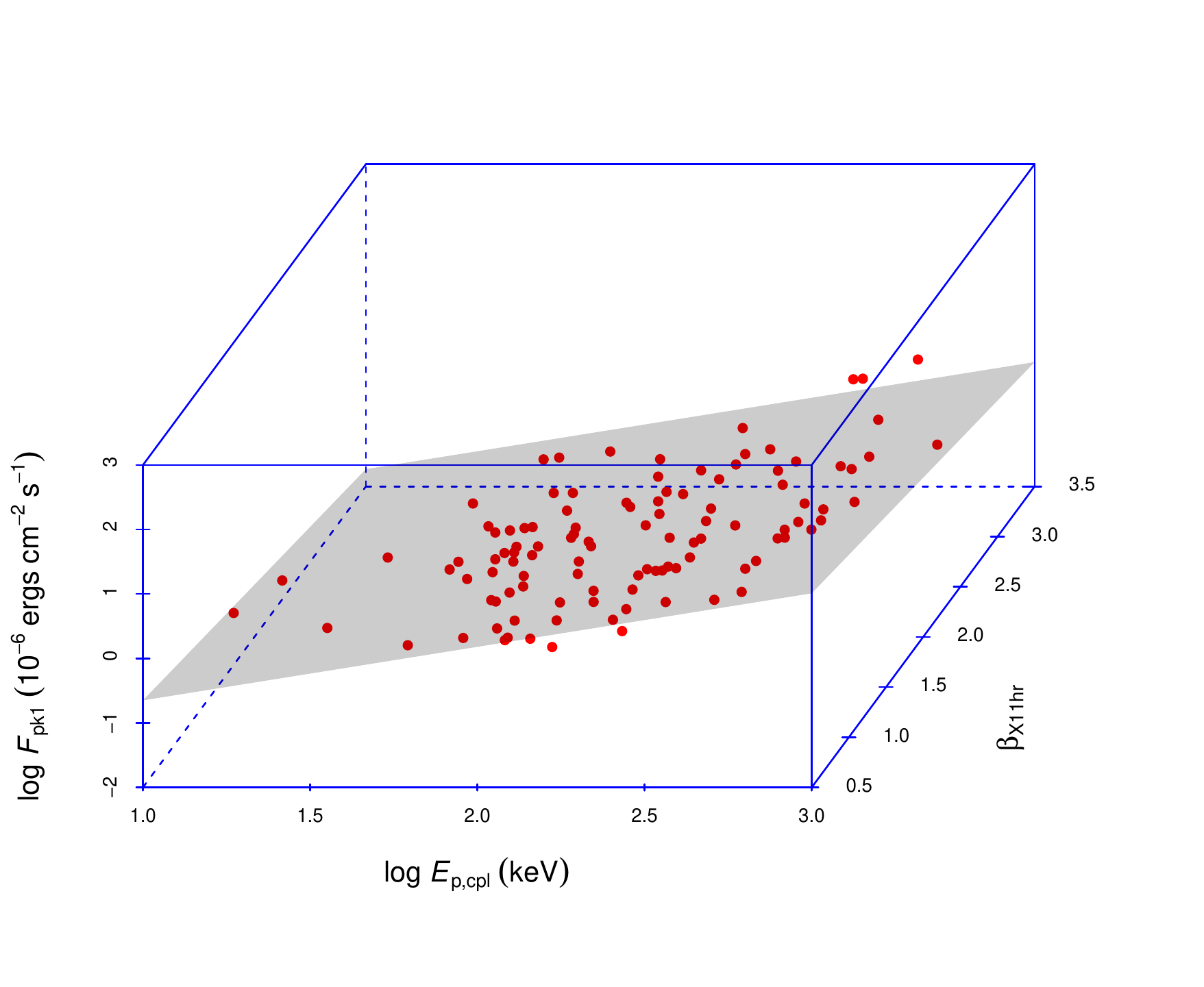}
\includegraphics[width=0.45\textwidth]{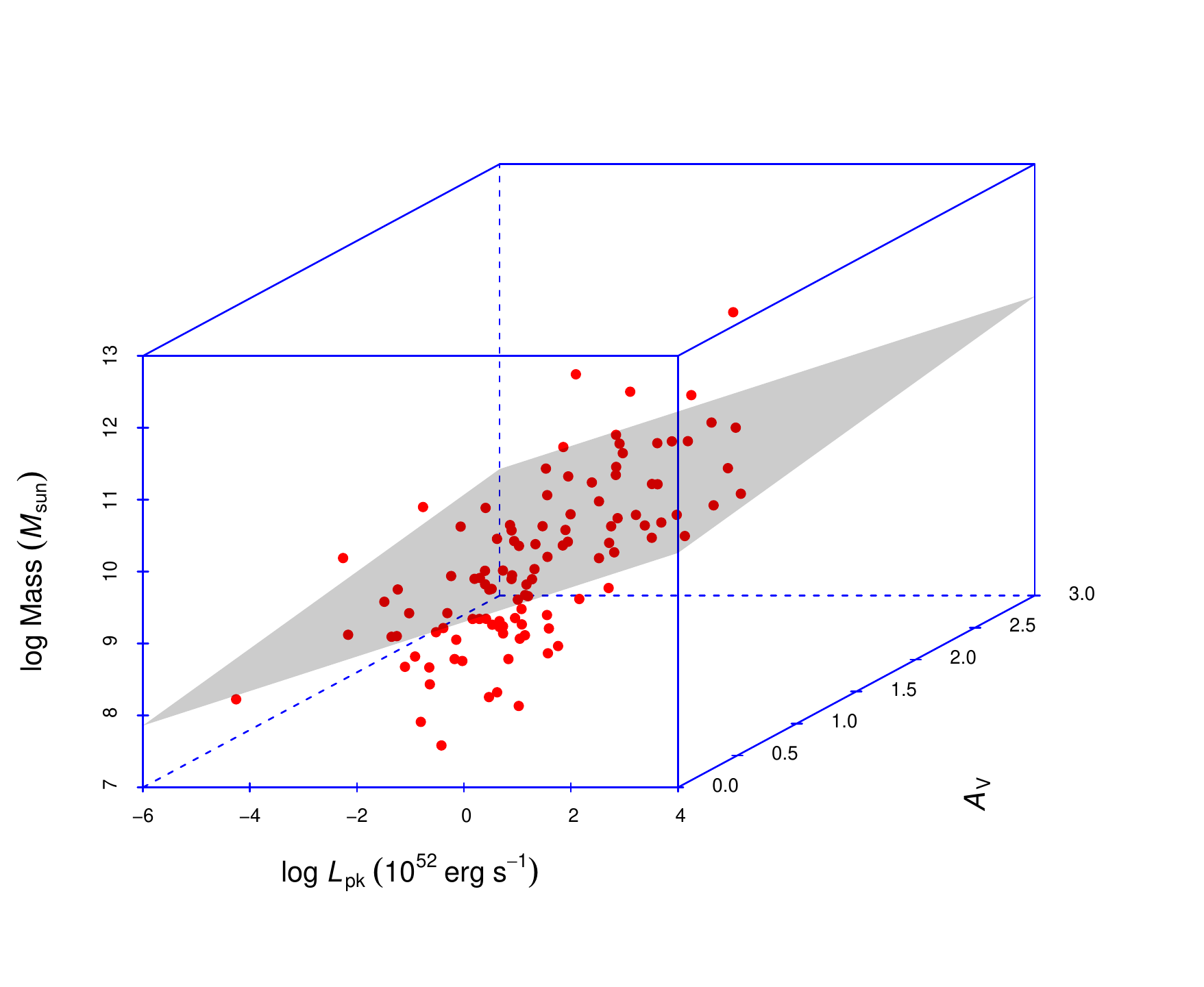}

\includegraphics[width=0.45\textwidth]{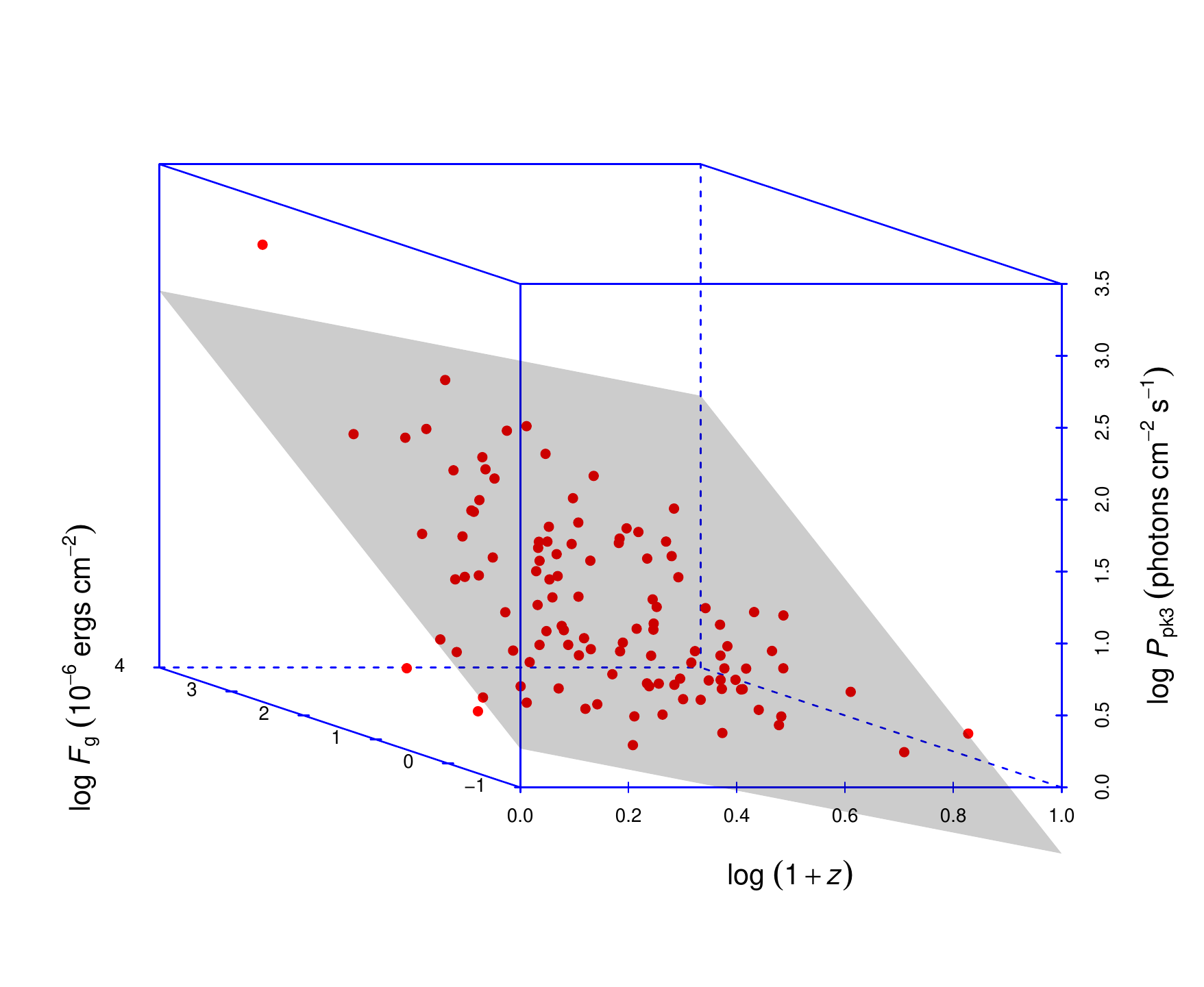}
\includegraphics[width=0.45\textwidth]{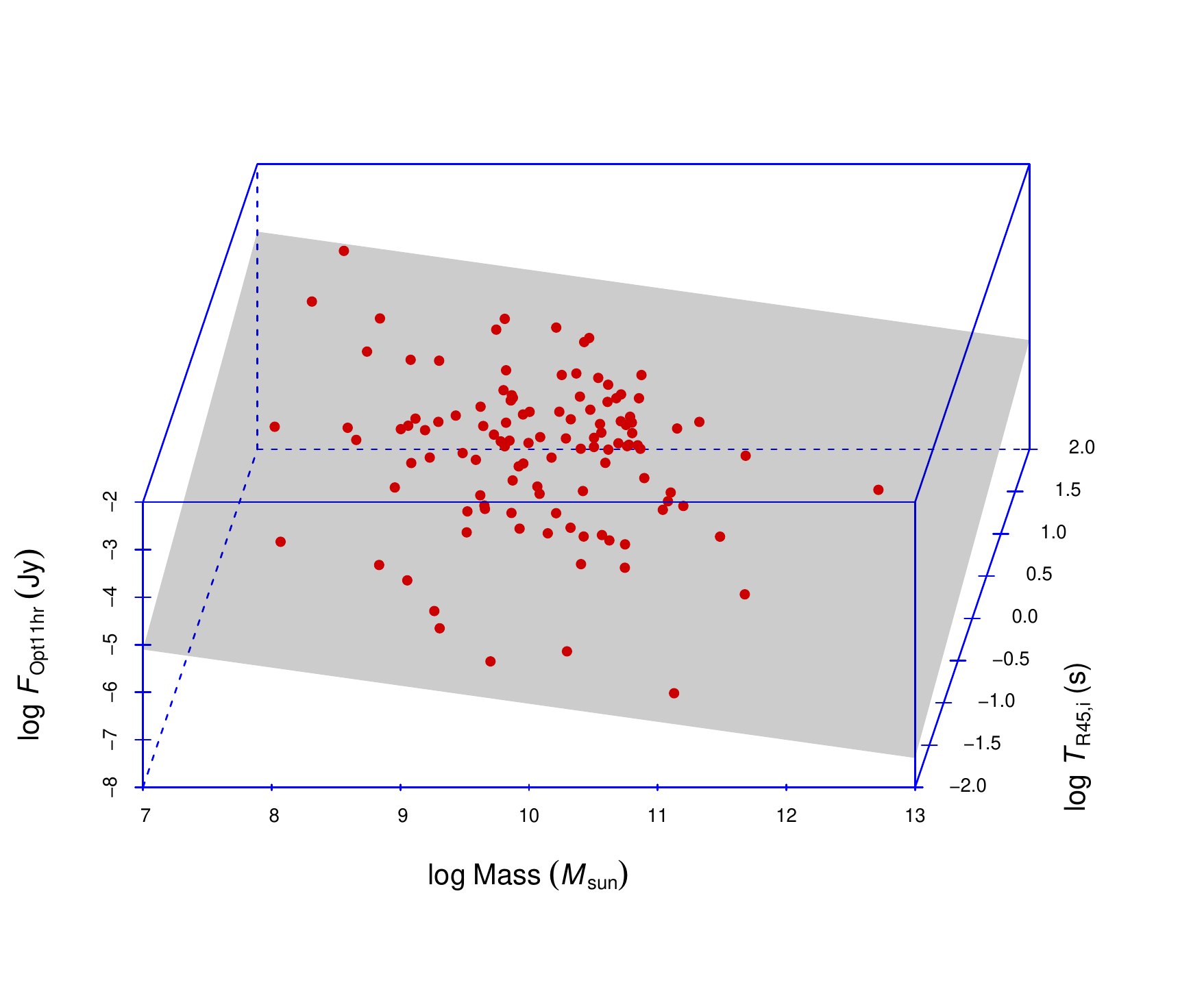}

\includegraphics[width=0.45\textwidth]{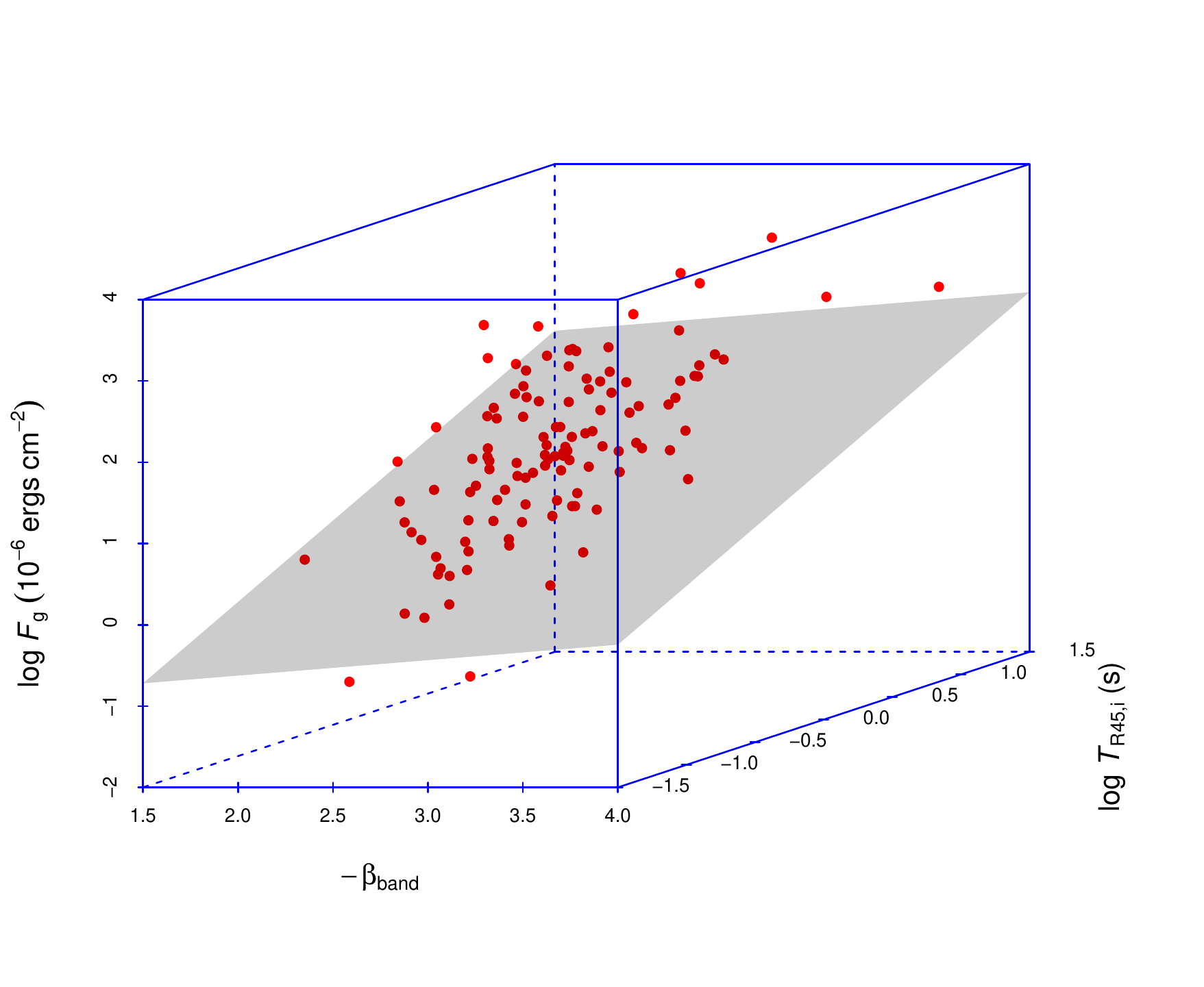}
\includegraphics[width=0.45\textwidth]{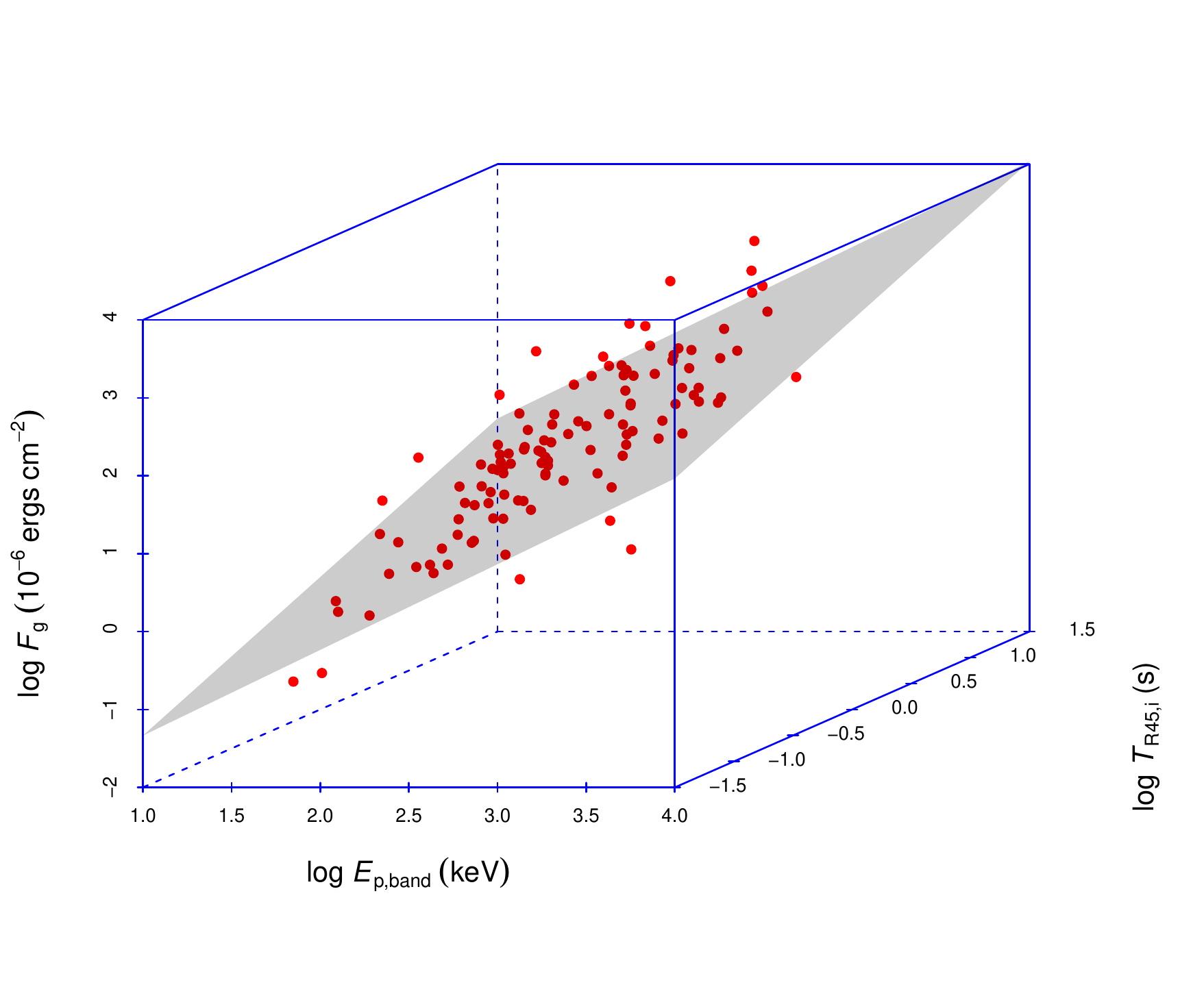}

\center{Fig. \ref{fig:three}---Continued}
\end{figure*}


\clearpage
\begin{figure*}

\includegraphics[width=0.45\textwidth]{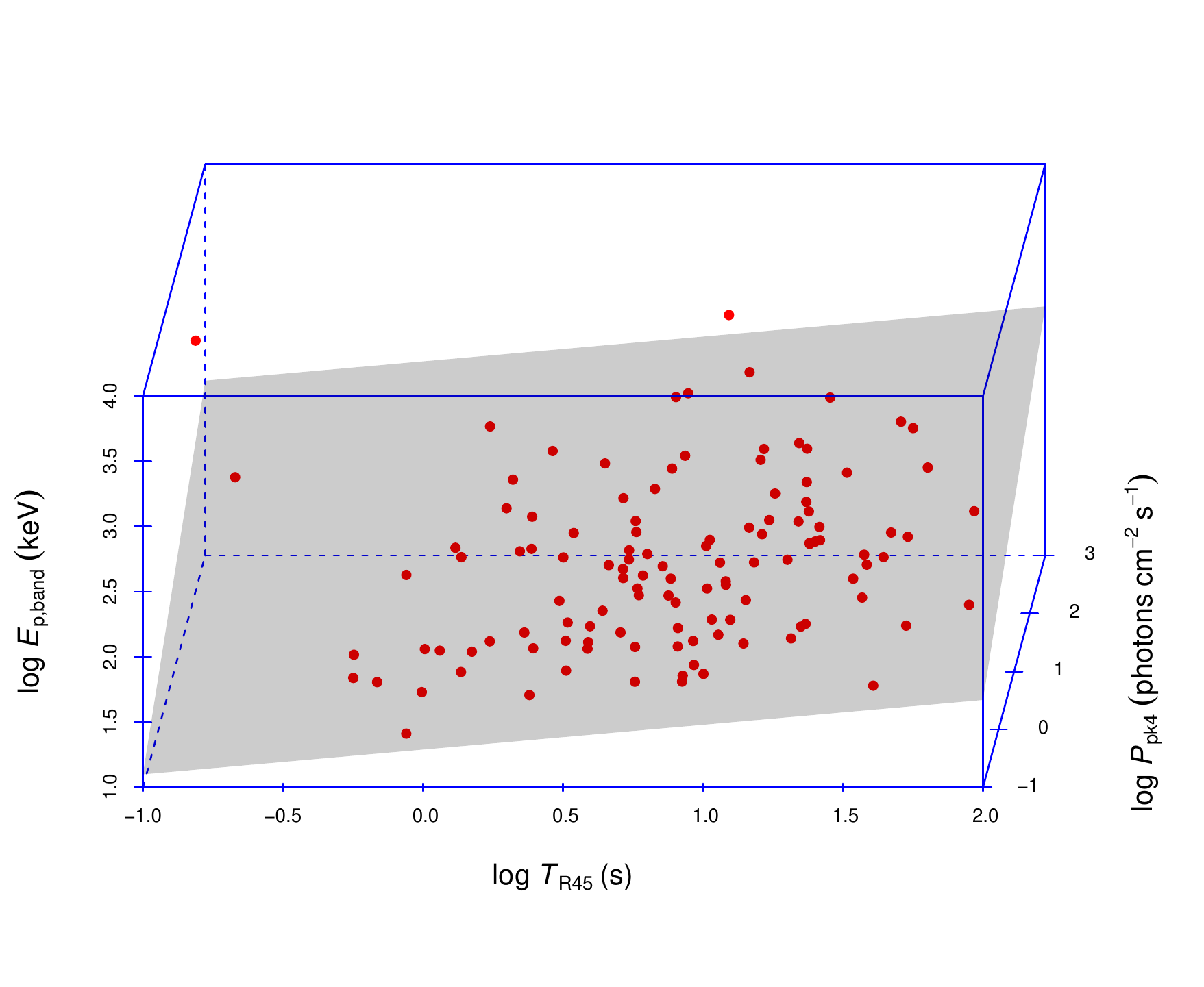}
\includegraphics[width=0.45\textwidth]{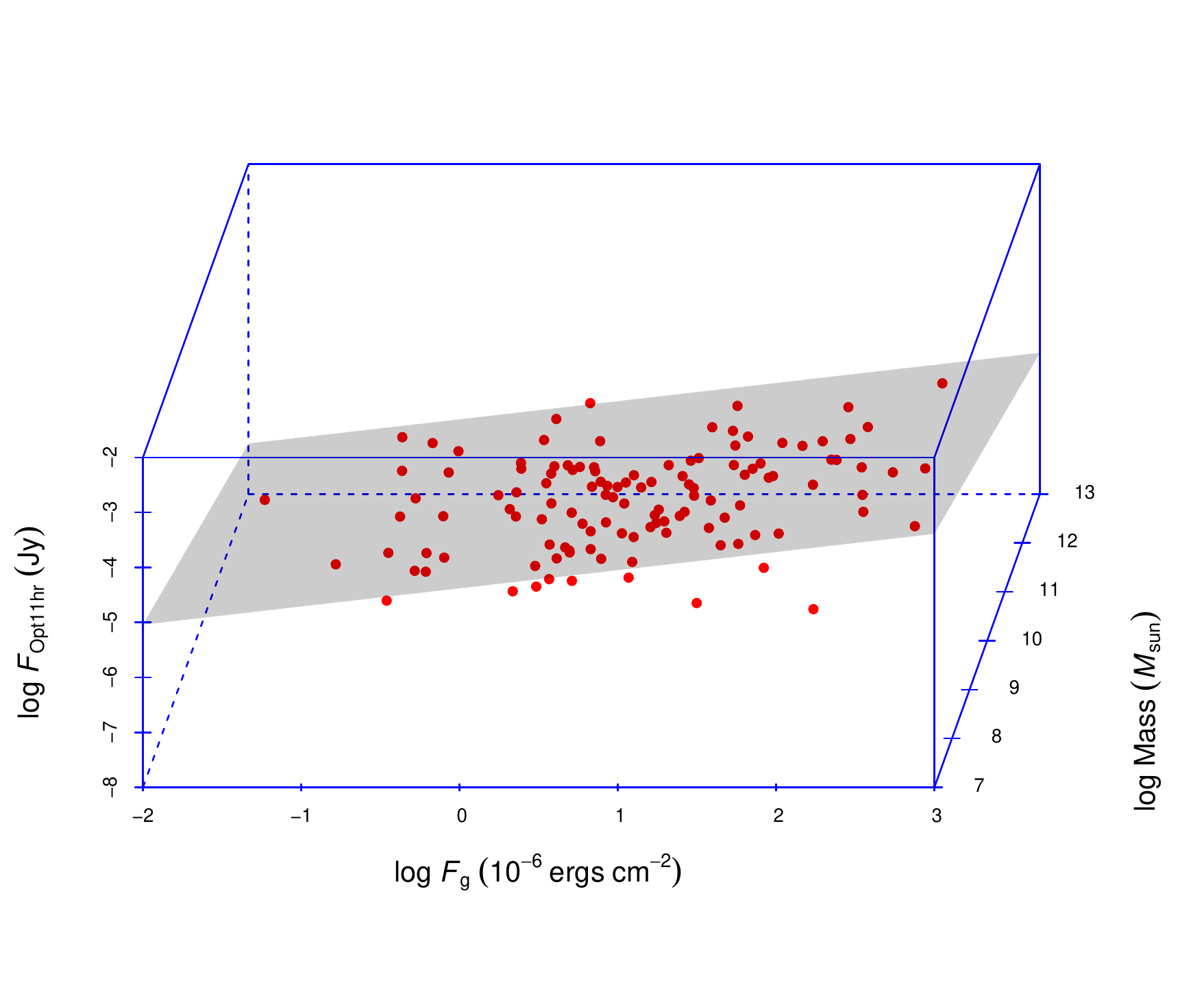}

\includegraphics[width=0.45\textwidth]{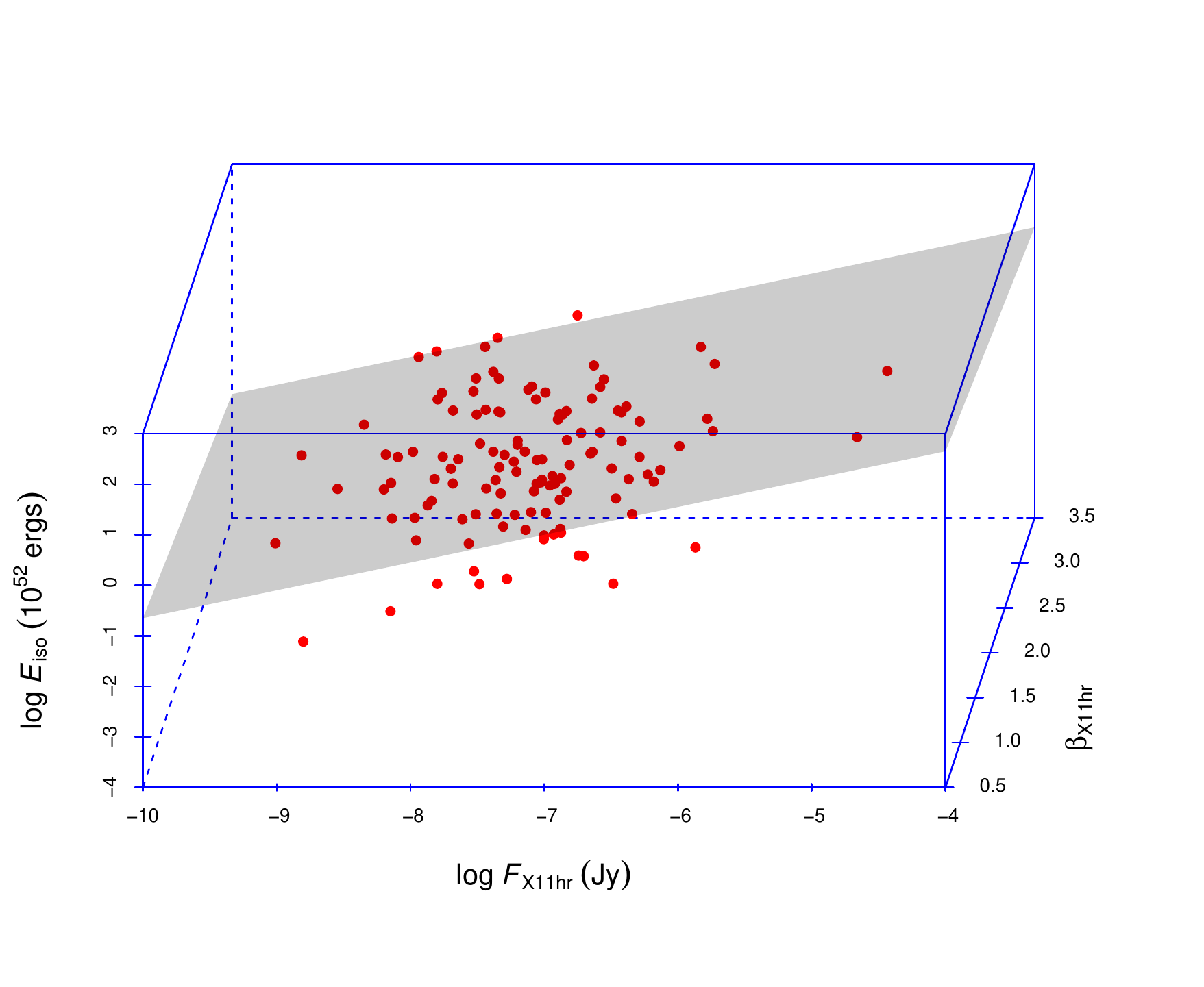}
\includegraphics[width=0.45\textwidth]{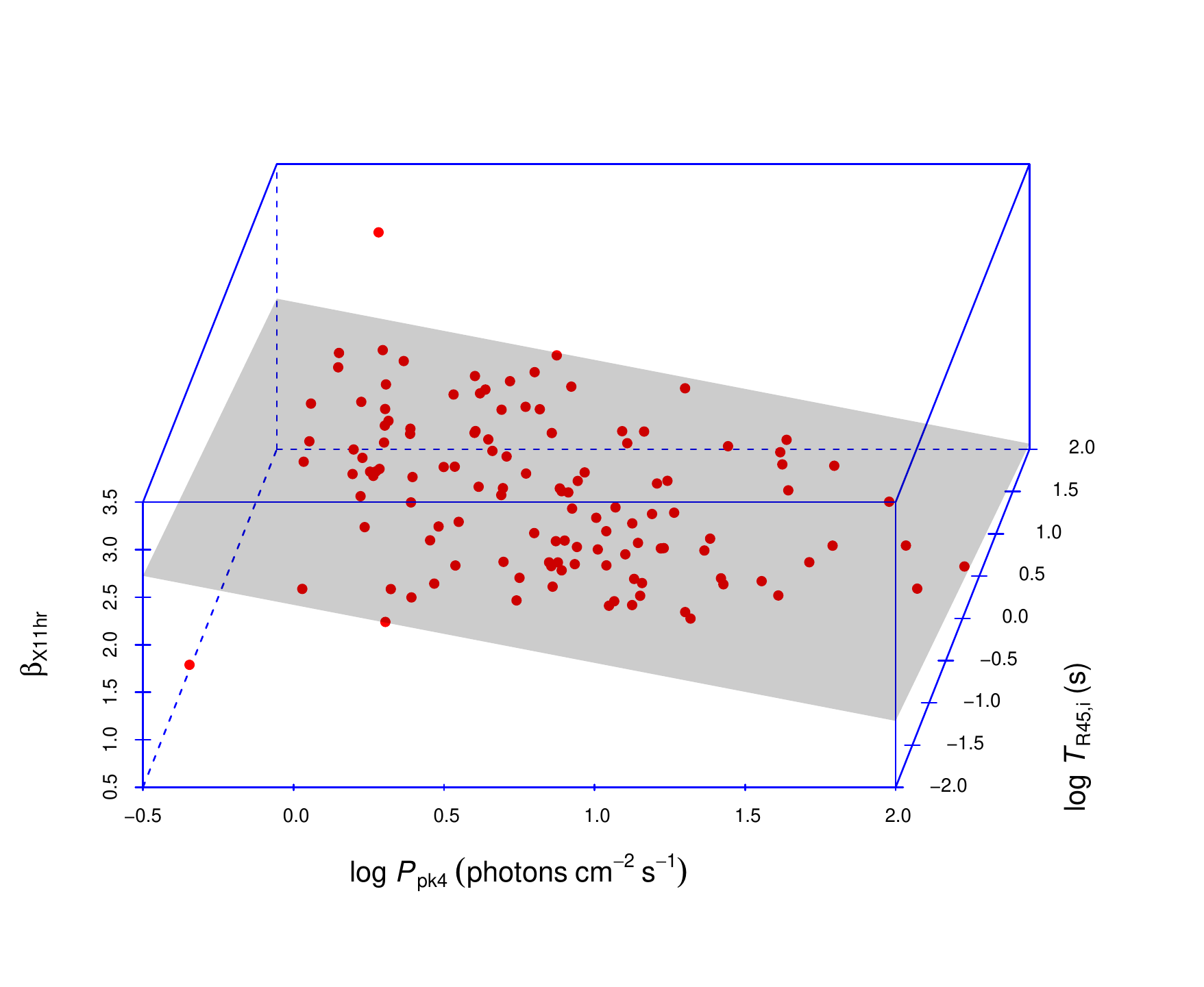}

\includegraphics[width=0.45\textwidth]{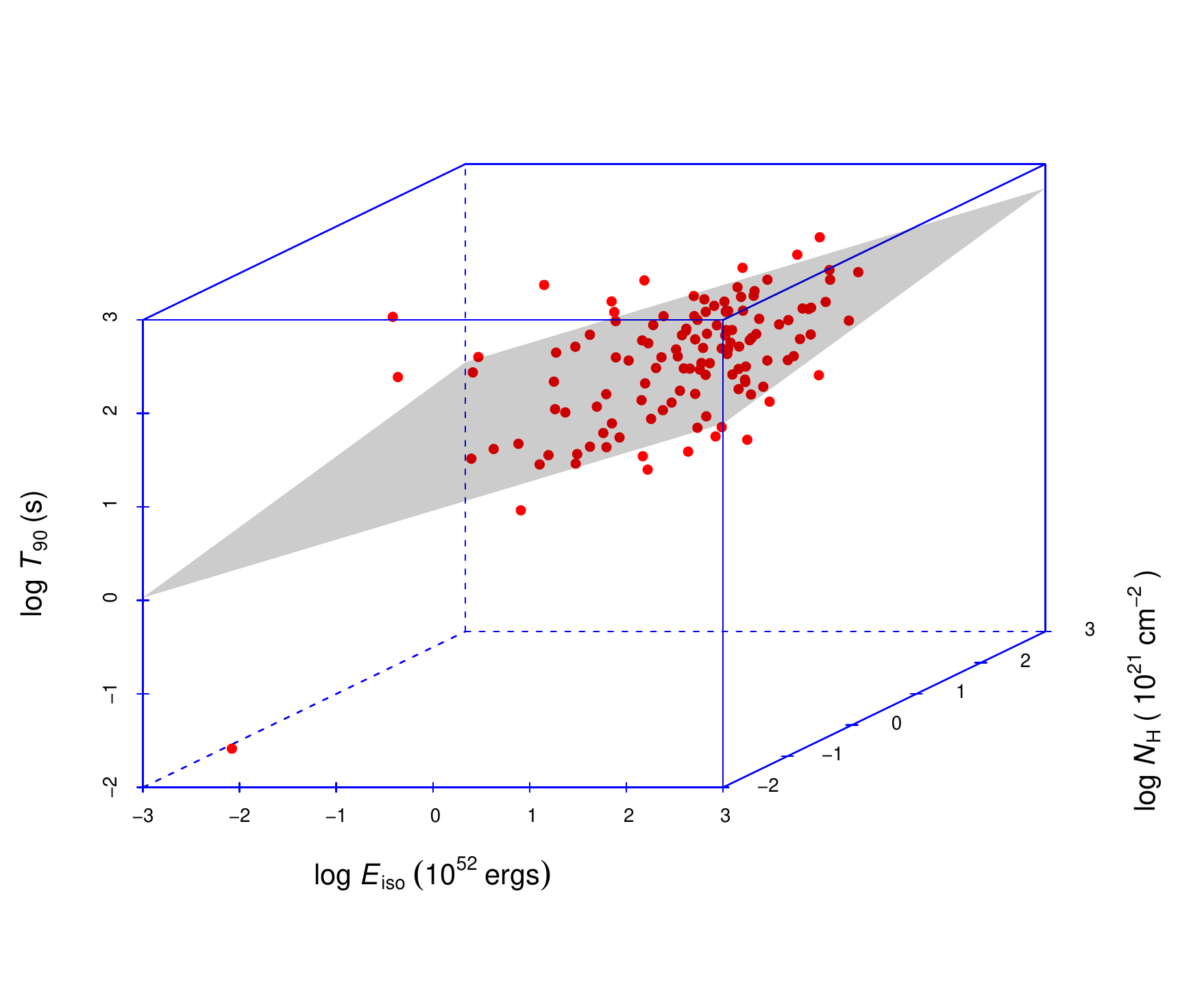}
\includegraphics[width=0.45\textwidth]{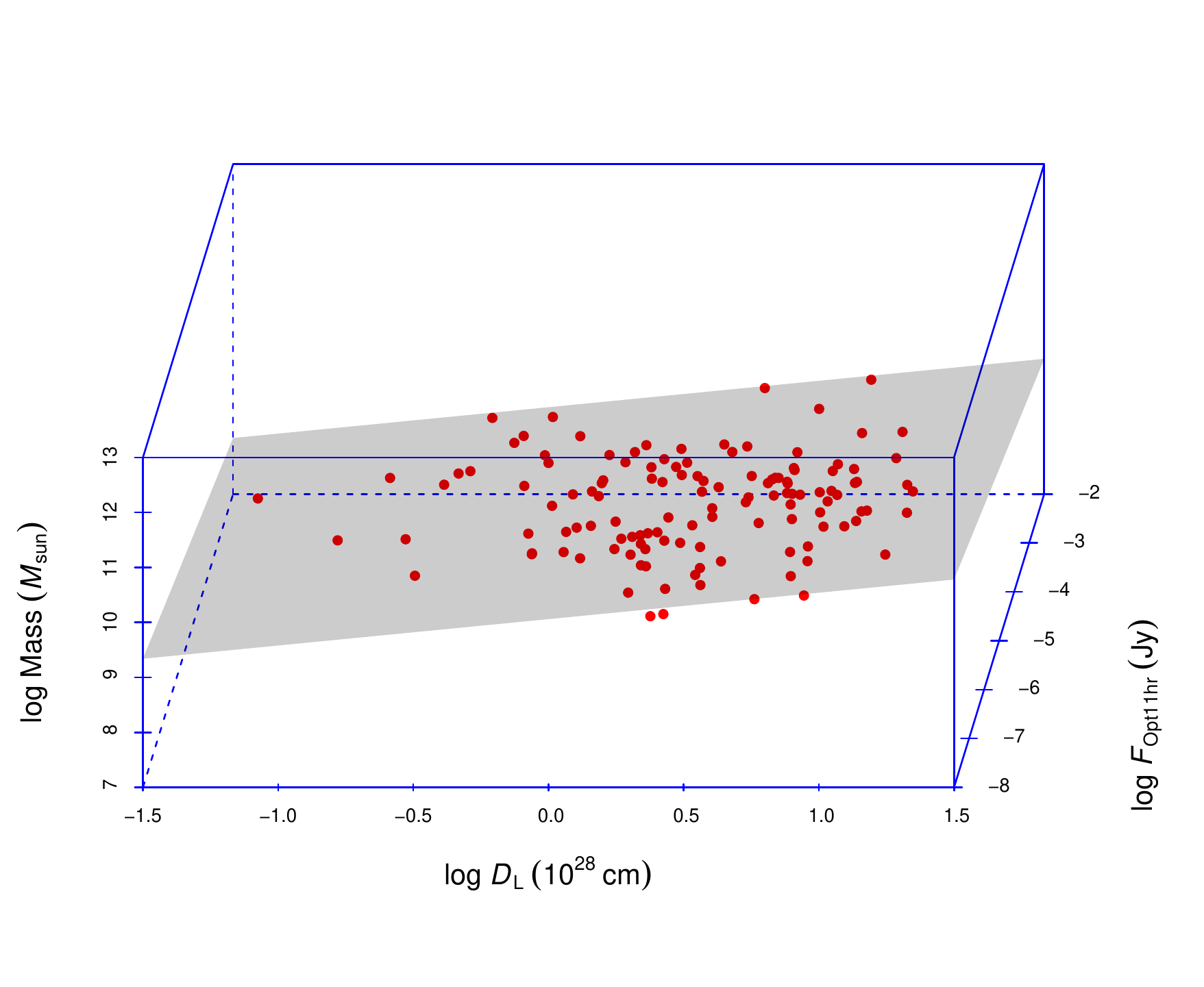}

\center{Fig. \ref{fig:three}---Continued}
\end{figure*}


\clearpage
\begin{figure*}

\includegraphics[width=0.45\textwidth]{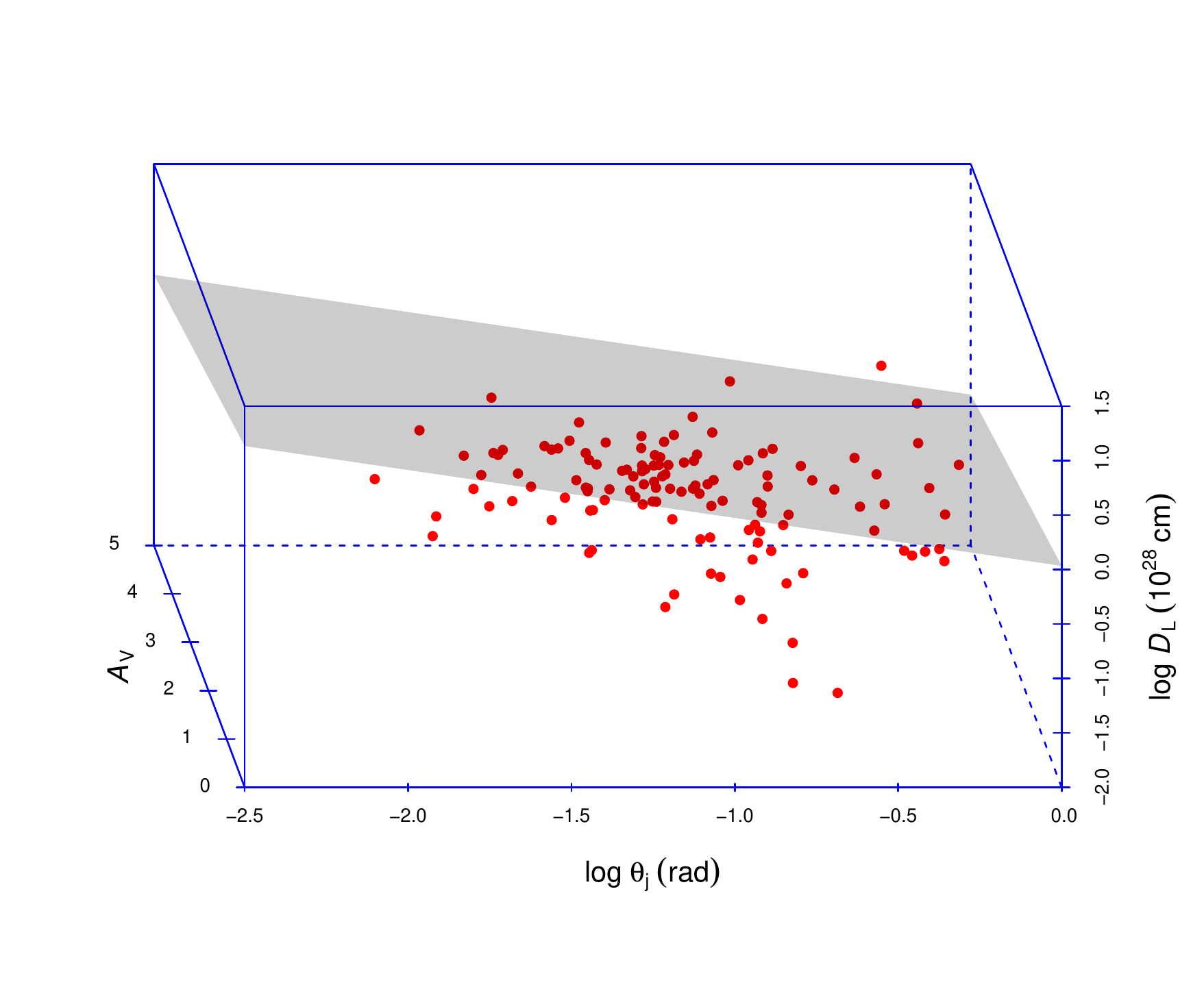}
\includegraphics[width=0.45\textwidth]{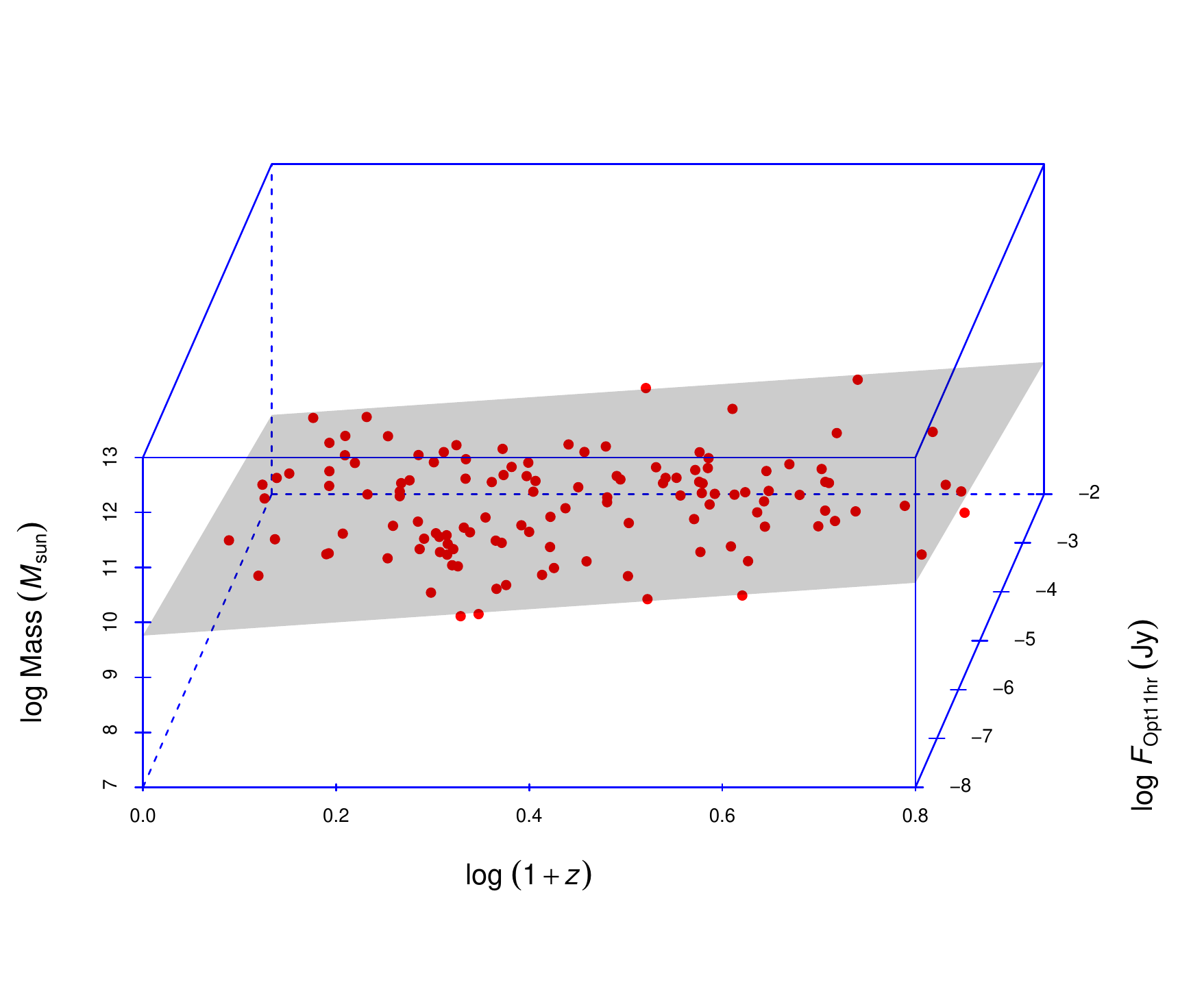}

\includegraphics[width=0.45\textwidth]{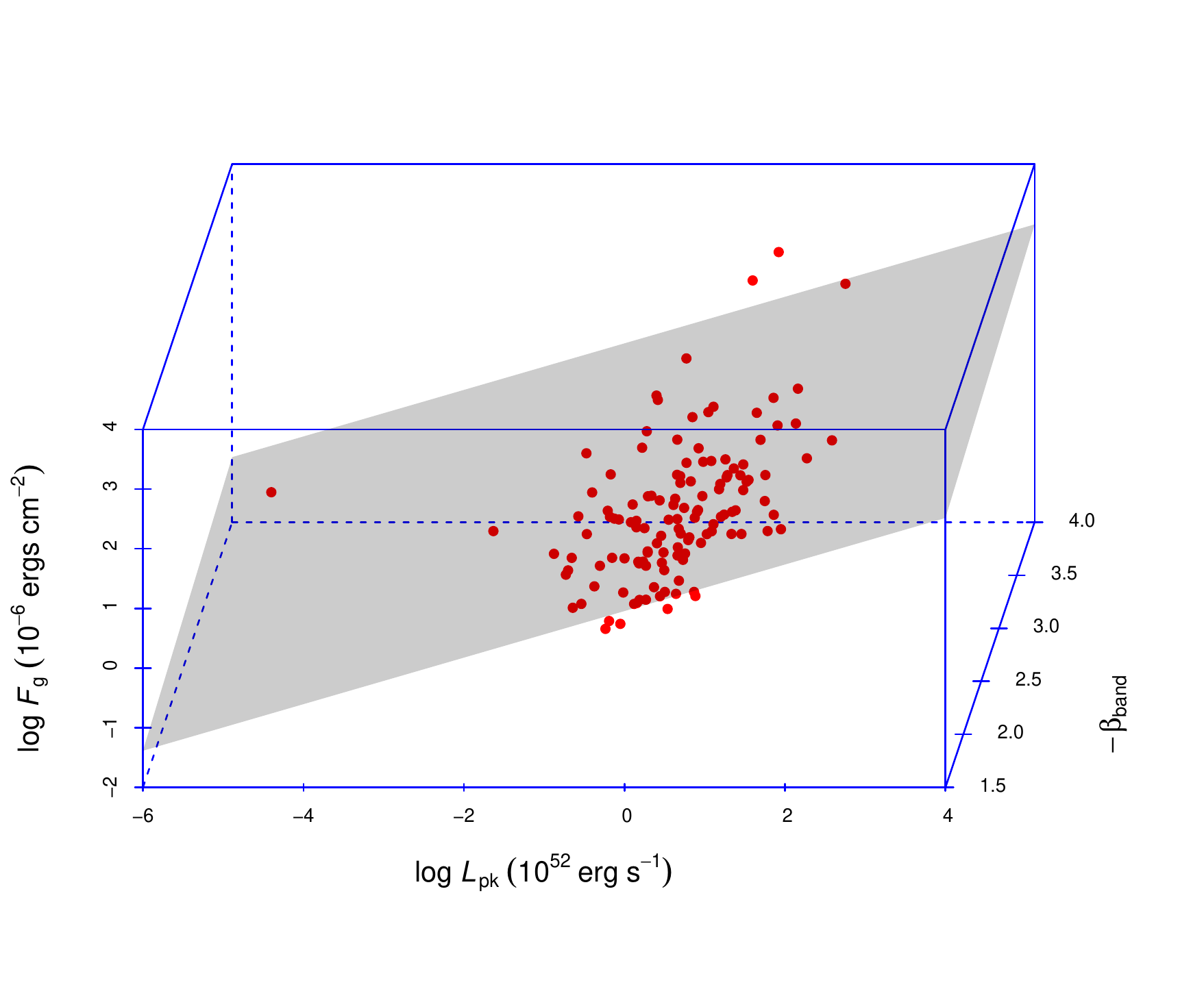}
\includegraphics[width=0.45\textwidth]{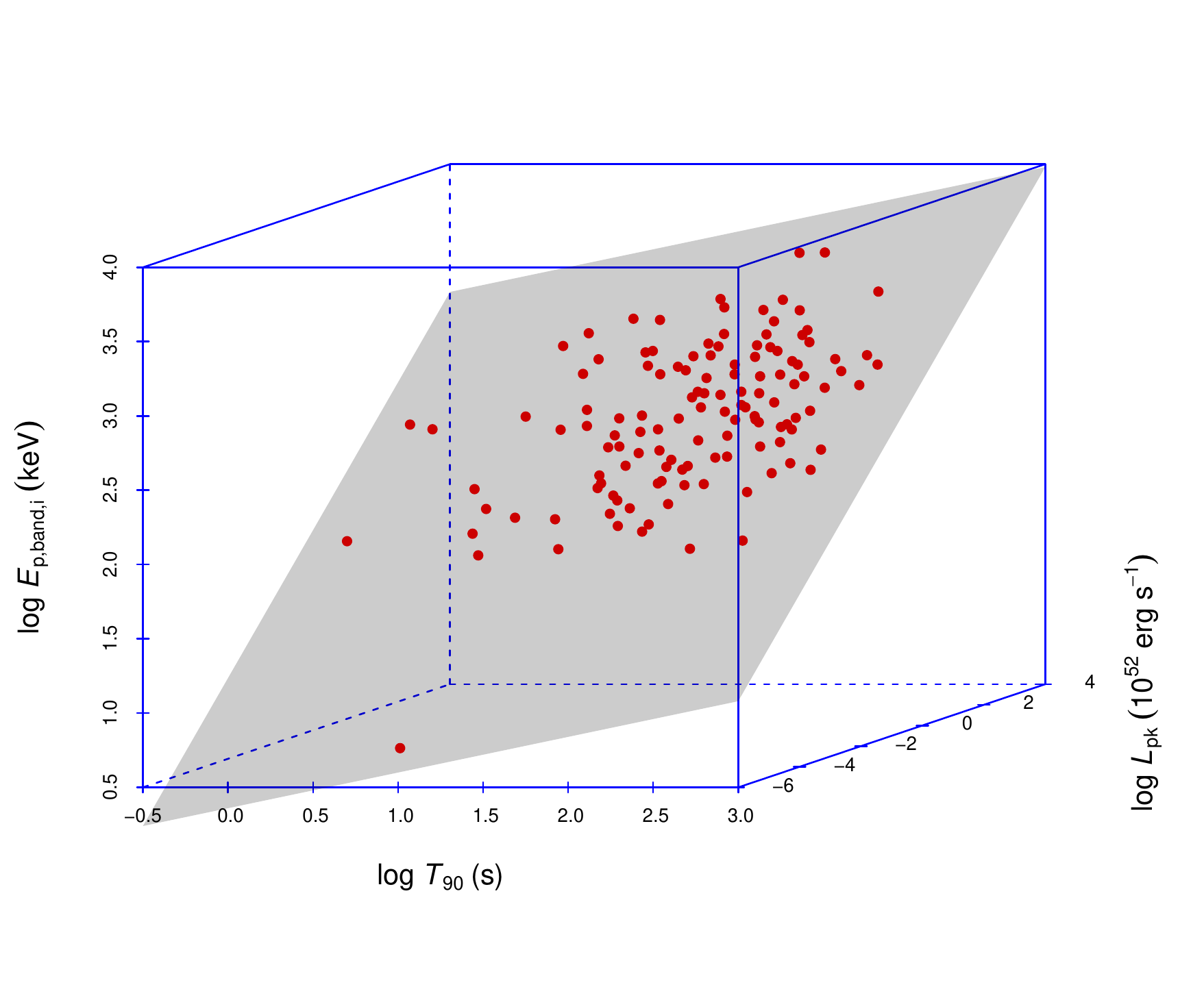}

\includegraphics[width=0.45\textwidth]{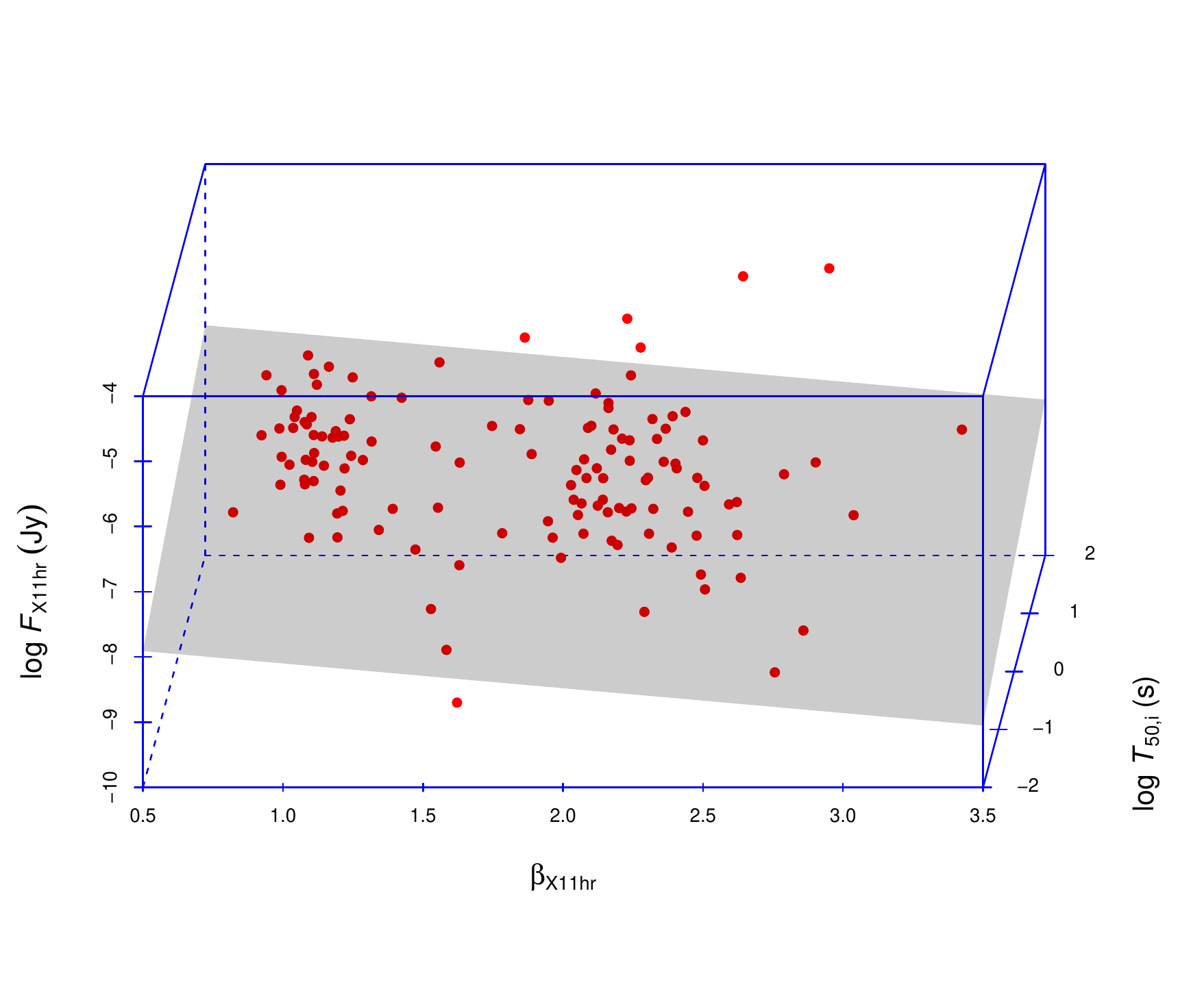}
\includegraphics[width=0.45\textwidth]{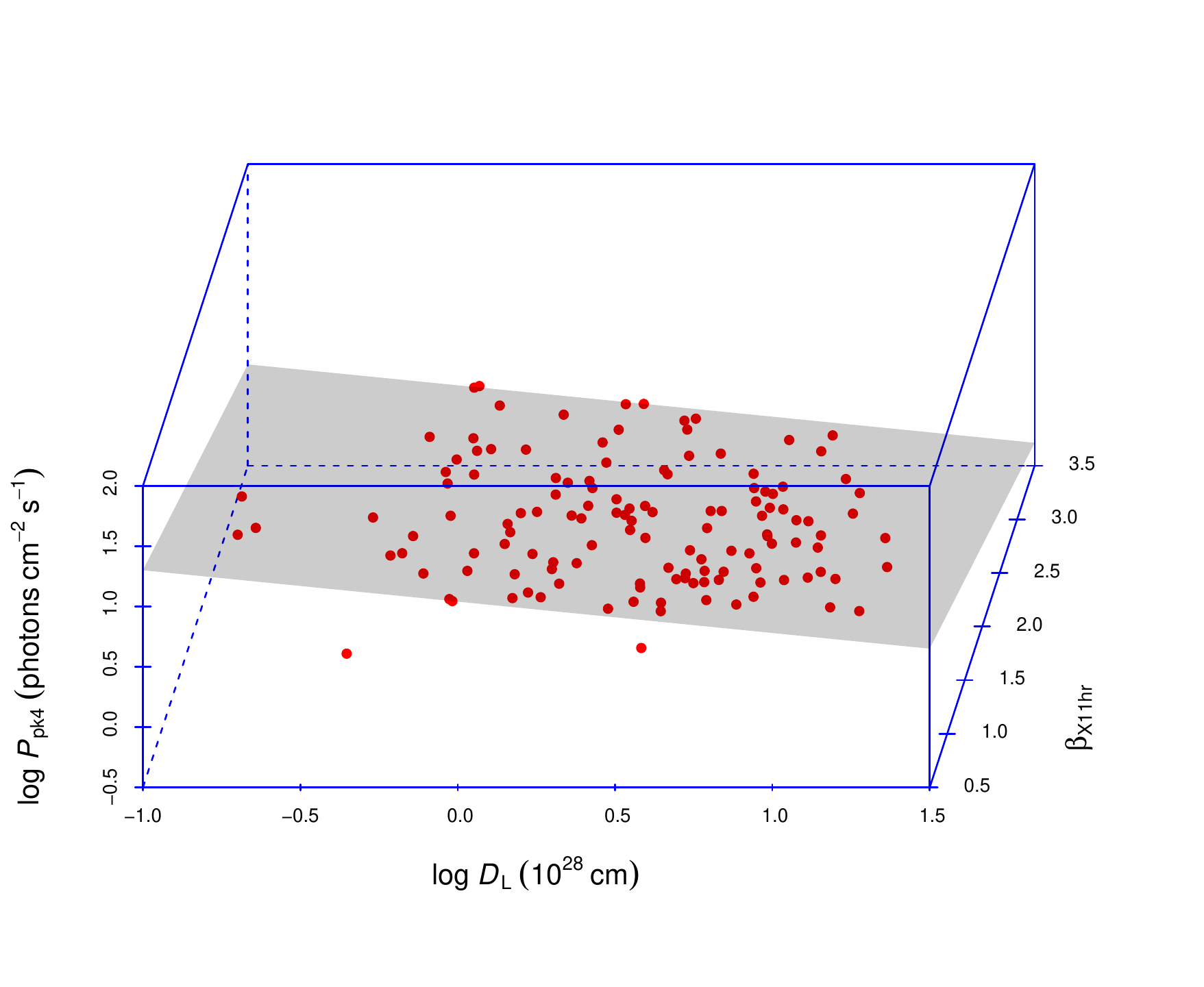}

\center{Fig. \ref{fig:three}---Continued}
\end{figure*}


\clearpage
\begin{figure*}

\includegraphics[width=0.45\textwidth]{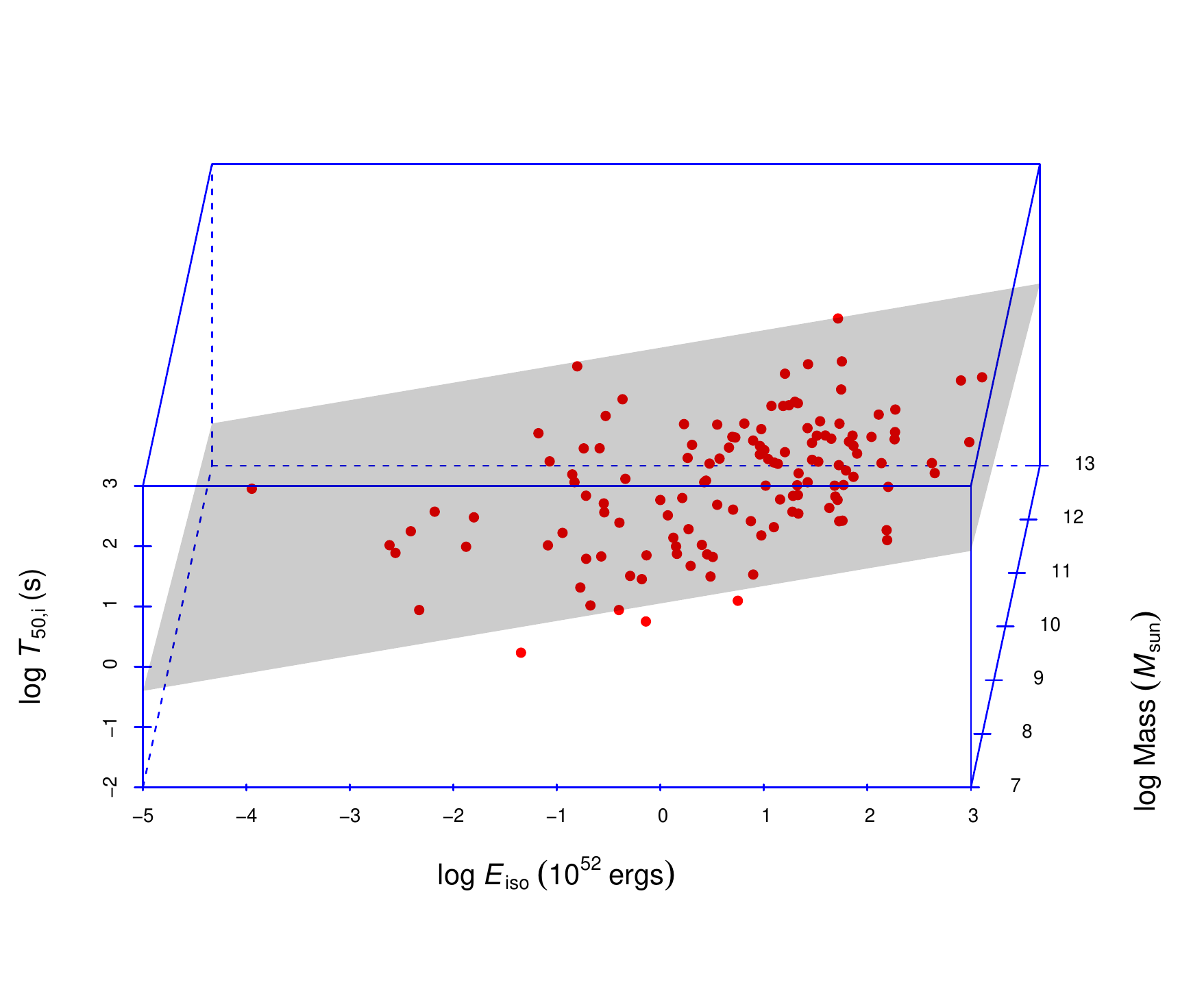}
\includegraphics[width=0.45\textwidth]{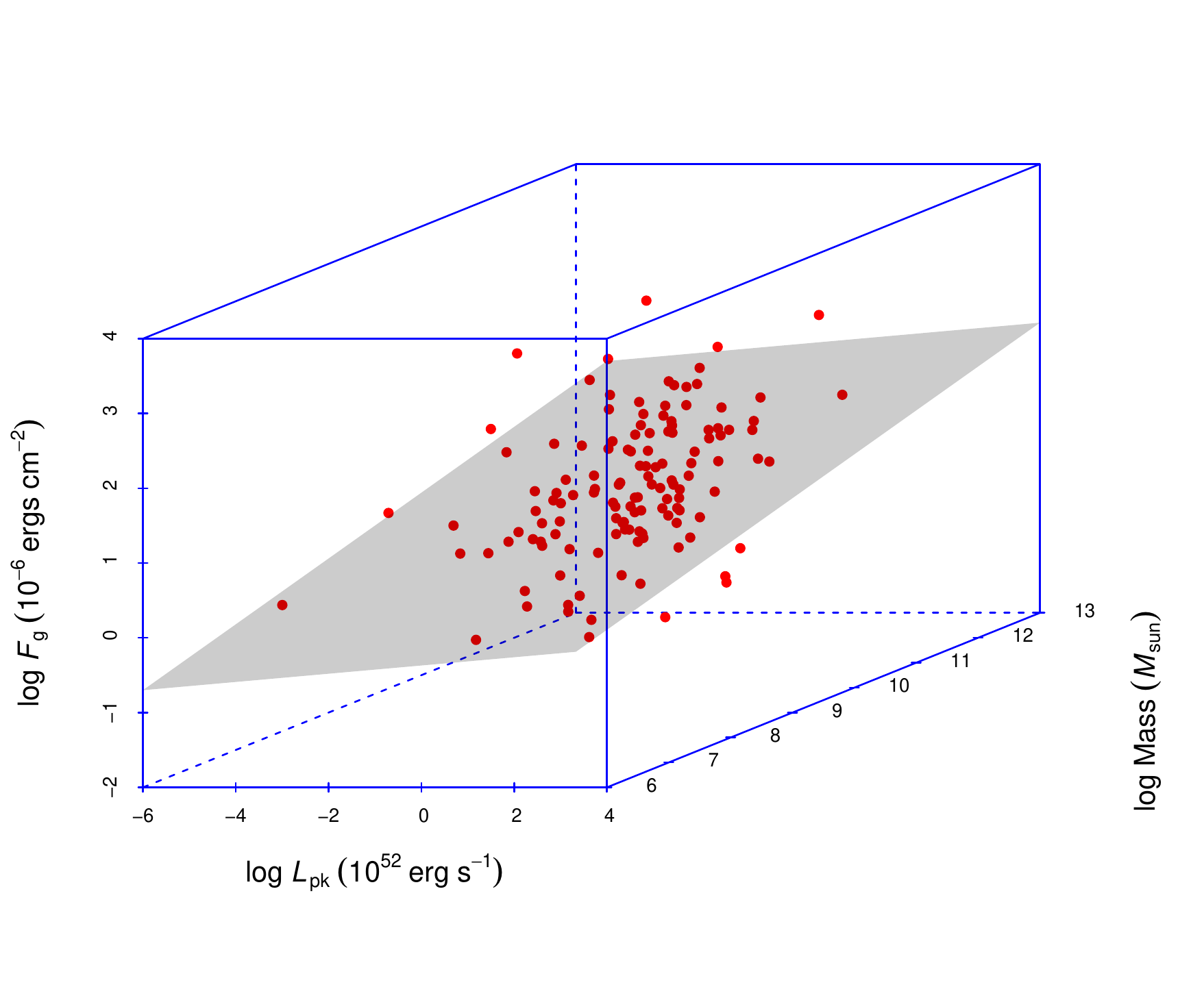}

\includegraphics[width=0.45\textwidth]{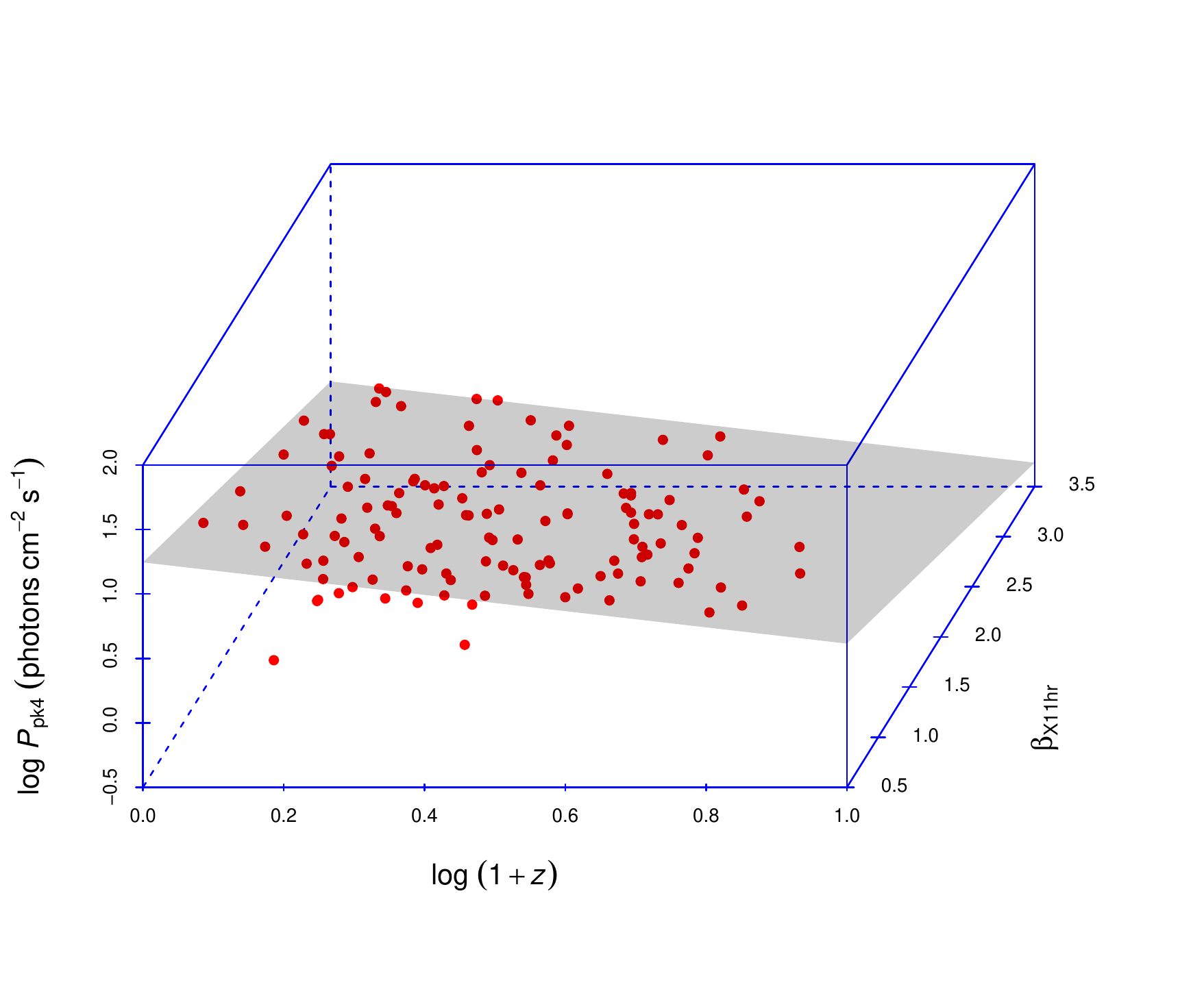}
\includegraphics[width=0.45\textwidth]{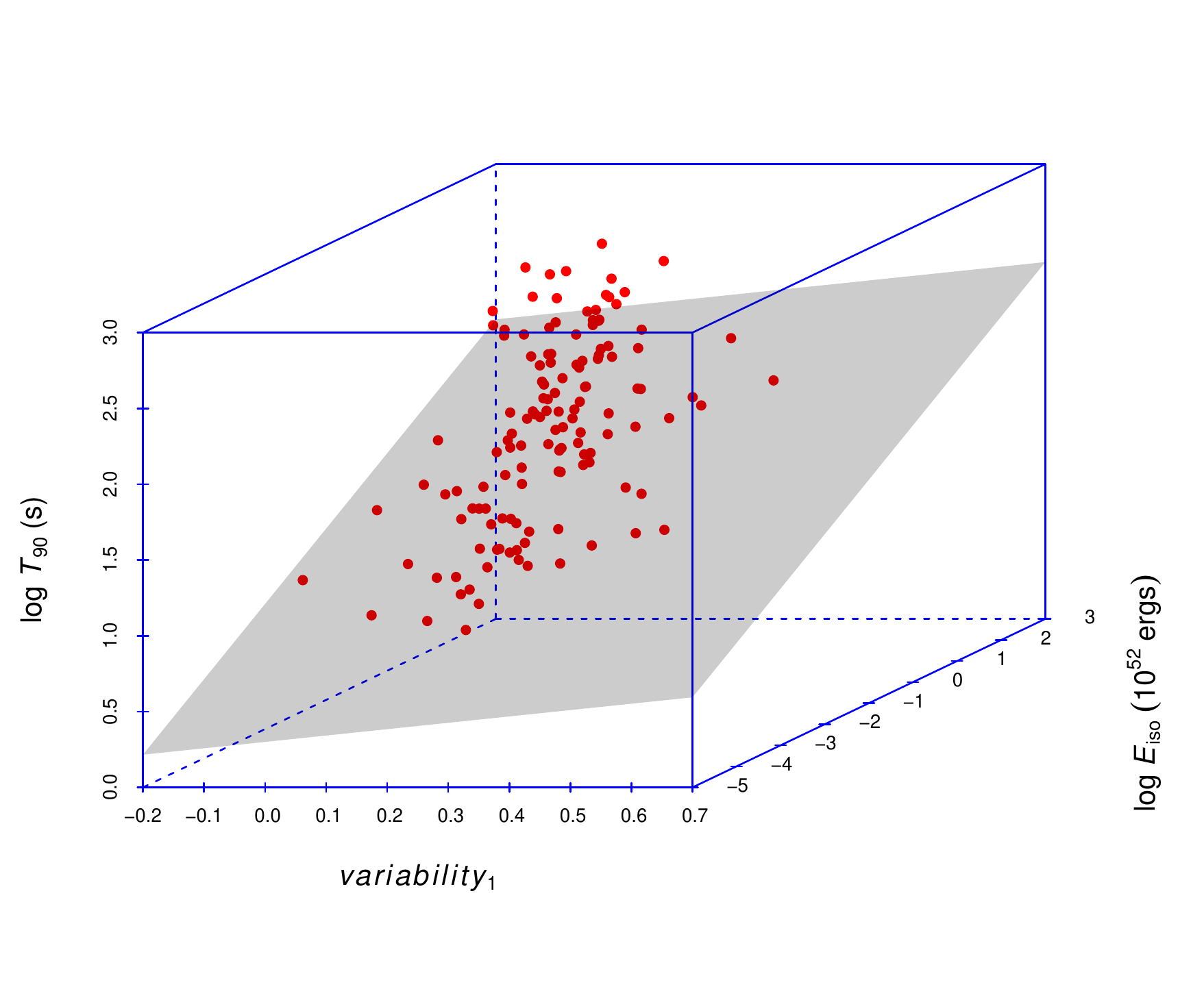}

\includegraphics[width=0.45\textwidth]{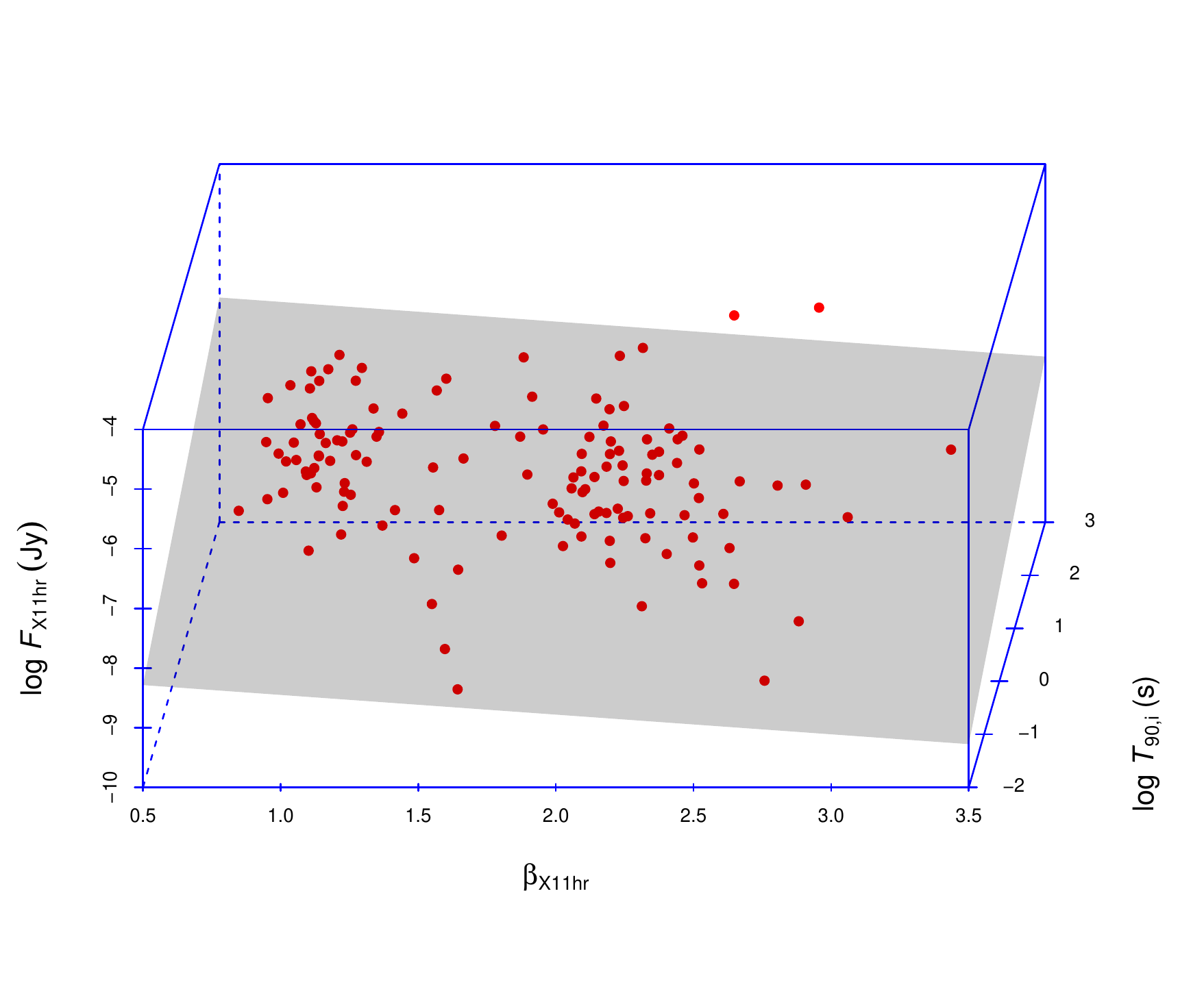}
\includegraphics[width=0.45\textwidth]{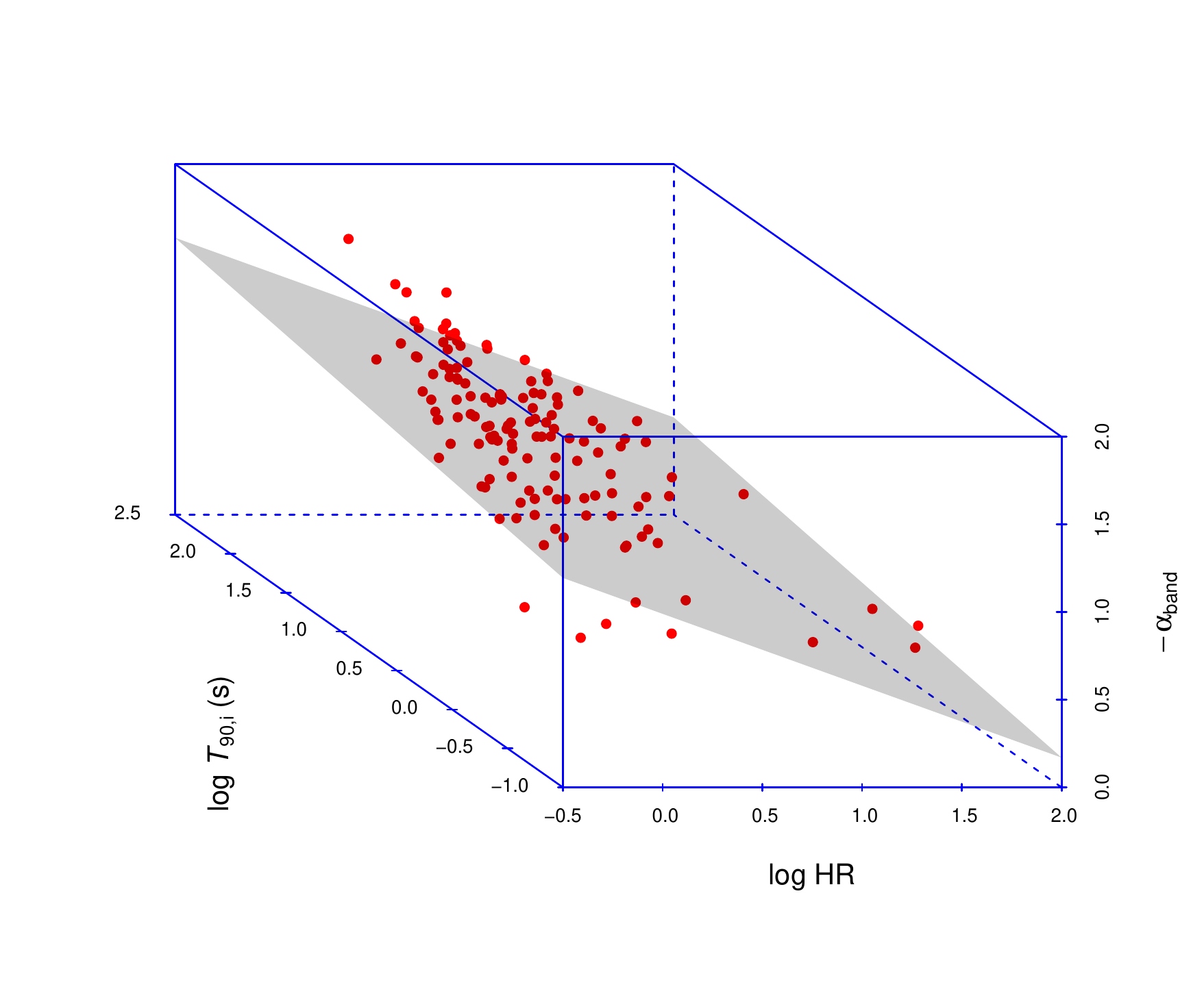}

\center{Fig. \ref{fig:three}---Continued}
\end{figure*}


\clearpage
\begin{figure*}

\includegraphics[width=0.45\textwidth]{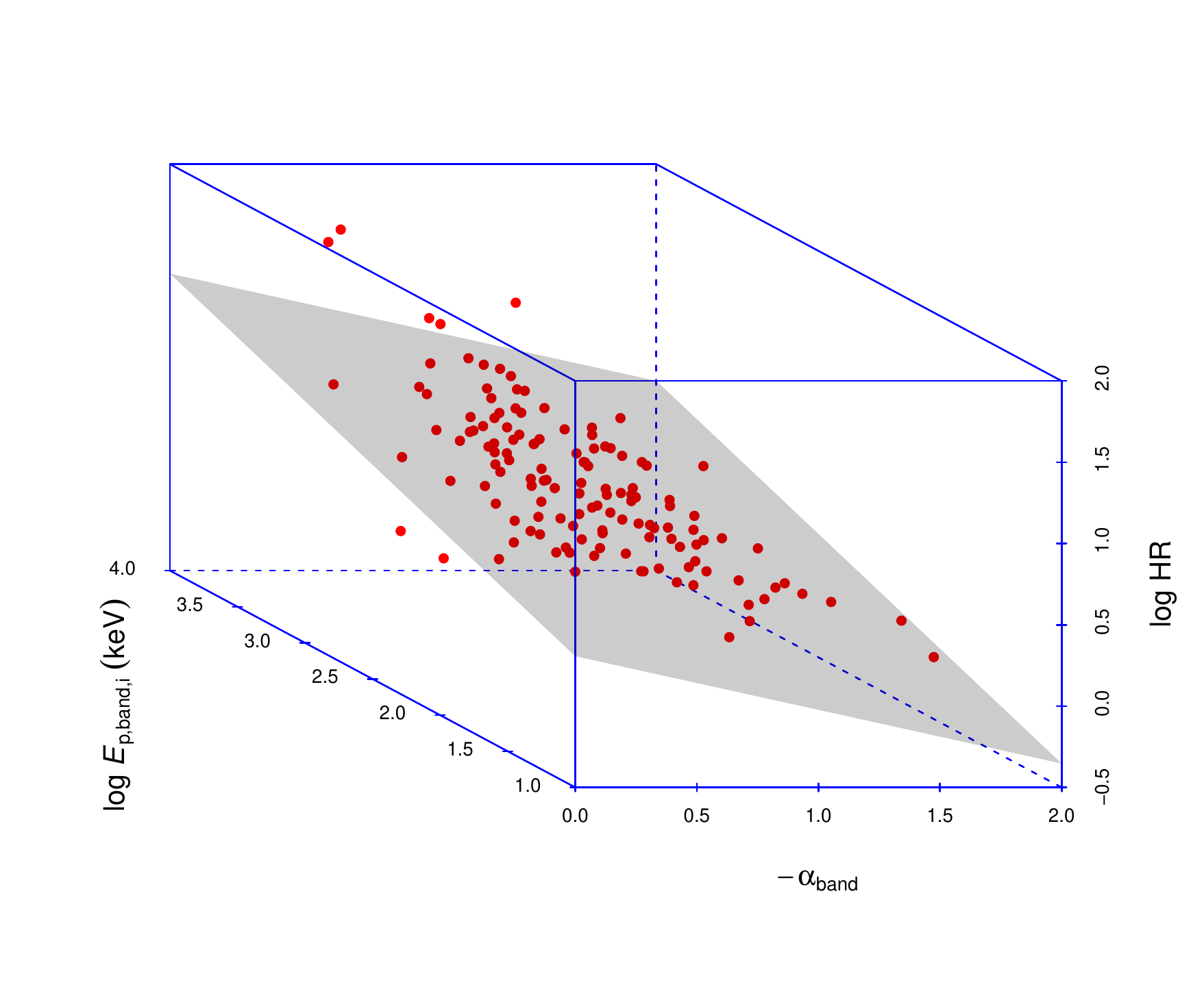}
\includegraphics[width=0.45\textwidth]{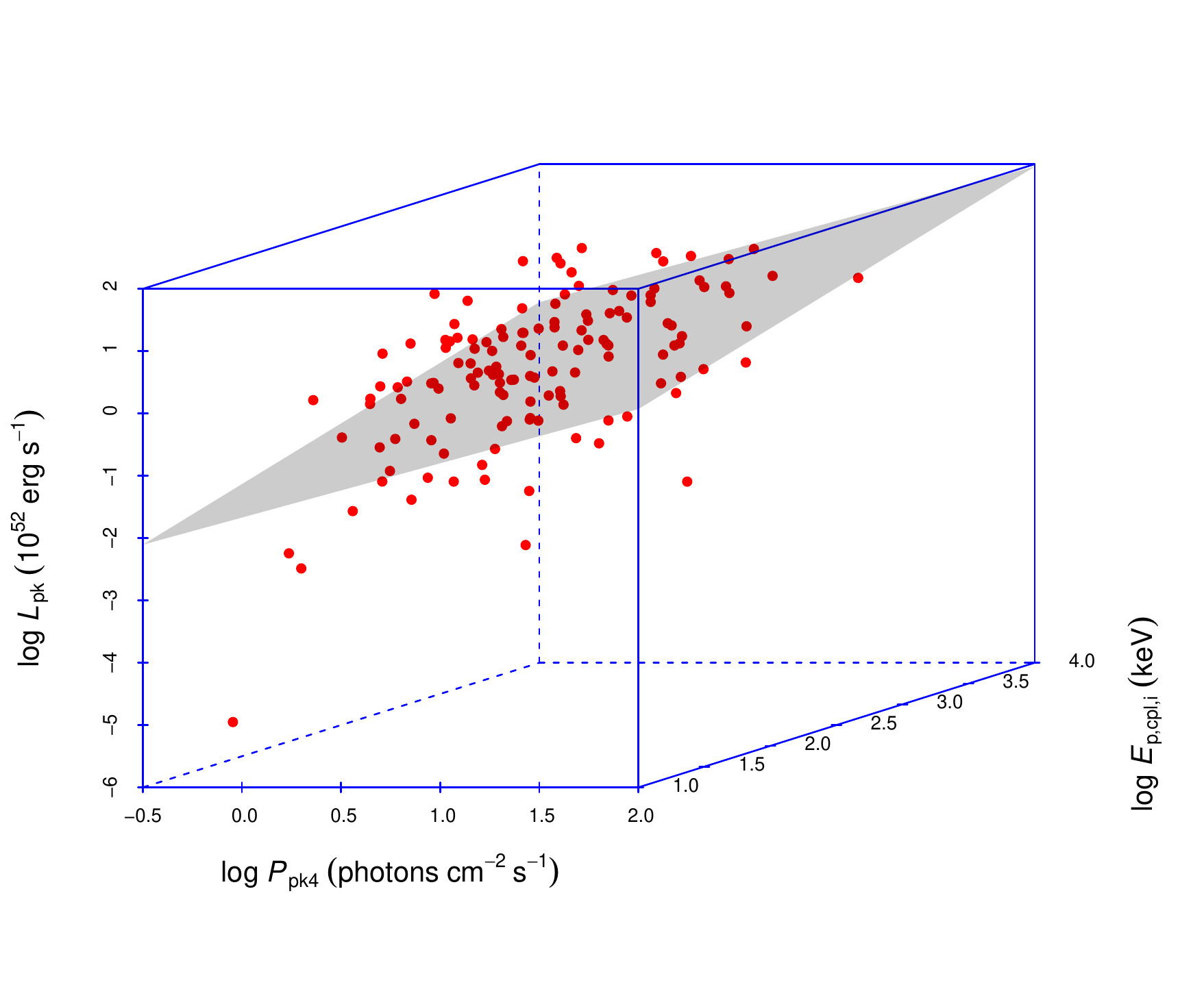}

\includegraphics[width=0.45\textwidth]{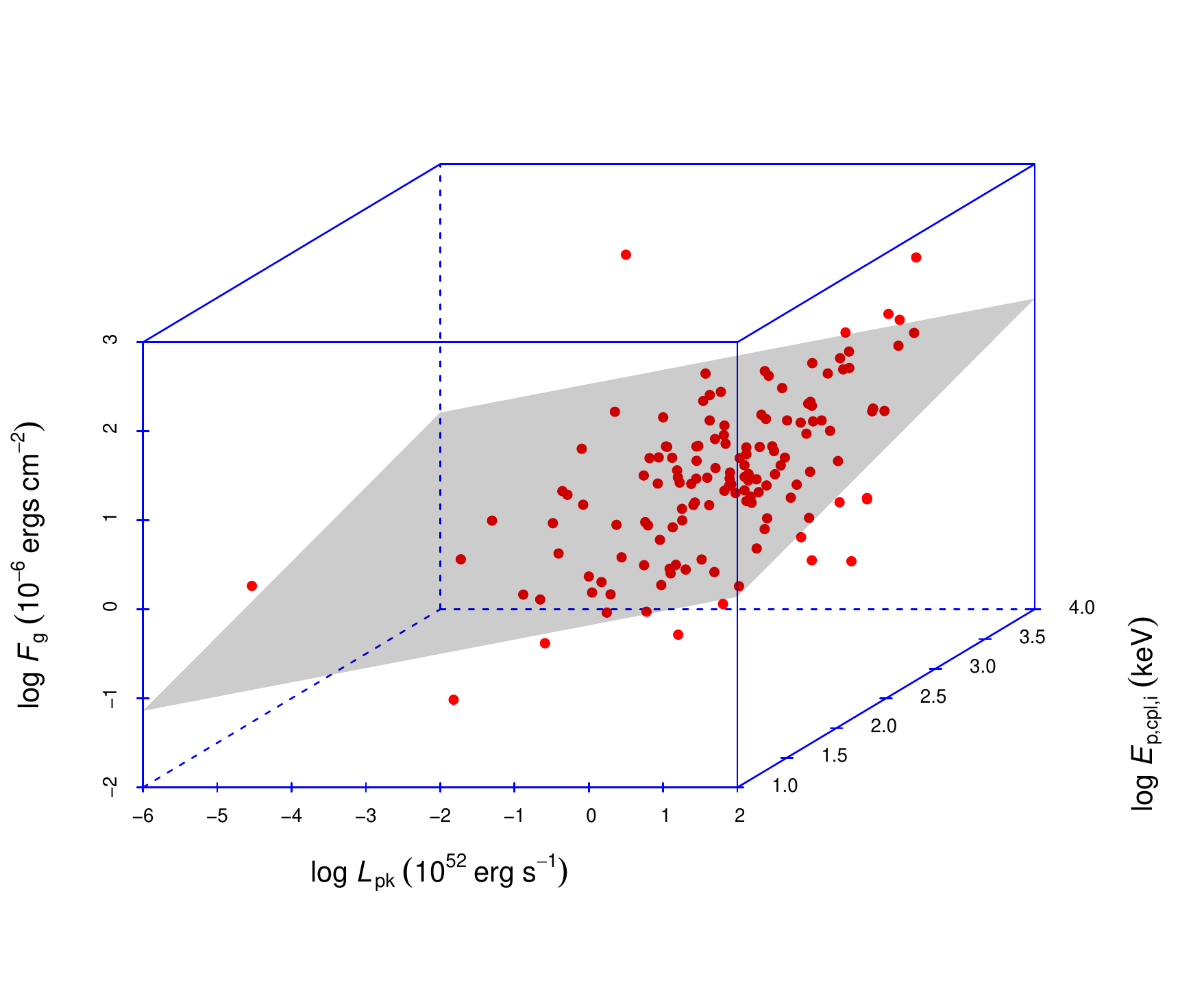}
\includegraphics[width=0.45\textwidth]{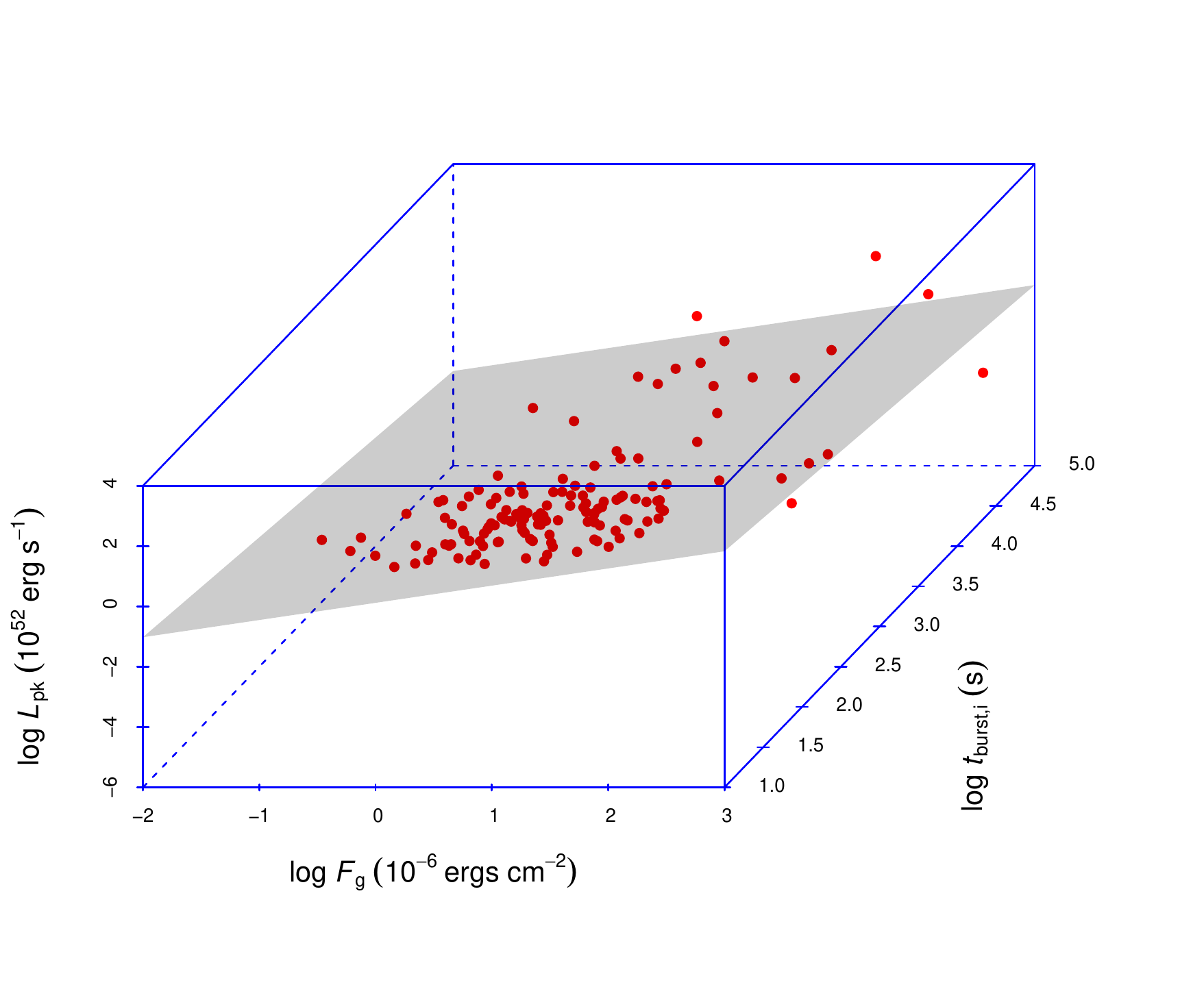}

\includegraphics[width=0.45\textwidth]{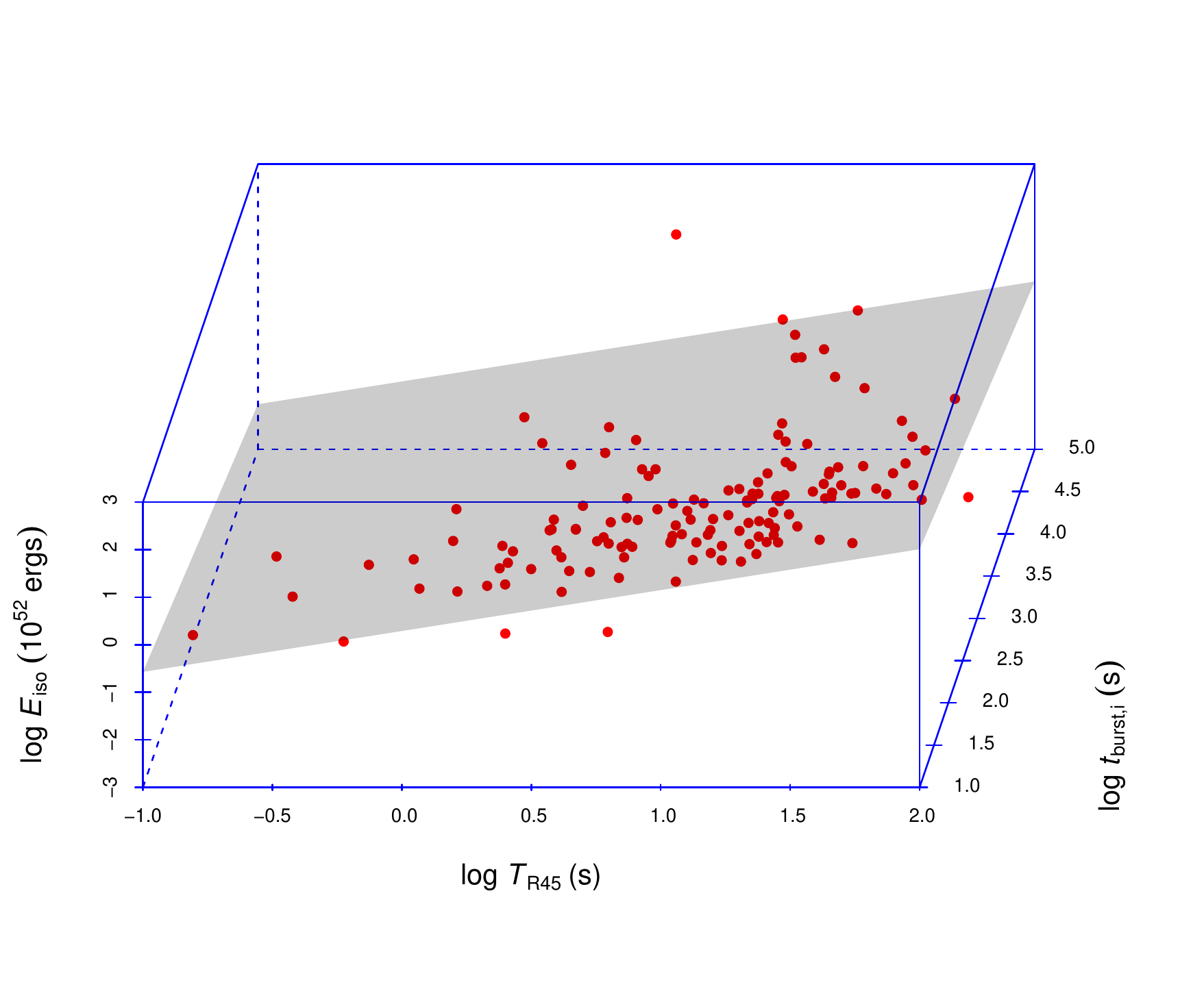}
\includegraphics[width=0.45\textwidth]{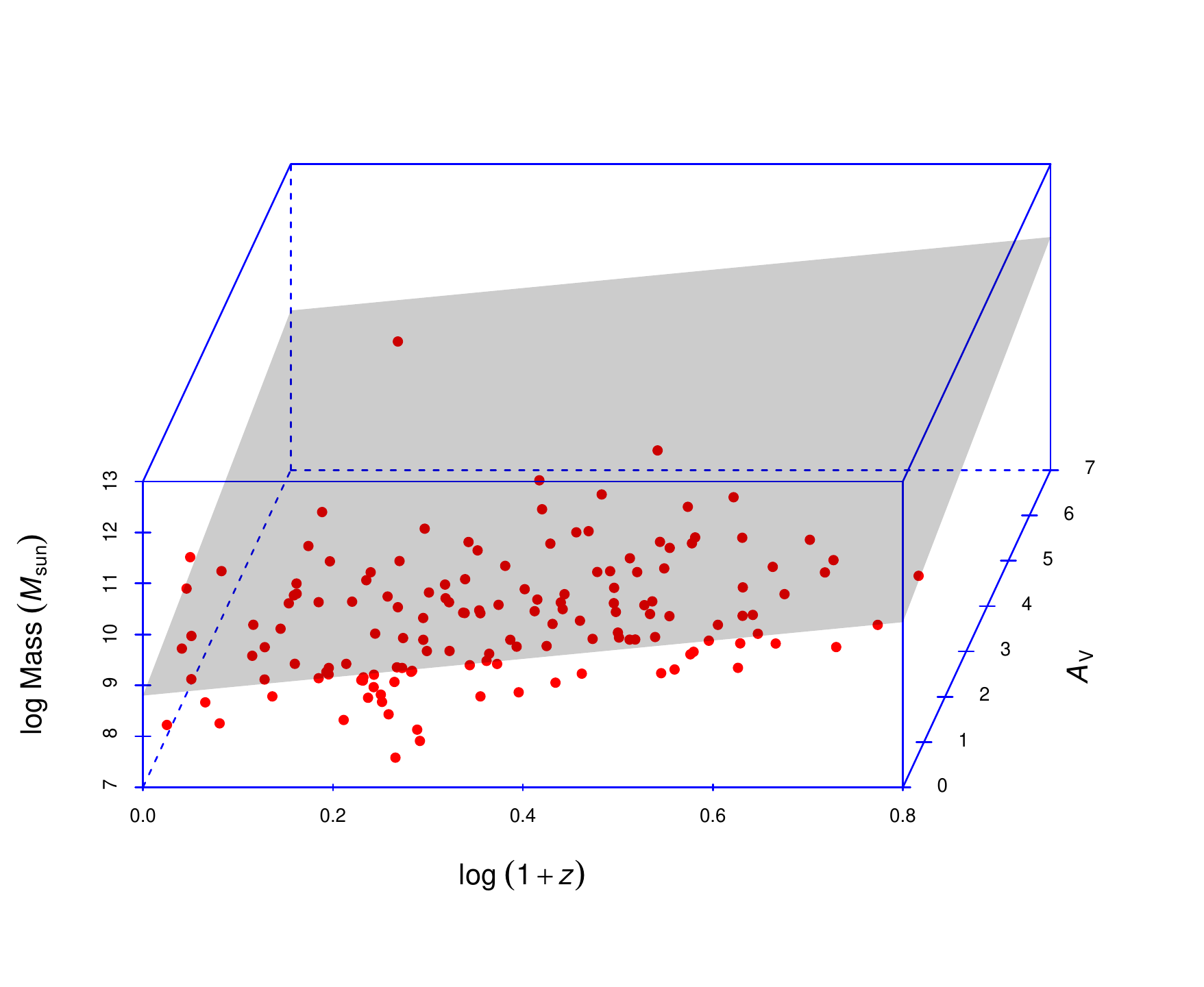}

\center{Fig. \ref{fig:three}---Continued}
\end{figure*}


\clearpage
\begin{figure*}

\includegraphics[width=0.45\textwidth]{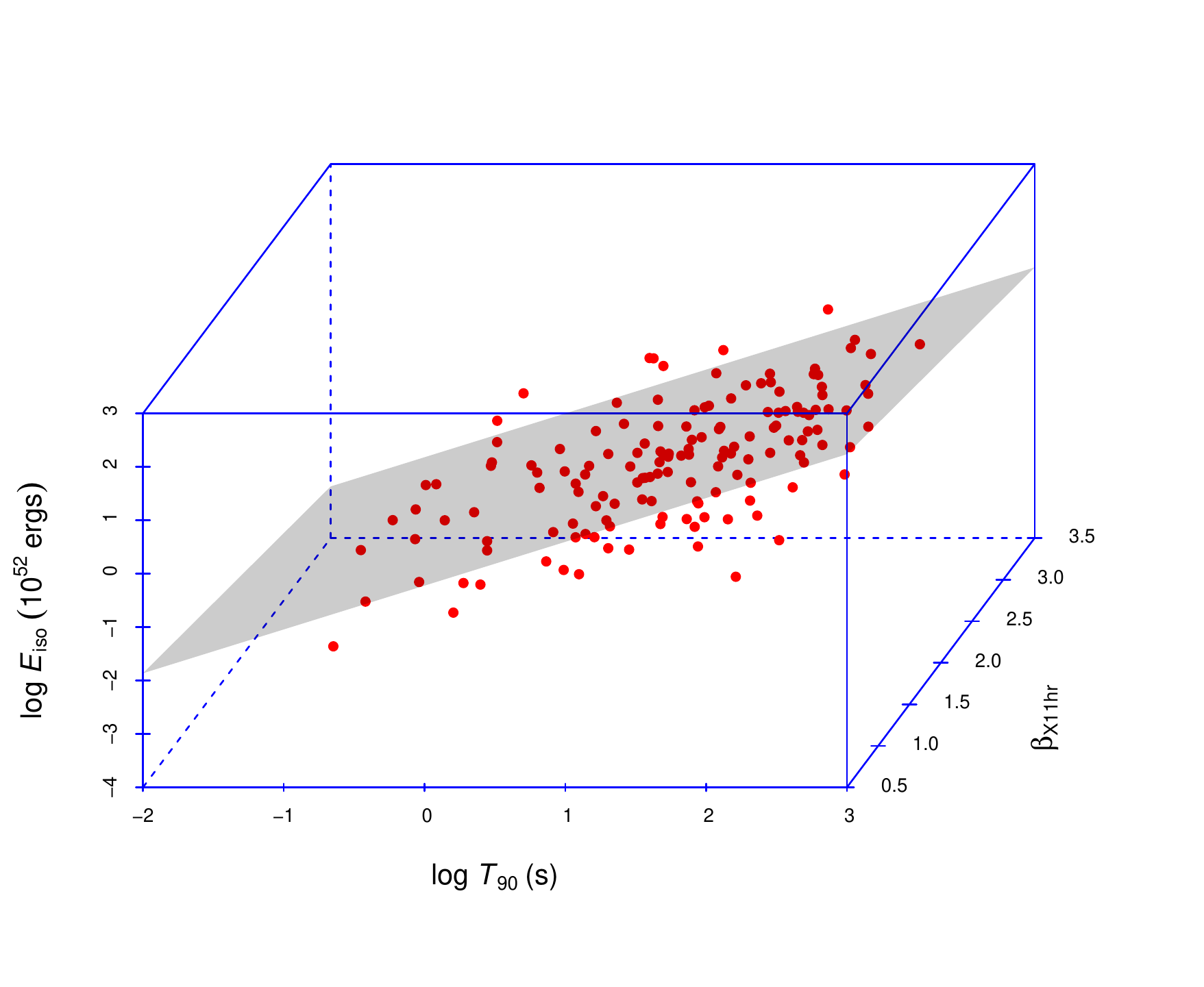}
\includegraphics[width=0.45\textwidth]{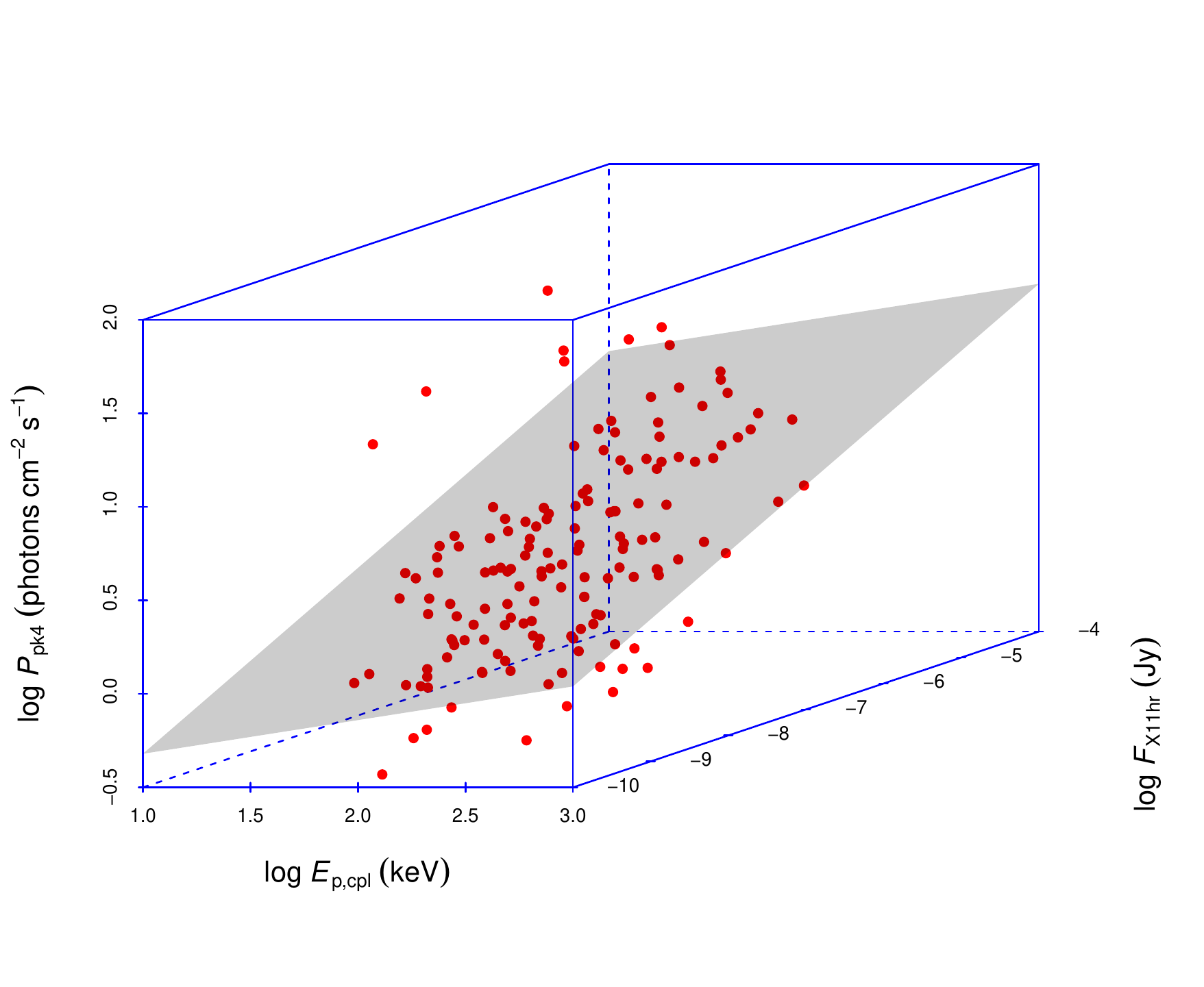}

\includegraphics[width=0.45\textwidth]{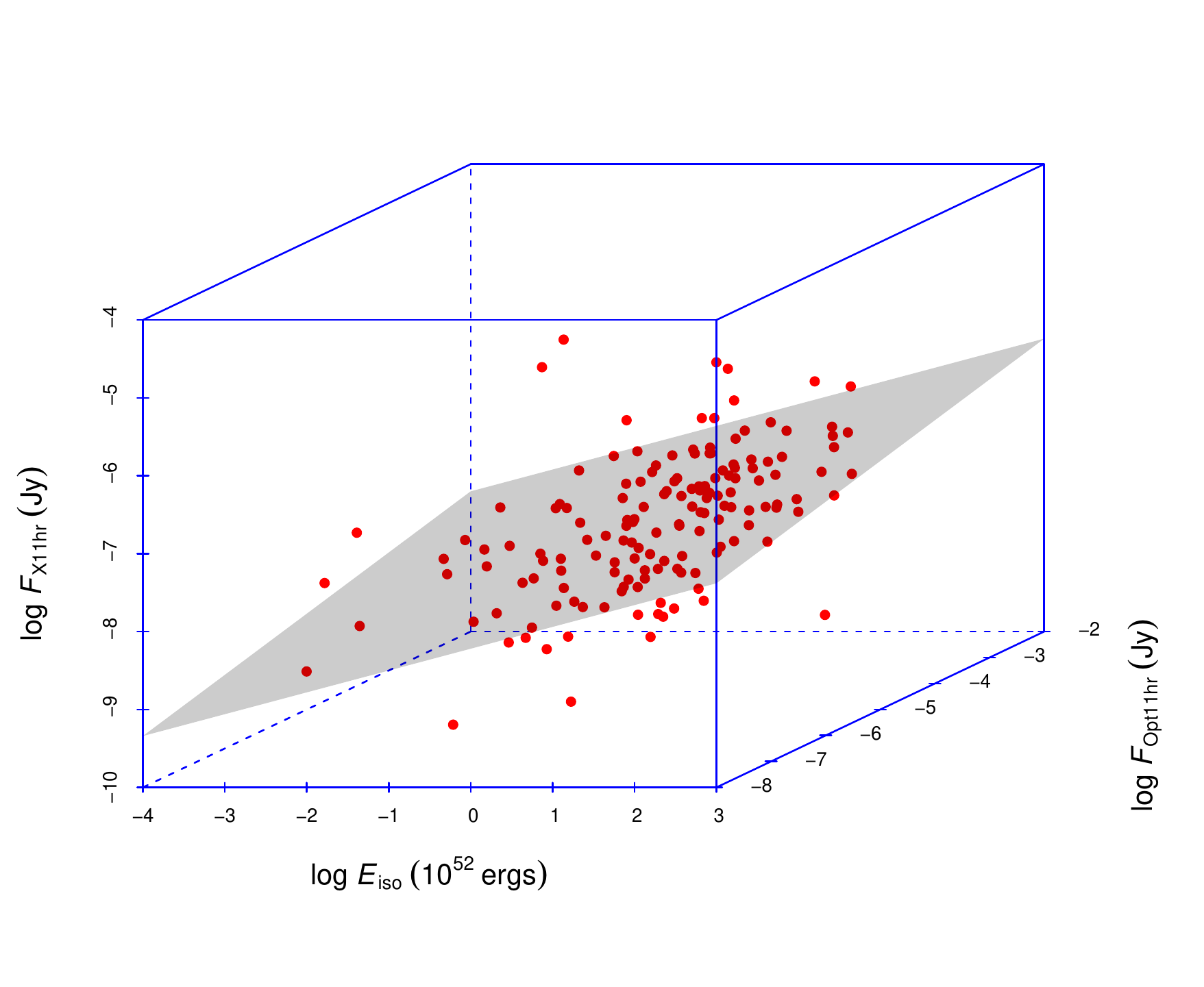}
\includegraphics[width=0.45\textwidth]{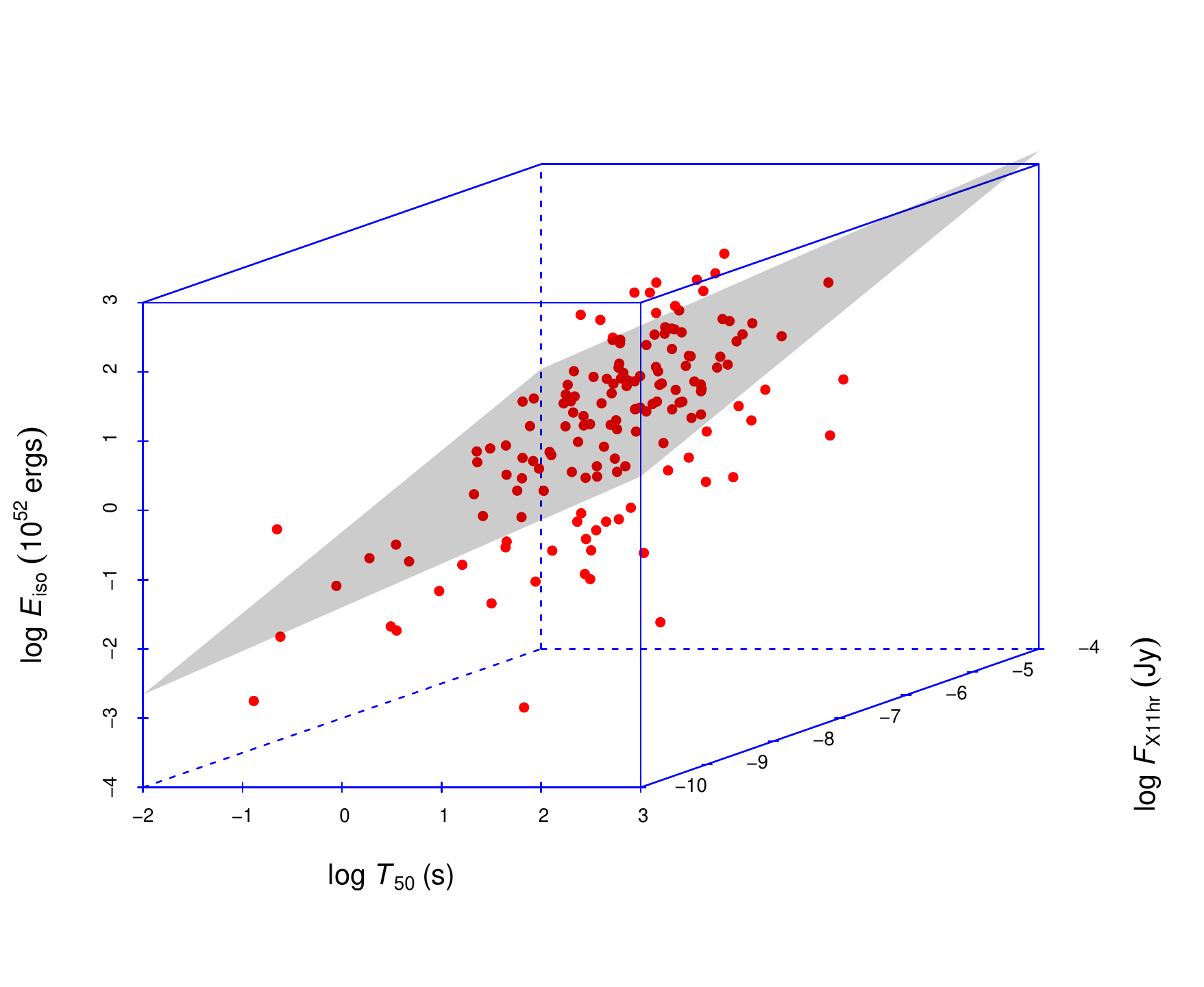}

\includegraphics[width=0.45\textwidth]{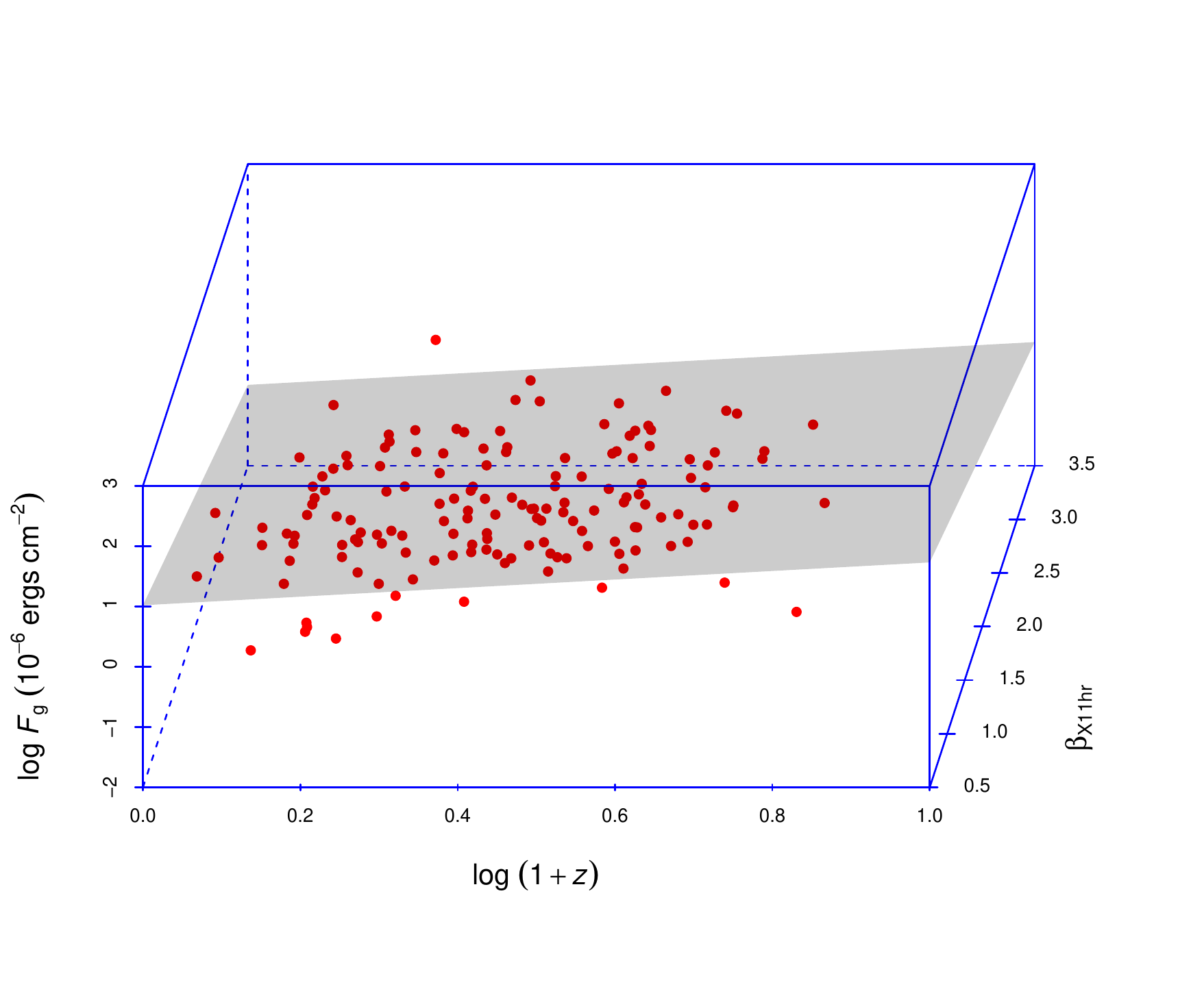}
\includegraphics[width=0.45\textwidth]{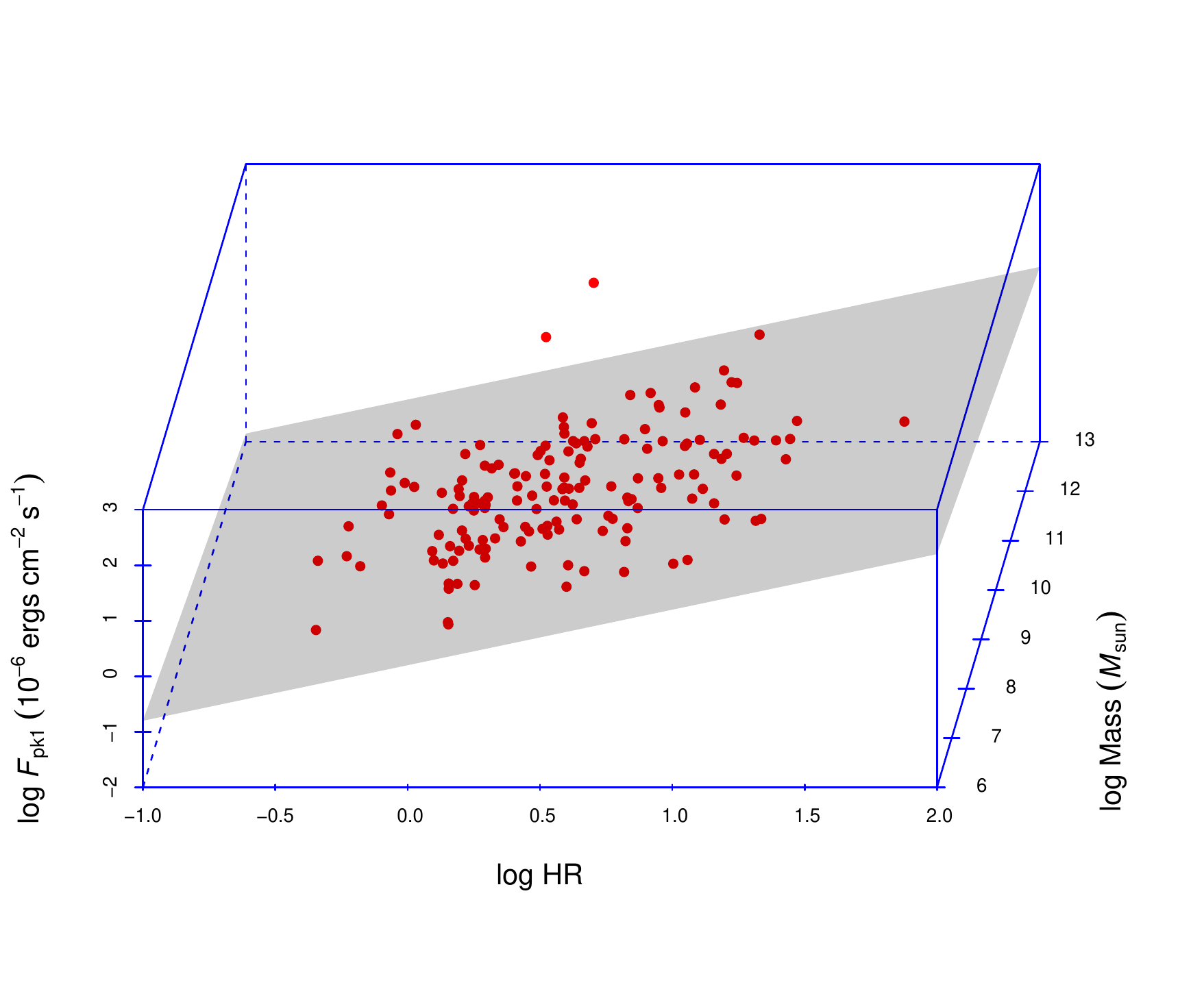}

\center{Fig. \ref{fig:three}---Continued}
\end{figure*}


\clearpage
\begin{figure*}

\includegraphics[width=0.45\textwidth]{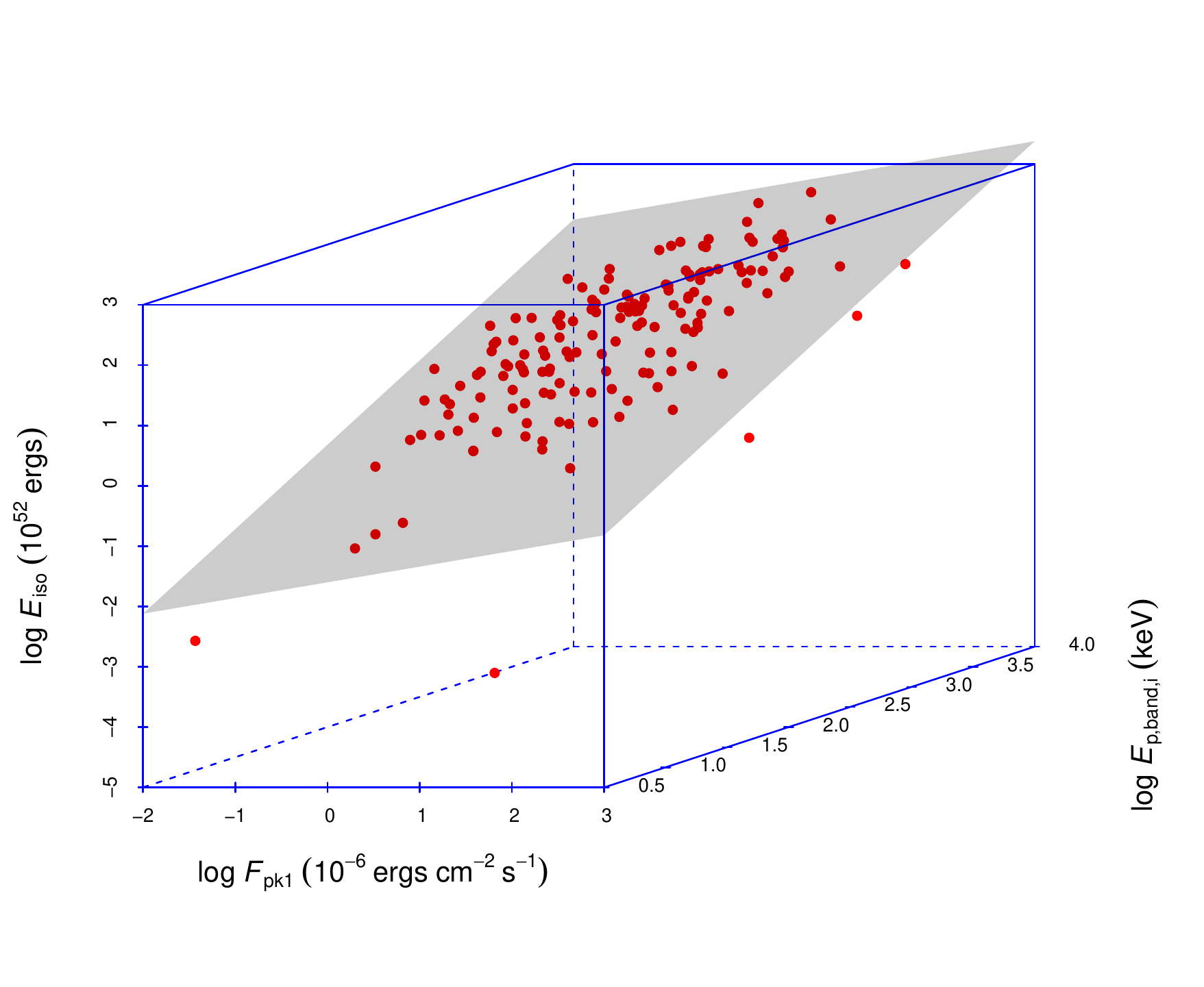}
\includegraphics[width=0.45\textwidth]{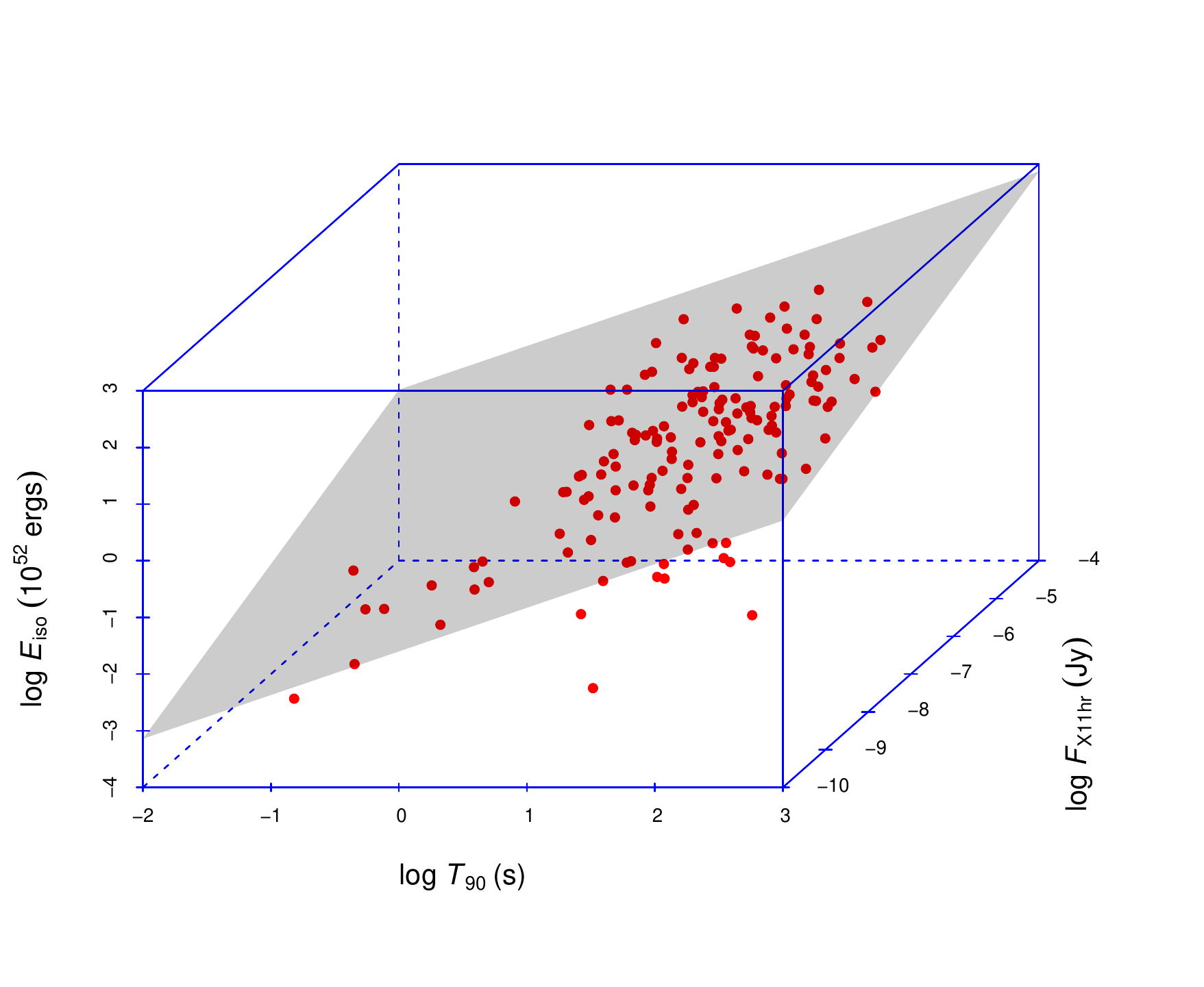}

\includegraphics[width=0.45\textwidth]{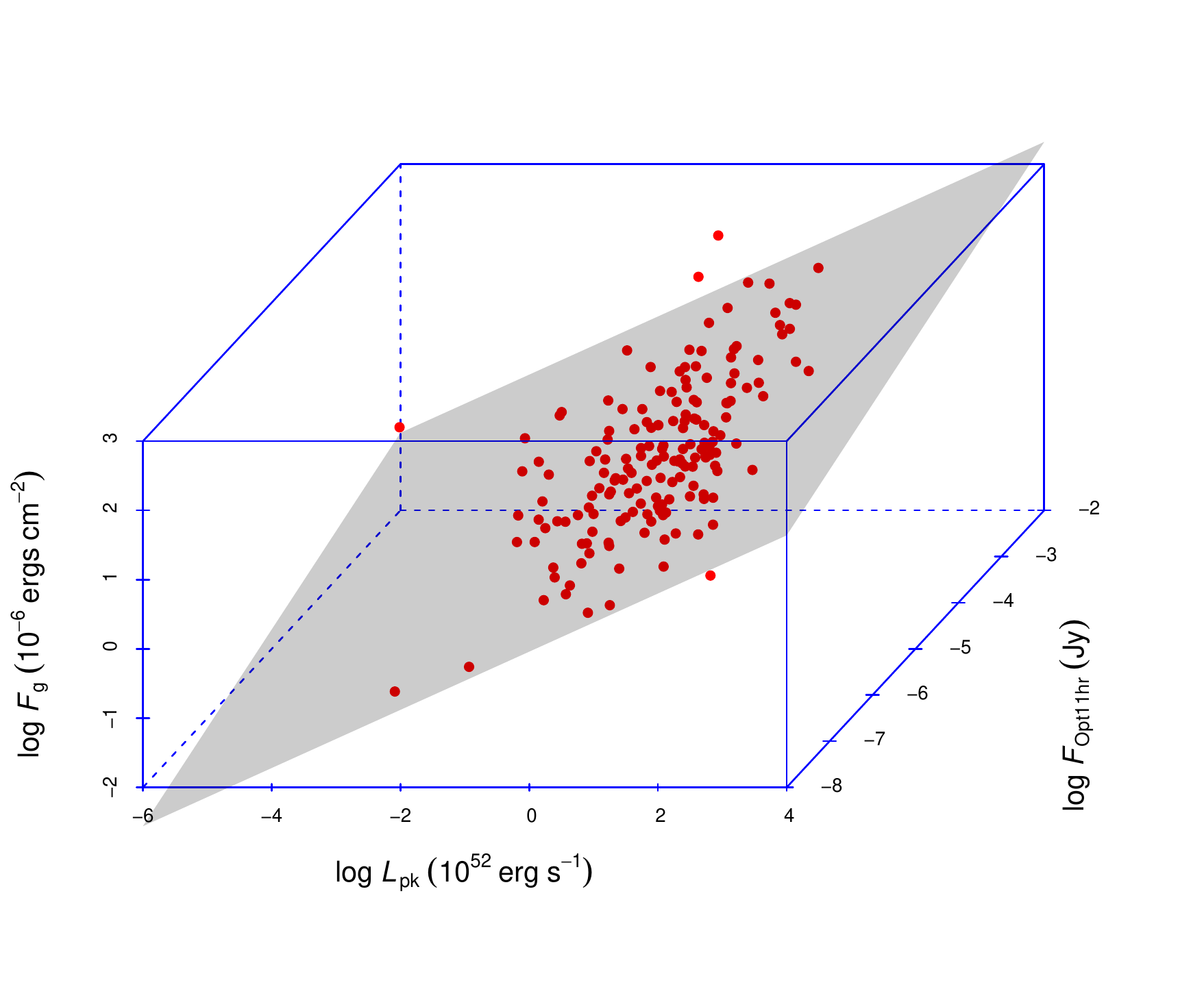}
\includegraphics[width=0.45\textwidth]{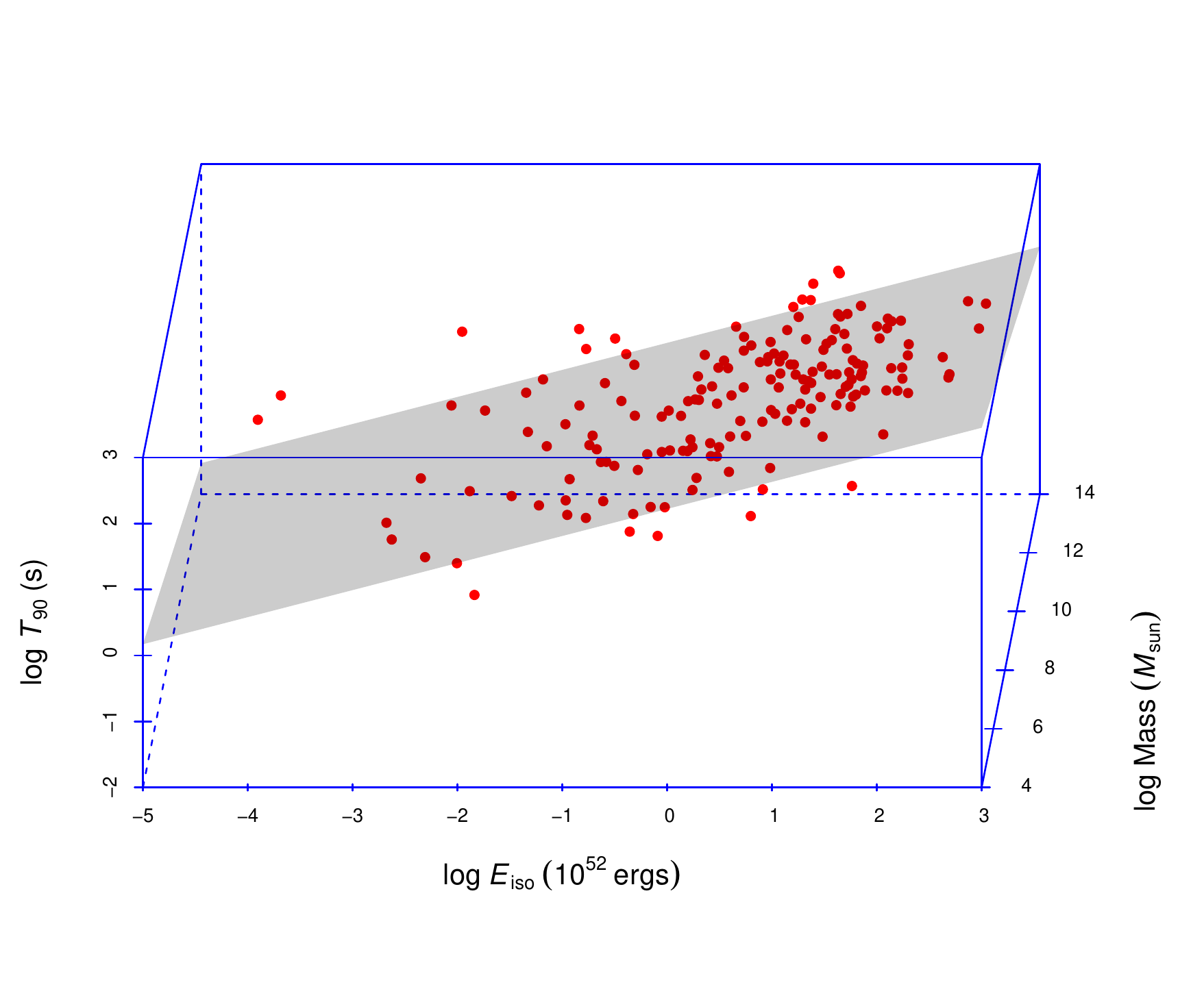}

\includegraphics[width=0.45\textwidth]{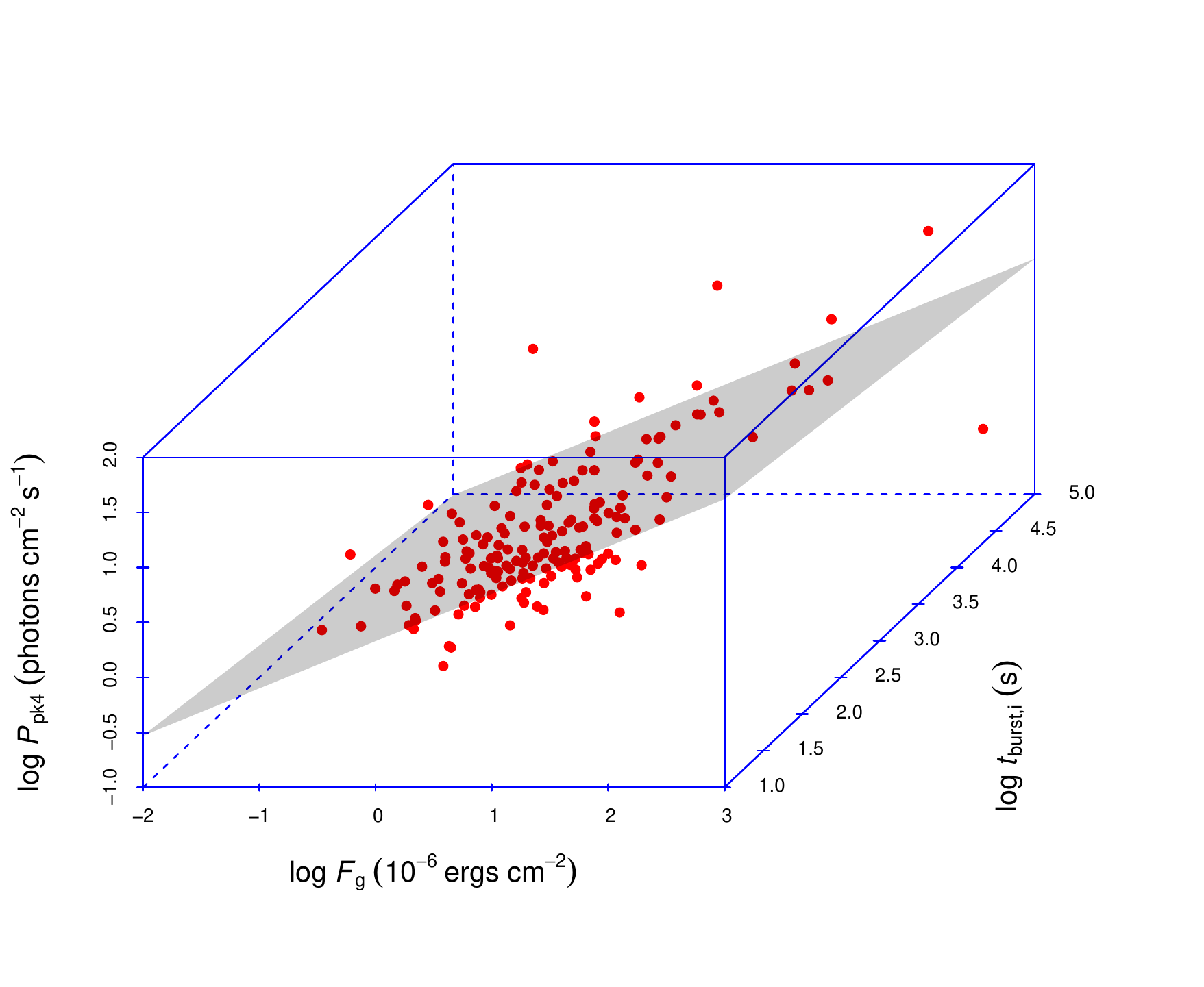}
\includegraphics[width=0.45\textwidth]{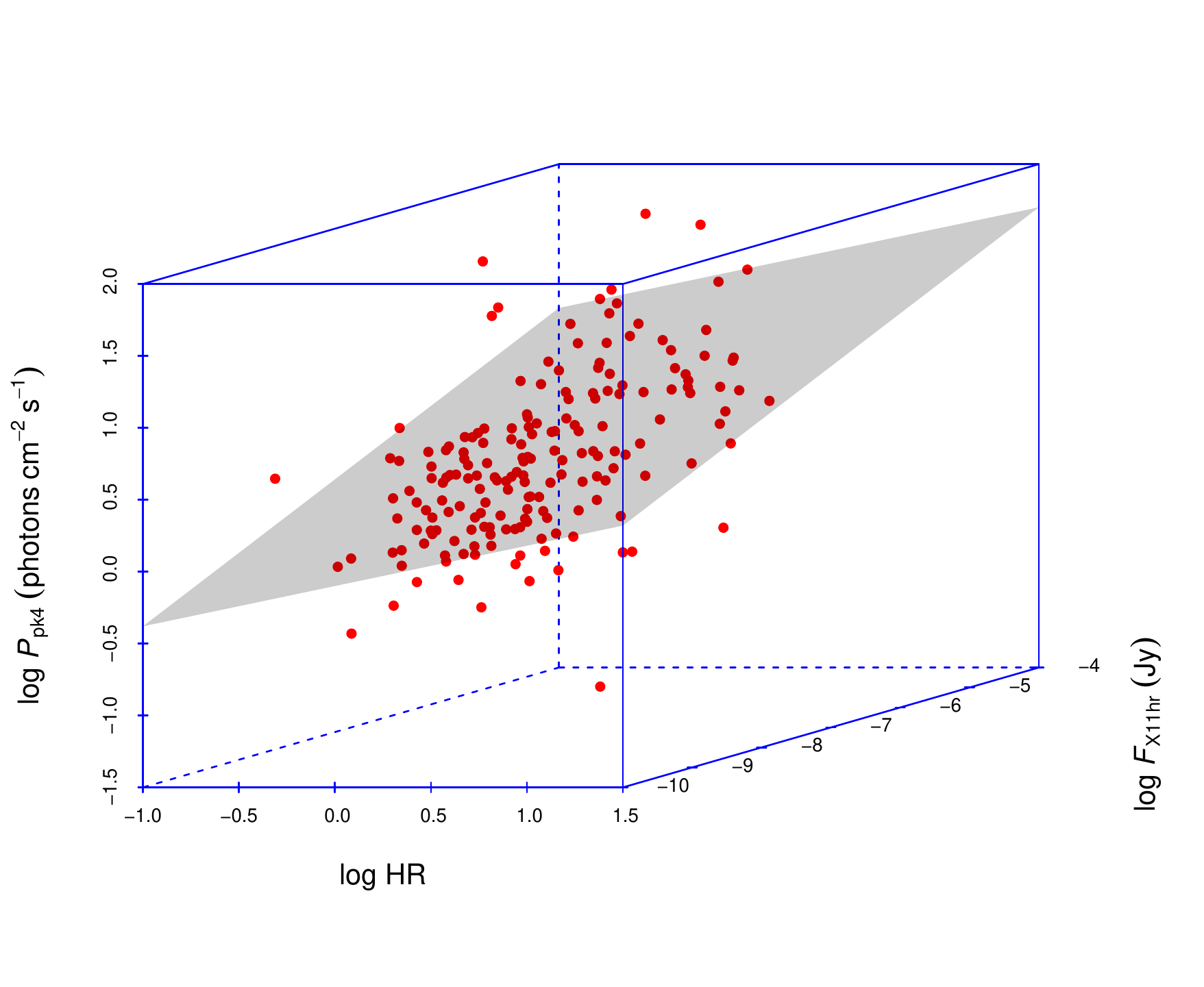}

\center{Fig. \ref{fig:three}---Continued}
\end{figure*}


\clearpage
\begin{figure*}

\includegraphics[width=0.45\textwidth]{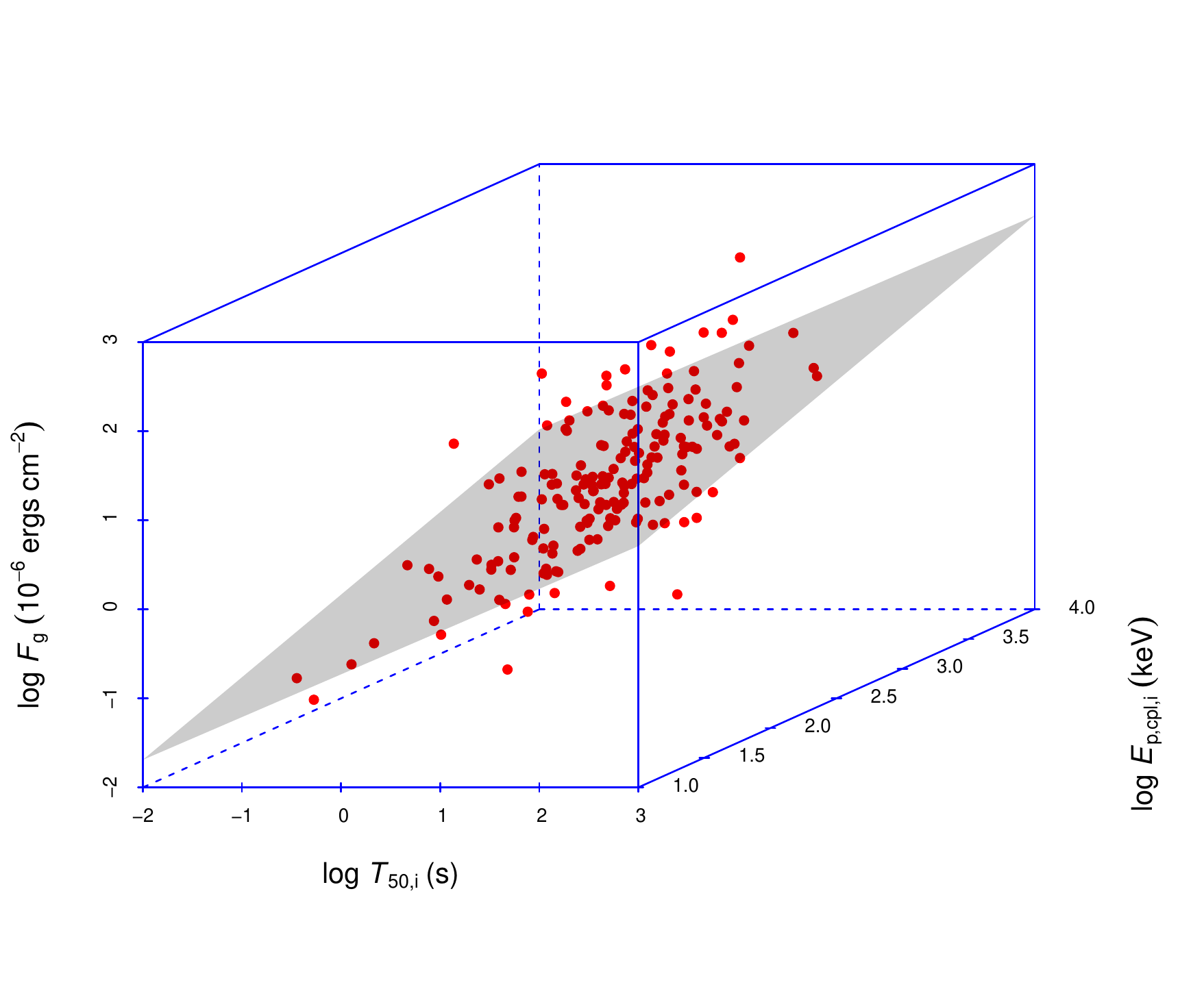}
\includegraphics[width=0.45\textwidth]{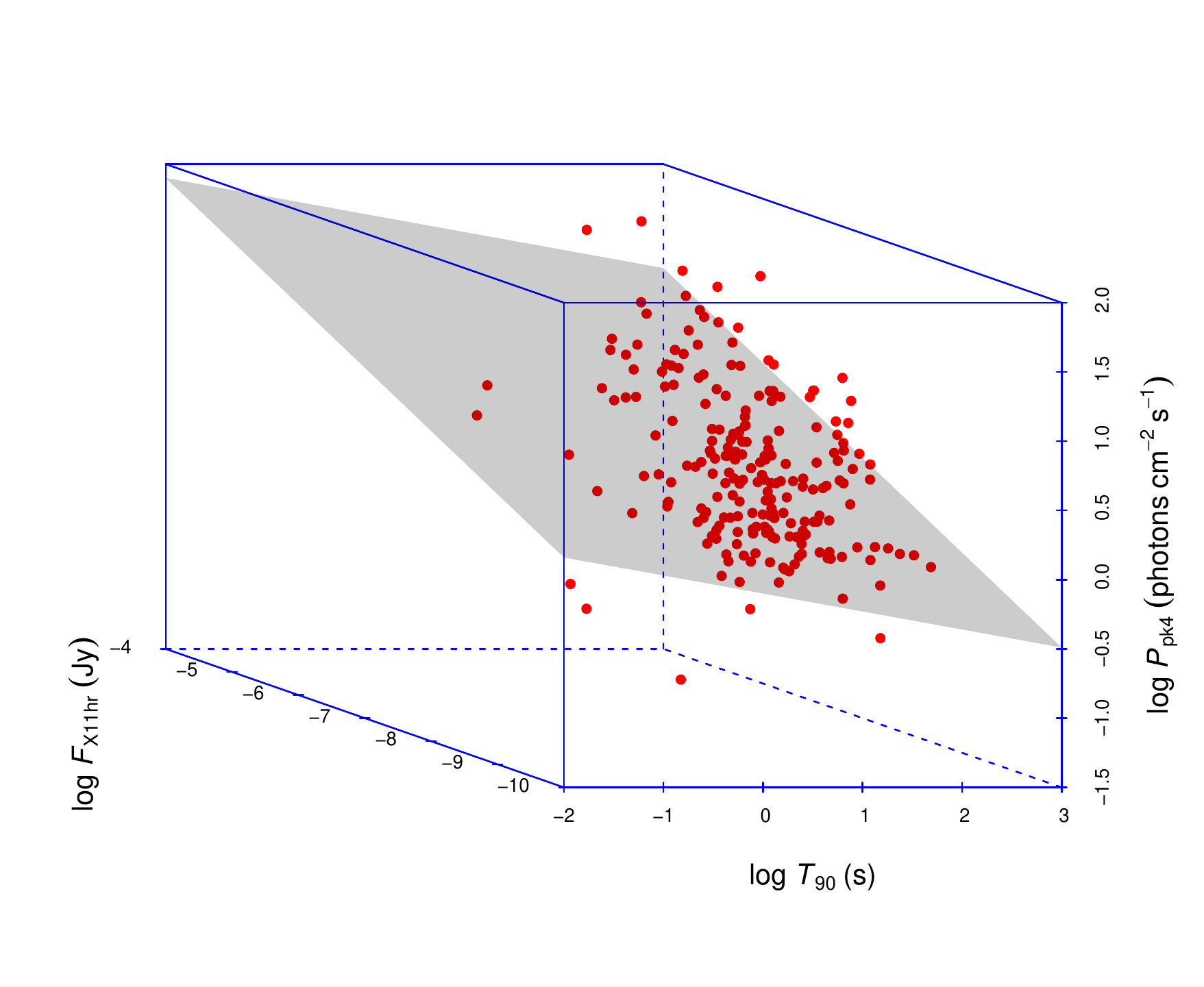}

\includegraphics[width=0.45\textwidth]{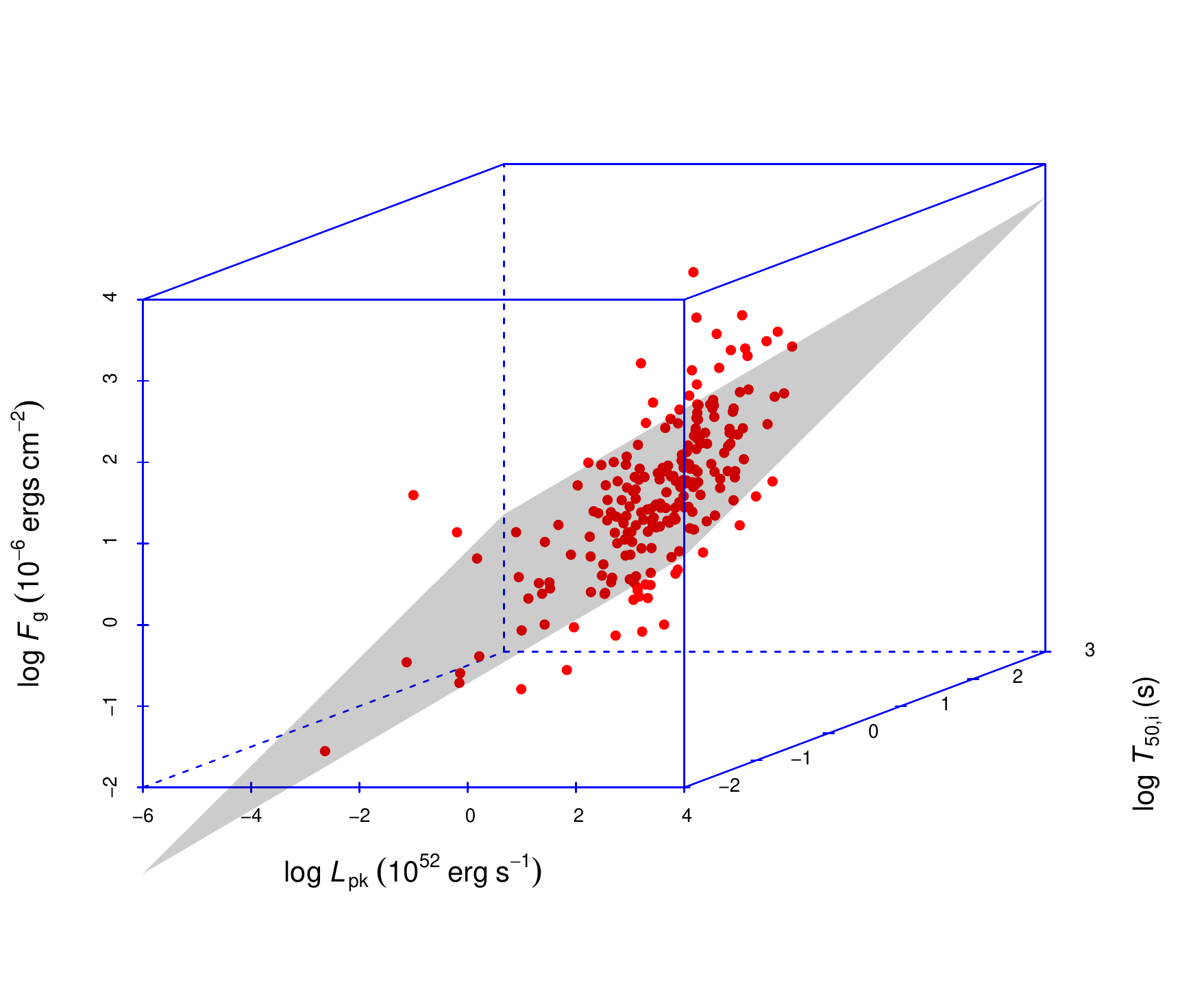}
\includegraphics[width=0.45\textwidth]{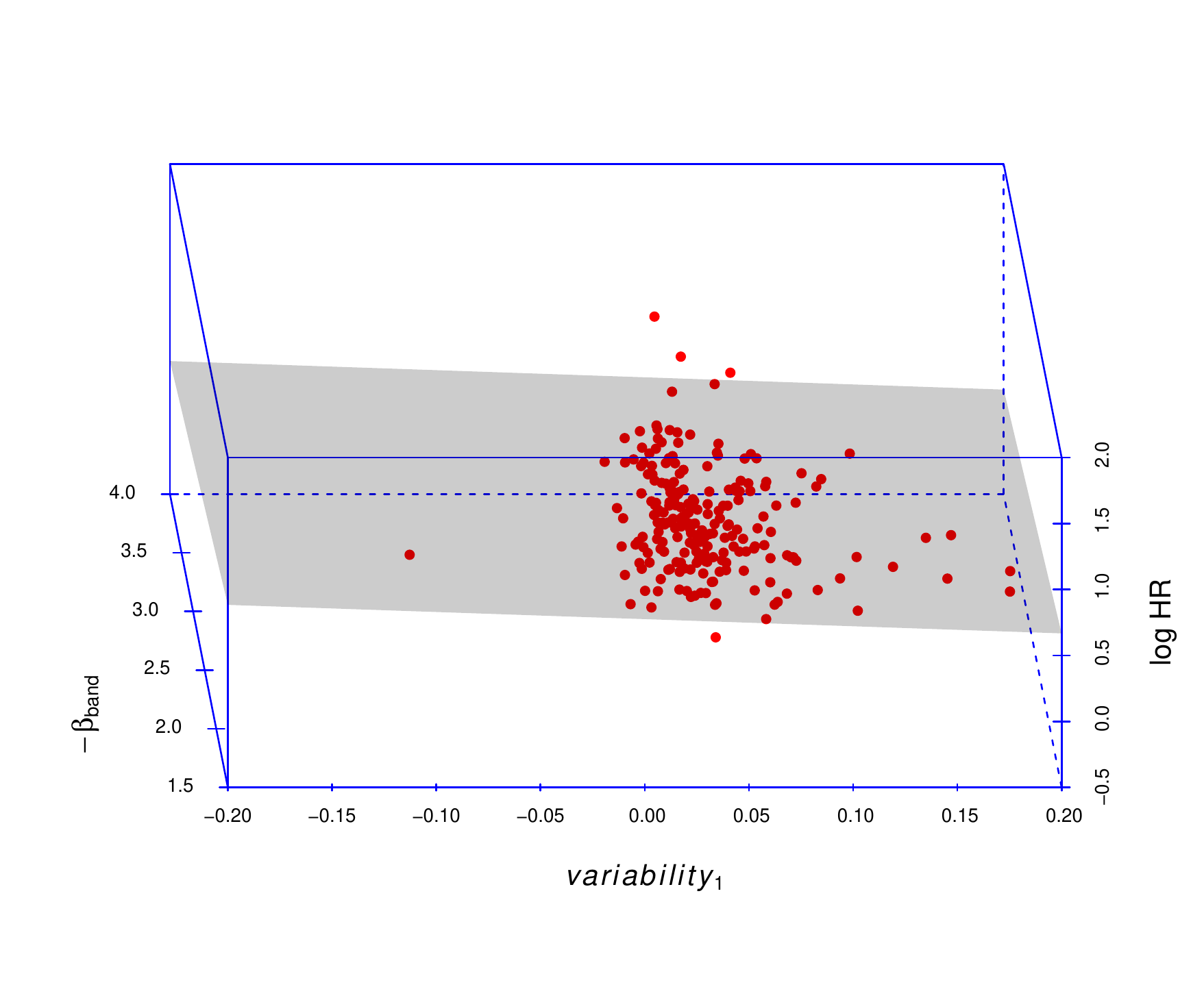}

\includegraphics[width=0.45\textwidth]{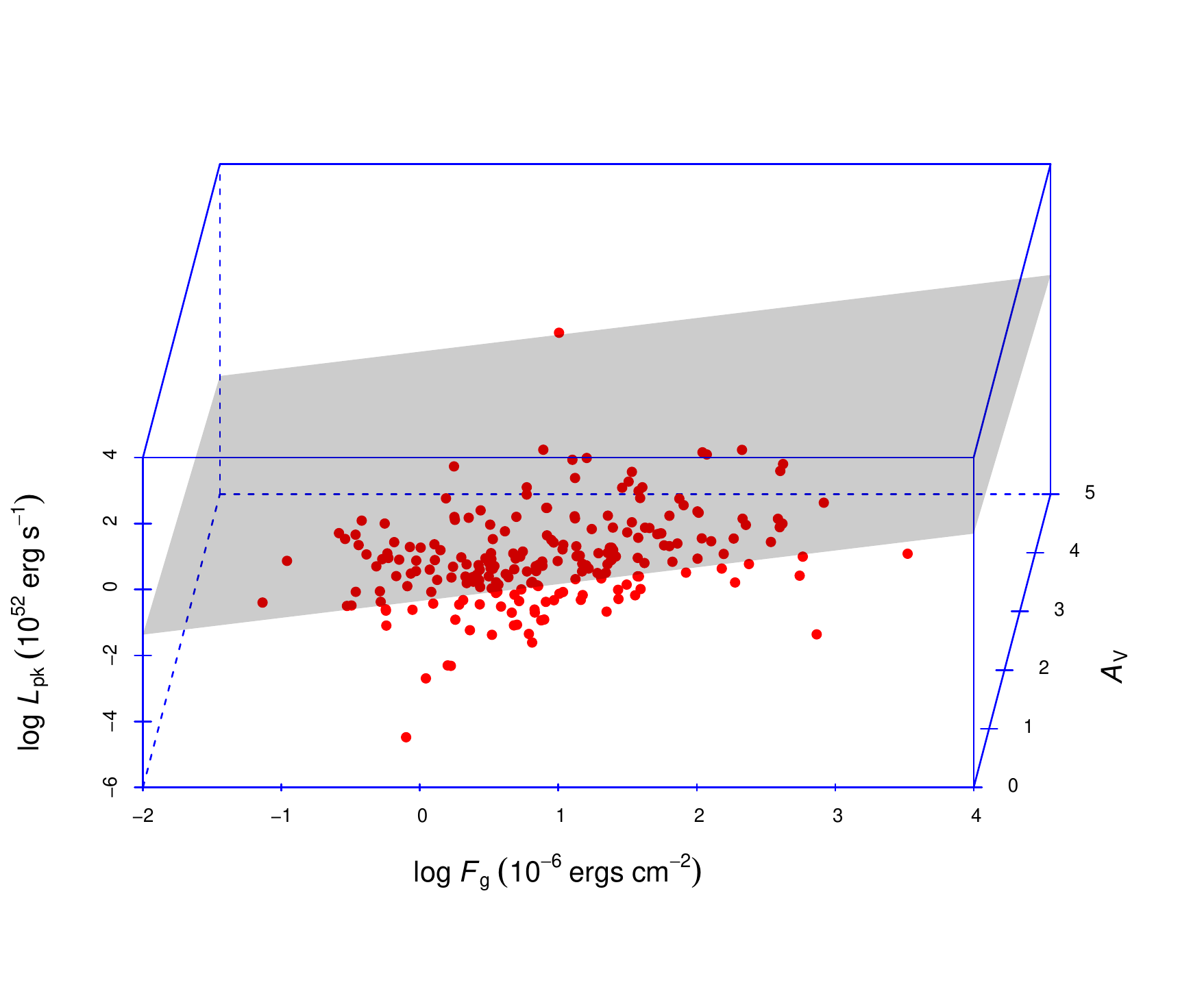}
\includegraphics[width=0.45\textwidth]{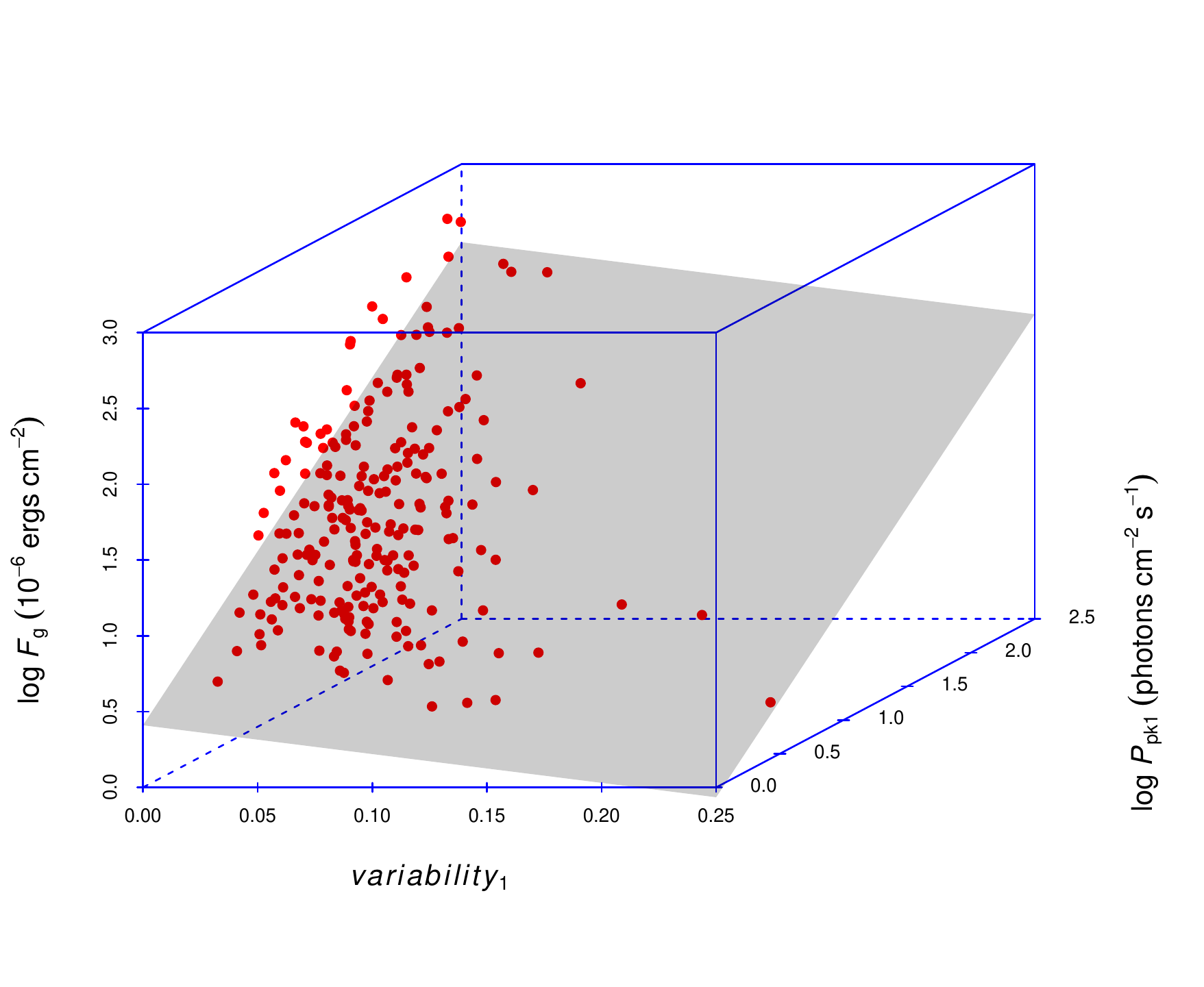}

\center{Fig. \ref{fig:three}---Continued}
\end{figure*}


\clearpage
\begin{figure*}

\includegraphics[width=0.45\textwidth]{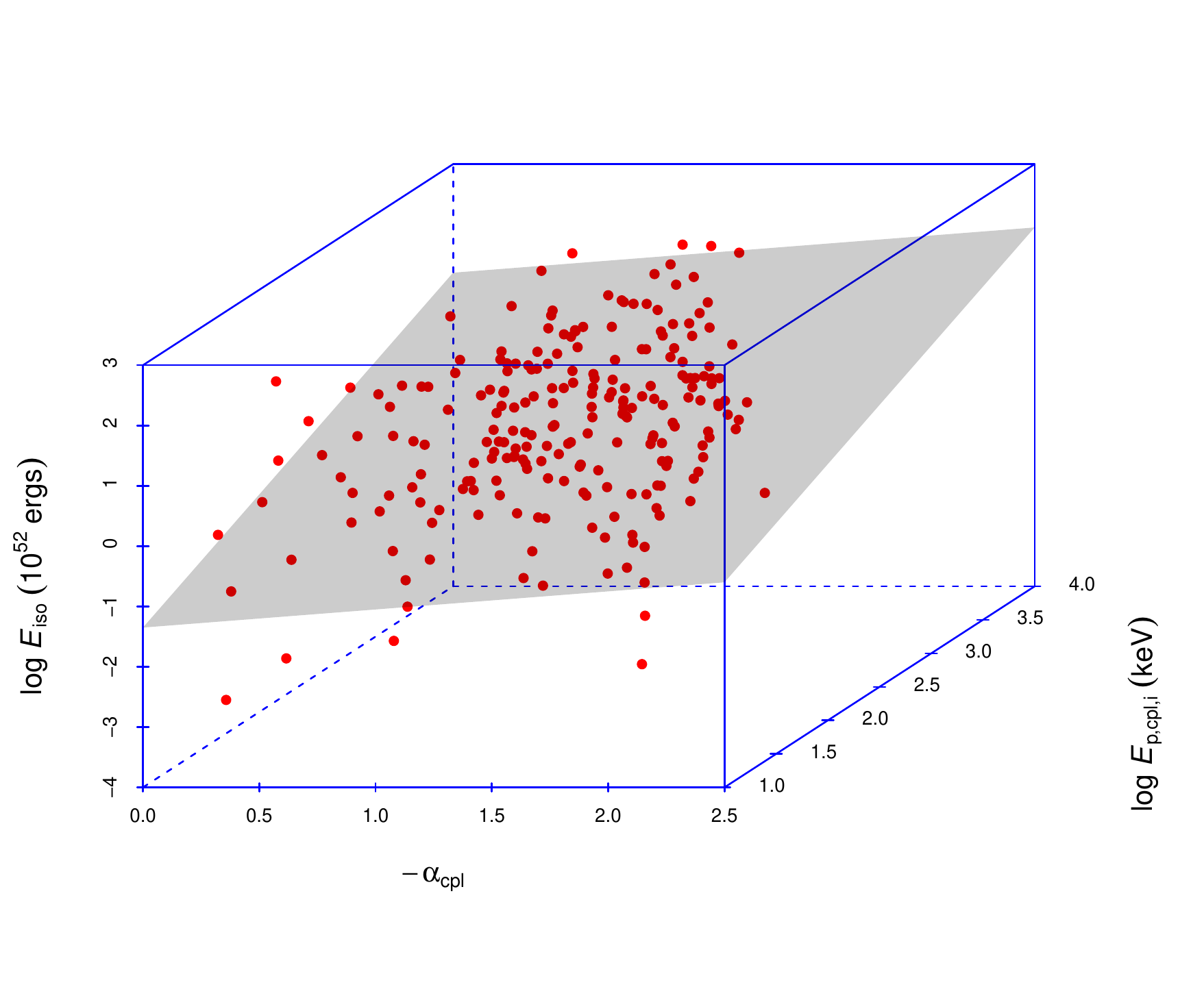}
\includegraphics[width=0.45\textwidth]{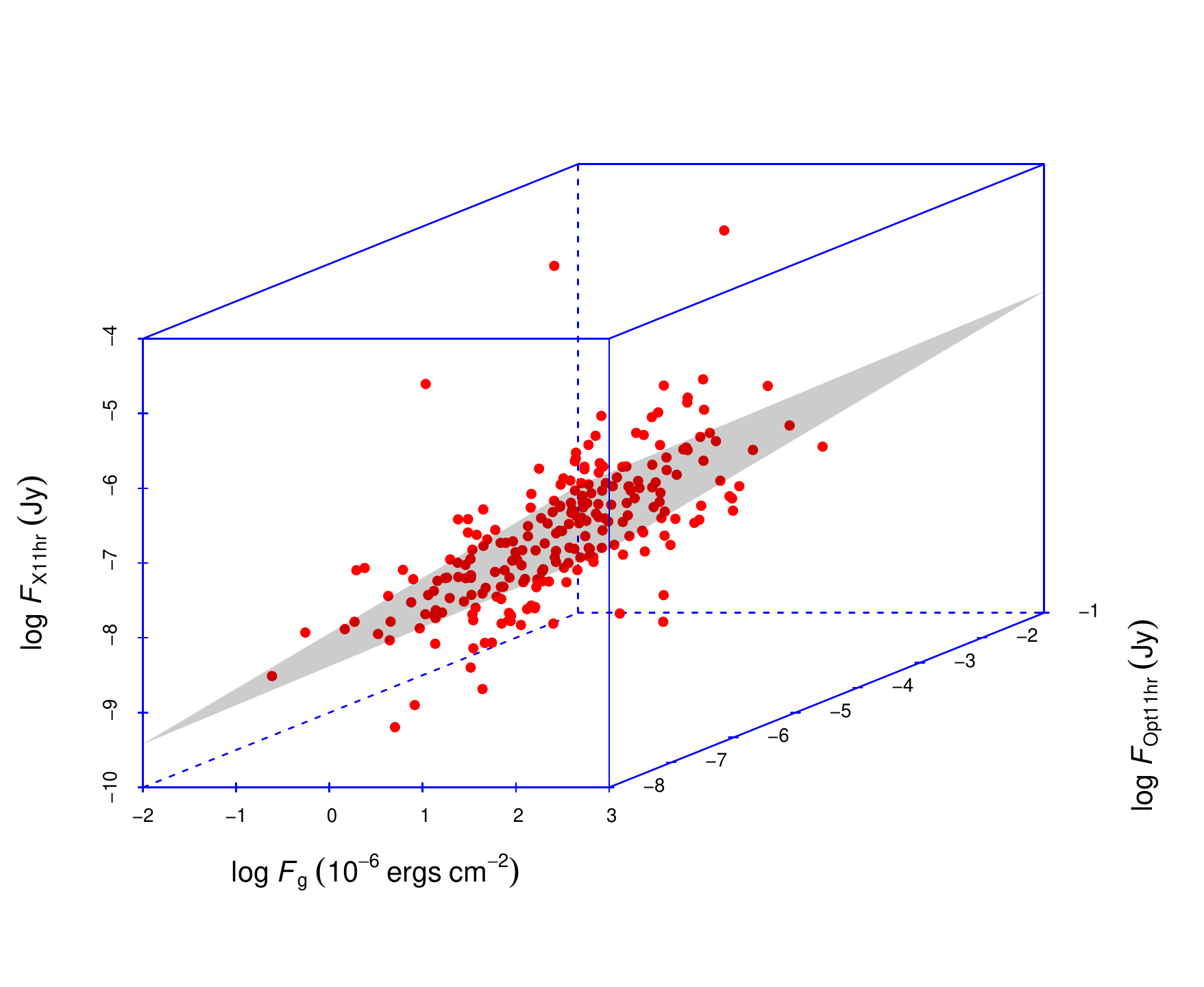}

\includegraphics[width=0.45\textwidth]{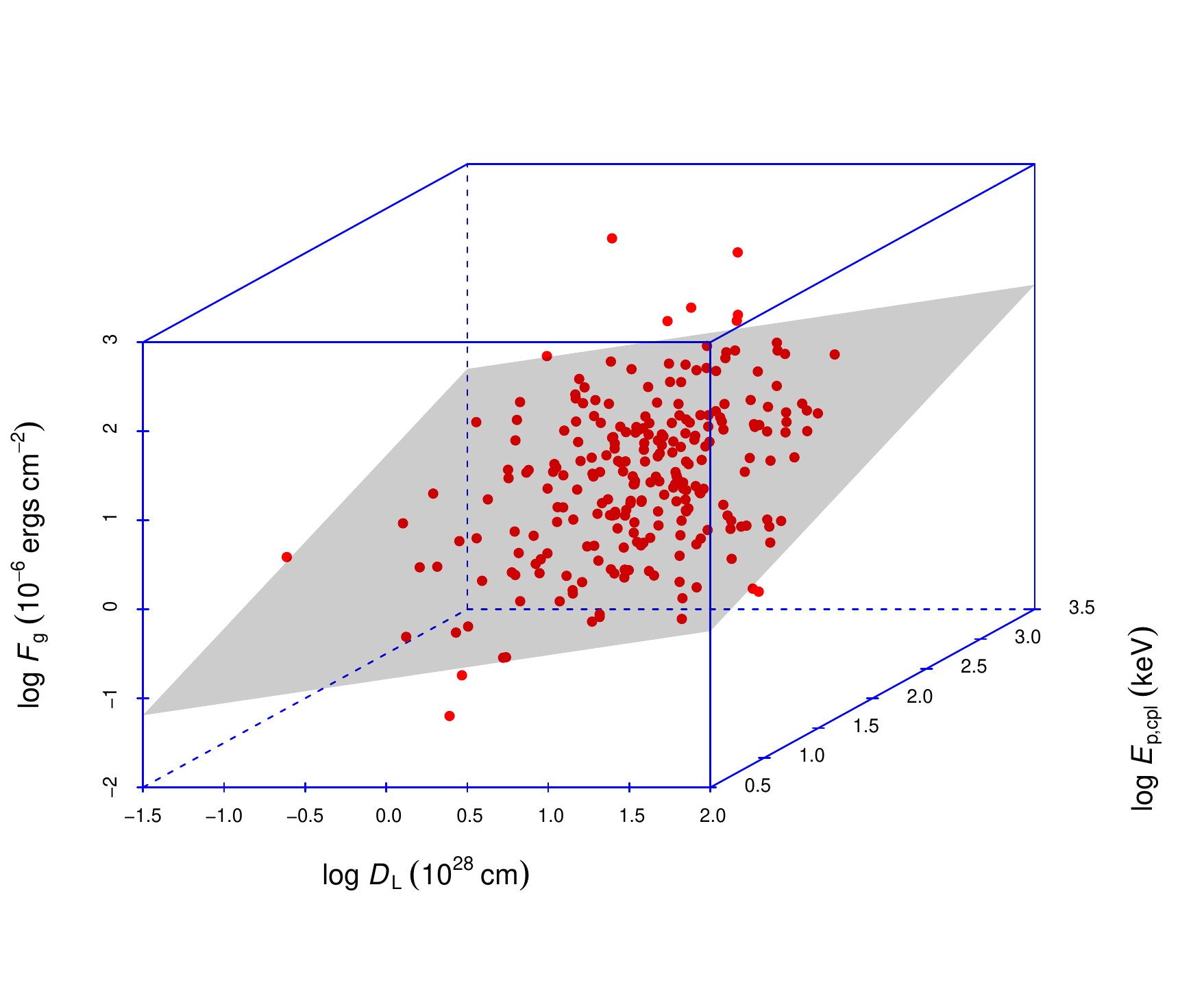}
\includegraphics[width=0.45\textwidth]{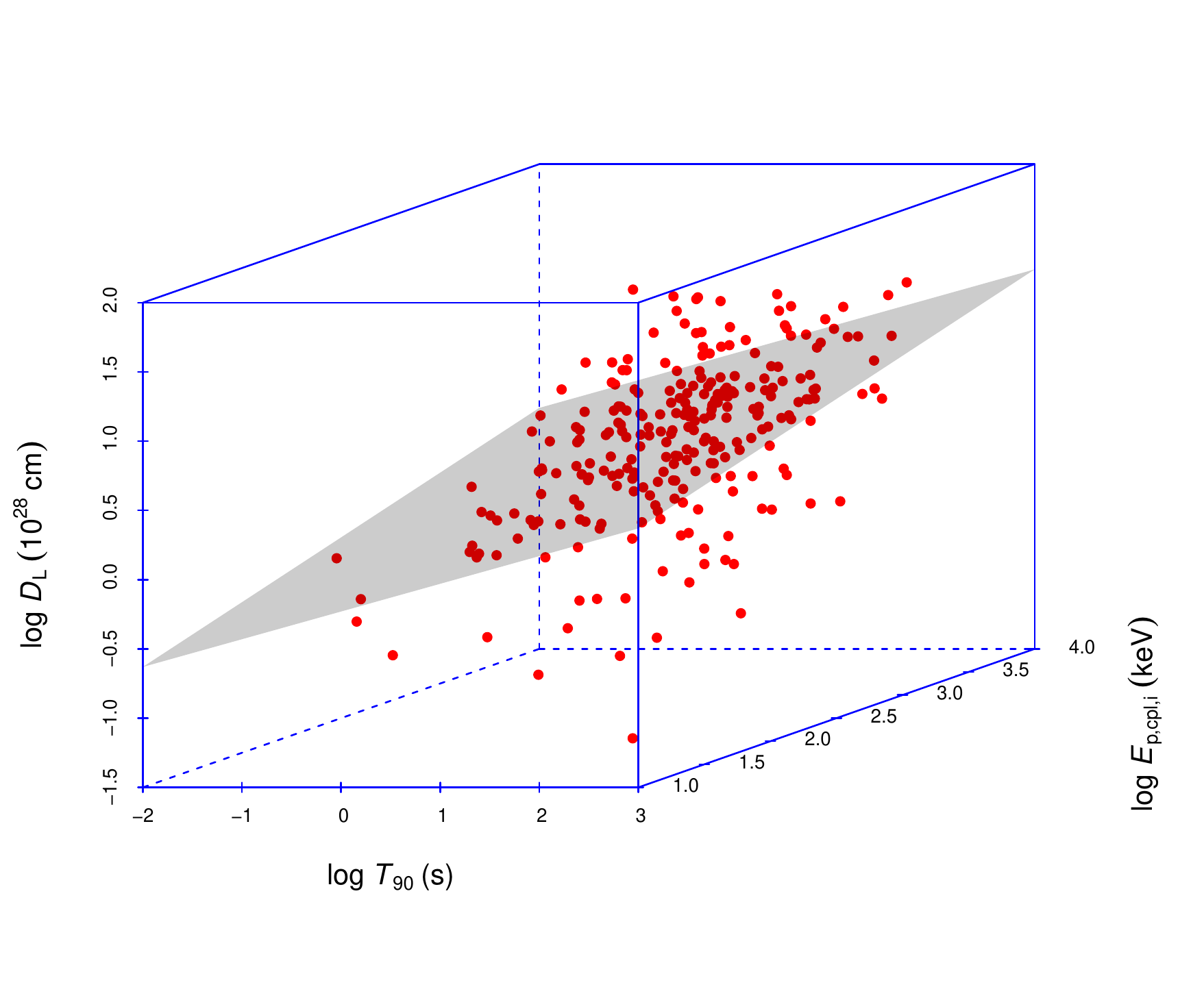}

\includegraphics[width=0.45\textwidth]{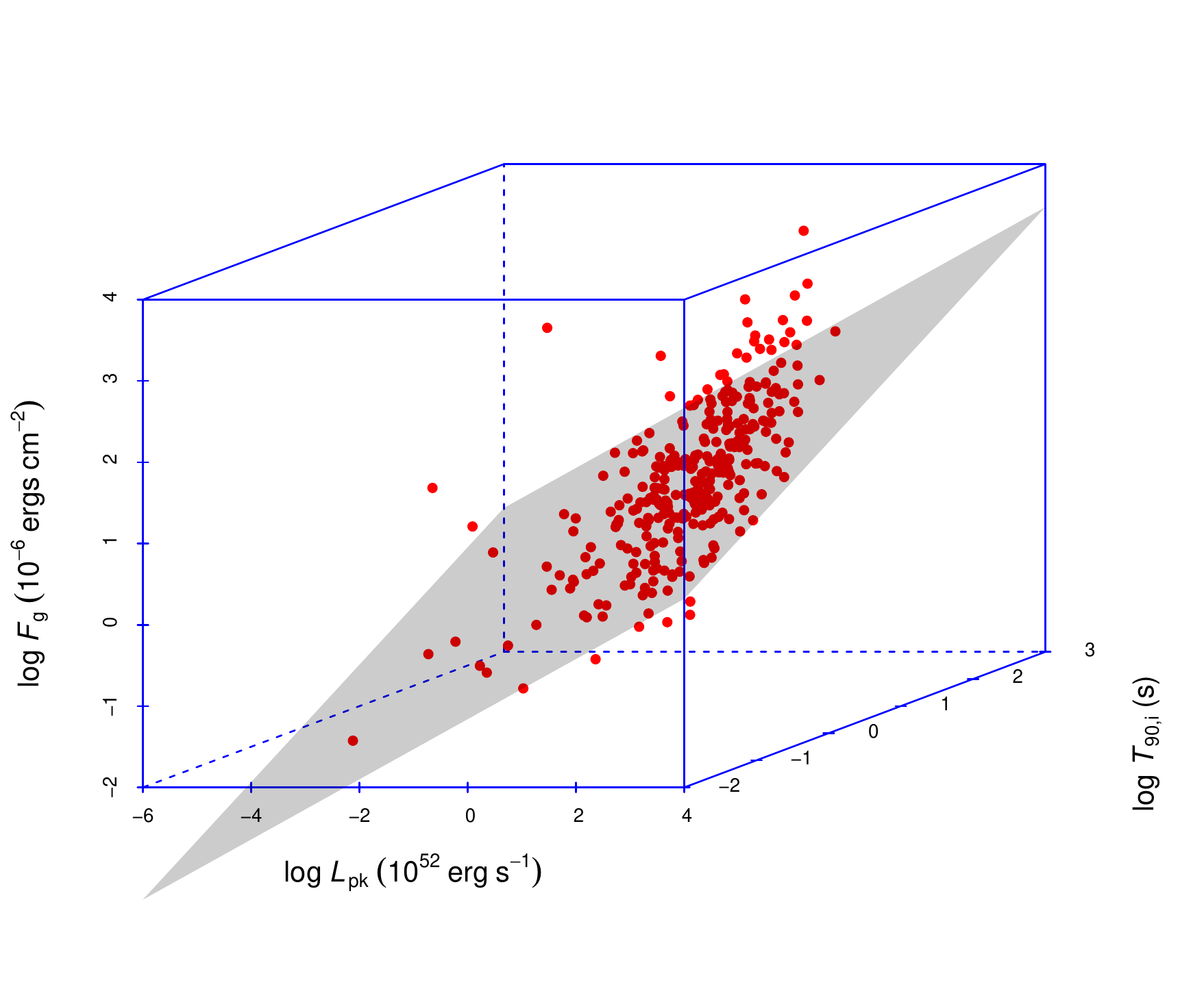}
\includegraphics[width=0.45\textwidth]{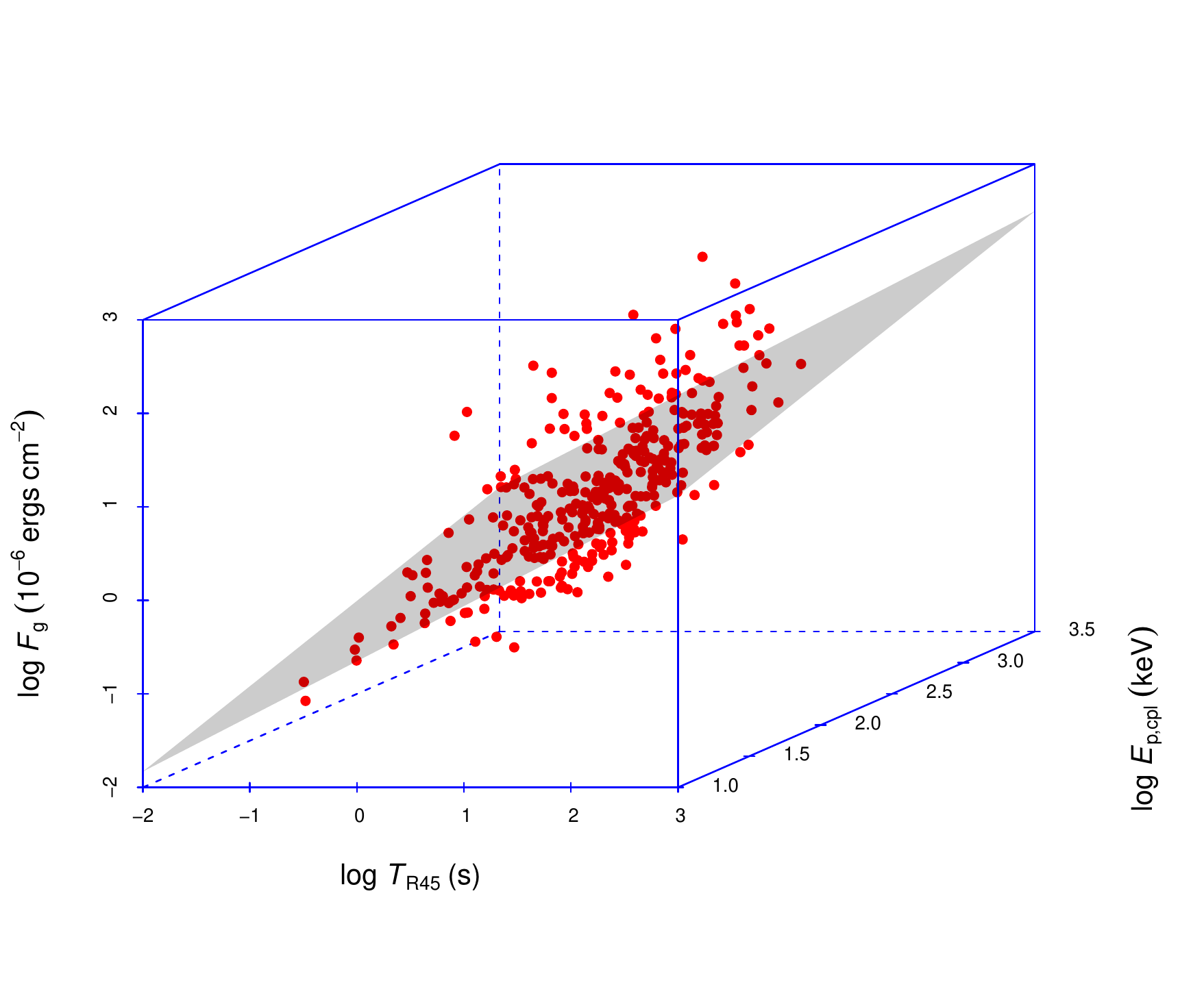}

\center{Fig. \ref{fig:three}---Continued}
\end{figure*}


\clearpage
\begin{figure*}

\includegraphics[width=0.45\textwidth]{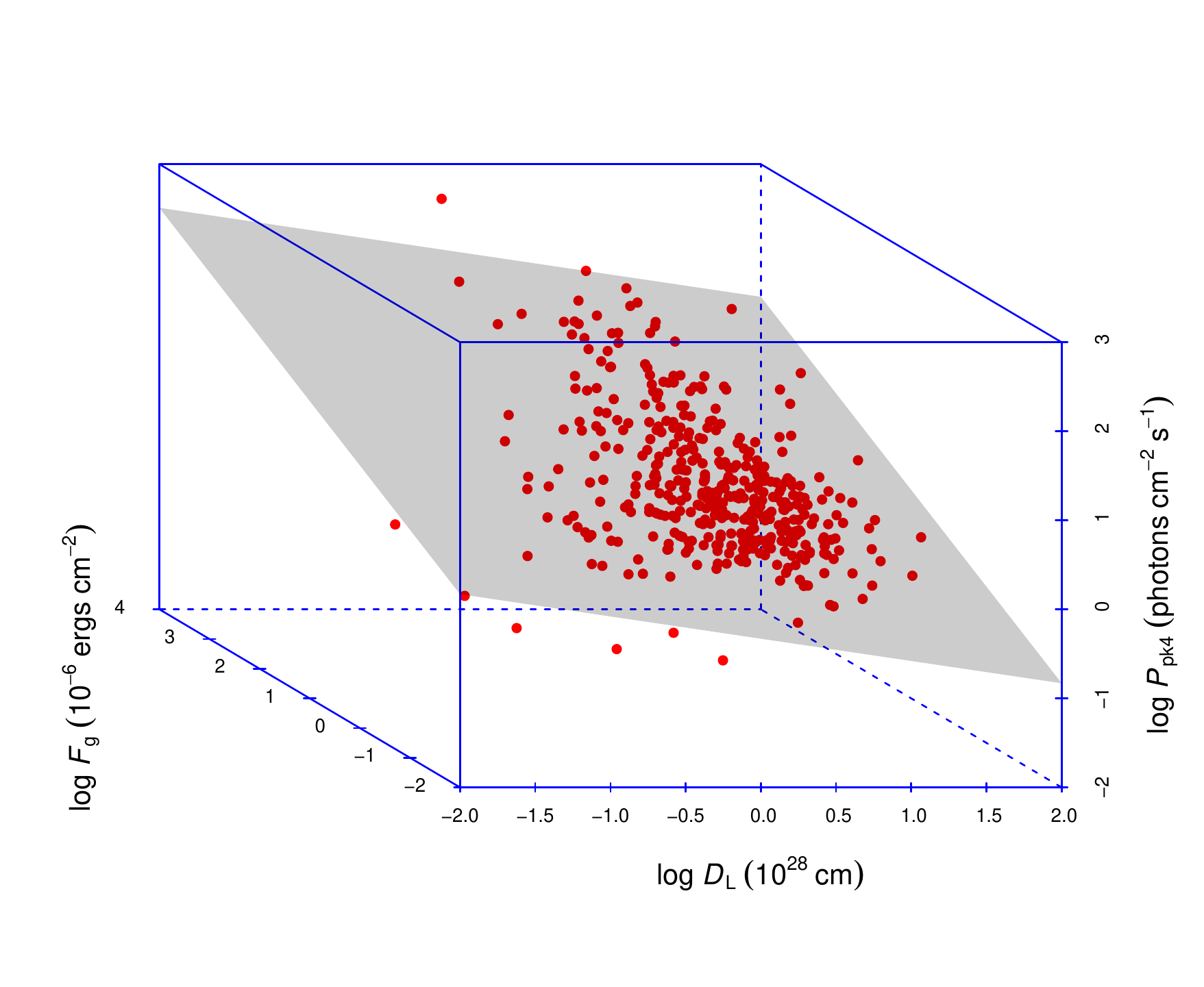}
\includegraphics[width=0.45\textwidth]{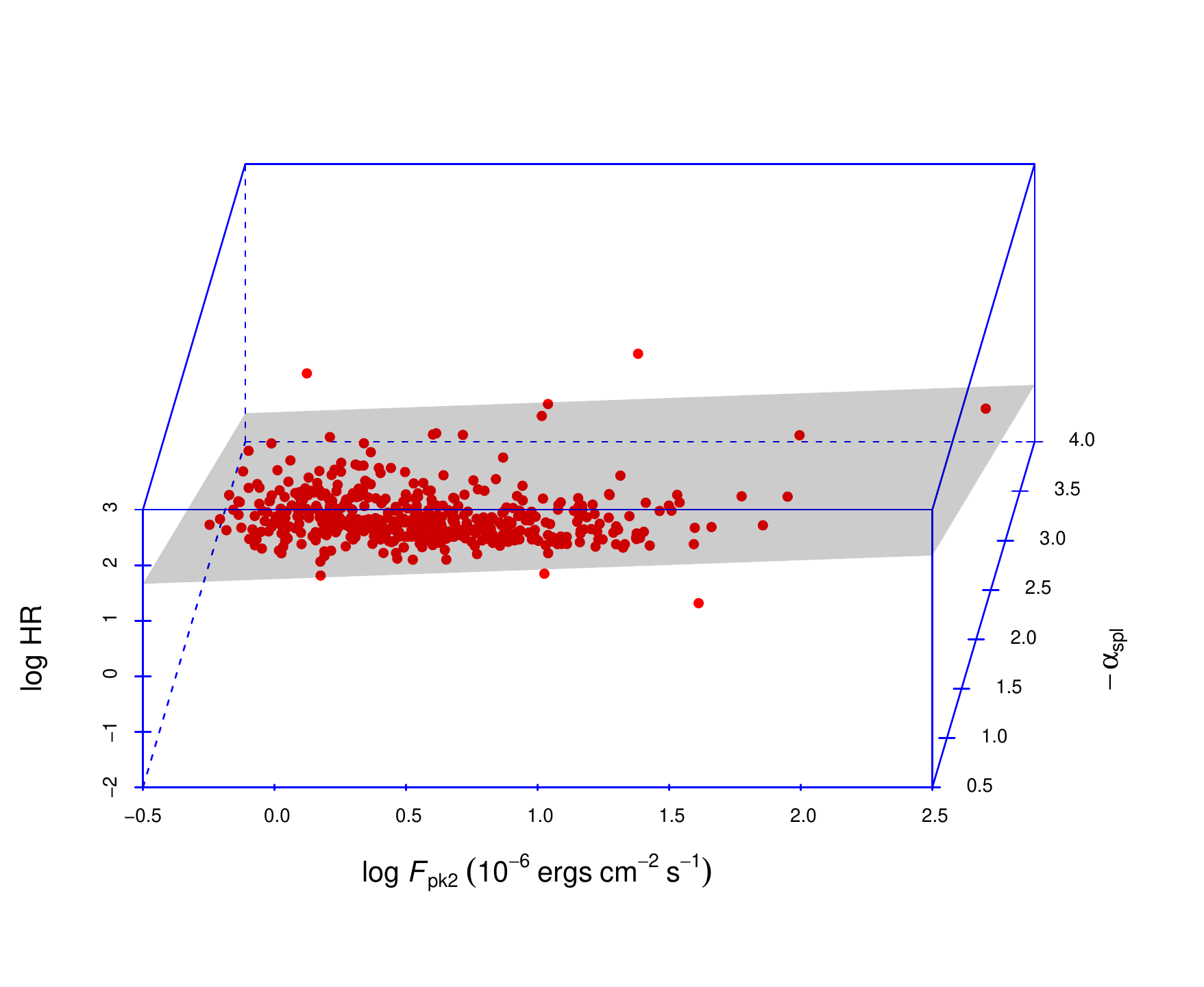}

\includegraphics[width=0.45\textwidth]{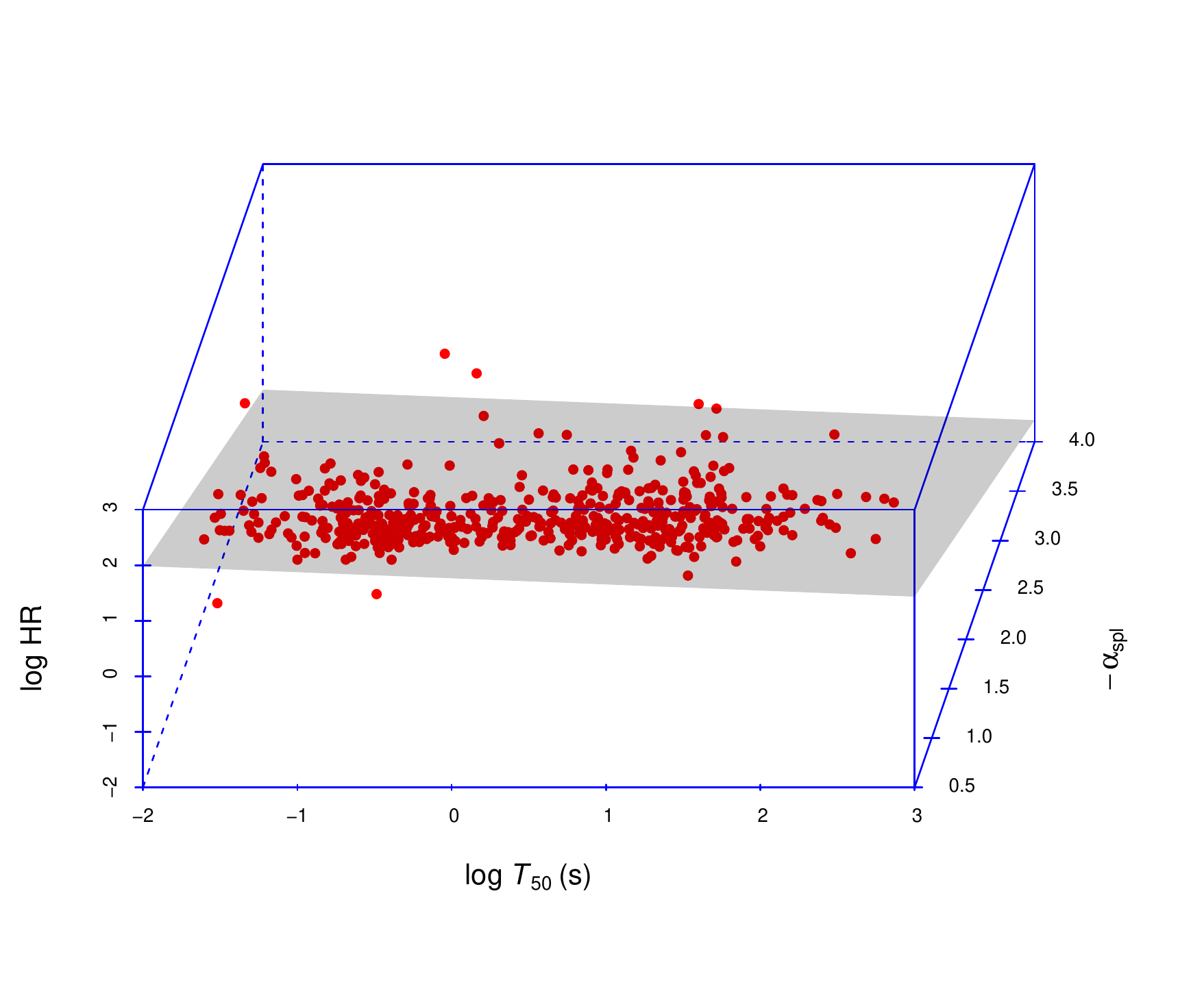}
\includegraphics[width=0.45\textwidth]{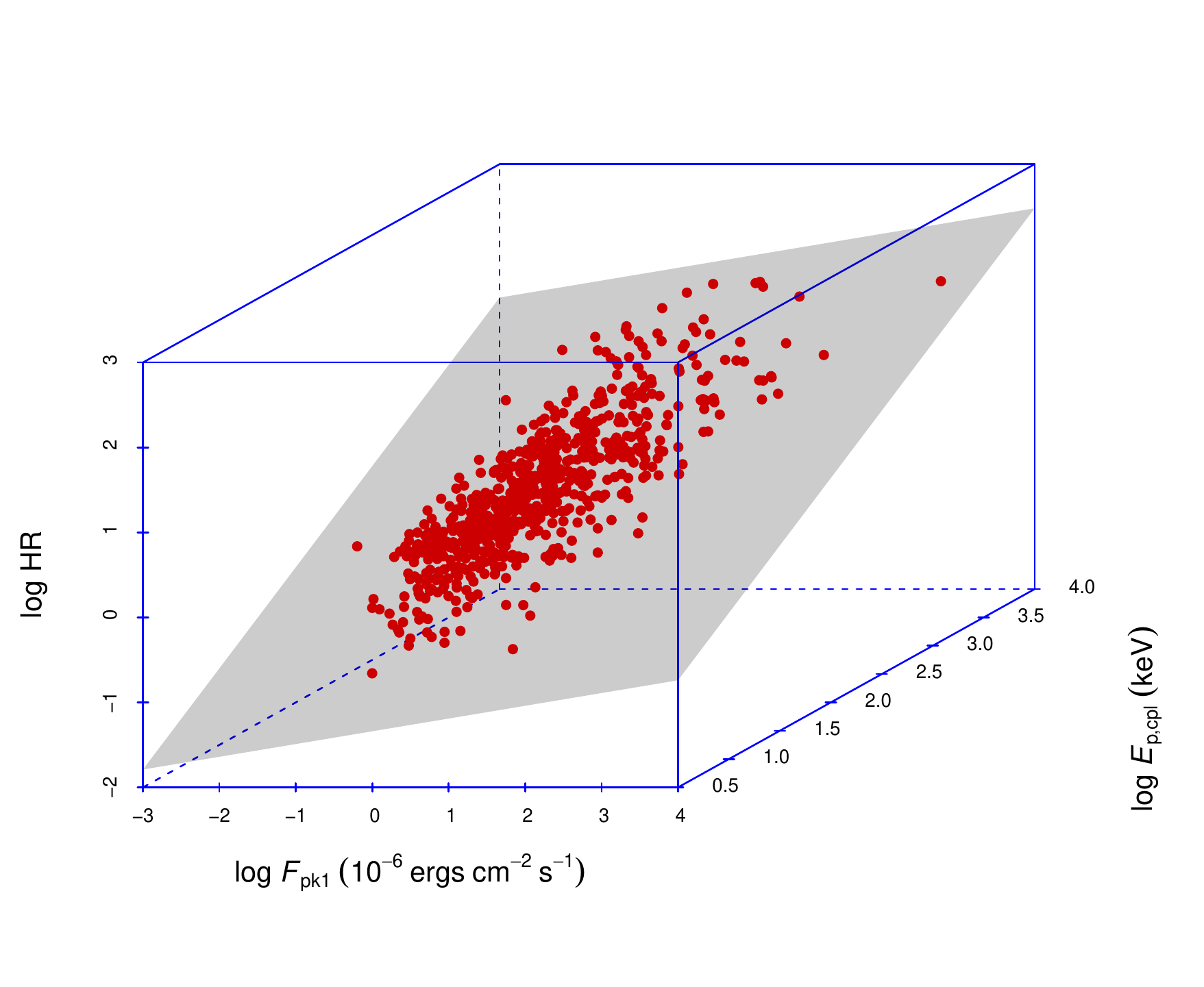}

\includegraphics[width=0.45\textwidth]{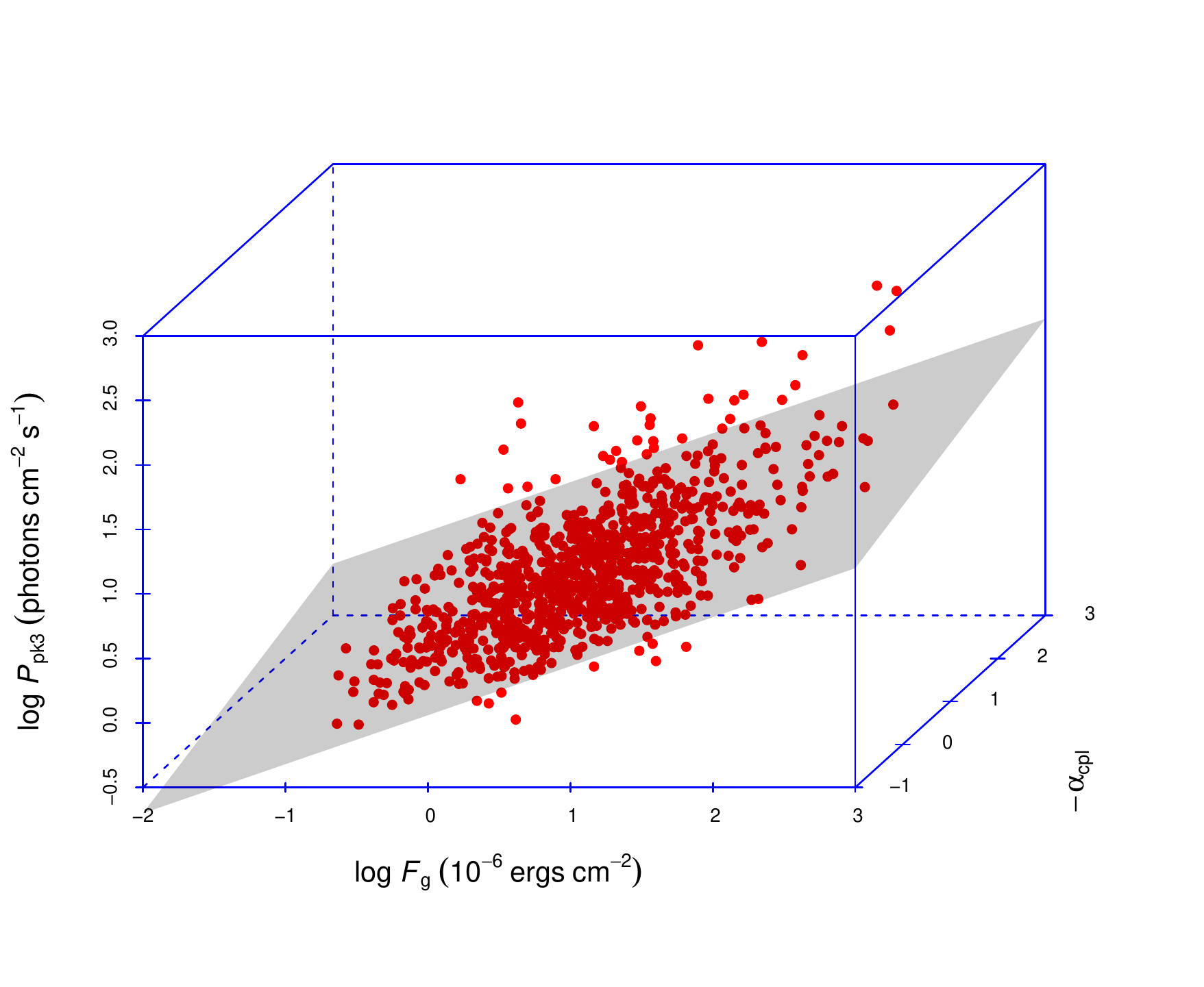}
\includegraphics[width=0.45\textwidth]{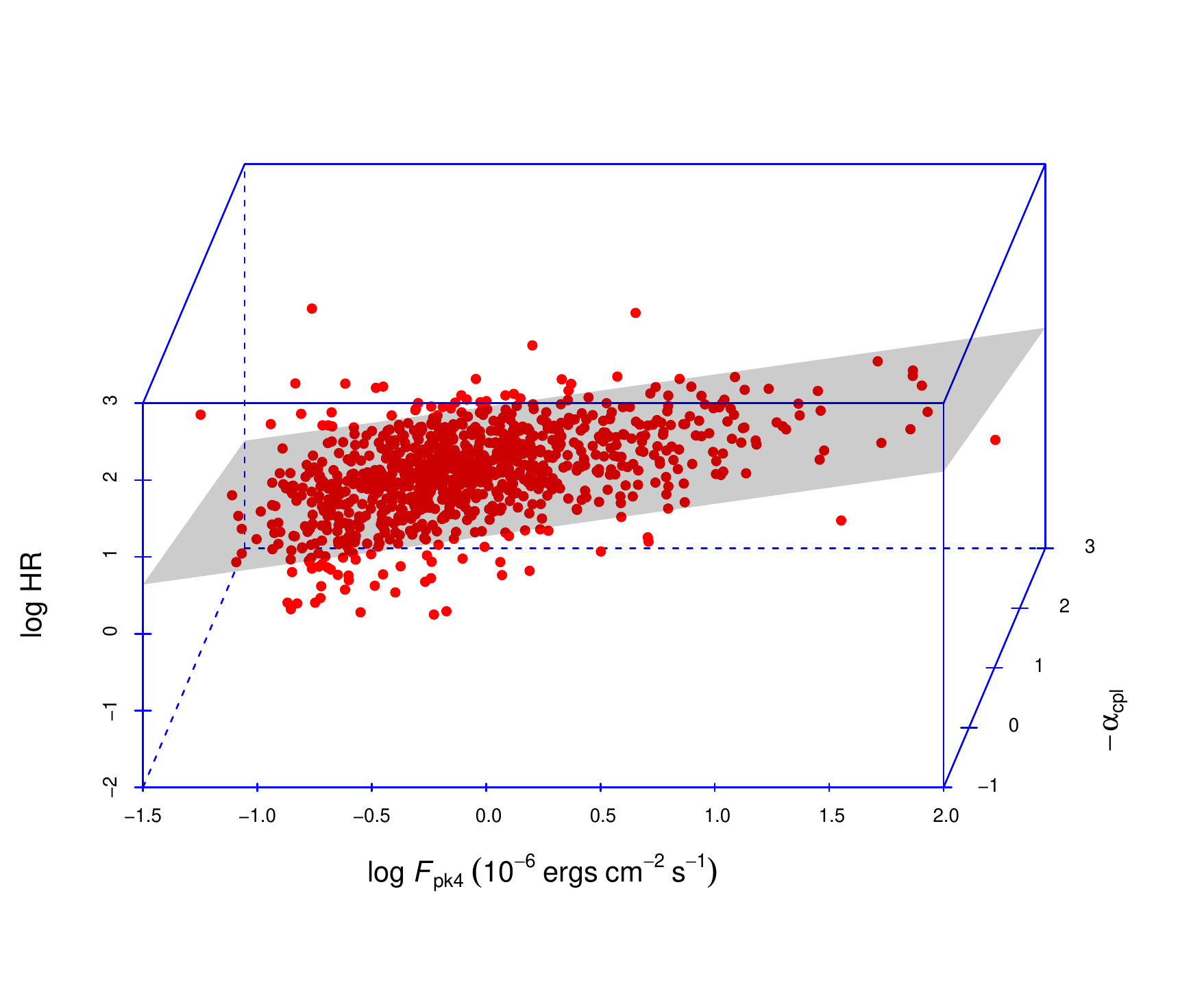}

\center{Fig. \ref{fig:three}---Continued}
\end{figure*}


\clearpage
\begin{figure*}

\includegraphics[width=0.45\textwidth]{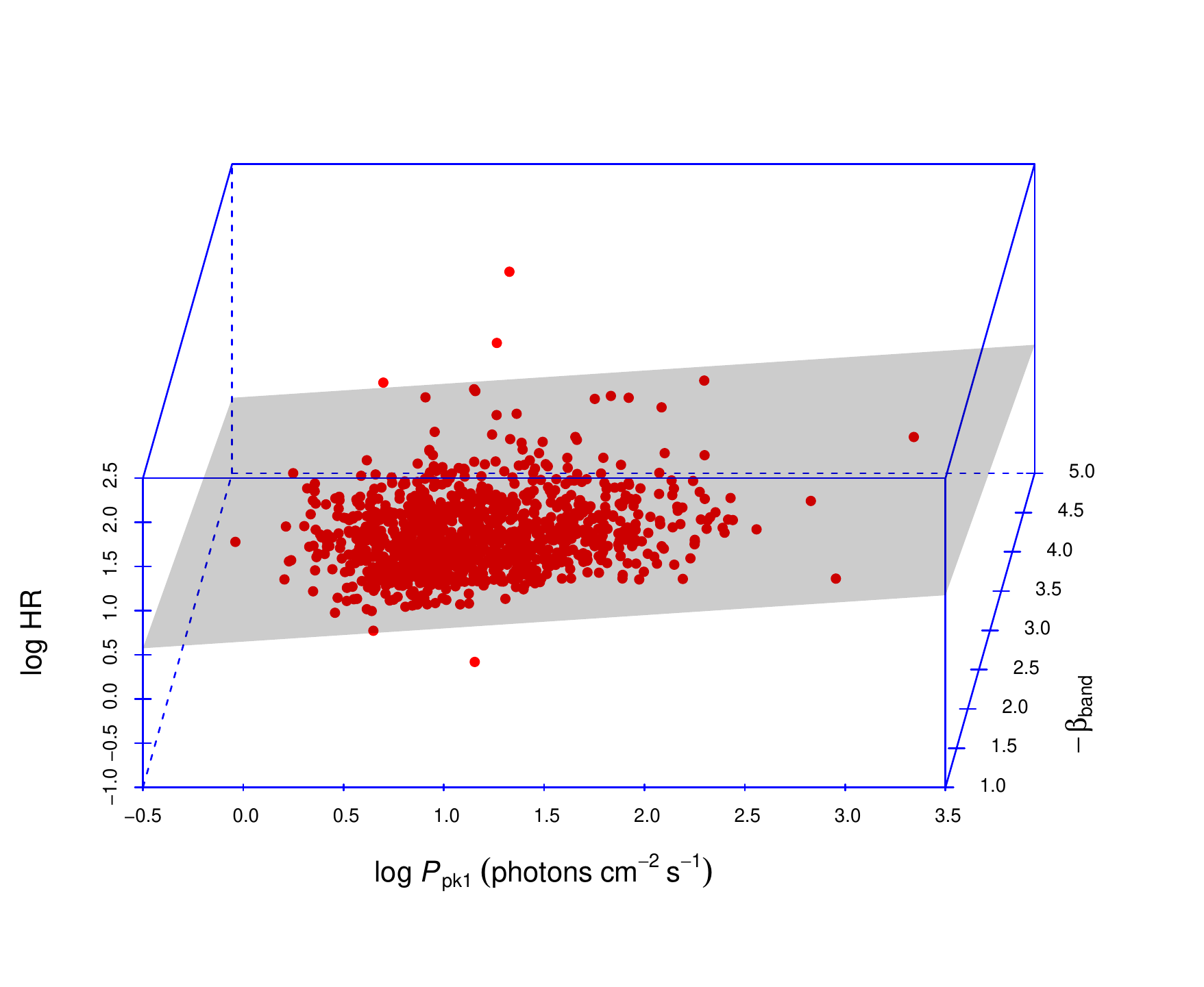}
\includegraphics[width=0.45\textwidth]{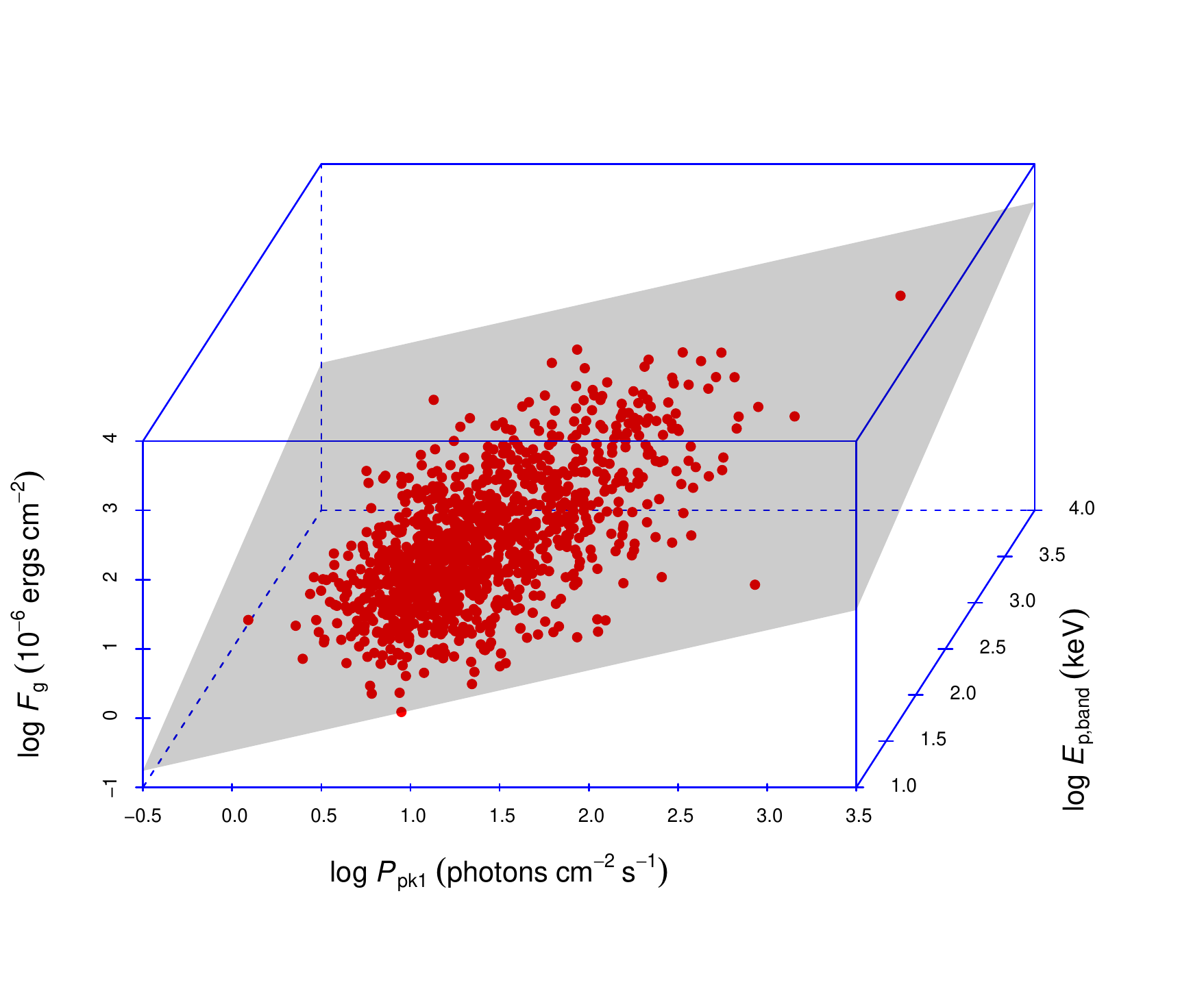}

\includegraphics[width=0.45\textwidth]{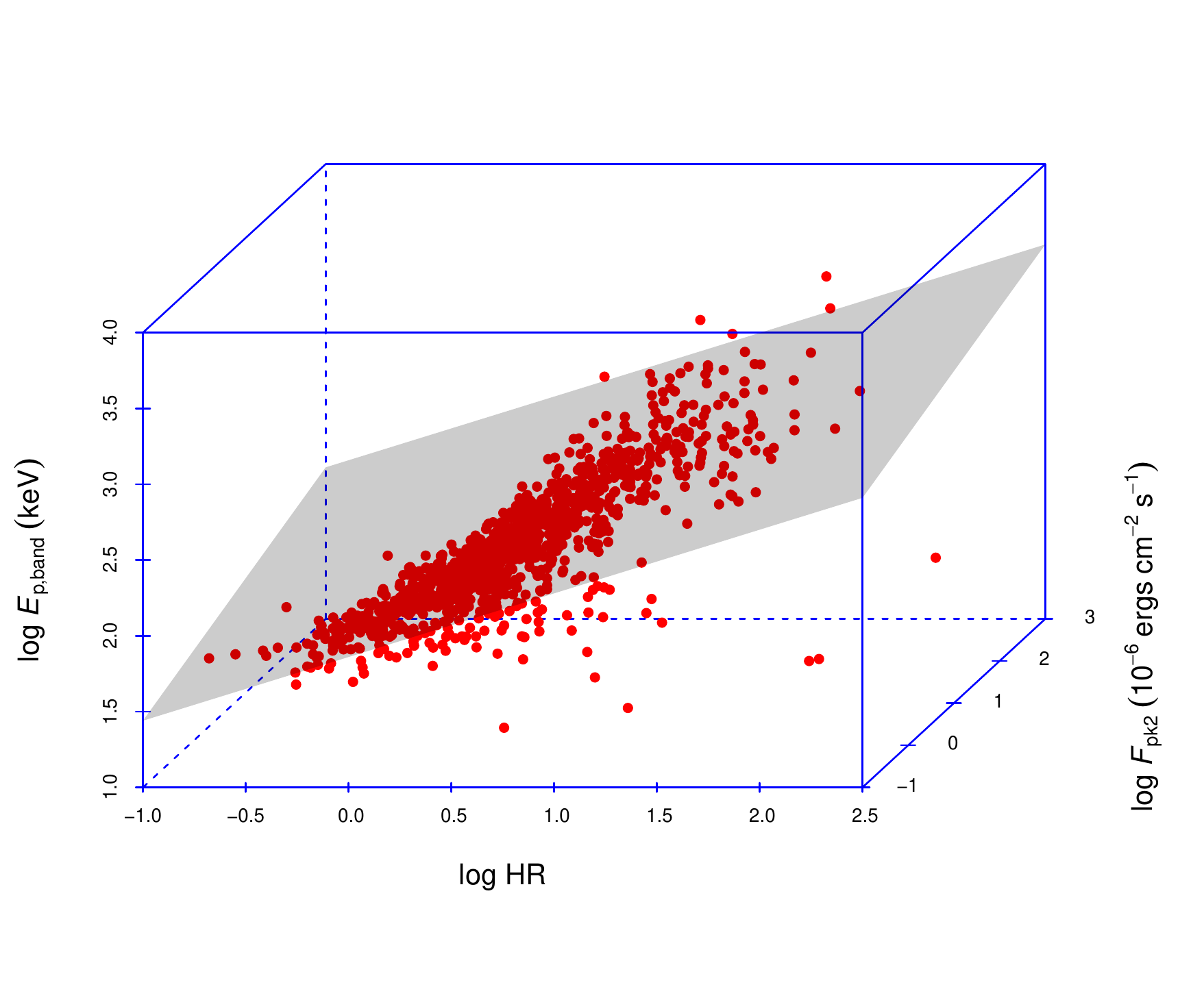}
\includegraphics[width=0.45\textwidth]{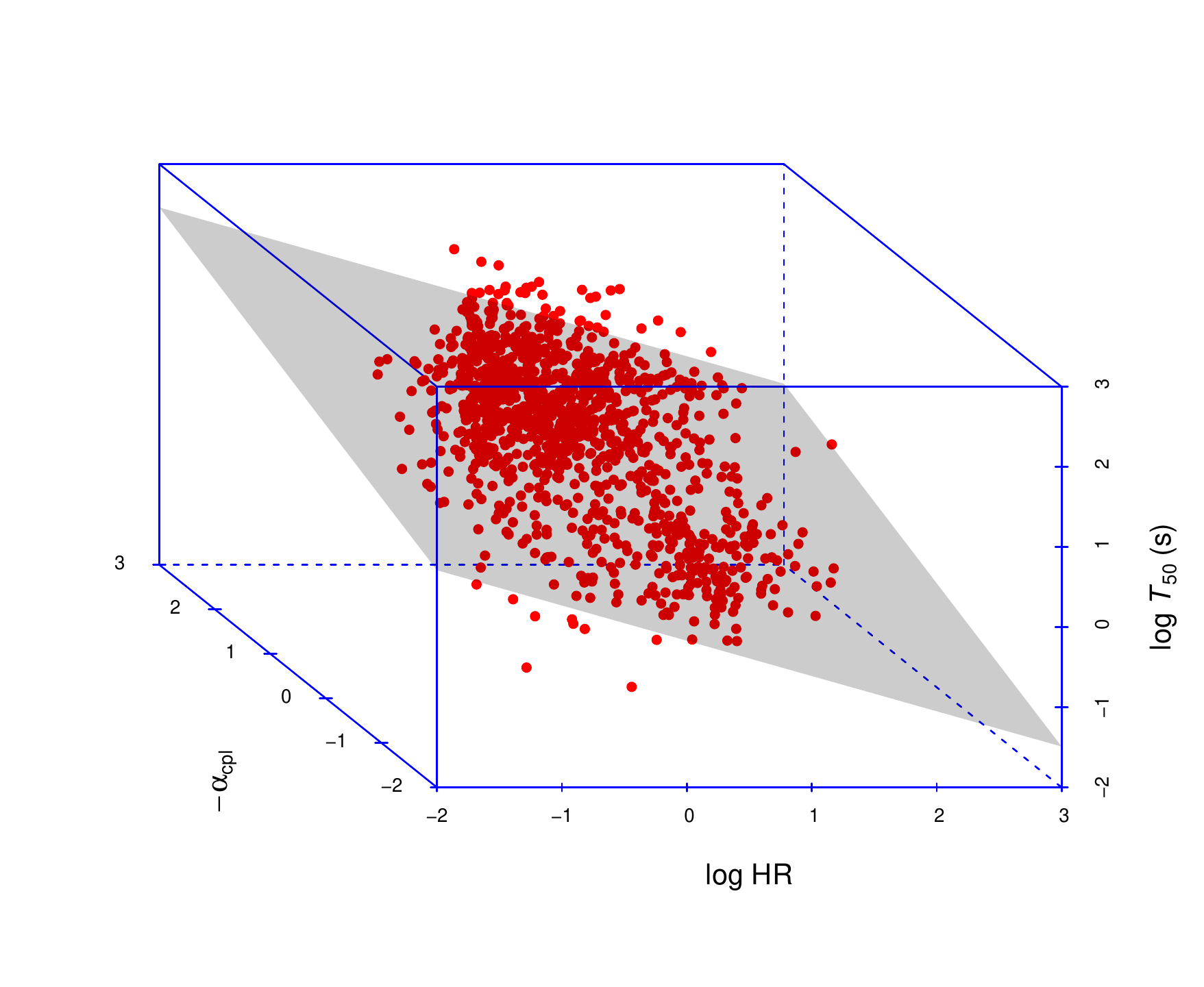}

\includegraphics[width=0.45\textwidth]{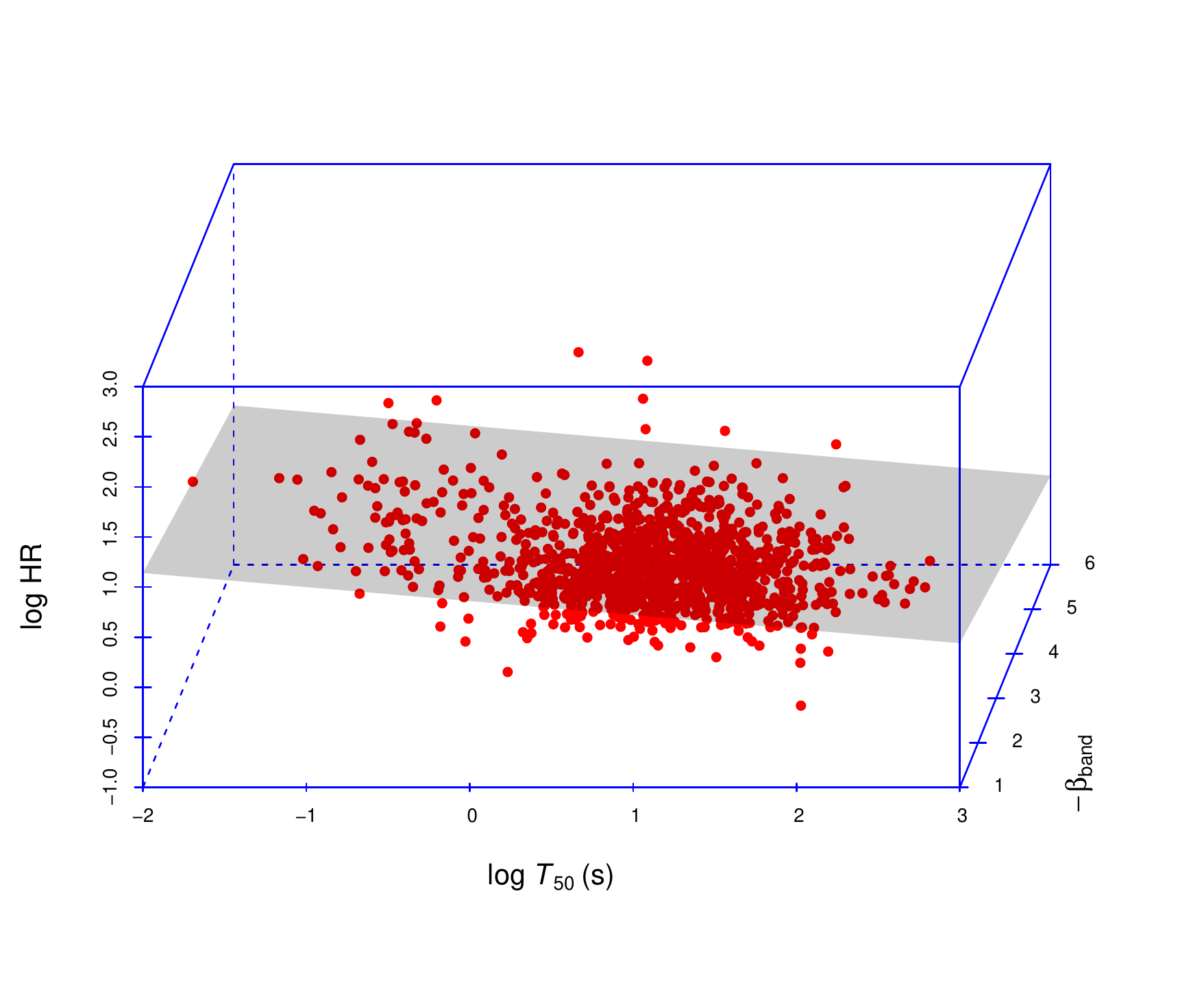}
\includegraphics[width=0.45\textwidth]{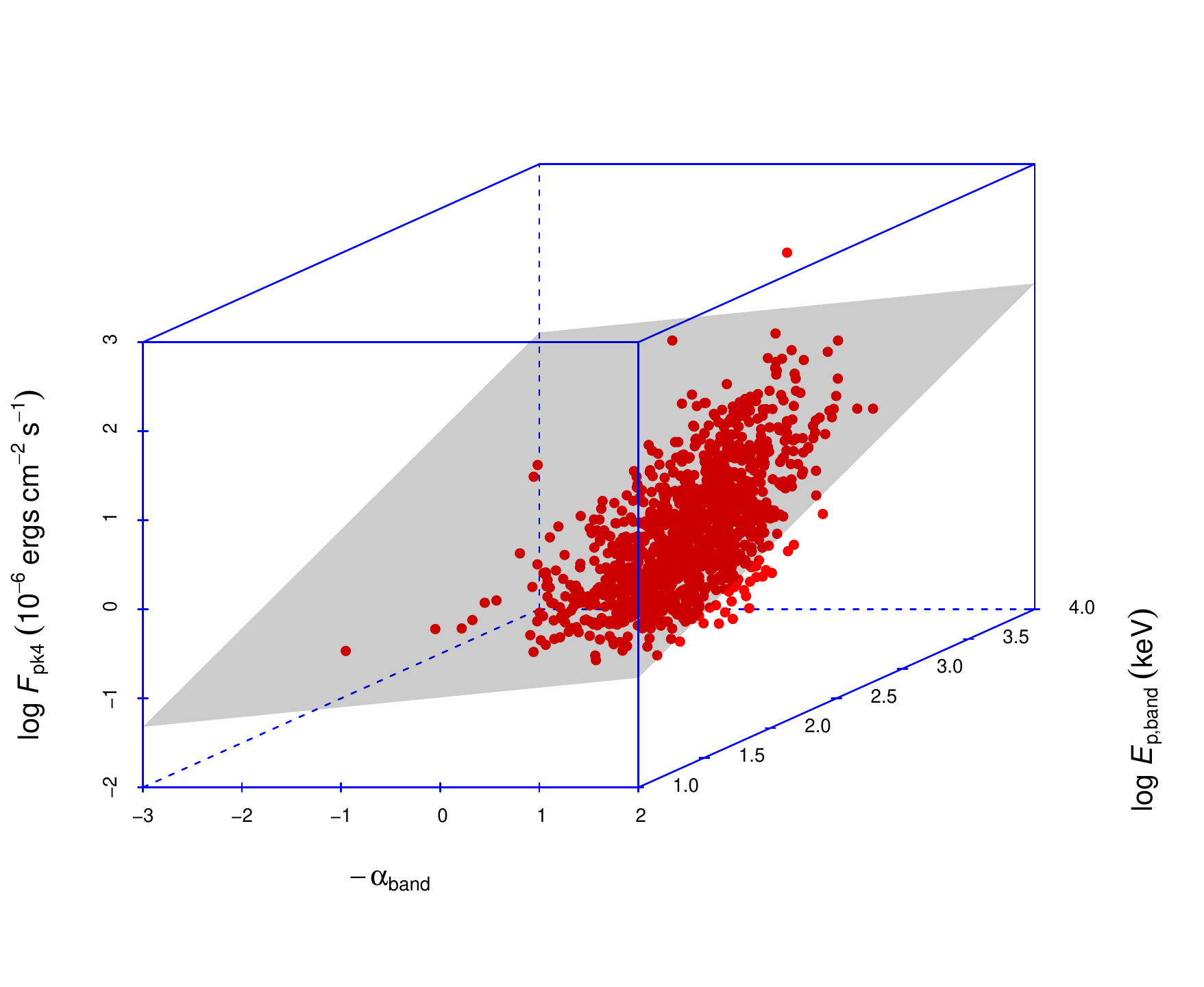}

\center{Fig. \ref{fig:three}---Continued}
\end{figure*}


\clearpage
\begin{figure*}

\includegraphics[width=0.45\textwidth]{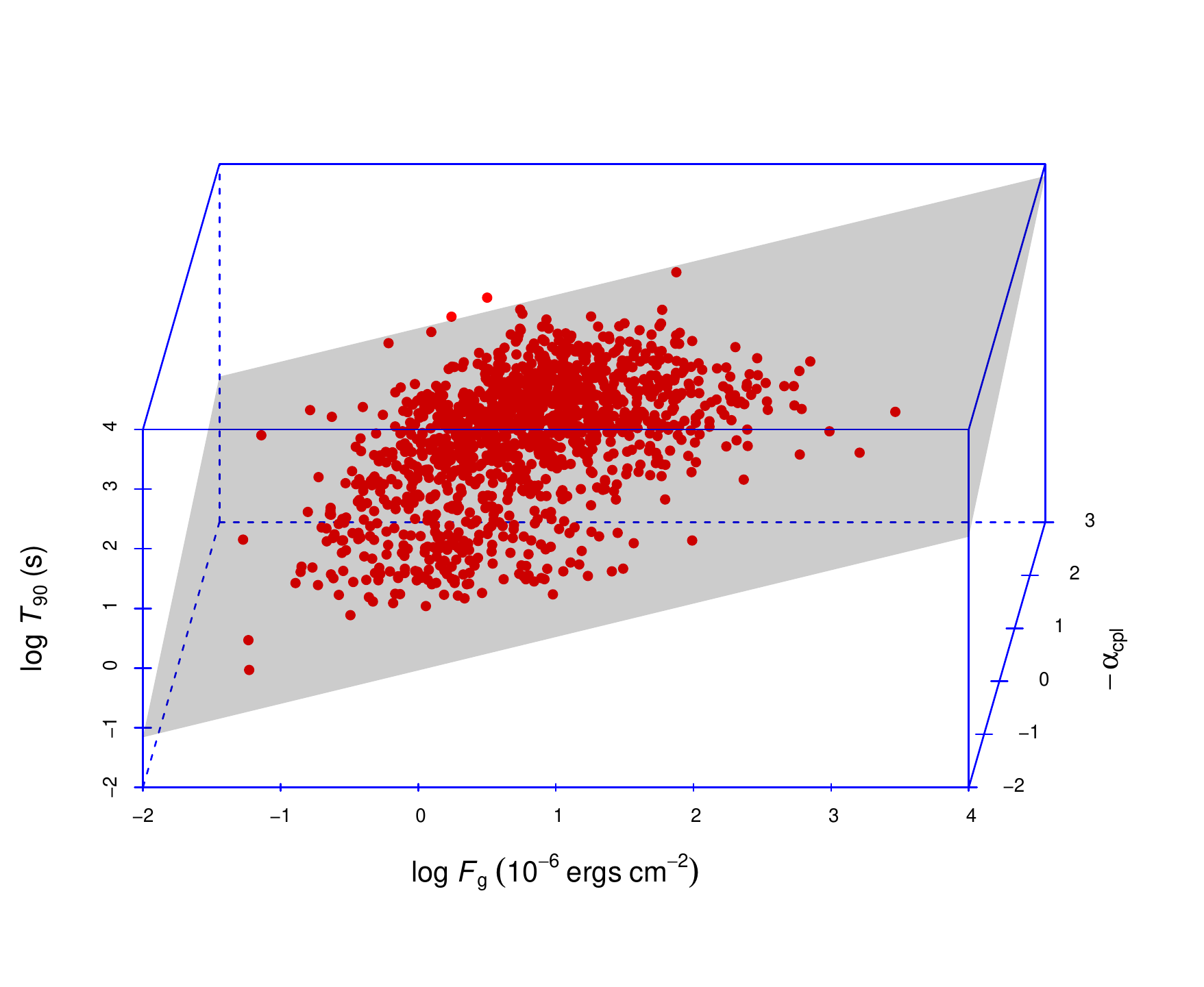}
\includegraphics[width=0.45\textwidth]{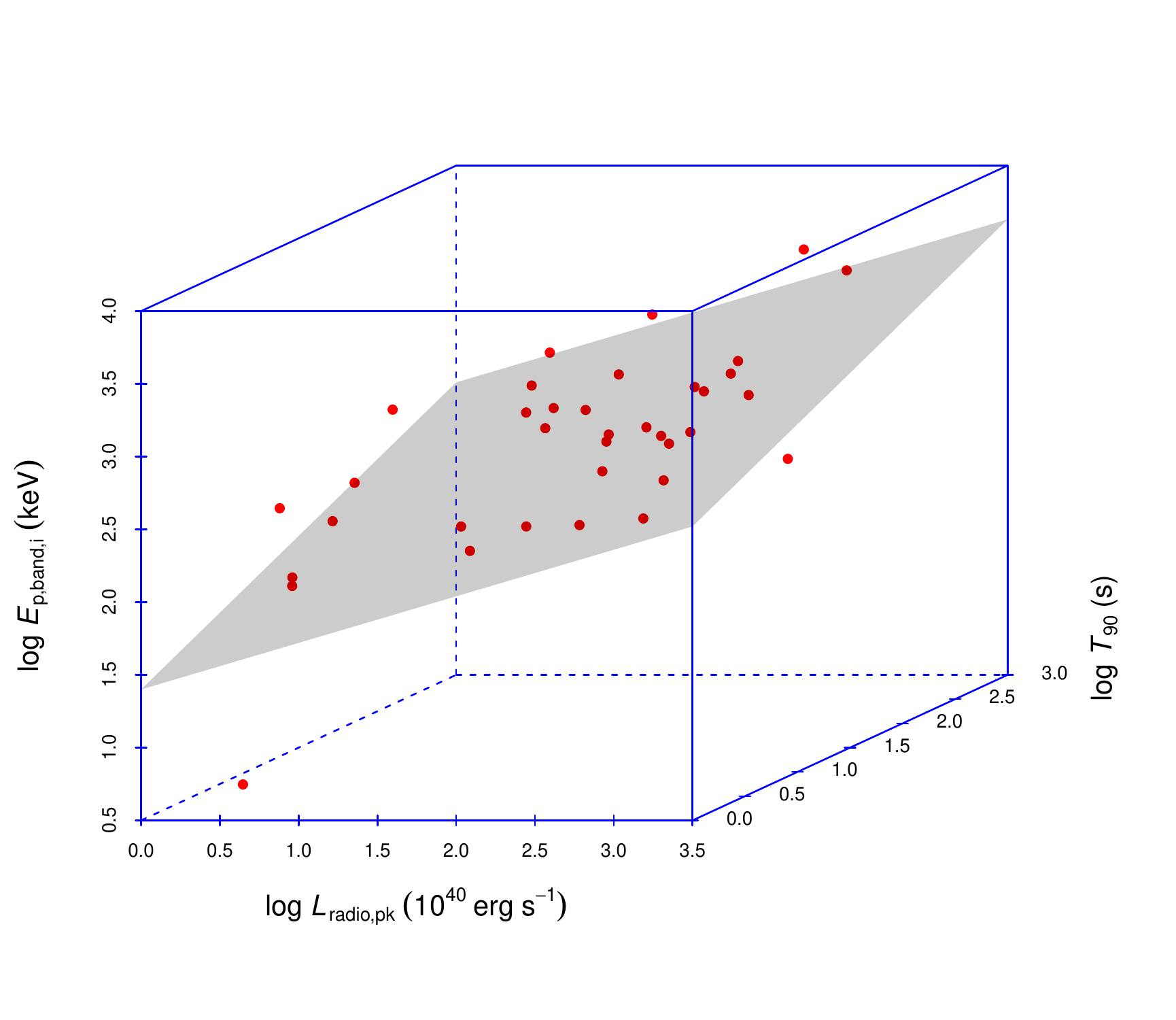}

\includegraphics[width=0.45\textwidth]{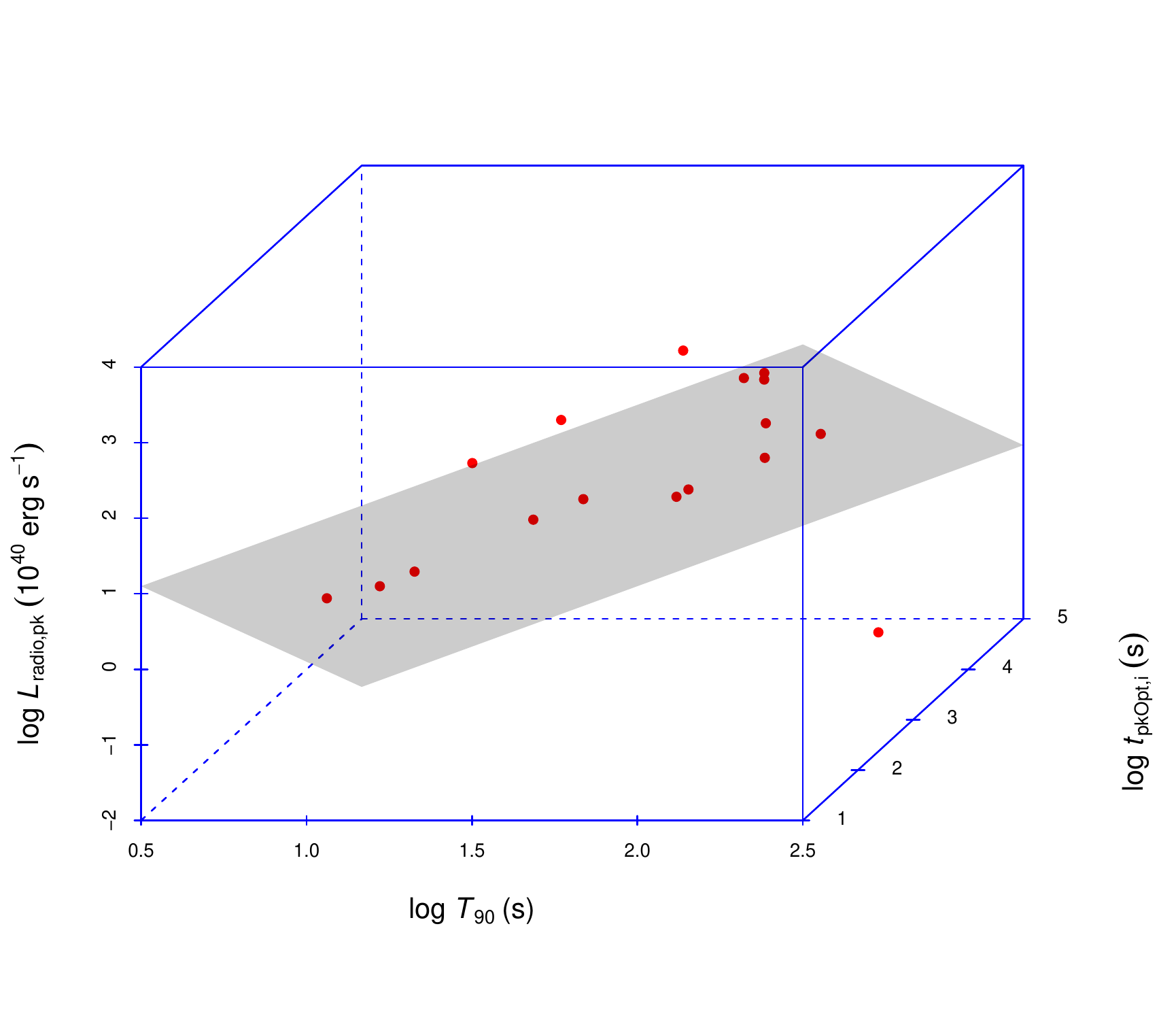}
\includegraphics[width=0.45\textwidth]{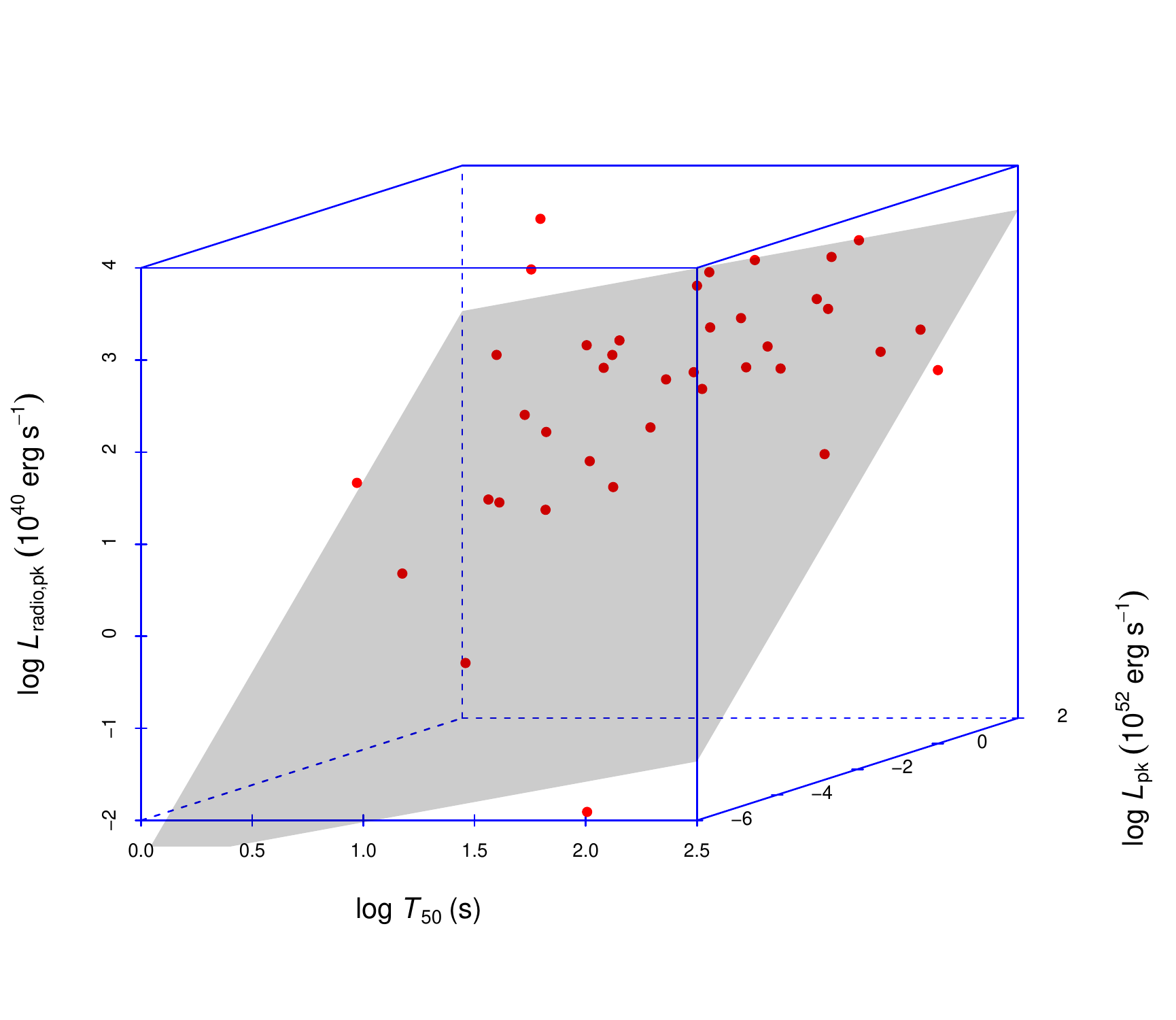}

\includegraphics[width=0.45\textwidth]{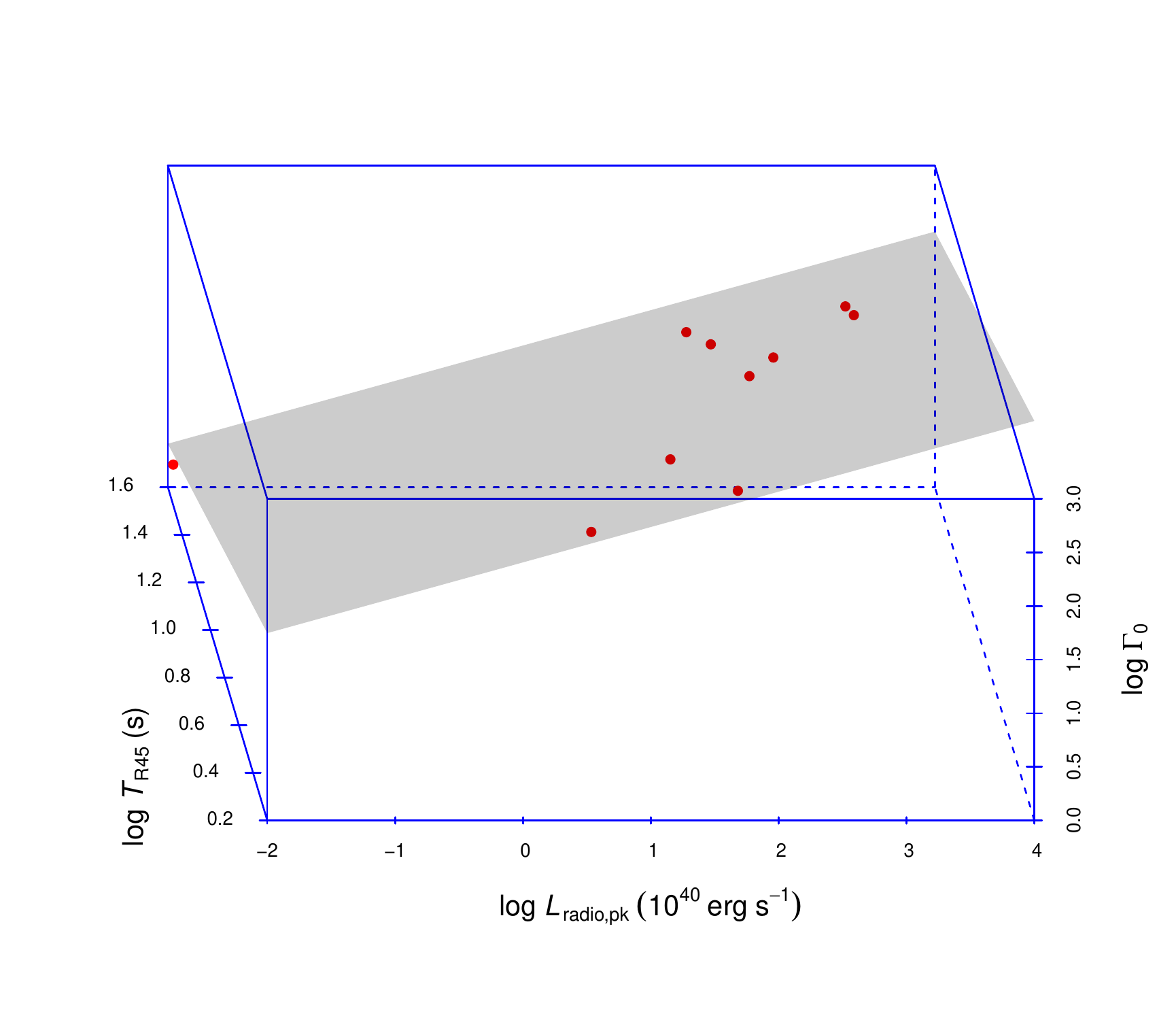}
\includegraphics[width=0.45\textwidth]{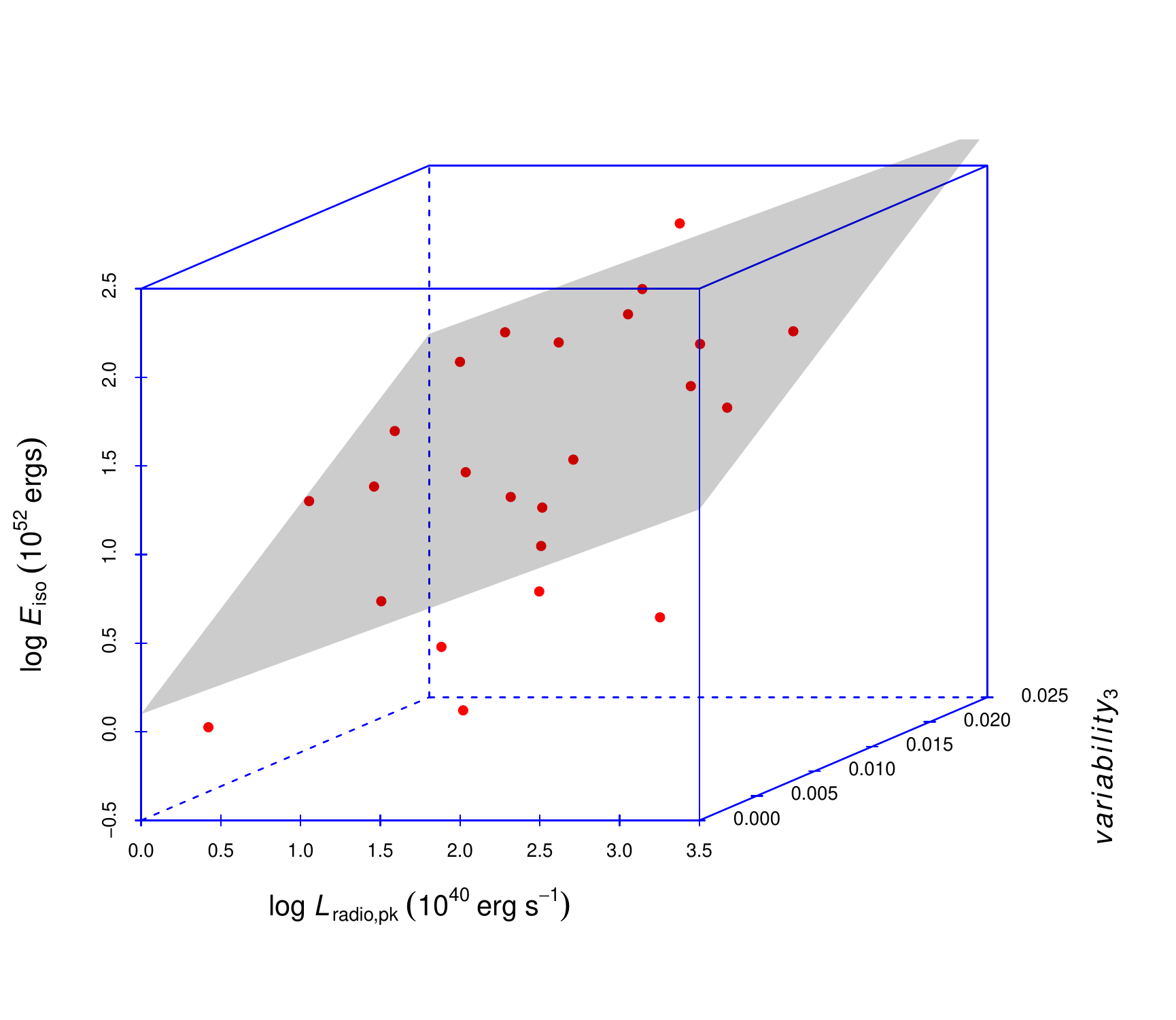}

\center{Fig. \ref{fig:three}---Continued}
\end{figure*}


\clearpage
\begin{figure*}

\includegraphics[width=0.45\textwidth]{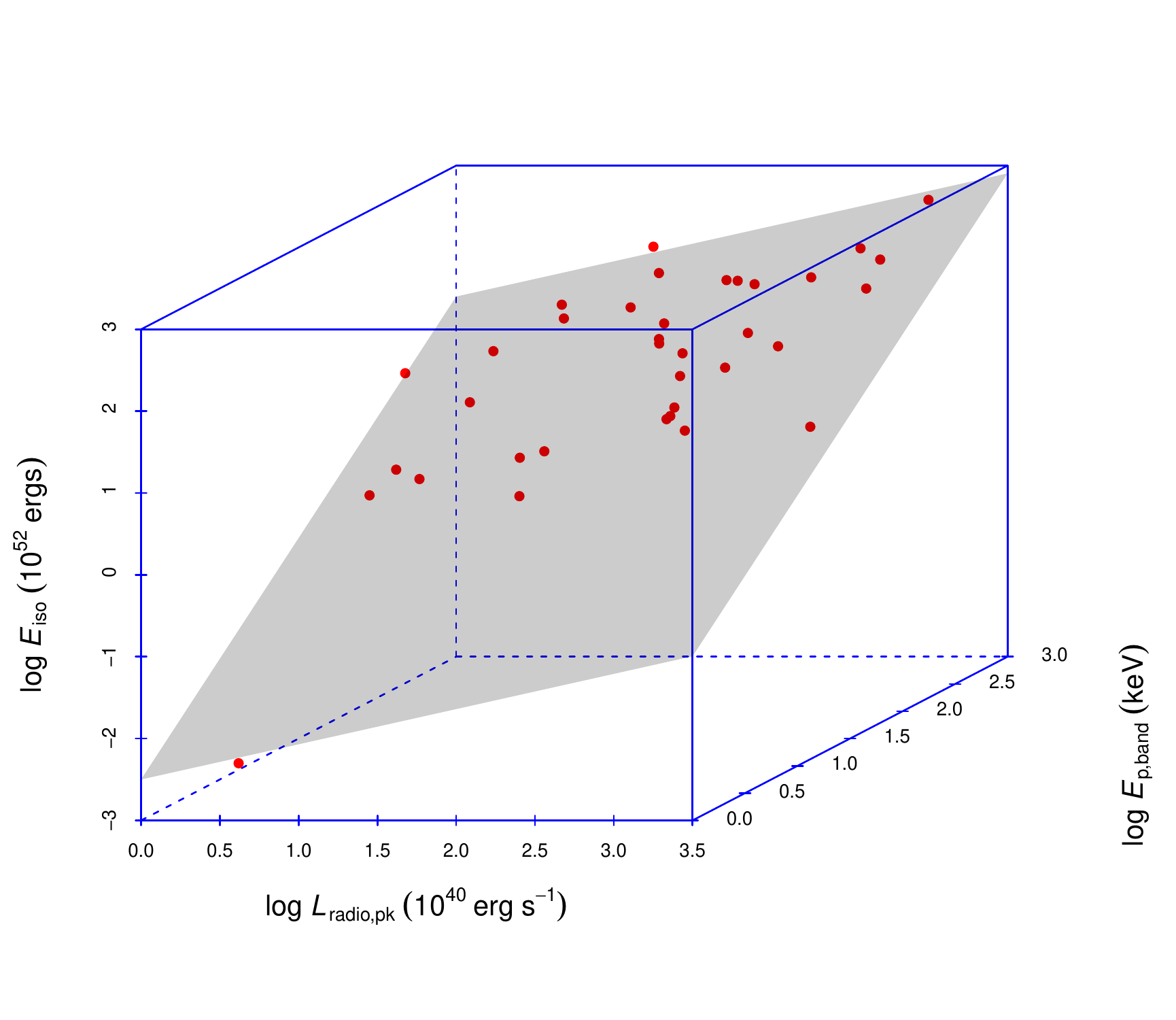}
\includegraphics[width=0.45\textwidth]{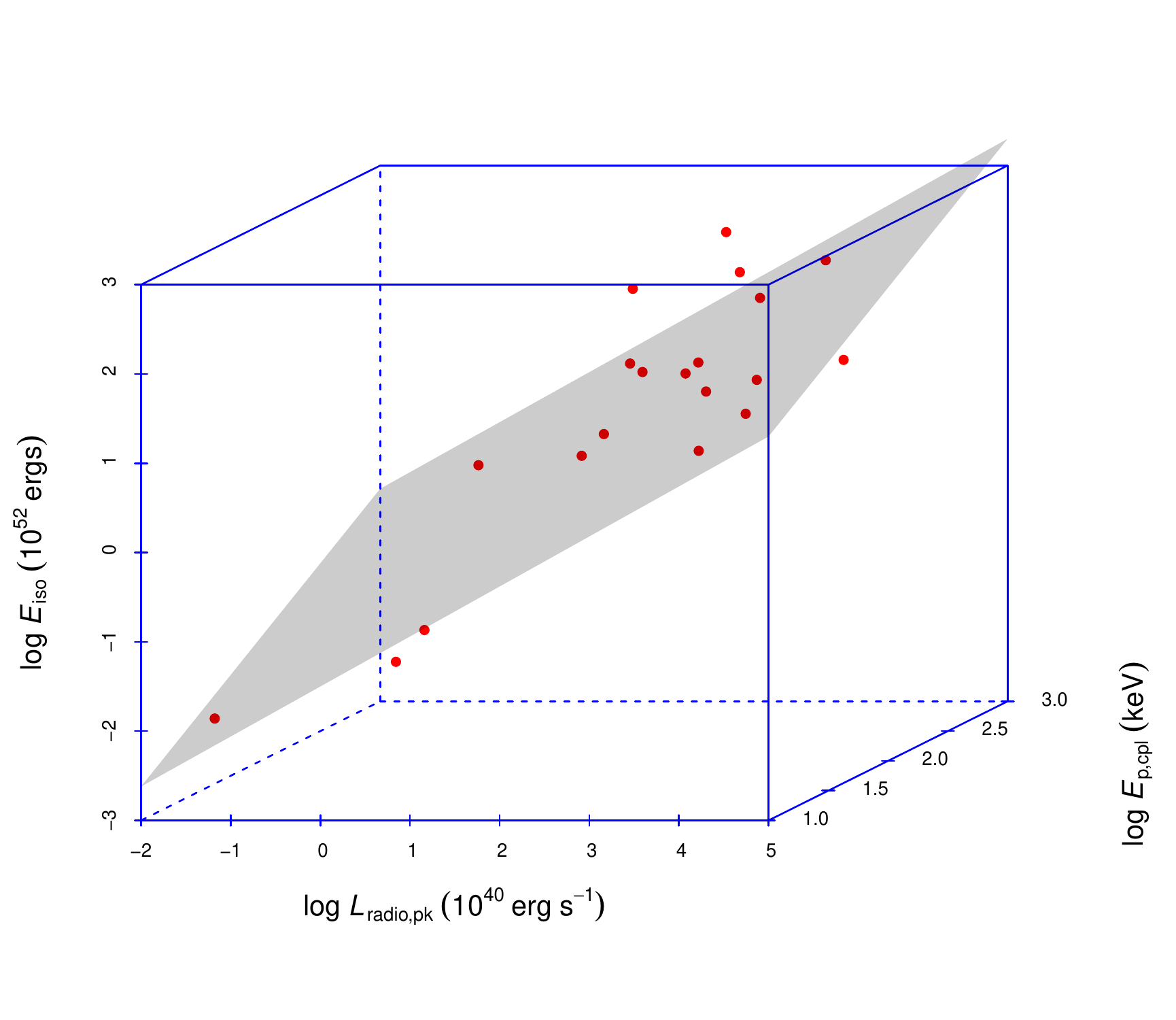}

\includegraphics[width=0.45\textwidth]{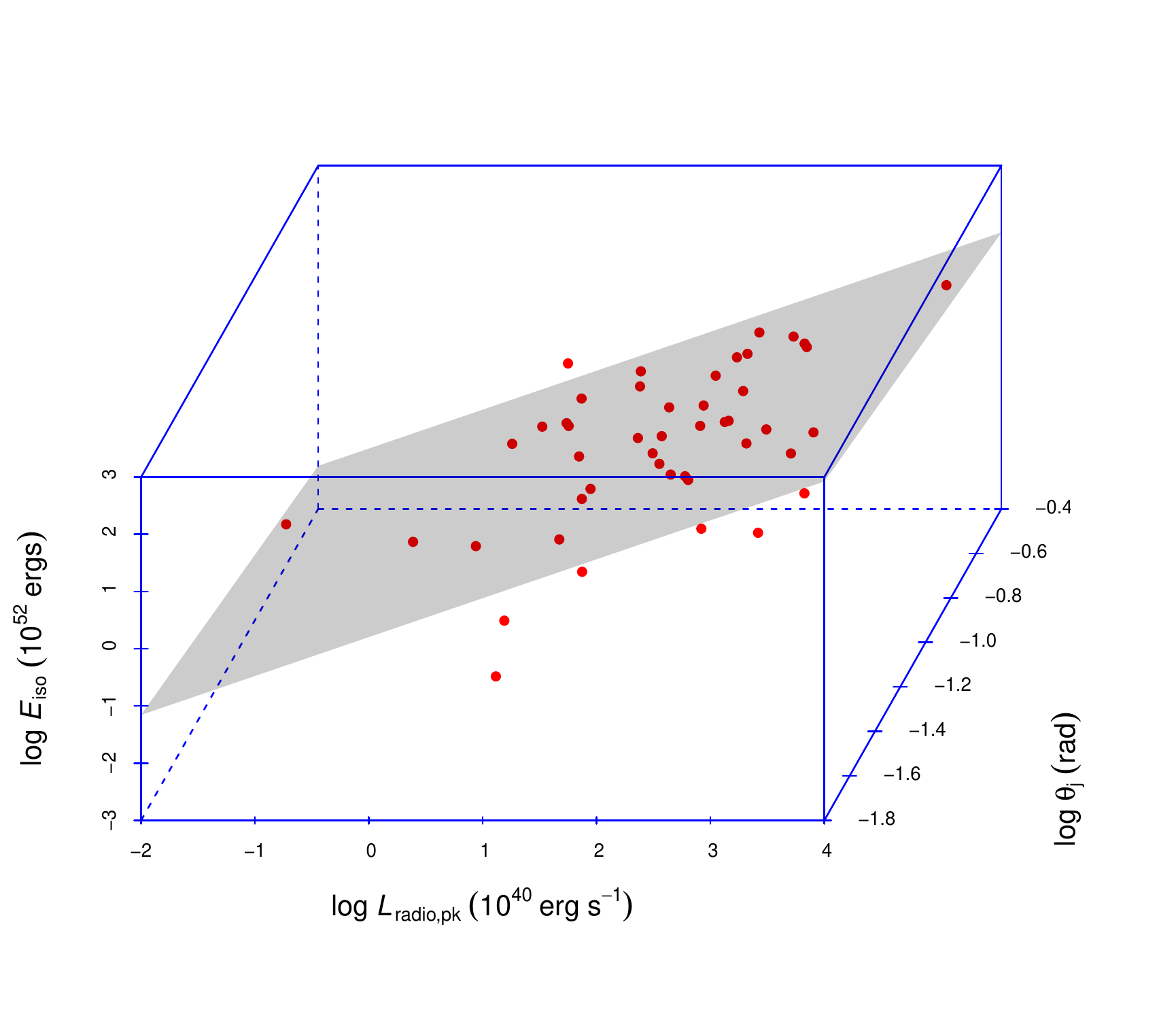}
\includegraphics[width=0.45\textwidth]{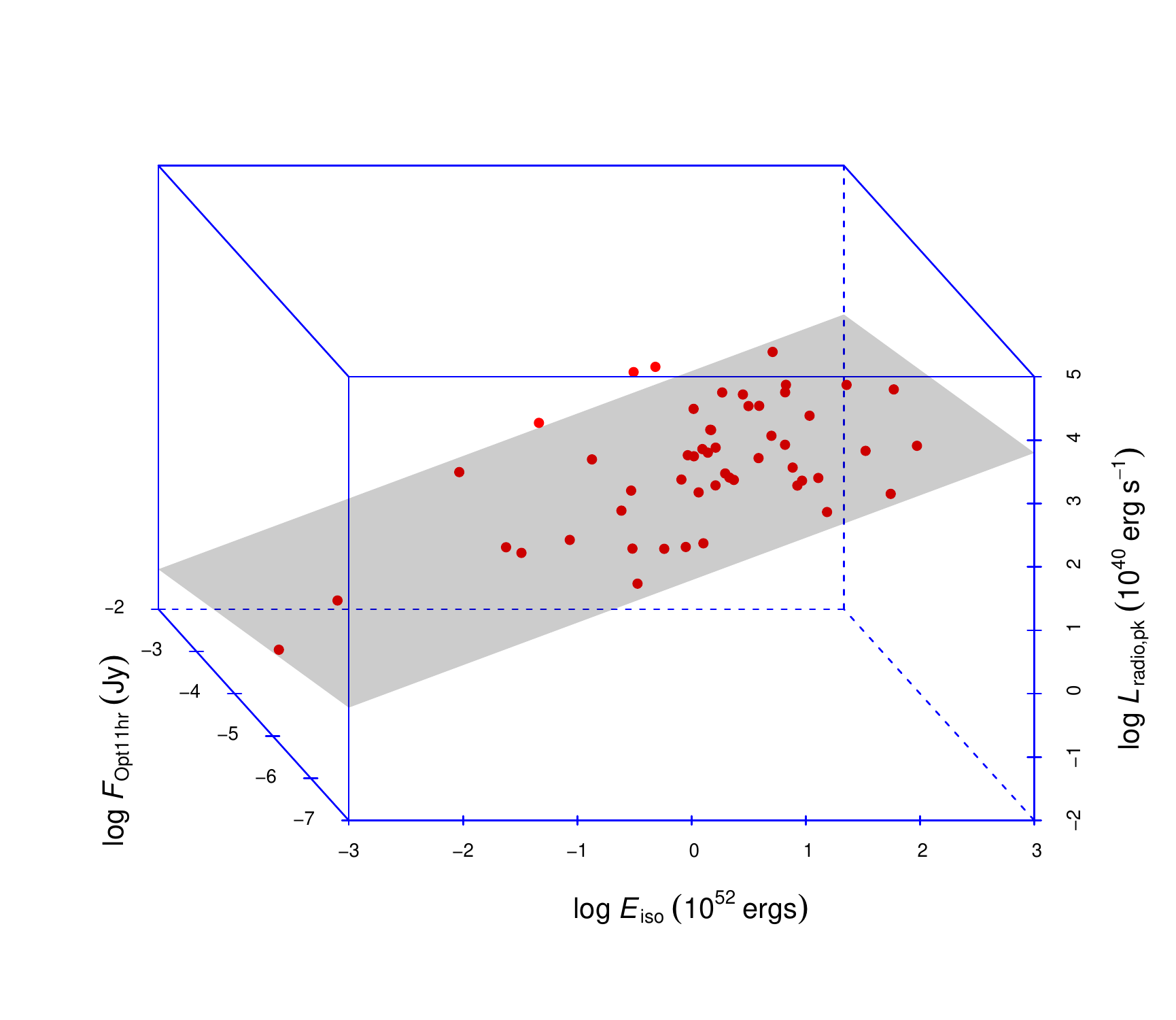}

\includegraphics[width=0.45\textwidth]{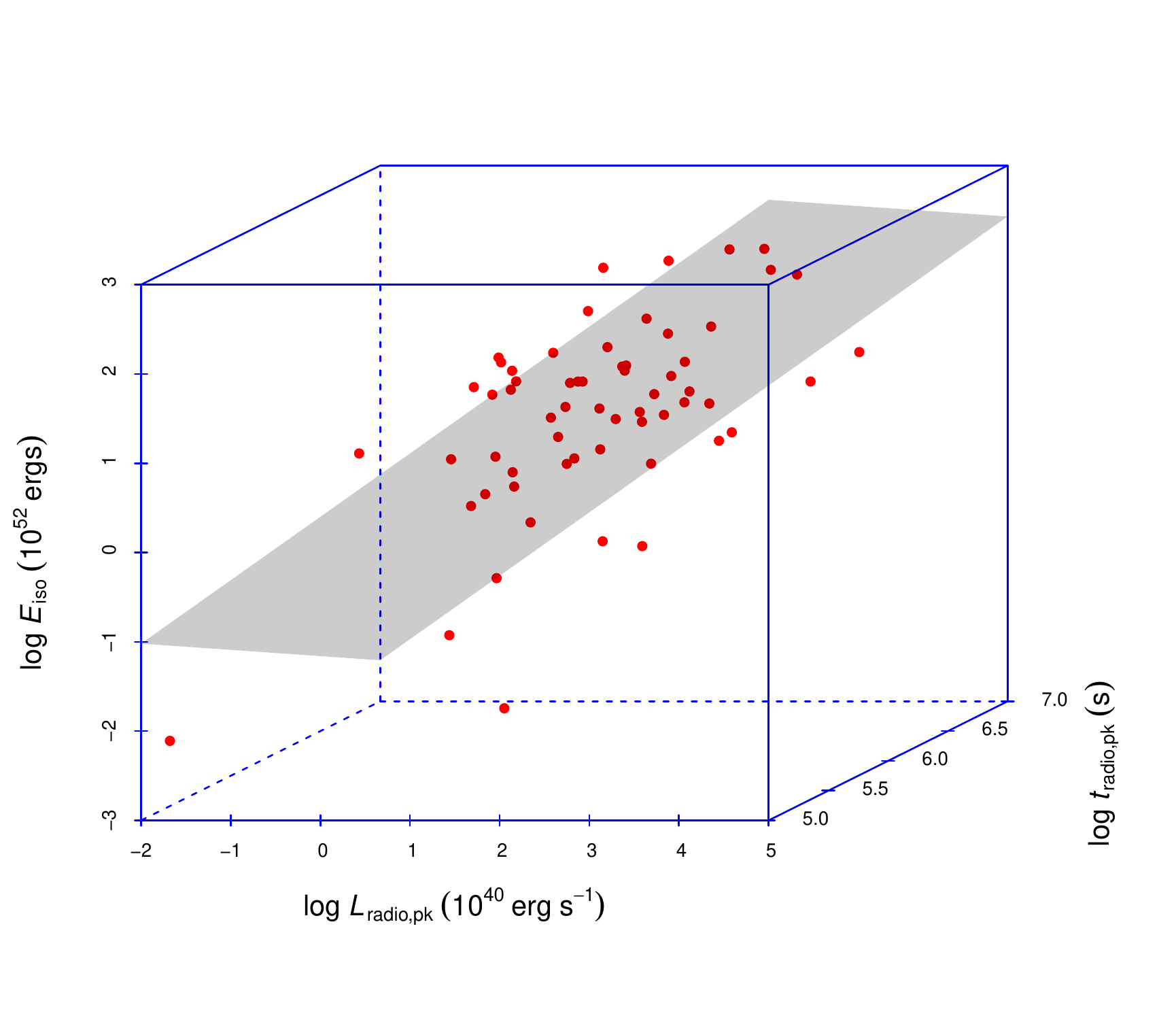}
\includegraphics[width=0.45\textwidth]{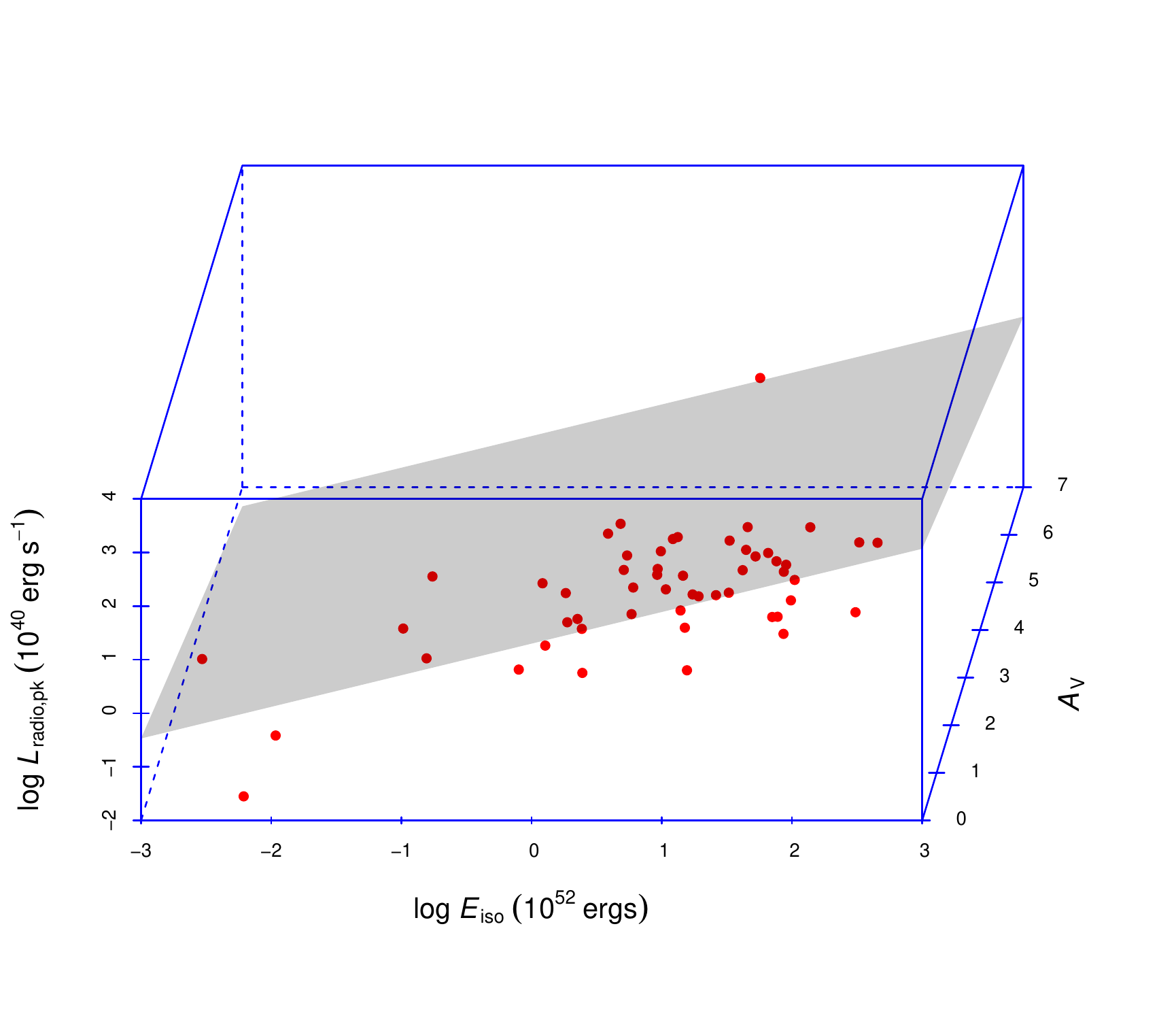}

\center{Fig. \ref{fig:three}---Continued}
\end{figure*}


\clearpage
\begin{figure*}

\includegraphics[width=0.45\textwidth]{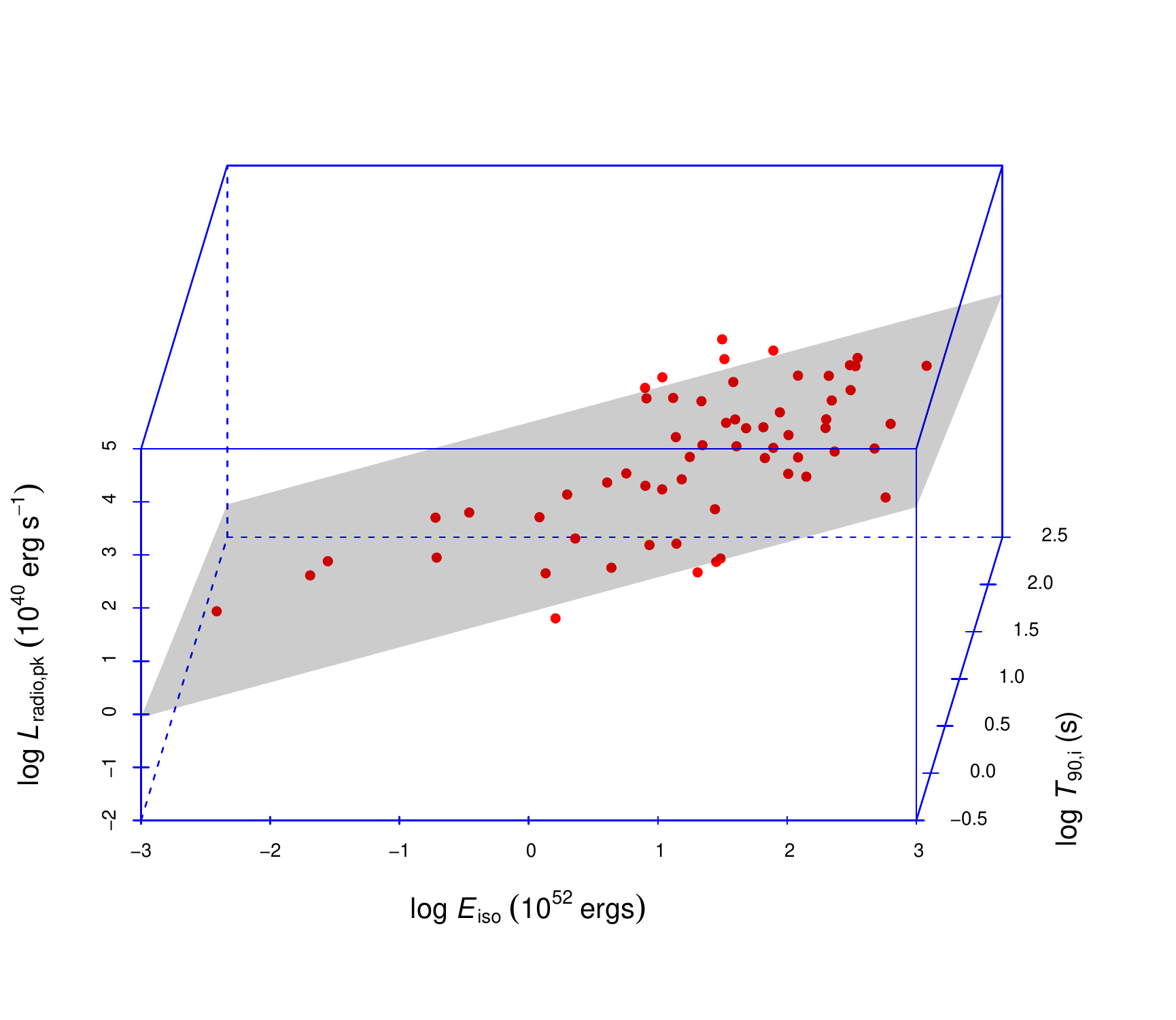}
\includegraphics[width=0.45\textwidth]{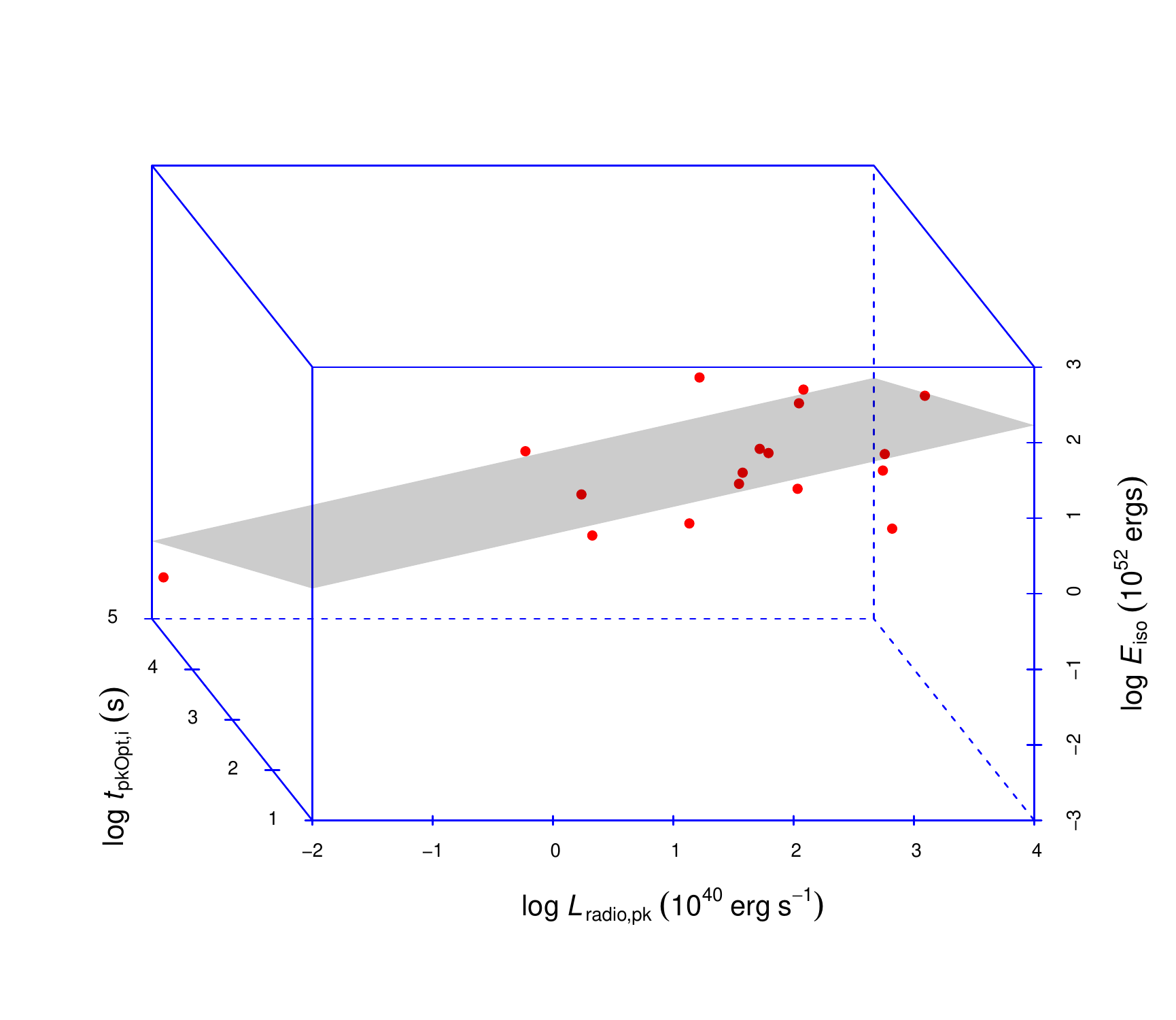}

\includegraphics[width=0.45\textwidth]{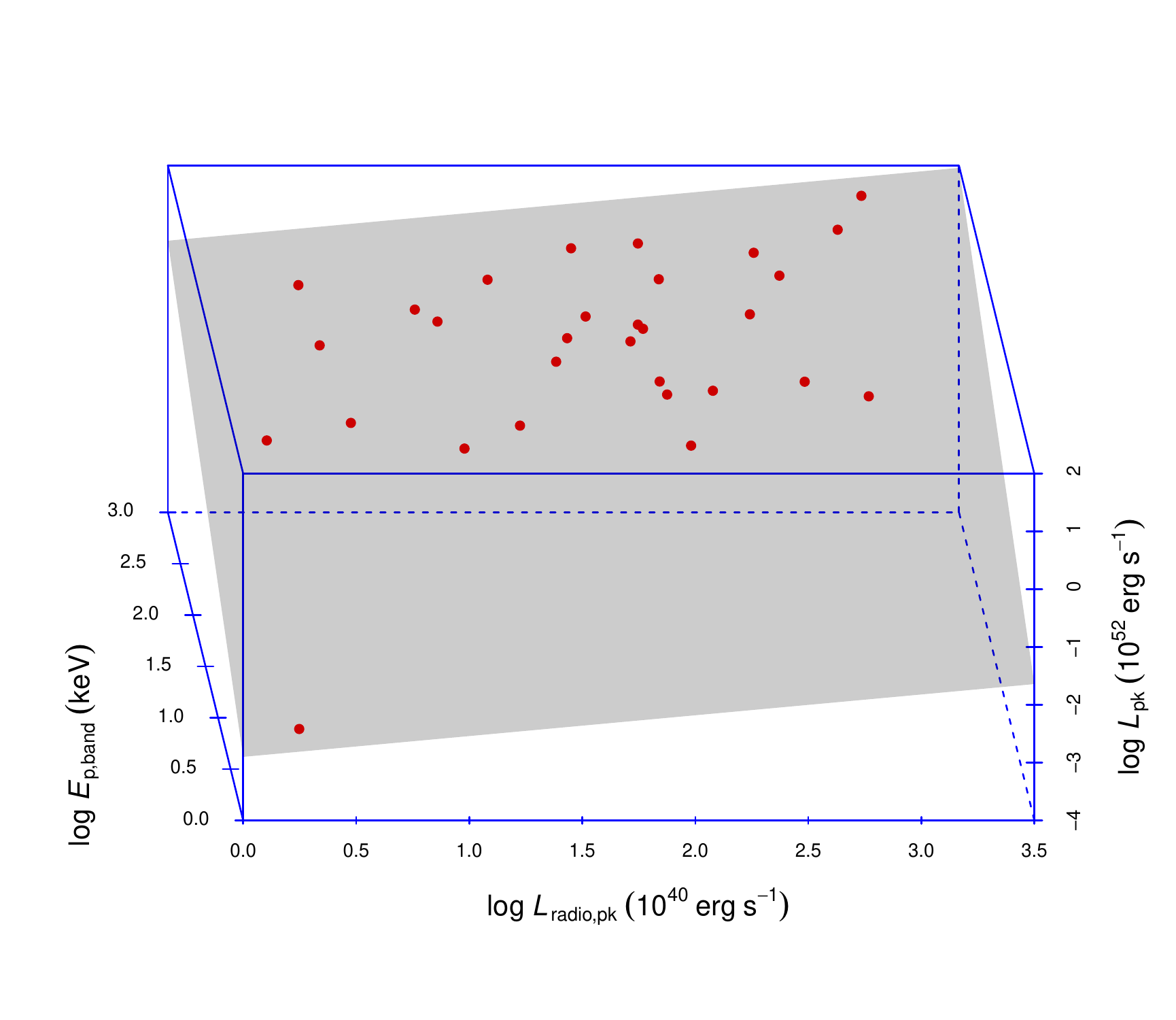}
\includegraphics[width=0.45\textwidth]{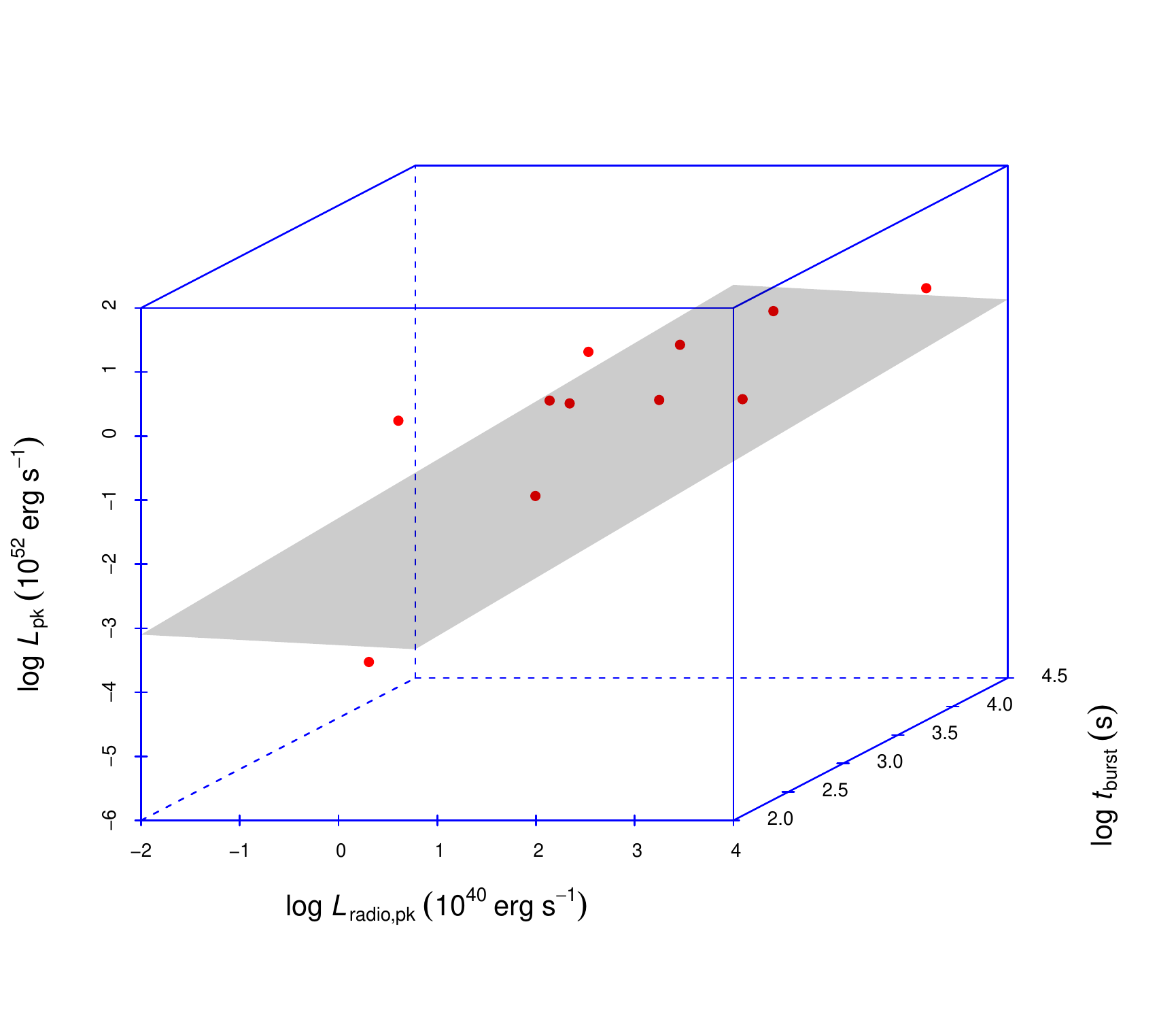}

\includegraphics[width=0.45\textwidth]{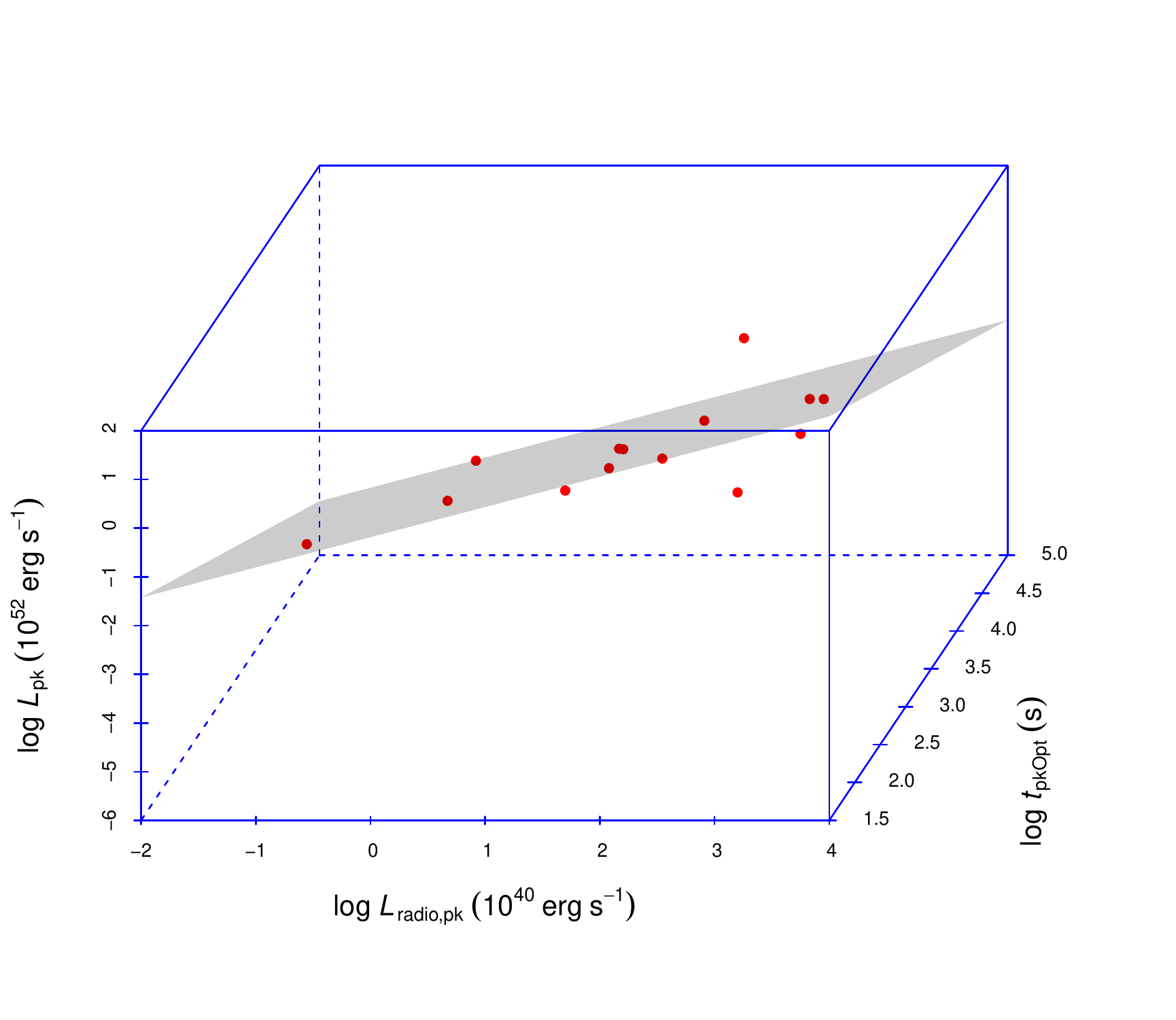}
\includegraphics[width=0.45\textwidth]{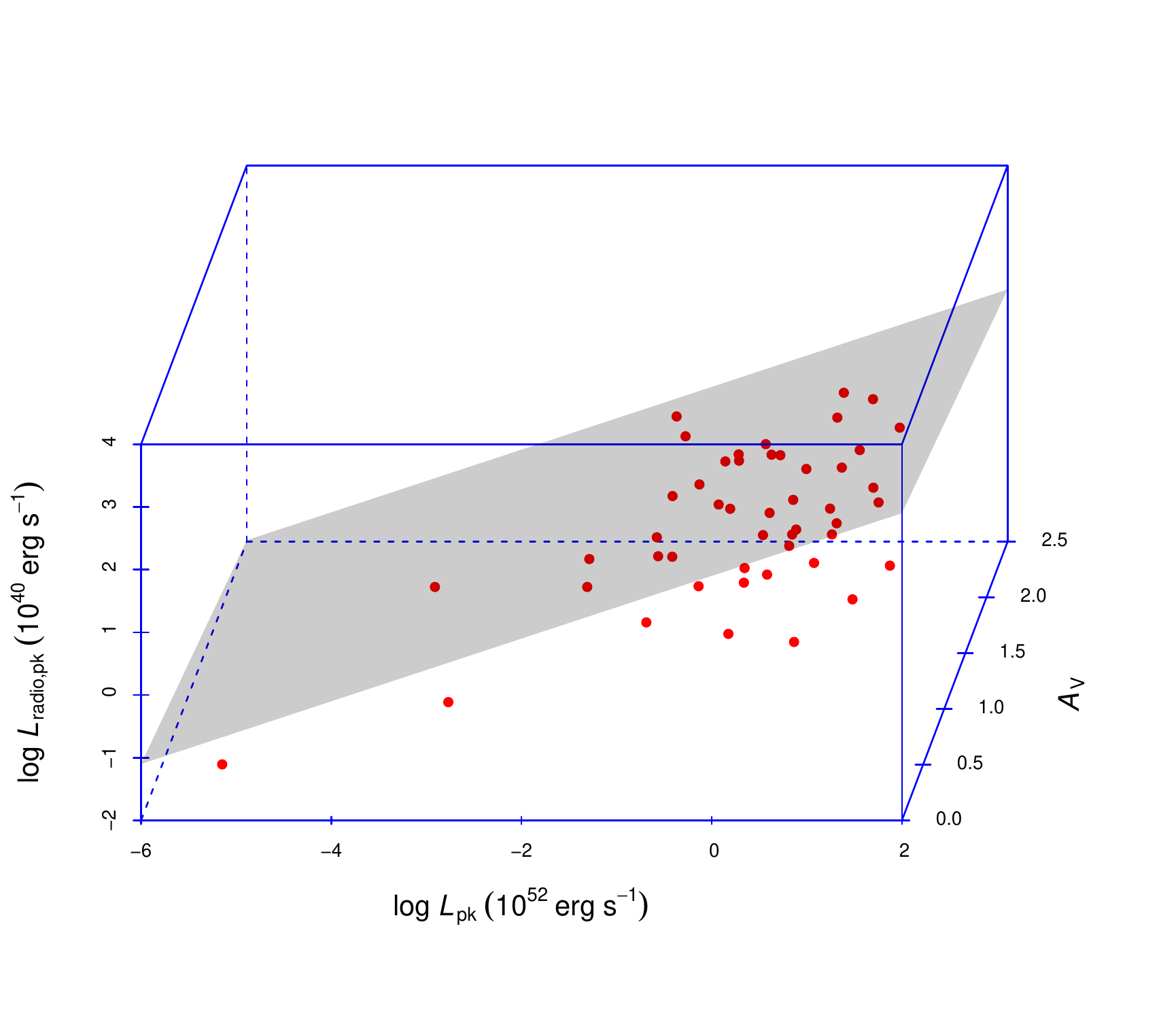}

\center{Fig. \ref{fig:three}---Continued}
\end{figure*}


\clearpage
\begin{figure*}

\includegraphics[width=0.45\textwidth]{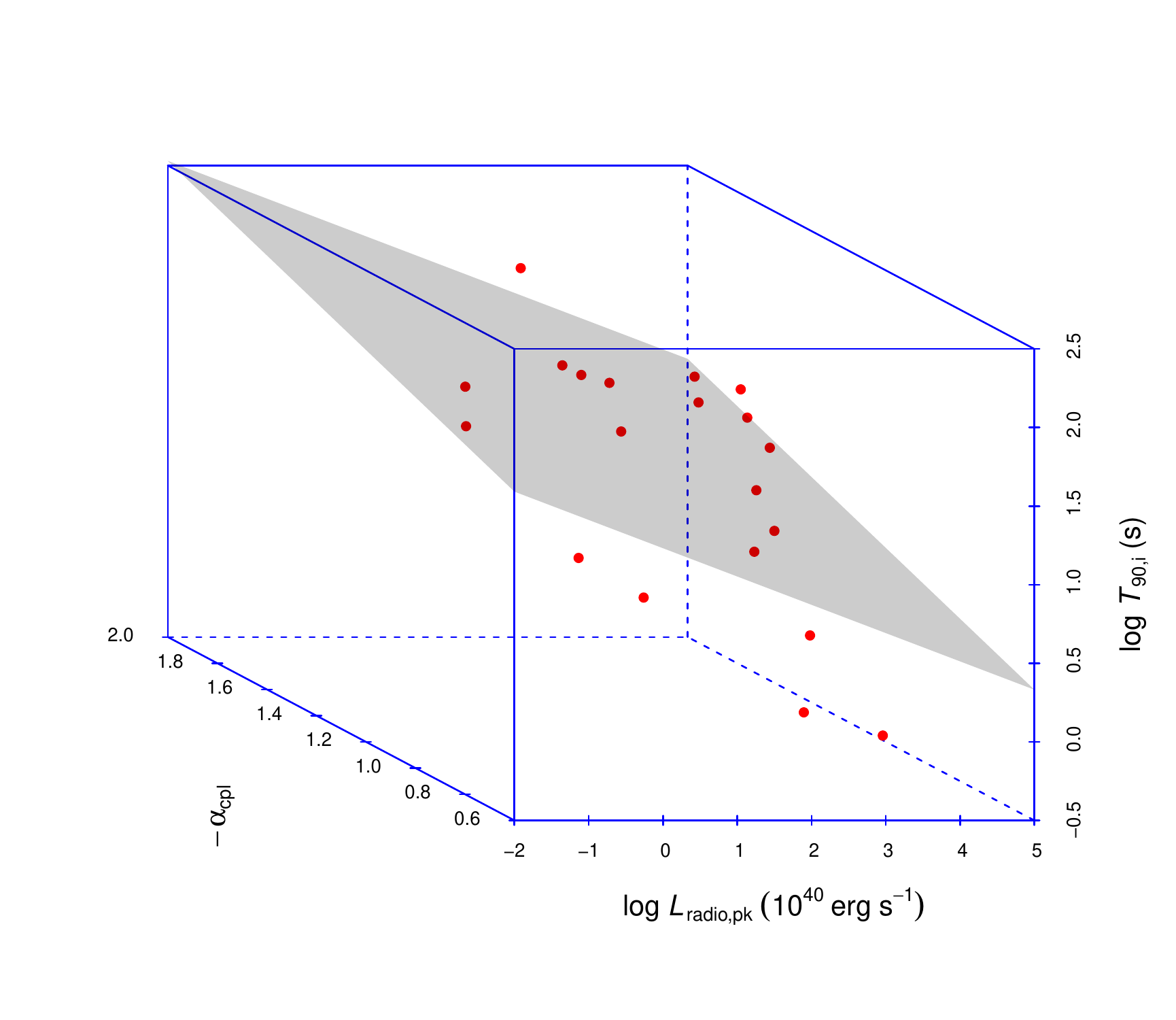}
\includegraphics[width=0.45\textwidth]{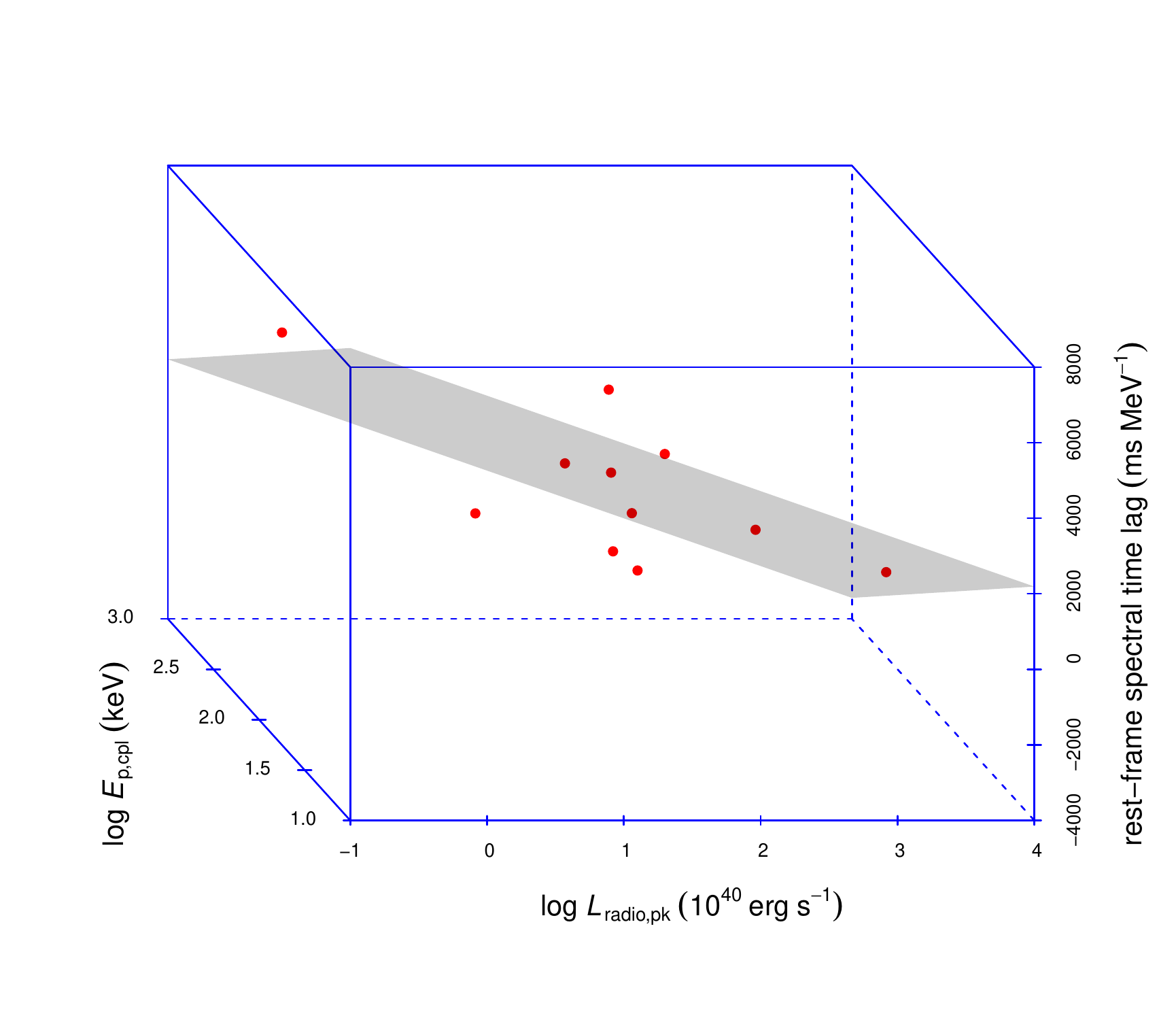}

\includegraphics[width=0.45\textwidth]{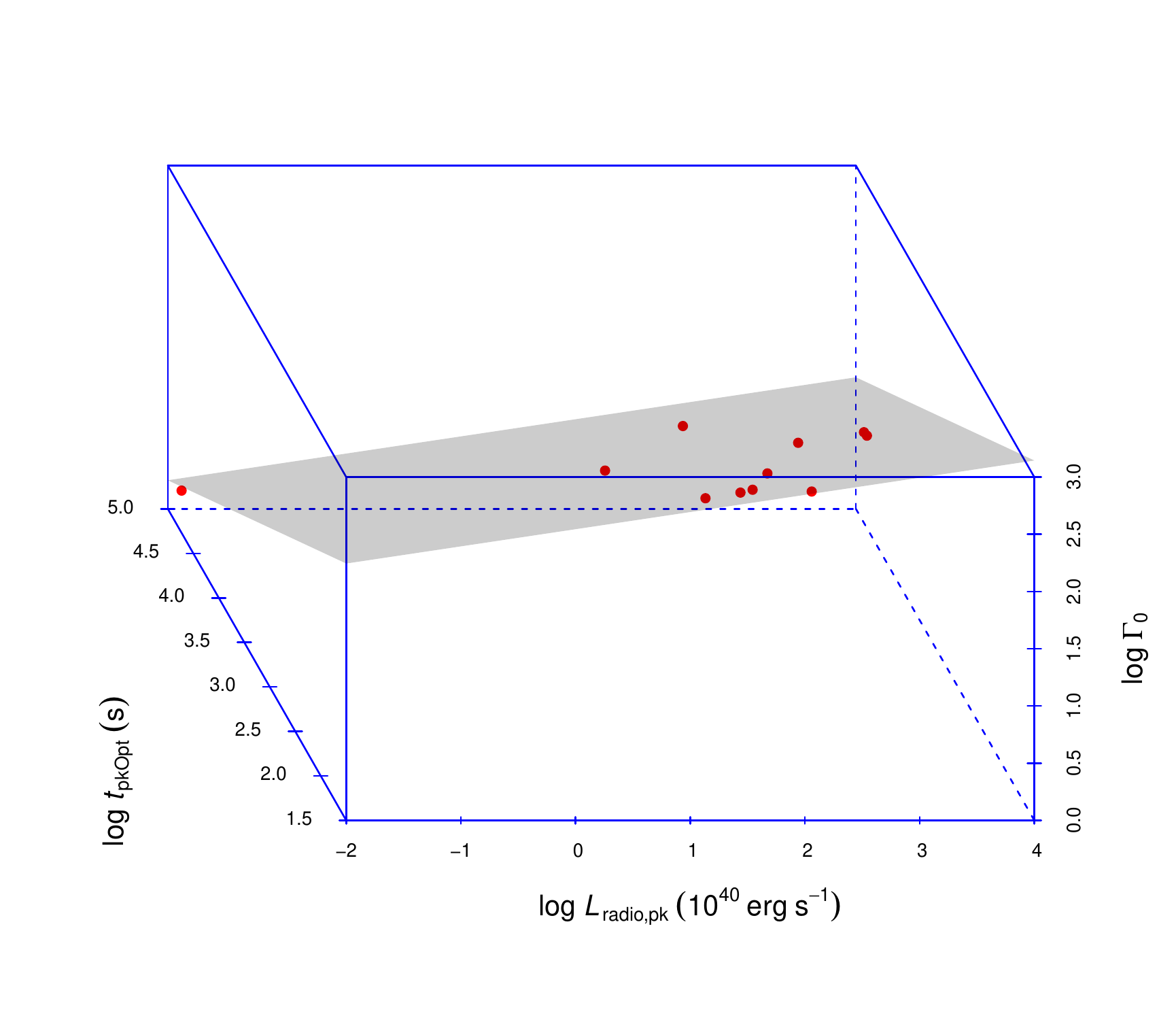}
\includegraphics[width=0.45\textwidth]{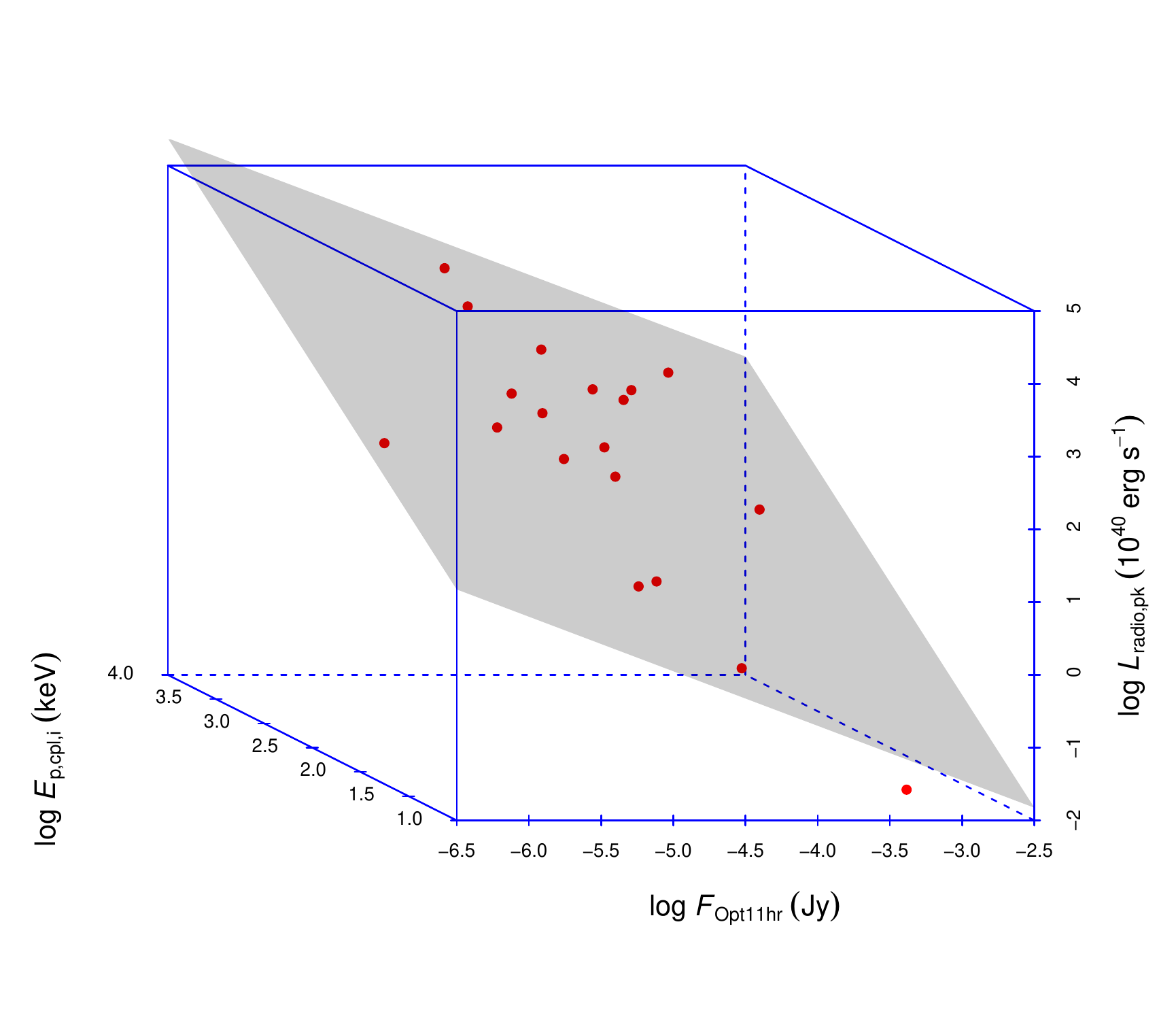}

\includegraphics[width=0.45\textwidth]{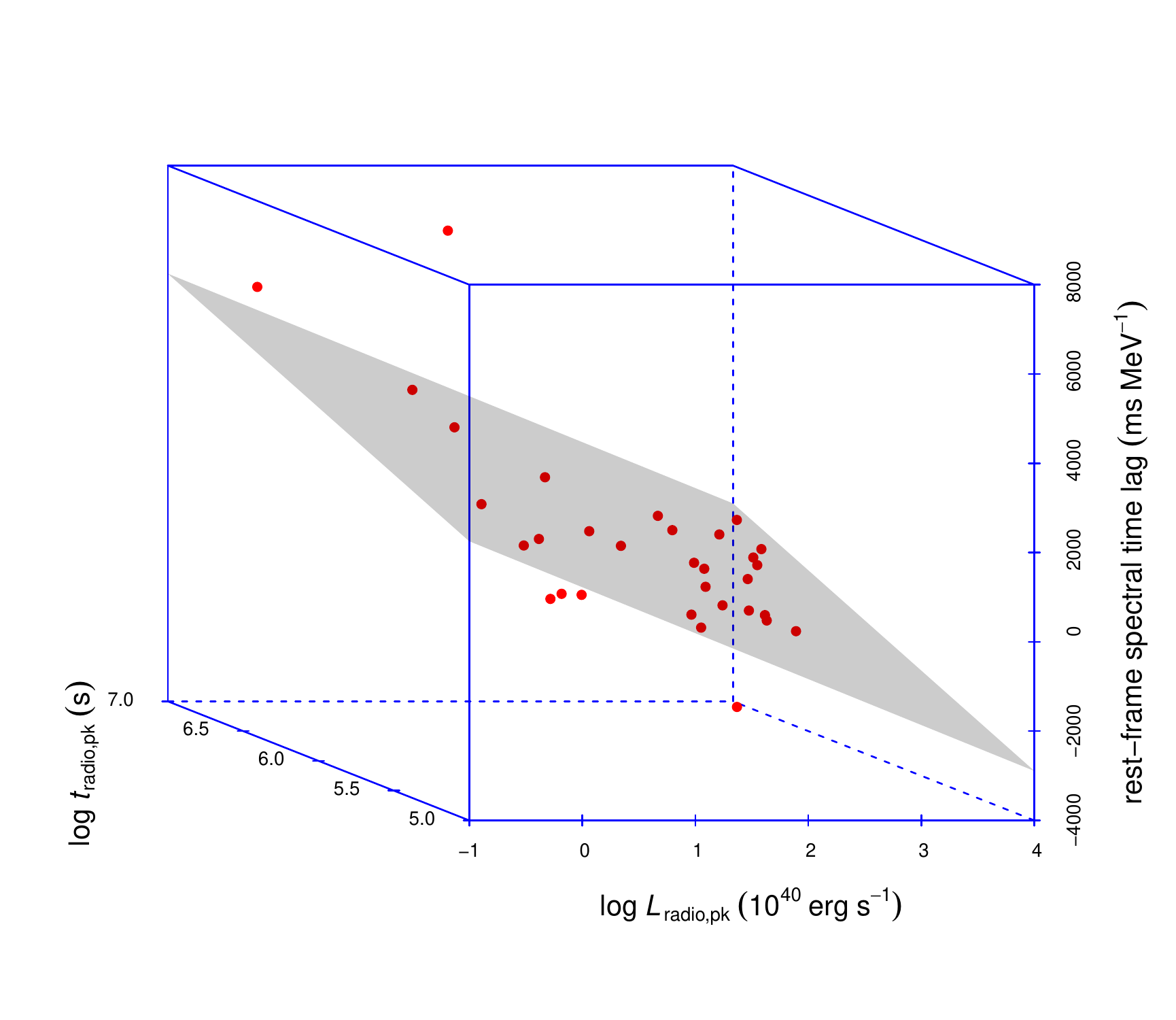}
\includegraphics[width=0.45\textwidth]{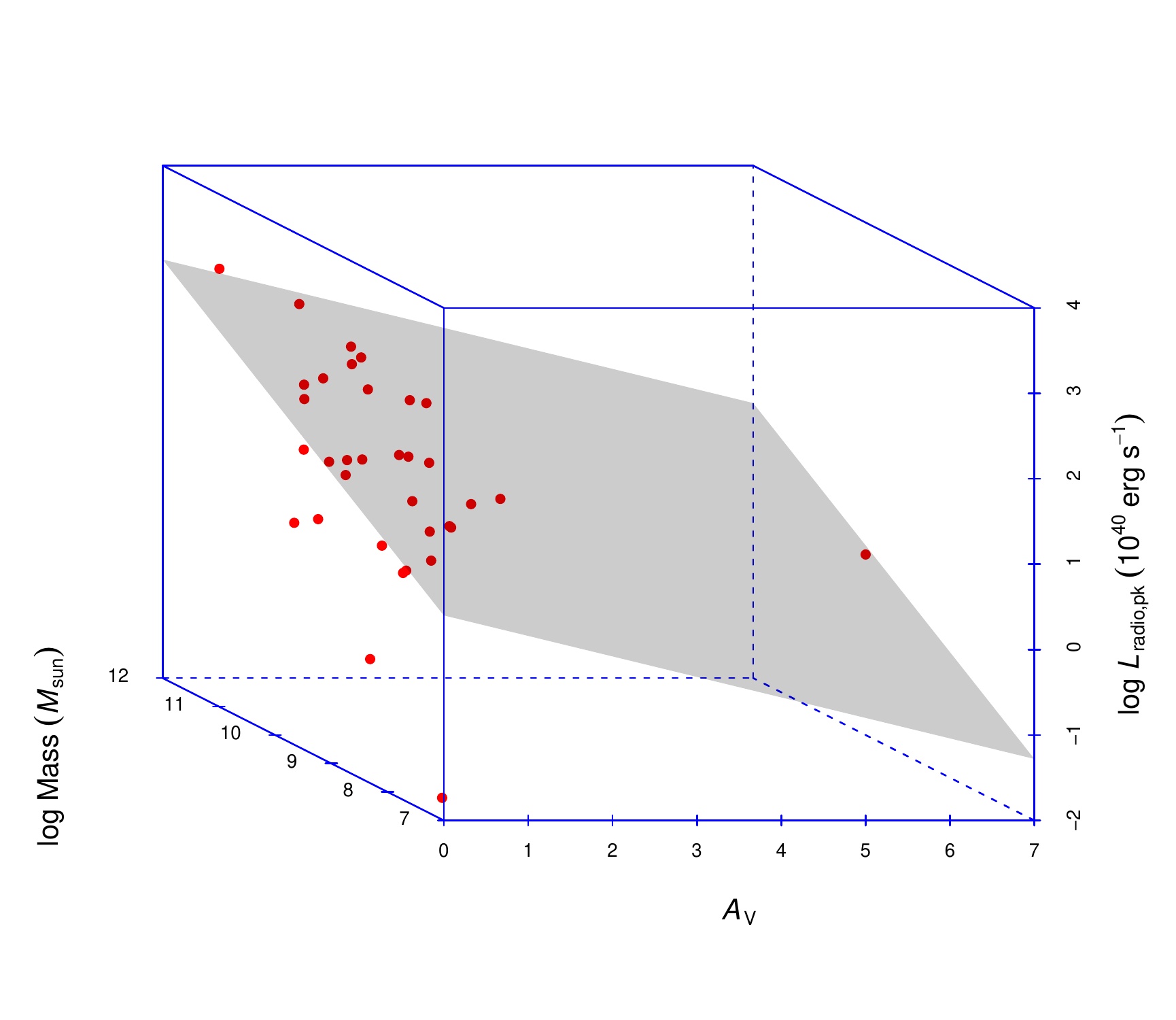}

\center{Fig. \ref{fig:three}---Continued}
\end{figure*}


\clearpage
\begin{figure*}

\includegraphics[width=0.45\textwidth]{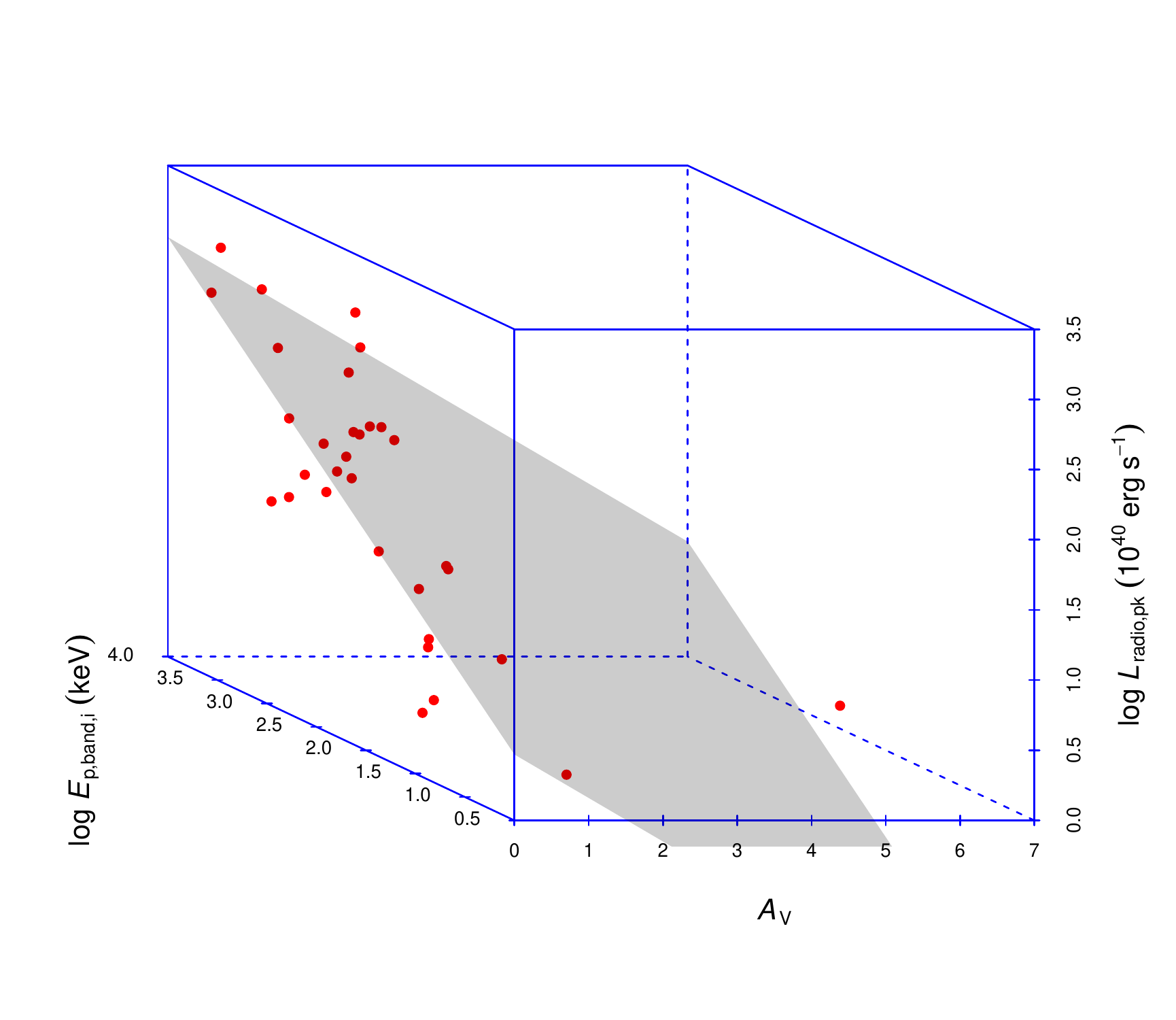}
\includegraphics[width=0.45\textwidth]{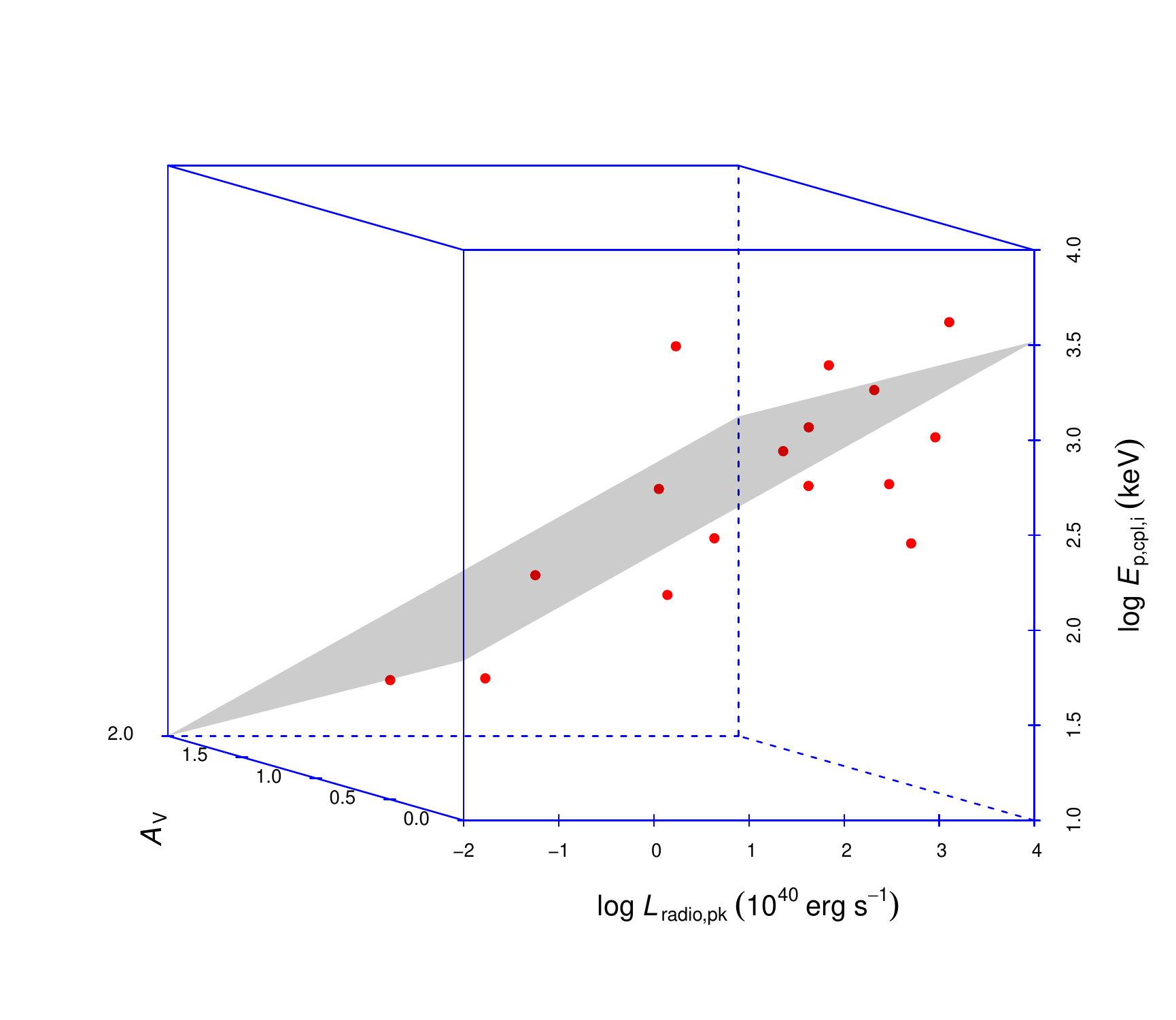}

\includegraphics[width=0.45\textwidth]{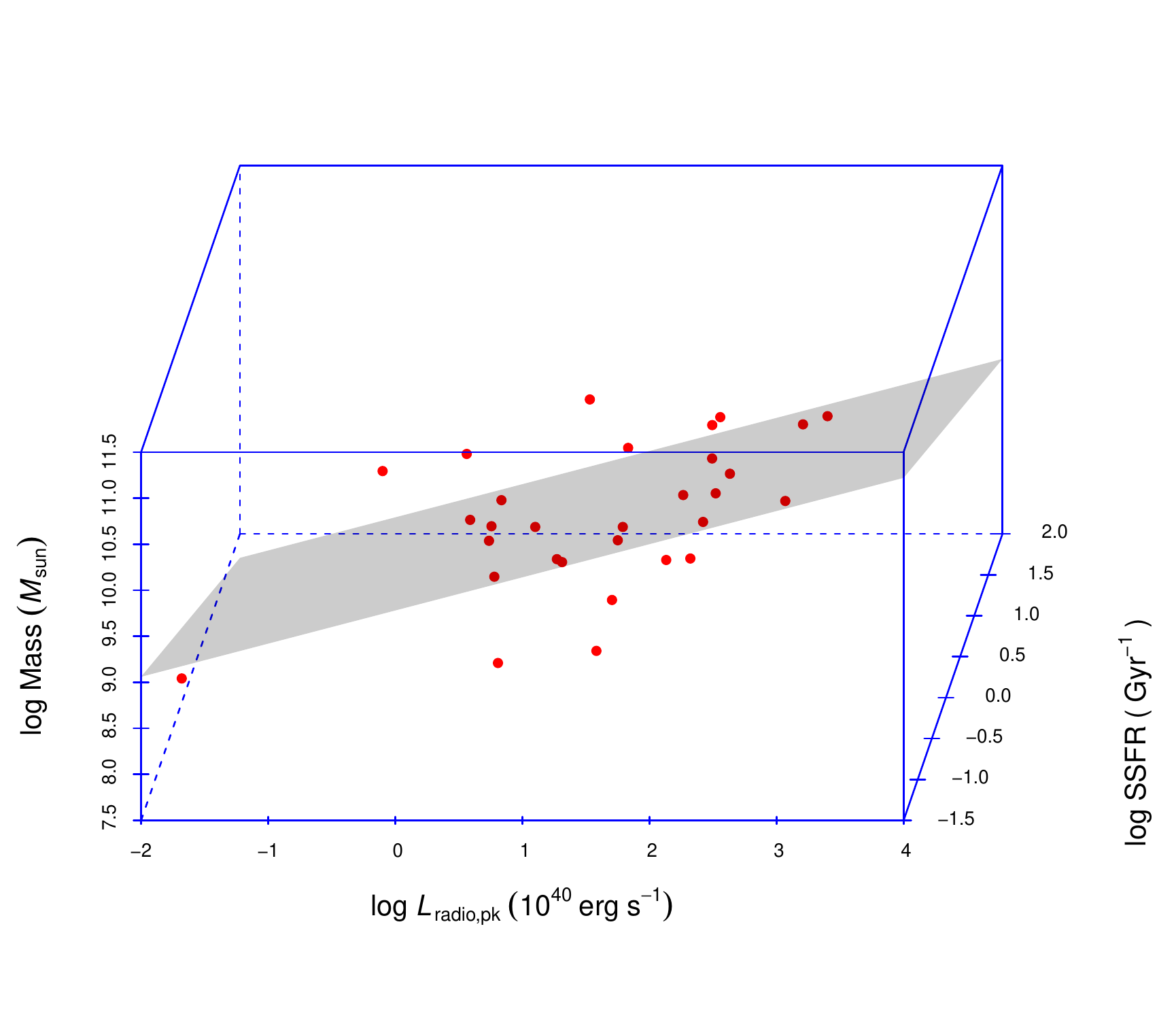}
\includegraphics[width=0.45\textwidth]{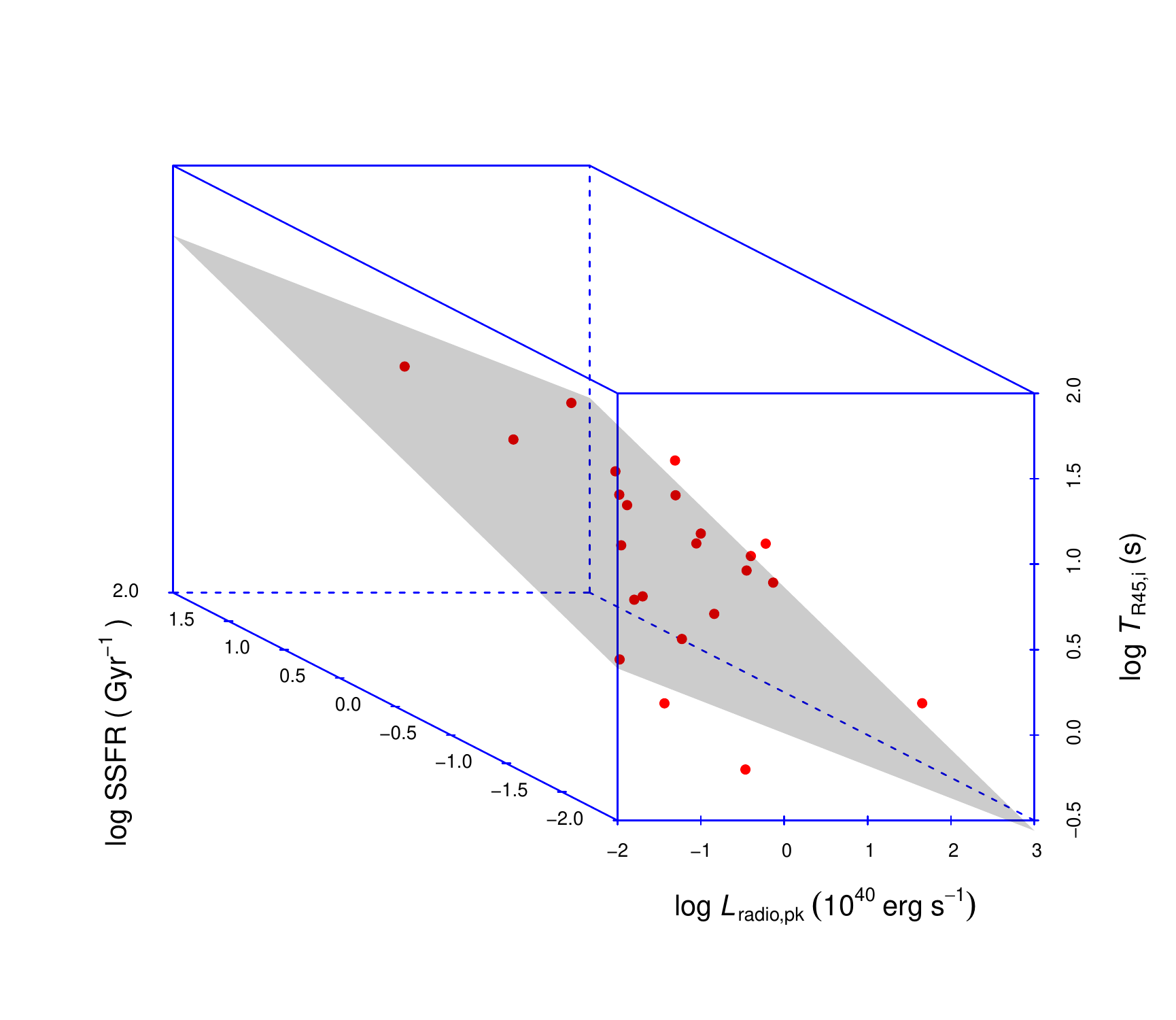}

\center{Fig. \ref{fig:three}---Continued}
\end{figure*}

\startlongtable
\begin{deluxetable}{cccccc}
\tablecaption{Big table samples \label{tab:bigtable}}
\tablehead{
\colhead{GRB} & \colhead{160509A} & \colhead{160422A} & \colhead{160419A} & \colhead{160417A} & \colhead{160412A}
}
\startdata
trigger time & 08:59:04.36 & 11:59:00.81 & 15:16:37 & 04:23:41 & 04:23:54 \\
 \hline
instrument & Fermi/MAXI & Fermi/CGBM & Fermi/Swift & Swift & Swift \\
 \hline
trigger number & \tabincell{c} {484477130 \\ 160509374} & \tabincell{c}{483019143 \\ 160422499} & \tabincell{c} {482771799 \\ 160419637 \\ 683383} & 683076 & 682522 \\
 \hline
coordinate & \tabincell{c}{$20h40m24s$ \\ $+76\arcdeg00\arcmin$} & \tabincell{c}{$02h48m12s$ \\ $-57\arcdeg54\arcmin$} & \tabincell{c}{$01h05m42s$ \\ $-27\arcdeg21\arcmin$} & \tabincell{c}{$08h00m58s$ \\ $+07\arcdeg38\arcmin$} & \tabincell{c}{$02h12m26s$ \\ $-67\arcdeg39\arcmin$} \\
 \hline
error & 30 & 3 & 3 & 3 & 3 \\
 \hline
$z$ & 1.17 & ... & ... & ... & ... \\
 \hline
reference & (1) &  &  &  &  \\
 \hline
\tabincell{c} {$D_{\rm L}$ \\ ($\rm 10^{\rm 28} ~ cm$)} & 2.27\tablenotemark{m} & ... & ... & ... & ... \\
\hline
reference &  &  &  &  &  \\
 \hline
\tabincell{c} {$T_{\rm 90}$ \\ ($\rm s$)} & $369.7^{\rm +0.8}_{\rm -0.8}$ & $8.7^{\rm +0.18}_{\rm -0.18}$\tablenotemark{b} & $8.8^{\rm +0.67}_{\rm -0.67}$\tablenotemark{b} & $15^{\rm +1.2}_{\rm -1.2}$\tablenotemark{b} & $31.6^{\rm +2.4}_{\rm -2.4}$\tablenotemark{b} \\
\hline
reference & (1) & (3) & (5) & (7) & (8) \\
 \hline
\tabincell{c} {$T_{\rm 50}$ \\ ($\rm s$)} & ... & ... & ... & ... & ... \\
\hline
reference &  &  &  &  &  \\
 \hline
\tabincell{c} {$T_{\rm R45}$ \\ ($\rm s$)} & ... & ... & ... & ... & ... \\
\hline
reference &  &  &  &  &  \\
 \hline
$variability_{\rm 1}$ & ... & ... & ... & ... & ... \\
\hline
reference &  &  &  &  &  \\
 \hline
$variability_{\rm 2}$ & ... & ... & ... & ... & ... \\
\hline
reference &  &  &  &  &  \\
 \hline
$variability_{\rm 3}$ & ... & ... & ... & ... & ... \\
\hline
reference &  &  &  &  &  \\
 \hline
\tabincell{c} {$F_{\rm g}$ \\ ($\rm 10^{\rm -6} ~ ergs ~ cm^{\rm -2}$)} & $218.1^{\rm +1.8}_{\rm -1.8}$\tablenotemark{c} & $95.15^{\rm +1.26}_{\rm -1.26}$\tablenotemark{c} & $4.6^{\rm +0.73}_{\rm -0.73}$\tablenotemark{bc} & $1.3^{\rm +0.4}_{\rm -0.4}$\tablenotemark{bc} & $4.56^{\rm +0.5}_{\rm -0.5}$\tablenotemark{bc} \\
\hline
reference & (1) & (4) & (6) & (6) & (6) \\
 \hline
HR & $5.853^{\rm +0.056}_{\rm -0.056}$\tablenotemark{j} & $3.383^{\rm +0.011}_{\rm -0.011}$\tablenotemark{j} & $0.88^{\rm +0.23}_{\rm -0.23}$\tablenotemark{j} & $2.19^{\rm +0.7}_{\rm -0.7}$\tablenotemark{j} & $2.46^{\rm +0.26}_{\rm -0.26}$\tablenotemark{j} \\
\hline
reference &  &  &  &  &  \\
 \hline
\tabincell{c} {$E_{\rm \gamma,iso}$ \\ ($\rm 10^{\rm 52} ~ ergs$)} & $57.6^{\rm +0.5}_{\rm -0.5}$ & ... & ... & ... & ... \\
\hline
reference & (1) &  &  &  &  \\
 \hline
\tabincell{c} {$L_{\rm pk}$ \\ ($\rm 10^{\rm 52} ~ ergs ~ s^{\rm -1}$)} & ... & ... & ... & ... & ... \\
\hline
reference &  &  &  &  &  \\
 \hline
\tabincell{c} {$F_{\rm pk1}$ \\ ($\rm 10^{\rm -6} ~ ergs ~ cm^{\rm -2} ~ s^{\rm -1}$)} & $21.86^{\rm +0.89}_{\rm -0.89}$\tablenotemark{j} & $19.29^{\rm +0.89}_{\rm -0.89}$\tablenotemark{j} & $1.06^{\rm +0.19}_{\rm -0.19}$\tablenotemark{j} & $0.27^{\rm +0.11}_{\rm -0.11}$\tablenotemark{j} & $0.474^{\rm +0.08}_{\rm -0.08}$\tablenotemark{j} \\
\hline
reference &  &  &  &  &  \\
 \hline
\tabincell{c} {$F_{\rm pk2}$ \\ ($\rm 10^{\rm -6} ~ ergs ~ cm^{\rm -2} ~ s^{\rm -1}$)} & ... & ... & ... & ... & ... \\
 \hline
\tabincell{c} {$F_{\rm pk3}$ \\ ($\rm 10^{\rm -6} ~ ergs ~ cm^{\rm -2} ~ s^{\rm -1}$)} & ... & ... & ... & ... & ... \\
 \hline
\tabincell{c} {$F_{\rm pk4}$ \\ ($\rm 10^{\rm -6} ~ ergs ~ cm^{\rm -2} ~ s^{\rm -1}$)} & ... & ... & ... & ... & ... \\
 \hline
\tabincell{c} {$P_{\rm pk1}$ \\ ($\rm photons ~ cm^{\rm -2} ~ s^{\rm -1}$)} & ... & ... & ... & ... & ... \\
\hline
reference &  &  &  &  &  \\
 \hline
\tabincell{c} {$P_{\rm pk2}$ \\ ($\rm photons ~ cm^{\rm -2} ~ s^{\rm -1}$)} & ... & ... & ... & ... & ... \\
\hline
reference &  &  &  &  &  \\
 \hline
\tabincell{c} {$P_{\rm pk3}$ \\ ($\rm photons ~ cm^{\rm -2} ~ s^{\rm -1}$)} & ... & ... & ... & ... & ... \\
\hline
reference &  &  &  &  &  \\
 \hline
\tabincell{c} {$P_{\rm pk4}$ \\ ($\rm photons ~ cm^{\rm -2} ~ s^{\rm -1}$)} & $71.27^{\rm +0.66}_{\rm -0.66}$\tablenotemark{c} & $93.8^{\rm +0.6}_{\rm -0.6}$\tablenotemark{c} & $11.51^{\rm +0.99}_{\rm -0.99}$\tablenotemark{bc} & $1.63^{\rm +0.33}_{\rm -0.33}$\tablenotemark{bc} & $2.59^{\rm +0.21}_{\rm -0.21}$\tablenotemark{bc} \\
\hline
reference & (2) & (4) & (6) & (6) & (6) \\
 \hline
$-\alpha_{\rm band}$ & $0.89^{\rm +0.01}_{\rm -0.01}$ & $0.95^{\rm +0.01}_{\rm -0.01}$ & ... & ... & ... \\
\hline
reference & (2) & (4) &  &  &  \\
 \hline
$-\beta_{\rm band}$ & $2.11^{\rm +0.02}_{\rm -0.02}$ & $2.42^{\rm +0.04}_{\rm -0.04}$ & ... & ... & ... \\
\hline
reference & (2) & (4) &  &  &  \\
 \hline
\tabincell{c} {$E_{\rm p,band}$ \\ ($\rm keV$)} & $370^{\rm +7}_{\rm -7}$ & $242^{\rm +4}_{\rm -4}$ & ... & ... & ... \\
\hline
reference & (2) & (4) &  &  &  \\
 \hline
$-\alpha_{\rm cpl}$ & ... & ... & $0.76^{\rm +0.2}_{\rm -0.2}$\tablenotemark{b} & ... & ... \\
\hline
reference &  &  & (5) &  &  \\
 \hline
\tabincell{c} {$E_{\rm p,cpl}$ \\ ($\rm keV$)} & ... & ... & $107.1^{\rm +21.29}_{\rm -21.29}$\tablenotemark{b} & ... & ... \\
\hline
reference &  &  & (5) &  &  \\
 \hline
$-\alpha_{\rm spl}$ & ... & ... & ... & $1.93^{\rm +0.15}_{\rm -0.15}$\tablenotemark{b} & $1.88^{\rm +0.048}_{\rm -0.048}$\tablenotemark{b} \\
\hline
reference &  &  &  & (7) & (8) \\
 \hline
\tabincell{c} {$\theta_{\rm j}$ \\ ($\rm rad$)} & ... & ... & ... & ... & ... \\
\hline
reference &  &  &  &  &  \\
 \hline
\tabincell{c} {spectral time lag \\ ($\rm ms ~ MeV^{\rm -1}$)} & ... & ... & ... & ... & ... \\
\hline
reference &  &  &  &  &  \\
 \hline
$\Gamma_{0}$ & ... & ... & ... & ... & ... \\
\hline
reference &  &  &  &  &  \\
 \hline
\tabincell{c} {$\log t_{\rm burst}$ \\ ($\rm s$)} & ... & ... & ... & ... & ... \\
\hline
reference &  &  &  &  &  \\
 \hline
\tabincell{c} {$t_{\rm pkX}$ \\ ($\rm s$)} & ... & ... & ... & ... & ... \\
\hline
reference &  &  &  &  &  \\
 \hline
\tabincell{c} {$t_{\rm pkOpt}$ \\ ($\rm s$)} & ... & ... & ... & ... & ... \\
\hline
reference &  &  &  &  &  \\
 \hline
\tabincell{c} {$F_{\rm X11hr}$ \\ ($\rm 10^{\rm -23} ~ ergs ~ cm^{\rm -2} ~ s^{\rm -1} ~ Hz^{\rm -1}$)} & ... & ... & ... & ... & ... \\
\hline
reference &  &  &  &  &  \\
 \hline
$\beta_{\rm X11hr}$ & ... & ... & ... & ... & ... \\
\hline
reference &  &  &  &  &  \\
 \hline
\tabincell{c} {$F_{\rm Opt11hr}$ \\ ($\rm 10^{\rm -23} ~ ergs ~ cm^{\rm -2} ~ s^{\rm -1} ~ Hz^{\rm -1}$)} & ... & ... & ... & ... & ... \\
\hline
reference &  &  &  &  &  \\
 \hline
\tabincell{c} {$t_{\rm radio,pk}$ \\ ($\rm s$)} & ... & ... & ... & ... & ... \\
\hline
reference &  &  &  &  &  \\
 \hline
\tabincell{c} {$F_{\rm radio,pk}$ \\ ($\rm 10^{\rm -23} ~ ergs ~ cm^{\rm -2} ~ s^{\rm -1} ~ Hz^{\rm -1}$)} & ... & ... & ... & ... & ... \\
\hline
reference &  &  &  &  &  \\
 \hline
\tabincell{c} {host galaxy offset \\ ($\rm kpc$)} & ... & ... & ... & ... & ... \\
\hline
reference &  &  &  &  &  \\
 \hline
metallicity & ... & ... & ... & ... & ... \\
\hline
reference &  &  &  &  &  \\
 \hline
Mag & ... & ... & ... & ... & ... \\
\hline
reference &  &  &  &  &  \\
 \hline
\tabincell{c} {$N_{\rm H}$ \\ ($\rm 10^{\rm 21} ~ cm^{\rm -2}$)} & ... & ... & ... & ... & ... \\
\hline
reference &  &  &  &  &  \\
 \hline
$A_{\rm V}$ & ... & ... & ... & ... & ... \\
\hline
reference &  &  &  &  &  \\
 \hline
\tabincell{c} {SFR \\ ($\rm M_{\bigodot} ~ yr^{\rm -1}$)} & ... & ... & ... & ... & ... \\
\hline
reference &  &  &  &  &  \\
 \hline
\tabincell{c} {$\log SSFR$ \\ ($\rm Gyr^{\rm -1}$)} & ... & ... & ... & ... & ... \\
\hline
reference &  &  &  &  &  \\
 \hline
\tabincell{c} {Age \\ ($\rm Myr$)} & ... & ... & ... & ... & ... \\
\hline
reference &  &  &  &  &  \\
 \hline
\tabincell{c} {$\log Mass$ \\ ($M_{\bigodot}$)} & ... & ... & ... & ... & ... \\
\hline
reference &  &  &  &  &  \\
 \hline
\enddata
\tablenotetext{A}{We changed the GRB name or the name is a little different in different papers and GCN, because some different GRBs have same name in different instruments. For an example, Fermi GRB 110916 and MAXI GRB 110916 are two different GRBs with different trigger time, but they have same GRB name, so we changed Fermi GRB 110916 to GRB 110916A, and MAXI GRB 110916 to GRB 110916B.}
\tablenotetext{a}{The errors are imputed by MICE algorithm.}
\tablenotetext{b}{The errors in the original papers are in 90\% confidence level, we changed the errors to 1 $\sigma$ confidence level by multiplying 0.995/1.645. }
\tablenotetext{c}{The values are calculated using the spectral values in order change the energy band.}
\tablenotetext{d}{The unit is different from the original papers.}
\tablenotetext{e}{The error is estimated as the central value multiplying 0.1.}
\tablenotetext{f}{We use the BASTE $\alpha$ peak value -1.1 as common when $\alpha$ is not well constrained.}
\tablenotetext{g}{We use the BASTE $\beta$ peak value -2.2 as common when $\beta$ is not well constrained.}
\tablenotetext{h}{The value in the original paper is in rest-frame, we change the value using rest-frame divided by $(1+z)$ to observer-frame.}
\tablenotetext{i}{We changed the values into logarithm or from logarithm into the normal form.}
\tablenotetext{j}{The values are calculated using other parameters.}
\tablenotetext{k}{We convert different metallicity calibrations into \citet{Kobulnicky2004} calibrator using the method in \citet{Kewley2008}.}
\tablenotetext{m}{The $D_{\rm L}$ is calculated using $z$ with cosmology parameters in \citet{Planck2016}.}
\tablecomments{All the results are in machine readable table. The description of every parameter is in Section \ref{sec:sample}.}
\end{deluxetable}
References. 
(1) \citet{Laskar2016};
(2) \citet{RobertsGCN19411};
(3) \citet{RicciariniGCN19345};
(4) \citet{BurnsGCN19331};
(5) \citet{CummingsGCN19328};
(6) \url{https://swift.gsfc.nasa.gov/archive/grb_table/index.php};
(7) \citet{BarthelmyGCN19323};
(8) \citet{UkwattaGCN19301};
(9) \citet{StamatikosGCN19302};
(10) \citet{SakamotoGCN19276};
(11) \citet{FrederiksGCN19312};
(12) \citet{PalmerGCN19263};
(13) \citet{RobertsGCN19265};
(14) \citet{MereghettiGCN19251};
(15) \citet{MarkwardtGCN19240};
(16) \citet{LienGCN19234};
(17) \citet{KrimmGCN19216};
(18) \citet{HuiGCN19198};
(19) \citet{CummingsGCN19188};
(20) \citet{BarthelmyGCN19181};
(21) \citet{ToelgeGCN19161};
(22) \citet{UkwattaGCN19148};
(23) \citet{StamatikosGCN19113};
(24) \citet{SakamotoGCN19106};
(25) \citet{PalmerGCN19091};
(26) \citet{OzawaGCN19085};
(27) \citet{vonGCN19061};
(28) \citet{MarkwardtGCN19066};
(29) \citet{HuiGCN19056};
(30) \citet{LienGCN19043};
(31) \citet{KrimmGCN19038};
(32) \citet{BarthelmyGCN19020};
(33) \citet{BarthelmyGCN18998};
(34) \citet{Dado2016};
(35) \citet{CummingsGCN18959};
(36) \citet{BarthelmyGCN18944};
(37) \citet{BarthelmyGCN18929};
(38) \citet{UkwattaGCN18919};
(39) \citet{StamatikosGCN18899};
(40) \citet{PalmerGCN18882};
(41) \citet{RobertsGCN18861};
(42) \citet{NakahiraGCN18845};
(43) \citet{VeresGCN18844};
(44) \citet{PalmerGCN18829};
(45) \citet{GolenetskiiGCN18867};
(46) \citet{YamadaGCN18814};
(47) \citet{VeresGCN18796};
(48) \citet{MoriyamaGCN18810};
(49) \citet{VeresGCN18787};
(50) \citet{LienGCN18751};
(51) \citet{KrimmGCN18752};
(52) \citet{BarthelmyGCN18754};
(53) \citet{BissaldiGCN18736};
(54) \citet{SenumaGCN18792};
(55) \citet{ToelgeGCN18727};
(56) \citet{KawakuboGCN18724};
(57) \citet{GolenetskiiGCN18837};
(58) \citet{YuGCN18715};
(59) \citet{CummingsGCN18699};
(60) \citet{BarthelmyGCN18683};
(61) \citet{UkwattaGCN18670};
(62) \citet{StamatikosGCN18668};
(63) \citet{SakamotoGCN18648};
(64) \citet{StanbroGCN18639};
(65) \citet{vonGCN18628};
(66) \citet{MarkwardtGCN18622};
(67) \citet{LienGCN18604};
(68) \citet{KrimmGCN18593};
(69) \citet{CummingsGCN18580};
(70) \citet{YoshidaGCN18605};
(71) \citet{StanbroGCN18570};
(72) \citet{MalesaniGCN18540};
(73) \citet{UkwattaGCN18542};
(74) \citet{TanvirGCN18524};
(75) \citet{StamatikosGCN18527};
(76) \citet{XuGCN18505};
(77) \citet{SakamotoGCN18514};
(78) \citet{Nappo2017};
(79) \citet{PalmerGCN18496};
(80) \citet{MarkwardtGCN18469};
(81) \citet{LienGCN18452};
(82) \citet{KrimmGCN18429};
(83) \citet{CummingsGCN18410};
(84) \citet{BarthelmyGCN18396};
(85) \citet{UkwattaGCN18382};
(86) \citet{StamatikosGCN18387};
(87) \citet{SakamotoGCN18368};
(88) \citet{RobertsGCN18358};
(89) \citet{GolenetskiiGCN18356};
(90) \citet{Augusto2016};
(91) \citet{PalmerGCN18328};
(92) \citet{BissaldiGCN18299};
(93) \citet{MarkwardtGCN18292};
(94) \citet{Zheng2009};
(95) \citet{LienGCN18268};
(96) \citet{KrimmGCN18256};
(97) \citet{GolenetskiiGCN18259};
(98) \citet{RobertsGCN18229};
(99) \citet{CummingsGCN18232};
(100) \citet{BarthelmyGCN18223};
(101) \citet{UkwattaGCN18214};
(102) \citet{BissaldiGCN18201};
(103) \citet{DeliaGCN18187};
(104) \citet{StamatikosGCN18196};
(105) \citet{RobertsGCN18190};
(106) \citet{YuGCN18178};
(107) \citet{SakamotoGCN18170};
(108) \citet{Cano2017};
(109) \citet{PalmerGCN18157};
(110) \citet{MarkwardtGCN18148};
(111) \citet{YuGCN18149};
(112) \citet{JenkeGCN18130};
(113) \citet{KrimmGCN18129};
(114) \citet{BarthelmyGCN18110};
(115) \citet{UkwattaGCN18091};
(116) \citet{TanvirGCN18080};
(117) \citet{StamatikosGCN18086};
(118) \citet{YounesGCN18081};
(119) \citet{GolenetskiiGCN18073};
(120) \citet{VeresGCN18066};
(121) \citet{PalmerGCN18055};
(122) \citet{MarkwardtGCN18048};
(123) \citet{BissaldiGCN18041};
(124) \citet{LienGCN18038};
(125) \citet{LienGCN18034};
(126) \citet{KrimmGCN18020};
(127) \citet{JenkeGCN18015};
(128) \citet{CummingsGCN18013};
(129) \citet{BarthelmyGCN18002};
(130) \citet{JenkeGCN17994};
(131) \citet{YuGCN17975};
(132) \citet{UkwattaGCN17973};
(133) \citet{StamatikosGCN17941};
(134) \citet{SakamotoGCN17930};
(135) \citet{GolenetskiiGCN17918};
(136) \citet{PalmerGCN17907};
(137) \citet{RobertsGCN17906};
(138) \citet{CummingsGCN17895};
(139) \citet{MarkwardtGCN17890};
(140) \citet{YuGCN17891};
(141) \citet{LienGCN17880};
(142) \citet{YuGCN17863};
(143) \citet{FujinumaGCN17875};
(144) \citet{DeGCN17822};
(145) \citet{RobertsGCN17819};
(146) \citet{LienGCN17814};
(147) \citet{YounesGCN17813};
(148) \citet{BurnsGCN17807};
(149) \citet{KrimmGCN17795};
(150) \citet{RobertsGCN17793};
(151) \citet{Zhang2014};
(152) \citet{CummingsGCN17776};
(153) \citet{BaumgartnerGCN17774};
(154) \citet{BarthelmyGCN17761};
(155) \citet{Siellez2016};
(156) \citet{UkwattaGCN17740};
(157) \citet{GolenetskiiGCN17727};
(158) \citet{Deng2016};
(159) \citet{StamatikosGCN17701};
(160) \citet{SakamotoGCN17675};
(161) \citet{Dichiara2016};
(162) \citet{ZhangGCN17674};
(163) \citet{StanbroGCN17658};
(164) \citet{PalmerGCN17637};
(165) \citet{vonGCN17623};
(166) \citet{MarkwardtGCN17628};
(167) \citet{LienGCN17604};
(168) \citet{KrimmGCN17596};
(169) \citet{CummingsGCN17581};
(170) \citet{YuGCN17579};
(171) \citet{BaumgartnerGCN17562};
(172) \citet{RobertsGCN17561};
(173) \citet{BarthelmyGCN17539};
(174) \citet{StamatikosGCN17516};
(175) \citet{Gorbovskoy2016};
(176) \citet{BurnsGCN17525};
(177) \citet{SakamotoGCN17519};
(178) \citet{ConnaughtonGCN17511};
(179) \citet{MarkwardtGCN17491};
(180) \citet{vonGCN17481};
(181) \citet{MereghettiGCN17476};
(182) \citet{KrimmGCN17471};
(183) \citet{CummingsGCN17457};
(184) \citet{BaumgartnerGCN17445};
(185) \citet{BurnsGCN17432};
(186) \citet{BarthelmyGCN17426};
(187) \citet{StamatikosGCN17406};
(188) \citet{BurnsGCN17408};
(189) \citet{UkwattaGCN17410};
(190) \citet{PelassaGCN17389};
(191) \citet{PelassaGCN17388};
(192) \citet{SakamotoGCN17401};
(193) \citet{PalmerGCN17374};
(194) \citet{JenkeGCN17364};
(195) \citet{StanbroGCN17353};
(196) \citet{GolenetskiiGCN17351};
(197) \citet{WisemanGCN17336};
(198) \citet{MarkwardtGCN17330};
(199) \citet{Sang2016};
(200) \citet{LienGCN17329};
(201) \citet{vonGCN17319};
(202) \citet{BurnsGCN17328};
(203) \citet{StanbroGCN17308};
(204) \citet{StanbroGCN17295};
(205) \citet{StanbroGCN17292};
(206) \citet{CummingsGCN17274};
(207) \citet{Xie2016};
(208) \citet{Fong2015};
(209) \citet{StanbroGCN17276};
(210) \citet{BaumgartnerGCN17266};
(211) \citet{CummingsGCN17256};
(212) \citet{RobertsGCN17249};
(213) \citet{BarthelmyGCN17239};
(214) \citet{JenkeGCN17241};
(215) \citet{JenkeGCN17220};
(216) \citet{UkwattaGCN17213};
(217) \citet{YuGCN17216};
(218) \citet{Ghirlanda2017};
(219) \citet{Lin2016};
(220) \citet{StamatikosGCN17202};
(221) \citet{JenkeGCN17189};
(222) \citet{SakamotoGCN17176};
(223) \citet{PalmerGCN17175};
(224) \citet{Liu2015};
(225) \citet{BurnsGCN17150};
(226) \citet{MarkwardtGCN17140};
(227) \citet{RobertsGCN17143};
(228) \citet{RobertsGCN17133};
(229) \citet{LienGCN17127};
(230) \citet{KrimmGCN17083};
(231) \citet{Cucchiara2015};
(232) \citet{JenkeGCN17094};
(233) \citet{CummingsGCN17046};
(234) \citet{BaumgartnerGCN17044};
(235) \citet{CummingsGCN17020};
(236) \citet{ZhangGCN17021};
(237) \citet{BarthelmyGCN17011};
(238) \citet{UkwattaGCN17010};
(239) \citet{RobertsGCN17001};
(240) \citet{RobertsGCN16987};
(241) \citet{XuGCN16983};
(242) \citet{RobertsGCN16971};
(243) \citet{Ruffini2016};
(244) \citet{StamatikosGCN16960};
(245) \citet{SakamotoGCN16942};
(246) \citet{PalmerGCN16935};
(247) \citet{MarkwardtGCN16927};
(248) \citet{Turpin2016};
(249) \citet{LienGCN16916};
(250) \citet{vonGCN16905};
(251) \citet{KrimmGCN16893};
(252) \citet{RobertsGCN16889};
(253) \citet{CummingsGCN16892};
(254) \citet{PelassaGCN16900};
(255) \citet{BaumgartnerGCN16870};
(256) \citet{vonGCN16850};
(257) \citet{BarthelmyGCN16845};
(258) \citet{UkwattaGCN16839};
(259) \citet{PelassaGCN16835};
(260) \citet{StamatikosGCN16827};
(261) \citet{SakamotoGCN16799};
(262) \citet{ZhangGCN16798};
(263) \citet{GolenetskiiGCN16807};
(264) \citet{PalmerGCN16768};
(265) \citet{Troja2016};
(266) \citet{ZhangGCN16775};
(267) \citet{JenkeGCN16762};
(268) \citet{MarkwardtGCN16756};
(269) \citet{vonGCN16754};
(270) \citet{GolenetskiiGCN16755};
(271) \citet{LienGCN16736};
(272) \citet{KrimmGCN16721};
(273) \citet{RobertsGCN16708};
(274) \citet{CummingsGCN16699};
(275) \citet{RobertsGCN16700};
(276) \citet{RobertsGCN16680};
(277) \citet{ZhangGCN16669};
(278) \citet{Lipunov2016};
(279) \citet{PelassaGCN16658};
(280) \citet{BaumgartnerGCN16652};
(281) \citet{StanbroGCN16636};
(282) \citet{BurnsGCN16626};
(283) \citet{BarthelmyGCN16615};
(284) \citet{UkwattaGCN16613};
(285) \citet{CummingsGCN16598};
(286) \citet{JenkeGCN16599};
(287) \citet{StamatikosGCN16584};
(288) \citet{ZhangGCN16590};
(289) \citet{Bhat2016};
(290) \citet{BurnsGCN16579};
(291) \citet{SakamotoGCN16573};
(292) \citet{Littlejohns2015};
(293) \citet{ZhangGCN16561};
(294) \citet{ZhangGCN16537};
(295) \citet{Pescalli2016};
(296) \citet{CummingsGCN16481};
(297) \citet{StamatikosGCN16462};
(298) \citet{YounesGCN16452};
(299) \citet{vonGCN16450};
(300) \citet{YounesGCN16447};
(301) \citet{ConnaughtonGCN16419};
(302) \citet{PalmerGCN16423};
(303) \citet{BarthelmyGCN16404};
(304) \citet{MarkwardtGCN16402};
(305) \citet{GolenetskiiGCN16389};
(306) \citet{StanbroGCN16385};
(307) \citet{KrimmGCN16370};
(308) \citet{Li2016ApJS};
(309) \citet{Cano2015};
(310) \citet{CummingsGCN16354};
(311) \citet{GolenetskiiGCN16351};
(312) \citet{CummingsGCN16346};
(313) \citet{StanbroGCN16347};
(314) \citet{GolenetskiiGCN16328};
(315) \citet{StanbroGCN16319};
(316) \citet{UkwattaGCN16306};
(317) \citet{StamatikosGCN16292};
(318) \citet{StamatikosGCN16284};
(319) \citet{Melandri2015};
(320) \citet{PalmerGCN16240};
(321) \citet{KruhlerAA2015};
(322) \citet{YuGCN16203};
(323) \citet{YuGCN16199};
(324) \citet{Kopac2015};
(325) \citet{YounesGCN16189};
(326) \citet{KrimmGCN16186};
(327) \citet{JenkeGCN16115};
(328) \citet{CummingsGCN16111};
(329) \citet{BarthelmyGCN16105};
(330) \citet{UkwattaGCN16103};
(331) \citet{JenkeGCN16084};
(332) \citet{YuGCN16081};
(333) \citet{CummingsGCN16073};
(334) \citet{StamatikosGCN16063};
(335) \citet{PelassaGCN16066};
(336) \citet{vonGCN16042};
(337) \citet{YuGCN16032};
(338) \citet{GolenetskiiGCN16025};
(339) \citet{YounesGCN16014};
(340) \citet{MarkwardtGCN15996};
(341) \citet{StanbroGCN15977};
(342) \citet{StanbroGCN15975};
(343) \citet{KrimmGCN15962};
(344) \citet{GolenetskiiGCN15943};
(345) \citet{FitzpatrickGCN15935};
(346) \citet{CummingsGCN15934};
(347) \citet{BarthelmyGCN15908};
(348) \citet{UkwattaGCN15906};
(349) \citet{ZhangGCN15866};
(350) \citet{BarthelmyGCN15847};
(351) \citet{StamatikosGCN15836};
(352) \citet{CummingsGCN15820};
(353) \citet{vonGCN15811};
(354) \citet{vonGCN15790};
(355) \citet{PalmerGCN15774};
(356) \citet{MarkwardtGCN15769};
(357) \citet{XiongGCN15751};
(358) \citet{KrimmGCN15738};
(359) \citet{Greiner2015};
(360) \citet{vonGCN15716};
(361) \citet{YounesGCN15709};
(362) \citet{XiongGCN15688};
(363) \citet{XiongGCN15687};
(364) \citet{BaumgartnerGCN15664};
(365) \citet{vonGCN15651};
(366) \citet{Racusin2016};
(367) \citet{BarthelmyGCN15620};
(368) \citet{StamatikosGCN15613};
(369) \citet{PelassaGCN15599};
(370) \citet{GolenetskiiGCN15754};
(371) \citet{vonGCN15591};
(372) \citet{SakamotoGCN15584};
(373) \citet{CollazziGCN15565};
(374) \citet{YuGCN15554};
(375) \citet{vonGCN15528};
(376) \citet{MarkwardtGCN15521};
(377) \citet{PelassaGCN15573};
(378) \citet{GolenetskiiGCN15549};
(379) \citet{LienGCN15508};
(380) \citet{CollazziGCN15503};
(381) \citet{KrimmGCN15499};
(382) \citet{BarthelmyGCN15457};
(383) \citet{FitzpatrickGCN15434};
(384) \citet{BarthelmyGCN15456};
(385) \citet{Urata2015};
(386) \citet{vonGCN15401};
(387) \citet{vonGCN15396};
(388) \citet{ZhangGCN15382};
(389) \citet{BarthelmyGCN15370};
(390) \citet{ZhangGCN15360};
(391) \citet{UkwattaGCN15354};
(392) \citet{FitzpatrickGCN15332};
(393) \citet{JenkeGCN15331};
(394) \citet{XiongGCN15315};
(395) \citet{Golkhou2014};
(396) \citet{SatoGCN15298};
(397) \citet{vonGCN15300};
(398) \citet{CummingsGCN15293};
(399) \citet{PalmerGCN15272};
(400) \citet{Greiner2014};
(401) \citet{LienGCN15234};
(402) \citet{ZhangGCN15219};
(403) \citet{XiongGCN15175};
(404) \citet{BaumgartnerGCN15163};
(405) \citet{Cano2014};
(406) \citet{Zhang2016};
(407) \citet{Beskin2015};
(408) \citet{CollazziGCN15129};
(409) \citet{UkwattaGCN15116};
(410) \citet{GolenetskiiGCN15125};
(411) \citet{StamatikosGCN15108};
(412) \citet{FitzpatrickGCN15104};
(413) \citet{GolenetskiiGCN15095};
(414) \citet{PalmerGCN15092};
(415) \citet{MarkwardtGCN15083};
(416) \citet{LienGCN15076};
(417) \citet{YuGCN15070};
(418) \citet{GoldsteinGCN15053};
(419) \citet{BarthelmyGCN15041};
(420) \citet{UkwattaGCN15031};
(421) \citet{GolenetskiiGCN15023};
(422) \citet{StamatikosGCN15016};
(423) \citet{FoleyGCN15011};
(424) \citet{CollazziGCN15005};
(425) \citet{FitzpatrickGCN14999};
(426) \citet{Wang2014};
(427) \citet{PelassaGCN14962};
(428) \citet{Contopoulos2014};
(429) \citet{StamatikosGCN14952};
(430) \citet{ByrneGCN14941};
(431) \citet{SakamotoGCN14942};
(432) \citet{PalmerGCN14925};
(433) \citet{ByrneGCN14940};
(434) \citet{BarthelmyGCN14899};
(435) \citet{XiongGCN14903};
(436) \citet{Wang2016};
(437) \citet{GolenetskiiGCN14872};
(438) \citet{FitzpatrickGCN14839};
(439) \citet{KrimmGCN14833};
(440) \citet{Yu2015A};
(441) \citet{YuGCN14801};
(442) \citet{Yu2015B};
(443) \citet{GolenetskiiGCN14809};
(444) \citet{CollazziGCN14765};
(445) \citet{Berger2014};
(446) \citet{BarthelmyGCN14736};
(447) \citet{KrimmGCN14726};
(448) \citet{Jeong2014};
(449) \citet{GolenetskiiGCN14698};
(450) \citet{BarthelmyGCN14693};
(451) \citet{KrimmGCN14694};
(452) \citet{JenkeGCN14663};
(453) \citet{CummingsGCN14659};
(454) \citet{UkwattaGCN14636};
(455) \citet{SakamotoGCN14613};
(456) \citet{BurgessGCN14583};
(457) \citet{vonGCN14560};
(458) \citet{PalmerGCN14554};
(459) \citet{vonGCN14530};
(460) \citet{Goldstein2016};
(461) \citet{Song2016};
(462) \citet{Gao2015B};
(463) \citet{Japelj2014};
(464) \citet{Niino2016};
(465) \citet{vonGCN14442};
(466) \citet{RauGCN14435};
(467) \citet{Zhang2015A};
(468) \citet{KrimmGCN14399};
(469) \citet{Wei2014};
(470) \citet{Serino2014};
(471) \citet{ChaplinGCN14346};
(472) \citet{BarthelmyGCN14343};
(473) \citet{TierneyGCN14329};
(474) \citet{GolenetskiiGCN14356};
(475) \citet{BarthelmyGCN14315};
(476) \citet{BarthelmyGCN14296};
(477) \citet{Davanzo2014};
(478) \citet{NorrisGCN14306};
(479) \citet{XiongGCN14283};
(480) \citet{PelassaGCN14271};
(481) \citet{GolenetskiiGCN14275};
(482) \citet{YuGCN14261};
(483) \citet{ChaplinGCN14235};
(484) \citet{ChaplinGCN14236};
(485) \citet{UkwattaGCN14197};
(486) \citet{GoldsteinGCN14189};
(487) \citet{PalmerGCN14163};
(488) \citet{PelassaGCN14154};
(489) \citet{BarthelmyGCN14146};
(490) \citet{GolenetskiiGCN14135};
(491) \citet{MarkwardtGCN14133};
(492) \citet{Lv2014A};
(493) \citet{BaumgartnerGCN14111};
(494) \citet{GolenetskiiGCN14104};
(495) \citet{Liang2015};
(496) \citet{BaumgartnerGCN14111};
(497) \citet{Heussaff2013};
(498) \citet{TierneyGCN14039};
(499) \citet{GolenetskiiGCN14022};
(500) \citet{GolenetskiiGCN14005};
(501) \citet{GolenetskiiGCN13979};
(502) \citet{BaumgartnerGCN13961};
(503) \citet{BaumgartnerGCN13942};
(504) \citet{GoldsteinGCN13951};
(505) \citet{Kaneko2015};
(506) \citet{SakamotoGCN13917};
(507) \citet{UkwattaGCN13880};
(508) \citet{BarthelmyGCN13869};
(509) \citet{PelassaGCN13872};
(510) \citet{XiongGCN13860};
(511) \citet{SakamotoGCN13836};
(512) \citet{PalmerGCN13828};
(513) \citet{MarkwardtGCN13807};
(514) \citet{GolenetskiiGCN13789};
(515) \citet{GolenetskiiGCN13787};
(516) \citet{BarthelmyGCN13784};
(517) \citet{GolenetskiiGCN13781};
(518) \citet{PelassaGCN13773};
(519) \citet{PelassaGCN13771};
(520) \citet{GolenetskiiGCN13758};
(521) \citet{GruberGCN13754};
(522) \citet{McGlynnGCN13741};
(523) \citet{GolenetskiiGCN13707};
(524) \citet{SakamotoGCN13689};
(525) \citet{GolenetskiiGCN13674};
(526) \citet{PalmerGCN13669};
(527) \citet{Xu2013};
(528) \citet{GolenetskiiGCN13676};
(529) \citet{BarthelmyGCN13659};
(530) \citet{GolenetskiiGCN13621};
(531) \citet{BarthelmyGCN13633};
(532) \citet{CummingsGCN13604};
(533) \citet{BarthelmyGCN13594};
(534) \citet{BaumgartnerGCN13581};
(535) \citet{BarthelmyGCN13572};
(536) \citet{UkwattaGCN13568};
(537) \citet{GolenetskiiGCN13552};
(538) \citet{MarkwardtGCN13535};
(539) \citet{Dereli2015};
(540) \citet{CummingsGCN13481};
(541) \citet{BaumgartnerGCN13472};
(542) \citet{Von2014};
(543) \citet{UkwattaGCN13450};
(544) \citet{GolenetskiiGCN13440};
(545) \citet{GolenetskiiGCN13445};
(546) \citet{BarthelmyGCN13404};
(547) \citet{SakamotoGCN13405};
(548) \citet{GolenetskiiGCN13378};
(549) \citet{PalmerGCN13360};
(550) \citet{GolenetskiiGCN13354};
(551) \citet{GolenetskiiGCN13351};
(552) \citet{GolenetskiiGCN13341};
(553) \citet{MarkwardtGCN13333};
(554) \citet{GruberGCN13339};
(555) \citet{CummingsGCN13310};
(556) \citet{GolenetskiiGCN13315};
(557) \citet{BaumgartnerGCN13291};
(558) \citet{GolenetskiiGCN13295};
(559) \citet{GolenetskiiGCN13272};
(560) \citet{GolenetskiiGCN13268};
(561) \citet{Laskar2015};
(562) \citet{YounesGCN13214};
(563) \citet{SakamotoGCN13195};
(564) \citet{CollazziGCN13194};
(565) \citet{PalmerGCN13186};
(566) \citet{GolenetskiiGCN13158};
(567) \citet{MarkwardtGCN13154};
(568) \citet{Delia2014};
(569) \citet{GolenetskiiGCN13100};
(570) \citet{GolenetskiiGCN13103};
(571) \citet{GolenetskiiGCN13074};
(572) \citet{BarthelmyGCN13052};
(573) \citet{UkwattaGCN13041};
(574) \citet{StamatikosGCN13039};
(575) \citet{PalmerGCN13007};
(576) \citet{NorrisGCN13015};
(577) \citet{PelassaGCN13013};
(578) \citet{MarkwardtGCN12998};
(579) \citet{GolenetskiiGCN12996};
(580) \citet{BarthelmyGCN12983};
(581) \citet{CummingsGCN12968};
(582) \citet{BarthelmyGCN12963};
(583) \citet{BarthelmyGCN12955};
(584) \citet{BaumgartnerGCN12946};
(585) \citet{RauGCN12950};
(586) \citet{UkwattaGCN12924};
(587) \citet{Bosnjak2014};
(588) \citet{BarthelmyGCN12889};
(589) \citet{Perley2016A};
(590) \citet{GolenetskiiGCN12824};
(591) \citet{BarthelmyGCN12815};
(592) \citet{RauGCN12806};
(593) \citet{GolenetskiiGCN12701};
(594) \citet{BarthelmyGCN12689};
(595) \citet{Wei2016};
(596) \citet{UkwattaGCN12671};
(597) \citet{StamatikosGCN12651};
(598) \citet{Gao2016};
(599) \citet{Kann2016};
(600) \citet{Stratta2013};
(601) \citet{GolenetskiiGCN12627};
(602) \citet{MarkwardtGCN12620};
(603) \citet{BarthelmyGCN12602};
(604) \citet{CummingsGCN12581};
(605) \citet{Margutti2012};
(606) \citet{BaumgartnerGCN12551};
(607) \citet{GolenetskiiGCN12532};
(608) \citet{TierneyGCN12529};
(609) \citet{PalmerGCN12504};
(610) \citet{BarthelmyGCN12507};
(611) \citet{MarkwardtGCN12485};
(612) \citet{KrimmGCN12476};
(613) \citet{SakamotoGCN12464};
(614) \citet{SakamotoGCN12477};
(615) \citet{CummingsGCN12457};
(616) \citet{GolenetskiiGCN12456};
(617) \citet{BarthelmyGCN12445};
(618) \citet{Sparre2014};
(619) \citet{Michalowski2015};
(620) \citet{Greiner2016};
(621) \citet{BarthelmyGCN12399};
(622) \citet{FitzpatrickGCN12386};
(623) \citet{Frederiks2013};
(624) \citet{SakamotoGCN12312};
(625) \citet{GolenetskiiGCN12301};
(626) \citet{PalmerGCN12292};
(627) \citet{GolenetskiiGCN12278};
(628) \citet{GolenetskiiGCN12249};
(629) \citet{Tsutsui2013B};
(630) \citet{Ackermann2013};
(631) \citet{CummingsGCN12201};
(632) \citet{BaumgartnerGCN12175};
(633) \citet{XiongGCN12073};
(634) \citet{BaumgartnerGCN12049};
(635) \citet{BarthelmyGCN12035};
(636) \citet{UkwattaGCN12030};
(637) \citet{StamatikosGCN12016};
(638) \citet{GolenetskiiGCN12019};
(639) \citet{Nava2012};
(640) \citet{Melandri2012};
(641) \citet{Allison2017};
(642) \citet{LinGCN11952};
(643) \citet{GolenetskiiGCN11951};
(644) \citet{CummingsGCN11937};
(645) \citet{Virgili2012};
(646) \citet{BarthelmyGCN11921};
(647) \citet{UkwattaGCN11902};
(648) \citet{GolenetskiiGCN11893};
(649) \citet{PalmerGCN11818};
(650) \citet{BarthelmyGCN11811};
(651) \citet{MarkwardtGCN11804};
(652) \citet{KrimmGCN11793};
(653) \citet{BarthelmyGCN11783};
(654) \citet{CummingsGCN11776};
(655) \citet{BarthelmyGCN11757};
(656) \citet{BaumgartnerGCN11764};
(657) \citet{Ukwatta2012};
(658) \citet{UkwattaGCN11703};
(659) \citet{StamatikosGCN11691};
(660) \citet{SakamotoGCN11677};
(661) \citet{vonGCN11671};
(662) \citet{Cucchiara2011};
(663) \citet{Chandra2012};
(664) \citet{BarthelmyGCN11557};
(665) \citet{CummingsGCN11546};
(666) \citet{StamatikosGCN11527};
(667) \citet{Fermi2012};
(668) \citet{Balazs2015};
(669) \citet{Liu2016};
(670) \citet{Kopac2012};
(671) \citet{Tunnicliffe2014};
(672) \citet{FoleyGCN11434};
(673) \citet{FrederiksGCN11439};
(674) \citet{BaumgartnerGCN11414};
(675) \citet{GolenetskiiGCN11408};
(676) \citet{BarthelmyGCN11388};
(677) \citet{GolenetskiiGCN11384};
(678) \citet{Gorbovskoy2012};
(679) \citet{SakamotoGCN11358};
(680) \citet{GolenetskiiGCN11350};
(681) \citet{KrimmGCN11312};
(682) \citet{CummingsGCN11289};
(683) \citet{BaumgartnerGCN11281};
(684) \citet{BarthelmyGCN11218};
(685) \citet{UkwattaGCN11207};
(686) \citet{StamatikosGCN11202};
(687) \citet{SakamotoGCN11169};
(688) \citet{Ghirlanda2012};
(689) \citet{Barniol2014};
(690) \citet{Geng2016};
(691) \citet{CummingsGCN11069};
(692) \citet{BarthelmyGCN11058};
(693) \citet{vonGCN11002};
(694) \citet{SakamotoGCN10993};
(695) \citet{PalmerGCN10988};
(696) \citet{Ahlgren2015};
(697) \citet{MarkwardtGCN10972};
(698) \citet{Piranomonte2015};
(699) \citet{KrimmGCN10964};
(700) \citet{BarthelmyGCN10896};
(701) \citet{Japelj2016};
(702) \citet{Volnova2014};
(703) \citet{Kruhler2011};
(704) \citet{Arcodia2016};
(705) \citet{CummingsGCN10803};
(706) \citet{BaumgartnerGCN10801};
(707) \citet{AfonsoGCN10782};
(708) \citet{StamatikosGCN10732};
(709) \citet{Ripa2011};
(710) \citet{CummingsGCN10660};
(711) \citet{Zhang2011};
(712) \citet{Nava2011};
(713) \citet{Levesque2010B};
(714) \citet{PalmerGCN10509};
(715) \citet{Greiner2011};
(716) \citet{BaumgartnerGCN10501};
(717) \citet{BarthelmyGCN10417};
(718) \citet{SakamotoGCN10371};
(719) \citet{PalmerGCN10351};
(720) \citet{Yonetoku2010};
(721) \citet{Sakamoto2011};
(722) \citet{Filgas2011};
(723) \citet{SakamotoGCN10180};
(724) \citet{Nemmen2012};
(725) \citet{Lu2012B};
(726) \citet{Wang2013};
(727) \citet{Galli2013};
(728) \citet{Robertson2012};
(729) \citet{Guetta2011};
(730) \citet{Butler2010};
(731) \citet{Kopac2013};
(732) \citet{Minaev2014};
(733) \citet{Liang2009};
(734) \citet{Qin2013B};
(735) \citet{Zhang2009};
(736) \citet{Krimm2009};
(737) \citet{Perley2013};
(738) \citet{Lv2012};
(739) \citet{Lien2016};
(740) \citet{Dichiara2013A};
(741) \citet{Nardini2011};
(742) \citet{Li2012B};
(743) \citet{Golkhou2015};
(744) \citet{Vianello2009};
(745) \citet{Kann2010};
(746) \citet{Huja2009};
(747) \citet{ThoeneGCN8135};
(748) \citet{Bhat2012};
(749) \citet{Xiao2011};
(750) \citet{Schady2011};
(751) \citet{Kruhler2009};
(752) \citet{Ukwatta2010};
(753) \citet{Ripa2009};
(754) \citet{Nava2008};
(755) \citet{Perley2009};
(756) \citet{Nardini2010};
(757) \citet{Zheng2009};
(758) \citet{Cenko2009};
(759) \citet{Minaev2010};
(760) \citet{Rossi2008};
(761) \citet{Hjorth2012B};
(762) \citet{Gehrels2008};
(763) \citet{Nysewander2009};
(764) \citet{Foley2008};
(765) \citet{Xiao2009B};
(766) \citet{Graham2013};
(767) \citet{Kasliwal2008};
(768) \citet{Schady2010};
(769) \citet{Butler2007};
(770) \citet{Perley2015};
(771) \citet{Contini2016};
(772) \citet{Ghisellini2009};
(773) \citet{Covino2016};
(774) \citet{Gao2010};
(775) \citet{Savaglio2009};
(776) \citet{Bellm2008};
(777) \citet{Rizzuto2007};
(778) \citet{Campisi2008};
(779) \citet{Mosquera2008};
(780) \citet{Willingale2007B};
(781) \citet{Schaefer2007};
(782) \citet{Rykoff2009};
(783) \citet{Melandri2008};
(784) \citet{Fong2010};
(785) \citet{De2015};
(786) \citet{Graham2017};
(787) \citet{Pelangeon2008};
(788) \citet{Leibler2010};
(789) \citet{Collazzi2008};
(790) \citet{Xiao2009A};
(791) \citet{Levesque2010A};
(792) \citet{Mannucci2011};
(793) \citet{Tsutsui2013A};
(794) \citet{Zaninoni2013};
(795) \citet{Ghirlanda2007};
(796) \citet{Guidorzi2016};
(797) \citet{TanvirGCN3031};
(798) \citet{Li2015A};
(799) \citet{Sakamoto2008};
(800) \citet{Rau2005};
(801) \citet{Guidorzi2005};
(802) \citet{Friedman2005};
(803) \citet{Sako2005};
(804) \citet{Sakamoto2005};
(805) \citet{Li2008};
(806) \citet{Mazets2002};
(807) \citet{Barraud2003};
(808) \citet{Frontera2004};
(809) \citet{Frontera2009};
(810) \citet{De2006};
(811) \citet{Guidorzi2011};
(812) \citet{Atteia2003};
(813) \citet{Christensen2004};
(814) \url{https://gammaray.nsstc.nasa.gov/batse/grb/catalog/current/index.html};
(815) \citet{Shahmoradi2010};
(816) \citet{Goldstein2013};
(817) \citet{Ashcraft2007};
(818) \citet{Ghirlanda2009};
(819) \citet{Bloom2002};
(820) \citet{Jimenez2001};
(821) \citet{Kaneko2006};
(822) \citet{Ramirez2002};
(823) \citet{Smith2002};
(824) \citet{Reichart2001};
(825) \citet{Frail2001};
(826) \citet{Schaefer2003};
(827) \citet{Sokolov2001};
(828) \citet{Simic2012};
(829) \citet{Fenimore2000};
(830) \citet{Svensson2010};
(831) \citet{Bosnjak2006};
(832) \citet{Postnov2000};
(833) \citet{Bloom2001};
(834) \citet{Berger2003};
(835) \citet{Fragile2004};
(836) \citet{Amati2008};
(837) \citet{Wei2010};
(838) \citet{Paciesas1999};
(839) \citet{Lloyd2002};
(840) \citet{Peng2012};
(841) \citet{Preece2000};
(842) \citet{Meegan1996};
(843) \citet{Sazonov1998};
(844) \citet{Hakkila2007};
(845) \citet{Fishman1994};
(846) \citet{Terekhov1994}
\startlongtable
\begin{deluxetable}{cccccccc}
\tablecaption{Imputation results \label{tab:imputation}}
\tablehead{
\colhead{Parameters} & \colhead{RIV} & \colhead{FMI} & \colhead{RE} & \colhead{Parameters} & \colhead{RIV} & \colhead{FMI} & \colhead{RE}
}
\startdata
$T_{\rm 901}$ & 0.00000358  & 0.00000358  &  1  & $T_{\rm 902}$ & 0.000126  & 0.000126  &  1 \\
\hline
$T_{\rm 501}$ &  0.000143 & 0.000143  & 1  & $T_{\rm 502}$ & 0.0000796  & 0.0000796  & 1   \\
\hline
$T_{\rm R451}$ & 0.0122  &  0.0122 &  0.998  & $T_{\rm R452}$ & 0.00172  &  0.00172 & 1  \\
 \hline
$variability_{\rm 11}$ &  0.0137 &  0.0136 &   0.997 & $variability_{\rm 12}$ & 0.00491  &  0.0049 & 0.999  \\
 \hline
$F_{\rm g1}$ &  0.0000144 &  0.0000144 &  1  & $F_{\rm g2}$ & 0.000317  & 0.000317  & 1  \\
 \hline
HR1 &  0.000454 &  0.000454 & 1  & HR2 &  0.000354 & 0.000354  &  1   \\
 \hline
$E_{\rm \gamma,iso1}$ & 0.00588  & 0.00587  &  0.999  & $E_{\rm \gamma,iso2}$ & 0.00071  & 0.00071  &  1 \\
 \hline
$F_{\rm pk11}$ & 0.000259  &  0.000259 &  1  & $F_{\rm pk12}$ &  0.000901 & 0.000901  &  1 \\
 \hline
$F_{\rm pk21}$ & 0.000324  &  0.000324 &  1  & $F_{\rm pk22}$ &  0.000849 &  0.000849 &    1\\
 \hline
$P_{\rm pk11}$ &  0.000205 &  0.000205 &  1  & $P_{\rm pk12}$ &  0.000274 &  0.000274 & 1  \\
 \hline
$-\alpha_{\rm band1}$ & 0.000911  &  0.000911 &  1  & $-\alpha_{\rm band2}$ & 0.000289  & 0.000289  & 1  \\
 \hline
$-\beta_{\rm band1}$ & 0.0017  &  0.0017 & 1  & $-\beta_{\rm band2}$ &  0.000599 & 0.000599  &  1  \\
 \hline
$E_{\rm p,band1}$ &  0.00176 &  0.00176 & 1  & $E_{\rm p,band2}$ &  0.000471 &  0.000471 & 1   \\
 \hline
$-\alpha_{\rm cpl1}$ & 0.00186  &  0.00185 & 1  & $-\alpha_{\rm cpl2}$ & 0.000343  &  0.000343 &  1  \\
 \hline
$E_{\rm p,cpl1}$ &  0.00018 &  0.00018 &  1 & $E_{\rm p,cpl2}$ &  0.000244 &  0.000244 &  1  \\
 \hline
$\theta_{\rm j1}$ & 0.00547  & 0.00546  &  0.999  & $\theta_{\rm j2}$ & 0.00924  & 0.0092  &   0.998 \\
 \hline
$\Gamma_{01}$ &  0.111 &  0.105 &  0.979  & $\Gamma_{02}$ &  0.0273 & 0.0269  & 0.995  \\
 \hline
$t_{\rm pkOpt1}$ &  0.044 & 0.043  &  0.991 & $t_{\rm pkOpt2}$ & 0.0147  & 0.0146  & 0.997   \\
 \hline
$F_{\rm X11hr1}$ & 0.0131  &  0.0131 &  0.997 & $F_{\rm X11hr2}$ & 0.0388  & 0.038  &  0.992   \\
 \hline
$t_{\rm radio,pk1}$ & 0.0197  &  0.0195 &  0.996  & $t_{\rm radio,pk2}$ & 0.0211  & 0.0209  & 0.996  \\
 \hline
$F_{\rm radio,pk1}$ & 0.111  &  0.104 &  0.98 & $F_{\rm radio,pk2}$ & 0.0227  &  0.0224 &  0.996   \\
 \hline
offset1 & 0.0161  &  0.016 &   0.997 & offset2 & 0.0119  & 0.0118  &  0.998  \\
 \hline
metallicity1 & 0.0569  &  0.0552 &  0.989  & metallicity2 &  0.013 & 0.0129  &  0.997  \\
 \hline
$N_{\rm H1}$ & 0.0016  &  0.0016 & 1  & $N_{\rm H2}$ & 0.00767  & 0.00764  &  0.998  \\
 \hline
$A_{\rm V1}$ & 0.00734  & 0.00731  &  0.999 & $A_{\rm V2}$ & 0.0319  & 0.0313  &  0.994  \\
 \hline
SFR1 & 0.0231  & 0.0228  &  0.995  & SFR2 &  0.0331 &  0.0325 & 0.994  \\
 \hline
$\log Mass1$ &  0.0157 &  0.0156 &  0.997  & $\log Mass2$ &  0.0082 & 0.00816  & 0.998  \\
 \hline
$T_{\rm 90,i1}$ &  0.00311 &  0.0031 &  0.999  & $T_{\rm 90,i2}$ & 0.0000847  &  0.0000847 & 1  \\
 \hline
$T_{\rm 50,i1}$ & 0.00634  & 0.00632  &  0.999  & $T_{\rm 50,i2}$ & 0.000355  & 0.000355  &  1 \\
 \hline
$T_{\rm R45,i1}$ & 0.00296  &  0.00296 &  0.999  & $T_{\rm R45,i2}$ & 0.00202  & 0.00202  &  1 \\
 \hline
$E_{\rm p,band,i1}$ & 0.00547  & 0.00546  &  0.999 & $E_{\rm p,band,i2}$ & 0.0104  & 0.0104  & 0.998    \\
 \hline
$E_{\rm p,cpl,i1}$ & 0.0061  & 0.00608  & 0.999  & $E_{\rm p,cpl,i2}$ & 0.00463  & 0.00462  &   0.999 \\
 \hline
$t_{\rm pkOpt,i1}$ &  0.0237 & 0.0234  &  0.995  & $t_{\rm pkOpt,i2}$ &  0.0196 & 0.0194  & 0.996  \\
 \hline
$t_{\rm radio,pk,i1}$ & 0.0416  &  0.0407 & 0.992  & $t_{\rm radio,pk,i2}$ & 0.0183  & 0.0182  & 0.996   \\
 \hline
 \enddata
\tablecomments{You can see Section \ref{sec:imputation} for the definations of RIV, FMI and RE. The description of every parameter is in Section \ref{sec:sample}}
\end{deluxetable}
\begin{deluxetable*}{cccccc}[b!]
\tablecaption{Correlation coefficient and correlation ratio results  \label{tab:coefficient}}
\tablehead{
\colhead{Parameters} & \colhead{ \tabincell{c} {$\log (1+z)$ \\ vs \\ $\log D_{\rm L}$} } & \colhead{ \tabincell{c} {$\log (1+z)$ \\ vs \\ $\log T_{\rm 90}$} } & \colhead{ \tabincell{c} {$\log (1+z)$ \\ vs \\ $\log T_{\rm 50}$} } & \colhead{ \tabincell{c} {$\log (1+z)$ \\ vs \\ $\log T_{\rm R45}$} } & \colhead{ \tabincell{c} {$\log (1+z)$ \\ vs \\ $variability_{1}$} }
}
\startdata
Pearson coefficient & 0.76$\pm$0.042 & 0.21$\pm$0.022 & 0.18$\pm$0.028 & 0.3$\pm$0.019 & 0.0019$\pm$0.05 \\
Pearson p-value & 1.7$\times 10^{-8}$ & 8.9$\times 10^{-7}$ & 7.6$\times 10^{-4}$ & 9$\times 10^{-9}$ & 9.8$\times 10^{-1}$ \\
Spearman coefficient & 0.91$\pm$0.012 & 0.2$\pm$0.016 & 0.17$\pm$0.015 & 0.24$\pm$0.013 & 0.066$\pm$0.059 \\
Spearman p-value & 4.8$\times 10^{-213}$ & 2.5$\times 10^{-6}$ & 1.5$\times 10^{-3}$ & 3.9$\times 10^{-6}$ & $4.1\times 10^{-1}$ \\
Kendall coefficient  & 0.83$\pm$0.011 & 0.13$\pm$0.011 & 0.11$\pm$0.01 &  0.17$\pm$0.0088 & 0.04$\pm$0.04 \\
Kendall p-value  & 3$\times 10^{-194}$ & 2.8$\times 10^{-6}$ & 1.2$\times 10^{-3}$ & 3.7$\times 10^{-6}$ & $4.5\times 10^{-1}$ \\
Correlation ratio  & 0.091$\pm$0.0067 & 0.66$\pm$0.0041 & 0.51$\pm$0.006 & 0.29$\pm$0.0039 & 0.71$\pm$0.044 \\
Cosine similarity  & 0.88$\pm$0.014 & 0.8$\pm$0.02 & 0.77$\pm$0.019  & 0.75$\pm$0.007 & 0.3$\pm$0.067 \\
   \hline
\enddata
\tablecomments{Correlation coefficients and correlation ratio between two different parameters. The definitions are in Section \ref{subsec:coefficient}. We considered all the errors using MC method \citep{Zou2017}. All the errors are in 1 $\sigma$ confidence level. All the results are in machine readable table. The description of every parameter is in Section \ref{sec:sample}.}
\end{deluxetable*}

\begin{deluxetable*}{ccccc}[b!]
\tablecaption{Linear regression results between two parameters  \label{tab:linear2}}
\tablehead{
\colhead{$y$} & \colhead{$x$} & \colhead{$b$} & \colhead{$a$} & \colhead{Adjusted $R^{2}$}
}
\startdata
$\log 1+z$  & $\log D_{\rm L}$ & 0.39$\pm$0.014 & 0.22$\pm$0.011 & 0.89 \\
$\log D_{\rm L}$  & $\log 1+z$ & 1.5 $\pm$0.19 & -0.13$\pm$0.083 & 0.89 \\
$\log 1+z$  & $\log T_{\rm 90}$ & 0.069$\pm$0.0066 & 0.31$\pm$0.012 & 0.064 \\
$\log T_{\rm 90}$  & $\log 1+z$ & 0.61$\pm$0.098 & 1.2$\pm$0.042 & 0.064 \\
$\log 1+z$  & $\log T_{\rm 50}$ & 0.058$\pm$0.0069 & 0.36$\pm$0.0078 & 0.049 \\
   \hline
\enddata
\tablecomments{$y$ is dependent variable, $x$ is independent variable, $b$ is the intercept of the linear model, $a$ is the linear coefficient of $x$. Adjusted $R^{2}$ is used to measure the goodness of our regression model, it means the percentage of variance explained considering the parameter freedom. All the errors are in 1 $\sigma$ error bars. All the results are in machine readable table. The description of every parameter is in Section \ref{sec:sample}.}
\end{deluxetable*}

\begin{deluxetable*}{ccccccc}[b!]
\tablecaption{Linear regression results between three parameters  \label{tab:linear3}}
\tablehead{
\colhead{$y$} & \colhead{$x_{1}$} & \colhead{$x_{2}$} & \colhead{$a_{1}$} & \colhead{$a_{2}$} & \colhead{$b$} & \colhead{Adjusted $R^{2}$}
}
\startdata
$\log HR$  & $-\alpha_{\rm spl}$ & $\log Age$ & -0.96$\pm$0.23 & -0.028$\pm$0.096 & 2.3$\pm$0.27 & 0.9997 \\
$-\alpha_{\rm spl}$  & $\log HR$ & $\log Age$ & -0.9$\pm$0.23 & 0.0037$\pm$0.098 & 2.2$\pm$0.4 & 0.9997 \\
$\log F_{\rm pk2}$  & $\log F_{\rm pk3}$ & $\log t_{\rm pkX,i}$ & 0.97$\pm$0.44 & 0.014$\pm$0.65 & 0.0071$\pm$0.59 & 0.9991 \\
$\log F_{\rm pk3}$  & $\log F_{\rm pk2}$ & $\log t_{\rm pkX,i}$ & 0.98$\pm$0.44 & 0.077$\pm$0.62 & -0.13$\pm$0.55 & 0.9991 \\
$\log F_{\rm pk2}$  & $variability_{2}$ &  $\log F_{\rm pk3}$& 0.25$\pm$0.71 & 0.96$\pm$0.08 & 0.025$\pm$0.16 & 0.9989 \\
   \hline
\enddata
\tablecomments{$x_{1}$ and $x_{2}$ are independent variables, $a_{1}$ and $a_{2}$ are the linear coefficients of $x_{1}$ and $x_{2}$ respectively. $y$, $b$ and adjusted $R^{2}$ are same meanings as in Table \ref{tab:linear2}. All the errors are in 1 $\sigma$ error bars. All the results are in machine readable table. The description of every parameter is in Section \ref{sec:sample}.}
\end{deluxetable*}

\begin{deluxetable*}{ccc}[b!]
\tablecaption{Differences between the current and previous analysis  \label{tab:comparison}}
\tablehead{
\colhead{reference} & \colhead{correlations} & \colhead{sample number}
}
\startdata
our result & $\log  E_{\rm p,band} = (0.24 \pm 0.011) \times \log E_{\rm iso} + (1.9 \pm 0.017)$ & 180 \\
\citet{Amati2002} & $E_{\rm p,band} \propto E_{\rm iso}^{0.52 \pm 0.06}$ & 12 \\
   \hline
our result & $\log E_{\rm p,band,i} = (0.35 \pm 0.011) \times \log E_{\rm iso} + (2.3 \pm 0.018)$ & 178 \\
\citet{Amati2008} & $E_{\rm p,band,i} \propto E_{\rm iso}^{0.57 \pm 0.01}$ & 70 \\
\citet{Yonetoku2010} & $E_{\rm iso} = 10^{53.00 \pm 0.045} \times [\frac{E_{\rm p,band,i}}{355keV}]^{1.57 \pm 0.099}$ & 101 \\
\citet{Demianski2017} & $\log (\frac{E_{\rm iso}}{1 erg}) = 1.75^{+0.18}_{-0.16} \times \log [\frac{E_{\rm p,band,i}}{300 keV}] + (52.53 \pm 0.02)$ & 162 \\
   \hline
our result & $\log L_{\rm pk} = (1.2 \pm 0.055) \times \log E_{\rm p,band,i} + (-2.9 \pm 0.14)$ & 127 \\
\citet{Yonetoku2004} & $\frac{L_{\rm pk}}{10^{\rm 52} ~ erg ~ s^{\rm -1}}=(2.34^{+2.29}_{-1.76}) \times 10^{-5} [\frac{E_{\rm p,band,i}}{1 \rm keV}]^{2 \pm 0.2}$ & 11 \\
\citet{Yonetoku2010} & $L_{\rm pk} = 10^{52.43 \pm 0.037} \times [\frac{E_{\rm p,band,i}}{355keV}]^{1.60 \pm 0.082}$ & 101 \\
   \hline
our result & $\log  F_{\rm g} = (0.85 \pm 0.057) \times \log E_{p,band,i} + (-1 \pm 0.15)$ & 179 \\
\citet{Lloyd2000} & $F_{\rm g} \propto E_{p,band,i}^{0.28 \pm 0.04}$ & 5 \\
   \hline
our result & $\log \Gamma_{0} = (0.35 \pm 0.014) \times \log E_{\rm iso} + (1.9 \pm 0.018)$ & 51 \\
\citet{Liang2010} & $ \log{\Gamma_{\rm 0}}=(0.269\pm0.002)\log{E_{\rm iso,52}} + (2.291\pm0.002)$ & 19 \\
\citet{Lv2012} & $ \log{\Gamma_{\rm 0}}=(0.29\pm0.002)\log{E_{\rm iso,52}} + (1.96\pm0.002)$ & 38 \\
   \hline
our result & $\log \Gamma_{0} = (-0.49 \pm 0.011) \times \log t_{\rm pkOpt,i} + (3.2 \pm 0.026)$ & 45 \\
\citet{Liang2010} & $\log \Gamma_{0} = (-0.63 \pm 0.04) \times \log t_{\rm pkOpt,i} + (3.69 \pm 0.09)$ & 19 \\
   \hline
\enddata
\tablecomments{Differences between the current and previous analysis of four relations. Our sample number is obviously bigger than previous analysis.}
\end{deluxetable*}

\end{document}